FERMILAB-FN-0992-E

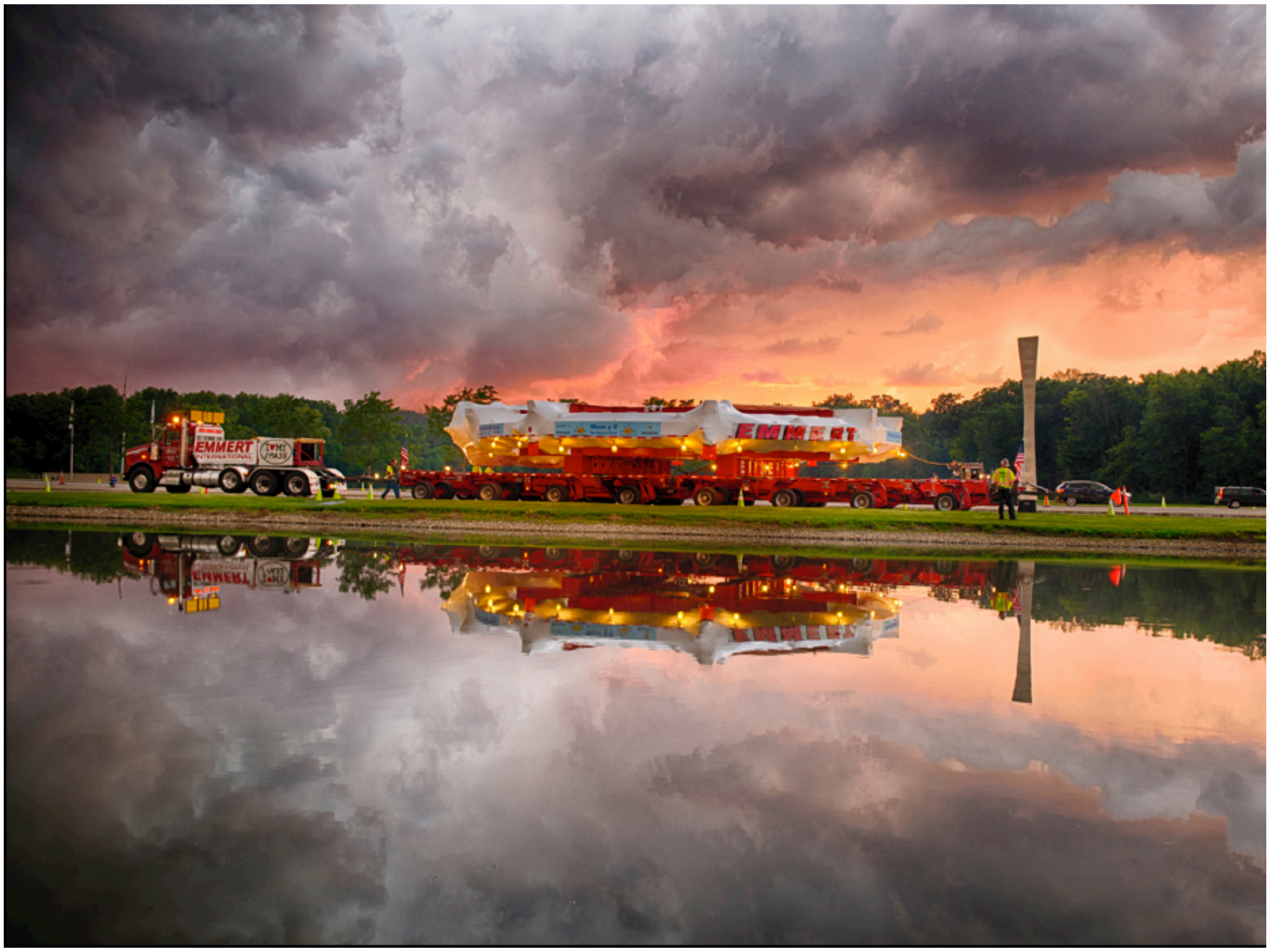

# Muon $g-2$

# Technical Design Report

15 June 2015
Fermi National Accelerator Laboratory
Batavia, IL 60510
www.fnal.gov
**Managed by**
Fermi Research Alliance, FRA
For the United States Department of Energy under
Contract No. DE-AC02-07-CH-11359

Contacts: C. Polly – Project Manager (polly@fnal.gov)
M.E. Convery – Deputy Project Manager (convery@fnal.gov)
D. Hertzog – Co-Spokesperson (hertzog@uw.edu)
B. L. Roberts – Co-Spokesperson (roberts@bu.edu)

U.S. DEPARTMENT OF ENERGY | Office of

This document was prepared for the CD2/3 Review on June 25 – 26, 2015

# Muon $(g-2)$ Technical Design Report

E989 Collaboration

J. Grange, V. Guarino, P. Winter, K. Wood, and H. Zhao
*Argonne National Laboratory, Lemont, IL, USA*

R.M. Carey, D. Gastler, E. Hazen, N. Kinnaird, J.P. Miller, J. Mott, A. Palladino, and B.L. Roberts[1]
*Boston University, Boston, MA, USA*

J. Benante, J. Crnkovic, W.M. Morse, H. Sayed, and V. Tishchenko
*Brookhaven National Laboratory, Upton, NY, USA*

V.P. Druzhinin, B.I. Khazin, I.A. Koop, I. Logashenko, Y.M. Shatunov, and E. Solodov
*Budker Institute of Nuclear Physics, Novosibirsk, Russia*

M. Korostelev, D. Newton, and A. Wolski
*The Cockcroft Institute, Daresbury and University of Liverpool, Liverpool, United Kingdom*

R. Bjorkquist, A. Chapelain, N. Eggert, A. Frankenthal, L. Gibbons, S. Kim, A. Mikhailichenko, Y. Orlov, D. Rubin, and D. Sweigart
*Cornell University, Ithaca, NY, USA*

D. Allspach, G. Annala, E. Barzi, K. Bourland, G. Brown, B.C.K. Casey, S. Chappa, M.E. Convery, B. Drendel, H. Friedsam, T. Gadfort, K. Hardin, S. Hawke, S. Hayes, W. Jaskierny, C. Johnstone, J. Johnstone, V. Kashikhin, C. Kendziora, B. Kiburg, A. Klebaner, I. Kourbanis, J. Kyle, N. Larson, A. Leveling, A.L. Lyon, D. Markley, D. McArthur, K.W. Merritt, N. Mokhov, J.P. Morgan, H. Nguyen, J-F. Ostiguy, A. Para, C.C. Polly[2], M. Popovic, E. Ramberg, M. Rominsky, D. Schoo, R. Schultz, D. Still, A.K. Soha, S. Strigonov, G. Tassotto, D. Turrioni, E. Villegas, E. Voirin, G. Velev, L. Welty-Rieger, D. Wolff, C. Worel, J-Y. Wu, and R. Zifko
*Fermi National Accelerator Laboratory, Batavia, IL, USA*

K. Jungmann and C.J.G. Onderwater
*University of Groningen, Groningen, The Netherlands*

P.T. Debevec, S. Ganguly, M. Kasten, S. Leo, K. Pitts, and C. Schlesier
*University of Illinois at Urbana-Champaign, Urbana, IL, USA*

K. Giovanetti
*James Madison University, Harrisonburg, VA, USA*

V.A. Baranov, V.N. Duginov, N.V. Khomutov, V.A. Krylov, N.A. Kuchinskiy, and V.P. Volnykh
*Joint Institute for Nuclear Research, Dubna, Russia*

M. Gaisser, S. Haciomeroglu, Y-I. Kim, S. Lee, M-J Lee, and Y.K. Semertzidis
*Institute for Basic Science and KAIST, Daejeon, Republic of Korea*

E. Won
*KAIST and Korea University, Seoul, Republic of Korea*

C. Crawford, R. Fatemi, W.P. Gohn, T.P. Gorringe, W. Korsch, and B. Plaster
*University of Kentucky, Lexington, KY, USA*

A. Anastasi[3], D. Babusci, S. Dabagov[4], C. Ferrari[5], A. Fioretti[5], C. Gabbanini[5], D. Hampai, A. Palladino, and G. Venanzoni
*Laboratori Nazionali di Frascati dell'INFN, Frascati, Italy*

T. Bowcock, J. Carroll, S. Charity, B. King, S. Maxfield, K. McCormick, J. Price, D. Sim, A. Smith, T. Teubner, W. Turner, M. Whitley, and M. Wormald
*University of Liverpool, Liverpool, United Kingdom*

R. Chislett, S. Kilani, M. Lancaster, E. Motuk, T. Stuttard, and M. Warren
*University College London, London, United Kingdom*

D. Flay, D. Kawall, and Z. Meadows
*University of Massachusetts, Amherst, MA, USA*

T. Chupp, R. Raymond, and A. Tewlsey-Booth
*University of Michigan, Ann Arbor, MI, USA*

M.J. Syphers and D. Tarazona
*Michigan State University, East Lansing, MI, USA*

J. Holzbauer, B. Quinn, and W. Wu
*University of Mississippi*

S. Catalonotti[6], R. Di Stefano[7], M. Iacovacci[6], and S. Mastroianni
*INFN, Sezione di Napoli, Napoli, Italy*

S. Chattopadhyay, M. Eads, M. Fortner, D. Hedin, and N. Pohlman
*Northern Illinois University, DeKalb, IL, USA*

A. de Gouvea
*Northwestern University, Evanston, IL, USA*

B. Abi, F. Azfar, and S. Henry
*University of Oxford, Oxford, United Kingdom*

F. Gray
*Regis University, Denver, CO, USA*

G. Di Sciascio and D. Moricciani
*INFN, Sezione di Roma Tor Vergata, Roma, Italy*

C. Fu, X. Ji, L. Li, and H. Yang
*Shanghai Jiao Tong University, Shanghai, China*

D. Stöckinger
*Technische Universität Dresden, Dresden, Germany*

J. Ritchie
*University of Texas, Austin*

G. Cantatore[8], D. Cauz[9], M. Karuza[10], G. Pauletta[9], and L. Santi[9]
*INFN, Sezione di Trieste e G.C. di Udine, Italy*

S. Baeßler, M. Bychkov, E. Frlez, and D. Pocanic
*University of Virginia, Charlottesville, VA, USA*

L.P. Alonzi, M. Fertl, A. Fienberg, N. Froemming, A. Garcia, D.W. Hertzog[1], J. Kaspar, P. Kammel, R. Osofsky, M. Smith, E. Swanson, and T. van Wechel
*University of Washington, Seattle, WA, USA*

K. Lynch
*York College, CUNY, Jamaica, NY, USA*

Monday 15$^{th}$ June, 2015

[1]Co-spokesperson
[2]Project Manager
[3]Dipartimento di Fisica e di Scienze della Terra dell'Università di Messina, Messina, Italy
[4]Lebedev Physical Institute and NRNU MEPhI, Moscow, Russia
[5]Istituto Nazionale di Ottica, UOS Pisa, Consiglio Nazionale delle Ricerche, Italy
[6]Università di Napoli, Napoli, Italy
[7]Università di Cassino, Cassino, Italy
[8]Università di Trieste, Trieste, Italy
[9]Università di Udine, Udine, Italy
[10]University of Rijeka, Rijeka, Croatia

DISCLAIMER

This work was prepared as an account of work sponsored by an agency of the United States Government. Neither the United States Government nor any agency thereof, nor any of their employees, nor any of their contractors, subcontractors, or their employees makes any warranty, express or implied, or assumes any legal liability or responsibility for the accuracy, completeness, or any third partys use or the results of such use of any information, apparatus, product, or process disclosed, or represents that its use would not infringe privately owned rights. Reference herein to any specific commercial product, process, or service by trade name, trademark, manufacturer, or otherwise, does not necessarily constitute or imply its endorsement, recommendation, or favoring by the United States Government or any agency thereof or its contractors or subcontractors. The views and opinions of authors expressed herein do not necessarily state or reflect those of the United States Government or any agency thereof.

## Executive Summary

The Muon $g-2$ Experiment, E989 at Fermilab, will measure the muon anomalous magnetic moment, $a_\mu \equiv (g-2)/2$, to unprecedented precision: the goal is 0.14 parts per million (ppm). The worth of such an undertaking is coupled to the fact that the Standard Model (SM) prediction for $a_\mu$ can also be determined to similar precision. As such, the comparison between experiment and theory provides one of the most sensitive tests of the completeness of the model. The Brookhaven-based E821 experiment, which completed data taking in 2001, determined $a_\mu$(Expt) to 0.54 ppm. Steady improvements in theory since that time have resulted in a present SM uncertainty on $a_\mu$(SM) of 0.42 ppm. The experimental measurement and SM predictions differ by 3.3 to 3.6 standard deviations, depending on which evaluation of the lowest-order hadronic contribution in the SM is used:

$$\Delta a_\mu(\mathrm{Expt} - \mathrm{SM}) = (286 \pm 80) \times 10^{-11} \quad (1)$$
$$= (260 \pm 78) \times 10^{-11} \quad (2)$$

(see Chapter 2 for details). This is a highly cited result, owing in part to the many natural SM extensions, from supersymmetry to dark photons, that could cause such an effect. The planned four-fold improvement in experimental precision compared to E821, could establish beyond doubt a signal for new physics—if the central value of the measurement remains unchanged. During the time it will require to mount, run and analyze the data, the SM hadronic predictions are expected to become even more precise; thus the comparison of experiment to theory will be quite powerful, no matter what final values are found. The Motivation for the new experiment and a detailed exposition on the SM theory is provided in Chapter 2 of this document.

The original E989 Proposal, and the additional design work now completed in preparation of this Technical Design Report (TDR), outline a credible plan to achieve the experimental goal in a timely and cost-efficient manner. The approach is anchored by the re-use of the existing precision muon storage ring, an efficient and parasitic deployment of the Fermilab proton complex and beamlines, and strategic upgrades or replacements of outdated or under-performing components from E821. The experiment will be carried out by a collaboration of accelerator, atomic, nuclear and particle physicists, drawing from domestic and international universities and national laboratories. The collaboration retains a strong core of experienced participants from BNL E821, augmented by many new groups selected for their expertise in areas that are required to mount a next-generation experiment.

In many ways, E989 is a unique, large-scale Project. Several core aspects involve proven elements from E821 that will be retained in whole or with minor upgrades. This is especially true for the storage ring elements and the magnetic field measuring tools, which will be relocated, re-assembled and restored to operation. Many of these items are well beyond a normal TDR stage in terms of design; indeed, they exist and often require no more than testing and minor repair. In contrast, several items have been identified as requiring a new approach to meet the demands of a higher rate experiment with lower systematic uncertainties. Chief among them is a new storage ring kicker and, possibly, a new inflector magnet. The storage-ring electrostatic quadrupoles will undergo an operational upgrade and one set will be redesigned to better allow for the beam passage through them as it enters the storage ring. The beam position mapping will employ a unique in-vacuum tracking system and the instrumentation for the precession frequency measurement—calorimeters, fast digitizers and modern data acquisition—will all be new. Naturally, the entire pion-to-muon beam path from target to storage ring is unique at Fermilab.

The BNL experiment was statistics limited. With a persistent and tantalizing hint of new physics, it has been recognized for many years that a next-generation effort is required to lead to a true discovery. A number of informal studies led to the realization that relocation of the storage ring to Fermilab would provide the ideal environment for the next generation experiment. The Booster, the Recycler, and the existing antiproton target station can be used to acquire a 20-fold increase in statistics in a timely manner. One can take direct advantage of the experience with the unique and well-understood storage ring developed for E821. The proposed beam environment and relatively modest experimental upgrades would provide a better measurement environment that will lead to reduced systematic uncertainties. The E989 Proposal was presented to the Fermilab PAC in March 2009. Cost evaluations by an independent committee followed, and beam delivery studies were initiated. Following the completion of the Proton Improvement Plan, Fermilab can service the NO$\nu$A experiment fully and provide excess proton cycles to adequately meet the unique needs of the $g-2$ experiment. The experimental technical approach described in this TDR is conservative. It is built on the foundation and lessons learned from several generations of $g-2$ experiments at CERN and then Brookhaven, and we retain key personnel that provide necessary overlap with the most recent effort.

The beam-use plan has evolved further such that it now largely overlaps with the needs of the Mu2e Experiment. Together, $g-2$ and Mu2e have become the first tenants of the new Muon Campus, which involves several buildings, beamlines and infrastructure support. A new general purpose building, MC-1, has been designed with specific attention to the needs of the $g-2$ experiment—e.g., stable floor, temperature control to $\pm 2^{\circ}$ F, and the necessary services. Ground-breaking for MC-1 occurred in May, 2013 and beneficial occupancy occured in May 2014.

Following the Project Overview, the TDR is organized as follows. We begin with chapters on the physics motivation—including the discussion standard-model and non-standard-model physics—and the experimental strategy. The measurement involves ambitious statistical and systematic uncertainty goals. Chapter 5 then provides a road map that summarizes our plans to meet the statistical and systematic uncertainty targets. The factors contributing to these estimates are distributed throughout the document, as they fall into different WBS categories, so this chapter also provides both short descriptions of the factors that underlie the uncertainty targets, and pointers to the specific sections that discuss the topics in detail. The bulk of the TDR then describes the experimental design that can reach those targets, from production and delivery of the muon beam through to the slow controls and monitoring of the data taking process.

# Contents

# Chapter 1

# Project Overview

## 1.1 Mission

The mission of the Muon $g-2$ experiment at Fermilab and its context in the High Energy Physics program have been officially summarized in excerpts from the DOE Mission Need Statement [1] approved in September 2012.

> The primary mission of the DOE Office of High Energy Physics (HEP) is to understand the universe at its most fundamental level through the study of matter and energy, space and time, and the forces governing basic interactions. The Standard Model of particle physics documents the current status of that understanding, which is known to be both stunningly robust, yet necessarily incomplete. An urgent mission of the particle physics community is to complete the Model. The tools required include high-energy colliders, which can directly produce the highest mass particles, and high-intensity accelerators, which can tease out the tiny effects of still unknown interactions in the data through high precision.
>
> One of the more persistent hints of new physics has been the deviation between the measured muon anomalous magnetic moment, $a_\mu = (g-2)/2$, and its Standard Model expectation, where both are determined to a precision of 0.5 parts per million. This fundamental measurement has been pursued for decades with increasing precision. The discrepancy has been interpreted to point toward several attractive candidates for Standard Model extensions: supersymmetry, extra dimensions, or a dark matter candidate. The Large Hadron Collider (LHC) is now delivering on its promise to explore physics at the highest mass ranges to date, although no new physics has yet been found. A new and more precise muon $g-2$ experiment offers a strategic opportunity to search for new physics through alternative means, which could lead to a fuller and more coherent picture of the underlying physics.
>
> The search for new physics can be carried out in complementary ways on the different frontiers of particle physics. The measurement of the muon anomalous magnetic moment is sensitive to interactions at the TeV scale, which is also the scale probed by the LHC. The capability gap filled by the new muon $g-2$ experiment derives from the ability of the measurement to elucidate the underlying physics we hope to discover at the LHC and probe areas of the newly discovered physics that are inaccessible to the LHC experiments.

The current muon $g-2$ measurement is used as a benchmark for new physics and has been used as input into the parameter space explored in almost all model dependent searches for new physics at the LHC, but the current discrepancy between the muon $g-2$ measurement and the theoretical prediction could be explained as a statistical fluctuation at the three-sigma level and has only been observed by one experiment. If the discrepancy is false, then this will cause serious confusion in interpreting LHC results. The discrepancy needs to be confirmed and established above the accepted discovery threshold of five standard deviations above a fluctuation.

Should LHC discover new physics at the TeV scale and the $g-2$ discrepancy is confirmed, a precise determination of $g-2$ is expected to provide direct measurements of the coupling constants of the new particles responsible for the discrepancy, fundamental parameters of the underlying theory and a window on the underlying symmetries of the new physics. In many cases, it is expected that these parameters will not be measured with adequate precision at the LHC alone.

There are no facilities, equipment, or services currently existing or being acquired within the Department of Energy, other government agencies, public organizations, private entities or international bodies that are sufficient to address these gaps.

The goal of the Muon $g-2$ experiment at Fermilab is a four-fold improvement in the experimental precision thereby reducing the error on $a_\mu$ to 140 ppb. If the discrepancy measured in E821 is truly an indication of new physics, then the difference with the current theoretical prediction will exceed the $5\sigma$ discovery threshold. Obtaining this precision requires observation of the muon spin precession with more than 20 times the statistics of the BNL E821 experiment while controlling systematics at the 100 ppb level.

## 1.2 Muon $g-2$ Project Scope

The scope of the Muon $g-2$ project can be easily visualized by looking at the work breakdown structures (WBS) shown in Figures 1.1 and 1.2.[1] Outside of the management portions, the project is divided into three major areas encompassing the accelerator modifications, storage ring, and detectors. The disassembly and transport of E821 equipment, now complete, has also been costed as part of the project. By agreement with DOE OHEP, the disassembly and transport proceeded as an OPC cost in advance of the schedule of CD reviews. A very substantial amount of the scope with the detector WBS is being funded via an NSF MRI which was awarded in mid-2013, a DOE Early Career Grant to Brendan Casey, or an in-kind contribution from INFN collaborators. The division of scope among the various funding sources is described in detail in the Detector Scope Document, GM2-DocDB-1313. However, the full scope appears in the DOE Project's resource loaded schedule of activities and is managed as a unified project, with all risks documented and controlled through the project office and progress toward the deliverables tracked using the Fermilab EVMS.

[1]Figures 1.1 and 1.2 also show the organization chart and institutional responsibilities at the time of writing the TDR.

### 1.2.1 Accelerator (WBS 476.2)

The accelerator portion of the project includes the upgrades and modifications required to convert the existing antiproton complex at Fermilab into a muon source and deliver a high-purity beam of 3.1 GeV/$c$ muons to the Muon $g-2$ experiment. A number of common accelerator components are needed for both the $g-2$ and Mu2e projects. These elements have been pulled out of both projects in order to better facilitate management of these components to simultaneously meet the specifications and schedule demands of both experiments. These common elements and their current status are discussed in Section 1.3.2. A brief description of the accelerator improvements required solely for $g-2$, corresponding to each L3 area in the WBS, is given in the list below.

- **Target Station (WBS 476.2.2)** The AP0 target hall, formerly used for antiproton production, will be utilized for the production of the muon beam. Protons from the Booster with 8 GeV kinetic energy will impact a target immediately upstream of a Li lens. The target and lens are the same as those used for the antiproton production. The Li lens has to be pulsed at a lower gradient, but higher repetition rate, thus requiring some modifications to the power supply. Pions with a momentum around 3.1 GeV/$c$ are focussed by the Li lens and directed out of the target hall through a 3 degree bend created by a single-turn pulsed magnet (PMAG). Protons that pass through the target are deposited in a water-cooled beam dump which requires replacement due to a leak that developed near the end of Tevatron operation.

- **Beamline (WBS 476.2.3)** The beamline portion of the WBS includes all modifications required to a multitude of beamlines. Starting with the primary proton beam before it impacts the target, the final focussing quadrupole needs to be replaced with a triplet to provide more flexibility in focussing the protons into a smaller spot-size on the production target. When the pion beam emerges from the target vault, it is captured into a new high-density FODO lattice that provides a better muon capture efficiency by tripling the number of quadrupoles currently in the line. The muon beam is then brought through another section of beamline and injected into the Delivery Ring, formerly called the Debuncher. The muons are circulated a few times around the Delivery Ring to allow protons to separate by time-of-flight before an abort kicker is fired to remove the protons.[2] The muons are then extracted from the Delivery Ring and brought through a new section of beamline connecting to the g-2 storage ring.

- **Controls and Instrumentations (WBS 476.2.4)** Standard accelerator control systems and interlocks will be required for beam delivery to the experiment. New instrumentation will be required in the new beamlines and instrumentation in existing portions of the accelerator complex will require modification due to the relatively low intensity of the pulsed muon beam. Safety systems including interlocks in the beamlines and the controls to the MC-1 building, the new experimental hall housing the storage ring, are also included in this area.

[2] The injection into the Delivery Ring and the abort system are also needed by Mu2e to inject 8 GeV protons into the Delivery Ring and provide a safe beam abort, therefore these systems are part of the general accelerator improvements discussed in the next section.

### 1.2.2 Ring (WBS 476.3)

The ring portion of the WBS includes all of the preparations needed to reassemble and install the E821 $g-2$ storage ring at Fermilab and connect it to a new cryogenic plant that will be constructed from Tevatron refrigerators.[3] A number of subsystems associated with the injection and storage of the muon beam will be upgraded. Field monitoring equipment is an integral part of storage ring and poses one of the largest challenges to the experiment. The NMR systems and calibration procedures must be capable of determining the magnetic field to better than 100 ppb, with an ultimate goal of achieving a 70 ppb uncertainty on the average field acting on the stored muon population.

- **Magnet (WBS 476.3.2)** The storage ring magnet must be reassembled holding very tight tolerances, starting with the foundation of base plates and jacks, followed by installation of more than 700 tons of return yoke and the superconducting coils. A total of 72 pole pieces line the top and bottom of the magnet gap and must then be installed with even tighter tolerances. Vacuum systems for the superconducting coils will be newly-provided since the older E821 equipment was absorbed into the RHIC complex.

- **Inflector (WBS 476.3.3)** The old inflector from E821 is currently used as the default plan for injecting into the storage ring. Reinstallation of the old inflector requires new vacuum systems and a connection to the cryogenic plant through the existing lead pot. However, an alternative inflector is still considered a technical opportunity, which would both mitigate a risk of delay to the project (if the old inflector fails to turn on or fails during commissioning) and also improve the muon yield. A new inflector with a larger opening and less material across the beam channel would allow for a better match into the storage ring, thus reducing beam oscillations from the unmatched dispersion, and increase the storage efficiency by as much as a factor of four.

- **Storage Ring Vacuum (WBS 476.3.4)** The vacuum chambers will require some modifications to accommodate relocated NMR probes and *in vacuo* straw trackers. The chambers will be installed with vacuum equipment that is either newly-purchased or recycled from the Tevatron wherever possible.

- **Kickers (WBS 476.3.5)** The electromagnetic kickers that place the injected muon beam onto a central orbit are being redesigned to provide a more powerful kick while sustaining the increased repetition rate of the new $g-2$ experiment. A Blumlein design is the preferred option to produce a kick with the correct pulse-width along with sharp rise and fall times.

[3] The cryogenic plant is being constructed off-project as part of an overall plan to provide cryogens to both $g-2$ and Mu2e experimental halls.

- **Quadrupoles (WBS 476.3.6)** The quadrupole system is being upgraded to run the ring at a higher $n$ value and provide a more massless quadrupole plate in the injection region where the muon beam has to pass.

- **Controls and Instrumentation (WBS 476.3.7)** The storage ring controls and instrumentation will need to be upgraded to be compatible with modern systems used at Fermilab. Instrumentation inside the superconducting coils will not be altered, but the read-out, monitoring, and controls all require updating.

- **Field (WBS 476.3.8)** The magnetic field portion of the WBS is disproportionately large compared to other level 3 areas due to the complexity associated with shimming the magnet to high uniformity and the many NMR systems required to determine the absolute field strength. A series of active and passive shimming techniques are used to produce an extremely uniform magnetic field when integrated azimuthally around the storage ring. In order to measure the field in the storage region, the NMR trolley from E821 will be reused with some minor upgrades. This device is pulled out of a garage every 2-3 days and travels around the ring making measurements of the field at the center of the magnet gap without ever having to break vacuum. In between trolley runs, a series of fixed NMR probes monitor the field changes at the edges of the storage volume to better interpolate field changes from one trolley run to the next. Finally, the NMR trolley must be absolutely calibrated in a region of the storage ring where the magnetic field has been shimmed with even higher uniformity. Plunging probes are used to determine the field at the location of the NMR trolley probes and are absolutely calibrated against a special, spherical probe that has been used for past muonium hyperfine and $g-2$ experiments. All of these NMR systems require updated readout and data acquisition systems.

### 1.2.3 Detector (WBS 476.4)

The detectors and electronics for the experiment will all be newly constructed to meet the demands of measuring the spin precession of the muon to a statistical error of 100 ppb, while controlling systematics on $\omega_a$ to the 70 ppb level. This is a substantial improvement over the E821 experiment, and better gain stability and corrections due to overlapping events in the calorimeters are crucial systematics addressed in the new design. A new tracking system will allow for better monitoring of the stored muon population, thus improving the convolution of the stored muon population with the magnetic field volume, and establishing corrections to $\omega_a$ that arise from the electric field and pitch corrections; see Section 4.4. The data acquisition must be able to handle the increased data rates and allow for the traditional T analysis of the data and the new Q method described in Section 16.1.2.

- **Calorimeters (WBS 476.4.2)** New calorimeters will be constructed using an array of $PbF_2$ crystals readout by SiPMs. Unlike in E821, where the calorimeters were read out as one monolithic block, the array of crystals will allow for spatial resolution of pileup. A stable voltage distribution is required to maintain the gain stability requirements, along with a calibration system capable of verifying that the stability requirements are being met.

- **Trackers (WBS 476.4.3)** New straw trackers will be installed at select positions inside the main ring vacuum chambers to allow precise reconstruction of the decay positrons. As with the calorimeters and any other materials close to the muon storage region, great care must be taken not to create any perturbations to the magnetic field. The trackers will be readout by ASDQ ASICs that provide amplification, shaping, and discrimination. The discriminated signals are digitized by a TDC implemented in a field-programmable gate array (FPGA). Also included is a collection of smaller, dedicated detectors that are installed around the storage ring. This category includes an entrance counter to mark the initial time when a muon bunch enters the ring, an extinction monitor to check for leakage protons between the muon bunches, and deployable fiber harp monitors within the storage ring. The fiber harps are strung with scintillating fibers and allow for a direct, but destructive, measurement of the distribution of stored muons and their associated beam dynamics parameters.

- **Backend Electronics (WBS 476.4.4)** Signals from the calorimeters will be digitized with new 800 MHz waveform digitizers which must be synchronized through a distributed clock system.

- **DAQ (WBS 476.4.5)** A new MIDAS-based data acquisition system will be developed to collect data from calorimeters and trackers, while also providing online monitoring of the data quality. For each injection of a muon bunch into the storage ring, the DAQ has to gather data from the various front-ends, package it through an event builder, and ship the data to mass storage for later offline analysis.

- **Slow Controls (WBS 476.4.6)** The slow controls system is new and encompasses an array of functionality including monitoring various environmental conditions to be stored and used if needed later for determining data quality, monitoring diagnostics for various subsystems, setting alarms to alert control room operators of problems, and providing automated controls to interface with various subsystems.

### 1.2.4 E821 Equipment Transfer (WBS 476.5)

The transport of the superconducting coils from Brookhaven to Fermilab was one of the highest risk elements of the project. In order to mitigate this risk as early as possible in the project timeline, the DOE authorized the equipment transfer to proceed on operating funds counted against the project TPC, but not part of the formal CD process. The transport has now been completed successfully, with all of the equipment which we are reusing from BNL E821 now on site at Fermilab.

## 1.3 Muon $g-2$ Dependencies Outside of the Project

In addition to the scope outlined in the previous section, there are a number of off-project components required for the success of Muon $g-2$ and the more global Fermilab program.

### 1.3.1 Proton Improvement Plan and NOvA Upgrades

The Proton Improvement Plan (PIP) at Fermilab is required to meet the demands of future proton economics and enable the 40 year old Linac and Booster machines to run reliably for another 20 years. Without this upgrade, the Booster can not reliably deliver the 9 Hz of beam required to feed the Main Injector for the NOvA program. Experiments like MicroBooNE (data in 2014), Muon $g-2$ (data in 2016), and Mu2e (data in 2019) all rely on an increase in the Booster repetition rate. The PIP is a staged set of improvements that eventually will increase the Booster to its maximum 15 Hz output. The Muon $g-2$ experiment requires 3 out of every 15 Booster batches that are then split into fourths and delivered at an average 12 Hz rate to the storage ring. Although the MicroBooNE experiment will be the first to suffer if the PIP goals are not recognized, it is important for these accelerator upgrades to stay on track in order to reduce conflicting beam demands since it is likely that MicroBooNE will continue to take data in parallel with Muon $g-2$. The Proton Improvement Plan is well underway with the Cockroft-Walton already replaced by a modern RFQ as an injector to the Linac, but the schedule has been affected by financial constraints.

Protons from the Booster need to be injected directly into the Recycler for $g-2$. This connection was recently completed as part of the NOvA project and will have been commissioned and in operation several years prior to the start of $g-2$. However, the kickers that enable the injection into the Recycler will have to operate at a higher repetition rate, but within their design specifications, for simultaneous operation of NOvA and $g-2$ (or eventually Mu2e). MicroBooNE directly uses the beam from the Booster and so does not place the same demands on injection into the Recycler.

### 1.3.2 The Muon Campus

The Muon $g-2$ and Mu2e experiments both reuse the anti-proton source to create individually customized muon sources. The initial plans that were developed independently for the experiments were fraught with conflicts. Over the last two years a plan has emerged to overcome those conflicts and replace them with synergies. Areas were identified where common equipment could be constructed to facilitate both experiments in a way that the overall cost of the muon program is minimized while compatibility is maximized. Furthermore, by treating the common pieces as more general civil construction and accelerator upgrades, the flexibility of the laboratory infrastructure increases and opportunities for future experiments beyond $g-2$ and Mu2e are enabled. In order to meet the combined specifications for $g-2$ and Mu2e, while also keeping an eye towards the future, these upgrades are separately managed in a series of General Plant Projects (GPPs) and Accelerator Improvement Projects (AIPs). The collection of upgrades has come to be known as the Muon Campus at Fermilab, and is broadly outlined in the list below.

- **MC-1 Building GPP:** This is the building that will house the $g-2$ storage ring in the high-bay, power supplies for large sections of the Muon $g-2$ and Mu2e beamlines in a central section, and the cryo facility for both experiments. This GPP is substantially complete, and we have received beneficial occupancy of the cryo room and the experimental hall.

- **Beamline Enclosure GPP:** This GPP provides a new tunnel to connect the former pbar source to provide beam to the MC-1 and Mu2e buildings.

- **Muon Campus Infrastructure GPP:** This GPP covers a few miscellaneous civil construction projects needed by both experiments including providing cooling for the He compressors reused from the Tevatron at the A0 building, an extension of the MI-52 building to provide room for extra power supplies and cooling skids, and possibly an additional electrical feeder.

- **Cryo Plant AIP:** This cryo plant will be constructed in the MC-1 Building reusing four refurbished refrigerators from the Tevatron to provide cooling to the Muon $g-2$ and Mu2e superconducting coils. The AIP is currently around 40% complete and installation in the MC-1 building has started.

- **Recycler RF AIP:** This AIP will add an RF system to the Recycler to allow protons from the Booster to be rebunched into the narrow $\approx 100$ ns pulses needed for Muon $g-2$ and Mu2e.

- **Beam Transport AIP:** This AIP will create a new extraction kicker and connection from the Recycler to transport the primary proton beam to the Muon Campus.

- **Delivery Ring AIP:** This AIP will provide the common modifications needed to transform the pbar source into a delivery ring capable of providing muons to Muon $g-2$ and slow-spill protons to the Mu2e target.

  A more detailed description of the accelerator components can be found in Chapter 7 with a summary given in Table 7.7.

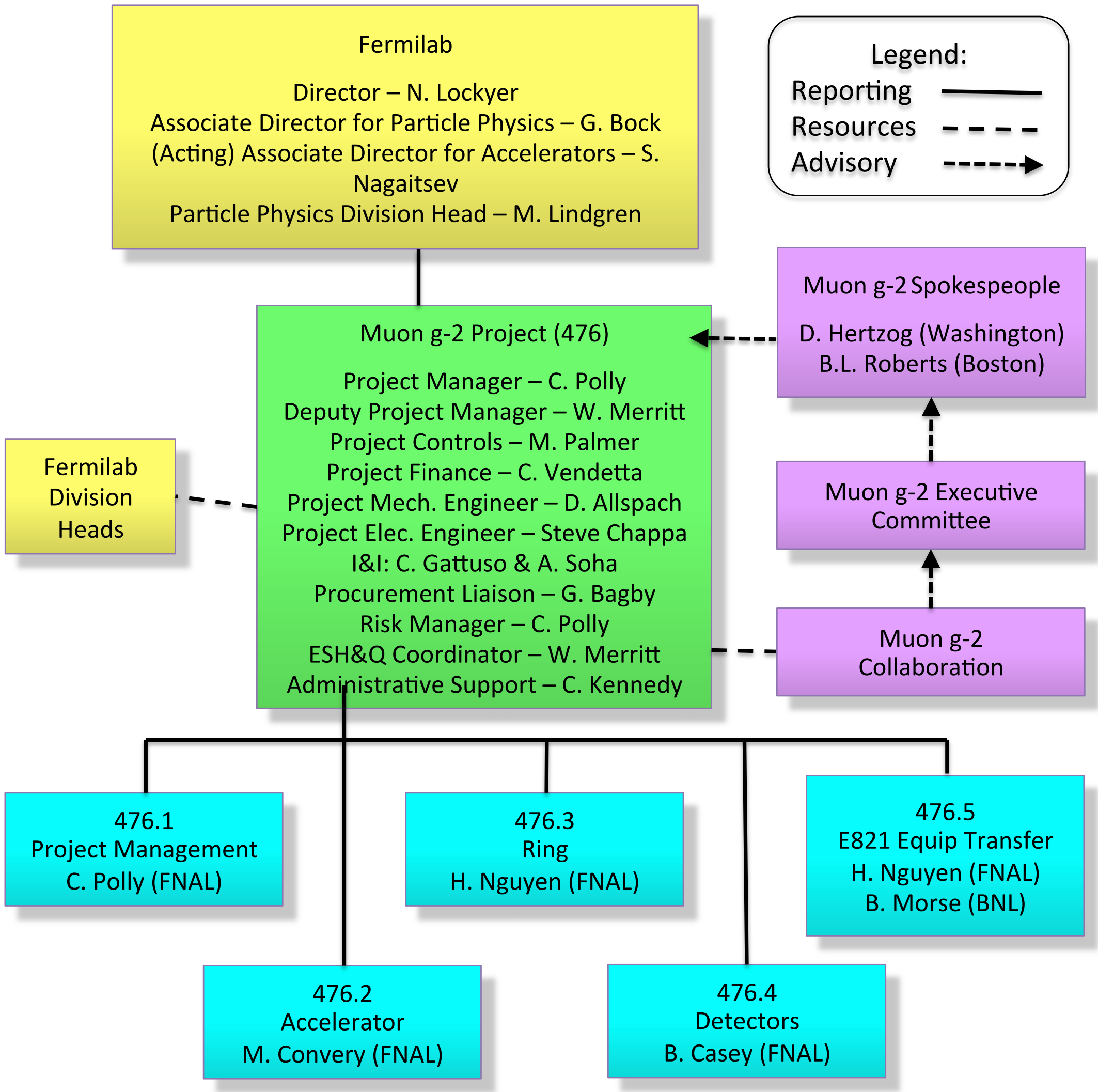

Figure 1.1: Organizational chart and WBS structure to Level 2.

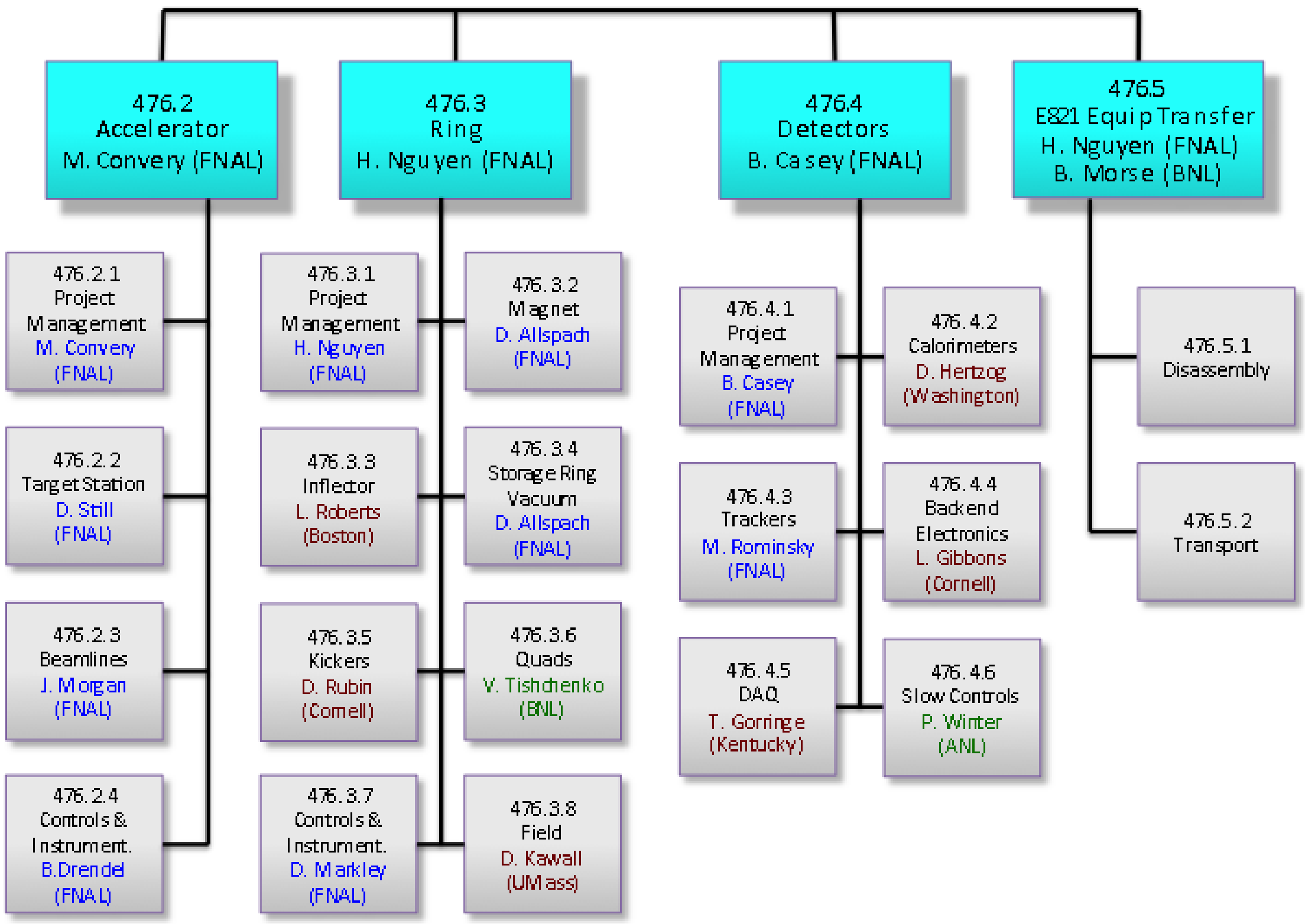

Figure 1.2: Organizational chart and WBS structure to Level 3.

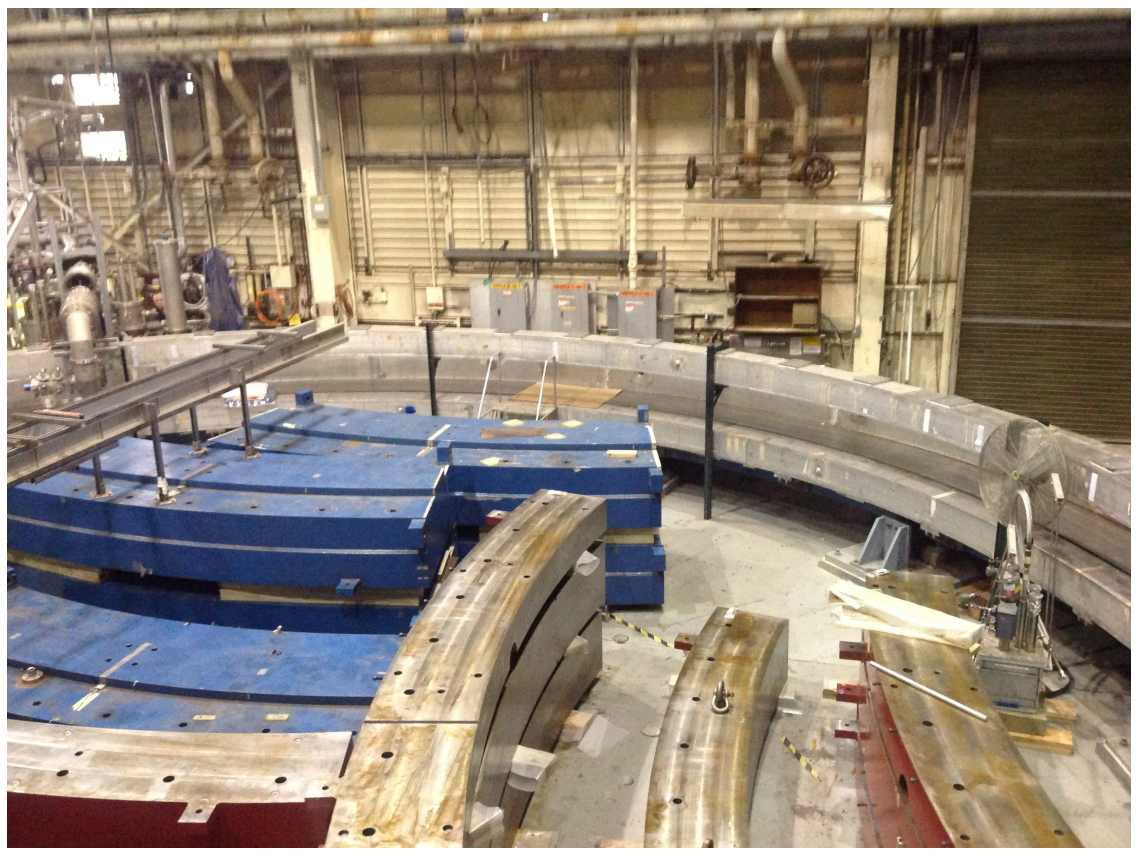

(a) February 2013 disassembly at Brookhaven

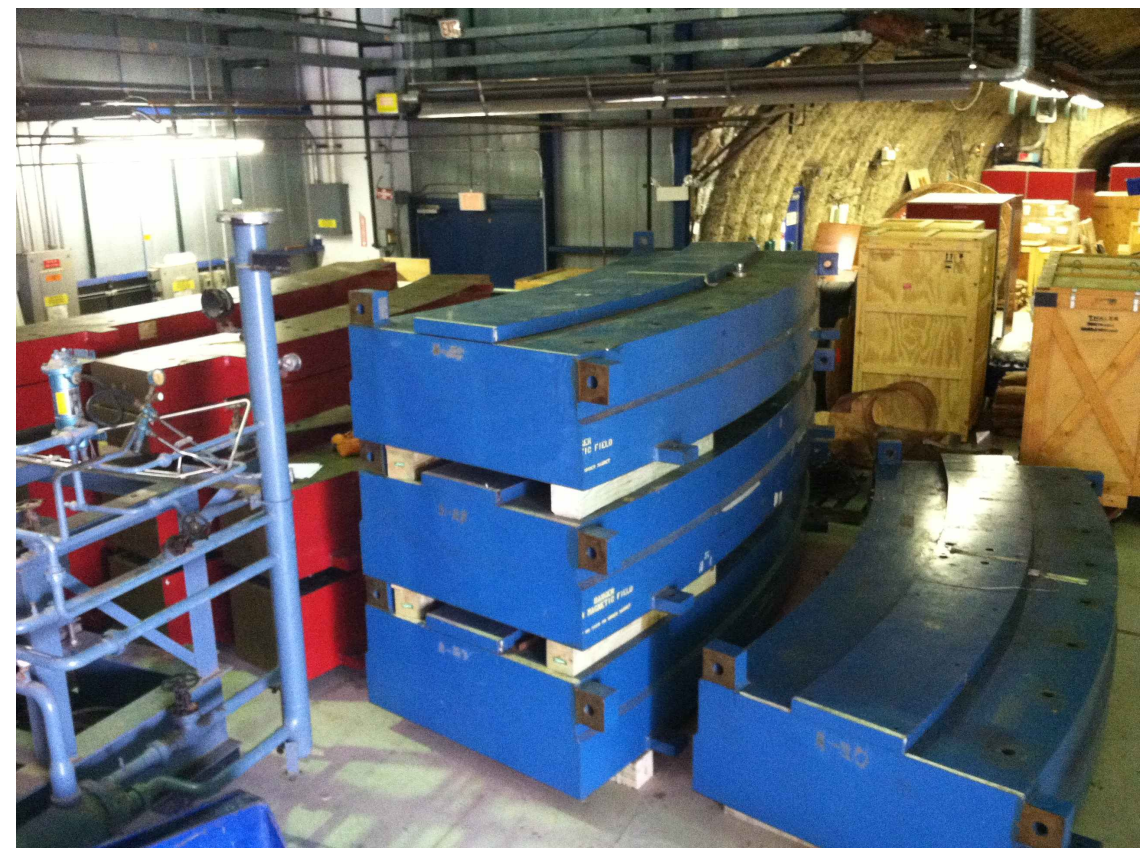

(b) May 2013 yoke steel stored at Fermilab

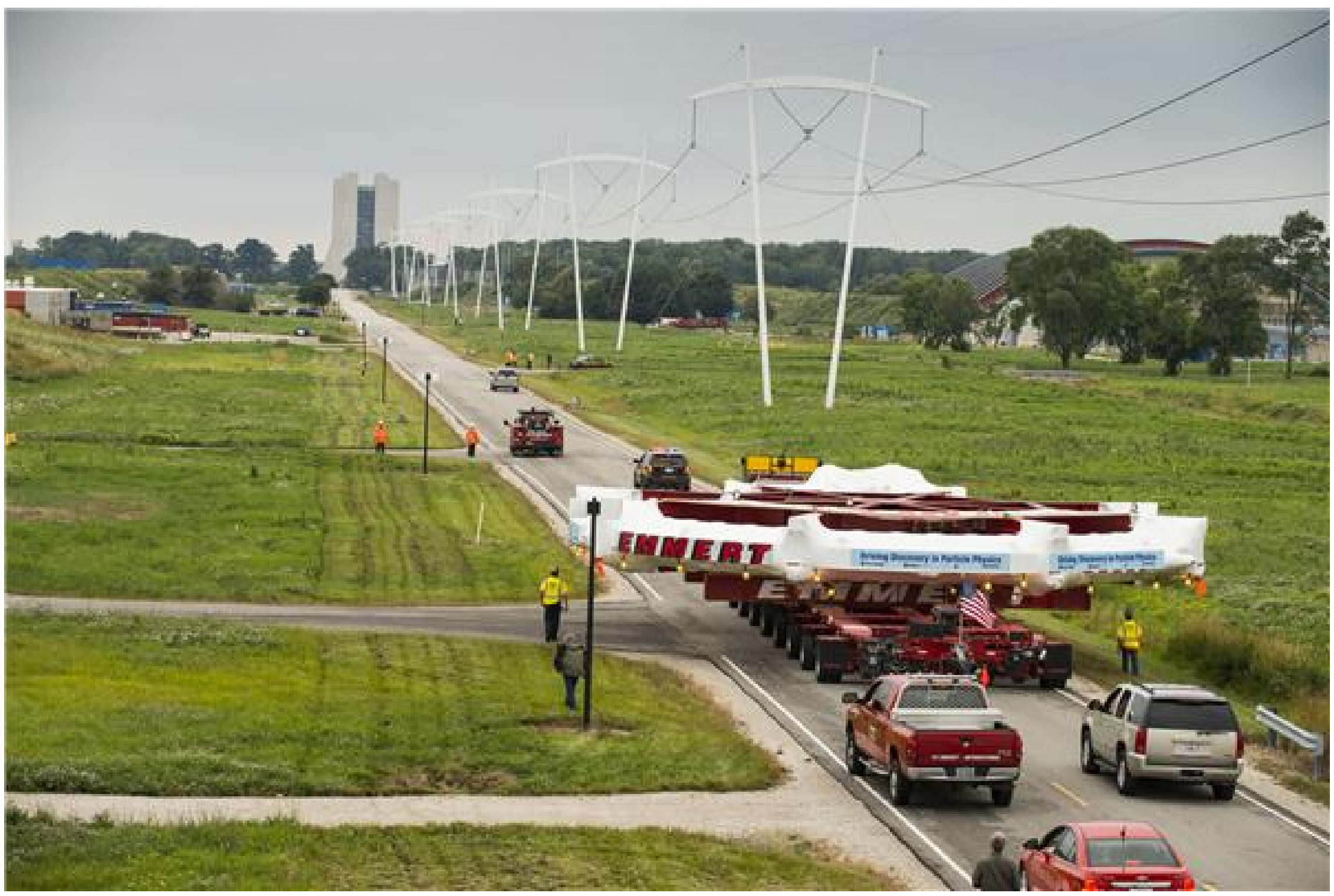

(c) July 2013 storage ring arrives at Fermilab

Figure 1.3: Pictures of disassembly and transport showing (a) storage ring at Brookhaven in February 2013, (b) a large portion of the yoke steel stored at Fermilab in May 2013, and (c) the storage ring arriving at Fermilab in July 2013.

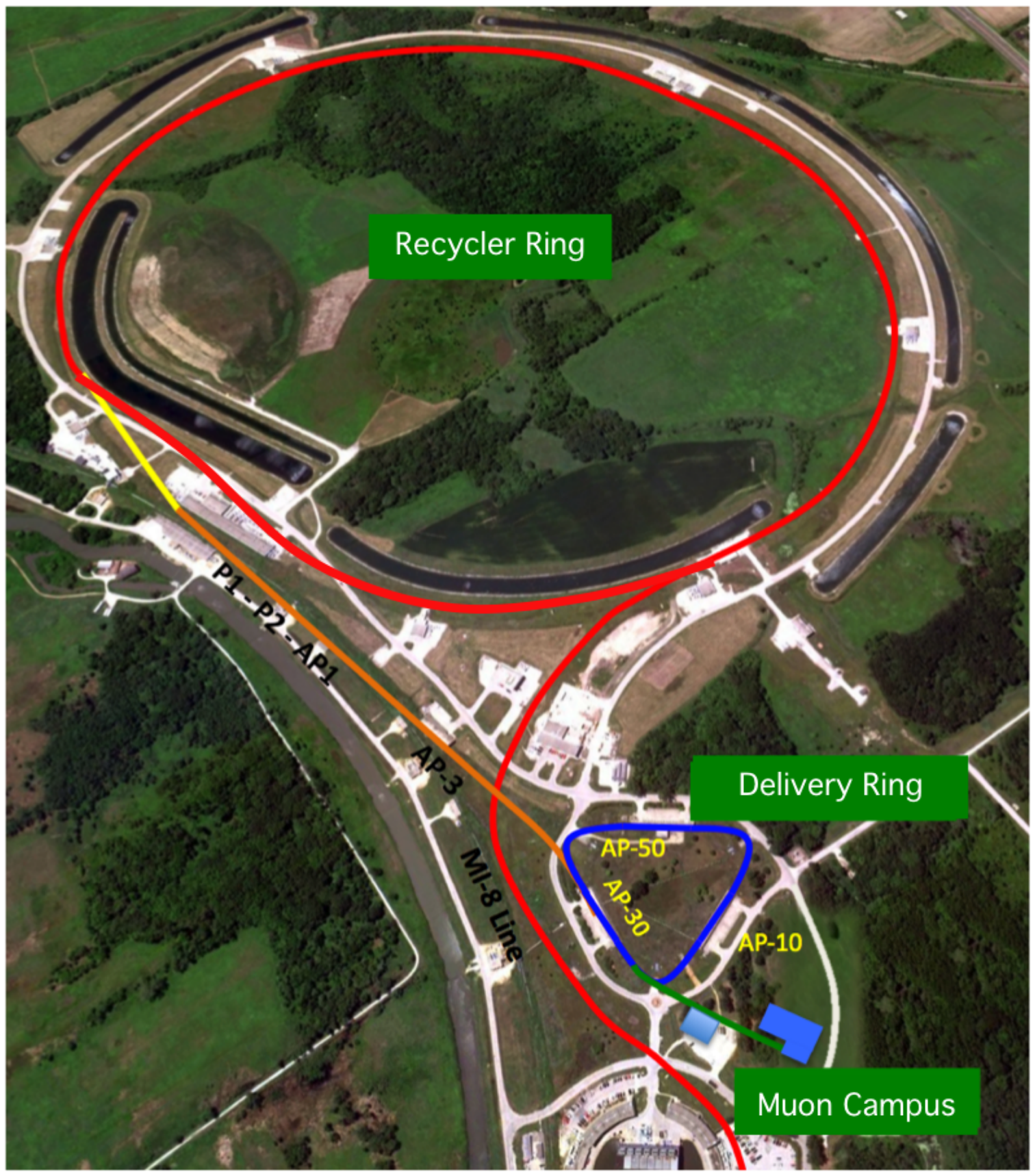

Figure 1.4: Aerial view of Muon Campus in relation to accelerator complex.

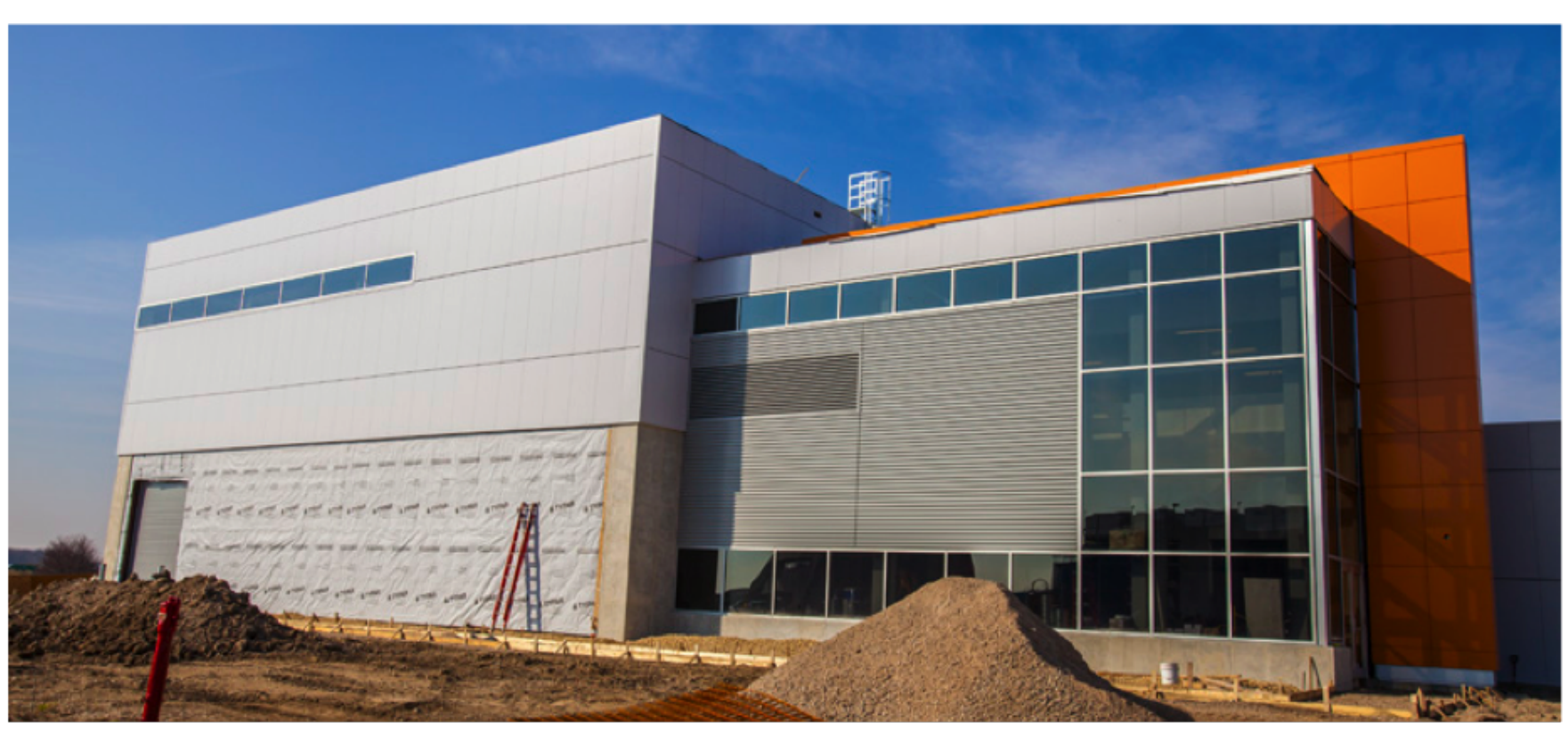

Figure 1.5: The completed MC-1 building in May 2014.

# Chapter 2

# Introduction and Physics Motivation

## 2.1 Introduction

This chapter gives the physics context of magnetic moment measurements, the Standard Model expectations, along with the reach of such experiments to identify and constrain physics beyond the Standard Model. Except for a broad-brush mention of the experimental technique, the details are left for later chapters. Chapter 3 gives an overview of the experimental method, and the subsequent chapters give the details. We attempt to follow the WBS structure in those later chapters.

## 2.2 Magnetic and Electric Dipole Moments

The study of magnetic moments of subatomic particles grew up with the development of quantum mechanics. For fermions the magnetic dipole moment (MDM) is related to the spin by

$$\vec{\mu} = g\frac{Qe}{2m}\vec{s}. \tag{2.1}$$

where $Q = \pm 1$ and $e > 0$. Our modern interpretation of the Stern-Gerlach experiments [1, 2] is that their observation: "to within 10% the magnetic moment of the silver atom is one Bohr magneton" was telling us that the $g$-factor of the un-paired electron is equal to 2. However, reaching this conclusion required the discovery of spin [3], quantum mechanics [4] along with with Thomas' relativistic correction [5]. Phipps and Taylor [6] repeated the Stern-Gerlach experiment in hydrogen, and mentioned the electron spin explicitly. One of the great successes of Dirac's relativistic theory [7] was the prediction that $g \equiv 2$.

For some years, the experimental situation remained the same. The electron had $g = 2$, and the Dirac equation seemed to describe nature. Then a surprising and completely unexpected result was obtained. In 1933, against the advice of Pauli who believed that the proton was a pure Dirac particle [8], Stern and his collaborators [9] showed that the $g$-factor of the proton was $\sim 5.5$, not the expected value of 2. Even more surprising was the discovery in 1940 by Alvarez and Bloch [10] that the neutron had a large magnetic moment.

In 1947, motivated by measurements of the hyperfine structure in hydrogen that obtained splittings larger than expected from the Dirac theory [11, 12, 13], Schwinger [51] showed that from a theoretical viewpoint these "discrepancies can be accounted for by a small additional electron spin magnetic moment" that arises from the lowest-order radiative correction to the Dirac moment[1],

$$\frac{\delta\mu}{\mu} = \frac{1}{2\pi}\frac{e^2}{\hbar c} = 0.001162. \tag{2.2}$$

It is useful to break the magnetic moment into two terms:

$$\mu = (1+a)\,\frac{e\hbar}{2m}, \quad \text{where} \quad a = \frac{(g-2)}{2}. \tag{2.3}$$

The first term is the Dirac moment, 1 in units of the appropriate magneton $e\hbar/2m$. The second term is the anomalous (Pauli) moment [14], where the dimensionless quantity $a$ (Schwinger's $\delta\mu/\mu$) is sometimes referred to as the *anomaly*.

### 2.2.1 The Muon

The muon was first observed in a Wilson cloud chamber by Kunze[15] in 1933, where it was reported to be "a particle of uncertain nature." In 1936 Anderson and Neddermeyer[16] reported the presence of "particles less massive than protons but more penetrating than electrons" in cosmic rays, which was confirmed in 1937 by Street and Stevenson[17], Nishina, Tekeuchi and Ichimiya[18], and by Crussard and Leprince-Ringuet[19]. The Yukawa theory of the nuclear force had predicted such a particle, but this "mesotron" as it was called, interacted too weakly with matter to be the carrier of the strong force. Today we understand that the muon is a second generation lepton, with a mass about 207 times the electron's. Like the electron, the muon obeys quantum electrodynamics, and can interact with other particles through the electromagnetic and weak forces. Unlike the electron which appears to be stable, the muon decays through the weak force predominantly by $\mu^- \to e^-\nu_\mu\bar{\nu}_e$. The muon's long lifetime of $\simeq 2.2\ \mu$s permits precision measurements of its mass, lifetime, and magnetic moment.

### 2.2.2 The Muon Magnetic Moment

The magnetic moment of the muon played an important role in the discovery of the generation structure of the Standard Model (SM). The pioneering muon spin rotation experiment at the Nevis cyclotron observed parity violation in muon decay [20], and also showed that $g_\mu$ was consistent with 2. Subsequent experiments at Nevis [22] and CERN [23] showed that $a_\mu \simeq \alpha/(2\pi)$, implying that in a magnetic field, the muon behaves like a heavy electron. Two additional experiments at CERN required that contributions from higher-order QED [24], and then from virtual hadrons [25] be included into the theory in order to reach agreement with experiment.

### 2.2.3 The Muon Electric Dipole Moment

Dirac [7] discovered an electric dipole moment (EDM) term in his relativistic electron theory. Like the magnetic dipole moment, the electric dipole moment must be along the spin. We can write an EDM expression similar to Eq. (2.1),

$$\vec{d} = \eta \left( \frac{Qe}{2mc} \right) \vec{s}, \tag{2.4}$$

where $\eta$ is a dimensionless constant that is analogous to $g$ in Eq. (2.1). While magnetic dipole moments (MDMs) are a natural property of charged particles with spin, electric dipole moments (EDMs) are forbidden both by parity and by time reversal symmetry.

[1] A misprint in the original paper has been corrected here.

The search for an EDM dates back to the suggestion of Purcell and Ramsey [26] in 1950, well in advance of the paper by Lee and Yang [27], that a measurement of the neutron EDM would be a good way to search for parity violation in the nuclear force. An experiment was mounted at Oak Ridge [28] soon thereafter that placed a limit on the neutron EDM of $d_n < 5 \times 10^{-20}$ $e$-cm, although the result was not published until after the discovery of parity violation.

Once parity violation was established, Landau [29] and Ramsey [30] pointed out that an EDM would violate both $P$ and $T$ symmetries. This can be seen by examining the Hamiltonian for a spin one-half particle in the presence of both an electric and magnetic field,

$$\mathcal{H} = -\vec{\mu} \cdot \vec{B} - \vec{d} \cdot \vec{E}. \tag{2.5}$$

The transformation properties of $\vec{E}$, $\vec{B}$, $\vec{\mu}$ and $\vec{d}$ are given in Table 2.2.3, and we see that while $\vec{\mu} \cdot \vec{B}$ is even under all three symmetries, $\vec{d} \cdot \vec{E}$ is odd under both $P$ and $T$. Thus the existence of an EDM implies that both $P$ and $T$ are not good symmetries of the interaction Hamiltonian, Eq. (2.5). The EDM is a $CP$-odd quantity, and if observed, would be the manifestation of a new source of $CP$ violation. The search for a muon EDM provides a unique opportunity to search for an EDM of a second-generation particle.

Table 2.1: Transformation properties of the magnetic and electric fields and dipole moments.

| | $\vec{E}$ | $\vec{B}$ | $\vec{\mu}$ or $\vec{d}$ |
|---|---|---|---|
| $P$ | - | + | + |
| $C$ | - | - | - |
| $T$ | + | - | - |

Concerning these symmetries, Ramsey states [30]:

> "However, it should be emphasized that while such arguments are appealing from the point of view of symmetry, they are not necessarily valid. Ultimately the validity of all such symmetry arguments must rest on experiment."

Fortunately this advice has been followed by many experimental investigators during the intervening 50 years. Since the Standard Model $CP$ violation observed in the neutral kaon and B-meson systems is inadequate to explain the predominance of matter over antimatter in the universe, the search for new sources of $CP$ violation beyond that embodied in the CKM formalism takes on a certain urgency. Searches for a permanent electric dipole moment of the electron, neutron, and of an atomic nucleus have become an important part of the search for physics beyond the Standard Model. The present limits on subatomic EDMs is given in Table 2.2.3.

Table 2.2: EDM Limits for various systems

| Particle | EDM Limit ($e$-cm) | SM value ($e$-cm) |
|---|---|---|
| $p$ [31] | $7.9 \times 10^{-25}$ | |
| $n$ [32] | $2.9 \times 10^{-26}$ | $\simeq 10^{-32}$ |
| $^{199}$Hg [31] | $3.1 \times 10^{-29}$ | $\simeq 10^{-32}$ |
| $e^-$ [33] | $1.05 \times 10^{-27}$ | $< 10^{-41}$ |
| $\mu$ [34] | $1.8 \times 10^{-19}$ | $< 10^{-38}$ |

## 2.3 Quick Summary of the Experimental Technique

Polarized muons are produced (see Chapter 7) and injected into the storage ring (see Chapter 12). The magnetic field is a dipole field, shimmed to ppm level uniformity. Vertical focusing is provided by electrostatic quadrupoles (see Chapter 13).

Two frequencies are measured experimentally: The rate at which the muon polarization turns relative to the momentum, called $\omega_a$, and the value of the magnetic field normalized to the Larmor frequency of a free proton, $\omega_p$.

The rate at which the spin[2] turns relative to the momentum, $\vec{\omega}_a = \vec{\omega}_S - \vec{\omega}_C$, where $S$ and $C$ stand for spin and cyclotron. These two frequencies are given by

$$\omega_S = -g\frac{Qe}{2m}B - (1-\gamma)\frac{Qe}{\gamma m}B; \tag{2.6}$$

$$\omega_C = -\frac{Qe}{m\gamma}B; \tag{2.7}$$

$$\omega_a = \omega_S - \omega_C = -\left(\frac{g-2}{2}\right)\frac{Qe}{m}B = -a\frac{Qe}{m}B \tag{2.8}$$

(where $e > 0$ and $Q = \pm 1$). There are two important features of $\omega_a$: (i) It only depends on the anomaly rather than on the full magnetic moment; (ii) It depends linearly on the applied magnetic field. In the presence of an electric field $\omega_a$ is modified

$$\vec{\omega}_a = -\frac{Qe}{m}\left[a_\mu \vec{B} - \left(a_\mu - \left(\frac{mc}{p}\right)^2\right)\frac{\vec{\beta}\times\vec{E}}{c}\right] \tag{2.9}$$

If operated at the "magic" momentum $p_{magic} = m/\sqrt{a_\mu} \simeq 3.09$ GeV/c the electric field contribution cancels in first order, and requires a small correction in second order.

[2]The term 'spin' is often used in place of the more accurate term 'polarization'

The magnetic field is weighted by the muon distribution, and also averaged over the running time weighed by the number of stored muons to determine the value of $\omega_p$ which is combined with the average $\omega_a$ to determine $a_\mu$. The reason for the use of these two frequencies, rather than $B$ measured in tesla can be understood from Eq. 2.9. To obtain $a_\mu$ from this relationship requires precise knowledge of the muon charge to mass ratio.

To determine $a_\mu$ from the two frequencies $\omega_a$ and $\omega_p$, we use the relationship

$$a_\mu = \frac{\omega_a/\omega_p}{\lambda_+ - \omega_a/\omega_p} = \frac{\mathcal{R}}{\lambda_+ - \mathcal{R}}\,, \tag{2.10}$$

where the ratio $\lambda_+ = \mu_{\mu^+}/\mu_p = 3.183\,345\,137\,(85)$ is the muon-to-proton magnetic moment ratio [43] measured from muonium (the $\mu^+e^-$ atom) hyperfine structure[45] (see Section 15.1.1 for futher details). Of course, to use $\lambda_+$ to determine $a_{\mu^-}$ requires the assumption of *CPT* invariance, *viz.* ($a_{\mu^+} = a_{\mu^-}$; $\lambda_+ = \lambda_-$). The comparison of $\mathcal{R}_{\mu^+}$ with $\mathcal{R}_{\mu^-}$ provides a *CPT* test. In E821

$$\Delta\mathcal{R} = \mathcal{R}_{\mu^-} - \mathcal{R}_{\mu^+} = (3.6 \pm 3.7) \times 10^{-9} \tag{2.11}$$

## 2.4 Results from E821

### 2.4.1 Measurement of $a_\mu$

The E821 Collaboration working at the Brookhaven Laboratory AGS used an electric quadrupole field to provide vertical focusing in the storage ring, and shimmed the magnetic field to ±1 ppm uniformity on average. The storage ring was operated at the "magic" momentum, $p_{magic} = 3.094$ GeV/c, ($\gamma_{magic} = 29.3$), such that $a_\mu = (m/p)^2$ and the electric field did not contribute to $\omega_a$.[3] The result is [36, 37]

$$a_\mu^{\rm E821} = 116\,592\,089(54)_{stat}(33)_{syst}(63)_{tot} \times 10^{-11} \quad (\pm 0.54\,\text{ppm}). \tag{2.12}$$

The results from E821 are shown in Fig. 2.1 (a) along with the Standard-Model value which is discussed below in Section 2.5. The importance of this result is illustrated in Fig. 2.1 (b) with a plot of the citations as a function of year.

## 2.5 The Standard-Model Value of $a_\mu$

The Standard-Model (SM) value of the muon anomaly can be calculated with sub-parts-per-million precision[4]. The comparison between the measured and the SM prediction provides a test of the completeness of the Standard Model. At present, there appears to be a three- to four-standard deviation between these two values, which has motivated extensive theoretical and experimental work on the hadronic contributions to the muon anomaly.

A lepton ($\ell = e,\, \mu,\, \tau$) has a magnetic moment which is along its spin, given by the relationship

$$\vec{\mu}_\ell = g_\ell \frac{Qe}{2m_\ell}\vec{s}, \qquad \underbrace{g_\ell = 2}_{\text{Dirac}}(1 + a_\ell), \qquad a_\ell = \frac{g_\ell - 2}{2} \tag{2.13}$$

[3] The magic momentum was first employed by the third CERN collaboration [25].

[4] This section is taken from Ref. [50]

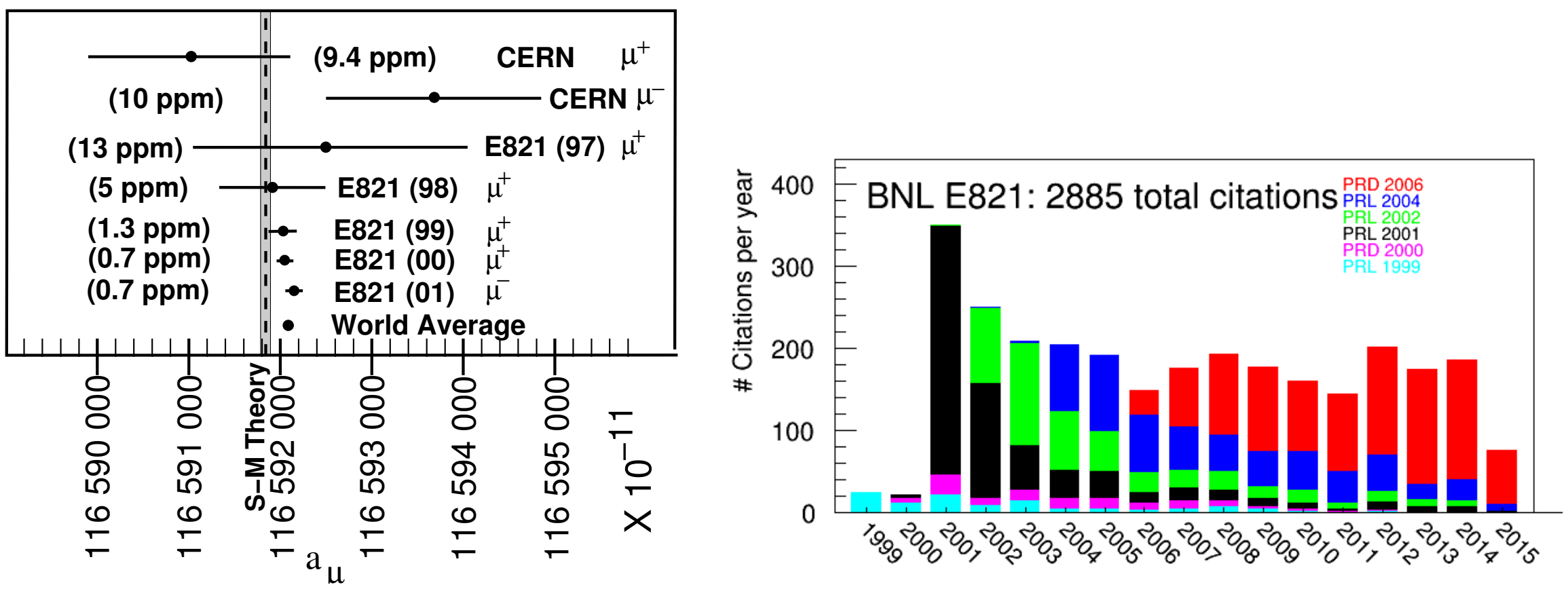

(a) (b)

Figure 2.1: (a)Measurements of $a_\mu$ from CERN and BNL E821. The vertical band is the SM value using the hadronic contribution from Ref. [71] (see Table 2.3). (b) Citations to the E821 papers by year as of April 2015: light blue [38] plus [39]; green [40]; red [41]; blue [36]; and yellow the Physical Review article [37].

where $Q = \pm 1$, $e > 0$ and $m_\ell$ is the lepton mass. Dirac theory predicts that $g \equiv 2$, but experimentally, it is known to be greater than 2. The small number $a$, the anomaly, arises from quantum fluctuations, with the largest contribution coming from the mass-independent single-loop diagram in Fig. 2.2(a). With his famous calculation that obtained $a = (\alpha/2\pi) = 0.00116\cdots$, Schwinger [51] started an "industry", which required Aoyama, Hayakawa, Kinoshita and Nio to calculate more than 12,000 diagrams to evaluate the tenth-order (five loop) contribution [52].

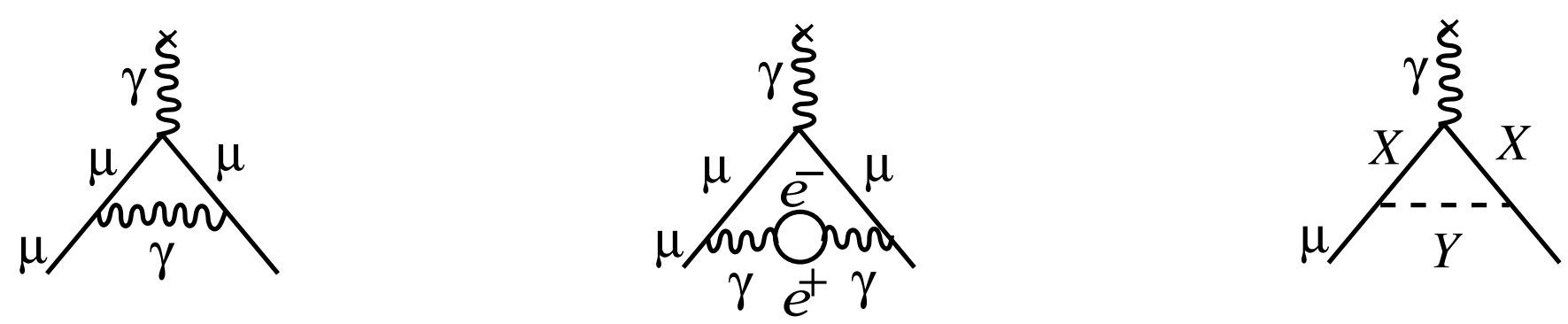

(a) (b) (c)

Figure 2.2: The Feynman graphs for: (a) The lowest-order (Schwinger) contribution to the lepton anomaly ; (b) The vacuum polarization contribution, which is one of five fourth-order, $(\alpha/\pi)^2$, terms; (c) The schematic contribution of new particles $X$ and $Y$ that couple to the muon.

The interaction shown in Fig. 2.2(a) is a chiral-changing, flavor-conserving process, which gives it a special sensitivity to possible new physics [53, 54]. Of course heavier particles can also contribute, as indicated by the diagram in Fig. 2.2(c). For example, $X = W^\pm$ and $Y = \nu_\mu$, along with $X = \mu$ and $Y = Z^0$, are the lowest-order weak contributions. In the Standard-Model, $a_\mu$ gets measureable contributions from QED, the strong interaction, and from the electroweak interaction,

$$a^{SM} = a^{QED} + a^{Had} + a^{Weak}. \tag{2.14}$$

In this document we present the latest evaluations of the SM value of $a_\mu$, and then discuss expected improvements that will become available over the next five to seven years. The uncertainty in this evaluation is dominated by the contribution of virtual hadrons in loops. A worldwide effort is under way to improve on these hadronic contributions. By the time that the Fermilab muon $(g-2)$ experiment, E989, reports a result later in this decade, the uncertainty should be significantly reduced. We emphasize that the existence of E821 at Brookhaven motivated significant work over the past thirty years that permitted more than an order of magnitude improvement in the knowledge of the hadronic contribution. Motivated by Fermilab E989 this work continues, and another factor of two improvement could be possible.

Both the electron [55] and muon [37] anomalies have been measured very precisely:

$$a_e^{exp} = 1\,159\,652\,180.73\,(28) \times 10^{-12} \quad \pm 0.24\,\text{ppb} \tag{2.15}$$

$$a_\mu^{exp} = 1\,165\,920\,89\,(63) \times 10^{-11} \quad \pm 0.54\,\text{ppm} \tag{2.16}$$

While the electron anomaly has been measured to $\simeq 0.3$ ppb (parts per billion) [55], it is significantly less sensitive to heavier physics, because the relative contribution of heavier virtual particles to the muon anomaly goes as $(m_\mu/m_e)^2 \simeq 43000$. Thus the lowest-order hadronic contribution to $a_e$ is [56]: $a_e^{\text{had,LO}} = (1.875 \pm 0.017)\ 10^{-12}$, 1.5 ppb of $a_e$. For the muon the hadronic contribution is $\simeq 60$ ppm (parts per million). So with much less precision, when compared with the electron, the measured muon anomaly is sensitive to mass scales in the several hundred GeV region. This not only includes the contribution of the $W$ and $Z$ bosons, but perhaps contributions from new, as yet undiscovered, particles such as the supersymmetric partners of the electroweak gauge bosons (see Fig. 2.2(c)).

### 2.5.1 Summary of the Standard-Model Value of $a_\mu$

#### QED Contribution

The QED contribution to $a_\mu$ is well understood. Recently the four-loop QED contribution has been updated and the full five-loop contribution has been calculated [52]. The present QED value is

$$a_\mu^{\text{QED}} = 116\ 584\ 718.951\ (0.009)(0.019)(0.007)(.077) \times 10^{-11} \tag{2.17}$$

where the uncertainties are from the lepton mass ratios, the eight-order term, the tenth-order term, and the value of $\alpha$ taken from the $^{87}$Rb atom $\alpha^{-1}(Rb) = 137.035\,999\,049(90)$ [0.66 ppb]. [57].

**Weak contributions**

The electroweak contribution (shown in Fig. 2.3) is now calculated through two loops [58, 59, 60, 61, 62, 63, 64, 65]. The one loop result

$$\begin{aligned} a_\mu^{\mathrm{EW}(1)} &= \frac{G_F}{\sqrt{2}} \frac{m_\mu^2}{8\pi^2} \left\{ \underbrace{\frac{10}{3}}_{W} + \underbrace{\frac{1}{3}(1-4\sin^2\theta_W)^2 - \frac{5}{3}}_{Z} \right. \\ &+ \left. \mathcal{O}\left(\frac{m_\mu^2}{M_Z^2}\log\frac{M_Z^2}{m_\mu^2}\right) + \frac{m_\mu^2}{M_H^2}\int_0^1 dx \frac{2x^2(2-x)}{1-x+\frac{m_\mu^2}{M_H^2}x^2} \right\} \\ &= 194.8 \times 10^{-11}\,, \end{aligned} \tag{2.18}$$

was calculated by five separate groups [66] shortly after the Glashow-Salam-Weinberg theory was shown by 't Hooft to be renormalizable. Due to the small Yukawa coupling of the Higgs boson to the muon, only the $W$ and $Z$ bosons contribute at a measurable level in the lowest-order electroweak term.

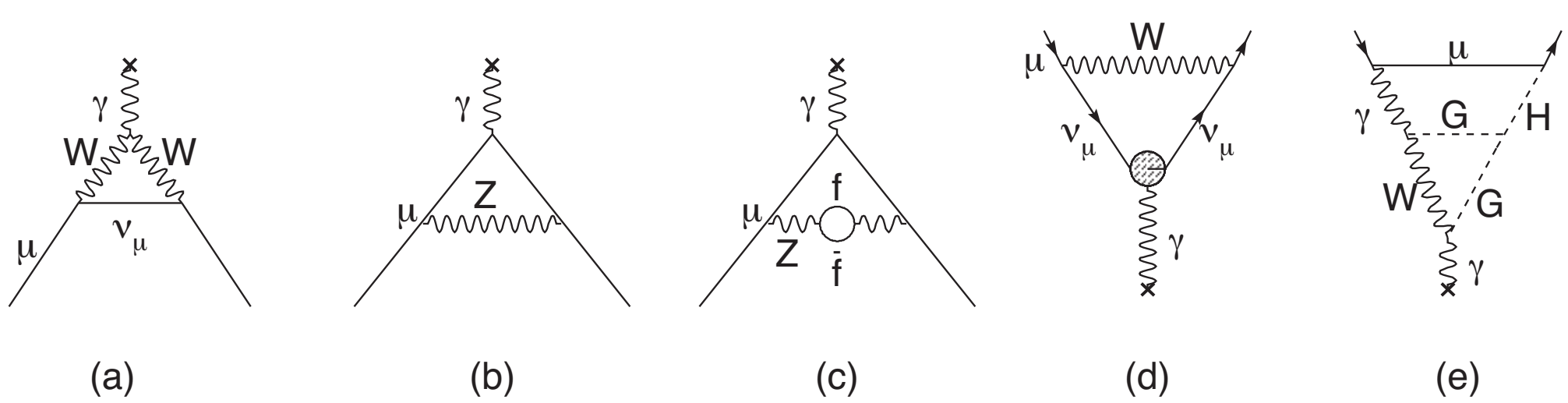

Figure 2.3: Weak contributions to the muon anomalous magnetic moment. Single-loop contributions from (a) virtual $W$ and (b) virtual $Z$ gauge bosons. These two contributions enter with opposite sign, and there is a partial cancellation. The two-loop contributions fall into three categories: (c) fermionic loops which involve the coupling of the gauge bosons to quarks, (d) bosonic loops which appear as corrections to the one-loop diagrams, and (e) a new class of diagrams involving the Higgs where $G$ is the longitudinal component of the gauge bosons. See Ref. [67] for details. The $\times$ indicates the photon from the magnetic field.

The two-loop electroweak contribution (see Figs. 2.3(c-e)), which is negative [60, 59, 58], has been re-evaluated using the LHC value of the Higgs mass and consistently combining exact two-loop with leading three-loop results [65]. The total electroweak contribution is

$$a_\mu^{\mathrm{EW}} = (153.6 \pm 1.0) \times 10^{-11} \tag{2.19}$$

where the error comes from hadronic effects in the second-order electroweak diagrams with quark triangle loops, along with unknown three-loop contributions [61, 68, 69, 70]. The leading logs for the next-order term have been shown to be small [61, 65]. The weak contribution is about 1.3 ppm of the anomaly, so the experimental uncertainty on $a_\mu$ of $\pm 0.54$ ppm now probes the weak scale of the Standard Model.

### Hadronic contribution

The hadronic contribution to $a_\mu$ is about 60 ppm of the total value. The lowest-order diagram shown in Fig. 2.4(a) dominates this contribution and its error, but the hadronic light-by-light contribution Fig. 2.4(e) is also important. We discuss both of these contributions below.

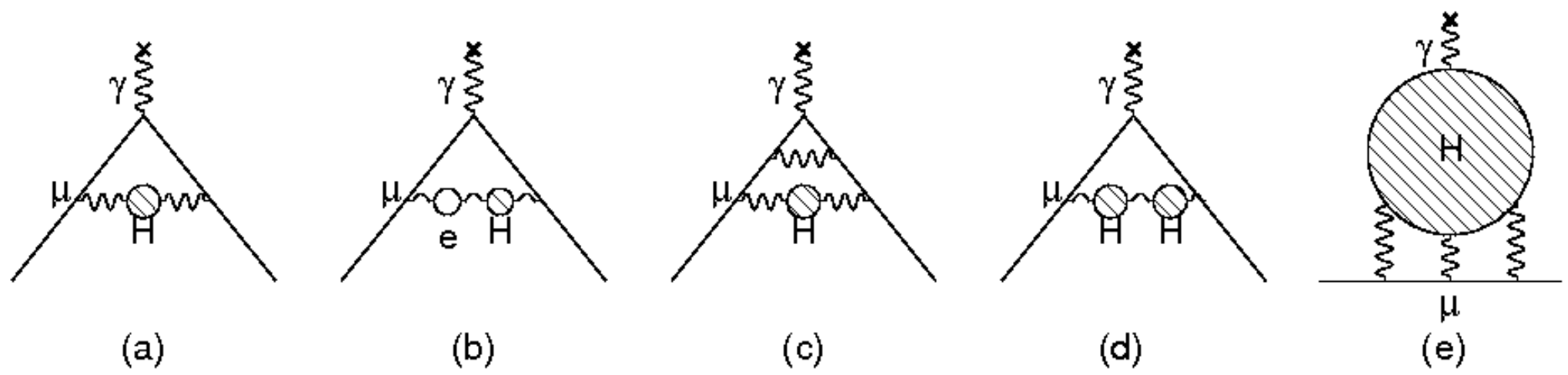

Figure 2.4: The hadronic contribution to the muon anomaly, where the dominant contribution comes from the lowest-order diagram (a). The hadronic light-by-light contribution is shown in (e).

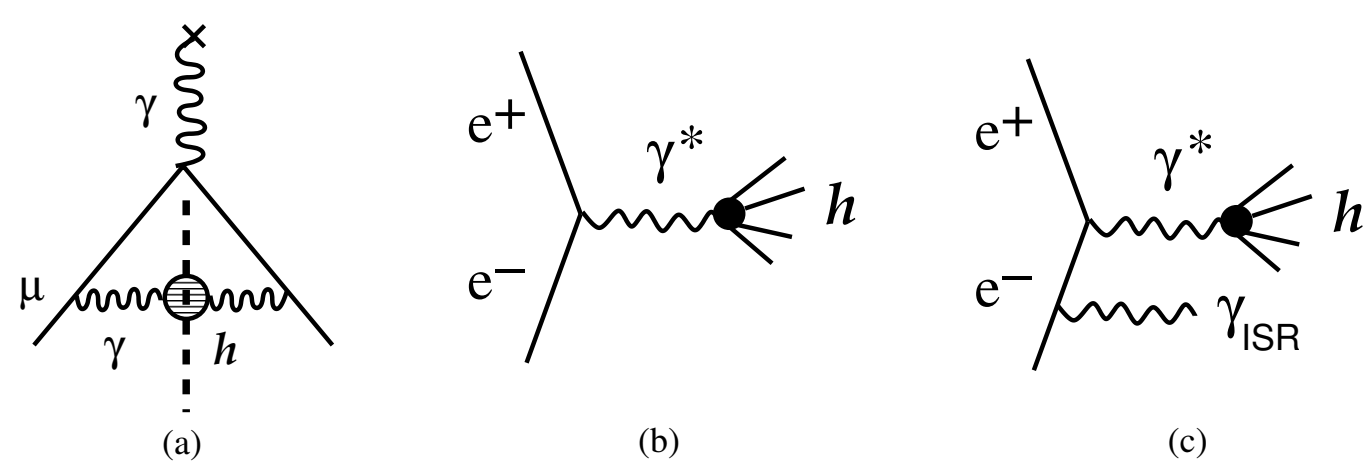

Figure 2.5: (a) The "cut" hadronic vacuum polarization diagram; (b) The $e^+e^-$ annihilation into hadrons; (c) Initial state radiation accompanied by the production of hadrons.

The energy scale for the virtual hadrons is of order $m_\mu c^2$, well below the perturbative region of QCD. However it can be calculated from the dispersion relation shown pictorially in Fig. 2.5,

$$a_\mu^{\mathrm{had;LO}} = \left(\frac{\alpha m_\mu}{3\pi}\right)^2 \int_{m_\pi^2}^{\infty} \frac{ds}{s^2} K(s) R(s), \quad \text{where} \quad R \equiv \frac{\sigma_{\mathrm{tot}}(e^+e^- \to \text{hadrons})}{\sigma(e^+e^- \to \mu^+\mu^-)}, \tag{2.20}$$

using the measured cross sections for $e^+e^- \to$ hadrons as input, where $K(s)$ is a kinematic factor ranging from 0.4 at $s = m_\pi^2$ to 0 at $s = \infty$ (see Ref. [67]). This dispersion relation relates the bare cross section for $e^+e^-$ annihilation into hadrons to the hadronic vacuum polarization contribution to $a_\mu$. Because the integrand contains a factor of $s^{-2}$, the values of $R(s)$ at low energies (the $\rho$ resonance) dominate the determination of $a_\mu^{\mathrm{had;LO}}$, however at the level of precision needed, the data up to 2 GeV are very important. This is shown in Fig. 2.6, where the left-hand chart gives the relative contribution to the integral for the different energy regions, and the right-hand gives the contribution to the error squared on the integral. The contribution is dominated by the two-pion final state, but other low-energy multi-hadron cross sections are also important. These data for $e^+e^-$ annihilation to hadrons are also important as input into the determination of $\alpha_{QED}(M_Z)$ and other electroweak precision measurements.

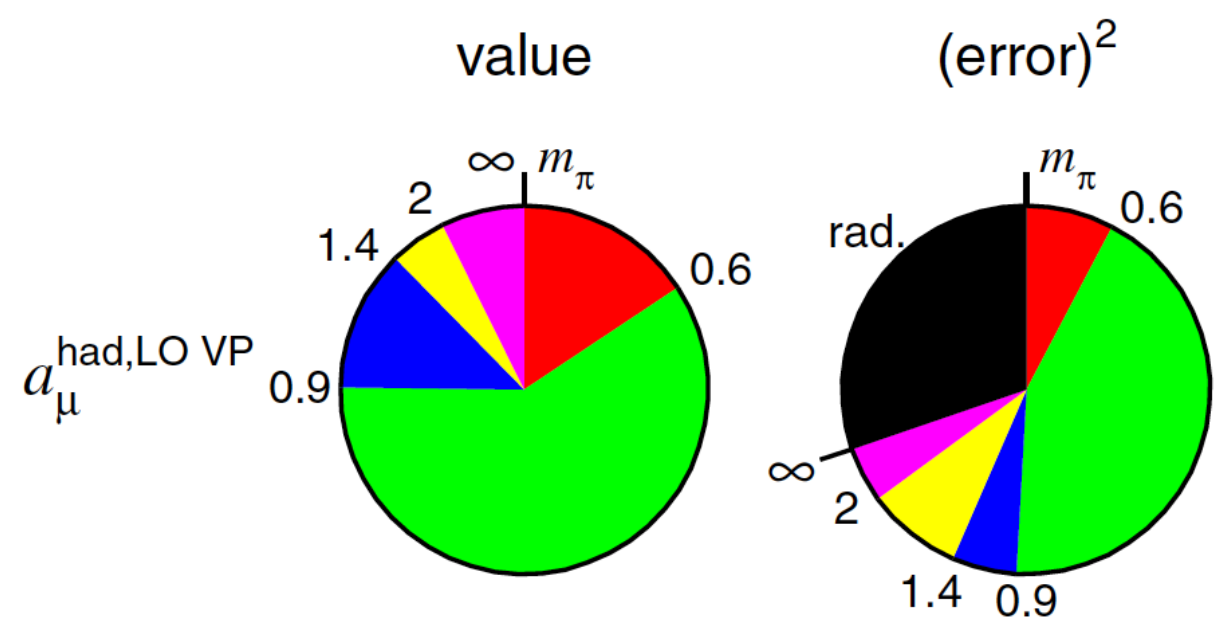

Figure 2.6: Contributions to the dispersion integral for different energy regions, and to the associated error (squared) on the dispersion integral in that energy region. Taken from Hagiwara et al. [72].

Two recent analyses [71, 72] using the $e^+e^- \to hadrons$ data obtained:

$$a_\mu^{\text{had;LO}} = (6\ 923 \pm 42) \times 10^{-11}\,, \tag{2.21}$$
$$a_\mu^{\text{had;LO}} = (6\ 949 \pm 43) \times 10^{-11}\,, \tag{2.22}$$

respectively. Important earlier global analyses include those of Hagiwara et al. [73], Davier, et al., [74], Jegerlehner and Nyffler [75].

In the past, hadronic $\tau$ spectral functions and CVC, together with isospin breaking corrections have been used to calculate the hadronic contribution [76, 71]. While the original predictions showed a discrepancy between $e^+e^-$ and $\tau$ based evaluations, it has been shown that after $\gamma$-$\rho$ mixing is taken into account, the two are compatible [77]. Recent evaluations based on a combined $e^+e^-$ and $\tau$ data fit using the Hidden Local Symmetry (HLS) model have come to similar conclusions and result in values for $a_\mu^{\text{HVP}}$ that are smaller than the direct evaluation without the HLS fit [78, 79].

The most recent evaluation of the next-to-leading order hadronic contribution shown in Fig. 2.4(b-d), which can also be determined from a dispersion relation, is [72]

$$a_\mu^{\text{had;NLO}} = (-98.4 \pm 0.6_{\text{exp}} \pm 0.4_{\text{rad}}\ ) \times 10^{-11}\,. \tag{2.23}$$

Very recently, also the next-to-next-to-leading order hadronic contribution has been evaluated [80], with a result of the order of the expected future experimental uncertainty. This result will be included in future evaluations of the full SM theory prediction.

### Hadronic light-by-light contribution

The hadronic light-by-light contribution (HLbL) cannot at present be determined from data, but rather must be calculated using hadronic models that correctly reproduce properties of QCD. This contribution is shown below in Fig. 2.7(a). It is dominated by the long-distance contribution shown in Fig. 2.7(b). In fact, in the so called chiral limit where the mass gap between the pseudoscalars ( Goldstone-like) particles and the other hadronic particles (the $\rho$ being the lowest vector state in Nature) is considered to be large, and to leading order in the $1/N_c$–expansion ($N_c$ the number of colors), this contribution has been calculated analytically [81] and provides a long-distance constraint to model calculations. There is also a short-distance constraint from the operator product expansion (OPE) of two electromagnetic currents which, in specific kinematic conditions, relates the light-by-light scattering amplitude to an Axial-Vector-Vector triangle amplitude for which one has a good theoretical understanding [82].

Unfortunately, the two asymptotic QCD constraints mentioned above are not sufficient for a full model independent evaluation of the HLbL contribution. Most of the last decade calculations found in the literature are compatible with the QCD chiral and large-$N_c$ limits. They all incorporate the $\pi^0$-exchange contribution modulated by $\pi^0\gamma^*\gamma^*$ form factors correctly normalized to the Adler, Bell-Jackiw point-like coupling. They differ, however, on whether or not they satisfy the particular OPE constraint mentioned above, and in the shape of the vertex form factors which follow from the different models.

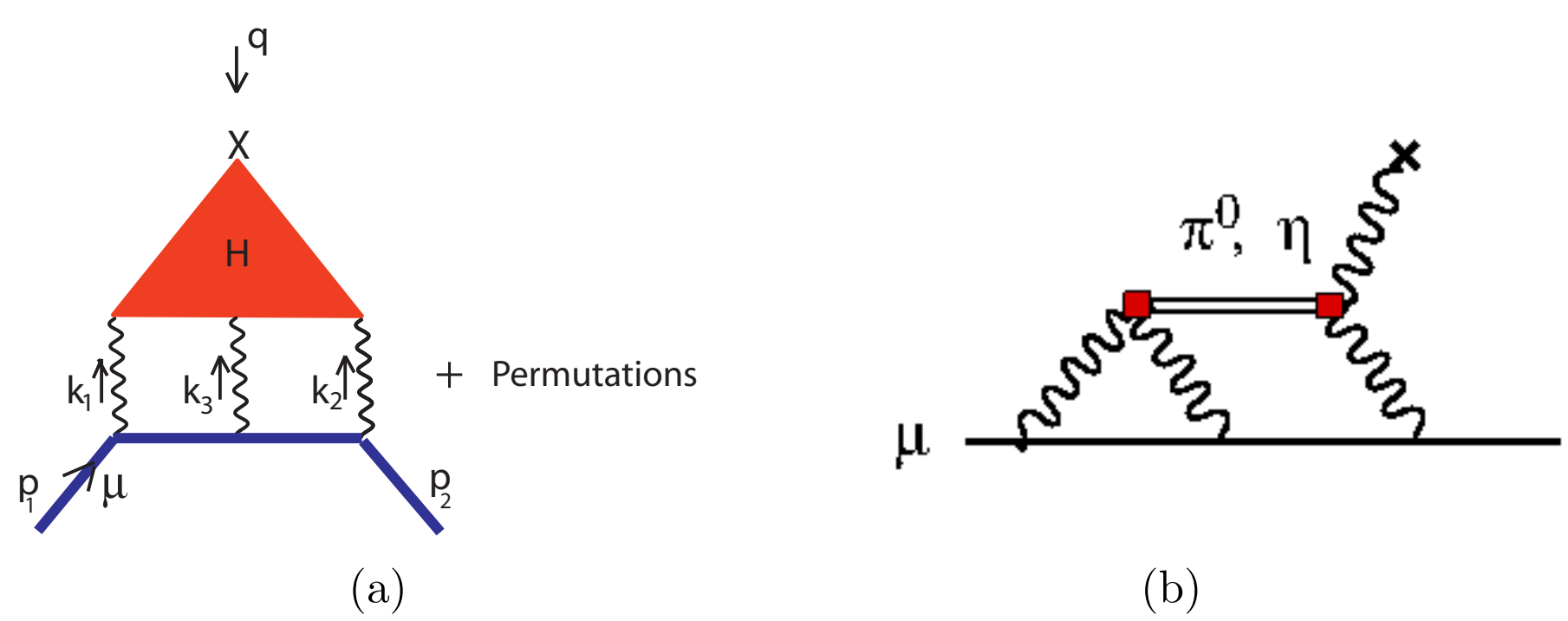

Figure 2.7: (a)The Hadronic Light-by contribution. (b) The pseudoscalar meson contribution.

A synthesis of the model contributions, which was agreed to by authors from each of the leading groups that have been working in this field, can be found in ref. [83][5]. They obtained

$$a_\mu^{\rm HLbL} = (105 \pm 26) \times 10^{-11} \,. \tag{2.24}$$

An alternate evaluation [75, 84] obtained, $a_\mu^{\rm HLbL} = (116 \pm 40) \times 10^{-11}$, which agrees well with the Glasgow Consensus [83]. Additional work on this contribution is underway on a number of fronts, including on the lattice. A workshop was held in March 2011 at the Institute for Nuclear Theory in Seattle [85] which brought together almost all of the interested experts. A second workshop followed at the Mainz Institute for Theoretical Physics in April 2014 [86].

[5]This compilation is generally referred to as the "Glasgow Consensus" since it grew out of a workshop in Glasgow in 2007. http://www.ippp.dur.ac.uk/old/MuonMDM/

One important point should be stressed here. The main physics of the hadronic light-by-light scattering contribution is well understood. In fact, but for the sign error unraveled in 2002, the theoretical predictions for $a_\mu^{\rm HLbL}$ have been relatively stable for more than ten years[6].

### 2.5.2 Summary of the Standard-Model Value and Comparison with Experiment

We determine the SM value using the new QED calculation from Aoyama [52]; the electroweak from Ref. [65], the hadronic light-by-light contribution from the "Glasgow Consensus" [83]; and lowest-order hadronic contribution from Davier, et al., [71], or Hagiwara et al., [72], and the higher-order hadronic contribution from Ref. [72]. A summary of these values is given in Table 2.3.

Table 2.3: Summary of the Standard-Model contributions to the muon anomaly. Two values are quoted because of the two recent evaluations of the lowest-order hadronic vacuum polarization.

| | Value ($\times 10^{-11}$) units |
|---|---|
| QED ($\gamma + \ell$) | $116\,584\,718.951 \pm 0.009 \pm 0.019 \pm 0.007 \pm 0.077_\alpha$ |
| HVP(lo) [71] | $6\,923 \pm 42$ |
| HVP(lo) [72] | $6\,949 \pm 43$ |
| HVP(ho) [72] | $-98.4 \pm 0.7$ |
| HLbL | $105 \pm 26$ |
| EW | $153.6 \pm 1.0$ |
| Total SM [71] | $116\,591\,802 \pm 42_{\text{H-LO}} \pm 26_{\text{H-HO}} \pm 2_{\text{other}}\ (\pm 49_{\text{tot}})$ |
| Total SM [72] | $116\,591\,828 \pm 43_{\text{H-LO}} \pm 26_{\text{H-HO}} \pm 2_{\text{other}}\ (\pm 50_{\text{tot}})$ |

This SM value is to be compared with the combined $a_\mu^+$ and $a_\mu^-$ values from E821 [37] corrected for the revised value of $\lambda = \mu_\mu/\mu_p$ from Ref [43],

$$a_\mu^{\rm E821} = (116\,592\,089 \pm 63) \times 10^{-11} \quad (0.54\,{\rm ppm}), \tag{2.25}$$

which give a difference of

$$\begin{aligned} \Delta a_\mu({\rm E821} - {\rm SM}) &= (287 \pm 80) \times 10^{-11}\ [71] &(2.26)\\ &= (261 \pm 80) \times 10^{-11}\ [72] &(2.27)\end{aligned}$$

depending on which evaluation of the lowest-order hadronic contribution that is used [71, 72].

This comparison between the experimental values and the present Standard-Model value is shown graphically in Fig. 2.1. The lowest-order hadronic evaluation of Ref. [79] using the hidden local symmetry model results in a difference between experiment and theory that ranges between 4.1 to 4.7$\sigma$.

[6] A calculation using a Dyson-Schwinger approach [87] initially reported a much larger value for the HLbL contribution. Subsequently a numerical mistake was found. These authors are continuing this work, but the calculation is still incomplete.

This difference of 3.3 to 3.6 standard deviations is tantalizing, but we emphasize that whatever the final agreement between the measured and SM value turns out to be, it will have significant implications on the interpretation of new phenomena that might be found at the LHC and elsewhere. Because of the power of $a_\mu$ to constrain, or point to, speculative models of New Physics, the E821 results have been highly cited, see Fig. 2.1 (b) and Section 2.7 below.

## 2.6 Expected Improvements in the Standard-Model Value

The present uncertainty on the theoretical value is dominated by the hadronic contributions [71, 72] (see Table 2.3). The lowest-order contribution determined from $e^+e^- \rightarrow$ hadrons data using a dispersion relation is theoretically relatively straightforward. It does require the combination of data sets from different experiments. The only significant theoretical uncertainty comes from radiative corrections, such as vacuum polarization (running $\alpha$), along with initial and final state radiation effects, which are needed to obtain the correct hadronic cross section at the required level of precision. This was a problem for the older data sets. In the analysis of the data collected over the past 15 years, which now dominate the determination of the hadronic contribution, the treatment of radiative corrections has been significantly improved. Nevertheless, an additional uncertainty due to the treatment of these radiative corrections in the older data sets has been estimated to be of the order of $20 \times 10^{-11}$ [72]. As more data become available, this uncertainty will be significantly reduced.

There are two methods that have been used to measure the hadronic cross sections: The energy scan (see Fig. 2.5(b)), and using initial state radiation with a fixed beam energy to measure the cross section for energies below the total center-of-mass energy of the colliding beams (see Fig. 2.5(c)). Both are being employed in the next round of measurements. The data from the new experiments that are now underway at VEPP-2000 in Novosibirsk and BESIII in Beijing, when combined with the analysis of existing multi-hadron final-state data from BaBar and Belle, should significantly reduce the uncertainty on the lowest-order hadronic contribution.

The hadronic-light-by-light contribution does not lend itself easily to determination by a dispersion relation, see however recent progress reported in Ref. [88] and in talks at the Mainz workshop [86]. Nevertheless there are some experimental data that can help to pin down related amplitudes and to constrain form factors used in the model calculations.

### 2.6.1 Lowest-order Hadronic Contribution

Much experimental and theoretical work is going on worldwide to refine the hadronic contribution. The theory of $(g-2)$, relevant experiments to determine the hadronic contribution, including work on the lattice, have featured prominently in the series of tau-lepton workshops and PHIPSI workshops which are held in alternate years. Over the development period of Fermilab E989, we expect further improvements in the SM-theory evaluation. This projection is based on the following developments:

### Novosibirsk

The VEPP2M machine has been upgraded to VEPP-2000. The maximum energy has been increased from $\sqrt{s} = 1.4$ GeV to 2.0 GeV. Additionally, the SND detector has been upgraded and the CMD2 detector was replaced by the much-improved CMD3 detector. The cross section will be measured from threshold to 2.0 GeV using an energy scan, filling in the energy region between 1.4 GeV, where the previous scan ended, up to 2.0 GeV, the lowest energy point reached by the BES collaboration in their measurements. See Fig. 2.6 for the present contribution to the overall error from this region. Engineering runs began in 2009, and data collection started in 2011. So far two independent energy scans between 1.0 and 2.0 GeV were performed in 2011 and 2012. The peak luminosity of $3 \times 10^{31}\text{cm}^{-2}\text{s}^{-1}$ was achieved, which was limited by the positron production rate. The new injection facility, scheduled to be commissioned during the 2013-2014 upgrade, should permit the luminosity to reach $10^{32}\text{cm}^{-2}\text{s}^{-1}$ . Data collection resumed in late 2012 with a new energy scan over energies below 1.0 GeV. The goal of experiments at VEPP-2000 is to achieve a systematic error 0.3-0.5% in the $\pi^+\pi^-$ channel, with negligible statistical error in the integral. The high statistics, expected at VEPP-2000, should allow a detailed comparison of the measured cross-sections with ISR results at BaBar and DAΦNE. After the upgrade, experiments at VEPP-2000 plan to take a large amount of data at 1.8-2 GeV, around the $N\bar{N}$ threshold. This will permit ISR data with the beam energy of 2 GeV, which is between the PEP2 energy at the $\Upsilon(4S)$ and the 1 GeV $\phi$ energy at the DAΦNE facility in Frascati. The dual ISR and scan approach will provide an important cross check on the two central methods used to determine the HVP.

### The BESIII Experiment

The BESIII experiment at the Beijing tau-charm factory BEPC-II has already collected several inverse femtobarns of integrated luminosity at various centre-of-mass energies in the range 3 - 4.5 GeV. The ISR program includes cross section measurements of: $e^+e^- \to \pi^+\pi^-$, $e^+e^- \to \pi^+\pi^-\pi^0$, $e^+e^- \to \pi^+\pi^-\pi^0\pi^0$ – the final states most relevant to $(g-2)_\mu$. Presently, a data sample of 2.9 fb$^{-1}$ at $\sqrt{s} = 3.77$ GeV is being analyzed, but new data at $\sqrt{s} > 4$ GeV can be used for ISR physics as well and will double the statistics. Using these data, hadronic invariant masses from threshold up to approximately 3.5 GeV can be accessed at BESIII. Although the integrated luminosities are orders of magnitude lower compared to the $B$-factory experiments BaBar and BELLE, the ISR method at BESIII still provides competitive statistics. This is due to the fact that the most interesting mass range for the HVP contribution of $(g-2)_\mu$, which is below approximately 3 GeV, is very close to the centre-of-mass energy of the collider BEPC-II and hence leads to a configuration where only relatively low-energetic ISR photons need to be emitted, providing a high ISR cross section. Furthermore, in contrast to the $B$ factories, small angle ISR photons can be included in the event selection for kinematic reasons which leads to a very high overall geometrical acceptance. Compared to the KLOE experiment, background from final state radiation (FSR) is reduced significantly as this background decreases with increasing center of mass energies of the collider. BESIII is aiming for a precision measurement of the ISR $R$-ratio $R_{\mathrm{ISR}} = N(\pi\pi\gamma)/N(\mu\mu\gamma)$ with a precision of about 1%. This requires an excellent pion-muon separation, which is achieved by training a multi-variate neural network. As a preliminary result, an absolute cross section measurement of the reaction $e^+e^- \to \mu^+\mu^-\gamma$ has been achieved, which agrees with the QED prediction within 1% precision.

Moreover, at BESIII a new energy scan campaign is planned to measure the inclusive $R$ ratio in the energy range between 2.0 and 4.6 GeV. Thanks to the good performance of the BEPC-II accelerator and the BESIII detector a significant improvement upon the existing BESII measurement can be expected. The goal is to arrive at an inclusive $R$ ratio measurement with about 1% statistical and 3% systematic precision per scan point.

### Summary of the Lowest-Order Improvements from Data

A substantial amount of new $e^+e^-$ cross section data will become available over the next few years. These data have the potential to significantly reduce the error on the lowest-order hadronic contribution. These improvements can be obtained by reducing the uncertainties of the hadronic cross-sections from 0.7% to 0.4% in the region below 1 GeV and from 6% to 2% in the region between 1 and 2 GeV as shown in Table 2.4.

| | $\delta(\sigma)/\sigma$ present | $\delta a_\mu$present | $\delta(\sigma)/\sigma$ future | $\delta a_\mu$future |
|---|---|---|---|---|
| $\sqrt{s} < 1$ GeV | 0.7% | 33 | 0.4% | 19 |
| $1 < \sqrt{s} < 2$ GeV | 6% | 39 | 2% | 13 |
| $\sqrt{s} > 2$ GeV | | 12 | | 12 |
| total | | 53 | | 26 |

Table 2.4: Overall uncertainty of the cross-section measurement required to get the reduction of uncertainty on $a_\mu$ in units $10^{-11}$ for three regions of $\sqrt{s}$ (from Ref. [93]).

### Lattice calculation of the Lowest-Order HVP:

With computer power presently available, it is possible for lattice QCD calculations to make important contributions to our knowledge of the lowest-order hadronic contribution. Using several different discretizations for QCD, lattice groups around the world are computing the HVP [94, 95, 96, 97, 98] (see also several recent talks at Lattice 2013 (Mainz)). The varied techniques have different systematic errors, but in the continuum limit $a \to 0$ they should all agree. Many independent calculations provide a powerful check on the lattice results, and ultimately the dispersive ones too.

Several groups are now performing simulations with physical light quark masses on large boxes, eliminating significant systematic errors. So called quark-disconnected diagrams are also being calculated, and several recent theory advances will help to reduce systematic errors associated with fitting and the small $q^2$ regime [99, 95, 100, 101, 102, 103]. While the HVP systematic errors are well understood, significant computational resources are needed to control them at the $\sim 1\%$ level, or better. Taking into account current resources and those expected in the next few years, the lattice-QCD uncertainty on $a_\mu(\mathrm{HVP})$, currently at the $\sim 5\%$-level, can be reduced to 1 or 2% within the next few years. This is already interesting as a wholly independent check of the dispersive results for $a_\mu(\mathrm{HVP})$. With increasing experience and computer power, it should be possible to compete with the $e^+e^-$ determination of $a_\mu(\mathrm{HVP})$ by the end of the decade, perhaps sooner with additional technical advances.

### 2.6.2 The Hadronic Light-by-Light contribution

There are two major approaches to improving the HLbL contribution, beyond theoretical work on refining the existing model calculations: Using experimental data from measurements of $\gamma^*$ physics at BESIII and KLOE; calculations on the lattice.

Any experimental information on the neutral pion lifetime and the transition form factor is important in order to constrain the models used for calculating the pion-exchange contribution (see Fig. 2.7(b)). However, having a good description, e.g. for the transition form factor, is only necessary, not sufficient, in order to uniquely determine $a_\mu^{\mathrm{HLbL};\pi^0}$. As stressed in Ref. [106], what enters in the calculation of $a_\mu^{\mathrm{HLbL};\pi^0}$ is the fully off-shell form factor $\mathcal{F}_{\pi^{0*}\gamma^*\gamma^*}((q_1+q_2)^2, q_1^2, q_2^2)$ (vertex function), where also the pion is off-shell with 4-momentum $(q_1+q_2)$. Such a (model dependent) form factor can for instance be defined via the QCD Green's function $\langle VVP \rangle$, see Ref. [84] for details. The form factor with on-shell pions is then given by $\mathcal{F}_{\pi^0\gamma^*\gamma^*}(q_1^2, q_2^2) \equiv \mathcal{F}_{\pi^{0*}\gamma^*\gamma^*}(m_\pi^2, q_1^2, q_2^2)$. Measurements of the transition form factor $\mathcal{F}_{\pi^0\gamma^*\gamma}(Q^2) \equiv \mathcal{F}_{\pi^{0*}\gamma^*\gamma^*}(m_\pi^2, -Q^2, 0)$ are in general only sensitive to a subset of the model parameters and do not permit the reconstruction the full off-shell form factor.

For different models, the effects of the off-shell pion can vary a lot. In Ref. [84] the off-shell lowest meson dominance (LMD) plus vector meson dominance (LMD+V) form factor was proposed and the estimate $a_{\mu;\mathrm{LMD+V}}^{\mathrm{HLbL};\pi^0} = (72 \pm 12) \times 10^{-11}$ was obtained (see also Ref. [107]). The error estimate comes from the variation of all model parameters, where the uncertainty of the parameters related to the off-shellness of the pion completely dominates the total error. In contrast to the off-shell LMD+V model, many other models, e.g. the VMD model or constituent quark models, do not have these additional sources of uncertainty related to the off-shellness of the pion. These models often have only very few parameters, which can all be fixed by measurements of the transition form factor or from other observables. Therefore, for such models, the precision of the KLOE-2 measurement can dominate the total precision of $a_\mu^{\mathrm{HLbL};\pi^0}$.

Essentially all evaluations of the pion-exchange contribution use for the normalization of the form factor, $\mathcal{F}_{\pi^{0*}\gamma^*\gamma^*}(m_\pi^2, 0, 0) = 1/(4\pi^2 F_\pi)$, as derived from the Wess-Zumino-Witten (WZW) term. Then the value $F_\pi = 92.4$ MeV is used without any error attached to it, i.e. a value close to $F_\pi = (92.2 \pm 0.14)$ MeV, obtained from $\pi^+ \to \mu^+\nu_\mu(\gamma)$ [108]. If one uses the decay width $\Gamma_{\pi^0\to\gamma\gamma}$ for the normalization of the form factor, an additional source of uncertainty enters, which has not been taken into account in most evaluations [109]. Until recently, the experimental world average of $\Gamma_{\pi^0\to\gamma\gamma}^{PDG} = 7.74 \pm 0.48$ eV [108] was only known to 6.2% precision. Due to the poor agreement between the existing data, the PDG error of the width average is inflated (scale factor of 2.6) and it gives an additional motivation for new precise measurements. The PrimEx Collaboration, using a Primakoff effect experiment at JLab, has achieved 2.8% fractional precision [110]. There are plans to further reduce the uncertainty to the percent level. Though theory and experiment are in a fair agreement, a better experimental precision is needed to really test the theory predictions.

### Impact of KLOE-2 measurements on $a_\mu^{\mathrm{HLbL};\pi^0}$

For the new data taking of the KLOE-2 detector, which is expected to start by the end of 2013, new small angle tagging detectors have been installed along DAΦNE beam line.

These "High Energy Tagger" detectors [111] offer the possibility to study a program of $\gamma\gamma$ physics through the process $e^+e^- \to e^+\gamma^* e^-\gamma^* \to e^+e^-X$.

Thus a coincidence between the scattered electrons and a $\pi^0$ would provide information on $\gamma^*\gamma^* \to \pi^0$ [104], and will provide experimental constraints on the models used to calculate the hadronic light-by-light contribution [105].

In Ref. [112] it was shown that planned measurements at KLOE-2 could determine the $\pi^0 \to \gamma\gamma$ decay width to 1% statistical precision and the $\gamma^*\gamma \to \pi^0$ transition form factor $\mathcal{F}_{\pi^0\gamma^*\gamma}(Q^2)$ for small space-like momenta, $0.01\ \mathrm{GeV}^2 \leq Q^2 \leq 0.1\ \mathrm{GeV}^2$, to 6% statistical precision in each bin. The simulations have been performed with the Monte-Carlo program EKHARA [113] for the process $e^+e^- \to e^+e^-\gamma^*\gamma^* \to e^+e^-\pi^0$, followed by the decay $\pi^0 \to \gamma\gamma$ and combined with a detailed detector simulation. The results of the simulations are shown in Figure 2.8. The KLOE-2 measurements will allow to almost directly measure the slope of the form factor at the origin and check the consistency of models which have been used to extrapolate the data from larger values of $Q^2$ down to the origin. With the decay width $\Gamma^{\mathrm{PDG}}_{\pi^0\to\gamma\gamma}$ $[\Gamma^{\mathrm{PrimEx}}_{\pi^0\to\gamma\gamma}]$ and current data for the transition form factor $\mathcal{F}_{\pi^0\gamma^*\gamma}(Q^2)$, the error on $a_\mu^{\mathrm{HLbL};\pi^0}$ is $\pm 4 \times 10^{-11}$ $[\pm 2 \times 10^{-11}]$, not taking into account the uncertainty related to the off-shellness of the pion. Including the simulated KLOE-2 data reduces the error to $\pm(0.7 - 1.1) \times 10^{-11}$.

### BESIII Hadronic light-by-light contribution

Presently, data taken at $\sqrt{s}$ =3.77 GeV are being analyzed to measure the form factors of the reactions $\gamma^*\gamma \to X$, where $X = \pi^0, \eta, \eta', 2\pi$.

BESIII has launched a program of two-photon interactions with the primary goal to measure the transition form factors (TFF) of pseudoscalar mesons as well as of the two-pion system in the spacelike domain. These measurements are carried out in the single-tag mode, i.e. by tagging one of the two beam leptons at large polar angles and by requiring that the second lepton is scattered at small polar angles. With these kinematics the form factor, which in general depends on the virtualities of the two photons, reduces to $F(Q^2)$, where $Q^2$ is the negative momentum transfer of the tagged lepton. At BESIII, the process $\gamma\gamma^* \to \pi^0$, which is known to play a leading contribution in the HLbL correction to $(g-2)$, can be measured with unprecedented precision in the $Q^2$ range between 0.3 GeV$^2$ and 4 GeV$^2$. In the future BESIII will also embark on untagged as well as double-tag measurements, in which either both photons are quasi-real or feature a high virtuality. The goal is to carry out this program for the final states $\pi^0, \eta, \eta', \pi\pi$. It still needs to be proven that the small angle detector, which recently has been installed close to the BESIII beamline, can be used for the two-photon program.

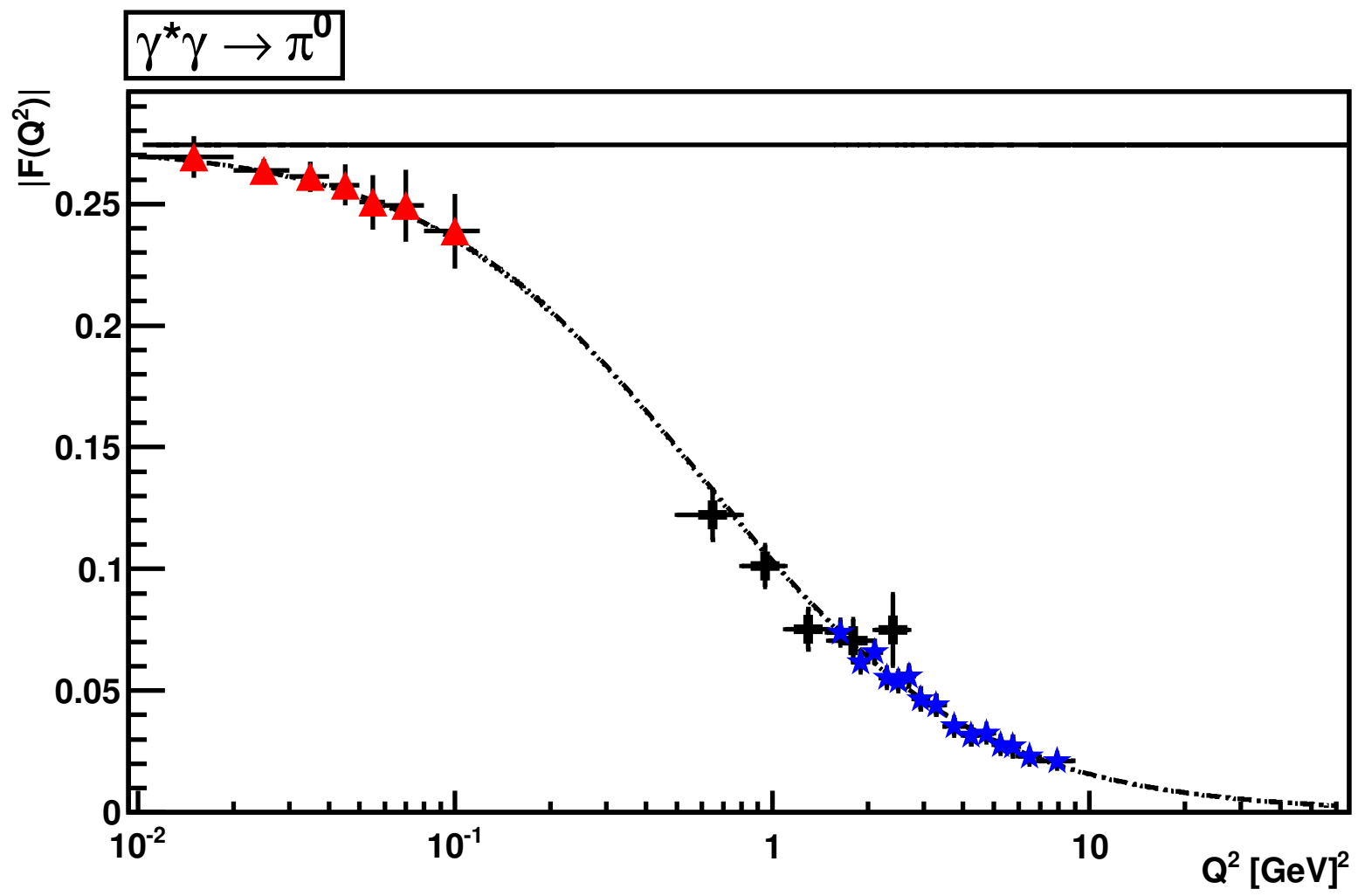

Figure 2.8: Simulation of KLOE-2 measurement of $F(Q^2)$ (red triangles) with statistical errors for 5 fb$^{-1}$, corresponding to one year of data taking. The dashed line is the $F(Q^2)$ form factor according to the LMD+V model [84, 107], the solid line is $F(0) = 1/(4\pi^2 F_\pi)$ given by the Wess-Zumino-Witten term. Data [114] from CELLO (black crosses) and CLEO (blue stars) at high $Q^2$ are also shown for illustration.

**Lattice calculation of Hadronic Light-by-Light Scattering:**

Model calculations show that the hadronic light-by-light (HLbL) contribution is roughly $(105 \pm 26) \times 10^{-11}$, $\sim 1$ ppm of $a_\mu$. Since the error attributed to this estimate is difficult to reduce, a modest, but first principles calculation on the lattice would have a large impact. Recent progress towards this goal has been reported [96], where a non-zero signal (statistically speaking) for a part of the amplitude emerged in the same ball-park as the model estimate. The result was computed at non-physical quark mass, with other systematic errors mostly uncontrolled. Work on this method, which treats both QED and QCD interactions non-perturbatively, is continuing. The next step is to repeat the calculation on an ensemble of gauge configurations that has been generated with electrically charged sea quarks (see the poster by Blum presented at Lattice 2013). The charged sea quarks automatically include the quark disconnected diagrams that were omitted in the original calculation and yield the complete amplitude. As for the HVP, the computation of the HLbL contribution requires significant resources which are becoming available. While only one group has so far attempted the calculation, given the recent interest in the HVP contribution computed in lattice QCD and electromagnetic corrections to hadronic observables in general, it seems likely that others will soon enter the game. And while the ultimate goal is to compute the HLbL contribution to 10% accuracy, or better, we emphasize that a lattice calculation with even a solid 30% error would already be very interesting. Such a result, while not guaranteed, is not out of the question during the next 3-5 years.

### 2.6.3 Summary of the Standard Model Contribution

The muon and electron anomalous magnetic moments are among, if not the most precisely measured and calculated quantities in all of physics. The theoretical uncertainty on the Standard-Model contribution to $a_\mu$ is $\simeq 0.4$ ppm, slightly smaller than the experimental error from BNL821. The new Fermilab experiment, E989, will achieve a precision of 0.14 ppm. While the hadronic corrections will most likely not reach that level of precision, their uncertainty will be significantly decreased. The lowest-order contribution will be improved by new data from Novosibirsk and BESIII. On the timescale of the first results from E989, the lattice will also become relevant.

The hadronic light-by-light contribution will also see significant improvement. The measurements at Frascati and at BESIII will provide valuable experimental input to constrain the model calculations. There is hope that the lattice could produce a meaningful result by 2018.

We summarize possible near-future improvements in the table below. Since it is difficult to project the improvements in the hadronic light-by-light contribution, we assume a conservative improvement: That the large amount of work that is underway to understand this contribution, both experimentally and on the lattice, will support the level of uncertainty assigned in the "Glasgow Consensus". With these improvements, the overall uncertainty on $\Delta a_\mu$ could be reduced by a factor 2. In case the central value would remain the same, the statistical significance would become 7-8 standard deviations, as it can be seen in Fig. 2.9.

| Error | [71] | [72] | Future |
|---|---|---|---|
| $\delta a_\mu^{\rm SM}$ | 49 | 50 | 35 |
| $\delta a_\mu^{\rm HLO}$ | 42 | 43 | 26 |
| $\delta a_\mu^{\rm HLbL}$ | 26 | 26 | 25 |
| $\delta(a_\mu^{\rm EXP} - a_\mu^{\rm SM})$ | 80 | 80 | 40 |

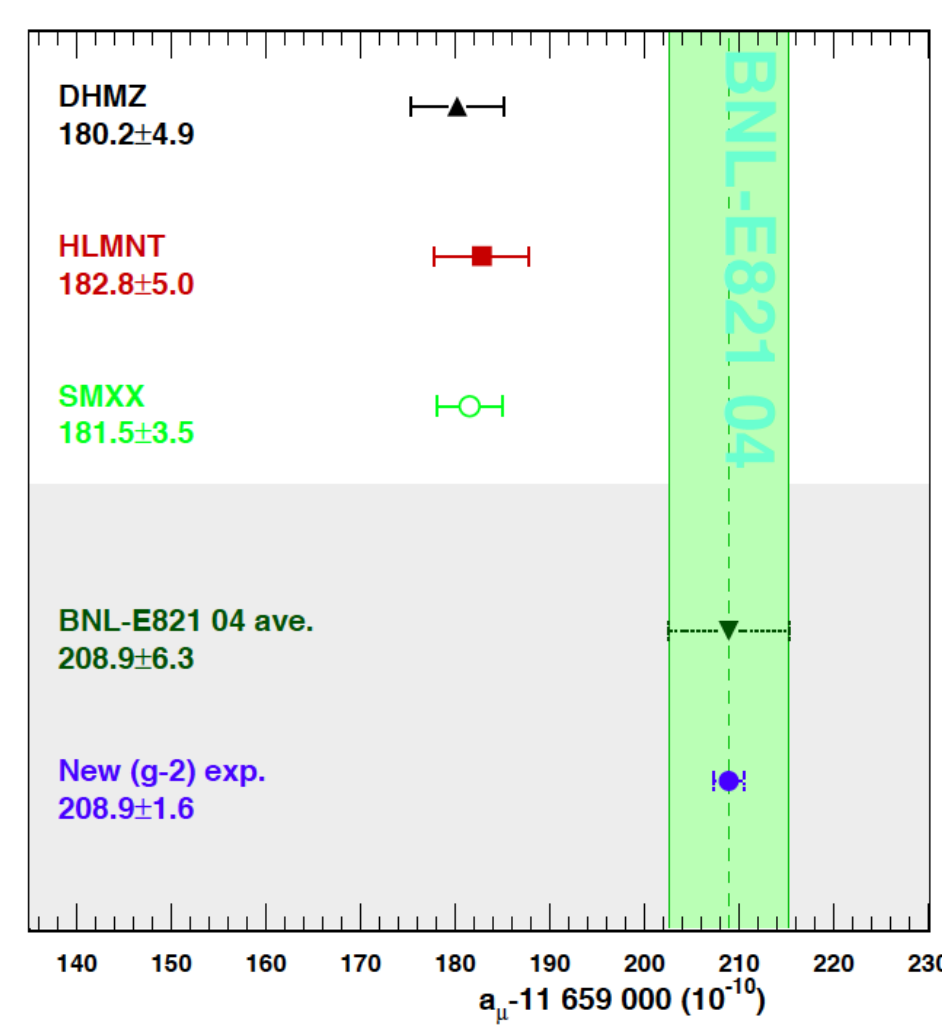

Figure 2.9: Estimated uncertainties $\delta a_\mu$ in units of $10^{-11}$ according to Refs. [71, 72] and (last column) prospects for improved precision in the $e^+e^-$ hadronic cross-section measurements. The final row projects the uncertainty on the difference with the Standard Model, $\Delta a_\mu$. The figure give the comparison between $a_\mu^{\rm SM}$ and $a_\mu^{\rm EXP}$. DHMZ is Ref. [71], HLMNT is Ref. [72]; "SMXX" is the same central value with a reduced error as expected by the improvement on the hadronic cross section measurement (see text); "BNL-E821 04 ave." is the current experimental value of $a_\mu$; "New (g-2) exp." is the same central value with a fourfold improved precision as planned by the future (g-2) experiments at Fermilab and J-PARC.

Thus the prognosis is excellent that the results from E989 will clarify whether the measured value of $a_\mu$ contains contributions from outside of the Standard Model. Even if there is no improvement on the hadronic error, but the central theory and experimental values remain the same, the significance of the difference would be over $5\sigma$. However, with the worldwide effort to improve on the Standard-Model value, it is most likely that the comparison will be even more convincing.

## 2.7 Physics Beyond the Standard Model

For many years, the muon anomaly has played an important role in constraining physics beyond the SM [47, 48, 53, 115, 54, 116]. The more than 2000 citations to the major E821 papers [37, 36, 41, 40], demonstrates that this role continues. The citations are shown as a function of year in Fig. 2.1 (b). It is apparent that with the LHC results available in 2012, interest in the BNL results has risen significantly. As discussed in the previous section, the present SM value is smaller than the experimental value by $\Delta a_\mu(\text{E821} - \text{SM})$. The discrepancy depends on the SM evaluation, but it is generally in the $> 3\sigma$ region; a representative value is $(261 \pm 80) \times 10^{-11}$, see Eq. (2.27).

In this section, we discuss how the muon anomaly provides a unique window to search for physics beyond the standard model. If such new physics is discovered elsewhere, e.g. at the LHC, then $a_\mu$ will play an important role in sorting out the interpretation of those discoveries. We discuss examples of constraints placed on various models that have been proposed as extensions of the standard model. Perhaps the ultimate value of an improved limit on $a_\mu$ will come from its ability to constrain the models that have not yet been invented.

#### Varieties of physics beyond the Standard Model

The LHC era has had its first spectacular success in summer 2012 with the discovery of a new particle compatible with the standard model Higgs boson. With more data, the LHC experiments will continue to shed more light on the nature of electroweak symmetry breaking (EWSB). It is very likely that EWSB is related to new particles, new interactions, or maybe to new concepts such as supersymmetry, extra dimensions, or compositeness. Further open questions in particle physics, related e.g. to the nature of dark matter, the origin of flavor or grand unification, indicate that at or even below the TeV scale there could be rich physics beyond the standard model.

Unravelling the existence and the properties of such new physics requires experimental information complementary to the LHC. The muon $(g-2)$, together with searches for charged lepton flavor violation, electric dipole moments, and rare decays, belongs to a class of complementary low-energy experiments.

In fact, the muon magnetic moment has a special role because it is sensitive to a large class of models related and unrelated to EWSB and because it combines several properties in a unique way: it is a flavour- and CP-conserving, chirality-flipping and loop-induced quantity. In contrast, many high-energy collider observables at the LHC and a future linear collider are chirality-conserving, and many other low-energy precision observables are CP- or flavour-violating. These unique properties might be the reason why the muon $(g-2)$ is the only among the mentioned observables which shows a significant deviation between the experimental value and the SM prediction, see Eq. (2.27). Furthermore, while $g-2$ is sensitive to leptonic couplings, $b$- or $K$-physics more naturally probe the hadronic couplings of new physics. If charged lepton-flavor violation exists, observables such as $\mu \to e$ conversion can only determine a combination of the strength of lepton-flavor violation and the mass scale of new physics. In that case, $g-2$ can help to disentangle the nature of the new physics.

The role of $g-2$ as a discriminator between very different standard model extensions is well illustrated by a relation stressed by Czarnecki and Marciano [48]. It holds in a wide range of models as a result of the chirality-flipping nature of both $g-2$ and the muon mass: If a new physics model with a mass scale $\Lambda$ contributes to the muon mass $\delta m_\mu(\text{N.P.})$, it also contributes to $a_\mu$, and the two contributions are related as

$$a_\mu(\text{N.P.}) = \mathcal{O}(1) \times \left(\frac{m_\mu}{\Lambda}\right)^2 \times \left(\frac{\delta m_\mu(\text{N.P.})}{m_\mu}\right). \tag{2.28}$$

The ratio $C(\text{N.P.}) \equiv \delta m_\mu(\text{N.P.})/m_\mu$ cannot be larger than unity unless there is fine-tuning in the muon mass. Hence a first consequence of this relation is that new physics can explain the currently observed deviation (2.27) only if $\Lambda$ is at the few-TeV scale or smaller.

In many models, the ratio $C$ arises from one- or even two-loop diagrams, and is then suppressed by factors like $\alpha/4\pi$ or $(\alpha/4\pi)^2$. Hence, even for a given $\Lambda$, the contributions to $a_\mu$ are highly model dependent.

It is instructive to classify new physics models as follows:

- Models with $C(\text{N.P.}) \simeq 1$: Such models are of interest since the muon mass is essentially generated by radiative effects at some scale $\Lambda$. A variety of such models have been discussed in [48], including extended technicolor or generic models with naturally vanishing bare muon mass. For examples of radiative muon mass generation within supersymmetry, see e.g. [117, 118]. In these models the new physics contribution to $a_\mu$ can be very large,
$$a_\mu(\Lambda) \simeq \frac{m_\mu^2}{\Lambda^2} \simeq 1100 \times 10^{-11} \left(\frac{1\ \text{TeV}}{\Lambda}\right)^2. \tag{2.29}$$
and the difference Eq. (2.27) can be used to place a lower limit on the new physics mass scale, which is in the few TeV range [119, 118].

- Models with $C(\text{N.P.}) = \mathcal{O}(\alpha/4\pi)$: Such a loop suppression happens in many models with new weakly interacting particles like $Z'$ or $W'$, little Higgs or certain extra dimension models. As examples, the contributions to $a_\mu$ in a model with $\delta = 1$ (or 2) universal extra dimensions (UED) [120] and the Littlest Higgs model with T-parity (LHT) [121] are given by
$$a_\mu(\text{UED}) \simeq -5.8 \times 10^{-11}(1 + 1.2\delta)S_{\text{KK}}, \tag{2.30}$$
$$a_\mu(\text{LHT}) < 12 \times 10^{-11} \tag{2.31}$$
with $|S_{\text{KK}}| \lesssim 1$ [120]. A difference as large as Eq. (2.27) is very hard to accommodate unless the mass scale is very small, of the order of $M_Z$, which however is often excluded e.g. by LEP measurements. So typically these models predict very small contributions to $a_\mu$ and will be disfavored if the current deviation will be confirmed by the new $a_\mu$ measurement.

Exceptions are provided by models where new particles interact with muons but are otherwise hidden from searches. An example is the model with a new gauge boson associated to a gauged lepton number $L_\mu - L_\tau$ [122], where a gauge boson mass of $\mathcal{O}(100\ \text{GeV})$ and large $a_\mu$ are viable; see however [123], which discusses a novel constraint that disfavors large contributions to $a_\mu$ in this model.

- Models with intermediate values for $C(\text{N.P.})$ and mass scales around the weak scale: In such models, contributions to $a_\mu$ could be as large as Eq. (2.27) or even larger, or smaller, depending on the details of the model. This implies that a more precise $a_\mu$-measurement will have significant impact on such models and can even be used to measure model parameters. Supersymmetric (SUSY) models are the best known examples, so muon $g-2$ would have substantial sensitivity to SUSY particles. Compared to generic perturbative models, supersymmetry provides an enhancement to $C(\text{SUSY}) = \mathcal{O}(\tan\beta \times \alpha/4\pi)$ and to $a_\mu(\text{SUSY})$ by a factor $\tan\beta$ (the ratio of the vacuum expectation values of the two Higgs fields). Typical SUSY diagrams for the magnetic dipole moment, the electric dipole moment, and the lepton-number violating conversion process $\mu \to e$ in the field of a nucleus are shown pictorially in Fig. 2.10. The shown diagrams contain the SUSY partners of the muon, electron and the SM $\text{U}(1)_Y$ gauge boson, $\tilde{\mu}$, $\tilde{e}$, $\tilde{B}$. The full SUSY contributions involve also the SUSY partners to the neutrinos and all SM gauge and Higgs bosons. In a model with SUSY masses equal to $\Lambda$ the SUSY contribution to $a_\mu$ is given by [124, 48, 125]

$$a_\mu(\text{SUSY}) \simeq \text{sgn}\,(\mu)\ 130 \times 10^{-11}\ \tan\beta\ \left(\frac{100\ \text{GeV}}{\Lambda}\right)^2 \tag{2.32}$$

which indicates the dependence on $\tan\beta$, and the SUSY mass scale, as well as the sign of the SUSY $\mu$-parameter. The formula still approximately applies even if only the smuon and chargino masses are of the order $\Lambda$ but e.g. squarks and gluinos are much heavier. However the SUSY contributions to $a_\mu$ depend strongly on the details of mass splittings between the weakly interacting SUSY particles (for details and the current status of the SUSY prediction for $a_\mu$ see e.g. [124, 125, 126, 127]). Thus muon $g-2$ is sensitive to SUSY models with SUSY masses in the few hundred GeV range, and it will help to measure SUSY parameters.

There are also non-supersymmetric models with similar enhancements. For instance, lepton flavor mixing can help. An example is provided in Ref. [129] by a model with two Higgs doublets and four generations, which can accommodate large $\Delta a_\mu$ without violating constraints on lepton flavor violation. In variants of Randall-Sundrum models [130, 131, 132] and large extra dimension models [133], large contributions to $a_\mu$ might be possible from exchange of Kaluza-Klein gravitons, but the theoretical evaluation is difficult because of cutoff dependences. A recent evaluation of the non-graviton contributions in Randall-Sundrum models, however, obtained a very small result [134].

Further examples include scenarios of unparticle physics [135, 136] (here a more precise $a_\mu$-measurement would constrain the unparticle scale dimension and effective couplings), generic models with a hidden sector at the weak scale [137] or a model with the discrete flavor symmetry group $T'$ and Higgs triplets [138] (here a more precise $a_\mu$-measurement would constrain hidden sector/Higgs triplet masses and couplings), or the model proposed in Ref. [139], which implements the idea that neutrino masses, leptogenesis and the deviation in $a_\mu$ all originate from dark matter particles. In the latter model, new leptons and scalar particles are predicted, and $a_\mu$ provides significant constraints on the masses and Yukawa couplings of the new particles.

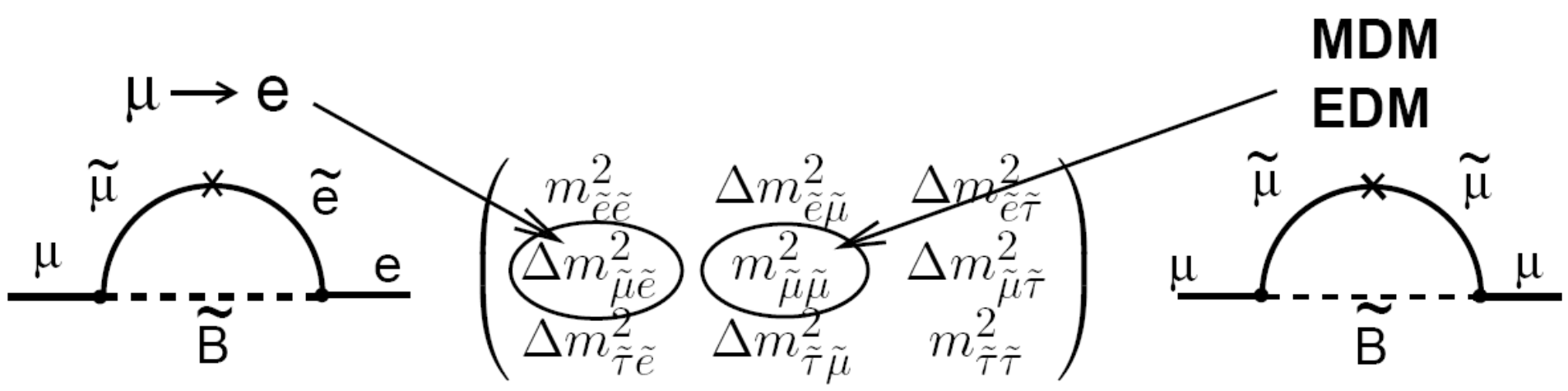

Figure 2.10: The SUSY contributions to the anomaly, and to $\mu \to e$ conversion, showing the relevant slepton mixing matrix elements. The MDM and EDM give the real and imaginary parts of the matrix element, respectively. The $\times$ indicates a chirality flip.

The following types of new physics scenarios are quite different from the ones above:

- Models with extended Higgs sector but without the $\tan\beta$-enhancement of SUSY models. Among these models are the usual two-Higgs-doublet models. The one-loop contribution of the extra Higgs states to $a_\mu$ is suppressed by two additional powers of the muon Yukawa coupling, corresponding to $a_\mu(\text{N.P.}) \propto m_\mu^4/\Lambda^4$ at the one-loop level. Two-loop effects from Barr-Zee diagrams can be larger [140], but typically the contributions to $a_\mu$ are negligible in these models.

- Models with additional light particles with masses below the GeV-scale, generically called dark sector models: Examples are provided by the models of Refs. [141, 142], where additional light neutral gauge bosons can affect electromagnetic interactions. Such models are intriguing since they completely decouple $g-2$ from the physics of EWSB, and since they are hidden from collider searches at LEP or LHC (see however Refs. [143, 144] for studies of possible effects at dedicated low-energy colliders and in Higgs decays at the LHC). They can lead to contributions to $a_\mu$ which are of the same order as the deviation in Eq. (2.27). Hence the new $g-2$ measurement will provide an important test of such models.

To summarize: many well-motivated models can accommodate larger contributions to $a_\mu$ — if any of these are realized $g-2$ can be used to constrain model parameters; many well-motivated new physics models give tiny contributions to $a_\mu$ and would be disfavored if the more precise $g-2$ measurement confirms the deviation in Eq. (2.27). There are also examples of models which lead to similar LHC signatures but which can be distinguished using $g-2$.

### Models with large contributions to $a_\mu$ versus LHC data

We first focus on two particularly promising candidate models which could naturally explain a deviation as large as Eq. (2.27): dark sector models and SUSY models.

Dark sector models involve very light new particles with very weak interactions. They are constrained by other low-energy observables, such as $(g-2)$ of the electron, but there is a natural window in parameter space, where they can accommodate large contributions to $a_\mu$. These models are hardly constrained by LHC data.

The situation is very different for SUSY models. SUSY searches are a central part of the LHC experiments and have not revealed any evidence for SUSY particles, so SUSY models are already strongly constrained by current LHC data. In the following we discuss why and how SUSY models are still compatible with large contributions to $a_\mu$.

At the one-loop level, the diagrams of the minimal supersymmetric standard model (MSSM) involve the SUSY partners the gauge and Higgs bosons and the muon-neutrino and the muon, the so-called charginos, neutralinos and sneutrinos and smuons. The relevant parameters are the SUSY breaking mass parameters for the 2nd generation sleptons, the bino and wino masses $M_2$, $M_1$, and the Higgsino mass parameter $\mu$. Strongly interacting particles, squarks and gluinos, and their masses are irrelevant on this level.

If all the relevant mass parameters are equal, the approximation (2.32) is valid, and the dominant contribution is from the chargino–sneutrino diagrams. If there are large mass splittings, the formula becomes inappropriate. For example, if $\mu$ is very large, the bino-like neutralino contribution of Fig. 2.10 is approximately linear in $\mu$ and can dominate. If there is a large mass splitting between the left- and right-handed smuon, even the sign can be opposite to Eq. (2.32), see the discussions in [124, 125, 126, 127].

On the two-loop level, further contributions exist which are typically subleading but can become important in regions of parameter space. For instance, there are diagrams without smuons or sneutrinos but with e.g. a pure chargino or stop loop [145, 146]. Such diagrams can even be dominant if first and second generation sfermions are very heavy, a scenario called effective SUSY.

Constraints from $a_\mu$ and LHC experiments and theoretical bias lead to the following conclusions:

- If supersymmetry is the origin of the deviation in $a_\mu$, at least some SUSY particles cannot be much heavier than around 700 GeV (for $\tan\beta = 50$ or less), most favorably the smuons and charginos/neutralinos.
- The negative results of the LHC searches for SUSY particles imply lower limits of around 1 TeV on squark and gluino masses. However, the bounds are not model-independent but valid in scenarios with particular squark and gluino decay patterns.
- The constraint that a SM-like Higgs boson mass is around 125 GeV requires either very large loop corrections from large logarithms or non-minimal tree-level contributions from additional non-minimal particle content.
- The requirement of small fine-tuning between supersymmetry-breaking parameters and the Z-boson mass prefers certain particles, in particular stops, gluinos and Higgsinos to be rather light.

A tension between these constraints seems to be building up, but the constraints act on different aspects of SUSY models. Hence it is in principle no problem to accommodate all the experimental data in the general minimal supersymmetric standard model, for recent analyses see Refs. [147, 148]; for benchmark points representing different possible parameter regions see [127].

The situation is different in many specific scenarios, based e.g. on particular high-scale assumptions or constructed to solve a subset of the issues mentioned above. The Constrained MSSM (CMSSM) is one of the best known scenarios. Here, GUT-scale universality relates SUSY particle masses, in particular the masses of colored and uncolored sfermions of all generations. For a long time, many analyses have used $a_\mu$ as a central observable to constrain the CMSSM parameters, see e.g. [149]. The most recent analyses show that the LHC determination of the Higgs boson mass turns out to be incompatible with an explanation of the current $\Delta a_\mu$ within the CMSSM [150, 151, 152]. Hence, the CMSSM is already disfavored now, and it will be excluded if the future $a_\mu$ measurement confirms the current $\Delta a_\mu$.

Likewise, in the so-called natural SUSY scenarios (see e.g. [156, 157]) the spectrum is such that fine-tuning is minimized while squarks and gluinos evade LHC bounds. These scenarios can explain the Higgs boson mass but fail to explain $g{-}2$ because of the heavy smuons.

On the other hand, the model of Ref. [153] is an example of a model with the aim to reconcile LHC-data, naturalness, and $g{-}2$. It is based on gauge-mediated SUSY breaking and extra vector-like matter, and it is naturally in agreement with FCNC constraints and the Higgs boson mass value. In this model the SUSY particles can be light enough to explain $g{-}2$, but in that case it is on the verge of being excluded by LHC data.

The rising tension between the constraints mentioned above, and further recent model-building efforts to solve it, are also reviewed in Refs. [154, 155]. In these references, more pragmatic approaches are pursued, and parameter regions within the general MSSM are suggested which are in agreement with all experimental constraints. All suggested regions have in common that they are split, i.e. some sparticles are much heavier than others. Ref. [154] suggests to focus on scenarios with light non-colored and heavy colored sparticles; Ref. [155] proposes split-family supersymmetry, where only the third family sfermions are very heavy. In both scenarios, $g{-}2$ can be explained, and the parameter space of interest can be probed by the next LHC run.

In the general model classification of the previous subsection the possibility of radiative muon mass generation was mentioned. This idea can be realized within supersymmetry, and it leads to SUSY scenarios quite different from the ones discussed so far. Since the muon mass at tree level is given by the product of a Yukawa coupling and the vacuum expectation value of the Higgs doublet $H_d$, there are two kinds of such scenarios. First, one can postulate that the muon Yukawa coupling is zero but chiral invariance is broken by soft supersymmetry-breaking $A$-terms. Then, the muon mass, and $a_\mu^{\rm SUSY}$, arise at the one-loop level and there is no relative loop suppression of $a_\mu^{\rm SUSY}$ [117, 118]. Second, one can postulate that the vacuum expectation value $\langle H_d \rangle$ is very small or zero [158, 159]. Then, the muon mass and $a_\mu^{\rm SUSY}$ arise at the one-loop level from loop-induced couplings to the other Higgs doublet. Both scenarios could accommodate large $a_\mu^{\rm SUSY}$ and TeV-scale SUSY particle masses.

Hence, in spite of the current absence of signals for new physics at the LHC, dark sector and SUSY models provide two distinct classes of models which are viable and can accom-modate large contributions to $a_\mu$. These examples of the CMSSM, natural SUSY, extended SUSY models, split MSSM scenarios, and radiative muon mass generation illustrate the model-dependence of $g-2$ within SUSY and its correlation to the other constraints. Clearly, a definitive knowledge of $a_\mu^{\rm SUSY}$ will be very beneficial for the interpretation of LHC data in terms of SUSY or any alternative new physics model.

### $a_\mu$ and model selection and parameter measurement

The LHC is sensitive to virtually all proposed weak-scale extensions of the standard model, ranging from supersymmetry, extra dimensions and technicolor to little Higgs models, unparticle physics, hidden sector models and others. However, even if the existence of physics beyond the standard model is established, it will be far from easy for the LHC alone to identify which of these — or not yet thought of — alternatives is realized. Typically LHC data will be consistent with several alternative models.

The previous subsection has given examples of qualitatively different SUSY models which are in agreement with current LHC data. Even worse, even if in the future the LHC finds many new heavy particles which are compatible with SUSY, these new states might allow alternative interpretations in terms of non-SUSY models. In particular universal-extra-dimension models (UED) [160], or the Littlest Higgs model with T-parity (LHT) [161, 162] have been called "bosonic SUSY" since they can mimick SUSY but the partner particles have the opposite spin as the SUSY particles, see e.g. [163]. The muon $g-2$ would especially aid in the selection since UED or Littlest Higgs models predict a tiny effect to $a_\mu$ [120, 121], while SUSY effects are often much larger.

On the other hand, a situation where the LHC finds no physics beyond the standard model but the $a_\mu$ measurement establishes a deviation, might be a signal for dark sector models such as the secluded U(1) model [141], with new very weakly interacting light particles which are hard to identify at the LHC [143, 142, 144].

Next, if new physics is realized in the form of a non-renormalizable theory, $a_\mu$ might not be fully computable but depend on the ultraviolet cutoff. Randall-Sundrum or universal extra dimension models are examples of this situation. In such a case, the $a_\mu$ measurement will not only help to constrain model parameters but it will also help to get information on the ultraviolet completion of the theory.

The complementarity between $a_\mu$ and LHC can be exemplified quantitatively within general SUSY, because this is a well-defined and calculable framework. Fig. 2.11 illustrates the complementarity in selecting between different models.

The red points in the left plot in Fig. 2.11 show the values for the so-called SPS benchmark points [167] and new benchmark points E1, E4, NS1. The points E1, E4 are the split scenarios defined in Endo et al, Ref. [154] (cases (a) and (d) with $M_2 = 300$ GeV and $m_L = 500$ GeV), the point NS1 is the natural SUSY scenario defined in Ref. [156]. These points span a wide range and can be positive or negative, due to the factor sign($\mu$) in Eq. (2.32). The discriminating power of the current (yellow band) and an improved (blue band) measurement is evident from Fig. 2.11(a).

Even though several SPS points are actually experimentally excluded, their spread in Fig. 2.11(a) is still a good illustration of possible SUSY contributions to $a_\mu$. E.g. the split scenarios of Refs. [154, 155] are comparable to SPS1b, both in their $g-2$ contribution and in the relevant mass spectrum. Natural SUSY is similar to SPS2, which corresponds to a heavy sfermion scenario. Similarly, the "supersymmetry without prejudice" study of Ref. [168] confirmed that the entire range $a_\mu^{\rm SUSY} \sim (-100 \ldots +300) \times 10^{-11}$ was populated by a reasonable number of "models" which are in agreement with other experimental constraints. Therefore, a precise measurement of $g-2$ to $\pm 16 \times 10^{-11}$ will be a crucial way to rule out a large fraction of models and thus determine SUSY parameters.

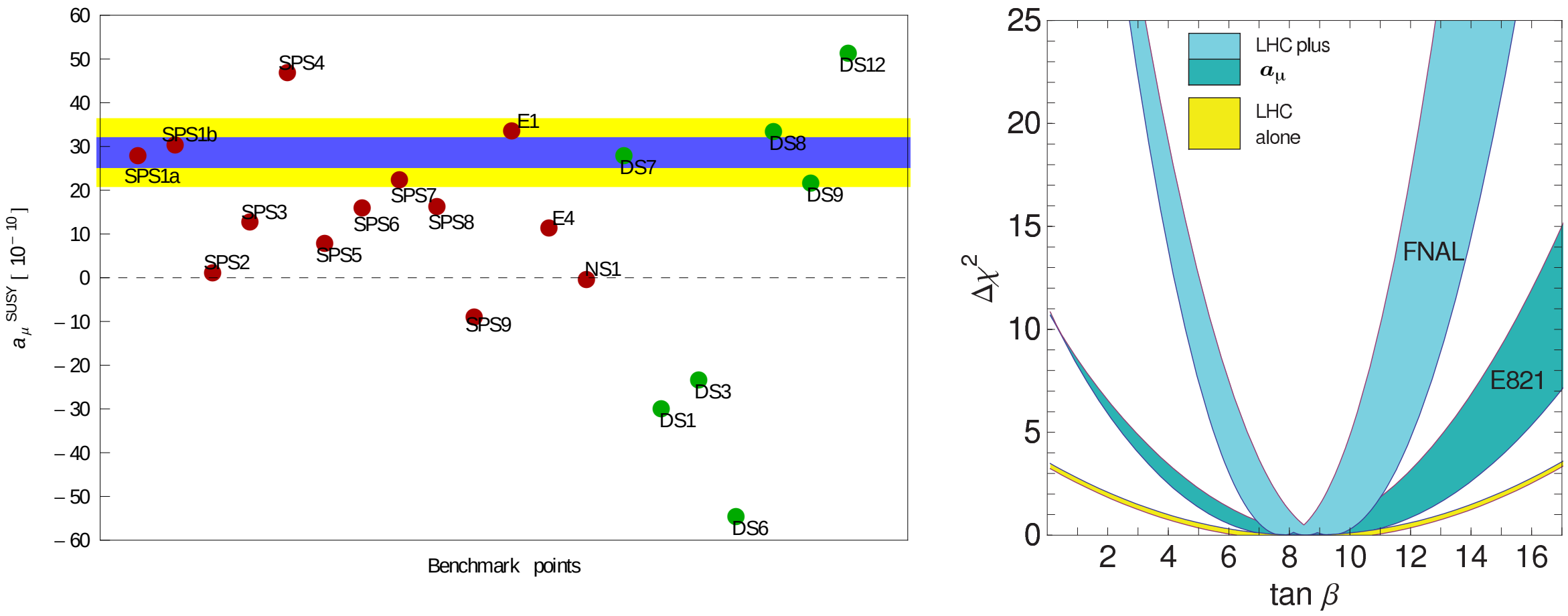

(a) (b)

Figure 2.11: (a) SUSY contributions to $a_\mu$ for the SPS and other benchmark points (red), and for the "degenerate solutions" from Ref. [164]. The yellow band is the $\pm 1\ \sigma$ error from E821, the blue is the projected sensitivity of E989. (b) Possible future $\tan\beta$ determination assuming that a slightly modified MSSM point SPS1a (see text) is realized. The bands show the $\Delta\chi^2$ parabolas from LHC-data alone (yellow) [166], including the $a_\mu$ with current precision (dark blue) and with prospective precision (light blue). The width of the blue curves results from the expected LHC-uncertainty of the parameters (mainly smuon and chargino masses) [166].

One might think that if SUSY exists, the LHC-experiments will find it and measure its parameters. Above it has been mentioned that SUSY can be mimicked by "bosonic SUSY" models. The green points in Fig. 2.11(a) illustrate that even within SUSY, certain SUSY parameter points can be mimicked by others. The green points correspond to "degenerate solutions" of Ref. [164] — different SUSY parameter points which cannot be distinguished at the LHC alone (see also Ref. [165] for the LHC inverse problem). Essentially the points differ by swapping the values and signs of the SUSY parameters $\mu$, $M_1$, $M_2$. They have very different $a_\mu$ predictions, and hence $a_\mu$ can resolve such LHC degeneracies.

The right plot of Fig. 2.11 illustrates that the SUSY parameter $\tan\beta$ can be measured more precisely by combining LHC-data with $a_\mu$. It is based on the assumption that SUSY is realized, found at the LHC and the origin of the observed $a_\mu$ deviation (2.27). To fix an example, we use a slightly modified SPS1a benchmark point with $\tan\beta$ scaled down to $\tan\beta = 8.5$ such that $a_\mu^{\rm SUSY}$ is equal to an assumed deviation $\Delta a_\mu = 255 \times 10^{-11}$.[7] Ref.

[7]The actual SPS1a point is ruled out by LHC; however for our purposes only the weakly interacting

[166] has shown that then mass measurements at the LHC alone are sufficient to determine $\tan\beta$ to a precision of $\pm 4.5$ only. The corresponding $\Delta\chi^2$ parabola is shown in yellow in the plot. In such a situation one can study the SUSY prediction for $a_\mu$ as a function of $\tan\beta$ (all other parameters are known from the global fit to LHC data) and compare it to the measured value, in particular after an improved measurement. The plot compares the LHC $\Delta\chi^2$ parabola with the ones obtained from including $a_\mu$, $\Delta\chi^2 = [(a_\mu^{\rm SUSY}(\tan\beta) - \Delta a_\mu)/\delta a_\mu]^2$ with the errors $\delta a_\mu = 80 \times 10^{-11}$ (dark blue) and $34 \times 10^{-11}$ (light blue). As can be seen from the Figure, using today's precision for $a_\mu$ would already improve the determination of $\tan\beta$, but the improvement will be even more impressive after the future $a_\mu$ measurement.

One should note that even if better ways to determine $\tan\beta$ at the LHC alone might be found, an independent determination using $a_\mu$ will still be highly valuable, as $\tan\beta$ is one of the central MSSM parameters; it appears in all sectors and in almost all observables. In non-minimal SUSY models the relation between $\tan\beta$ and different observables can be modified. Therefore, measuring $\tan\beta$ in different ways, e.g. using certain Higgs- or $b$-decays at the LHC or at $b$-factories and using $a_\mu$, would constitute a non-trivial and indispensable test of the universality of $\tan\beta$ and thus of the structure of the MSSM.

In summary, the anomalous magnetic moment of the muon is sensitive to contributions from a wide range of physics beyond the standard model. It will continue to place stringent restrictions on all of the models, both present and yet to be written down. If physics beyond the standard model is discovered at the LHC or other experiments, $a_\mu$ will constitute an indispensable tool to discriminate between very different types of new physics, especially since it is highly sensitive to parameters which are difficult to measure at the LHC. If no new phenomena are found elsewhere, then it represents one of the few ways to probe physics beyond the standard model. In either case, it will play an essential and complementary role in the quest to understand physics beyond the standard model at the TeV scale.

particles are relevant, and these are not excluded. The following conclusions are neither very sensitive to the actual $\tan\beta$ value nor to the actual value of the deviation $\Delta a_\mu$.

# Chapter 3

# Overview of the Experimental Technique

In this chapter we give an overview of how the experiment is done. This is followed by a number of chapters that give the details of the specific hardware being developed for E989. The order of those chapters follows the WBS as closely as possible.

The experiment consists of the following steps:

1. Production of an appropriate pulsed proton beam by an accelerator complex.
2. Production of pions using the proton beam that has been prepared.
3. Collection of polarized muons from pion decay $\pi^+ \to \mu^+ \nu_\mu$
4. Transporting the muon beam to the $(g-2)$ storage ring.
5. Injection of the muon beam into the storage ring.
6. Kicking the muon beam onto stored orbits.
7. Measuring the arrival time and energy of positrons from the decay $\mu^+ \to e^+ \bar{\nu}_\mu \nu_e$
8. Precise mapping and monitoring of the precision magnetic field

Central to the determination of $a_\mu$ is the spin equation[1]

$$\vec{\omega}_a = -\frac{Qe}{m}\left[a\vec{B} - \left(a - \frac{1}{\gamma^2 - 1}\right)\frac{\vec{\beta}\times\vec{E}}{c}\right] = -\frac{Qe}{m}\left[a_\mu\vec{B} - \left(a_\mu - \left(\frac{mc}{p}\right)^2\right)\frac{\vec{\beta}\times\vec{E}}{c}\right], \quad (3.1)$$

that gives the rate at which the muon spin turns relative the momentum vector, which turns with the cyclotron frequency. The electric field term is there since we use electrostatic vertical focusing in the ring. At the magic momentum, $p_{\rm m} = 3.09$ GeV/c, the effect of the motional magnetic field (the $\vec{\beta}\times\vec{E}$ term) vanishes.

Measurement of $a_\mu$ requires the determination of the muon spin frequency $\omega_a$ and the magnetic field $\vec{B}$ averaged over the muon distribution.

[1]See Section 3.3 for the details.

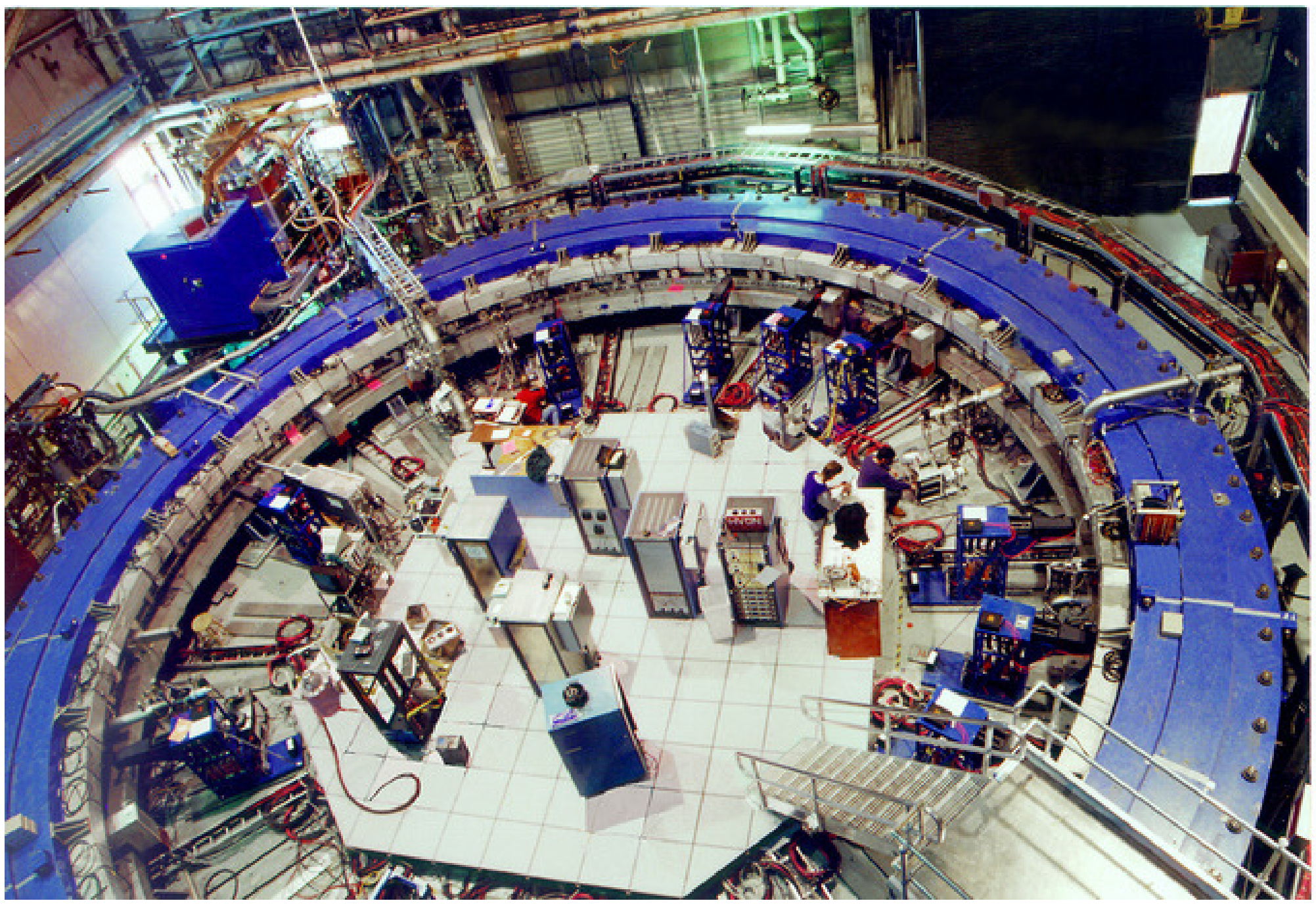

Figure 3.1: The E821 storage-ring magnet at Brookhaven Lab.

## 3.1 Production and Preparation of the Muon Beam

E989 will bring a bunched beam from the 8 GeV Booster to a pion production target located where the antiproton production target was in the Tevatron collider program (see Chapter 7). Pions of 3.11 GeV/c ±5% will be collected and sent into a large-acceptance beamline. Muons are produced in the weak pion decay

$$\pi^{\mp} \to \mu^{\mp} + \bar{\nu}_{\mu}(\nu_{\mu}). \tag{3.2}$$

The antineutrino (neutrino) is right-handed (left-handed) and the pion is spin zero. Thus the muon spin must be antiparallel to the neutrino spin, so it is also right-handed (left-handed). A beam of polarized muons can be obtained from a beam of pions by selecting the highest-energy muons (a "forward beam") or by selecting the lowest-energy muons (a "backward beam"), where forward or backward refers to whether the decay is forward or backward in the center-of-mass frame relative to the pion momentum. Polarizations significantly greater than 90% are easily obtained in such beams. The pions and daughter muons will be injected into the Delivery Ring (the re-purposed $\bar{p}$ debuncher ring), where after several turns the remaining pions decay. The surviving muon beam will be extracted and brought to the muon storage ring built for E821 at Brookhaven.

## 3.2 Injection into the Storage Ring

A photograph of the E821 magnet is shown in Figure 3.1. It is clear from the photo that this "storage ring" is very different from the usual one that consists of lumped elements. The storage ring magnet is energized by three superconducting coils shown in Fig 3.2(b). The continuous "C" magnet yoke is built from twelve 30° segments of iron, which were designed to eliminate the end effects present in lumped magnets. This construction eliminates the large gradients that would make the determination of the average magnetic field, $\langle B \rangle$, very difficult. Furthermore, a small perturbation in the yoke can effect the field at the ppm level at the opposite side of the ring. Thus every effort is made to minimize holes in the yoke, and other perturbations. The only penetrations through the yoke are to permit the muon beam to enter the magnet as shown in Fig 3.2(a), and to connect cryogenic services and power to the inflector magnet and to the outer radius coil (see Fig. 3.2(b)). Where a hole in the yoke is necessary, extra steel was placed around the hole on the outside of the yoke to compensate for the missing material.

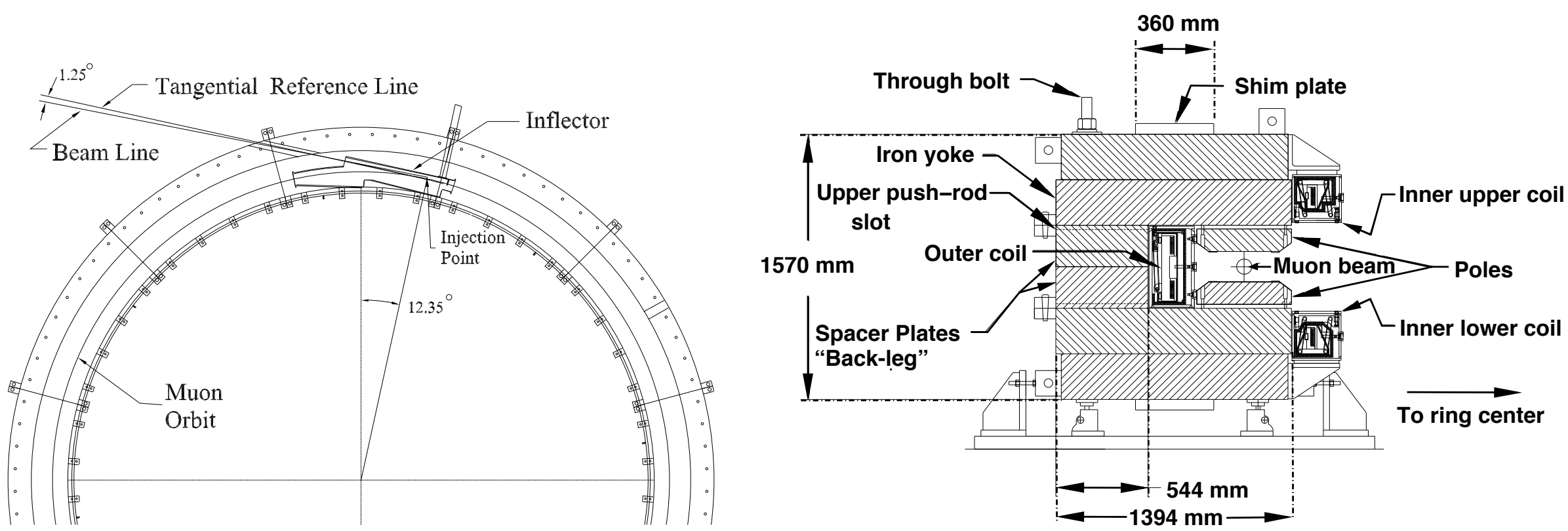

(a) (b)

Figure 3.2: (a) Plan view of the beam entering the storage ring. (b) Elevation view of the storage-ring magnet cross section.

The beam enters through a hole in the "back-leg" of the magnet and then crosses into the inflector magnet, which provides an almost field free region, delivering the beam to the edge of the storage region. The geometry is rather constrained, as can be seen in Fig. 3.3(a). The injection geometry is sketched in Fig. 3.3(b). The kick required to put magic momentum muons onto a stable orbit centered at magic radius is on the order of 10 mrad.

The requirements on the muon kicker are rather severe:

1. Since the magnet is continuous, any kicker device has to be inside of the precision magnetic field region.

2. The kicker hardware cannot contain magnetic elements such as ferrites, because they will spoil the uniform magnetic field.

3. Any eddy currents produced in the vacuum chamber, or in the kicker electrodes by the kicker pulse must be negligible by 10 to 20 $\mu$s after injection, or must be well known and corrected for in the measurement.

4. Any new kicker hardware must fit within the real estate that was occupied by the E821 kicker. The available space consists of three consecutive 1.7 m long spaces; see Fig. 3.5

5. The kicker pulse should be shorter than the cyclotron period of 149 ns.

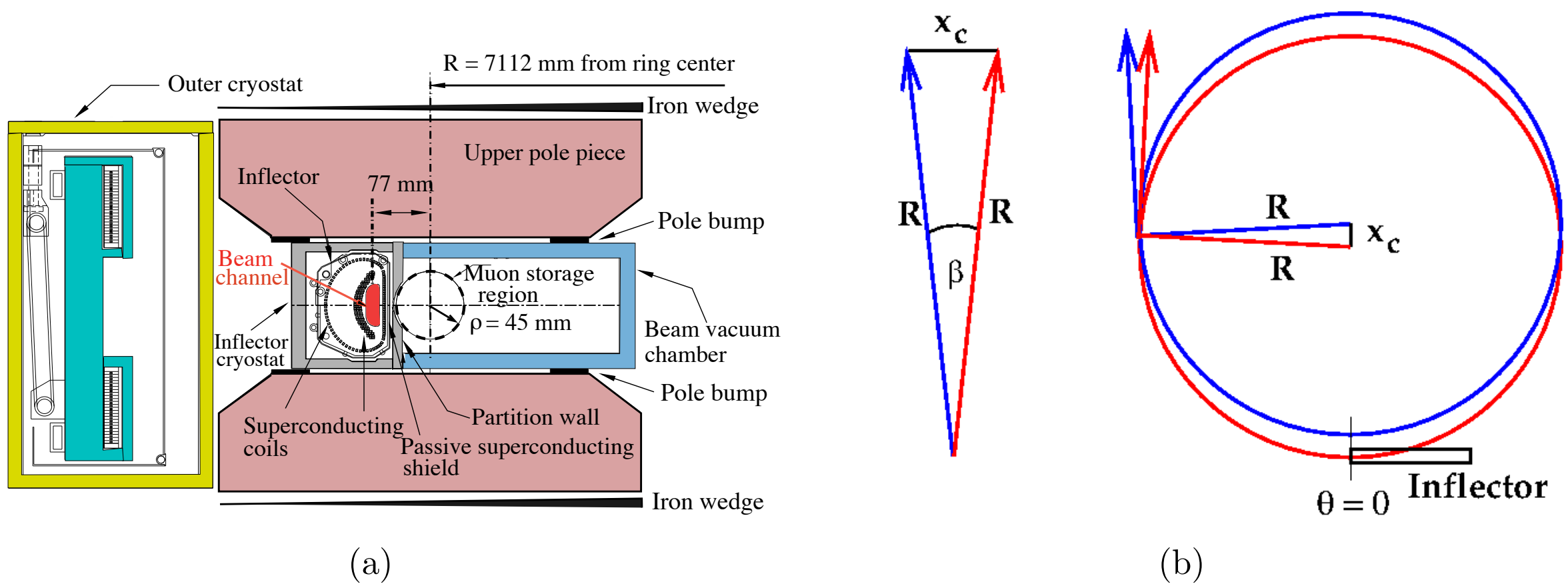

Figure 3.3: (a) The inflector exit showing the incident beam center 77 mm from the center of the storage region. The incident muon beam channel is highlighted in red. (b) The geometry of the necessary kick. The incident beam is the red circle, and the kick effectively moves the red circle over to the blue one.

## 3.3 The Spin Equations

Measurements of magnetic and electric dipole moments make use of the torque on a dipole in an external field:

$$\vec{\tau} = \vec{\mu} \times \vec{B} + \vec{d} \times \vec{E}, \tag{3.3}$$

where we include the possibility of an electric dipole moment ($\vec{d}$). Except for the original Nevis spin rotation experiment, the muon magnetic dipole moment experiments inject a beam of polarized muons into a magnetic field and measure the rate at which the spin turns relative to the momentum, $\vec{\omega}_a = \vec{\omega}_S - \vec{\omega}_C$, where $S$ and $C$ stand for spin and cyclotron, respectively. These two frequencies, in the absence of any other external fields, are given by

$$\omega_S = -g\frac{Qe}{2m}B - (1-\gamma)\frac{Qe}{\gamma m}B; \tag{3.4}$$

$$\omega_C = -\frac{Qe}{m\gamma}B; \tag{3.5}$$

$$\omega_a = \omega_S - \omega_C = -\left(\frac{g-2}{2}\right)\frac{Qe}{m}B = -a_\mu\frac{Qe}{m}B \tag{3.6}$$

(where $e > 0$ and $Q = \pm 1$). There are two important features of $\omega_a$:

- It only depends on the anomaly rather than on the full magnetic moment.
- It depends linearly on the applied magnetic field.

To measure the anomaly, it is necessary to measure $\omega_a$, and to determine the magnetic field $B$. The relevant quantity is $\langle B \rangle$, which is the magnetic field convolved with the muon beam distribution,

$$\langle B \rangle = \int M(r,\theta) B(r,\theta) r dr d\theta, \tag{3.7}$$

where the magnetic field $B(r,\theta)$ is expressed in terms of a multipole expansion

$$B(r,\theta) = \sum_{n=0}^{\infty} r^n \left[c_n \cos n\theta + s_n \sin n\theta\right], \tag{3.8}$$

and the muon distribution is expressed in terms of moments

$$M(r,\theta) = \sum_{m=0}^{\infty} [\xi_m(r)\cos m\theta + \sigma_m(r)\sin m\theta]. \tag{3.9}$$

Because the harmonics $\sin n\theta \sin m\theta$, etc., are orthogonal and vanish for $m \neq n$ when integrated over one period, non-vanishing integrals come from products of the same moment/multipole, in the expression for $\langle B \rangle$. To determine $\langle B \rangle$ to sub-part-per-million (ppm) precision, one either needs excellent knowledge of the multipole and moment distributions for $B$ and $M$; or care must be taken to minimize the number of terms, with only the leading term being large, so that only the first few multipoles are important. This was achieved in the most recent experiment [1] by using a circular beam aperture, and making a very uniform dipole magnetic field.

However there is one important issue to be solved: How can the muon beam be confined to a storage ring if significant magnetic gradients cannot be used to provide vertical focusing? The answer to this question was discovered by the third CERN collaboration [2], which used an electric quadrupole field to provide vertical focusing. Of course, a relativistic particle feels a motional magnetic field proportional to $\vec{\beta} \times \vec{E}$, but the full relativistic spin equation contains a cancellation as can be seen below. Assuming that the velocity is transverse to the magnetic field ($\vec{\beta} \cdot \vec{B} = 0$), one obtains [3, 4]

$$\vec{\omega}_a = -\frac{Qe}{m}\left[a_\mu \vec{B} - \left(a_\mu - \left(\frac{mc}{p}\right)^2\right)\frac{\vec{\beta}\times\vec{E}}{c}\right] = -\frac{Qe}{m}\left[a_\mu \vec{B} - \left(a_\mu - \frac{1}{\gamma^2-1}\right)\frac{\vec{\beta}\times\vec{E}}{c}\right]. \tag{3.10}$$

For the "magic" momentum $p_{\rm magic} = m/\sqrt{a} \simeq 3.09$ GeV/c ($\gamma_{\rm magic} = 29.3$), the second term vanishes, and the electric field does not contribute to the spin motion *relative* to the momentum.[2] If $g = 2$, then $a_\mu = 0$ and the spin would follow the momentum, turning at the cyclotron frequency.

If an electric dipole moment were to be present (see Eq. 2.4), it would modify the spin equation to

$$\vec{\omega}_{a\eta} = \vec{\omega}_a + \vec{\omega}_\eta = -\frac{Qe}{m}\left[a\vec{B} - \left(a - \frac{1}{\gamma^2-1}\right)\frac{\vec{\beta}\times\vec{E}}{c}\right] - \eta\frac{Qe}{2m}\left[\frac{\vec{E}}{c} + \vec{\beta}\times\vec{B}\right] \tag{3.11}$$

[2]Small corrections to the measured frequency must be applied since $\vec{\beta} \cdot \vec{B} \simeq 0$ and not all muons are at the magic momentum. These are discussed in Chapter 4.

where $\eta$ plays the same role for the EDM as $g$ plays for the MDM.

To good approximation, $\omega_a$ is directed parallel to the $\vec{B}$ field, and $\omega_\eta$ is directed radially since the motional electric field proportional to $\vec{\beta} \times \vec{B}$ dominates over the quadrupole electric field. The net effects of the EDM are to tip the plane of polarization precession out of the ring plane by the angle $\delta = tan^{-1} \frac{\eta\beta}{2a_\mu}$ (see Fig. 3.4), and to increase the magnitude of the precession according to $\omega = \sqrt{\omega_a^2 + \omega_\eta^2} = \sqrt{\omega_a^2 + \left(\frac{e\eta\beta B}{2m}\right)^2}$. This tipping causes the average vertical component of the momentum of the decay positrons to oscillate with frequency $\omega_a$, but out of phase with the number oscillation (Eq. 3.18) by $\pi/2$.

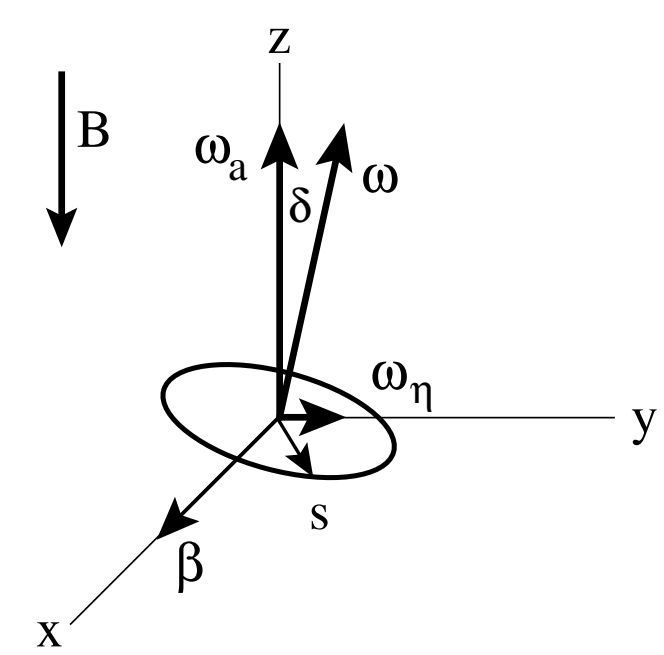

Figure 3.4: (b) The vectors $\vec{\omega}_a$ and $\vec{\omega}_\eta$ showing the tipping of the precession plane because of the presence of an electric dipole moment.

Since E989 will be equipped with three tracking stations that are useful for determining the properties of the stored muon beam, the up-down oscillating EDM signal comes for free. E989 should be able to improve on the E821 muon EDM limit [5] of

$$d_\mu < 1.8 \times 10^{-19}\ e \cdot \text{cm (95\%C.L.)} \tag{3.12}$$

two or more orders of magnitude. The most recent measurement of the electron EDM obtained [6] $d_e < 8.7 \times 10^{-29}$ $e$·cm (90% C.L). While a naive scaling between the electron and muon EDM goes linearly with mass, there are SUSY models that predict a much larger scaling [7].

## 3.4 Vertical Focusing with Electrostatic Quadrupoles

The storage ring acts as a weak-focusing betatron, with the vertical focusing provided by electrostatic quadrupoles. The ring is operated at the magic momentum, so that the electric field does not contribute to the spin precession. However there is a second-order correction to the spin frequency from the radial electric field, which is discussed below. There is also a correction from the vertical betatron motion, since the spin equations in the previous section were derived with the assumption that $\vec{\beta} \cdot \vec{B} = 0$.

A pure quadrupole electric field provides a linear restoring force in the vertical direction, and the combination of the (defocusing) electric field and the central (dipole) magnetic field ($B_0$) provides a net linear restoring force in the radial direction. The important parameter is the field index, $n$, which is defined by

$$n = \frac{\kappa R_0}{\beta B_0}, \tag{3.13}$$

where $\kappa$ is the electric quadrupole gradient and $R_0$ is the storage ring radius. For a ring with a uniform vertical dipole magnetic field and a uniform quadrupole field that provides vertical focusing covering the full azimuth, the stored particles undergo simple harmonic motion called betatron oscillations, in both the radial and vertical dimensions. The beam motion is discussed in more detail in the following chapter.

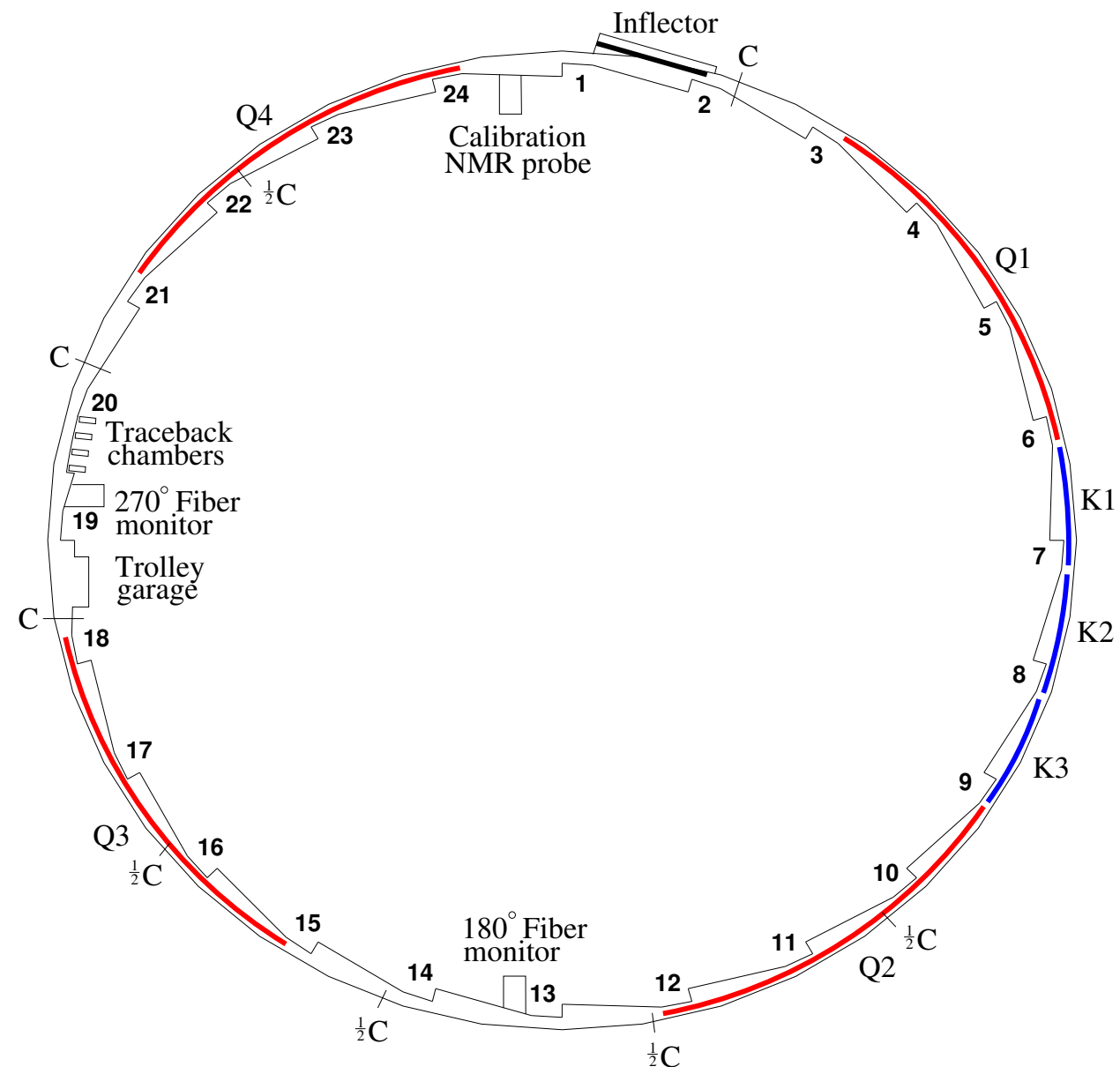

Figure 3.5: The layout of the storage ring, as seen from above, showing the location of the inflector, the kicker sections (labeled K1-K3), and the quadrupoles (labeled Q1-Q4). The beam circulates in a clockwise direction. Also shown are the collimators, which are labeled "C", or "$\frac{1}{2}$C" indicating whether the Cu collimator covers the full aperture, or half the aperture. The collimators are rings with inner radius: 45 mm, outer radius: 55 mm, thickness: 3 mm. The scalloped vacuum chamber consists of 12 sections joined by bellows. The chambers containing the inflector, the NMR trolley garage, and the trolley drive mechanism are special chambers. The other chambers are standard, with either quadrupole or kicker assemblies installed inside. An electron calorimeter is placed behind each of the radial windows, at the position indicated by the calorimeter number.

## 3.5 Muon Decay

The dominant muon decay is

$$\mu^{\mp} \rightarrow e^{\mp} + \nu_\mu(\bar{\nu}_\mu) + \bar{\nu}_e(\nu_e) \tag{3.14}$$

which also violates parity.

Since the kinematics of muon decay are central to the measurements of $a_\mu$, we discuss the general features in this section. Additional details are given in Ref. [8]. From a beam of pions traversing a straight beam-channel consisting of focusing and defocusing elements (FODO), a beam of polarized, high energy muons can be produced by selecting the "forward" or "backward" decays. The forward muons are those produced, in the pion rest frame, nearly parallel to the pion laboratory momentum and are the decay muons with the highest laboratory momenta. The backward muons are those produced nearly anti-parallel to the pion momentum and have the lowest laboratory momenta. The forward $\mu^-$ ($\mu^+$) are polarized along (opposite) their lab momenta respectively; the polarization reverses for backward muons. The E821 experiment used forward muons, as will E989, the difference being the length of the pion decay line, which in E989 will be 1,900 m, compared with 80 m in E821.

The pure $(V-A)$ three-body weak decay of the muon, $\mu^- \to e^- + \nu_\mu + \bar{\nu}_e$ or $\mu^+ \to e^+ + \bar{\nu}_\mu + \nu_e$, is "self-analyzing", that is, the parity-violating correlation between the directions in the muon rest frame (MRF) of the decay electron and the muon spin can provide information on the muon spin orientation at the time of the decay. When the decay electron has the maximum allowed energy in the MRF, $E'_{\rm max} \approx (m_\mu c^2)/2 = 53$ MeV, the neutrino and anti-neutrino are directed parallel to each other and at 180° relative to the electron direction. The $\nu\bar{\nu}$ pair carry zero total angular momentum; the electron carries the muon's angular momentum of 1/2. The electron, being a lepton, is preferentially emitted left-handed in a weak decay, and thus has a larger probability to be emitted with its momentum *anti-parallel* rather than parallel to the $\mu^-$ spin. Similarly, in $\mu^+$ decay, the highest-energy positrons are emitted *parallel* to the muon spin in the MRF.

In the other extreme, when the electron kinetic energy is approaches zero in the MRF, the neutrino and anti-neutrino are emitted back-to-back and carry a total angular momentum of one. In this case, the electron spin is directed opposite to the muon spin in order to conserve angular momentum. Again, the electron is preferentially emitted with helicity -1; however, in this case its momentum will be preferentially directed *parallel* to the $\mu^-$ spin. The positron, in $\mu^+$ decay, is preferentially emitted with helicity +1, and therefore its momentum will be preferentially directed *anti-parallel* to the $\mu^+$ spin.

With the approximation that the energy of the decay electron $E' >> m_e c^2$, the differential decay distribution in the muon rest frame is given by[9],

$$dP(y',\theta') \propto n'(y')\,[1 \pm \mathcal{A}(y')\cos\theta']\,dy'd\Omega' \tag{3.15}$$

where $y'$ is the momentum fraction of the electron, $y' = p'_e/p'_{e\ \rm max}$, $d\Omega'$ is the solid angle, $\theta' = \cos^{-1}(\hat{p}'_e \cdot \hat{s})$ is the angle between the muon spin $\hat{s}$ and $\vec{p}\,'_e$, $p'_{e\ \rm max}c \approx E'_{max}$, and the $(-)$ sign is for negative muon decay. The number distribution $n(y')$ and the decay asymmetry $\mathcal{A}(y')$ are given by

$$n(y') = 2y'^2(3-2y') \quad \text{and} \quad \mathcal{A}(y') = \frac{2y'-1}{3-2y'}. \tag{3.16}$$

Note that both the number and asymmetry reach their maxima at $y' = 1$, and the asymmetry changes sign at $y' = \frac{1}{2}$, as shown in Figure 3.6(a).

The CERN and Brookhaven based muon $(g-2)$ experiments stored relativistic muons of the magic momentum in a uniform magnetic field, which resulted in the muon spin precessing with constant frequency $\vec{\omega}_a$, while the muons traveled in circular orbits. If *all* decay electrons were counted, the number detected as a function of time would be a pure exponential; therefore we seek cuts on the laboratory observable to select subsets of decay electrons whose numbers oscillate at the precession frequency. The number of decay electrons in the MRF varies with the angle between the electron and spin directions, the electrons in the subset should have a preferred direction in the MRF when weighted according to their asymmetry as given in Equation 3.15. At $p_\mu \approx 3.094$ GeV/c the directions of the electrons resulting from muon decay in the laboratory frame are very nearly parallel to the muon momentum regardless of their energy or direction in the MRF. The only practical remaining cut is on the electron's laboratory energy. An energy subset will have the desired property: there will be a net component of electron MRF momentum either parallel or antiparallel to the laboratory muon direction. For example, suppose that we only count electrons with the highest laboratory energy, around 3.1 GeV. Let $\hat{z}$ indicate the direction of the muon laboratory momentum. The highest-energy electrons in the laboratory are those near the maximum MRF energy of 53 MeV, and with MRF directions nearly parallel to $\hat{z}$. There are more of these high-energy electrons when the $\mu^-$ spins are in the direction opposite to $\hat{z}$ than when the spins are parallel to $\hat{z}$. Thus the number of decay electrons reaches a maximum when the muon spin direction is opposite to $\hat{z}$, and a minimum when they are parallel. As the spin precesses the number of high-energy electrons will oscillate with frequency $\omega_a$. More generally, at laboratory energies above $\sim 1.2$ GeV, the electrons have a preferred average MRF direction parallel to $\hat{z}$ (see Figure 3.6). In this discussion, it is assumed that the spin precession vector, $\vec{\omega}_a$, is independent of time, and therefore the angle between the spin component in the orbit plane and the muon momentum direction is given by $\omega_a t + \phi$, where $\phi$ is a constant.

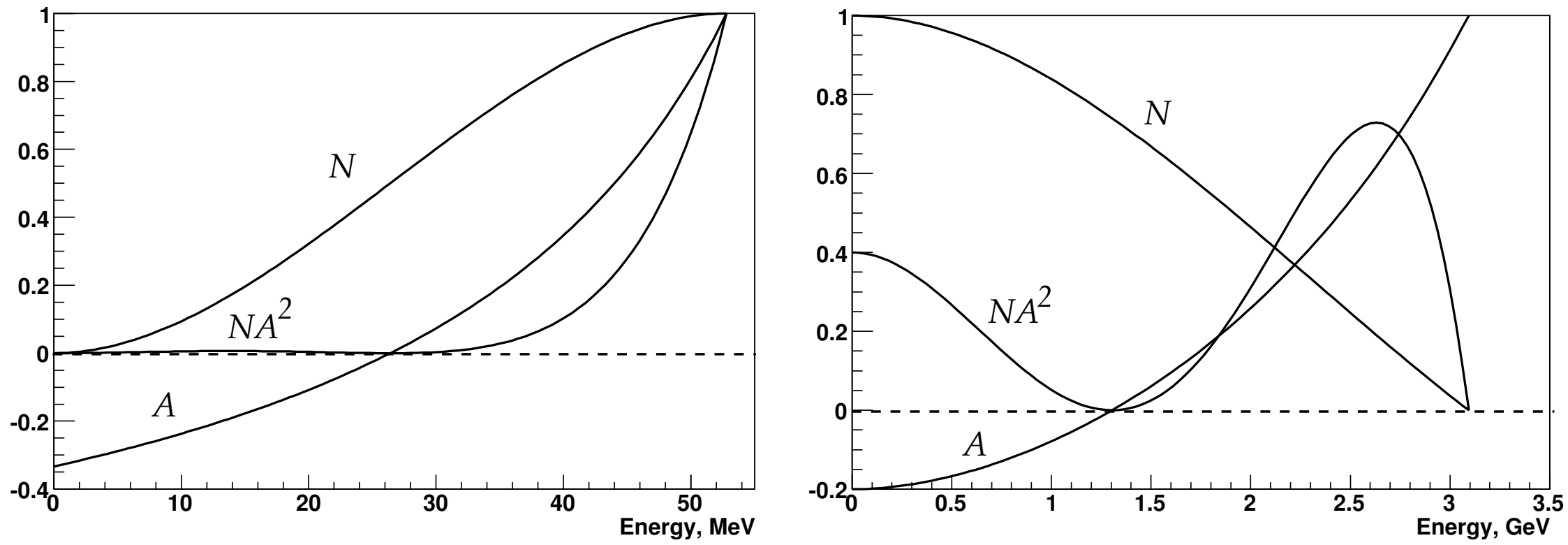

(a) Muon Rest Frame (b) Laboratory Frame

Figure 3.6: Number of decay electrons per unit energy, N (arbitrary units), value of the asymmetry $A$, and relative figure of merit $NA^2$ (arbitrary units) as a function of electron energy. Detector acceptance has not been incorporated, and the polarization is unity. For the third CERN experiment and E821, $E_{max} \approx 3.1$ GeV ($p_\mu = 3.094$ GeV/c) in the laboratory frame.

Equations 3.15 and 3.16 can be transformed to the laboratory frame to give the electron number oscillation with time as a function of electron energy,

$$N_d(t,E) = N_{d0}(E)e^{-t/\gamma\tau}[1 + A_d(E)\cos(\omega_a t + \phi_d(E))], \tag{3.17}$$

or, taking all electrons above threshold energy $E_{th}$,

$$N(t,E_{th}) = N_0(E_{th})e^{-t/\gamma\tau}[1 + A(E_{th})\cos(\omega_a t + \phi(E_{th}))]. \tag{3.18}$$

In Equation 3.17 the differential quantities are,

$$A_d(E) = \mathcal{P}\frac{-8y^2 + y + 1}{4y^2 - 5y - 5}, \quad N_{d0}(E) \propto (y-1)(4y^2 - 5y - 5), \tag{3.19}$$

and in Equation 3.18,

$$N(E_{th}) \propto (y_{th} - 1)^2(-y_{th}^2 + y_{th} + 3), \qquad A(E_{th}) = \mathcal{P}\frac{y_{th}(2y_{th} + 1)}{-y_{th}^2 + y_{th} + 3}. \tag{3.20}$$

In the above equations, $y = E/E_{max}$, $y_{th} = E_{th}/E_{max}$, $\mathcal{P}$ is the polarization of the muon beam, and $E$, $E_{th}$, and $E_{max} = 3.1$ GeV are the electron laboratory energy, threshold energy, and maximum energy, respectively.

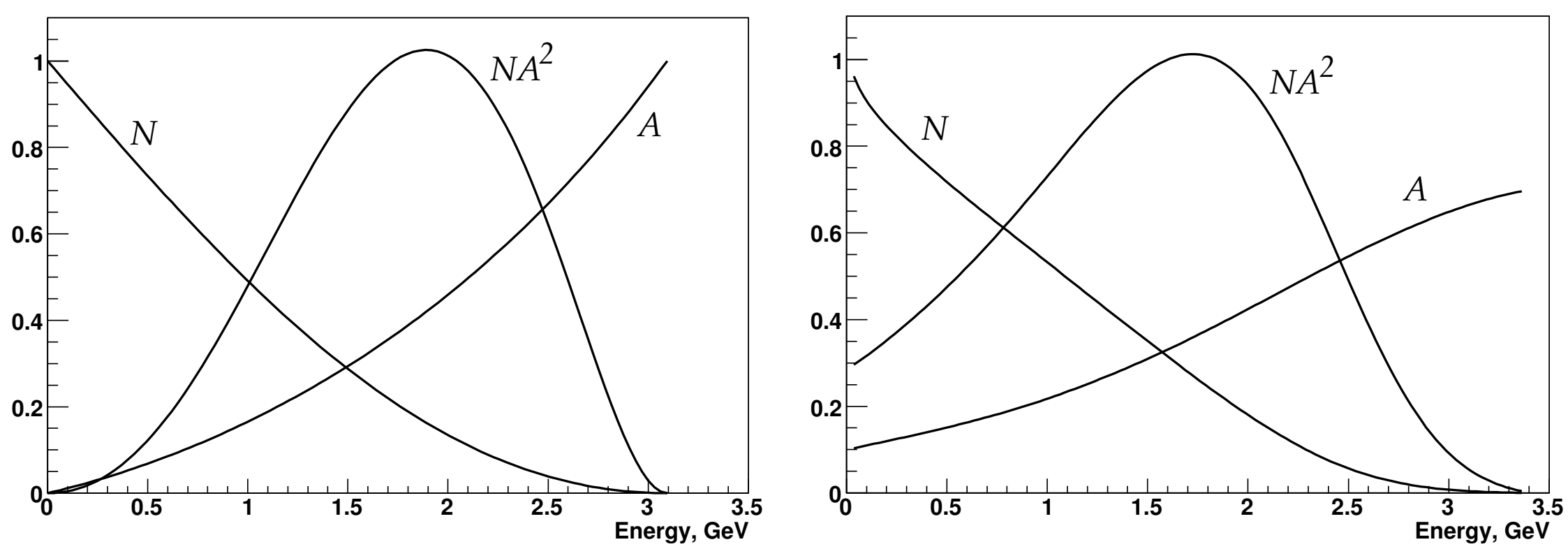

(a) No detector acceptance or energy resolution included (b) Detector acceptance and energy resolution included

Figure 3.7: The integral $N$, $A$, and $NA^2$ (arbitrary units) for a single energy-threshold as a function of the threshold energy; (a) in the laboratory frame, not including and (b) including the effects of detector acceptance and energy resolution for the E821 calorimeters. For the third CERN experiment and E821, $E_{max} \approx 3.1$ GeV ($p_\mu = 3.094$ GeV/c) in the laboratory frame.

The fractional statistical error on the precession frequency, when fitting data collected over many muon lifetimes to the five-parameter function (Equation 3.18), is given by

$$\delta\epsilon = \frac{\delta\omega_a}{\omega_a} = \frac{\sqrt{2}}{2\pi f_a \tau_\mu \sqrt{NA^2}}. \tag{3.21}$$

where $N$ is the total number of electrons, and $A$ is the asymmetry, in the given data sample. For a fixed magnetic field and muon momentum, the statistical figure of merit is $NA^2$, the quantity to be maximized in order to minimize the statistical uncertainty.

The energy dependencies of the numbers and asymmetries used in Equations 3.17 and 3.18, along with the figures of merit $NA^2$, are plotted in Figures 3.6 and 3.7 for the case of E821. The statistical power is greatest for electrons at 2.6 GeV (Figure 3.6). When a fit is made to all electrons above a single energy threshold, the optimal threshold energy is about 1.7-1.8 GeV (Figure 3.7).

The resulting arrival-time spectrum of electrons with energy greater than 1.8 GeV from the final E821 data run is shown in Fig. 3.8. While this plot clearly exhibits the expected features of the five-parameter function, a least-square fit to these 3.6 billion events gives an unacceptably large chi-square. A number of small effects must be taken into account to obtain a reasonable fit, which will be discussed in Chapter 5.

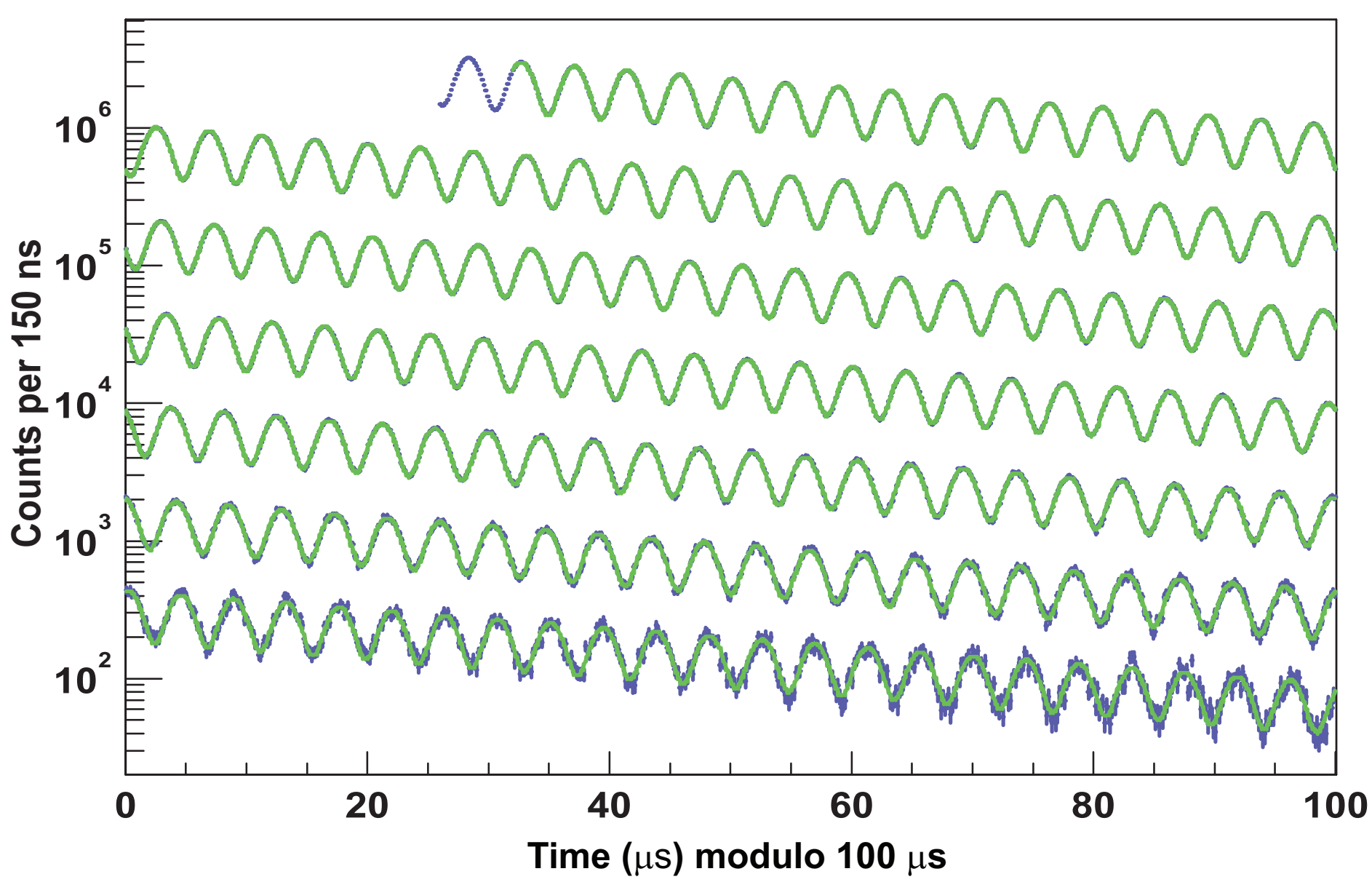

Figure 3.8: Histogram, modulo 100 $\mu$ s, of the number of detected electrons above 1.8 GeV for the 2001 data set as a function of time, summed over detectors, with a least-squares fit to the spectrum superimposed. Total number of electrons is $3.6 \times 10^9$. The data are in blue, the fit in green.

## 3.6 The Magnetic Field

The rate at which the muon spin turns relative to its momentum (Eq. 3.10) depends on the anomaly $a_\mu$ and on the average magnetic field given by Eq. 3.7. Thus the determination of $a_\mu$ to sub-tenths of a ppm requires that both $\omega_a$ and $\langle B \rangle$ be determined to this level. The muon beam is confined to a cylindrical region of 9 cm diameter, which is 44.7 m in length. The volume of this region is $\simeq 1.14$ m$^3$ or $\simeq 40$ ft$^3$, which sets the scale for the magnetic field measurement and control. The E989 goal is to know the magnetic field averaged over running time and the muon distribution to an uncertainty of $\pm 70$ parts per billion (ppb).

The problem breaks into several pieces:

1. Producing as uniform magnetic field as possible by shimming the magnet.
2. Stabilizing $B$ in time at the sub-ppm level by feedback, with mechanical and thermal stability.
3. Monitoring $B$ to the 20 ppb level around the storage ring during data collection.
4. Periodically mapping the field throughout the storage region and correlating the field map to the monitoring information without turning off the magnet between data collection and field mapping. It is essential that the magnet not be powered off unless absolutely necessary.
5. Obtaining an absolute calibration of the $B$-field relative to the Larmor frequency of the free proton.

The only magnetic field measurement technique with the sensitivity needed to measure and control the $B$-field to the tens of ppb is nuclear magnetic resonance (NMR). As in E821, E989 will implement a pulsed NMR setup. In this configuration a $\pi/2$ RF pulse is used to rotate the proton spin and the resulting free-induction decay (FID) will be detected by a pick-up coil around the sample. The E821 baseline design used the NMR of protons in a water sample with a $CuSO_4$ additive that shortened the relaxation time, with the probes tuned to operate in a 1.45 T field. When the water evaporated from a few of the probes, the water was replaced with petroleum jelly, which has the added features of a smaller sensitivity to temperature changes and no evaporation.

Special nuclear magnetic resonance (NMR) probes [10, 1] were used in E821 to measure and to monitor the magnetic field during the experimental data collection.[3] Three types of probes exist: a spherical water probe that provides the absolute calibration to the free proton; cylindrical probes that monitor the field during data collection, and also in an NMR trolley to map the field; and a smaller spherical probe which can be plunged into the muon storage region by means of a bellows system to transfer the absolute calibration to the trolley probes. A collection of 378 cylindrical probes placed in symmetrically machined grooves on the top and bottom of the muon beam vacuum chamber provide a point-to-point measure of the magnetic field while beam is in the storage ring. Probes at the same azimuthal location but different radii gave information on changes to the quadrupole component of the field at that location.

[3]The probes are described in Chapter 15

The field mapping trolley contains 17 cylindrical probes arranged in concentric circles as shown in Figure 3.9. At several-day intervals during the running periods, the beam will be turned off, and the field mapping trolley will be driven around the inside of the evacuated beam chamber measuring the magnetic field with each of the 17 trolley probes at 6,000 locations around the ring. One of the resulting E821 field maps, averaged over azimuth, is shown in Figure 3.9(b) for reference.

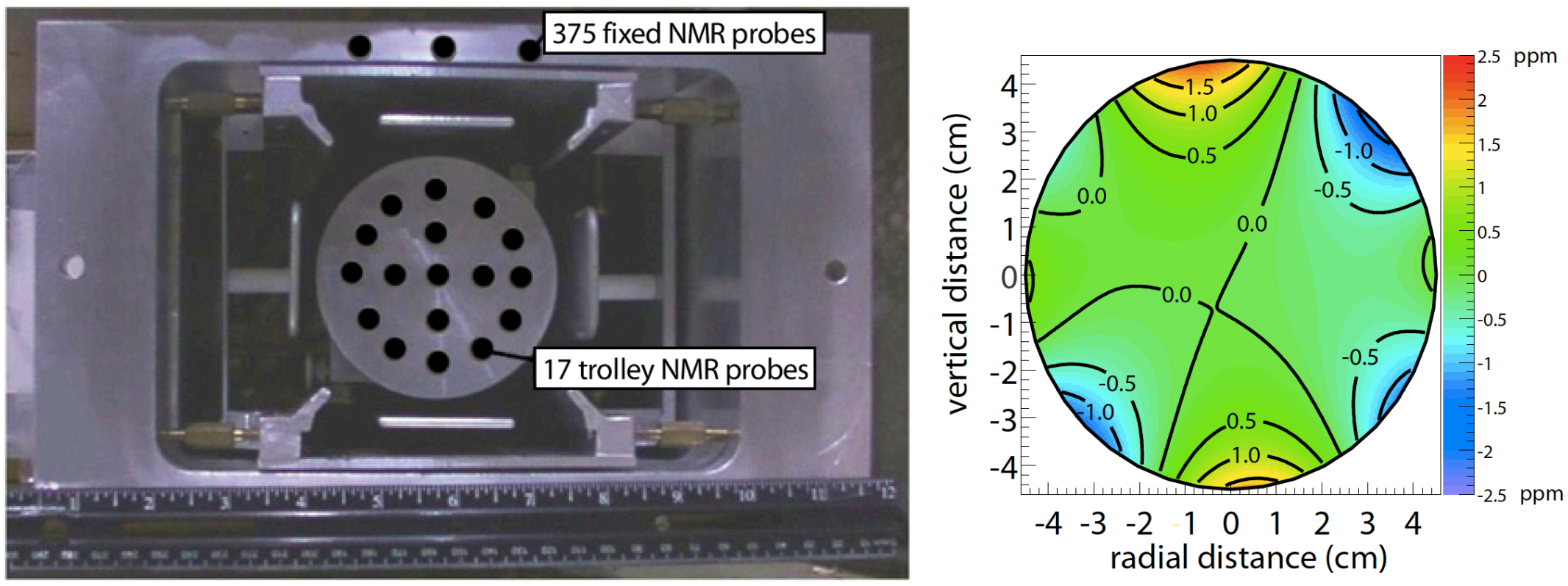

(a) (b)

Figure 3.9: (a) The electrostatic quadrupole assembly inside a vacuum chamber showing the NMR trolley sitting on the rails of the cage assembly. Seventeen NMR probes are located just behind the front face in the places indicated by the black circles. The inner (outer) circle of probes has a diameter of 3.5 cm (7 cm) at the probe centers. The storage region has a diameter of 9 cm. The vertical location of three of the 180 upper fixed probes is also shown. An additional 180 probes are located symmetrically below the vacuum chamber. (Reprinted with permission from [1]. Copyright 2006 by the American Physical Society.) (b) A contour plot of the magnetic field averaged over azimuth, 0.5 ppm intervals.

The absolute calibration utilizes a probe with a spherical water sample [11]. The Larmor frequency of a proton in a spherical water sample is related to that of the free proton through $f_{\rm L}({\rm sph}-{\rm H_2O},T) = [1-\sigma({\rm H_2O},T)]\, f_{\rm L}({\rm free})$, [12, 13] where $\sigma({\rm H_2O}, 34.7^\circ\ {\rm C}) = 25.790(14)\times 10^{-6}$ is from the diamagnetic shielding of the proton in the water molecule, determined from [14]

$$\sigma({\rm H_2O}, 34.7^\circ C) = 1 - \frac{g_p({\rm H_2O}, 34.7^\circ C)}{g_J(H)}\frac{g_J(H)}{g_p(H)}\frac{g_p(H)}{g_p({\rm free})}\,. \tag{3.22}$$

The terms are: the ratio of the $g$-factors of the proton in a spherical water sample to that of the electron in the hydrogen ground state ($g_J(H)$) [14]; the ratio of electron to proton $g$-factors in hydrogen [15]; the bound-state correction relating the $g$-factor of the proton bound in hydrogen to the free proton [16, 17]. The temperature dependence is from Reference [18]. An alternate absolute calibration would be to use an optically pumped $^3$He NMR probe [19]. This has several advantages: the sensitivity to the probe shape is negligible, and the temperature dependence is also negligible. This option is being explored for E989.

The calibration procedure described above permits the magnetic field to be expressed in terms of the Larmor frequency of a free proton, $\omega_p$. The magnetic field is weighted by the muon distribution, and also averaged over the running time weighed by the number of stored muons to determine the value of $\omega_p$ which is combined with the average $\omega_a$ to determine $a_\mu$. The reason for the use of these two frequencies, rather than $B$ measured in tesla can be understood from Eq. 3.10. To obtain $a_\mu$ from this relationship requires precise knowledge of the muon charge to mass ratio.

To determine $a_\mu$ from the two frequencies $\omega_a$ and $\omega_p$, we use the relationship

$$a_\mu = \frac{\omega_a/\omega_p}{\lambda_+ - \omega_a/\omega_p} = \frac{\mathcal{R}}{\lambda_+ - \mathcal{R}}\,, \tag{3.23}$$

where the ratio

$$\lambda_+ = \mu_{\mu^+}/\mu_p = 3.183\,345\,137\,(85) \tag{3.24}$$

is the muon-to-proton magnetic moment ratio [20] measured from muonium (the $\mu^+e^-$ atom) hyperfine structure[21]. Of course, to use $\lambda_+$ to determine $a_{\mu^-}$ requires the assumption of *CPT* invariance, *viz.* ($a_{\mu^+} = a_{\mu^-}$; $\lambda_+ = \lambda_-$). The comparison of $\mathcal{R}_{\mu^+}$ with $\mathcal{R}_{\mu^-}$ provides a *CPT* test. In E821

$$\Delta\mathcal{R} = \mathcal{R}_{\mu^-} - \mathcal{R}_{\mu^+} = (3.6 \pm 3.7) \times 10^{-9} \tag{3.25}$$

# Chapter 4

# Beam Dynamics and Beam Related Systematic Errors

## 4.1 Introduction

In this chapter we discuss the behavior of a beam in a weak-focusing betatron, and the features of the injection of a bunched beam that are important in the determination of $\omega_a$. We also discuss the corrections to the measured frequency $\omega_a$ that come from the vertical betatron motion, and the fact that not all muons are at the magic momentum (central radius) in the storage ring. The final section of this chapter discusses the systematic errors that come from the pion and muon beamlines.

## 4.2 The Weak Focusing Betatron

The behavior of the beam in the $(g-2)$ storage ring directly affects the measurement of $a_\mu$. Since the detector acceptance for decay electrons depends on the radial coordinate of the muon at the point where it decays, coherent radial motion of the stored beam can produce an amplitude modulation in the observed electron time spectrum. Resonances in the storage ring can cause particle losses, thus distorting the observed time spectrum, and must be avoided when choosing the operating parameters of the ring. Care is taken in setting the frequency of coherent radial beam motion, the "coherent betatron oscillation" (CBO) frequency, which lies close to the second harmonic of $f_a = \omega_a/(2\pi)$. If $f_{\rm CBO}$ is too close to $2f_a$, the beat frequency, $f_- = f_{CBO} - f_a$, complicates the extraction of $f_a$ from the data, and can introduce a significant systematic error.

A pure quadrupole electric field provides a linear restoring force in the vertical direction, and the combination of the (defocusing) electric field and the central magnetic field provides a linear restoring force in the radial direction. The $(g-2)$ ring is a weak focusing ring[1, 2, 3] with the field index

$$n = \frac{\kappa R_0}{\beta B_0}, \tag{4.1}$$

where $\kappa$ is the electric quadrupole gradient, $B_0$ is the magnetic field strength, $R_0$ is the magic radius $\equiv 7112$ mm, and $\beta$ is the relativistic velocity of the muon beam. For a ring with a uniform vertical dipole magnetic field and a uniform quadrupole field that provides vertical focusing covering the full azimuth, the stored particles undergo simple harmonic motion called betatron oscillations, in both the radial and vertical dimensions.

The horizontal and vertical motion are given by

$$x = x_e + A_x \cos(\nu_x \frac{s}{R_0} + \delta_x) \quad \text{and} \quad y = A_y \cos(\nu_y \frac{s}{R_0} + \delta_y), \tag{4.2}$$

where $s$ is the arc length along the trajectory. The horizontal and vertical tunes are given by

$$\nu_x = \sqrt{1-n} \quad \text{and } \nu_y = \sqrt{n}. \tag{4.3}$$

Several $n$ - values were used in E821 for data acquisition: $n = 0.137,\ 0.142$ and $0.122$. The horizontal and vertical betatron frequencies are given by

$$f_x = f_C\sqrt{1-n} \simeq 0.929 f_C \quad \text{and} \quad f_y = f_C\sqrt{n} \simeq 0.37 f_C, \tag{4.4}$$

where $f_C$ is the cyclotron frequency and the numerical values assume that $n = 0.137$. The corresponding betatron wavelengths are $\lambda_{\beta_x} = 1.08(2\pi R_0)$ and $\lambda_{\beta_y} = 2.7(2\pi R_0)$. It is important that the betatron wavelengths are not simple multiples of the circumference, as this minimizes the ability of ring imperfections and higher multipoles to drive resonances that would result in particle losses from the ring.

Table 4.1: Frequencies in the $(g-2)$ storage ring, assuming that the quadrupole field is uniform in azimuth and that $n = 0.137$.

| Quantity | Expression | Frequency [MHz] | Period [$\mu$s] |
|---|---|---|---|
| $f_a$ | $\frac{e}{2\pi mc} a_\mu B$ | 0.228 | 4.37 |
| $f_C$ | $\frac{v}{\pi R_0}$ | 6.7 | 0.149 |
| $f_x$ | $\sqrt{1-n} f_c$ | 6.23 | 0.160 |
| $f_y$ | $\sqrt{n} f_c$ | 2.48 | 0.402 |
| $f_{\text{CBO}}$ | $f_c - f_x$ | 0.477 | 2.10 |
| $f_{\text{VW}}$ | $f_c - 2f_y$ | 1.74 | 0.574 |

As a reminder, the muon frequency, $\omega_a$ is determined by the average magnetic field weighted by the muon distribution and the magnetic anomaly:

$$\vec{\omega}_a = -\frac{Qe}{m}\left[a_\mu \vec{B} - \left(a_\mu - \left(\frac{mc}{p}\right)^2\right)\frac{\vec{\beta}\times\vec{E}}{c}\right]. \tag{4.5}$$

The field index also determines the angular acceptance of the ring. The maximum horizontal and vertical angles of the muon momentum are given by

$$\theta^x_{\max} = \frac{x_{\max}\sqrt{1-n}}{R_0}, \quad \text{and} \quad \theta^y_{\max} = \frac{y_{\max}\sqrt{n}}{R_0}, \tag{4.6}$$

where $x_{\max}, y_{\max} = 45$ mm is the radius of the storage aperture. For a betatron amplitude $A_x$ or $A_y$ less than 45 mm, the maximum angle is reduced, as can be seen from the above equations.

## 4.3 Weak Focusing with Discrete Quadrupoles

For a ring with discrete quadrupoles, the focusing strength changes as a function of azimuth, and the equation of motion looks like an oscillator whose spring constant changes as a function of azimuth $s$. The motion is described by

$$x(s) = x_e + A\sqrt{\beta(s)}\cos(\psi(s) + \delta), \tag{4.7}$$

where $\beta(s)$ is one of the three Courant-Snyder parameters.[2]

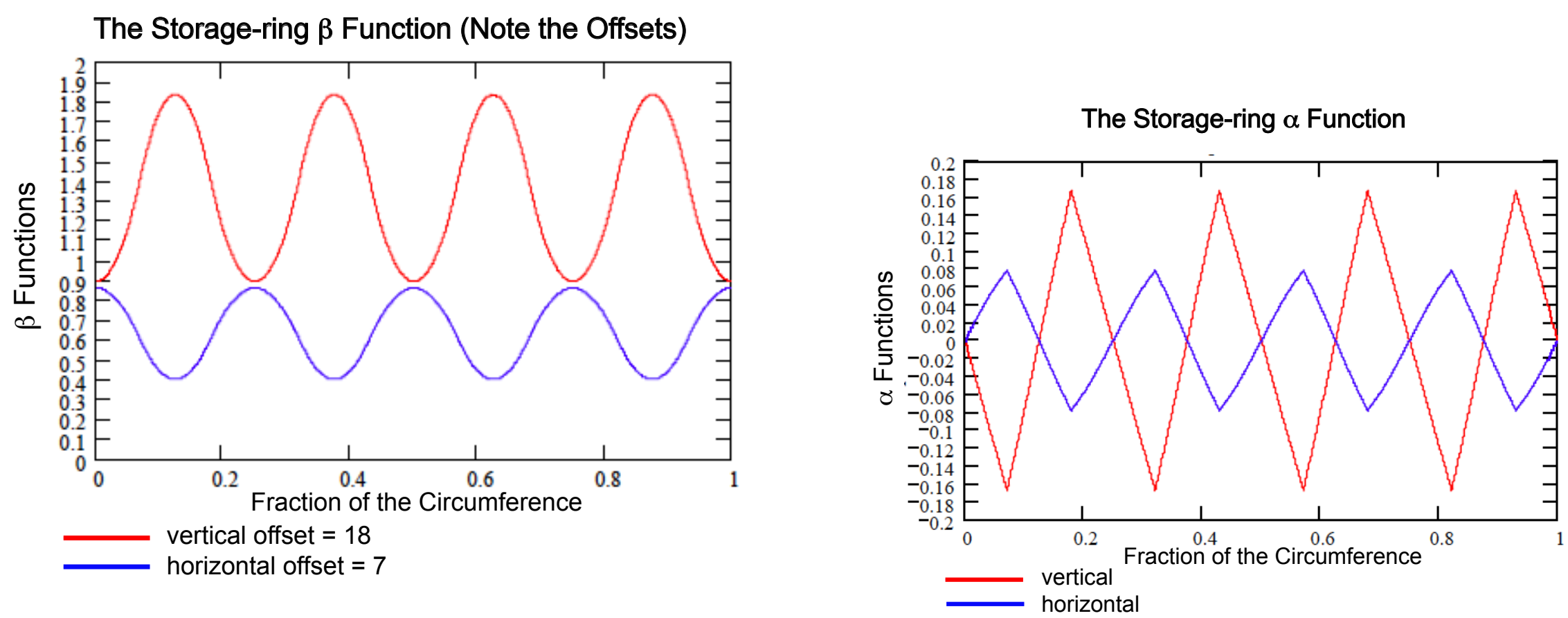

(a) (b)

Figure 4.1: (a) The horizontal (radial) and vertical beta functions for the E821 lattice. Note the scale offset. (b) The horizontal (radial) and vertical alpha functions for the E821 lattice. The $n$-value is 0.134 for both. (From Ref. [9]

The layout of the storage ring is shown in Figure 4.2(a). The four-fold symmetry of the quadrupoles was chosen because it provided quadrupole-free regions for the kicker, tracking chambers, fiber monitors, and trolley garage; but the most important benefit of four-fold symmetry is to reduce the peak-to-peak betatron oscillation amplitudes, with $\sqrt{\beta_{\max}/\beta_{\min}} = 1.03$. The beta and alpha functions for the $(g-2)$ storage ring [9] are shown in Fig. 4.1.

Resonances in the storage ring will occur if $L\nu_x + M\nu_y = N$, where $L$, $M$ and $N$ are integers, which must be avoided in choosing the operating value of the field index. These resonances form straight lines on the tune plane shown in Figure 4.2(b), which shows resonance lines up to fifth order. The operating point lies on the circle $\nu_x^2 + \nu_y^2 = 1$.

The detector acceptance depends on the radial position of the muon when it decays, so that any *coherent* radial beam motion will amplitude modulate the decay $e^{\pm}$ distribution. This can be understood by examining Fig. 4.3. A narrow bunch of muons starts its radial betatron oscillation at the point $s = 0$. The circumference of the ring is $2\pi\rho$ so the $x$-axis shows successive revolutions around the ring. The radial betatron wavelength is longer than the circumference $2\pi\rho$. The rate at which the muon bunch moves toward and then away from the detector is given by $f_{CBO} = f_C - f_x$. The CBO wavelength is slightly over 14 revolutions of the ring.

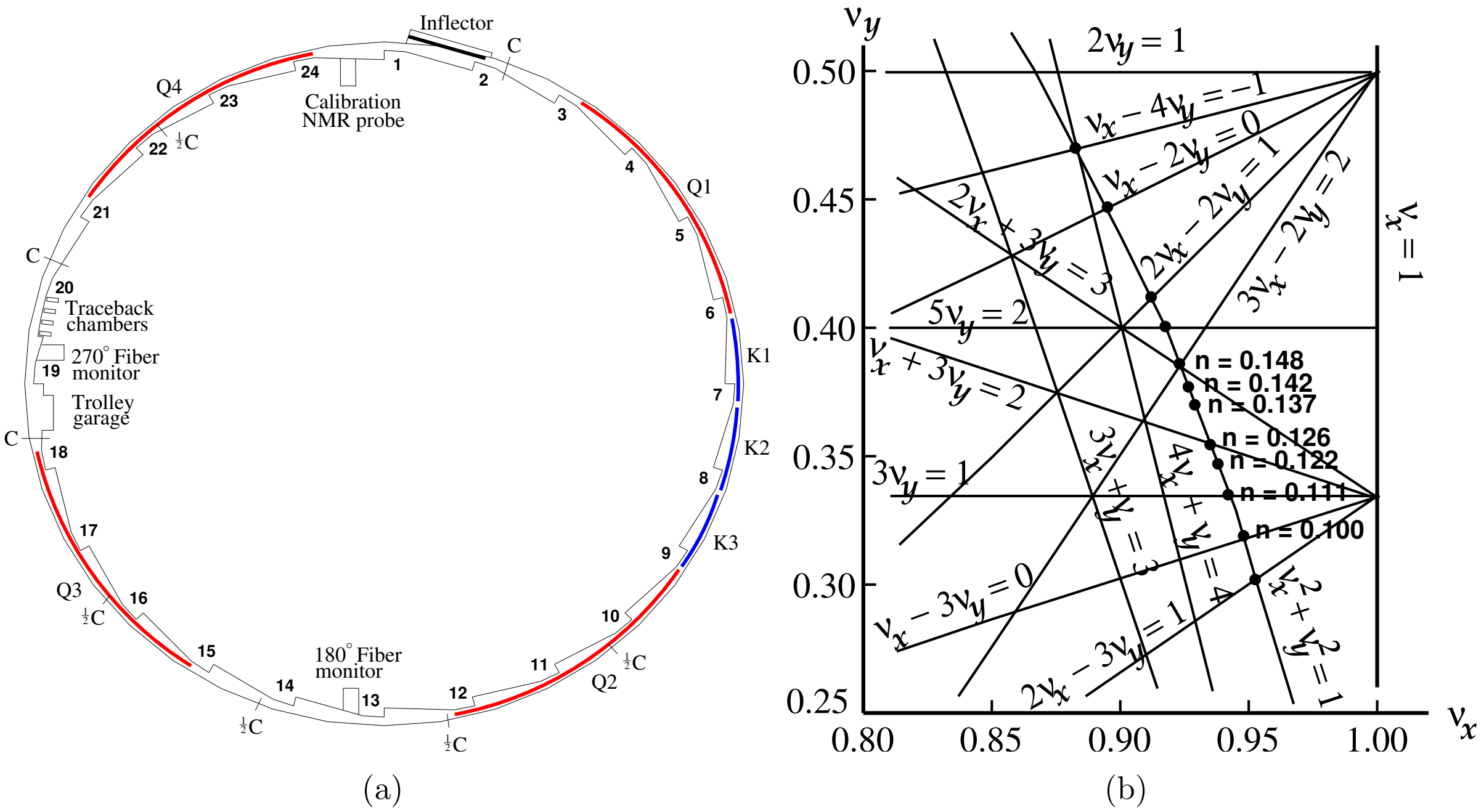

Figure 4.2: (a) The layout of the storage ring. (b)The tune plane, showing the three operating points used during our three years of E821 running.

The presence of the CBO was first discovered in E821 from a plot that showed an azimuthal variation in the value of $a_\mu$ shown in Fig. 4.4(a). When the CBO is included, this azimuthal dependence disappears. Because the CBO wavelength is only slightly greater than the circumference, its effect almost washes out when all detectors are added together. Adding all detectors together was one of the techniques used in E821 to eliminate CBO effect. However, the four-fold symmetry of the ring was broken by the kicker plates that covered one section of the ring, so the cancellation was not perfect, but good enough. This will most likely not be true in E989, so it is important to minimize the CBO effects. See Chapter 13 for further discussion. Since some detectors saw more injection flash than others, this meant that data at times earlier than around 40 $\mu$s was discarded in those analyses. Other analyzers included the CBO and were able to use data from the "quiet" detectors at earlier times.

The principal frequency will be the "Coherent Betatron Frequency,"

$$f_{\rm CBO} = f_C - f_x = (1 - \sqrt{1-n}) f_C \simeq 470 \text{ kHZ}, \tag{4.8}$$

which is the frequency at which a single fixed detector sees the beam coherently moving back and forth radially. This CBO frequency is close to the second harmonic of the $(g-2)$ frequency, $f_a = \omega_a/2\pi \simeq 228$ Hz.

An alternative way of thinking about the CBO motion is to view the ring as a spectrometer where the inflector exit is imaged at each successive betatron wavelength, $\lambda_{\beta_x}$. In principle, an inverted image appears at half a betatron wavelength; but the radial image is spoiled by the $\pm 0.3\%$ momentum dispersion of the ring. A given detector will see the beam move radially with the CBO frequency, which is also the frequency at which the horizontal

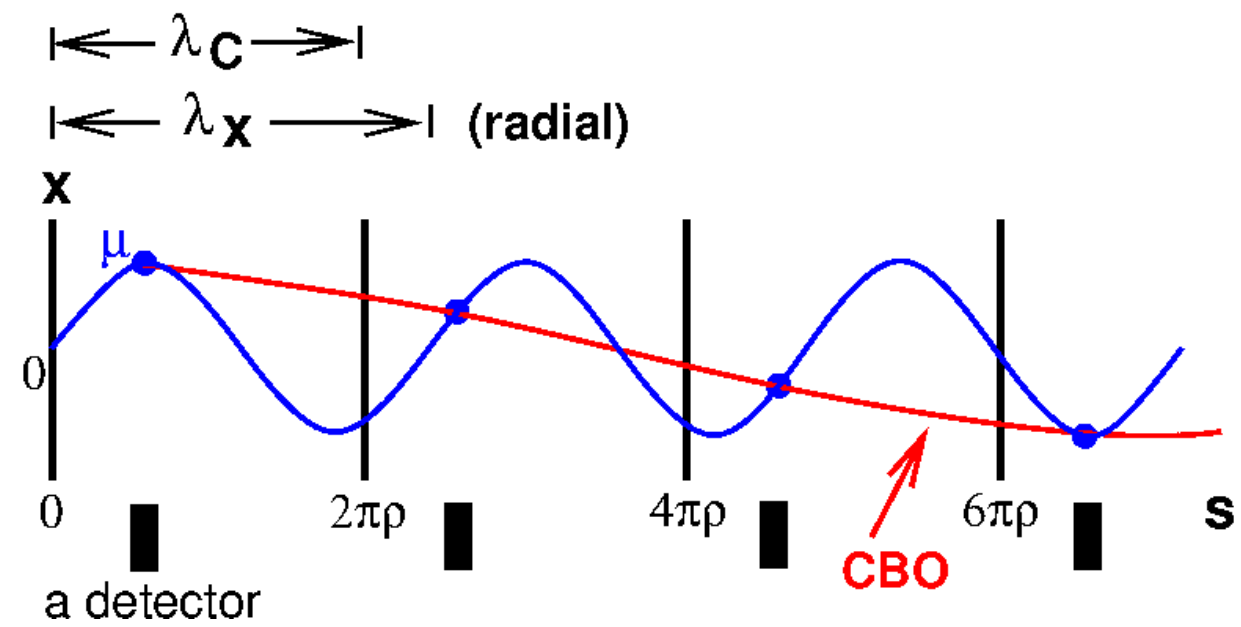

Figure 4.3: A cartoon of the coherent betatron motion (CBO). The radial CBO oscillation is shown in blue for 3 successive betatron wavelengths, the cyclotron wavelength (the circumference) is marked by the black vertical lines. One detector location is shown. Since the radial betatron wavelength is larger than the circumference, the detector sees the bunched beam slowly move closer and then further away. The frequency that the beam appears to move in and out is $f_{CBO}$ .

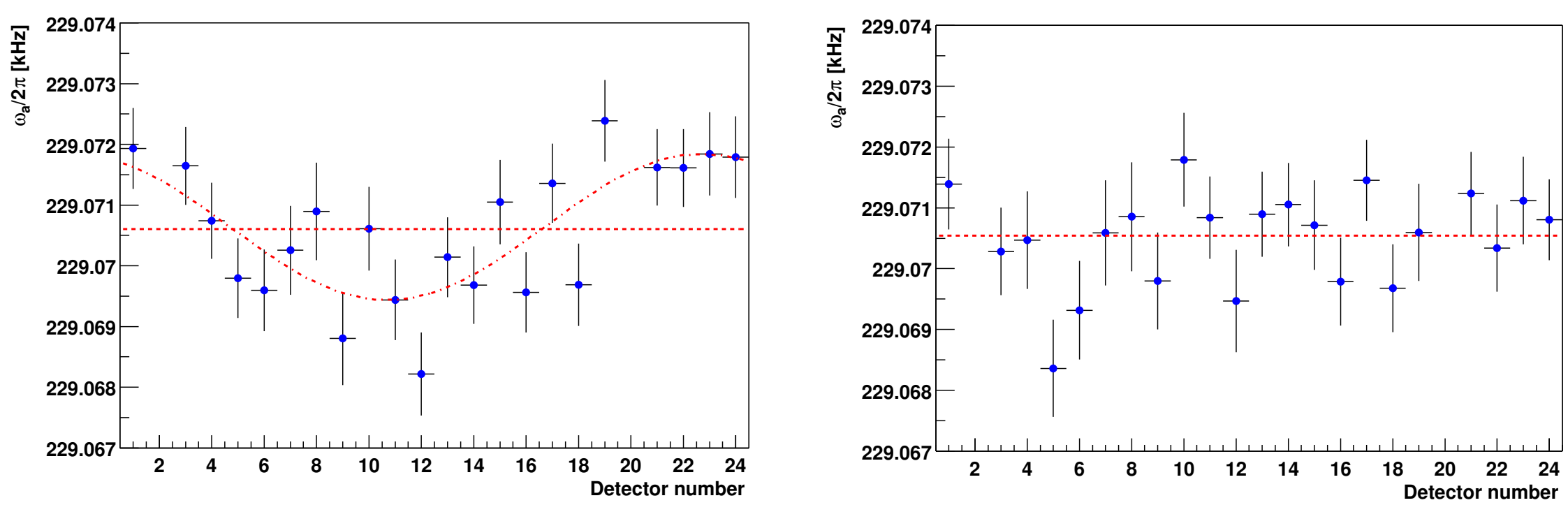

(a) No CBO in Fit

(b) CBO in Fit

Figure 4.4: The dependence of the extracted value of $a_\mu$ vs. detector number. (a)With no CBO in the fit function. (b) With CBO included in the fit function.

waist precesses around the ring. The vertical waist betatron wavelength is only 2.7 turns, and disappears rather quickly. A number of frequencies in the ring are tabulated in Table 4.1

The CBO frequency and its sidebands are clearly visible in the Fourier transform to the residuals from a fit to the five-parameter fitting function Equation 3.18, and are shown in Figure 4.5. The vertical waist frequency is barely visible. In 2000, the quadrupole voltage was set such that the CBO frequency was uncomfortably close to the second harmonic of $f_a$, thus placing the difference frequency $f_- = f_{CBO} - f_a$ next to $f_a$. This nearby sideband forced us to work very hard to understand the CBO and how its related phenomena affect the value of $\omega_a$ obtained from fits to the data. In 2001, we carefully set $f_{CBO}$ at two different values, one well above, the other well below $2f_a$, which greatly reduced this problem.

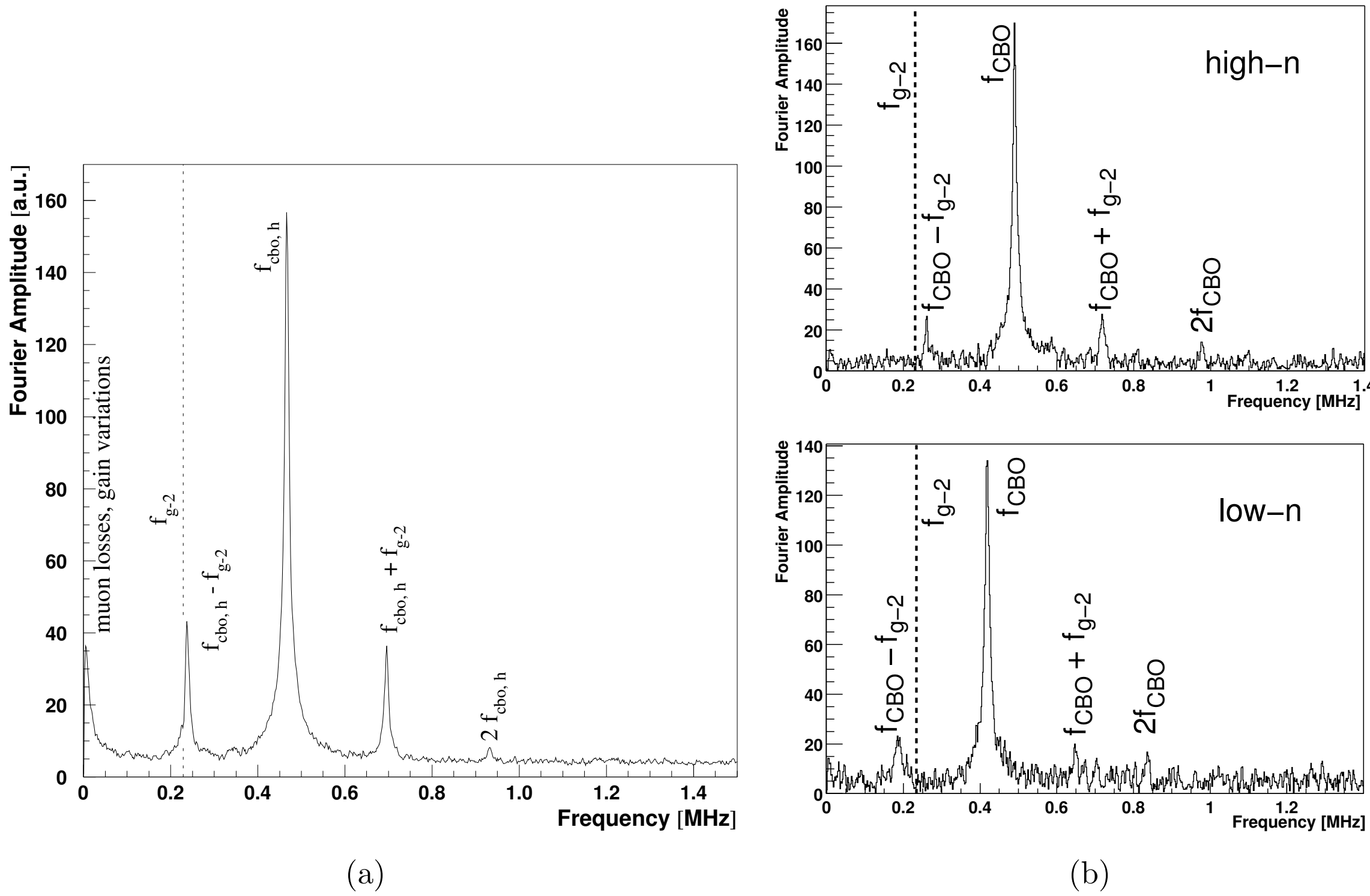

Figure 4.5: The Fourier transform to the residuals from a fit to the five-parameter function, showing clearly the coherent beam frequencies. (a) is from 2000, when the CBO frequency was close to $2\omega_a$, and (b) shows the Fourier transform for the two n-values used in the 2001 run period.

### 4.3.1 Monitoring the Beam Profile

Knowledge of the distribution of stored beam in the storage ring is necessary for several important corrections to the measured muon spin rotation frequency. There are three tools available to determine this distribution:

1. Tracking chambers (see Chapter 19) that measure the trajectories of the decay positrons, and reconstruct the vertical and horizontal spacial distribution of stored muons.

2. Measurement of the beam de-bunching after injection into the ring; called the "fast rotation analysis", which is discussed below.

3. Fiber beam monitors, which consist of $x$ and $y$ arrays of 0.5 mm scintillating fibers that can be inserted into the storage region to measure the central part of the muon distribution (see Chapter 20).

Because of the limited momentum acceptance of the Recycler Ring, the minimum proton bunch width is 120 ns, as is shown in Fig. 7.5. In E821 the beam had an rms $\sim$ 25 ns. These beam widths should be compared to the cyclotron period of the storage ring of 149 ns. We first discuss the E821 case, with its narrow beam. The momentum distribution of stored muons produces a corresponding distribution in radii of curvature. The distributions depend on the phase-space acceptance of the ring, the phase space of the beam at the injection point, and the kick given to the beam at injection. The narrow 18 mm horizontal aperture of the E821 inflector magnet restricts the stored momentum distribution to about $\pm 0.15\%$. As the muons circle the ring, the muons at smaller radius (lower momentum) eventually pass those at larger radius repeatedly after multiple transits around the ring, and the bunch structure largely disappears after 60 $\mu$s . This de-bunching can be seen in the E821 data in Figure 4.6 where the signal from a single detector is shown at two different times following injection. The bunched beam is seen very clearly in the left figure, with the 149 ns cyclotron period being obvious. The slow amplitude modulation comes from the $(g-2)$ precession. By 36 $\mu$s the beam has largely de-bunched.

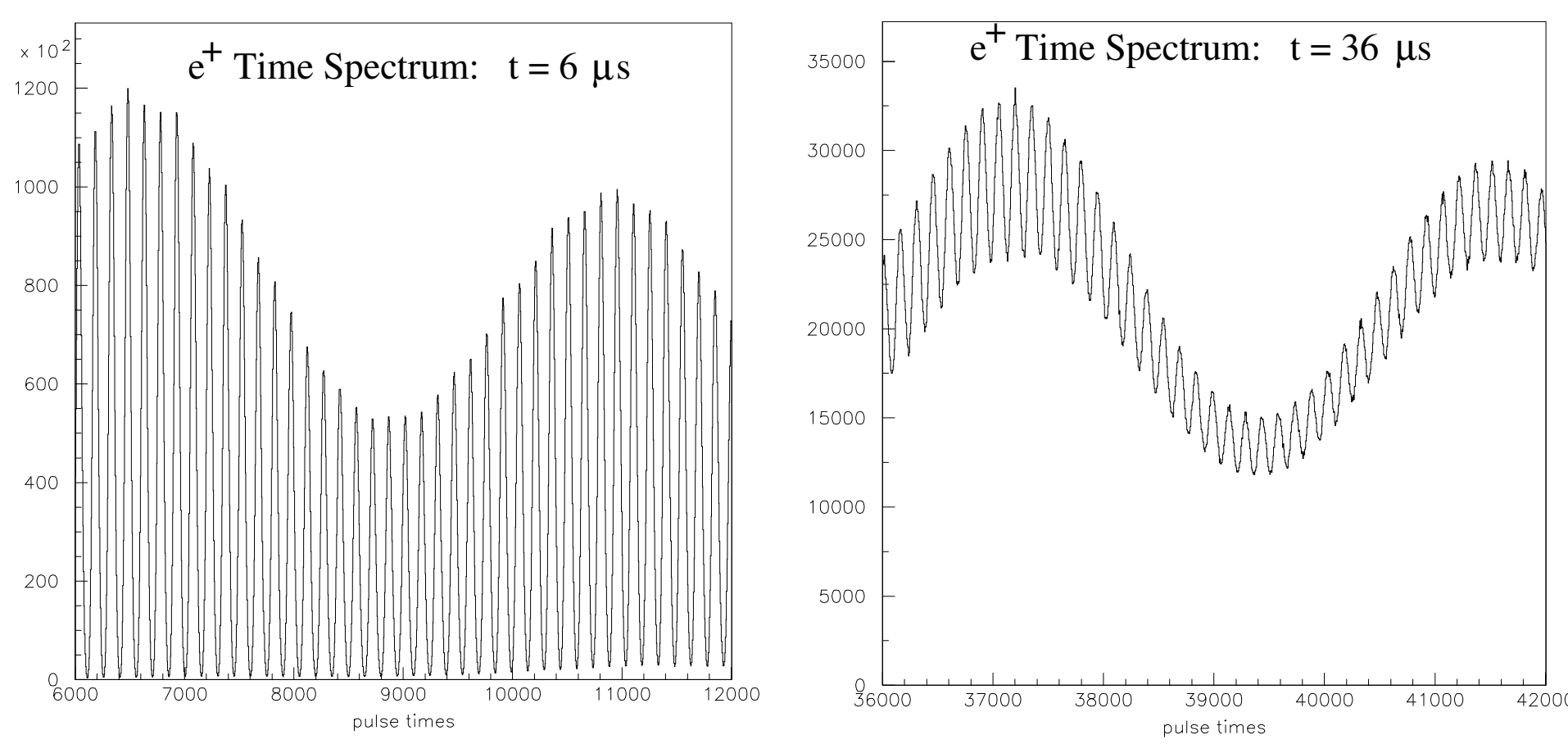

Figure 4.6: The time spectrum of a single E821 calorimeter soon after injection. The spikes are separated by the cyclotron period of 149 ns. The time width of the beam at injection was $\sigma \simeq 23$ ns.

Only muons with orbits centered at the central radius have the "magic" momentum, so knowledge of the momentum distribution, or equivalently the distribution of equilibrium radii, is important in determining the correction to $\omega_a$ caused by the radial electric field used for vertical focusing. Two methods of obtaining the distribution of equilibrium radii from the beam debunching were employed in E821. One method uses a model of the time evolution of the bunch structure. A second, alternative procedure uses modified Fourier techniques[8].

We discuss the former method, which was descended from the third CERN experiment, and show a preliminary study that demonstrates the ability to use this method to determine the distribution of equilibrium radii in E989. The initial bunched beam is modeled as an ensemble of particles having an unknown frequency distribution and a narrow time spread. The model assumes that every time slice of the beam has the same frequency profile but the time width is left as a fit parameter, as is the exact injection time. The distribution of angular frequencies will cause the bunched beam to spread out around the ring over time, in a manner that depends uniquely on the momentum distribution. In particular, the time evolution of any finite frequency slice is readily specified. A given narrow bin of frequencies contributes linearly to the time spectrum. The total time spectrum is a sum over many of these frequency components, with amplitudes that can be determined using $\chi^2$ minimization. The momentum distribution is then determined from the frequency distribution (or equivalently, from the radial distribution) by

$$\frac{p - p_0}{p_0} = (1 - n)\left(\frac{R - R_0}{R_0}\right). \tag{4.9}$$

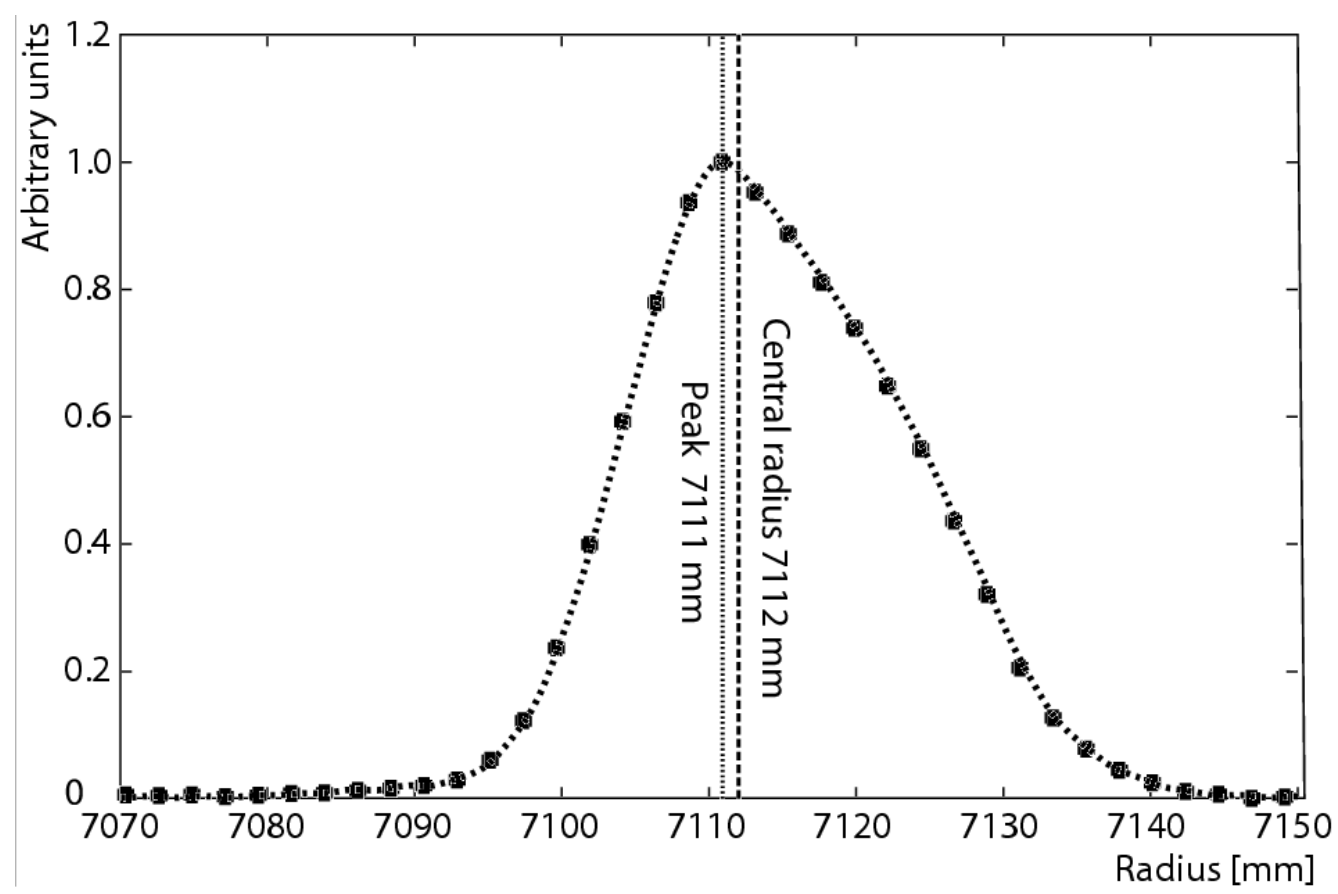

Figure 4.7: The distribution of equilibrium radii obtained from the beam de-bunching. The solid circles are from a de-bunching model fit to the data, and the dotted curve is obtained from a modified Fourier analysis.

The result of the fast-rotation analysis from one of the E821 running periods is shown in Fig. 4.7. The smooth curve is obtained from the modified Fourier transform analysis. The peak of the distribution lies below the nominal magic radius of 7112 mm but the mean is somewhat larger, $7116 \pm 1$ mm for this run period. The rms width is about 10 mm, and the two methods give equivalent results.

Early in the planning for E989, it became clear that the Recycler beam would be much wider than that produced by the BNL AGS. A preliminary beam profile, shown in Fig. 4.8(a), was used to determine whether the fast rotation analysis could be used for such a wide beam. The equilibrium distribution for the simulation was chosen to be Gaussian, with a mean of 7112 mm and width 14.2 mm. The time structure seen by a single detector is shown in Fig. 4.8(b), which can be compared to Fig. 4.6. The distribution of equilibrium radii obtained from the analysis of the debunching is shown in Fig. 4.8(c). The input mean was recovered in the analysis. Several questions will be addressed in future studies: What is the connection between the $t_0$ phase and the distribution of equilibrium radii? What happens if the equilibrium radius is changed significantly by beam scraping after injection? Would this be easier to detect and correct for with a narrower pulse?

While the scintillating-fiber monitors were not that useful in measuring the beam profile, they were extremely useful in measuring the various frequencies in the muon beam motion. The pulse height from a single fiber varies as the beam oscillates across it, and show clearly the vertical and horizontal tunes as expected. In Figure 4.9, the horizontal beam centroid motion is shown, with the quadrupoles powered asymmetrically during scraping, and then symmetrically after scraping. A Fourier transform of the latter signal shows the expected frequencies, including the cyclotron frequency of protons stored in the ring.

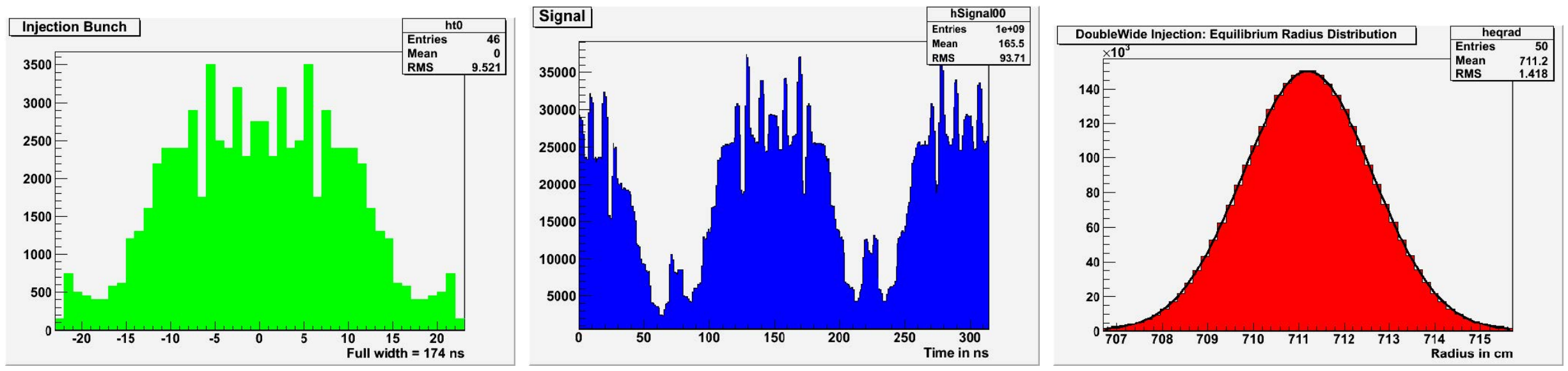

(a) (b) (c)

Figure 4.8: Simulations of a temporally wide beam. (a) An early version of the Recycler output beam. (b) The time spectrum shortly after injection, which can be compared with the left-hand E821 calorimeter after injection shown Fig. 4.6. (c) The distribution of equilibrium radii extracted from the debunching in these simulated data.

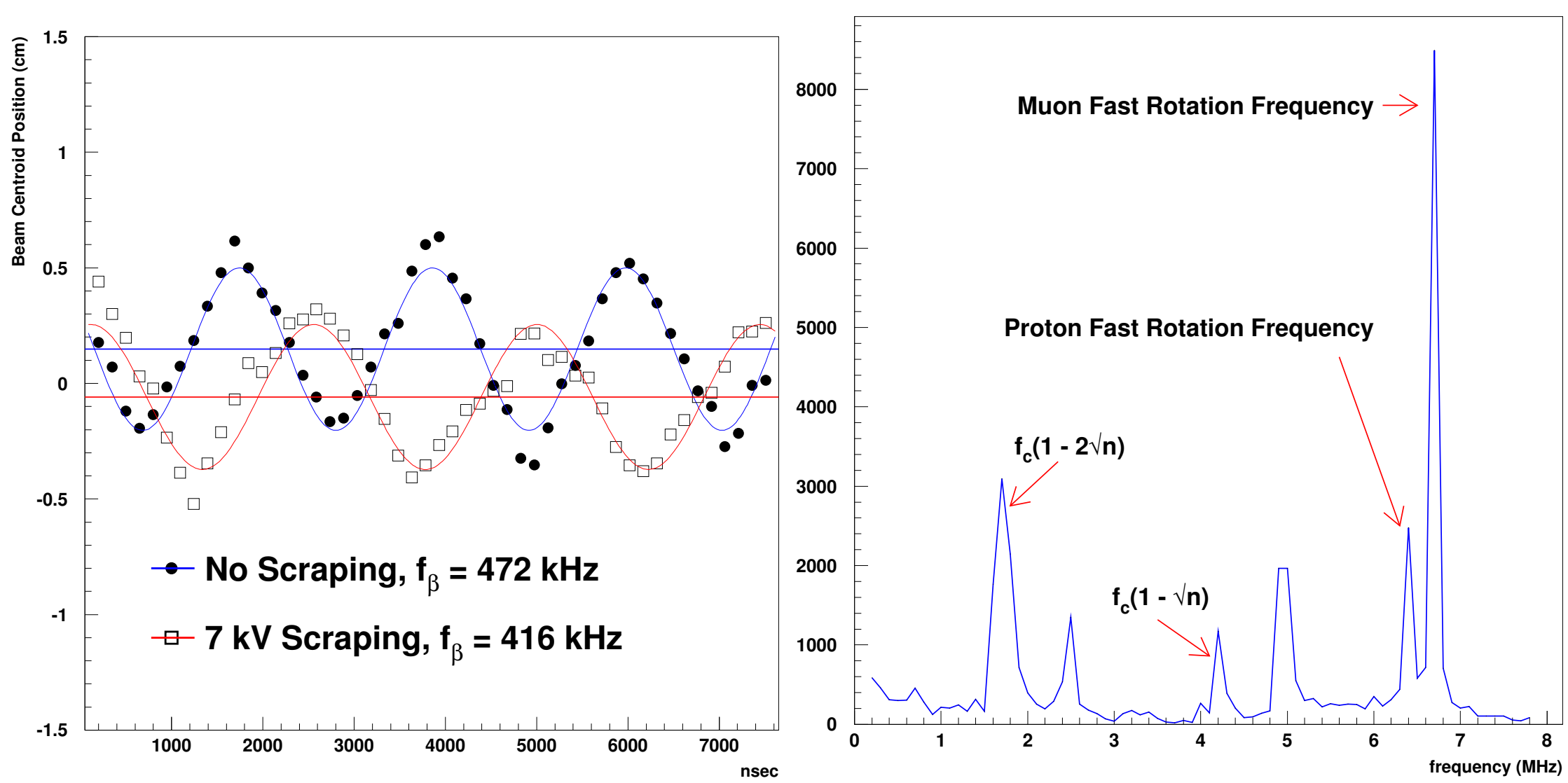

(a) (b)

Figure 4.9: (a) The horizontal beam centroid motion with beam scraping and without, using data from the scintillating fiber hodoscopes; note the tune change between the two. (b) A Fourier transform of the pulse from a single horizontal fiber, which shows clearly the vertical waist motion, as well as the vertical tune. The presence of stored protons is clearly seen in this frequency spectrum.

## 4.4 Corrections to $\omega_a$: Pitch and Radial Electric Field

In the simplest case, in the absence of an electric field and when the velocity is perpendicular to the magnetic field, the rate at which the spin turns relative to the momentum is given by

$$\omega_a = \omega_S - \omega_C = -\left(\frac{g-2}{2}\right)\frac{Qe}{m}B = -a\frac{Qe}{m}B \tag{4.10}$$

The spin equation modified by the presence of an electric field was introduced earlier, with the assumption that the velocity is transverse to the magnetic field. In the approximation that all muons are at the magic momentum, $\gamma_{\rm magic} = p_{\rm magic}/m\beta$, the electric field does not affect the spin rotation.

At the current and proposed levels of experimental precision, corrections for the approximations that the velocity is perpendicular to the field and that all muons are at the magic momentum must be made; the vertical betatron motion must be included, and the storage ring momentum acceptance of $\pm 0.5\%$ means that the muons have a range of momenta not quite at the magic momentum. Corrections to the measured value for $\omega_a$ from these two effects were made to the data in E821 after the data were un-blinded. In the 2001 data set, the electric field correction for the low $n$-value data set was $+0.47 \pm 0.05$ ppm. The pitch correction was $+0.27 \pm 0.04$ ppm. These are the only corrections made to the $\omega_a$ data.

We sketch the derivation for E821 and E989 below[4]. For a general derivation the reader is referred to References [6, 7].

For the more general case where $\vec{\beta} \cdot \vec{B} \neq 0$ and $\vec{E} \neq 0$, the cyclotron rotation frequency becomes:

$$\vec{\omega}_C = -\frac{Qe}{m}\left[\frac{\vec{B}}{\gamma} - \frac{\gamma}{\gamma^2-1}\left(\frac{\vec{\beta}\times\vec{E}}{c}\right)\right], \tag{4.11}$$

and the spin rotation frequency becomes[5]

$$\vec{\omega}_S = -\frac{Qe}{m}\left[\left(\frac{g}{2}-1+\frac{1}{\gamma}\right)\vec{B} - \left(\frac{g}{2}-1\right)\frac{\gamma}{\gamma+1}(\vec{\beta}\cdot\vec{B})\vec{\beta} - \left(\frac{g}{2}-\frac{\gamma}{\gamma+1}\right)\left(\frac{\vec{\beta}\times\vec{E}}{c}\right)\right]. \tag{4.12}$$

Substituting for $a_\mu = (g_\mu - 2)/2$, we find that the spin difference frequency is

$$\vec{\omega}_{diff} = \vec{\omega}_S - \vec{\omega}_C = -\frac{Qe}{m}\left[a_\mu\vec{B} - a_\mu\left(\frac{\gamma}{\gamma+1}\right)(\vec{\beta}\cdot\vec{B})\vec{\beta} - \left(a_\mu - \frac{1}{\gamma^2-1}\right)\frac{\vec{\beta}\times\vec{E}}{c}\right]. \tag{4.13}$$

Strictly speaking, the rate of change of the angle between the spin and the momentum vectors, $|\vec{\omega}_a|$='precession frequency', is equal to $|\vec{\omega}_{diff}|$ only if $\vec{\omega}_S$ and $\vec{\omega}_C$ are parallel. For the E821 and E989 experiments, the angle between $\vec{\omega}_S$ and $\vec{\omega}_C$ is always small and the rate of oscillation of $\vec{\beta}$ out of pure circular motion is fast compared to $\omega_a$, allowing us in the following discussion the make the approximation that $\vec{\omega}_a \simeq \vec{\omega}_{diff}$. More general calculations, where this approximation is not made, are found in References [6, 7]. In the E821 and E989 limits, the results presented here are the same as in the References.

If $\vec{\beta} \cdot \vec{B} = 0$, the angle between the momentum and spin reduces to the previously introduced expression

$$\vec{\omega}_a \simeq \vec{\omega}_{diff} = -\frac{Qe}{m}\left[a_\mu \vec{B} - \left(a_\mu - \frac{1}{\gamma^2 - 1}\right)\frac{\vec{\beta} \times \vec{E}}{c}\right]. \tag{4.14}$$

For $\gamma_{\rm magic} = 29.3$ ($p_\mu = 3.09$ GeV/$c$), the second term vanishes and the electric field does not contribute to the spin precession. In that case, the spin precession is independent of muon momentum; *all* muons precess at the same rate. Because of the high uniformity of the B-field, a precision knowledge of the stored beam trajectories in the storage region is not required.

First we calculate the effect of the electric field due to muons not exactly at $\gamma_{\rm magic} = 29.3$, for the moment neglecting the $\vec{\beta} \cdot \vec{B}$ term. If the muon momentum is different from the magic momentum, the precession frequency is given by

$$\omega_a' = \omega_a\left[1 - \beta\frac{E_r}{cB_y}\left(1 - \frac{1}{a_\mu \beta^2 \gamma^2}\right)\right], \tag{4.15}$$

where $\omega_a = -a\frac{Qe}{m}B$. Using $p = \beta\gamma m = (p_m + \Delta p)$, after some algebra one finds

$$\frac{\omega_a' - \omega_a}{\omega_a} = \frac{\Delta\omega_a}{\omega_a} = -2\frac{\beta E_r}{cB_y}\left(\frac{\Delta p}{p_m}\right). \tag{4.16}$$

Thus the effect of the radial electric field reduces the observed frequency from the simple frequency $\omega_a$ given in Equation 4.13. Now

$$\frac{\Delta p}{p_m} = (1-n)\frac{\Delta R}{R_0} = (1-n)\frac{x_e}{R_0}, \tag{4.17}$$

where $x_e$ is the muon's equilibrium radius of curvature relative to the central orbit. The electric quadrupole field is

$$E = \kappa x = \frac{n\beta c B_y}{R_0}x. \tag{4.18}$$

We obtain

$$\frac{\Delta\omega}{\omega} = -2n(1-n)\beta^2\frac{x x_e}{R_0^2}, \tag{4.19}$$

so clearly the effect of muons not at the magic momentum is to lower the observed frequency. For a quadrupole focusing field plus a uniform magnetic field, the time average of $x$ is just $x_e$, so the electric field correction is given by

$$C_E = \frac{\Delta\omega}{\omega} = -2n(1-n)\beta^2\frac{\langle x_e^2\rangle}{R_0^2}, \tag{4.20}$$

where $\langle x_e^2\rangle$ is determined from the fast-rotation analysis (see Figure 4.6). The uncertainty on $\langle x_e^2\rangle$ is added in quadrature with the uncertainty in the placement of the quadrupoles of $\delta R = \pm 0.5$ mm ($\pm 0.01$ ppm), and with the uncertainty in the mean vertical position of the beam, $\pm 1$ mm ($\pm 0.02$ ppm). For the low-$n$ 2001 sub-period, $C_E = 0.47 \pm 0.054$ ppm.

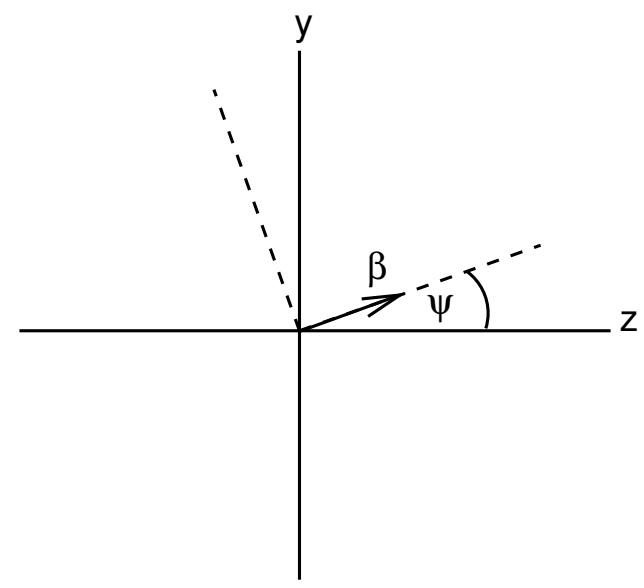

Figure 4.10: The coordinate system of the pitching muon. The angle $\psi$ varies harmonically. The vertical direction is $\hat{y}$ and $\hat{z}$ is the azimuthal (beam) direction.

The betatron oscillations of the stored muons lead to $\vec{\beta} \cdot \vec{B} \neq 0$. Since the $\vec{\beta} \cdot \vec{B}$ term in Equation 4.12 is quadratic in the components of $\vec{\beta}$, its contribution to $\vec{\omega}_S$ will not generally average to zero. Thus the spin precession frequency has a small dependence on the betatron motion of the beam. It turns out that the only significant correction comes from the vertical betatron oscillation; therefore it is called the pitch correction (see Equation 4.13). As the muons undergo vertical betatron oscillations, the "pitch" angle between the momentum and the horizontal (see Figure 4.10) varies harmonically as $\psi = \psi_0 \cos \omega_y t$, where $\omega_y$ is the vertical betatron frequency $\omega_y = 2\pi f_y$, given in Equation 4.4. In the approximation that all muons are at the magic $\gamma$, we set $a_\mu - 1/(\gamma^2 - 1) = 0$ in Equation 4.13 and obtain

$$\vec{\omega}_a \simeq \vec{\omega}_{diff} = -\frac{Qe}{m}\left[a_\mu \vec{B} - a_\mu \left(\frac{\gamma}{\gamma+1}\right)(\vec{\beta} \cdot \vec{B})\vec{\beta}\right]. \tag{4.21}$$

We adopt the (rotating) coordinate system shown in Figure 4.10, where $\vec{\beta}$ lies in the $yz$-plane, $z$ being the direction of propagation, and $y$ being vertical in the storage ring. Assuming $\vec{B} = \hat{y}B_y$, $\vec{\beta} = \hat{z}\beta_z + \hat{y}\beta_y = \hat{z}\beta\cos\psi + \hat{y}\beta\sin\psi$, we find

$$\vec{\omega}'_a = -\frac{Qe}{m}[a_\mu \hat{y} B_y - a_\mu \left(\frac{\gamma}{\gamma+1}\right)\beta_y B_y(\hat{z}\beta_z + \hat{y}\beta_y)]. \tag{4.22}$$

The small-angle approximation $\cos\psi \simeq 1$ and $\sin\psi \simeq \psi$ gives the component equations

$$\omega'_{ay} = \omega_a \left[1 - \left(\frac{\gamma-1}{\gamma}\right)\psi^2\right] \tag{4.23}$$

and

$$\omega'_{az} = -\omega_a \left(\frac{\gamma-1}{\gamma}\right)\psi. \tag{4.24}$$

It is seen that the direction of $\vec{\omega}'_a$ in Figure 4.10 oscillates at the pitch frequency. We are interested in the overall precession rate about the y-axis, which can be obtained in terms of the period between the times that $\psi = 0$, or the average rate of precession during the pitch period. To facilitate obtaining this average, we project $\vec{\omega}'_a$ onto axes parallel and perpendicular to $\vec{\beta}$, using a standard rotation. Using the small-angle expansions $\cos\psi \simeq 1 - \psi^2/2$, and $\sin\psi \simeq \psi$, we find the transverse component of $\omega'_a$ is given by

$$\omega_\perp = \omega'_{ay}\cos\psi - \omega'_{az}\sin\psi \simeq \omega_a\left[1 - \frac{\psi^2}{2}\right]. \tag{4.25}$$

As can be seen from Table 4.1, the pitching frequency $\omega_y$ is more than an order of magnitude larger than the frequency $\omega_a$, so that $\omega_\parallel$ changes sign rapidly, thus averaging out its effect on $\omega'_a$. Therefore $\omega'_a \simeq \omega_\perp$,

$$\omega'_a \simeq -\frac{Qe}{m} a_\mu B_y \left(1 - \frac{\psi^2}{2}\right) = -\frac{q}{m} a_\mu B_y \left(1 - \frac{\psi_0^2 cos^2 \omega_y t}{2}\right). \tag{4.26}$$

Taking the time average yields a pitch correction

$$C_p = -\frac{\langle \psi^2 \rangle}{2} = -\frac{\langle \psi_0^2 \rangle}{4} = -\frac{n}{4}\frac{\langle y^2 \rangle}{R_0^2}, \tag{4.27}$$

where we have used Equation 4.6 $\langle \psi_0^2 \rangle = n\langle y^2 \rangle / R_0^2$. The quantity $\langle y_0^2 \rangle$ was both determined experimentally and from simulations. For the 2001 period, $C_p = 0.27 \pm 0.036$ ppm, the amount the precession frequency is lowered from that given in Equation 4.5 because $\vec{\beta} \cdot \vec{B} \neq 0$.

We see that both the radial electric field and the vertical pitching motion *lower* the observed frequency from the simple difference frequency $\omega_a = (e/m)a_\mu B$, which enters into our determination of $a_\mu$ using Equation 3.23. Therefore our observed frequency must be *increased* by these corrections to obtain the measured value of the anomaly. Note that if $\omega_y \simeq \omega_a$ the situation is more complicated, with a resonance behavior that is discussed in References [6, 7].

## 4.5 Systematic Errors from the Pion and Muon Beamlines

Systematic effects on the measurement of $\omega_a$ occur when the muon beam injected and stored in the ring has a correlation between the muon's spin direction and its momentum. For a straight beamline, by symmetry, the averaged muon spin is in the forward direction for all momenta muons. However, muons born from pion decay in a bending section of the beamline will have a spin-momentum correlation, especially when the bend is used to make a momentum selection. This is illustrated in Fig. 4.11. For E821 we had a 32 degree bend with D1/D2 to select the pion momentum, and a 21 degree bend with D5 to select the muon momentum. 57% of the pions were still left at the latter bend. A plot of the simulated muon radial spin angle vs. momentum for the E821 beamline is shown in Fig. 4.12. The FNAL experiment beamline bends are given in Table 4.2.

Table 4.2: FNAL beamline horizontal bends.

| **Bend** | **Pions left** | *dp/p* | **Purpose** |
|---|---|---|---|
| 3 degree | 96% | ±10% | Pion momentum selection |
| 19 degree | 41% | ±2% | M2 to M3 |
| Delivery Ring (DR) | 18% | ±2% | Remaining pions decay |
| After DR | $< 10^{-3}$ | ±1% | Muon momentum selection |

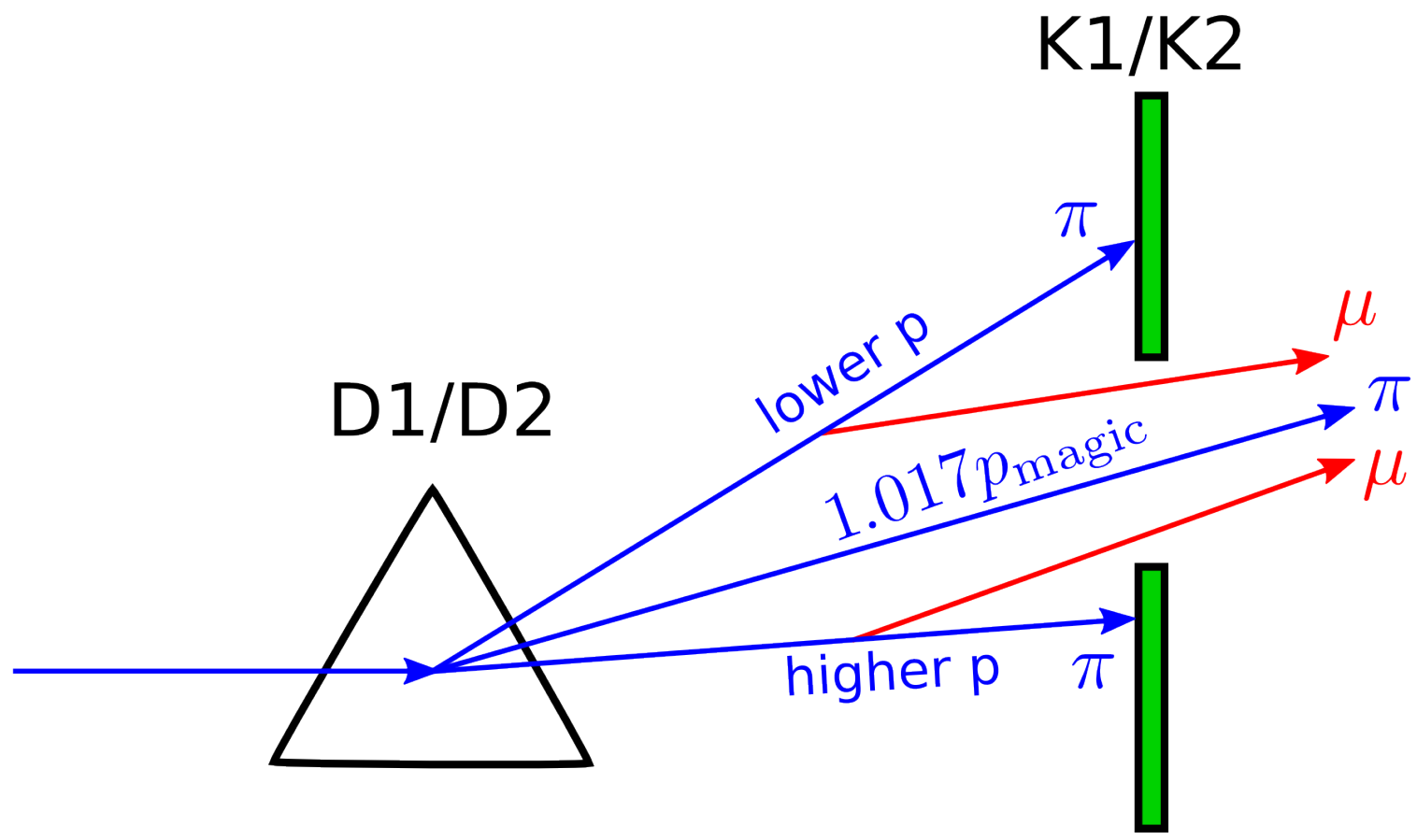

Figure 4.11: Cartoon of the E821 pion/muon beam going through D1/D2. The pions (blue arrows) with momentum (1.017±0.010) times the magic momentum pass through the K1/K2 collimator (green rectangles) slits. Some pions decay after the D1/D2 bend and the decay muons (red arrows) pass through the collimator slit. These muons may have approximately magic momentum, and finally are stored in the muon storage ring. The muon spin direction will then be correlated with it's momentum.

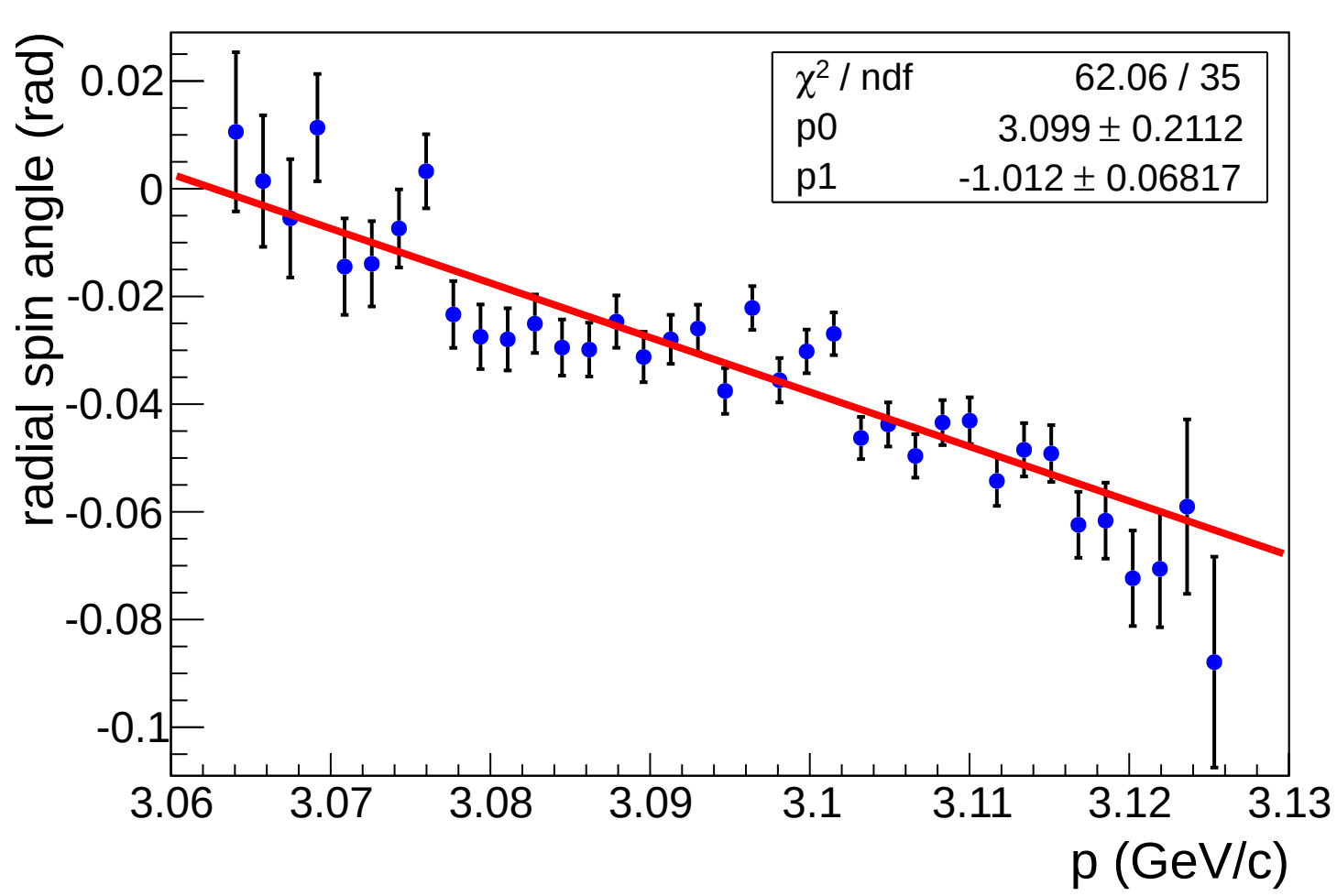

Figure 4.12: Simulation from Hugh Browns BETRAF program of the spin-momentum correlation of muons entering the E821 storage ring, i.e., at the end of the inflector magnet (symbols). The red line is linear fit to data points.

The systematic effect is calculated from:

$$\left\langle \frac{d\Theta_{\text{spin}}}{dt} \right\rangle = \left\langle \frac{d\Theta_{\text{spin}}}{dp} \frac{dp}{dt} \right\rangle \tag{4.28}$$

where $dp/dt$ occurs because the muon lifetime in the lab frame is gamma times the rest frame lifetime. This gave an E821 beamline "differential decay" systematic effect on the measurement of $\omega_a$ of 0.05ppm, which was sufficiently small for E821 that we didn't need to correct for it.

The design philosophy for the FNAL beamline is significantly different from that of E821. For E821 we had a beamline whose length was about the pion $\beta\gamma c\tau$, so to minimize the pion "flash" we selected $(1.017 \pm 0.010)$ times the magic momentum pions after the target and then selected $(1.0 \pm 0.005)$ times the magic momentum just before the muon storage ring. For the FNAL beamline, effectively all the pions will have decayed before the muon storage ring. The pion momentum selection right after the target is only a 3 degree bend and selects ±10% in momentum. The capture probability $Y_{\mu\pi}$ for the long straight section of the beamline is shown in Fig. 4.13. With ±10% momentum acceptance, the pions which are headed for the low momentum side of the beamline acceptance (see Fig. 4.11) can not give a magic momentum muon. The pions which are headed for the high momentum side of the beamline acceptance will be very inefficient in giving a magic momentum muon. Note that this is suggested by Fig. 4.13, but we haven't yet done the FNAL beamline simulation in the bending regions. For later bends, a larger fraction of the pions will have decayed prior to the bend compared to E821 (see Table 4.2). We believe this bending section of the beamline systematic error will be less or equal the E821 error, but we haven't properly simulated it yet. The time line for the simulation calculation is given in the next section.

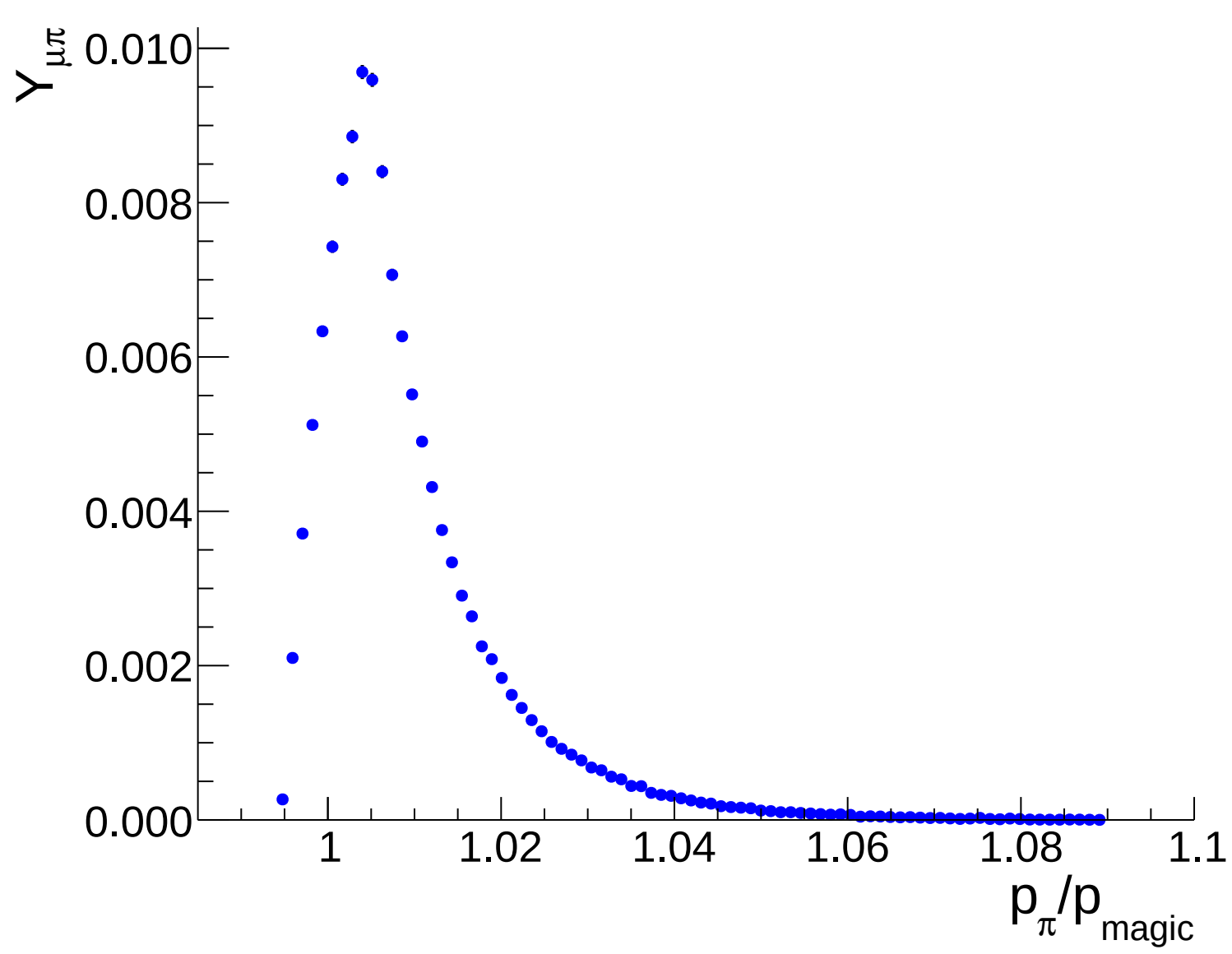

Figure 4.13: Parametric phase space calculation of the $\pi$-$\mu$ capture probability in the straight section of the FNAL pion decay channel. The muons have the magic momentum ±0.5%.

Another systematic effect comes when the muons go around the delivery ring (DR). The cyclotron and anomalous magnetic moment frequencies are:

$$\omega_c = \frac{eB}{m\gamma} \qquad \omega_a \approx \frac{eaB}{m} \tag{4.29}$$

The former is exact while the latter is good to the sub-ppm level. The "spin tune" is then:

$$Q_{\text{spin}} = \frac{\omega_a}{\omega_c} \approx a\gamma \tag{4.30}$$

The spin-momentum correlation after seven turns in the DR, is shown in Fig. 4.14. The slope is less than the slope shown in Fig. 4.12. Of course, Fig. 4.14 is exact, but the energies of the muons in the storage ring are different from their energies in the DR due to the material the beam passes through between the DR and the storage ring. Once the simulation is complete, we will correct our measured value of $\omega_a$ for the beamline differential decay effect.

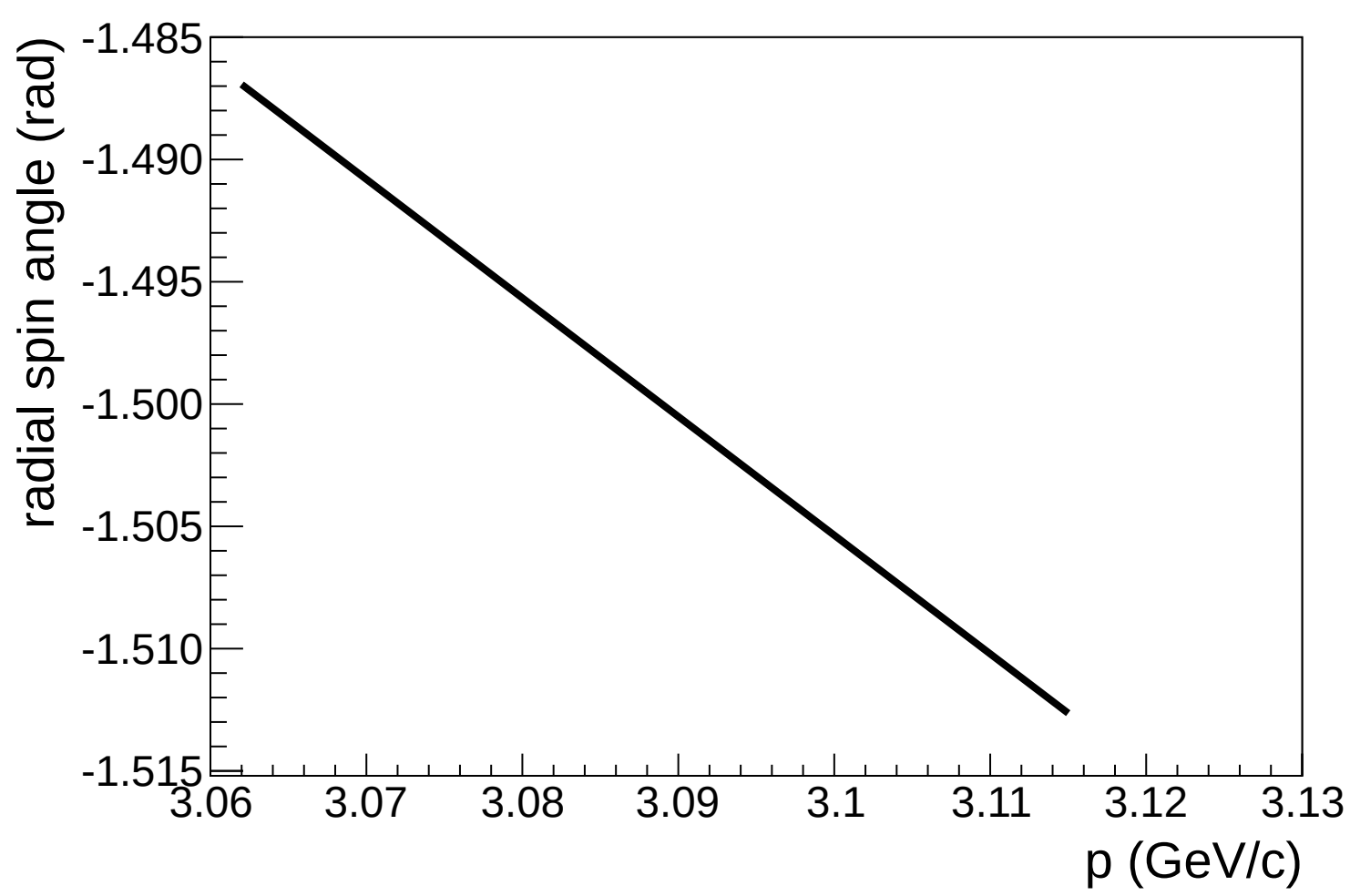

Figure 4.14: Radial spin angle vs. momentum after seven turns in the DR.

Such correlations also couple to the lost muon systematic error. For E821, the differential lost muon rate was about $10^{-3}$ per lifetime, while the differential decay rate was $1.2 \times 10^{-3}$ per lifetime. As discussed above, the FNAL differential lost muon rate will be less than $10^{-4}$ per lifetime.

### 4.5.1 Simulation plan and time line

We are planning to study the beamline systematic errors independently in two ways, using phase-space calculations and tracking. The phase-space calculations were first used by W.M. Morse for E821 [10]. In E989 the phase-space calculation were used to guide the design of the beamline [11] and to estimate the muon capture probability in the straight section for this document. While the phase-space method is approximation, it gives quick insight into the problem and allows to make studies of an idealized beamline with required characteristics without having the actual design of the beamline.

For tracking calculations several off-the-shelf accelerator packages have been considered, `TRANSPORT`, `TURTLE`, `DECAY TURTLE`, `MAD`, `TURTLE with MAD input`. Suitable tracking program for $(g-2)$ must be capable of *i)* describing decay of primary particles (pions) into secondary particles (muons) and transporting the secondary particles and *ii)* transporting spin through the beamline. It turned out that none of the existing programs can be used "as is" for the studies of systematic errors in $(g-2)$. Some modification are needed of any of the existing programs. Lack of the source code in some cases (`DECAY TURTLE`) makes implementation of the missing features impossible. Our current plan for tracking simulations is to use the program `G4Beamline` for the following reason *i)* the program is well-supported and is under active development, *ii)* it is based on `Geant4` toolkit which is widely used in physics simulations, *iii)* spin tracking has been recently implemented in `Geant4`, *iv)* the accelerator team is planning to use `G4Beamline` for beamline simulations, therefore the input configuration file for the $(g-2)$ beamline will be provided by the experts, *v)* the common ground between `G4Beamline` and the downstream simulation program `g2RingSim` for the $(g-2)$ storage ring will simplify the task of combining the two programs together for back-to-back simulations.

Recently, a preliminary version of the `G4Beamline` for $(g-2)$ was released with significant boost in performance and bug fixes. The construction of the $(g-2)$ beamline model for `G4Beamline` is in progress. Basing on our experience, we expect to get the results from `G4Beamline` simulations in six months.

`G4Beamline` simulations for the straight section will be confronted with the phase space simulation to cross-check the two codes. In parallel, we are planning to extend the phase space method to the bending sections of the beamline (beamline elements with dispersion).

Finally, the production and collection of pions in the target station was simulated by `MARS` (see section 7.4.1). We are planning to confront `MARS` and `G4Beamline` simulations of the target station to cross-check the two codes.

### 4.5.2 Coherent Betatron Oscillation Systematic Error Simulations

The theory of coherent betatron oscillations (CBO) is given in the Beam Dynamics Section. Briefly, the E821 inflector was not well matched to the storage ring [12]. Furthermore, the E821 kicker did not provide the optimal kick. Large coherent betatron oscillations were observed in E821. These affect both the spin motion of the muon and the decay positron acceptance. Fig. 4.15 shows the spin precession from just the muon $g-2$, and the additional spin precession due to a fully coherent betatraon oscillation. The latter amplitude is $10^{-4}$ times the former. The CBO amplitude within the muon $dN/dt$ plot for each detector station is shown in Fig. 4.16. The E821 kicker had thicker plates than the E821 electric quads. The detectors shadowed by the kicker plates (detectors 7-9) had about twice the CBO amplitude compared to the detectors shadowed by the quad plates (4-6, 10-12, 16-18, and 22-24). The dominant CBO modulation effect seen in the muon $dN/dt$ plot is due to the decay positron acceptance. The E989 inflector and kicker teams are studying upgrades to the E821 design. The simulated CBO mean and width from a E989 kicker study with the E821 inflector is shown in Fig. 4.17. The E989 mean CBO modulation in this figure is about three times less than observed in E821. The E989 width modulation is about the same as observed in E821.

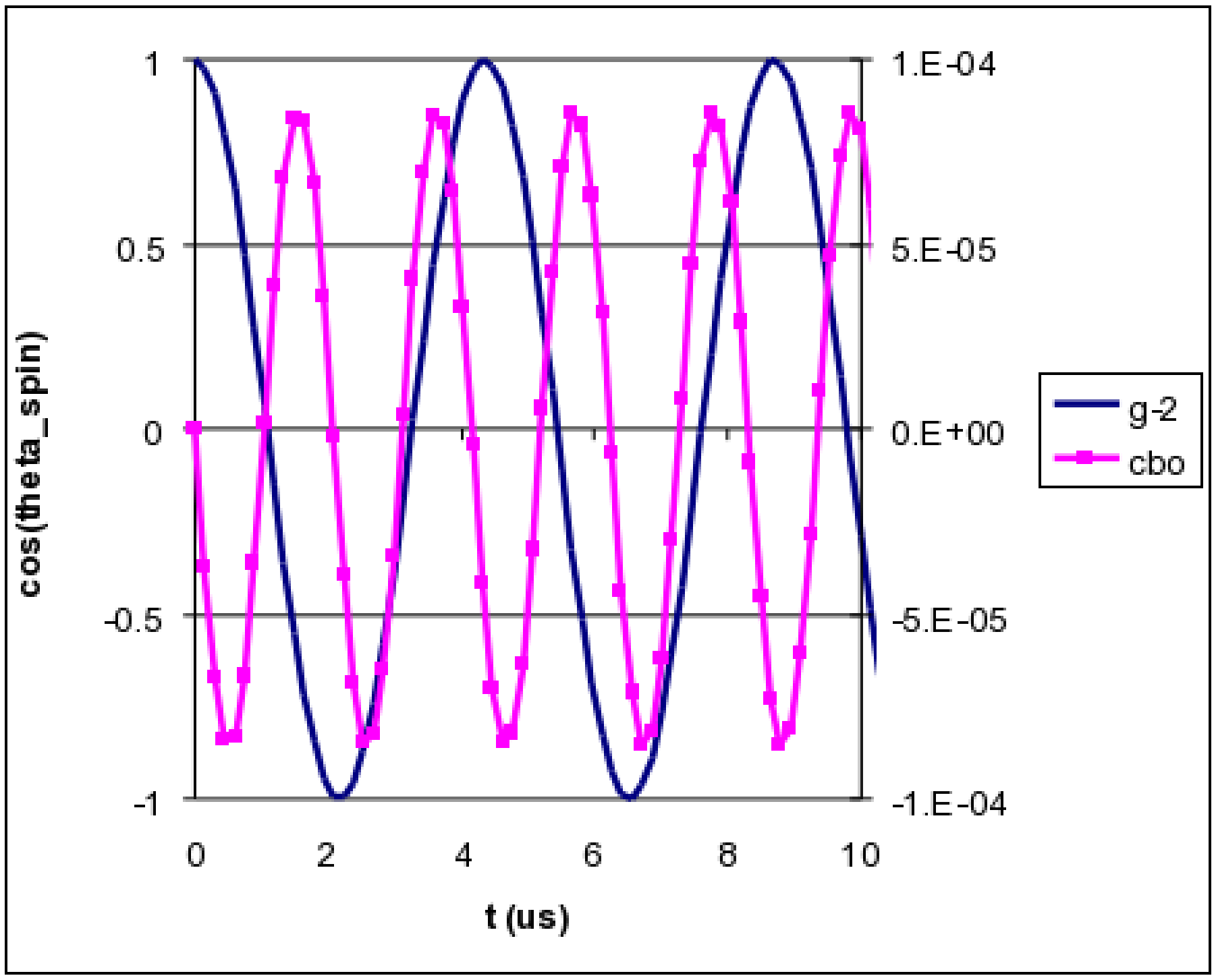

Figure 4.15: Spin precession from just the muon $g-2$ (lhs), and the additional spin precession due to a fully coherent betatraon oscillation (rhs).

The muon $dN/dt$ multi-parameter function [12] is shown below. Fig. 4.18 show the effect of the CBO on the fitted muon $g-2$ frequency *vs.* the CBO frequency when equ. 4.31 was used to generate the data with the E821 parameters, but the fit was done with only the $V-A$ theory five parameters: $N_0$, $\tau_\mu$, $A$, $\omega_a$, and $\phi$. The E821 and E989 CBO frequencies are indicated. E989 will be 30% less sensitive to a given CBO modulation compared to E821 (2001 data). Since one never really understands systematic errors, our goal is to make all the beam dynamics systematic errors as low as reasonably achievable, i.e., zero, if possible.

The E821 muon $dN/dt$ analyzers found that the CBO de-coherence when fit to an exponential gave $\tau_{\rm CBO} \approx 0.1 - 0.14$ ms [12] for different running periods. Fig. 4.19 shows the fit to one of the running periods from 2001.

$$N(t) = \frac{N_0}{\gamma\tau_\mu} e^{-t/\gamma\tau_\mu} \cdot \Lambda(t) \cdot V(t) \cdot B(t) \cdot C(t) \cdot [1 - A(t)\cos(\omega_a t + \phi(t))] \qquad (4.31)$$

$$\Lambda(t) = 1 - A_{\rm loss} \int_0^t L(t) e^{-t/\gamma\tau_\mu} dt$$

$$V(t) = 1 - e^{-t/\tau_{\rm VW}} A_{\rm VW} \cos(\omega_{\rm VW} t + \phi_{\rm VW})$$

$$B(t) = 1 - A_{\rm br} e^{-t/\tau_{\rm br}}$$

$$C(t) = 1 - e^{-t/\tau_{\rm CBO}} A_1 \cos(\omega_{\rm CBO} t + \phi_1)$$

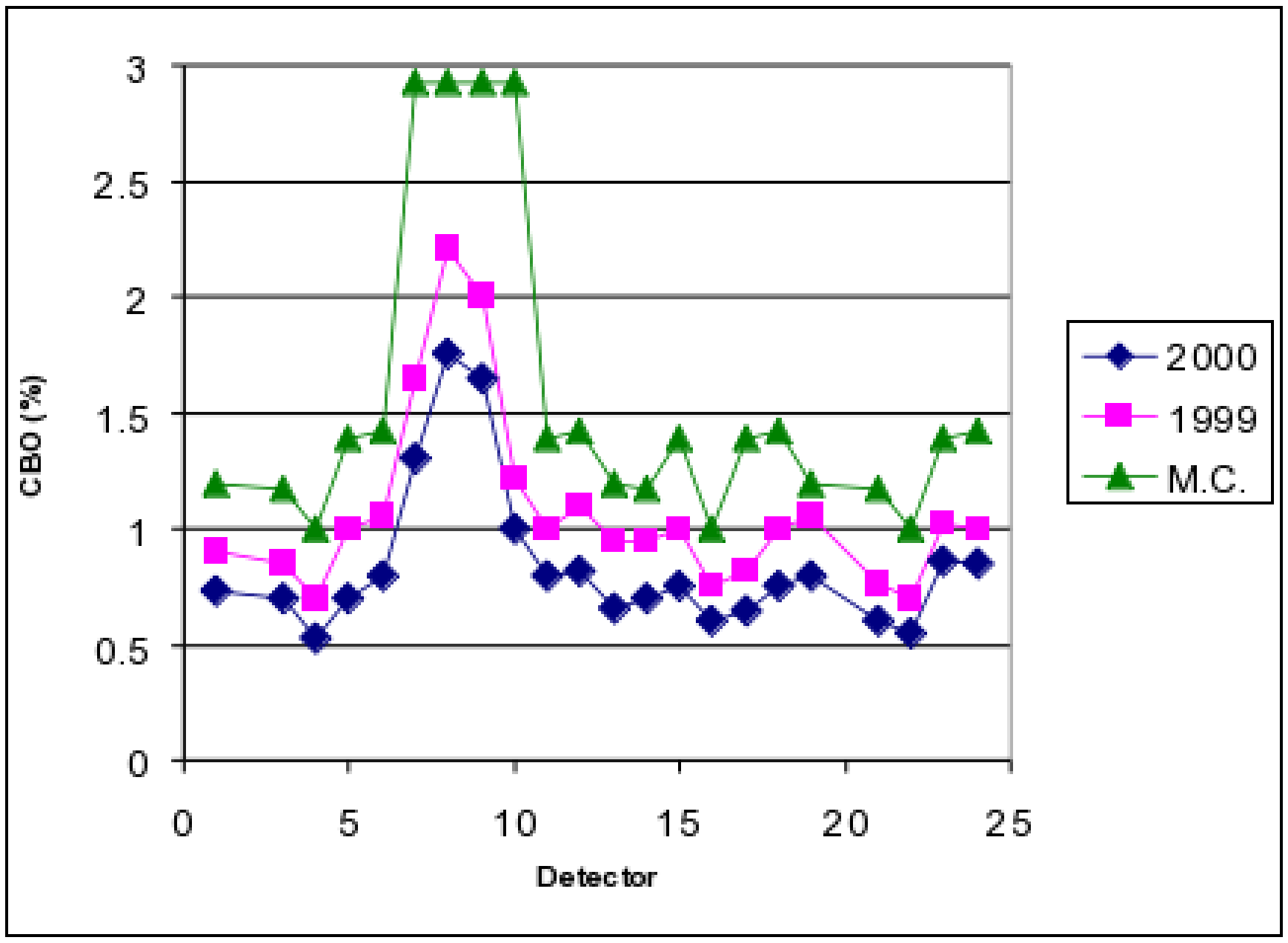

Figure 4.16: E821 observed and positron acceptance simulated CBO $A_1$ amplitudes vs. detector number. Detectors 7-9 were shadowed by the E821 kicker plates. The inflector angle was changed between the 1999 and 2000 runs.

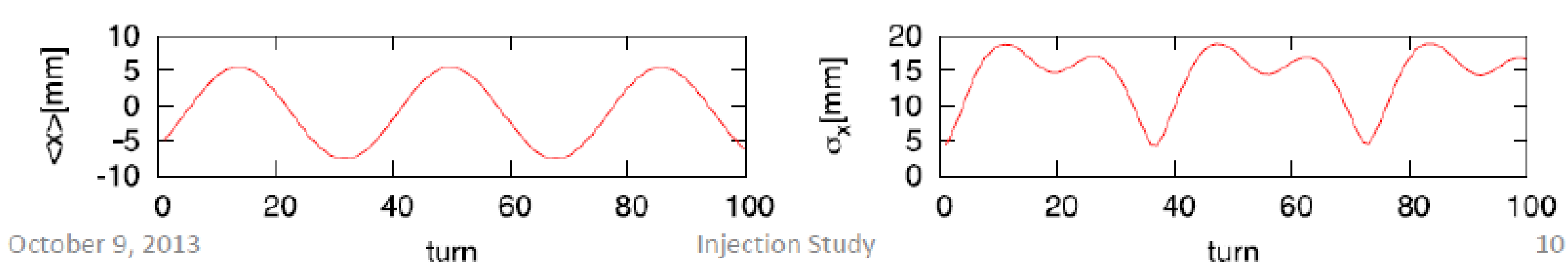

Figure 4.17: E989 kicker simulation showing the CBO modulation of the mean, and the width of the muon distribution vs. turn number [13]. The modulation of the mean is three times less than E821. The width modulation is about the same as E821.

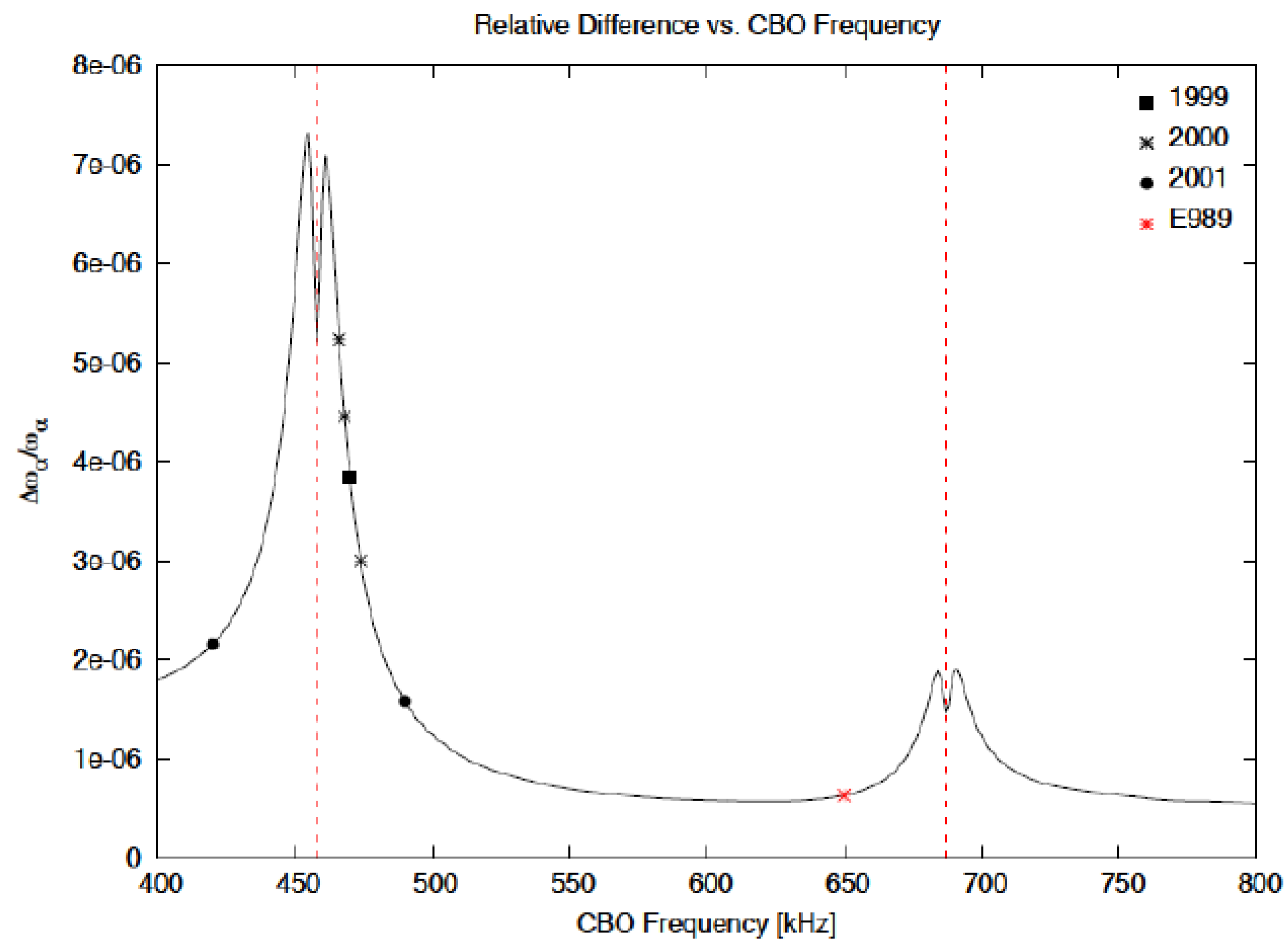

Figure 4.18: Effect of the CBO on the fitted muon $g-2$ frequency vs. CBO frequency (see text for discussion). The E821 CBO frequencies and the planned E989 frequency are indicated.

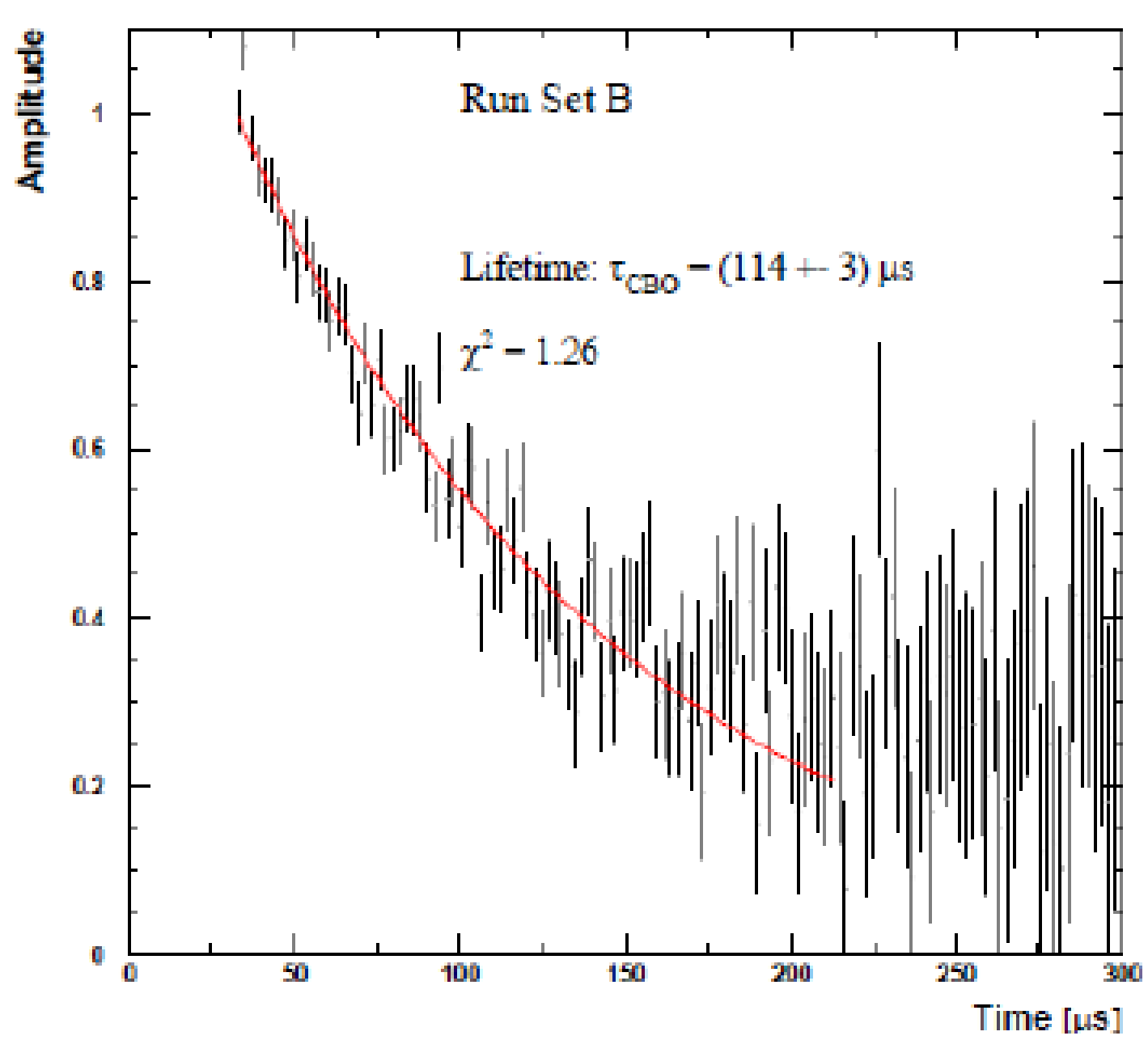

Figure 4.19: Time distribution of residuals from the 5-parameter fit at the CBO frequency for one set of E821 2001 run.

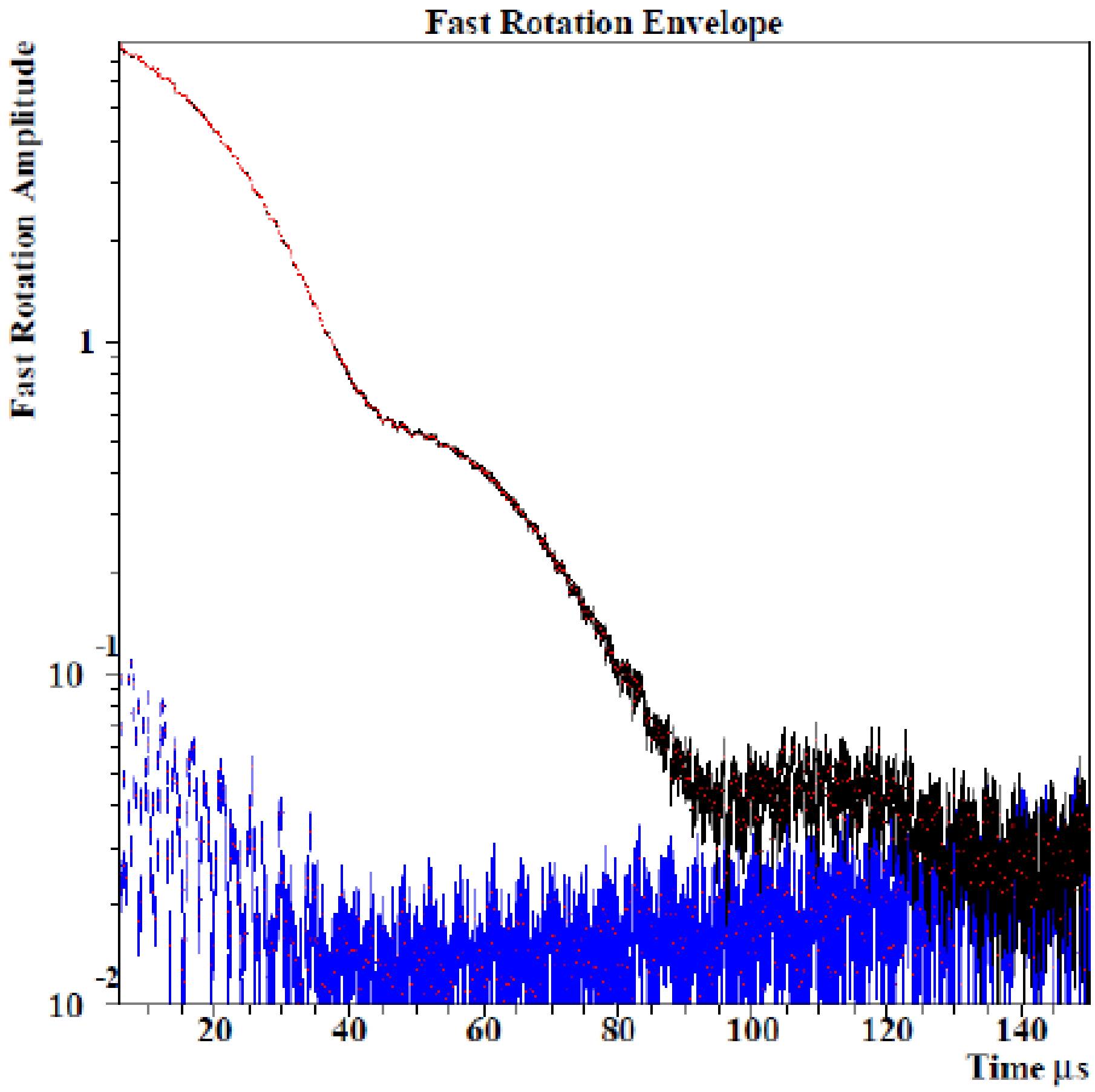

Figure 4.20: E821 de-coherence of the fast rotation envelope (red points with black statistical errors) from the 2000 run. Blue is after binning at the revolution period, before accidental overlaps were corrected.

$$A(t) = A\left(1 - e^{-t/\tau_{\rm CBO}} A_2 \cos(\omega_{\rm CBO} t + \phi_2)\right)$$
$$\phi(t) = \phi_0 + e^{-t/\tau_{\rm CBO}} A_3 \cos(\omega_{\rm CBO} t + \phi_3) \quad (4.32)$$

Next we discuss the calculation of CBO de-coherence, due to the muons having different beam dynamics frequencies.

$$f_{\rm CBO} = f_{\rm rev}\,(1 - Q_x) \quad (4.33)$$
$$\frac{df_{\rm CBO}}{f_{\rm CBO}} = \frac{df_{\rm rev}}{f_{\rm rev}} \oplus \frac{dQ_x}{1 - Q_x} \quad (4.34)$$

The E821 de-coherence of the revolution frequency is shown in Fig. 4.20. $df_{\rm rev}/f_{\rm rev} \approx 1.5{\times}10^{-3}$. From the muon $dN/dt$ plot fits, $df_{\rm CBO}/f_{\rm CBO} \approx 8{\times}10^{-3}$. This gives $dQ_x \approx 5{\times}10^{-4}$. This is the main source of the de-coherence of the coherent betatron oscillations.

Ref. [15] simulated the E821 2000 run CBO de-coherence due to the tune spread from the electric quadrpoles [14]. The simulated mean and width at the CBO frequency is shown in Fig. 4.21. The mean had only the CBO frequency, but the width had both the CBO frequency, and twice the CBO frequency [12]. The muon $dN/dt$ plot had both the CBO frequency, and twice the CBO frequency. Ref. [15] then concluded that: "The beam width contributes 20-30% to the observed CBO signal for detectors 10-24 and starting at 25 $\mu$s". However, this logic is incorrect, as the CBO can have only the first harmonic, for example, but if the detector acceptance is non-linear vs. the betatron $x$-$x'$ oscillations, the other harmonics will appear in the muon $dN/dt$ plot. Nevertheless, Fig. 4.22 shows 70% mean and 30% width modulation (points labeled sum), and an exponential with $\tau = 114$ $\mu$s, which matches the Fig. 4.19 time distribution quite well. The fraction of the CBO modulation in the muon $dN/dt$ plot due to the mean, and the fraction due to the width will be determined in the simulation study. The CBO simulation will determine the E989 $C(t)$, $A(t)$, and $\phi(t)$ parameters for the kicker, quad, and free detectors, and the CBO systematic error for the $Q$ and $T$ methods of analysis.

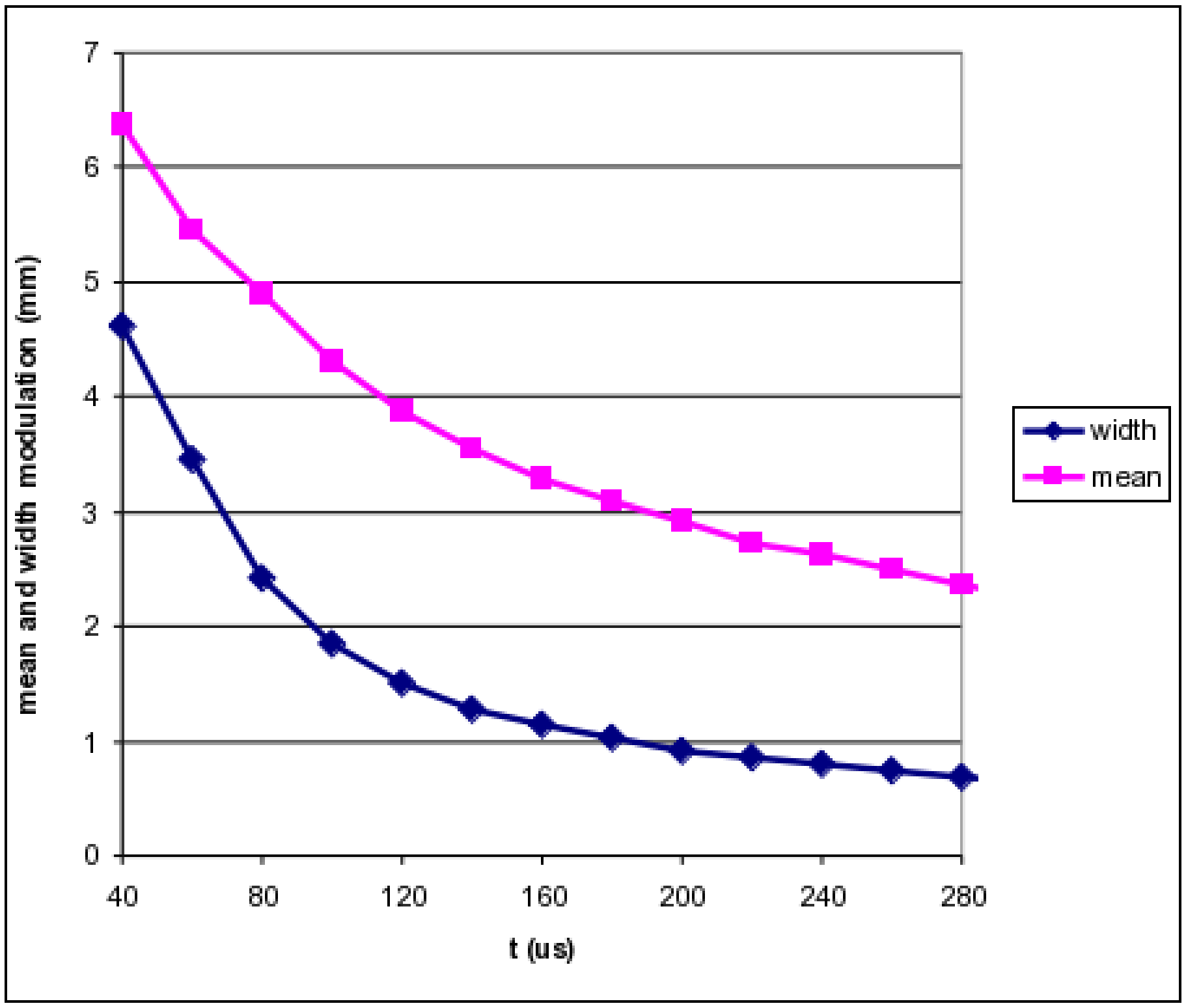

Figure 4.21: Ref. [15] horizontal mean and width modulation at the CBO frequency.

### 4.5.3 Lost Muons

A systematic error occurs if the muons lost from the storage ring at late times have a different average spin direction compared to the stored muons. This difference in the spin direction occurs due to the production and storage processes. The E821 storage ring injection capture efficiency was $(4 \pm 1)\%$ [12]. Thus about 96% of the injected muons were lost. The E821 muon loss rate after 30 $\mu$s was lost/stored $\approx 10^{-3}$, or lost/injected $\approx 4 \times 10^{-5}$. Our goal is to reduce the lost muon rate after 30 $\mu$s by at least an order of magnitude compared to the E821 rate.

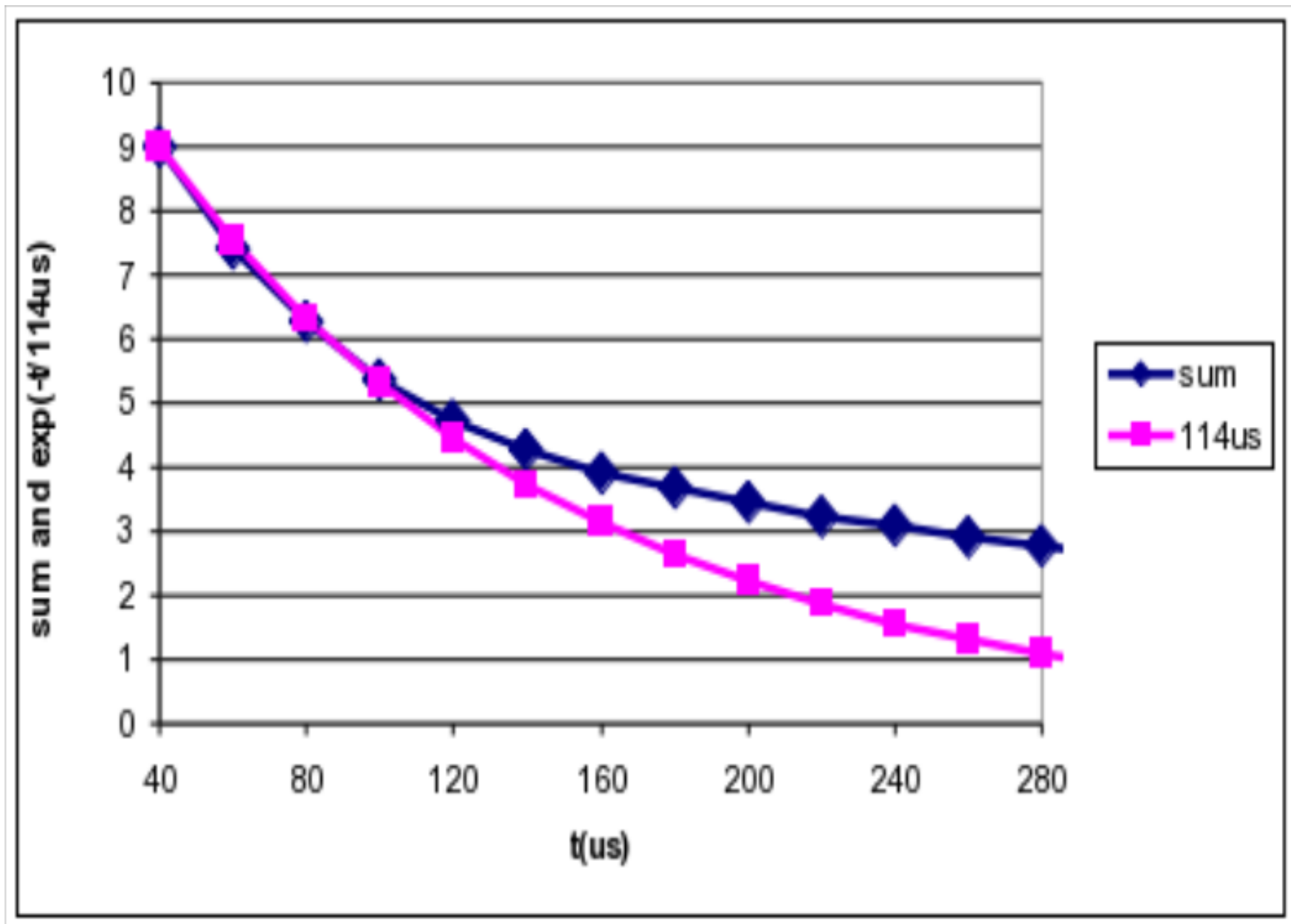

Figure 4.22: 70% mean and 30% width modulation (sum), and exponential with $\tau = 114$ $\mu$s.

A schematic drawing of the E821 muon storage ring vacuum chamber is shown in Fig. 4.23. Some of the E821 full collimators were changed to 1/2-collimators, since the E821 kicker did not give an adequate kick for the first turn (see Kicker Section). For E989, all the collimators will be full collimators. The distortion of the closed orbit due to non-perfect magnetic fields, for uniform and perfect electric quads, is:

$$\Delta X_e(\Theta) \approx \frac{R_0}{B_0} \sum_{N=1}^{\infty} \frac{B_{yNC}\cos(N\Theta) + B_{yNS}\sin(N\Theta)}{-N^2 + Q_x^2} \quad (4.35)$$

$$\Delta Y_e(\Theta) \approx \frac{R_0}{B_0} \sum_{N=0}^{\infty} \frac{B_{RNC}\cos(N\Theta) + B_{RNS}\sin(N\Theta)}{-N^2 + Q_y^2} \quad (4.36)$$

$$(4.37)$$

Fig. 4.24 shows the lost muon results of the E989 phase space model by a BNL high school summer student. One can readily see the effect a non-uniform magnetic field has on the muon losses. This study was limited by statistics, but the zero values for lost muons have at least a factor of ten fewer lost muons than E821 after 30 $\mu$s. However, this study assumed infinitely thick collimators, i.e., the muon was lost as soon as it hit a collimator. We next need a tracking study, including finite electric quads with non-perfect fields, following the muon after it first strikes the collimator, etc.

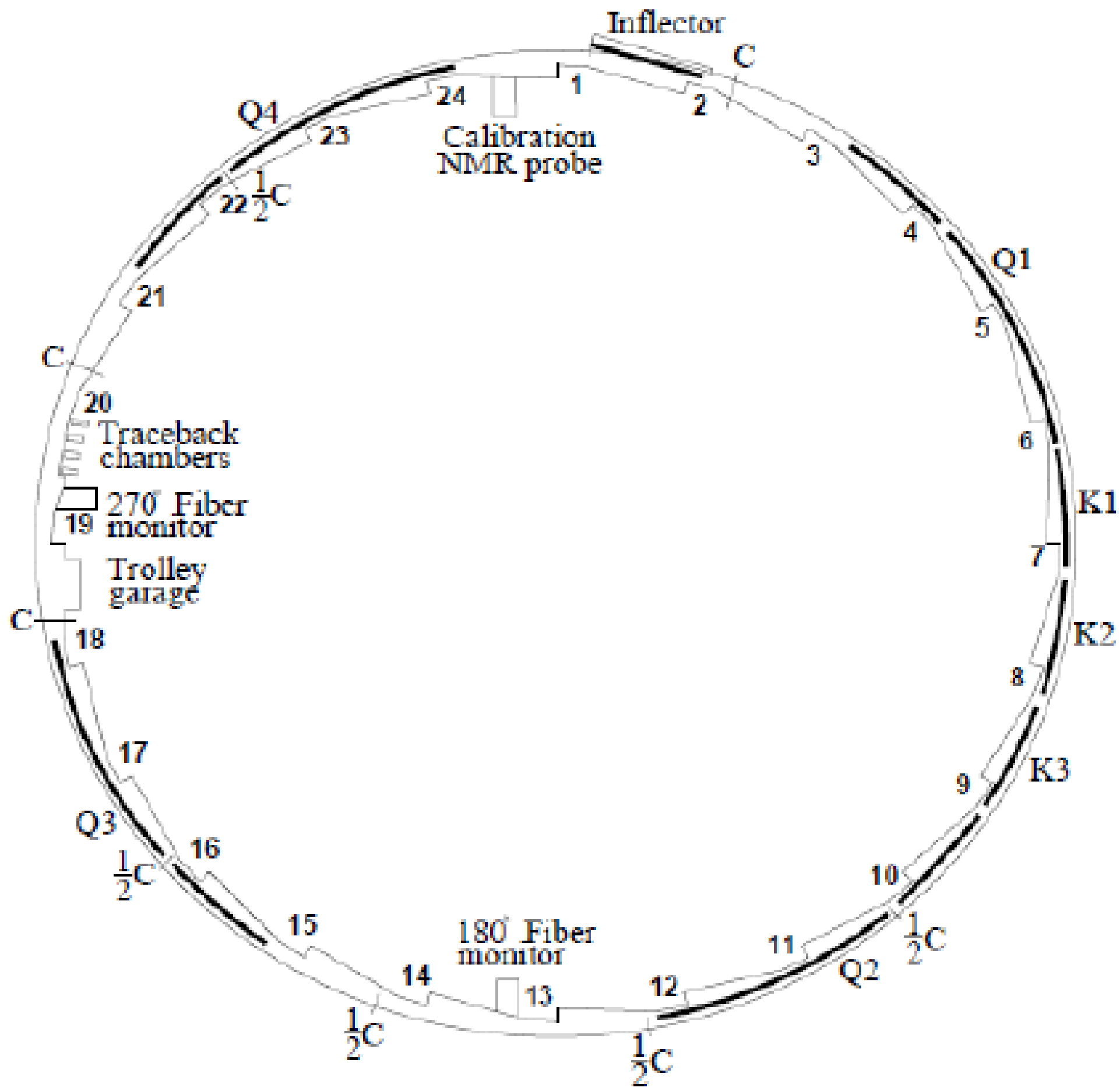

Figure 4.23: E821 vacuum chambers showing the locations of the electric quads and collimators.

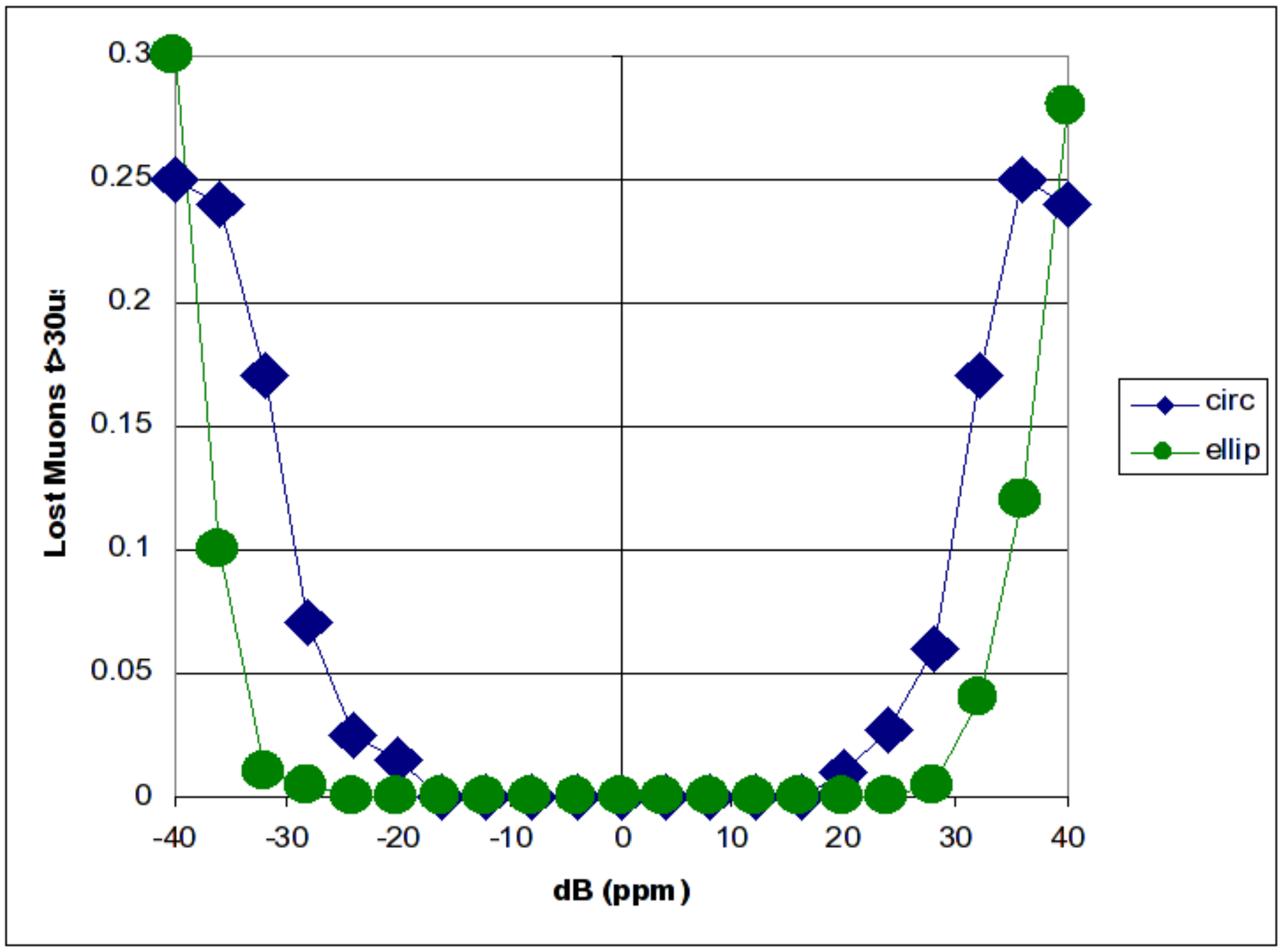

Figure 4.24: Lost muons after 30s from a phase space study vs. magnetic field uniformity, for both circular and elliptical collimators. Elliptical collimators follow $\sqrt{\beta_{x,y}}$.

### 4.5.4 Electric Field and Pitch Corrections

The theory of the electric and pitch corrections is given in the Beam Dynamics Section. The E821 electric field and pitch corrections to the anomaly were $(0.47 \pm 0.05)$ ppm and $(0.27 \pm 0.04)$ ppm, respectively. The table below gives more detail on the systematic errors.

| Systematic Effect | $E$ correction | Pitch |
|---|---|---|
| Difference Between Data and Simulation | ±50 ppb | ±30 ppb |
| Beam *vs.* Quad Electrode Position Uncertainty | ±20 ppb | ±20 ppb |

Our E989 goal is $< 30$ ppb for the electric field and pitch corrections combined. The electric field correction requires a precise knowledge of the momentum distribution of the stored muons. This is obtained by the so-called "fast rotation analysis", where the beam is observed to de-bunch as it rotates around the ring. The equilibrium closed orbit is given by:

$$x_{\rm e} = D\frac{dp}{p} \tag{4.38}$$

For E821 the head and the tail of the incoming bunch had identical momentum distributions. This will not be the acse for the E989 beam, since the beam goes around the Delivery Ring (DR) a number of times. The DR has $\langle D \rangle = 2$ m. Some pions decay to muons in the DR, so we have to track the pion momentum, which is higher than the muon momentum, and then the muon momentum. For muons which have five turns around the DR, for example:

$$\delta L \approx 10\pi(2\text{m})\frac{dp}{p} \approx 63\text{m}\frac{dp}{p} \tag{4.39}$$

$$\delta t \approx \frac{63\text{m}}{c}\frac{dp}{p} \approx 210\text{ns}\frac{dp}{p} \tag{4.40}$$

Putting in $dp/p$ of several parts per thousand shows that this will be a small effect, but it will be studied in the simulation.

### 4.5.5 Collimator Study

The E821 collimators were IR = 45 mm, OR = 55 mm, and thickness 3 mm Cu [12]. The E821 collimator design was based on a “back of the envelope” calculation. For E989 we need a real simulation to minimize the lost muon systematic error, maximize the positron detection, and allow adequate space for supplementary detectors.

### 4.5.6 Simulation Responsibilities and Schedule

Each calculation listed needs to be done independently by at least two different people, or by one person but with a different method, i.e., analytical calculation, phase space simulation, tracking simulation, etc. BDT = Beam Dynamics Team (BNL, FNAL, Univ. Mississippi, and CAST, Korea). The dates shown are estimates of when the simulation studies will be completed by calendar year and quarter.

**Differential decay Q2 2015**

1. Kicker – BDT
2. Muon spin in DR – BDT
3. Pion decays in bends – BDT
4. Straw system – Detector Team

**CBO Q3 2015**

1. $F(t)$ for the de-coherence – BDT
2. Kicker plate study – BDT
3. Effect on E989 $\omega_a$ for $Q$ method – BDT
4. Effect on E989 $\omega_a$ for $T$ method – BDT
5. Straw system/fiber beam monitor system – Detector Team
6. Hardware CBO damping – BDT

**Lost muons Q1 2015**

1. Phase space – BDT

2. Tracking – BDT
3. New "scraping" hardware? – BDT

**Collimators (analytical, phase space, tracking) Q1 2015**

1. Number and Thickness – lost muon study – BDT
2. Number and Thickness – decay positron study – Detector Team

**Pitch Correction Q2 2015**

1. Straw system/fiber beam monitor system – Detector Team
2. Beam Dynamics – BDT

**E field Correction Q2 2015**

1. Fast rotation – BDT
2. Beam Dynamics – BDT

**Distortion of closed orbit due to non-perfect electric quad fields Q2 2014- Finished [17]**

1. Analytical calculation – BDT
2. Tracking – BDT

**Distortion of closed orbit due to non-perfect magnetic fields Q2 2014 - Finished [18]**

1. Analytical calculation of effect on average magnetic field – BDT/Magnetic Field Team
2. Tracking – BDT

**Geometric Phase Q2 2014 - Finished [16]**

1. Analytical calculation – BDT
2. Tracking – BDT

# Chapter 5

# Statistical and Systematic Errors for E989

E989 must obtain twenty-one times the amount of data collected for E821. Using the $T$ method (see Section 16.1.2) to evaluate the uncertainty, $1.5 \times 10^{11}$ events are required in the final fitted histogram to realize a 100 ppb statistical uncertainty. The systematic errors on the anomalous precession frequency $\omega_a$, and on the magnetic field normalized to the proton Larmor frequency $\omega_p$, are each targeted to reach the $\pm 70$ ppb level, representing a threefold and twofold improvement, respectively, compared to E821. E989 will have three main categories of uncertainties:

- **Statistical.** The least-squares or maximum likelihood fits to the histograms describing decay electron events vs. time in the fill will determine $\omega_a$, the anomalous precession frequency. The uncertainty $\delta\omega_a$ from the fits will be purely statistical (assuming a good fit). A discussion of the fitting sensitivity using various weighting schemes is given in Chapter 16, Section 16.2. The final uncertainty depends on the size of the data set used in the fit, which in turn depends on the data accumulation *rate* and the *running time*. These topics are discussed here.

- $\omega_a$ **Systematics.** Additional systematic uncertainties that will affect $\delta\omega_a$ might be anything that can cause the extracted value of $\omega_a$ from the fit to differ from the true value, beyond statistical fluctuations. Categories of concern include the detection system (e.g., gain stability and pileup immunity discussed in Chapter 16), the incoming beamline (lost muons, spin tracking), and the stored beam (coherent betatron oscillations, differential decay, $E$ and pitch correction uncertainties). These latter topics are discussed in Chapter 4.

- $\omega_p$ **Systematics.** The magnetic field is determined from proton NMR in a procedure described in Chapter 15. The uncertainties are related to how well known are the individual steps from absolute calibration to the many stages of relative calibration and time-dependent monitoring. The"statistical" component to these measurements is negligible.

The purpose of this chapter is twofold. First, we summarize the event-rate calculation from initial proton flux to fitted events in the final histograms in order to determine the running time required to meet the statistical goals of the experiment. We also gather the results of many systematic uncertainty discussions that are described in various chapters throughout this document and roll up the expected systematic uncertainty tables for E989.

## 5.1 Event Rate Calculation Methodologies

The E989 Proposal [1] event-rate estimate was made by taking a **relative comparison approach** using like terms with respect to the known situation for rates in the E821 BNL experiment. Many factors allowed for trivial adjustments (proton fills per second, kinematics of the decay line length, kinematics of the decay line capture), while others relied on expected improvements in specific hardware components (optimized storage ring kicker pulse shape and magnitude, open-ended inflector, thinner or displaced Q1 outer plate and standoffs). In E821, the transmission through the closed-ended inflector and subsequently through the Q1 outer plates, followed by an imperfect kick, combined to give a sub-optimal storage ring efficiency factor, but individually the contributions from each element were not known as well as their product.

The E989 Conceptual Design Report [2] used that approach to estimate the need for a run duration of $17 \pm 5$ months, which included 2 months of overall commissioning and 2 months of systematic studies. The CDR also provided a bottom-up estimate, although at the time of the document, key simulations were just beginning. That approach suggested 18 months, perfectly in agreement with the relative calculation. Here, we present our estimate based on full **End-to-End Simulation** of the data accumulation rate. Many technical improvements since the CDR have tended to increase the overall data rate. However, the default use of the existing E821 inflector eliminates an anticipated gain. We have increased considerably from 2 to 6 the number of months that will be required to commission the entire accelerator chain and experiment.

### 5.1.1 Bottom-Up Event Rate Calculation

Table 5.1 contains a sequential list of factors that affect the event rate based on a bottom-up, full simulation approach. We assume the Proton Improvement Plan delivery of 4 batches of $4 \times 10^{12}$ protons to the Recycler per 1.33 s supercycle with the Booster operating at 15 Hz. Each proton batch is split into four proton bunches of intensity $10^{12}$; thus, the experiment will receive 16 proton bunches per supercycle, or a rate of 12 Hz. Each bunch corresponds to a "fill" of the Storage Ring. Four sequential stages of the simulation result in the estimates of positrons recorded by detectors per fill, and thus provide an estimate of the required operation of the experiment to achieve the statistical precision of 100 ppb stated in the Proposal and, importantly, the instantaneous rates on the many detector systems used in the experiment. The major simulation stages are:

1. Pion production on the target
2. Muon capture from pion decay, and subsequent transport to the storage ring entrance
3. Muon transmission into, and subsequent capture in, the storage ring

4. Muon decay positron acceptance by the detectors

The tools used include `MARS` for particle production, `G4beamline` and `BMAD` for beam transport and optimization, and `g2ringsim`, which is a `GEANT-4`-based full description of the storage ring and detector systems built in the `ART` framework. They are described in expert respective Chapters that follow. Here we present a linear narrative that will guide the reading of Table 5.1. The Table is then further justified with a sequence of Notes that pertain to each entry.

The particle flow is as follows. A burst of $10^{12}$ 8-GeV kinetic energy protons is focussed in the final stages of the M1 beamline to a spot size of 0.15 mm as it strikes the pion production target. The time distribution of the protons in the burst has an unusual "W-shaped" intensity profile with a maximum width of approximately 110 ns and a concentrated peak in the center. The target and lithium lens system was used previously for antiproton production. It is repurposed and optimized for the production and capture of 3.1 GeV/$c$ positive particles in a fairly broad momentum bite. These are bent with the pulsed BMAG into the newly optimized M2 FODO lattice, which evolves following a short horizontal bend to the M3 beamline. The length of these sections, which is where the majority of muons are collected and captured, is approximately 270 m, where approximately 80% of the pions have decayed to muons.

The now mostly muon beam enters the Delivery Ring (DR) where it will make a variable number of revolutions (we anticipate 3 - 5) before being extracted by a fast kicker to the M5 beamline which delivers the beam to the $g-2$ Storage Ring entrance. The combination of beamlines so assembled admits at least a $40\pi$ mm-mrad phase space and has a momentum width $\delta p/p \sim 2\%$. The muon distribution retains the time profile described above. The purpose of this nearly 2 km path is to allow essentially all pions to decay to muons and to allow a time separation between muons and protons in the DR such that the protons can be removed by a kicker safely out of time from the passing muon burst. Thus, an essentially pure muon beam arrives at the Storage Ring at the magic muon momentum of 3.094 GeV/$c$. We assume that after a period of up to 6 months of steady commissioning and optimization, one can achieve $> 90\%$ transmission to the ring.

These muons must enter the Storage Ring through a hole in the back leg of the magnet yoke. They next enter a superconducting inflector magnet whose purpose is to null the strong return field flux that passes through the steel; it cancels the 1.4 T storage ring field over a 1.7 m path. This device is non-trivial. It has a small aperture, and includes coils covering both ends that introduce multiple scattering. The residual (non-canceled) fringe field along the path from the outside of the yoke to the exit of the inflector bends the beam left and right, the effect being to further restrict the transmission fraction. The beam emerges into the Storage Ring volume at an angle that is corrected by a $\sim 12$ mrad transverse outward kick during the first quarter turn. The newly designed magnetic kicker field profile in both space and time affects the storage efficiency. To reduce the muon loss rate for "stored" muons, the quadrupole system is used to scrape the beam along fixed collimaters and then return it to center. The transverse stored beam profile is reduced at the cost of $\sim 13\%$ of the muon flux.

Once stored—typically defined as a muon that remains in the storage volume for at least 100 turns—the muon decays can be studied using standard GEANT-based tools. To enhance the statistics in the subsequent simulations, we start by modeling the stored distribution with a polarized "muon gas" where we can then study the decays and detector acceptance and response.

Combined, the above sequence nets a yield of nearly 1100 recorded positrons per fill, each having an energy above the nominal threshold cut of 1.86 GeV, which maximizes the experimental sensitivity figure of merit. The yield is not more than $1.1\times10^{-9}$/pot. Therefore, each stage of the simulation requires optimized tools and an interface to subsequent phases through intermediate files. We note that full spin tracking is included. The following notes

Table 5.1: Event rate calculation using a bottom-up approach.

| **Item** | **Factor** | **Value per fill** | **Note** |
|---|---|---|---|
| Protons on target | | $10^{12}$ p | 1 |
| Positive pions captured in FODO, $\delta p/p = \pm0.5\%$ | $1.2\times10^{-4}$ | $1.2\times10^{8}$ | 2 |
| Muons captured and transmitted to SR, $\delta p/p = \pm2\%$ | 0.67% | $8.1\times10^{5}$ | 3 |
| Transmission efficiency after commissioning | 90% | $7.3\times10^{5}$ | 4 |
| Transmission and capture in SR | $(2.5\pm0.5)\%$ | $1.8\times10^{4}$ | 5 |
| Stored muons after scraping | 87% | $1.6\times10^{4}$ | 6 |
| Stored muons after 30 $\mu$s | 63% | $1.0\times10^{4}$ | 7 |
| Accepted positrons above E = 1.86 GeV | 10.7% | $1.1\times10^{3}$ | 8 |
| Fills to acquire $1.6\times10^{11}$ events (100 ppb) | | $1.5\times10^{8}$ | 9 |
| Days of good data accumulation | 17 h/d | 202 d | 10 |
| Beam-on commissioning days | | 150 d | 11 |
| Dedicated systematic studies days | | 50 d | 12 |
| Approximate running time | | $402\pm80$ d | 13 |
| Approximate total proton on target request | | $(3.0\pm0.6)\times10^{20}$ | 14 |

explain entries in Table 5.1:

1. We assume a 0.15 mm spot size at the final focus of the M1 line on the target and an average proton pulse flux of $10^{12}$ from the Recycler, after a 4-fold split of the injected batch from the Booster.

2. MARS calculation. Assumes the (improved) proton spot size on target of 0.15 mm, which increased the yield compared to the measured rates at 0.5 mm spot size. Assumes $40\pi$-mm-mrad emittance. Measurement verifies yield of positive particles. Simulation shows that 45% are pions. The target yield is assumed to be optimized by adjustments of the geometry compared to that in the CDR. Combined optimizations increased yield by the factor 1.35 compared to the CDR.

3. This is a multistep, full G4beamline simulation including all elements from the beginning of the M2 FODO, the bend to M3, three revolutions of the Delivery Ring, and transport along M5 to the last quad prior to the Storage Ring. Spin tracking gives a muon polarization average of 95%. Pions are assumed to have decayed; protons are kicked away in the DR from their time-of-flight lag.

4. After commissioning period of up to 6 months, estimate a 90% transmission from M2 to the end of M5, including losses in the DR kickers and accumulated misalignments of magnets. This is an expert opinion based on experience with the antiproton complex.

5. BMAD and g2ringsim calculations starting with muons from the output of G4beamline, which are transported through the back leg of the magnet, through the inflector (multiple scattering included), into the ring. They are kicked with a 20 ns rise time, 20 ns fall time, 80 ns flat top magnetic field. Losses occur from the apertures, the fringe fields, the non-ideal kicker pulse width and the natural Storage Ring acceptance. Both simulations suggest a storage fraction of $\sim (2.5 \pm 0.5)\%$.

6. We take the simple geometrical ratio of $(4.2/4.5)^2 = 0.87$ to establish a 2 mm annulus, given a position uncertainty of the quads of 0.5 mm.

7. Factor $exp(-t/\tau_\mu)$ with $t = 30$ $\mu$s and $\tau_\mu = 64.4$ $\mu$s.

8. Monte Carlo acceptance of the 24 calorimeters of 10.7% for events with energy above 1.86 GeV and striking the front face of one of the 24 calorimeter stations. Estimate includes all losses owing the material (quads, kicker plates, vacuum chambers).

9. With $T$ method analysis, resolution of calorimeters folded in, and the polarization of 0.95 from the simulation, the asymmetry is $A = 0.38$ and the number of required events in the fit is $1.6 \times 10^{11}$ for a 100 ppb statistical uncertainty.

10. Assume uptime data collection of 17 hours per day obtained as follows. One 3-h duration trolley run per 2 days loses 1.5 h/d. Accelerator uptime average is estimated at 85% and experiment livetime (including any functional downtime) is 90%.

11. Estimate of time to commission the new experiment and machine operation sequence. This is based, in part, on past experience at BNL and FNAL.

12. Generous estimate of dedicated systematic studies throughout the full measurement period.

13. Net data taking estimate. The range of $\pm 20\%$ is based on uncertainty in the storage fraction. Other factors may increase the uncertainty range.

14. Total proton request for the delivered beam to the experiment.

## 5.2 $\omega_a$ systematic uncertainty summary

Our plan of data taking and hardware changes addresses the largest systematic uncertainties and aims to keep the total combined uncertainty below 70 ppb. Experience shows that many of the "known" systematic uncertainties can be addressed in advance and minimized, while other more subtle uncertainties appear only when the data is being analyzed. Because we have devised a method to take more complete and complementary data sets, we anticipate the availability of more tools to diagnose such mysteries should they arise. Table 5.2 summarizes this section.

Table 5.2: The largest systematic uncertainties for the final E821 $\omega_a$ analysis and proposed upgrade actions and projected future uncertainties for data analyzed using the $T$ method. The relevant Chapters and Sections are given where specific topics are discussed in detail.

| Category | E821 [ppb] | E989 Improvement Plans | Goal [ppb] | Chapter & Section |
|---|---|---|---|---|
| Gain changes | 120 | Better laser calibration | | |
| | | low-energy threshold | 20 | 16.3.1 |
| Pileup | 80 | Low-energy samples recorded | | |
| | | calorimeter segmentation | 40 | 16.3.2 |
| Lost muons | 90 | Better collimation in ring | 20 | 13.10 |
| CBO | 70 | Higher $n$ value (frequency) | | |
| | | Better match of beamline to ring | $< 30$ | 13.9 |
| $E$ and pitch | 50 | Improved tracker | | |
| | | Precise storage ring simulations | 30 | 4.4 |
| Total | 180 | Quadrature sum | 70 | |

## 5.3 $\omega_p$ systematic uncertainty summary

The magnetic field is mapped by use of NMR probes. A detailed discussion is found in Chapter 15. In Table 5.3 we provide a compact summary of the expected systematic uncertainties in E989 in comparison with the final achieved systematic uncertainties in E821. The main concepts of how the improvements will be made are indicated, but the reader is referred to the identified text sections for the details.

Table 5.3: Systematic uncertainties estimated for the magnetic field, $\omega_p$, measurement. The final E821 values are given for reference, and the proposed upgrade actions are projected. Note, several items involve ongoing R&D, while others have dependencies on the uniformity of the final shimmed field, which cannot be known accurately at this time. The relevant Chapters and Sections are given where specific topics are discussed in detail.

| Category | E821 [ppb] | Main E989 Improvement Plans | Goal [ppb] | Chapter |
|---|---|---|---|---|
| Absolute field calibration | 50 | Special 1.45 T calibration magnet with thermal enclosure; additional probes; better electronics | 35 | 15.4.1 |
| Trolley probe calibrations | 90 | Plunging probes that can cross calibrate off-central probes; better position accuracy by physical stops and/or optical survey; more frequent calibrations | 30 | 15.4.1 |
| Trolley measurements of $B_0$ | 50 | Reduced position uncertainty by factor of 2; improved rail irregularities; stabilized magnet field during measurements* | 30 | 15.3.1 |
| Fixed probe interpolation | 70 | Better temperature stability of the magnet; more frequent trolley runs | 30 | 15.3 |
| Muon distribution | 30 | Additional probes at larger radii; improved field uniformity; improved muon tracking | 10 | 15.3 |
| Time-dependent external magnetic fields | – | Direct measurement of external fields; simulations of impact; active feedback | 5 | 15.6 |
| Others † | 100 | Improved trolley power supply; trolley probes extended to larger radii; reduced temperature effects on trolley; measure kicker field transients | 30 | 15.7 |
| Total systematic error on $\omega_p$ | 170 | | 70 | 15 |

*Improvements in many of these categories will also follow from a more uniformly shimmed main magnetic field.

†Collective smaller effects in E821 from higher multipoles, trolley temperature uncertainty and its power supply voltage response, and eddy currents from the kicker. See 15.7.

# Chapter 6

# Civil Construction Off-Project

The experimental hall is funded as a General Plant Project (GPP), as part of the Muon Campus Program. The beamline and tunnel from the delivery ring to the hall are separate GPP and Accelerator Improvement Projects (AIP). The locations of the buildings on the muon campus is shown in Fig. 6.1.

## 6.1 The MC1 Building

The muon storage ring will be located in the MC-1 Building on the Muon Campus, which is shown in Fig. 6.2. While it is a general purpose building, the design and features are extremely important to the success of E989. The principal design considerations are a very stable floor, and good temperature stability in the experimental hall. Both of these features were absent at Brookhaven, and presented difficulties to the measurement of the precision field. This design serves E989, and subsequent experiments well. One portion of the MC1 building will house beamline power supplies and cryo facilities for the two initial experiments on the muon campus: $(g-2)$ and Mu2e.

The floor in the experimental area is constructed from reinforced concrete $2'\ 9''$ (84 cm) thick. The floor is $12'$ below grade. Core samples show that the soil at the location is very compacted, the floor settling is expected to be about $0.25''$ fully loaded.

This floor will be significantly better than the floor in Building 919 at Brookhaven, where the ring was housed for E821. That floor consisted of three separate pieces: a concrete spine down the middle of the room, with a concrete pad on each side of the spine. Thus the foundation of the ring will be much more mechanically stable than it was at BNL.

Even more important is the temperature stability available in MC-1. The HVAC system will hold the temperature steady to $\pm 2^\circ$ F during magnet operation and data collection. This stability, combined with thermal insulation around the magnet will minimize the changes in the field due to temperature changes in the experimental hall.

A floor plan of MC-1 is shown in Fig. 6.4. The experimental hall is $80' \times 80'$ with a 30 ton overhead crane. The loading dock in the lower left-hand corner is accessed through the roll-up door labeled in Fig. 6.2 . Unlike in BNL 919, the crane coverage is significantly larger than the storage-ring diameter, simplifying many tasks in assembling the ring.

A detailed MC-1 document is available from FESS, titled "MC-1 Building", dated March

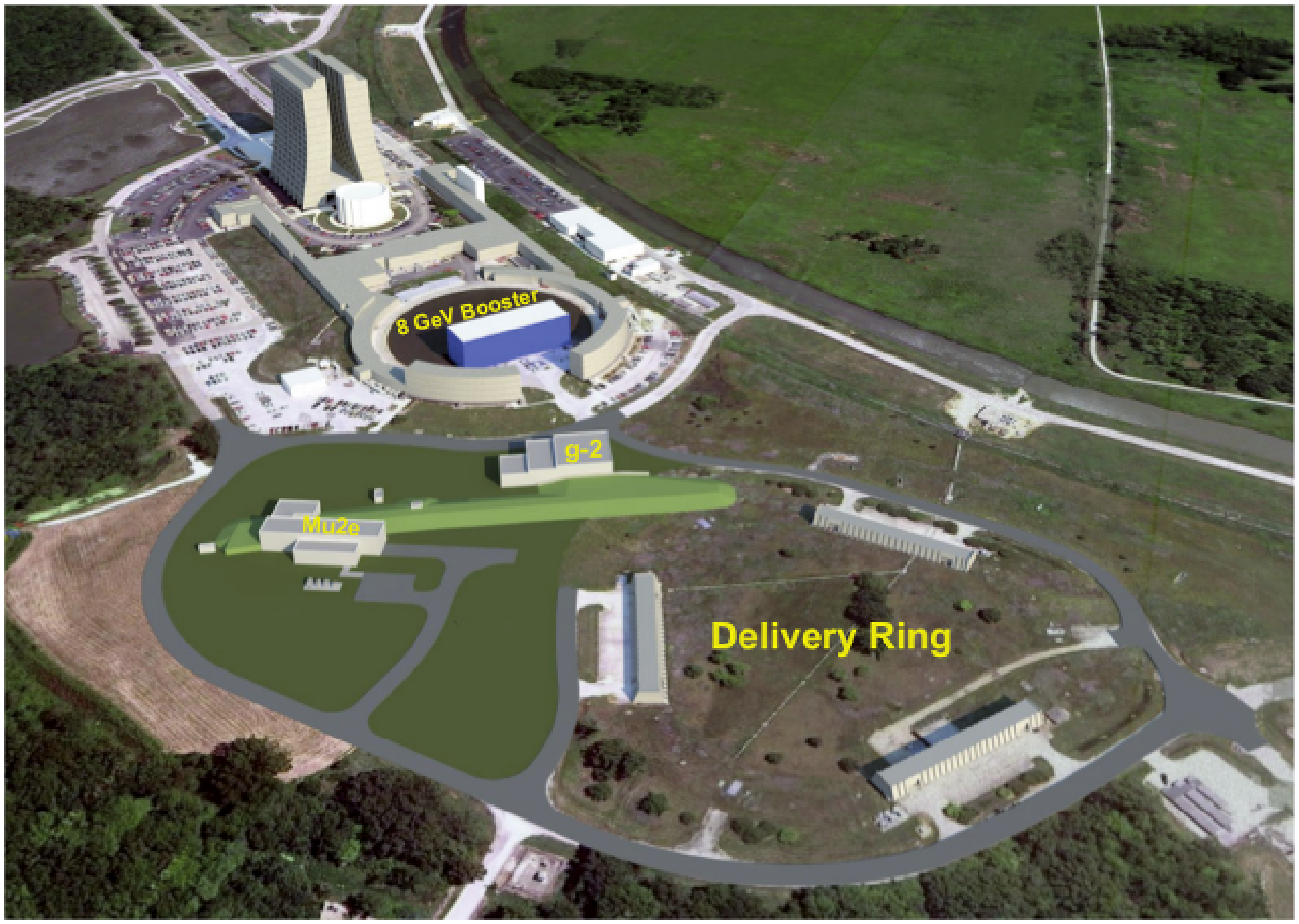

Figure 6.1: The layout of the Muon Campus, which lies between the former Antiproton Rings and the Booster Accelerator. The locations of the $(g-2)$ and Mu2e experiments are labeled.

2012.

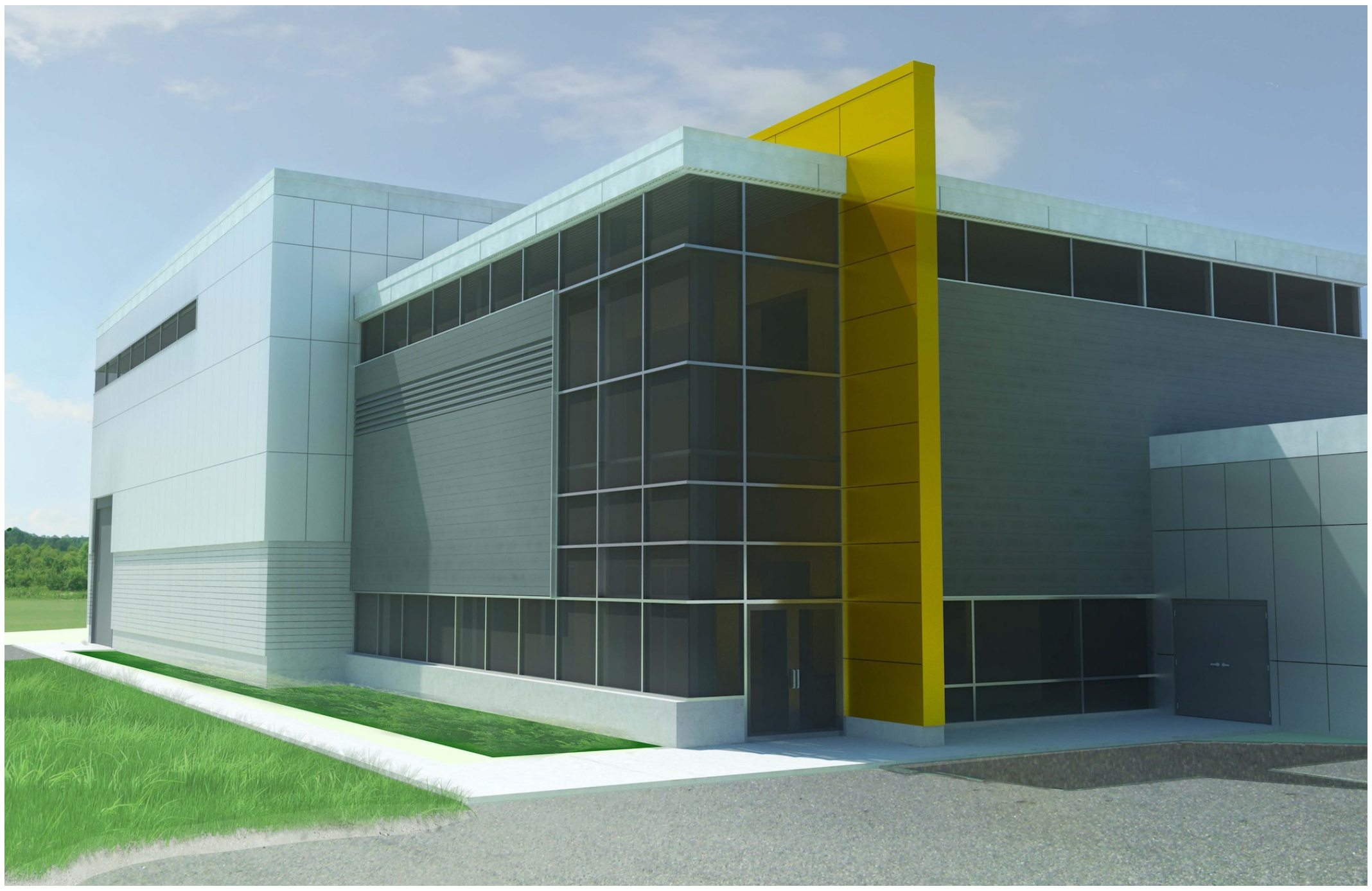

Figure 6.2: A rendering of the MC-1 building.

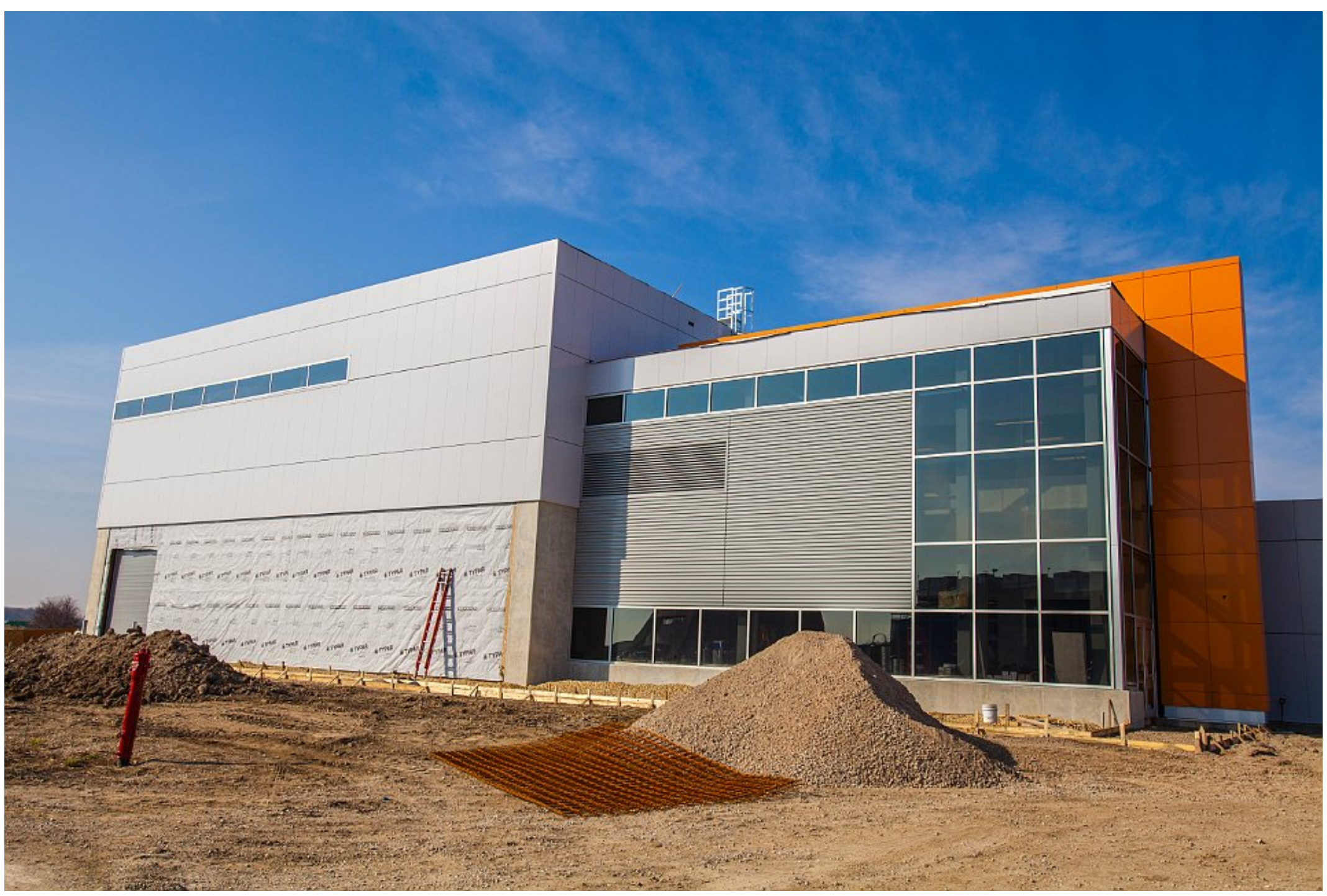

Figure 6.3: A photograph of the MC-1 building on April 18, 2014. Installation of various ring-related components began in the spring of 2014 and will continue throughout the summer.

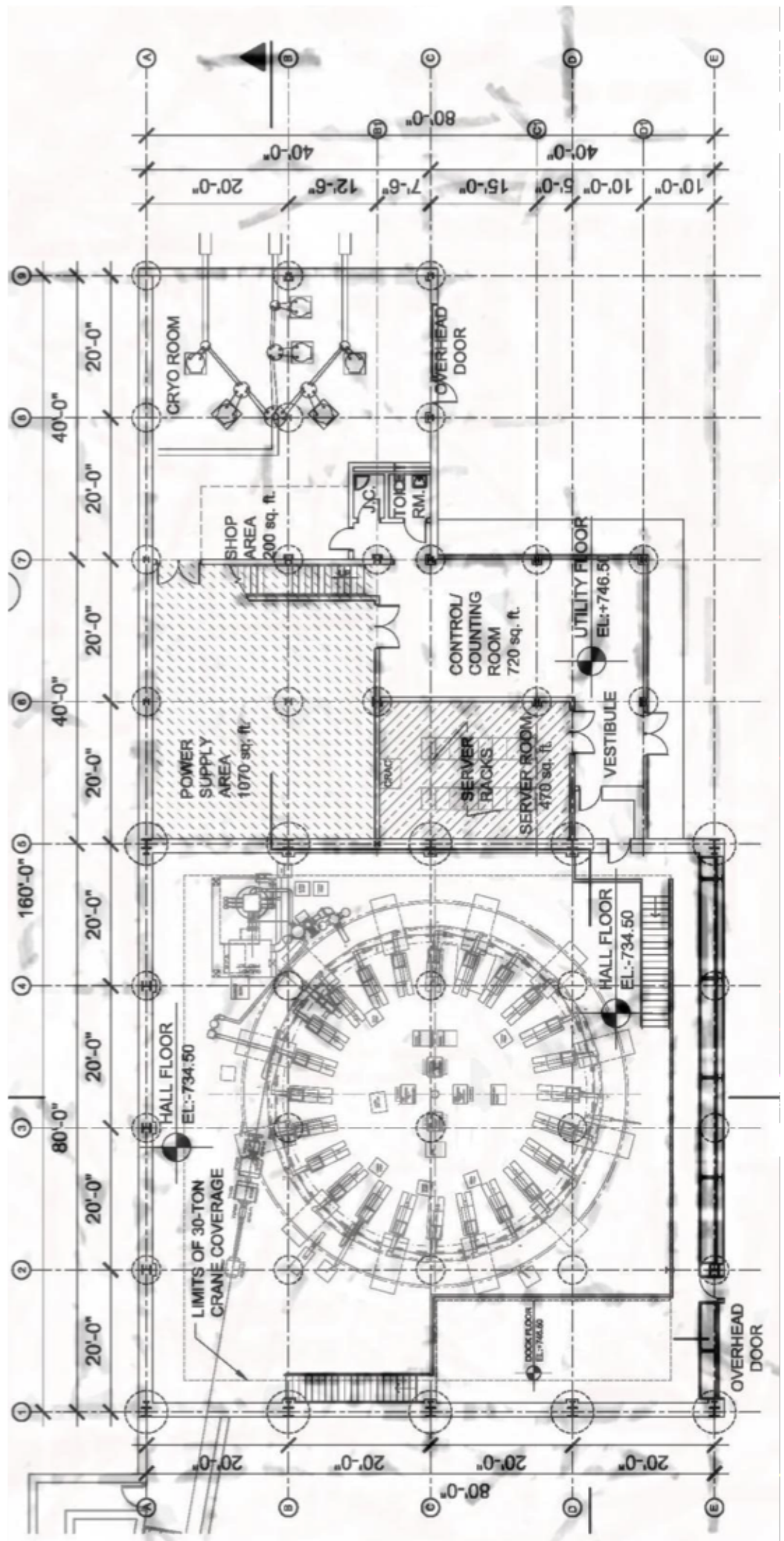

Figure 6.4: The first-floor layout of the MC1 building.

# Chapter 7

# Accelerator and Muon Delivery

In order to achieve a statistical uncertainty of 0.1 ppm, the total $(g-2)$ data set must contain at least $1.8 \times 10^{11}$ detected positrons with energy greater than 1.8 GeV, and arrival time greater than 30 $\mu$s after injection into the storage ring. This is expected to require $4 \times 10^{20}$ protons on target including commissioning time and systematic studies. For optimal detector performance, the number of protons in a single pulse to the target should be no more than $10^{12}$ and the number of secondary protons transported into the muon storage ring should be as small as possible. Data acquisition limits the time between pulses to be at least 10 ms. The revolution time of muons around the storage ring is 149 ns, and therefore the experiment requires the bunch length to be no more than ∼100 ns. Systematic effects on muon polarization limit the momentum spread $dp/p$ of the secondary beam. Requirements and general accelerator parameters are given in Table 7.1.

| Parameter | Design Value | Requirement | Unit |
|---|---|---|---|
| Total protons on target | $2.3 \times 10^{20}$/year | $4 \times 10^{20}$ | protons |
| Interval between beam pulses | 10 | $\geq 10$ | ms |
| Max bunch length (full width) | 120 (95%) | $< 149$ | ns |
| Intensity of single pulse on target | $10^{12}$ | $10^{12}$ | protons |
| Max Pulse to Pulse intensity variation | $\pm 10$ | $\pm 50$ | % |
| $\lvert dp/p \rvert$ of pions accepted in decay line | 2-5 | 2 | % |
| Momentum of muon beam | 3.094 | 3.094 | GeV/c |
| Muons to ring per $10^{12}$ protons on target | $(0.5 - 1.0) \times 10^5$ | $\geq 6000$ stored | muons |

Table 7.1: General beam requirements and design parameters.

## 7.1 Overall Strategy

The $(g-2)$ experiment at Fermilab is designed to take advantage of the infrastructure of the former Antiproton Source, as well as improvements to the Proton Source and the conversion of the Recycler to a proton-delivery machine. It is also designed to share as much infrastructure as possible with the Mu2e experiment in order to keep overall costs low.

The Antiproton Accumulator will no longer be in use, and many of its components will be reused for the new and redesigned Muon beamlines. Stochastic cooling components and other infrastructure no longer needed in the Debuncher ring will be removed in order to improve the aperture, proton abort functionality will be added, and the ring will be renamed the Delivery Ring (DR). The former AP1, AP2, and AP3 beamlines will be modified and renamed M1, M2, and M3. The DR Accelerator Improvement Project (AIP) will provide upgrades to the Delivery Ring. The Beam Transport AIP will provide aperture improvements to the P1, P2, and M1 lines needed for future muon experiments using 8 GeV protons, including $(g-2)$. The layout of the beamlines is shown in Fig. 7.1.

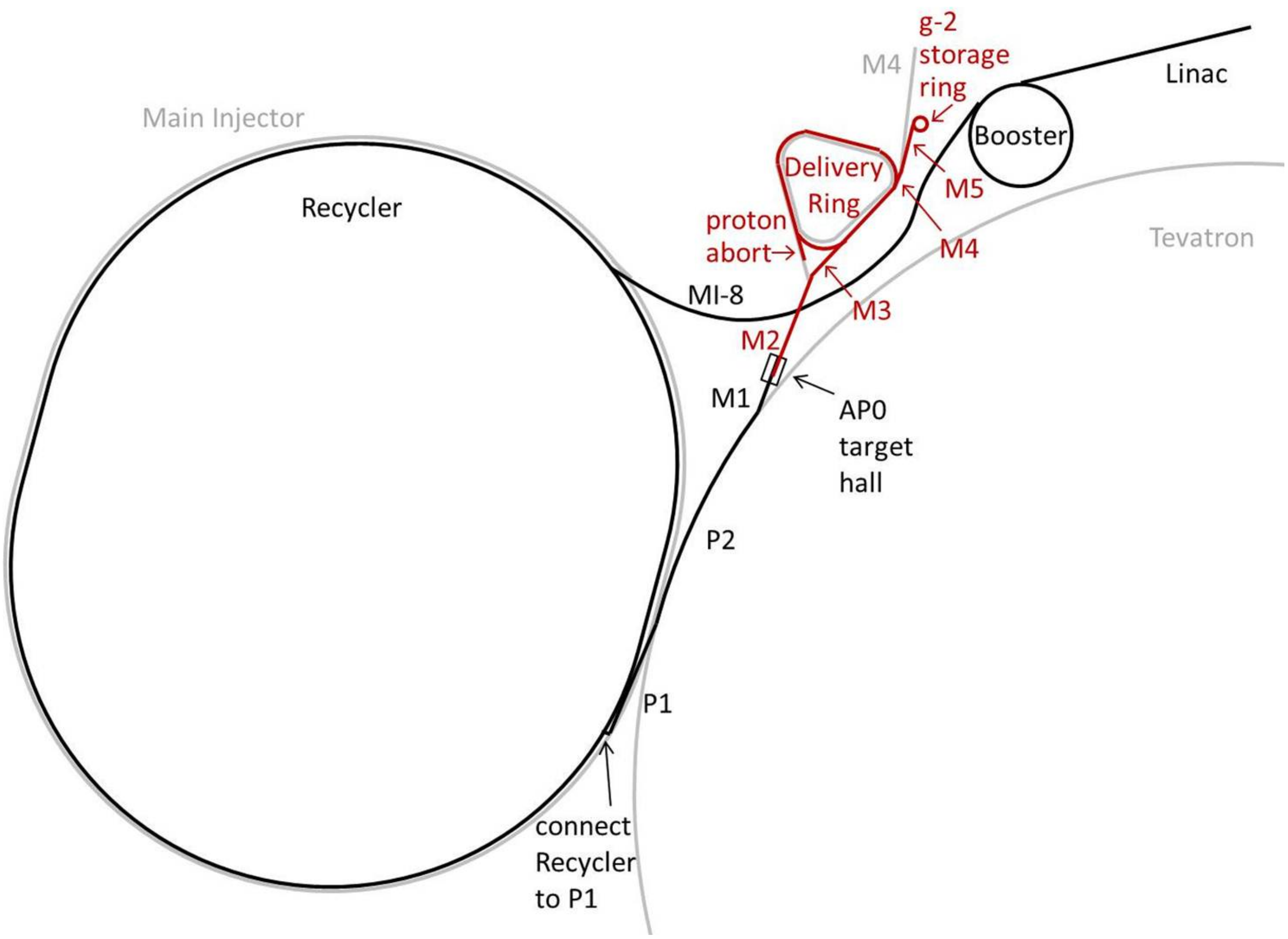

Figure 7.1: Path of the beam to $(g-2)$. Protons (black) are accelerated in the Linac and Booster, are re-bunched in the Recycler, and then travel through the P1, P2, and M1 lines to the AP0 target hall. Secondary beam (red) then travels through the M2 and M3 lines, around the Delivery Ring, and then through the M4 and M5 lines to the muon storage ring.

The Proton Improvement Plan [1], currently underway, will allow the Booster to run at 15 Hz, at intensities of $4 \times 10^{12}$ protons per Booster batch. The Main Injector (MI) will run with a 1.333 s cycle time for its neutrino program (NO$\nu$A), with twelve batches of beam from the Booster being accumulated in the Recycler and single-turn injected into the MI at the beginning of the cycle. While the NO$\nu$A beam is being accelerated in the MI, eight Booster batches will be available for experimental programs such as $(g-2)$ which use 8 GeV protons. Extraction from the Recycler to the P1 beamline, required for $(g-2)$, will be implemented in the Beam Transport AIP.

Protons from the Booster with 8 GeV kinetic energy will be re-bunched into four bunches in the Recycler and transported one at a time through the P1, P2, and M1 beamlines to a target at AP0. Secondary beam from the target will be collected using a lithium lens, and positively-charged particles with a momentum of 3.11 GeV/c ($\pm \sim 10\%$) will be selected using a bending magnet. Secondary beam leaving the Target Station will travel through the M2 and M3 lines which are designed to capture as many muons with momentum 3.094 GeV/c from pion decay as possible. The beam will then be injected into the Delivery Ring. After several revolutions around the DR, essentially all of the pions will have decayed into muons, and the muons will have separated in time from the heavier protons. A kicker will then be used to abort the protons, and the muon beam will be extracted into the new M4 line, and finally into the new M5 beamline which leads to the $(g-2)$ storage ring. Note that the M3 line, Delivery Ring, and M4 line are also designed to be used for 8 GeV proton transport by the Mu2e experiment.

The expected number of muons transported to the storage ring per $10^12$ protons on target, based on target-yield simulations using the antiproton-production target and simple acceptance assumptions, is $(0.5-1.0)\times 10^5$. Beam tests were conducted using the existing Antiproton-Source configuration with total charged-particle intensities measured at various points in the beamline leading to the Debuncher, which confirmed the predicted yields to within a factor of two [2]. More details are given in Sec. 7.4.1.

## 7.2 Protons from Booster

During the period when $(g-2)$ will take data, the Booster is expected to run with present intensities of $4\times 10^{12}$ protons per batch, and with a repetition rate of 15 Hz. In a 1.333 s Main-Injector NO$\nu$A cycle, twelve Booster batches are slip-stacked in the Recycler and then accelerated in the MI and sent to NO$\nu$A. While the Main Injector is ramping, the Recycler is free for a period of seven Booster cycles to send 8 GeV (kinetic energy) protons to $(g-2)$. The RF manipulations of beam for $(g-2)$ in the Recycler (Sec. 7.3.1) allow $(g-2)$ to take three Booster batches in a 1.33 s NO$\nu$A cycle, or four in a 1.40 s cycle. Figure 7.2 shows a possible time structure of beam pulses to $(g-2)$.

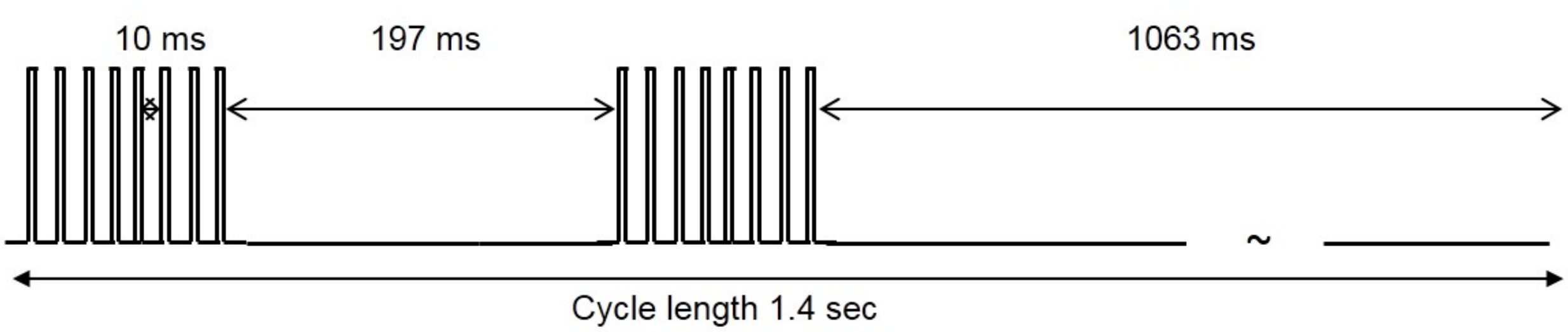

Figure 7.2: Possible time structure of beam pulses to $(g-2)$.

The following section describes improvements needed to run the proton source reliably at 15 Hz.

### 7.2.1 Proton Improvement Plan

The Fermilab Accelerator Division has undertaken a Proton Improvement Plan (PIP) [1] with the goals of maintaining viable and reliable operation of the Linac and Booster through 2025, increasing the Booster RF pulse repetition rate, and doubling the proton flux without increasing residual activation levels.

The replacement of the Cockroft-Walton pre-accelerator with a radio-frequency quadrupole (RFQ) to increase reliability of the pre-accelerator and to improve beam quality was completed in 2012.

The Booster RF solid-state upgrade necessary for reliable 15 Hz RF operations involved the replacement of 40-year-old electronics that are either obsolete, difficult to find, or unable to run at the required higher cycle-rate of 15 Hz, and allows for easier maintenance, shorter repair times, and less radiation exposure to personnel. The solid-state upgrade was completed in 2013.

Refurbishment of the Booster RF cavities and tuners, in particular, cooling, is also necessary in order to operate at a repetition rate of 15 Hz and is expected to be complete in 2015.

Other upgrades, replacements, and infrastructure improvements are needed for viable and reliable operation. Efforts to reduce beam loss and thereby lower radiation activation include improved methods for existing processes, and beam studies, e.g., aimed at finding and correcting aperture restrictions due to misalignment of components.

The proton flux through the Booster over the past two decades and projected into 2016 based on expected PIP improvements is shown in Fig. 7.3.

The new PIP flux goal will double recent achievements and needs to be completed within five years. The goal of doubling the proton flux will be achieved by increasing the number of cycles with beam. The intensity per cycle is not planned to increase.

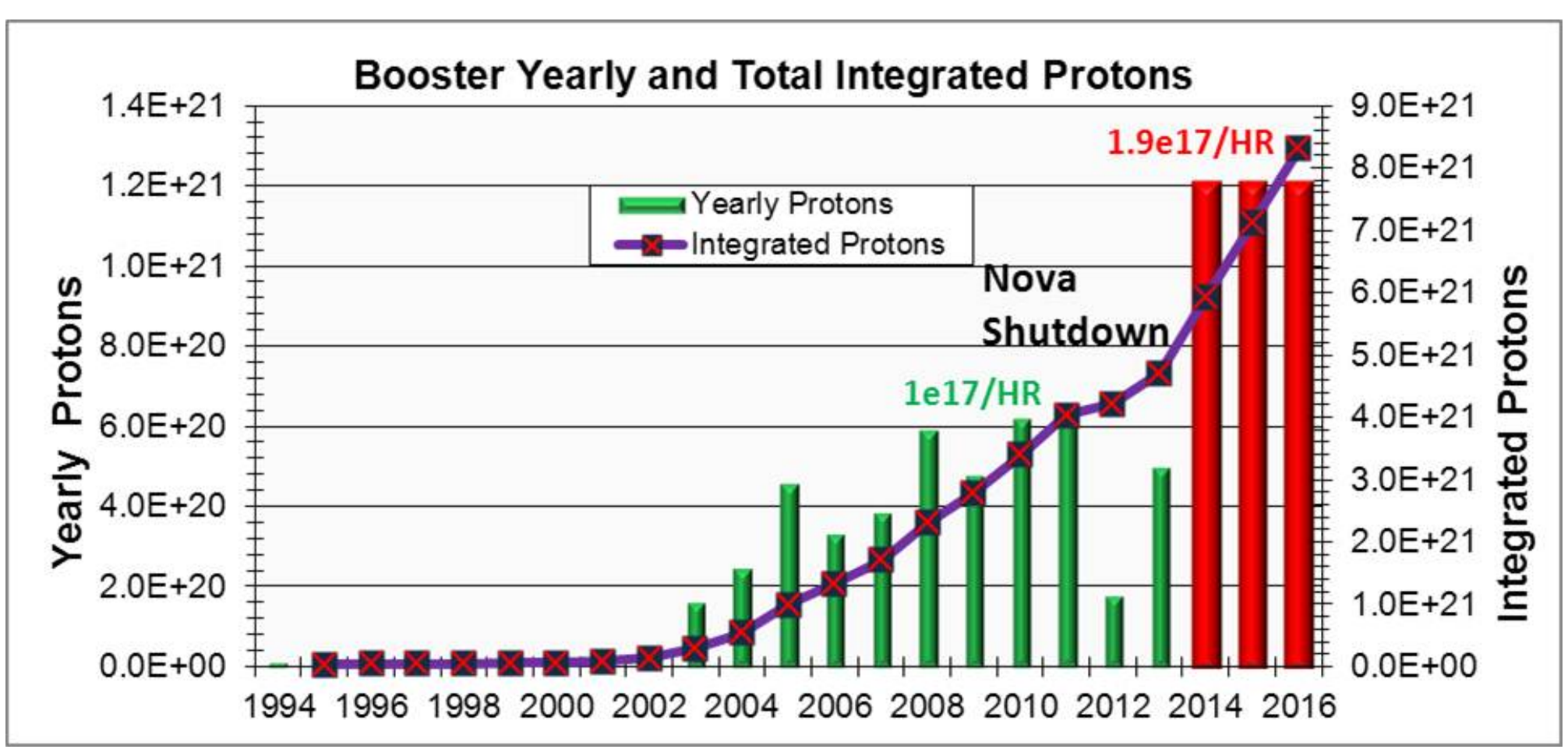

Figure 7.3: Yearly and integrated proton flux (including PIP planned flux increase).

## 7.3 Recycler

The $(g-2)$ experiment requires a low number of decay positrons in a given segment of the detector, and therefore requires that the full intensity of a Booster pulse ($4 \times 10^{12}$ protons) be redistributed into four bunches of $1 \times 10^{12}$ protons each. These bunches should be spaced no closer than 10 ms to allow for muon decay and data acquisition in the detector. Because the revolution time of muons in the $(g-2)$ ring is 149 ns, the longitudinal extent of the bunches should be no more than 120 ns. The Recycler modifications needed to achieve these requirements will be made under the Recycler RF AIP, and are described below.

### 7.3.1 Recycler RF

The proposed scheme for $(g-2)$ bunch formation [3] uses one RF system, 80 kV of 2.5 MHz RF. The design of the RF cavities will be based on that of existing 2.5 MHz cavities which were used in collider running, but utilizing active ferrite cooling. The ferrites of the old cavities and the old power amplifiers will be reused in the new system.

In order to avoid bunch rotations in a mismatched bucket, the 2.5 MHz is ramped "adiabatically" from 3 to 80 kV in 90 ms. Initially the bunches are injected from the Booster into matched 53 MHz buckets (80 kV of 53 MHz RF), then the 53 MHz voltage is turned off and the 2.5 MHz is turned on at 3 kV and then ramped to 80 kV. The first 2.5 MHz bunch is then extracted and the remaining three bunches are extracted sequentially in 10 ms intervals. The formation and extraction of all four bunches takes three Booster cycles, or 2 Booster batches can be rebunched into eight bunches and extracted to $(g-2)$ in four Booster cycles. This limits the $(g-2)$ experiment to using three of the available eight Booster batches in every 1.33 s Main-Injector NO$\nu$A cycle, or four in a 1.4 s Main-Injector cycle..

Simulated 2.5 MHz bunch profiles are shown in Fig. 7.4. The 53 MHz voltage was ramped down from 80 to 0 kV in 10 ms and then turned off. The 2.5 MHz voltage was snapped to 3 kV and then adiabatically raised to 80 kV in 90 ms. The maximum momentum spread is $dp/p = \pm 0.28\%$. The overall efficiency is 95%, and 95% of the beam captured is contained within 120 ns. Roughly 75% of the beam is contained in the central 90 ns and 60% in 50 ns.

Although the Recycler is not yet configured to do such RF manipulations, by using the 2.5 MHz coalescing cavities in the Main Injector, the proposed bunch-formation scheme was tested with beam. In general, the agreement between simulations and data is very good. For illustration, the comparison between the beam measurements and the simulations for the case in which the 2.5 MHz voltage is ramped adiabatically from 3 to 70 kV in 90 ms is shown in Fig. 7.5.

Extraction from the Recycler and primary proton beam transport will be described in the beamline section, Sec. 7.5.

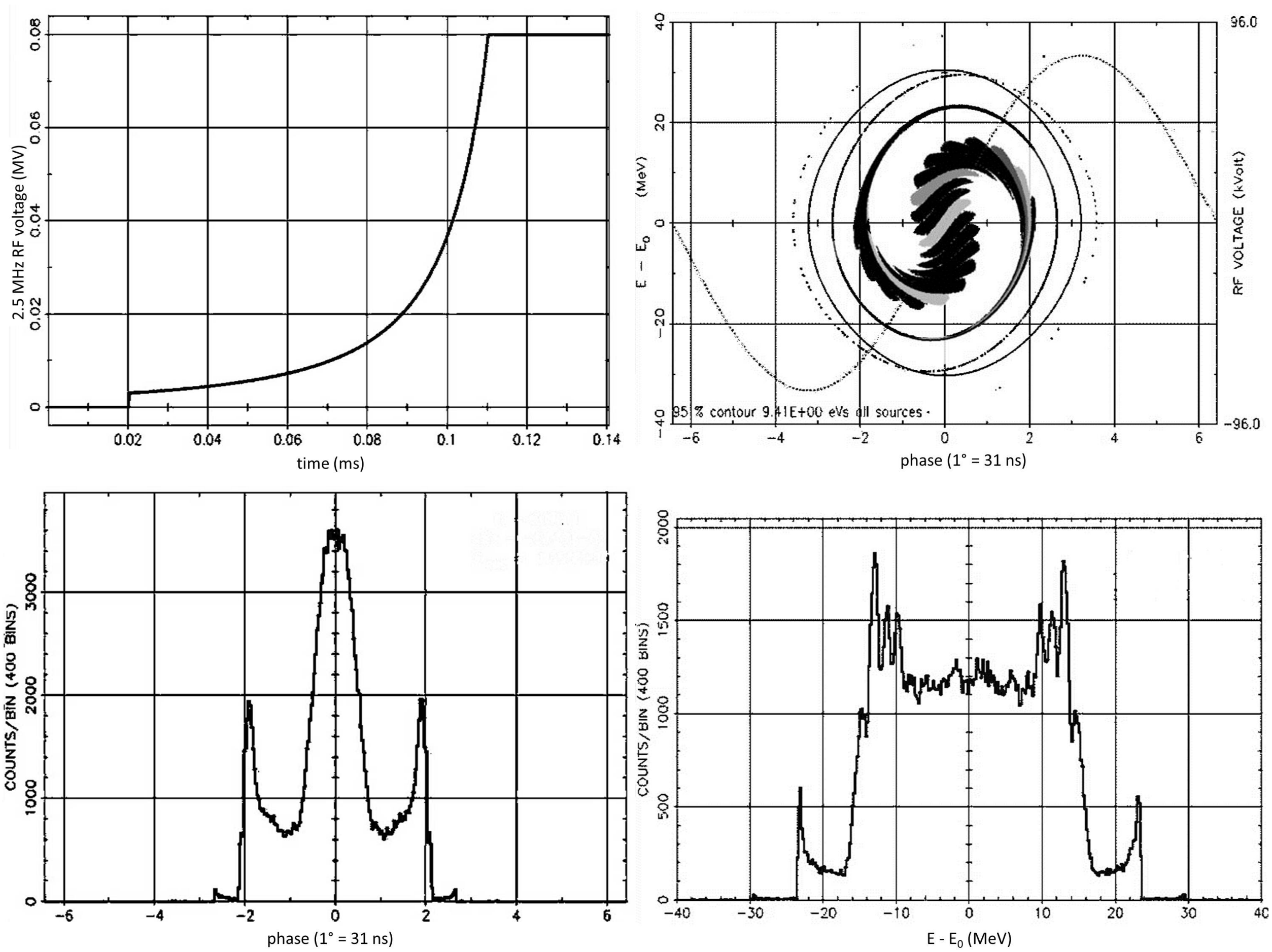

Figure 7.4: Results of RF simulations: 2.5 MHz voltage curve (upper left), phase space distribution (upper right), phase projection (lower left) and momentum projection (lower right).

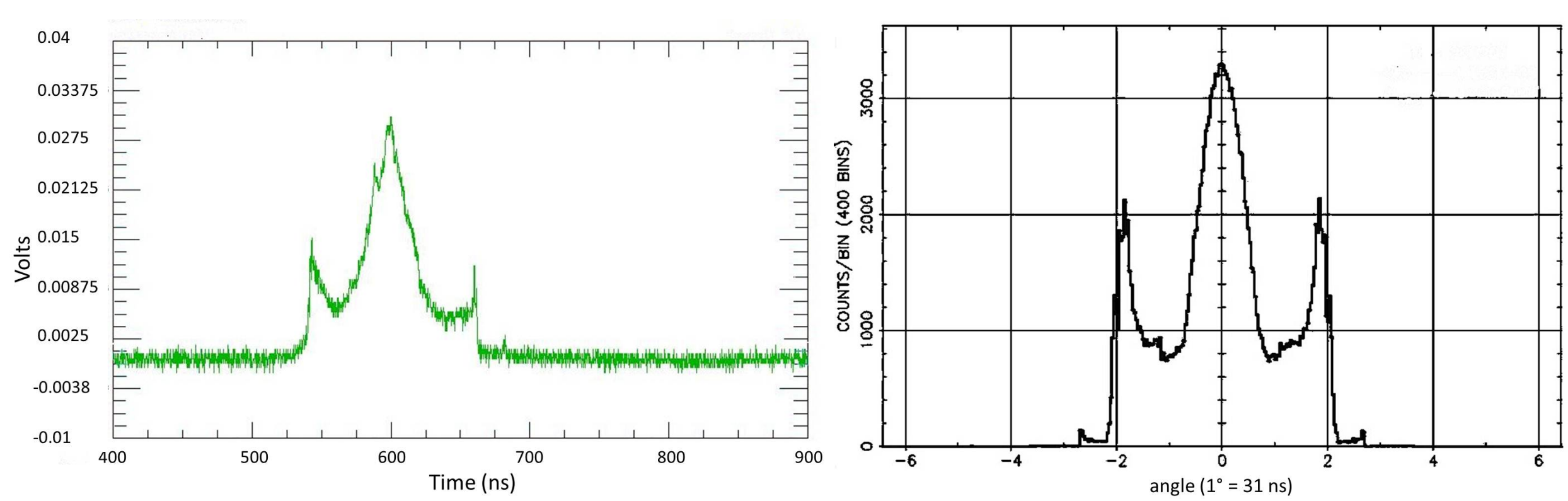

Figure 7.5: Comparison of beam profile (left) with simulation (right) for the case in which the 2.5 MHz voltage is ramped "adiabatically" from 3-70 kV in 90 ms. In both profiles, 95% of the particles captured are contained within 120 ns.

## 7.4 Target station

The $(g-2)$ production target station will reuse the existing target station that was operated for antiproton production for the Tevatron Collider since 1985, while incorporating certain modifications. The $(g-2)$ target station will be optimized for maximum $\pi^+$ production per proton on target (POT) since the experiment will utilize muons from pion decay. Repurposing the antiproton target station to a pion production target station takes full advantage of a preexisting tunnel enclosure and service building with no need for civil construction. Also included are target vault water cooling and air ventilation systems, target systems controls, remote handling features with sound working procedures and a module test area. Figure 7.6 shows the current target-station (vault) layout. The overall layout of the target-vault modules will be unchanged from that used for antiproton production. The major differences in design will include different primary and secondary beam energies, polarity of the selected particles, and pulse rate. Upgrades to pulsed power supplies are required.

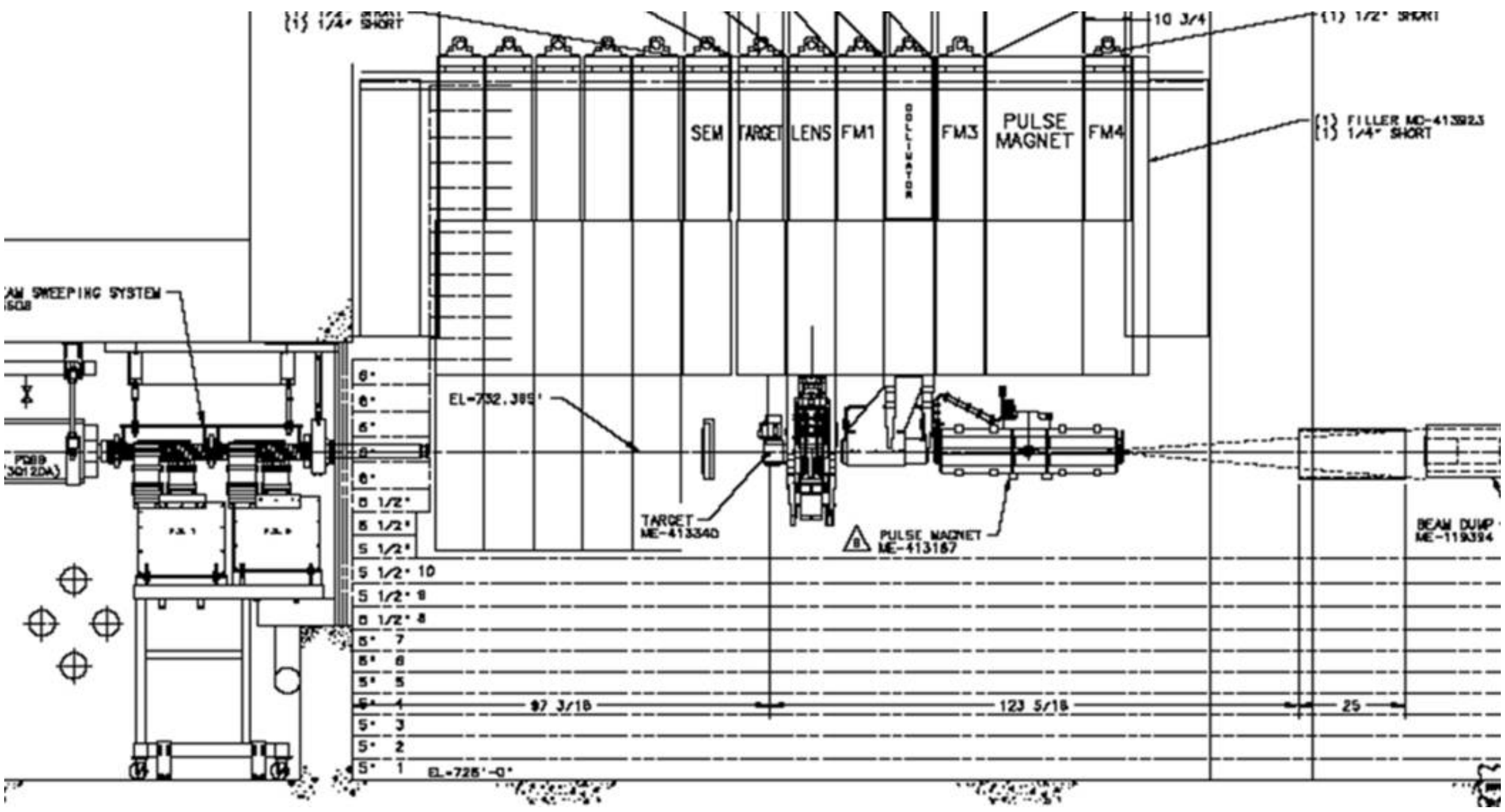

Figure 7.6: Layout of the $(g-2)$ target station.

The production target station consists of five main devices: the pion production target, the lithium lens, a collimator, a pulsed magnet, and a beam dump. Once the primary beam impinges on the target, secondaries from the proton-target interaction are focused by the lithium lens and then momentum-selected, centered around a momentum of 3.11 GeV/c, by a pulsed dipole magnet (PMAG). This momentum is slightly above the magic momentum needed to measure the muon anomalous magnetic moment in the downstream muon ring. The momentum-selected particles are bent 3° into a channel that begins the M2 beamline. Particles that are not momentum-selected will continue forward and are absorbed into the target-vault beam dump. An overview of some of the required beam design parameters for the $(g-2)$ target system can be found in Table 7.2.

| Parameter | FNAL $(g-2)$ 12 Hz |
|---|---|
| Intensity per pulse | $10^{12}$ p |
| Total POT per cycle | $16 \times 10^{12}$ p |
| Number of pulses per cycle | 16 |
| Cycle length | 1.33 s |
| Primary energy | 8.89 GeV |
| Secondary energy | 3.1 GeV |
| Beam power at target | 17.2 kW |
| Beam size $\sigma$ at target | 0.15-0.30 mm |
| Selected particle | $\pi^+$ |
| $\lvert dp/p\rvert$ (PMAG selection) | 10% |

Table 7.2: Beam parameters for the target station.

One significant difference the $(g-2)$ production target station will have from the antiproton production target station is the pulse rate at which beam will be delivered to the target station. The $(g-2)$ production rate will need to accommodate 16 pulses in 1.33 s with a beam pulse-width of 120 ns. This is an average pulse rate of 12 Hz. The antiproton production pulse rate routinely operated at 1 pulse in 2.2 s or 0.45 Hz. This is a challenging factor that drives the cost of the design since the lithium lens and pulsed magnet will need to pulse at a significantly higher rate. Figure 7.2 shows a possible $(g-2)$ pulse scenario for pulsed devices and timing for proton beam impinging on the target.

### 7.4.1 The $(g-2)$ production target and optimization of production

The target to be used for the $(g-2)$ experiment is the antiproton production target used at the end of the Tevatron Collider Run II. This target is expected to produce a suitable yield of approximately $10^{-5}$ $\pi^+$/POT within $|dp/p| < 2\%$ based on simulations. This target design has a long history of improvements for optimization and performance during the collider run. The target is constructed of a solid Inconel 600 core and has a radius of 5.715 cm with a typical chord length of 8.37 cm. The center of the target is bored out to allow for pressurized air to pass from top to bottom of the target to provide internal cooling to the Inconel core. It also has a cylindrical beryllium outer cover to keep Inconel from being sputtered onto the lithium lens from the impinging protons. The target has a motion control system that provides three-dimensional positioning with rotational motion capable of 1 turn in 45 s. This target and the target motion system need no modifications or enhancements to run for the $(g-2)$ experiment. Figure 7.7 shows a drawing and a photo of the current target.

Beam tests were performed to measure the yield from this target in 2012 [2]. The instrumentation measured total number of charged particles and did not differentiate between particle species. Measurements were recently repeated using a Cerenkov counter to measure the particle composition of the beam; data analysis is still in progress. The yield of positive 3.1-GeV secondaries from $10^{12}$ 8-GeV protons on target measured in the beam tests was about 85% of the $9.3 \times 10^8$ particles predicted [4] using a G4beamline [5] simulation at the

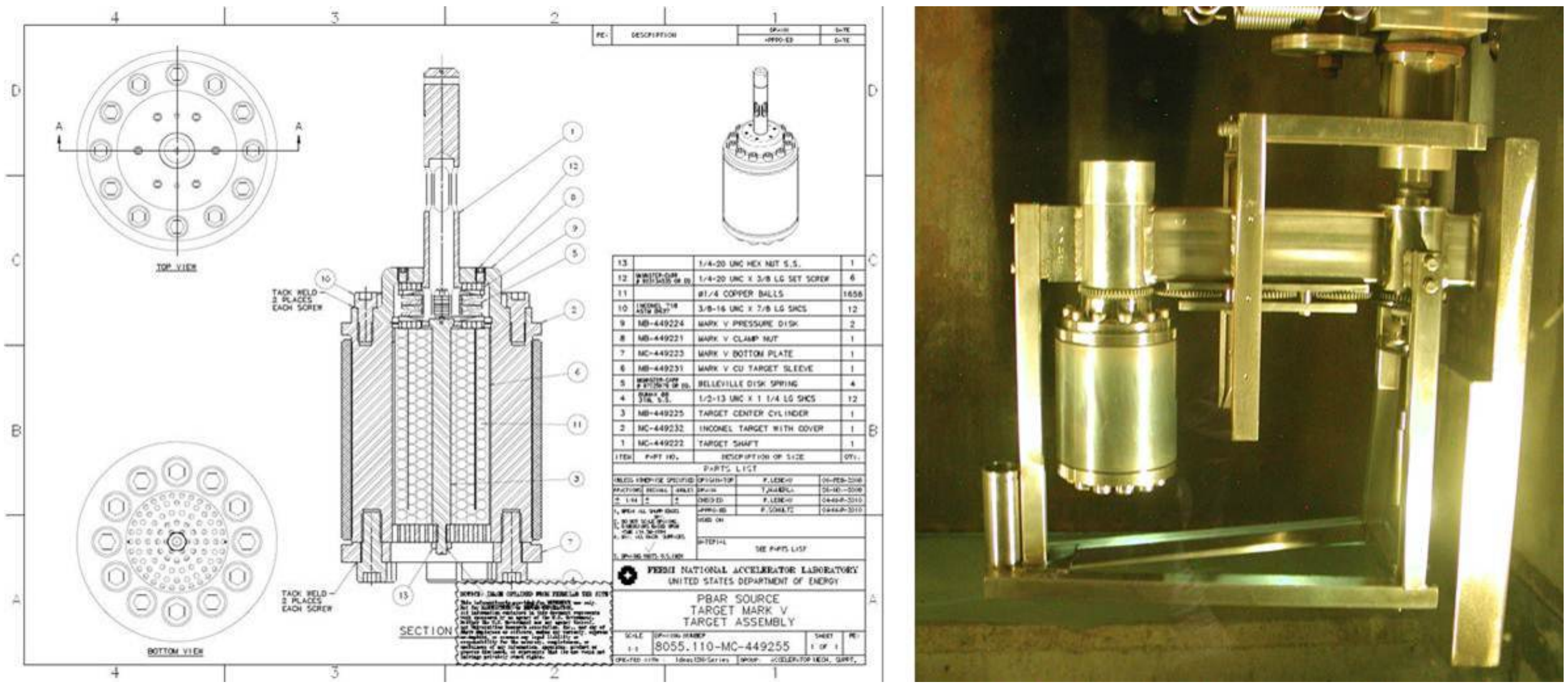

Figure 7.7: Current default target to be used for the $(g-2)$ target station.

704 location in the AP2 beamline shown in Fig. 7.8, and about 70% of the $4.3\times10^7$ particles at the 728 location [2]. The spot size of the beam on target was $\sigma_x \approx \sigma_y >\sim 0.5$ mm. As discussed in the beamlines section, we plan to reduce the spot size to 0.15 mm.

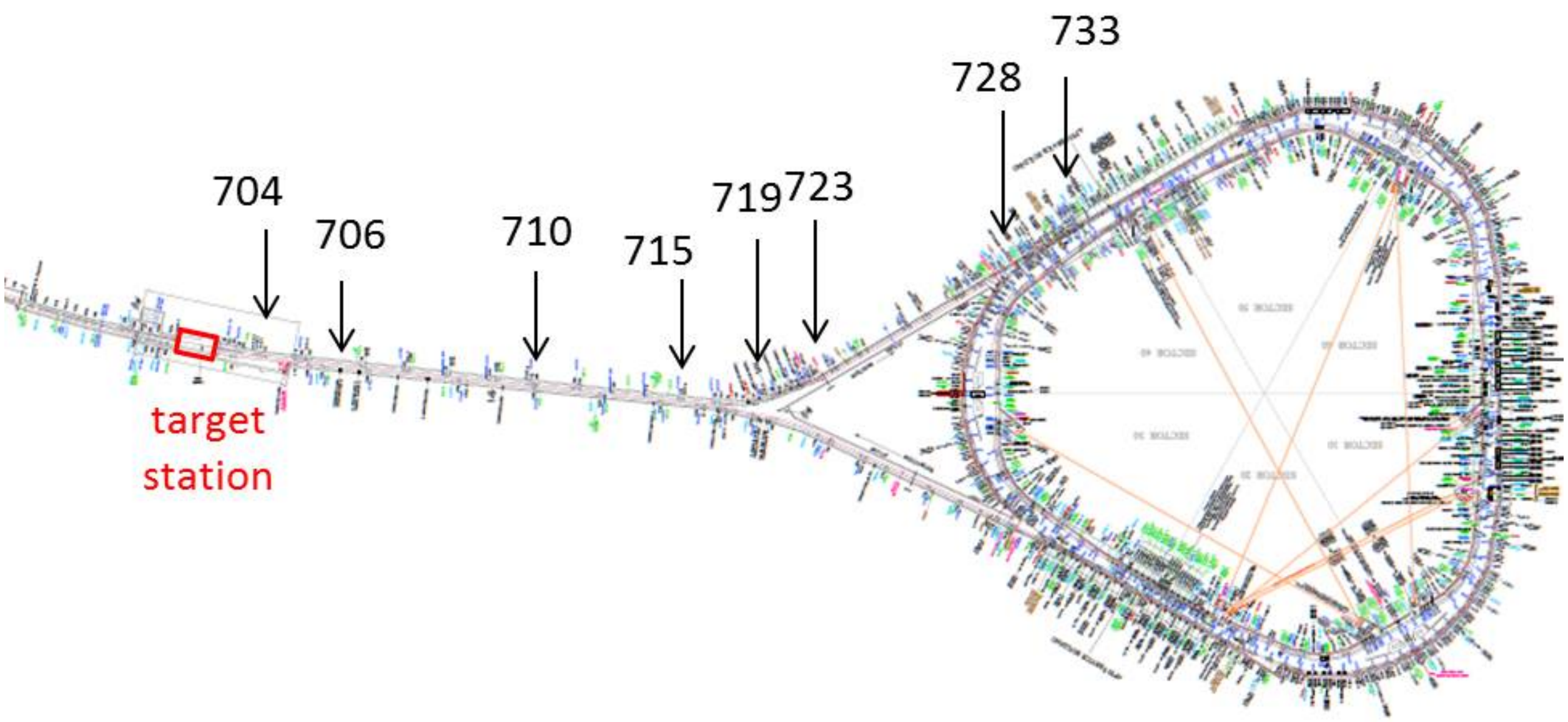

Figure 7.8: Locations in AP2 line.

The spot size of the beam on the target is an important parameter in determining the pion yield. Initial values for the spot size were simply scaled from the $\sigma_x = \sigma_y = 0.15$ mm size of the beam for 120-GeV antiproton production to $\sigma_x = \sigma_y = 0.55$ mm for 8.9 GeV. Optimized results from MARS [6] simulations (Fig. 7.9) for the impinging-proton spot size can be seen in Fig. 7.10. The simulation result demonstrates that if the spot size is reduced from the original 0.55 mm to 0.15 mm, a ∼15% increase in pion production can be achieved for the current target-to-lens distance of 28 cm. These modifications are not directly made to the target station or target components but to the beamline just upstream of the target. Details of the beamline optics incorporating this optimization for pion yield can be found in Sec 7.5.4. Combining the spot size with optimization of the target-to-lens distance discussed in Sec. 7.4.2 gives an improvement of ∼30%.

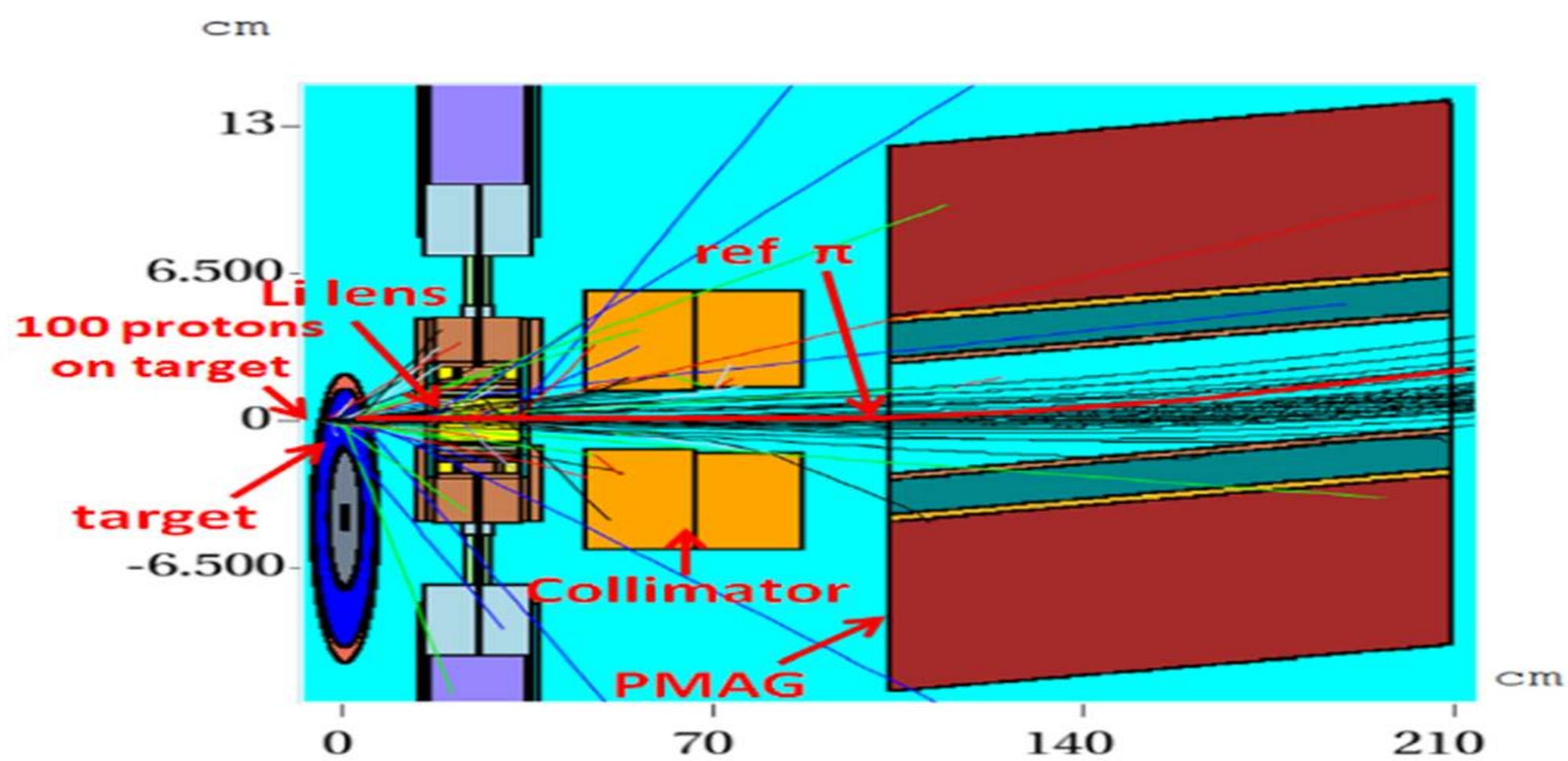

Figure 7.9: Graphical representation of target system used in MARS for simulated yield results.

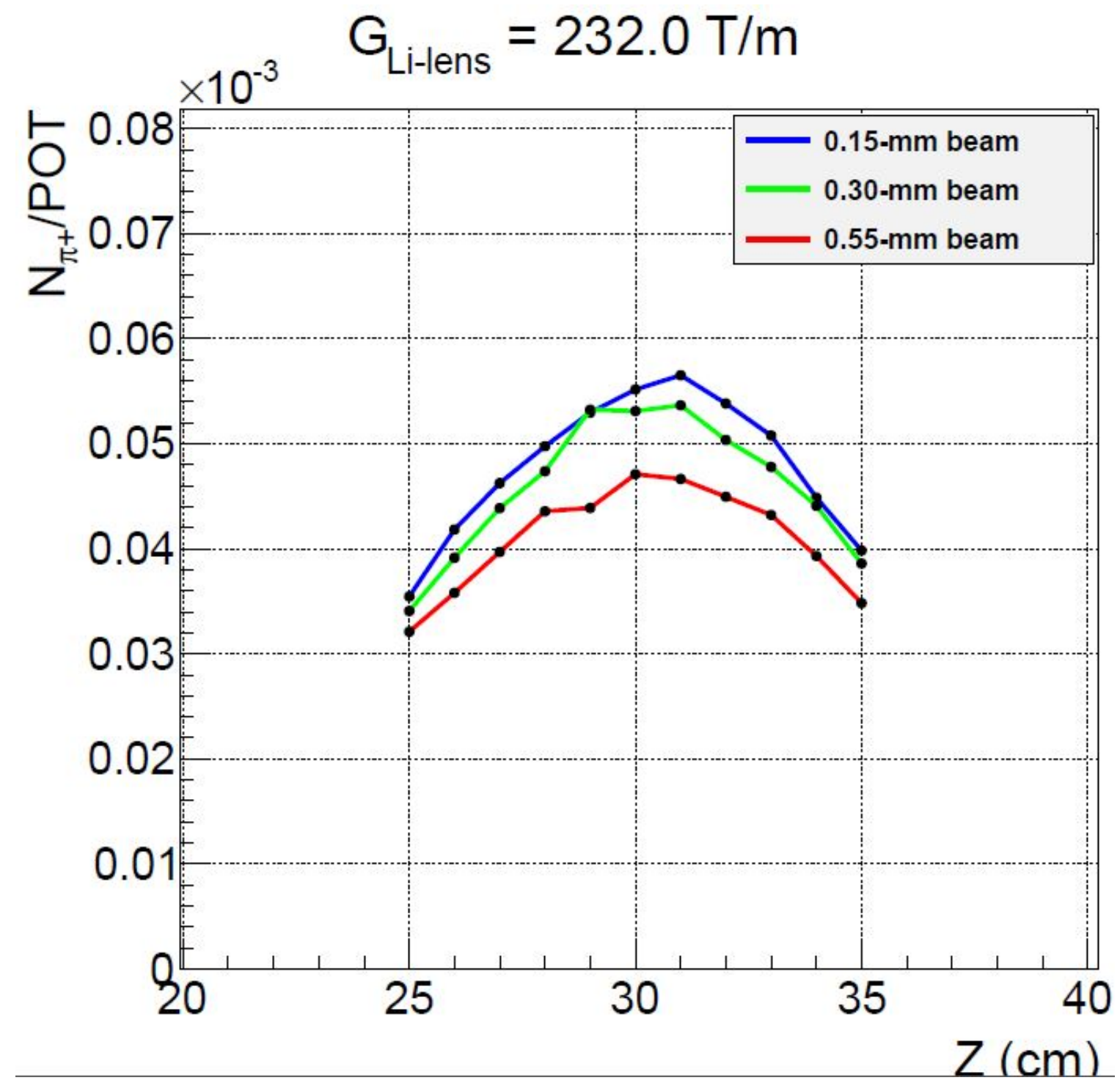

Figure 7.10: MARS simulation results for optimization of the distance between the target and the lithium lens for three beam spot sizes on target.

### 7.4.2 Focusing of secondaries from the target

The lithium collection lens is a 1 cm radius cylinder of lithium that is 15 cm long and carries a large current pulse that provides a strong isotropically focusing effect to divergent incoming secondaries after the initial interaction of impinging particles with the target [11]. The lithium lens cylinder is contained within a toroidal transformer, and both lens and transformer are water cooled. The peak current produced by the secondary of the transformer is a factor 8 larger that the primary peak current. Figure 7.11 is a drawing of the lithium lens depicting (a) the transformer and lens body, and (b) details of the lithium cylinder.

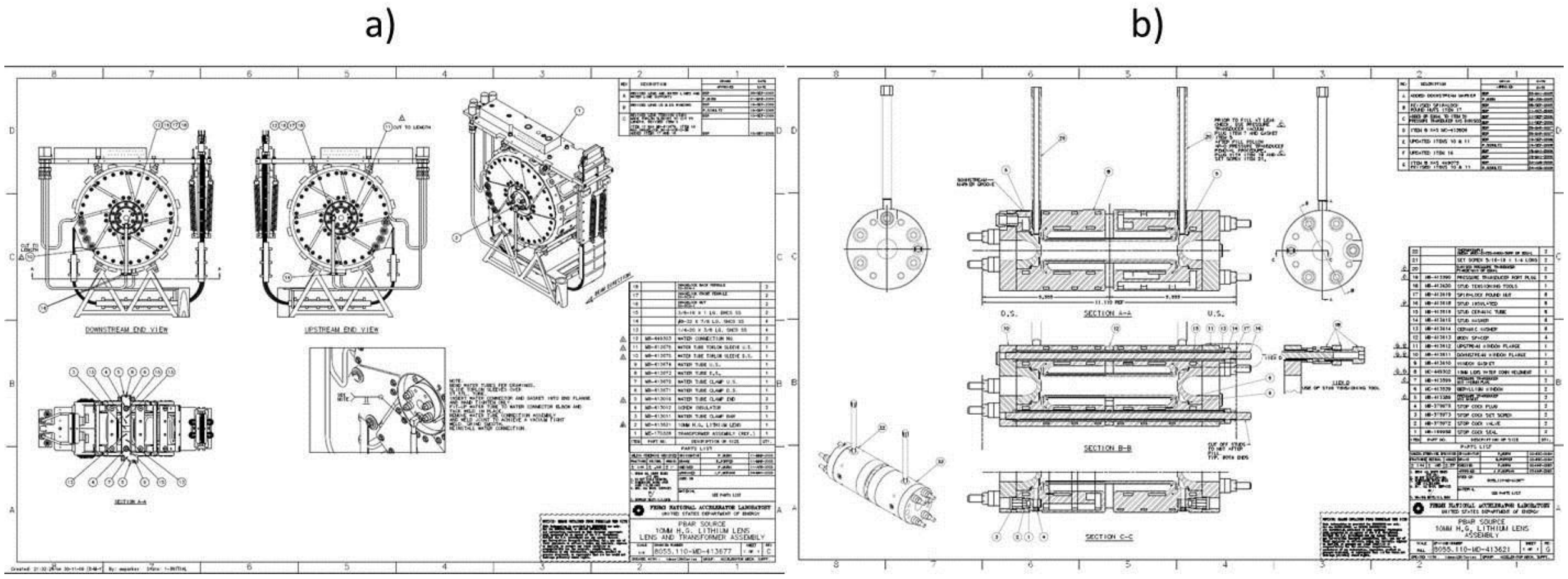

Figure 7.11: Drawing of the lithium lens and transformer (a) and the lithium cylinder body (b).

During antiproton production for the Collider Run II, the lens pulsed at a peak secondary current of 450 kA, which is equivalent to a gradient of 670 T/m at 8.9 GeV/c with a base pulse width of 400 $\mu$s. Scaling the lens gradient for use at 3.11 GeV/c for $(g-2)$, the gradient required will be 232 T/m at a pulsed secondary current of 155 kA with the same 400 $\mu$s pulse width. (The peak current produced by the secondary of the transformer is a factor of eight larger that the primary peak current.) The gradient for $(g-2)$ will still accommodate the same range of focal lengths from the target to the lens with the nominal distance taken to be 28 cm. The range of distances from the target to the lens is limited by the design of the target vault area for Colliding beam and is costly and difficult to change. Table 7.3 provides an overview of required operating parameters.

Accommodating the $(g-2)$ 12-Hz average pulse rate for the lithium lens is one of the biggest challenges and concerns for repurposing the antiproton target station for $(g-2)$. Even though the peak current and gradient will be reduced by a factor of about 3, the pulse rate will increase by a factor of 24 compared to the operation for antiproton production. Resistive and beam heating loads, cooling capacity, and mechanical fatigue are all concerns that are warranted for running the lithium lens at the $(g-2)$ repetition rate.

Therefore, in order to gain confidence that the lens will be able to run under these conditions, a preliminary ANSYS [7] analysis has been conducted. This analysis simulated thermal and mechanical fatigue for the lens based on the pulse timing scenario in Fig 7.2 and at a gradient of 230 T/m. These results were compared to results from a similar analysis for the lens operating under the antiproton-production mode of a gradient of 670 T/m at a pulse rate of 0.5 Hz [8]. Figure 7.12 (left) shows the ANSYS output thermal profile of a cutaway of the lens operating at 12 Hz. The lithium body corner is a temperature-sensitive location and should avoid lithium melting temperatures of 453.75 K. The corner temperature was found to reach a maximum temperature of 376 K. The plot on the right of Fig. 7.12 is the increase in maximum temperature of the lithium over the 16 pulses, depicting a change in temperature of 22 K when the operating temperature has come to equilibrium. We conclude from this analysis that the lithium lens is adequately cooled to operate at the nominal $(g-2)$ pulse rate.

| Lens operation | Pulse width ($\mu$s) | Peak secondary current (kA) | Gradient (T/m) | Pulses per day |
|---|---|---|---|---|
| Antiproton production | 400 | 450 | 670 | 38,880 |
| $(g-2)$ pion production | 400 | 155 | 232 | 1,036,800 |

Table 7.3: Comparison of lithium lens parameters for $(g-2)$ operations and antiproton production.

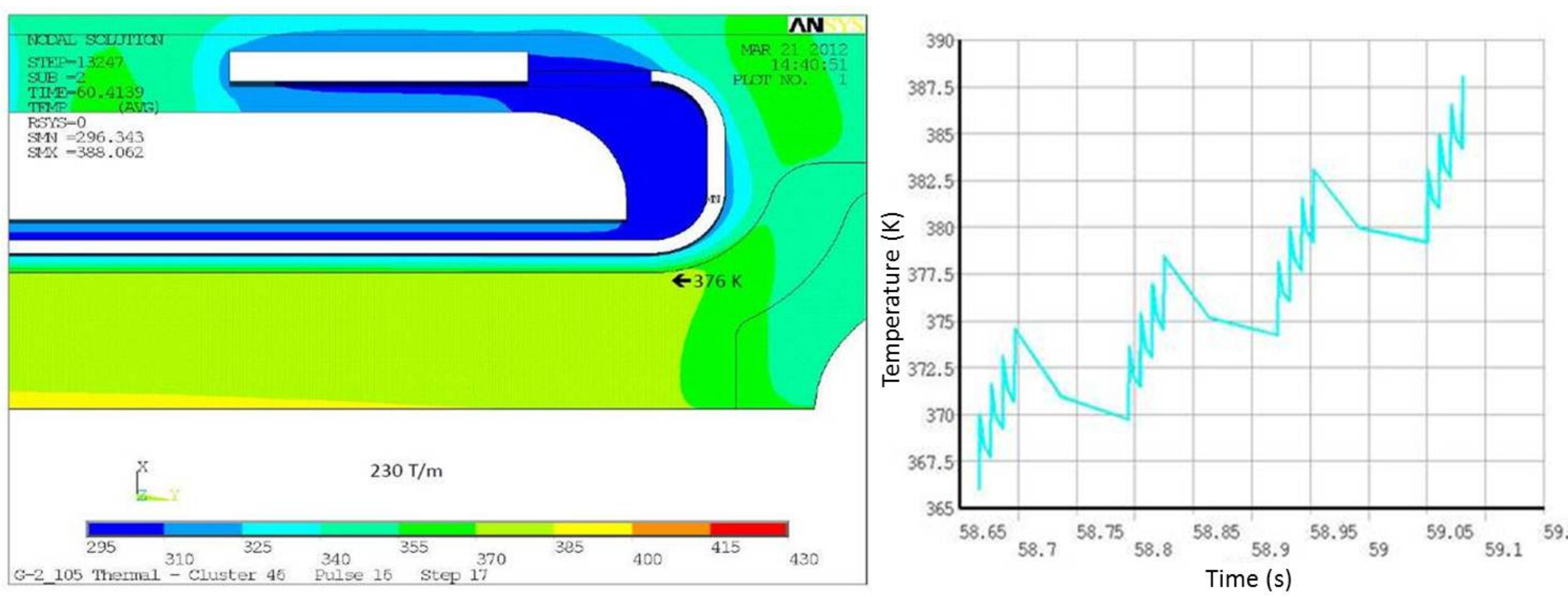

Figure 7.12: Simulated thermal profile from ANSYS for the lens operating at an average pulse rate of 12 Hz *(left)* depicting little beam heating and a corner temperature of 376 K. *(right)* Plots showing lens temperature increase over the 16 pulses.

Mechanical fatigue was also assessed for the lithium lens. Figure 7.13 depicts a constant life fatigue plot developed for the lens from the ANSYS analysis. The two red lines represent upper and lower estimates of fatigue limits for the lens material. The red data points represent fatigues for gradients of 1000 T/m, 670 T/m, and two points at 230 T/m for a lithium preload pressure of 3800 and 2200 psi, respectively. For the lens operating in the antiproton production conditions of 670 T/m, the mechanical fatigue was a large concern in the lens design. It appears that for the $(g-2)$ case, the mechanical fatigue will be a comparatively small concern.

This initial assessment of the lithium lens suggests that is should be able to operate at the $(g-2)$ repetition rate. However, since the operation of the lithium lens at the average 12 Hz rate is crucial, testing of the lens at 12 Hz was needed. The lens has been pulsed in a test station at a 12 Hz rate in order to confirm that 1M pulses per day can be achieved and sustained over many months. The lens was pulsed 80 million times without problems, and data from these tests were used to confirm predictions of the ANSYS model.

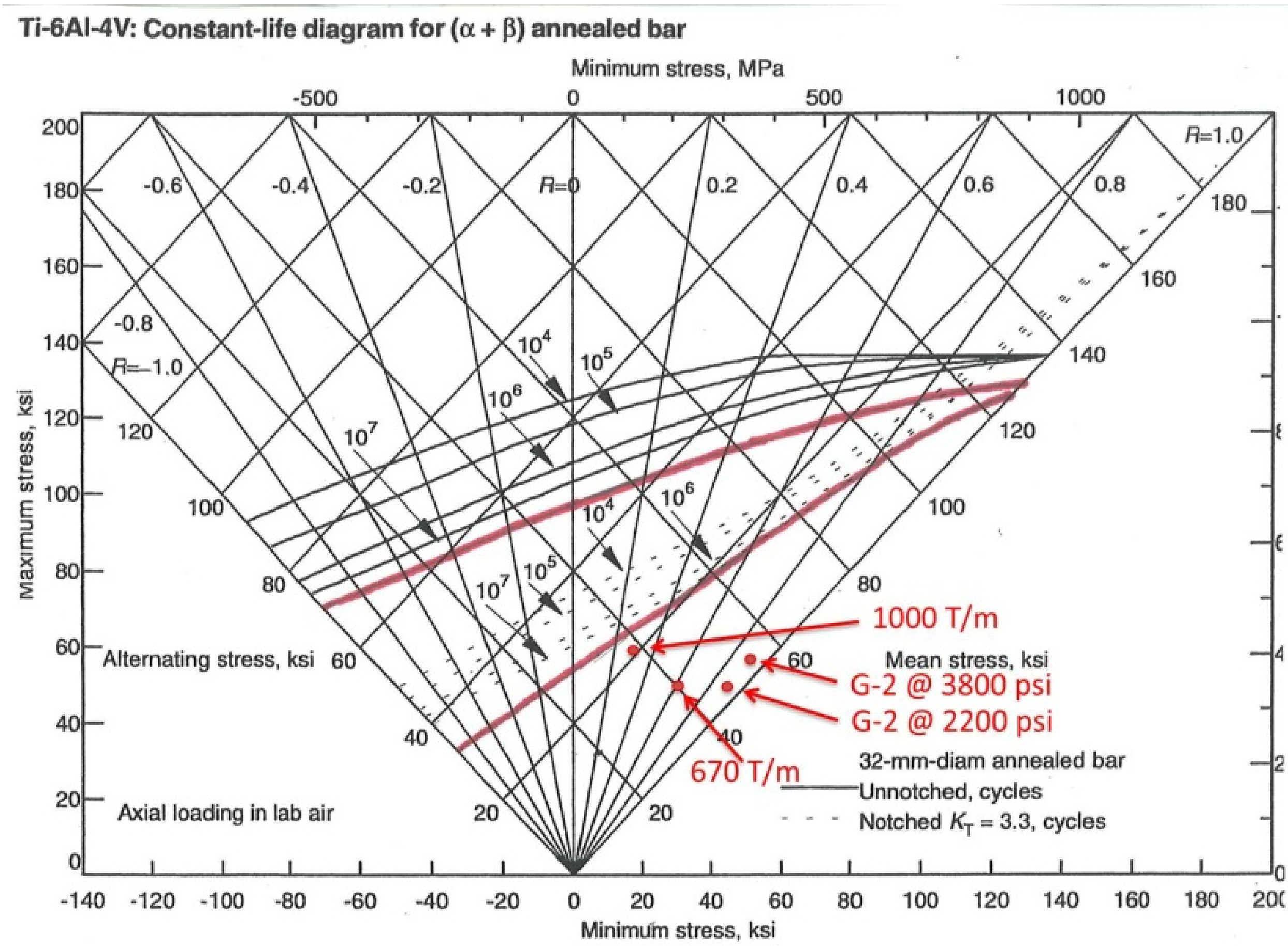

Figure 7.13: Constant-life fatigue plot of the lithium lens for antiproton and $(g-2)$ modes showing that mechanical fatigue for the $(g-2)$ pulse rate is a small concern.

The same ANSYS analysis was also used to determine if the repetition rate of the lens pulsing could be increased above 12 Hz (or 16 pulses in 1.33 s) for operational periods that may allow an increased repetition rate [9]. Lens thermal estimates for rates of 20, 24 and 28 pulses in the 1.33-s time period were conducted and analyzed. Figure 7.14 shows the results of this analysis. There are two concerns with increasing the repetition rate above 12 Hz. First, as the temperature rises and approaches the lithium melting temperature there is an increased risk of lithium leakage due to increased plasticity. The plasticity this close to the melting temperature is incredibly hard to model or predict. Pulsing at 15 Hz seems possible, but 18 Hz and above seems risky in this regard. The second concern is that as the temperature increases, so does the stress on the septum wall. In addition, the yield temperature of a material decreases as the temperature increases. While 20, 24 and 28 pulses per cycle have been modeled thermally, a structural analysis has not been done as of this writing. Simple scaling suggests that 15 Hz again seems possible but 18 Hz seems risky. This needs to be modeled to say with more certainty.

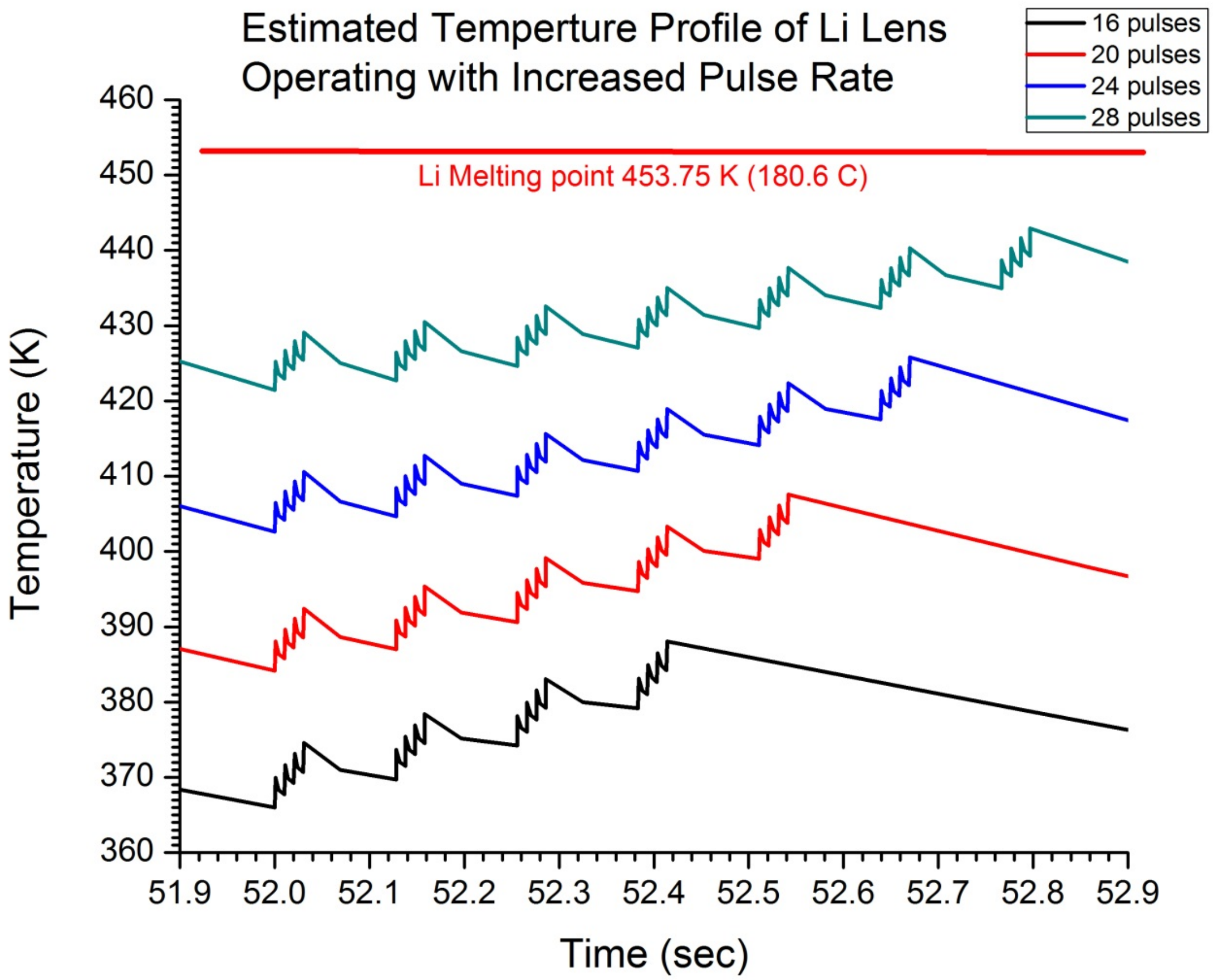

Figure 7.14: Temperature profile of the Li lens operating with increased pulse rate as modeled using ANSYS.

Temperature and stress effects with increased lens gradient were also considered. This has the same risks as increasing the pulse rate because the ultimate effect of increased thermal loading is the same. Due to the fact that the beam heating is so low compared to the joule heating, it is possible to ignore the beam heating and simply scale the joule heating for a first-order approximation. For example, a 25% increase in the repetition rate (from 12 to 15 Hz) corresponds to a 25% increase in thermal loading. Understanding that thermal power is $I^2R$, a 12% increase in current and gradient would correspond to a 25% increase in thermal loading. So 260 T/m would be equivalent to a repetition rate of 15 Hz.

In order to optimize the yield of pions from the target system as a function of lens gradient and target-to-lens distance, MARS simulations were conducted [10]. Figure 7.10 shows the results of the MARS simulation. For a nominal gradient of 232 T/m, pion production peaks at a target-to-lens distance of 30 cm to 31 cm. If we can increase the lens gradient, then a larger target-to-lens distance would be optimal, and a larger increase in yield could be possible as shown in Fig. 7.15.

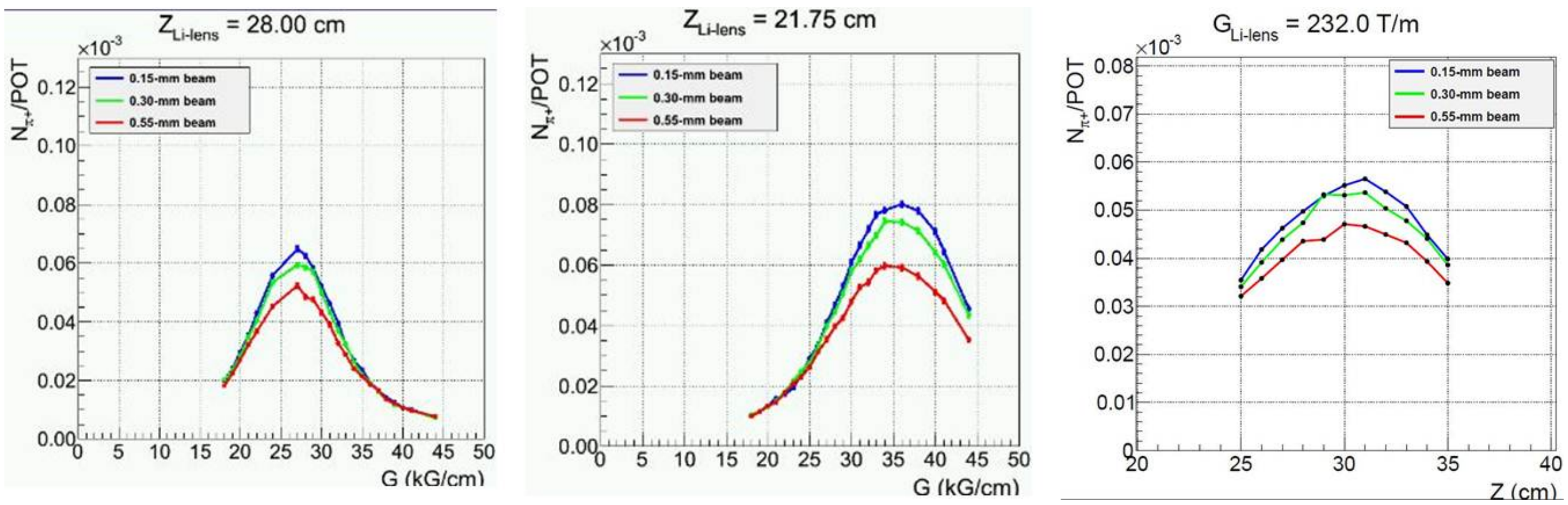

Figure 7.15: MARS simulation results for optimization of target-to-lens distance and lens gradient shown for three beam spot sizes on target.

Table 7.4 is an overview of the nominal operating parameters for the Li lens which will be the used for $(g-2)$ operation. Much attention has been given to the Li lens to confirm that it will be able to operate under the $(g-2)$ pulse rate scenario. There are three suitable lens spares that can be used during the $(g-2)$ data-taking period. The major upgrade to the Li lens system will be to modify the power supply to operate at the $(g-2)$ pulse rate. Details of the power supply design are presented in Sec 7.4.4.

| Lens parameters | Value |
|---|---|
| Nominal gradient | 232 T/m |
| Nominal peak lens current | 155 kA |
| Nominal target to lens distance | 30 cm |
| Min target to lens distance | 21.75 cm |
| Max target to lens distance | 33.4 cm |
| Max Power Supply Current | 25 kA |
| Max gradient (for max PS output) | 301 T/m |

Table 7.4: Operating parameters for the lithium lens.

### 7.4.3 Pulsed magnet (PMAG) and collimator

The pulsed magnet, shown in Fig. 7.16, selects 3.115 GeV/c positive particles and bends them 3° into the channel that begins the M2 beamline. The magnet will operate with a field of 0.53 T and is 1.07 m long with an aperture of 5.1 cm horizontally and 3.5 cm vertically. It is a single-turn magnet that has incorporated radiation-hard hardware such as ceramic insulation between the magnet steel and the single-conductor bars, as well as Torlon-insulated bolts [11]. The pulsed magnet has a typical pulse width of 350 $\mu$s and similarly to the lithium lens, will need to accommodate the $(g-2)$ pulse rate shown in Fig. 7.2. The pulsed magnet is water cooled. In addition to the magnet currently in the target vault, there are three spares.

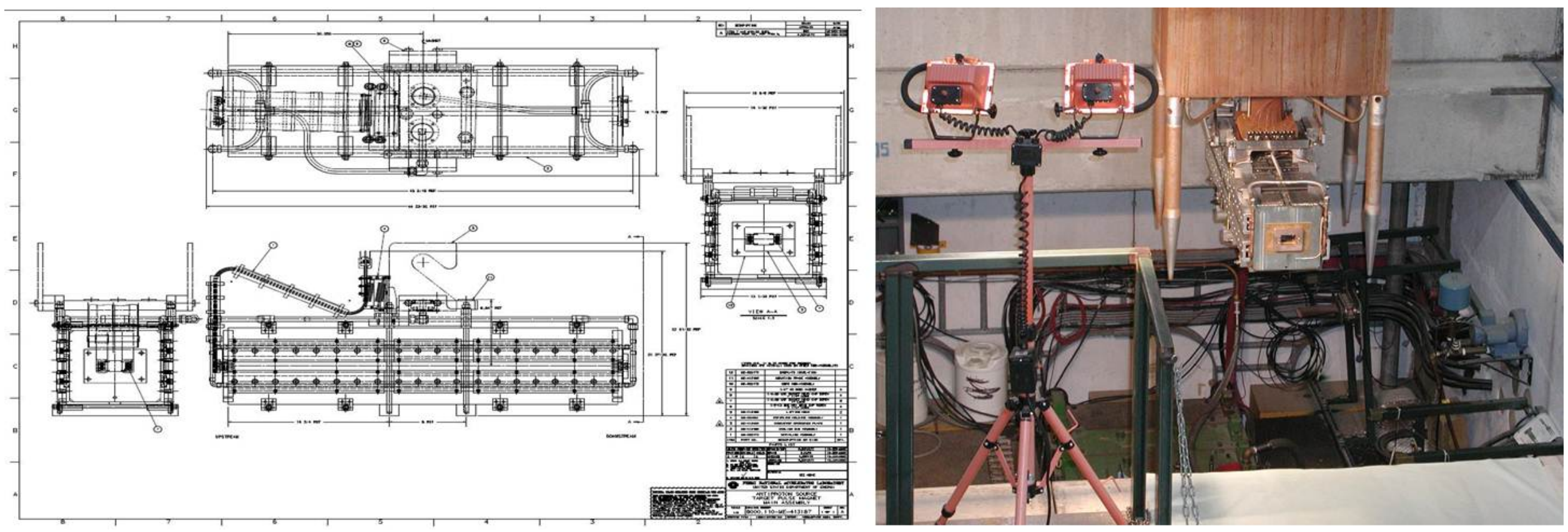

Figure 7.16: Pulsed magnet (PMAG) used for momentum-selection of pions.

One initial concern regarding the pulsed magnet was that while operating in the polarity needed to collect positive secondaries, the magnet would have an increase in energy deposited in the downstream end of the magnet compared to antiproton production where negative secondaries were collected. An increase in energy deposition could potentially lead to magnet failures, and therefore running with positive polarity might require a redesign of the magnet. A MARS simulation was conducted to look at the energy deposition across the entire pulsed magnet compared to the antiproton production case. The simulated magnet was segmented in order to highlight sensitive areas. The simulation concluded that although the map of energy deposition for the positive particle polarity with 8-GeV protons on target was different than for the antiproton production case (120-GeV protons on target), there were no locations where the deposited energy was higher, and the total was an order of magnitude lower [12]. The negative particle polarity case was more than two times lower for 8-GeV primary beam than for 120-GeV. Therefore a new pulsed magnet design was not needed and the plan is to use the device currently installed.

In order to accommodate the $(g-2)$ pulse rate, the pulsed magnet power supply will also need to be modified into one similar to the new supply for the lithium lens with improved charging capability. Details of the power supply design are presented in Sec. 7.4.4.

A collimator is located directly upstream of the pulsed magnet. The purpose of the collimator is to provide radiation shielding to the pulsed magnet to improve its longevity. It is a water-cooled copper cylinder 12.7 cm in diameter and 50.8 cm long. The hole through the center of the cylinder is 2.54 cm diameter at the upstream end, widening to a diameter of 2.86 cm at the downstream end. The existing collimator is currently planned to be used without modification.

### 7.4.4 Lithium-lens and pulsed-magnet (PMAG) power supplies

The lithium-lens and pulsed-magnet power supplies will both need to be upgraded in order to meet the $(g-2)$ pulse rate scenario shown in Fig. 7.2. The requirements and specifications for the lens and pulsed-magnet power supply systems can be seen in Table 7.5. During the design process, it was found to be cost-beneficial to use the same power supply design for both supplies since their load characteristics and power supply output are similar. Both systems currently have existing power supplies that will be modified to produce the $(g-2)$ power supplies. Modifications to the existing supplies will include new larger charging supply systems, additional enclosures to house a large capacitor bank for bulk energy storage, and new power supply controls [13].

| **Power Supply Specification** | | |
|---|---|---|
| | **Lens** | **PMAG** |
| **device** | | |
| type | transformer-lens | 1-turn magnet |
| location | AP0 | AP0 |
| inductance | 3.54 $\mu$H,<br>from transformer primary | 2.56 $\mu$H |
| resistance | 12.31 $m\Omega$ | 2.61 $m\Omega$ |
| **current program** | | |
| pulsed | 1/2 sinewave | 1/2 sinewave |
| peak nominal current | 20 kA | 15.3 kA |
| peak maximum current | 25 kA | 18 kA |
| pulse base | 400$\mu$s (same as existing) | 355$\mu$s (same as existing) |
| maximum rep rate | 100 Hz | 100 Hz |
| average rep rate | 12 Hz | 12 Hz |
| maximum ave rep rate | 18 Hz | 18 Hz |
| **regulation** | | |
| drift and stability | ±0.1% of maximum | ±0.1% of maximum |
| **other** | | |
| AC input | 480 VAC, 3-phase | 480 VAC, 3-phase |
| cooling | air and/or LCW | air and/or LCW |
| controls | accelerator timing system | accelerator timing system |
| power supply location | AP0, must fit within present power-supply footprint | AP0, must fit within present power-supply footprint |

Table 7.5: Lithium-lens and pulsed-magnet power supply requirements and specifications.

Fig. 7.17 shows a high-level schematic diagram that will be used for the design of the power supplies. The lens and PMAG power supplies both require similar half-sinewave pulsed currents. These high-power pulses are provided through a solid-state switch (pulsed SCRs) that connects a charged capacitor bank (Pulsed Cap Bank) to the load. After the pulse, the capacitor-bank voltage is brought back to the correct polarity through a "Charge Recovery" circuit. The design of the power supply includes a "Charge Transfer" mechanism to minimize the pulsed loading on the 480-VAC distribution systems. Without this feature, a costly dedicated 13.8 kV/480 VAC transformer would have been necessary to operate the power supplies. The Charge Transfer hardware consists of a 12-pulse rectified power supply, a large capacitor bank, and a high-voltage solid-switch (IGBT). Between load pulses, the IGBT switch closes and transfers energy from the capacitor bank to the pulsed capacitor bank to make up for the energy lost during the pulse. Since the capacitor bank is much larger (by a factor >30) than the pulsed capacitor bank, this helps reduce the effect on the AC line of the 10 ms burst of pulses. The "Filter Choke" in series with the capacitor bank further evens out the pulsing. Given the age of the existing power supply controls and the added Charge Transfer feature required, a completely new control system will be designed and built. The re-designed pulsed power supply will also incorporate a "deQ-ing" system to maintain the required 0.1% regulation of the Pulsed Cap Bank energy.

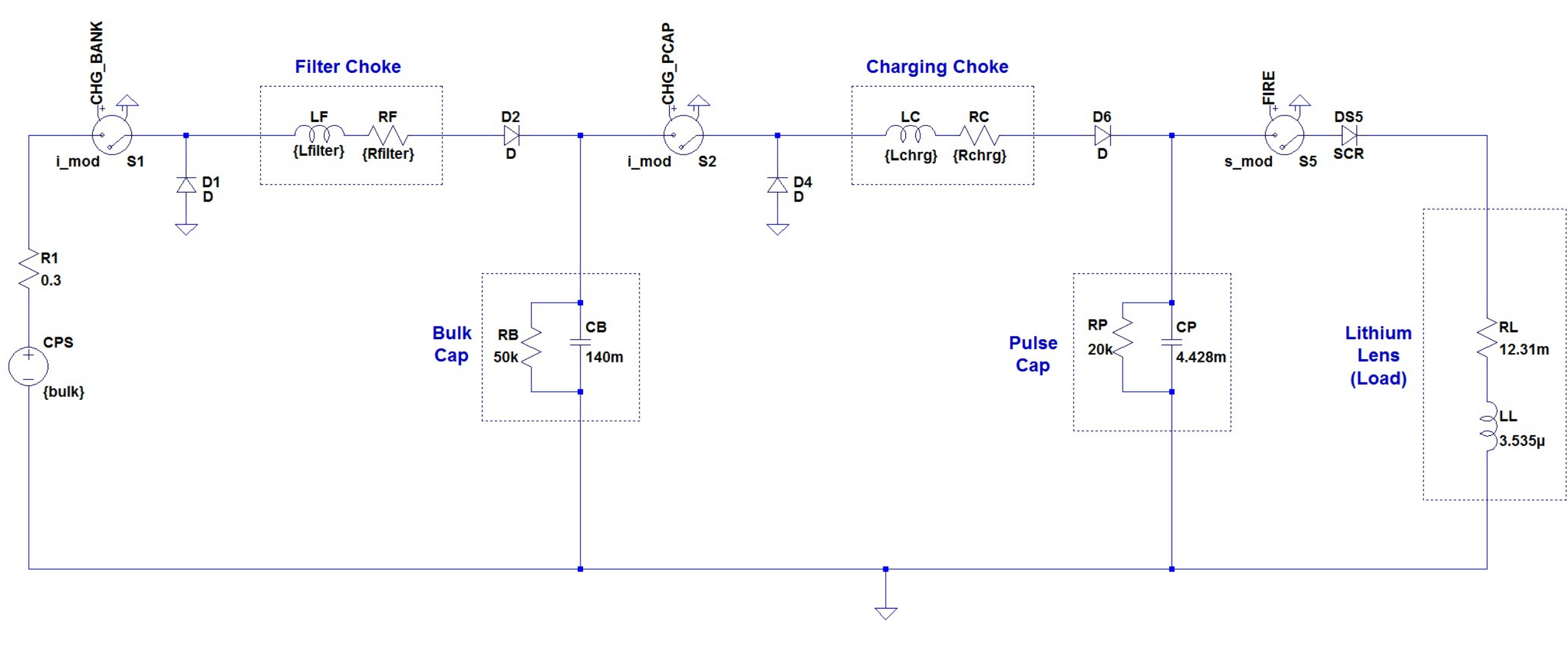

Figure 7.17: Lithium-lens and pulsed-magnet power supply high-level schematic.

In order to accommodate all of the modifications to the existing power supplies, the size of their enclosures must increase and the layout of the components will be changed. Figure 7.18 shows the existing power supplies that are located at the AP0 target hall.

Figure 7.19 shows the layout of the power supplies including the changes to incorporate the new components. An additional enclosure will be added to hold the charging inductors and the new power supply controls. The energy storage section, made up of a parallel array of dry-film capacitors, will be housed in the middle enclosure above the large pulse capacitors which will be reused. Two new phase-controlled DC charging supplies, one each for the lens and PMAG will replace the existing 480-VAC line transformers as shown in Fig. 7.20.

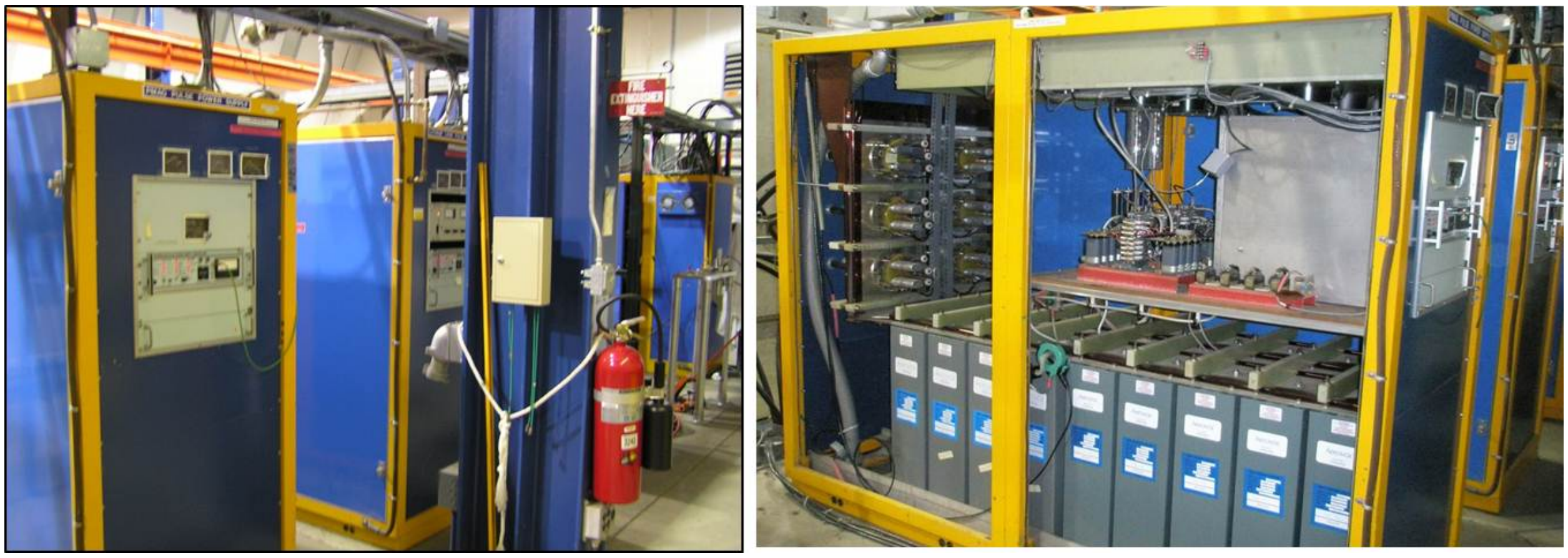

Figure 7.18: Existing lens and PMAG power supplies at AP0.

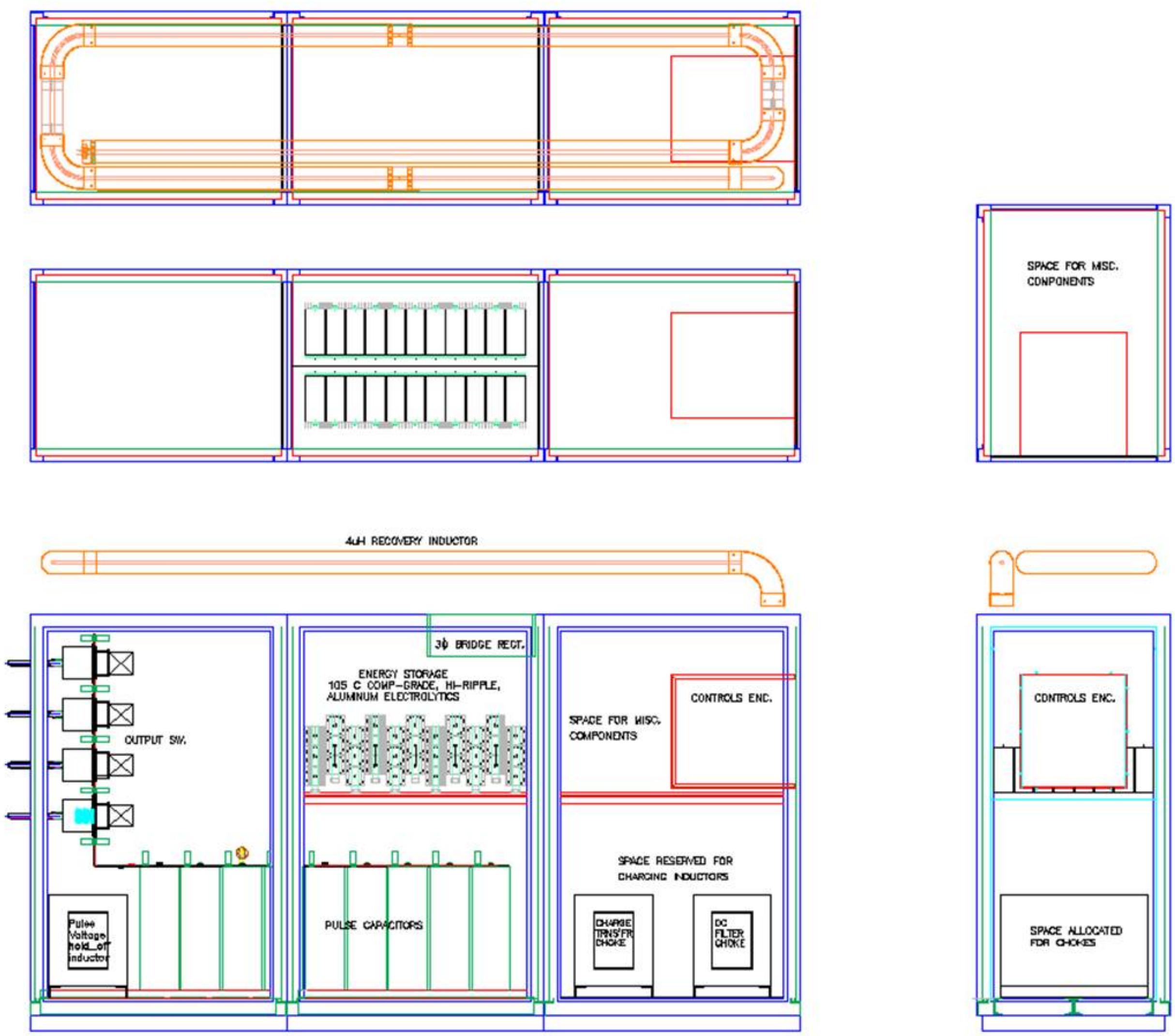

Figure 7.19: Power supply layout including new components. Note that the proposed bulk cap bank will utilize dry film types rather than the electrolytics shown, due to cost and reliability issues.

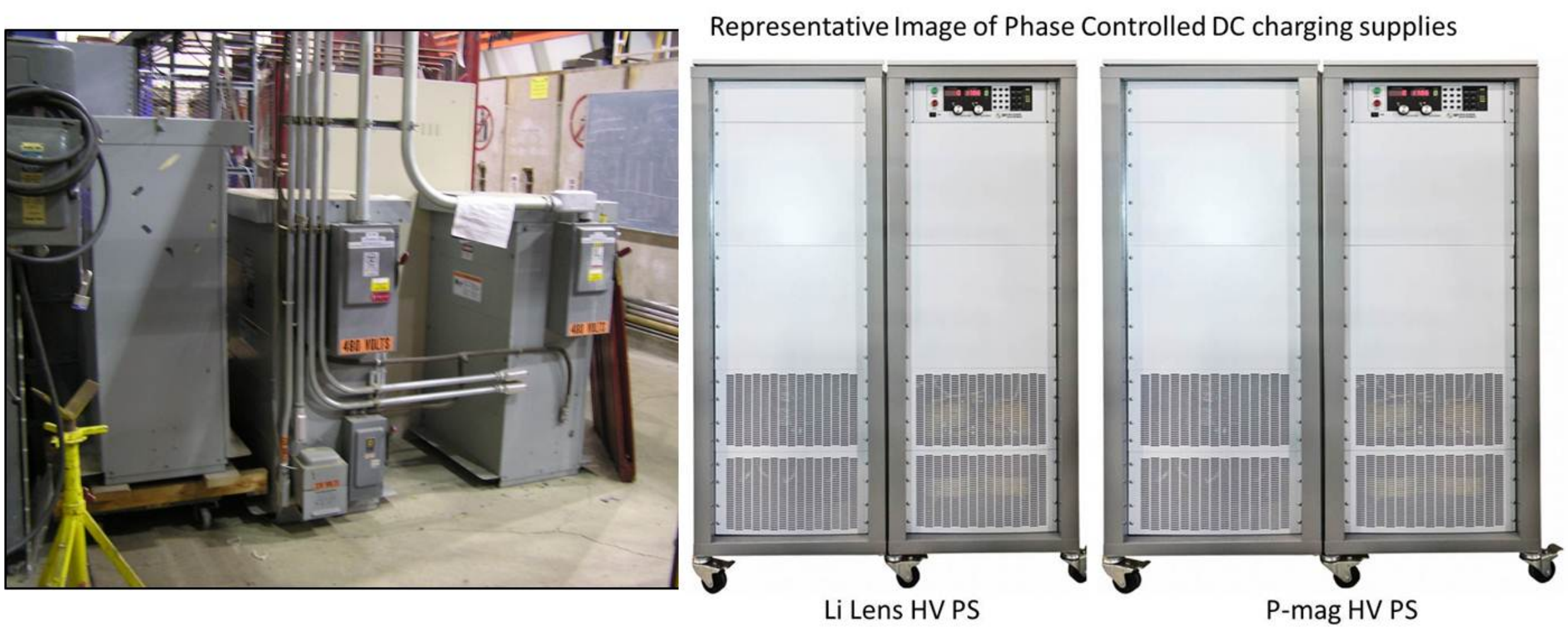

Figure 7.20: Existing 480 VAC transformers *(left)* that will be replaced by new phase controlled DC charging supplies *(right)*.

### 7.4.5 Target station beam dump

The target-station beam dump absorbs particles which are not momentum-selected by the pulsed dipole magnet and continue straight ahead. The location of the beam dump can be seen in Fig. 7.21. The current beam dump has a graphite and aluminum core which is water cooled, surrounded by an outer steel box. The graphite core is 16 cm in diameter and 2 m in length, and is designed to handle a beam power of 80 kW [14]. The existing dump has a known water leak that developed at the end of the collider run, and will be replaced with an updated copy of the 80 kW beam dump, shown in Fig. 7.22. The maximum beam energy load for $(g-2)$ would occur if $(g-2)$ takes advantage of extra cycles, for example if the NO$\nu$A experiment were not able to run. At a rate of 18 Hz, the beam energy load would be 25 kW, which is easily accommodated with the current dump design.

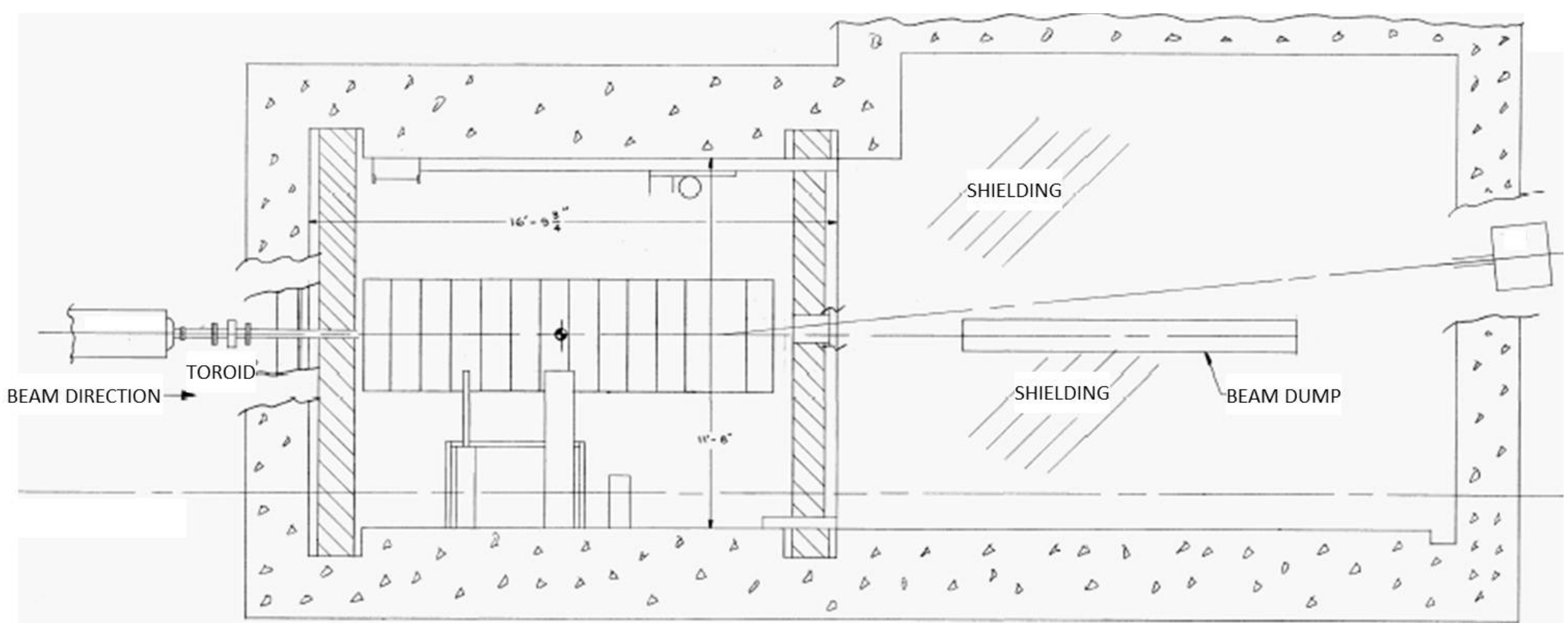

Figure 7.21: Layout of the target-station beam dump.

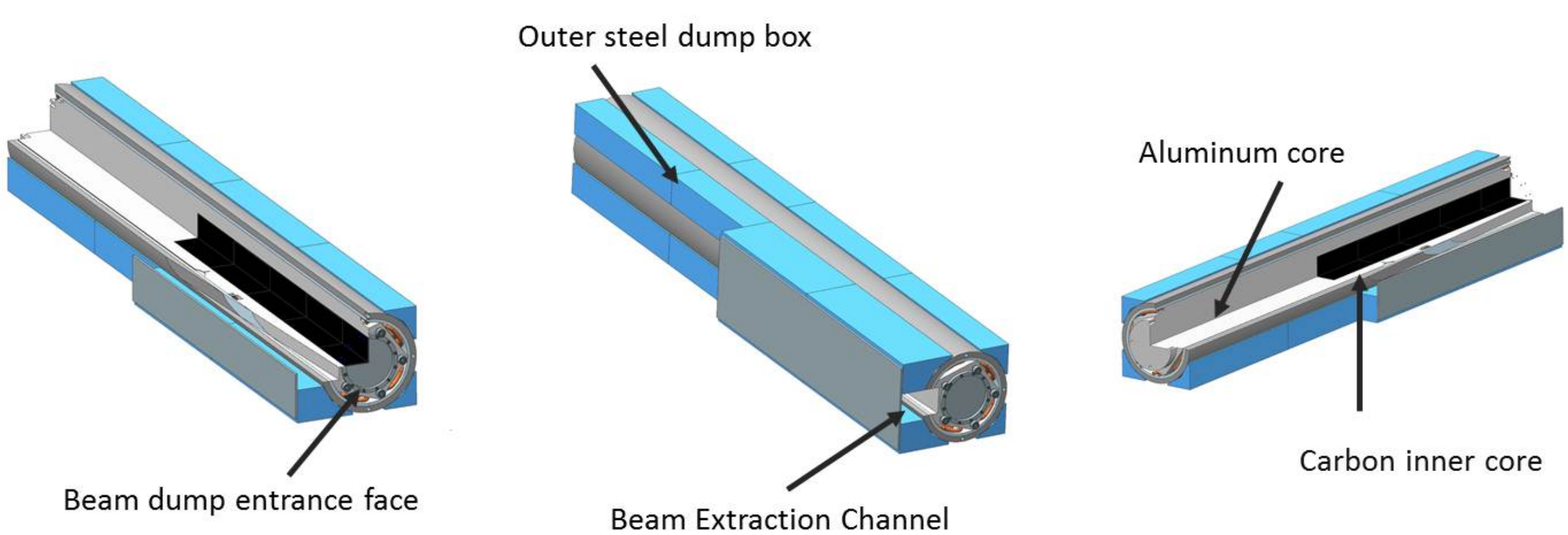

Figure 7.22: Design of the AP0 beam dump core.

### Target station beam dump removal and installation

The AP0 beam dump has been in operation for 27 years and upon removal of the beam dump core from the dump shield it will be found to be highly activated. Therefore, MARS simulations have been conducted on the existing beam dump in order to estimate the residual radioactive dose [15]. An estimate of the residual dose is needed in order to develop a detailed plan for the removal of the existing beam dump and the installation of the new beam dump. A detailed plan for removing the dump will be essential for controlling worker radiation exposure levels and preventing the spread of radioactive contamination. Table 7.6 shows the results of the MARS simulation which estimates the radiation dose rates for different parts of the beam dump core. The sum column represents the upper limit of peak dose rate. Rates at the upstream core on contact are estimated to be 427 rem/h.

| Location | 2001 | 2004 | 2007 | 2011 | 2014 | sum |
|---|---|---|---|---|---|---|
| beam-right dump core | 0.03 | 0.23 | 1.30 | 23.00 | 0.02 | 24.6 |
| beam-left dump core | 0.03 | 0.18 | 1.10 | 18.00 | 0.02 | 19.3 |
| bottom of dump core | 0.03 | 0.20 | 1.20 | 20.00 | 0.02 | 21.4 |
| top locator plate | 0.01 | 0.07 | 0.54 | 12.00 | 0.01 | 12.6 |
| upstream core | 1.30 | 15.00 | 61.00 | 350.00 | 0.06 | 427.4 |
| downstream core | 1.10 | 12.00 | 52.00 | 290.00 | 0.04 | 355.1 |
| right side plug | 0.01 | 0.03 | 0.20 | 4.70 | 0.01 | 4.9 |
| left side plug | 0.00 | 0.02 | 0.18 | 4.10 | 0.01 | 4.3 |
| upstream lower plug face | 0.00 | 0.00 | 0.03 | 0.56 | 0.00 | 0.6 |
| downstream lower plug face | 0.01 | 0.03 | 0.24 | 5.20 | 0.01 | 5.5 |

Table 7.6: Summary of partial peak radiation dose rates in rem/h at contact with various surfaces of the beam dump and beam dump plug. The peak dose rates are taken from the MARS histogram results for each irradiation/cooling period. The upper limit of peak dose rate is indicated in the sum column.

Based on the MARS radiation dose rate results, the plan for removing the beam dump will include constructing a steel coffin that the beam dump will be placed in once removed. The coffin design is shown in Fig. 7.23. The approximate weight of the coffin is 10,200 lb, and the approximate weight of the coffin, dump assembly, and dump locator plate assembly at 13,100 lbs. The coffin is made of 4-in plates of 1018 Steel, cold drawn, UNS G10180 with top or side lift points and a detachable lid.

The beam dump will be removed and placed inside the coffin and then lowered and stored in the bottom of the hot rack storage located at AP0. It is possible that the coffin and beam dump will be transported to the on-site Neutrino Target Service Building (NTSB) in the future for long term storage. Figure 7.24 shows a photo of the front face of the beam dump as taken in 2003 and also of the top of the dump plug below the surface of the shielding blocks.

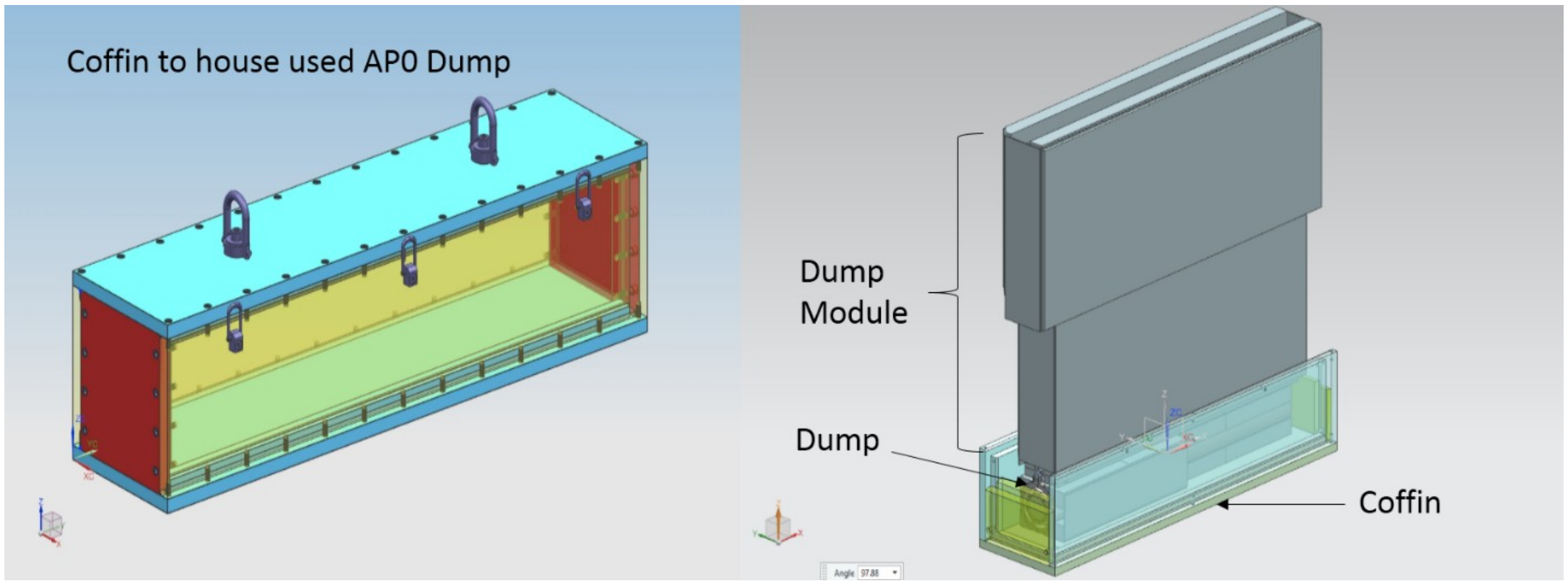

Figure 7.23: Steel coffin that will be used to house the removed beam dump to shield and mitigate radiation concerns of the beam dump.

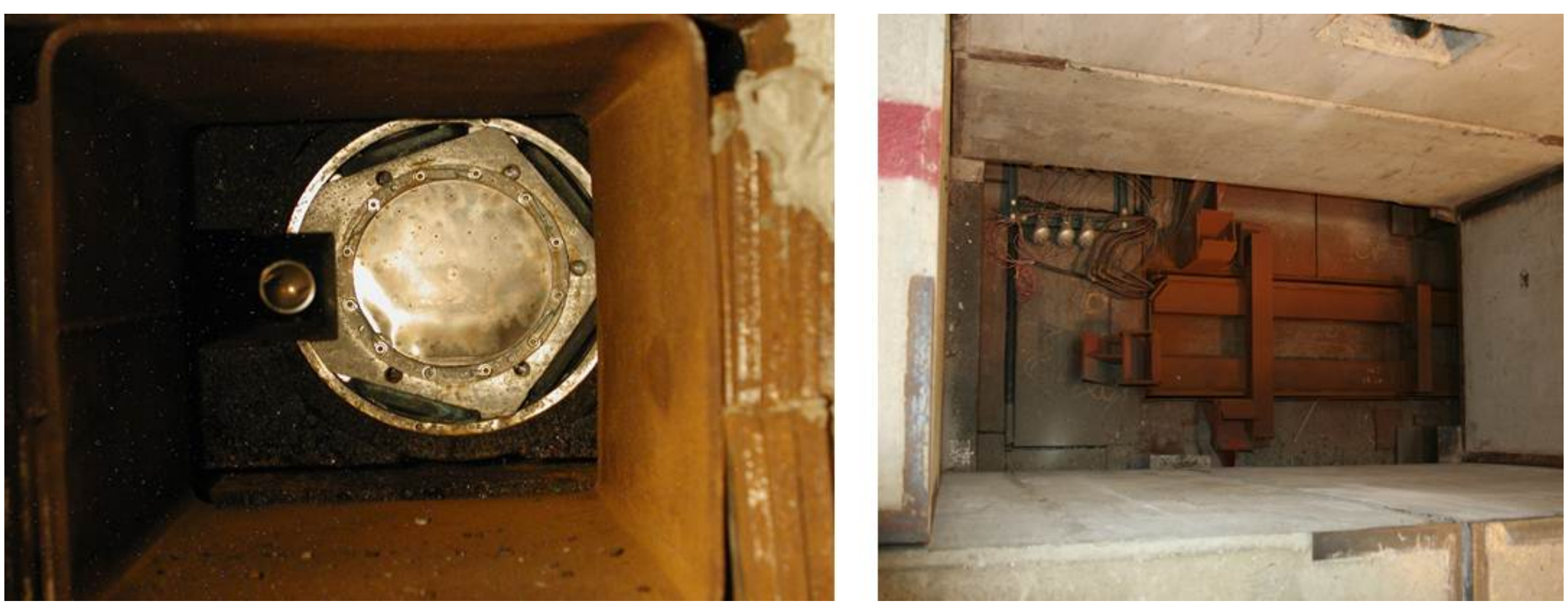

Figure 7.24: *(left)* Photo of the front face of the beam dump as taken in 2003. *(right)* Looking down on the top of the dump plug below the surface of the shielding blocks .

## 7.5 Beam Transport Lines

### 7.5.1 Overview of $(g-2)$ beamlines

The existing tunnel enclosures and beamlines connecting the Recycler Ring to the Delivery Ring will be largely reused for $(g-2)$ operation. However, there are fundamental differences between the way the Rings and beamlines were operated for Collider Operation and how they will be used to support the Muon Campus. A high-intensity, 8 GeV kinetic energy proton beam will be transported to the AP0 Target Station in $(g-2)$ operation and to the Delivery Ring for the Mu2e experiment. The increase in intensity from Collider Operation in conjunction with the beam size of the 8 GeV beam will present challenges for efficient beam transfer. The beamlines downstream of the AP0 Target Station will need to be reconfigured to connect to the D30 straight section of the Delivery Ring. New extraction lines will be constructed to transport beam from the D30 straight section to the $(g-2)$ and Mu2e experiments. Careful planning is required for the D30 straight section of the Delivery Ring due to the presence of both the injection and extraction points. The extraction line will also need to support both single-turn extraction for $(g-2)$ and resonant extraction for Mu2e.

### 7.5.2 Beamline Changes from Collider Operation

During antiproton ("Pbar") operation in Collider Run II, the P1 line connected to the Main Injector at the MI 52 location. The P1 line supported operation with three different beam energies, 150 GeV for protons to the Tevatron, 120 GeV for Pbar production and SY120 operation, and 8 GeV for protons and antiprotons to and from the Antiproton Source. (SY120 refers to the "Switchyard" of beamlines used for the 120-GeV fixed-target program.) The junction between the P1 and P2 lines occurs at F0 in the Tevatron enclosure. The P2 line ran at two different beam energies, 120 GeV for antiproton production and SY120 operation and 8 GeV for protons and antiprotons to and from the Antiproton Source. The P2, P3 (for SY120 operation), and AP1 lines join at the F17 location in the Tevatron enclosure. The AP1 line also operated at 120 GeV and 8 GeV, but is not used for SY120 operation. The AP3 line only runs at a kinetic energy of 8 GeV. The AP3 line connects with the AP1 line in the Pre-Vault beam enclosure near the Target Vault and terminates at the Accumulator.

After the conversion from collider to NO$\nu$A and $(g-2)$ operation, the Recycler will become part of the proton transport chain and will connect directly with the Booster. There will be a new beamline connection between the Recycler Ring and the P1 line. The P1 line will become a dual energy line, with no further need to deliver 150 GeV protons with the decommissioning of the Tevatron. The P2 line will continue to operate at both 8 GeV for the Muon experiments and 120 GeV for SY120 operation. The AP2 and AP3 lines will need to be completely dismantled and reconfigured to support both the transport of muon secondaries via the Target Station for $(g-2)$ and protons via the target bypass for Mu2e. The $(g-2)$ 3.1 GeV secondary beamline emanating from the Target Station and the Mu2e 8 GeV primary beamline bypassing the Target Station will merge and follow a single line to the Delivery Ring. The new injection line will connect to the Delivery Ring in the D30 straight section. The extraction line also originates in the D30 straight section and has to be capable of supporting both resonant and single-turn extraction.

The beamlines that made up the Antiproton Source, those that have an “AP” prefix, will be modified, reconfigured and renamed prior to $(g-2)$ operation. The AP1 line will only operate at an energy of 8 GeV and will be renamed M1. The AP1 line will be largely unchanged, with the exception of the replacement of some magnets to improve aperture and the addition of a Final Focus quadrupole triplet at the end of the line. The AP2 line will become two separate beamlines and no longer be continuous. The upstream end of the line will be part of the pion decay channel for the $(g-2)$ experiment and will be renamed M2. It will provide a connection from the Pbar AP0 Target Station to the M3 line, which will continue the pion decay channel to the Delivery Ring. The downstream section of AP2 will become the abort and proton removal line from the Delivery Ring. The reconfigured AP3 line will be required to transport both 8 GeV beam for the Mu2e experiment and also a 3.1 GeV secondary beam for the $(g-2)$ experiment and will be renamed M3. The 18.5° right bend will be changed from a three to a two dipole configuration in order to avoid higher beta functions in this region. The M3 line will also be modified to connect to the Delivery Ring (formerly Debuncher) instead of the Accumulator. The extraction line connecting the Delivery Ring to the experiments will be called M4. The M5 line will vertically branch from the M4 line shortly after leaving the Delivery Ring and continue to the $(g-2)$ storage ring in the MC-1 building. Figure 7.25 compares the Pbar beamline configuration with that to be used for $(g-2)$ and Mu2e operation. In general, the AP1, AP2 and AP3 lines will refer to the old Pbar beamline configuration and M1, M2, M3, M4 and M5 will refer to the beamline configuration for $(g-2)$ operation.

Figure 7.26 shows another view of the Muon Campus beamlines, including the experimental halls.

Most of the common improvements to the beamlines and Delivery Ring that benefit Mu2e, $(g-2)$, and future experiments will be incorporated into several Accelerator Improvement Projects (AIPs). They are the Recycler RF AIP, Cryo AIP, Beam Transport AIP, and Delivery Ring AIP. The Recycler RF AIP, as the name implies, will provide an RF system in the Recycler that is capable of forming the short 2.5 MHz bunches required by the experiments. The Cryo AIP provides cryogenics for the $(g-2)$ storage ring and to the Mu2e solenoids. The Beam Transport AIP will complete a connecting beamline between the Recycler and P1 Line and will be responsible for replacing magnets at key locations to improve aperture. The Delivery Ring AIP has numerous improvements that are of common benefit to both $(g-2)$ and Mu2e, such as the injection and abort/proton removal systems. Table 7.7 summarizes which improvements are contained in the various AIPs, as well as those that will be managed as part of the Mu2e and $(g-2)$ projects. Project Managers for the various projects will work closely together to ensure they interface properly. Virtually all of the work that is incorporated into the AIPs must be completed prior to beam operation to $(g-2)$.

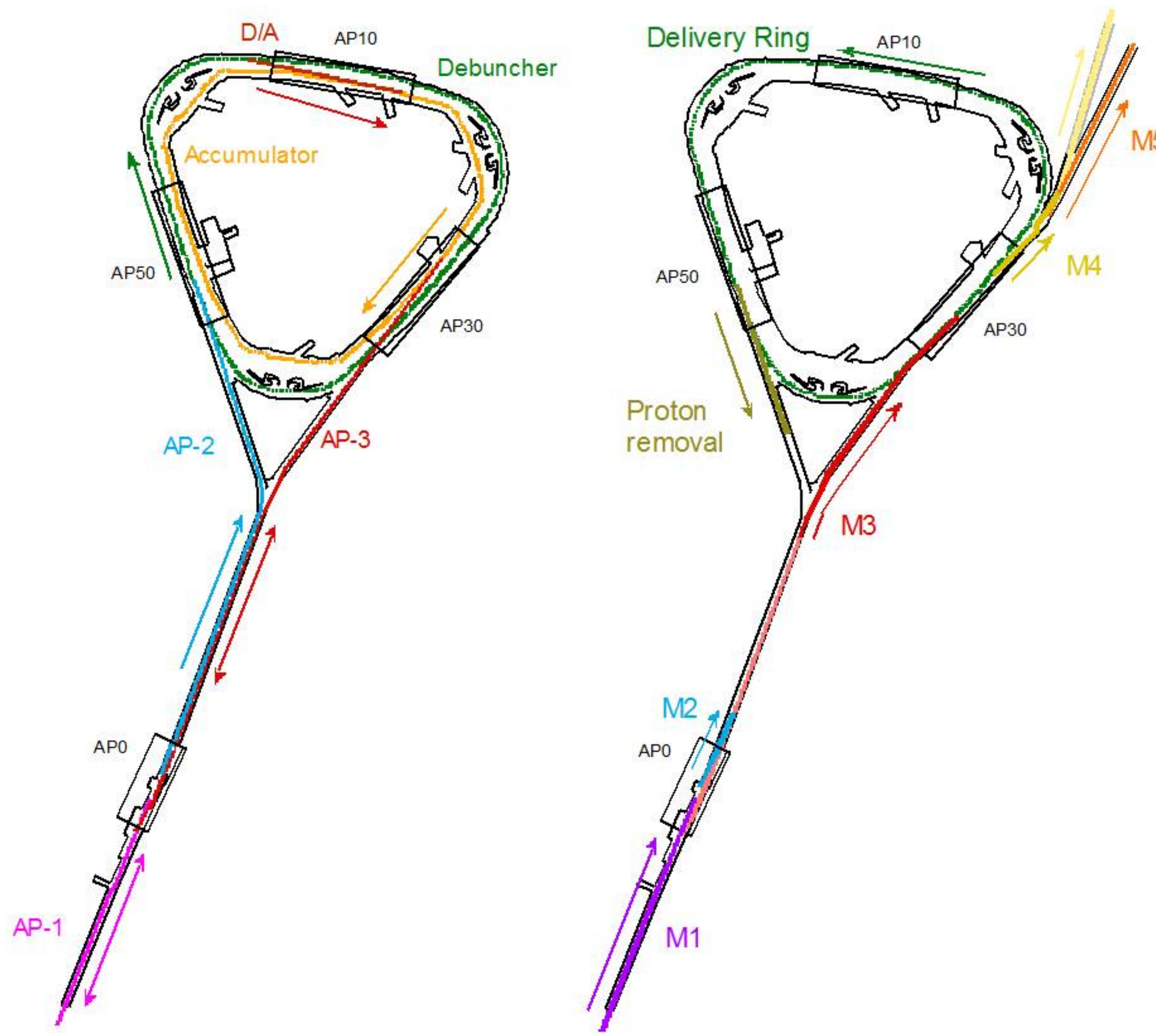

Figure 7.25: Layout of the Antiproton Source beamlines (left) and the reconfigured beamlines for $(g-2)$ operation (right).

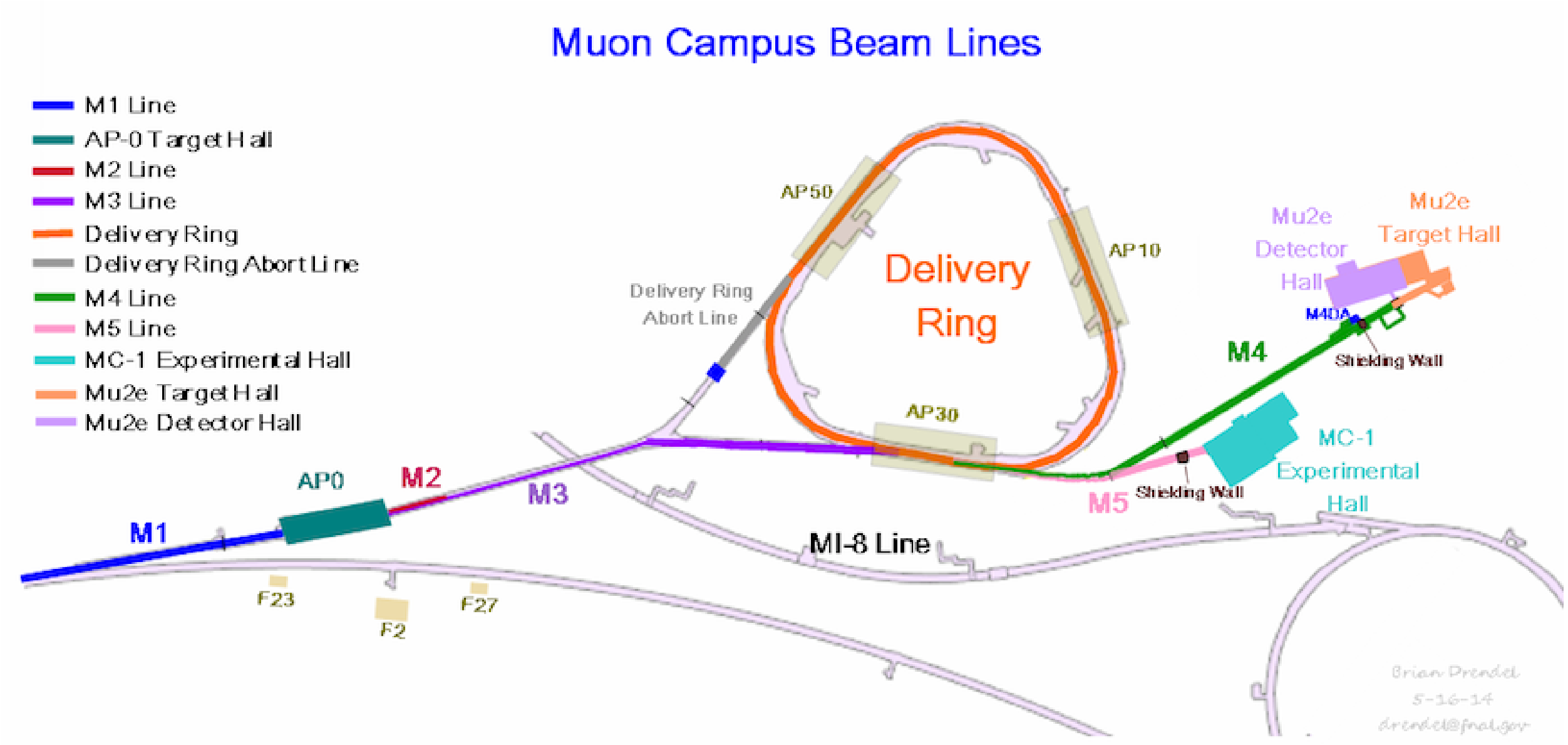

Figure 7.26: The Muon Campus beamlines and experimental halls.

| Description | Project | Comment |
|---|---|---|
| Cryogenics | CR AIP | |
| Recycler RF upgrade | RR AIP | |
| Recycler extraction/P1 stub line | BT AIP | |
| P1,P2 and M1 aperture upgrade | BT AIP | **M1 final focus quadrupoles on $(g-2)$** |
| **Reconfigure AP2 and AP3** | $(g-2)$ | **New lines are called M2 and M3** |
| **Final focus to AP0 Target Station** | $(g-2)$ | |
| **AP0 Target Station upgrades** | $(g-2)$ | |
| Beam transport instrumentation | BT AIP | |
| Beam transport infrastructure | BT AIP | |
| Delivery Ring injection | DR AIP | |
| **D30 straight section preparation** | $(g-2)$ | |
| Delivery Ring modification | DR AIP | |
| DR abort/proton removal | DR AIP | |
| *Delivery Ring RF system* | *Mu2e* | |
| Delivery Ring controls | DR AIP | |
| Delivery Ring instrumentation | DR AIP | *DCCT and Tune measure are Mu2e* |
| *Resonant extraction from DR* | *Mu2e* | |
| **Fast extraction from DR** | $(g-2)$ | |
| Delivery Ring infrastructure | DR AIP | |
| **Extraction line to split** | $(g-2)$ | **Upstream M4 line** |
| *Extraction line from split to Mu2e* | *Mu2e* | *Downstream M4, including extinction* |
| **Extraction line from split to $(g-2)$** | $(g-2)$ | **Beamline to MC-1 building** |

Table 7.7: Beamline, Delivery-Ring, and other upgrades and associated project: $(g-2)$ project, Mu2e project, Delivery Ring Accelerator Improvement Project (DR AIP), Beam Transport (BT) AIP, Recycler RF (RR) AIP, and Cryo (CR) AIP.

### 7.5.3 Proton Beam Transport to the Target Station

Beam transport of the 8 GeV primary beam from the Recycler Ring (RR) to the Target Station closely resembles the scheme used to transport 120 GeV protons for antiproton production in Collider operation. The most notable differences are the change in beam energy and the switch from the Main Injector to the RR as the point of origin for the P1 line. The beamlines will be modified to 1) provide a connection between the RR and P1 line, 2) improve aperture to accommodate the larger beam size and intensity, and 3) reconfigure the final focus region in order to reach the desired spot size on the production target. Table 7.8 lists the beamlines connecting the RR with the Target Station and their respective lengths.

| Beam Line | Length (m) |
|---|---|
| RR to P1 | 43 |
| P1 | 182 |
| P2 | 212 |
| AP1 (M1) | 144 |
| RR to Target Total | 581 |

Table 7.8: Recycler Ring to Target beamline lengths.

#### Recycler Ring to P1 line stub

Operation of $(g-2)$ and Mu2e requires the transport of protons from the RR rather than the Main Injector. A new transfer line from the RR to the P1 beamline will be constructed to facilitate proton beam transport from the RR to the Delivery Ring. This new beamline provides a way to deliver 8 GeV kinetic energy protons to the Delivery Ring, via the RR, using existing beam transport lines and without the need for new civil construction.

**Beamline Design** The P1 line is lower in elevation than the RR, thus the beam will be extracted downward. This will be accomplished with a horizontal kicker that will displace beam into the field region of a Lambertson magnet that will bend beam down. The kickers are located immediately downstream of the RR 520 location and the Lambertson will be just downstream of the RR 522 location. Due to space limitations, only two vertical bend centers made up of the Lambertson and a dipole are used in the new line. An integer multiple of 360° in betatron phase advance between the two bending centers is required to cancel the vertical dispersion from the bends. The new beamline needs to intercept the existing P1 line in a location that doesn't disturb the extraction trajectory from the Main Injector, which will be retained for SY120 operation. That junction point will be located near quadrupole Q703.The angles of both the Lambertson and the vertical bending magnet (V903, VBEND in Fig. 7.27) were obtained by matching the site coordinates from the RR to P1 line using TRANSPORT [16] code. Figure 7.27 shows the layout of the new line, with the existing P1 line drawn in red.

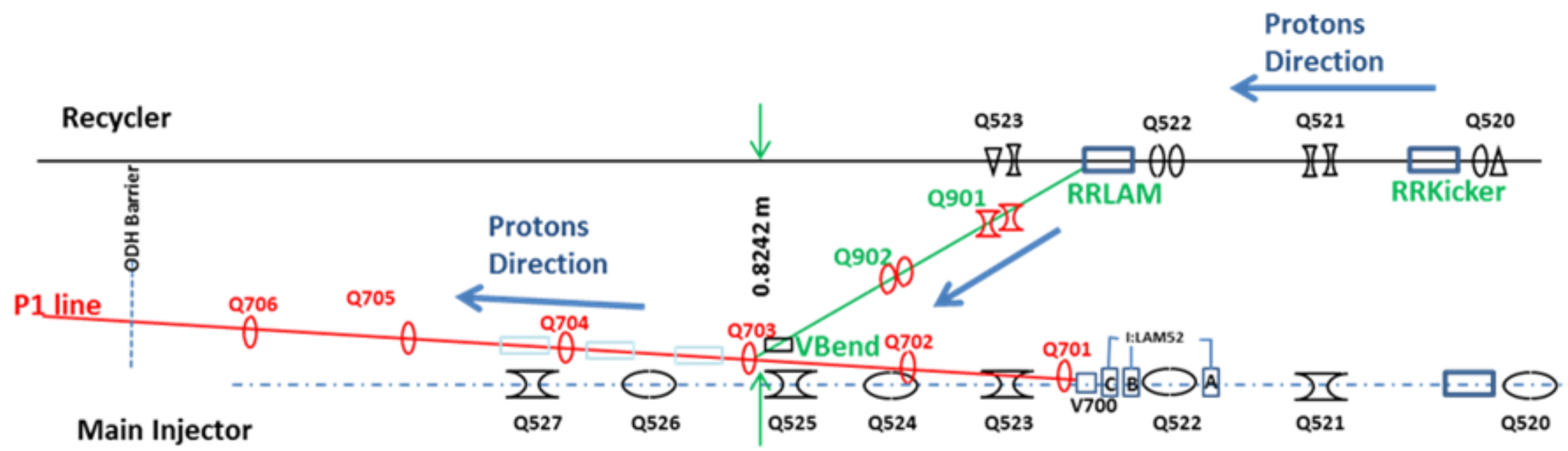

Figure 7.27: The new Recycler Ring to P1 connecting beamline.

**Kickers** The $(g-2)$/Mu2e extraction kicker will be of the same design as the kickers used during collider operation, but will be potted instead of using Fluorinert for electrical insulation. The physical dimensions and properties of the kickers are listed in Table 7.9. The plan is to reuse the ceramic vacuum chamber from old RR kicker magnets, which are slightly smaller than the standard RR vacuum chamber. The kicker system will be made up of two magnets producing 0.79 mr bend each for a total kick of 1.58 mr. The new kicker power supplies will be located in the MI-52 service building. Power supplies for the new beamline magnets will also be located at MI-52. This service building will be expanded to accommodate the new power supplies.

| Recycler Extraction Kicker RKB-25 | |
|---|---|
| Parameter | Value |
| Ferrite length | 46.6 in |
| Case length | 64.0 in |
| Insert length | 67.78 in |
| Print number | ME-481284 |
| Maximum strength (each) | 0.279 kG m |
| Maximum kick (each) | 0.94 mr @ 8 GeV/$c^2$ |
| Required kick (each) | 0.79 mr @ 8 GeV/$c^2$ |
| Rise time, 3% - 97% | 140 ns |

Table 7.9: RR extraction kicker parameters.

**Lambertson** The Lambertson magnet will be rolled 2.7° from the vertical and V903 rolled -4.0° to provide a small horizontal translation in order to create the proper horizontal trajectory required to match to the P1 line. The V903 dipole magnet is a 1.5-m long "ADCW"-type that will provide a 21 mr bend, matching the bend of the Lambertson. There will be two permanent quadrupoles and two electromagnetic trim quadrupoles located between the Lambertson and vertical dipole magnets that make up the dogleg between the RR and P1 line. Due to space constraints, the permanent quadrupoles are shifted downstream from their ideal locations by 0.25 m. A more detailed technical description of the design features of the new beam line stub can be found in Ref. [17]. Figure 7.28 shows the lattice functions from the Recycler to the AP0 Target Station.

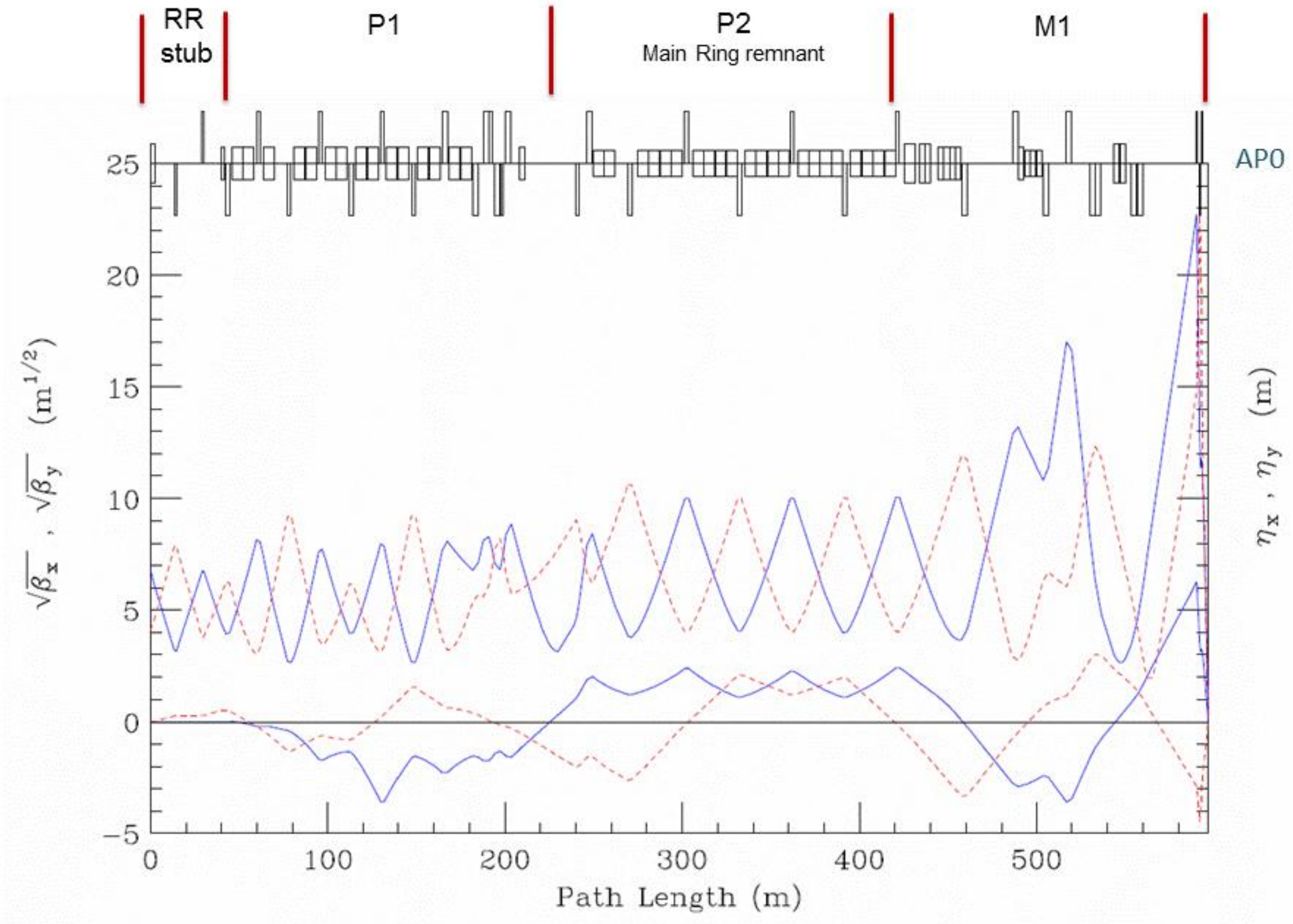

Figure 7.28: Lattice functions for primary beamlines from the Recycler to the Target Station.

**Recycler orbit** The RR extraction scheme incorporates a permanent horizontal 3-bump in the RR to improve aperture. The bump displaces the circulating beam outward 25 mm at the upstream end of the Lambertson (RLAM). Figure 7.29 shows the trajectories of the circulating and extracted beams, including the horizontal bump at the Lambertson. The bump is created by horizontal trim dipoles at the 524, 522 and 520 locations. The extraction kickers displace the extracted beam inward 25 mm at the same location. This creates a separation of the RR circulating beam and extracted beam at the front face of the Lambertson of 50 mm.

**Apertures** The Recycler extraction Lambertson has an adequate aperture for both the circulating and extracted beams. Figure 7.30 shows the footprint of both beams at the Lambertson for both a $10\sigma$ and $6\sigma$ beam size. The vertical bend magnet has a relatively small physical horizontal aperture, but is located where the horizontal beta functions are also small. The horizontal acceptance of the vertical dipole is actually larger than that of the Lambertson, despite the smaller physical aperture. The quadrupole and trim magnets are modeled after those in the Recycler and have good apertures.

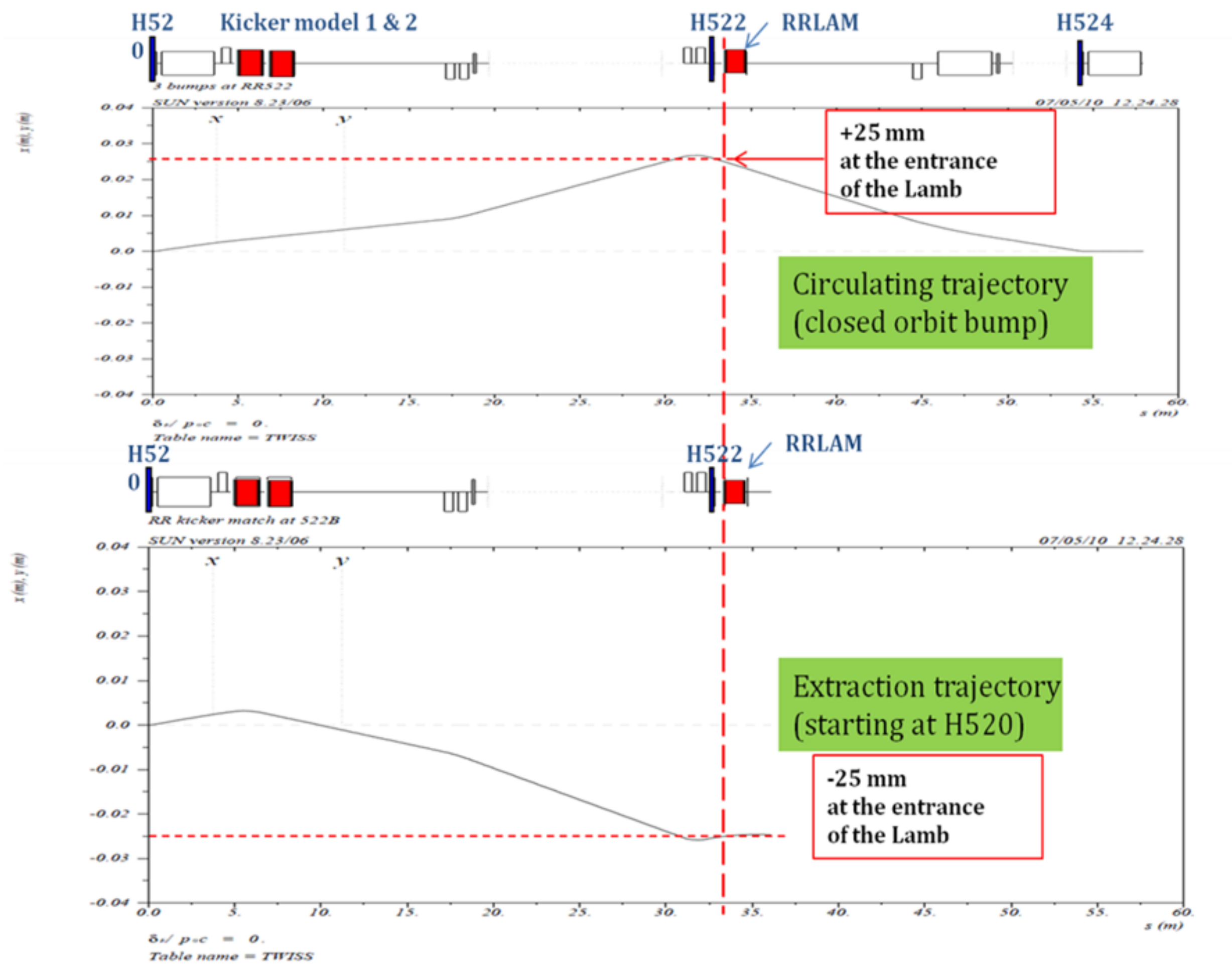

Figure 7.29: Horizontal trajectories for circulating and extracted beam from the Recycler.

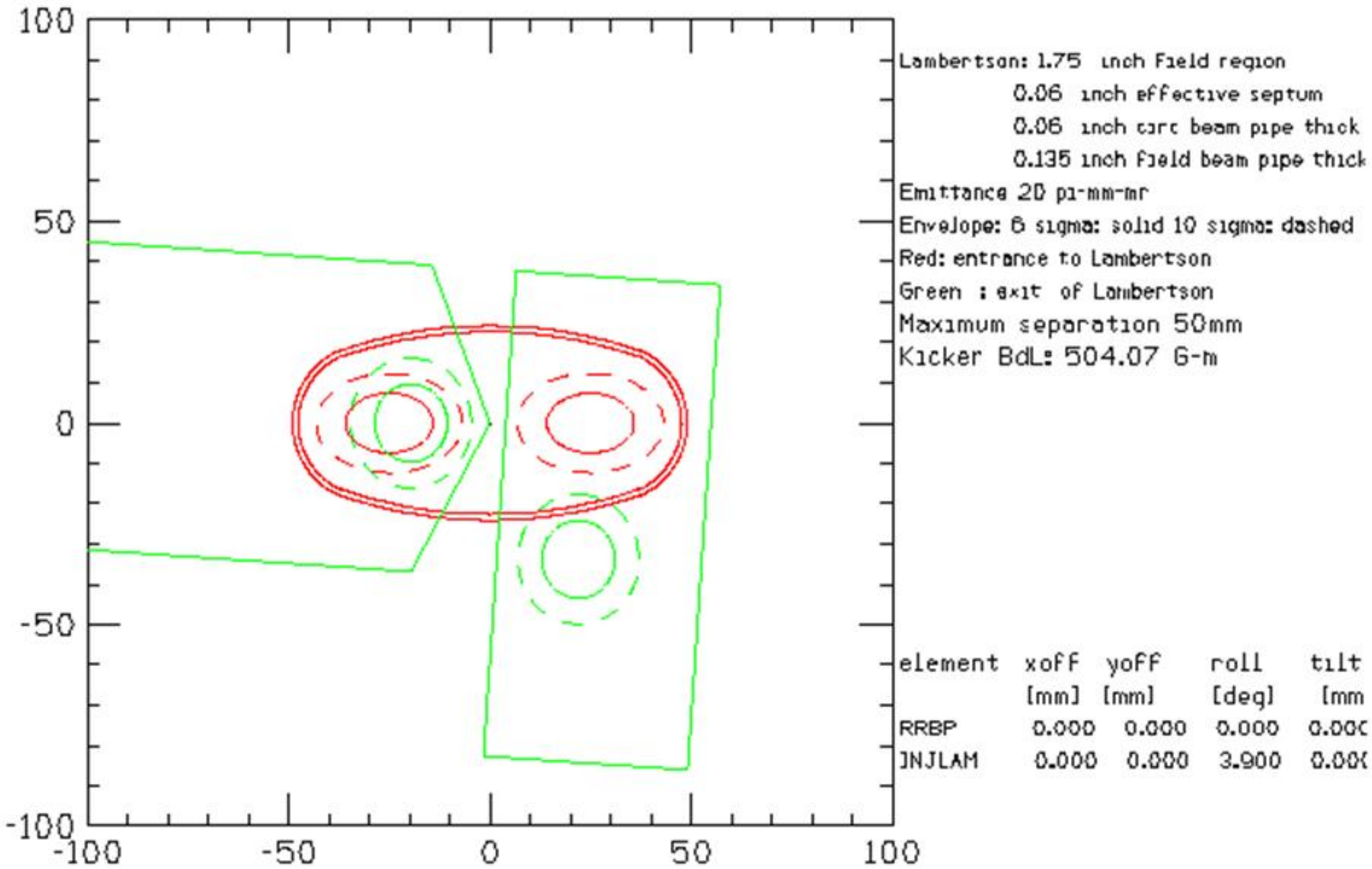

Figure 7.30: Beam sizes at the entrance (red) and exit (green) of the extraction Lambertson. The dashed outline represents $10\sigma$ and the solid outline $6\sigma$ beam for a normalized emittance of $18\pi$ mm-mr.

### 7.5.4 P1, P2 and M1 Aperture Improvements

The increased intensity and beam size planned for muon operation will lead to unacceptably high beam loss unless apertures are improved in the P1, P2 and M1 lines. Limiting apertures in these beamlines were identified during Collider Run II which simplified the process of identifying locations to improve for Muon operation. The elimination of M1 120 GeV operation for antiproton stacking provides an opportunity to improve the aperture utilizing weaker magnets that previously were not practical for use as replacements.

The introduction of the P1-line stub has eliminated several aperture restrictions that were associated with Main Injector extraction. In particular, the vertical C-magnets that follow the MI-52 Lambertson will be avoided with the new stub line. Most of the P1 line after the P1-line stub has good aperture, until the former junction area with the now decommissioned Tevatron. The vertical dipole at the 714 location was installed as a C-magnet because of its proximity with the Tevatron and has a small horizontal aperture. The decommissioning of the Tevatron allows the replacement of this magnet with a conventional dipole that will increase the horizontal acceptance by more than 50%. The new magnet must also be capable of producing enough field strength to operate at 120 GeV and support SY120 operation. The four Tevatron F0 Lambertsons will no longer be needed to inject protons into the Tevatron and can be removed to improve the aperture, also in the horizontal plane.

The P2 line will remain a dual-energy line supporting $(g-2)$ and SY120 operation, so the junction between the P2, M1, and P3 beamlines at F17 will remain. The aperture for both $(g-2)$ and SY120 operation will substantially improve with the proposed replacement of the F17 C-magnets with a large aperture CDA dipole magnet that both beams will pass through. The B3 dipole at the F-17 location has good aperture and will remain; the B3 and CDA will be bused together and run from one power supply. The B3 and CDA are powered to bend beam into the M1 line and not powered for SY120 beam that will continue from the P2 line into the P3 line. Figure 7.31 shows a comparison of the existing and planned magnet layouts at F17.

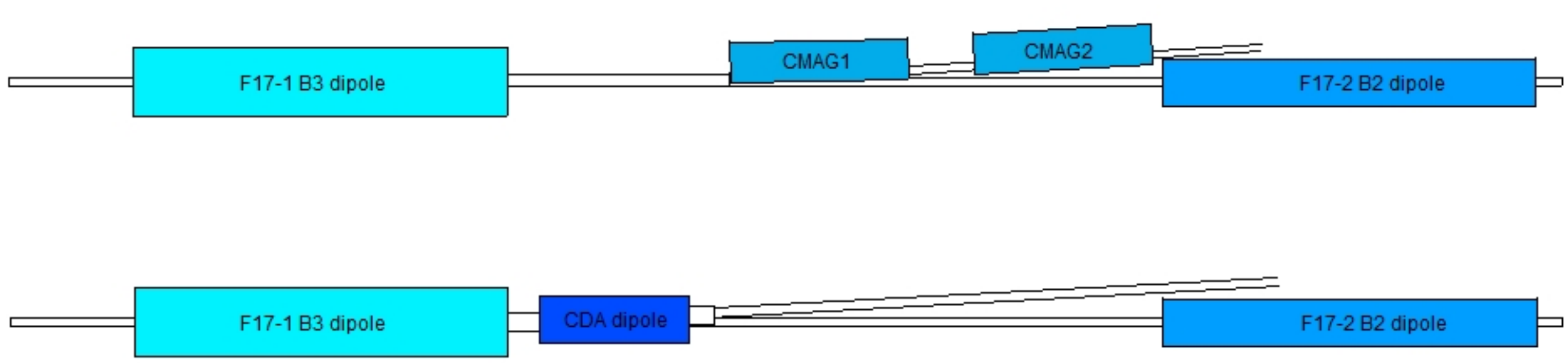

Figure 7.31: Existing *(upper)* magnet layout with two small aperture C-magnets and the planned *(lower)* layout with a single large aperture CDA dipole.

M1 will only operate at 8 GeV for $(g-2)$ operation, so the eight small-aperture EPB dipole magnets that make up the HV100 and HV102 strings can be replaced with larger-aperture, weaker dipoles. The number of dipoles can be reduced from four to two in each string. The 1.5 m MDC dipole magnets have a pole gap that is 2.25 in instead of 1.5 in and provides more than a factor of two increase in acceptance. Several trims will also be replaced or relocated to complete the aperture upgrade. Replacement trims will be repurposed from the Accumulator. The final-focus region at the end of M1 is described separately in the next section. Table 7.10 is a listing of M1 Line magnets and highlights the magnets that have been changed to improve the physical apertures in the RR to Target Station lines. Reference [17] has a more detailed explanation of the devices used to improve the aperture and how the improvements will be implemented.

| Magnet | Type | Current (A) | Power Supply |
|---|---|---|---|
| F17B3 | B3 | 280.0 | I:F17B3 |
| **F17CDA** | **CDA** | **280.0** | **I:F17B3** |
| HT100 | SY Bump | 25.0 | M:HT100 |
| **HV100A** | **MDC** | **92.3** | **M:HV100** |
| **HV100B** | **MDC** | **92.3** | **M:HV100** |
| Q101 | 3Q120 | 4.8 | M:Q101 |
| VT11A | SY Bump | 25.0 | M:VT101A |
| **VT101** | **NDA** | **25.0** | **M:VT101** |
| Q102 | 3Q120 | 3.2 | M:Q102 |
| **HV102A** | **MDC** | **87.3** | **M:HV102** |
| **HV102B** | **MDC** | **87.3** | **M:HV102** |
| Q103 | 3Q120 | 7.2 | M:Q103 |
| Q104 | 3Q120 | 10.3 | M:Q104 |
| Q105A | 3Q120 | 2.6 | M:Q105 |
| Q105B | 3Q120 | 2.6 | M:Q105 |
| **M:HT105** | **NDA** | **25.0** | **M:HT105** |
| V105A | EPB | 56.1 | M:V105 |
| V105B | EPB | 56.1 | M:V105 |
| Q106A | 3Q120 | 0.9 | M:Q106 |
| Q106B | 3Q120 | 0.9 | M:Q106 |
| Q107A | 3Q120 | 0.0 | M:Q107 |
| Q107B | 3Q120 | 0.0 | M:Q107 |
| **VT108A** | **NDA** | **25.0** | **M:VT108** |
| **HT108A** | **NDA** | **25.0** | **M:HT107** |
| **VT108B** | **NDB** | **25.0** | **M:VT108** |
| **HT108B** | **NDB** | **25.0** | **M:HT107** |
| **Q108A** | **SQD** | **369.2** | **M:Q108A** |
| **Q108B** | **SQE** | **421.9** | **M:Q108B** |
| **Q108C** | **SQD** | **372.7** | **M:Q108C** |

Table 7.10: M1-line dipoles, quadrupoles, and trims (HT and VT prefix). Magnets that were changed to improve aperture are shown in bold (the old Q108A&B and Q109A&B 3Q120 quadrupoles were also removed).

### Final Focus Region

The desired spot size on the production target, a proton beam $\sigma$ in both planes of 0.15 mm, is the same as what was used in antiproton production during collider operation. Because the beam momentum is 8.89 GeV/c for $(g-2)$ operation instead of the 120 GeV/c that was used for antiproton production, much smaller beta functions are required to achieve this spot size (0.068 m vs. 0.878 m, respectively). The existing quadrupole configuration in AP1 cannot produce the desired spot size and will need to be changed in order to achieve the desired spot size with good beam transmission. Figure 7.32 shows a modified version of the scheme proposed in Ref. [18], where a quadrupole triplet replaces the last quadrupole, PQ9B, in the AP1 line. Figure 7.32 shows the optics of the entire M1 line, with the final focus occurring on the far right. The quadrupoles making up the triplet need to be as short as possible while concurrently producing a very strong integrated gradient. The PQ8A&B and PQ9A magnets are not powered and can be removed to improve aperture. Larger aperture NDA and MDB trim magnets from surplus Pbar inventory will replace HT107 and VT108 to provide adequate aperture.

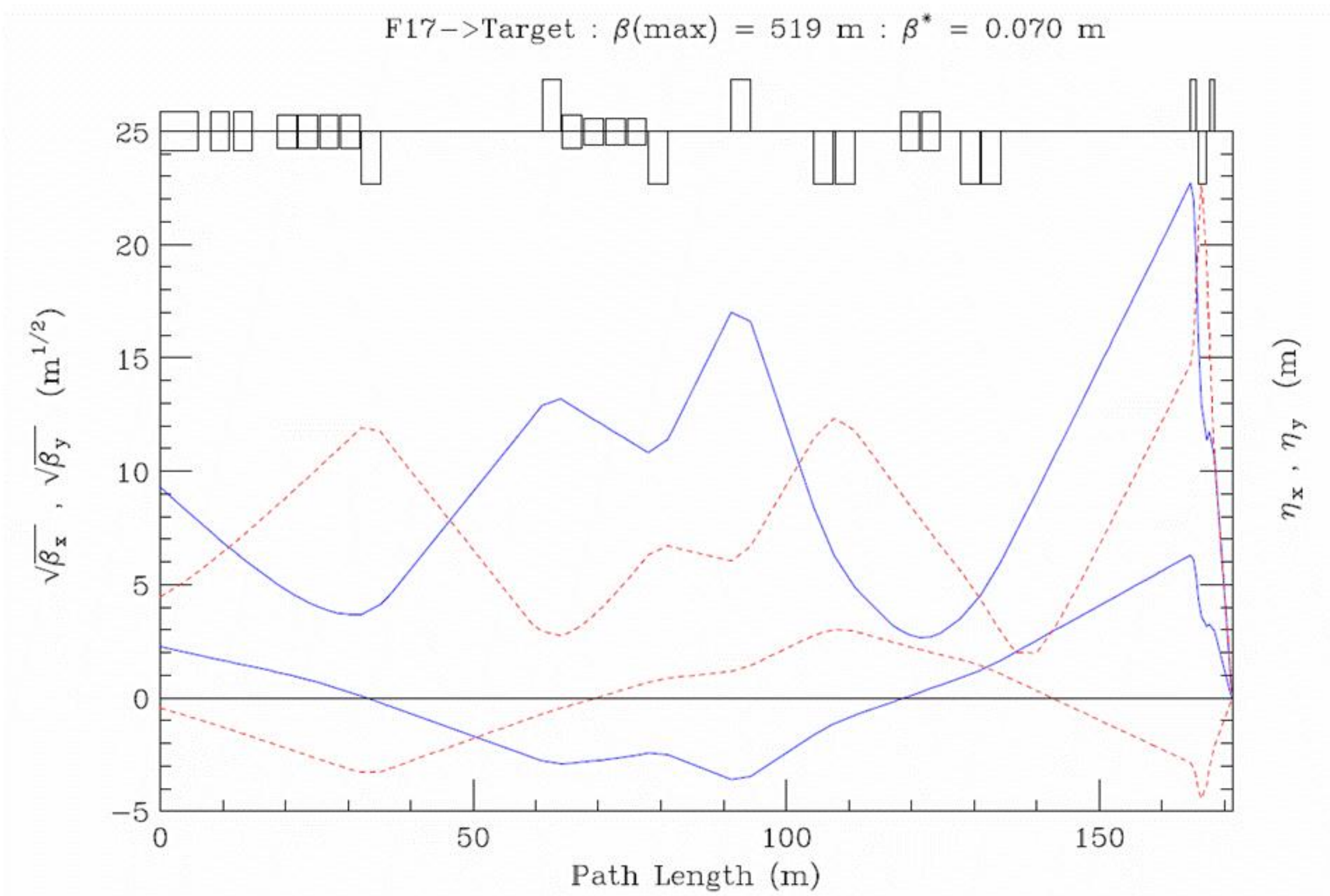

Figure 7.32: Beta functions (horizontal is blue, vertical is red in upper traces) and dispersion functions (horizontal is blue, vertical is red in lower traces) for the M1 line. The Final Focus quadrupole triplet is on the far right side of the plot.

The best compromise between maximizing integrated field, minimizing quadrupole length and providing adequate aperture, from available magnets, is to use a triplet made of an SQD – SQE – SQD combination. The three magnets will all be repurposed from the decommissioned Pbar AP2 and AP3 lines. The quadrupoles are required to run up to 425 A in order to achieve the desired 0.15-mm spot size, which is close to the highest currents these types of magnets have ever operated at. The SQE magnet in the middle of the triplet is the strongest Pbar quadrupole available and operates at the highest current of the triplet quadrupoles.

### 7.5.5 M2 and M3: Pion to muon decay beamlines

The M2 and M3 lines are designed to capture as many magic-momentum muons from pion decay as possible. The M2 line will be rebuilt from parts of the former AP2 line, which transports secondary beam from the Target Station. The M3 line, primarily rebuilt from the former AP2 and AP3 lines, begins as a target-bypass which will be used by the Mu2e experiment to transport primary 8-GeV protons. For $(g-2)$, the M2 line crosses the tunnel into the M3 line at the upstream end of the Transport Enclosure. Focusing of the secondary beam within the Target Station is limited by available space in the target vault. Immediately following the Target Station, the M2 line starts with an existing series of four quadrupoles, followed by eight more quadrupoles and a dipole, which then match into the lattice of the M3 line. Figure 7.33 shows the existing Pbar beamlines versus the new configuration with the M2 line merging into the M3 line.

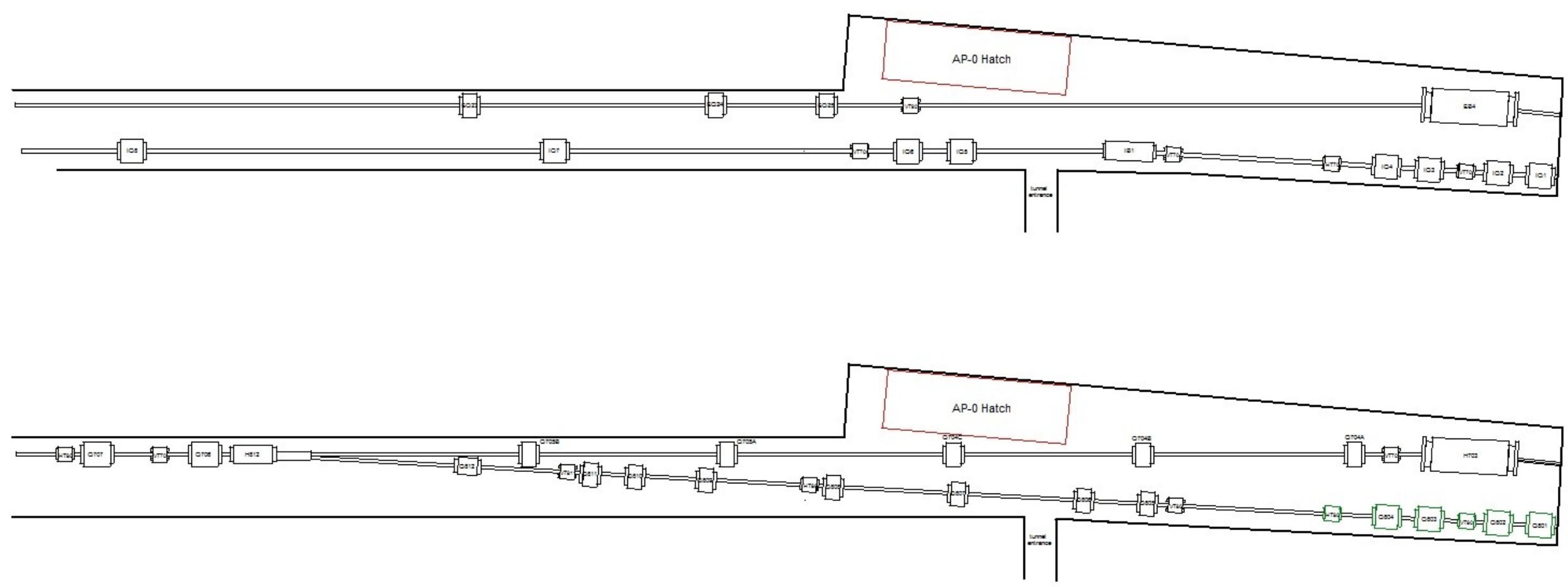

Figure 7.33: Present AP2 and AP3 configuration downstream of Target Vault *(top)* and M2 line merging into M3 line *(bottom)*.

#### Design layout

With the exception of a few specialized insertions, the upstream M2 and M3 lines largely track the trajectories of the former AP2 and AP3 lines. The first 22 m of M2, including the Target Station, remains unchanged from the AP2 configuration. Pions collected from the lithium lens are transported into the M2 line via the existing PMAG dipole, which bends the beam through 3° (52 mr). A second dipole from AP2, which had bent the beam another 3° to align it with the left side of the tunnel, has been removed. The M2 line continues across the tunnel, intersecting the M3 line 28 m further downstream (50 m total M2-line length from the lithium lens to the M3 line). At the intersection point, a large aperture 5D32 dipole provides the second 3° bend required for beam to match the M3-line trajectory. The magnet layout in the vicinity of the M2/M3 merge is constrained by the physical dimensions of the magnets, summarized in Fig. 7.34.

To cancel the horizontal dispersion created by PMAG, there is 540° of horizontal phase advance between the PMAG dipole in the Target Station and the 5D32 switch dipole at the M2/M3 merge. The 5D32 dipole has an unusually large pole width that can accommodate both the beams from the M2 line and the upstream M3 line. The magnet is simply switched on and off to change between the $(g-2)$ and Mu2e modes of operation. The 5D32 dipole also has a 4.125-in pole gap that can provide more than $40\pi$-mm-mr acceptance with the 24-m vertical beta function at the downstream end of the magnet. The optics of the M2 line and the M3 line immediately after the merge is shown in Fig. 7.35.

The M3 line between the M2/M3 merge and Right Bend insertion has short matching sections on each end with a long transport FODO section making up most of that part of the line. The FODO cells are characterized by a 90° phase advance and 5.50-m half-cell length. The M3 line upstream of the Right Bend follows the path of the old AP3 line and can be built with existing AP2 and AP3 devices. This part of M3 must maintain small beta functions to serve as a continuation of the muon decay channel. Since the M3 line must also operate with 8-GeV protons for Mu2e operation, scaled magnet currents must be within power-supply and magnet operating limits. Figure 7.36 shows the beta and dispersion functions of M2 and M3 upstream of the Right Bend.

A specialized insertion created by two SDB dipoles bends the beam horizontally 18.5° (323 mr) to the right, aligning with the existing AP3 line. The 18.5° horizontal bend has the two bend centers separated by a quadrupole triplet of SQEs to generate the 180° of betatron phase advance needed to locally cancel the horizontal dispersion. There are short matching sections on either side of this insertion to transition in and out of FODO lattices. The beam continues for 63.0 m to the beginning of the geometric and optical matching section between the M3 line and the Delivery Ring (DR) injection point in the D30 straight section. The FODO cells in this region are characterized by 72° of phase advance and a half-cell length of 5.613 m.

This final injection section satisfies multiple, interleaved design constraints:

- Providing the optical match between the lattice functions of the M3 line and those of the DR;
- A 86 mr horizontal right bend to align with the D30 straight section, and;
- An overall 4-ft elevation drop from M3 to the DR, performed in two steps.

The first step of the drop in elevation uses two MDC dipoles bending through 67.8 mr. The second down-bend is provided by a SDD dipole bending downward at 102.3 mr.

Embedded in the level beamline section between the first and second elevation step changes, two MDC dipoles bend horizontally, each through 43.1 mr to align the trajectory with the D30 straight section. The dipoles form an achromatic bend embedded in the achromatic vertical descent to the Delivery Ring. The final nine quadrupoles in the line perform the optical match between the 72° FODO cells and the Delivery Ring. Figure 7.37 shows the path of the M2 and M3 lines from the Target Station to the Delivery Ring.

The final stages of injection occur entirely in the vertical plane, with the final up-bend produced by a combination of a C-magnet (also called a septum dipole) in the beamline, followed by a large-aperture focusing quadrupole Q303 and a pulsed magnetic septum dipole in the Delivery Ring. The C-magnet bends upward 35 mr, and steers the beam 11.6-cm high off-axis through Q303, generating another 30 mr of upward vertical kick. The septum adds 35 mr of bend up. Three kicker modules upstream of quad Q202 close the trajectory onto the orbit of the Delivery Ring. Figure 7.38 shows the optics for this part of the M3 line.

The total beamline length from the face of the target-station lithium lens to mid-quad Q202 in the Delivery Ring is 296 m. Parameters of the main magnets are listed in Table 7.11. Figure 7.39 shows the complete path of the $(g-2)$ beam through M2 and M3 from the Target Station to the Delivery Ring.

| Magnet | Type | Current (A) | Power Supply |
|---|---|---|---|
| Q801 | SQC | 95.2 | D:Q801 |
| Q802 | SQC | 107.0 | D:Q802 |
| VT802 | NDB | 25.0 | D:VT802 |
| Q803 | SQC | 107.0 | D:Q802 |
| Q804 | SQC | 132.3 | D:Q804 |
| HT804 | NDB | 25.0 | D:HT804 |
| VT804 | NDB | 25.0 | D:VT804 |
| Q805 | SQA | 45.7 | D:Q805 |
| Q806 | SQA | 260.9 | D:Q806 |
| Q807 | SQA | 260.9 | D:Q806 |
| Q808 | SQA | 260.9 | D:Q806 |
| HT808 | NDB | 25.0 | D:HT808 |
| Q809 | SQA | 260.9 | D:Q806 |
| Q810 | SQA | 77.7 | D:Q810 |
| Q811 | SQA | 279.9 | D:Q811 |
| VT811 | NDB | 25.0 | D:VT811 |
| Q812 | 4Q24 | 143.3 | D:Q812 |
| H812 | 5D32 | 903.6 | D:H812 |
| M2/M3 merge | | | |
| Q706 | SQD | 82.7 | D:Q706 |
| VT706 | NDB | 25.0 | D:VT706 |
| Q707 | SQD | 104.4 | Q707 |
| HT707 | NDB | 25.0 | D:HT707 |
| Q708 | SQB | 83.8 | D:Q708 |
| Q709 | SQC | 94.7 | D:Q709 |
| HT709 | NDB | 25.0 | D:HT709 |
| Q710 | SQC | 94.7 | D:Q709 |
| Q711 | SQC | 94.7 | D:Q709 |
| Q712 | SQC | 94.7 | D:Q709 |
| Q713 | SQC | 94.7 | D:Q709 |
| Q714 | SQC | 94.7 | D:Q709 |
| Q715 | SQC | 94.7 | D:Q709 |
| Q716 | SQC | 94.7 | D:Q709 |
| Q717 | SQC | 94.7 | D:Q709 |
| Q718 | SQC | 94.7 | D:Q709 |
| HT718 | NDB | 25.0 | D:HT718 |
| Q719 | SQC | 94.7 | D:Q709 |
| Q720 | SQC | 94.7 | D:Q709 |
| Q721 | SQC | 94.7 | D:Q709 |
| Q722 | SQC | 94.7 | D:Q709 |
| VT722 | NDB | 25.0 | D:VT722 |
| Q723 | SQC | 94.7 | D:Q709 |
| Q724 | SQC | 87.3 | D:Q724 |
| VT724 | NDB | 25.0 | D:VT724 |
| HT724 | NDB | 25.0 | D:HT724 |

| Magnet | Type | Current (A) | Power Supply |
|---|---|---|---|
| Q725 | SQD | 74.8 | D:Q725 |
| Q726 | SQB | 88.5 | D:Q726 |
| H726 | SDB | 364.1 | D:H726 |
| Q727 | SQE | 84.7 | D:Q727 |
| Q728 | SQE | 81.4 | D:Q728 |
| VT728 | NDB | 25.0 | D:VT728 |
| D:Q728 | SQE | 84.7 | D:Q727 |
| H729 | SDB | 364.1 | D:H726 |
| Q730 | SQC | 77.0 | D:Q730 |
| Q731 | SQC | 77.0 | D:Q730 |
| HT732 | NDA | 25.0 | D:HT732 |
| Q732 | SQC | 77.0 | D:Q730 |
| Q733 | SQC | 77.0 | D:Q730 |
| Q734 | SQC | 77.0 | D:Q730 |
| Q735 | SQC | 77.0 | D:Q730 |
| Q736 | SQC | 77.0 | D:Q730 |
| VT736 | NDA | 25.0 | D:VT736 |
| Q737 | SQC | 77.0 | D:Q730 |
| Q738 | SQC | 77.0 | D:Q730 |
| Q739 | SQC | 77.0 | D:Q730 |
| HT739 | NDA | 25.0 | D:HT739 |
| Q740 | SQC | 77.0 | D:Q730 |
| Q741 | SQC | 77.0 | D:Q730 |
| Q742 | SQC | 90.0 | D:Q742 |
| V742 | MDC | 307.7 | D:V742 |
| HT742 | NDA | 25.0 | D:HT742 |
| Q743 | SQB | 73.3 | D:Q743 |
| VT743 | NDB | 25.0 | D:VT743 |
| V743 | MDC | 307.7 | D:V742 |
| Q744 | SQB | 72.7 | D:Q744 |
| HT744 | NDA | 25.0 | D:HT744 |
| H744 | MDC | 192.4 | D:H744 |
| Q745 | SQE | 100.7 | D:Q745 |
| Q746 | SQD | 90.0 | D:Q746 |
| Q747 | SQD | 90.0 | D:Q746 |
| Q748 | SQE | 100.7 | D:Q745 |
| H748 | MDC | 192.4 | D:H744 |
| Q749 | SQC | 89.6 | D:Q749 |
| Q750 | SQD | 219.8 | D:Q750 |
| V750 | SDD | 428.1 | D:V750 |
| Q751 | 4Q16 | 57.5 | D:Q751 |
| Q752 | 4Q16 | 57.5 | D:Q751 |
| ICMAG | MSDA | 248.9 | D:ICMAG |
| Q303 (DR) | LQE | 456.5 | D:QT303 |
| ISEP | SEPT | 8,750.0 | D:ISEP |

Table 7.11: M2- and M3-line dipoles, quadrupoles and trims (HT and VT prefix) from the Target Station to the Delivery Ring.

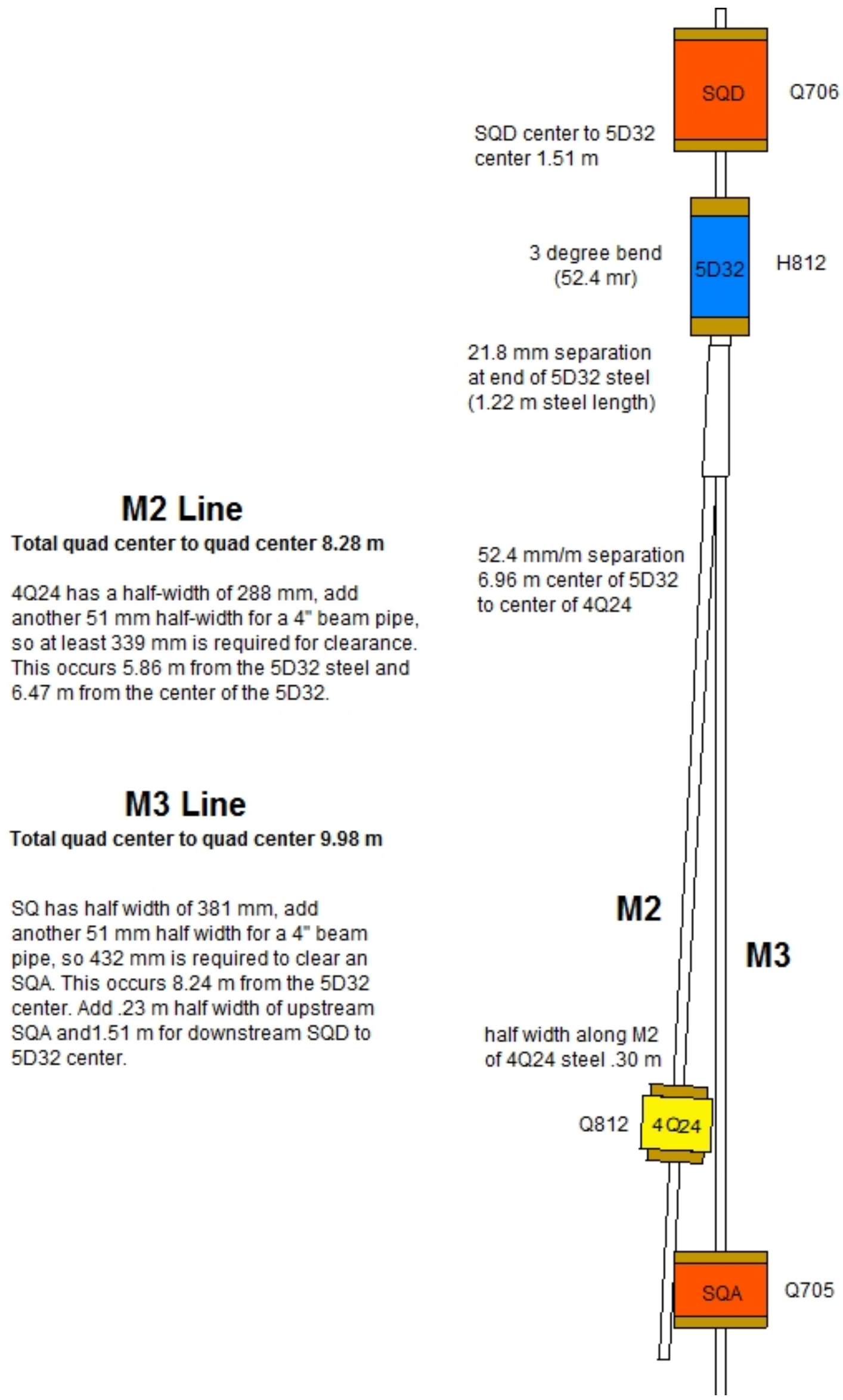

Figure 7.34: M2/M3 merge geometry.

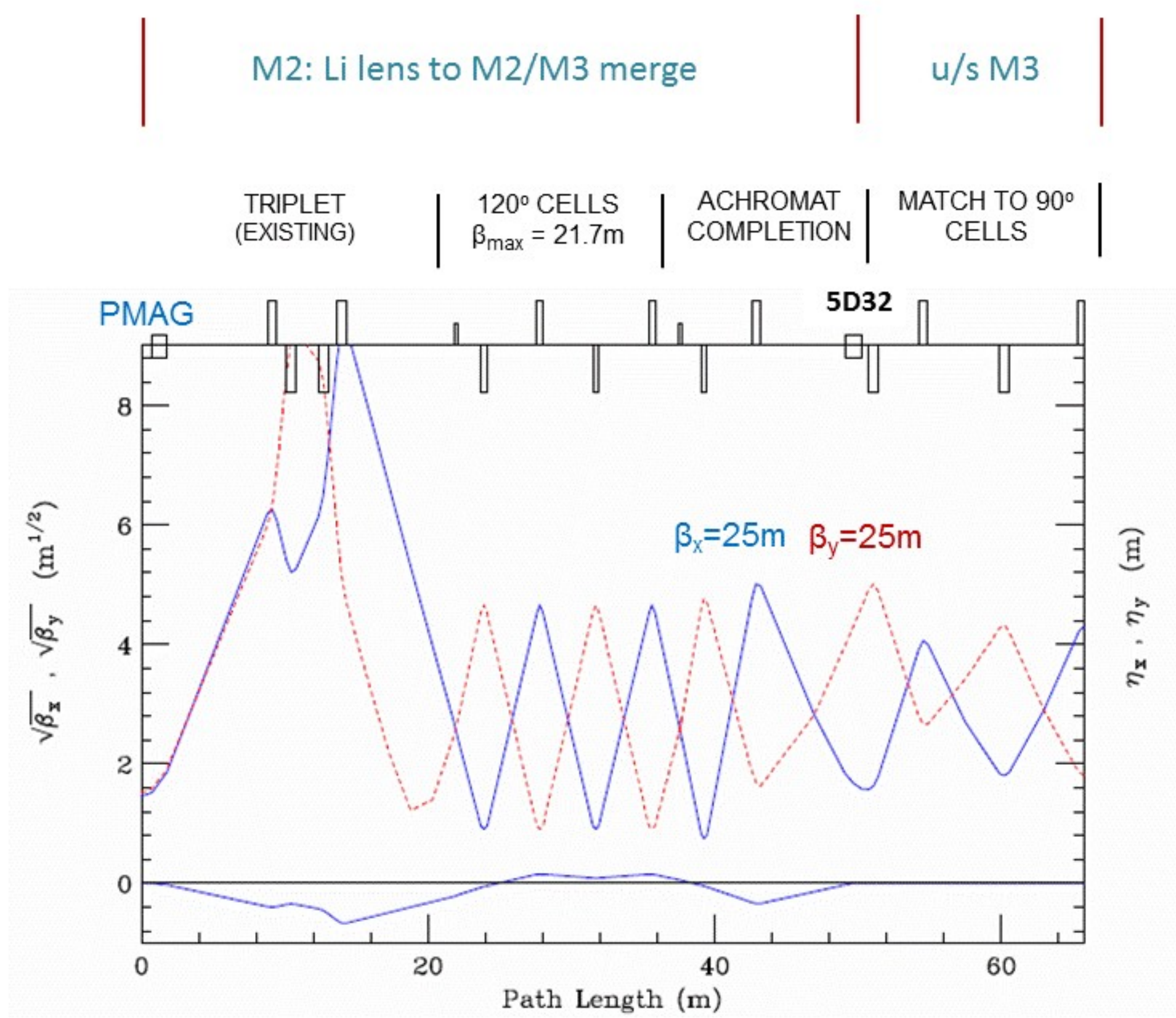

Figure 7.35: M2-line beta functions above and dispersion functions below (horizontal blue and vertical red), including the 5D32 dipole at the M2/M3 merge, and match into M3.

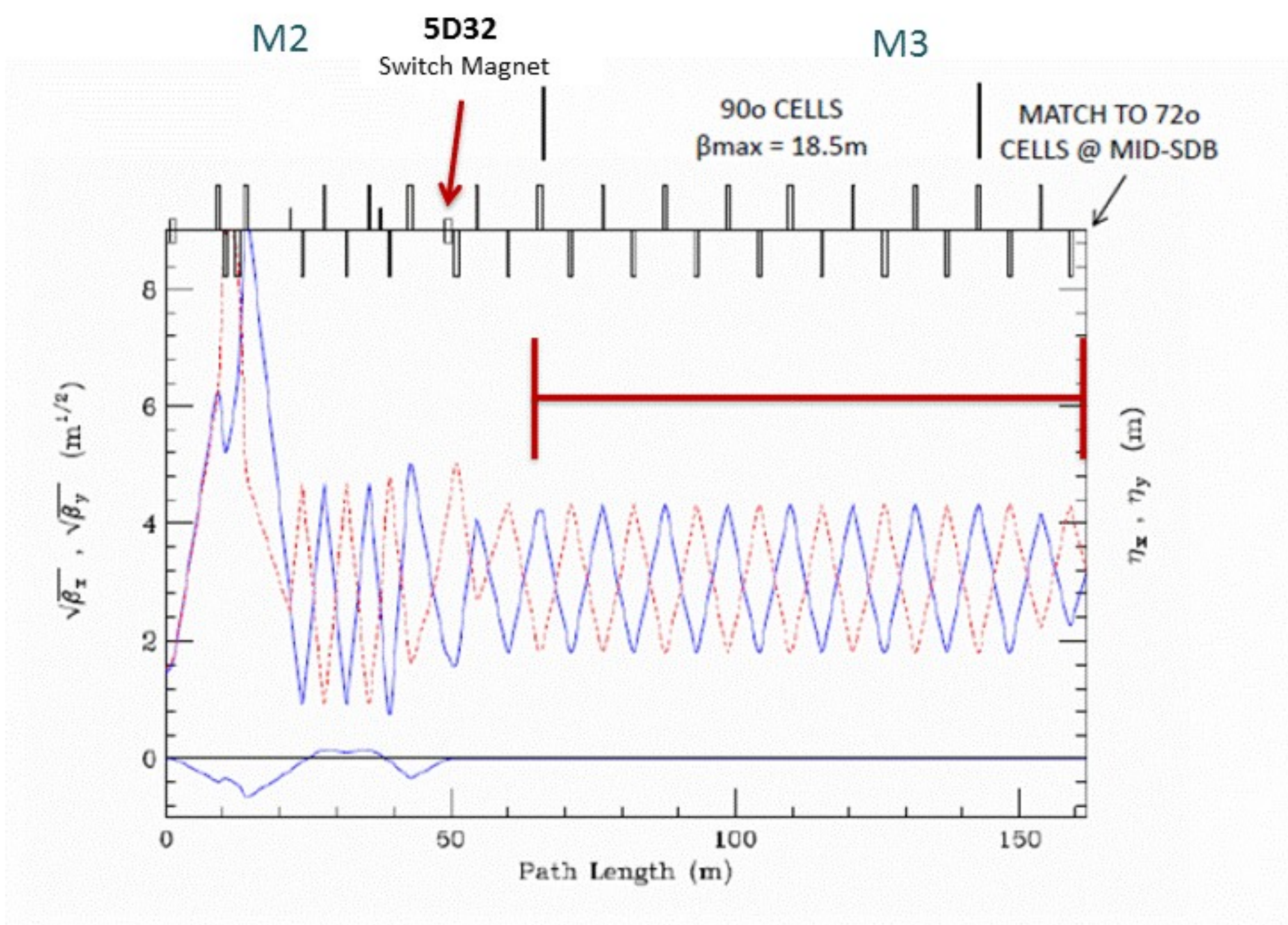

Figure 7.36: M2- and M3-line beta functions above and dispersion functions below (horizontal blue and vertical red), ending at the first "Right Bend" dipole.

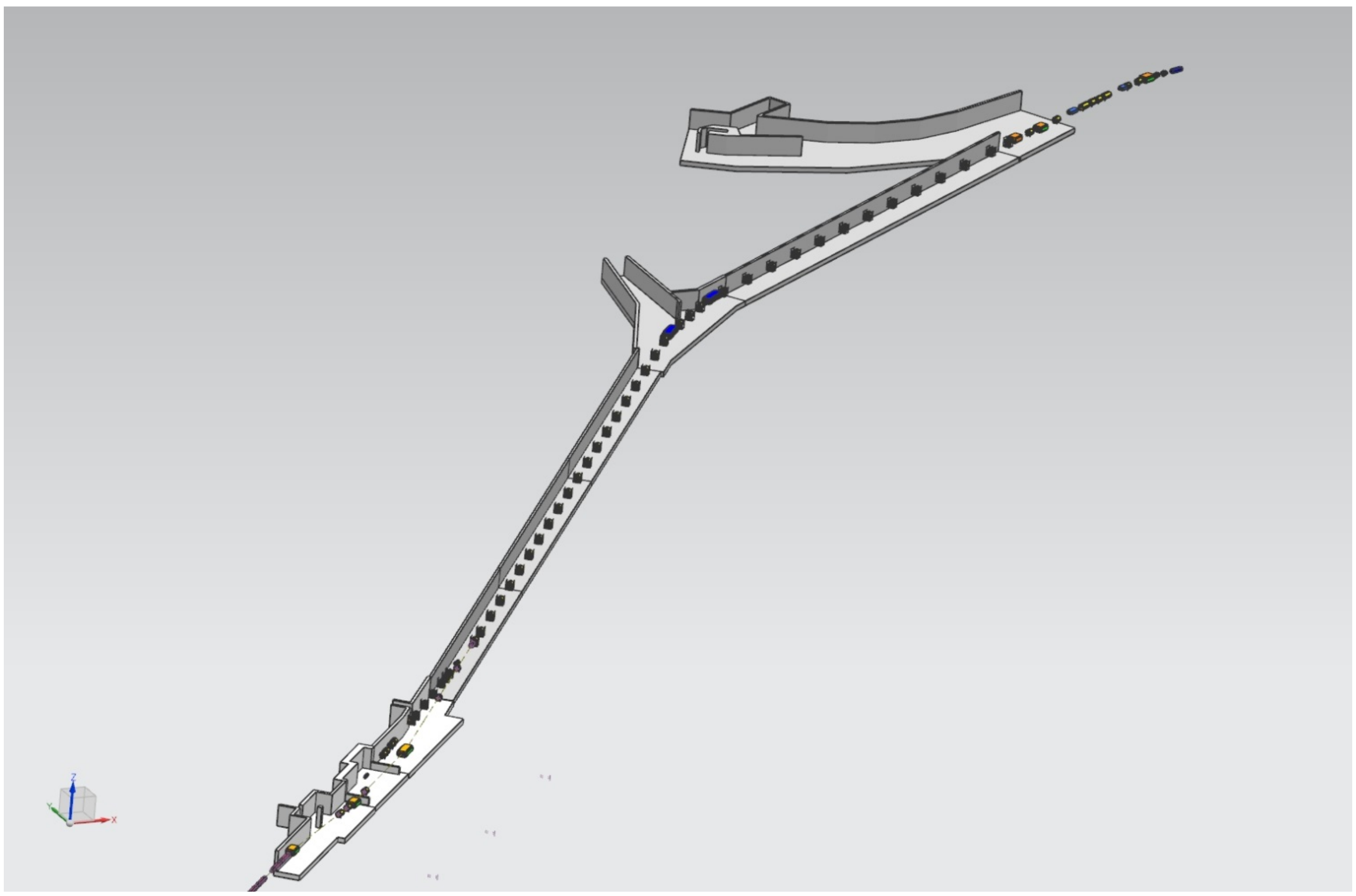

Figure 7.37: M2 and M3 lines from Target Station to Delivery Ring.

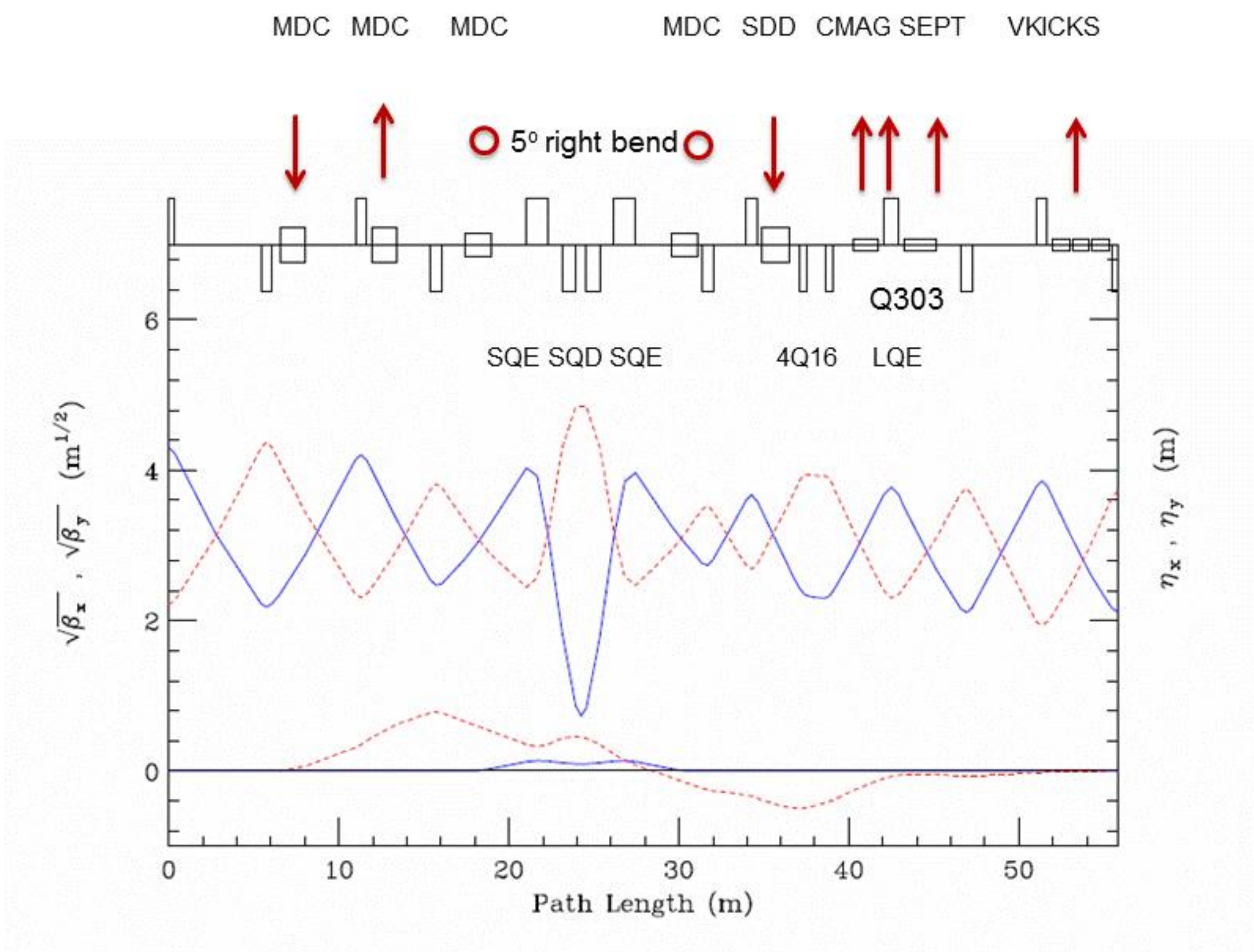

Figure 7.38: Downstream section of the M3 line, through the vertical and horizontal translations to match into the Delivery Ring, beta functions above and dispersion functions below (horizontal blue and vertical red).

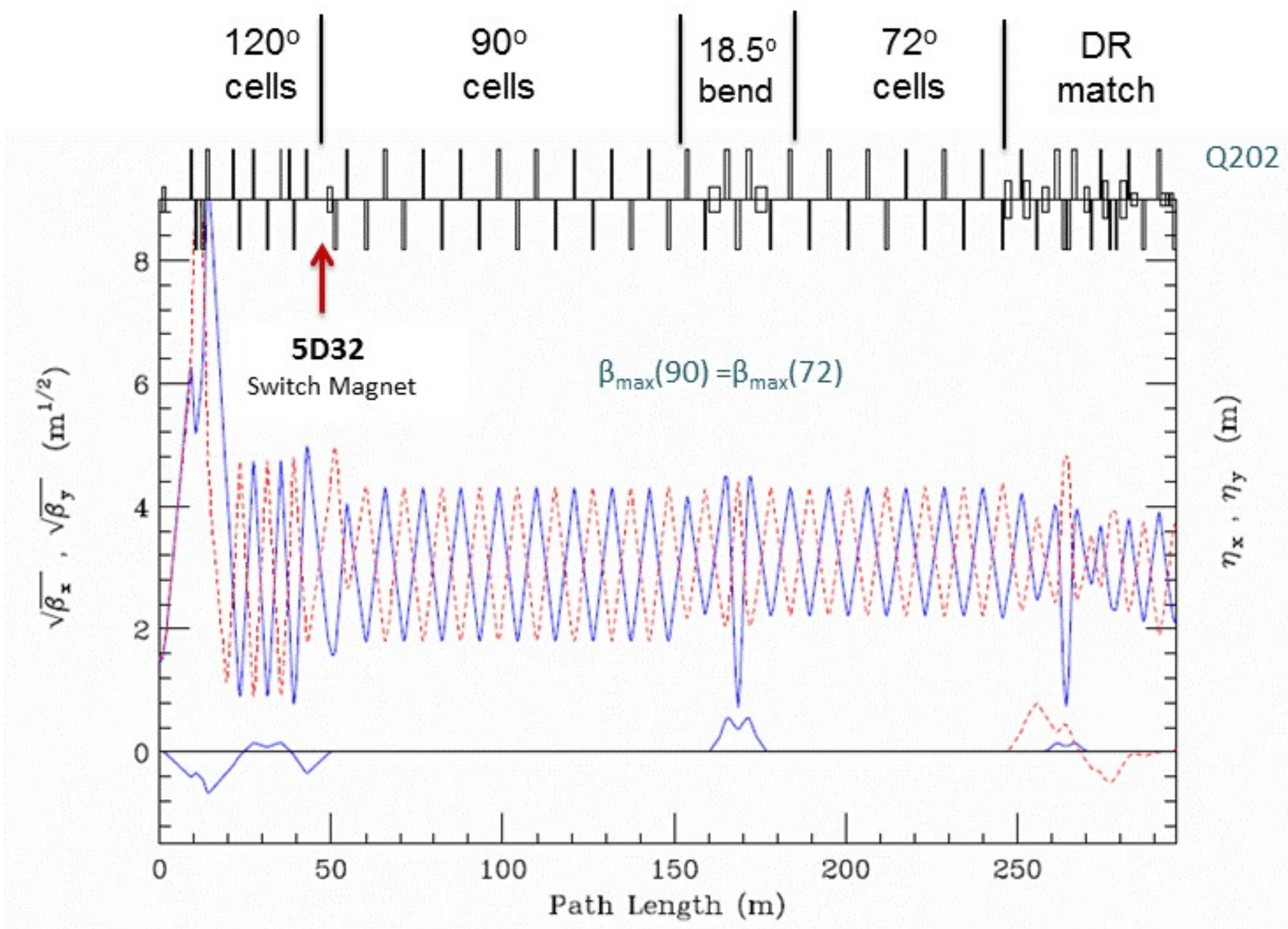

Figure 7.39: M2 and M3 lines from the Target Station to the Delivery Ring.

### 7.5.6 Delivery Ring

The Pbar Debuncher ring will largely remain intact for $(g-2)$ operation and will be renamed the Delivery Ring for its new role in providing muons to the experiment. A considerable amount of equipment left over from Pbar operation will need to be removed from the Debuncher. Most of the equipment targeted for removal was used for stochastically cooling the antiproton beam during collider operation and is not needed for $(g-2)$. Some of these devices also have small apertures, so the ring acceptance will be improved with their removal. The cooling tanks in the D30 straight section also need to be removed to provide room for the new injection and extraction devices.

The Pbar Accumulator ring will not be needed for $(g-2)$ and Mu2e operation and will become a source of magnets, power supplies and other components for use in the reconfigured beamlines. In particular, the downstream section of M3 and the M4 (extraction) line will be largely made up of former Accumulator components. Some larger-aperture magnets will also be needed in the injection and extraction regions and will come from the Accumulator or other surplus sources.

#### Rings Lattice and Acceptance

The original design lattice for the Debuncher will be used for the Delivery Ring with few modifications. The lattice has a 3-fold symmetry with additional mirror symmetry in each of the three periods, with three zero-dispersion straight sections: D10, D30 and D50. The original lattice parameters were largely dictated by the requirements for Pbar stochastic cooling and the RF systems. The Debuncher was designed with a large transverse and longitudinal momentum acceptance in order to efficiently RF-debunch and stochastically cool antiprotons from the production target. This lattice design is also well suited for $(g-2)$ operation. During Collider Run II, the original lattice was distorted somewhat in order to reduce the beam size in the stochastic cooling tanks that had limiting apertures. Since these tanks will be removed, the lattice that will be used for $(g-2)$ will revert back to an earlier Debuncher optics that incorporated improvements over the original design lattice. Figure 7.40 shows the lattice functions for one period of the Debuncher.

It should be noted that the design acceptance of the Debuncher was $20\pi$ mm-mr, while the (g-2) acceptance requirement is $40\pi$ mm-mr. During the 25 years of Pbar operation, numerous aperture improvements were undertaken to boost the acceptance of the Debuncher. After the final Collider Run II aperture improvements were put in place in 2007, the measured acceptance of the Debuncher was as high as $33\pi$ mm-mr in both transverse planes. The $(g-2)$ design goal of a $40\pi$ mm-mr acceptance for the Delivery Ring, while reusing as much of the original equipment as possible, presents a difficult challenge.

The transverse acceptances of the Debuncher dipole, quadrupole, sextupole, and trim magnets are quite large. The smallest magnet acceptance is in the vertical plane of the dipoles and is approximately $54\pi$ mm-mr on one end, growing to $79\pi$ mm-mr on the other end. The dipoles have a $90\pi$-mm-mr or larger horizontal acceptance ($90\pi$-mm-mr for the $\pm 2\%$ momentum spread and locations with the largest dispersion) and the other magnets have a 100 $\pi$-mm-mr or larger acceptance in both planes. Since the original Debuncher lattice will not be significantly changed for $(g-2)$ operation, the main Delivery-Ring magnets will not be limiting apertures. In general, devices with a physical aperture of 50 mm or greater provide an acceptance of over $40\pi$ mm-mr in the Debuncher, and select locations can provide that acceptance for devices that have an aperture of 40 mm, as long as they are relatively short in length.

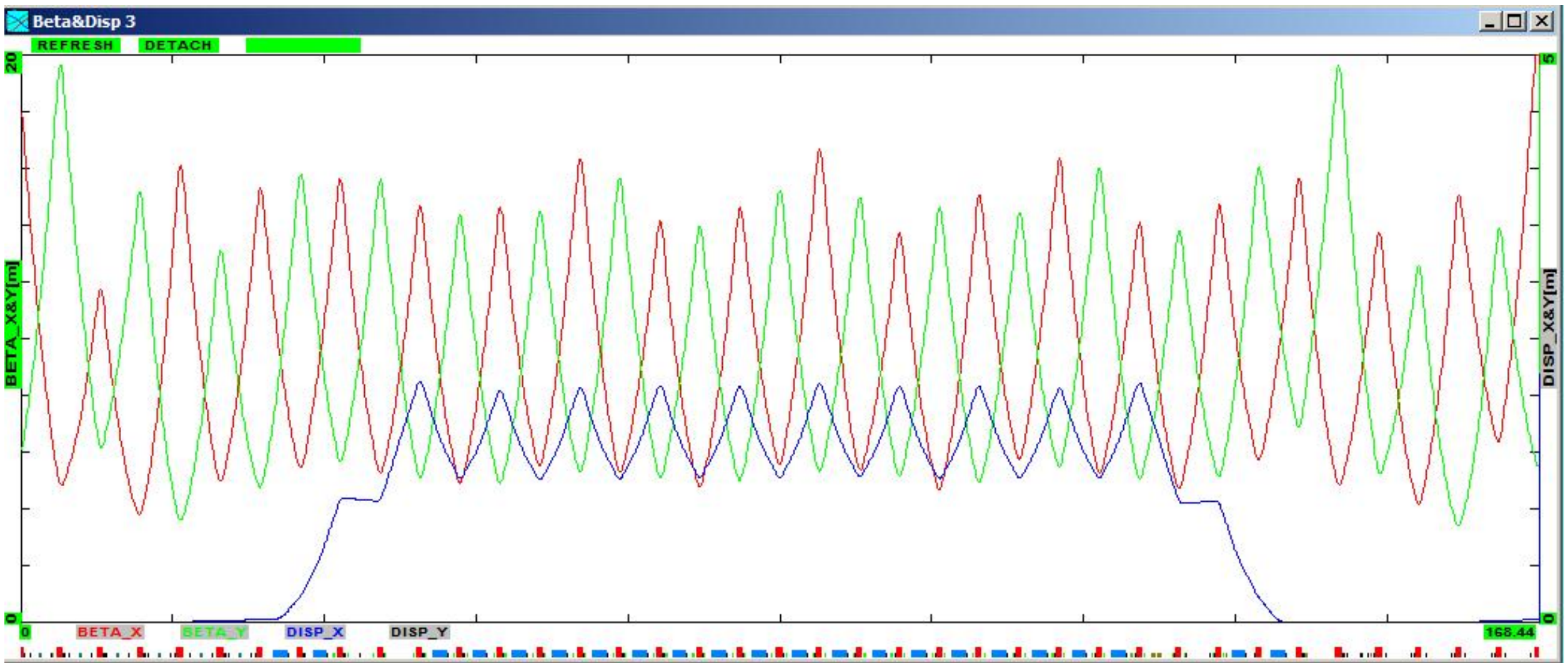

Figure 7.40: Debuncher/Delivery Ring lattice functions through 1/3 of the ring. $\beta_x$ is in red, $\beta_y$ in green, and horizontal dispersion in blue.

During Collider operation, the smallest physical apertures in the Debuncher came from stochastic cooling tanks, RF cavities, instrumentation, and devices used for injecting and extracting beam. Many of these devices will be removed as part of the repurposing of the Debuncher for the muon experiments. Some of these devices, most notably the kickers, will be retained in the interest of economy and/or complexity and lead-time of manufacture. Other devices, such as the injection septum, will be new devices with necessarily small physical apertures in order to provide enough bend strength.

There is only one RF cavity planned for the Delivery Ring, which is needed to support Mu2e operation and will have an aperture similar to the Debuncher rotator cavities (approximately 80 mm). Since the rotator cavities had an acceptance that was greater than $100\pi$ mm-mr, the new cavity will have ample aperture for use in both Mu2e and $(g-2)$ operations. RF cavities used for antiproton production will be removed prior to $(g-2)$ operation. A diagnostic RF system, DRF-3, will remain to facilitate closed orbit measurements in the Delivery Ring.

Many of the beam detectors used during Pbar operation had small physical apertures in order to improve sensitivity. Since the beam intensities when running $(g-2)$ are expected to be even smaller than those seen during Pbar operation, designers will need to be mindful of the aperture needs of the $(g-2)$ experiment. Similarly, when instrumentation is being considered for reuse in the Delivery Ring, the physical aperture and proposed tunnel location will be analyzed for adequate acceptance.

Both injection from the M3 line and extraction to the M4 line take place in the D30 straight section. Injection will be located in the upstream half of the straight section, and the pulsed magnetic septum and kicker magnets will have small apertures in order to provide adequate bending strength. The septum has a small aperture in both planes, while the kicker is primarily limited in the horizontal plane. The septum is a modified Booster-style (BSE) magnetic septum magnet. The septum modifications involve increasing the pole gap from 28 mm to 47 mm in order to greatly improve the horizontal acceptance, and reducing the septum thickness from 14 mm to 9 mm to increase the vertical acceptance. The injection kicker system will be made up of two surplus Pbar AP4 injection kicker magnets. The horizontal aperture is only 41 mm and will likely be one of the limiting apertures of the Delivery Ring. The extraction kicker system will be made up of two Pbar extraction kicker magnets. They have a vertical aperture of 41 mm and will also be one of the limiting apertures of the Delivery Ring.

### Kickers and Septa

The kickers and septa required for $(g-2)$ operation will need to operate at a much higher frequency than that used for antiproton production, with peak rates increasing as much as a factor of 30. In an effort to make the new kicker systems more economical and eliminate a long lead-time device, existing Pbar kicker magnets will be reused. Kickers will be required for injection and extraction from the Delivery Ring as well as for proton removal. Table 7.12 compares kicker parameters for existing Pbar systems to the specifications for the $(g-2)$ injection and proton-removal kickers. The rise and fall time specifications for $(g-2)$ are generally less strict than what was needed for antiproton production, due to the short bunch length of the muons (and protons). Decreasing the rise time of the proton removal kicker, however, will reduce the number of turns required in the Delivery Ring to adequately separate the protons from the muons. Although the Pbar kicker magnets are suitable for reuse, new power supplies will be needed to operate at the increased rate. Resistive loads for the kickers will need to be cooled with Fluorinert. A single Fluorinert distribution system is planned, with piping bridging the distance between the load resistors from kickers in the D30 and D50 straight sections.

| Kicker (modules) | Integrated Field (kG-m) | Kick Angle (mr) | Rise Time 95%/5% (ns) | Fall Time 95%/5% (ns) | Flat Top Time (ns) |
|---|---|---|---|---|---|
| Debuncher Extraction (3) | 1.34 | 4.6 | 150 | 150 | 1500 |
| Debuncher Injection (3) | 1.81 | 6.1 | 185 | 185 | 1500 |
| Delivery-Ring Injection (2) | 0.64 | 6.2 | n/a | 800 | 300 |
| Delivery-Ring Extraction (2) | 0.77 | 5.8 | 450 | n/a | 200 |
| Delivery-Ring Proton Removal (1) | 0.52 | 6.2 | 180 | n/a | 270 |

Table 7.12: Existing Pbar (top) and future $(g-2)$ (bottom) kicker strength and waveform specifications.

The septa and pulsed power supplies used during Pbar operation are not suitable for rapid cycling and cannot be used for $(g-2)$. The septa have no internal cooling to handle the increased heat load from the planned high duty cycle, and the power supplies are not able to charge quickly enough. The Booster-style septum magnet design will be modified to have the necessary size and field strength required for use in the injection and proton removal systems. The power supplies used in the Booster to power the septum magnets also appear to be a good fit. Although they are designed to operate at a lower frequency (15 Hz) than the peak needed for $(g-2)$, the lower operating current (for 3.1 GeV/c versus 8.89 GeV/c momentum) should more than compensate for changes to the heat load and mechanical stresses due to the increased pulse rate. The new septa can be the same length as their Pbar counterparts so they can comfortably fit between quadrupoles in the injection and proton removal regions.

### Delivery Ring D30 straight section

The Delivery-Ring injection and extraction regions will both be located in the D30 straight section. Due to the physical constraints of the tunnel, the M3 line will trace a path above the Delivery Ring before descending. Similarly, the M4 line will be located above the Delivery Ring until the ring bends away at the edge of the straight section. In both cases, the tight quadrupole spacing in the Delivery Ring leaves little room for the descending and ascending beamlines. The extraction line will closely follow the trajectory of the decommissioned AP4 (Booster to Debuncher) line. The tunnel in this region has an existing stub region that the extraction line will pass through, eliminating the need for civil construction to widen and strengthen the tunnel. Figure 7.41 shows the layout of injection and extraction devices in the D30 straight section.

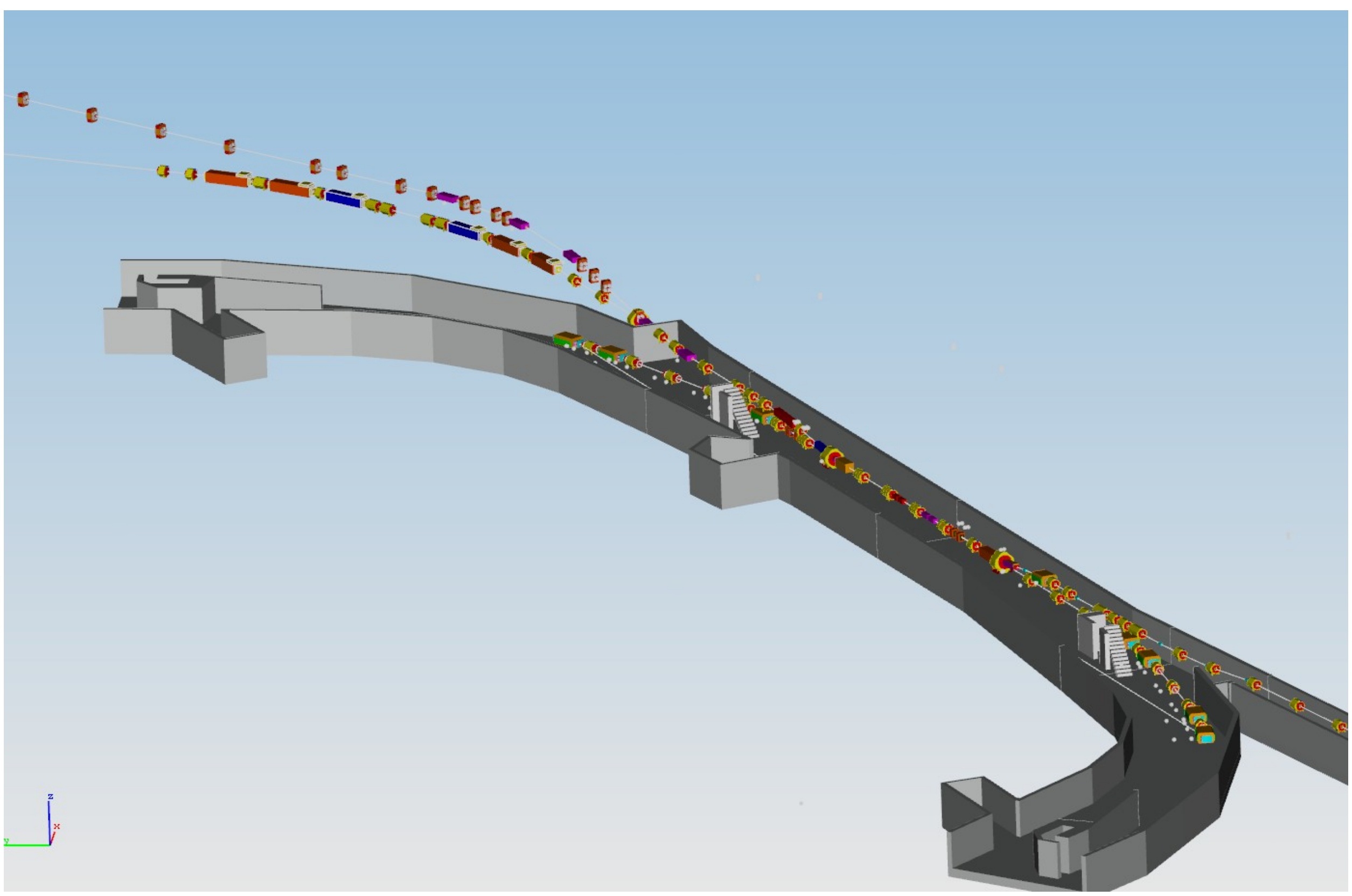

Figure 7.41: D30 straight section, injection on right, extraction on left.

The work required to prepare the D30 straight section for the new beamlines is considerable. Without modification, both the M3 and M4 lines would have physical conflicts with existing utilities and ring devices in the areas of elevation change to and from ring level. The existing cable trays on the Debuncher side of the ring are also where magnet hangers will be required for the new beamlines. The cable trays are full of cables, many of which need to be retained or replaced in order to operate the beamlines. The Debuncher, cable trays, utilities, lighting, cable routing, and numerous other subsystems will need to be relocated as part of the D30 straight section reconfiguration. Although not required for $(g-2)$ operation, extensive radiation shielding will be installed in the tunnel for Mu2e, so the reconfiguration must accommodate the future shielding.

The main features of the reconfiguration are as follows

- The AP3 beamline must be removed and devices temporarily stored for future use in the new beamlines. Accumulator magnets must be removed to allow room for tunnel activities.
- Debuncher magnets in and adjacent to the D30 straight section must be temporarily removed to allow access to the equipment that needs to be relocated or replaced. Stochastic cooling tanks that are no longer needed will be removed.
- Removal of existing Debuncher cable trays, relocated to the center of the tunnel, addition of cross-over trays from the Accumulator side to augment central trays.
- Removal of existing cabling to make way for new cabling to support the reconfigured beamlines and Delivery Ring. In addition to magnet cables, there are also safety system, instrumentation, network, abort link, shunt and motion control cables to be maintained or replaced.
- Relocation of tunnel utilities, primarily cooling water, electrical power infrastructure and tunnel lights.
- Reconfiguration of the main power supply buses that bypass all or part of the straight section (Main Bend bus, QF bus, QD bus, QSS bus, SF bus, SD bus).
- Installation of M3 and M4 lines, including injection and extraction devices.
- Reinstallation of Delivery Ring, including relocation of magnets to accommodate the reconfigured ring elements (motion controlled quad stands, large aperture quadrupoles at D3Q3 and D2Q5)

Figures 7.42 and 7.43 provide an overview of the D30 straight section reconfiguration, showing blocks of devices for removal and installation.

### Injection

The M3 line runs above the Delivery Ring in the upstream end of the D30 straight section and ends with a vertical translation into the ring. M3 injection will be achieved with a combination of a C-magnet, D3Q3 quadrupole, magnetic septum, and kicker magnets, which will all provide vertical bends. The septum and C-magnet are both based on existing designs, which reduces overall costs, but modified to improve the aperture. Both magnet designs required modifications in order to attain the $(g-2)$ acceptance goal of $40\pi$ mm-mr.

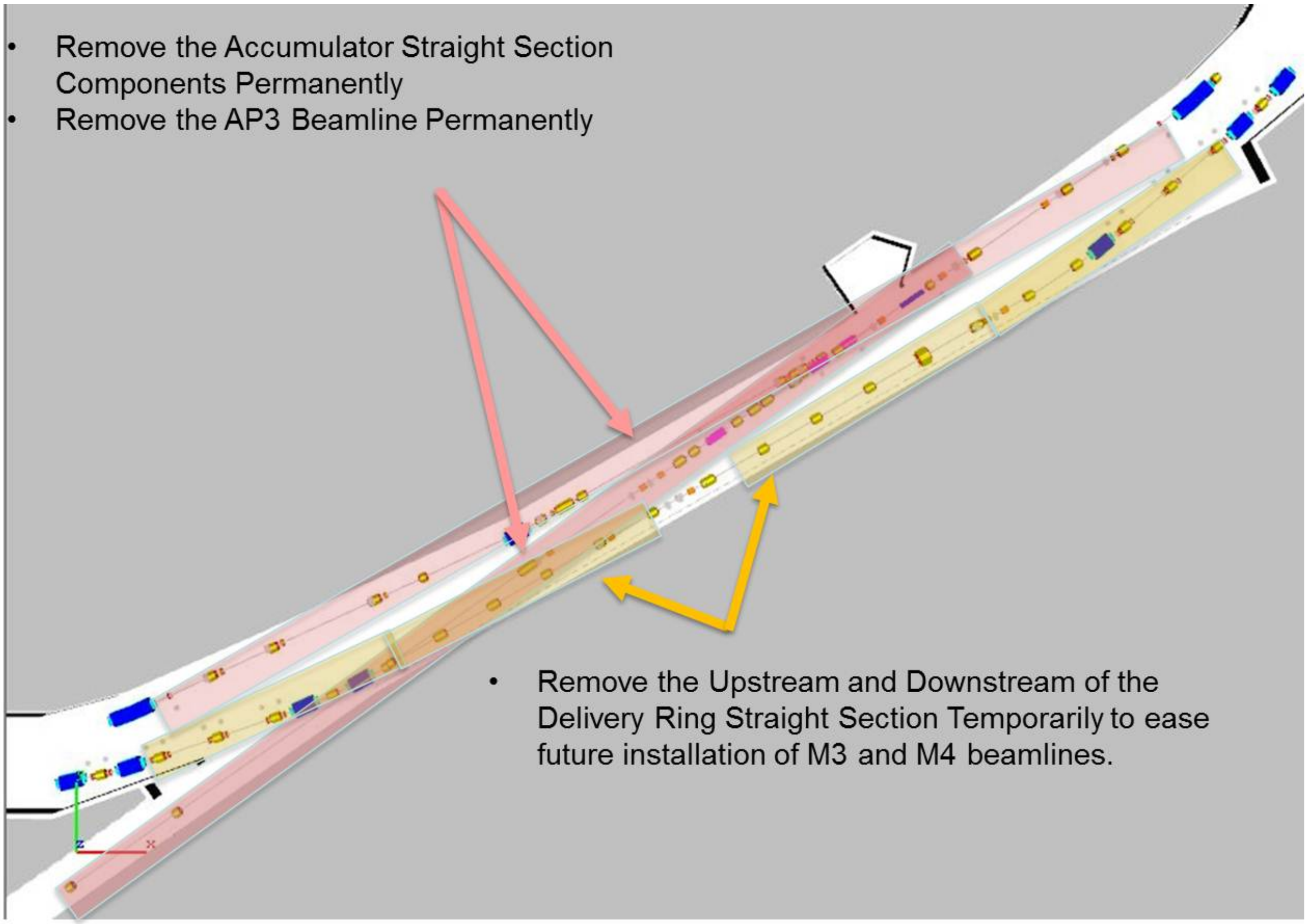

Figure 7.42: Devices in the D30 straight section to be removed.

The magnetic septum is a modified Booster-style (BSE) magnet, with an increased pole gap and a thinner septum to improve aperture. The BSE magnet has a 1.1-in pole gap, which will be increased to 1.85 in for the new septum. Similarly, the C-magnet is a larger aperture (2.1 in instead of 1.6 in) and shorter (2.2 m instead of 3.0 m) version of C-magnet designs already in use and has been designated an MSDA magnet. An identical C-magnet is used in the extraction region, but with different vacuum pipe geometry. The descending beam in M3 will pass through the C-magnet first and will be bent upward by 35 mr. The beam will continue well above the center of the D3Q3 quadrupole and receive a 30-mr upward kick. Since the beam is up to 140 mm above the centerline of the quadrupole, a large-bore quadrupole magnet is required in order to provide adequate aperture. The large quadrupole at D3Q3 will be the LQE magnet from the D2Q5 location, which will be replaced by an 8-in quadrupole, as described below. The LQx magnets were designed to have a substantial good-field region that extends between the poles. Similar arrangements with LQ magnets can be found in Pbar at D4Q5 (former AP2 injection, planned proton removal) and D6Q6 (former Debuncher extraction). The injected beam then passes through the field region of the septum magnet and receives a 39-mr upward bend as required for the necessary trajectory entering the injection kicker magnets. The beam leaves the septum with a 4-mr residual angle, because the septum and kicker are not 90° apart. The kicker magnets provide a final 6.1-mr vertical bend to place the injected beam on the closed orbit of the Delivery Ring.

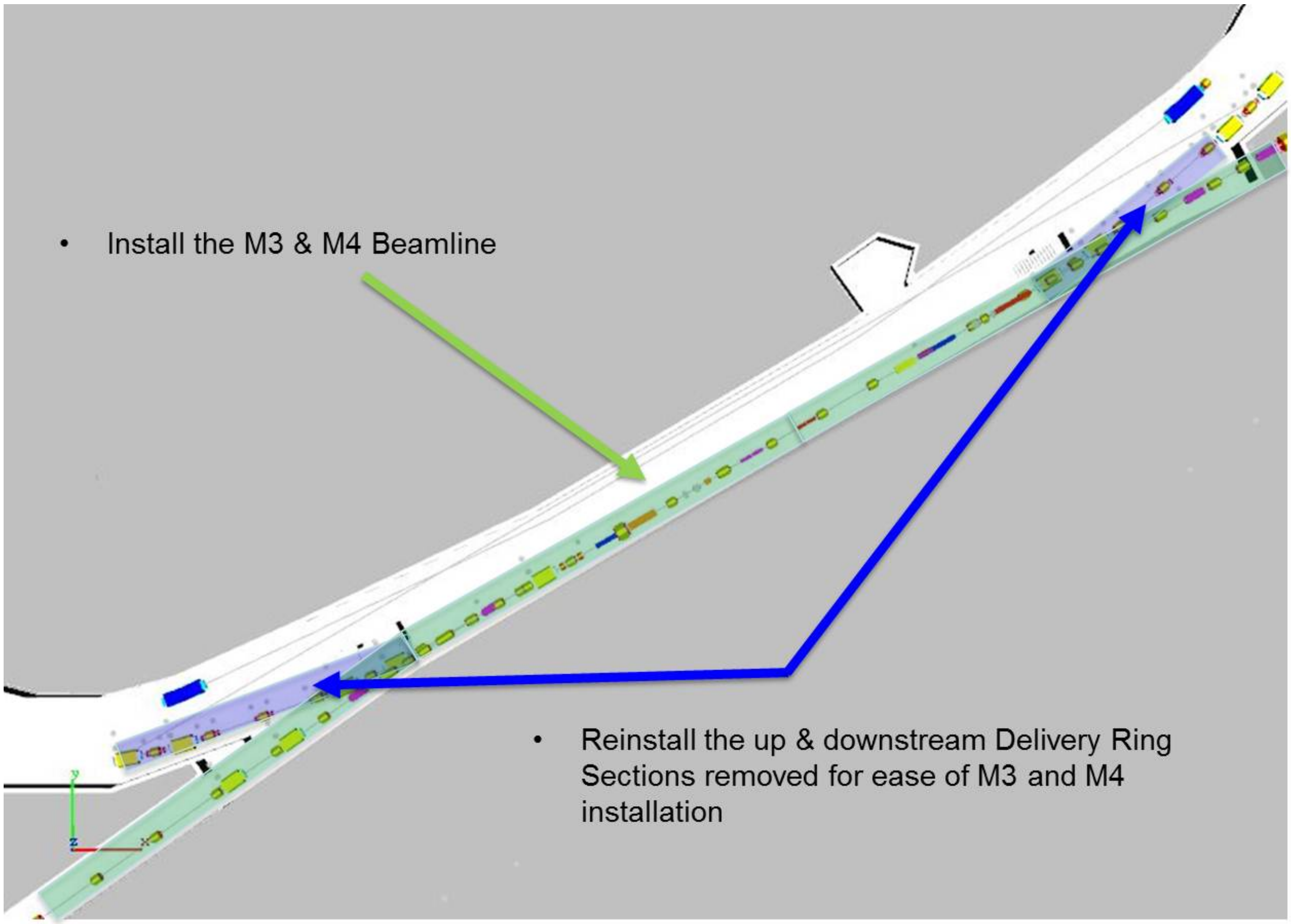

Figure 7.43: Devices in the D30 straight section to be installed.

The two-module injection kicker system is located between the D30Q and D2Q2 magnets. To minimize the horizontal $\beta$ function and maximize acceptance, the kickers will be located as close to the D2Q2 quadrupole as possible. Spare Pbar injection kicker magnets will be refurbished and reused for injection. The magnets are already designed to be oriented vertically, so little additional effort will be required to convert them to their new application. They will require a new motion-controlled stand, based on the existing Debuncher injection kicker stand. Kicker rise and fall time specifications and power supply information was provided in Table 7.12 and the accompanying text. Figure 7.44 shows the injection devices and their location in the Delivery Ring, along with their bend angles. Due to the large vertical excursion through the top of the D3Q2 magnet, a vertical bump across the injection region will be incorporated to lower the beam and improve the aperture. The quadrupole magnets at D2Q2, D30Q and D3Q4 will be displaced to create the bump by generating steering due to the beam passing off-center through the magnets. To create a 15-mm downward displacement at D3Q2, the magnets will be lowered by 8.1, 11.0, and 4.2 mm respectively.

It would be beneficial, but not necessary for $40\pi$-mm-mr acceptance, to install an existing "extended star chamber" quadrupole at the D3Q2 location. SQC-312, in magnet storage, was previously located at D4Q4 in the Pbar AP2 injection area and has an extended top lobe in its star chamber. This magnet is slated for installation at D3Q2.

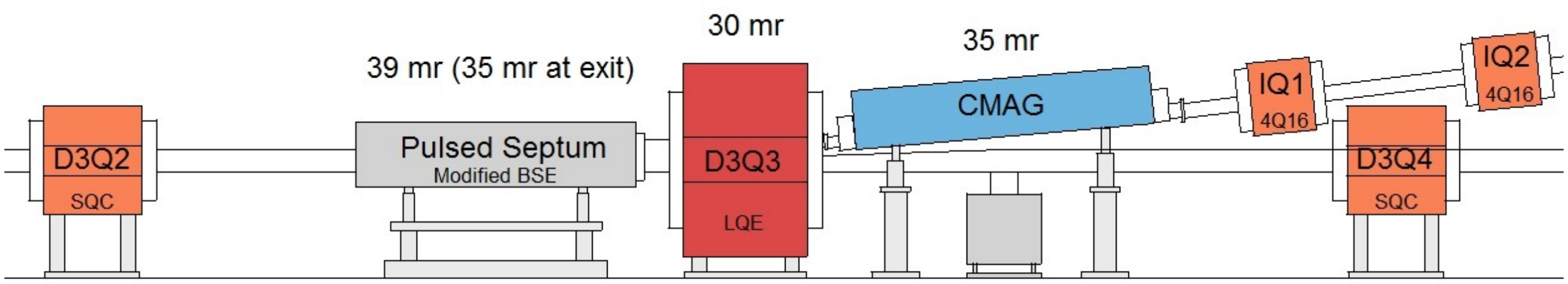

Figure 7.44: Delivery-Ring injection devices.

**Extraction**

Extraction from the Delivery Ring takes place in the downstream half of the D30 straight section. The extraction channel and the first 30 m of the M4 line will be used for both Mu2e resonant extraction and $(g-2)$ single-turn extraction. This arrangement avoids the complexity and additional expense of dual extraction lines in the limited available space. It also eliminates the need to remove potentially highly radioactive objects from the ring when switching between experiments. The ideal extraction configuration will provide enough aperture for both the Mu2e resonantly-extracted proton beam and the $(g-2)$ muon beam to be transported efficiently through the M4 line.

A Lambertson and C-magnet pair will be used, in conjunction with the intervening D2Q5 quadrupole, to bend the beam upward out of the Delivery Ring. In the interest of compatibility between $(g-2)$, Mu2e, and future muon experiments, a Lambertson magnet is required for extraction. The resonant-extraction process used for Mu2e is very restrictive on the size, strength, and location of the electrostatic septa that are required to split the extracted beam. The electrostatic septa will be located on either side of the D2Q3 quadrupole, and are expected to be about 1.5 m long for the upstream and 2.0 m long for the downstream septum. However, they will not be in place until after the $(g-2)$ run is completed. In order to achieve the goal of a combined extraction channel and beamline, the $(g-2)$ extraction kickers must be located in a lattice location that is $\sim n\pi/4$ radians from the Lambertson, where $n$ is an odd integer, and in an area not already occupied by injection or extraction devices.

The $(g-2)$ extraction kickers will be located between the D2Q2 and D2Q3 quadrupoles. There will be two kicker modules of approximately 0.85 m length each. During the dedicated period of $(g-2)$ operation, the kickers will be located as close to the D2Q3 quadrupole as possible in order to minimize the vertical $\beta$ function and maximize acceptance. The kicker magnets will be repurposed Pbar extraction kicker magnets that have a vertical aperture of 41 mm. The kicker magnets will be powered in series from a single power supply. There is also an alternative layout concept that would allow $(g-2)$ to operate after the Mu2e electrostatic septa are installed. The septum would need to be short enough to leave room for a single kicker near the D2Q2 quadrupole in this arrangement. Also, the kicker magnet would need to be modified in order to provide enough bending strength. The relocation of the kicker would also reduce aperture unless the $\beta$ functions in this region could be suppressed by about 20%.

The Lambertson is a newly designed magnet, based on the NO$\nu$A MLAW Lambertson, but is shorter and has a larger aperture. This Lambertson design was based on an insertion length of 2 m or less and a larger pole gap of 2.2 in and has been designated as an MLG magnet. Figure 7.45 shows the solid model of the Delivery-Ring Extraction Lambertson. The Lambertson magnetic fields, including the fields in the "field free" region, have been extensively modeled [19].

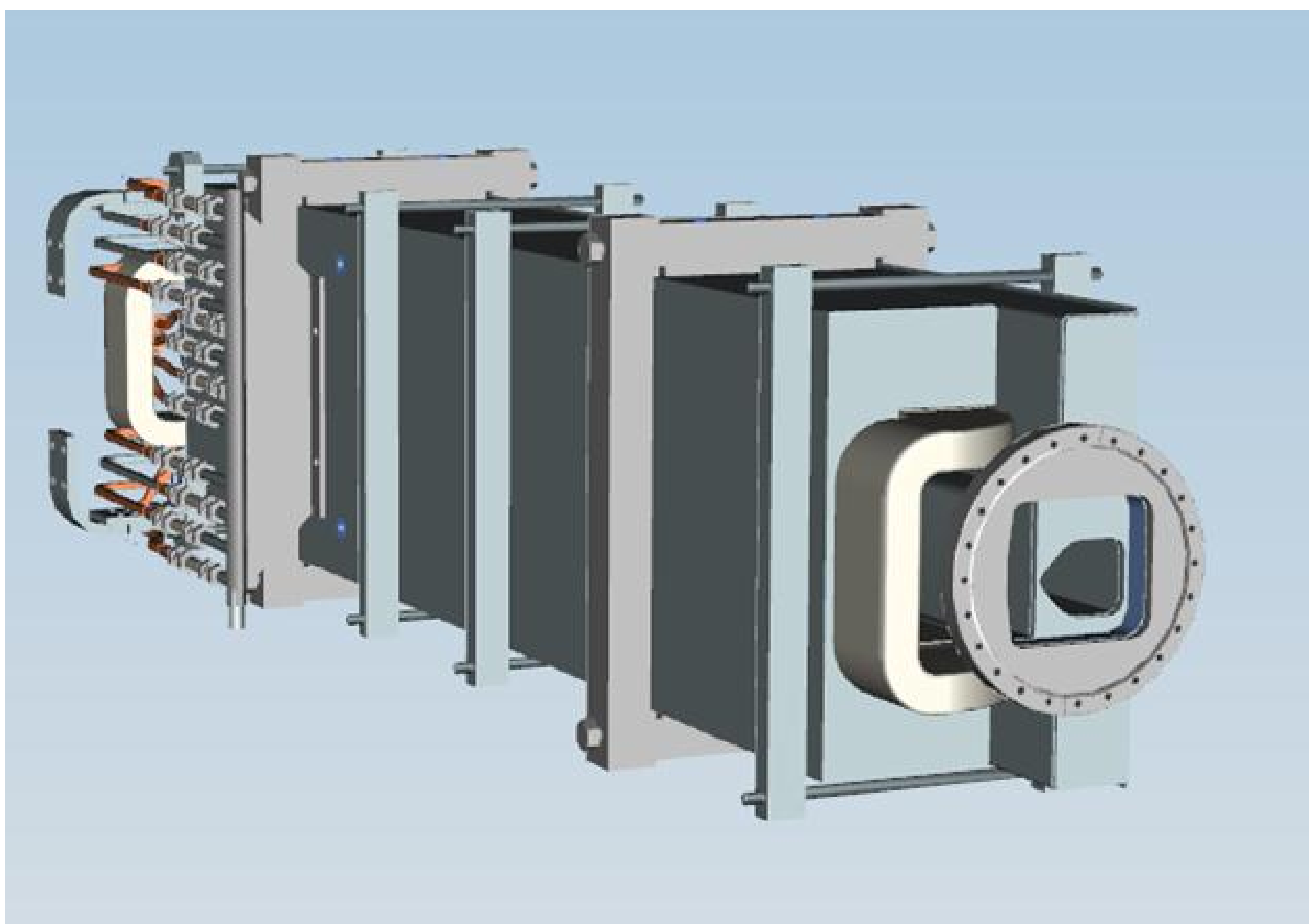

Figure 7.45: Delivery-Ring Extraction Lambertson, looking downstream.

The beam passing through the D2Q5 quadrupole has large offsets in both planes, due to the use of a Lambertson in the extraction process. The beam is kicked horizontally into the field region of the Lambertson, then bent upwards. There are two significant implications to the extraction design in order to achieve a $40\pi$-mm-mr acceptance. The first is that a larger-aperture quadrupole is needed than the available Pbar LQ series. A surplus BNL 8-in quadrupole (8Q24) will be used to provide the additional aperture. Even with the increased aperture, a large horizontal 4-bump across the extraction region is required for $(g-2)$ beam to fit within the available physical aperture (the bump is not required for Mu2e operation). The quadrupole magnets at D2Q3, D2Q4, D2Q6 and D2Q7 will be displaced horizontally to create the bump by generating steering due to the beam passing off-center through the magnets. To create a 40 mm outward displacement at D2Q5, the quadrupoles will be offset by 15.6, 11.3, 14.8 and 20.1 mm, respectively. They will all be displaced towards the right (wall) side so that beam will be bumped further to the right side through the Lambertson. Figure 7.46 shows the layout of the extraction devices for $(g-2)$ operation and $40\pi$-mm-mr acceptance.

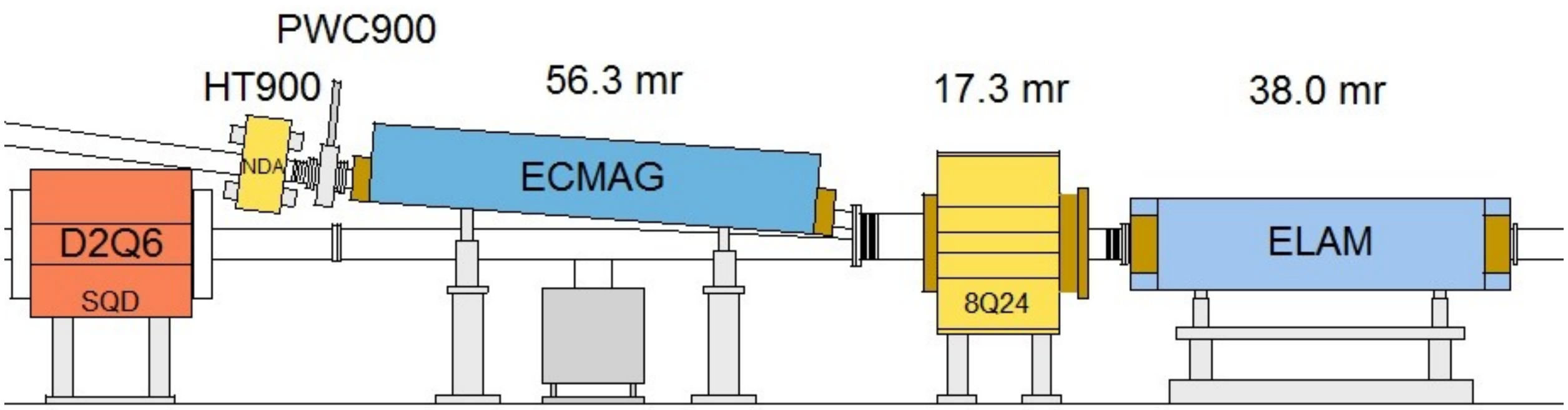

Figure 7.46: Delivery-Ring extraction devices.

**Proton Removal (Abort) System**

The proton removal system is an example of both repurposing an otherwise unneeded part of the Antiproton Source and implementing a dual function system that can be used by both the $(g-2)$ and Mu2e experiments. During Mu2e operation, an abort is needed to minimize uncontrolled proton beam loss and to "clean up" beam left at the end of resonant extraction. The proton beam must be removed quickly, by means of kicker magnets, in order to minimize losses in the ring. The $(g-2)$ experiment can benefit from the removal of protons before they reach the storage ring. The abort system can serve this purpose, as long as the protons sufficiently slip in time to create a gap for the kickers to rise through.

The old Debuncher injection point from the AP2 line in the D50 straight section will be used for the abort and proton removal systems. Recall that most of the AP2 line will be removed and replaced with the new M2 line that will merge with the M3 line upstream of the right bend. The downstream end of AP2, where antiprotons were formerly injected into the Debuncher, can now be used to extract protons from the Delivery Ring. This is made possible by the change in beam direction (as viewed from above) from clockwise to counterclockwise. The existing Pbar injection kicker magnets can be reused, although a new power supply will be needed to operate at the frequency needed to support Mu2e and $(g-2)$. The septum magnet and power supply will also need to be upgraded for the same reason. The new larger-aperture septum magnet will be identical to what was previously described for injection into the Delivery Ring. The section of the AP2 beamline being repurposed will require the addition of a vertical bending magnet to steer beam into the abort dump located in the middle of the Transport tunnel. Figure 7.47 shows the layout of the abort line.

The most economical plan to minimize the number of turns necessary to separate protons from muons is to only power the first kicker magnet, which provides the shortest rise time, a strong enough kick and requires only a single power supply. The rise time of the kickers with this configuration is about 180 ns. The kickers will be reconfigured for Mu2e operation, because all three kicker magnets are required to provide enough strength due to the higher beam momentum for Mu2e. For $(g-2)$ proton removal, the 180-ns rise time requires several revolutions around the Delivery Ring to provide enough gap between the muons and protons for the kicker to rise through. Table 7.13 lists the separation between the beams and the gap size for different numbers of turns. Four turns around the Delivery Ring would be required to cleanly remove all of the protons without disturbing the muons. All of the protons could be removed in three turns, but some of the muons would also be deflected. The table is based on the assumptions already stated: that the kicker rise time is 180 ns, the proton and muon bunch lengths are 120 ns and that the kicker should not disturb any of the muons.

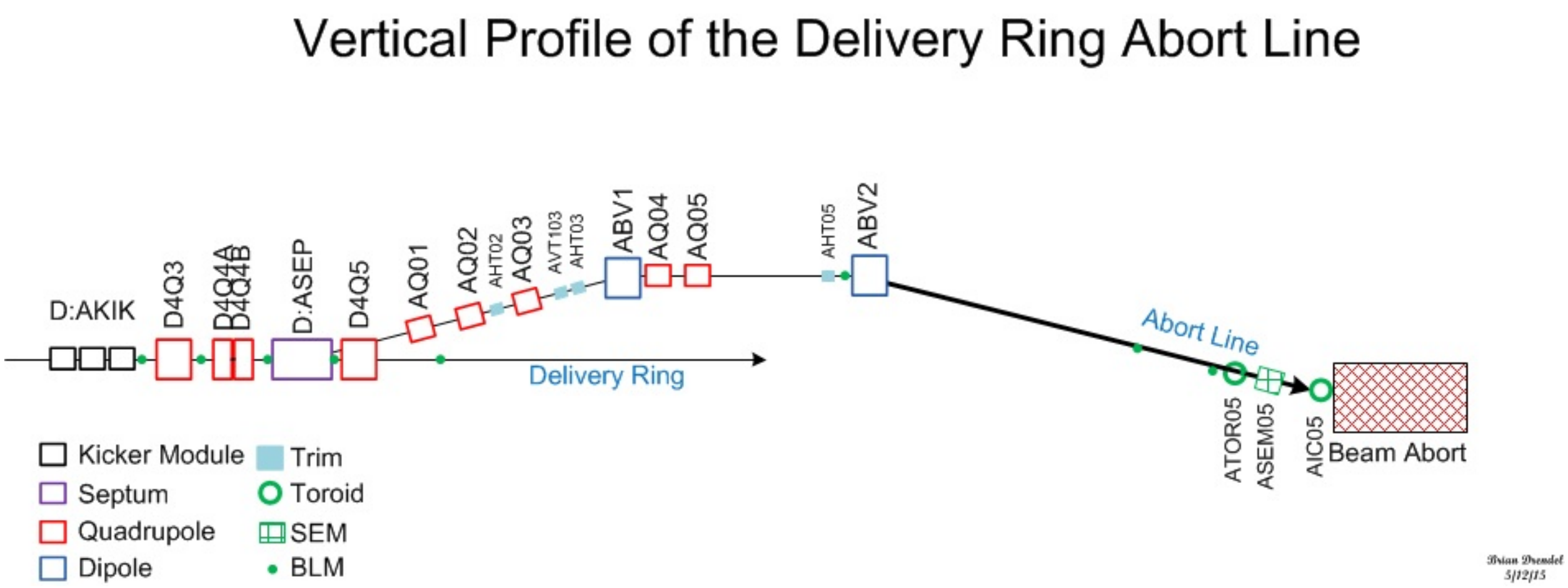

Figure 7.47: Side view of the Delivery Ring Abort/Proton Removal line.

| | Muon vs. Proton | | |
|---|---|---|---|
| | Centroid time difference (ns) | Gap size (ns) | Impact of proton removal kickers |
| Injection | 40 | None | Unable to kick protons only |
| $1^{st}$ turn at Abort | 91 | None | Unable to kick protons only |
| $2^{nd}$ turn at Abort | 161 | 41 | 25% of protons removed |
| $3^{rd}$ turn at Abort | 231 | 111 | 85% of protons removed |
| $4^{th}$ turn at Abort | 301 | 181 | Protons cleanly removed |
| $5^{th}$ turn at Abort | 371 | 251 | Protons cleanly removed |

Table 7.13: Efficiency of proton-removal system for different number of turns in the Delivery Ring, based on a 120-ns bunch length and 180-ns kicker rise time.

As the kicker magnets "fill" during the rising current waveform, the kicker magnetic field and bending strength increase proportionally. Protons are completely removed from the Delivery Ring when the kicker strength is about 85% of what is needed to center beam in the abort channel. Between 85% and 100% of the nominal kicker strength, some of the protons will be lost on the Abort Septum instead of traveling to the abort. When the kicker strength is rising and below 85%, some of the protons remain in the Delivery Ring. In addition to separating the beams to improve removal efficiency, the percentage of protons removed can also be increased by firing the kicker earlier and disturbing part of the muons.

A side benefit of the muons taking multiple turns around the Delivery Ring is that virtually all of the pions will have decayed before the muons reach the storage ring. The primary potential problem with this proton removal concept is due to differential decay systematic errors caused by the different muon path lengths as they travel through the Delivery Ring. An analysis has been done that indicates that this will not be a significant problem [20].

### 7.5.7 Muon transport to storage ring: M4 and M5 lines

There are a number of physical constraints that dictate the design and geometry of the M4 and M5 beamlines.

- Beginning with the vertical extraction trajectory from the Delivery Ring, transition to the M4 line elevation of 48 in above the Delivery Ring.
- Vertically separate the M5 line from the M4 line and set the final elevation of M5 to that of the $(g-2)$ ring, 49 in above the MC-1 service building floor.
- The tunnel enclosures housing the downstream M4 line to Mu2e and M5 line to MC-1 are separated through two independent horizontal bend strings of 40.2° and 27.1°, respectively
- The horizontal bearing and final location of the M5 line from the horizontal bend center to the $(g-2)$ ring is set by the ring center and azimuthal orientation (injection point into the ring) which has now been fixed in site coordinates.

After extraction from the Delivery Ring is complete, beam passes through a series of vertical steering magnets through part of the M4 line, then bends upward into the M5 line and continues to the $(g-2)$ Storage Ring. The 30-m long upstream section of the M4 line, between the Delivery Ring and the beginning of the M5 line, must be capable of operating at 8 GeV/c momentum for operation to the Mu2e experiment. The large differences in beam size and energy place difficult, sometimes conflicting, demands on the optics and magnet selection for this part of the M4 line. The downstream part of M4, making up the bulk of the line, continues another 215 m to the Mu2e production target.. The M5 line is 100 m long and includes a horizontal bend string to provide the proper entry position and angle into the $(g-2)$ Storage Ring. The civil constraints of the local geography and proximity of the two beamlines and respective enclosures further complicate and restrict the layout of both external beamlines.

**Civil Layout** The local geography for much of the Muon Campus is shown in Fig. 7.48. Civil and geographical constraints (avoidance of wetlands, for example) dictate a ∼40° bend after extraction from the D30-straight-section to optimize the location of the Mu2e experimental hall. Another civil engineering constraint is the location of the MC-1 building and $(g-2)$ storage ring. The $(g-2)$ experiment was positioned to avoid even low-level stray magnetic fields from Mu2e components on the one side (maximal distance from the strong Mu2e experimental solenoids) and Booster fields on the other. There are also utility corridors on the Booster side that further constrained the location of MC-1. Extending the M5 line would cause a conflict with the existing South Booster Road and reduce maneuvering room for delivering equipment. These factors set the minimum amount of left bend required for the M5 line at $\sim 27.1^{\circ}$.

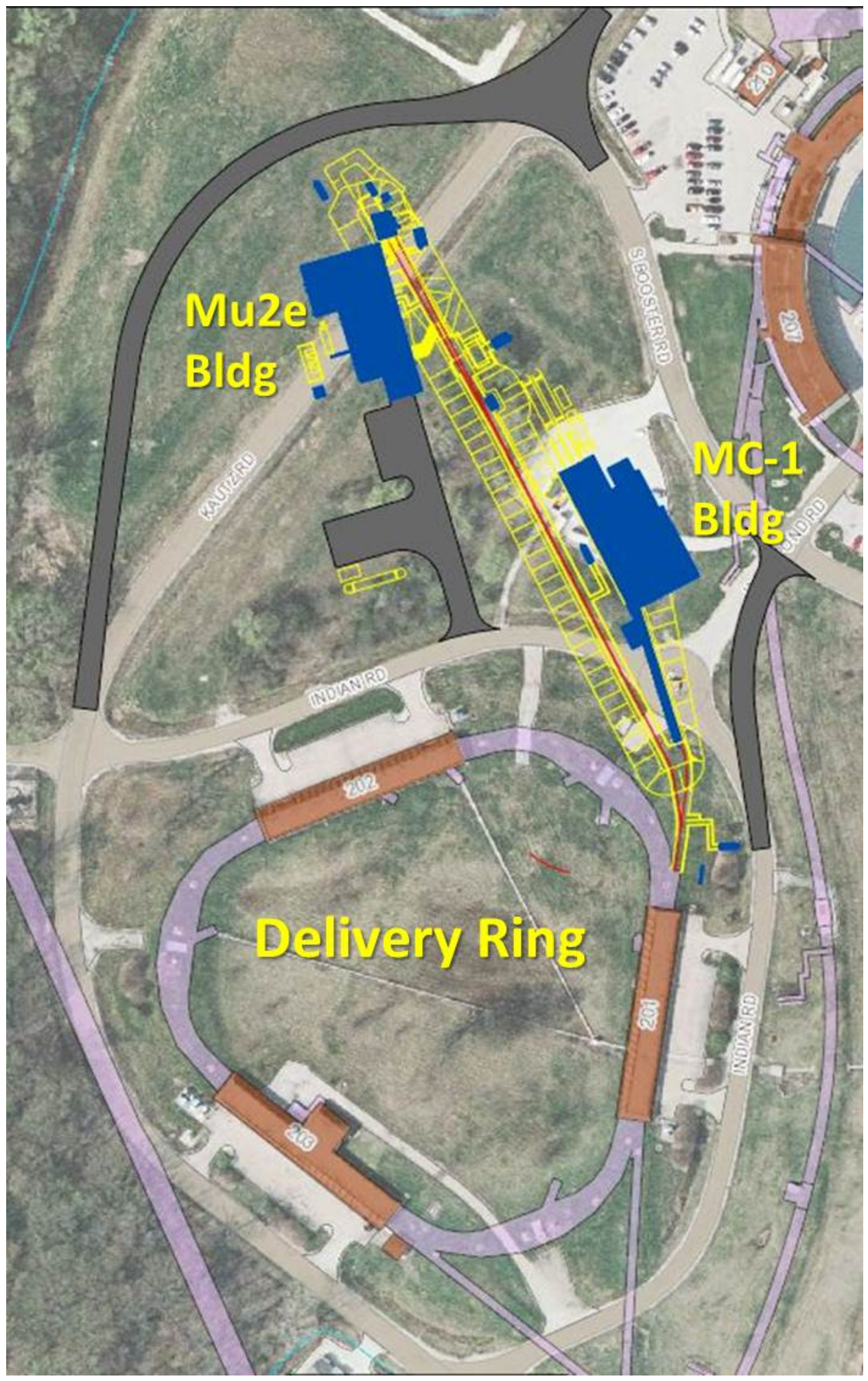

Figure 7.48: Tunnel enclosures and service buildings for the Delivery Ring, M4 and M5 lines, and experiments

The very short distance ($\sim$120 m) from the common extraction Lambertson to the $(g-2)$ storage ring mandates efficient, space-conserving separation of the M4 and M5 lines. Since physical separation from the Delivery Ring must occur vertically, the most efficient separation of the two lines is also vertical. This is accomplished by reversing a vertical-bend dipole in this section. Strong, independent horizontal left-bend dipole strings then direct beam to either the Mu2e or $(g-2)$ experiment. Final separation into independent civil enclosures is achieved by utilizing a large difference in the strengths of the left bends between the M4 and M5 lines. These strong horizontal left-bend strings must immediately follow the vertical separation stage in both lines. Rapid separation is particularly important for the M5 line given the short distance to the experiment and the need for bending and matching sections.

In summary, the M4 and M5 Lines must be designed with the following physical features:

- Horizontal kick into a Lambertson for vertical extraction from the D30 straight
- Vertical separation from the Delivery Ring magnetic components [section common to Mu2e/$(g-2)$ that takes advantage of existing tunnel civil construction]
- Vertical separation from Mu2e through a reversed vertical dipole. This section cleanly derives a separate beamline for $(g-2)$ by changing the bend strength and polarity of a single dipole between $(g-2)$ and Mu2e operation. Another dipole is added to the M5 line to level the beamline off at the storage ring elevation
- The final elevation of the M5 line (@225.1223 m above sea level) is 49 in above the projected civil elevation of the MC-1 experimental hall floor
- The final elevation of the M5 line is also 6.2 ft above the M4 line elevation (@223.2245 m) and 10.2 ft above the Delivery Ring elevation (@222.005097 m)
- A 27.09° horizontal bend string fixes the direction of the beamline from the D30 straight towards the geographic location chosen for the $(g-2)$ storage ring. (Mu2e has a 40.2° bend). The difference in bend eventually separates the two experimental beamline enclosures.

**Beam Properties and Capability**

- A horizontal kick into the field region of a Lambertson from the D30 straight section of the Delivery Ring towards the inside of the ring
- Horizontal Delivery Ring "bump" to move $(g-2)$ extracted beam away from the edge of the D2Q5 quadrupole aperture
- Beamline must transport a $40\pi$ mm-mrad acceptance and a momentum spread up to $\pm 0.5\%$ with small changes in beta functions for off-momentum particles (5% or less)
- Variable matching conditions at injection to the $(g-2)$ ring to accommodate the aperture restrictions of the current inflector and a possible new inflector.
- Beam position and angle scan capability in the final focus region of $\pm 1$ cm, with no angle change, and $\pm 3$ mr, with no position offset, both vertical and horizontal, to optimally tune injection into the storage ring
- 0.3 m reserved from last beamline element to the entrance of the $(g-2)$ ring backleg to avoid interference with fringe fields and to provide room for instrumentation for the experiment

### Optics Insertions

- **M4/M5 $(g-2)$ vertical achromat:** The $(g-2)$ vertical achromat is a complex 7-bend achromat. The vertical bends include the extraction Lambertson, quadrupole steering from D2Q5, the C magnet, the first leveling bend (EDWA), two upward-bending MDC dipoles (one is reversed polarity from the Mu2e configuration) and a final reverse-bend CDC dipole. Achromatic optics are required in the M4/M5 lines to suppress vertical dispersion from the D30 vertical extraction system. Dispersion must be suppressed upstream of the horizontal left-bend string to avoid coupling between the two planes. Independent vertical dispersion cancellation must be implemented in the M4 and M5 lines to $(g-2)$ and in the M4 line to Mu2e.

- **M4/M5 separation of Mu2e and $(g-2)$:** To separate the two lines physically and optically, the separation must occur vertically due to space constraints. Although combined, the M4 line between the Delivery Ring and M5 line must be independently tunable in the vertical-bend section in order for both $(g-2)$ and Mu2e. The beamline tune of both lines must satisfy conditions for a vertical achromat.

- **M5 horizontal dispersion module:** An achromatic module of 3 equal-strength dipoles producing a 27.1° horizontal bend to the left. The dipole string is located after the final vertical bend.

- **M5 FODO cell transport section:** A FODO section between the horizontal bend string and the final focus section is designed to transport beam with minimal losses.

- **M5 final focus section:** A strong-focusing and tunable final focus telescope is designed to adjust and optimize optical parameters of injection into the ring. The design of the final focus must accommodate both the aperture restrictions in the inflector and overcome the strong-focusing fringe fields along the injection beam trajectory. The section must have the flexibility to accommodate a larger aperture Inflector, that requires larger beta functions in the Final Focus quadrupoles.

- **M5 matching:** The beamline also requires the necessary matching sections between the custom insertions and achromatic modules.

Only ∼85 m is available for the M5 beamline insertions after accomplishing the vertical elevation change. Space restrictions do not permit momentum collimation to be incorporated into the external beamline.

### Beamline Sections

As stated above, the M5 beamline is best described in terms of its modular functionality. Correspondingly, the following descriptions detail the important sections, and discuss the design approach for each section, including a review of the extraction process. The location of each section in the overall external beamline layout is shown in Fig. 7.49.

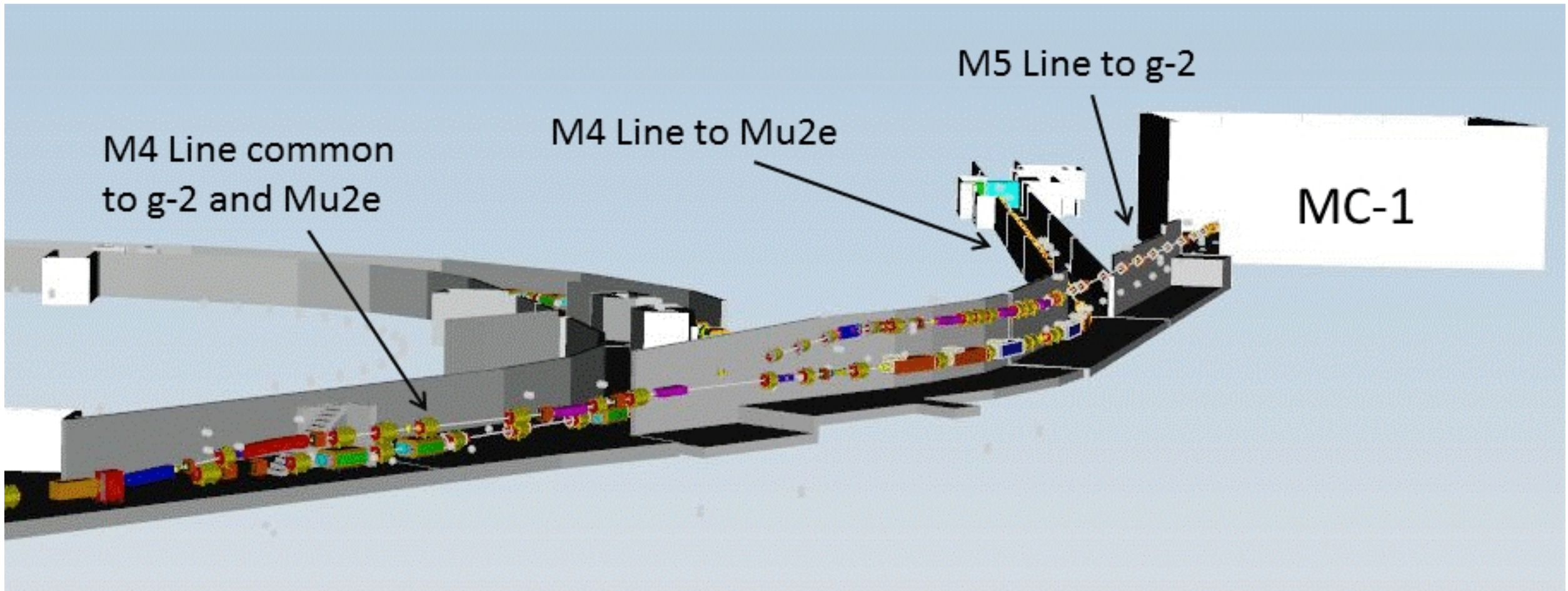

Figure 7.49: An overview of the Delivery-Ring extraction region, the shared upstream M4 line, and the downstream M4 and M5 lines.

**The Vertical M4/M5 Section** Once the beam clears the Delivery-Ring components, it can be steered onto a centered mid-plane trajectory. Steering trim magnets have been strategically placed to correct for any differences between the $(g-2)$/Mu2e and kicker/septa forms of extraction. The exact extraction orbit depends sensitively on the D30 quadrupole strengths, and these depend on the Delivery-Ring tunes established for resonant extraction or muon beam delivery for Mu2e and $(g-2)$, respectively. It is unlikely the quad strengths will be identical, but they are expected to be similar to each other. For the optics design described here,the 2004 Run II Debuncher operational strengths were used (a symmetric lattice ideally suited for $(g-2)$ operation).

The initial bend upwards is so strong (to clear the Delivery-Ring components), that the beamline must be leveled before the final M4-line elevation. This is necessary to allow sufficient space to implement a vertical achromat, which requires significant phase advance generated by quadrupoles. Leveling the beamline reference trajectory at an intermediate elevation allows a straight section to be inserted with sufficient space for a sequence of quadrupoles that generate the needed phase advance to cancel vertical dispersion after the final set of vertical bends. Given the still-limited vertical clearance and the need to have a large bend angle, an EDWA type dipole, which has small core dimensions, can be installed after D2Q6 with a bend equal and opposite to the combined bends of the Lambertson, C magnet, and D2Q5 focusing quadrupole. Leveling the line at 0.8128 m, or ~32 in, above the Delivery-Ring centerline provides for a long elevated "straight" that allows SQ series quadrupoles to be installed without conflicts with the Delivery Ring below. The only conflicts are with the extended saddle coils of the DR dipoles, and these must be avoided.

Downstream of the vertical leveling bend, an achromat is implemented using four quadrupoles. This straight section is followed by two MDC dipoles for Mu2e with reverse bends (up/down) that elevate the Mu2e extracted beam to a final elevation of 1.22 m (4 ft) above the Delivery Ring. For $(g-2)$ operation, the last vertical dipole in the M4/M5 section reverses polarity and increases in strength to switch beam delivery from the M4 to the M5 line. For $(g-2)$, therefore, three dipoles are required (the last Mu2e vertical dipole is reversed), sending the beam steeply upward to achieve rapid separation of the M5 line from the M4 line. This rapid separation proves critical in order to position the strong horizontal bend section; otherwise the ring location would have moved eastward into a utility corridor. The common M4/M5 part of the beamline thus extends from the C magnet to the vertical dipole, V907, (the last vertical dipole in the Mu2e configuration) after which the two beamlines are completely separate.

Figure 7.50 displays the achromatic optics of Delivery-Ring extraction from the center of the first quadrupole upstream of the Lambertsons to the end of the achromat. These optical functions are predicated on an assumed matched beam distribution extracted from the Delivery Ring. This may not be the case, and extracted beam properties may differ significantly between $(g-2)$ and Mu2e. Therefore it is important that the two vertical achromats have been separated between the M5 line and the M4 line and can be independently tuned. The physical layout of this section is shown in Fig. 7.51.

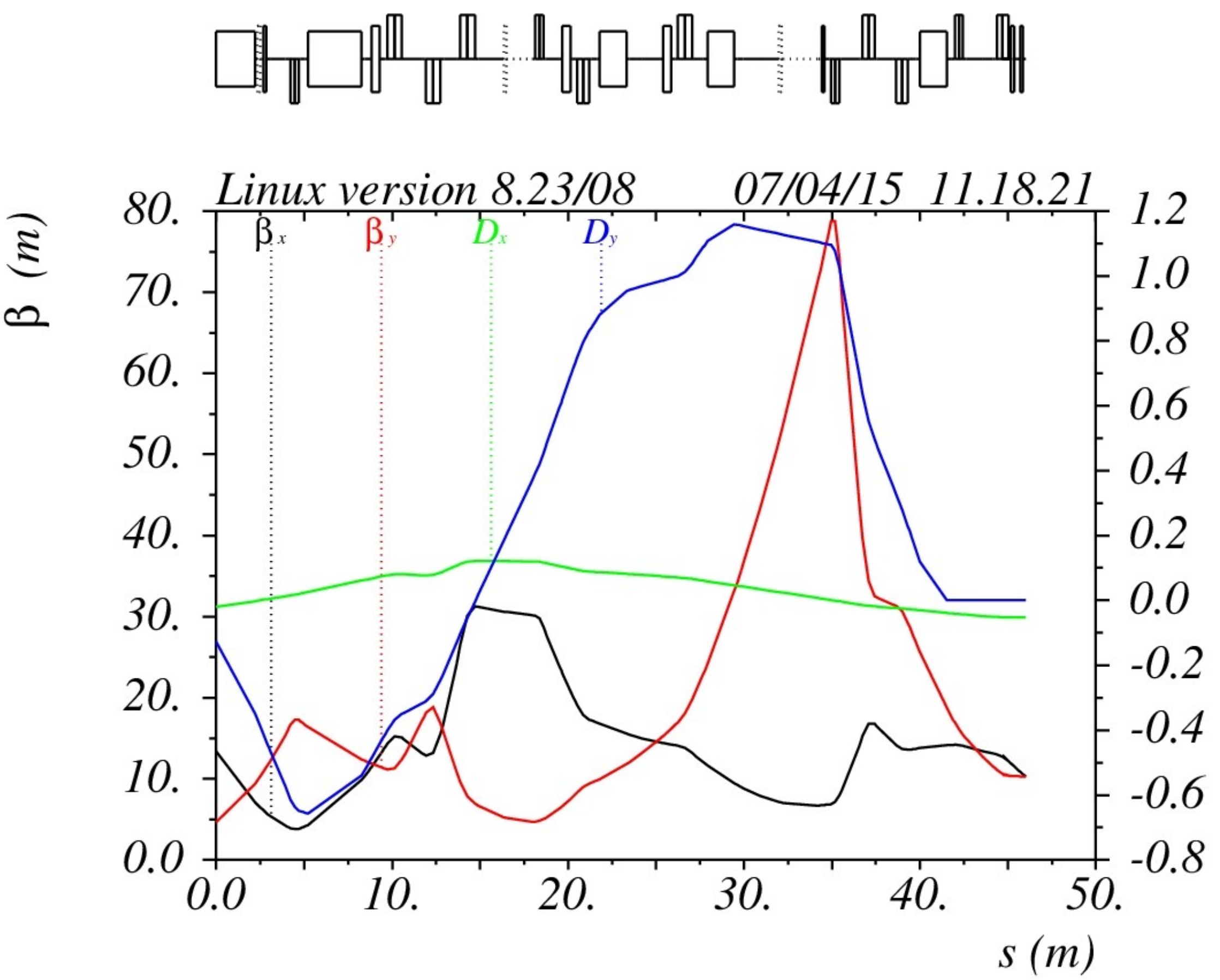

Figure 7.50: The extraction optics showing the Lambertson and C magnet (at left) followed by an opposite-sign vertical bend and quadrupoles to form the achromat. This is followed by a final bend up and then level again to the elevation of the beamline.

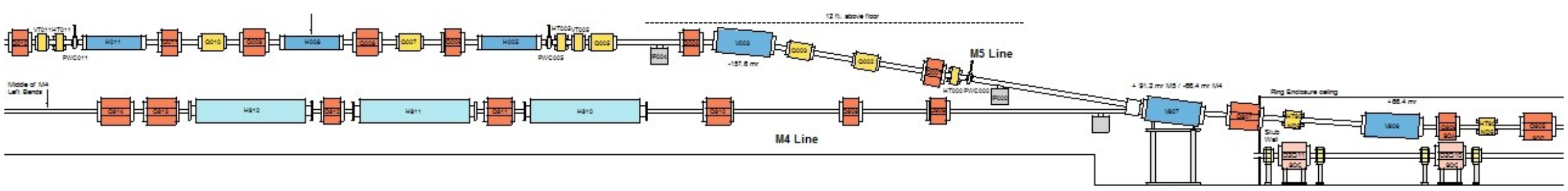

Figure 7.51: Layout of the extraction section showing the vertical bends and the separation of the beamlines to Mu2e and $(g-2)$.

**Horizontal Bend String** Immediately downstream of the vertical section, a strong westerly bend is required in order to meet the constraints of the location of the $(g-2)$ ring, particularly the critical bore coordinates through the yoke and inflector position. The horizontal separation of the M4 and M5 lines after the extended vertical separation requires the horizontal bend module to be located as close to the end of the vertical section and as compact as possible. The bend increases significantly with any further downstream translation of the bend center or rotation of the $(g-2)$ storage ring. The M5 horizontal bend string, as designed, is already close to the maximum bend that can be achieved in the limited space available. Thus, maintaining a feasible bend center location is central to an efficient beam transport design.

The location of the storage ring and MC-1 building predicates a westerly bend of 27.1°. The bearing of the bend must be physically implemented within the existing enclosure location and exactly match to the injection trajectory of the $(g-2)$ storage ring. Figure 7.52 shows the present optimized beamline location in red as determined by a) the ring position, b) the injection alignment requirements, and c) as derived from the bend center to the upstream 1.25° bend center (relative to the ring tangent at the exit of the inflector) to injection center coordinates approximating the fringe-field effect. The green circle represents the upstream end of the inflector. Alignment is discussed in detail in the Final Focus section.

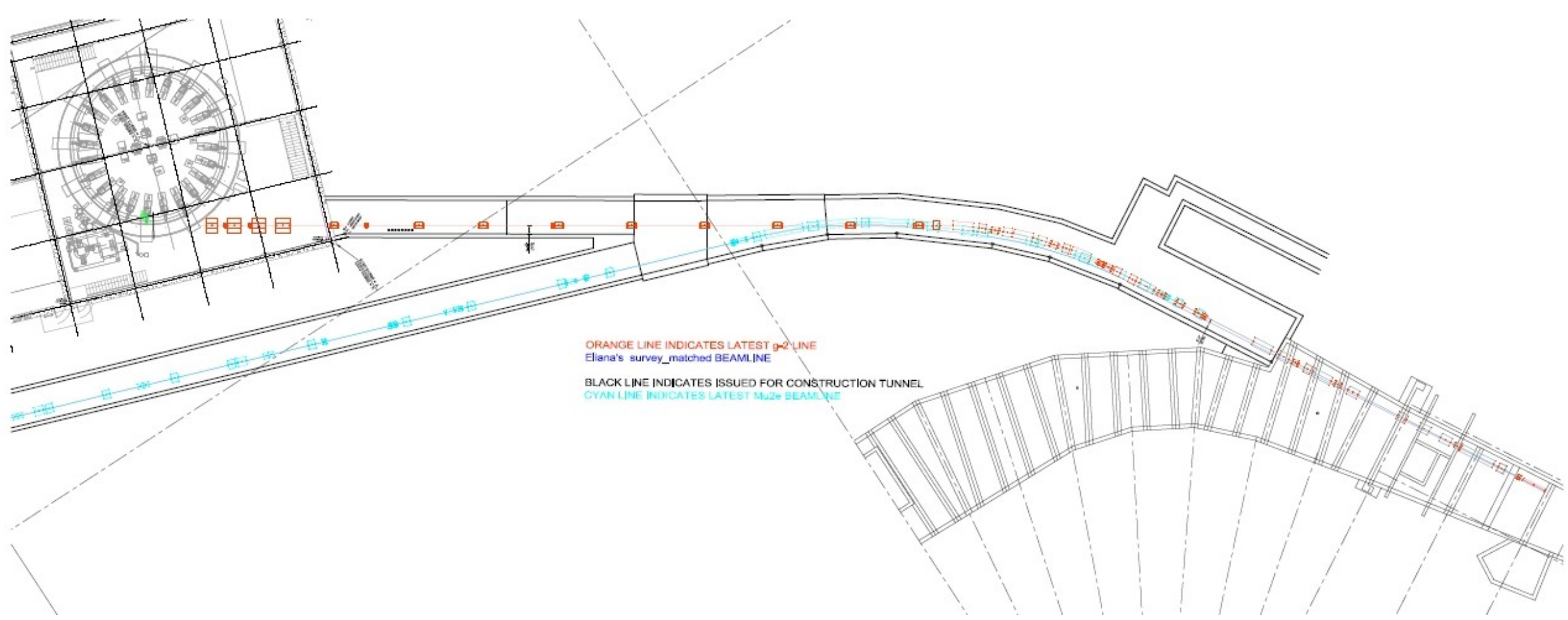

Figure 7.52: Layout of the horizontal bend section showing the horizontal separation of the M4 and M5 lines. The red line represents the optimized upstream-M4 and M5 beamline layout and the green circle the injection point at the upstream end of the inflector. The cyan line shows the beam trajectory through the downstream M4 line to Mu2e.

The horizontal bend design employs a 3-bend module comprised of three MDC dipoles in series as shown in Fig. 7.53 with each MDC delivering 1/3 of the total bend. Quadrupoles in this module supply 120° of phase advance between each dipole, with a symmetry point at the center $(D = 0)$ to cancel horizontal dispersion, fulfilling conditions for a linear achromat. Upstream of this module a three-quadrupole matching section connects the optics of the vertical section with the closed optics of the horizontal bend module.

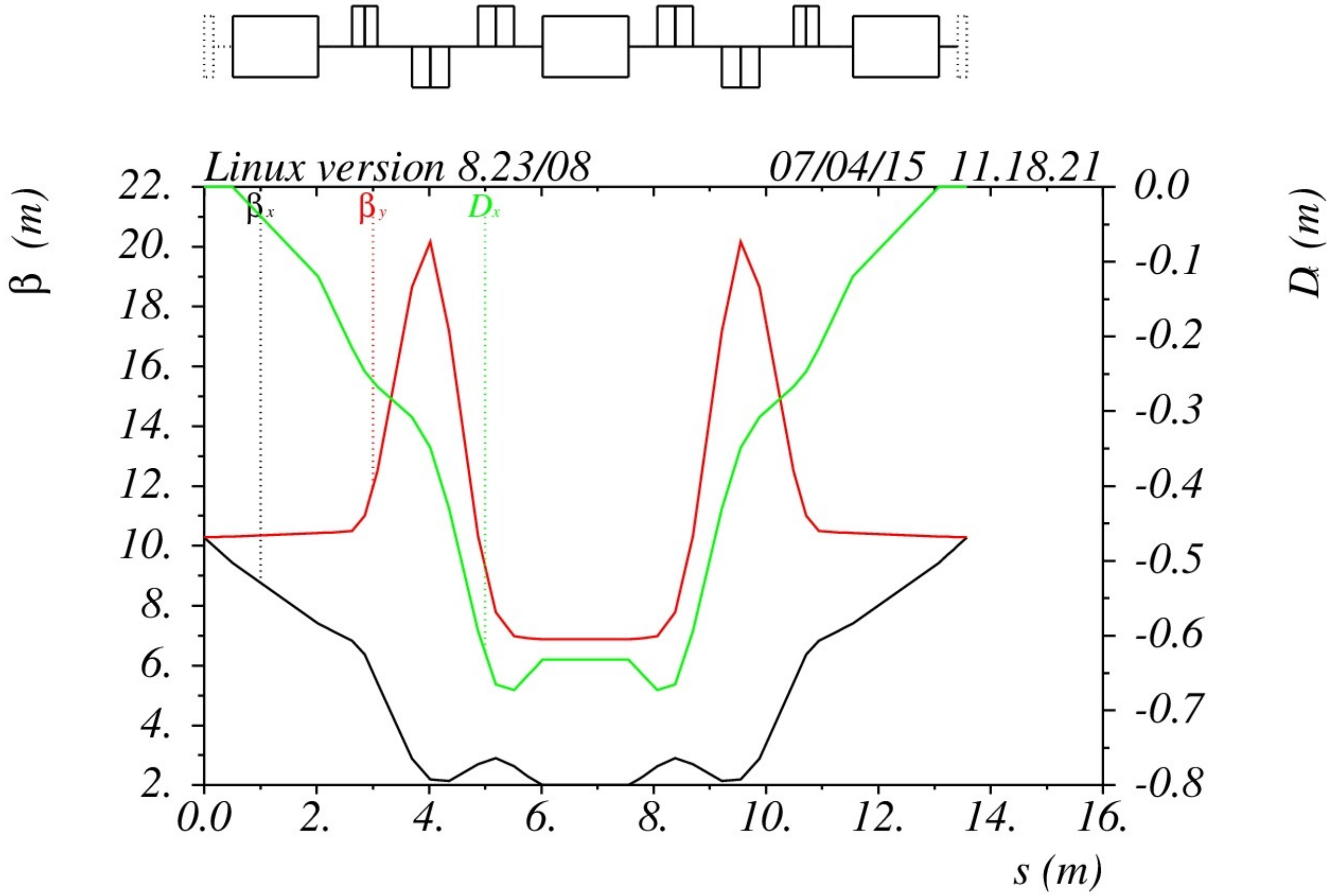

Figure 7.53: The optics of the $(g-2)$ horizontal bend insert.

### The FODO-Cell Transport Section

The basic FODO-cell optics transports beam most efficiently with the lowest losses and maximum acceptance. FODO cells are the simplest magnetic lens configuration consisting of alternating horizontally- and vertically-focusing quadrupole elements. Therefore this type of module was implemented to transfer beam from the horizontal bend string to the M5-line final-focus quadrupoles. The FODO cell structure has 90° of phase advance per cell. The current half-cell length (distance between quadrupoles) is 6-7 m and the peak beta value is 22 m, giving a beam size of ±3 cm through this section of the line. What is convenient about this type of interface is that the integrated length of the FODO insertion can be varied by 10-20% without significantly impacting the optics or the matching to upstream and downstream sections. Two consecutive FODO cells are used between the horizontal bend and final-focus modules as shown in Fig. 7.54.

### Injection into the $(g-2)$ Ring

The M5 beamline height is set to the $(g-2)$ storage-ring elevation of 49 in above the MC-1 floor elevation of 734.5 ft, or 738.58875 ft (225.1223 m). Horizontally, the bearing of the M5 line and the final-focus optics is dictated by the storage ring Inflector and the steering effect of the strong fringe fields as beam crosses these fields into the main field of the storage ring,

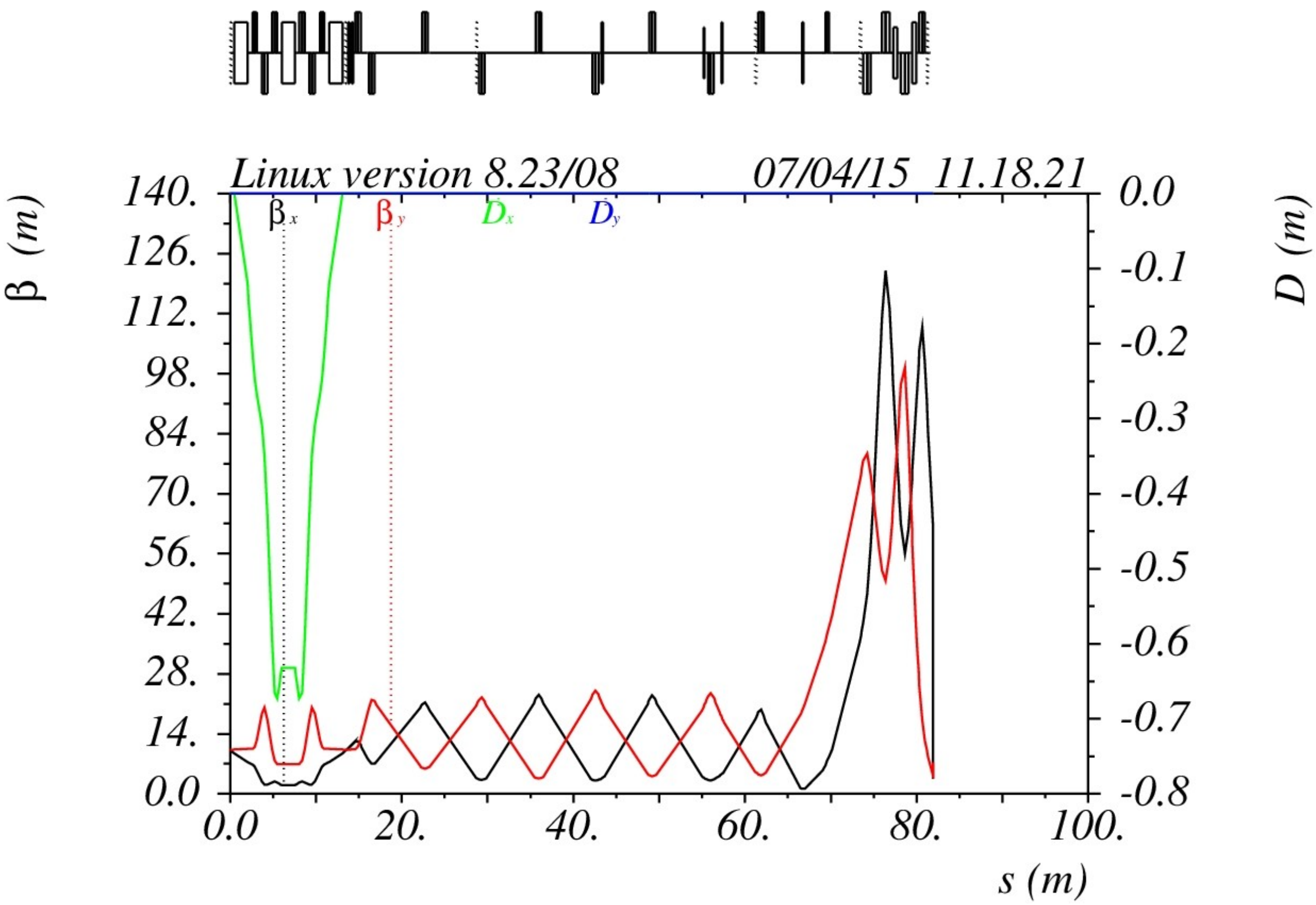

Figure 7.54: The M5 beamline starting at the beginning of the horizontal bend to the point of injection on the right.

77 mm offset tangentially from the reference orbit of the $(g-2)$ ring for 3.1 MeV/c muons as shown in Fig. 7.55 (right). The ring location and rotational orientation within the hall are therefore very critical and completely determine the bearing of the external beamline downstream of the horizontal bend string, as there is insufficient remaining distance to implement another horizontal steering section. The optical matching horizontally is further complicated by the restricted horizontal aperture of the inflector, which is 18 mm at its maximum as shown in Fig. 7.55.

The orientation of the inflector within the ring is given in Fig. 7.56 *(left)*. The linear optical functions of the ring are well established and the exit of the inflector where beam enters the ring is essentially in the center of an "open" section between the electrostatic quadrupoles so that the following are the required matching conditions, with a plot of the $(g-2)$ storage-ring lattice functions in Fig. 7.56.

$$\beta_x \approx 7.9\text{m}$$

$$\alpha_x \approx 0$$

$$\beta_y \approx 18.9\text{m}$$

$$\alpha_y \approx 0$$

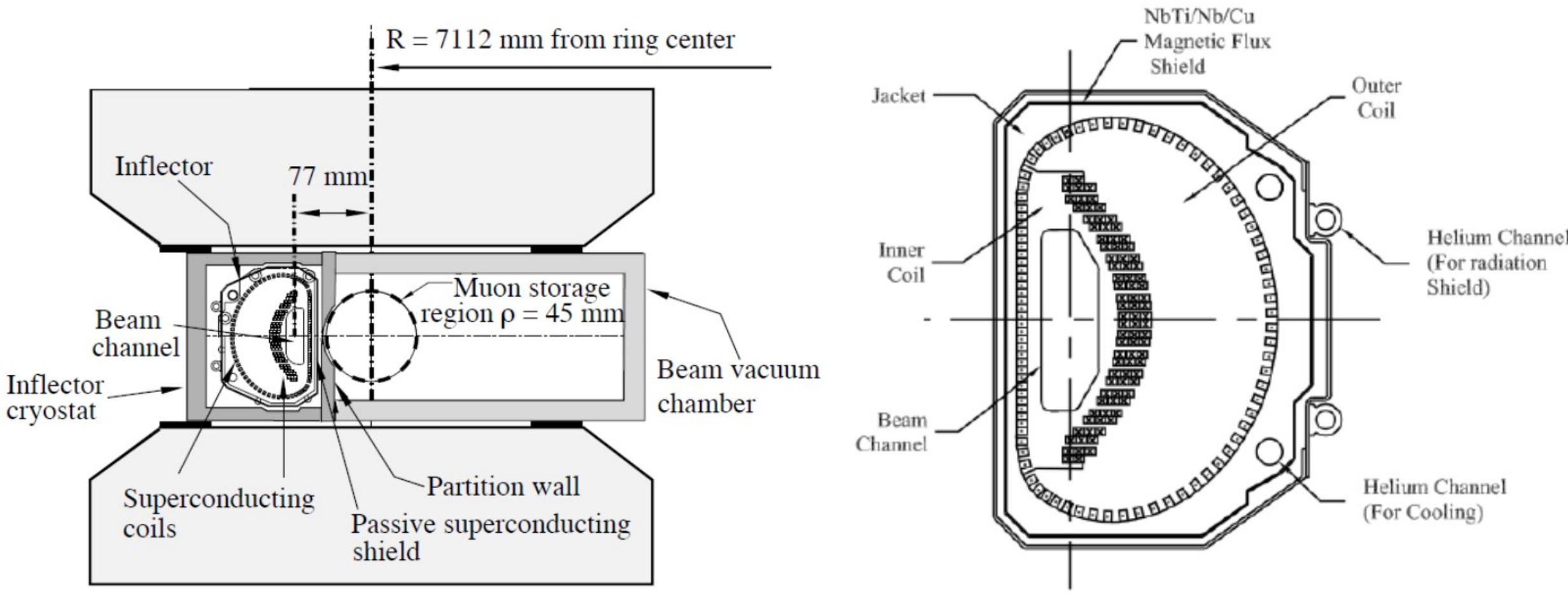

Figure 7.55: The inflector cross section which has a horizontal aperture of ∼9-18 mm and vertical of 36-56 mm.

$$D_x \approx 8.3\text{m}$$

$$D'_x \approx 0$$

However, the restricted horizontal aperture of the inflector means that the horizontal optical functions and dispersion cannot be matched at injection for a $40\pi$-mm-mr geometric emittance without losing a large fraction of the incoming beam – the inflector is the dominant aperture restriction and optical constraint for injection. The largest betatron function that can be transmitted efficiently is only 2 m and is achieved by a waist at the center of the inflector of 1.5 m (The inflector length is 1.715262 m) as in the design of the original experiment at BNL. There is no remaining aperture for increased beam size due to dispersion, so horizontal dispersion is suppressed after the horizontal bend string. This leads to the following requirements for injection:

- $\beta_x \sim 1.5$ m at the center of the inflector leading to an increased mismatched beta function of 2 m and at the point of injection and a mismatched $\alpha_x$ of about -0.6. Vertical functions can be matched with the current inflector.

- No dispersion is possible with efficient transmission through the inflector causing a dispersion wave through the machine shown in Fig. 7.57, cutting the momentum acceptance by a factor of two.

The optical matching and steering of the external beamline is further complicated by the extreme effects of the fringe fields which will be discussed in the following section.

**Fringe Field Effects** The large fringe fields of the main 1.45 T field depicted in Fig. 7.58 have a major impact on the optics and on the angle of injection. Even the weak field in the yoke has a significant deflection (in the opposite direction as the main fringe field) because of the extended length of the trajectory in this section. For the storage ring, the main field and poletip configuration produce extended fringe fields with a strong vertically focusing component due to the tangential crossing of this field (edge focusing effect). A more detailed description of the fringe fields entering the $(g-2)$ storage ring can be found in Ref. [23].

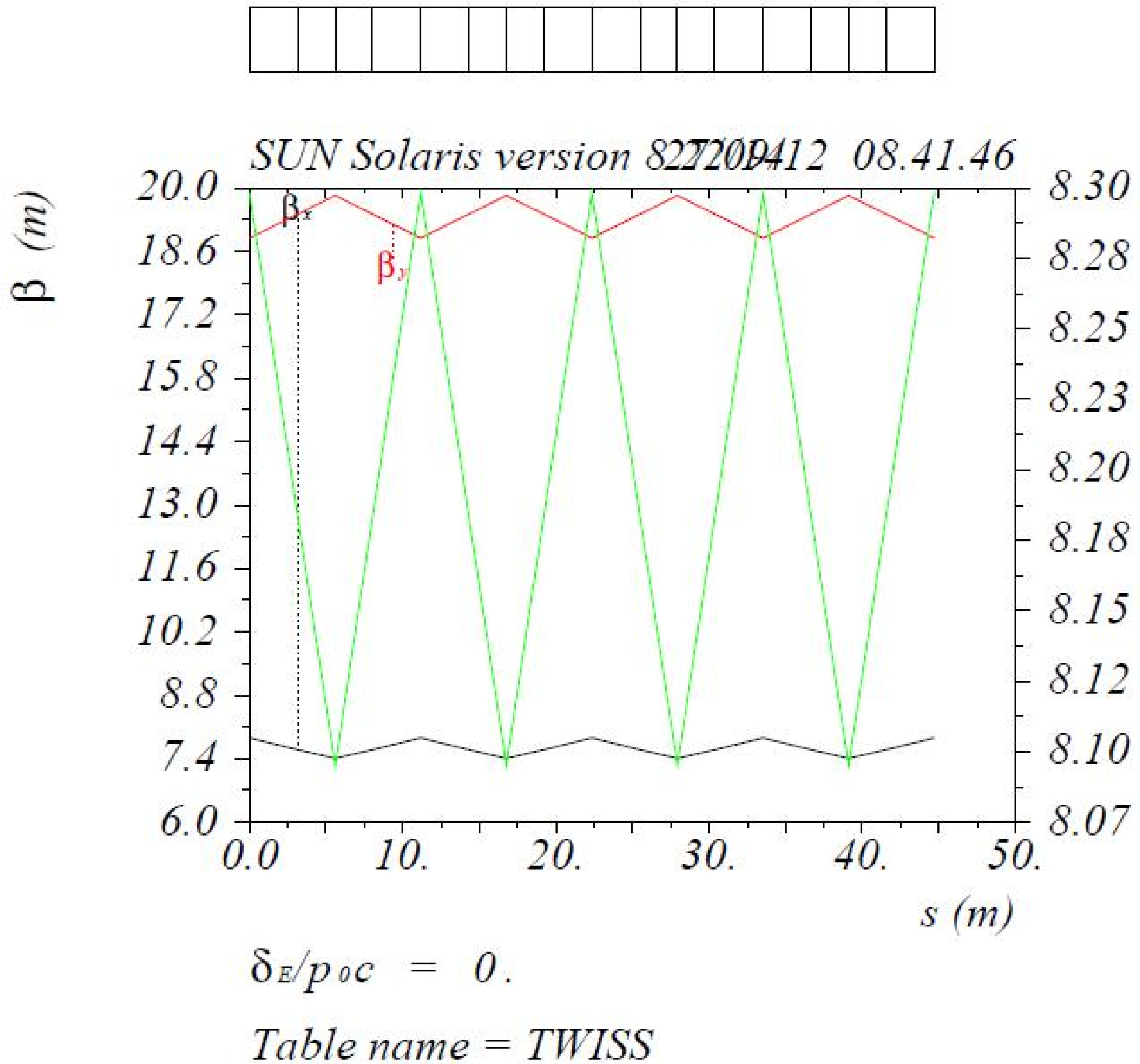

Figure 7.56: The linear ring optical functions.

The actual trajectory through this section is complicated by the field-nullifying impact of the inflector. The combined fields [24] have been computed and shown in Fig. 7.59 with the actual trajectory shown. The complicated trajectory was also initially computed for the original experiment [25]. Note that the superconducting inflector has a sharp field fall-off at the ends, but is inserted crossing the fringe fields giving rise to a reversed bend with respect to the main field in the upstream half of the inflector. The deviation from tangential injection is therefore a complicated convolution of the different fields from the yoke, main ring field, and inflector. Since such a complicated trajectory cannot be accommodated in the physical design, only the net effect is compensated for, which is 1.25° from tangential injection.

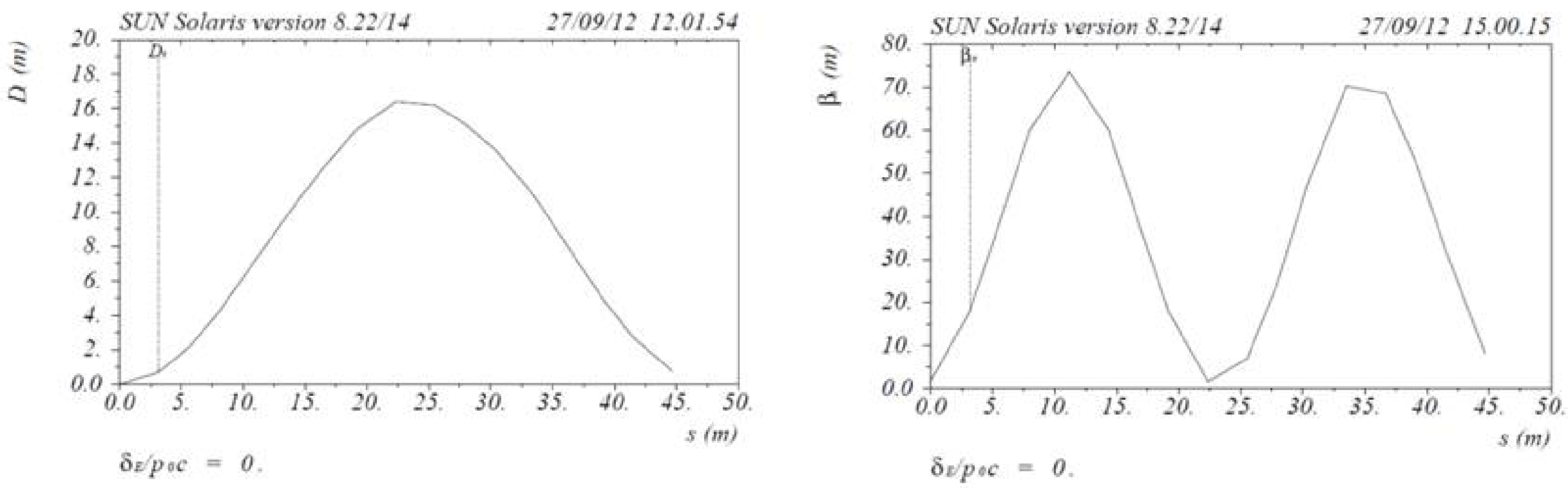

Figure 7.57: The dispersion wave around the ring created by zero dispersion at the inflector and the horizontal betatron mismatch due to the inflector.

**Final Focus** The strong vertical focusing and horizontal defocusing of the fringe field must be compensated and overcome by a strong final-focus telescope with a focusing quadrupole as close to the iron yoke as possible. Since there are weak but extended fringe fields from the bore through the yoke, a match point is set 30 cm outside of the yoke entrance for optical functions. Final-focus quadrupoles are upstream of this optical matching point to avoid any interfering interaction of the magnet with the fringe field of the yoke. Both matrices were inserted appropriately into the line to test and understand if a single design could accommodate both transfer matrices. The lattice functions at injection are $\beta_x = 2.0$ m, $\alpha_x = -0.6$, and $\beta_y = 2.0$ m, $\alpha_y = 0$. These are back propagated to the 4.3 m upstream match point using both $R$ matrices, which produce very different matching conditions: BNL $R$ matrix: $\beta_x = 2.0$ m, $\alpha_x = -0.6$, $\beta_y = 2.0$ m, $\alpha_y = 0$, and Rubin $R$ matrix $\beta_x = 2.0$ m, $\alpha_x = -0.6$, $\beta_y = 2.0$ m, $\alpha_y = 0$. The differences in the optics of the final focus are shown in Fig. 7.60. Note that steering trims are located within the final-focus quadrupoles. Corresponding trims are located further upstream in 90° phase advance locations in order to implement independent position and angle steering.

The optics of the complete line is shown in Fig. 7.61. Table 7.14 lists magnet types and $(g-2)$ operating currents for the dipoles, quadrupoles and trims in the upstream M4 and M5 lines.

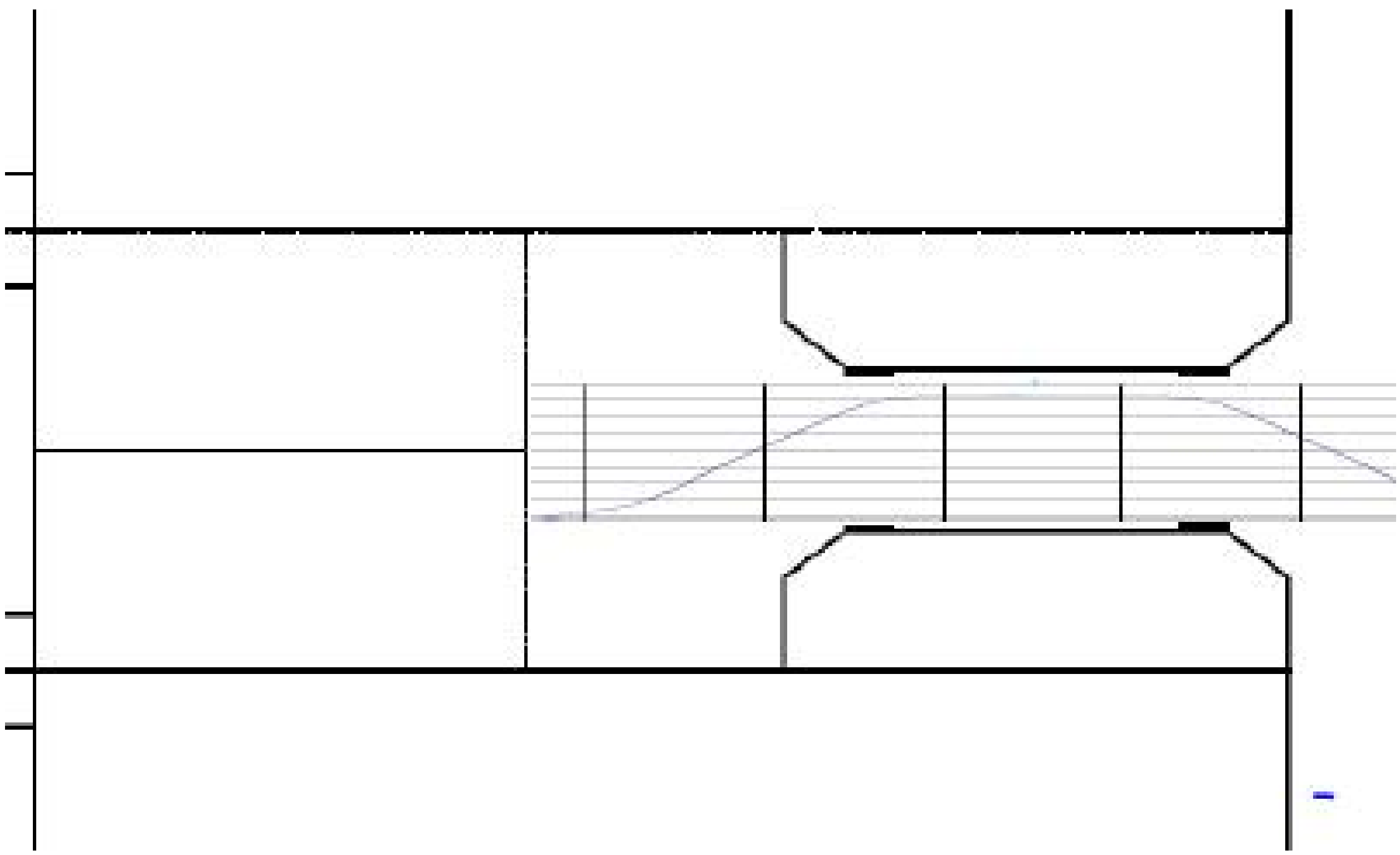

Figure 7.58: Storage ring magnetic field profile superimposed over an outline of the poletip.

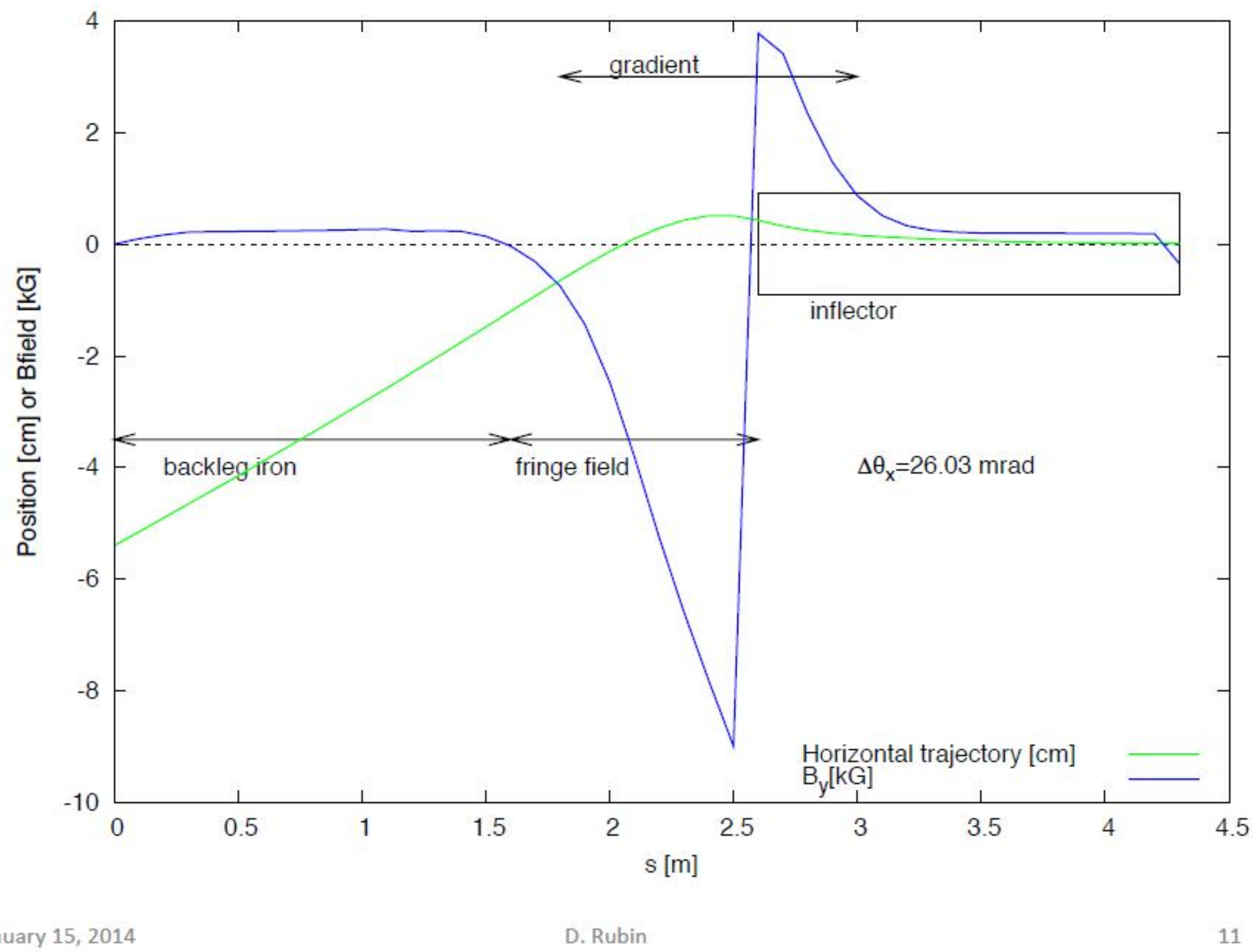

Figure 7.59: Superimposed fields from the yoke, main ring and inflector in blue. The green line is the beam trajectory.

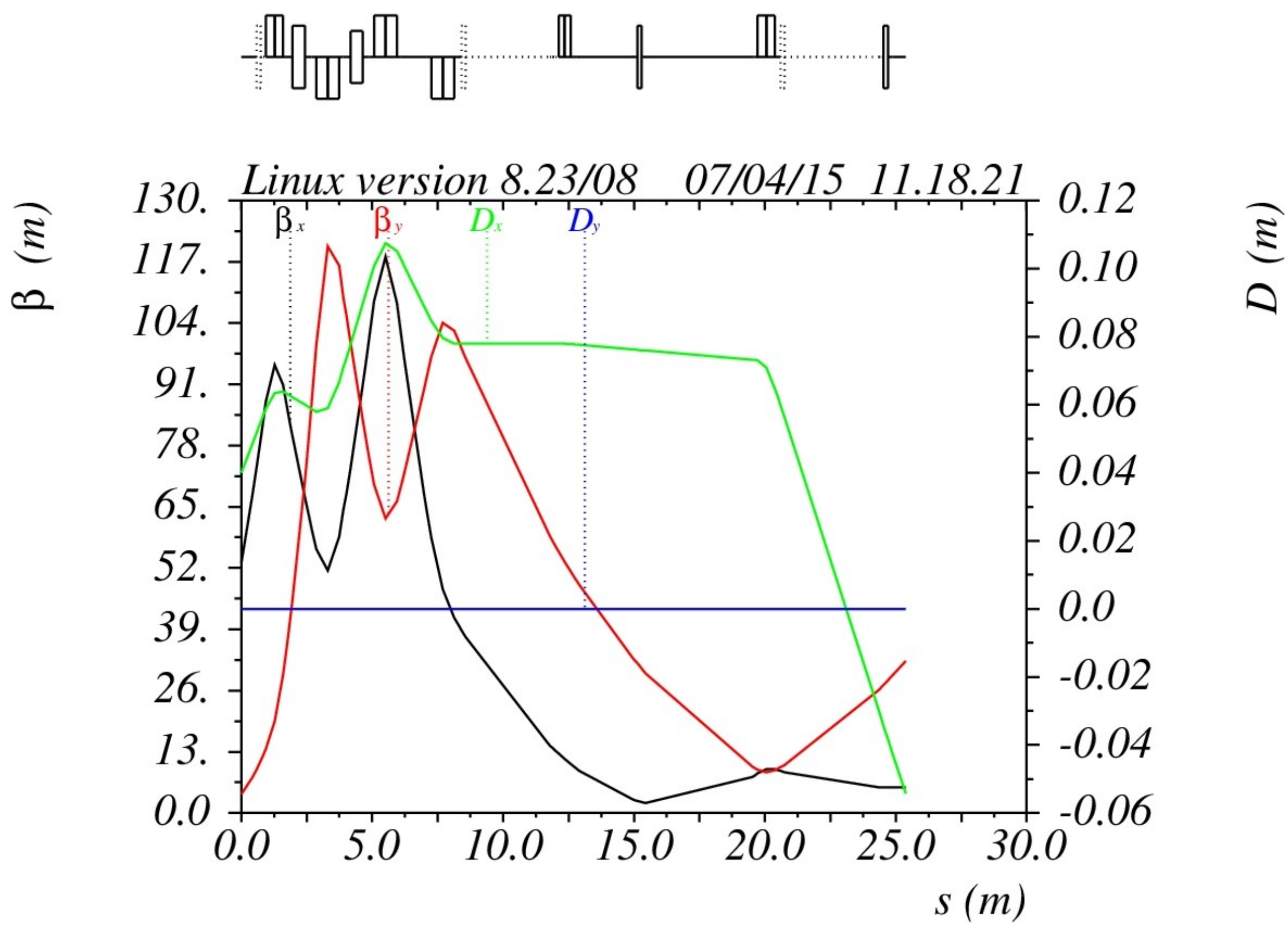

Figure 7.60: Final focus optics with matched injection conditions to the g-2 ring in all but dispersion using the BNL matrix description. The plots are left to right, downstream to upstream with the "4.3 m" match point as the origin.

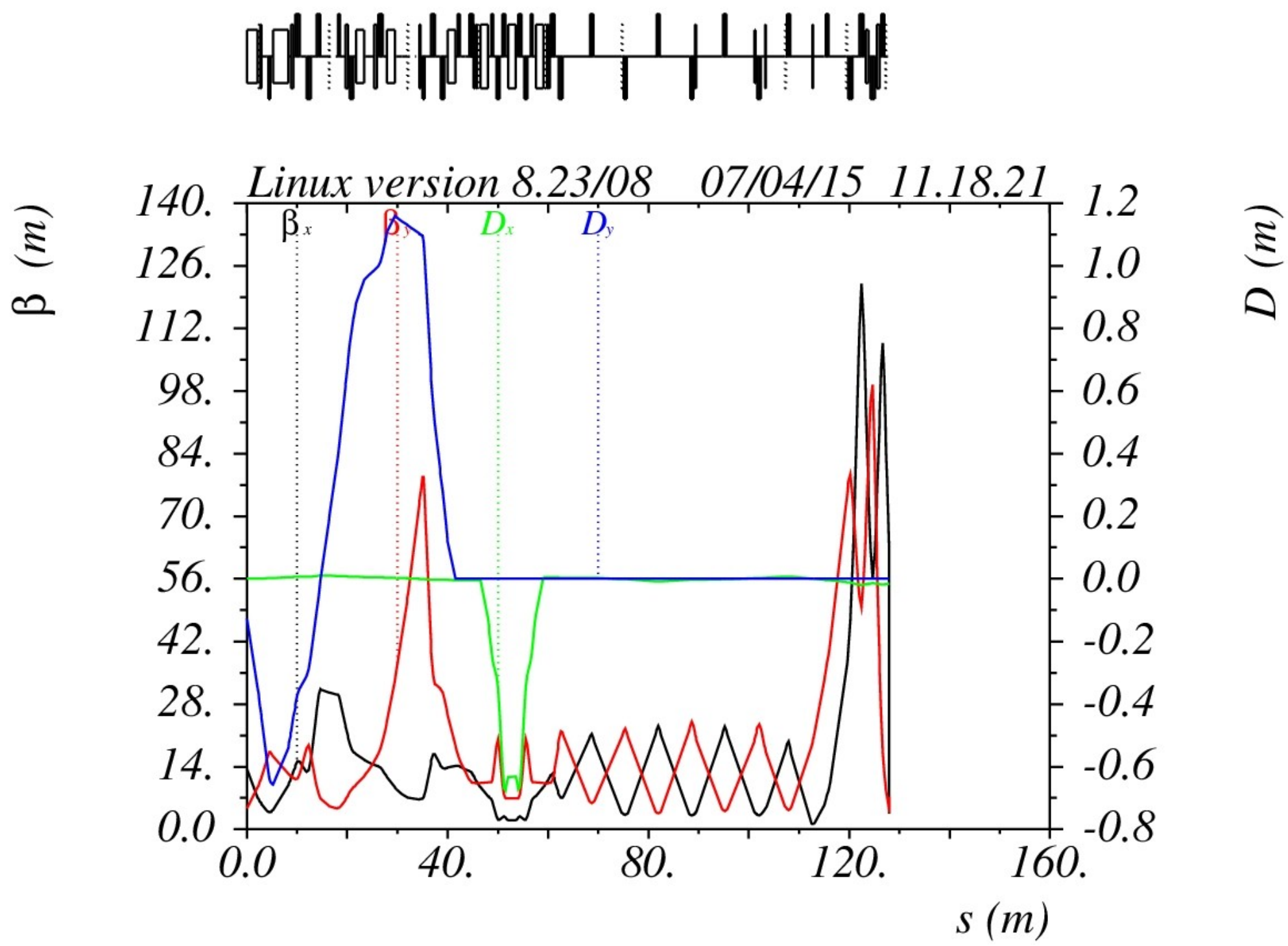

Figure 7.61: Optics of the $(g-2)$ external beamline from the extraction C-magnet to storage-ring injection.

| Magnet | Type | Current (A) | Power Supply |
|---|---|---|---|
| ELAM | MLG | 487.3 | D:ELAM |
| Q205 (DR) | 8Q32 | 796.0 | D:QT205 |
| ECMAG | MSDA | 354.6 | D:ECMAG |
| HT900 | NDA | 25.0 | D:HT900 |
| Q901 | SQA | 90.2 | D:Q901 |
| V901 | EDWA | 173.8 | D:V901 |
| HT901 | NDB | 25.0 | D:HT901 |
| Q902 | SQD | 48.9 | D:Q902 |
| Q903 | SQD | 102.1 | D:Q903 |
| Q904 | SQD | 63.6 | D:Q904 |
| Q905 | SQA | 58.1 | D:Q905 |
| HT905 | NDB | 25.0 | D:HT905 |
| Q906 | SQC | 33.8 | D:Q906 |
| HT906 | NDB | 25.0 | D:HT906 |
| V906 | MDC | 302.4 | D:V906 |
| Q907 | SQD | 16.0 | D:Q907 |
| V907 | MDC | 415.8 | D:V907 |
| M4/M5 split | | | |
| HT000 | NDA | 25.0 | D:HT000 |
| Q001 | SQA | 146.7 | D:Q001 |
| Q002 | 4Q24 | 103.9 | D:Q002 |
| Q003 | 4Q24 | 23.8 | D:Q003 |
| V003 | CDC | 993.4 | D:V003 |
| Q004 | SQA | 25.7 | D:Q004 |
| Q005 | 4Q24 | 19.3 | D:Q005 |
| VT005 | NDA | 25.0 | D:VT005 |
| HT005 | NDA | 25.0 | D:HT005 |
| H005 | MDC | 713.3 | D:H005 |
| Q006 | SQA | 132.1 | D:Q006 |
| Q007 | 4Q24 | 262.9 | D:Q007 |
| Q008 | SQB | 217.1 | D:Q008 |
| H008 | MDC | 713.3 | D:H008 |
| Q009 | SQB | 217.1 | D:Q008 |
| Q010 | 4Q24 | 262.9 | D:Q007 |
| Q011 | SQA | 132.1 | D:Q006 |
| H011 | MDC | 713.3 | D:H005 |
| HT011 | NDA | 25.0 | D:HT011 |
| VT011 | NDA | 25.0 | D:VT011 |
| Q012 | 4Q24 | 104.8 | D:Q012 |
| Q013 | 4Q24 | 116.7 | D:Q013 |
| Q014 | 4Q24 | 73.9 | D:Q014 |
| Q015 | 4Q24 | 73.9 | D:Q015 |
| Q016 | 4Q24 | 89.0 | D:Q016 |
| Q017 | 4Q24 | 83.0 | D:Q017 |
| VT017 | NDA | 25.0 | D:VT017 |
| Q018 | 4Q24 | 89.0 | D:Q016 |
| VT018 | NDA | 25.0 | D:VT018 |
| Q019 | 4Q24 | 83.0 | D:Q017 |
| HT019 | NDA | 25.0 | D:HT019 |
| Q020 | 4Q24 | 124.0 | D:Q020 |
| Q021 | SQA | 6.0 | D:Q021 |
| Q022 | LQD | 320.0 | D:Q022 |
| Q023 | LQD | 640.0 | D:Q023 |
| VT023 | NDBW | 25.0 | D:VT023 |
| Q024 | LQD | 815.0 | D:Q024 |
| HT024 | NDBW | 25.0 | D:HT024 |
| Q025 | LQB | 840.0 | D:Q025 |

Table 7.14: Magnet locations, type, operating current and power supply configurations for the upstream M4 and M5 lines.

### Alignment of the Ring and Injection Beamline

The alignment of the downstream section of the M5 line is critical given the accuracy required to target the inflector correctly and enter the ring tangential to the circulating beam reference orbit. The site coordinates of the MC-1 building were specified initially to place the experimental hall in a location optimal from a shielding and civil-engineering perspective and in consideration of the proximate Mu2e external beamline and experiment. This was further complicated by the fact that the MC-1 service building construction was accelerated to accommodate reassembly and cool down of the $(g-2)$ storage ring. This meant that the storage-ring position and rotation was further constrained because the MC-1 building location was fixed. The rotational orientation of the storage ring is critical given the strict tolerances of the injection system and the fact that a switchyard or injection-bump configuration is not spatially permitted by the external beamline enclosure (at least not without significant cost). The site coordinates now specified for the $(g-2)$ storage ring, and extrapolation of the required incoming injection trajectory and beamline coordinates can be found in Ref. [23] and are shown in Fig. 7.62.

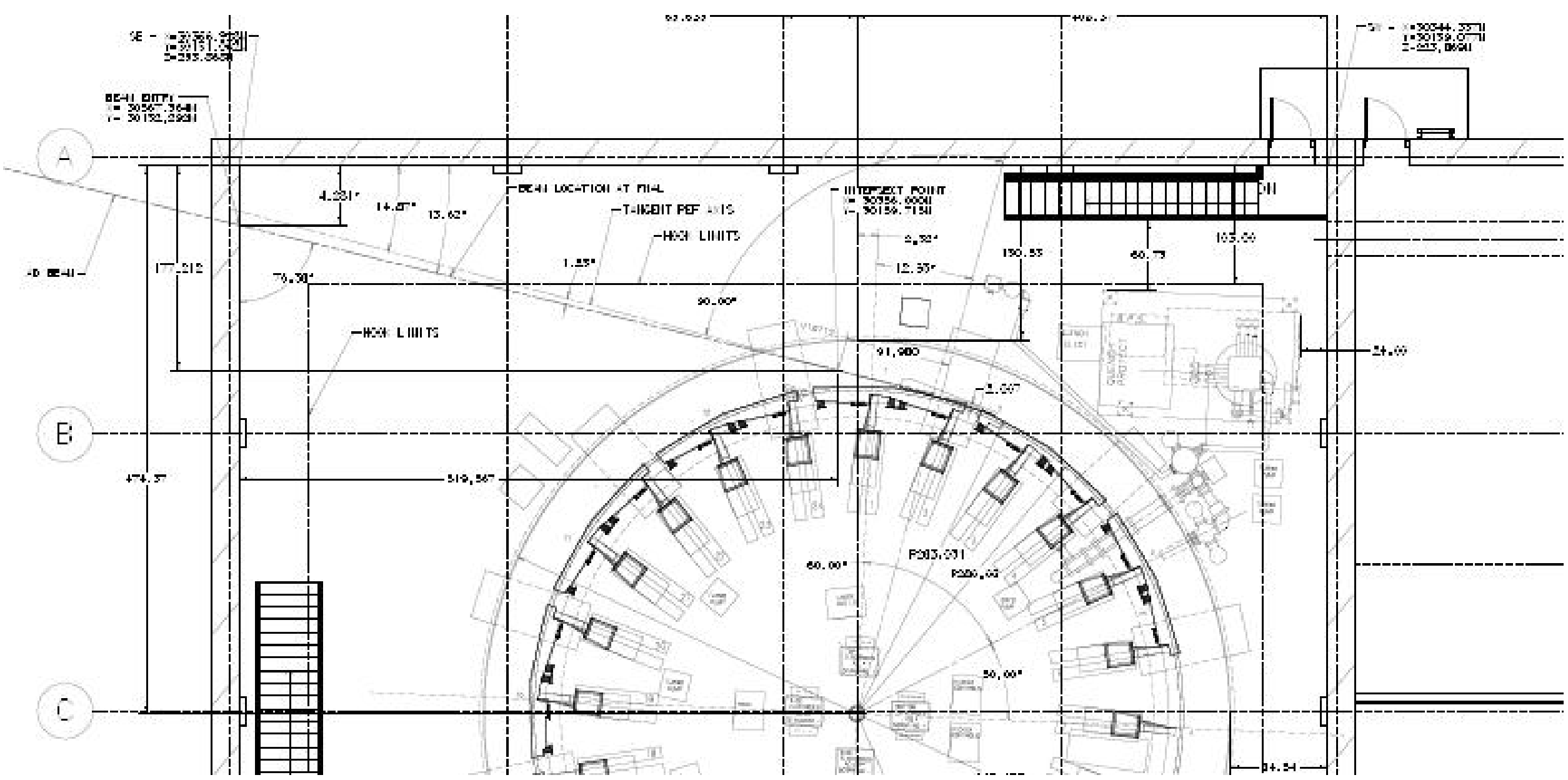

Figure 7.62: Alignment coordinates of the MC-1 hall, beam entry and the preliminary 1.25° point coordinates which have been cross checked or recomputed.

### 7.5.8 Vacuum Systems

The existing vacuum systems in the rings and transport lines have performed very well during Pbar operation. Typical vacuum readings in the Debuncher and transport lines were approximately $1 \times 10^{-8}$ Torr. The Debuncher has good ion-pump coverage that should generally be adequate for $(g-2)$ operation. Stochastic cooling tanks, kickers and septa that will be removed during the conversion have built-in ion pumps, so some of these pumps will need to be installed in the vacated spaces. Injection and extraction devices will either have ion pumps integrated into the design, or have additional pumping capacity added to the surrounding area. Vacuum components from the AP2 and AP3 lines should provide most of the needs for the reconfigured M2 and M3 lines. The Accumulator has enough surplus ion pumps available to cover part of the needs for the extraction beamlines. Most of the vacuum pipe for the M4 and M5 lines will need to be purchased. The ion-pump density in the new beamlines will be at least as great as what was used during Pbar operation. Vacuum controls from Pbar will be repurposed for the new beamlines.

The Delivery Ring will retain the present scheme with six vacuum sectors separated by pneumatic isolation valves that can be controlled remotely. The vacuum valves in the D30 straight section will be rearranged slightly to accommodate the new component layout. There will also be manually controlled valves, particularly in the D30 straight section, to allow smaller sections to be isolated. The beamlines will also have isolation valves, both pneumatic and manual, to facilitate repairs and reduce pump-down times.

### 7.5.9 Infrastructure Improvements

Electrical power for the Antiproton Source is provided by Feeder 24, which operated with a power level of about 4.4 MW during Pbar operation. Although the $(g-2)$ power load is expected to be considerably less than what was used in Pbar by virtue of the reduced beam momentum, the Mu2e experiment must also be able to operate the same magnets at 8.89 GeV/c. For Mu2e, most service buildings are expected to use approximately the same amount of power as they did in Pbar operation. The exception is the AP-30 service building, where there will be an increase in power load from the injection- and extraction-line power supplies. A power test was performed on the individual service building transformers to aid in predicting the power needs for Mu2e [26]. Also, since the Accumulator will no longer be used, approximately 1.4 MW will be available for new loads.

Presently, Pbar magnets and power supplies receive their cooling water from the Pbar 95° Low Conductivity Water (LCW) system. The cooling requirements for $(g-2)$ are expected to be lower than for Pbar operation. However, Mu2e will operate at 8.89 GeV/c and create a substantially larger heat load than $(g-2)$. Fortunately, the removal of the heat load from decommissioning the Accumulator and the D/A line should be enough to offset the increase from the extraction line and other new loads. The M4 and M5 lines will have an LCW branch that will run the length of the new tunnels and connect to the Debuncher header in the D30 straight section. The LCW will also continue into the MC-1 building to be used in the power supply room. If necessary, it is also possible to design smaller closed-loop systems that heat-exchange with the Chilled Water system. This strategy has been used to cool some of the loads in the Target Station. The Chilled Water system has adequate capacity and is already distributed to the existing Pbar service buildings as well as to the new MC-1 building.

### 7.5.10 Power Supplies

The magnet power supply systems will provide the necessary current and regulation to transport beam through the Muon Campus beam lines. The transport beam line power supplies will be operated in DC current mode. The power supply system will build on existing power system designs for voltage, current and controls that have been developed by the FNAL Accelerator Division Electrical Engineering Support Department in order to keep the maintenance cost as low as possible.

The Muon Campus beamlines are expected to use two different formats for power supplies. The first type is switch-mode commercial power supplies that can be procured from multiple vendors and operated in voltage mode. Most commercial power supplies are unstable on inductive loads (such as magnets) and do not have built-in compensation correction, which must be added for them to work properly. The switch-mode power supplies have the advantage of being efficient and compact and are more cost effective for lower power applications. The second type of beam line power supplies will be phase-controlled supplies commonly called SCR-type supplies. These supplies will be semi-custom designed and built to FNAL specification. SCR-style supplies will be required to use the AD E/E Support designed voltage regulator and then use a current regulator to close the current loop. New SCR supplies will only be purchased for high-power applications. Some existing Pbar SCR supplies will also be reused in order to save money. For instance, the Delivery Ring power supplies, which are mostly the SCR type, will be reused for Muon Campus operation.

#### System Layout

AD E/E Support has designed and developed a current regulation and controller system that is used in the DC application. The control portion of the system uses a Programmable Logic Controller (PLC) to manage input power as well as status readback and control through the Accelerator Control NETwork (ACNET) system. A variety of commercial power supplies are used in voltage mode to provide the current to the magnets. This requires an interface circuit to convert the status and control to signal levels that can be provided to the controls system. Two versions of the interface chassis exist and can support 8-16 power supplies.

Each current regulator uses a PC-104 embedded micro controller to provide current regulation for up to four power supplies. This controller also collects the status and control information from the power supplies via the PLC and converts this information into a format that the ACNET controls systems can present to operations.

**Current Regulation** There are two critical parts of the current regulation system; one is the current-measuring device and the other is the stability of the current reference. Each system will have a total current monitoring DCCT (Direct Current Current Transformer) installed that is used to provide accurate and stable current feedback to the regulator. The current regulation system will be a recent design constructed for the last Main Injector and Linac Ion source upgrades. This system is a Digital/Analog combined regulator built using a PC-104 embedded processor system that provides the current regulation by providing a total voltage reference to the power supply. This regulator supports four power supplies in a single chassis and provides all of the voltage drive through the power supply for the main current, including any correction needed. The DCCTs used for the feedback will be commercial devices procured at the level that meets the long term stability requirements of the experiment. This regulator system can support operation at the ±4 ppm level as designed but can be improved to the ±0.25 ppm level by procuring a high-performance DCCT. The plan is to use ±300 ppm for the beamline power supplies, which is more economical due to the lower cost current feedback devices. This is a typical regulation specifications for beamlines at FNAL. This system is not intended for fast ramping power supplies and has a $dI/dt$ limiter used during startup. The existing power supplies for the Delivery Ring have much tighter regulation tolerances.

The current regulator has one PC-104 processor that sends an 18-bit digital reference to each of four Sigma Delta DACs in temperature-regulated modules (see Fig. 7.63). Each DAC module receives the analog current from a current output DCCT/HAL probe and converts the current to a voltage using burden resistors in the temperature-regulated module. The reference and current signals are subtracted to generate an error signal, which is then amplified 100 times. This amplified error is then sent to the PC-104 module that adjusts the drive to the power supply to minimize the error signal. The reason for the amplification is to reduce the sensitivity of the PC-104s AD converter. The magnet parameters are loaded into the PC-104 along with the current loop bandwidth and maximum gain limit. It then uses this information to provide the correct correction to each power supply.

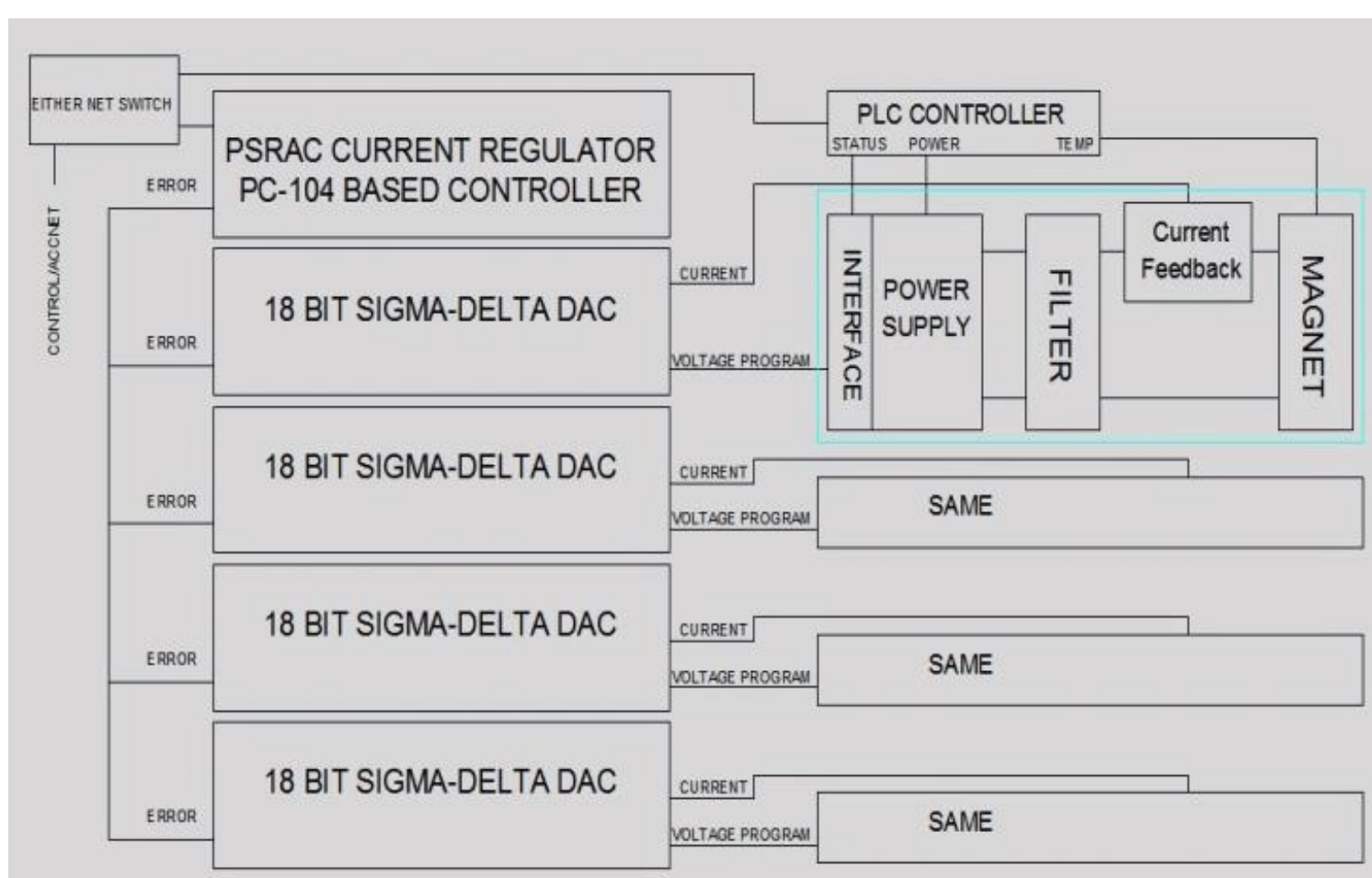

Figure 7.63: Current regulator.

An additional feature of this system is that the PC-104 has a transient recorder built in that will record trip events and provide data for analysis, which is very useful during single-event trips that happen very infrequently. A second option built in this regulator has a window detector that can be set up to monitor current, DAC settings, and the current error signals to ensure that they are within a set range. These limits are set up using an independent path into the processor. The PC-104 monitors and uses four analog signals: current reference, current, voltage, and current error. These signals are stored in the transient recorder during a trip and can be plotted at 1440 Hz.

**PLC Controller** All of the control and status read backs are provided through the PC-104 current regulator using a single E-net connection to the ACNET system. The power-supply system uses a PLC to collect data from four power supplies via an E-net connection and passes it through the PC-104 to ACNET. The information is collected in the PC-104 that converts it to ACNET format. The PC-104 also provides the On, Off, and Reset functions to the power supplies through the PLC that manages things that are common to all four supplies. The PLC is a device that allows for the level shifting of signals from the many different types of power supplies that will be used. All of the signals and controls become local to the power supply location with only an E-net cable back to the controls system. This has the benefit of reducing the amount of controls cards needed for collecting data, and the cable that is needed to connect to the cards.

All of the power supplies will need monitoring and control of the 480 VAC input power. The PLC is used to manage the control of this power and to monitor signals common to all supplies: safety system, door interlocks, smoke detectors, and magnet over-temperature. The PLC interfaces the 480 VAC input power to the supplies using a custom-built starter panel that we have chosen to have a limit of 40 kW. Two of these 40 kW starter panels can be installed in a standard relay rack to power two supplies or groups of supplies. This choice to use 40 kW as a power limit to each group of supplies is based on providing reasonable wire for the input power, size of starter on the panel, and a reasonable amount of space for the high-current cables to exit the rack to the magnets. This allows enough room to install up to four 10-kW power supplies in one half of a rack to reduce the amount of floor space needed for racks of power supplies. The starter panel provides a place for this connection; the main contactor on the starter panel uses a +24 VDC coil that can be directly connected to the Electrical Safety System.

**Interface Chassis** We will be using the PC-104 embedded current-regulation system with many different sizes and manufacturers of power supplies. Some use TTL for status and control, while others use +24 VDC, so the interface chassis and cards provide a place to convert signals to useful levels that can be sent back to the controls system. Using the PLC and interface chassis allows the PC-104 code to be identical for all systems, with the need for only some of the PLC code to be unique.

**Switch-Mode Power Supply** We plan to group switch-mode (SM) power supplies by size in order to reduce the number of different sizes we need to procure and the number of spares to store. The specification for the SM-style power supplies will define the voltage, current, and power level for each size, as well as a voltage regulation and ripple. The plan is to share a common line voltage for all supplies so that supplies with different power levels can be used in the same rack. There are only three or four manufacturers in the US that can meet all of the Muon Campus needs.

**SCR-style power supply** The specification for the SCR-style power supplies will be based on the present design of the 75 kW power supplies used in the Main Injector. This specification requires the use of the FNAL Accelerator Division E/E Support designed voltage regulator. E/E Support will have these voltage regulators constructed, and two copies are provided to the manufacturer to use for testing. The reason for this is to reduce the maintenance load on the engineering staff caused by unique regulation electronics of many different manufacturers. SCR supplies in general support two quadrant operation; this will not be needed for the Muon Campus beamlines. However, to minimize the variety of supplies requiring support in the Accelerator Division, this requirement will be maintained.

Manufacturers of modern SCR power supplies use PLCs internal to the equipment rather than constructing custom circuit boards to provide control connections. The specification for the power supply will include the detailed information needed to ensure that any PLCs used are compatible with maintenance tools we have on hand.

SCR supplies will need LCW cooling for at least the SCRs and possibly the magnetics. The LCW cooling water will be sourced from the M5-line tunnel to the MC-1 power supply room. The power supplies will need 50 kW of cooling with a minimum pressure range of 60-100 psi and a flow of 18 GPM.

### Power supply locations

Power Supplies for the Muon Campus will be housed in a combination of existing and newly constructed buildings. Power supply changes for the M1 line are relatively minor and can be incorporated into the existing F23 and AP-0 service buildings. Although the changes to convert the AP2 and AP3 lines into the M2 and M3 lines are significant, there will be enough room in the existing F27, AP0, and AP-30 service buildings to accommodate them. The upstream M4 line will have power supplies located in the AP-30 service building. The M5 line will be powered by supplies located in the MC-1 power supply room. In addition, the MC-1 power supply room will also house supplies for a large part of the downstream M4 line, including the critical A/C Dipole system, which is not needed for $(g-2)$. The sharing of the MC-1 power supply room saved considerable cost for the Mu2e experiment by eliminating the need for an additional service building or a larger power supply area in the Mu2e service building with extremely long cable runs. Figures 7.64 and 7.65 show the power supply layouts in the AP-30 and MC-1 service buildings.

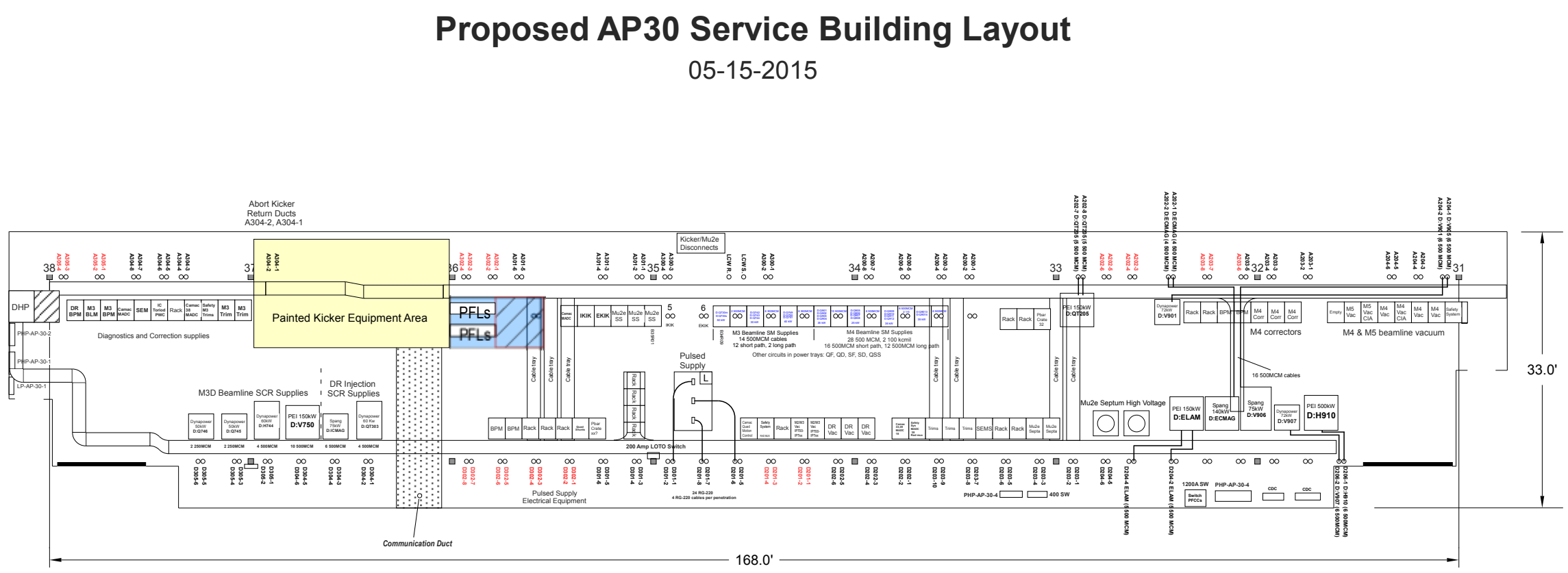

Figure 7.64: Power Supply layout in the AP-30 Service Building.

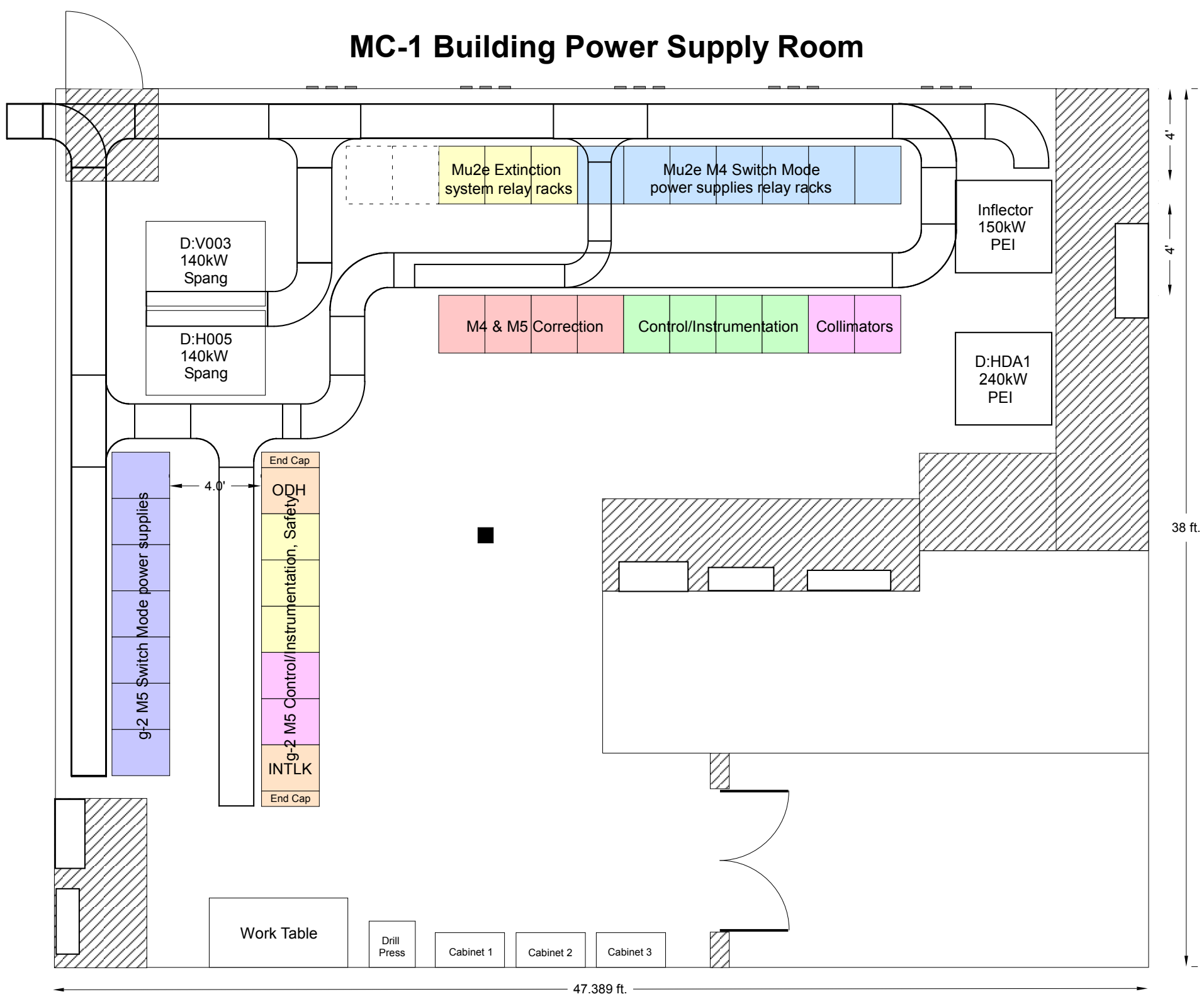

Figure 7.65: Power Supply layout in the MC-1 Power Supply Room.

## 7.6 Controls and beam monitoring

### 7.6.1 Accelerator controls

A well-established controls system allows devices in the former Antiproton-Source ("Pbar"), now Muon, service buildings and tunnel enclosures to receive information such as synchronization signals and to communicate back to other accelerator systems. A map of the service buildings, labeled "AP" for former Antiproton-Source buildings, and "F" for buildings which are part of the F-sector of the Tevatron, is shown in Fig. 7.66. Devices in the new extraction beamlines and MC-1 building will also need to be connected to the controls system.

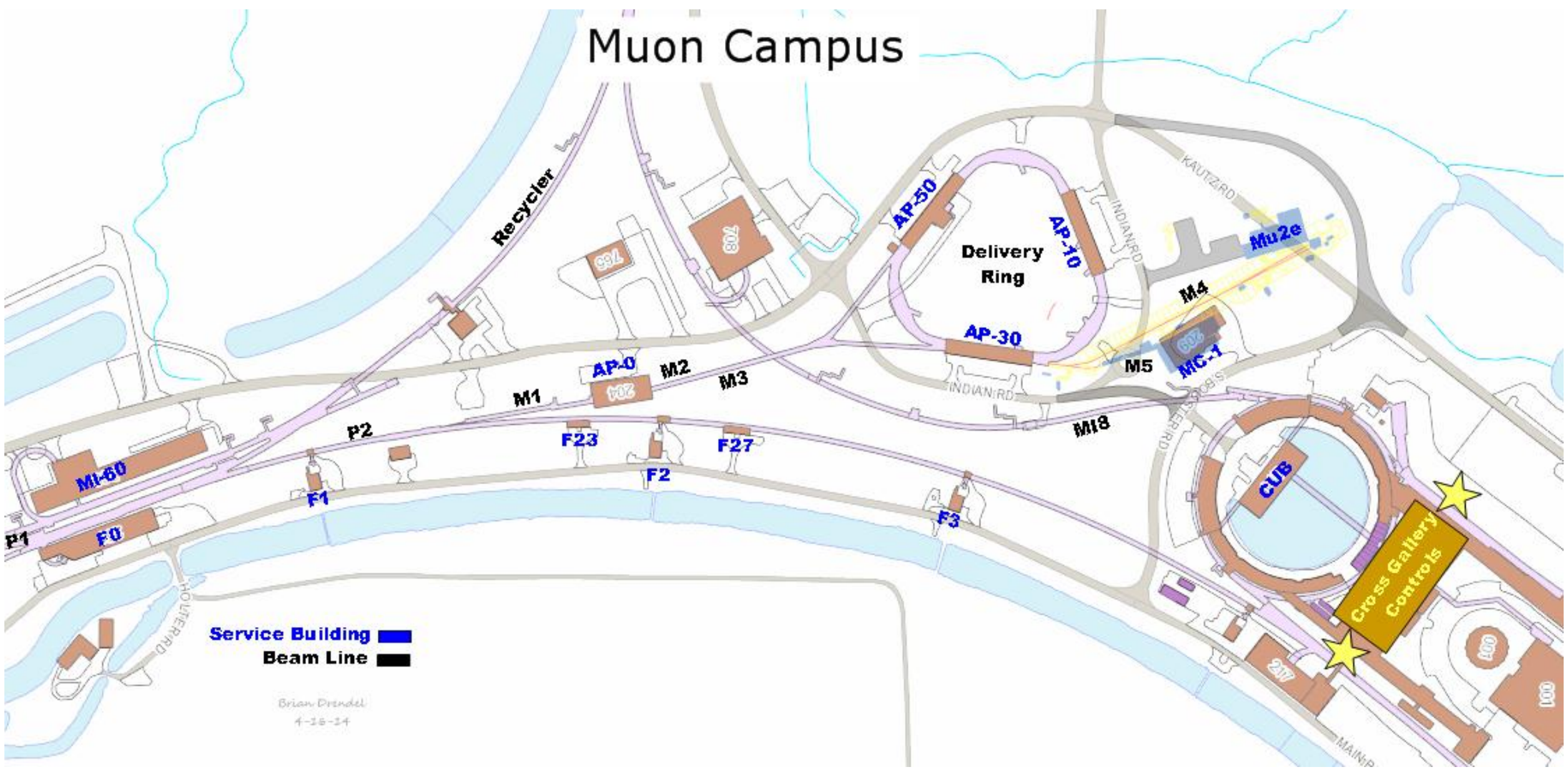

Figure 7.66: Muon Campus service buildings.

For completeness, all changes to the controls system needed for the Muon Campus are described here. The re-routing of controls in the Delivery Ring is covered by the Delivery Ring AIP, and work required only for Mu2e is covered by the Mu2e project. Cable pulls to the MC-1 building and work required to establish network, phone, and the Fire and Utility System in the MC-1 building is covered by the MC-1 Building GPP. Other work related to establishing accelerator controls in the MC-1 building is on the $(g-2)$ project.

#### CAMAC and links

The existing accelerator service buildings will continue to use the existing legacy controls infrastructure. These service buildings include all of the Main Injector service buildings, as well as F0, F1, F2, F23, F27, AP0, AP10, AP30 and AP50. Future Muon Campus service buildings, including MC-1, will be upgraded to a more modern controls infrastructure which will be discussed later in this document.

**CAMAC** Computer Automated Measurement and Control (CAMAC) crates exist in each service building and communicate with the control system through a VME-style front-end computer over a 10 MHz serial link as shown in Fig. 7.67. Both digital and analog status and control of many accelerator devices occur through the CAMAC front ends. There should be ample CAMAC-crate coverage for $(g-2)$ operation in the existing Muon service buildings, as there is excess capacity in most of the existing crates, and very few crates will need to be added or moved.

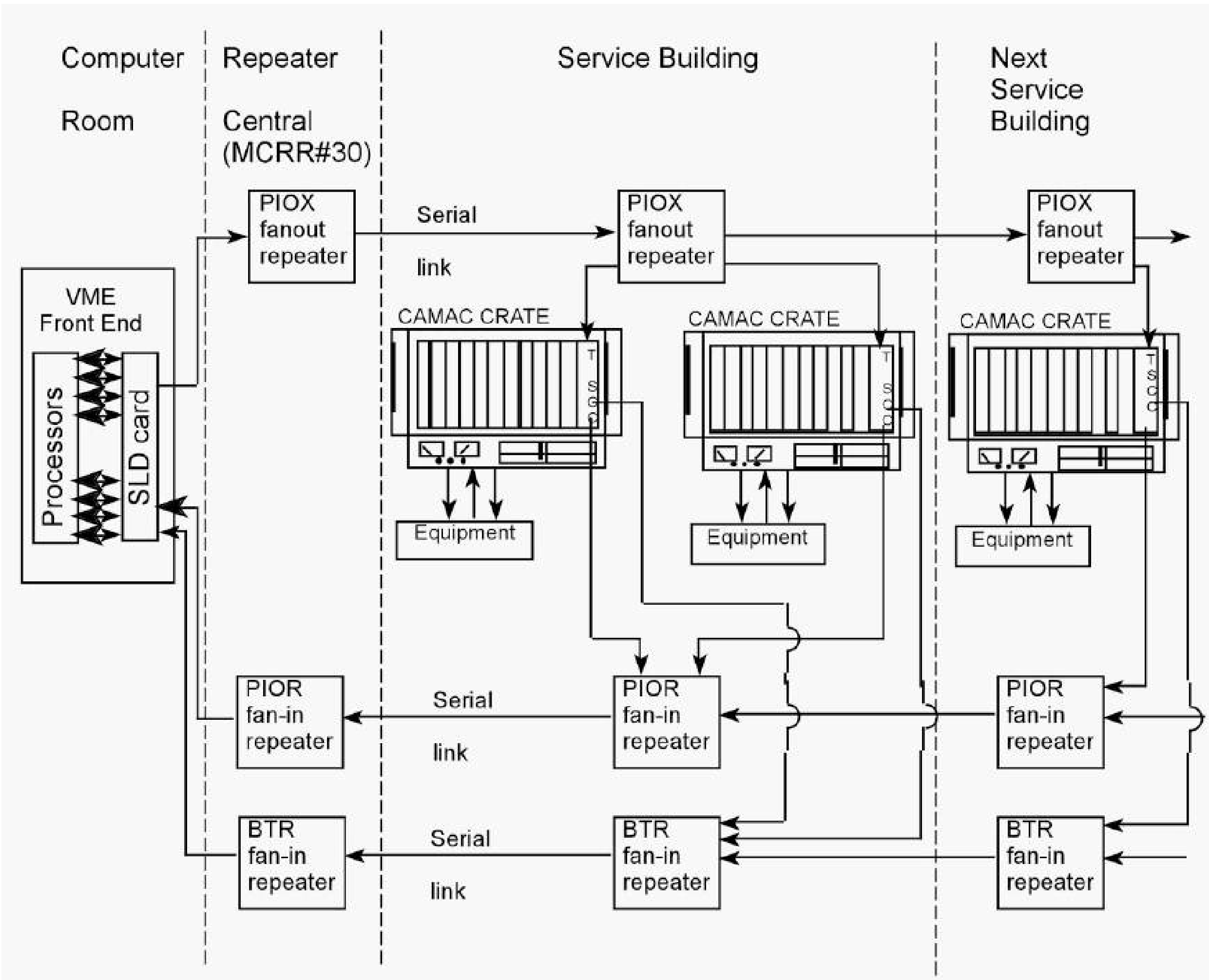

Figure 7.67: Legacy CAMAC crates interfacing VME front ends via serial links provide both analog and digital status and control of accelerator devices, and will continue to be used in existing Muon service buildings [27]. Drawing courtesy of the AD Operations Controls Rookie Book [28].

**Serial Links** There are serial links that are distributed through and between the service buildings via the accelerator enclosures that provide the necessary communications paths for CAMAC as well as other necessary signals such as clock signals, the beam permit loop, and the Fire and Utilities System (FIRUS). Controls serial links can be run over multimode fiber-optic cable or copper Heliax cable. Most Muon links that run through accelerator enclosures are run over Heliax, which should function normally in the radiation environment expected for $(g-2)$ operations [29].

**TCLK** Accelerator device timing that does not require synchronization to the RF buckets will remain on the existing 10 MHz Tevatron Clock (TCLK) system. The existing TCLK infrastructure will remain in existing service buildings and new TCLK link feeds will be run via multimode fiber optic cable from the Mac Room to the MC-1 service building [29].

**Beam Synch** Accelerator device timing for devices that require synchronization to the RF buckets will continue to be handled through the Beam Synch Clocks; however, a few changes will be required to maintain functionality. The F0, F1 and F2 service buildings will need both 53-MHz Main Injector beam synch (MIBS) for SY120 operations and 2.5-MHz Recycler beam synch (RRBS) for $(g-2)$ and Mu2e operations. These buildings already support multiple beam synch clocks, so the addition of RRBS will require minimal effort. An obsolete 53-MHz Tevatron beam synch (TVBS) feed in the MI60 control room will be replaced with a 2.5-MHz RRBS feed in order to provide the necessary functionality. The remaining Muon service buildings currently use 53-MHz MIBS, but will require 2.5-MHz RRBS for $(g-2)$ and Mu2e operations. This functionality can be obtained by replacing the MIBS feed at F0 with RRBS and using the existing infrastructure. Further upgrades and cable pulls will only be required if it is later determined that both MIBS and RRBS are required in these service buildings. New beam synch feeds to the MC-1 building were run via multimode fiber-optic cable from the Mac Room as part of the MC-1 Building GPP [29].

**Beam Permit** The Delivery-Ring permit loop provides a means of inhibiting incoming beam when there is a problem with the beam delivery system. The Pbar beam permit infrastructure will be used in the existing buildings. The CAMAC 201 and 479 cards, which provide the 50-MHz abort loop signal and monitor timing, will need to be moved from the Mac Room to AP50 to accommodate the addition of the abort kicker at AP50. Existing CAMAC 200 modules in each CAMAC crate can accommodate up to eight abort inputs each. If additional abort inputs are required, spare CAMAC 200 modules will be repurposed from the Tevatron and will only require a minor modification. The permit loop will be extended to the MC-1 building via multimode fiber-optic cable from the Mac Room. Implementation of a Hot-Link Rack Monitor abort card is not expected to be completed by the time of $(g-2)$ operations. As a result the abort inputs from devices in the MC-1 building will be transported to existing CAMAC 200 modules in the AP-30 service building via a Heliax cable that will be pulled through the accelerator enclosures [29].

Operational and permit scenarios are under development. The capability of running beam to the Delivery-Ring dump when Mu2e and $(g-2)$ are down will be needed, as well as the ability to run to either experiment while the other is down.

### Hot-Link Rack Monitor

New controls installations will use Hot-Link Rack Monitors (HRMs) in place of CAMAC. A HRM runs on a VME platform that communicates with the control system over Ethernet as shown in Fig. 7.68. Unlike CAMAC, no external serial link is required, minimizing the need for cable pulls between buildings. Each HRM installation provides 64 analog input channels, 8 analog output channels, 8 TCLK timer channels, and 8 bytes of digital I/O. This incorporates the features of multiple CAMAC cards into a single, compact chassis. Like CAMAC, when additional functionality or controls channels are needed, additional units can be added. Two HRMs will be installed in the MC-1 building and should provide ample controls coverage for both accelerator and experimental devices.

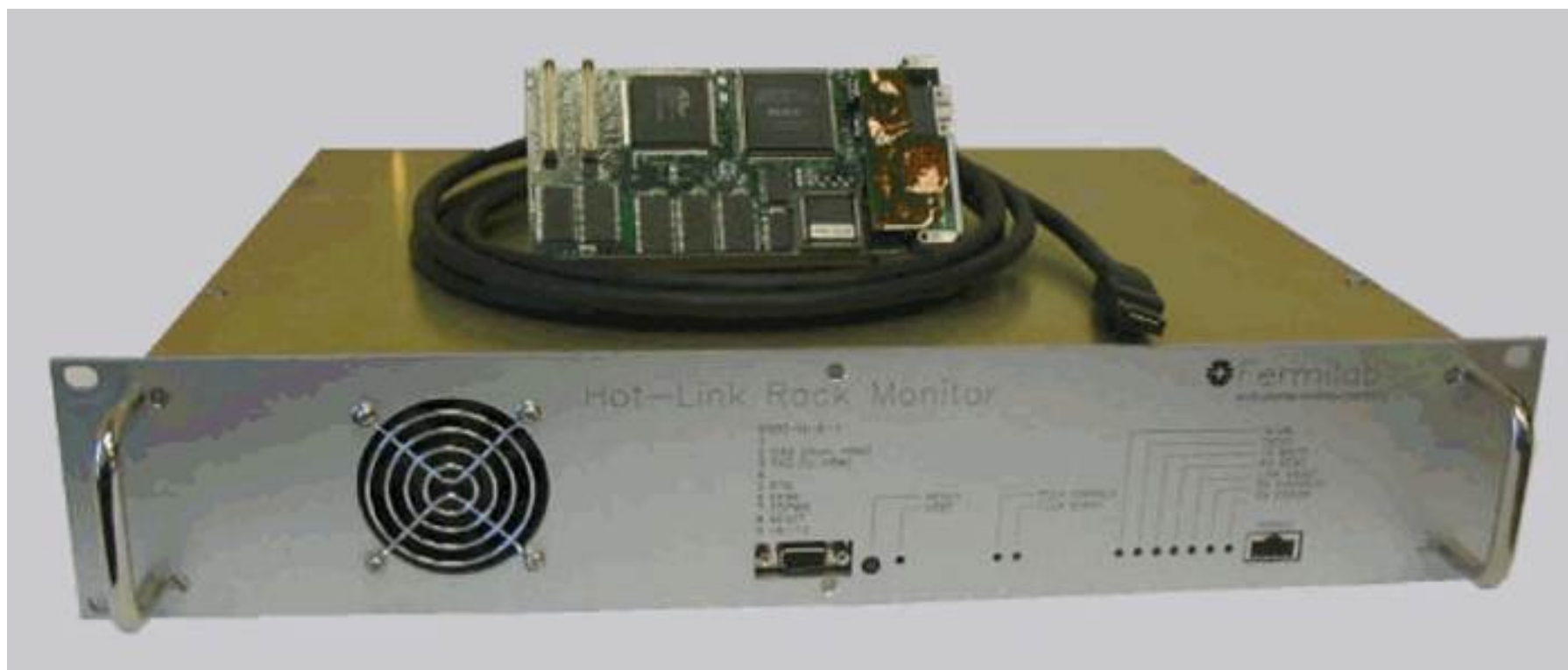

Figure 7.68: A Hot-Link Rack Monitor is a flexible data acquisition system composed of a remote unit and a PCI Mezzanine card that resides in a VME crate. Each HRM provides provides sixty-four 16-bit analog input channels, 8 analog output channels, 8 TCLK timer channels and 8 bytes of digital I/O. HRMs will eventually replace all of the functionality of CAMAC [30].

**Ethernet**

Many modern devices have some form of Ethernet user-interface. In addition, many devices and remote front-ends use Ethernet to interface with the control system, instead of using the traditional CAMAC. The results are an increasing demand on the Controls Ethernet. Figure 7.69 is a map of the Muon Controls network. All of the current Muon Ring service buildings have Gigabit fiber-optic connections from the Cross-Gallery computer room to Cisco network switches centrally located in each service building. These will provide ample network bandwidth and connections after the reconfiguration for $(g-2)$ and Mu2e. A central Ethernet switch that fans out to the other Muon buildings is currently located in AP10, but will need to be moved to AP30, as will be discussed later in this document [31].

Ethernet connects between the Muon-Ring service buildings via multimode fiber-optic cable paths that traverse the Rings enclosure on the Accumulator side. The multimode fiber currently in place will be replaced by single-mode fiber under the Delivery Ring AIP as needed for the high-radiation environment of Mu2e.

Most beamline service buildings have gigabit fiber connected to centrally located network switches that provide ample network bandwidth and connections. AP0, F23, and F27 are the only three buildings that do not have this functionality. AP0 runs off a 10 Mbps hub that connects to 10Base5 "Thicknet" that runs through the Transport and Rings enclosures back to AP10, while F23 and F27 run off 802.11b wireless from MI60. Both are 10 Mbps shared networks with limited bandwidth and connectivity. It is anticipated that the network in these three buildings will be sufficient for $(g-2)$ operations.

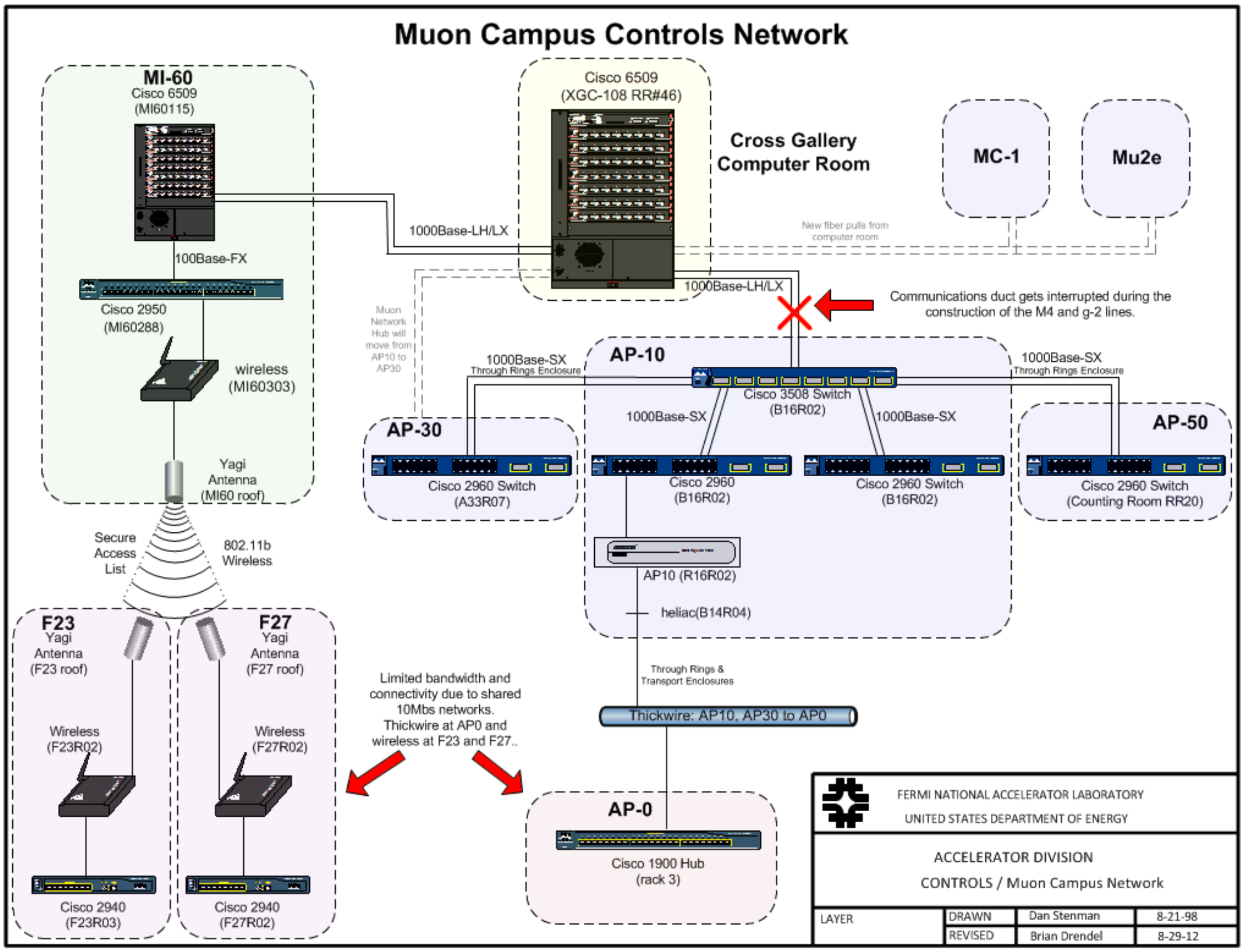

Figure 7.69: Controls Ethernet to the Muon service buildings is expected to be adequate for $(g-2)$ operations. The central switch at AP10 will be moved to AP30. Legacy networks at AP0, F23, and F27 have limited bandwidth and connectivity, but should be sufficient for $(g-2)$ operations.

### Controls connectivity

Civil construction of the M4 and M5 beamline enclosures will result in the removal of the underground controls communication duct that provides the connectivity between the Accelerator Controls NETwork (ACNET) and the Muon Campus [32]. Included in this communication duct is the fiber-optic cable that provides Ethernet connectivity, as well as 18 Heliax cables that provide the controls serial links and other signals including the FIRUS [29]. These cables currently connect from this communications duct to the center of the 20 location in the Rings enclosure, and travel through cable trays on the Delivery Ring side to the AP10 service building. New communications ducts from the existing manholes are being constructed as part of a General Plant Project. These communications ducts go directly to AP30, MC-1 and Mu2e service buildings without going through accelerator enclosures. See Fig. 7.70 for drawings of the current and future controls connectivity paths.

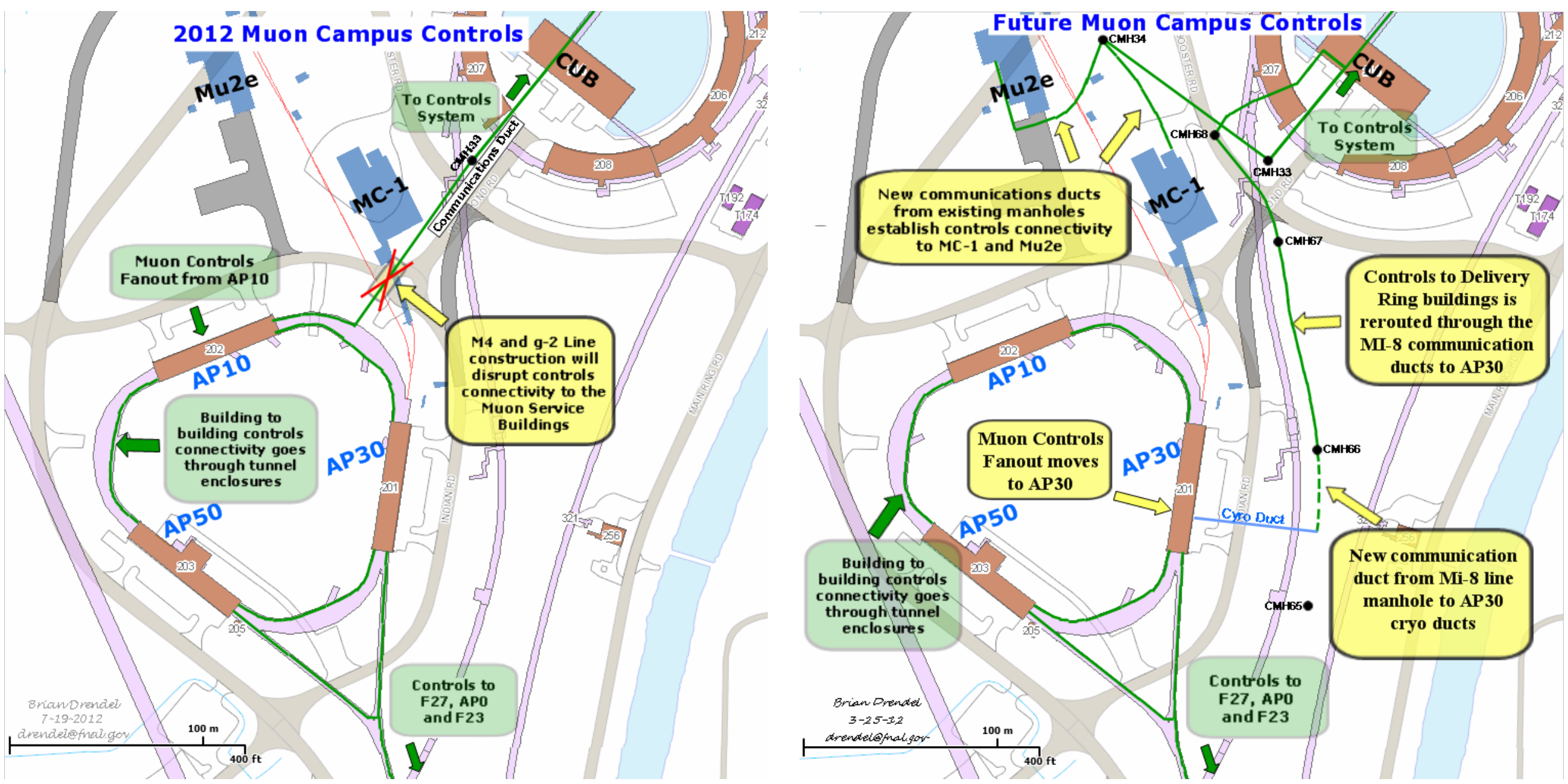

Figure 7.70: *(left)* Communication paths prior to Muon Campus operations. During construction of the M4 and M5 beam line enclosures, the communications duct that provides controls connectivity to the Muon Campus will be interrupted and controls will need to be restored via a different path. *(right)* Controls signals will be rerouted through the MI-8 manholes to a newly constructed manhole near AP30. From this manhole, communication ducts will connect to existing and unused cryo duct work to get to the AP30 service building. New controls will need to be established at the MC-1 and Mu2e building via new communications ducts that connect to an existing manhole [33].

**Restoring connectivity** When the Heliax and fiber-optic cables are cut during the removal of the above-mentioned communications duct, controls connectivity will be lost. New fiber optic cable has been pulled from the cross gallery, through the MI-8 line communications ducts to AP30. As a result, the Ethernet and controls links will fan out from AP30 instead of AP10. This will require some additional controls hardware configuration and labor. Efforts will be made to minimize the disruption by staging the new hardware at AP30 before the communication duct is cut. This is especially important for FIRUS which is necessary for monitoring building protection. This work is being done as part of the Delivery Ring AIP. More details can be found in Refs. [33] and [34].

**Establish connectivity to MC-1** New fiber-optic cable will be pulled from the Cross Gallery to the MC-1 service building. Single-mode fiber is needed for Ethernet and FIRUS, and multimode fiber is needed for the timing links and the abort-permit loop. A bundle of 96-count single-mode and a bundle of 36-count multimode fiber-optic cable will be pulled to MC-1. The fiber bundles will share a common path with the fiber bundles headed toward Mu2e from the Cross Gallery to the manhole by Booster West Tower. Both fiber bundles will travel through a single inner duct to the manhole. The Mu2e and MC-1 fiber bundles will then branch off to a second manhole inside a common inner duct, and then separate into the new communication ducts to the Mu2e and MC-1 service buildings. The fiber bundles to the MC-1 building were pulled by the MC-1 Building GPP, and will be pulled to the Mu2e building by the Mu2e project. The fiber will provide ample connectivity for all Ethernet and controls signals for both the accelerator and experiment. The $(g-2)$ experiment anticipates requiring network rates approaching 100 MB/s during production data taking which can be handled easily with the proposed infrastructure.

### Safety system

The existing safety system enclosure interlock hardware installed in the Pre-Target, Pre-Vault, Vault, Transport and Delivery Rings will remain in place. The tunnel egress between the Delivery Ring and Transport enclosures on the AP2 side will be blocked as a result of the new beam abort dump. A safety system mini loop will be created on each side of the abort dump to satisfy ES&H requirements. Reset boxes will be repurposed from the Tevatron for these mini loop areas [35].

The Delivery Ring enclosure will be separated from the new extraction line enclosure under AP30 using a gate. The Delivery Ring side of the gate will use a reset box repurposed from the Tevatron. The Extraction enclosures area will be defined using interlocked gates. One gate is the Delivery Ring / Extraction Enclosure gate; a second gate will separate the Extraction Enclosure from the M4 Enclosure (beam to Mu2e). The third gate separates the Extraction Enclosure from the MC-1 experimental hall. The Extraction Enclosure and the MC-1 experimental hall will each use the Rack Mounted Safety System (RMSS) chassis for their safety system interlocks. These chassis will be mounted in a rack dedicated for safety system equipment. The Extraction Enclosure RMSS will be located in the AP-30 service buildings safety system relay rack and the MC-1 RMSS will be located in the MC-1 buildings safety system relay rack which is located in the power supply room. The RMSS chassis uses a reset box similar to the Main Injector [35].

The three existing Pbar area Critical Device Controllers (CDCs) will function much as they presently do, with one to bring beam into the AP-0 area, one to bring the beam on target for $(g-2)$ operation, and one to take the beam around the AP0 target for Mu2e operation. These three CDCs will remain in the existing AP0 safety-system relay rack. Three new CDCs will be installed in the AP30 safety system relay rack to accommodate Delivery Ring extraction for $(g-2)$ and Mu2e beam operations. One CDC will be called the Extraction CDC to bring beam out of the Delivery Ring and into the Extraction Enclosure. This CDC will be repurposed from the Recycler CDC. The second CDC will bring beam from the Extraction Enclosure to the MC-1 Experimental Hall and is named the MC-1 CDC. This CDC will be repurposed from the Tevatron CDC. The third CDC will bring beam from the Extraction Enclosure to the M4 Enclosure for Mu2e operation and will be named the M4 CDC. The M4 CDC will be repurposed from the Pelletron CDC. The Extraction CDC can only be permitted when the MC-1 CDC or the M4 CDC is permitted. A Safety System Logic Module will be installed in the AP30 safety system relay rack to accommodate the "OR" function needed for the Extraction CDC. This Logic Module will be repurposed from the Tevatron Logic Module. Existing interlocked radiation detectors may be moved if needed and the system modified to include Total Loss Monitors (TLMs). The key trees from Pre-Vault, Pre-Target, and Transport will remain in the Main Control Room (MCR), while the remote AP10 key tree will likely be moved from AP10 to the MCR [35].

Cryogenics will be used in the MC-1 Refrigerator Room and the experimental hall, so an Oxygen Deficiency Hazard (ODH) system will be implemented using a safety-rated PLC system. This PLC will be located in the MC-1 buildings Power Supply Room in a dedicated relay rack.

**Cable Path** Copper safety system cables will be pulled to AP30, MC-1 and Mu2e. The existing Safety System signal trunk lines, which consist of seven 20-conductor #18 AWG cables that run from the safety system vault room XGC-005, through the Central Utility Building (CUB) to AP10, will be interrupted due to the Muon Campus installation. These trunk lines will need to be spliced at CUB and replaced with new cables from CUB to the AP30 building. These cables will be pulled at the same time the Control System fiber in order to minimize contract electrician costs. Below we will outline how we will establish the Safety System signals for the Transport and Delivery Rings, as well as the new MC-1 and Mu2e areas. Figure 7.71 gives a pictorial representation of each of the required cable pulls [35] [33].

**Interlocks** The safety system will need to be reestablished to the existing Muon Campus areas when the seven 20-conductor cables are interrupted. New junction boxes will be installed at CUB and at AP30, new cables will be pulled as shown in Fig. 7.71, and a new safety-system end rack will be installed on the existing safety system relay rack at AP30 to accommodate three critical device controllers and a Logic Module for $(g-2)$ and Mu2e operations [35] [33].

**Radmux** The Multiplexed Radiation Monitoring Data Collection System (MUX) is operated by the ESH&Q / Radiation Protection / Instrumentation Team. The MUX system is used to collect data from connected radiation monitors throughout the accelerator areas, beamline areas, and test areas at the Laboratory. The system provides an interface between the radiation monitors and the hardware network, provides real-time data for its various users, logs the raw data, and processes the data into formatted reports for users and for archive purposes. Additionally the archived data serves as the legal record of radiation levels throughout the laboratory [36]. Radmux connectivity will be restored to the existing muon buildings and established to the MC-1 Experimental Hall via new cable pulls [33].

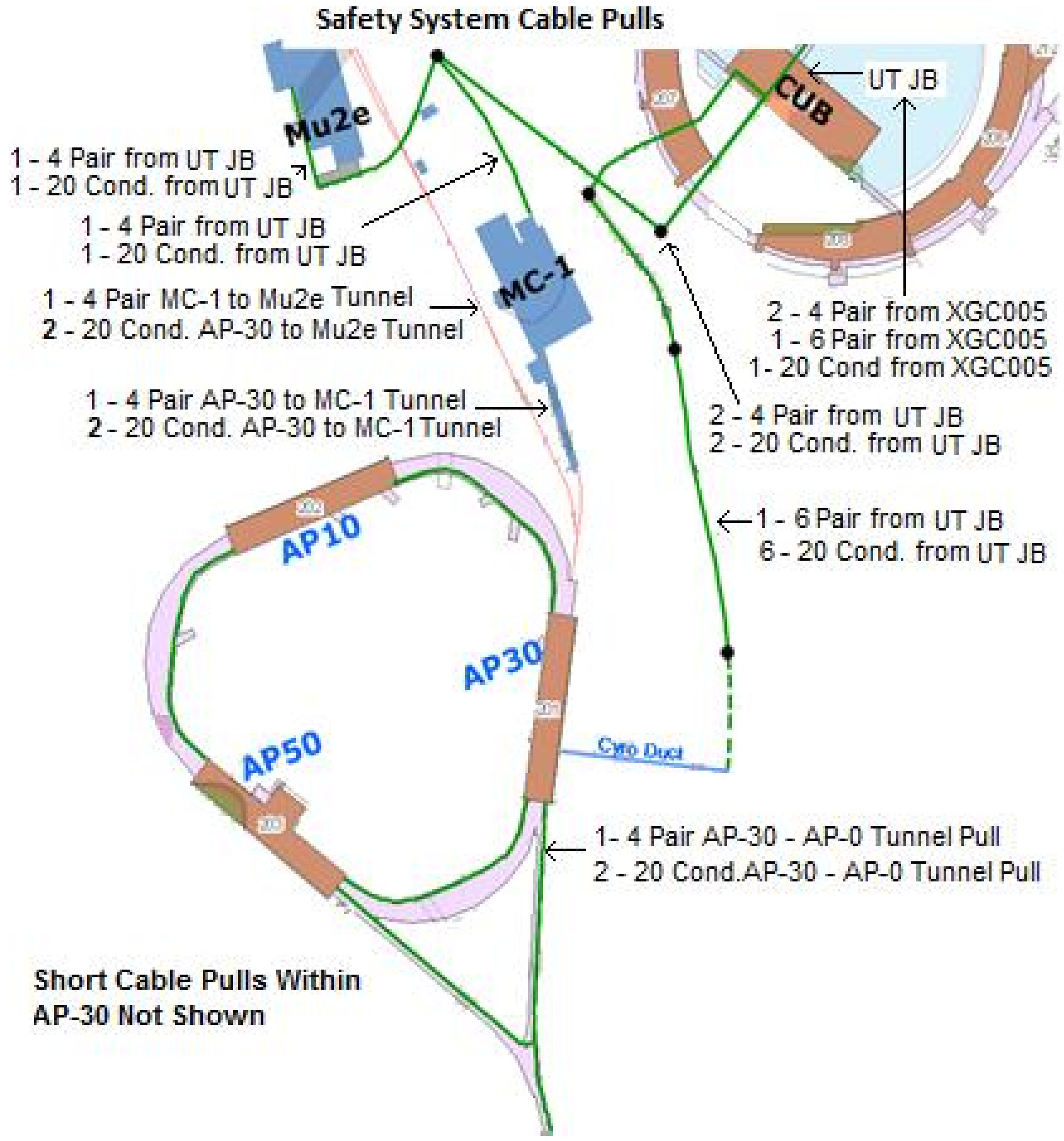

Figure 7.71: Safety system interlock cable pulls [35] [33].

**Phone** Phone connections to the existing Muon service buildings will be reestablished by splicing into the 400-pair cable in the MI-8 communications duct. A new section of 100-pair cable will be run from the splice via a new communications duct path established by the Delivery Ring AIP to the AP30 service building. Phone connections to the MC-1 Experimental Hall will be established by splicing into 400-conductor pair phone line in the CMH33 Manhole and running new 100-conductor pair phone line to the MC-1 and Mu2e Experimental Halls [37].

**Site Emergency Warning System** The Site Emergency Warning System (SEW) currently runs to the Muon Rings buildings over the CATV system. When the communications duct is cut, the CATV system will not be reestablished to the Muon Rings buildings. Instead, the SEWs will be run over single mode fiber optic cable to AP30 and then through the Muon Rings enclosure to AP10, where a connection will be made to the existing system. The fiber will be fusion-spliced to make one continuous fiber path all of the way to AP10. No cabling infrastructure will be needed for the SEWs in the Mu2e and MC-1 service buildings. A paging system will be internal to each building and will be tied to a radio receiver (called a TAR) which receives the SEWS radio broadcast system. The messages will be broadcast over the paging system [38].

### 7.6.2 Accelerator instrumentation

Beam monitoring can be divided into distinct zones: primary protons, mixed secondaries, proton secondaries, and muons. The locations of each of these areas are shown in Fig. 7.72. The expected beam properties in each of these areas are shown in Table 7.15.

| Beam Type | Particle Species | Beam Momentum (GeV/c) | Number of Particles per pulse | RF Bucket (MHz) | Bunch Length (ns) | Transverse Emittance (mm-mr) |
|---|---|---|---|---|---|---|
| Primary protons | p | 8.9 | $10^{12}$ | 2.515 | 120 | $18\pi$ |
| Mixed secondaries | $\mu^+$, $\pi^+$, p, $e^+$ | 3.1 | $10^7$ to $2 \times 10^8$ | 2.515 | 120 | $40\pi$ |
| Proton secondaries | p | 3.1 | $10^7$ | 2.515 | 120 | $40\pi$ |
| Muons | $\mu^+$ | 3.1 | $< 10^5$ | 2.515 | 120 | $40\pi$ |

Table 7.15: Expected properties of primary proton beam, secondary beam off the target, and muon beam from pion decay relevant to instrumentation designed to measure beam. Transverse emittances are 95% normalized.

#### Primary proton beam

Instrumentation for the primary proton beam in the Recycler, P1 stub, P1, P2 and M1 lines is covered by the Beam Transport AIP. Much of the instrumentation needed to measure the primary proton beam during $(g-2)$ operation already exists, but needs to be modified for use with the faster cycle times and 2.5-MHz RF beam structure. The overall beam intensity is similar to that seen in Pbar stacking operations, and in many cases requires only small calibration changes be made to the instrumentation. Toroids will be used to monitor beam intensity and will be used in conjunction with Beam Loss Monitors (BLMs) to maintain good transmission efficiency in the beamlines. Multiwires and Secondary Emission Monitors (SEMs) will provide beam profiles in both transverse planes. Beam Position Monitors (BPMs) will provide real-time orbit information and will be used by auto-steering software to maintain desired beam positions in the beamlines.

#### Mixed secondaries

Mixed-secondary beam will traverse the M2 and M3 lines, as well as the Delivery Ring. Changes to existing instrumentation are required in these areas as a result of the secondary beam being approximately two orders of magnitude lower in intensity than it was during the former Antiproton-stacking operations. In addition, 2.515 MHz bunch structure and a faster pulse rate must be taken into consideration. Mu2e beam will have beam intensities four to five orders of magnitude higher than $(g-2)$ operations in the M3 line and Delivery Ring, so design upgrades take into account the vastly different beam intensities required for both experiments. Beam studies have been conducted in order to help determine what instrumentation best suits the low-intensity secondaries of $(g-2)$ operations [2].

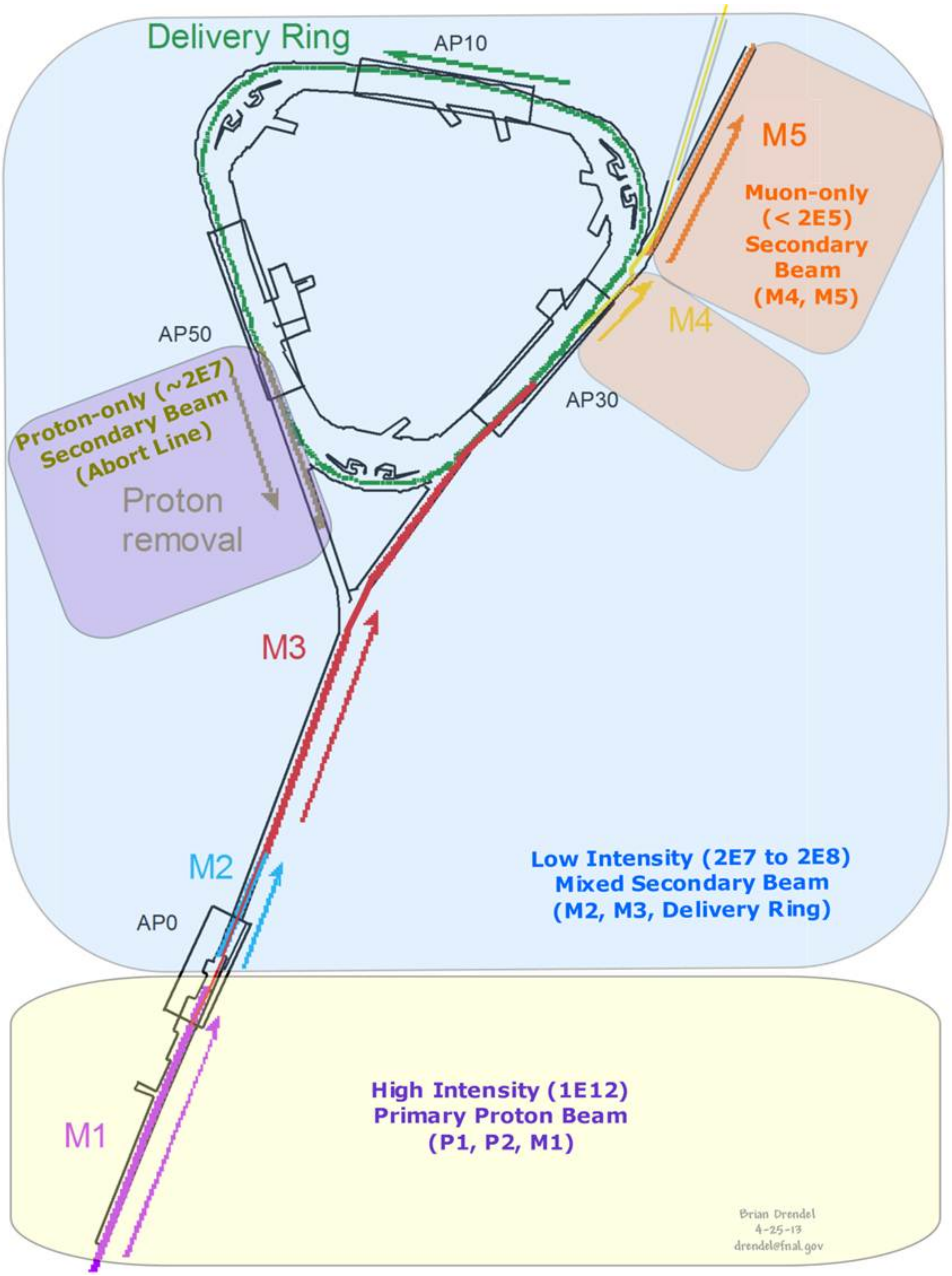

Figure 7.72: Beam monitoring can be divided into four different zones, each with different instrumentation schemes. High-intensity proton beam will be monitored with Toroids, BPMs and BLMs. Low-intensity secondary and proton-only secondary beam will be monitored with Ion Chambers, BLMs and SEMs. Muon-only beam will be monitored with Ion Chambers and PWCs.

**Toroids** Four toroids are available for use in the secondary beamlines and were the primary intensity-measurement device in these lines during Antiproton operations. These will be used for Mu2e operations; however, beam studies show that even with high gain and careful filtering, they are not able to measure beam at $(g-2)$ operational intensities [2]. As a result, toroids will not be used during normal $(g-2)$ operations, but will still be used with higher-intensity beams during commissioning and studies periods.

**Ion chambers** Ion chambers will become the primary beam-intensity measurement device for mixed-secondary beam. They are relatively inexpensive devices that can measure beam intensities with an accuracy of $\pm 5\%$ with as little as $10^5$ particles. Ion chambers were used in the AP2 line in the past, and work was done during beam studies to recommission the ion chamber that used to be operational near the end of the AP2 line [2]. For $(g-2)$ operations, ion chambers will be implemented in the M2 line, M3 line, and the Delivery Ring; these ion chambers will be installed in a vacuum can with motor controls to allow them to be pulled out of the beam. Figure 7.73 shows the ion chamber design. The vacuum can with motor controls that the ion chamber sits in will be described in the PWC section below.

Each ion chamber consists of one signal foil interleaved between 24 high-voltage foils. The foils are sealed in an aluminum chamber continuously purged with an 80% argon - 20% carbon dioxide gas mix. The standard ion chamber is shown in Fig. 7.73. Protons passing through the $ArCO_2$ gas generate 96 e/ion pairs or about $1.6 \times 10^{-17}$ charges/cm which equals about 1.6 pC for $10^5$ protons [39].

The ion chambers are made retractable because the beam going through those ion chambers and the vacuum windows required to separate beam tube vacuum from the $ArCO_2$2 gas required for the chamber would result in excessive Coulomb scattering during high-intensity Mu2e operations [40]. The solution is to make the ion chamber retractable much like what will be discussed in more detail in the Proportional Wire Chamber section below. The ion chamber will be installed inside of an anti-vacuum chamber with two titanium vacuum windows to provide a barrier between the gas needed for the ion chamber and the beamline vacuum. The entire anti-vacuum chamber would be mounted inside of a vacuum can that is common to beam tube vacuum. The ion chamber will be on a motorized drive that would allow it to be moved in or out of the path of beam [39].

Beam studies were completed to check the effectiveness of using ion chambers in the range of intensities expected during $(g-2)$ operation [2]. One ion chamber was installed in the upstream portion of the AP2 beam line at the 704 location, while the other ion chamber was located at the downstream portion of the AP2 line at the 728 location (Fig. 7.8. Both ion chambers were shown to integrate beam charge as expected over the normal range of $(g-2)$ operational intensities for the M2 and M3 lines as can be seen in Fig. 7.74.

**Wall Current Monitors** Wall Current Monitors (WCMs) are non-destructive intensity-measurement devices that could be used for the mixed-secondary beam. These devices have the advantage of being completely passive and not requiring a break in the vacuum, which may make them a better fit in the M3 line where we need to minimize beam losses during the higher intensities of Mu2e operations, and in the Delivery Ring where beam circulates multiple times for $(g-2)$ operations and for approximately 56 ms during Mu2e operations. A new WCM design has been developed that would provide accurate intensity measurements for secondary beam during $(g-2)$ operations. The design is based on that of a WCM for Mu2e extraction. Each slice of the slow-spilled Mu2e beam is approximately $2 \times 10^7$, which is consistent with the intensity that we would expect in the M3 line and Delivery Ring during $(g-2)$ operations. The prototype WCM is currently installed in the Delivery Ring and will remain in place during $(g-2)$ operations. If additional funding becomes available, additional WCMs could be built for other areas.

**Secondary Emission Monitors** SEMs will be used to measure beam profiles in the M2 and M3 lines. There are 24 SEMs in the former Antiproton-source beamlines available for use. SEM tunnel hardware will require some maintenance, and locations where SEMs are moved will require new cable pulls. Beam studies showed that special high-gain preamps are required to measure the low-intensity secondary beam during $(g-2)$ operations [2]; the design is described below.

SEMs will provide profile information via Fermilab generation-3 profile-monitor scanners and Fermilab standard profile-monitor software. Each scanner connects to one SEM and communicates to the control system via an Ethernet connection. The scanner is at the center of all profile monitor installations. It collects the charge from each of the detector wires and converts the values of the charges to a set of digital numbers. The data are transferred to the Accelerator Control System for analysis and display. Example profiles are shown in Fig. 7.75.

The scanner consists of five printed circuit boards, one controller board and four analog integrator boards. It has a set of 96 integrator circuits, 48 for horizontal and 48 for vertical. The integrators collect the charge from each of the detector titanium strips and converts it to a voltage value proportional to the total charge collected. The basic integration capacitor value for most scanners is 100 pf; this value provides the most sensitivity. Larger capacitors are used in higher-intensity beams to minimize the possibility of overloading the integrators. The integrators collect charge until they reach the end of the integration duration set by the user or until at least one wire reaches the preset threshold voltage. At the end of the integration period, the integrators are switched from the sample mode to the hold mode. The integrated voltages on each channel are measured one by one and converted to digital values.

The third-generation scanner (Fig. 7.76) is an evolution of the previous design. The SEM interfaces to the scanner through the integrator boards as in the previous version. The control board centers around an Altera Cyclone III FPGA, which handles sequence control, ADC conversion, TCLK decoding, and timing. Communications and data handling are performed by a Rabbit Semiconductor RCM3209 module. The Rabbit module includes the microprocessor and Ethernet interface. New features include Ethernet communications, advance triggering options, and background subtraction [41].

A new high-gain preamp has been designed to enable the existing SEMs to measure the low-intensity secondary beam [41]. The new preamp consists of two amplification stages. The first stage is a transimpedence amplifier with a gain of approximately $2 \times 10^7$ and an integration capacitor to slow down and widen the incoming pulse. The second stage reduces the DC offset of the first stage by a factor of about 100, then amplifies the remaining signal by about 100, with a low-pass roll-off of about 16 kHz. There is a DC blocking capacitor at the output of the amplifier to prevent any offset voltage of the amplifier from washing out our signal. The 1.5 k$\Omega$ resistance of the integrator in the scanner is accounted for in the gain equation ($R_{INT}$ in Fig 7.77). After integration in the scanner, the integrator output can be amplified by another factor of 10 or 100 if necessary.

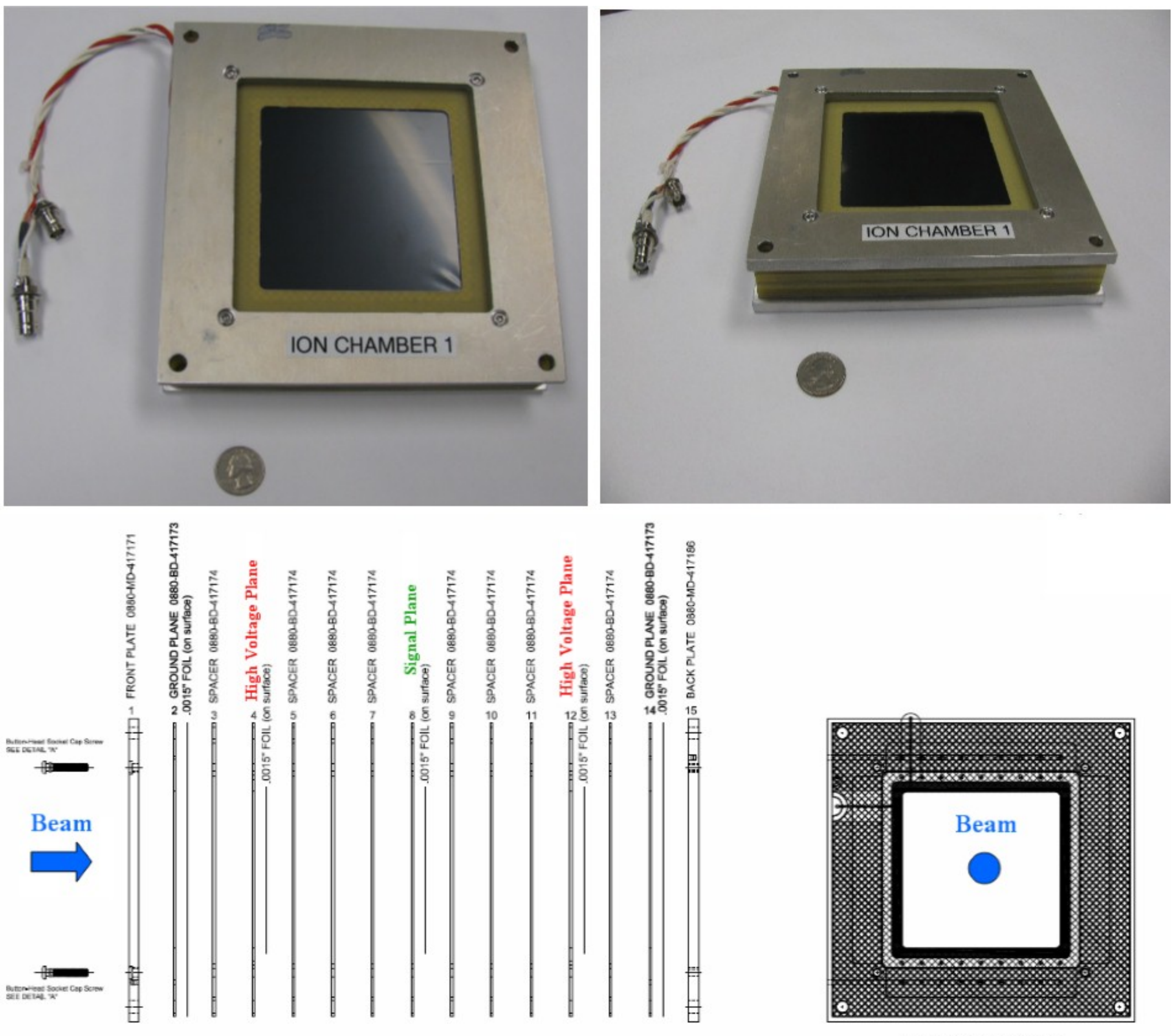

Figure 7.73: Finished ion chamber assembly (top). Ion Chamber layout (bottom). Ion Chamber assembly is made up of a single signal plane sandwiched between two high voltage foils with a ground plane and end plate on each end.

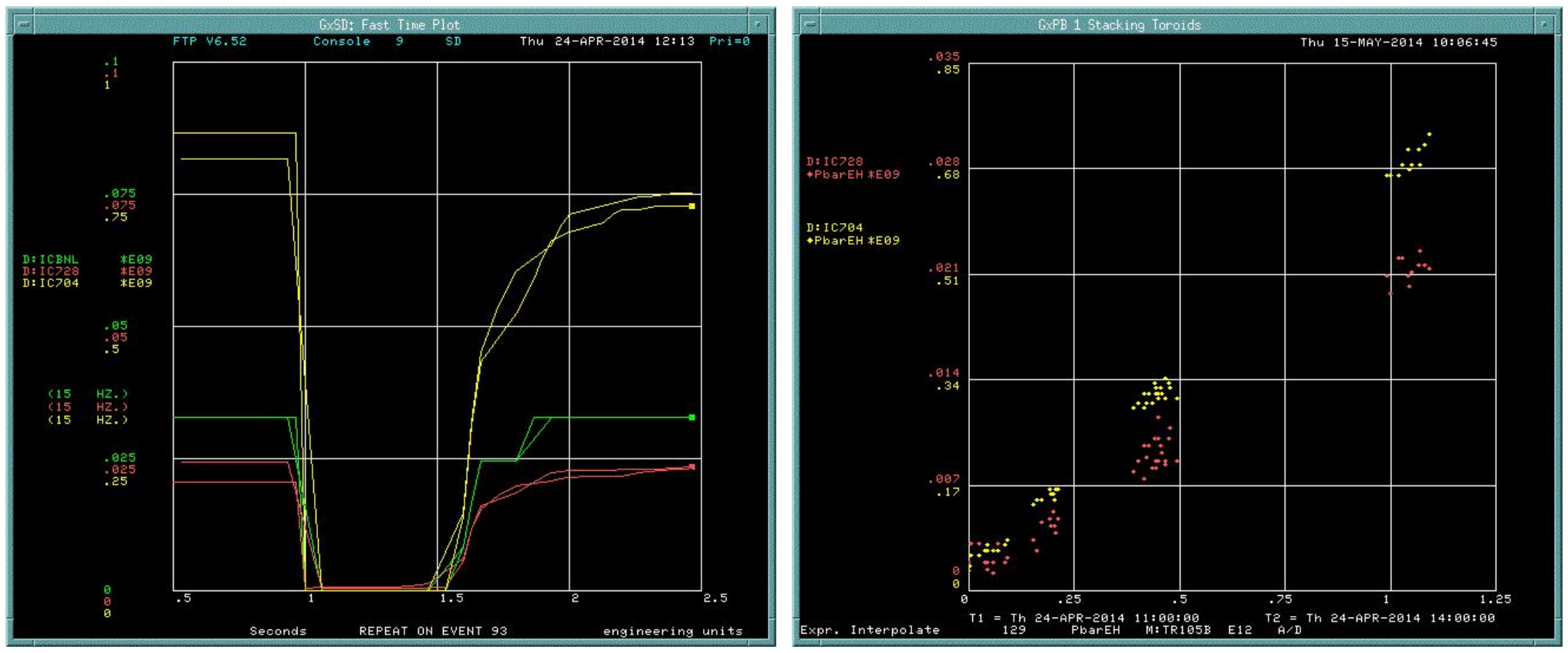

Figure 7.74: AP2 ion chamber performance was measured during beam studies. *(left)* Ion chamber integration over time. The signal is reset at 1.0 s and samples at beam time just after 1.5 s. The yellow trace is the intensity reported by the ion chamber at the 704 location and the red trace is that from the ion chamber at the 728 location. This plot was taken with $1 \times 10^{12}$ protons on target and shows an intensity of $7.5 \times 10^{8}$ particles at the 704 location and $2.5 \times 10^{7}$ at the 728 location. *(right)* The output of the same two ion chambers over varied intensities of beam on target. The response is linear through a wide range of beam intensities.

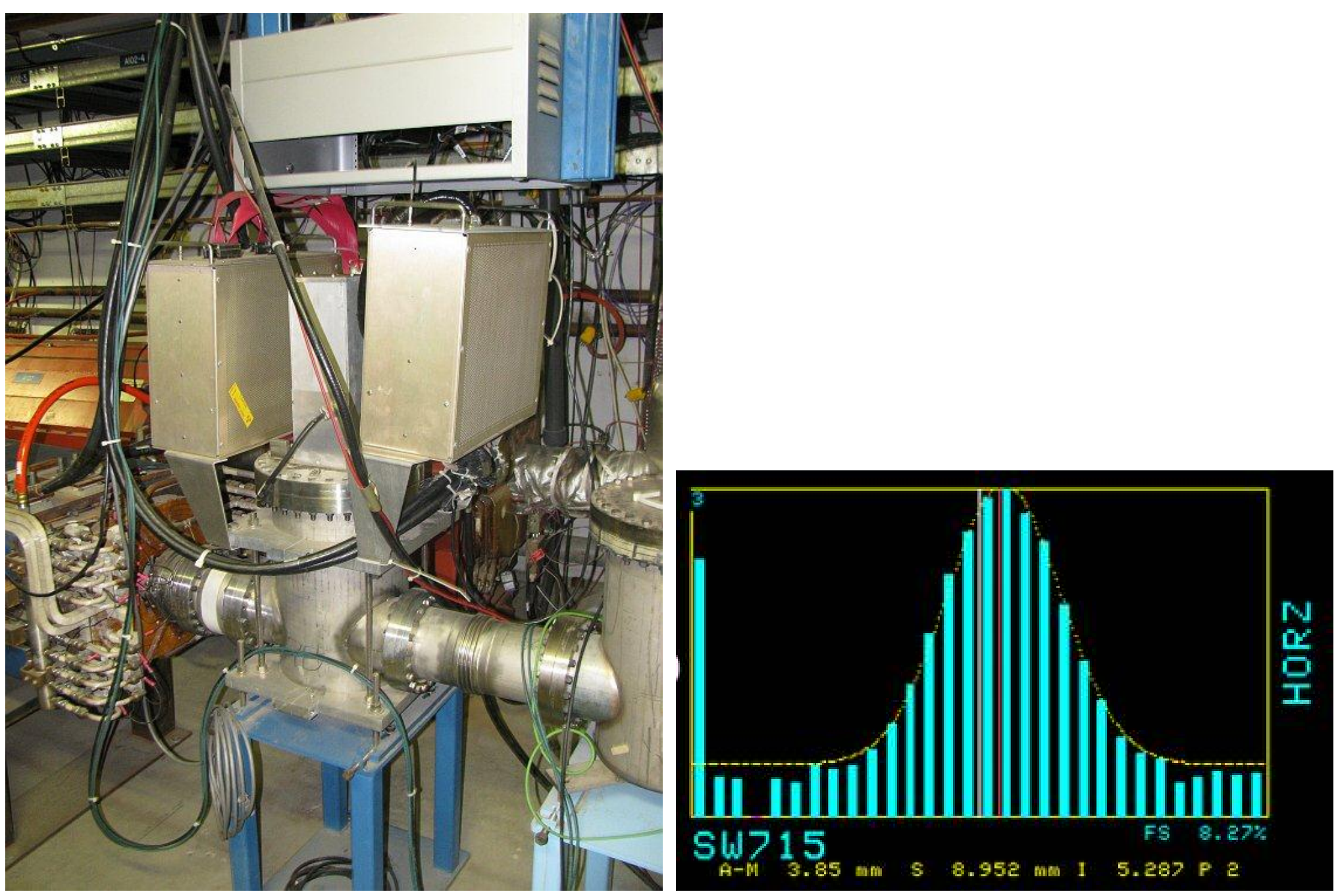

Figure 7.75: SEMs will be used to measure mixed secondary beam profiles. SEM tunnel hardware (left) is pictured. Preamp boxes are mounted next to the vacuum can. The SEM wires can be pulled out of the beam when not in use. SEMs can be used to measure beam profiles, positions and intensities (right).

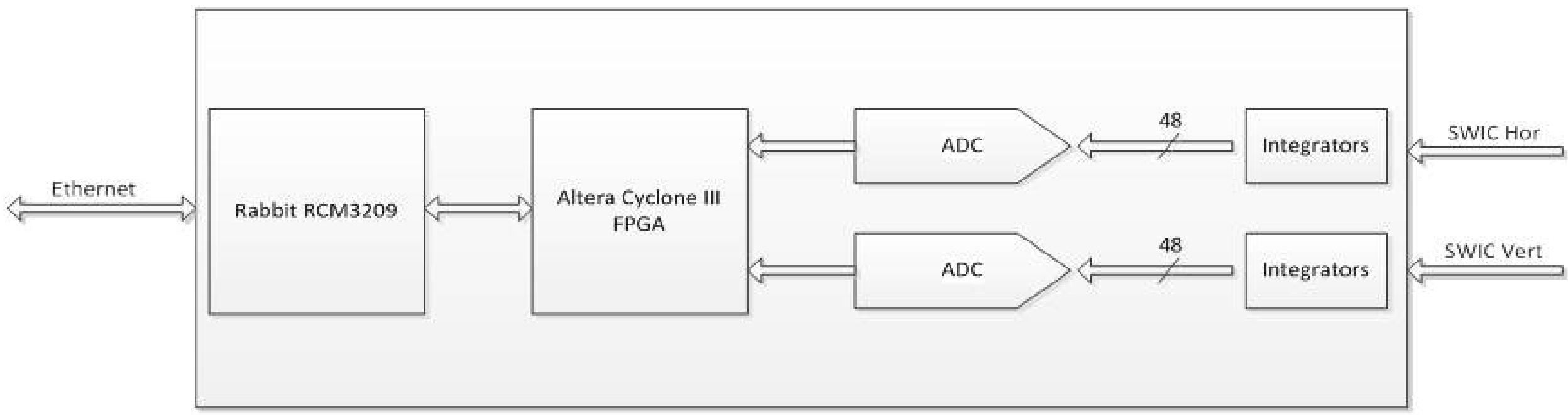

Figure 7.76: Scanner block diagram [41].

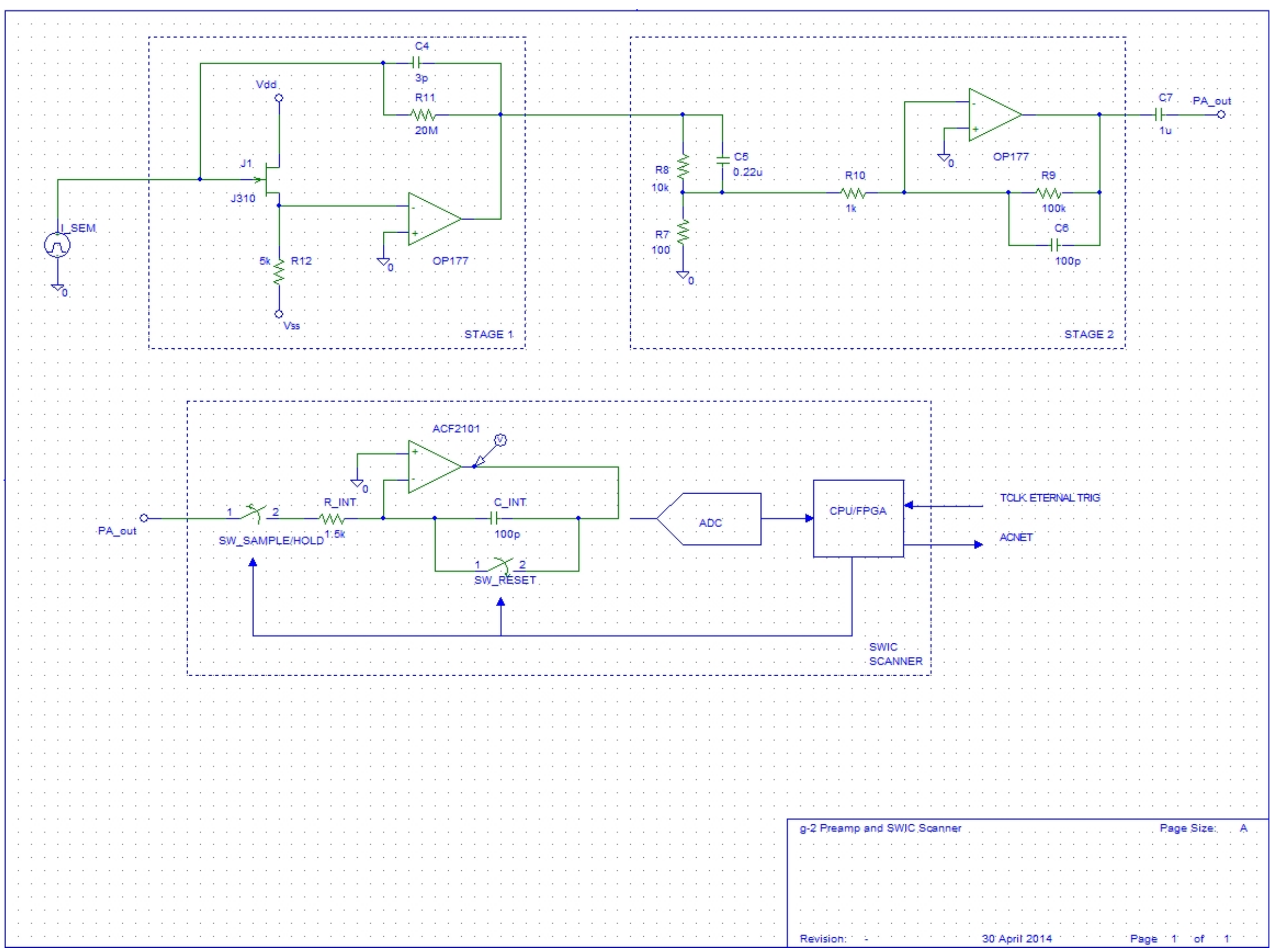

Figure 7.77: Profile monitor preamp design [41].

Prototype high-gain preamps were tested during beam studies in the AP2 line with 8-GeV beam on target and 3.1-GeV secondary beam with positive charge. Beam intensities were varied through the range expected for $(g-2)$ operations. Figure 7.78 shows SEM profiles at two locations in the AP2 line using the new high-gain preamps with the nominal $(g-2)$ intensity of $10^{12}$ protons on target.

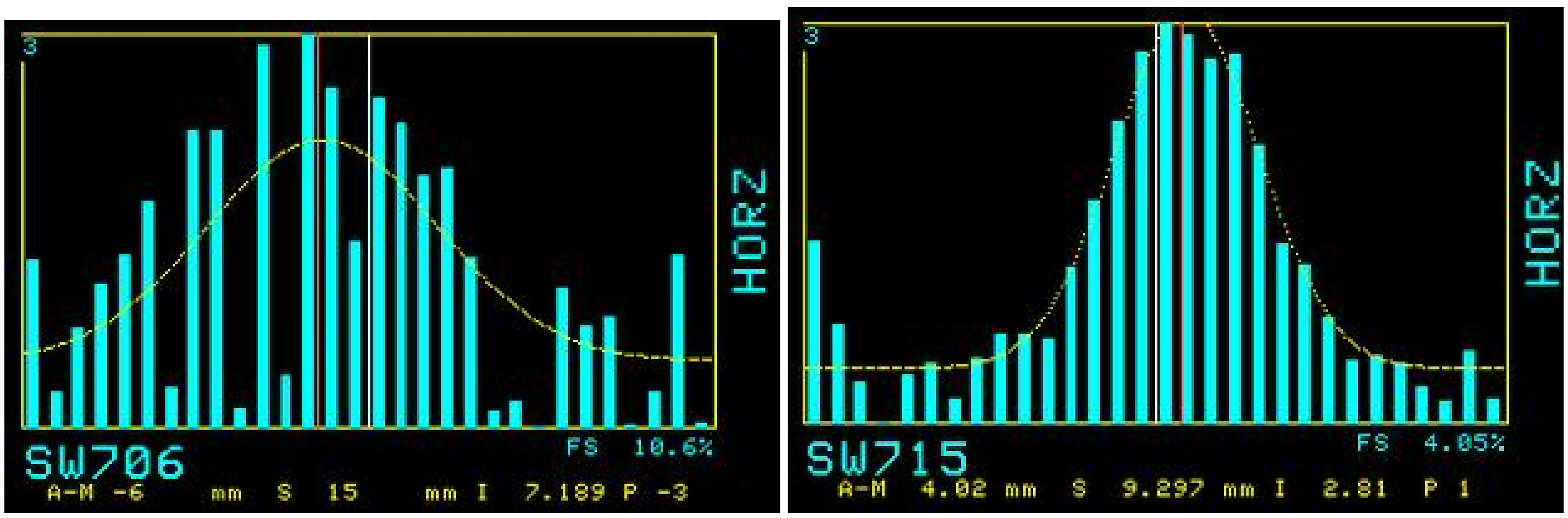

Figure 7.78: Demonstration that SEM wire profiles can be obtained at $(g-2)$ operational intensities and energies. Shown are profiles of low-intensity secondary beam collected with $10^{12}$ protons on target and an intensity of $10^9$ mixed secondary beam as measured by the ion chamber at the 704 location. The SEM at the 706 location has some bad wires which will be repaired during maintenance periods.

Figure 7.79 shows profiles with $10^{11}$ protons on target so that intensity at the 706 location approximates that expected at the end of the M3 line during $(g-2)$ operations in order to test the range of the SEM high-gain preamp.

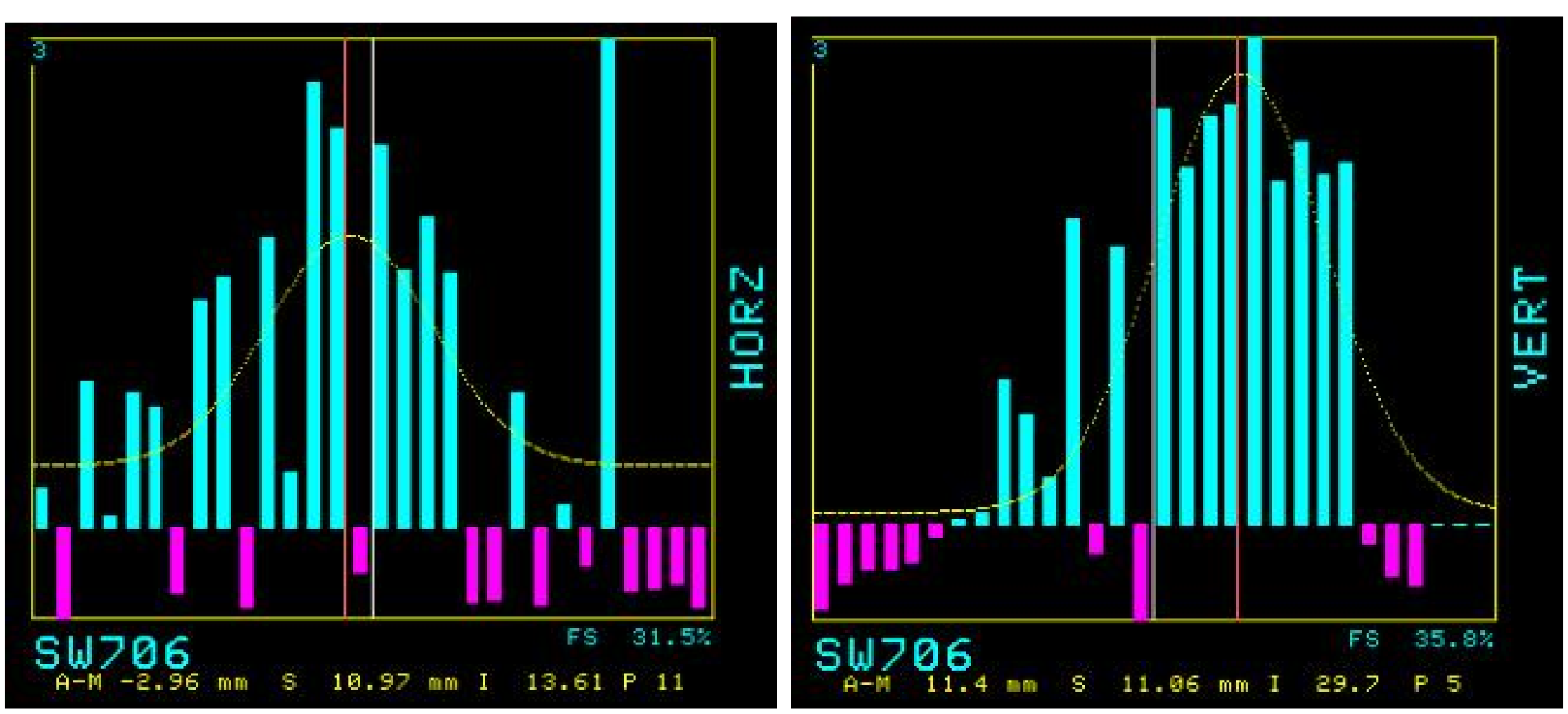

Figure 7.79: Profiles are still visible on SEM706 with $10^{11}$ protons on target and an intensity of $5 \times 10^7$ mixed secondary beam as measured by the ion chamber at the 704 location.

Large pulse-to-pulse noise variation was observed with the AP2-line SEMs, so a third-generation scanner was tested during beam studies. This scanner implements a hardware pulse-by-pulse background noise subtraction. For each beam cycle, the background noise is subtracted before the beam pulse arrives, and that signal is subtracted from a second sample taken at beam time. The results were very promising, giving us clean-looking profiles, as seen in Fig. 7.80.

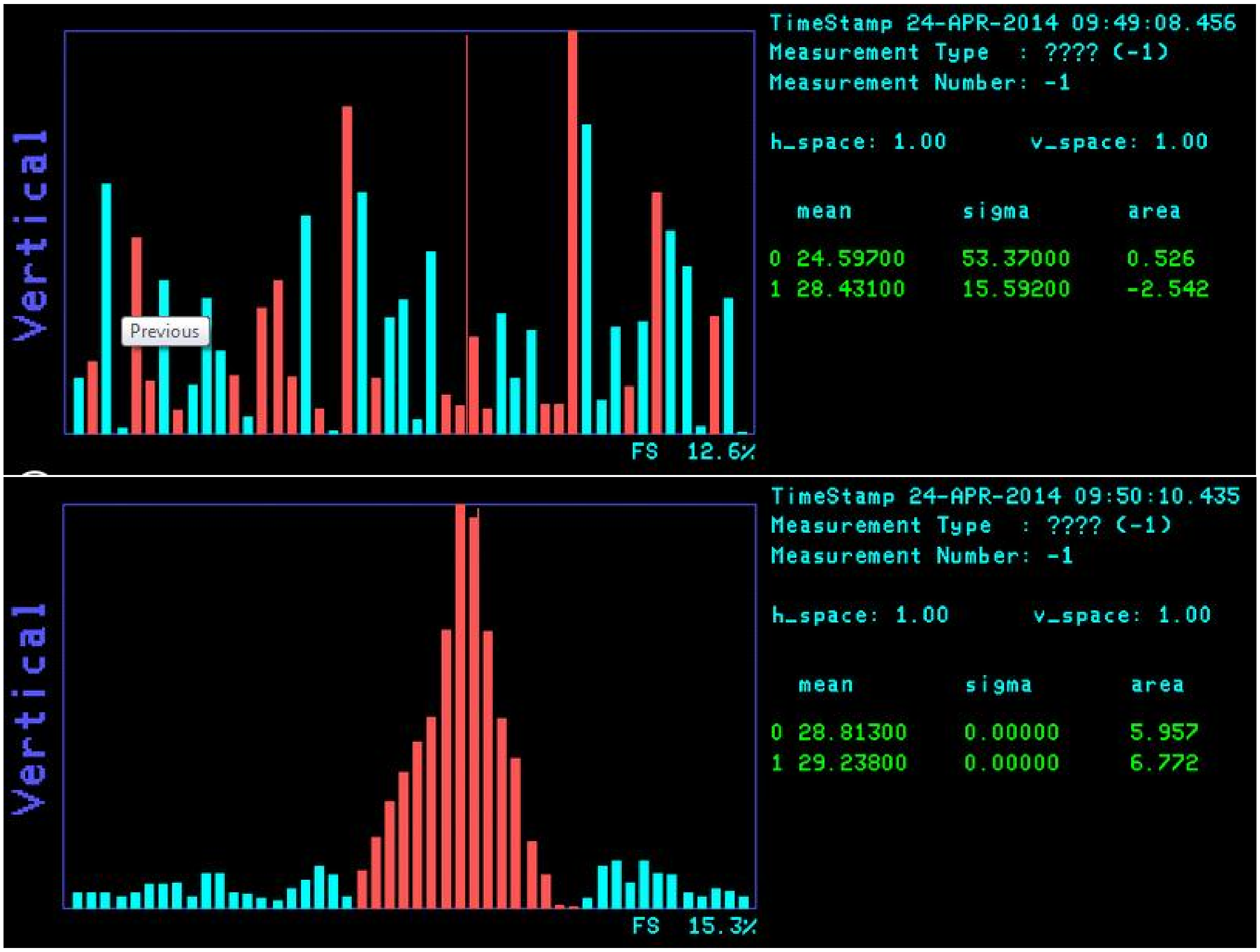

Figure 7.80: Wire profile at the 728 location for $10^{11}$ 120-GeV protons on target with $2.5\times10^{8}$ low-intensity secondary beam measured using an ion chamber at at 728. The top plot is the noise sample and the bottom plot shows the results of subtracting the noise sample from the raw beam signal.

**Beam Loss Monitors** BLMs (Fig. 7.81) will be used to help maintain good transmission efficiency through the beamlines. Both Delivery-Ring and AP3 loss monitors will use the existing hardware and electronics for $(g-2)$ operations, but will be replaced for the higher-intensity Mu2e operations. The BLM design allows for switching back and forth between the two separate BLM systems with minimal effort.

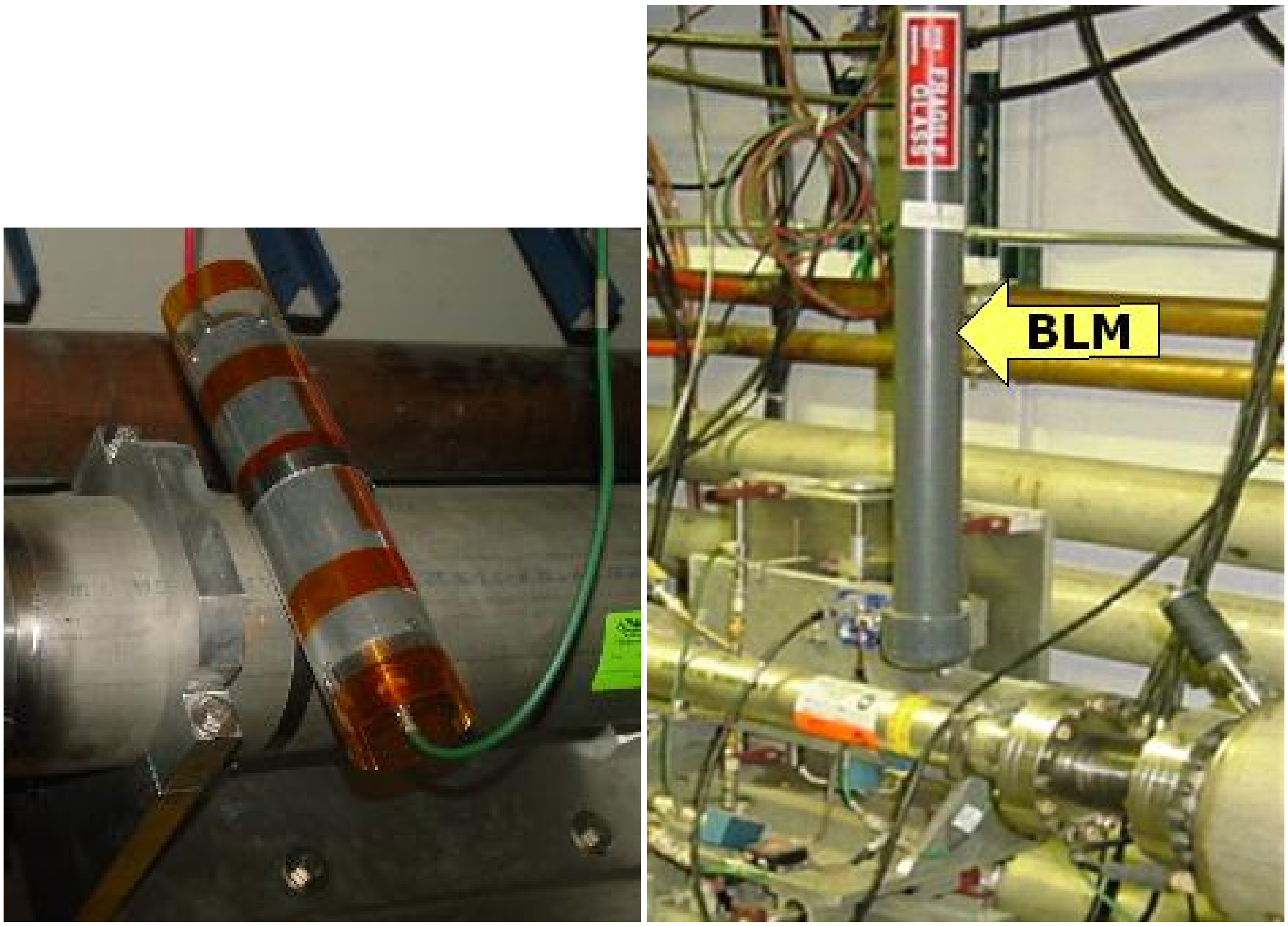

Figure 7.81: Two styles of BLMs will be used. Tevatron-style ion chamber loss monitors (left) will be used in areas of primary beam, and also in the Delivery Ring for Mu2e operations. The Pbar-style ion chamber, which consists of a plastic scintillator and a long light guide connected to a photomultiplier tube shielded from light in PVC, will be used in the Delivery Ring during $(g-2)$ operations.

The plastic-scintillator type BLM is sensitive to a small number of particles, making it ideal for Delivery Ring $(g-2)$ operations. The loss monitors are made up of a 4 in $\times$ 2 in $\times$ 1/2 in piece of plastic scintillator glued to a 36-in long Lucite light guide (see Fig. 7.82). At the end of the light guide, a small Lucite coupling attaches it to an RCA 4552 photomultiplier tube (PMT). The intent of the light guide is to keep the scintillator near the magnets but to extend the phototubes up and away from the region of beam loss. This assembly is mounted in a housing made up of PVC pipe and has feed-throughs for the high voltage and signal cables, as shown in Fig. 7.82.

The BLM output is processed through a series of three cards located in one or more NIM crates. Each service building has a single BLM rack to process loss signals for two sectors. The signals are passed from card to card via LIMO connections in the front panels of the cards. The BLM output first goes to an amplifier card, which handles twelve BLMs and amplifies each BLM signal by a factor of $\sim 10$. Each amplified signal is next sent to a quad or octal discriminator, which handles four or eight BLMs, respectively. This card levels the signal spike from the PMT caused by the lost particle and sends a NIM-level pulse to a Jorway quad scalar which handles four BLMs. The quad scalar is really a pulse counter that counts pulses during the gated period defined by the gate module. A CAMAC 377 card provides start, stop and clear times to the gate module for the gate pulse. Output from the Jorway quad scalar card is sent to the control system.

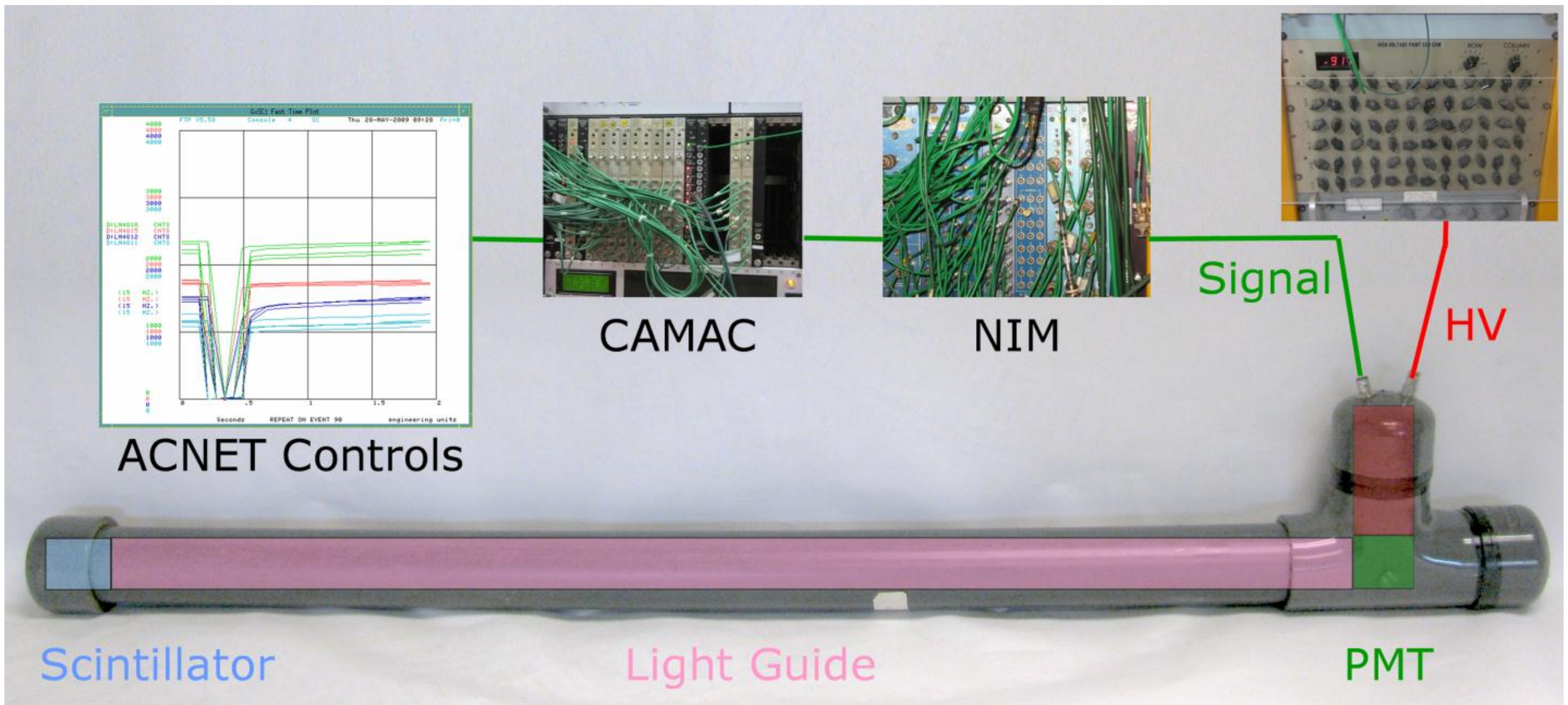

Figure 7.82: Delivery-Ring PMT BLM system.

**Mixed Secondaries Instrumentation Summary** Table 7.16 summarizes the instrumentation installation locations in the M2 beamline, M3 beamline, and Delivery Ring.

| Name | Device | Beam Line | Specific Location |
|---|---|---|---|
| SEM804 | SEM | M2 | Use existing SEM704 location |
| Tor804 | Toroid | M2 | Use existing Tor704 location |
| IC804 | Ion Chamber | M2 | Use existing IC704 location |
| SEM810 | SEM | M2 | Immediately downstream of Q811 |
| SEM702 | SEM | M3 | Use existing SEM926 location |
| SEM703 | SEM | M3 | Immediately downstream of H703 |
| SEM706 | SEM | M3 | Immediately downstream of Q706 |
| SEM711 | SEM | M3 | Immediately downstream of Q711 |
| SEM719 | SEM | M3 | Immediately downstream of Q719 |
| SEM726 | SEM | M3 | Immediately downstream of Q725 |
| SEM729 | SEM | M3 | Immediately downstream of Q730 |
| SEM740 | SEM | M3 | Immediately downstream of Q740 |
| IC740 | Ion Chamber | M3 | Immediately downstream of SEM740 |
| SEM744 | PWC | M3 | Immediately downstream of Q744 |
| SEM748 | PWC | M3 | Immediately downstream of Q749 |
| SEM204 | PWC | DR | Immediately upstream of ELAM |
| SEM302 | PWC | DR | Immediately downstream of ISEP |
| IC209 | Ion Chamber | DR | Immediately downstream of D2Q9 |
| SEM607 | PWC | DR | Use existing SEM607 location |
| SEM105 | PWC | DR | Near D1Q5 |
| IC102 | Ion Chamber | DR | Immediately upstream of D1Q2 |
| SEM403 | PWC | DR | Use existing SEM403 location |
| SEM506 | PWC | DR | Near D5Q6 |
| WCM503 | WCM | DR | Between D5Q3 and D5Q4. Use WCM until needed for Mu2e in M4 and then replace with ion chamber or another WCM. |

Table 7.16: Mixed secondary beam instrumentation in the M2 beamline, M3 beamline, and Delivery Ring [42].

### Proton Secondaries

Proton secondaries will extracted to the Delivery-Ring abort line and will have a similar beam intensity to that of the Delivery Ring. Instrumentation already located in that region will be used. A toroid will be used to measure beam intensity for Mu2e operations, but will be out of its operational range for $(g-2)$. Ion chambers, SEMs and BLMs will be used for $(g-2)$ in the same way they are for the mixed secondary lines.

### Muon Secondaries

Muons will traverse the upstream portion of the M4 line and the M5 line. The largest technical challenge will be measuring the low-intensity muon beam, which models show should be on the order of $10^5$ muons per pulse. This is two to three orders of magnitude smaller than the upstream mixed-secondary beam. Most of our standard diagnostics will not work at these beam intensities.

**Ion Chambers** Beam intensity will be measured with ion chambers as described in the ion chamber section above. This design will allow beam intensity measurements down to $10^5$ particles. The ion chamber in the M4 line will need to be retractable in order to be compatible with Mu2e operations, while the M5-line ion chambers can be permanently in the beam path.

**Proportional Wire Chambers** Beam profiles in the upstream M4 and M5 beamlines will be measured using Proportional Wire Chambers (PWCs). Other proposed solutions, such as the BNL Segmented Wire Ion Chambers (SWICs), would have required design of vacuum bypass systems as well as permanent vacuum windows in the path of the beam that would create significant losses due to Coulomb scattering effects [40]. PWCs are more sensitive than SWICs, with the capability of measuring beam intensities down to the $10^3$ particle range. When mounted inside refurbished Switchyard bayonet vacuum cans, the PWCs can be pulled out of the beam path when not in use. This eliminates the need for permanent vacuum windows and vacuum bypasses. A new design that will be used for the Switchyard beamlines was recently developed and provides the measuring capabilities needed to measure low-intensity muon-only beam for the $(g-2)$ experiment [43]. Using this existing design makes using PWCs even more cost effective.

The PWC has two planes of signal wires, one plane for horizontal and one for vertical. There are 48 signal wires in each plane which are 10-$\mu$m diameter gold-plated tungsten and can be configured with either 1 mm or 2 mm spacing. The wire planes are sandwiched between Aluminum high-voltage bias foils where negative voltage is applied. In addition to the bias foils, there are two more grounded foils on the outermost surfaces over the outer bias foils. These grounded foils balance the electrostatic field on the bias foil and prevent the bias foil from deflecting towards the sense wires. They also provide a degree of safety by covering the bias foils with a grounded conductive shield. Two end plates hold the entire assembly together. See Fig. 7.83 for a detailed view of the assembly.

The PWC assembly is filled with an 80% Argon and 20% Carbon Dioxide gas mixture. Ions are created when beam passes through the gas in the chamber. The positive ions are drawn toward the negatively charged high voltage foils, where they are neutralized. The electrons are drawn toward the signal wires. As the electrons get within close proximity of the sense wires the electrostatic field around the wires causes the electrons to accelerate, creating an electron cascade in the gas. The collected negative charge on the wires is then processed by the same type scanner as is used for the SEMs.

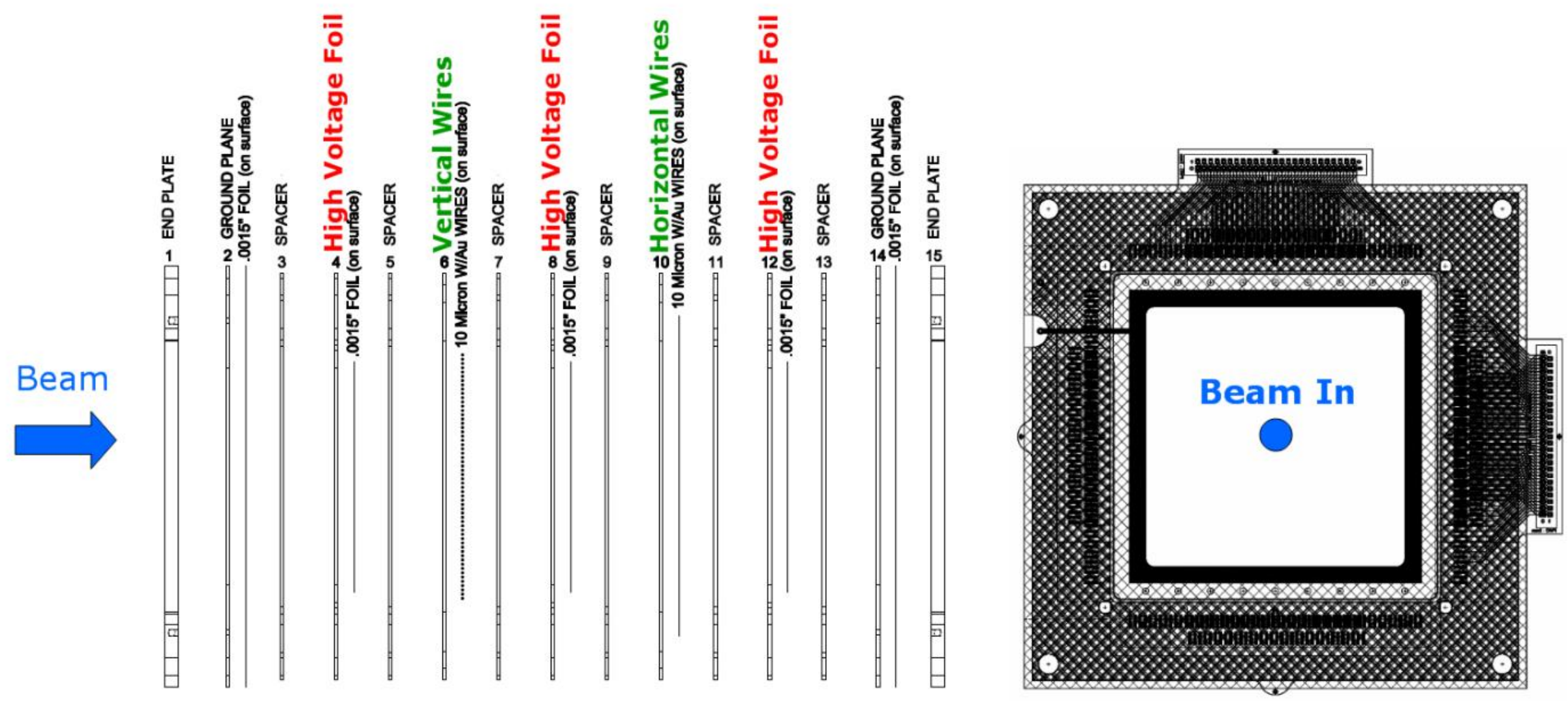

Figure 7.83: The Proportional Wire Chamber (PWC) assembly is made up of horizontal and vertical signal planes separated by high-voltage foils with a ground plane and end plate on each end. On the left is a side-view showing each layer of the assembly and on the right is a front (beam in) view showing the entire assembly.

As with the previously mentioned ion chambers, the gas filled PWCs must be isolated from beam tube vacuum. The PWCs will be packaged in an anti-vacuum box. The anti-vacuum box is a sturdy machined aluminum shell with a 0.003-in thick titanium foil window mounted on each side for the beam to pass through. The anti-vacuum box allows the detector to be mounted in a beamline vacuum chamber while the PWC inside the box remains at atmospheric pressure. A vacuum-tight duct attached to the box allows the gas tubing, signal and high-voltage cables to be routed from the PWC to outside the vacuum chamber.

In order to save engineering and assembly costs, the anti-vacuum boxes will be installed inside of bayonet vacuum vessels that are being repurposed from Switchyard. The bayonet-type drive slides the PWC linearly into and out of the beam with a screw drive system. Bayonet drives use a 72-RPM Superior Electric Slo-Syn AC synchronous stepping motor coupled directly to the screw shaft. The detector linear drive shaft is housed in a collapsible bellows that seals it from atmosphere. Figure 7.84 shows the PWC assembly, the anti-vacuum box, and the bayonet vacuum can. The same configuration is being used for the earlier-mentioned retractable ion chambers. In that case, the PWC assembly is modified to hold a single foil plane.

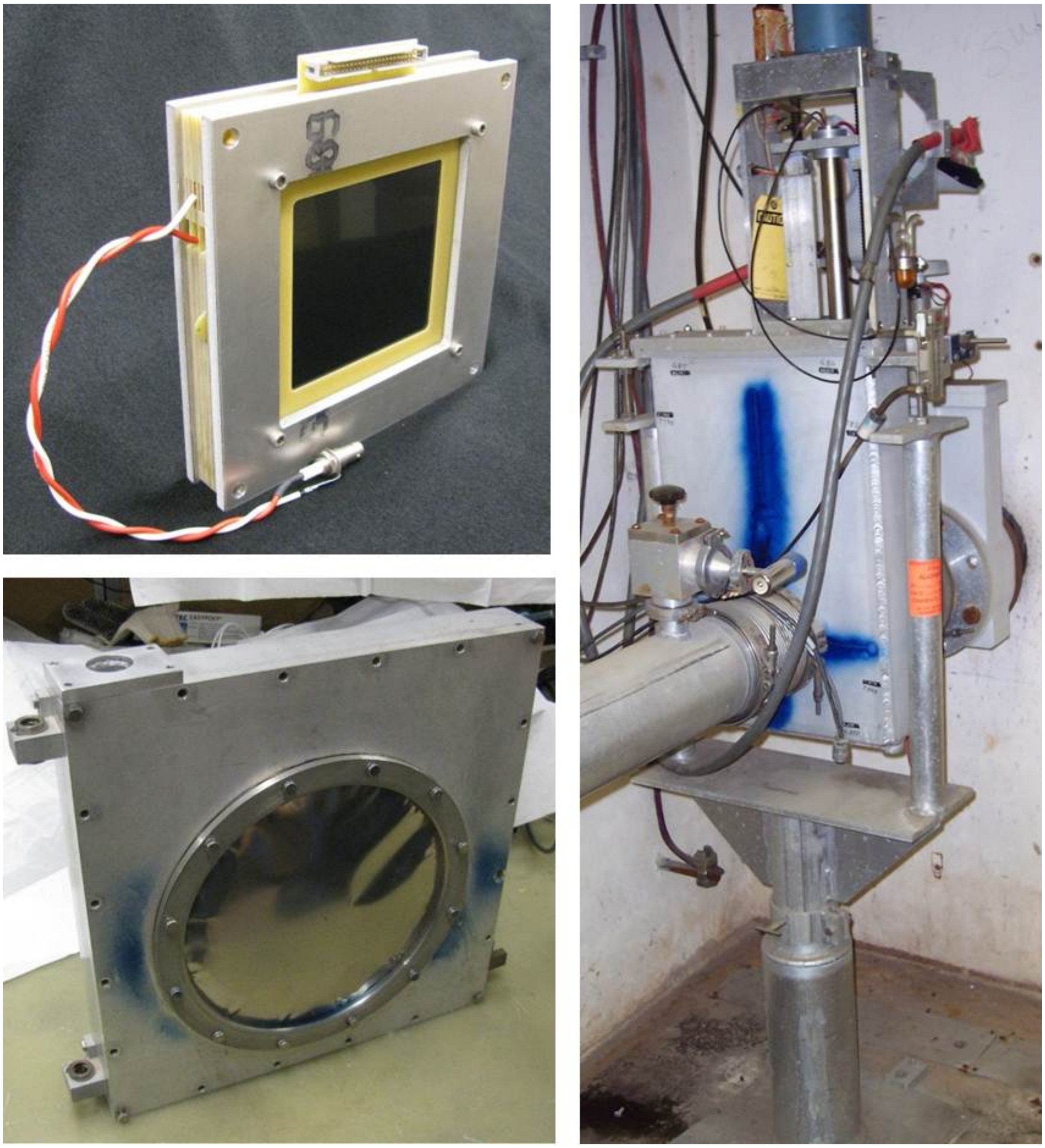

Figure 7.84: The first completed PWC prototype (top left). The signal connection is at the top and the high voltage connection comes out the left side. The PWC is installed in an anti-vacuum box (lower left). ArCO2 gas is pumped into this chamber, and there is a vacuum window on both front and back of this module. The anti-vacuum chamber is installed inside of the bayonet can (right) which is pumped down to beam tube vacuum. The PWC wires can be lowered into the beam or raised out of the beam via a motor drive.

**Cerenkov Counter** The BNL experiment E821 used a Cerenkov counter to measure the particle composition entering the $(g-2)$ ring. The detector was used to distinguish the relative particle compositions of $\pi+$, $e+$, and $\mu+$, but could not be used to measure protons [44].

The E821 Cerenkov detector was shipped to FNAL in 2012. The tank was refurbished and brought up to current ESH&Q standards [45] before being installed in the AP2 beamline for studies in 2014 (Fig. 7.85). The flammable isobutane gas was replaced by nonflammable Octafluorotetrahydrofuran ($C_4F_8O$), which simplified ESH&Q requirements. $C_4F_8O$ has been used in other Cerenkov detectors at FNAL, and calculations showed that the Cerenkov light angle $\theta_c$ and pressure thresholds were compatible with repeating the E821 particle composition measurements in the AP2 line [2].

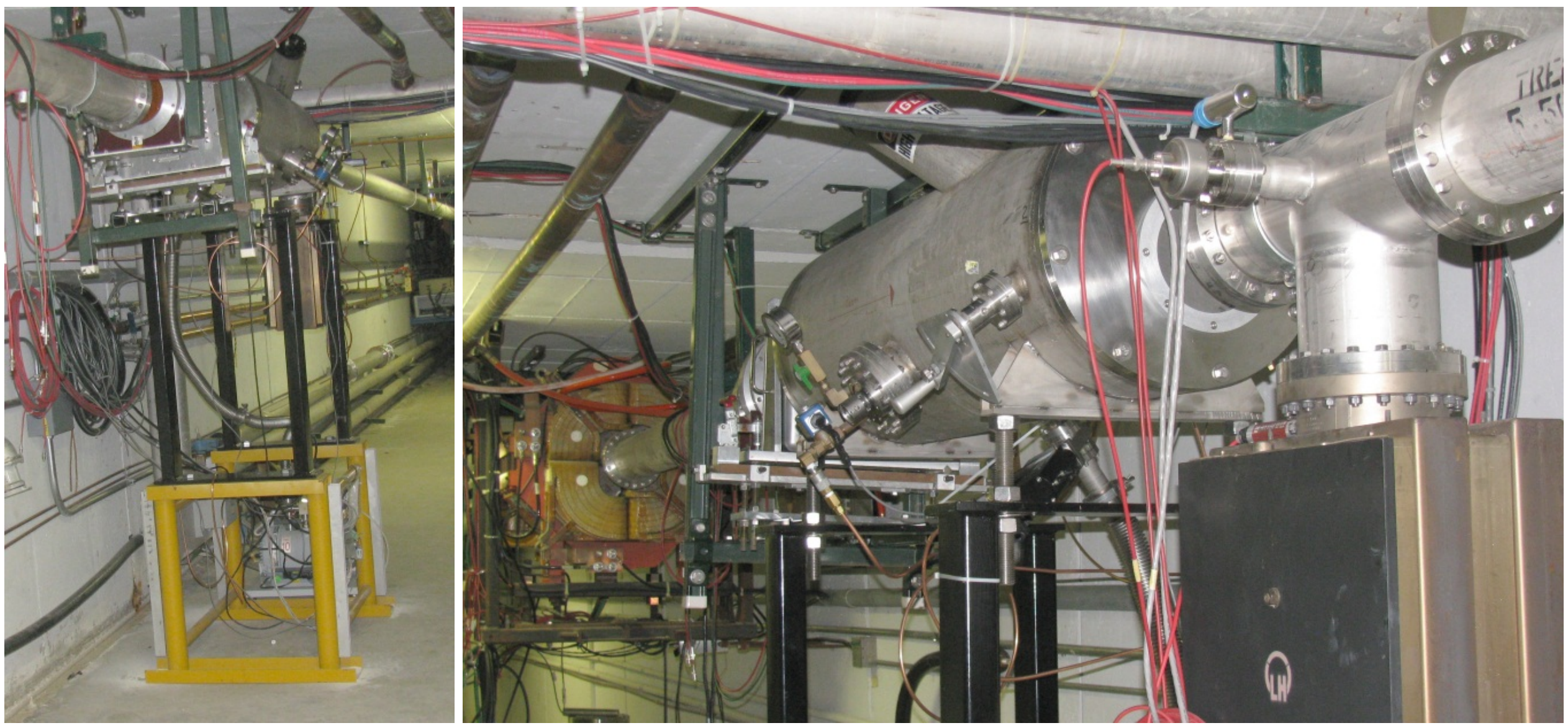

Figure 7.85: Cerenkov detector installed in the AP2 line.

A new controls system interface was designed and is shown in Fig. 7.86 [46], [2]. An Automation Direct DL405 series Programmable Logic Controller (PLC) in the AP50 service building handles most of the control and monitoring for the Cerenkov detector. The PLC has input and output capability for 24 VDC signals, relay contacts, and 0-10 V analog signals used by the pump cart and valves.

Monitoring and controls are through ACNET, with status bits for pumps and valves, analog readings for turbo pump speed and tank pressure, and control bits for pumps, valves and the test LED. The gate valve connects the pumps to the chamber. The solenoid valve allows the introduction of gas to the chamber. PLC logic prevents the gate valve from moving from the closed position to the open position if the turbo pump speed is above 10% in order to prevent damage to the turbo if there is gas in the chamber.

A small additional microcontroller board is used to communicate over a serial link with the Setra chamber pressure gauge. The microcontroller queries the gauge once per second, parses the response and writes the value into the PLC memory for presentation to ACNET [46].

Data were collected from the Cerenkov detector during the spring 2014 beam studies.

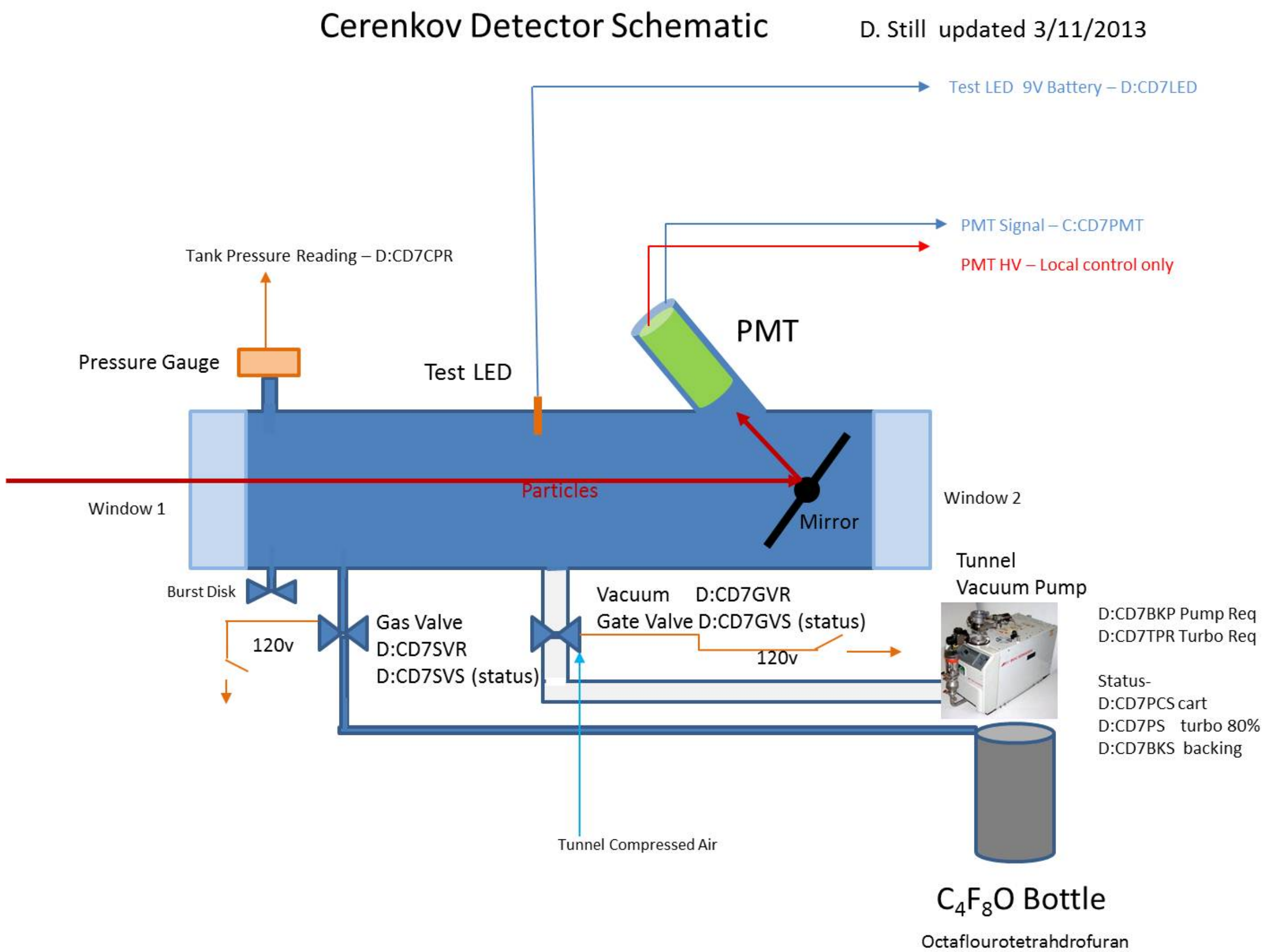

Figure 7.86: Cerenkov detector schematic showing basic PLC controls [2]

Beamlines were configured with 8 GeV beam on target and 3.1 GeV positive secondary beam in the AP2 line. Figure 7.87 shows the results of these studies with pulses of $2.7\times10^{12}$ protons on target. The intensity of secondary beam in the upstream AP2 line was approximately $2\times10^{9}$, while the beam intensity in the downstream portion of the line near the Cerenkov detector was approximately $6\times10^{7}$. Detector response was measured as the gas pressure was slowly raised to about 14 psi and again as gas pressure was lowered. As expected, Fig. 7.87 shows three distinct slopes representing the positrons, muons and pions [2].

With the successful implementation of the Cerenkov Detector in the AP2 line, the next step will be to move the electronics and controls to the MC-1 service building and the detector to the M5 beamline for commissioning. The Cerenkov detector will be installed between Q023 and Q024 in the M5 line as shown in Fig. 7.88, and modifications will be made to the detector stand to match the beampipe elevation. The Cerenkov detector is a significant source of Compton scattering, so during normal $(g-2)$ operations, the detector and vacuum windows will be replaced with a spool piece.

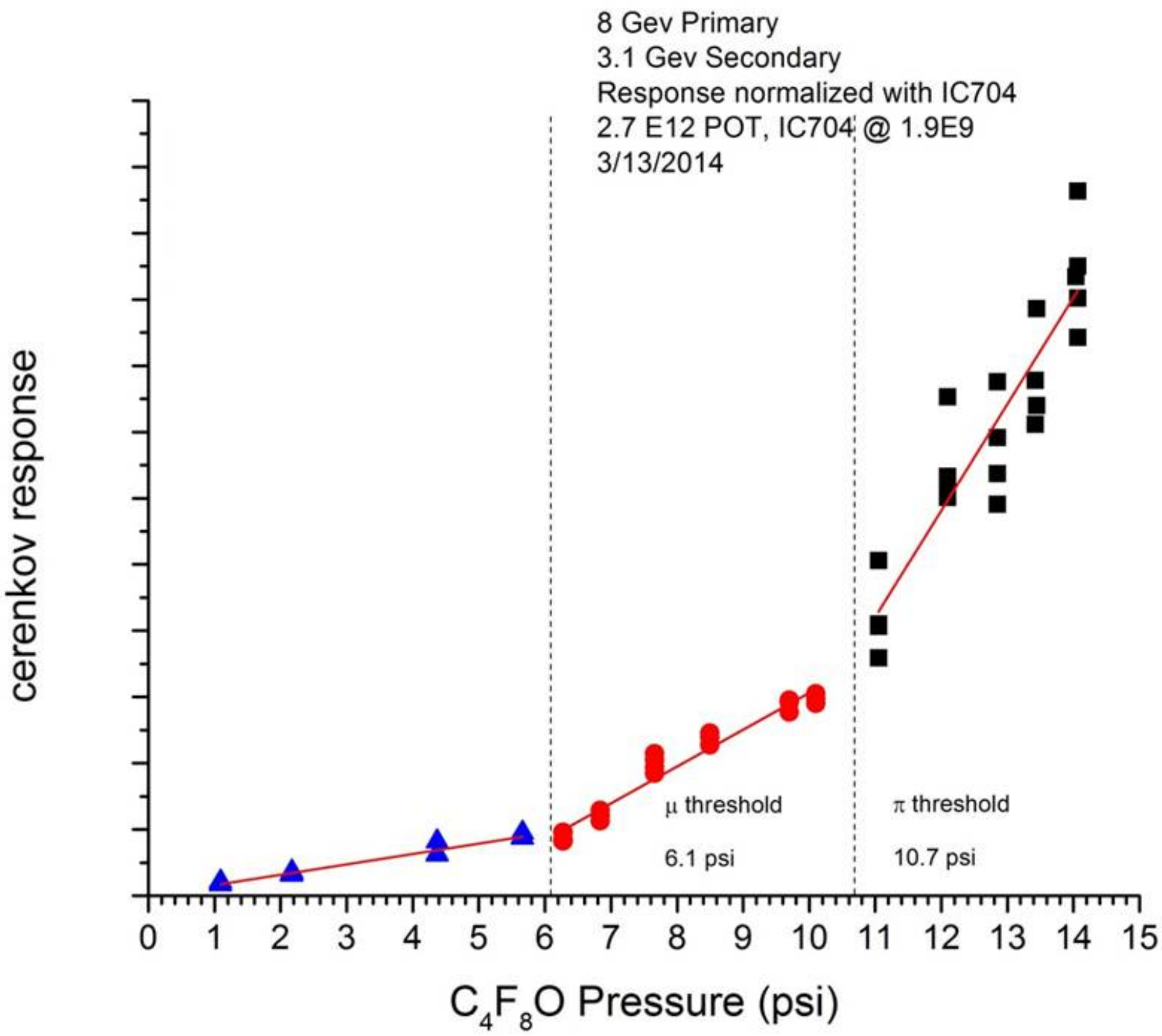

Figure 7.87: Cerenkov detector response (in arbitrary units) as a function of gas pressure. The three distinct slopes represent positrons, muons and pions [2].

Figure 7.88:

**Muon Instrumentation Summary** There will be one retractable ion chamber and two PWCs in the M4 beam line before the split. There will be an additional ion chamber and six PWCs in the M5 beam line. In addition, there will be a Cerenkov detector in the M5 line during beamline commissioning. Specific locations of these devices are outlined in Table 7.17.

| Name | Device | Beam Line | Specific Location |
|---|---|---|---|
| PC900 | PWC | M4 | Immediately downstream of c-magnet |
| IC901 | Ion Chamber | M4 | Immediately upstream of Q902 |
| PC903 | PWC | M4 | Immediately downstream of Q903 |
| PC000 | PWC | M5 | Half-way between V907 and Q001 |
| PC006 | PWC | M5 | Immediately upstream of H006 |
| PC012 | PWC | M5 | Immediately downstream of H012 |
| PC020 | PWC | M5 | Immediately downstream of Q020 |
| CD024 | Cerenkov | M5 | Immediately upstream of Q024 (commissioning only) |
| PC026 | PWC | M5 | Immediately downstream of Q026 |
| PC027 | PWC | M5 | Immediately downstream of Q027 |
| IC027 | Ion Chamber | M5 | Immediately downstream of PWC027 |

Table 7.17: Muon beam instrumentation in the M4 and M5 beamlines [47].

### Accelerator instrumentation summary

A summary of instrumentation devices which will potentially be used for $(g-2)$ is shown in Table 7.18.

| Beamline | Beam type | Intensity | Position | Profile | Loss |
|---|---|---|---|---|---|
| Primary protons | P1, P2, M1 | toroids | BPMs | multiwires, SEMs | BLMs |
| Mixed secondaries | M2, M3, DR | ion chambers | SEMs/PWCs | SEMs/PWCs | BLMs |
| Proton secondaries | DR abort | ion chambers | SEMs | SEMs | BLMs |
| Muons | M4, M5 | ion chambers | PWCs | | |

Table 7.18: Instrumentation to be used in the beamlines for $(g-2)$ operations.

## 7.7 ES&H, Quality Assurance, Value Management, Risk

### 7.7.1 ES&H

The Accelerator Division ES&H Department has the responsibility for providing Environmental, Safety, and Health coordination and oversight of ES&H for all accelerator work on the project. As with all Fermilab projects, attention to ES&H concerns will be part of the project management, and Integrated Safety Management will be incorporated into all processes. Line management responsibility for ES&H will be maintained on this project. Safe coordination of installation activities will be accomplished through the Project Management team, Project ES&H Coordinator, Project Engineer, and Task Manager. During installation, the Subcontractors, T&M Crafts, and all Fermilab personnel will utilize Job Hazard Analyzes to plan all work and to mitigate hazards. The Project Manager and Project ES&H Coordinator will audit compliance with all applicable ES&H requirements.

The handling and installation of magnets, vacuum systems, power supplies, and other accelerator components are common tasks within the Accelerator Division, and standard safety practices will be used. If any work falls outside of common practices, job hazard analyses will be conducted in order to ensure that the tasks are performed safely. Detailed procedures exist for handling components in the radioactive target vault, and the activation will be lower after years of not running beam than it was during antiproton production.

### 7.7.2 Quality Assurance

All aspects of the accelerator work will be periodically reviewed with regard to Quality Assurance issues from Conceptual Design through completion. The following elements will be included in the design and construction effort: an identification of staff assigned to each task with clear definition of responsibility levels and limit of authority as well as delineated lines of communication for exchange of information; requirements for control of design criteria and criteria changes and recording of standards and codes used in the development of the criteria; periodic review of the design process, drawings, and specifications to insure compliance with accepted design criteria.

### 7.7.3 Value Management

Significant cost savings have been incorporated into the $(g-2)$ accelerator design by utilizing the existing infrastructure from the Antiproton Source. This includes 1 km of tunnel complete with electrical infrastructure, cable trays, a cooling water distribution system, and safety interlocks. Service buildings with HVAC, cooling water, controls communication infrastructure, extensive electrical infrastructure, electronics racks, access roads and parking lots are also already in place.

The existing Target Station and its components will be reused: target, lens, collimator, momentum-selection magnet, target vault, cooling systems, a "hot" work cell, and tunnel access points with overhead crane coverage. A new target-station dump to replace the current one which has an internal water leak will be constructed using the existing design.

As many existing components as possible will be reused for the beamlines, including approximately 250 Antiproton-Source magnets plus about 30 beamline magnets from the previous $(g-2)$ experiment at BNL. New magnets will be based on existing designs, where practical. Power supplies will also be repurposed where practical, although modern switch-mode power supplies will be purchased which have high efficiency and power factor near unity, which will save operating costs, and which are also smaller in size and save substantial building space.

Much of the beamline instrumentation will also be recycled, including Secondary Emission Monitors and Beam Loss Monitors, with upgraded readout electronics where necessary to see the low-intensity $(g-2)$ secondary beam.

### 7.7.4 Risk

The largest risks to the cost and schedule of the accelerator work are delays of funding and lack of engineering support when it is needed.

Another large risk depends on Mu2e shielding needs in the Delivery-Ring D30 straight section, which have not yet been fully determined. Shielding may need to be placed in areas which would obstruct current plans for reconfiguration of beamlines and cable trays. Magnets may need to be made radiation-hard.

The external beamline depends on a new tunnel enclosure being built under a General Plant Project. If that project is delayed or if construction costs rise, there may be a burden on $(g-2)$.

Conflicts and difficulty of work in the congested area of the D30 straight section and the M3 line which joins the DR in that area are a schedule risk on the order of a month or two.

There is also an opportunity that the M2/M3 crossover design may be simplified and be made to cost up to $500k less.

Magnets which need to be built new and those which have been taken from the BNL beamline carry a risk on the order of $200k.

The possibility that existing accelerator controls infrastructure is not able to support $(g-2)$ is low, but carries risks on the order of $100-200k. The risk that various types of instrumentation cannot be refurbished or upgraded to see the low-intensity $(g-2)$ secondary beam would require new instrumentation to be built at a cost of roughly $200-400k and a 4-month delay.

The biggest technical risk was that the lithium lens used for focusing secondaries off the target would not be able to pulse at the $(g-2)$ rate. However, a lens has been pulsed in a test stand at the average 12-Hz rate for 80 million pulses without any sign of lens failure, confirming ANSYS simulations which predicted that mechanical fatigue should be less than it was during antiproton production.

There is an opportunity to save $100k if a new transformer will not be required in order to support the lens power supply.

The risk that the Target Station does not provide the desired yield may be handled by running the experiment for a longer period, or additional cooling may be needed for the final focus system, or a new target may be designed and constructed.

# Chapter 8

# Beam Rate Simulations

The ultimate goal of the beam delivery simulations is a complete "End-to-End" study from pion production on the target to stored muons in the ring. This goal will be achieved by using a sequence of specialized tools. They include a `MARS` calculation for pion production, `G4Beamline` and `MAD8` calculations of the pion-to-muon decay line, the Delivery Ring, and the final beamline into the storage ring, a detailed `GEANT4` and `BMAD` simulations of the transmission into the ring and the final storage fraction. This is ongoing work by many collaborators. In this chapter we present the status of the individual pieces, the results of Preliminary Design simulation studies of pion production, muon capture, beam transport and injection into the $(g-2)$ storage ring, and discuss future plans.

## 8.1 Pion Production at the Target

The description of the pion production target and lithium lens are given in Chapter 7. In this section we describe the software model of the target and lithium lens, give details of the simulations, summarize results and discuss future plans.

### 8.1.1 `MARS` Model and Beam Initial Conditions

Pion production in the target and focusing of the secondary beam by the lithium lens were simulated using `MARS` [1]. A graphical representation of the `MARS` model of the target and the lens is shown in Fig. 8.1. The target consists of a 11.43-cm-diameter 25.4-cm-long Inconel cylinder (1) enclosed into a 6-mm-thick Be container (red ring in Fig. 8.1). The target is off-centered horizontally relative to the beam axis by 4.31 cm. The intersection of the beam with the Inconel cylinder is therefore about 7.5 cm long.

Secondary beam particles produced in target are focused by the lithium lens. The most relevant part for our simulations is the 16-cm-long 2-cm-diameter Li cylinder (2) with a 232 T/m focusing magnetic field produced by the electric current flowing along the axis of the cylinder. We assume ideal focusing field (i.e., no attempts have been made so far to model a more realistic or time-dependent field).

In Ref. [2] a study was done of the pion collection efficiency dependence on the focusing field and position of the lithium lens relative to the production target. As expected, stronger

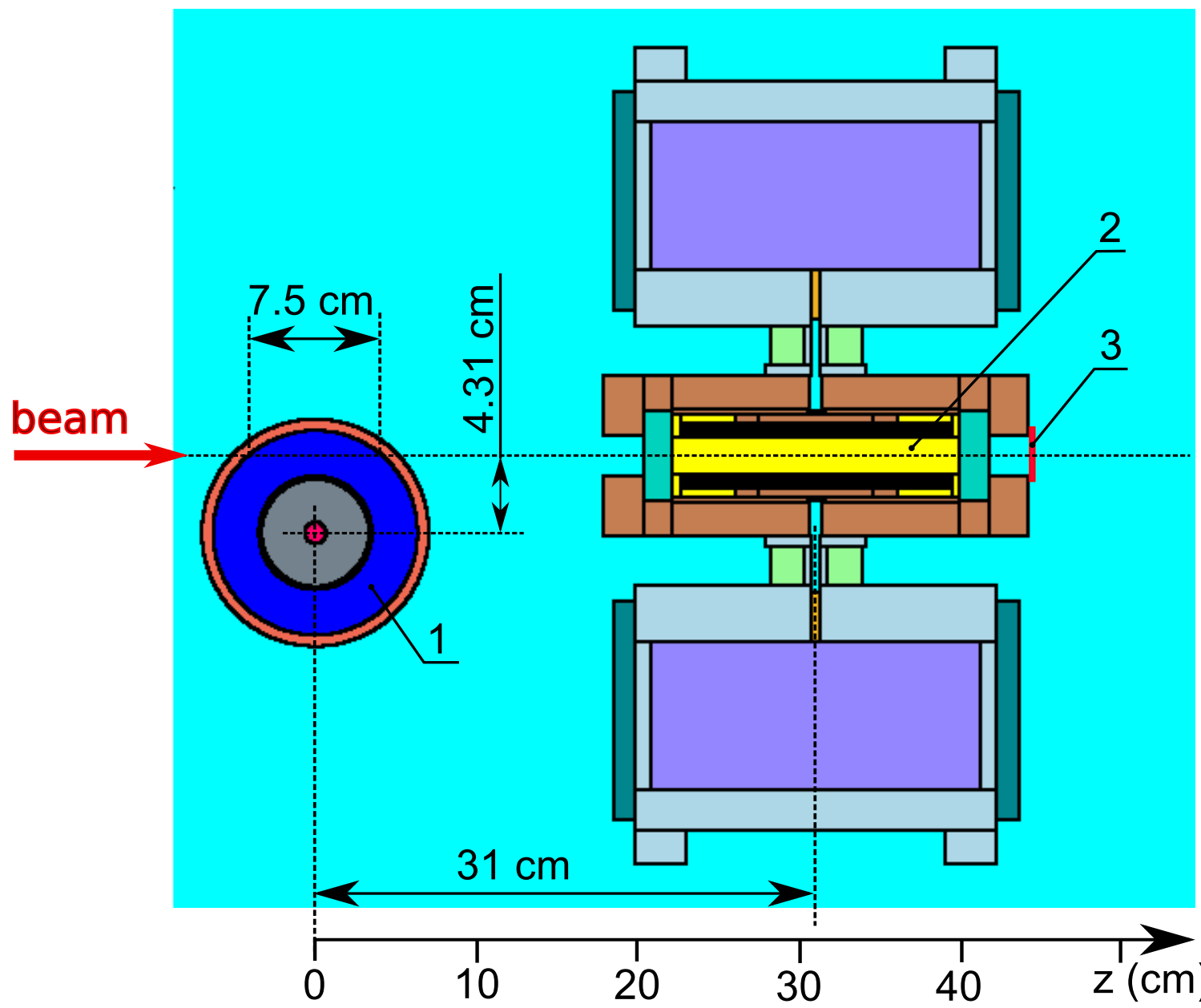

Figure 8.1: MARS model of the E989 target station (top view): Inconel target (1), 16-cm-long 2-cm-diameter Li cylinder with magnetic field (2), virtual detector, $z = 43$ cm (3).

focusing fields result in better pion collection.

For our final simulations we use the baseline design value of 232 T/m recommended by the experts. The chosen field strength results in high pion collection efficiency and allows for long-term reliable operation of the lithium lens at 12 Hz repetition rate in a pulsed mode with irregular time structure of the beam (see section 7.2). The position of the lithium lens with 232 T/m focusing field was also optimized for pion collection efficiency. The optimization was done by maximizing the number of pions inside a $40\pi$ mm-mrad acceptance ellipse at virtual detector (3) as seen in Fig. 8.1. Twiss parameters defining the phase-space ellipses were taken from the beamline lattice design in Ref. [3] ($\beta_x = 2.105$ m, $\alpha_x = 0.033$, $\beta_y = 2.274$ m, $\alpha_y = 0.001$). The pion momentum range was limited to a 2.7-3.5 GeV/c band. The simulated number of accepted pions per proton on target as function of lithium lens position is shown in Fig. 8.2. The highest number of pions was observed when the lithium lens is placed at $z = 31$ cm, which is within its operation range. In future, the position of the lithium lens can be refined to maximize the number of magic-momentum muons entering the $(g-2)$ storage ring.

The primary proton beam with $0.3\pi$-mm-mrad emittance has a kinetic energy of 8 GeV. As it was shown in Ref. [4] and also seen in Fig. 8.2, smaller proton beam sizes lead to higher pion fluxes within the beamline admittance. According to the baseline design, the size of the primary proton beam on target will be 0.15 mm (see Section 7.4.1). For our baseline calculations we assumed Gaussian proton beam with $\sigma_x = \sigma_y = 0.15$ mm spot size at $z = 0$ and $\sigma_{x'} = \sigma_{y'} = 2$ mrad angular divergence[1].

The physics information (momentum, coordinates, spin, etc.) of beam particles entering the virtual detector (3) with $|x|, |y| < 20$ mm, $|x'|, |y'| < 20$ mrad and $2.7 \leq p \leq 8.0$ GeV/c were recorded for the second simulation step using the `G4beamline` program. Thus, the phase space of recorded particle was chosen to be larger than the admittance of the beamline so as to not introduce any bias.

[1]The specified size of the beam spot at $z = 0$ is assumed in the absence of the target. Multiple scattering in the target material slightly broadens the primary beam.

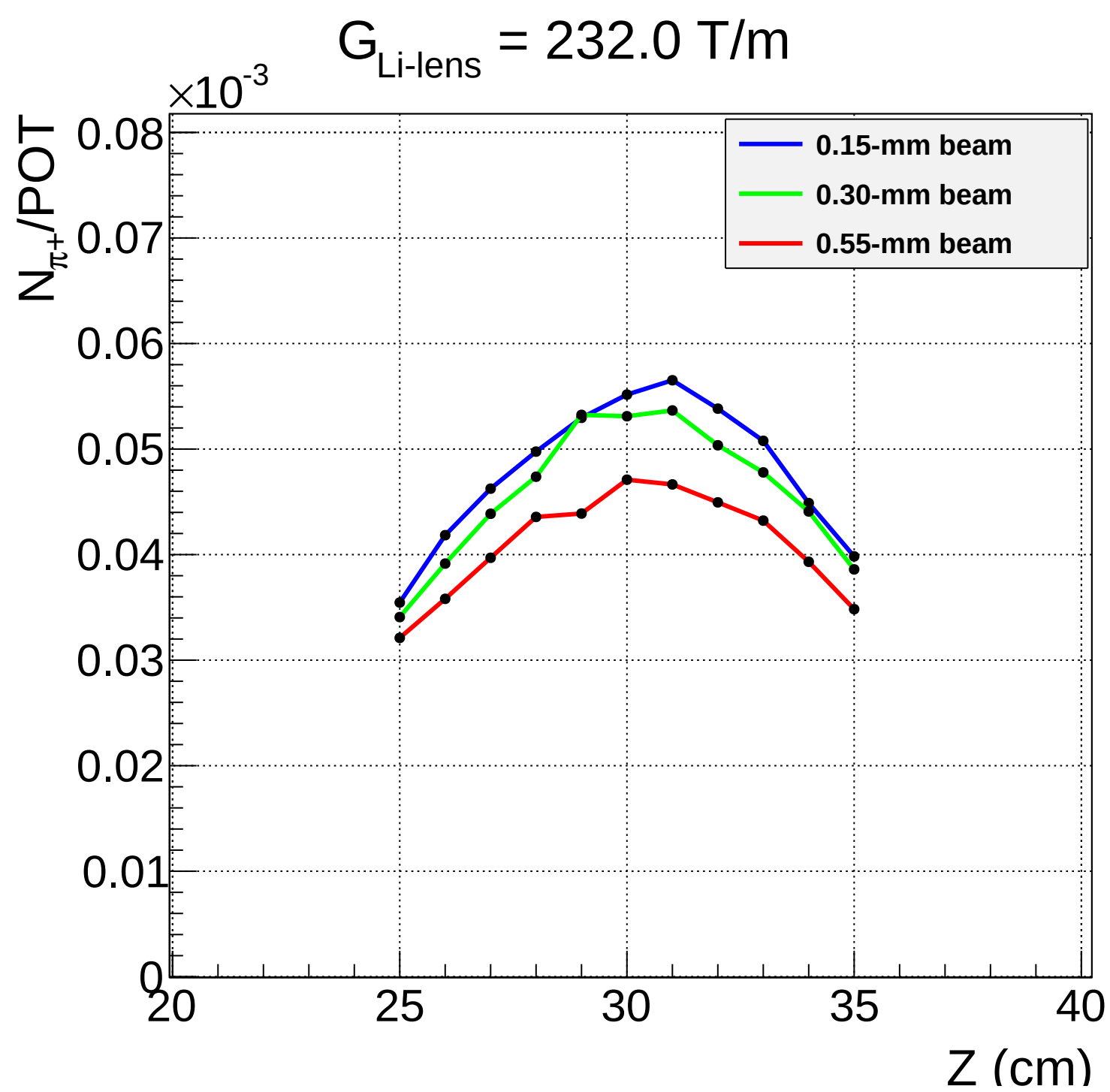

Figure 8.2: Number of pions at the downstream face of the lithium lens within $40\pi$ mm-mrad acceptance ellipse ($\beta_x = 2.105$ m, $\alpha_x = 0.033$, $\beta_y = 2.274$ m, $\alpha_y = 0.001$) as function of lithium lens position relative to the production target (center-to-center). Software cut on pion momentum was limited to $\pm 2.5\%$ around magic momentum. Three lines correspond to three different spot sizes of the proton beam as indicated in the insert.

### 8.1.2 Polarization of Muons off the Target

Magic-momentum muons originate from in-flight decays of pions with momenta in the range from 3.1 to 5.3 GeV/c. The longitudinal polarization dependence (along pion momentum) and emission angle (relative to pion momentum) of magic-momentum muons as a function of pion momentum are shown in Fig. 8.3. In the pion decay channel, the momentum distribution of pions is relatively narrow, $\Delta p_\pi / p_{\text{magic}} \approx \pm 10\%$ , producing highly-polarized magic-momentum muons. In contrast, the momentum distribution of pions in the target station is very broad, resulting in magic-momentum muons originate with a wide range of polarizations.

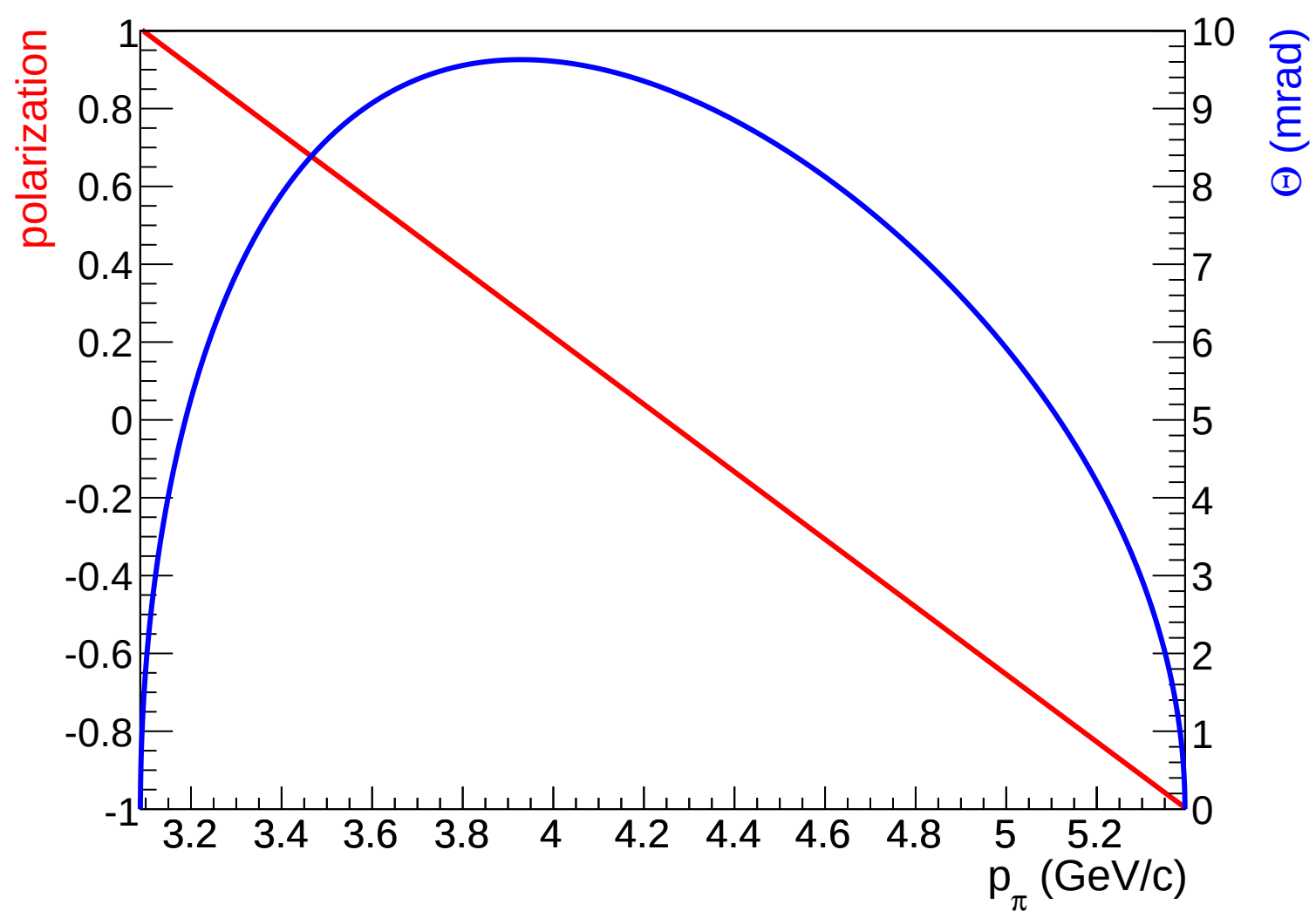

Figure 8.3: Horizontal polarization (red) or emission angle relative to pion momentum (blue) of magic-momentum muons as function of pion momentum. The components of muons polarization vector are plotted in a so-called *centerline* coordinate system where the Z axis is running down the center of the beamline, the X axis is beam left, and the Y axis is up.

The `MARS` simulation does not track polarization. Thus, we had to introduce spin tracking code into `MARS` ourselves. The distribution of horizontal components of polarization of magic-momentum muons ($\Delta p/p = \pm 0.5\%$) originating from pion decays upstream to virtual detector (3) (see Fig. 8.1) is shown in Fig. 8.4. The average polarization of these muons is about 0.55.

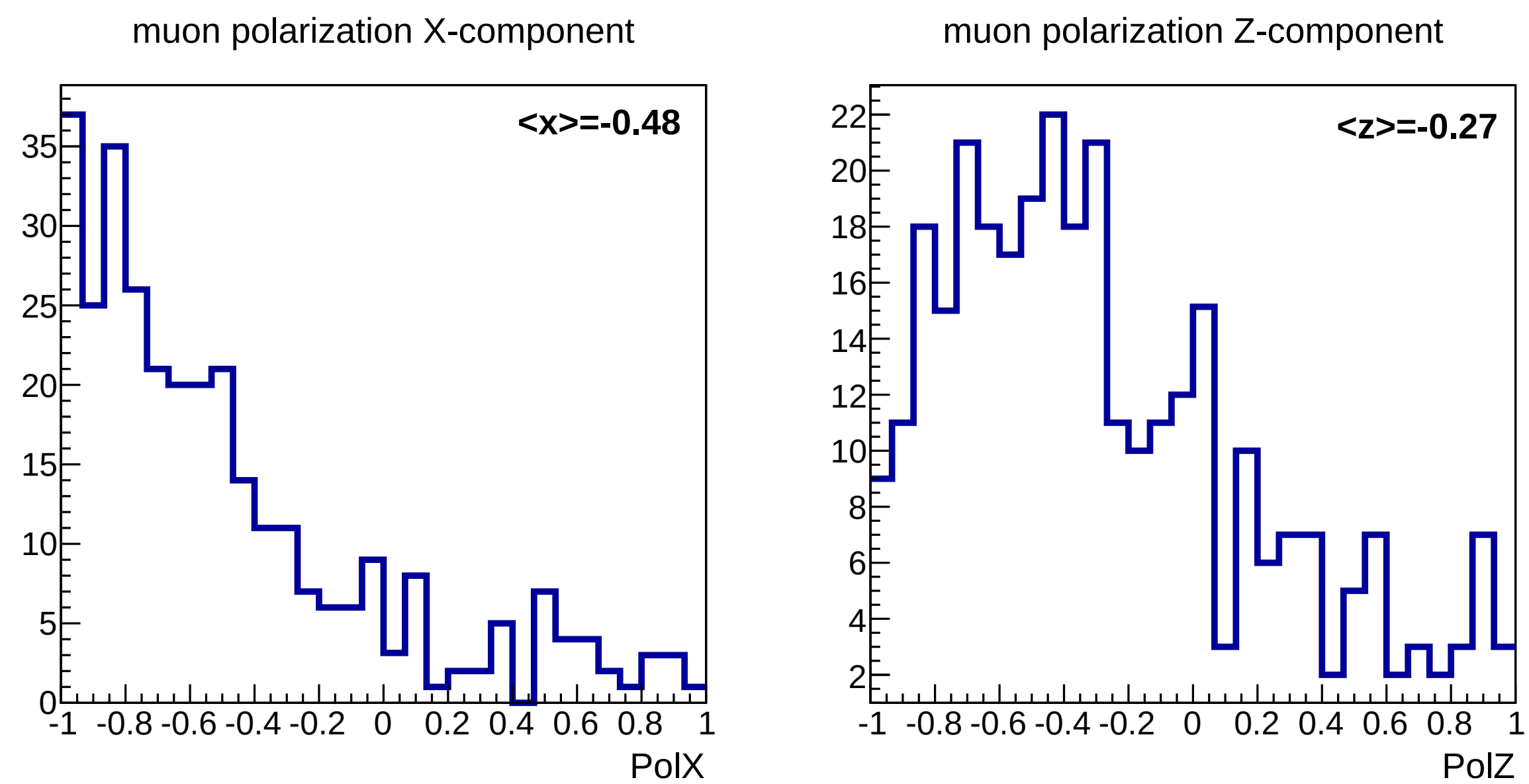

Figure 8.4: Horizontal components of polarization of magic-momentum muons ($\Delta p/p = \pm 0.5\%$).

## 8.2 Target–to–Storage-Ring Transport

Simulation of pion beam transport starting from virtual detector (3) (See Fig. 8.1), including pion decay and muon capture, and muon beam transport to the $(g-2)$ storage ring was performed using `G4beamline` program. In this section we present the `G4beamline` model of the beamline, describe difficulties we encountered in the process of building a software model of the beamline, present the results of simulations and discuss future plans.

### 8.2.1 `G4Beamline` Model and Optics Validation

`G4Beamline` is a particle tracking and simulation program based on the `GEANT4` toolkit. `GEANT4` was originally conceived for modeling detectors; `G4Beamline` extends `GEANT4` simulations to beamline elements, beam transport lines and decay beam lines. Concretely this means that `G4Beamline` provides standard beamline elements defined in terms of `GEANT4` primitives. In addition, it allows the user to work in beam coordinates, either to specify beamline geometry or to analyze and interpret tracking results. Note that all trajectory integrations are ultimately performed in absolute coordinates using time as the integration variable by `GEANT4`. This also implies that it is as accurate and realistic as the `GEANT4` toolkit implements.

`G4Beamline` allows simulation of important aspects of the $(g-2)$ experiment in an integrated manner, including muon production in the decay line, muon capture by the beamline and muon transport to the $(g-2)$ ring. Most importantly, in combination with the simulation program `gm2RingSim` – which is also based on `GEANT4` – the impact of beam-related systematic errors can be studied.

The $(g-2)$ delivery beamlines are designed using the standard optics code `MAD8`. Like virtually all beam optics codes, MAD8 does not have an explicit notion of three-dimensional space. Rather, a beamline is represented as an ordered sequence of elements which in turn, uniquely defines a design trajectory. Since the trajectory of a particle through a magnet depends on its initial conditions at the magnet input face, *a-priori* assumptions about the latter must be made. For all focusing elements, MAD8 assumes that the trajectory is a straight line entering and exiting at $(x = 0, x' = 0, y = 0, y' = 0)$ in the element local transverse coordinate system. For bending elements it is assumed that the particle entrance and exit angles are of equal magnitudes and opposite signs. In the case of an ideal sector magnet (SBEND) the entrance and exit angles with respect to the magnet faces are trivially zero. For a rectangular bend (RBEND) these angles are equal to $\theta/2$, $\theta$ being the net bending angle through the magnet.

The G4Beamline models of the delivery lines are constructed by direct translation of the design lattice in MAD8 format (Fig. 8.5). To maximize translation reliability, the latter is performed in an automated manner, using a program written specifically for that purpose.

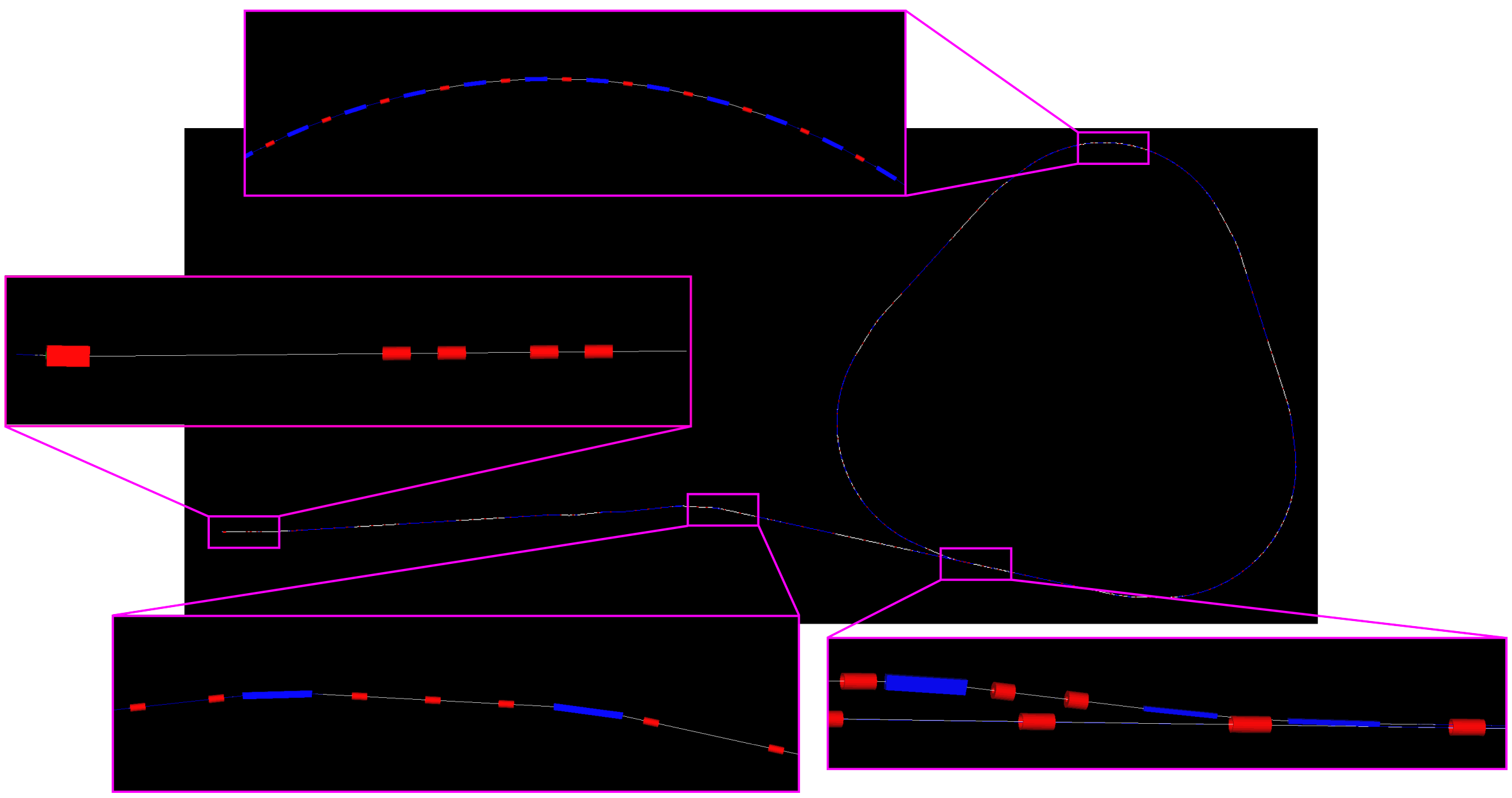

Figure 8.5: G4Beamline model of the *M2-M3-delivery ring* section of the $(g-2)$ delivery beamline.

Early on, a number of deficiencies in G4Beamline were identified. In collaboration with the code developer (who has been responsive to our needs and concerns), appropriate fixes were developed and incorporated into G4Beamline official releases.

Even though G4Beamline has been used with success by a number of groups, the code had never been exercised much or at all to model moderately complex non-planar lattices. Thus, when the absolute positions of the G4Beamline magnets were compared to the 3-dimensional site coordinates produced with the MAD8 survey command, some small but nevertheless noticeable discrepancies were observed.

We ultimately identified and addressed a number of issues

1. Rectangular bend magnets were effectively always rotated in space with respect to a

point located in the center of the magnet. To replicate MAD8 positioning, such magnets need to be rotated by half the net bend angle with respect to an axis passing through the origin of the upstream face. Note in passing this issue adversely affected the geometric acceptance.

2. The beam reference coordinate system was not transformed properly in rotated bending dipole magnets, resulting in incorrect transverse beam coordinates.

3. The polarization state was ignored when saving or reading a particle distribution. A related issue was that it was not possible to read an arbitrary initial polarization distribution. Uniform polarization was the initial condition supported by the program.

4. $N$ being the number of elements in the lattice, we observed an $O(N^2)$ scaling of the time required to integrate trajectories. The inefficiency was traced to the technique used to determine, at each time step, the position of a particle in local element coordinates for purpose of evaluating the local electromagnetic field. After discussions, an improved algorithm based on voxelization techniques was developed and implemented by the code author. Near $O(N)$ scaling is now achieved.

G4Beamline still has shortcomings (as of time of this writing) that are relevant to our simulations.

- it generally does not efficiently handle accurate integration through sharp-edged fields. This issue principally affects idealized bending elements where the magnetic field exhibits an abrupt jump at the upstream and downstream faces. The immediate effect is a small error on the particle angle which translates into a noticeable position error further downstream. To perform accurate numerical integration, the integrator needs to precisely anticipate the time at which a particle will cross a magnet face boundary. In practice, this turns out not to be an easy problem to deal with and the best strategy often ends up being the imposition of a relatively small limit on the maximum allowed step size. While the integration accuracy is improved, the cost in terms of computational efficiency tends to be high.

- G4Beamline does not correctly handle dipole edge focusing. This usually results in the introduction of a modest amount of beam envelope modulation which may affect the accuracy of computed acceptance. Work on a fix for this issue is on-going. Note that in principle, edge focusing effects are completely accounted for when a detailed fringe field description can be included. Currently this is supported only for rectangular bends with poles faces at nominal orientation.

G4Beamline offers two methods to position elements: (1) the sequential method where the position of each element is expressed relatively in a local coordinate system referred to as centerline coordinates, and (2) the absolute method where each element is positioned absolutely in a fixed global coordinate system. Since the sequential method most closely follows the default mode of operation of an optics code, we have relied on this approach to produce our initial G4Beamline geometry models.

While very convenient, the sequential technique of positioning beamline elements lacks generality. This is due to the implicit assumption that the design trajectory through each element is uniquely defined. This is not always the case for a number of reasons, including :

- Some elements are time-dependent e.g. kicker magnets.
- Some magnets are positioned in such a way that the design trajectory traverses them more than once. For example, to relax the requirements on the time-dependent kicker, the beam injected into the delivery ring enters at an offset through ring quadrupoles before being kicked to become tangential to the ring central ring orbit.

To simulate the entire $(g-2)$ beam transport process including transport to and injection into the delivery ring followed by extraction after a few turns it became necessary to split the simulation into a few steps. A distinct beamline model is constructed for each step and the input particle distribution is assumed to be the output from the previous step. Using this approach, kickers can be modeled as (different) time-independent devices while quadrupoles where the beam nominal trajectory is off-center can be represented by combined function bends.

To confirm that the `G4Beamline` geometry is equivalent to `MAD8`, a number of sanity checks are performed.

- For both models the path lengths measured along the design trajectories (as reported by each code) should be in agreement at the mm level.
- The current version of `G4Beamline` generates an output equivalent to that of the `MAD8 survey` (i.e. the absolute positions of the center of the aperture at the entrance and exit faces of every element, in absolute coordinates). For a correct translation, all these positions should agree at the mm level.
- Finally, an appropriately matched particle distribution is tracked through the `G4beamline` model. The lattice functions are then extracted from the beam size and emittances and compared to the output generated by the `MAD8 TWISS` command.

For future simulations, we intend to use a single absolute coordinate model of the entire $(g-2)$ delivery line (M2-M3-delivery-ring-M4-M5) geometry. This will eliminate the bookkeeping involved in breaking a simulation into multiple steps and allow the use of truly time dependent kicker elements. We should point out that an automated procedure to generate such an absolute description directly from a collection of `MAD8` optics files has recently been developed. In this case, while the other checks remain useful, there is obviously be no need to compare the absolute position of the elements with that of the `MAD8 survey` command. On the other hand, absolute positioning introduces some additional complexity mainly related ascertain that both the field and element orientations reported by the `MAD8 survey` are correctly translated into equivalent `G4Beamline` element rotations.

### 8.2.2 Beam Transmission

To record the beam particles for offline analysis we installed several virtual detectors in the `G4beamline` model: a `START` detector at $z = 5$ mm downstream to the lithium lens; virtual detector `VD5` in the injection region in the Delivery Ring (DR); virtual detector `DRR` in the DR near the point where the beam completes one revolution around DR; virtual detector `DRE` near the extraction point from the DR; and `STOP` detector after the last magnet of the entire beamline. An example placement of these detectors is shown in Fig. 8.6.

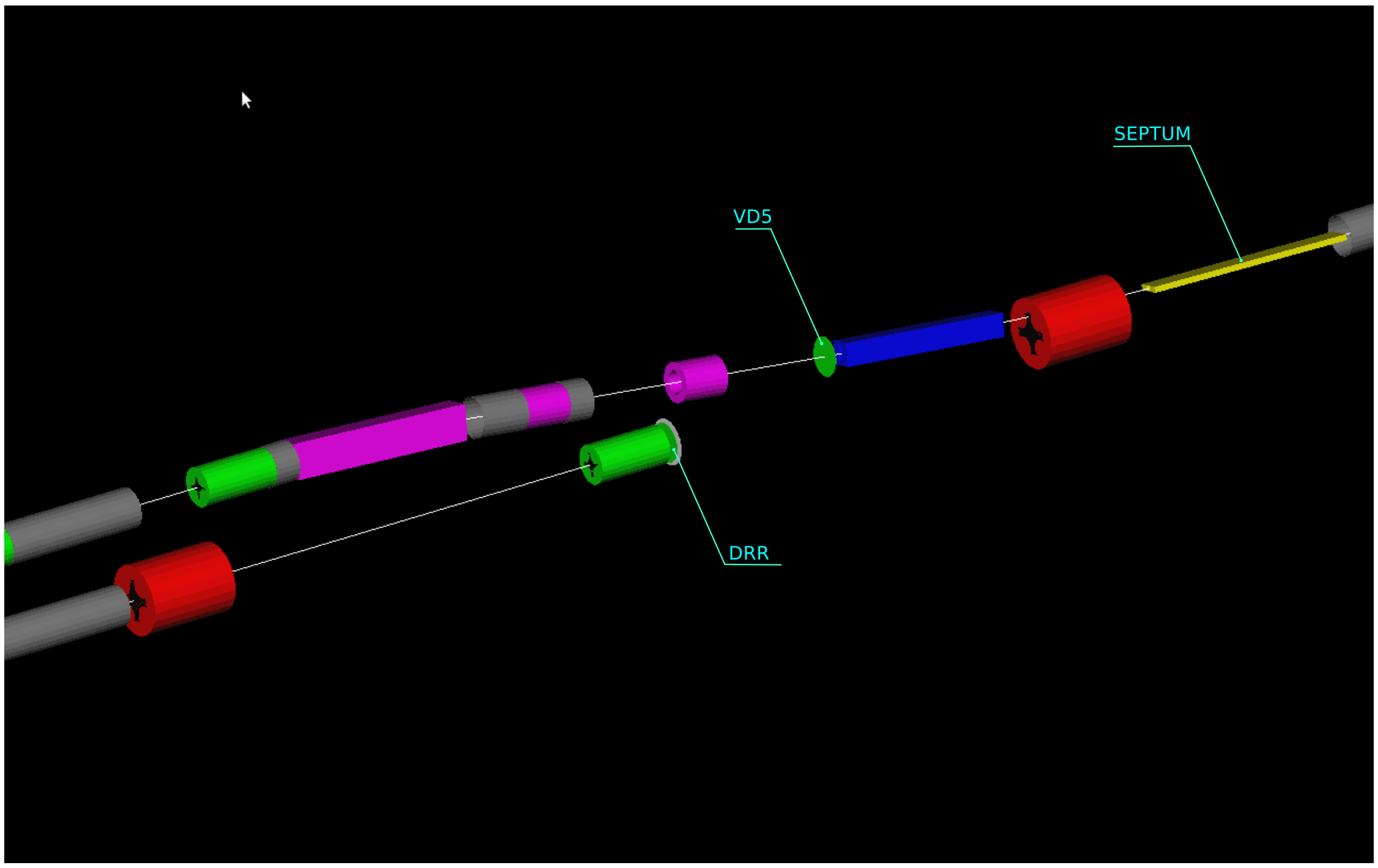

Figure 8.6: Location of virtual detectors near the injection point into the Delivery Ring in `G4beamline` model of the E989 beamline. `VD5` is located near the point where the beam is injected into the Delivery Ring (DR); `DRR` is located near the point where the beam completes one revolution around the DR.

To study the transmission of the beam through the beamline, we disabled decays of unstable particle in the `G4Beamline` physics list. In such conditions a beam particle can be lost only if it hits a beamline element. In our simulations, only natural apertures of beamline elements determine the beam transmission, no additional software cuts have been applied.

The horizontal ($x$) and vertical ($y$) phase space distributions of particles in the `START` detector are shown in Fig. 8.7. The same distributions accumulated using only those particles which propagate all the way through the beamline into the `STOP` detector are shown in Fig. 8.8. Thus, the emittance of the transmittable pion beam is about $40\pi$ mm-mrad in agreement with the `MAD` design of the beamline.

The momentum distributions of pions in the three virtual detectors are shown in Fig. 8.9. We remind the reader that we used relaxed phase space cuts to select `MARS` data for tracking in `G4beamline`. Therefore, the number of hits in the `START` detector is relatively large. Due to magnet apertures only a fraction of the initial beam can be transmitted through the entire beamline. We also start with a relatively broad momentum cut (blue line). As discussed in the previous chapter, the initial momentum range is selected using the PMAG magnet. The width of the momentum distribution of the beam entering the DR is about 3.5% (RMS). The final momentum selection takes place in the DR as the last beamline section M4-M5 has no momentum-selecting elements. At the end of the beamline the width of the momentum distribution is about 1.2% (RMS).

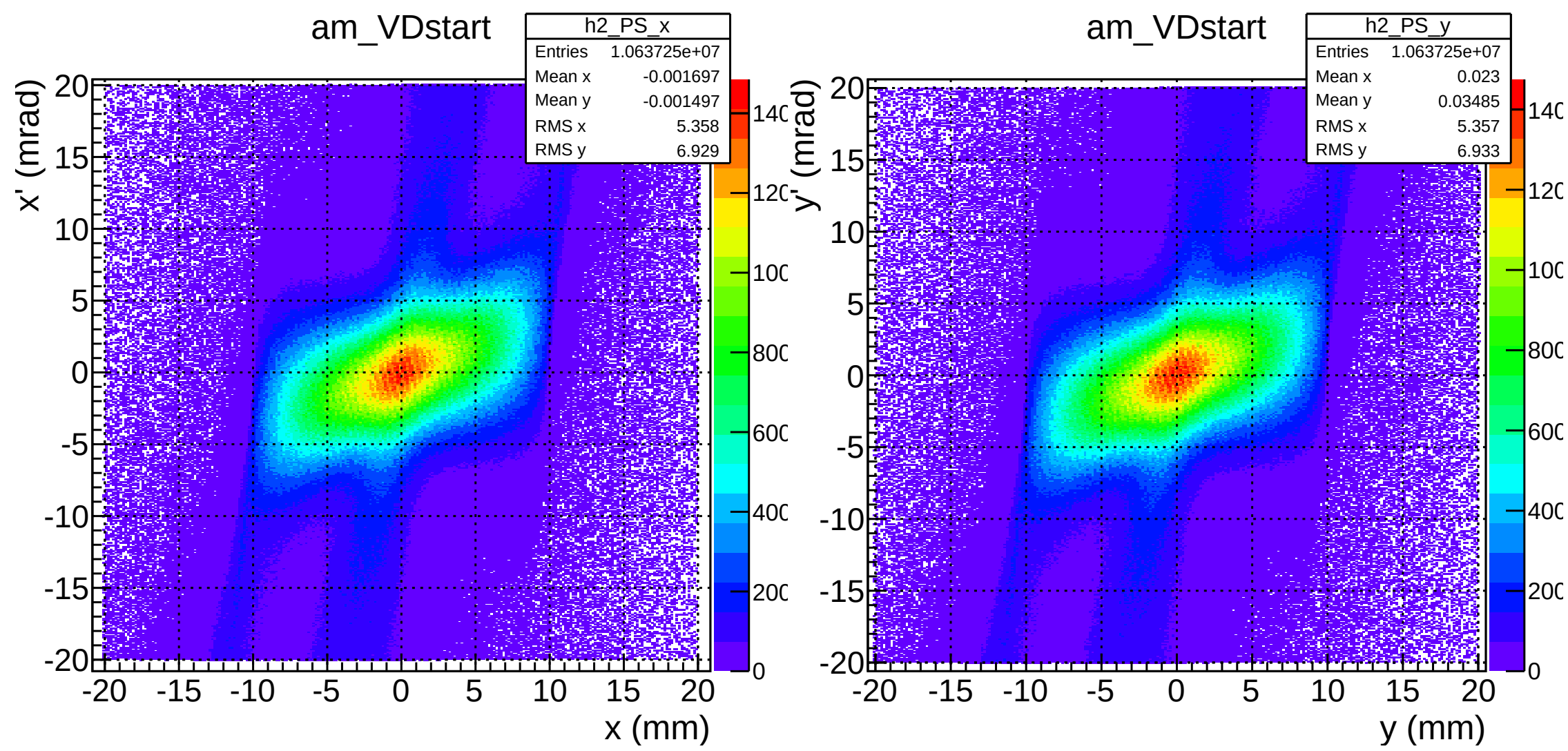

Figure 8.7: Horizontal (left) or vertical (right) phase space distributions of all particles in the START detector.

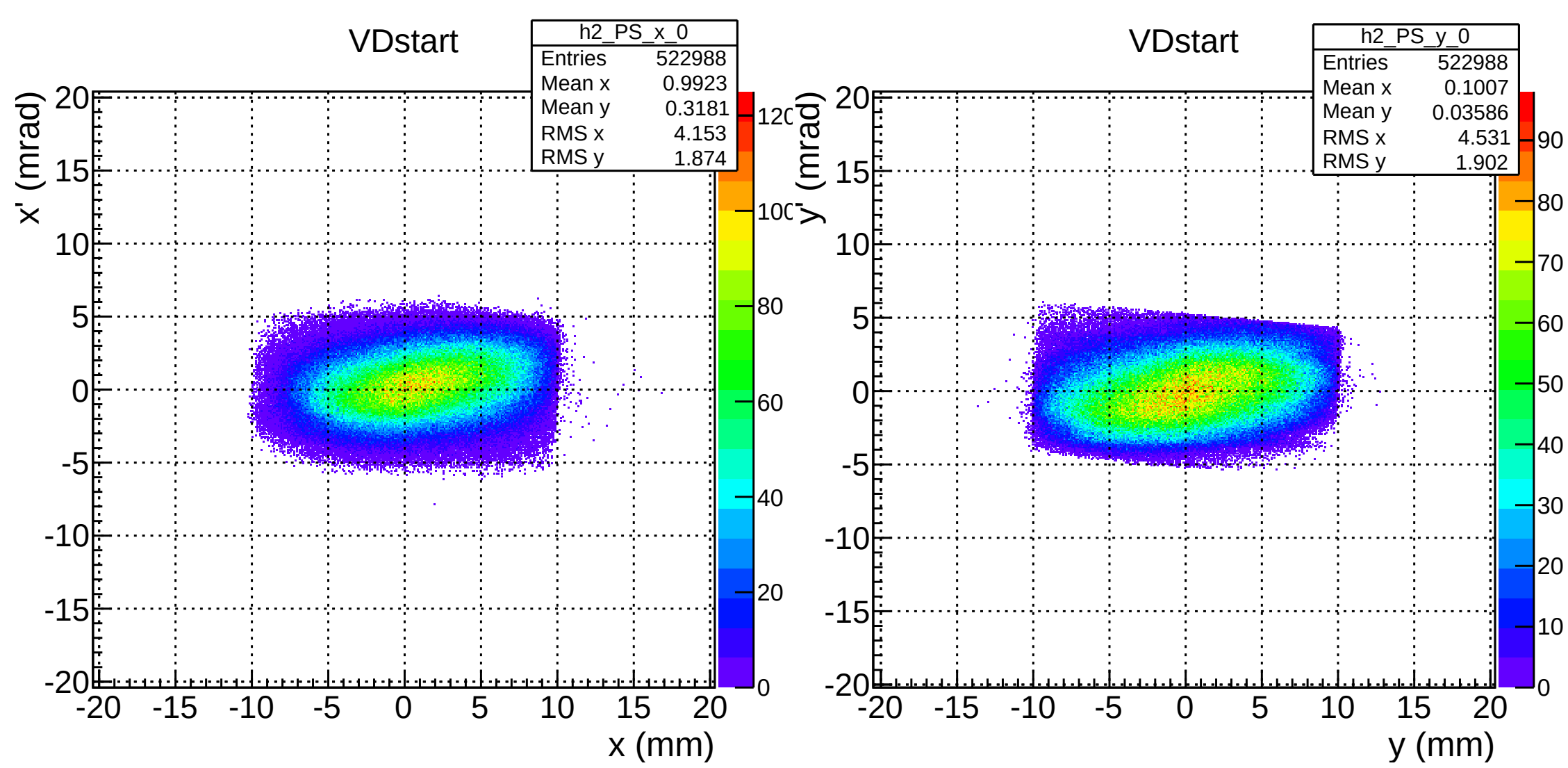

Figure 8.8: Horizontal (left) or vertical (right) phase space distributions in the START detector accumulated using only those pions which transmit through the entire beamline into the STOP detector. Decay of unstable particles disabled.

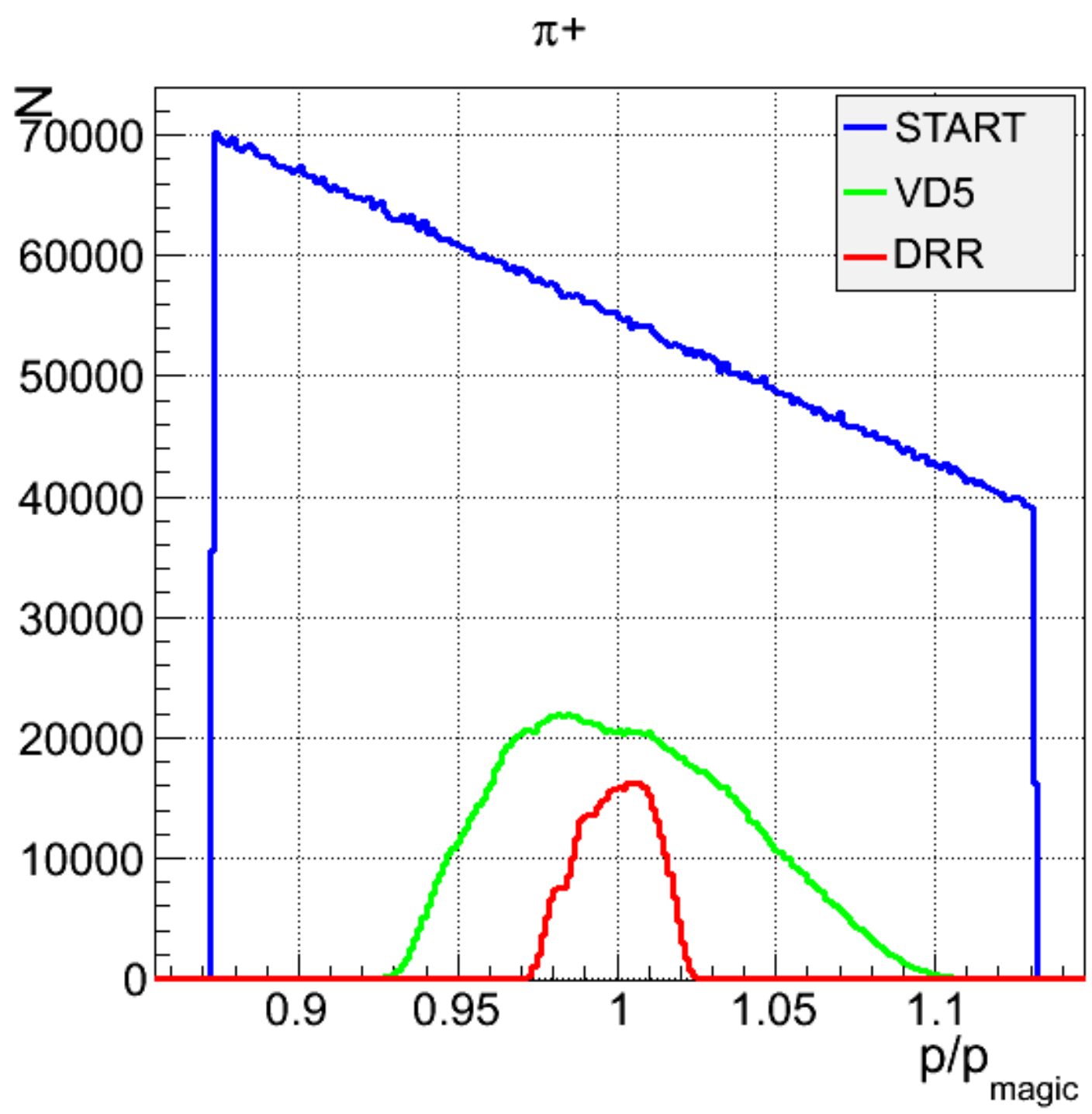

Figure 8.9: Momentum distribution of pions in START (blue), VD5 (green) and DRR (red) virtual detectors. Pion decay disabled.

### 8.2.3 Muon Collection

For the studies in this and the following sections we enable stochastic processes in G4beamline including decays of unstable particles. The momentum distributions of muons in virtual detectors START, VD5 and DRR are shown in Fig. 8.10. Sharp edges in the momentum distribution of muons in the START detector are due to a software cut of approximately 0.85-1.13 GeV/c applied to the MARS data to pre-select the relevant momentum range for tracking simulations. The momentum distribution of muons entering the DR (green histogram) has a complex shape because it includes muons originating before and after momentum-selecting beamline elements. After several revolutions in the DR almost all pions have decayed and the momentum distribution of muons resembles the distribution of beam particles in Fig. 8.9.

Since there is no momentum selection in the M4/M5 beamline section, the distribution of muon momenta at the end of the beamline (not shown in Fig. 8.10) has a similar shape to the one in the DRR detector. Currently, the design of the magnets for beam extraction from the DR has not been finalized yet, therefore in our model we assume 100% beam transmission through the apertures of magnets in the extraction optics. Therefore, the intensity of the muon beam reduces merely due to muon decays. Assuming three revolutions of the beam in the DR, the total number of magic-momentum muons ($\Delta p_\mu/p_{\rm magic} = \pm 0.5\%$) per proton on target at the downstream face of the last beamline magnet is $2.3 \times 10^{-7}$. However, as it will be discussed below, only a fraction of these muons will survive injection into the $(g-2)$ storage ring.

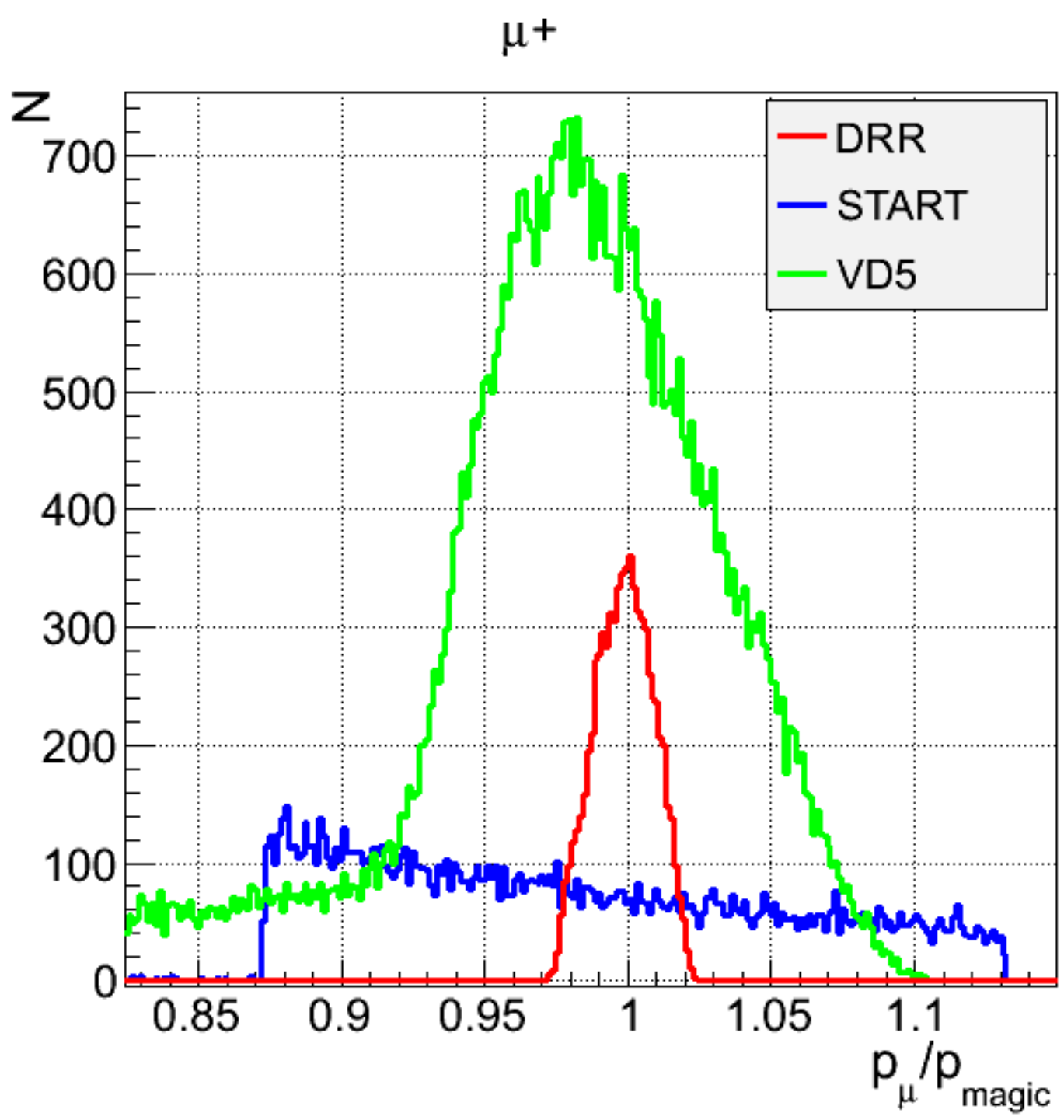

Figure 8.10: Momentum distribution of muons in `START` (blue), `VD5` (green), and `DRR` (red, after five turns around DR) virtual detectors.

Using $1.2 \times 10^{10}$ incident protons on target, `MARS` and `G4beamline` tracking simulations were performed to obtain a sample of about 10,000 muons in the `STOP` detector[2]. Physics parameters of these muons were recorded into an output file to be used as input for the studies of muon injection into the $(g-2)$ storage ring (see Section 8.3.1).

### 8.2.4 Beam Polarization and Distribution

The distribution of horizontal components ($x$ and $z$) of the polarization of magic-momentum muons ($\Delta p/p = \pm 0.5\%$) in the `STOP` detector is shown in Fig. 8.11. The average polarization is approximately 95%. This includes approximately 5% of muons with low polarization origintating upstream of the `START` detector (see Fig. 8.4).

[2]For this we simulated three turns around the Debuncher Ring

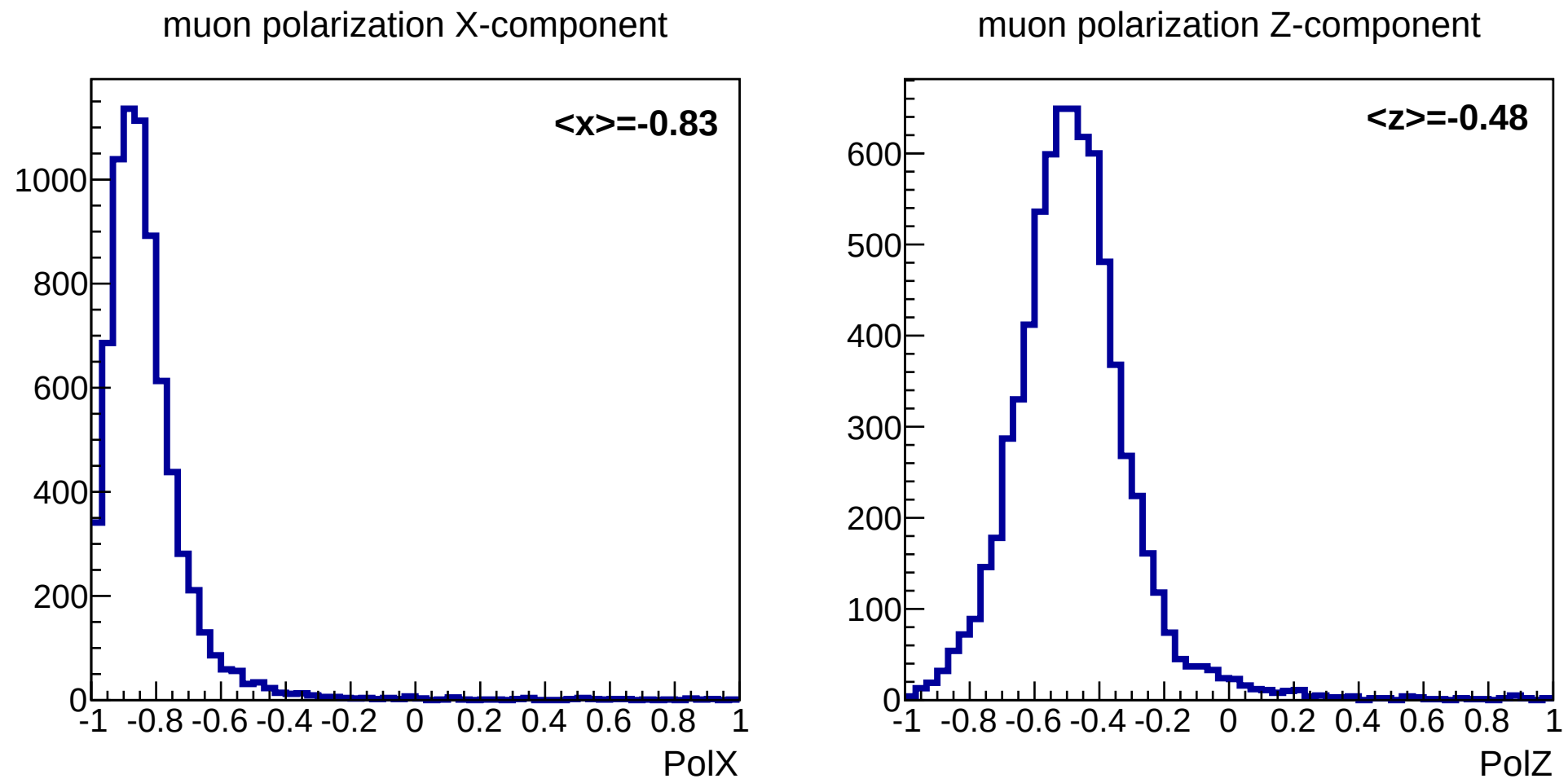

Figure 8.11: Horizontal components of polarization of magic-momentum muons ($\Delta p/p = \pm 0.5\%$) in the STOP detector.

### 8.2.5 Hadron Clean-up

The beam after the lithium lens is a mixture of mainly protons, neutrons and pions. The M2M3 beam line and Delivery Ring effectively selects all particles with a similar rigidity to the reference beam (predominantly protons and pions with momentum, $p \sim 3.1$ GeV/c) but the velocity difference between different species causes the mixed beam to separate into discrete populations. At the end of the M2M3 beamline, $\sim 82\%$ of the pions have decayed, so it is a mixed beam of mostly protons and muons that enters the Delivery Ring. After a number of turns round the Delivery Ring, the trailing proton beam can be extracted using the abort fast kicker. The rise-time of this kicker ($\sim$ 180 ns) determines the minimum gap required between the proton and muon beams to ensure a clean extraction of protons, without significant muon losses.

A MARS simulation of $10^9$ protons on target was used to calculate the expected particle phase-space distribution at the end of the lithium lens. This gave $\sim 12.7 \times 10^6$ protons, $\sim 1.1 \times 10^6$ positive pions and a small number of muons ($\sim$ 3000) at the start of the M2 line. The longitudinal distribution of all particles was assumed to follow that of the incident proton beam. The MAD8 tracking code [6] was used to track protons and positive pions separately through the M2M3 line and for several turns round the Delivery Ring. In both cases the lattice was tuned to a nominal momentum of 3.1 GeV/c. The apertures within the lattice were modeled in a variety of ways. The apertures of the quadrupole star chambers have a complicated geometry (see Fig. 8.12) with different sizes used in the M2M3 beamline and in the Delivery Ring. Both types were approximated with a tilted square aperture with side length of 108 mm in the Delivery Ring and 84 mm in the M2M3 lines. The arc dipoles are modeled with a rectangular aperture with dimensions $128 \times 60$ mm$^2$. At the M2M3 merging location an off-set quadrupole (Q733) is used to impart a 25 mrad dipole kick. This was modeled within the MAD8 lattice by a bending dipole with a superimposed quadrupole field component. A similar technique is used to model the D3Q3 (injection) and D2Q5 (extraction) quadrupoles which impart a 30 mrad vertical kick to aid the transition into and out of the Delivery Ring. All these quadrupoles are assigned circular apertures with a diameter of 200 mm.

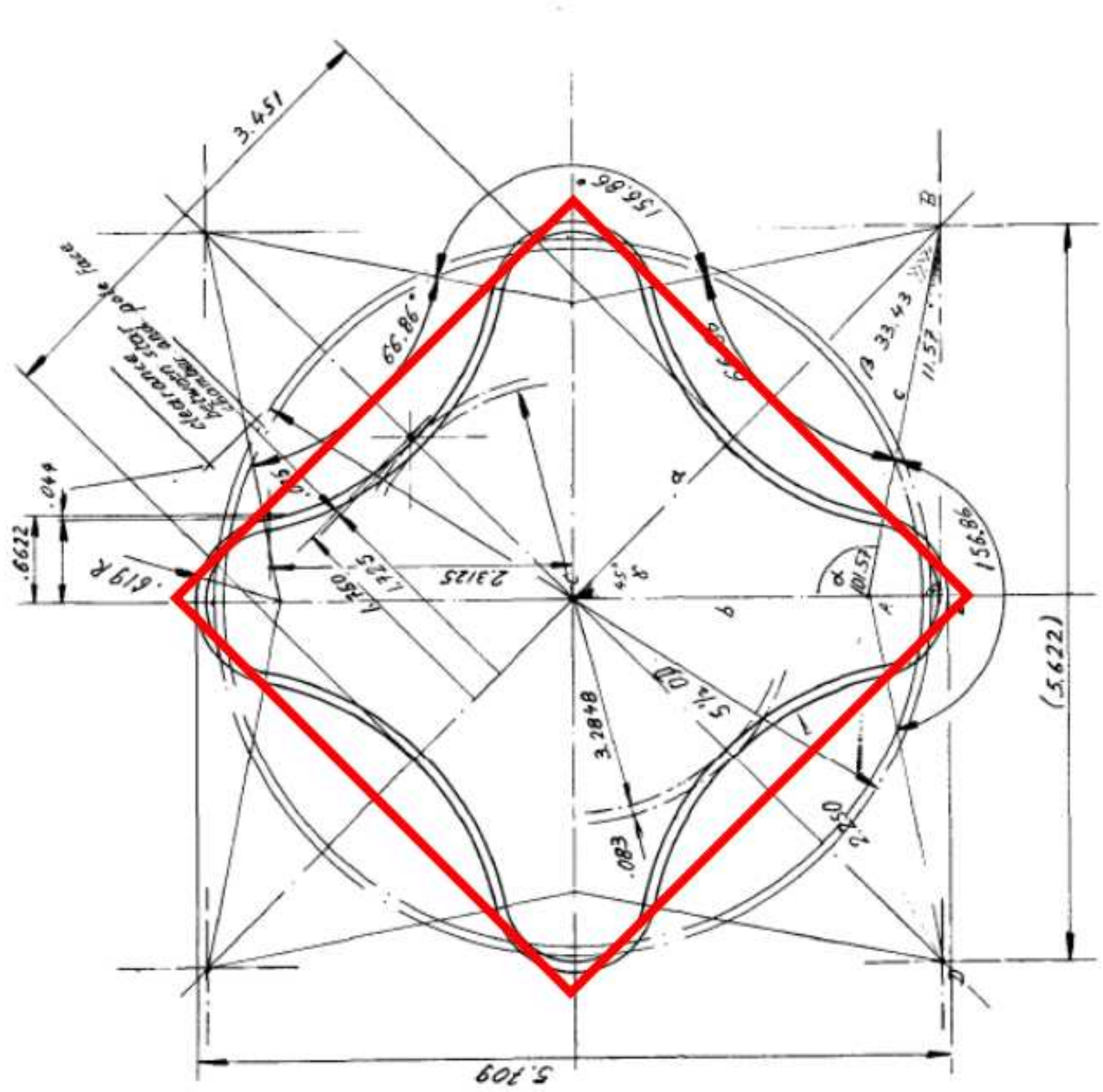

Figure 8.12: The approximation (red line) of the quadrupole star-chamber aperture used in the MAD8 model. The side length of the rectangular aperture was set to 108 mm for the quadrupoles in the Delivery Ring and 84 mm for those in the M2M3 beamline.

After tracking $\sim 12.7 \times 10^6$ protons from the end of the lithium lens, 113,930 survived to the middle of quadrupole Q202 in the Delivery Ring. The pions were also tracked from the beginning of the M2M3 line, with 90,038 surviving to the Delivery Ring. For these surviving pions, the probability of decay was calculated and a population of muons was randomly drawn from the parent pion population. The muon population was then tracked through the Delivery Ring. The separation between the beam centroids of the proton and muon populations was found to be 43.6 ns on injection to the Delivery Ring.

The populations of muons and protons were tracked separately round the Delivery Ring and their longitudinal distributions were calculated as a function of the number of turns. These longitudinal distributions are shown in Fig. 8.13 and the time interval between the two populations are summarised in Table 8.1. The rise time of the proton removal kicker magnets is $\sim$ 180 ns, so a clean separation of the two populations can be achieved on the fourth turn. These results are in broad agreement with survival studies made with `G4beamline` and analytic calculations of the separation between the two particle bunches.

Approximately 35% of all particles were lost in the first turn (mainly in the quadrupole apertures in the arcs), with the particle losses dropping below 1%, for both species on subsequent turns. The particle loss rates and a measure of the longitudinal distributions are summarised in Tables 8.2 (protons) and 8.3 (muons).

The work in this section has focused on modeling the M2M3/Delivery Ring lattice and the longitudinal distributions of the proton and muon populations. Future work will be concerned with a careful calculation of the absolute number of muons (and their distribution in phase-space) that are extracted from the Delivery Ring. It is expected that the current `MAD8` model will need improving to achieve this – in particular:

- The momentum deviation (from the reference 3.1 GeV/c) of some particles at the beginning of the M2M3 line is very large, which raises the question of how accurately the `MAD8` tracking model describes the motion of these particles. Modelling the lattice in other codes (`BMAD`, `PTC`) will enable the `MAD8` tracking to be benchmarked.

- The quadrupole apertures are described as rectangles rather than using an accurate description. Changing the aperture size in `MAD8` to intentionally under or overestimate the true aperture suggests that this approximation is reasonably good, but it may be useful to use a code that can realistically model the true aperture shape.

- The decay of pions to muons was approximated and was forced to occur at the entrance or exit of a lattice element, so the decay step length was no greater than the betatron wavelength. It is planned to implement an exact model of the decay dynamics, with the decay step length reduced to a distance much smaller than the betatron wavelength.

- It is planned to include muon spin tracking in the Delivery Ring simulations.

- All pions should be used in the decay calculation, rather than just the pions that survive at the end of the M2M3 line.

- The injection/proton separation/extraction kickers more will be modelled in more detail to better determine the number of muons delivered to the storage ring.

| | Centroid time difference | Gap size |
|---|---|---|
| Injection | 43.6 ns | None |
| 1st turn at abort | 94.6 ns | None |
| 2nd turn at abort | 169.1 ns | 36.1 ns |
| 3rd turn at abort | 243.6 ns | 108.4 ns |
| 4th turn at abort | 318.1 ns | 180.3 ns |
| 5th turn at abort | 392.6 ns | 252.2 ns |
| 6th turn at abort | 467.1 ns | 324.1 ns |

Table 8.1: Separation between protons and muons in the Delivery Ring.

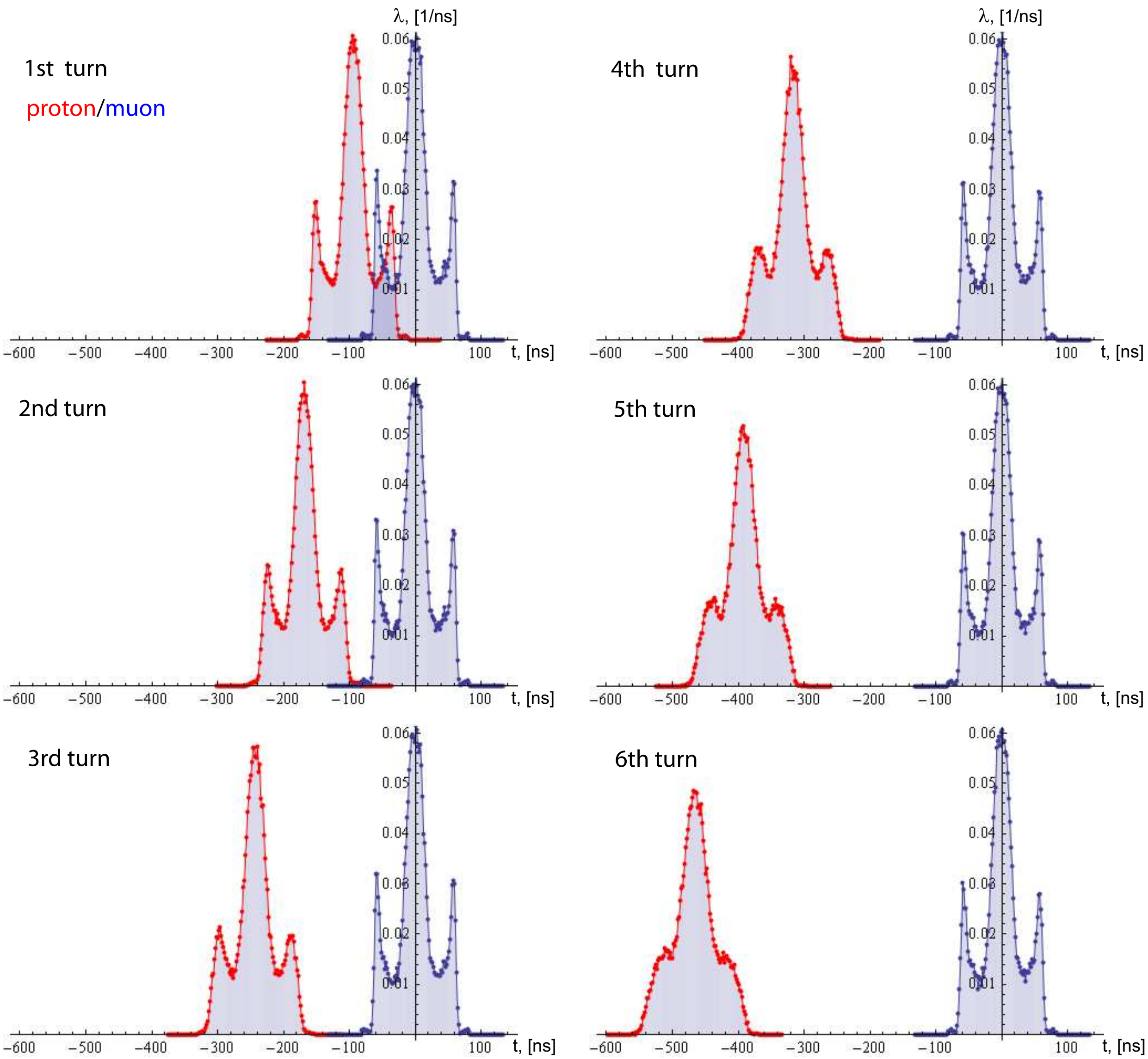

Figure 8.13: The normalised longitudinal distribution of the proton and muon populations as a function of turn number in the Delivery Ring

| | Particle loss per turn | 95% of protons within | 99% of protons within | RMS $\Delta p/p$ |
|---|---|---|---|---|
| 1st turn | 36.80 % | ± 60.8 ns | ± 66.5 ns | 0.01390 |
| 2nd turn | 0.78 % | ± 61.7 ns | ± 68.6 ns | 0.01381 |
| 3rd turn | 0.31 % | ± 63.0 ns | ± 70.6 ns | 0.01377 |
| 4th turn | 0.35 % | ± 64.2 ns | ± 72.8 ns | 0.01374 |
| 5th turn | 0.26 % | ± 65.6 ns | ± 75.0 ns | 0.01371 |
| 6th turn | 0.23 % | ± 67.2 ns | ± 77.2 ns | 0.01368 |

Table 8.2: Proton longitudinal distribution in the Delivery Ring

| | Particle loss per turn | 95% of muons within | 99% of muons within | RMS $\Delta p/p$ |
|---|---|---|---|---|
| 1st turn | 35.12 % | ± 60.1 ns | ± 64.2 ns | 0.01457 |
| 2nd turn | 0.75 % | ± 60.1 ns | ± 64.4 ns | 0.01447 |
| 3rd turn | 0.31 % | ± 60.2 ns | ± 64.6 ns | 0.01443 |
| 4th turn | 0.31 % | ± 60.3 ns | ± 65.0 ns | 0.01440 |
| 5th turn | 0.27 % | ± 60.4 ns | ± 65.4 ns | 0.01436 |
| 6th turn | 0.23 % | ± 60.5 ns | ± 65.8 ns | 0.01433 |

Table 8.3: Muon longitudinal distribution in the Delivery Ring

## 8.3 Storage Ring Simulations

### 8.3.1 Simulation of Muon Injection into the Ring

We have developed a model of the injection line and storage ring in order to simulate the injection process. The model is based on routines from the `BMAD` [5] accelerator modeling library. The simulation is a tool for evaluating dependencies on the kicker parameters, (as well as many other beam line and ring parameters). The model includes both the injection line and storage ring. Here we define the injection line as the portion of the beam line extending from the last quadrupole in the M5 line to the good field region of the storage ring dipole as shown in Fig. 8.14.

Just beyond the final M5 quadrupole the muons enter the storage ring through a hole in the backleg iron, emerging between the coils of the ring dipole and finally passing through the superconducting inflector magnet. Muons exit the downstream end of the inflector and enter the good field region of the storage ring. Our model of the storage ring, in addition to the uniform vertical B-field of the dipole, includes the electrostatic quadrupoles, collimators, and of course the kicker magnets. The magnetic field in the hole through the iron and the region from the inside of the backleg iron and through the inflector is based on maps computed with `Opera 3D`. Fig. 8.15 shows the fields along the trajectory of incoming muons.

As the vertical field is increasing in this region, from zero at the edge of the backleg iron, to 1.45T in the gap, the traversing muons experience a very significant horizontal defocusing, in addition to nonnegligible steering in the horizontal plane. Fig. 8.16 (Left) shows the muon trajectory with initial offset and angle chosen so that the beam exits the inflector tangent to the 7189 mm circle concentric with the closed orbit of the storage ring. The horizontal defocusing is evident in the $\beta$-function (Fig. 8.16-Right) propagated along the trajectory in Fig. 8.16(Left). The initial $\beta_{x/y}$ and $\alpha_{x/y}$ just upstream of the hole in the iron are chosen so that there will be a waist with $\beta_x \sim 1.5$ m and $\beta_y \sim 14$ m halfway through the 1.7 m long inflector. These $\beta$-values found to yield maximum transmission through the inflector and capture in the storage ring. Note that in order to compensate for the horizontal defocusing of the main dipole fringe field, $\beta_x$ will necessarily be large in the final quadrupole in the M5 line, and the quadrupole horizontally focusing and as near to the iron as possible.

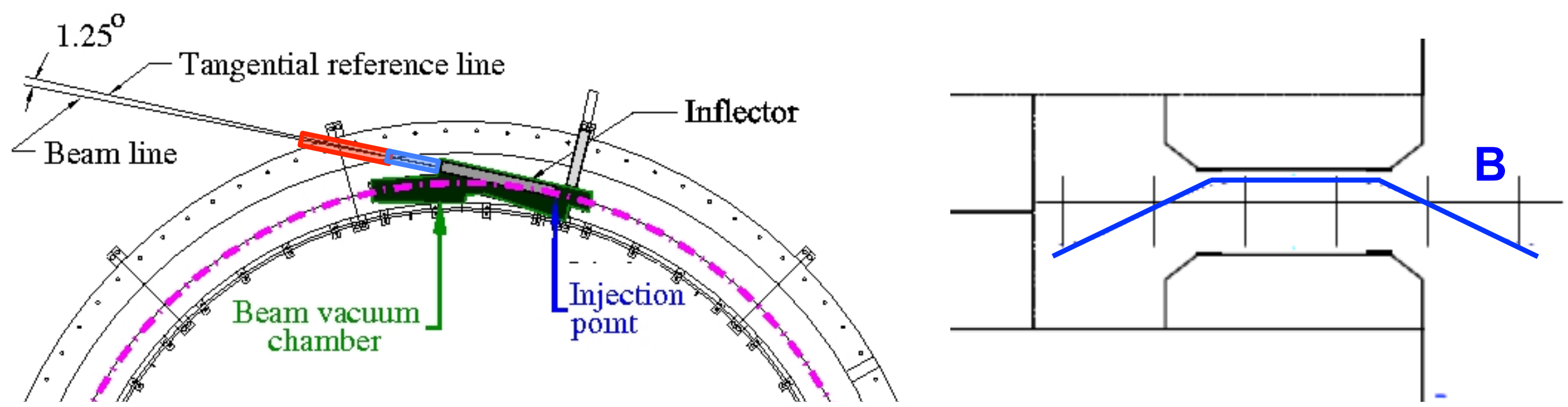

Figure 8.14: (Left) Injection line begins 30 cm upstream of the iron yoke. The hole through the iron is red. The hole through the outer cryostat is blue. Our $z$-axis is the tangential reference line. (Right) The vertical B-field increases from near zero at the inner edge of the iron to $\sim$ 1.4 T between the poles.

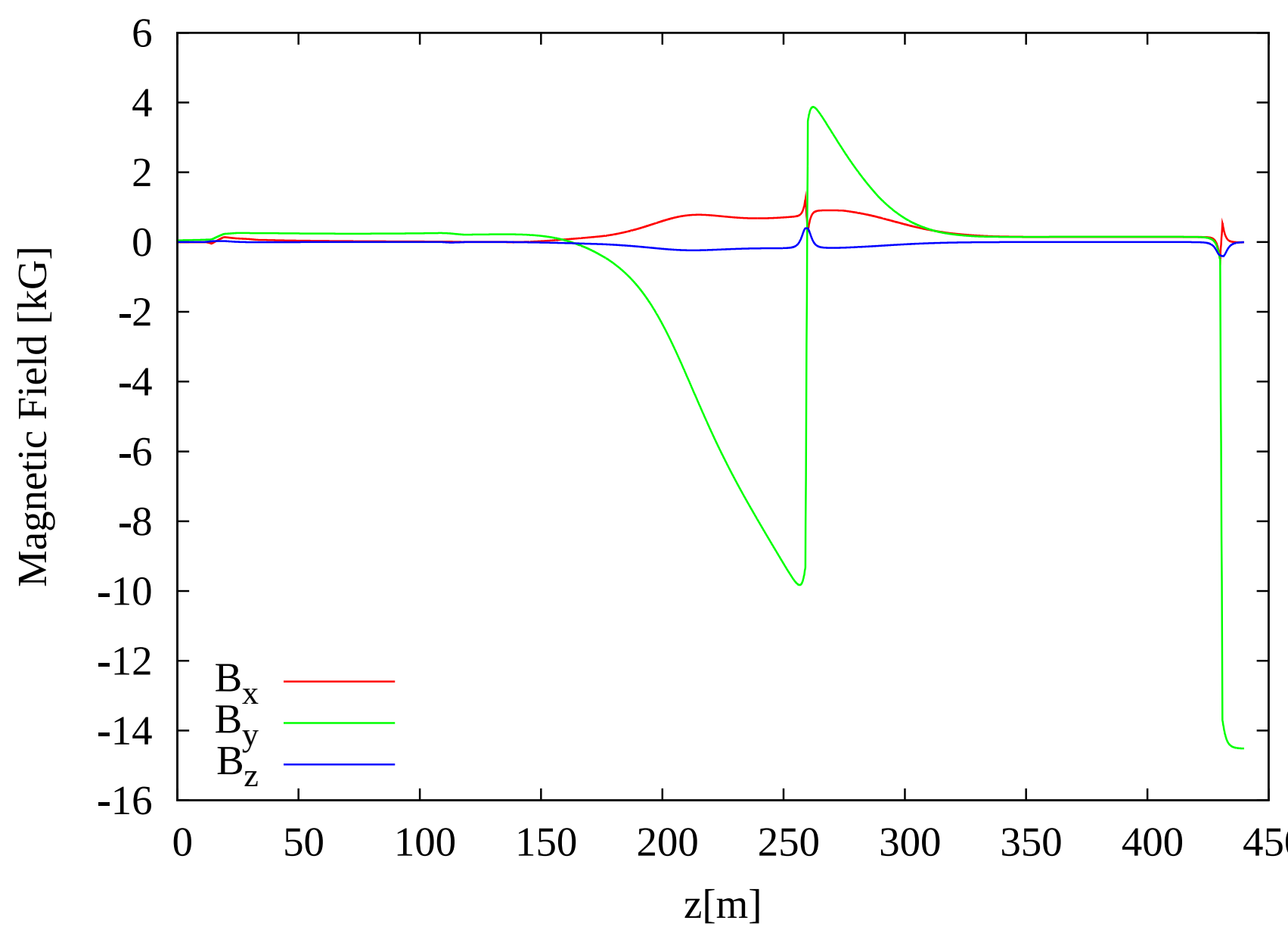

Figure 8.15: Sum of main magnet fringe field and inflector field along the injection line. The origin is 30 cm upstream of the yoke iron. The inflector exit is at 430 cm.

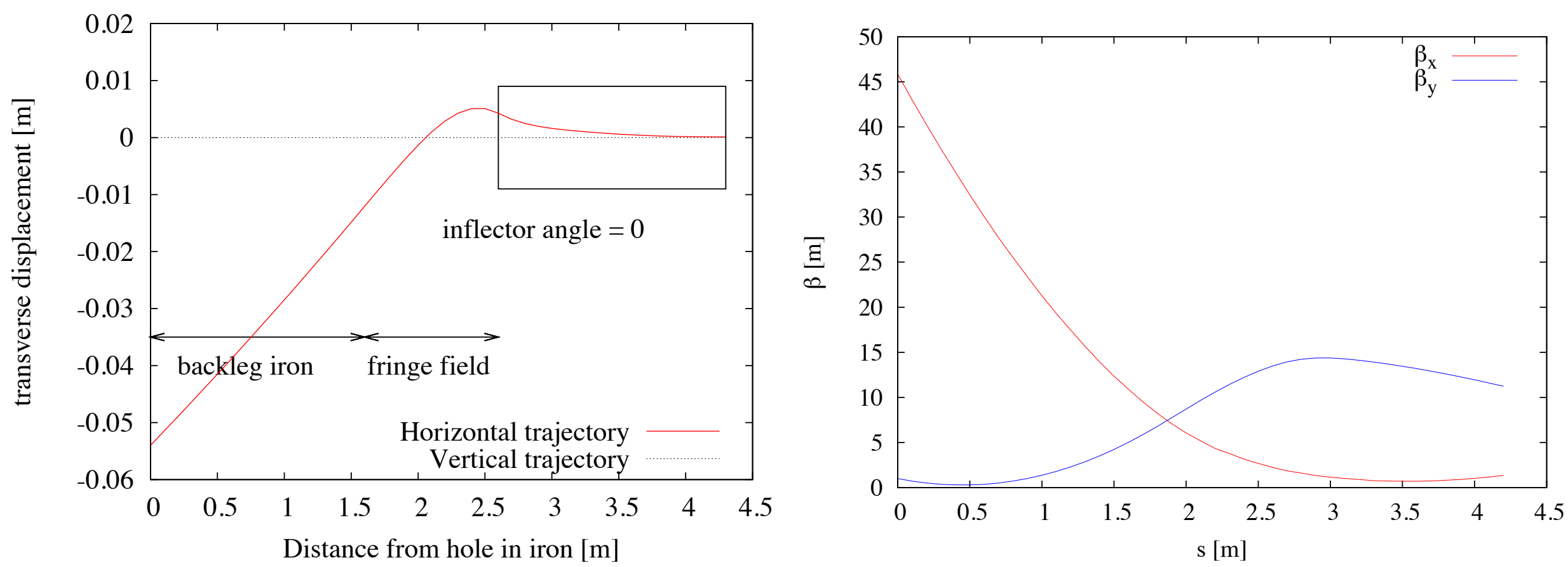

Figure 8.16: (Left) Trajectory through the injection channel. (Right) Horizontal and vertical $\beta$-functions with initial values at the upstream end of the line to yield $\alpha_x = \alpha_y \sim 0$ midway through the inflector, and $\beta_x = 1.5$ m. The limiting horizontal aperture is the inflector. The small value of $\beta_x$ optimizes transmission into the ring. The origin of the coordinate system in these plots is 30 cm upstream of the hole in the magnet iron.

The inflector coils overlap the entrance and exit of the inflector magnet and the phase space volume of the muon beam increases with multiple scattering and energy loss in the coils and cryostat windows.

For the study of capture efficiency we use the muon distribution in the 3D phase space generated by a delta function proton bunch on the conversion target (see Section 8.2.3). A temporal distribution is introduced to correspond with the profile of the incident proton bunch (see Section 7.3.1 ). The resulting distribution has an emittance $\epsilon_x \equiv \sigma_x \sigma_{px} \sim 11$ mm-mrad, and $\epsilon_y \sim 15$ mm-mrad. The energy spread is $\sim \pm 2\%$ and the length of the bunch $\tau = 120$ ns. Most of the beam fails to get through the injection line. The fraction that does survive is sensitive to

1. Horizontal angle and offset at the entrance to the injection line.
2. Effective $\beta$ and $\eta$ in the inflector. The inflector is the limiting aperture. Transmission is optimized when there is a $\beta_x$- waist midway through the inflector.

3. Kicker field; overall strength and uniformity.
4. Kicker pulse width, shape and timing.
5. Betatron tunes of storage ring.
6. Storage ring aperture as defined by collimators.
7. Inflector tilt.

We find maximum capture efficiency with the following twiss parameters midway through the inflector $\beta_x = 1.5$ m, $\beta_y = 9$ m and $\eta_x = 0$ m, and $\alpha_x = \alpha_y = \eta'_x = 0$. The incident offset and angle of the injected muons, 30 cm upstream of the hole in the magnet iron, are -5.5 cm and 24 mrad respectively. (The origin of the local coordinate system is the center of the downstream end of the inflector. The $z$-axis is the perpendicular to the radial line from the center of the storage ring to the center of the inflector aperture. The incident offset and angle are with respect to the $z$-axis, approximately 4.3 m upstream of the origin.) The optimum kicker field is found to be 200 G. For the purposes of the simulation we assume a kicker pulse with 80 ns flat top and 20 ns rise and fall time. The collimators limit the aperture of the storage ring to a circle with 45 mm radius. The betatron tunes are $Q_x = 0.9264$ and $Q_y = 0.3773$. The closed ring values $\beta_x = 7.99$ m and $\beta_y = 18.43$ m. The chromaticities, that is the energy dependence of the betatron tunes, are $dQ_x/d\delta = -0.104$ and $dQ_y/d\delta = 0.307$ where $\delta$ is the fractional energy offset.

The inflector magnet can be rotated (tilted) in the plane of the storage ring about its downstream end by a couple of mrad. This degree of freedom provides some ability to compensate for the fact the net magnetic field in the inflector is non-zero and the muon trajectory is not a straight line. We find best transmission with an inflector tilt angle of $\theta \sim 1.5$ mrad.

Dependence of some of the parameters of the stored beam on the kicker field is shown in Fig. 8.17 (Left), including number of captured muons and the coherent betatron oscillation amplitude of the distribution of stored muons. Fig. 8.17 (Right) shows the dependence on the dispersion at the inflector. If the dispersion is zero there is maximum transmission and capture. The number of muons captured depends very weakly on the dispersion, but the number of muons that are transmitted through the inflector and ultimately lost in the ring falls rapidly with dispersion. Finite dispersion in the inflector will reduce the background in the calorimeters and trackers by reducing the number of muons that will inevitably be lost in the first few turns.

Fig. 8.18 (Left) shows the fate of muons as the distribution proceeds through the fringe fields of the main dipole yoke, the inflector (with its scattering in the end coils and limiting aperture), and finally the fast kicker and around the ring. We find that muons that survive 20 turns in the storage ring are there to stay. Therefore for the purposes of study of dependence of capture efficiency and the like, the capture is defined as survival for 20 turns.

We note that a large fraction of the muon distribution delivered to the storage ring is outside the acceptance of the ring because,

- Temporal extent. The length of the muon bunch is 120 ns. For the purposes of the simulation we have assumed a kicker with 80 ns flattop, thus excluding a significant fraction of the bunch.

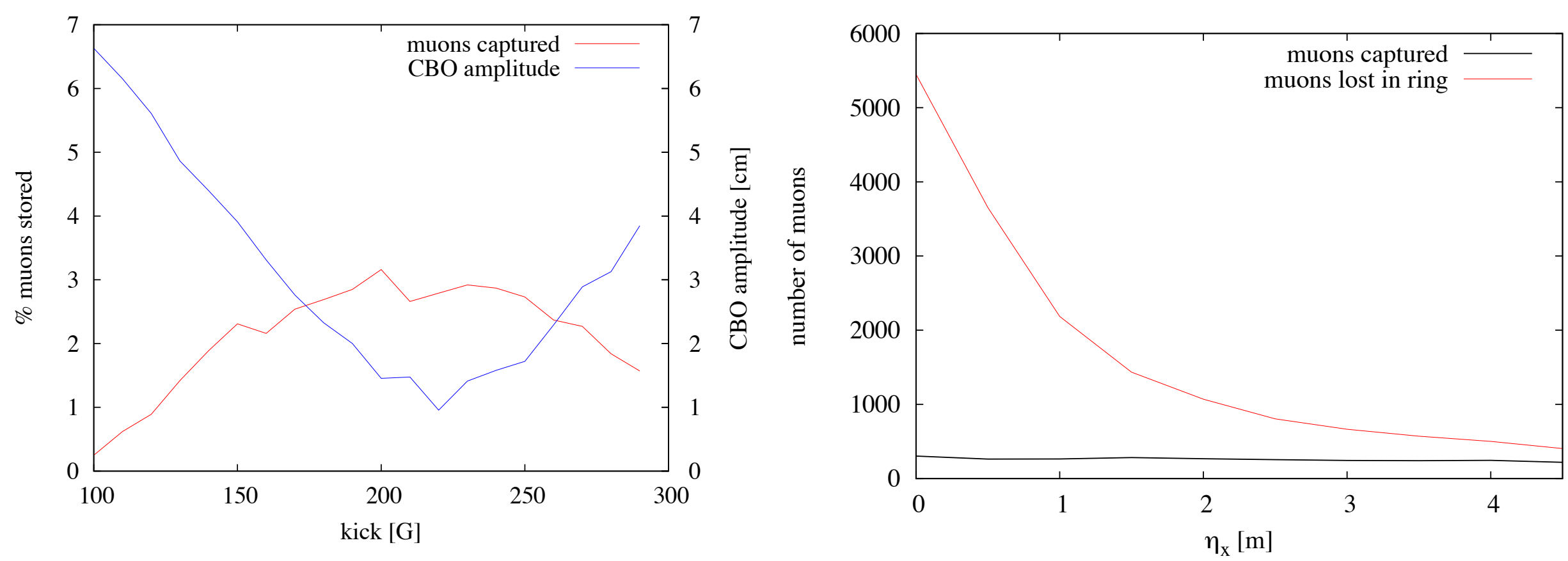

Figure 8.17: (Left) Percentage of muons stored, and amplitude of coherent betatron amplitude of the stored particles versus kicker field. (Right) Number of muons stored and number of muons lost in the ring as a function of dispersion ($\eta_x$) in the inflector. The incoming distribution of 10,000 muons is generated and propagated from target to the end of the M5 line. The temporal distribution is taken to correspond to that of the proton bunch.

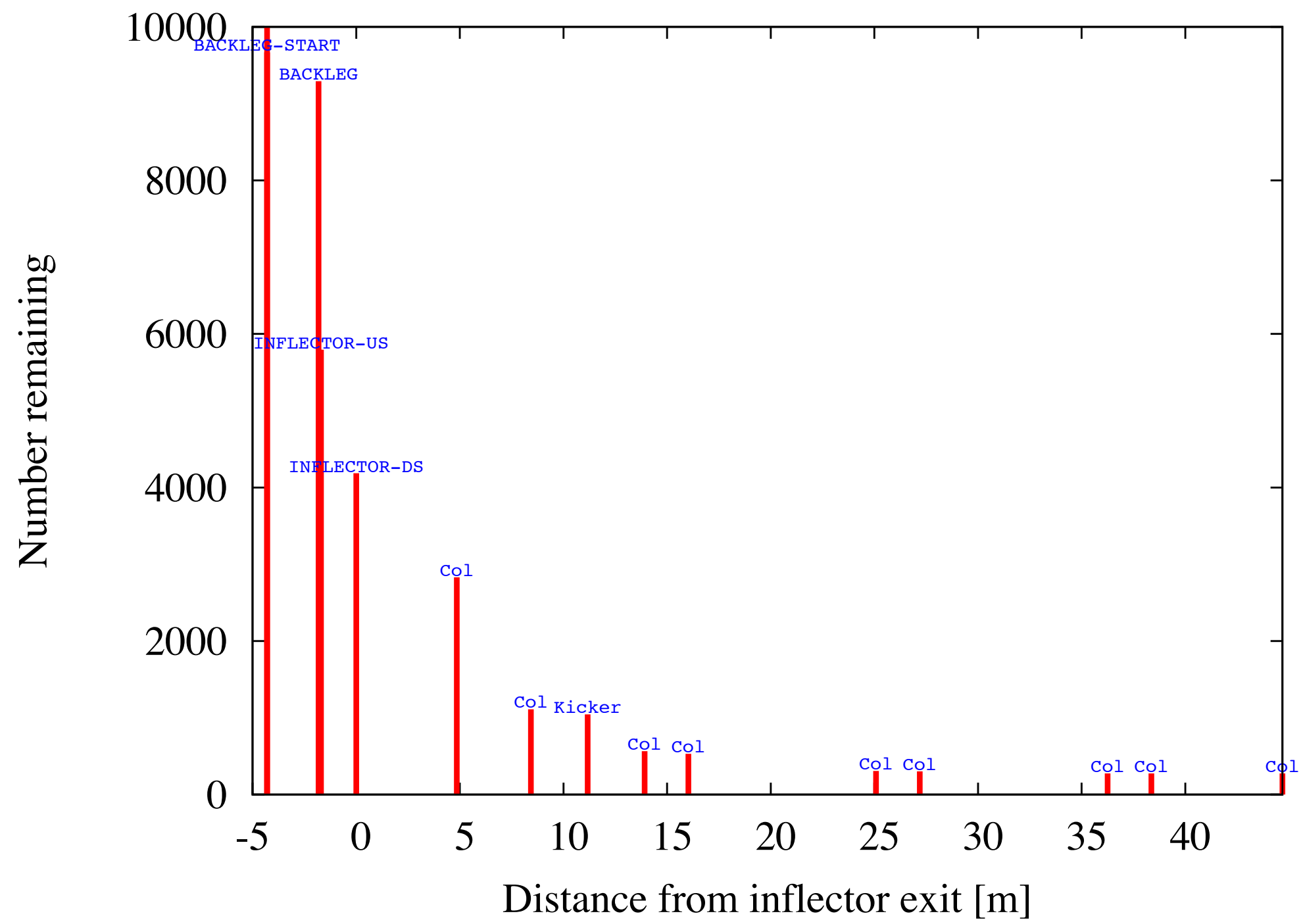

Figure 8.18: Number of muons remaining at points along the injection channel and the first turn around the ring. We find that approximately 80% of the muons that survive the first turn will be stored. The incoming distribution of 10,000 muons is generated and propagated from target to the end of the M5 line. The temporal distribution is taken to correspond to that of the proton bunch.

- Energy spread. The energy width of the distribution is 2% and the energy acceptance of the ring about 0.15%.

Fig. 8.19 shows how the distribution of muons evolves with transport through the injection line and then around the ring. The first of the three plots (Left) shows the horizontal, vertical, energy, and temporal distribution of the distribution projected to conform with the twiss parameters $\beta_x = 1.5$ m, $\beta_y = 9$ m, and $\eta_x = 0$ that are determined to yield maximum capture. Fig. 8.19 (Center) is the distribution of particles after passage through the injection channel including the scattering in the end coils. Particles outside the inflector aperture are lost. Note that when the particles exit the inflector their average horizontal offset is 77 mm. Finally, Fig. 8.19 shows the distribution of muons that survive 20 turns.

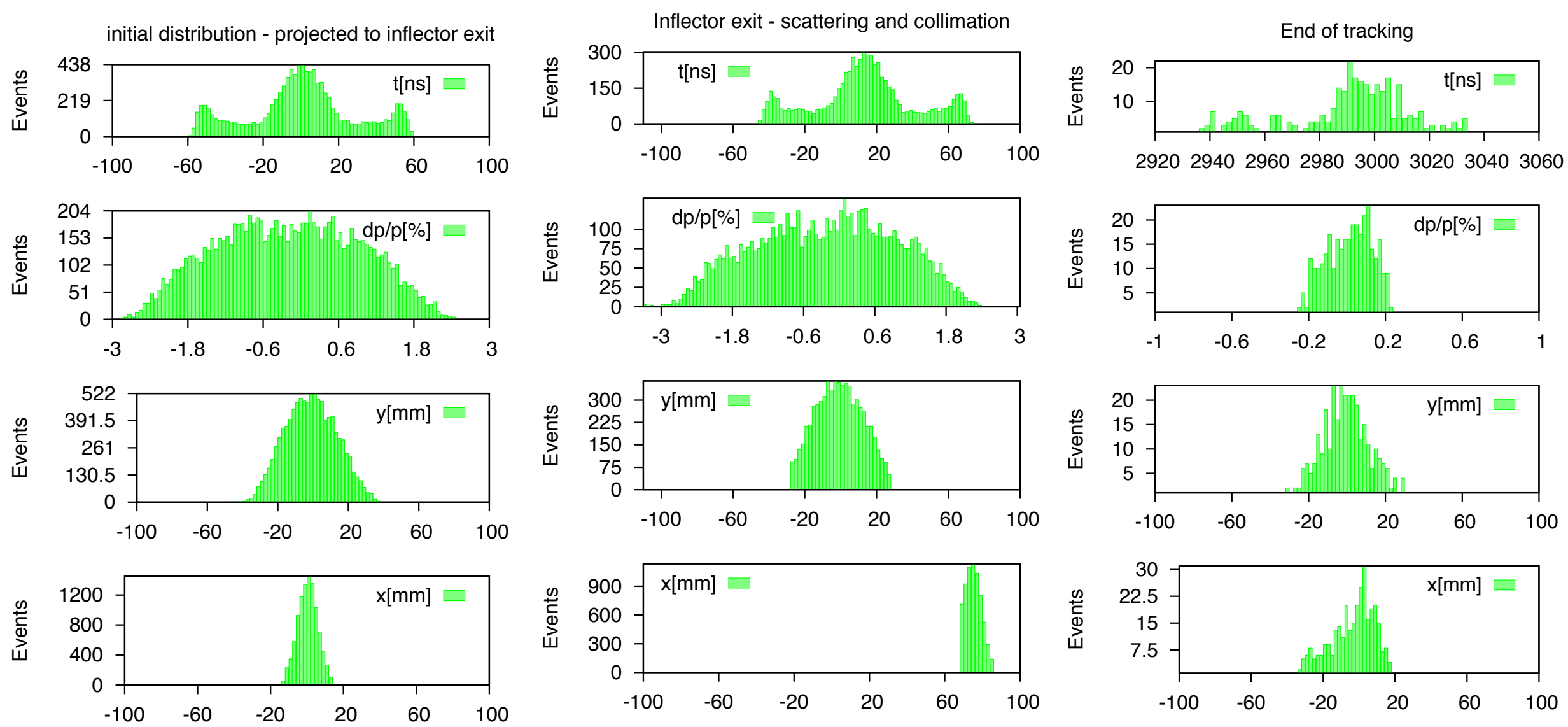

Figure 8.19: (Left) Distribution of of 10,000 muons from pions generated at the target, propagated through the decay line, delivery ring, and M5 line and projected to conform to the twiss parameters ($\beta_x = 1.5$ m, $\beta_y = 9$ m, $\eta_x = 0$) at the inflector exit. The temporal distribution is consistent with the temporal distribution of protons on the conversion target. (Center) Muons that emerge at the inflector exit after propagation through the injection line including scattering in the coil ends. Particles outside the inflector aperture are lost. (Right) Distribution of the 316 muons that survive 20 turns.

### 8.3.2 Muon Transmission and Storage Simulations

Muon transmission into the ring and the storage fraction are studied using a detailed `GEANT4` simulation of the E821 $g-2$ experiment, together with substitutions for certain elements as proposed in this TDR. The storage rate depends strongly on the amount of material the muon beam must traverse, as well as the intrinsic momentum spread ($dp/p$) of the muon beam. Common to the studies presented here is the assumption that an ideal storage ring kick will be provided to the incoming muon bunch, see Chapter 12. A baseline storage rate of 6.5% for a $40\pi$ mm-mrad muon beam with $|dp/p| < 0.5\%$ is predicted assuming the muon beam must scatter through the two closed ends of the *existing* E821 inflector and the *existing* outer Q1 plate and support. Under an ideal setting of a fully open inflector and no Q1 scattering, a storage of 22% is expected. A summary of a much larger set of studies is presented here.

### Simulation Overview

The $g-2$ muon storage region is a torus with central radius 7112 mm and a $\pm 45$ mm inner and outer radius as seen in Fig. 8.20. The $+x$ axis is directed toward the inflector where the muons enter the ring, $+z$ is aimed to the right 90° downstream of the $x$ axis, and the $y$ axis is oriented in and out of the page with the positive direction defined as outward. This coordinate system is useful to describe the ring as a whole (e.g., where is the inflector in relation to some other system), but a different beam-centric coordinate is used when describing beam dynamics. This coordinate system places the muon beam at the origin with the $+x$ direction defined as radially outward, the $+z$ direction aligned with the muon momentum direction or more commonly the azimuth direction in a cylindrical coordinate system, and the $y$ direction remains unchanged from the previous coordinates. This system is shown schematically in Fig. 8.20. The latter coordinate system will be used in this document unless otherwise specified.

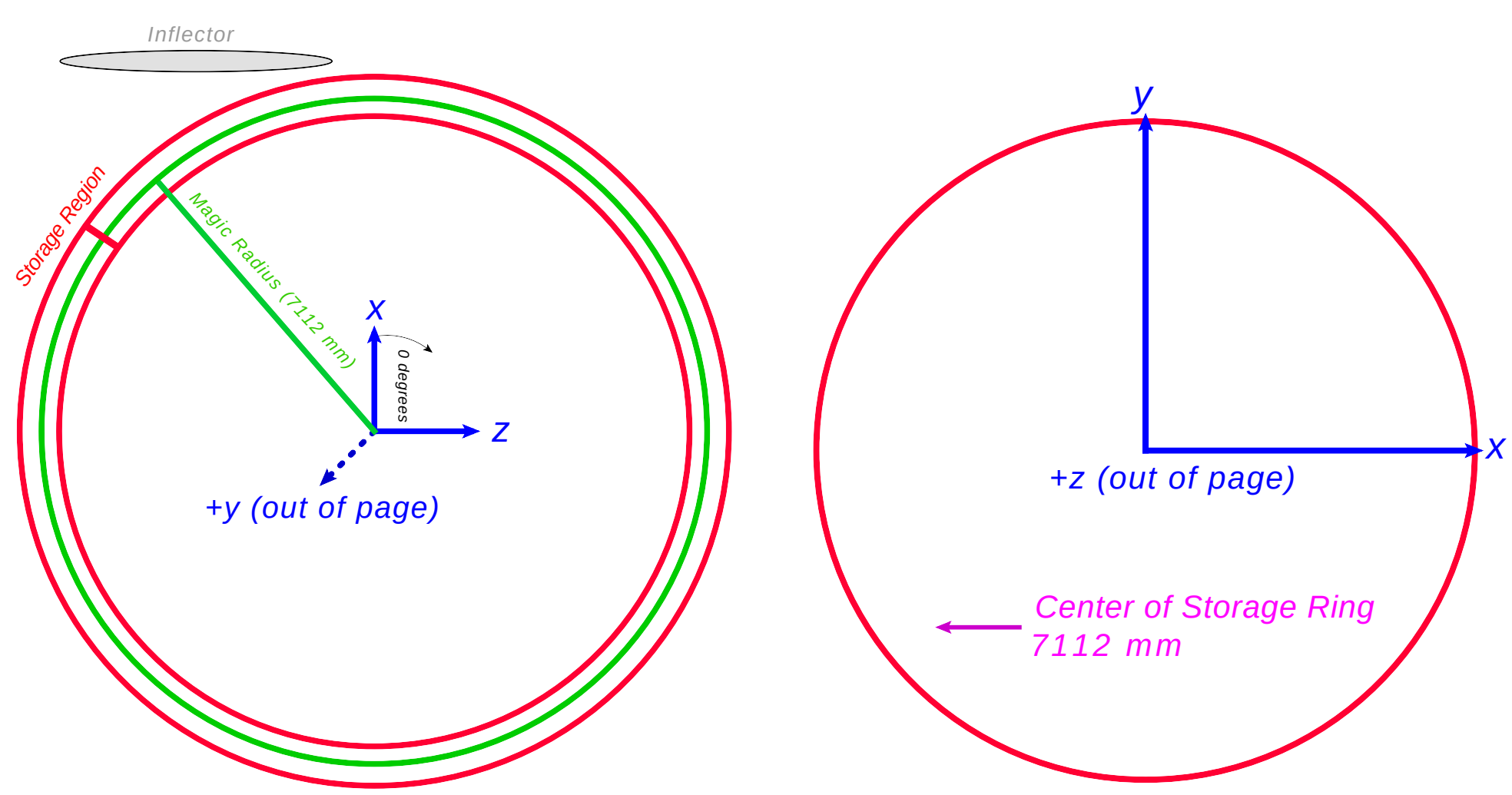

Figure 8.20: Left: Schematic of the $g-2$ muon storage region viewed from above with the associated coordinate system. The magic radius orbit is shown in green and the inner and outer boundaries of the muon storage region are shown in red. The inflector is shown in grey for orientation. Right: Schematic of the muon beam coordinate system viewed head-on inside the $g-2$ storage region

The muon beam used at the start of the simulation is created by an "inflector gun," a GEANT4 particle gun that allows the user to sample a particle phase space ($x$, $p_x/p_z$, $y$, $p_y/p_z$) given a set of beam emittances ($\epsilon_{x,y}$) and Twiss parameters ($\alpha_{x,y}$, $\beta_{x,y}$, $\gamma_{x,y}$). The beam emittance ellipse is defined such that 95% of the beam phase space is contained with the bounded region. This is represented by Eq. 8.1 below, with $x' \equiv p_x/p_z$ and $y' \equiv p_y/p_z$.

$$\gamma_x x^2 \; + 2\alpha_x x x' \; + \beta_x (x')^2 < \; \epsilon_x \tag{8.1}$$
$$\gamma_y y^2 \; + 2\alpha_y y y' \; + \beta_y (y')^2 < \; \epsilon_y \tag{8.2}$$

Only two of the three Twiss parameters are required since the third can be computed using the Courant-Snyder invariant relationship shown in Eq. 8.3. In practice, $\gamma$ is the derived quantity.

$$\beta\gamma - \alpha^2 = 1 \tag{8.3}$$

Fig. 8.21 is a schematic diagram indicating the relationship between the Twiss parameters and physical degrees of freedom $(x, x')$. It can be seen in this diagram that the maximum extent of the beam is given by $\sqrt{\varepsilon\beta}$ and the maximum $x'$ is given by $\sqrt{\varepsilon\gamma}$.

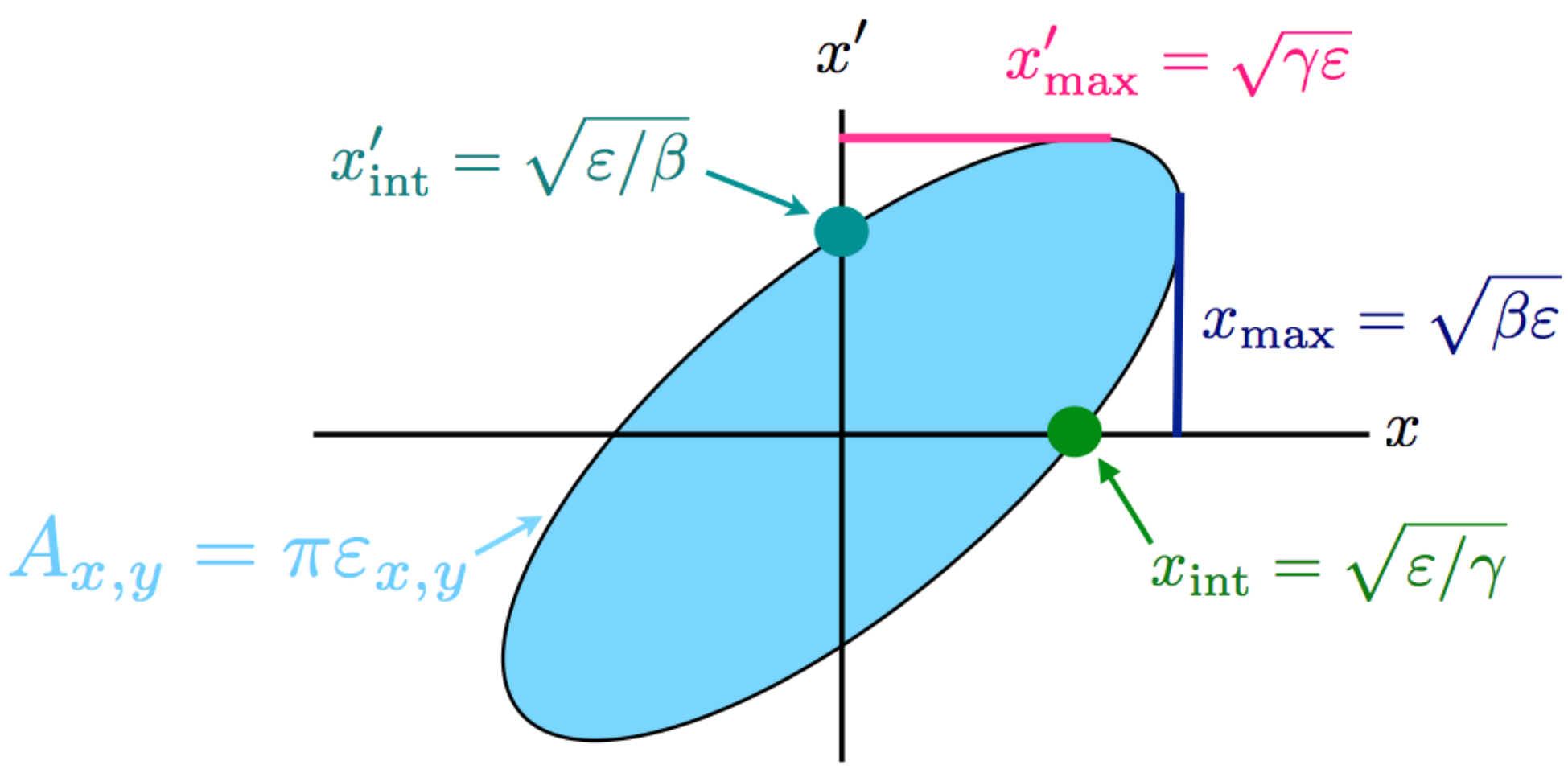

Figure 8.21: Relationship between the Twiss parameters and the physical degrees of freedom $x$ and $x'$.

The magnitude of the beam momentum is computed by sampling a Gaussian centered at the magic momentum ($p_{\mathrm{m}} \equiv m_\mu/a_\mu = 3.094$ GeV/$c$) and a width ($dp/p$). Typical values for $|dp/p|$ range between $10^{-4}$ and $10^{-1}$ for this study.

The beam is generated at a fixed $z$ position either along the inflector main axis or along the azimuthal direction within the $g-2$ storage region. Typically, the Twiss parameters are defined at the "downstream" end of the inflector one millimeter before the beam must traverse inflector coils. A transport matrix is employed to recompute the Twiss parameters when the beam originates at the "upstream" end of the inflector, which is defined as one millimeter before the beam must enter the outer inflector cryostat. A drift space approximation is used for the transport matrix. In all studies, the muon storage is computed as the ratio of muons remaining in the ring after 100 revolutions vs. the incoming flux, with muon decay turned off. The storage ring kicker magnetic field is assumed to be a square pulse applied to the first turn only, and at an ideal magnitude (typically 220 G), tuned to maximize the storage rate for the given conditions. The E821 LCR pulse was also studied for comparison. This non-ideal pulse shape (and magnitude) were not considered for E989. Variations studied and optimized in the following include the beam entrance "launch" angle into the inflector, the geometrical inflector angle with respect to a tangent to the storage ring central radius, and the momentum spread $|dp/p|$ of the incoming beam. Here we report only on the storage rate for $|dp/p| < 0.5\%$; the intrinsic momentum acceptance of the ring is much smaller.

A number of discrete variations were explored. They include:

- **Inflector Field:** *Mapped* means the computed, true inflector magnetic field is loaded and vectorially added to the main magnet fringe field. *Vanish* means the field within the inflector is identically zero (idealized).

- **Inflector Geometry:** *Closed-Closed* represents the existing E821 inflector with the magnetic coils covering both the upstream and downstream ends. It also includes the aluminum cryostat materials. *Open-Open* is a hypothetical new inflector with both upstream and downstream ends open. Intermediate cases have also been studied.

- **Quad Geometry:** *Full* is the existing E821 geometry for the outer Q1 quadrupole plate and the mechanical Macor standoffs that hold the plate in position. The trajectory of the incoming muon beam passes through these materials at a small glancing angle. The energy loss and multiple scattering have an impact on the storage fraction. *No Quads* represents the proposed E989 geometry where the Q1 outer plate is displaced radially such that no muons pass through these materials. Intermediate geometries—e.g., existing Q1 and removed standoffs—were also studied.

- **Incoming Beam Tune:** *The E821-match* parameters were determined by minimizing the beam amplitude within the inflector volume. *Ideal-match* parameters are derived by assuming ideal phase space matching into the storage ring with uniform quadrupole coverage.

The generated phase space for an $A = \epsilon\pi = 40\pi$ mm-mrad beam starting in the Downstream position with the two beam types, E821-match and Ideal-match, are shown in Fig. 8.22.

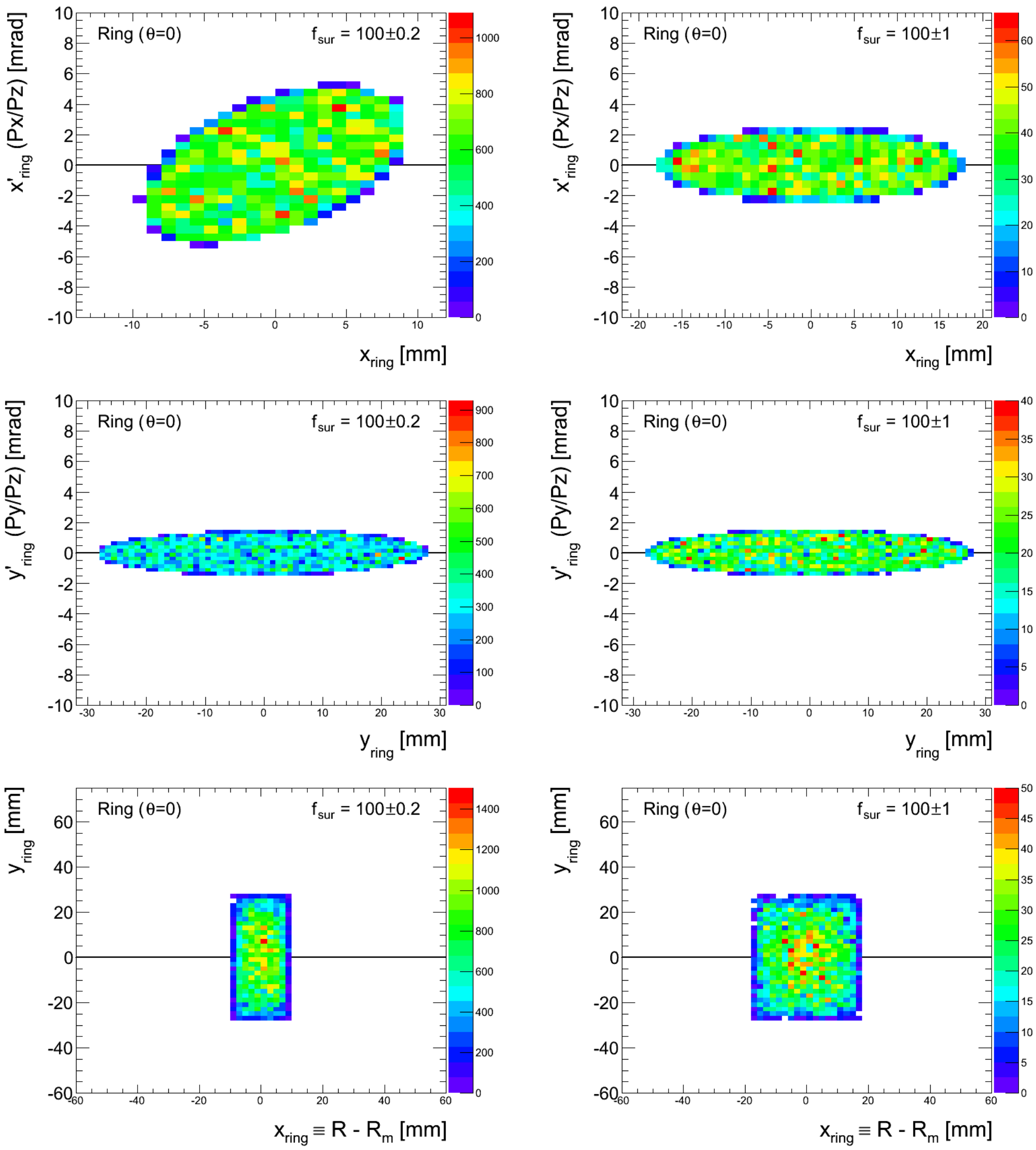

Figure 8.22: Generated Phase Space for the E821-match (left column) and Ideal-match (right column) Beams. The left column shows $x-x'$ and the right column shows $y-y'$. In all plots the origin intersects with the main inflector axis. The most noticeable difference between these two beams is the horizontal width.

**Results for Beam Starting at Upstream Entrance of the Inflector**

Here we present the main findings of the studies in which the incoming beam is launched into the upstream entrance of the inflector. It must cross both ends of the inflector (whether "open" or "closed," through the 1.7 m "D"-shaped inflector beam channel physical limitations, and through, if applicable, the outer Q1 plate/support system before entering the storage region. Fig. 8.23 is a schematic of the magic radius (red line) and the starting location of the muon momentum vector (blue arrow) and their relationship to other systems in the ring. The region indicated by "Kicker Plates" provides the idealized horizontal deflection appropriate to the given situation. There are two degrees of freedom, $\delta$ and $\ell$, as the beam enters the storage region. If the inflector is oriented such that it is tangent to the ring, then the maximum storage rate occurs when $\ell = 0$ and $\delta = -6$ mrad in the case of a fully vanishing magnetic field within the inflector. In the case of a fully mapped magnetic field, the optimal inflector angle $\delta$ is between $[-2, -4]$ mrad and the optimal launch angle $\ell$ is between $[-12, -14]$ mrad.

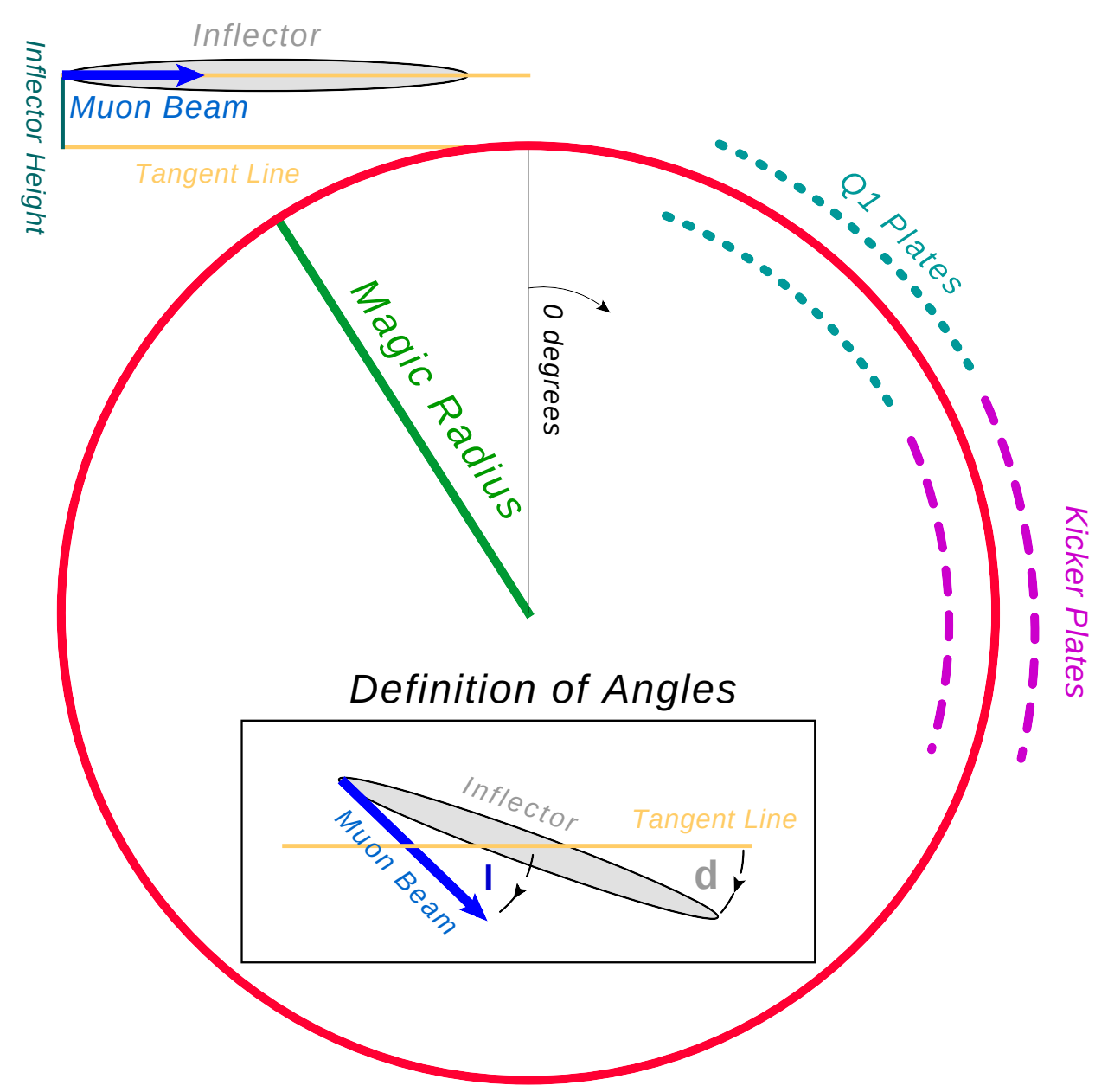

Figure 8.23: Schematic of the $g-2$ storage ring as viewed from above. The starting location of the muon momentum vector in relation to the magic radius and other detector elements is shown by the blue arrow. The beam originates at the inflector entrance.

The muon storage fraction approaches a maximum when the inflector is fully open and the outer Q1 plate and support are massless only for the E821-match beam tune. The Ideal-match beam storage rates are noticeably lower than the E821-match because more of the beam is lost while traversing the inflector beam channel. Similarly, closing the inflector will drop the storage rate by approximately a factor of two and is nearly equivalent to making the Q1 plate and support massless. Fig. 8.24 shows curves of storage fraction for several benchmark scenarios of inflector and Q1 plate/support geometries. The plots show the fraction vs. the momentum width of the incoming beam, assuming the beam flux is *common* for all scenarios; thus, the fraction is reduced as the beam width increases. It is instructive to compare performance at the common $|dp/p| < 0.5\%$ point to tie into the calculation of the muon flux described in Section 8.2.3. Table 8.4 shows the storage rate for the two inflector geometry options combined with the two Q1 geometries for the E821-match beam with $|dp/p| < 0.5\%$.

Table 8.4: Muon Storage Rates in % for 4 combinations of inflector and Q1 geometries for an E821-match muon beam with $|dp/p| < 0.5\%$. The underlined value for the open inflector and massless quads represents the maximum storage fraction obtainable for an incoming beam having $|dp/p| < 0.5\%$. The bold entry for closed inflector, massless quads represents the best estimate of the starting geometry for E989, prior to installation of a new inflector. All statistical uncertainties are well below 0.1%

| Q1 Geometry → Inflector Geometry ↓ | Massless Q1 Plate and Support | Massive Q1 Plate and Support |
|---|---|---|
| Fully Open | 22.0 | 13.0 |
| Fully Closed | **10.0** | 6.5 |

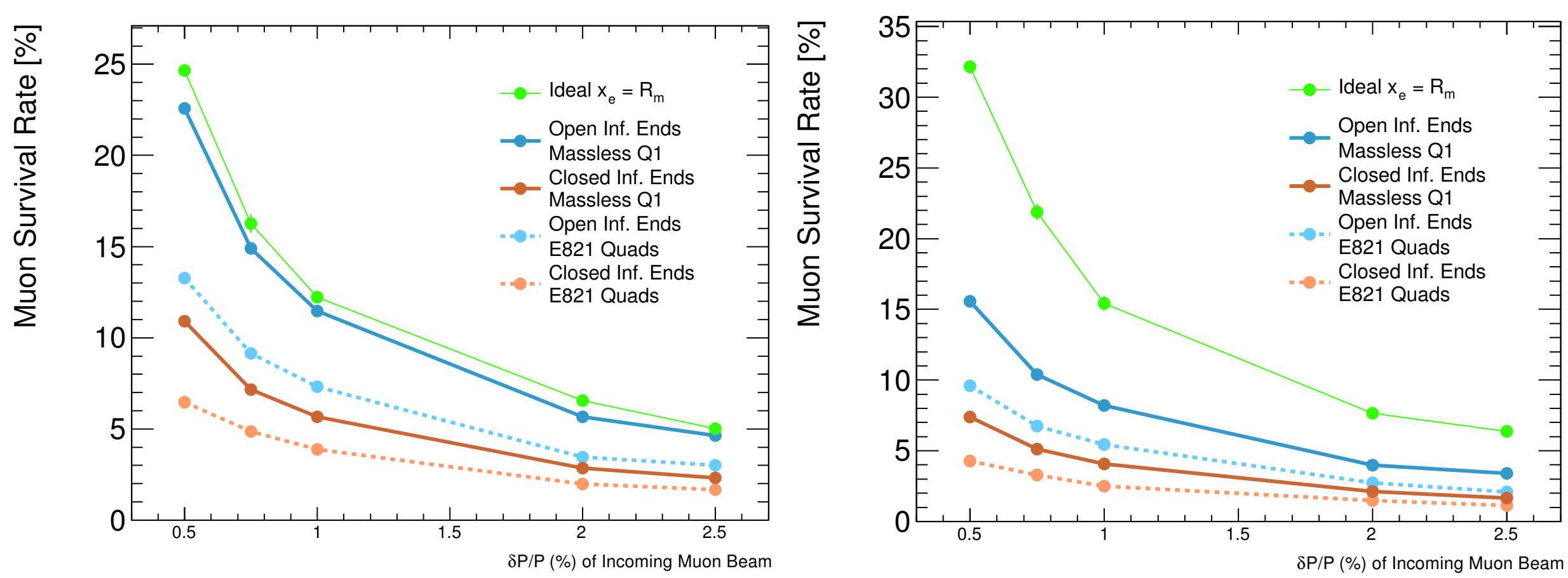

Figure 8.24: Comparison of the muon storage rates for a wide range of $dp/p$ with a variation of inflector and outer Q1 plate/support geometries assuming $A = 40\pi$ mm-mrad beam with the E821-inflector Twiss parameters (left) and the Ideal-match beam parameters (right). The "Ideal" entry represents a pencil beam launched at the magic radius inside the storage ring.

## 8.4 Summary

Using a sequence of specialized tools we developed a software model of the $(g-2)$ beamline, which allow us study various aspects of the beam transport and delivery for the E989 muon $(g-2)$ experiment at Fermilab. A complete "end-to-end" simulation has been performed to estimate the number of stored muons per proton on target, generate the expected distributions of stored muons, estimate the polarization of the muon beam and build a foundation for future systematic studies. Using the software models, we performed optimization studies of pion capture by the lithium lens, muon capture by the decay line, beam injection and transmission into the $(g-2)$ storage ring. According to our simulations, we expect to have approximately $8.1 \times 10^{-7}$ muons per proton on target at the end of the beamline (at the entrance into the $(g-2)$ inflector) with $\Delta p/p_{\mathrm{magic}} \approx 1.2\%$ RMS. Assuming baseline design parameters of the inflector, quadrupole Q1 and muon kicker, we expect to store approximately $2 \times 10^{-8}$ muons per proton on target in the $(g-2)$ storage ring. The polarization of the stored muons is about 95%.

As with any simulation, especially of a system that has not yet been built, these estimates can only yield expectations of performance and provide guidance going forward. Many effects have yet to be included into the simulations, such as random misalignment errors and powering/setting errors as set by the expected tolerances, though work will continue on these fronts and others. However there will always be unknown sources of errors, particle loss, emittance growth, etc., which can lead to a reduction in overall particle transmission. The beam line systems being developed at Fermilab for $(g-2)$ are flexible enough to provide many options for further optimization of beam performance after commissioning. The tact taken in the end-to-end simulation has been to create a modeling system that is as flexible as possible to adequately simulate the as-built system and to allow further modeling developments throughout the life of the experiment as more details of the beam delivery system are understood.

# Chapter 9

# The Muon Storage Ring Magnet

## 9.1 Introduction

As emphasized in Chapter 2, the determination of the muon anomaly $a_\mu$ requires a precise measurement of the muon spin frequency in a magnetic field $\omega_a$, and an equally precise measurement of the average magnetic field felt by the ensemble of precessing muons, $\langle B \rangle$. We repeat the spin equation given in Eq. 3.10, since it is central to the design of the storage-ring magnet.

$$\vec{\omega}_a = -\frac{Qe}{m}\left[a_\mu \vec{B} - \left(a_\mu - \left(\frac{mc}{p}\right)^2\right)\frac{\vec{\beta}\times\vec{E}}{c}\right]. \tag{9.1}$$

As explained in Chapter 2, the need for vertical focusing and exquisite precision on $\langle B \rangle$ requires that: either the muon trajectories be understood at the tens of parts per billion level, and the magnetic field everywhere be known to the same precision; or the field be as uniform as possible and well-measured, along with "reasonable knowledge" of the muon trajectories. This latter solution was first employed at CERN [1] and significantly improved by E821 at Brookhaven [2]. The uniformity goal at BNL was $\pm 1$ ppm when averaged over azimuth, with local variations limited to $\leq 100$ ppm.

Fermilab E989 will use the storage-ring magnet designed and built for Brookhaven E821, with additional shimming to further decrease the local variations in the magnetic field. This requires the relocation of the ring from BNL to Fermilab, which is described in detail in the following chapter. While the magnet steel comes apart and can be moved by conventional trucks, the 14.5 m diameter superconducting coils will need to be moved as a package, on a custom designed fixture that can be pulled by a truck to travel by road, and put on a barge to travel by sea, and then again by road to get it to the Fermilab site.

The storage ring is built as one continuous superferric magnet, an iron magnet excited by superconducting coils. A cross-section of the magnet is shown in Fig. 9.1. The magnet is C-shaped as dictated by the experiment requirement that decay electrons be observed inside the ring. The field, and hence its homogeneity and stability, are determined dominantly by the geometry, characteristics, and construction tolerances of the iron. Although both copper and superconducting coils were considered, the use of superconducting coils offered the following advantages: thermal stability once cold; relatively low power requirements; low voltage, and hence use of a low-voltage power supply; high $L/R$ time constant value and hence low ripple currents; and thermal independence of the coils and the iron. The main disadvantage was that the coils would have a much larger diameter and smaller height than any previously built superconducting magnet. However, since the E821 magnet team could not identify any fundamental problems other than sheer size, they decided to build superconducting coils.

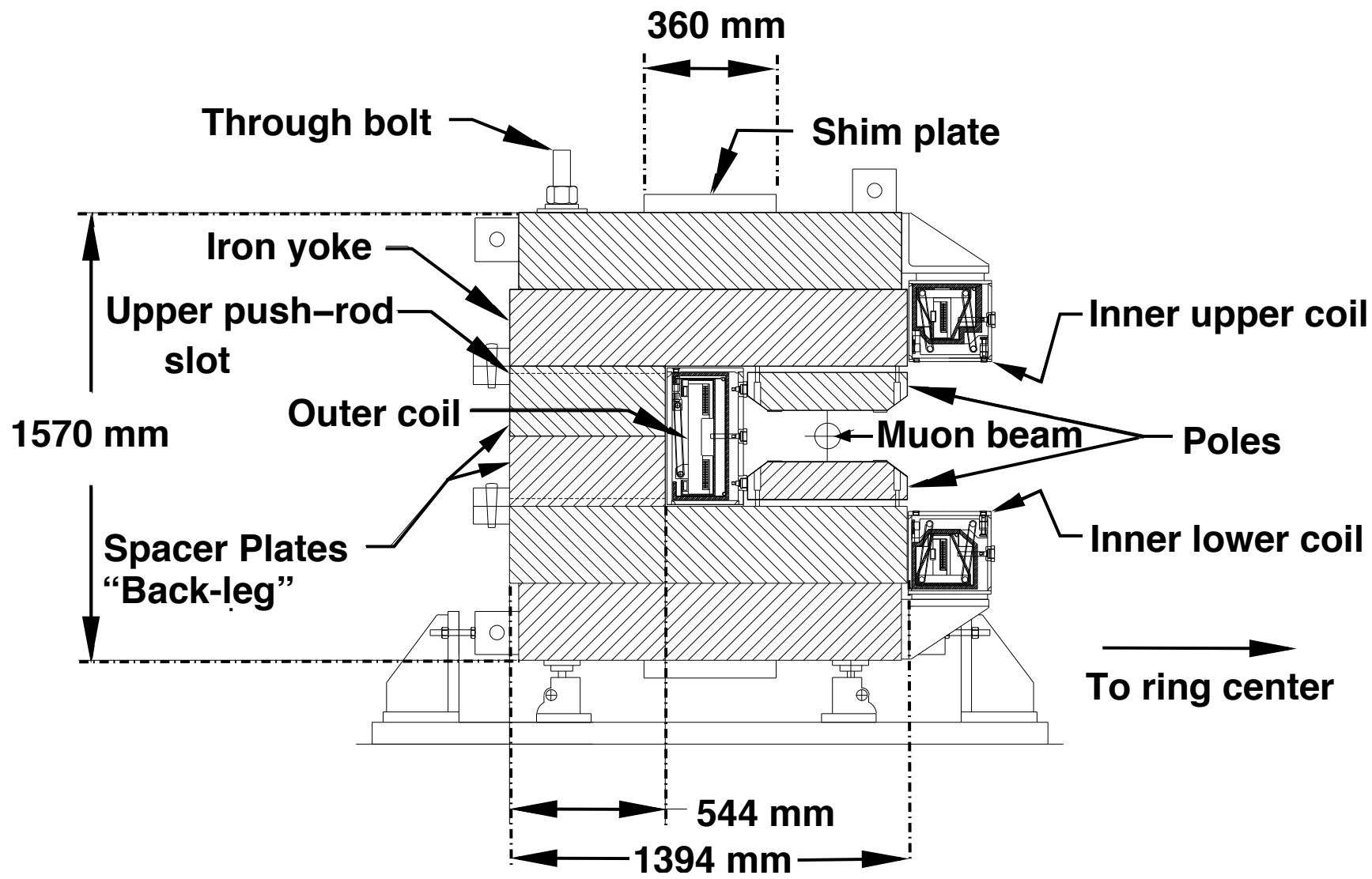

Figure 9.1: Cross section of the E821 storage-ring magnet. The yoke is made up of 12 azimuthal sections, each of which consists of six layers of high quality magnet steel provided by Lukins Steel Corporation. The pole pieces were provided by Nippon Steel Corporation.

To obtain the required precision in such a large diameter magnet with an economical design is an enormous challenge. The magnet had to be a mechanical assembly from subpieces because of its size. With practical tolerances on these pieces, variations up to several thousand ppm in the magnetic field could be expected from the assembled magnet. To improve this result by the required two to three orders of magnitude required a shimming kit.

Because of the dominant cost of the yoke iron, it was an economic necessity to minimize the total flux and the yoke cross-section. This led to a narrow pole, which in turn conflicts with producing an ultra-uniform field over the 9 cm good field aperture containing the muon beam.

A simple tapered pole shape was chosen which minimized variations in the iron permeability and field throughout the pole. The ratio of pole tip width to gap aperture is only 2/1. This results in a large dependence of the field shape with the field value B. However, since the storage ring is to be used at only one field, $B = 1.45$ T, this is acceptable. Because of dimensional and material property tolerance variation, the compact pole piece increases the necessity for a simple method of shimming.

Experience with computer codes, in particular with `POISSON` [4], had demonstrated that, with careful use, agreement with experiment could be expected at a level of $10^{-4}$ accuracy. `POISSON` is a two-dimensional (2D) or cylindrically symmetric code, appropriate for the essen-tially continuous ring magnet chosen for the $(g-2)$ experiment. Computational limitations, finite boundary conditions, and material property variations are all possible limitations on the accuracy of paper calculations of the design.

We will briefly discuss the design features that are relevant to E989, especially to moving the ring, but not repeat all the details given in Danby et al. [3], and in the E821 Design Report [5]. The parameters of the magnet are given in Table 9.1.

Table 9.1: Magnet parameters

| | |
|---|---|
| Design magnetic field | 1.451 T |
| Design current | 5200 A |
| Equilibrium orbit radius | 7112 mm |
| Muon storage region diameter | 90 mm |
| Inner coil radius - cold | 6677 mm |
| Inner coil radius - warm | 6705 mm |
| Outer coil radius - cold | 7512 mm |
| Outer coil radius - warm | 7543 mm |
| Number of turns | 48 |
| Cold mass | 6.2 metric tons |
| Magnet self inductance | 0.48 H |
| Stored energy | 6.1 MJ |
| Helium-cooled lead resistance | 6 $\mu\Omega$ |
| Warm lead resistance | 0.1 m$\Omega$ |
| Yoke height | 157 cm |
| Yoke width | 139 cm |
| Pole width | 56 cm |
| Iron mass | 682 metric tons |
| Nominal gap between poles | 18 cm |

## 9.2 Yoke Steel

E989 will reuse the yoke steel manufactured for the E821 experiment. The yoke pieces have been surveyed and disassembled at Brookhaven and have been shipped to Fermilab. The design and construction of the magnet has been documented and published in NIM [3] as well as the final report in Phys. Rev. D [2]. We summarize the main design features and issues here, with a discussion of potential improvements in Section 9.4.

Ideally, the $g-2$ magnet would be azimuthally symmetric. To ease the fabrication and assembly processes, the magnet was built with twelve 30° sectors. Each sector consists of an upper and lower yoke separated by a spacer plate as shown in Fig. 9.1. Due to the large thickness of the yoke (54 cm), the individual plates were fabricated separately and welded together after machining. The spacer plate is also split at the midplane to allow for the installation of beam pipes and other services after the lower section is in place but prior to the installation of the upper yoke. The yoke plates and spacers in each sector are all fastened together with eight long high-strength steel bolts that cover the full 1.57 m tall yoke. The total sector mass is $\approx$ 57,000 kg, which results in a total magnet mass of $\approx$ 680,000 kg.

Significant quality control efforts were taken during the manufacturing process to ensure that the magnet had sufficiently uniform permeability and the appropriate geometric shape. Both of these parameters have strong effects on the uniformity of the magnetic field in the storage region.

High-quality plates were manufactured by hot-rolling AISI 1006 iron to minimize magnetic voids in the material. These plates were manufactured with $< 0.08\%$ of carbon and other impurities. The finished plates were inspected ultrasonically to detect voids and inclusions, and analyzed chemically to understand the composition.

Although the yoke steel is partially magnetically isolated from the storage region by an air gap near the pole pieces, strict machining specifications are required to minimize non-uniformities in the storage region field. The surfaces of the yoke plates closest to the storage region were milled flat within 130 $\mu$m and 1.6 $\mu$m finish. Similarly, the spacer plate surfaces were milled flat within $\pm$130 $\mu$m, with a thickness accurate to $\pm$130 $\mu$m. These surfaces are parallel within 180 $\mu$m. The radial tolerance for each yoke plate and the spacer plates was $\pm$130 $\mu$m. When constructed, the vertical yoke gap had an rms deviation of $\pm$90 $\mu$m, or 500 ppm of the total air gap of 20 cm, and a full-width spread of $\pm$200 $\mu$m.

Each of the 12 sectors need to be connected smoothly to achieve azimuthal symmetry. To achieve azimuthal continuity, each sector end has four radial projections for bolts to fasten adjacent sector ends to each other. When the sectors are fitted to each other, shimmed, and the bolts tightened, relative motion of adjacent sectors is minimized. The average azimuthal gap between sectors was 0.8 mm, with an rms deviation of $\pm$0.2 mm.

When we begin to reconstruct the storage ring, we will clean the yoke steel and remove any rust that has developed. It will be important to do this in a non-destructive manner that maintains the high-level of precision achieved during manufacturing.

## 9.3 Poles and Wedges

E989 will reuse the pole pieces and wedge shims that were manufactured for the E821 experiment. The pole pieces and wedges have been removed from the storage ring at Brookhaven and have already been shipped to Fermilab where they are awaiting reassembly.

### 9.3.1 Poles

More stringent quality requirements are placed on the machining of the pole pieces than the yoke steel. The air gap between the yoke and pole pieces decouples the field region from non-uniformities in the yoke. Thus, irregularities in the pole pieces dominate the field aberrations. Ultra-pure continuous vacuum cast steel with $< 0.004\%$ carbon impurities is used for the pole pieces. The fabrication process greatly minimizes impurities such as ferritic inclusions or air bubbles.

A dimensioned view of the pole pieces is shown in Figure 9.2. Each 30° yoke sector contains three pole pieces (azimuthally). The pole pieces are 56 cm wide (radially), with a tolerance of 50 $\mu$m. The thickness (vertical) of each piece is 13.3 cm with a tolerance of 40 $\mu$m. The pole faces which define the storage ring gap have tight machining tolerances. Each face has a flatness tolerance of 25 $\mu$m, leading to upper and lower faces being parallel within a 50 $\mu$m tolerance. The surface finish is 0.8 $\mu$m. These machining tolerances are so stringent due to the large quadrupole moment introduced by non-parallel surfaces. An `OPERA-2D` simulation of the magnet has determined that a 100 $\mu$m tilt of the pole piece over its width corresponds to $>$ 100 ppm. This is in good agreement with the 2D `POISSON` calculations performed for the E821 simulations.

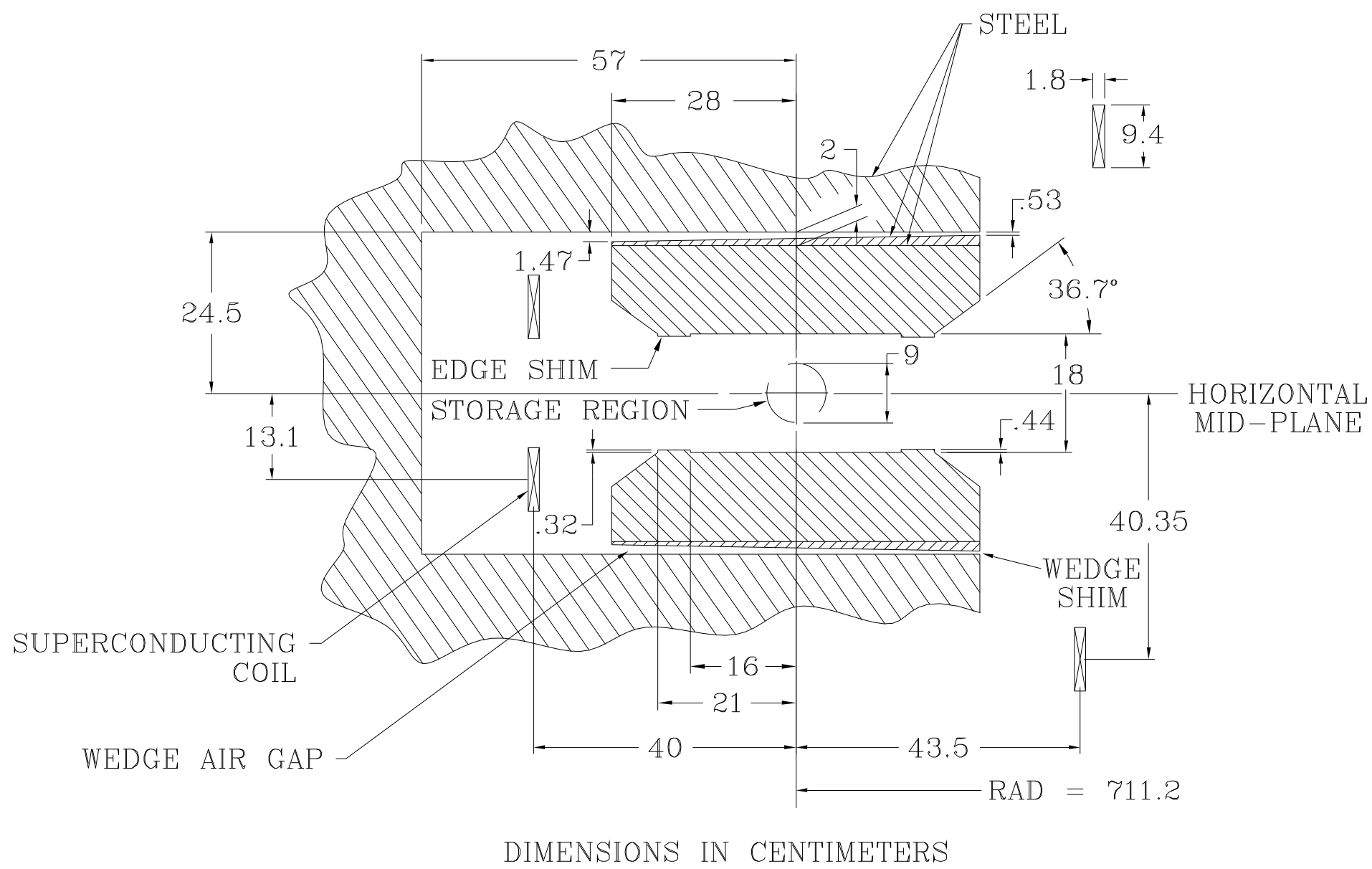

Figure 9.2: Cross section view of the magnet gap region.

Each yoke sector contains three pole pieces. Vertically, the pole pieces are mounted to the yoke plates with steel bolts. The outer two pieces are each machined radially, parallel to the yoke sector. The middle pole piece in each sector is interlocking, with an angle of 7° with respect to the radial direction. The pole pieces were isolated azimuthally by 80 $\mu$m kapton shims, which served two purposes. First, the kapton shims helped position the pole pieces at the correct azimuth. Second, the kapton electrically isolated the poles from each other, allowing small reproducible eddy currents. If the poles were all in contact with each other, large eddy currents would develop around the entire circumference of the ring during field ramping and energy extraction.

The pole gap distance was measured using a capacitive sensor, as described in Section 15.8.2. The gap was 18 cm with an rms variation of $\pm$23 $\mu$m, and a full range of 130 $\mu$m. As the magnet is powered, the induced torque causes the open side of the C-magnet (inner radius) to close slightly. Thus, during the installation, the poles were aligned with an opening angle of 80 $\mu$rad. A precise bubble level was used to achieve 50 $\mu$m precision over the length of the pole piece. Pole realignment will be part of the shimming process described in Section 15.8.2.

### 9.3.2 Wedges

The gaps between the yoke and poles isolate the yoke steel from the poles and provide a region where shims can be inserted to fine-tune the magnetic field. Steel wedges that are sloped radially (see Fig 9.2) are inserted to compensate for the intrinsic quadrupole moment produced by the C-magnet. There are 72 wedges in each 30° yoke sector. The induced quadrupole term depends on the slope of the wedge, which was calculated to be 1.1 cm over the 53 cm width for E821. This wedge angle was verified empirically, and no additional grinding was needed. The radial position of the wedges can be adjusted to change the total material in the gap, affecting only the dipole moment (see Section 15.8.3).

During the ramping of the main coil current, the thick end of the wedge attracts more field lines, leading to a torque. To prevent the wedges from deflecting vertically, an aluminum "anti-wedge" is used to fill the air gap between the wedge and the pole piece.

E989 will reuse the wedge-spacer combination as is. Fine tuning of the quadrupole moment can be achieved with active current shims, and pole bumps, as discussed in Section 15.8.3.

## 9.4 Thermal Effects

Temperature variations in the experimental hall are expected to be controlled within ±1 °C during the course of data taking. This will change the shape of the magnet, which will in turn change the magnetic field. We produced thermal simulations with `ANSYS` to quantify the geometric distortions, which are then input into the `OPERA-2D` model of the storage ring.

E821 used 3.5" of fiber glass insulation around the bulk of the yoke and 3/8" foam rubber insulation near the poles pieces, as shown in Figure 9.3 (a). Reasonable thermal film coefficients in the range of 5-25 W/m$^2C$ were used at the surfaces of the magnet. Thermal oscillations based on day-night temperature cycles are imposed on the $g-2$ magnet system and modeled with `ANSYS`. The air temperature is assumed to be spatially uniform throughout the hall. The model indicates that this will lead to thermal fluctuations at the yoke and pole pieces of a few tenths of a degree, as shown in Figure 9.3 (b). The pole pieces are constrained mechanically to prevent sliding, thus, in response to the thermal variations, they bend.

Figure 9.4 shows the response of the magnet under the 1 °C hall fluctuations. The contours show the maximum extent of the deflection for both radial shifts (Figure 9.4 (a)) and vertical shifts (Figure 9.4 (b)). The deflections are on the order of 1 $\mu$m per degree C change in the hall temperature. The parallelism of the pole faces is known to affect the higher-order multipole components of the magnetic field. Figure 9.5 plots the relative change in the pole gap as a function of radius for the thermal changes described above.

Two different thermal contact resistances of the pole foam rubber insulation were modeled. In both cases, the gap distortion leads to a change of about 1 $\mu$m. The pole gap distortions were input into the OPERA-2D magnetic field simulation. Distortions on the order of a few tenths of a ppm were observed in the sextupole and octupole moment with a change of 1 $\mu$m in the pole gap. Because the monitoring of the higher order multipole moments is done primarily with the trolley runs, extrapolation of the field map from the fixed probes during the main data collection will rely on stable magnet geometry.

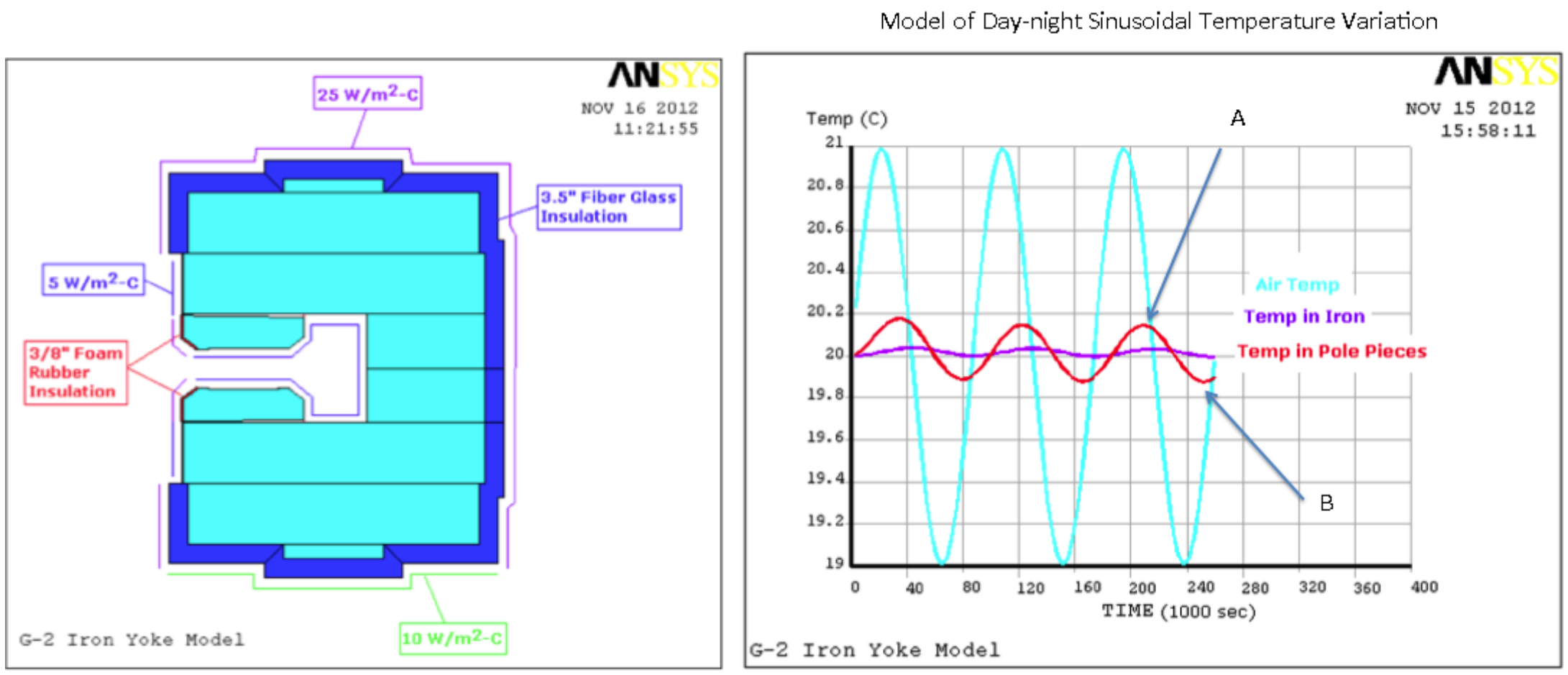

(a) (b)

Figure 9.3: (a) An ANSYS model of the $g-2$ storage ring includes the thermal insulation used in E821. (b) Thermal oscillations based on day-night temperature cycles are imposed on the $g-2$ magnet system assuming a $\pm 1\,^{\circ}$C. The temperature variations of the yoke (purple) and pole (red) are overlaid.

The ANSYS and OPERA tools nicely complement each other and allow us to understand the effects of magnet deflections in E989. We plan to repeat these studies with varied insulation thickness and with additional insulation around the inner superconducting coils. With a high quality temperature control system stabilizing the experimental hall and better thermal isolation of the steel, E989 will have significantly smaller time-dependent magnet distortions than E821. This will lead to more stable multipole components.

## 9.5 The Superconducting Coil System

### 9.5.1 Overview

The coil design was based on the TOPAZ solenoid at KEK [6]. TOPAZ conductor was used, with pure aluminum stabilizer and niobium-titanium superconductor in a copper matrix. Conductor characteristics are given in Table 9.5.1. At full field the critical temperature of the outer coil is 6.0 K. The magnet typically operates at 5.0 K. This represents 76% of the superconductor limit. Each coil block is effectively a very short solenoid with 24 turns, and one layer. The coils are wound from the inside of the ring so that, when powered, the coils push out radially against a massive aluminum mandrel. Cooling is indirect with helium pipes attached to the mandrels. The coil turns, coil stack and insulation are epoxied together, forming a monolithic block. The coils hang from the cryostat with low heat load straps, and the shrinkage and expansion of the coils is taken by the straps. The coils are located using radial stops on the inner radius. For the outer coil the stops transfer the force from the coil to the cryostat box, and push rods from the iron yoke transfer the force from the box to the iron (see Fig. 9.7). For the inner coils, pins replace the pushrods.

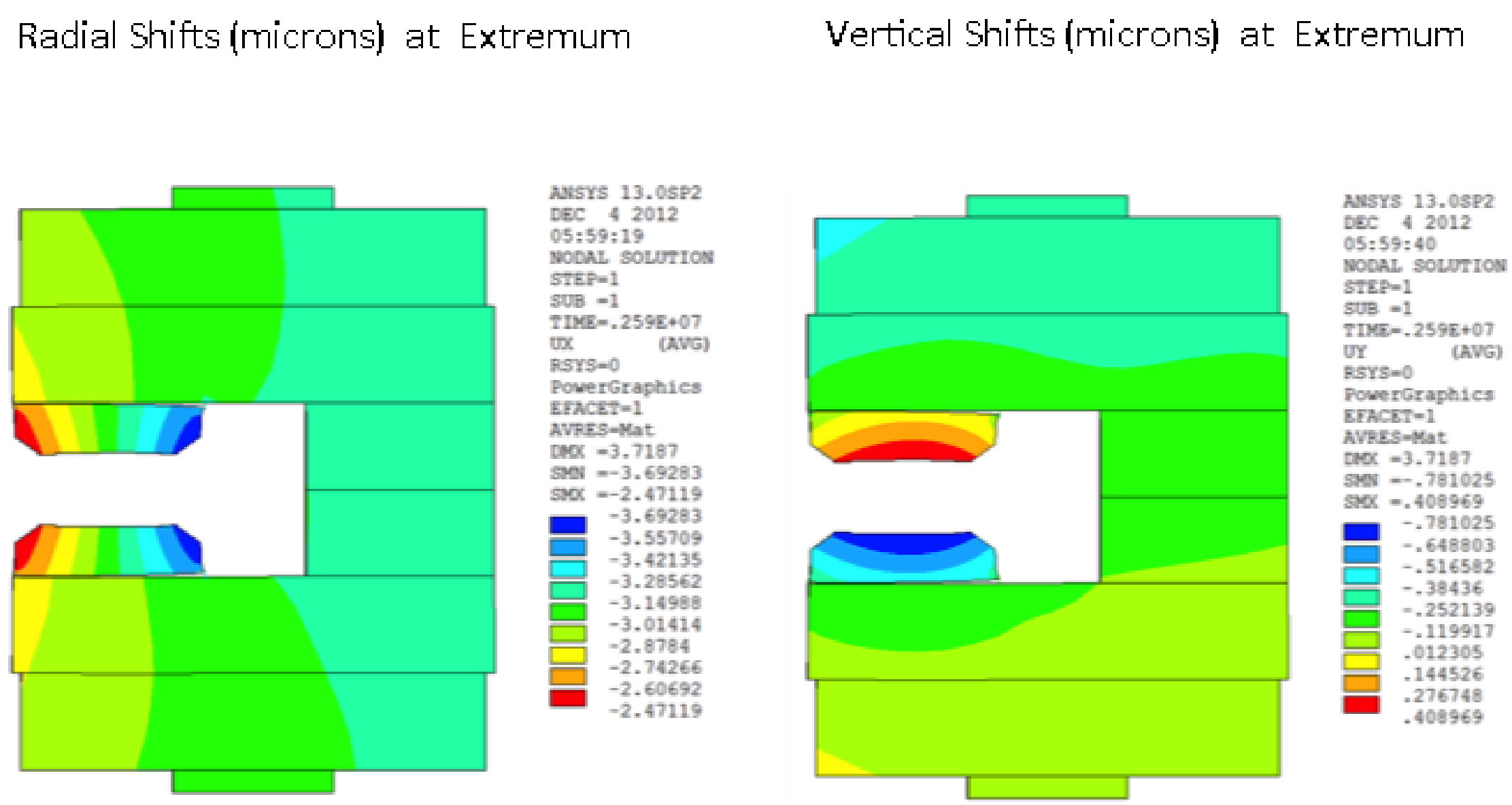

(a) (b)

Figure 9.4: The thermal fluctuations depicted in Figure 9.3 are imposed on the magnet, causing distortion of the magnet, as modeled in ANSYS. The deflections are decomposed in (a) the radial and (b) the vertical dimensions for the worst-case scenario.

When the coils are cooled, they contract down onto the radial stops into a scalloped shape. When powered, the Lorentz force pushes the coils outward, increasing the force against the mandrel, which provides cooling. This feature, the result of winding on the inside of the mandrel, reduces the risk of cooling problems even if the coil were to separate from the mandrel during transport [7].

A ground plane insulation band of 0.3 mm thickness was built from a sandwich of three layers of 50 $\mu$m kapton, epoxy coated, between two layers of epoxy-filled fiberglass. The insulation assembly was fully cured and placed into the mandrel. A 0.1-mm layer of B-stage epoxy film was placed between the mandrel and kapton laminate, and between the kapton laminate and the conductor block after winding. A 4.8-mm thick G-10 piece was placed on the winding ledge, and on top and on the inner radius of the completed coil block. The epoxy-filled fiberglass in the ground plane insulation sandwich improved heat transfer between coil and mandrel.

The coil was then wound using a machine that wrapped the superconductor with three overlapping layers of 25 $\mu$m of kapton and fiberglass filled with B-stage epoxy, 19 mm in width, laying the conductor into the mandrel with a compressive load as described in Ref. [3].

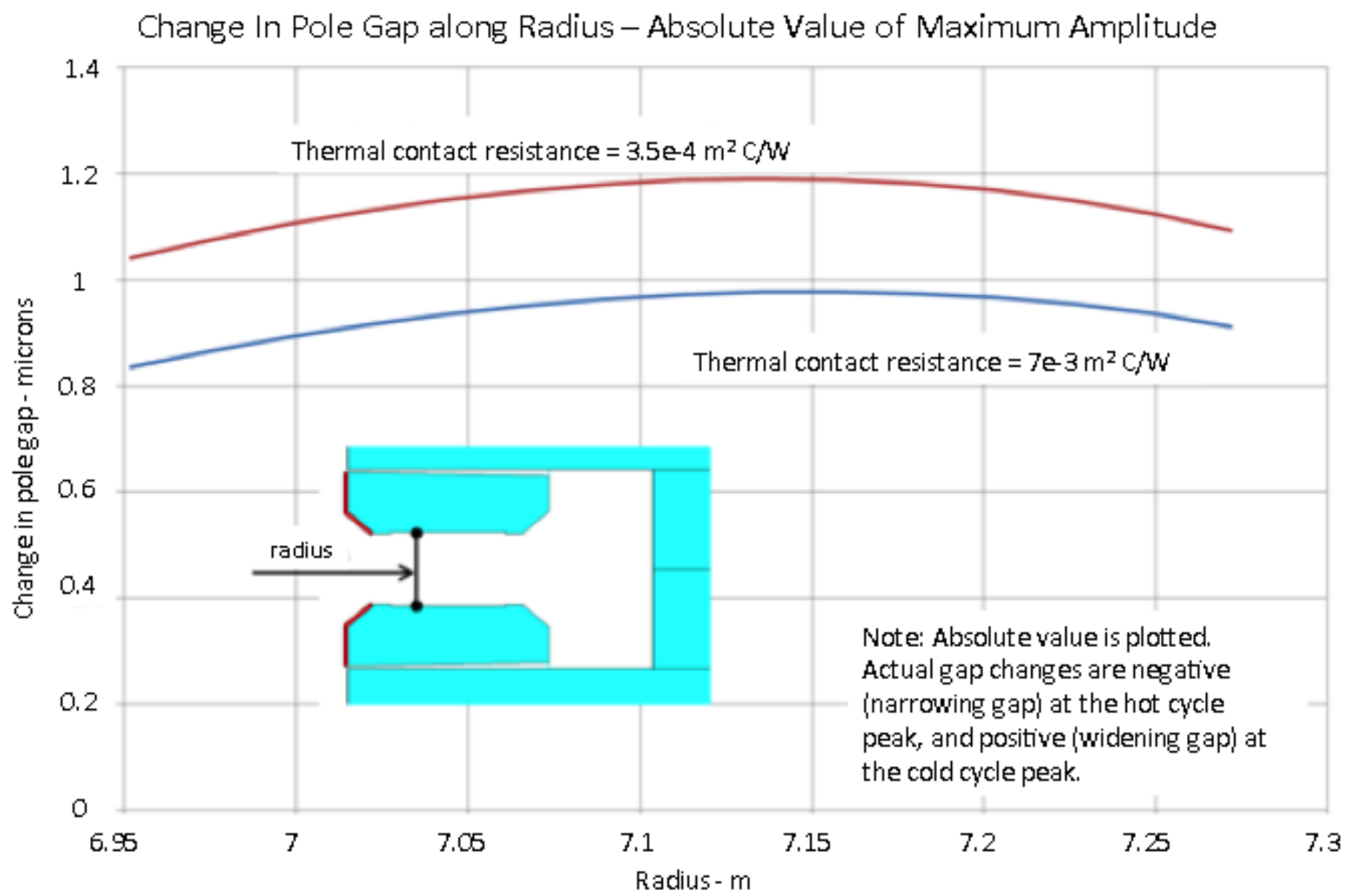

Figure 9.5: The deflections of the pole pieces under thermal variations are quantified in ANSYS simulations as a function of the radial coordinate. Typical fluctuations of 1 °C will produce micron scale distortions. Two different thermal contact resistances are shown.

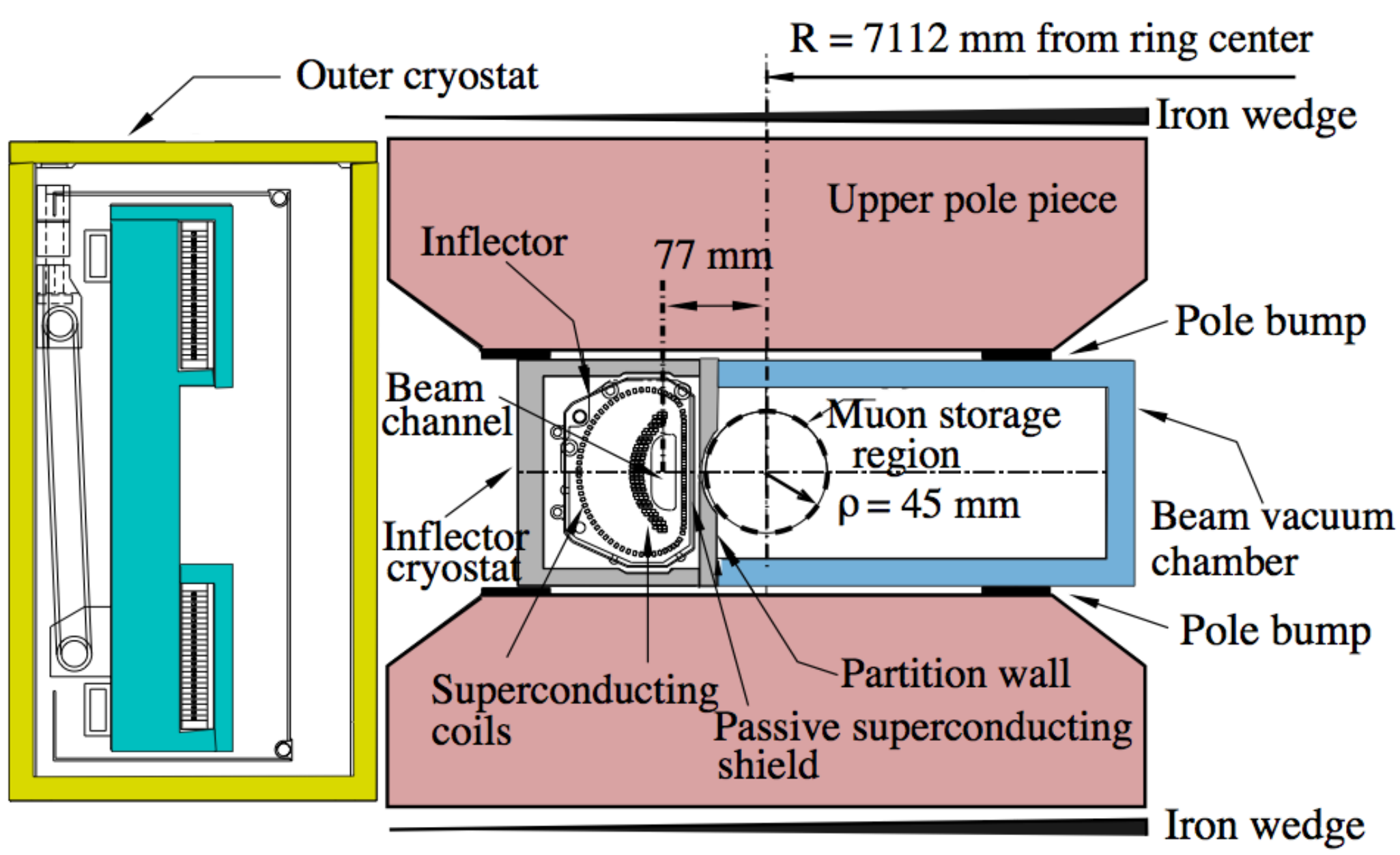

Figure 9.6: The arrangement of the pole pieces, shimming wedges and the inflector cryostat, showing the downstream end of the inflector where the beam exits. The beam is going into the page, and the ring center is to the right.

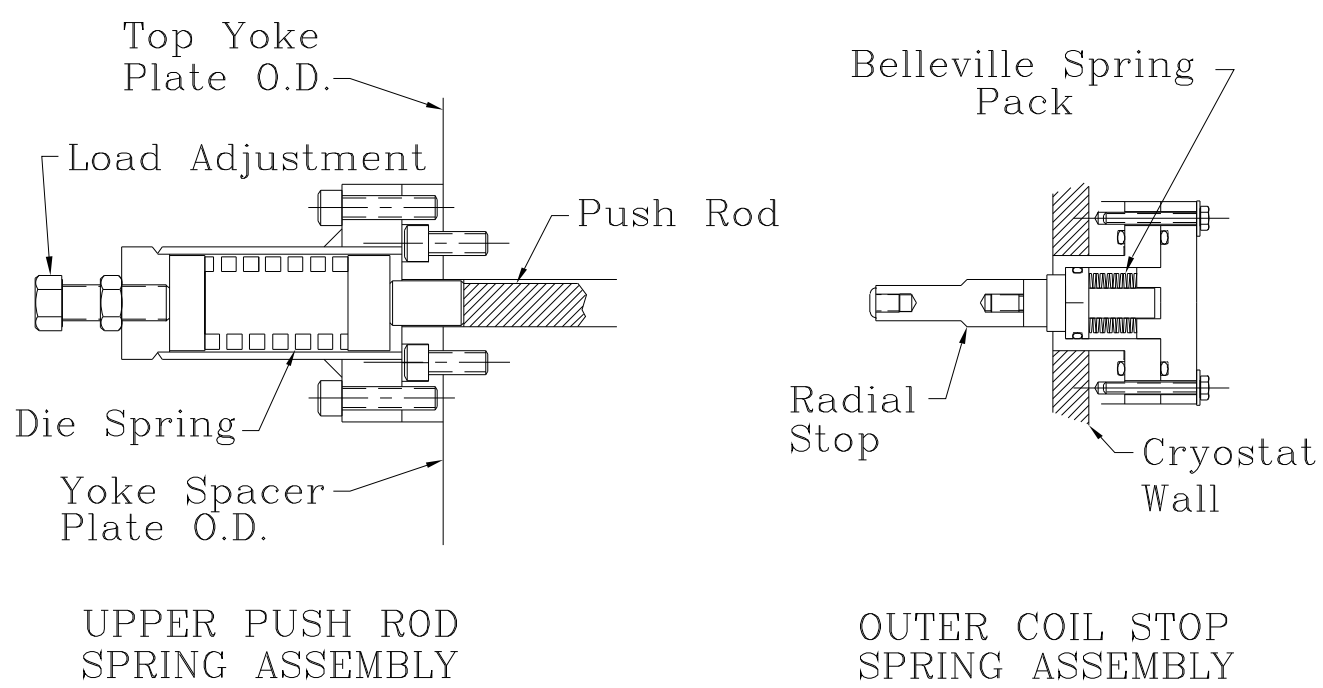

Figure 9.7: The spring-loaded radial stop and push rod. The stops are attached to the cryostat inner wall. The push rods preload the outer cryostat, attaching to the yoke at the outer radius, passing through a radial slot in the yoke to the outer cryostat.

The wrap was tested at 2000 V DC during the wind. Aluminum covers were added after the coil was wound, and the entire assembly heated to 125 °C to cure the epoxy. See Fig. 9.8.

The outer coil contains two penetrations, one to permit the beam to enter the ring, and one which could have permitted high voltage to be fed to a proposed electrostatic muon kicker. It was decided at the time to make this "kicker penetration" in the outer coil, but not to make a hole through the magnet yoke until it was shown that this kicker could be built (which was not demonstrated).

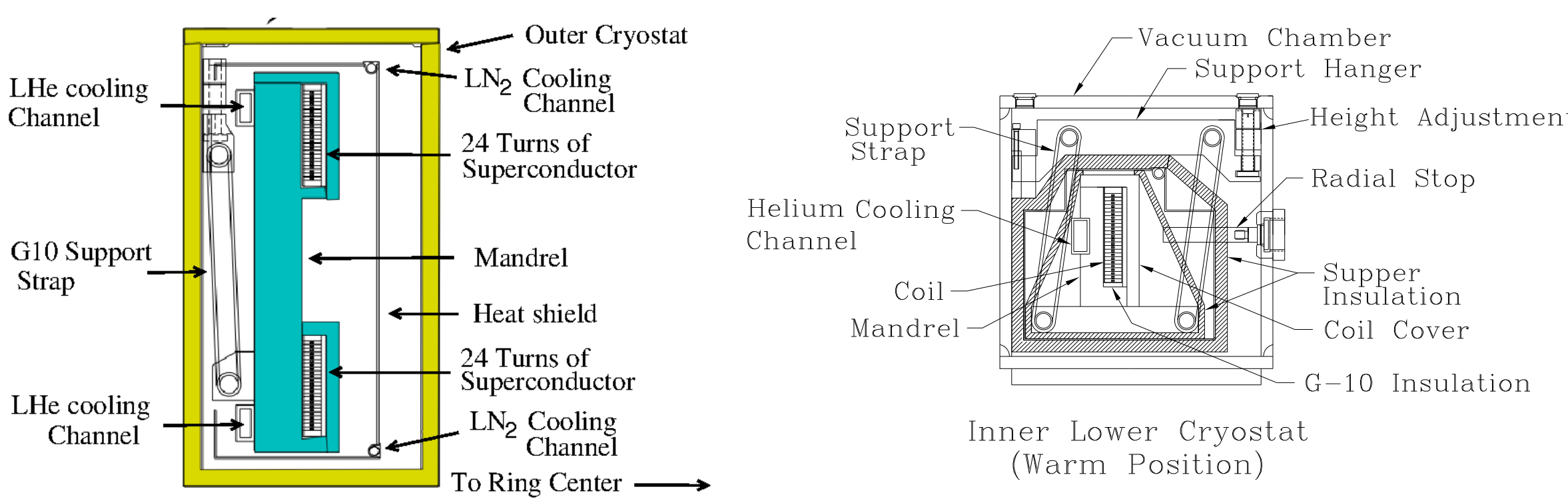

(a) Outer Coil (b) Inner Coil

Figure 9.8: The outer and inner coil structures. Both are shown in their warm configuration.

The coils are indirectly cooled with two-phase He flowing through channels attached to the mandrel, as shown in Fig. 9.8. The two-phase helium cooling avoids the increase in temperature that would occur in a circuit cooled with single-phase helium. The operating temperature of the coils is within 0.2 K of the coldest temperature in the cooling circuit. The advantages of two-phase cooling are: (1) the helium flows in well-defined flow circuits; (2) the total amount of helium that can be flashed off during a quench is limited to the mass of helium in the magnet cooling tubes; and (3) the location of the helium input and output from the cryostat and the location and orientation of the gas cooled leads are not affected by the cooling system [8].

Table 9.2: Superconductor parameters

| | |
|---|---|
| Superconductor type | NbTi/Cu |
| Nominal dimensions | 1.8 mm × 3.3 mm |
| NbTi/Cu ratio | 1:1 |
| Filament 50 $\mu$m | |
| Number of filaments | 1400 |
| Twist pitch | 27 mm |
| Aluminum stabilizer type | Al extrusion |
| Ni/Ti composite dimensions | 3.6 mm × 18 mm |
| Al/(NbTi + Cu) ratio | 10 |
| RRR (Al) | 2000-2500 |
| RRR (Cu) | 120-140 |
| $I_c$ | 8100 A (2.7 T, 4.2 K) |

The key to the operation of a two-phase helium cooling circuit is a helium dewar (the control dewar) that contains a heat exchanger. This heat exchanger sub-cools the helium from the J-T circuit before it enters the magnet cooling circuits. This isobaric cooling provides a higher ratio of liquid to gas with a higher pressure and lower temperature than the refrigerator J-T circuit alone would provide. This feature is important for the long cooling channels in the magnet cooling circuits. The use of a heat exchanger in the control dewar reduces the helium flow circuit pressure drop by a factor of two or more. The control dewar and heat exchanger also have the effect of damping out the oscillations often found in two-phase flow circuits. The helium in the control dewar acts as a buffer providing additional cooling during times when the heat load exceeds the capacity of the refrigerator.

The $(g-2)$ cooling system was originally designed to have three separate cooling circuits: a 218 m long cooling circuit that cools all three mandrels in series, the lead and coil interconnect circuits that are 32 m long (the gas-cooled leads are fed off of this circuit), and a 14 m long cooling circuit for the inflector magnet. Later the cooling system was modified to permit each of the mandrels to be cooled separately. Ultimately, the $(g-2)$ cooling system operates with parallel cooling circuits for the coils, inflector, and lead cooling. Electrically, the three coils are connected in series so that the two inner coils are in opposition to the outer coil to produce a dipole field between the inner and outer coils. The magnet is powered through a pair of tubular gas-cooled leads developed for this application. Each lead consists of a bundle of five tubes. Each tube in the bundle consists of three nested copper tubes with helium flow between the tubes. The copper tubes used in the leads are made from an alloy with a residual resistance ratio of about 64. The lead length is 500 mm. A typical cool down from 300 to 4.9 K takes about 10 days. Once the control dewar starts to accumulate liquid helium, it takes another day to fill the 1000 l dewar. In operation, the pressure drop across the magnet system is about 0.02 MPa (3.0 psi). We initiated several test quenches and had one unintentional quench when the cooling water was shut off to the compressors. The peak measured pressure during a 5200 A quench was 0.82 MPa (105 psi). Other places in the cooling circuit could have a pressure that is 40% higher. The quench pressure peak occurs 11 s after the start of the quench. The quench pressure pulse is about 12 s long compared to current discharge time constant at 5200 A of 31 s. The outer coil mandrel temperature reaches 38 K after the quench is over. Re-cooling of the magnet can commence within 5 min of the start of the quench. After a full current quench, it takes about 2 hours for the outer coil to become completely superconducting. The inner coils recover more quickly.

Table 9.3: Estimates of cryogenic heat leaks

| | | 4.9 K load (W) | 80 K load (W) |
|---|---|---|---|
| Magnet system heat load | Outer coil cryostat | 52 | 72 |
| | Two inner coils | 108 | 77 |
| | Inflector | 8 | 5 |
| | Interconnects | 11 | 46 |
| | **Magnet subtotal** | **179** | **200** |
| Distribution | Helium piping | 19 | |
| | Control dewar | 5 | |
| | Interconnects/valves | 33 | 32 |
| | Nitrogen piping | | 34 |
| | **Distribution subtotal** | **57** | **66** |
| Lead gas (1.1 g/s) | Equivalent refrigeration | 114 | |
| Total refrigeration | | 351 | 266 |
| Contingency | | 70 | 51 |
| **Cryogenic design** | **Operating point** | **421** | **308** |

Both persistent mode and power supply excitation were considered. The total flux, $\int \vec{B} \cdot d\vec{s}$, is conserved in persistent mode. However, room temperature changes would result in changes in the effective area. Thus although the flux, is conserved, the magnetic field in the muon storage region is not. Persistent mode would also require a high-current superconducting switch. Power supply excitation with NMR feedback was chosen, although no feedback was used for the 1997 run. This method gives excellent control of the magnetic field and allows the magnet to be turned off and on easily. The power supply parameters are shown in Table 9.5.1.

The quench protection design parameters were determined by the requirements of magnetic field stability and protection of the magnet system in case of a quench. When the energy is extracted, eddy currents are set up in the iron which oppose the collapse of the field. This can cause a permanent change in the magnetic field distribution [9]. This is sometimes called the ‘umbrella effect, since the shape of the change over a pole resembles an umbrella. The eddy currents are minimized if the energy is extracted slowly. There will also be eddy currents in the aluminum mandrels supporting the coils. Electrically, this can be represented by a one turn shorted transformer. These eddy currents will heat the mandrels and can cause the entire coil to become normal. This is called quench-back. This has several beneficial effects. The part of the stored energy that is deposited in the coil is deposited uniformly over the entire coil and mandrel assembly. Also, once quench-back occurs, the energy extraction process is dominated by the quenchback and not by the specifics of where the quench occurred. Therefore, the effects of a quench on the reproducibility of the magnetic field should be minimal.

Table 9.4: Power supply parameters

| | | |
|---|---|---|
| Rating | 5 V, 6500 A | |
| Rectifier | 480 VAC input, 12 pulse (Two ±15°, 6 pulse units in parallel) | |
| Output filter | 0.4 F | |
| Regulator | Low-level system | 0.1 ppm stability with 17 bit resolution |
| | Power section | Series regulator with 504 pass transistors |
| Cooling | Closed loop water system with temperature regulation | |
| Regulation | Current-internal DCCT | ±0.3 ppm over minutes to several hours |
| | Field-NMR feedback (current vernier) | ±0.1 ppm (limited by the electronics noise floor) |
| Manufacturer | Bruker, Germany | |

The energy extraction system consists of a switch, resistor, and quench detection electronics. An energy extraction resistor of 8 mΩ was chosen. Including the resistor leads, the room temperature resistance is 8.8 mΩ. This gives an 1/RC time constant of 1 minute. The actual time constant varies due to the temperature increase of the coil and dump resistor and the effect of eddy currents in the mandrels during the energy extraction (see below). This resistance value was calculated to cause quenchback in the outer mandrel within 2 seconds at full current. The quench protection circuit for E821 is shown in Figure 9.9. [1] The energy extraction trigger for a quench which originates in one of the coils is the voltage difference between matching coils; for example, $V(\text{outer} - \text{upper}) - V(\text{outer} - \text{lower})$. Since the inductance is effectively the same, the voltages should be equal even while charging the magnet, unless a quench develops in one coil. This quench threshold is set at 0.1 V. However, the coil interconnects are thermally coupled together with the helium tubes. It is possible that a quench in an interconnect could propagate to both coils almost simultaneously. Therefore, a voltage threshold of 10 mV was chosen for each interconnect. The outer upper to lower interconnect is only 1 m long. This threshold was set to 5 mV. The thresholds were determined by the requirement that the quench be detected within 0.2 s. The gas-cooled leads develop a voltage of typically 15 mV at full current. If the lead voltage exceeds 30 mV, the energy is extracted.

[1] These items are at the end of service life. The replacement system is described below.

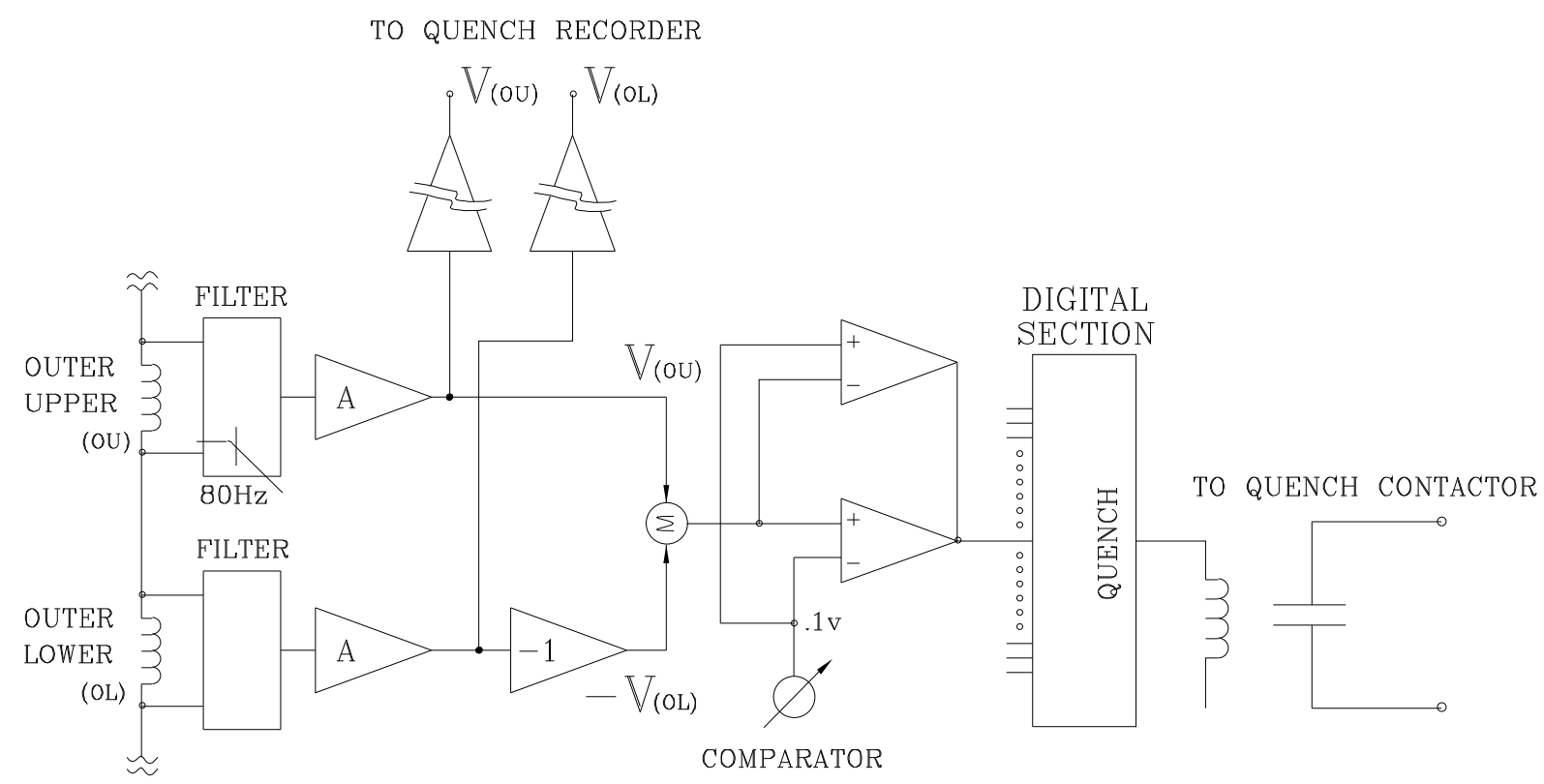

Figure 9.9: Diagram of the quench protection circuit for E821.

### 9.5.2 Preparations Prior to Transportation

No significant changes will be made to the design, and nearly all components are reused from E821. The WBS sections below describe the steps to reassemble and recommission the items above. We will not need to fabricate any parts, other than to replace old components or to build spares.

Prior to the coil transportation, room temperature tests were performed to verify as much as possible the working state of the system. These were:

- Electrical verification of the instruments connected to the coil and/or mandrel. These refer to the temperature probes, voltage taps for quench detection, and strain gauges. The instruments connected to the cryostat consists of thermometers, voltage taps, and strain gauges. These are indicated in figures 9.10 and 9.11. The strain gauges are attached to the straps, which counteract the Lorentz forces, and the radial stops, which counteract the shrinking of the radial stops due to cooling.

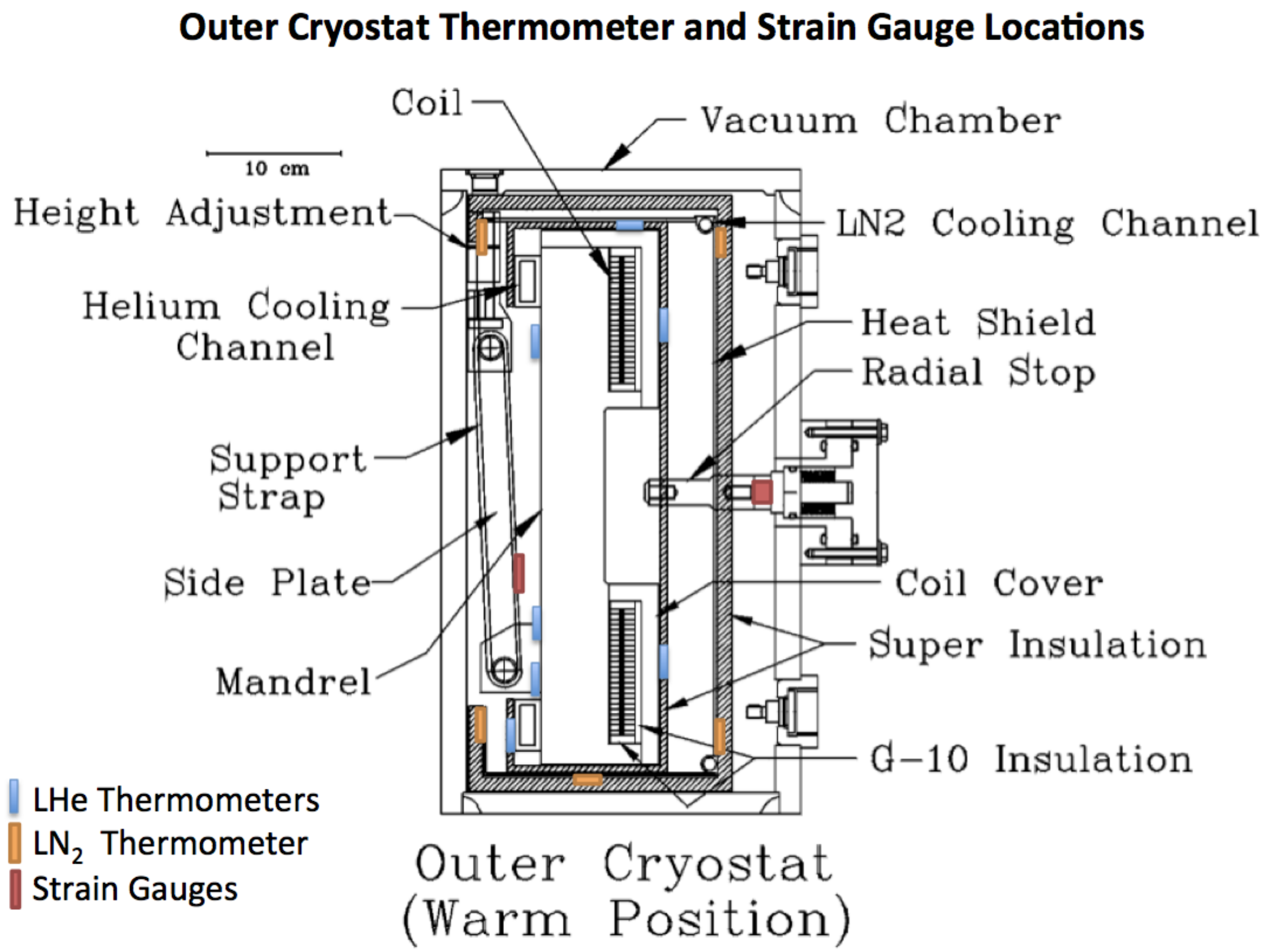

Figure 9.10: Location of outer coil instrumentation, showing the LHe $LN_2$ thermometers, and strain gauges. There are typically eight thermometers placed at each azimuthal location, with the positions indicated above.

- Resistance measurements of the coil at room temperature, which agree with measurements performed in 1995 (see table 9.5).
- The resistance between the coil leads and ground was measured to be a few kohms, where as an open resistance was expected. Further tests showed the 'short-to-ground' occurring at the connection between the inner lower coil and the power supply (see figure 9.16), and is a straight-forward repair. There is no short within the coils themselves. This short was likely present during E821 running, and would have contributed a 0.01 mA current-to-ground, out of a total of 5200 A. This is 2 ppb effect and would not have been seen in E821 (see references [11] and [12]).

Following this verification, the interconnections between the three coils (see figure 9.12) were separated by a grinding wheel. The temperature was monitored and kept well below 100 °C during the process to minimize degradation to the Aluminum resistance and the NbTi current-carrying capacity. Figures 9.13 and 9.14 show the details of the welds that were cut in this region.

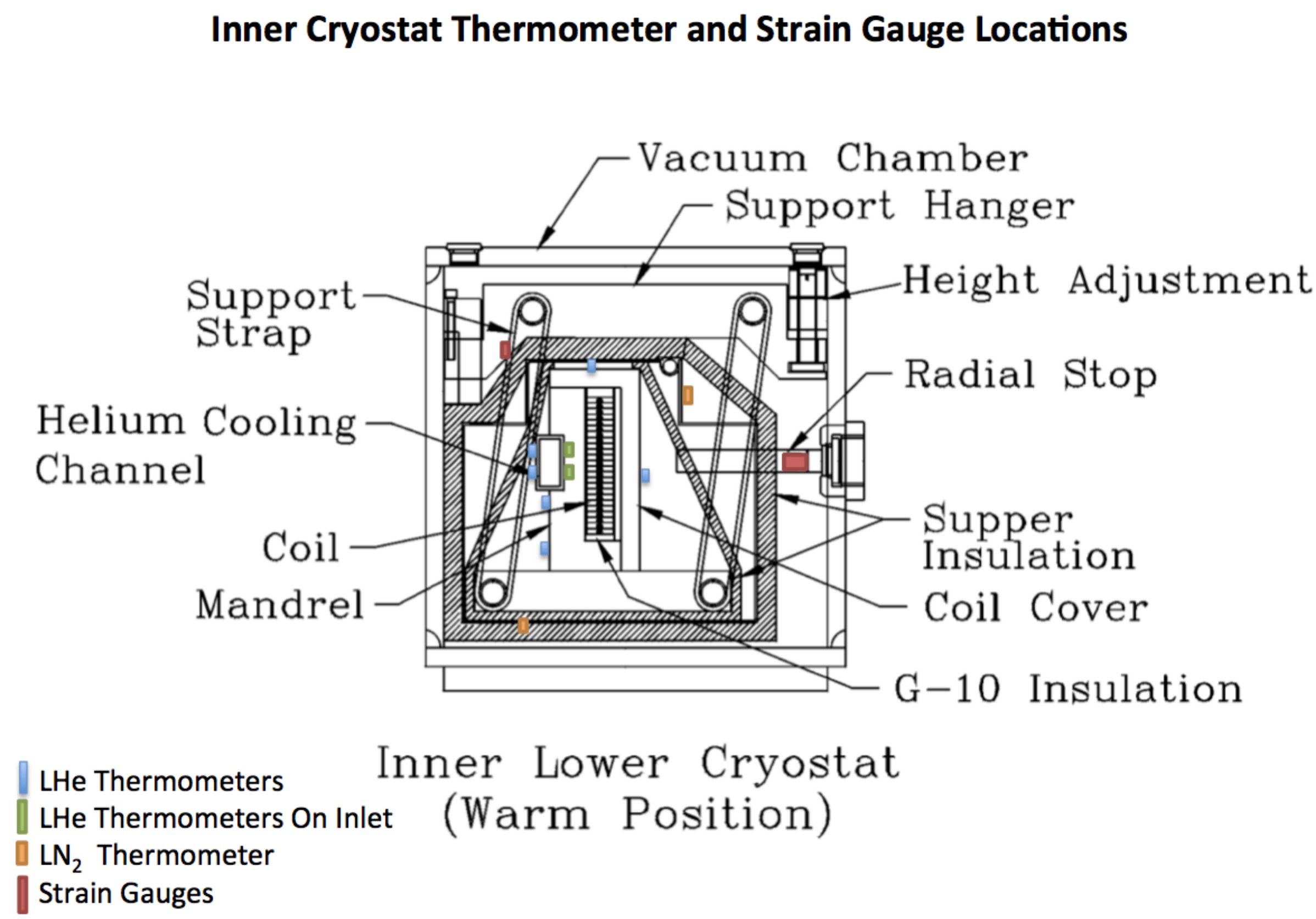

Figure 9.11: Location of the lower inner coil instrumentation, showing the LHe and $LN_2$ thermometers, and strain gauges. The instrument locations are symmetrically placed for the upper coil instruments. There are typically eight thermometers placed at each azimuthal location, with the positions indicated above.

### 9.5.3 Cryostat Vacuum Chambers

This WBS consists of the vacuum chambers that provide the thermal insulation for the coils. After the interconnects have been rewelded (see section 9.5.9), the vacuum flanges enclosing that region will be reconnected.

For transportation, a vacuum port connected to the outer cryostat was cut in order to gain clearance. Therefore, this pipe will be rewelded upon reassembly at Fermilab.

### 9.5.4 Vacuum Pumps

New or refurbished 'dry' vacuum pumps will be used to pump down the cryostat vacuum chambers. The pumps must remain a few meters away from the storage beam region in order to not perturb the precision magnetic field. The vacuum quality needed is about $10^{-4}$ Torr.

E821 implemented a mechanism described below to deal with a potential failure mode called the "cold cryostat problem". In the event of loss of cryostat vacuum while the coils are cold, the cryostat walls will also become cold and therefore will shrink. Such could happen if the cryogenic lines leaked cryogens into the vacuum. The outer cryostat will shrink until it hits the pole pieces. Therefore, the cryostat wall could experience stresses exceeding the allowable value for Aluminum.

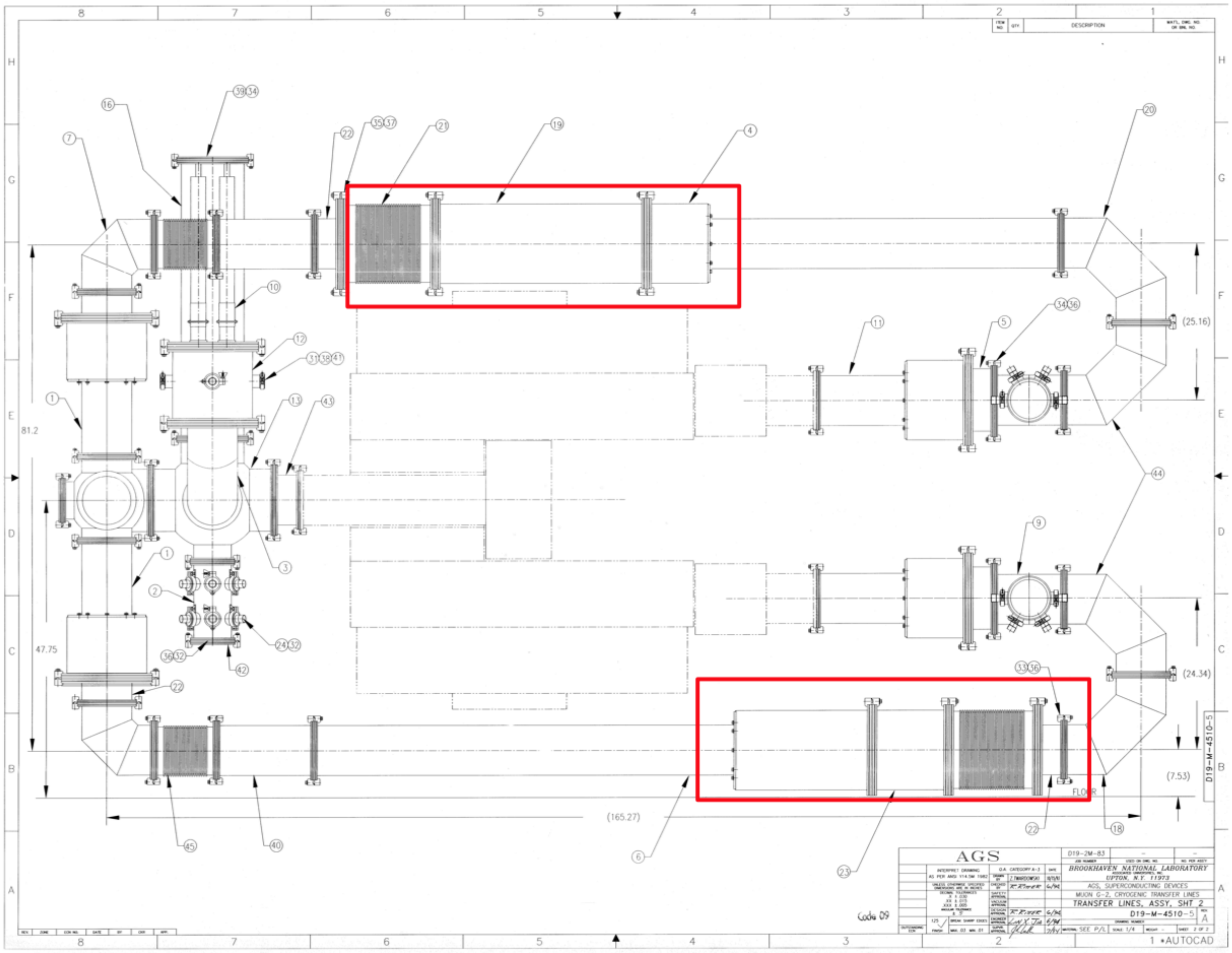

Figure 9.12: The connections between the three coils are indicated. The upper(lower) red box is the connection between the outer-upper (outer-lower) coil and the inner-upper (inner-lower) coil.

For this potential failure scenario, E821 implemented a scheme to trigger a large Roots blower vacuum pump to rapidly evacuate the vacuum chamber.

### 9.5.5 Power Supply and Quench Protection

The power supply, a Bruker type B-MN5/6500 linear supply, for the main ring will be the same unit used in Brookhaven and as described in reference [3]. Therefore, the design and specifications, and regulation mechanisms for operation will also be the same. The power supply and quench protection cabinet was moved from Brookhaven to Fermilab in mid-November of 2013, specifically to the Dzero Assembly Building (DAB) pit.

Various aspects of the Bruker supply, the quench protection cabinet, and quench detection electronics have been rexamined and recommissioned at Fermilab DAB. The findings are described in reference [37].

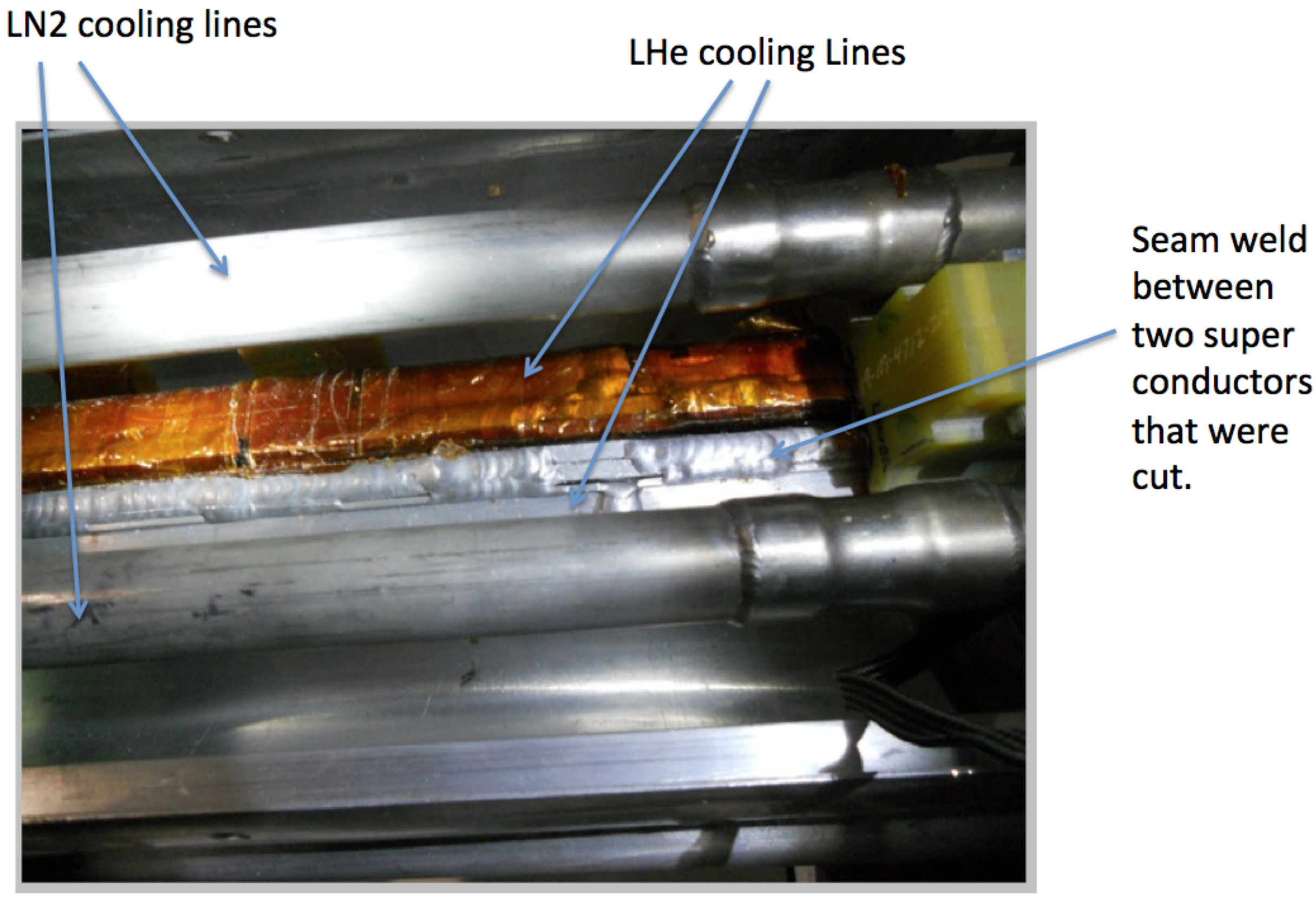

Figure 9.13: Photograph of the region of the interconnection, indicating welds and cooling lines.

### Power Supply Refurbishment and Modifications

Based on the findings, there are four areas of the Bruker power supplys implementation that needs to be modified or upgraded: cooling, controls, pass bank monitoring, and repair or replacement during operation. Overall, the assembly of the Bruker components within the cabinet is quite integrated inside of a relatively small cabinet space. This means that to get to a series of components, a major disassembly process will be required. Therefore, disassembly, where it is not absolutely required, shall be kept to a minimum. However, the following tasks should be implemented:

- Cooling: The cooling supply and the return for the Bruker cannot be directly connected to the chilled water supply system due to the relatively low maximum operating pressure of 8 bar (about 120 psi) and the large cooling flow requirement of 70 Liters/minute. In addition, the cooler the electronic components can be kept during operation, the better given the age of these components. Therefore, a separate regulated cooling system, as described in reference [33], will be implemented.

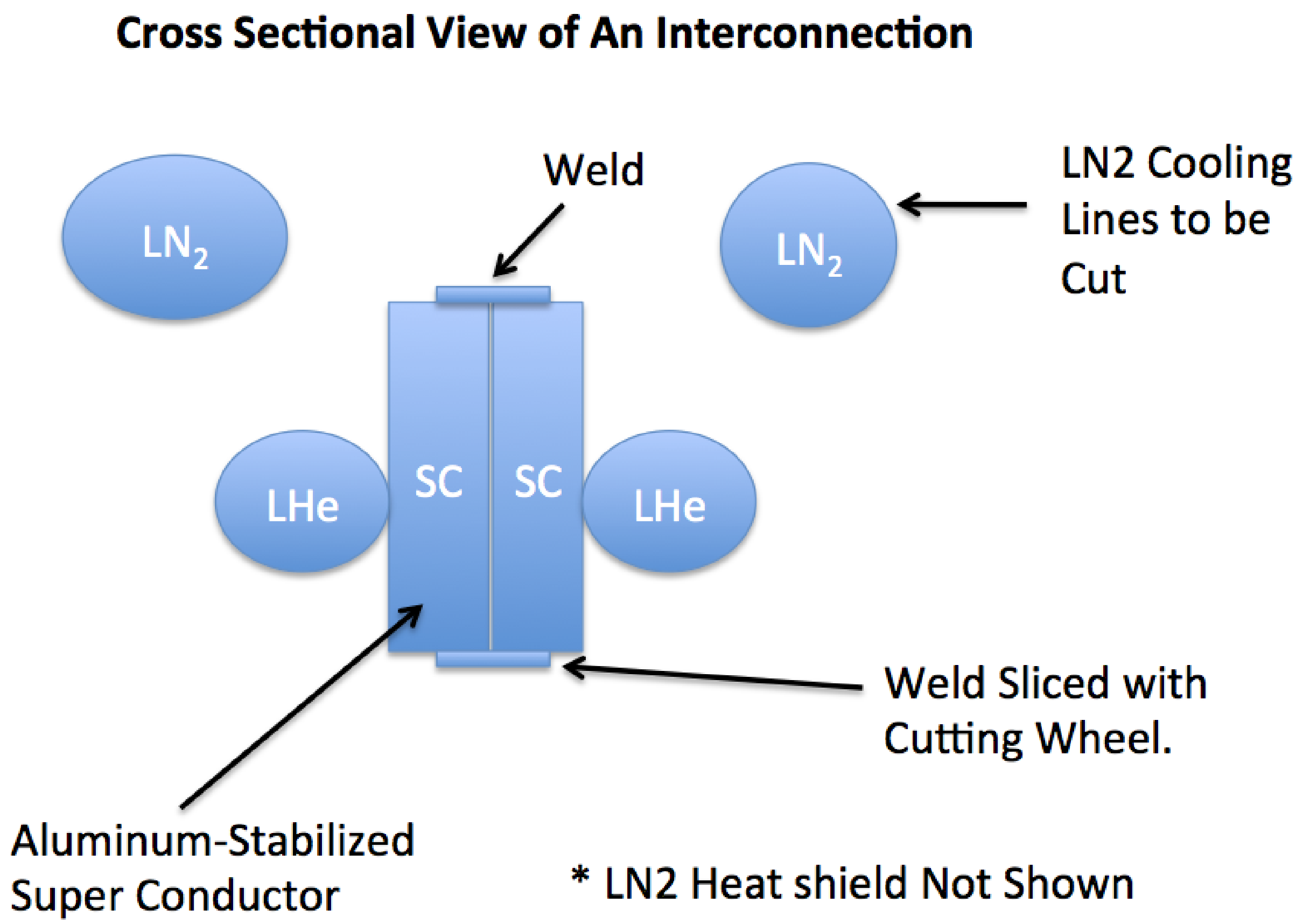

Figure 9.14: Diagram of the region of the interconnection, indicating welds and cooling lines that were cut, in order to facilitate the transportation.

- Controls: The original controls for the Bruker uses RS232 command words as described in reference [34] for remote control of the power supply. Local/remote control is implemented using a keyed selector switch. Therefore, it does not appear that front panel operation and remote control can occur at the same time. To implement the control and operation of the Bruker within the Muon g-2 control system, the RS232 interface must be connected to an emulator, either using LabView or a Beckhoff controls unit that can be accessed using Ethernet.

- Pass Bank: The output current is a summation of the outputs of 28 pass bank circuits, each containing 18 power transistors. As examined, there is no protection or monitoring of each pass bank to indicate a failed pass bank. Therefore, if one pass bank should fail, the current requirements for the other remaining ones increase. This could lead to a zipper-effect failure that can occur rather rapidly and thus take out the entire supply. Some sort of monitoring will be required on each pass bank to detect a failure and thus to shut down the power supply before this zipper effect can occur.

- Repair: As stated, replacement of a pass bank, or the transistors within a pass bank, will not be an easy undertaking and therefore, downtime can be problematic. In addition, there are no spare parts that came with the Bruker. To reduce downtime in the event of a pass bank failure, several spare pass banks (10% or three units) should be fabricated and the transistors matched to the existing pass banks characteristics. Other components, for which spares should be available, will be identified. These include the controls power supplies and printed circuit board components.

| Coil | Resistance (ohms) Sept-1995 | Resistance (ohms) Dec-2011 |
|---|---|---|
| Inner Upper | 0.429 | 0.430 |
| Inner Lower | 0.426 | 0.430 |
| Outer Upper | - | 0.483 |
| Outer Lower | - | 0.476 |
| Outer Upper + Lower | 0.952 | 0.958 |

Table 9.5: Room temperature coil resistance (ohms) measurements, showing consistency between Sept-1995 and Dec-2011.

### 9.5.6 Quench Protection System

For the Main Ring, a quench protection system is illustrated in the simplified diagram below (see figure 9.15). This differs with the diagram from the conceptual design report in some important aspects. First, the trip cabinet employs just a SPST DC breaker. It is not a selector switch. Once a quench is detected, the DC breaker disconnects one power supply output lead from the Main Ring lead.

Second, the blocking diode, located in the trip cabinet, conducts only during a quench due to the reverse direction of the quench current from the collapsing field and this current is then dissipated in the dump resistor. Thus, the end result of the quench detection system is a signal that trips the DC breaker. As shown, this signal is generated by any one of four conditions: an interlocks drop or some safety condition, one or more of the threshold levels are exceeded, an intentional trip due to the active pressing of one or more of the crash switches, or the failure of the main power supply.

### 9.5.7 Quench Protection Circuitry

As the E821 quench protection electronics have reached the end-of-service life, we will adopt the system that was used for the Dzero Solenoid quench protection [32]. The Dzero system is available for E989 use, and has the ability (without redundancy) to serve both the Main Ring and Inflector magnets. In case there is insufficient channel count, a second (backup) option is to construct a LabView based system similar to that used at Fermilabs superconducting magnet test facility, as described in reference [36]. Whichever specific system is used, figure 9.16 illustrates the overall circuitry to be implemented. The voltage thresholds for the generation of a quench trip signal will be set to the levels as used in the Brookhaven system for E821 and as described in reference [3].

- Coil Voltage: 100 mV

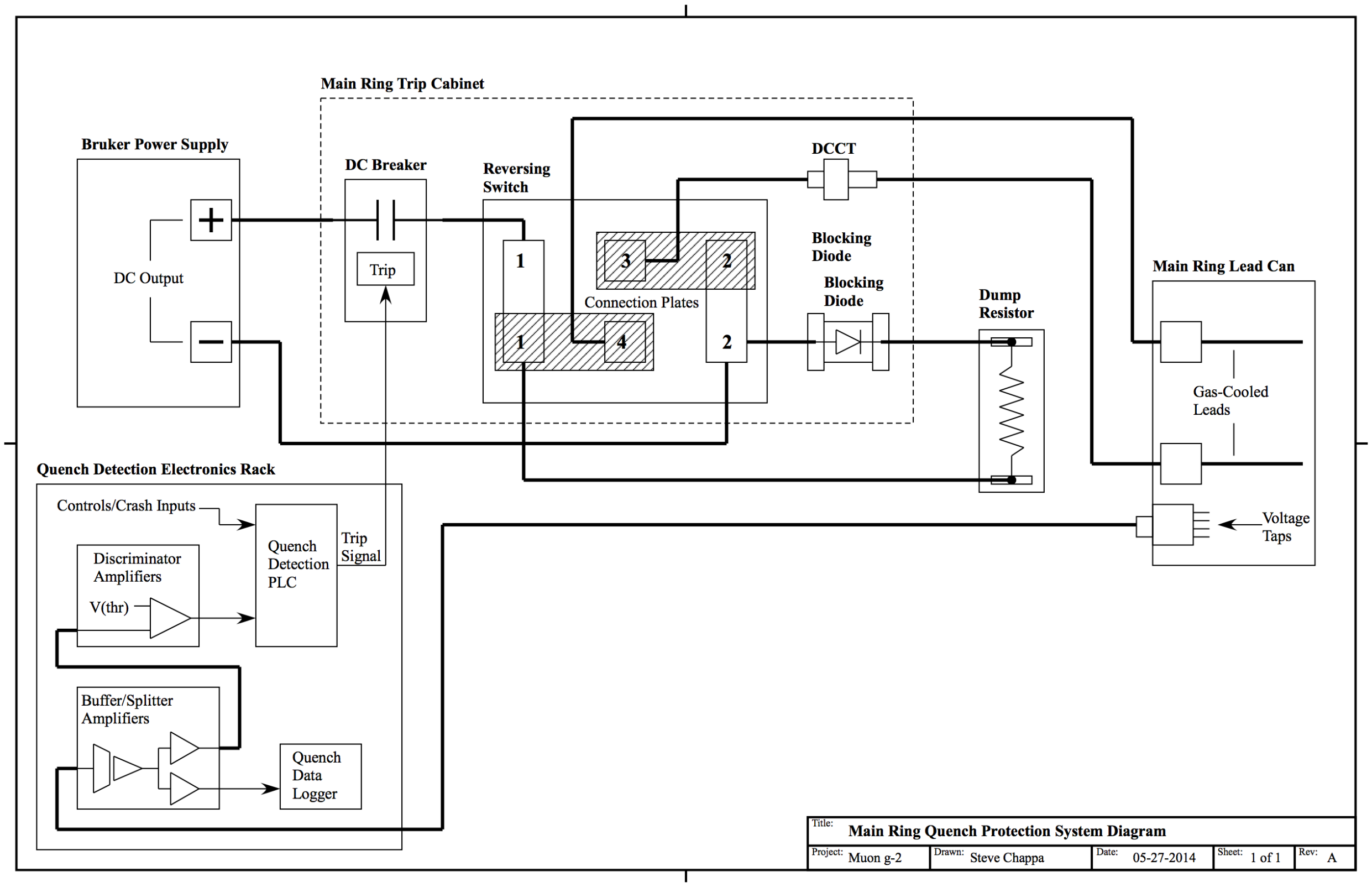

Figure 9.15: The Quench Protection System Diagram.

- Gas-cooled lead: 30 mV
- Outer/Inner coil interconnects: 10 mV
- Upper/Lower coil interconnects: 5 mV

In order to document the specific wiring and connector information of the Main Ring, a wiring schematic is being developed. According to the information at present, a draft of this schematic is shown in reference [35]. Once verified, this schematic will be used to diagram the specific voltage taps to the connectors.

### Power Supply and Quench Component Placement

As of now, the plan is to locate the Main Ring power supply and quench protection components within the main hall in MC-1 in a similar manner to that used in E821. The exact location on or near the platform will depend on the exact space and orientation of the components (trip cabinet, quench protection, electronics rack, etc.). The dump resistor will be placed outside of the main hall so as to not adversely affect the temperature within the hall. For runs of the DC output bus that exceed 10 or so feet, water-cooled 10KA bus work can be implemented to save on cost associated with the use of 535 cm locomotive cable.

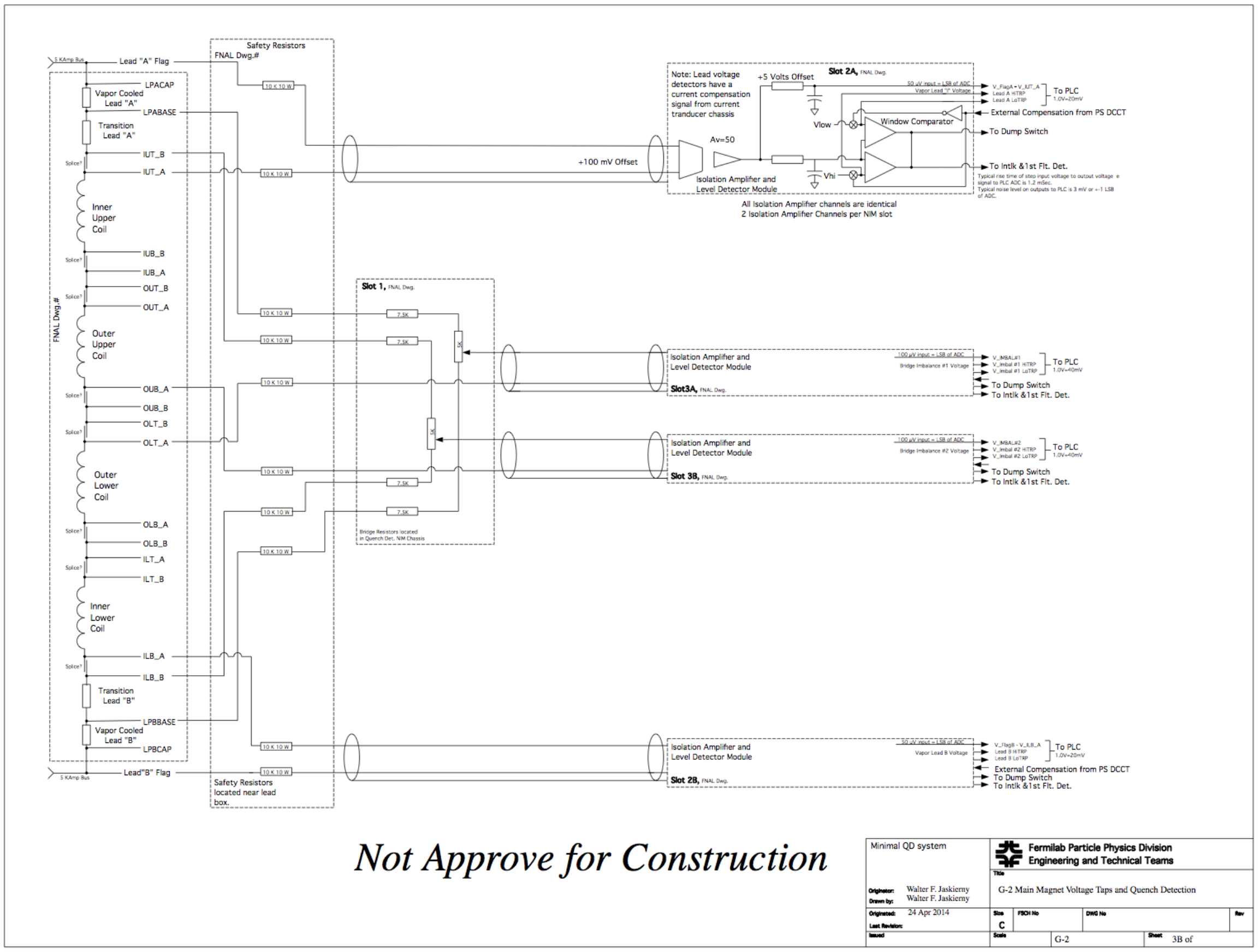

Figure 9.16: Main Magnet Voltage Tap and The Quench Protection System Diagram.

### 9.5.8 Cryogenics

This WBS refers to the cryogens (LHe and $LN_2$) required to cool the coils to 4.9 K, the cooling lines, the heat shields cooled to $LN_2$ temperatures, and the flow control valves. The E821 cryosystem will be reused as much as possible, especially the 1000 liter dewar. The E821 cryogenic flow diagram is shown in figure 9.17.

While no design changes will be made, this WBS requires considerable verification, reanalysis, and documentation due to the significant hazard and stored energy, and the potential for the 'cold cryostat' problem as described in the section above.

The documentation were generated for E821, and will be reaccessed for E989. Attention will be given to:

- Heat load of the system. There will be a slight increase due to the rewelding of the coils at the interconnection (see section 9.5.9) and a slightly longer run between the cryogenic plant and the LHe dewar.

- Flow diagram, pressure drops, and flow rates. These are specified to be identical to that of E821.

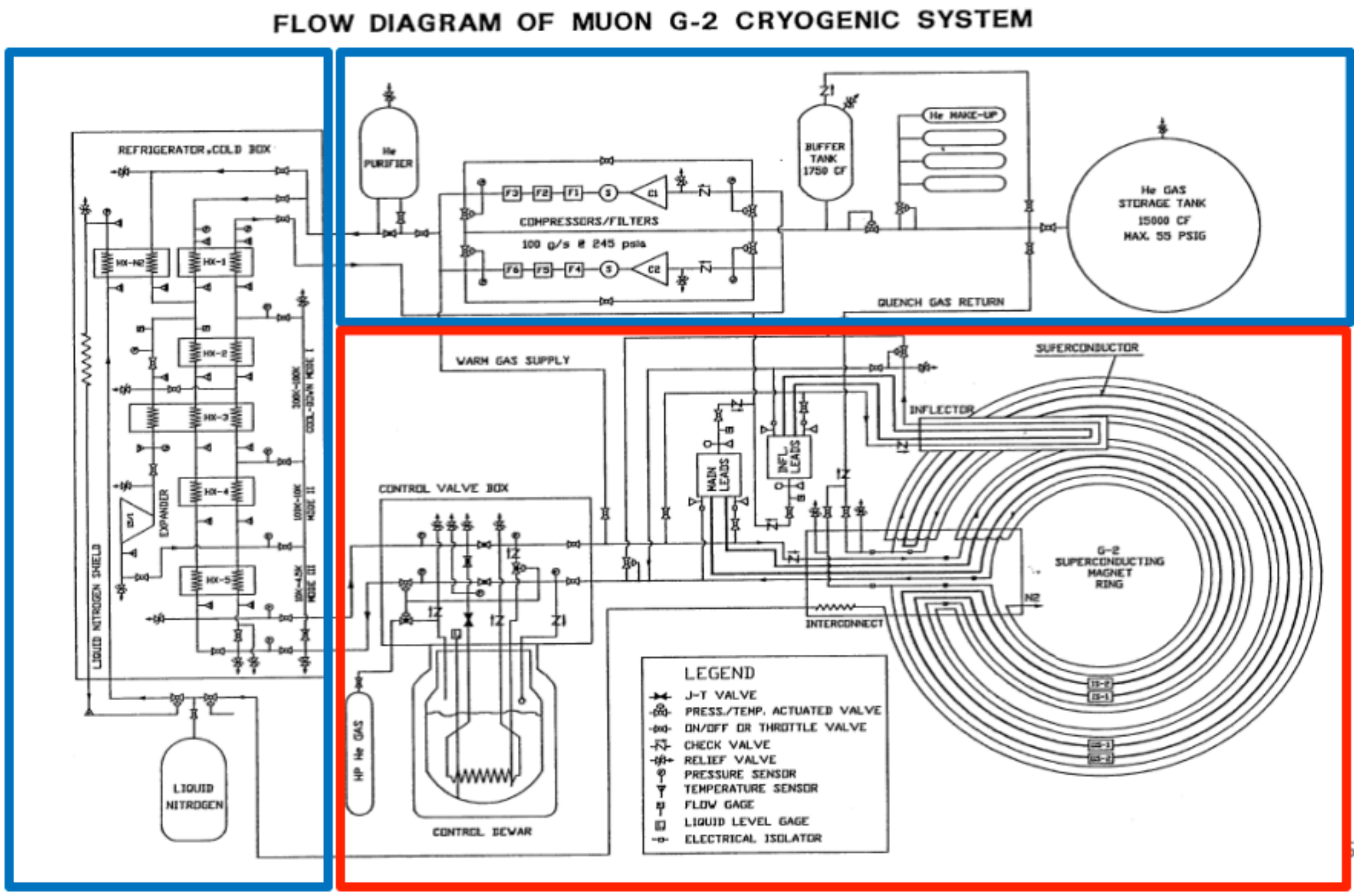

Figure 9.17: The cryogenic plant and its connection with E821 (G-2). The red box outlines the flow within building 919 at BNL. The upper (left) blue box outlines the LHe ($LN_2$) cryogenic plant.

- Cool-down and warm-up procedure.
- Connection to the Controls and Instrumentation WBS.

The Fermilab Accelerator Division (AD) will provide two dedicated refrigeration systems for E989. An important difference between E821 and E989 is that the latter will share aspects of the cryogenic plant with the Mu2e experiment[13]. Mu2e and g-2 helium gasses are mixed together, and therefore share a common compressor. The AD design includes a cryo adsorber to trap contamination.

### 9.5.9 Super Conducting Coils

After the steel yoke pieces and coils have been reinstalled into the correct position, the recommissioning activities can begin. The key recommissioning activities for this WBS consists of:

- Performing electrical continuity tests of the instruments such as thermometers and strain gauges.
- Performing electrical continuity tests on the coils. This also verifies unwanted thermal shorts.

- Rewelding the coil interconnects using pure Aluminum filler using the TiG (Tungsten inert Gas) welding technique.

- Rewelding the LHe and $LN_2$ cooling lines in the interconnection region.

During the interconnect cutting and the TiG rewelding process, a small amount of degradation to the pure Aluminum stabilizer is to be expected due to work hardening. Work hardening will cause a resistance increase of the Aluminum, therefore adding to the heat load at 4.9 K and a small local heating at the interconnection region.

It also reduces the critical current capacity ($I_c$) before the superconductors become normal. For the welding that took place during the E821 construction, the SC coil heating due to welding was modeled. The maximum temperature seen by the SC coils due to welding is 350°C[3]. Degradation of NbTi critical current of $< 5\%$ was measured for a 2 T field for an annealing time of 10s at 400°C[9].

For E821, the coil current and temperature were approximately 5200 A and 4.9 K respectively. The magnetic field at the coil was approximately 2 T. The critical temperature was estimated to be 6.2 K, giving a safety margin of 1.2 K[10]. However, the magnetic field at the interconnect is estimated to be $< 1$ T, and so the safety margin at the interconnect is even greater.

E821 also welded a test overlap joint, and measured a cold resistance of $16 \cdot 10^{-9}\Omega$ at 2 T. At the current of 5200 A, the heat load is 0.43 Watts at 4.9 K, as compared to the cooling capacity of 351 W at 4.9 K (see table 5 of reference [3]).

In summary, the coils were designed with a rather large safety margin. No quenches were observed to have taken place at the interconnects. To be conservative, we will measure the resistance properties of an overlap weld, cut it and reweld, and remeasure the resistance properties. Finally, work hardening of the Aluminum is strongly anti-correlated with yield strength[14]. Therefore, room temperature tests can be performed to gauge the level of work hardening.

We have studied the process of cutting and rewelding the superconductors. Several samples of 20 cm long overlap joints using g-2 super conductors have been produced, and tested at 4.2K and up to 8000 Amps. The results are described in reference [15]. Two 20 cm long overlap joints, made using ER4043 as the aluminum welding rod[2] were measured having a 7 nOhm and 10 nOhm joint resistance, respectively. They were cut and then rewelded (again using ER4043). The rewelded joints were remeasured to have a resistance of 21-25 nOhm. Extrapolating to the case of a 1 meter overlap joint, we expect the joint to have a 4-5 nOhm resistance. For the actual rewelding, we will use ER1100[3] as the welding rod, which is the purest aluminum available.

## 9.6 Simulations

The E989 collaboration have developed their own Opera2D and Opera3D models of the magnet, which are useful for understanding:

[2]A general purpose aluminum welding rod with 4.5% to 6% silicon, and 1145 °F melting temperature.
[3]An aluminum rod with lowest percentage of alloy agents, and 1215°F melting temperature.

- the large magnetic forces on the pole, yoke, coils, and wedge shims. These magnetic forces are input for calculations of the deflections and stresses [16, 17, 18].
- the transients (eddy) currents from magnet quenching. These currents warm up the cryostats, and cause small voltage differences across the ring [21, 20, 19].
- the dependence of the field quality on alignment mismatches, which are input to the alignment tolerances [22, 23, 24, 25].

The E989 models confirm E821's Opera2D and Poisson calculations. Figure 9.18 and 9.19 show force calculations on poles and coils.

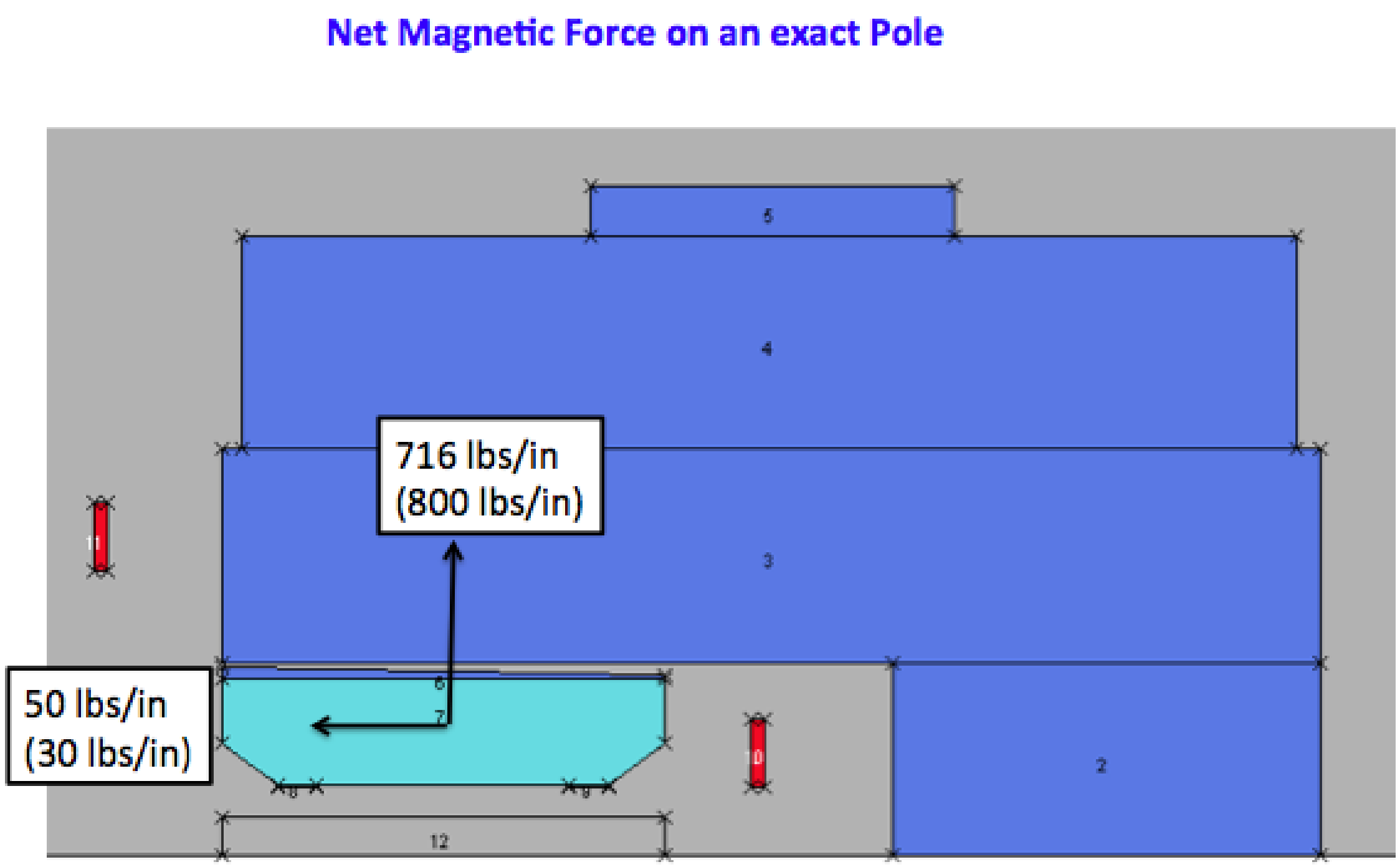

Figure 9.18: Magnetic force on the pole from an Opera2D calculation. The number in parenthesis is a calculation from BNL E821 g-2.

## 9.7 Alignment

This section of the TDR describes the necessary steps for the alignment of the g-2 magnet ring yokes and poles. Unlike the case of positioning a new accelerator system, this installation requires positioning the components to an existing location derived from BNL E821 measurements. During the construction of the g-2 ring at BNL the designers had the freedom to shape every sixth pole piece to the existing gap. In our case we do not have that degree-of-freedom. So we need to place the components as close to their as-found components as possible in order to close the circle of poles within their tolerances. While we have historical data on these gaps, there is a chance that we need to iterate to the final location in several steps. The following paragraph summarizes the required tolerances.

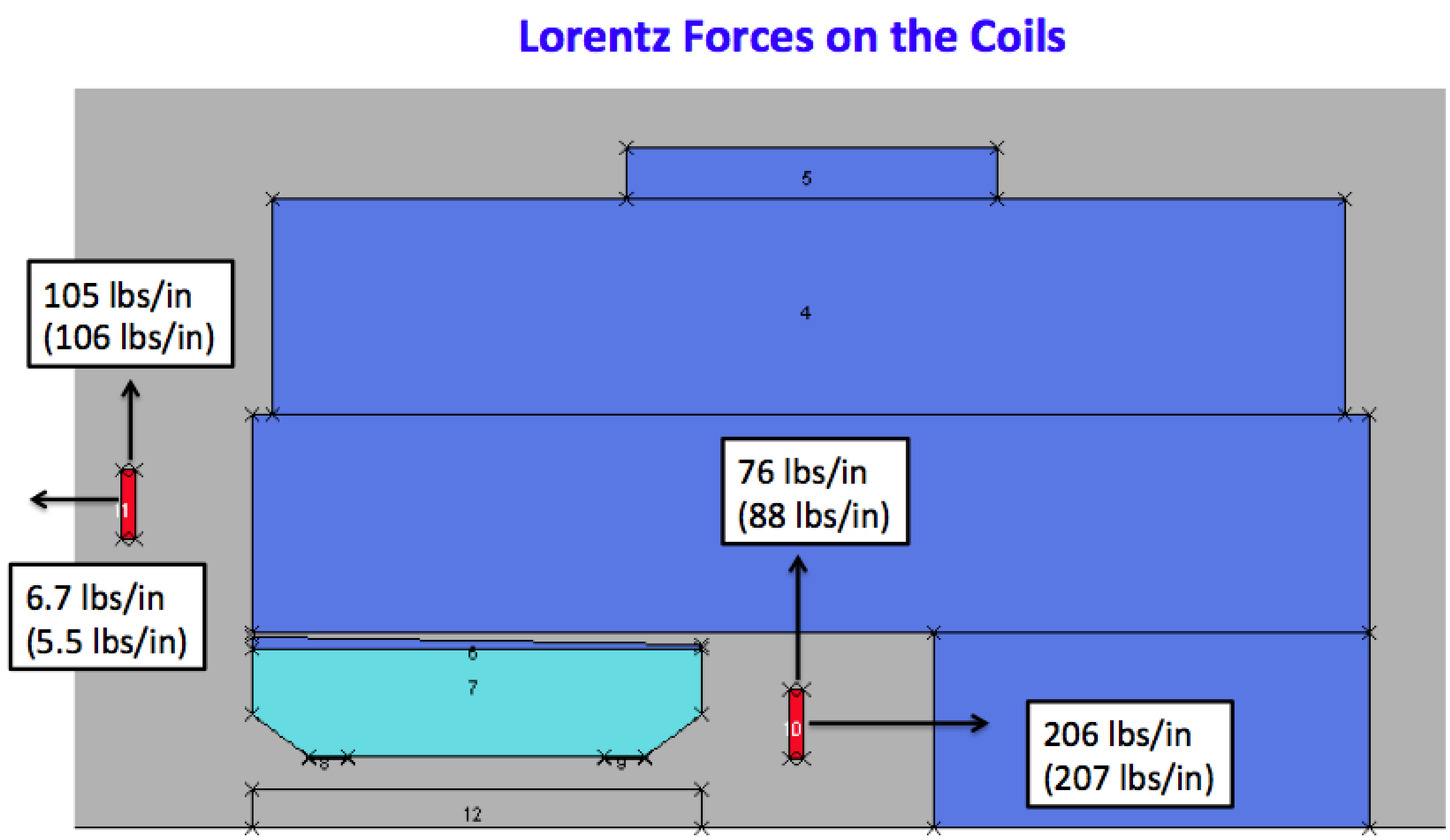

Figure 9.19: Magnetic force on the coils from an Opera2D calculation. The number in parenthesis is a calculation from BNL E821 g-2.

### 9.7.1 Required Position Tolerances

For E989, we aim for the tolerance achieved by E821. We fiducialized a subset of the yokes and poles, to verify whether they match the engineering drawings. The following paragraph summarizes the relevant alignment tolerances, and our QC results [26].

#### Yokes

The machining tolerances for the gap-facing surfaces of the upper and lower yokes were specified to be flat within ±0.13 mm, and this was confirmed by our QC check of two lower yokes. Both yokes show an RMS of ±0.03 mm with a peak to peak amplitude of ±0.1 mm. Similarly, the inner radial tolerance of all yoke pieces is specified to be ±0.13 mm. Again our QC measurements for two lower yokes confirm that these tolerances have been met for a best fit cylinder radius. However, the radii derived from these measurements deviate by ±2.5 mm on average from the design radius.

The parallelism for the top and bottom surfaces of the middle yoke (a.k.a. backleg) are specified with an RMS value of ±0.2 mm, and the thickness tolerance for these parts is ±0.13 mm. Considering the QC results for the lower yokes, these values will be accepted as face values without performing QC measurements. The yoke end faces need to be perpendicular to the gap facing surface to ±0.3 mrad and ±0.2 mrad in the radial direction. Both of these requirements have been met according to our QC measurements. Similarly, the chord length at the inner and outer yoke radii intersecting the end planes have been machined with an RMS of ± 0.25 mm. Here too our QC measurements confirm this machining tolerance.

The design azimuthal gap between individual yokes was specified to be 0.5 mm. The average reported gap spacing for the inner and outer yoke gaps is 0.8 mm with an RMS of ±0.2 mm. The lower, middle and upper yokes were matched to equalize the effective azimuthal gap for the three pieces. In the re-assembly this will automatically be achieved as we intend to use the same pieces at the same location as in the BNL setup and the provided pins provide the alignment of these pieces to each other.

### Poles

The machining tolerances for the flatness of the pole top surfaces creating the gap between the upper and lower poles were specified to be $\pm 25 \mu m$ with a ground surface finish of $\pm 0.8 \mu m$. Errors in pole thickness are small compared to the expected variation in gap distance. QC measurements of 12 poles show a pole thickness variation on the order of $\pm 47 \mu m$ which only exceeds slightly the manufacturing specification of $\pm 40 \mu m$.

Naturally, at that level of accuracy any temperature effects need to be taken into account. These effects will play a major role during the re-assembly of the massive steel components of the g-2 magnet ring. Very good temperature stability will be required when we fine tune the position of these components. The machining tolerance for the pole width is specified as 0.56 m $\pm 50 \mu m$. The QC measurements of the pole subset show that the average pole width deviates from the nominal value by 20 $\mu$m with an RMS of $\pm 27 \mu m$. For the azimuthal pole position, Kapton shims of thickness varying from 40 to 80 $\mu$m were used. The nominal inner yoke and pole radii are set to 6.83199 m from the ring center. Similarly, the yokes will be placed at that same distance.

### Magnet Gap

The nominal gap distance between the lower and upper poles is 180 mm. The gaps between upper and lower poles should be uniform and not tilted when the magnet is energized. Consequently, the upper poles need to be positioned tilted in the radial direction by about 200 $\mu$rad to account for yoke deformation when the magnet is energized. In other words the gap distance at the inner pole radius needs to be 100 $\mu$m larger than the nominal gap. Assuming that the lower poles will be set level with respect to gravity at the center of the ring, the upper poles need to be tilted radially to accomplish this requirement. The placement of the gaps in absolute space is less important than the pole-to-pole gap uniformity. No particular value for the absolute placements of the poles has been specified. However, from the historical BNL 821 data, it is obvious that the achieved pole placement stayed within a tolerance band of 0.5 mm while the relative pole to pole positioning achieved an average value of 12 $\mu$m with an RMS of $\pm 50 \mu m$.

### 9.7.2 Superconducting Coils

The superconducting coils operate at 5200 Amps producing a magnetic field of 1.45 T. There are two inner coils, supported by the upper and lower yokes respectively. The third, outer coil, resides inside the C-magnet close to the vacuum chamber. When cooled down to 5°K, the required coil radial position repeatability is ±0.5 mm, in order to maintain the required field quality. In the cool down process all coils shrink radially by 30 mm while the outer coil expands by 3 mm once powered. The outer coil hangs from the cryostat supported by low thermal-conductivity straps that take up the dynamic changes of the coil during the temperature cycling. The coils are located using radial stops on the inner most radius. For the outer coil the stops transfer the force from the coil to the cryostat box, and push rods from the iron yoke transfer the force from the box to the iron. For the inner coils, pins replace the pushrod. The outer coil is supported at 16 azimuthal locations. Due to the shrinking of the coils when cooled, the outer coil rises by 3.3 mm and needs to be positioned 1.5 mm below the magnet mid-plane under powered conditions. Any vertical adjustments are only possible as long as the upper yokes have not been installed. Consequently this sets the installation sequence. For the positioning of the outer coil, eight radial stops and their associated push-rods are utilized, while eight radial stops are used for the inner coils with the inner cryostat boxes locked at four azimuthal locations.

### 9.7.3 Alignment Concept

Fermilab operates a multitude of accelerators that are monitored utilizing a site wide survey control network. The origin of this network is congruent with the lattice design programs. The major difference between the lattice information provided by the Accelerator Physics Department and the survey control network is the required transformation between the planar rectangular coordinate system of the lattice to a curvilinear geodetic coordinate system that takes the approximate earth curvature into account as most of our survey equipment references to earths gravity. This setup ensures that the transport line feeding into the g-2 ring magnet and the ring structure are properly located to each other. The control network has been established and the 3D coordinates of the control points are known to ±0.5 mm. Based on this information we provided the layout of the incoming beam line and ring support plates.

#### Referencing

The re-assembly of the g-2 experiment at FNAL specifies very tight positioning tolerances, requiring the referencing of all lower yoke steel before placement in the ring. This step calls for the permanent installation of four fiducial markers per yoke. Utilizing a laser tracker system all pertinent yoke features required for the placement of the yokes need to be measured. Using this information we can then best fit the placement of the injection yoke on the incoming injection line and the proper ring location using the control network information. Measuring the locations of the connecting surfaces between yokes one can then determine the necessary shim size between yokes according to the provided BNL historic data. The only variable in this case is the measured nominal radius that needs to be held to a tolerance of $\pm 50 \mu m$.

### Instrumentation

Depending on environmental conditions, the laser tracker system (LTS) [27] provides 3D point accuracies on the order of $\pm 50\mu m$ in a spherical volume with radius up to 10 m. This system will be used for all referencing and positioning tasks of g-2 components (see figure 9.20).

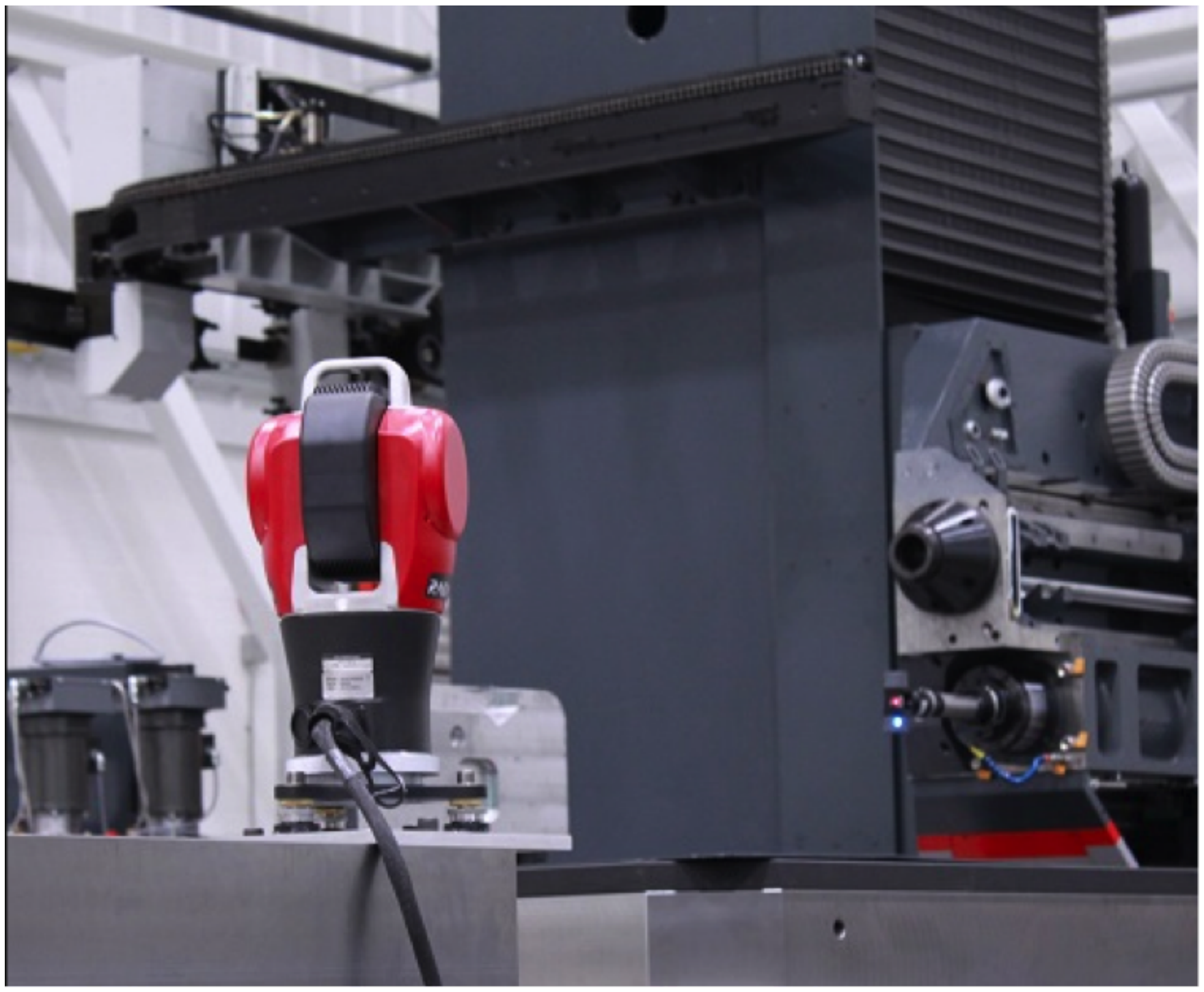

Figure 9.20: The Radian API Laser Tracker System.

Although the LTS is sufficient to provide the demanding position accuracies, it is not capable of fulfilling the elevation requirements. Therefore a separate, ultra-precision HAMAR [27] laser level system will be implemented (see 9.21). Depending on the environmental conditions, this system can provide a level laser plane within $\pm 2.5\mu$rad over a distance of 10 m. It is fast and easy to use and will be handy when frequently checking the levelness of the yokes and poles in between installation steps that may affect that dimension. The laser source will be placed close to the center of the ring on a stable Brunson support, while multiple sensors can be used on ring components. In order to capture any drifts or sudden changes in the laser plane three permanent monitoring stations at the ring in-field constantly monitor the level plane and provide information in case the system exceeds the tolerance range. Both of these systems cannot be utilized under operating conditions when the magnet is powered. At that point the BNL shimming trolley will be utilized. This trolley operates without the vacuum chamber installed and features four capacitance sensors at the four corners of the support frame. The Capacitec sensors measure distances to a fraction of a micron over a very limited range and are used to provide the ultimate gap measurement under operating conditions.

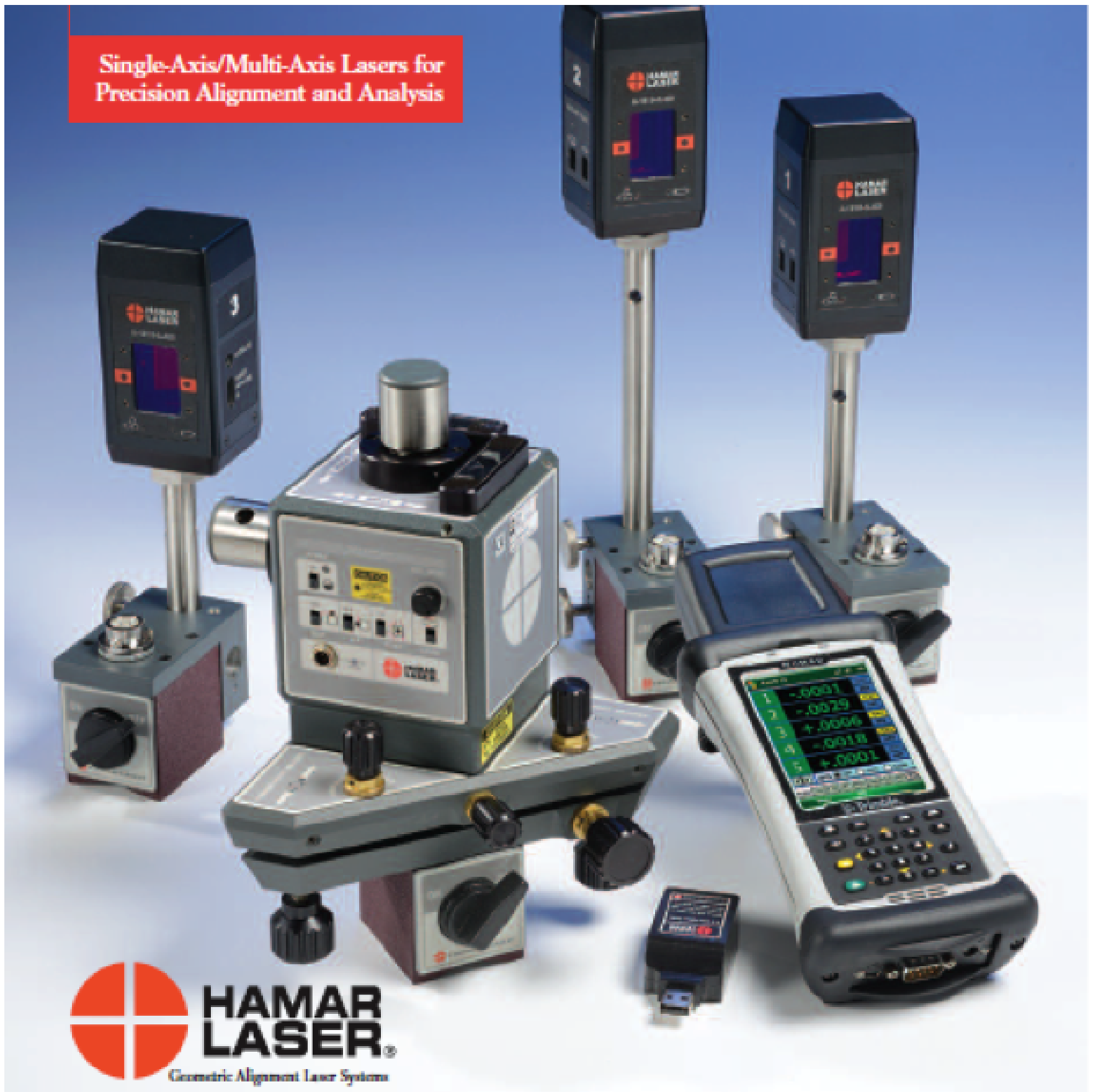

Figure 9.21: The HAMAR Ultra Precision L740 leveling laser system.

**Position Tools**

BNL E821 used primarily the lifting crane, radial adjusters, and jackstand adjustment screws to move the yoke and poles into the final location. For E989, additional tools have been built to adjust the yoke and azimuthal position. A new tool to adjust the elevation of the poles has also been built. These should improve the speed and accuracy of the reassembly.

## 9.8 Main Ring AC Power and Grounding Plan

The AC power distribution and the grounding for the Main Ring within the MC-1 hall are closely intertwined since the requirements for each function requires the interconnection of components for each system. How these components are connected and where they are connected defines the grounding plan for the main ring and can determine if the system will minimize or propagate noise throughout the ring and its individual detectors. Some of these detectors will require a clean AC power source and a clean and stable grounding system.

### 9.8.1 Building AC Power Distribution

The AC power within the MC-1 building is divided primarily into four main functions: 1) Building utilities, 2) Accelerator and Beamline power supplies, and 3) Main Ring detector and power supply loads, and 4) Computer and telecomm loads. AC power enters the MC-1 building through two 750 KVA transformers. One of the transformers, 1A, supplies power for the building utilities, the Main Rings clean power, and computer rooms AC power. The other transformer, 1B, supplies power mainly for the Accelerator power supplies and for the Main Ring and Inflector power supplies. Reference [29] illustrates a very simplified and partial Single-Line Electrical Drawing (SLED) that depicts how these circuits are divided up between the two transformers. Reference [28] contains the contractor or construction drawings used to specifically define all aspects of the electrical installation. However, these drawings do not include the distribution circuits used for the Main Ring since these AC power distribution circuits were defined after the MC-1 contracts were issued.

### 9.8.2 Main Ring and Detector Clean Power

The AC power entering the MC-1 building is usually assumed to meet certain standards as far as distortion from the ideal is concerned. However, there is no set standard to define this distortion in the US. Standards committees, such as IEEE and the IEC, are moving towards developing a universal standard. There already exists a series of guidelines and limitations set forth in several IEEE publications. These standards define limits of any 60 Hz sine-wave deviations such as changes in voltage, frequency, swell, harmonics, etc. If one assumes these standards are met, then the AC power, as seen by the building's feeder transformers, will be pretty much free of these distortions or clean. AC power can be made dirty (distortion) by what the individual loads can put back on the AC line. Thus AC power for certain loads can be kept clean by segmenting them from loads that are known to cause line distortions. Segmenting the loads is done primarily by assigning specific loads to specific distribution transformers. As seen in reference [29], the AC power for the Bruker, for instance, is supplied by a 480 VAC to 480 VAC, delta-to-wye, shielded transformer. This is commonly referred to as transformer isolation. Since only the Bruker power supply is connected to the transformers output, noise and distortion connected upstream on the transformers primary will have little effect on the transformers output and thus the Bruker.

Three other transformers are used in like manner to distribute power to 1) the detector electronic loads, 2) the main ring utilities such as computers and cryo and vacuum pumps, and 3) Main Ring components located outside of the ring such as the Blumlein equipment and various 120/208 VAC outlets. Since each of these transformers are shielded and rated to handle nonlinear loads, such as computers and power supplies, distortion on the secondary of one transformer will not significantly affect the power as seen on the secondary of another transformer. In addition to the separation provided by transformers, there will be transient suppressors installed in each distribution panel to protect the load bus from transients into the panel or transients caused by noisy or faulty loads. Line filtering can also be installed at the load location when incoming power to a piece of equipment or equipment within a rack requires additional protection from AC line noise.

### 9.8.3 Main Ring Grounding Plan

A grounding plan or strategy is designed to specifically define the interconnections to the earth, AC power distribution, and the MC-1 building. Ground connections are used to perform one or more of the three basic functions: 1) Provide electrical safety ground connections, 2) Provide a low-impedance path for HF noise currents to return to their source, 3) Minimize any voltage potential differences between different metal structures and the earth ground potential and thus minimize ground currents through these metal structures.

#### Electrical Safety Connections

Connections for electrical safety are governed by the National Electrical Code (NFPA 70) and the provisions and requirements must be followed. If done properly, most, if not all, safety ground connections will have no effect on an electrical systems noise performance.

#### Low-Impedance Path for HF Currents

HF ground currents, those above 1000 Hz, can be conducted or capacitively coupled between component metal and conductors or can be directly conducted from one metal structure to another. The higher the frequency, the more likely these currents can be coupled. Low-impedance ground connections can provide a path for these currents that is controlled and thus reduce the probability that they will couple into an input to a high-impedance amplifier or some other sensitive circuit component.

#### Minimize Voltage Potentials

To minimize DC or LF (those below 1000 Hz) currents from flowing through metal structures, grounding connections are specifically made between these structures and to the earth grounding electrode or electrodes. Ideally, the potential between the earth, the building, the electrical distribution, and Main Ring and Detector metal structures would all be zero volts. No potential, no ground currents. However, this is not possible. To implement these connections, two configurations are used: 1) a star or point ground configuration, 2) a grid grounding configuration. Each has their advantages.

### 9.8.4 Basic Grounding Connection Rules

The basic or fundamental set of rules must accompany a grounding plan. For the Main Ring and MC-1 Hall grounding connections, the following set of rules were chosen:

- Connect the Main Ring metal structures to the MC-1 building structures at one point.
- No floating pieces of metal. All metal or chassis components shall be grounded.
- Follow specific routing of cables along grounded structures. No point-to-point cabling.
- Most cable shield connections are to be made on a case-by-case basis. Generally, shields for cables that exit or enter the Main Ring are connected only at one point or to only one ground structure (coax cables being a special case).

- Avoid the creation of ground loops by adhering rules 3 and 4 listed above.
- Avoid the use of continuous metal or conductors that follow the perimeter of the Main Ring and form a circular path.

Keep in mind that the details of how these rules are implemented and where the grounding connections are made determine how successful the plan can be in creating a low-noise, solidly-referenced grounding environment.

### 9.8.5 Main Ring Grounding Configuration

The implementation of the Main Ring grounding plan starts with a basic diagram and the connection of four main components: 1) the Main Ring structure, 2) the Main Ring grounding structure arranged in a wheel spoke pattern, 3) the MC-1 building ground structure, and 4) the cable and interconnecting bridge that serves as the start-point grounding connection between the Main Ring and building ground. This star ground configuration prevents building ground currents from being conducted through the Main Rings metal components and ground structure from different sectors of the buildings ground structure. The main disadvantage with this configuration is that effort must be taken to determine and control all connections between the Main Ring and the building.

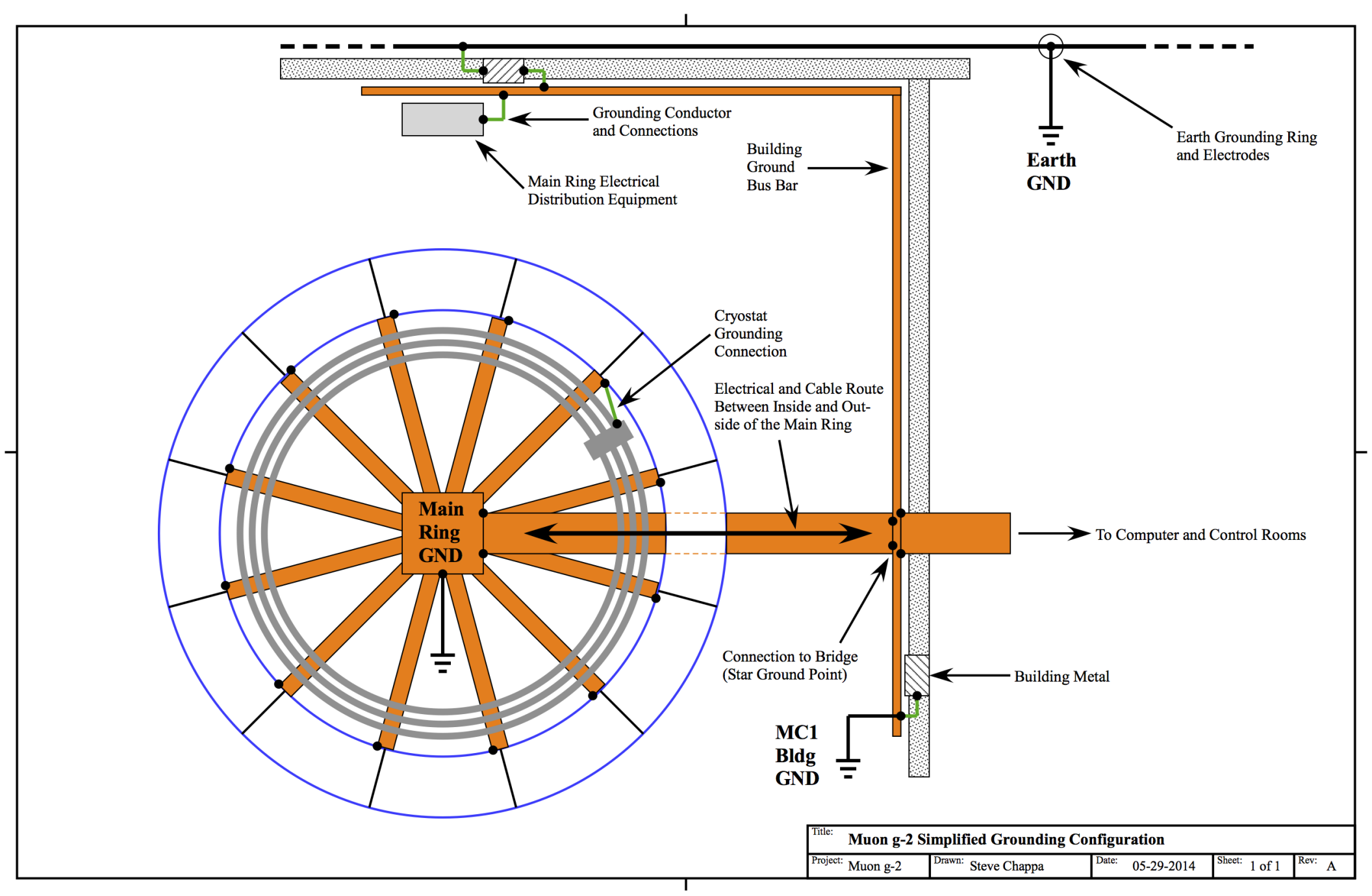

Figure 9.22: A simplified drawing of the grounding configuration. See reference [31] for detailed drawings.

### Main Ring Structure

This structure consists of the cryostat rings, the yoke and pole pieces, and ring interconnect. Each of these metal pieces need to be kept at the same potential to limit any current paths between them and to keep the eddy currents minimized by having relatively high-impedance breaks between these yoke sections. Notice that the common connection point for the three cryostat rings is connected to the Main Ring grounding structure only at one point (near the bridge).

### Main Ring Grounding Structure

To achieve the grounding connections for the different yoke sectors and for the pole pieces, plus various chassis or metal components on each sector, the grounding path cannot be a circular copper bus as would be easy to implement. Instead, in order to adhere to rule 6, a spoke pattern of flat copper plates (wide copper for HF low-impedance) is used to provide a solid grounding connection for each yoke sector and the metal pole pieces and components within that sector.

### Building Grounding Structure

A good building ground structure has all metal parts of that building, the electrical distributions grounding connections, and the earth grounding electrode connections arranged so that there is as much uniformity of ground potential throughout the building. However, this is not the case for the MC-1 hall. The electrical grounding connection to the building is a rather long distance from the area where the Main Rings star ground connection needs to be made. Because of this distance and the limited building connections to the electrical system, additional ground connections between the building and electrical system needs to be made using the wide copper bus bar that runs along the south and west walls (see [31]). This bus also provides a solid, low-impedance grounding connection for the Main Ring. Notice also since most of the cables exiting the Main Ring Needs to go to the computer or the control room, a cable tray with the flat copper plate is used to provide a low-impedance ground connection between the computer and control rooms and the Main Ring star ground point.

### Connection Bridge or Star Point

The center part of the spoke grounding pattern is then connected to the MC-1 building ground by the use of a wide copper bridge (wide copper for minimizing impedance at high frequencies). This bridge actually consists of all components necessary for electrical distribution, signal cables, cable tray, and kicker HV coax cables. The reason for this bridge is to provide a controlled path for all cables that connect between the building and Main Ring. By keeping these cable paths next to the common ground structure, this minimizes any cable loop area that might develop between the Main Ring and MC-1 building grounds and thus minimize the likelihood of creating ground loops that can cause noise problems for systems in and outside of the Main Ring.

## 9.9 ES&H, Quality Assurance, Value Management, Risks

### 9.9.1 Yoke, Pole, and Shims

The hazards will be in the stored mechanical energy of these very heavy items during the use of the crane during installation. While the magnet is powered, the super bolts will be stressed (stretched) to counteract the magnetic forces' tendency to close the pole gaps. The stored electro-mechanical energy is approximately 6 MJoules. The storage region magnetic field of 1.45T is also a hazard.

Mitigation would be to train the collaboration in proper procedures, controlling access to the area, and implementing a mechanism for detecting high magnetic susceptibility materials so that they are kept away from the high field area.

Quality assurance, value management, and risks concerns are minimal since these items have been built and worked to specifications for E821. They are also passive materials and have high mechanical strength. Since considerable E989 machinery has already developed to model the magnetic field and temperature, we can simulate all alignment requirements.

### 9.9.2 Power Supply and Quench Protection System

#### Hazards

The hazards during the testing processes will be primarily electrical. There will be an arc-flash hazard based on the incoming 480 VAC power feed and at 60-amps. A complete hazard analysis can be made once the units are received and initially examined. The mitigations for these hazards are:

- In the staging area, a safety disconnect switch will be used to provide emergency shut-off of power to the power supply.
- No operation of any component will be done unattended. There will always be 2 persons minimum present during any operation and testing.
- Only properly trained persons, Electrical NFPA-70E and LOTO II at a minimum, will be allowed to work on the equipment.
- Proper PPE and distances will be observed during operation and testing, especially during initial power up and during full-load operation. The level of PPE will be determined once the staging area is set up.

#### Risks

Risks to the successful completion of the tasks:

- Use of Obsolete Components: Level-Medium. This risk involves the use of relatively old electronics. Therefore, component failure probability (based on hours of previous operation and power cycles, will need to be assessed for critical components of the power supply. Mitigation, at a minimum, will be to list these components and acquire spares in case of failure. To minimize the effect this risk will have on the schedule, the component list will be generated in parallel with the other operations. Initial costs will be incurred to acquire these spare components.

- Catastrophic Failure During Testing: Level-Low to Medium. The units have operated before and they operated as initially designed. Therefore, testing of the components and of the units in stages should provide early indications of failure and allow replacement before the start of connection and commissioning.

- Difficulty in Acquiring the Proper Testing Equipment: Level-Low. At this time, most equipment needed for testing, is, or should be, readily available on site. Therefore, purchases of new equipment will be at a minimum.

- Catastrophic Failure During Commissioning: Level-Low. The risk here would be the most critical since failure of any component while connected to the ring coils would also present risk to the coils themselves. Mitigation here would be the exhaustive testing of components prior to connection to the ring and the powering up of the ring in stages of load and cooling. In addition, specific procedures for connection to the rings leads will be repeatedly and continuously reviewed.

**Quality Assurance**

The assurances that the units will operate as required are two-fold: First, the design is already done. Redesign should be at a minimum. Second, there will be exhaustive testing of each component prior to re-assembly. These testing procedures will be continuously reviewed during and after each stage of testing. Again, the primary process for all this is the initial examination followed by testing in stages as opposed to a one-time, massive final test. The staged testing allows deficiencies or problem areas to be identified at the earliest possible point in the task schedule. This again stresses the importance of the initial examination for the results of this will determine, in large part, the testing schedule.

**Value Management**

As much reuse of components will be implemented throughout these processes. Also, preliminary examination will serve to spot questionable components. Overall, the refurbishment process is used to minimize the need to purchase or redesign components and systems. Since the design is already done and it operated as designed, the risk and extra effort in developing a new design can be kept to a minimum.

### 9.9.3 The Cryogenic and Related Systems

The hazards are related to the use of LHe and $LN_2$ cryogenics: thermal energy and ODH. As described above, if the cooling lines in the vacuum cryostat leak, there is potential for the 'cold-cryostat' scenario (see above). These can be mitigated in the same fashion as E821. The cooling lines and vacuum chambers can be pressure tested at room temperature, following delivery from BNL and prior to use.

New quality assurance and value management requirements are minimal since these items have been built and worked to specifications for E821. Should the items fail during recommissioning, they can be easily replaced since they are commercially available items of reasonable cost. These are vacuum parts and cryogenic lines. The only outstanding technical skills required are Aluminum welding and vacuum leak-testing.

There are inherent risks in nearly all cryogenic systems since these are usually very complex and have long time scales. Even though E821 operated a successful system, significant engineering is required for recertification. A mitigating factor is that similar systems have been built at Fermilab, and is inline with the expertise of the project mechanical engineer.

### 9.9.4 The Superconducting Coil System

The coils systems do not present a significant health hazard. Quality assurance and value management concerns are minimal since these items have been built and worked to specifications for E821.

The risks are not considerable. The risk of damage to the coils, straps, heat shields, glue joints are minimal since the stresses expected during transportation are at least 4x smaller than the maximum allowable stress. The expected stresses and deflections due to transportation have been simulated by FEA.

However, we itemize them here since the replacement cost of the coils is beyond the scope of the project:

- The coil windings are on the inside radial surface of the mandrel, rather than on the outside. During power up, the coils push against the mandrel, thereby enhancing the thermal cooling. Therefore, the system can tolerate failures of the glued interface, which is designed to enhance thermal conductivity.

- Failure of the straps, though unexpected, can be detected as we slowly energize the coils. If a strap fails, it can be detected as a shift in the coil vertical position. It can be repaired by cutting an access hole in the vacuum cryostat. The downtime would be of order 2 weeks.

- Failure in the heat shield can be detected as the system taking too long for cool down. Failure of the cooling lines can be detected as loss of vacuum. These are repaired via cutting access holes into the vacuum cryostat. However, it will be difficult to locate the point of failure.

- There is a very slight risk of the Aluminum resistance at the interconnection becoming too high during the reweld. We are currently prototyping the rewelding process and study the correlation of resistance with yield strength. We can anneal the interconnection to improve the resistance.

# Chapter 10

# The Superconducting Inflector Magnet

In this chapter we first introduce the E821 inflector magnet, which is our baseline starting option. We then describe the shortcomings of this magnet, as well as the characteristics and the benefits that an improved inflector would have. During the period between the CD1 Review and the present, an Inflector Task Force (ITF) [1] has been set up to develop a design for a new inflector. Two options are being explored: A direct wind magnet using the technique developed at Brookhaven Lab by Brett Parker, and a double cosine theta magnet following the E821 design, but with open ends and a wider beam channel.

At present the new inflector is outside of the scope of the project budget cap. Because of the real benefits to the experiment, the new inflector is being studied as an 'optional scope contingency', to be brought into the baseline if we prove contingency is available. We will continue the design work, so that if/when adequate contingency becomes available, we can move forward with the production.

## 10.1 Introduction to Inflection Challenges

The typical storage ring is composed of lumped beamline elements such as dipoles, quadrupoles, sextapoles, etc., which leaves space for injection, extraction, and other beam manipulation devices. For the measurement of $a_\mu$, the requirement of $\pm 1$ ppm uniformity on the magnetic field, which in E989 must be known to $\leq \pm 70$ ppb, prohibits this usual design. Instead, as described in Chapter 9 the $(g-2)$ storage ring is designed as a monolithic magnet with no end effects. The "C"-magnet construction shown in Fig 9.1 presents several obstacles to transporting a beam into the storage ring: There must be holes through the back-leg of the magnet and through the outer coil cryostat and mandrel for the beam to enter the experiment. These holes must come through at an angle, rather than radially, which complicates the design, especially of the outer-coil cryostat.

A plan view of the beam path entering the storage ring is given in Fig. 10.1. Since the beam enters through the fringe field of the magnet, and then into the main 1.5 T field, it will be strongly deflected unless some magnetic device is present that cancels this field. This device is called the inflector magnet. In reality, there is a fringe field that grows approximately linearly as the beam moves radially inward from the hole in the outer cryostat to the location of the inflector entrance. This is sketched in Fig. 10.2.

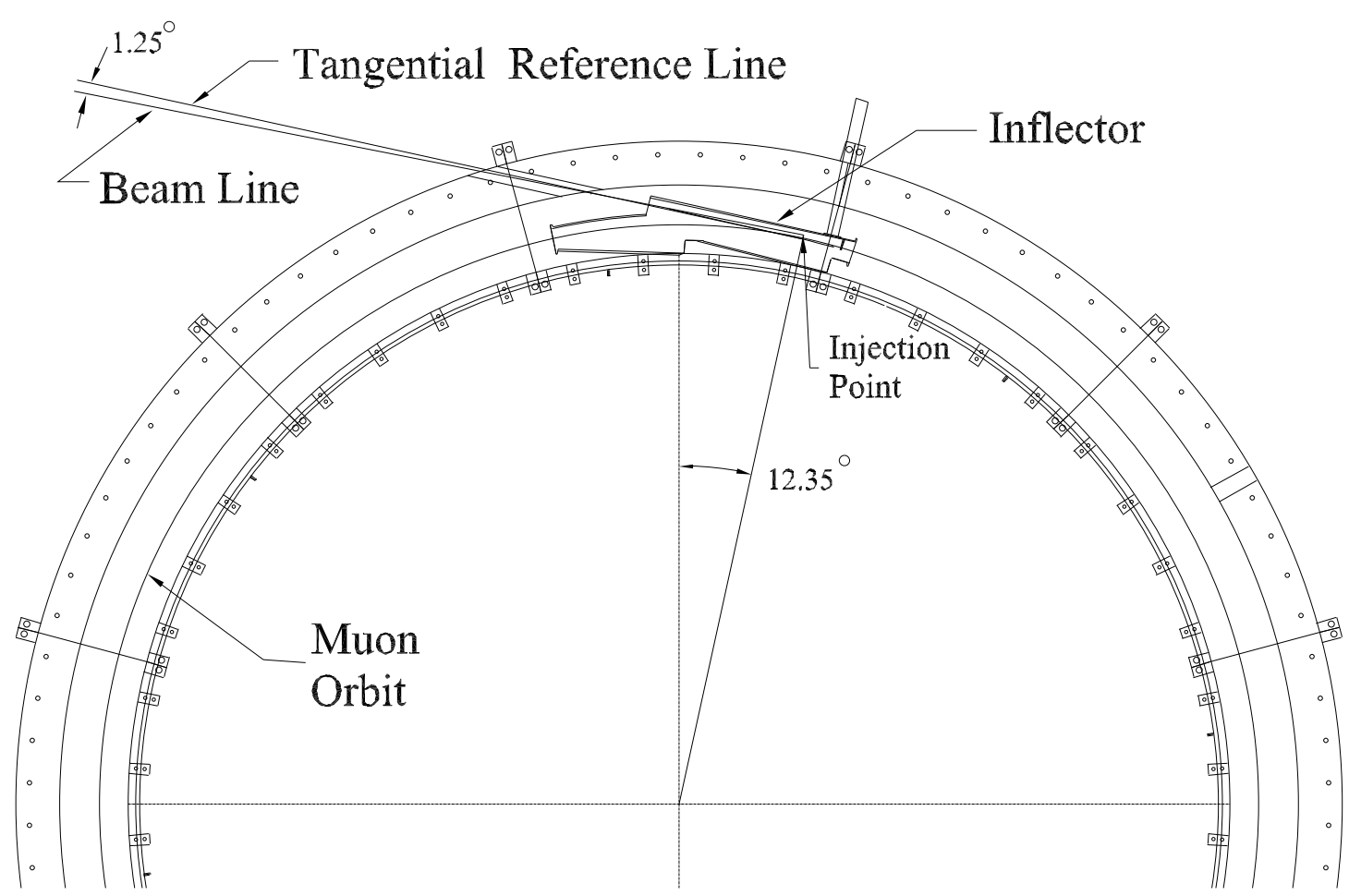

Figure 10.1: Plan view of the beam entering the storage ring.

The injection beam line is set to a 1.25° angle from the tangential reference line (Fig. 10.1). The inflector is aligned along this reference line and its downstream end is positioned at the injection point, which is tangent to the ring. The point where the reference line is tangent to the storage ring circumference is offset 77 mm radially outward from the muon central orbit. The main magnet fringe field, upstream of the inflector, bends the incoming beam by about 1.25°, so that the beam enters the inflector nearly parallel to the inflector axis.

The requirements on the inflector magnet are very restrictive:

1. To a good approximation it should null the storage ring field such that the muons are not deflected by the main 1.5 T field.

2. It should be a static device *to prevent time-varying magnetic fields correlated with injection*, which could affect $\int \vec{B} \cdot d\vec{\ell}$ seen by the stored muons and produce an "early to late" systematic effect.

3. It cannot "leak" magnetic flux into the precision shimmed storage-ring field that affects $\int \vec{B} \cdot d\vec{\ell}$ at the sub-ppm level.

4. It cannot contain any ferromagnetic material, which would distort the uniform magnetic field.

5. The inflector should have a "reasonable" aperture to match the beamline to the ring acceptance.

6. The inflector angle in the cryostat should be variable over the full range permitted by the constraints of the space available.

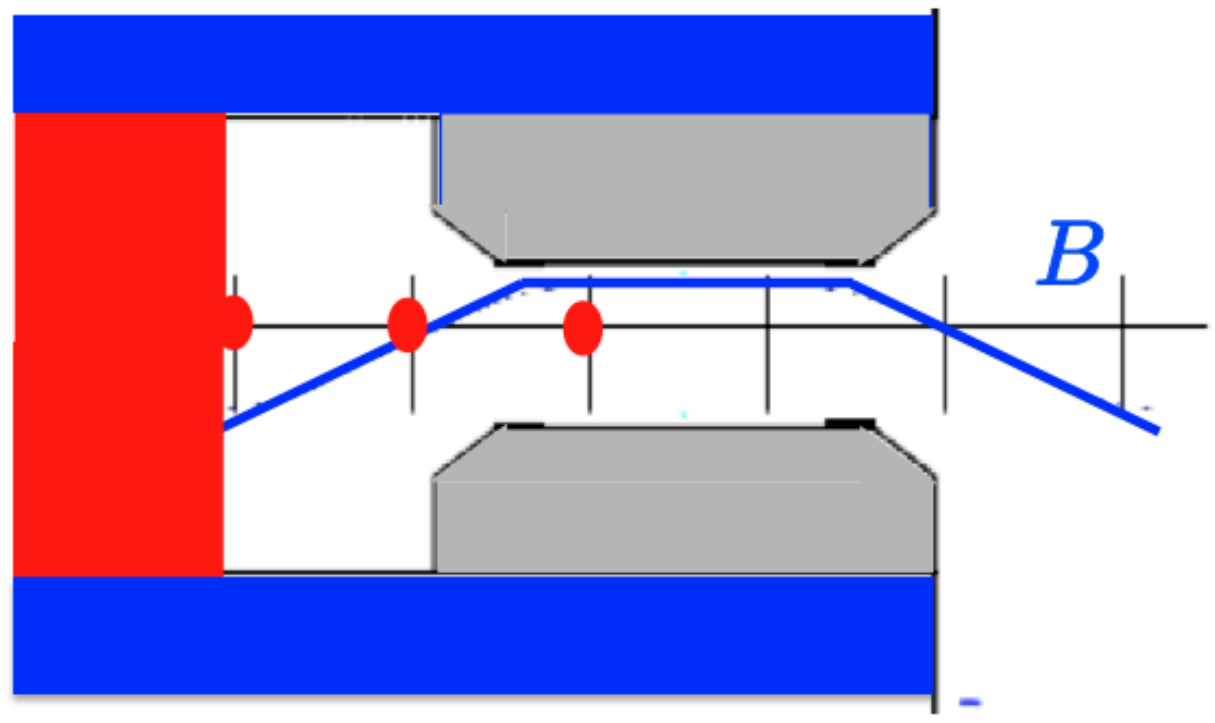

Figure 10.2: The fringe field of the main magnet over the radial range traversed by the beam. The left-hand red dot is where the beam exits hole in the outer coil. The center dot is where the beam enters the inflector. The right-hand dot is where the beam exits the inflector. The field inside of the inflector is not constant until part way down.

## 10.2 The E821 Inflector Design and Operation

Three possible solutions were considered in E821: A pulsed inflector, a superconducting flux exclusion tube, and a modified double $\cos\theta$ magnet. The pulsed inflector proved to be technically impossible at the repetition rate necessary at BNL. Furthermore it violates item 2 above. Naively one could imagine that a superconducting flux exclusion tube would work for this application. However, an examination of Fig. 10.3 shows that in the vicinity of the tube, the magnetic field is perturbed on the order of 10%, or 100,000 ppm [2], an unacceptable level. Attempts to figure out how to mitigate this problem were unsuccessful. This is because the large eddy currents needed to shield the 1.45 T field are large enough to affect the uniformity of the field seen by the muons contained in the red semicircle. However, this principle will re-appear in the discussion of how to shield the 200 G (20 mT) residual magnetic field from the truncated double $\cos\theta$ design employed in the E821 inflector. The properties of the E821 Inflector are summarized in Table 10.1

Table 10.1: Properties of the E821 Inflector.

| | |
|---|---|
| Overall dimension | 110(W)× 150(W)×2025(L) mm$^3$ |
| Magnetic length | 1700 mm |
| Beam aperture | 18 mm (W) × 56 mm (H) |
| Design current | 2850 A (with 1.45 T main field) |
| Number of turns | 88 |
| Channel field | 1.5 T (without main field) |
| Peak field | 3.5 T (at design current, with main dipole field) |
| Inductance | 2.0 mH |
| Resistance | 1.4 Ω (at 300 K) |
| Cold mass | 60 kg |
| Stored energy | 9 kJ (at design current) |

Table 10.2: Properties of the inflector superconductor.

| | |
|---|---|
| Configuration (NbTi:Cu:Al) | 1:0.9:3.7 |
| Stabilizer | Al (99.997% RRR = 750 |
| Process | Co-extrusion |
| NbTi/Cu composite | Diameter 1.6 mm monolith |
| NbTi filament | Diameter 0.02 mm |
| Number of filaments | 3050 |
| Twist pitch | 31 mm |
| Conductor dimension | 2 × 3 mm$^2$ |
| Insulated conductor dimension | 2.3 × 3.3 mm$^2$ |

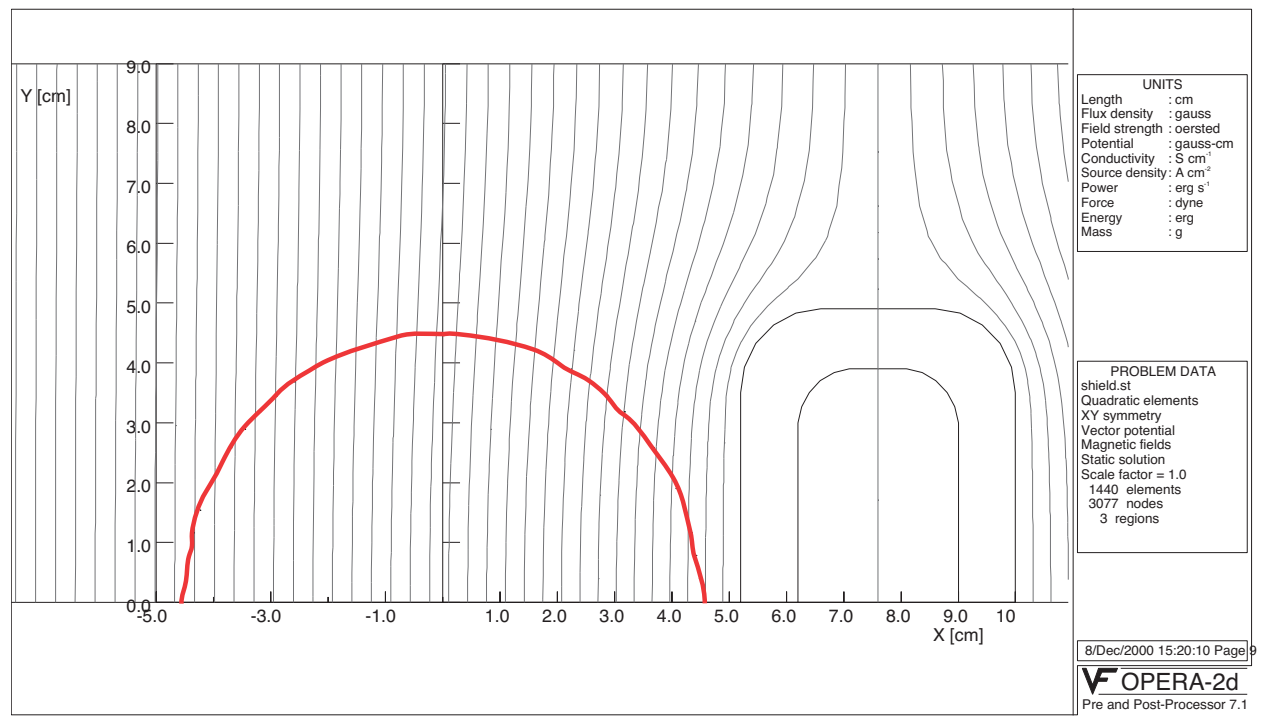

Figure 10.3: The calculated magnetic field outside of a superconducing flux exclusion tube placed in a 1.45 T magnetic field. The red circle is the muon beam storage region. (From Ref. [2])

### 10.2.1 Magnetic Design of the E821 Inflector

Only the double $\cos\theta$ design[4] satisfied the three criteria listed above. The double $\cos\theta$ design has two concentric $\cos\theta$ magnets with equal and opposite currents, which outside has negligible field from Ampère's law. A double $\cos\theta$ design provides a 1.5 T field close to the storage region, and traps its own fringe field, with a small residual fringe field remaining. However, what is needed for the $(g-2)$ beam channel is a septum magnet. This is achieved by truncating the two $\cos\theta$ distributions along a line of constant vector potential $A$ [4]. The truncation method is shown in Fig. 10.4, taken from Ref. [4], which should be consulted for additional details.

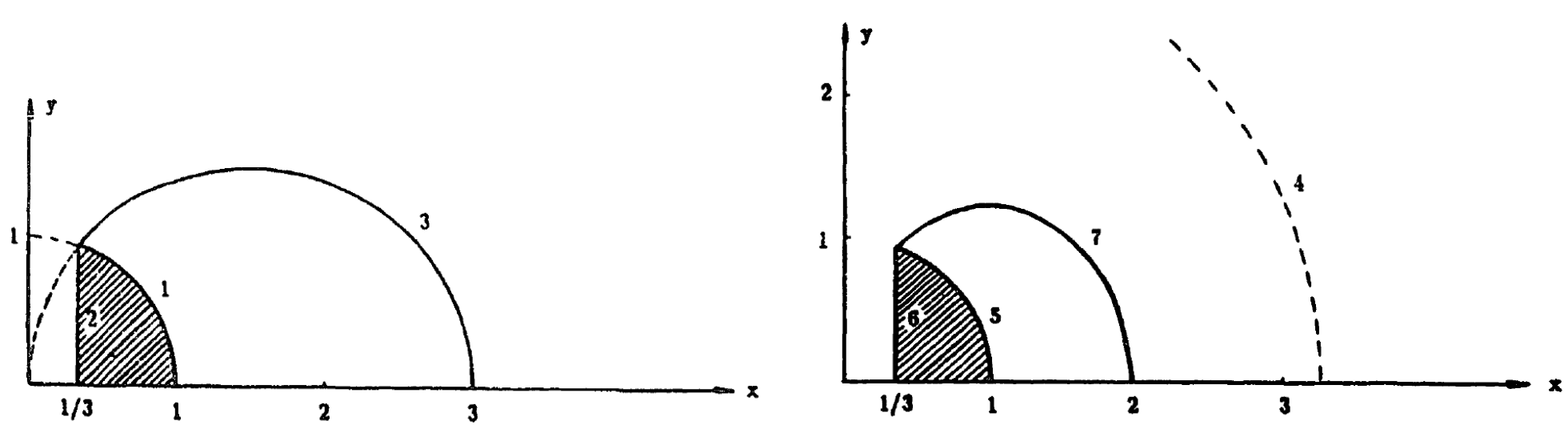

(a) Truncated Single $\cos\theta$ (b) Double Truncated $\cos\theta$

Figure 10.4: (a) The principle of the truncated single $\cos\theta$ magnet. (b) The principle of the truncated double $\cos\theta$ magnet.

Aluminum-stabilized superconductor was chosen for the BNL $(g-2)$ inflector: (a) to minimize the interactions of the incoming pion/muon beam at both upstream and downstream ends of the coil with no open apertures for the beam, and (b) to make the coils and cryostat design compact, so that the conductive cooling (without liquid helium containers surrounding the coils) can be achieved effectively. An existing Al-stabilized superconductor was supplied by Japan KEK (fabricated by Furukawa Co.). This conductor was developed for ASTROMAG (Particle Astrophysics Magnet Facility) [5, 6]. Fig. 10.5 shows the cross-section of this conductor. The basic parameters are listed in Table 10.2. From computer calculations, which include the self-field effect [7], show that the peak field seen by the inflector conductor filaments reaches 3.5 T. This is due to the superposition of the return flux and the main field. Short sample tests showed that the critical current of this superconductor is about 3890 A at 4.6 K and 3.5 T. In the $(g-2)$ storage ring, the inflector sees 1.45 T field (from the main magnet) even at zero operating current. From the conductor characteristics, the inflector operates at around 73% of the full load (at 4.6 K). The short sample test data and the inflector load line (in the storage ring field environment) are shown in Fig. 10.5(b).

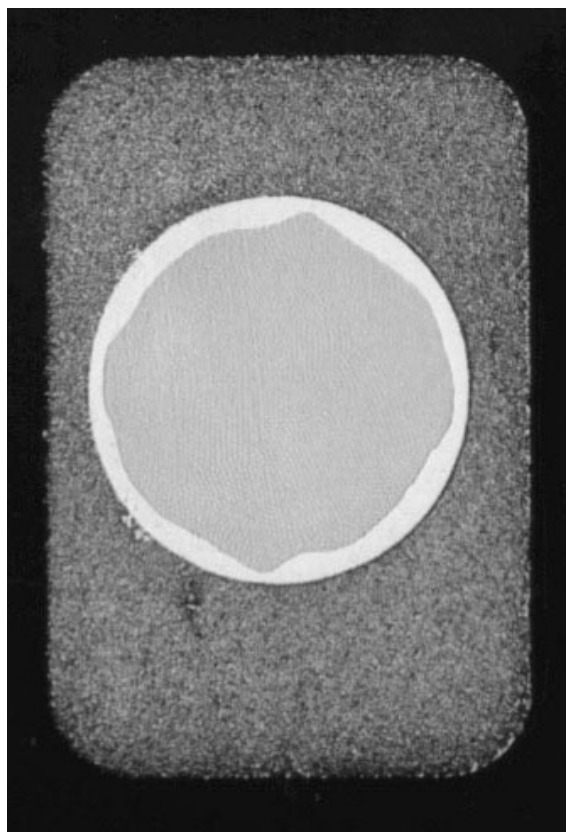

(a) SC cross-section

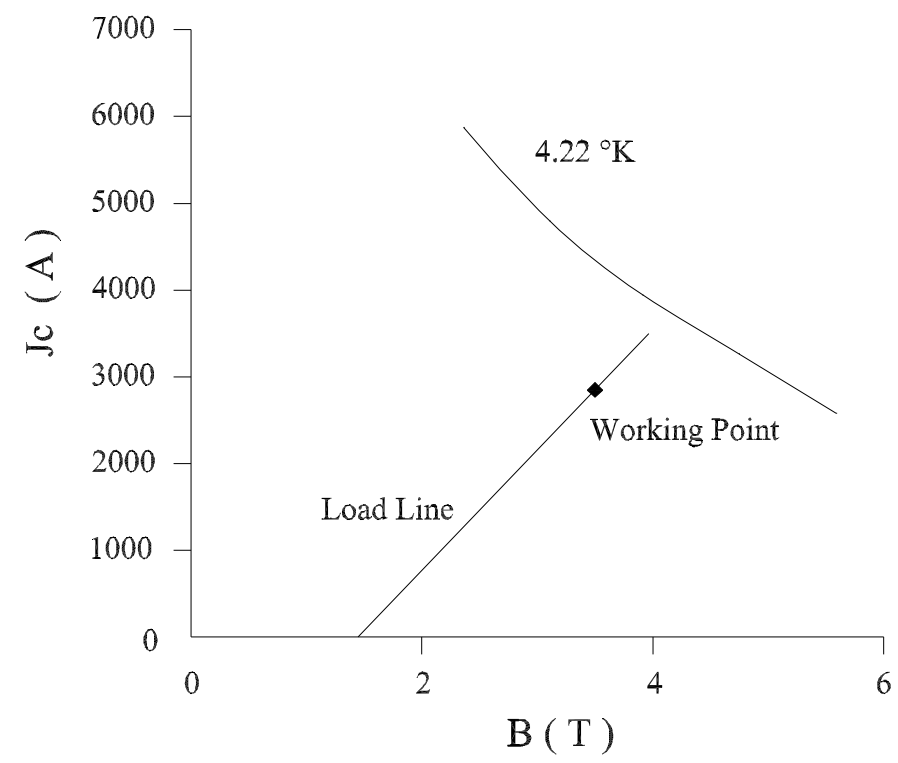

(b) Inflector Load Line

Figure 10.5: (a) The inflector superconductor cross-section. (b) Superconductor characteristics and the inflector load line in the environment of 1.45 T magnetic field.

The result is a magnet with conductors arranged as shown in Fig. 10.6(a). The conductors are connected in series, with an equal number with current into and out of the page. In Fig. 10.6(a) the current is flowing out of the page in the backward “D” shaped pattern of conductors, and into the page in the “C” shaped arrangement of conductors. The field from the inflector magnet is vertical up in the beam channel and downward in the return area, as shown in Fig 10.6(a). With the main storage ring field vertical, there is no field in the beam channel and $\simeq 3$ T field in the return area. With this design and the ASTROMAG conductor, it is difficult to enlarge the beam channel very much because moving the “C” arrangement of conductors to the left would quickly exceed their critical current.

There are two sources of magnetic flux from the inflector that can leak into the storage region. Because the field is produced by discrete conductors, rather than a continuous current distribution, some flux does leak out of this arrangement of conductors, see Fig. 10.6(b). The inflector lead configuration is also important, and when it was necessary to produce a second inflector, the lead configuration was changed to reduce this effect.

The coil was wound in two different pieces indicated by “inner” and “outer” coils in Fig. 10.6(a). One end of the coil is shown in Fig. 10.7(a) The choice was made to wind the coil over the beam channel, because this configuration would have less flux leakage, and was thought to be more stable from quenches. However, a 0.5 m prototype was constructed with one open and one closed end, which are shown in Fig. 10.7. This prototype inflector was operated in the earth's field, and then in an external 1.45 T field without incident.

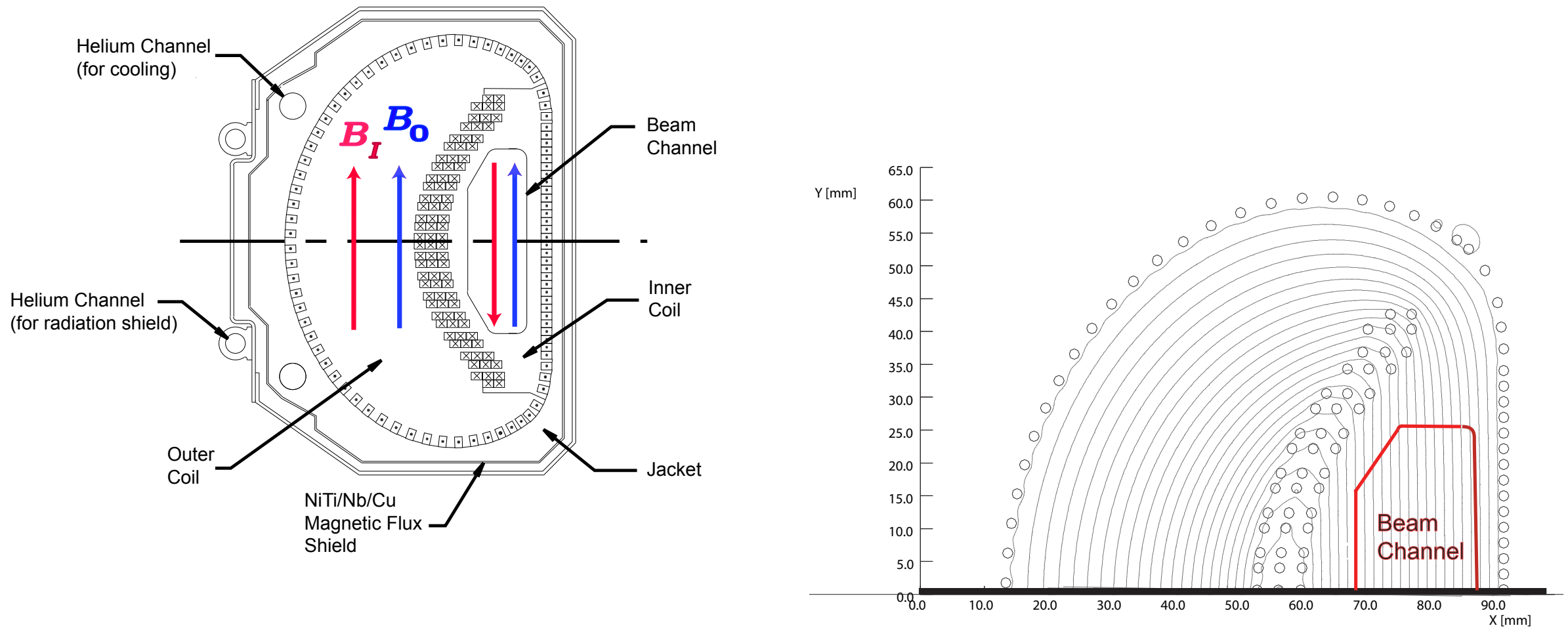

(a) Inflector Conductor Arrangement (b) Calculated Field

Figure 10.6: (a) The arrangement of conductors in the inflector magnet, showing the direction of the inflector field $B_I$ and the main field $B_0$ for a beam of positive muons going into the page. The current in the inner "C" is into the page and is out of the page in the backward "D". (b) Magnetic field lines generated by this arrangement of conductors. The beam aperture is $18 \times 56$ mm$^2$.

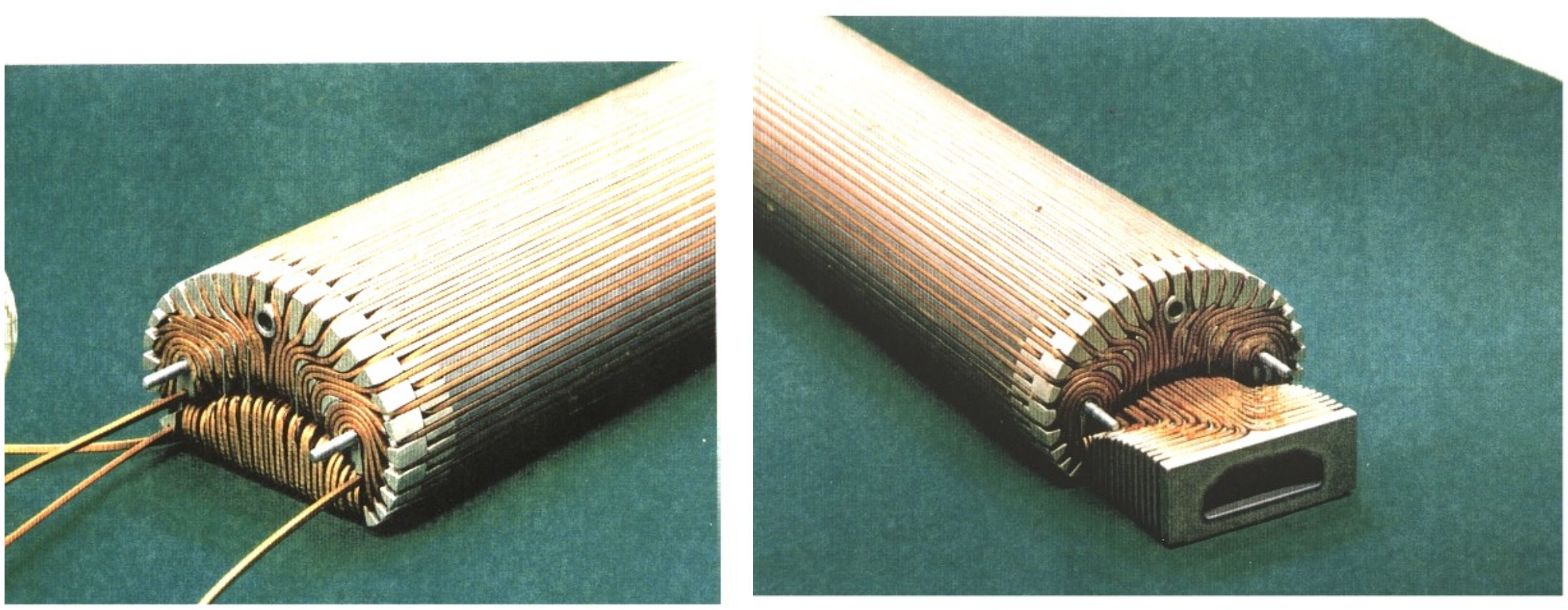

(a) Closed Inflector End (b) Open Inflector End

Figure 10.7: (a) The prototype closed inflector end. (b) The prototype open inflector end.

The inner coil and the outer coil are connected in series. The joint is located inside the downstream end of the coils; and is made by soldering the superconductors without removing the aluminum stabilizer. The joined leads were placed inside a U-shaped groove, as shown in Fig. 10.8, attached to the coil end structure. Cooling tubes run through the extender (aluminum block). One temperature sensor was mounted near the joint to monitor the local ohmic heating.

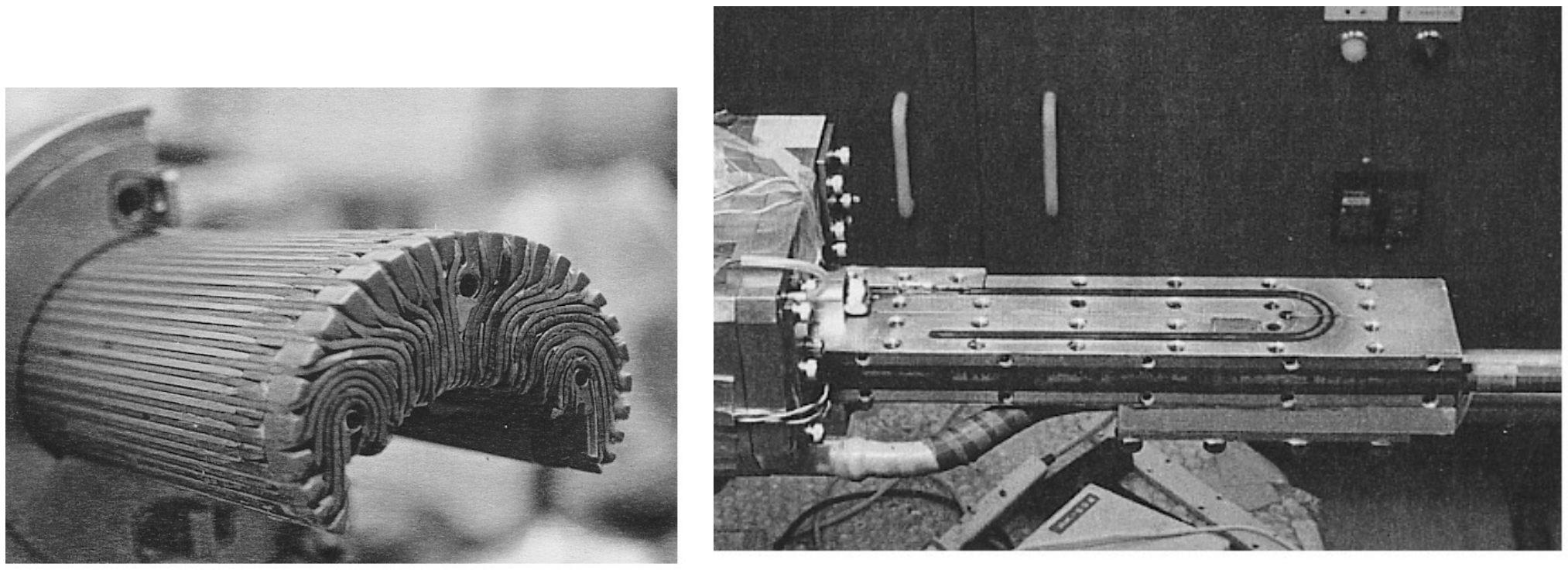

(a) Outer Inflector Coil (b) Coil Interconnect

Figure 10.8: (a) The arrangement of conductors in the inflector magnet.(b) The joint and lead holder for the interconnect.

The geometry of the inflector cryostat is complicated by the proximity of the outer-coil cryostat, the pole pieces and the muon beam. A sketch of the beam path through the outer coil is shown in Fig. 10.9(a). The complicated arrangement where the inflector entrance nests into the concave wall of the inflector cryostat is shown in Fig. 10.9(b). Fig. 10.11 shows the combined inflector cryostat and beam vacuum chamber. The cryostat region and beam region have different vacuums, so the inflector can be cooled, independent of whether the beam vacuum chamber is evacuated.

The exit of the inflector magnet is shown in Fig. 10.14, which clearly indicates the accelerator physics issue. The incident beam is contained in the red 18 mm $\times$ 56 mm "D"-shaped channel, while the stored beam is confined to a 90 mm diameter circular aperture. Thus it it impossible to match the $\beta$ or $\alpha$ functions between the ring and the muon beamline without unacceptable losses in the injection channel The result is a "$\beta$ wave" that reduces the acceptance of the ring.

### 10.2.2 Shielding the residual fringe field

At the design current, the maximal fringe field within the muon storage region was calculated to be about 200 G (1.4%) near the outer edge. The fringe field behaves in such a way that it is a rapidly varying function along the transverse direction, i.e. the radial direction of the storage ring, and essentially gives a negative disturbance. The fringe field of the inflector is opposite to the main field at the outer radius of the storage ring, and changes sign while crossing the central orbit.

The consequence of such a fringe field is severe. The high gradient of the field is beyond the working range of the NMR probes, so that the magnetic field map of the storage region would be incomplete, directly impacting the error of the measurement precision of the muon magnetic moment.

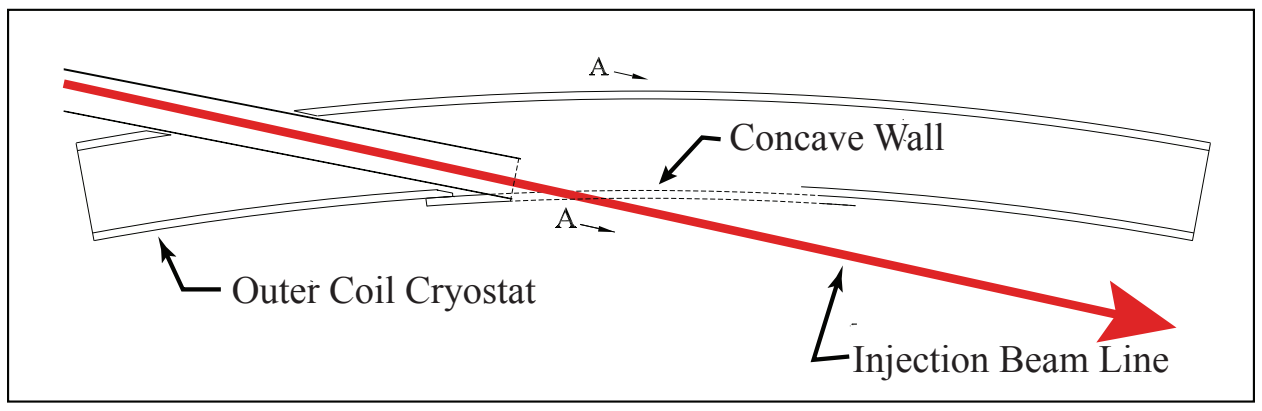

(a) Outer Coil Penetration

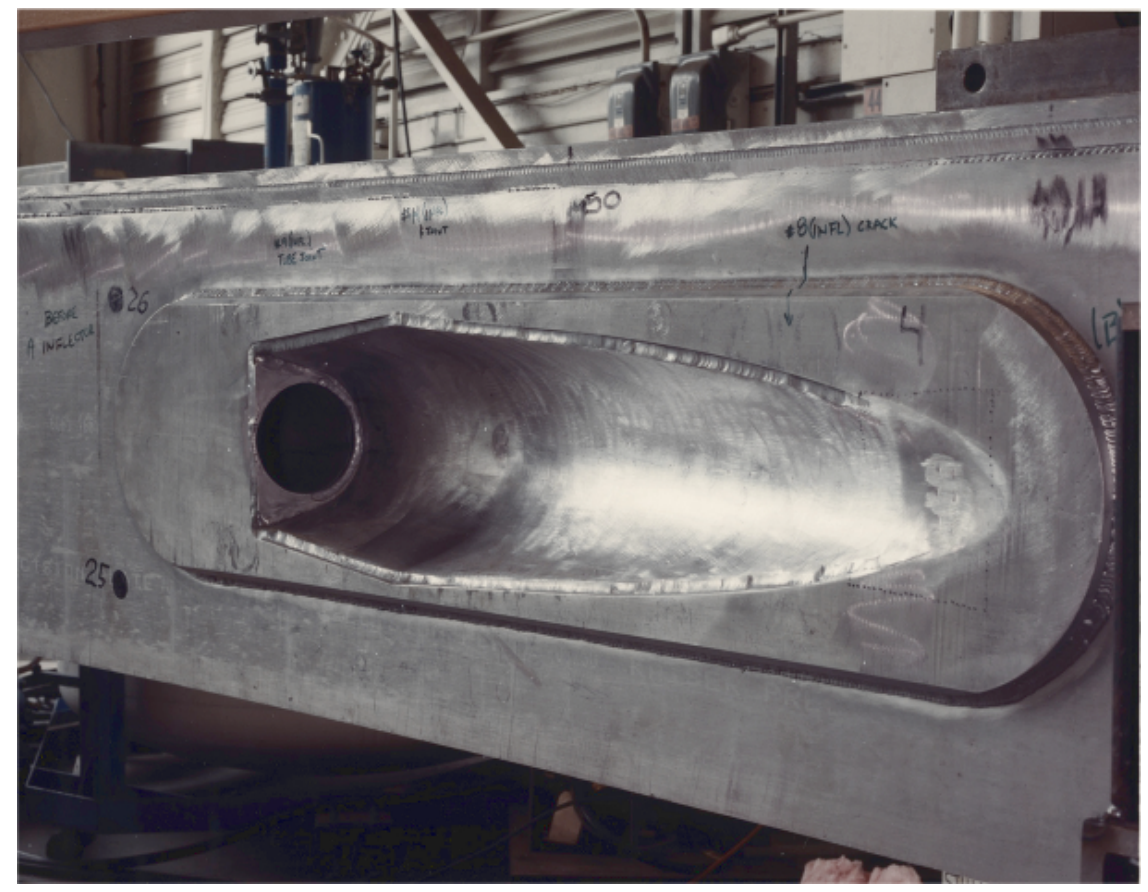

(b) Outer Coil Penetration

Figure 10.9: (a) A plan view of the beam penetration through the outer coil and cryostat. (b) A photo of the hole in the outer cryostat that the muon beam passes through.

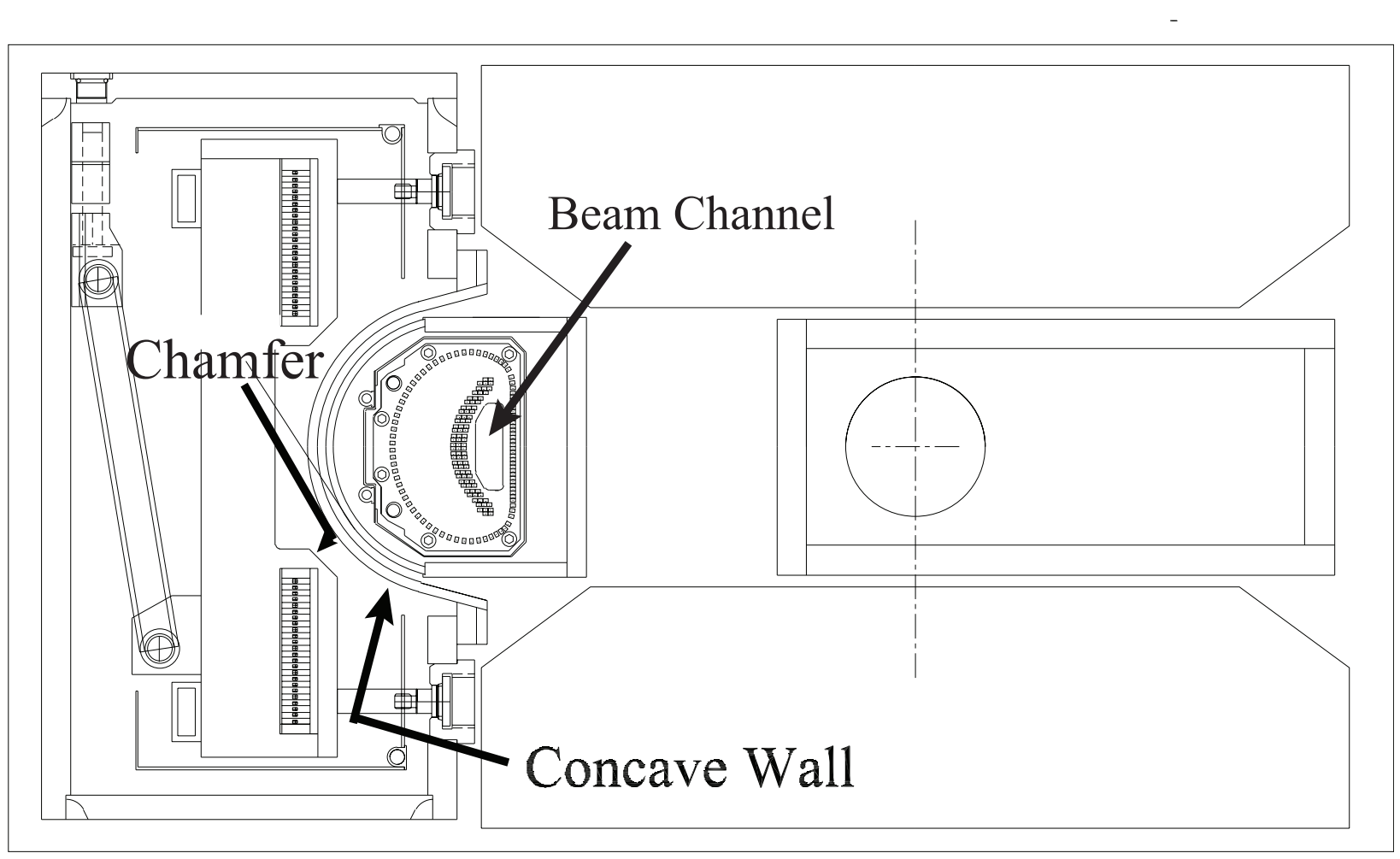

Figure 10.10: An elevation view of the inflector entrance showing the concave wall of the outer-coil cryostat where the beam exits the outer coil-cryostat.

Conventional magneto-static shimming studies to reduce this fringe field using computer simulations were carried out. The iron compensation must be located outside the muon storage region, far from the disturbance it is trying to shield. Thus its contribution to the central field would be a slowly varying function in this space (long wavelength), which is not able to cancel the larger gradient fringe field to an acceptable level [12].

The best way to eliminate a multipole fringe field is to create an opposite multipole current source with the same magnitude. The best such current source is the super-current generated inside a superconducting material due to the variation of the surrounding field.

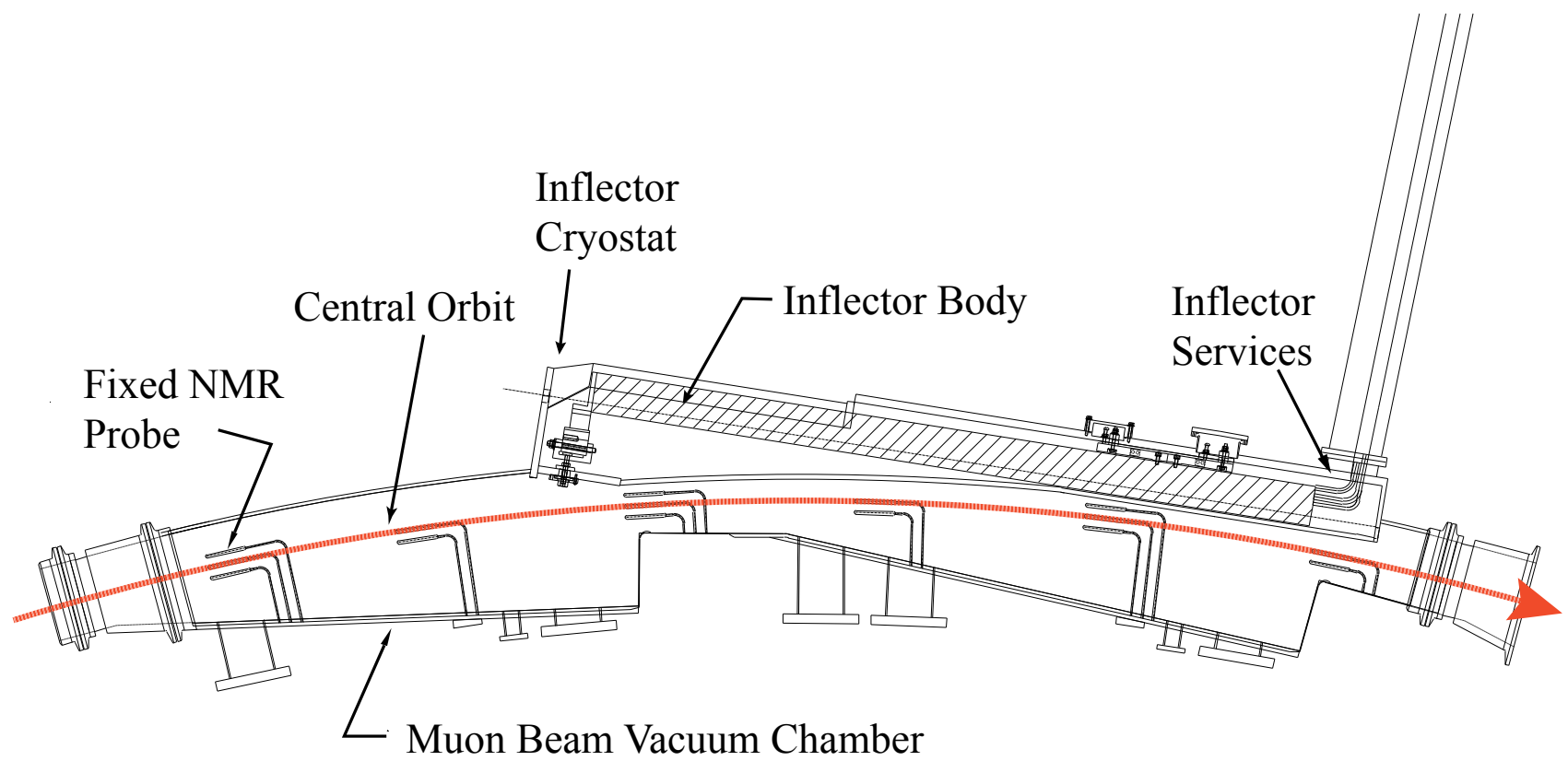

Figure 10.11: Plan view of the combined inflector cryostat-beam vacuum chamber arrangement. The inflector services (power, LHe and sensor wires) go through a radial hole in the back-leg outside of the storage-ring magnet. The NMR fixed probes are in grooves on the outside of the vacuum chambers, above and below the storage region. The red arrow shows the muon beam central orbit.

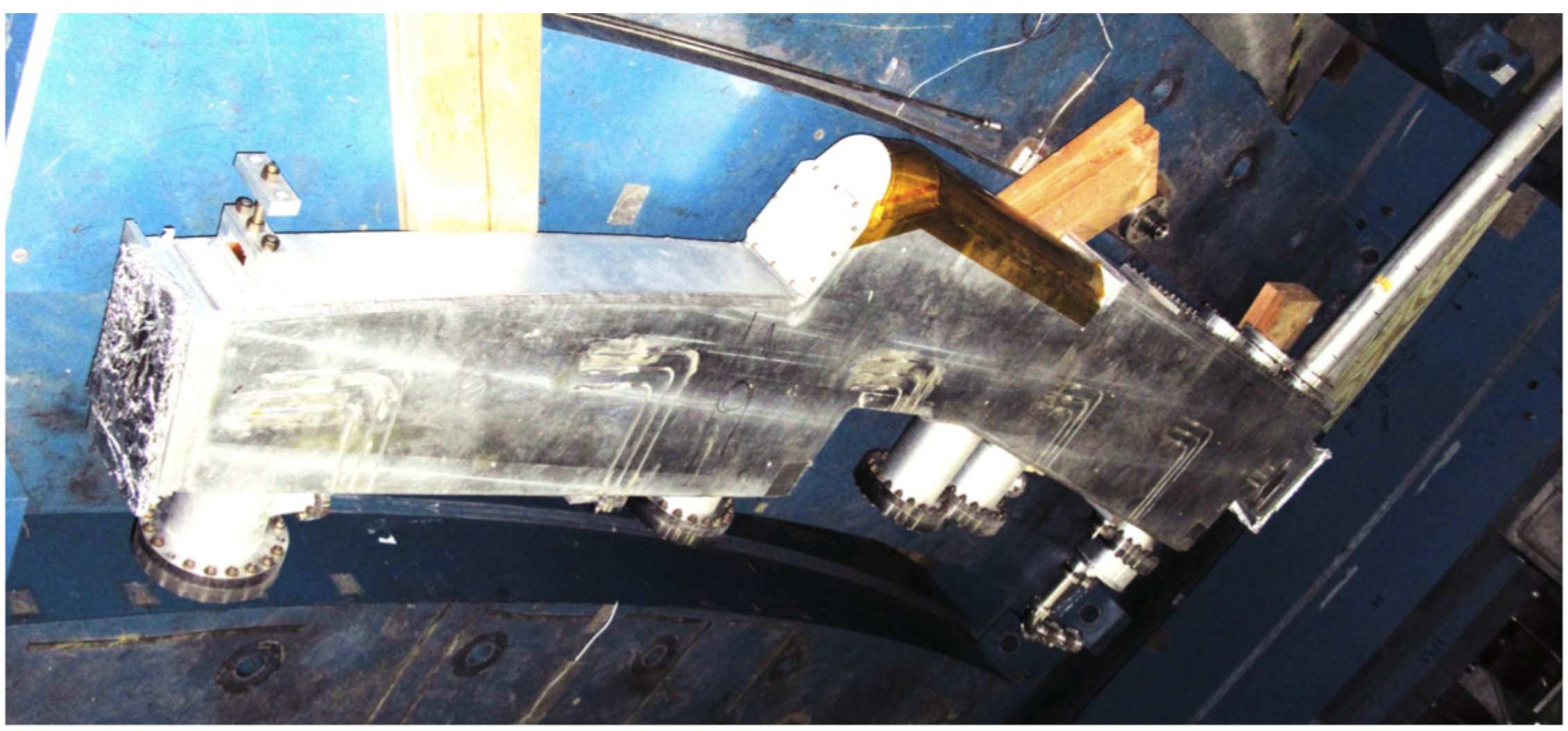

Figure 10.12: A photo of the inflector cryostat.

A method of using SC material to shield the inflector residual fringe field was studied and developed. The fringe field specification was then satisfied.

A test sheet of a superconducting shield was developed that contained 30 layers NbTi, 60 layers Nb, and 31 layers Cu. The Cu layers greatly improved the dynamic stability against flux jumping [9]. The Nb layers act as barriers, which prevent the diffusion of Ti into Cu. The diffusion could form hard inter-metallic layers and create difficulties for the rolling process. Fig. 10.15 shows the typical cross section of the sheet. Based on successful tests, Nippon Steel Corp. developed large, thin pieces of sheet especially for the $(g-2)$ inflector, to cover

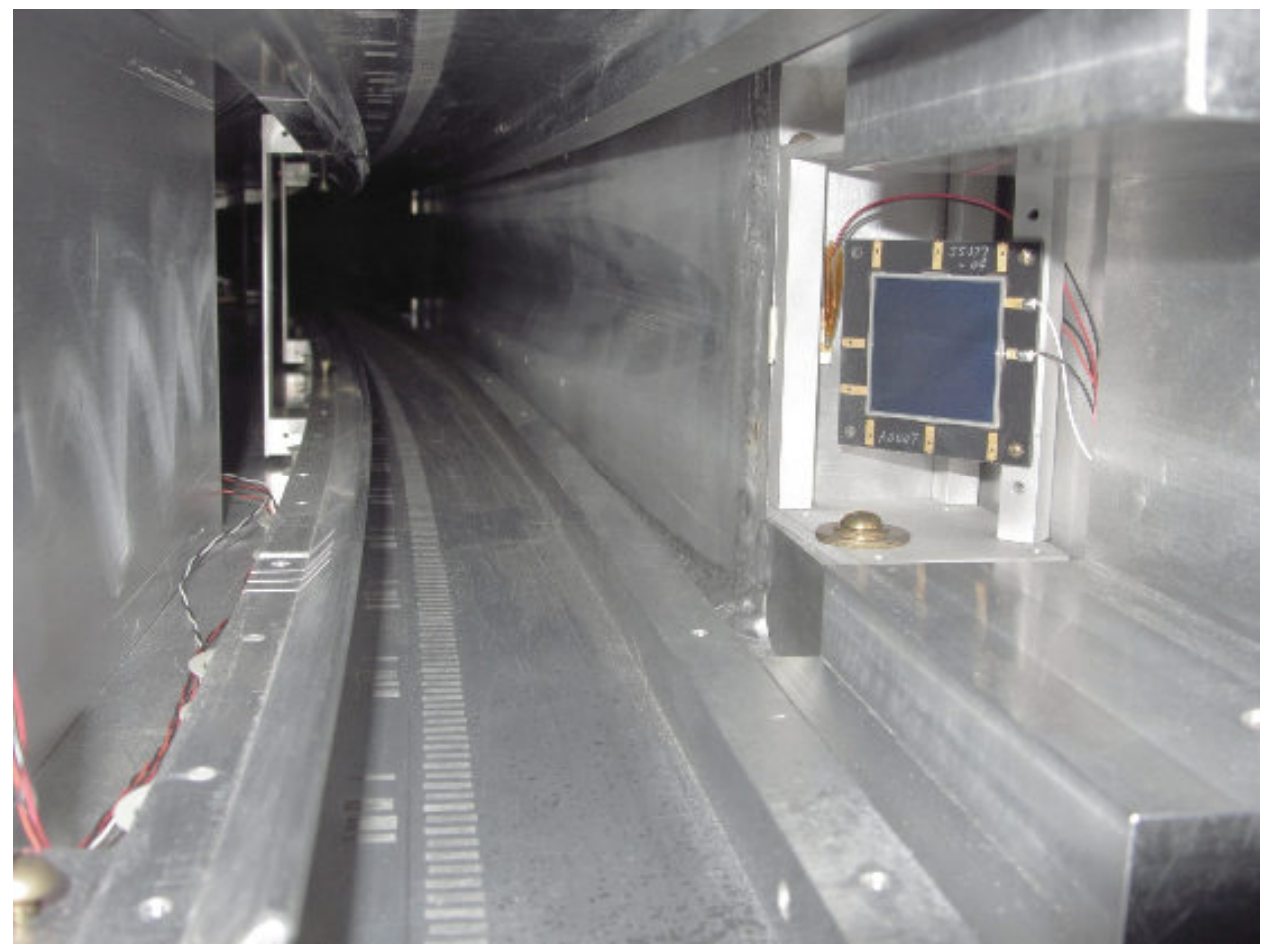

Figure 10.13: A photo of the inflector cryostat exit. A silicon detector (in blue) measured the beam profile just downstream of the inflector cryostat exit

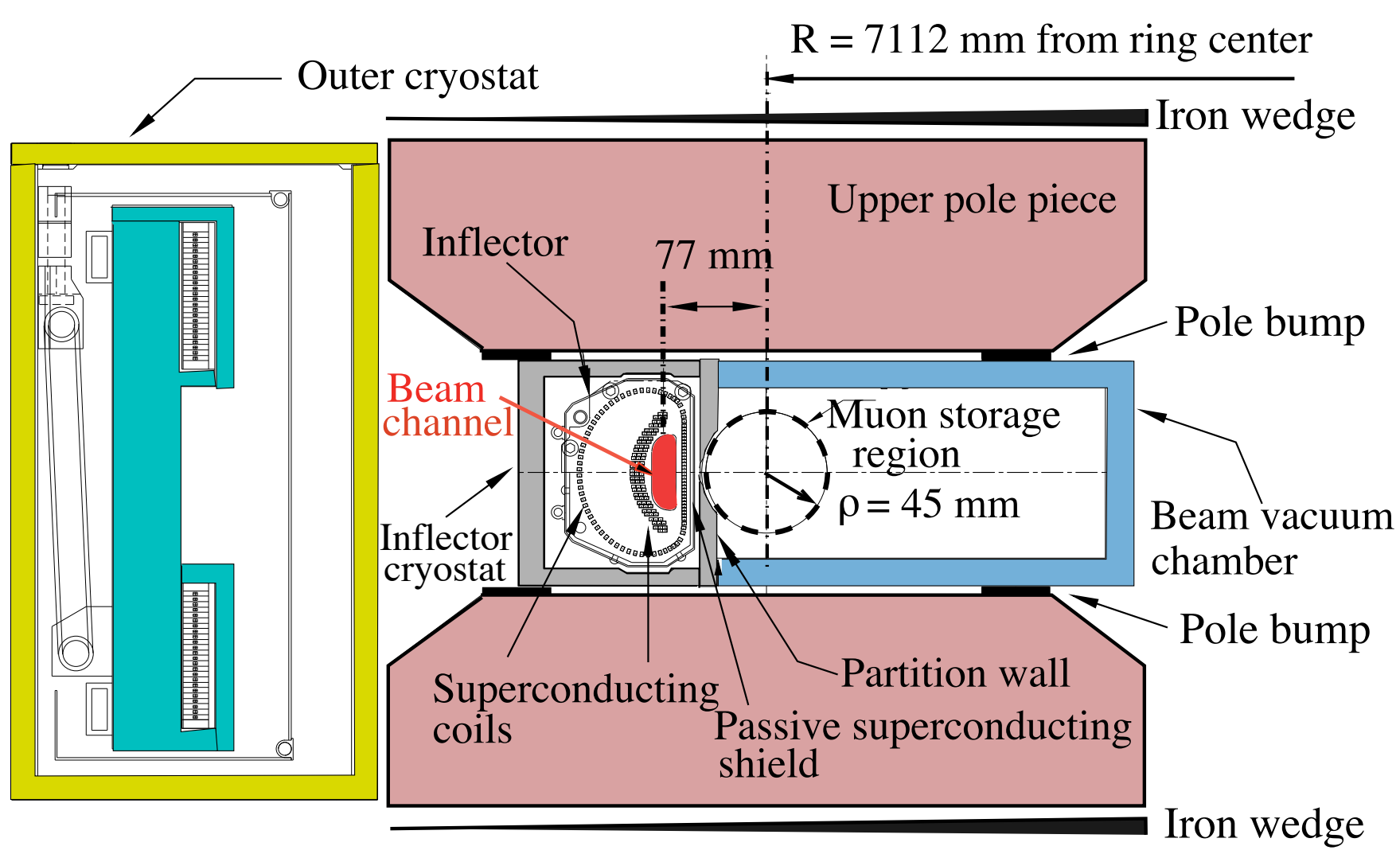

Figure 10.14: The inflector exit showing the incident beam center 77 mm from the center of the storage region. The incident muon beam channel is highlighted in red. (Modified from Fig. 9.6)

its $2 \times 0.5$ m$^2$ surface and to fit into the limited space between the storage region and main magnet coil. The shielding result was extremely satisfactory.

The steps in using the shield are as follows:

1. With the inflector warm ($\sim 20$ K) the storage ring magnet is powered and allowed to reach equilibrium.

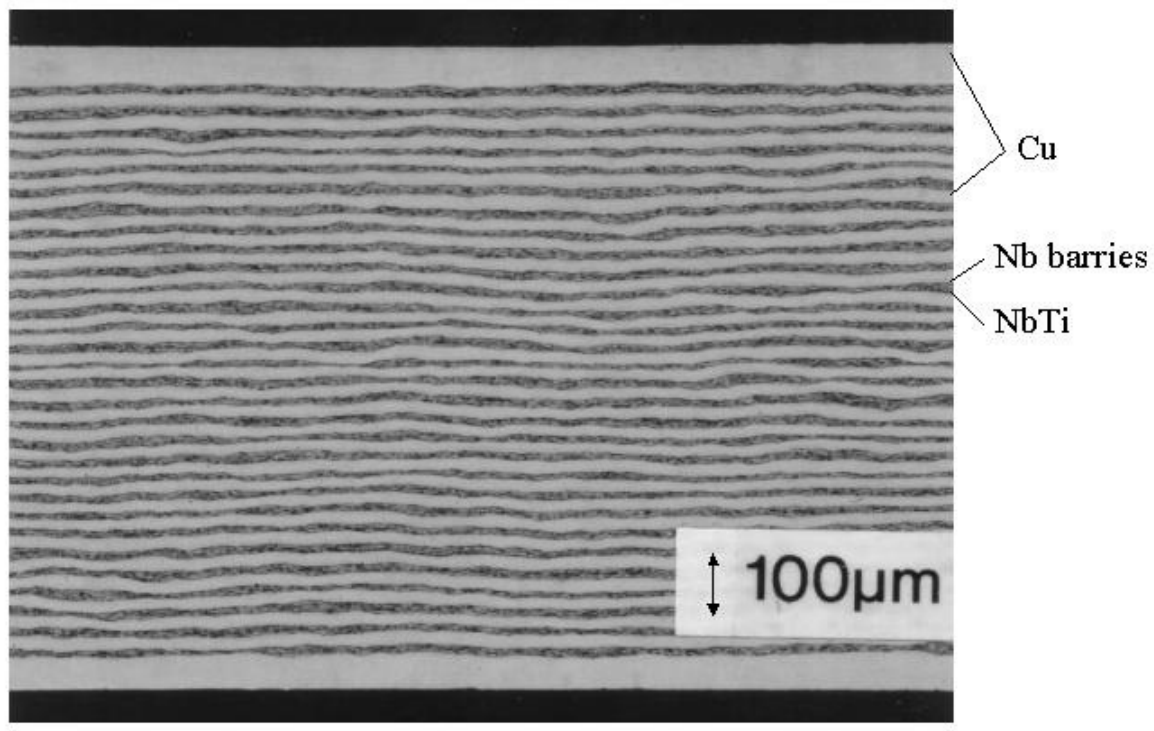

(a) SC shield X-section

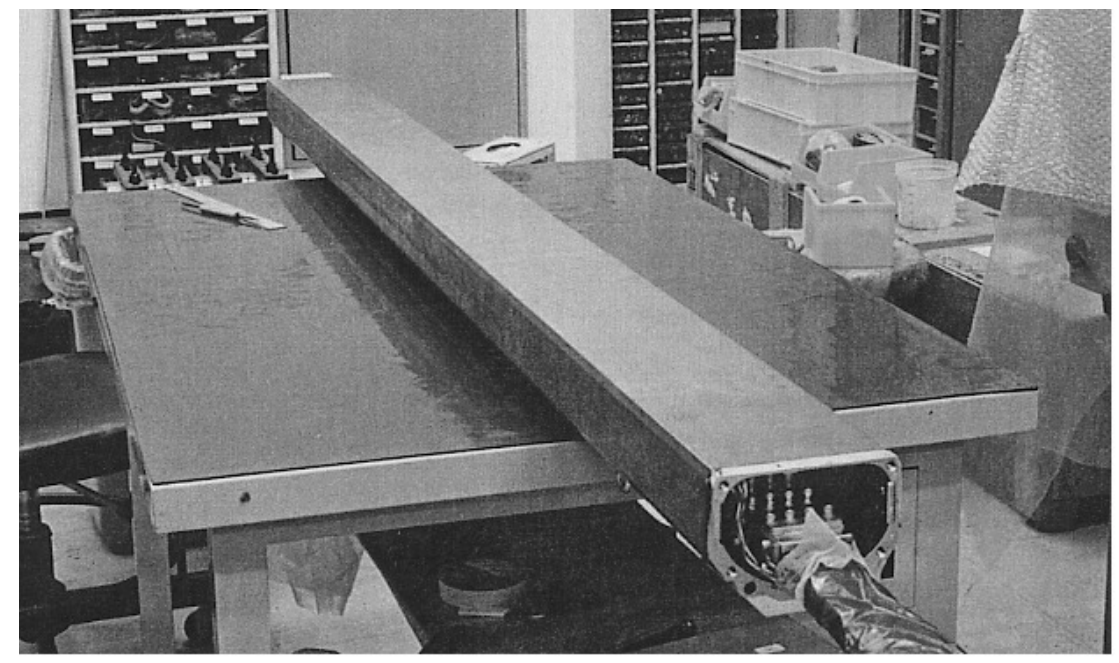

(b) SC shield installed

Figure 10.15: (a)Cross section of the multi-layer superconducting shield sheet. (b)The superconducting shield installed around the body of the inflector.

2. The inflector is then cooled to superconducting temperatures. The shield material is a Type-II super conductor, where $H_{C1} = 0.009$T for NbTi is the maximum field for the Meissner effect to occur. Therefore, as it is cooled down to the superconducting state, the shield is not able to expel the external field. Rather, the external field will fully penetrate the shield. and the shield traps the main field.

3. The inflector is then powered. In this superconducting state, the shield will exhibit perfect diamagnetism, and will resist any change in the flux penetration through its surface.

### 10.2.3 Performance of the E821 Inflector

Two full-size inflectors were produced. To emphasize the importance of the superconducting shield, we relate what happened when the shield on the first inflector was damaged. In the testing of the first inflector, an accident occurred, where the interconnect shown in Fig. 10.8(b) was melted, leaving a few centimeters of undamaged cable outside of the inflector body. In order to repair it, the superconducting shield was cut to give access to the damaged superconductor. After the repair, an attempt was made to apply a patch to the shield. Unfortunately this attempt was not completely successful. The resulting fringe field reduced the storage-ring field by 600 ppm (8.7 G) over a 1° azimuthal angle, resulting in unacceptable magnetic-field gradients for the NMR trolley probes closest to the inflector body. It was also realized that a significant fringe field came from the inflector leads. A field map, averaged over azimuth, from the 1999 run using the damaged inflector, and one from the 2001 run using the new inflector are shown in Fig. 10.16. The field in the region with large gradients had to be mapped by a special procedure following data taking. This large fringe field introduced an additional uncertainty into the average field of ±0.20 ppm (200 ppb) in the result [14]. The 1999 result had a total error of ±1.3 ppm, so the additional 0.2 ppm uncertainty introduced by the damaged shield was small compared to the statistical error of ±1.2 ppm. Had this error not be eliminated, its effect would have been quite serious for the 2000 and 2001 results, both of which had a total error of ±0.7 ppm.

The damaged inflector was replaced in mid 1999, well before the 2000 running period. Two modifications were made to the new inflector design: The superconducting shield was extended further beyond the downstream end; The lead geometry was changed to reduce the fringe field due to the inflector leads. Both of these improvements were essential to the excellent shielding obtained from the second inflector.

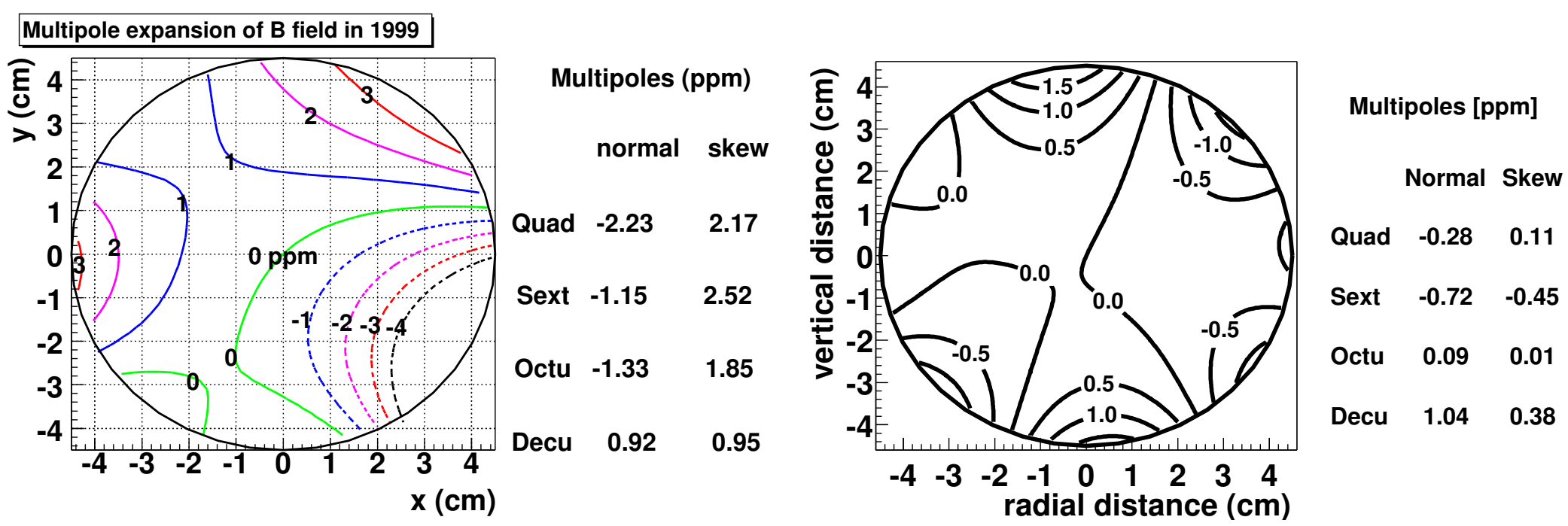

(a) 1999 Data Set

(b) 2001 Data Set

Figure 10.16: The average magnetic field $\langle B \rangle_{azimuth}$ (a) with the damaged inflector (1 ppm contours) (b) and with the second inflector (0.5 ppm contours). Note that the large disturbance in the average field was from a 600 ppm disturbance in the field over 1° in azimuth.

As with any system placed in or near the storage ring, the inflector must not disturb the magnetic field, the measurement of the magnetic field nor the distribution of stored muons at an appreciable level. The E821 inflector shield performed well but did leak some measurable amount of magnetic flux, even after the damaged inflector was replaced. Figure 10.17 shows the disturbance only minimally affects the edge of the storage volume and falls off rapidly. Except for probe 9, which is closest to the inflector, the local effect is at the $\simeq$ 100 ppm level.

This disturbance must couple to a distortion of the trolley guiding rails to produce a difference between the measured field and that experienced by the stored muons. To test a worst-case scenario, we assume a local distortion of the trolley rails perfectly in phase with the field deviation due to the inflector. Fig. 10.18 shows the resultant difference in measured field compared to the true field at the position of each trolley probe, given for various amplitudes of the rail distortion. E821 achieved a precision of 0.5 mm for the trolley rails and E989 will meet this or do better (see Chapter 11). Therefore, even for an amplitude of 1 mm perfectly in-phase with the field distortion due to the inflector, the resultant systematic error is below 10 ppb across the storage volume and so the E821 inflector exceeds the magnetic field requirements of E989.

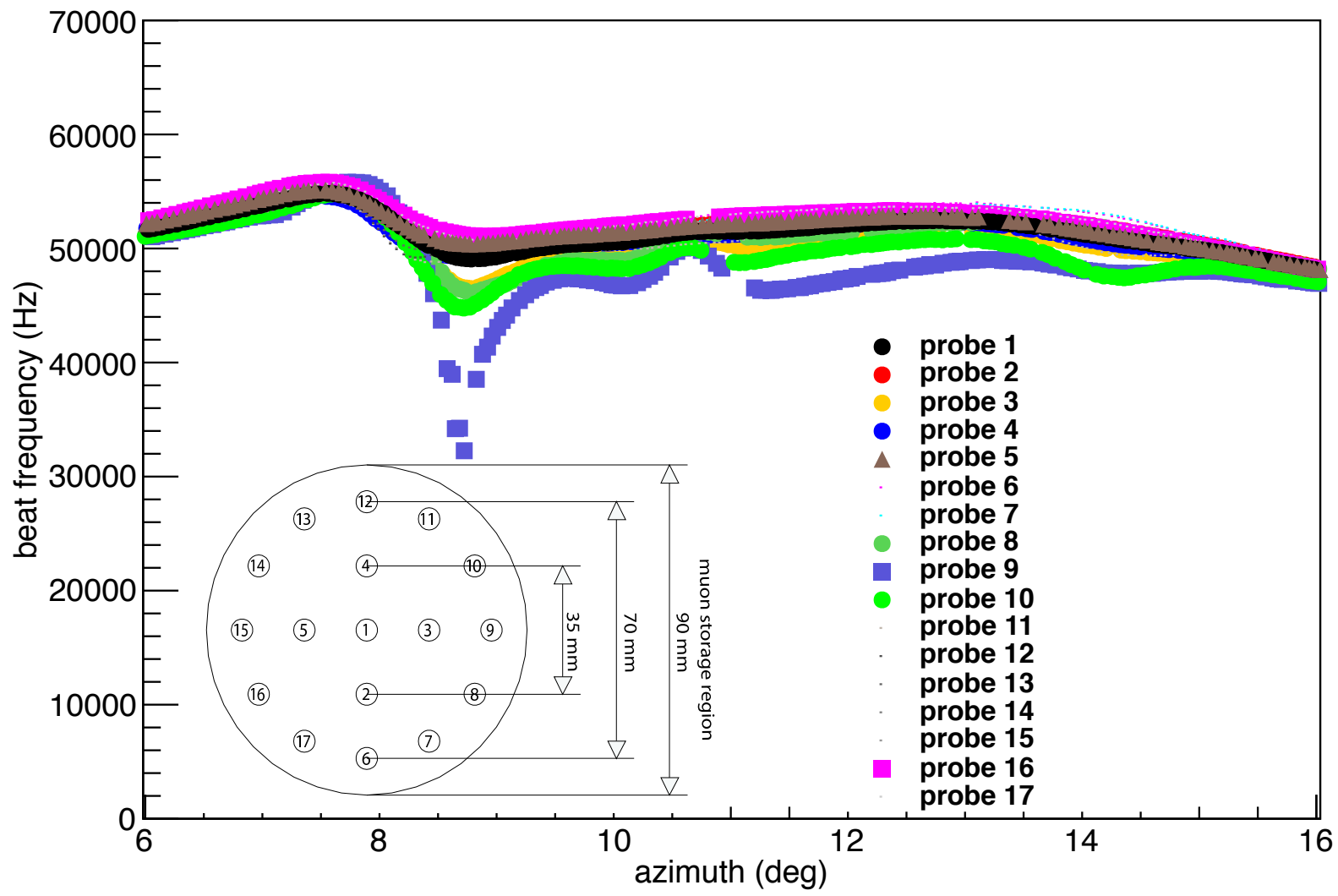

Figure 10.17: The response of the trolley probes to the second inflector fringe field during the 2001 data collection period. The insert shows the location of the trolley probes.

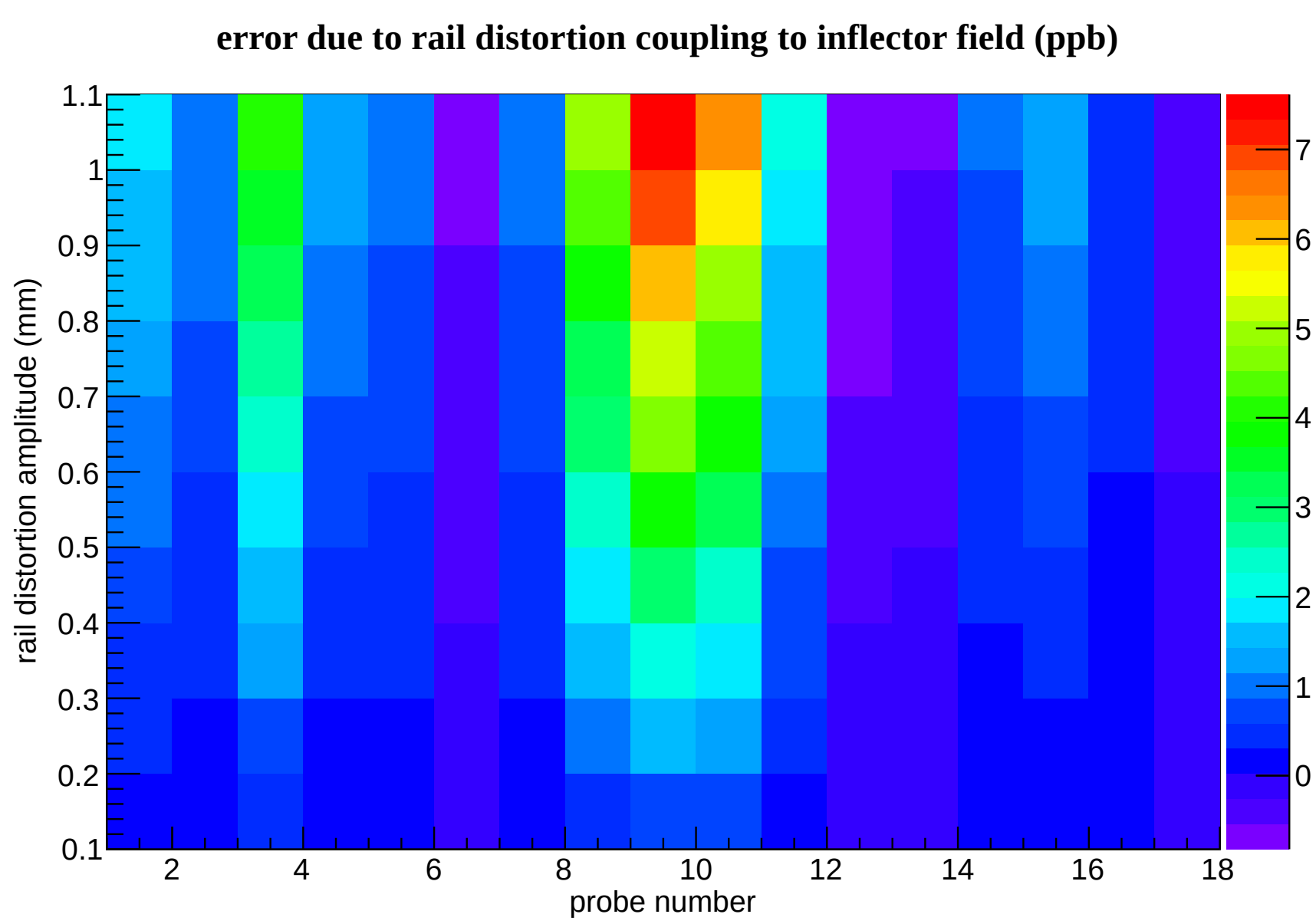

Figure 10.18: Shift in magnetic field experienced by each trolley probe for various distortions of the trolley rails perfectly in phase with the distortion of the magnetic field due to the E821 inflector. As even a 1 mm in-phase distortion of the rails causes a shift of $< 10$ ppb at the edge of the muon storage volume, the E821 inflector more than meets the requirements for E989.

## 10.3 Inflector Power Supply and Quench Protection

To power the existing inflector magnet, none of the equipment used in E821 at Brookhaven could be re-purposed. The original Transrex power supply is obsolete, as are its controls. The scenario with respect to the inflector quench detector electronics rack was the same as that for the Main Ring quench detector electronics: the components used were obsolete; no documentation exists, the PLC is obsolete; there are no spare PC boards; and there are missing cables and their documentation. We have identified a different, more modern power supply and quench protection circuit at Fermilab to re-purpose for E989.

This power supply, a PEI-150 (which is similar to the original Transrex but considerably updated in function and usability) was used as the power supply for the Dzero detector solenoid. It is presently located in the Dzero Assembly Building (DAB) and is connected to the controls, the AC power distribution, the quench protection system, and the cooling water system. Therefore, the preliminary testing was performed in its present location. The power supply and its control panel are shown below.

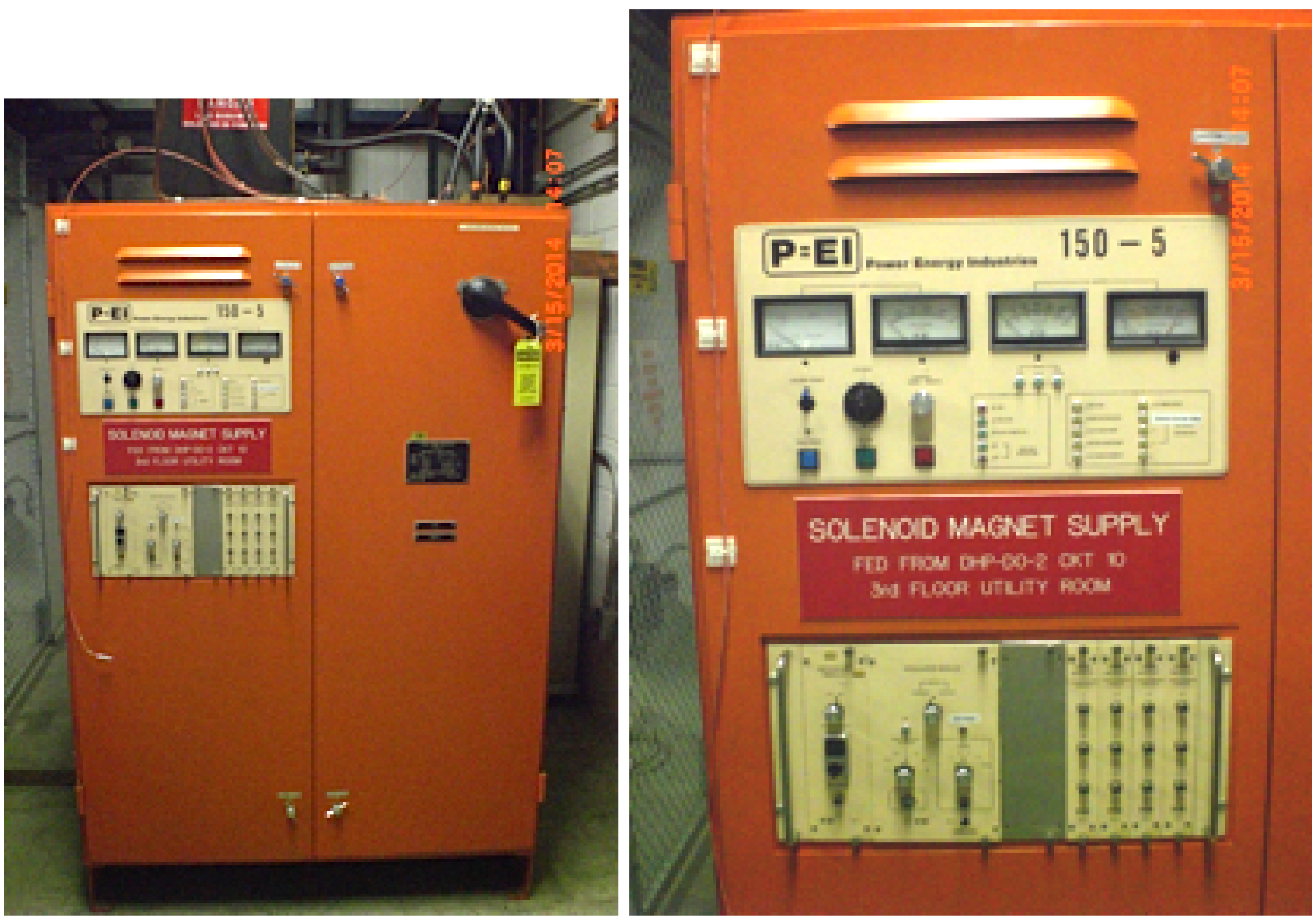

(a) The Power Supply (b) Control Panel

Figure 10.19: (a) The inflector power supply. magnet. (b) Close-up of the controls.

### 10.3.1 Power Supply Requirements

There were two requirements given for operation of the inflector power supply: 1) the maximum DC current output must be 3500A, 2) the maximum ramp rate be limited to 2A/sec. These requirements can be met by the PEI-150 unit. Using the proper output voltage taps, the configuration chosen can provide up to 5000A at 7.5 VDC output. Using the external reference voltage input, the ramp rate can be directly controlled. In addition, this ramp rate can be limited to 2A/sec by changing some components within the PEI's control module.

Current regulation given for the unit is 0.05% (500 ppm) when operated between 20% and 100% of rated output current. Since this is a newer and redesigned version of the originally used Transrex, this regulation specification is assumed to be adequate for the inflector's operation. Like the Transrex, there is sufficient ripple at 720 Hz ripple and this ripple can approximate 20% peak-to-peak into a resistive load. This ripple is the result of the 12-phase SCR firing operation. Therefore, an external 720 Hz notch filter is connected to the output of the supply prior to being connected to the quench protection trip cabinet.

The cooling system inside of the PEI power supply is much more robust than that used for the Main Ring. Therefore, a separate regulated cooling system is not required. However, the water/coolant must be low conductivity.

### 10.3.2 Preliminary Testing

Once the AC service was inspected and the cooling water system reconnected, the PEI-150 power supply was powered ON. The output current was manually ramped up to 300A into the existing dump resistor. Operational voltages and controls all appear to be normal. Next, the output bus was shorted together to allow for the full operational current of 3500 A. The output current was ramped up and held at 3500A for 45 minutes. Operation appeared normal and thus completing the preliminary checkout. The power supply will remain in its present location until some operational testing using an inductive load can be performed to determine the notch filter's performance, and to determine the power supply's regulation performance. Then the power supply and components will be disassembled and moved to the final testing area were the configuration and quench protection system can be reassembled and prepared for a series of final tests for certification.

### 10.3.3 Inflector Quench Protection

The quench protection originally used for E821 is not documented and there was no indication of the trip cabinet arrangement. The quench energy stored in the inflector magnet (about 9 KJ), as described in Ref. [2], was not dissipated using a dump resistor but the energy was dissipated within the magnet itself. This was accomplished by turning off the power supply and using the power supply's output impedance to provide a path for the current from the collapsing magnetic field. However, given that there is only one inflector magnet available for the operation at MC-1, there was a desire to provide a more precise control of the energy dissipation in order to provide a higher degree of protection for the inflector magnet in E989.

To provide this added degree of protection for the inflector during a quench, the quench protection system is almost identical to that used for the Main Ring magnet. The only difference is the number of tap voltages that need to be monitored. Also, a quench condition of the Main Ring magnet will be used to initiate an energy extraction from the inflector magnet.

The quench protection components, used for the Dzero solenoid, can be directly re-purposed for the inflector operation. This includes the trip cabinet with a remote-controlled reversing switch (if needed), the air-cooled 720 Hz notch filter, and the quench detector electronics rack. The quench protection circuitry, used for the Dzero solenoid and now being re-purposed for the inflector, is fully documented in ref [19]. In addition, there are existing spare modules for the controls and voltage tap input amplifiers. Only the coil voltage threshold for the generation of a quench trip signal was briefly described in ref [2]. The interconnect voltage will be set to 10 mV in consultation with Wuzheng Meng at BNL.

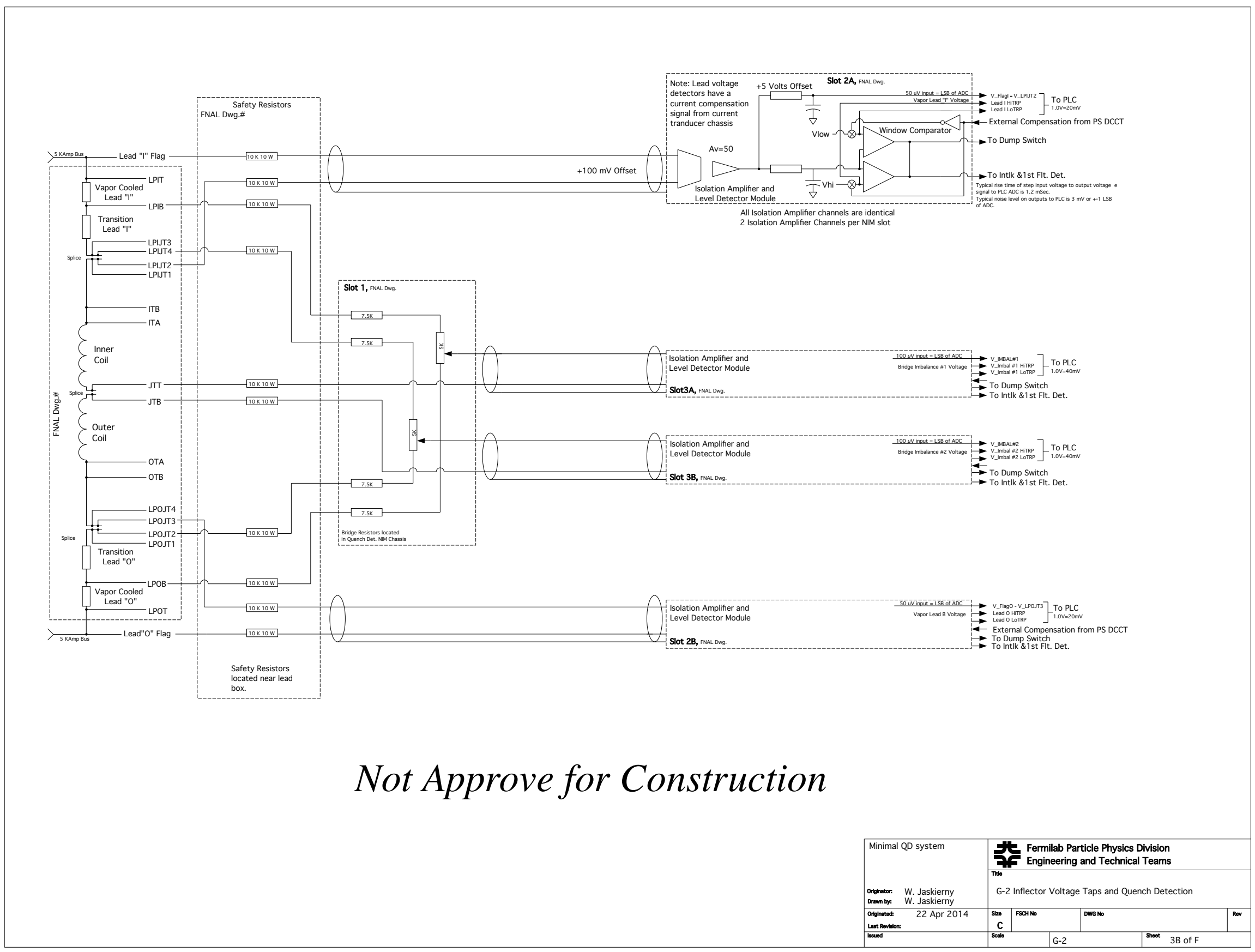

Figure 10.20: Inflector quench protection circuitry and tap locations.

In order to document the specific wiring and connector information of the inflector, a wiring schematic is being developed. According to the information at present, a draft of this schematic is shown in ref. [20]. Once verified, this schematic will be used to diagram the specific voltage taps to the connectors.

Table 10.3: Quench Protection Parameters.

| | |
|---|---|
| Coil Voltage | 120 mV |
| Gas-cooled lead | 30 mV |
| Coil interconnect | 10 mV |

### 10.3.4 Power Supply and Quench Components Placement

At present, we plan to locate the inflector power supply and quench protection components within the main hall in MC-1 and also on/under the platform used for the Main Ring power supply. In E821 at Brookhaven, the power supply was located in a different room outside of the ring hall. The exact location on or near the platform will depend on the space and orientation of the components (trip cabinet, quench protection, electronics rack, etc.). The dump resistor can be placed inside or near one of the component cabinets since the estimated energy to be dissipated is quite small and should not adversely affect the MC-1 hall environmental temperature. For runs of the DC output bus that exceed 10 feet, water-cooled 10 kA bus work can be implemented to save on cost associated with the use of 535 cm locomotive cable.

One concern of locating the PEI power supply inside the main hall is the effect of 60 Hz (or multiples) magnetic fields, emanating from the power supply and the notch filter, on the Main Ring's magnetic field. An initial set of measurements were taken when the PEI power supply was operating at full output of 3500A. These measurements are currently being evaluated. If needed, magnetic shielding can be placed around the power supply to help mitigate this adverse effect, or the location of the power supply will need to be re-evaluated.

## 10.4 Lessons for E989 from the E821 Inflector

The most important single lesson from the E821 inflector came from the flux leakage from the damaged inflector, and the realization that the first design of the inflector leads also contributed to this problem (see Fig.10.16). The $\pm 0.2$ ppm systematic error from this problem is three times the E989 magnetic field error budget of $\pm 0.07$ ppm. The highly localized 600 ppm perturbation at the location of the "repaired" superconducting shield could not be shimmed away. The second issue that must be addressed is the mismatch of the E821 inflector aperture and the storage ring acceptance. The third issue is to open the ends.

The guiding principles going forward are:

- *The flux inside of the inflector must be confined inside of the inflector and not permitted to leak into the storage region.*

- *Any new inflector design must have a horizontal (radial) aperture significantly larger than 18 mm; as close to 40 mm as possible. This will facilitate beam matching into the storage ring, and should reduce the coherent betatron oscillations, and the systematic error associated with it.*

- *The ends of the inflector need to be open, rather than have coil windings across them, again to facilitate matching and to eliminate multiple scattering.*

The latter two conditions could increase the number of stored muons by almost a factor of $\simeq 2 - 3$, which would permit the measurement of $a_\mu^-$ as well as $a_\mu^+$.

The muon injection efficiency achieved in E821 was around 2%. Early simulations predicted that it should be 5 - 7%. Simulations by Hugh Brown showed that opening the ends of the inflector in E821 would have increased the number of stored muons by a factor of $\times 1.8$. A new open-ended inflector with a larger horizontal aperture, as large as 30 to 40 mm diameter is desirable. We are working very hard to determine how much larger the aperture can be made, given the constraints of the vacuum system and magnet geometry.

## 10.5 Progress Toward A New Inflector

The E821 inflector magnet was a *tour de force* in magnet design, introducing the truncated double cosine theta magnet [4] and a superconducting shield [2] that provided a constant dipole field inside of the beam channel, with flux leakage of less that 100 ppm in the region populated by the majority of the muon beam. It worked perfectly for E821, and permitted a measurement of $a_\mu$ at the 0.5 ppm level. Nevertheless, issues that would be desirable to improve in the next-generation experiment were discussed in the previous section.

Several concepts have been considered to replace the existing inflector. Any new design is constrained by the injection geometry shown in Figs. 10.1, 10.9, 10.10, 10.11 and 10.14. A passive superconducting shield to remove any leakage flux from the new inflector will be essential.

The small aperture of the E821 inflector, and the coil windings over the beam channel make matching the beamline to the storage ring impossible. Since E989 plans to accumulate 21 times the data of E821, it is necessary to revisit the inflector aperture issue. Opening the radial aperture to a 30 to 40 mm would come close to matching with the incoming beam, and permit more muons to be stored. It would also reduce the amplitude of the coherent betatron oscillations, which cause one of the significant systematic errors. As the aperture gets larger and the centroid of the injected beam is displaced radially outward, a larger kick is needed to place the beam on orbit. Shielding the flux leakage from a larger open end will also be challenging.

In E989 the knowledge of the average magnetic field needs to be improved from $\pm 170$ ppb in E821 to $\pm 70$ ppb. While the plan to improve the magnetic field measurement and control is discussed in Chapter 15, this plan is meaningless if any device in the experiment spoils the field by introducing extraneous magnetic flux into the storage region. The damaged inflector in E821 demonstrated how a 200 ppb problem can easily be introduced.

Two possible suggestions have been proposed for a new inflector:

- A double-cosine $\theta$ design with a larger aperture with open ends.

- A multiple coil magnet using the direct wind technology.

### 10.5.1 A New Double Cosine $\theta$ Magnet

We have been studying this option with our colleagues in the Fermilab Technical Division. The truncated double cosine $\theta$ design encased in a multi-layer superconducting shield worked well in E821, albeit with the limitations discussed above.

The muon beam injection into the g-2 storage ring and the storage ring operation strongly depends on the inflector magnet performance. The magnetic design for the E821 inflector was carried out by W. Meng [3], and the engineering was done by the KEK-Tokin Company collaboration in Japan led by Akira Yamamoto. This inflector will be used during the first phase of experiments at FNAL. Nevertheless, it is very desirable to build the new inflector magnet with improved performance. The new magnet should have open both ends for the muon beam, and with an increased horizontal aperture. It should be noted that it is very difficult to improve the E821 magnet design, which was based on number of models, and cold tests. The old magnet geometry and parameters with closed ends are shown in Fig. 10.21 and Table 10.4

The E821 inflector used a unique NbTi superconducting shield made by Nippon Steel, Japan [2]. It is unclear at this time whether such multilayer superconducting shield production technology still exists. Other issues are that there is no full set of inflector magnet drawings because the Tokin Company went out of business some years ago. The available space for the magnet cold mass is very constrained and cannot not be increased without major changes in vacuum vessels, thermal shields, supports, etc. So, the following action items were chosen to proceed with the first step of the magnet design and fabrication:

- Disassembly of the damaged and repaired E821 inflector magnet with the intention to use most of magnet parts.
- Measure the parts' dimensions and produce the set of cold mass drawings.
- Wind inner and outer magnet coils using copper or superconducting cable into slots of old magnet aluminum mandrels. The coil ends should have an open configuration
- Make fringe field magnetic measurements of inner and outer coils assembly at room temperature.

Now the magnet is fully disassembled (See Fig. 10.22) and the magnet components are shown in Figs.10.23 - 10.25. After disassembly, most of the parts can be used again for the magnet modeling. The main goal for this model is to verify the coil winding technology with open ends obtaining reasonable end fringe fields. The first 3D magnetic field simulation results showed that for the open ends (See Fig. 10.26) the peak field on the superconducting shield could be larger than specified shield peak value 0.1 T obtained for the magnet with closed ends.

The main reason for the larger field on the shield is the vertical bends of superconductor on both ends which generate the large normal field component in the shield area. The optimization of the superconductor bends is the most critical issue for the open end magnet approach.

The second step in the new magnet design will be the magnet cross-section optimization to increase the magnet aperture in the horizontal direction. First simulations showed that the 10 mm aperture width increase at the same cold mass outer dimensions increases the peak field at the inner coil (the leftmost layer of the “C” shaped conductor arrangement in Fig. 10.6(a)) from 3.3 T to 5 T during the magnet operation in the 1.45 T background field from the main storage ring magnet. This is why the superconductor should have a much larger current carrying capacity at larger fields than for the E821 design. Because the E821 magnet stored energy of 9 kJ is rather low, it is possible to increase the volume of superconductor in the cable relative to the volume of copper stabilizer.

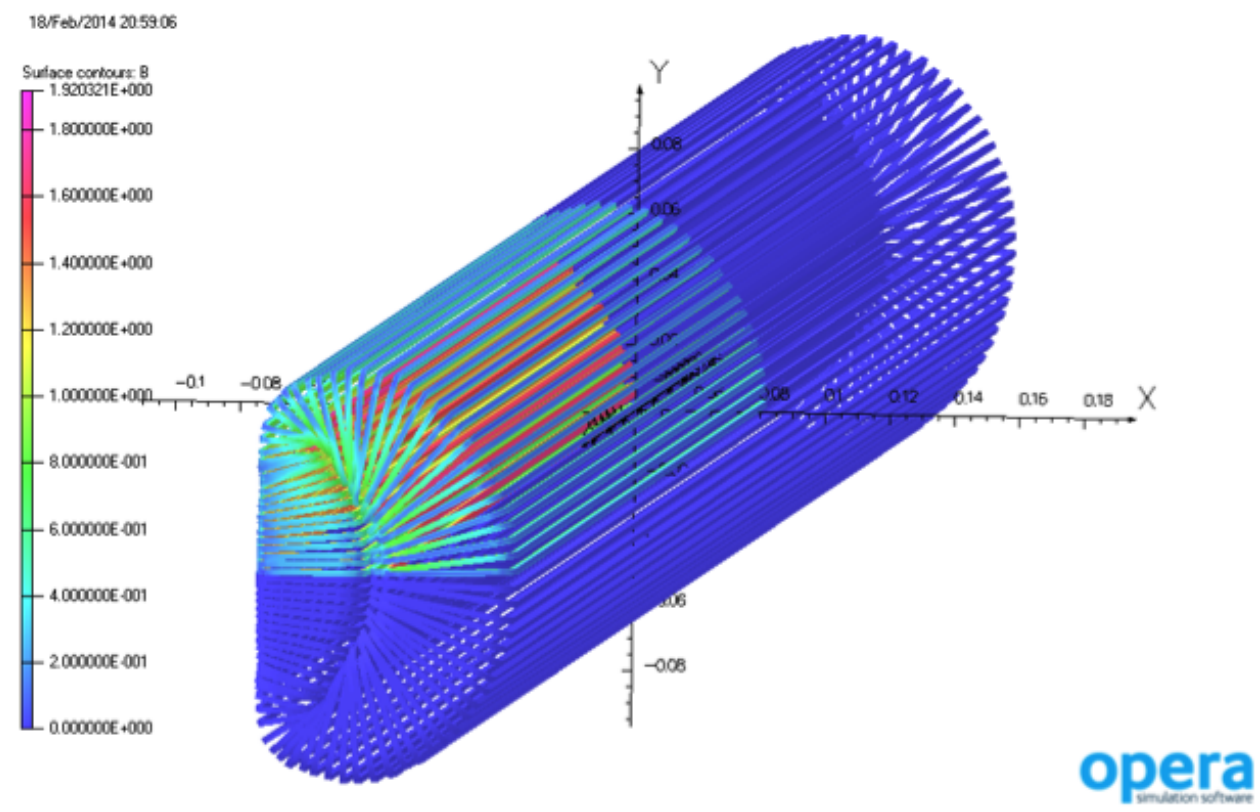

Figure 10.21: The E821 inflector coil geometry. $B_{\max} = 1.92$ T without an external magnetic field of $B = 1.45$ T

Table 10.4: Parameters of E821 inflector.

| Parameter | Units | Value |
|---|---|---|
| Dipole magnetic field in the beam pipe center | T | 1.45 |
| Magnet effective length | mm | 1696.4 |
| Coil length | mm | 1700 |
| Total length | mm | 2045 |
| Beam pipe width x height | mm | 18 x 56 |
| NbTi superconducting screen width x height | mm | 103.2 x 154.6 |
| NbTi superconducting screen length | mm | 1931 |
| Coil current | A | 2850 |
| Superconductor with Al stabilizer bare dimensions | mm | 2 x 3 |
| Superconductor ratio (NbTi:Cu:Al) | | 1.0:0.9:3.7 |
| Superconductor critical current at 3.5 T and 4.6 K | A | 3890 |
| Inductance | mH | 2.0 |
| Stored energy | kJ | 9.0 |
| Overall cold mass dimensions | mm | 110x150x2025 |

In the next section, we discuss the development of a new NbTi cable “six around one” with six NbTi strands around the central copper strand. This cable is capable of carrying a much larger current than the aluminum stabilized ASTROMAG conductor used in the present inflector. The superconducting cable grading could reduce the total volume of su-perconductor when the outer coil will be wound with the larger cable than the inner one. Two different cables have been investigated: six around one with 0.806 mm strand diameter for the inner coil and 1.0 mm strand for the outer.

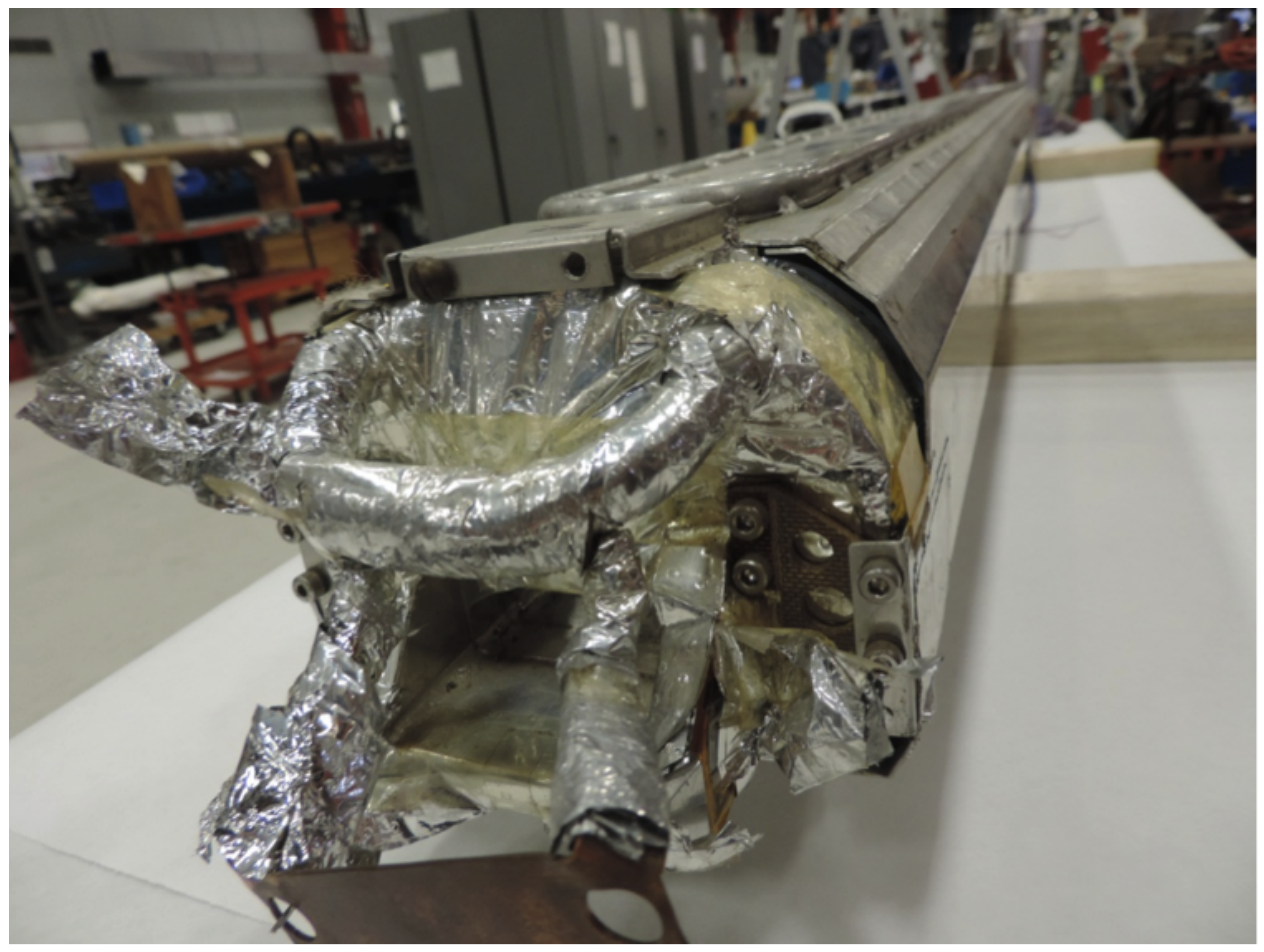

Figure 10.22: Inflector magnet cold mass with the thermal shield before disassembly.

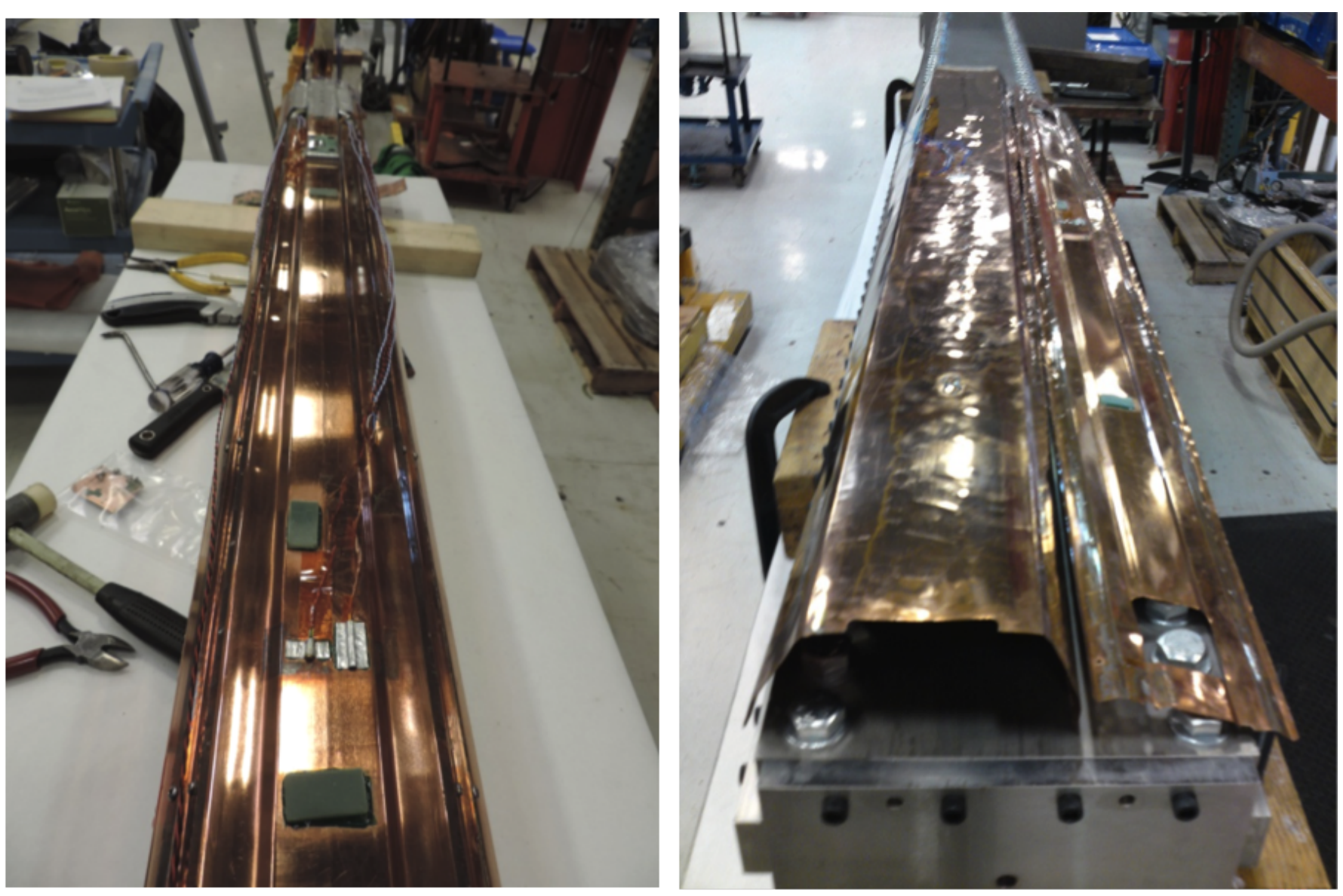

(a) Before removal (b) After removal

Figure 10.23: (a)The superconducting shield before (a) and after (b) removing it from the magnet.

The inflector magnet has a very tight specification for the fringe field leakage into the storage ring aperture. The main ring correction system would be designed to correct the average integrated field but is not capable of correcting local field distortions. To correct these local effects we propose to use correction coils made on the base of printed board technology. For example to correct 100 ppm field distortion, requires a printed circuit board with 10 A total current. These correctors could be mounted on the pole tips or on the vacuum vessel walls.

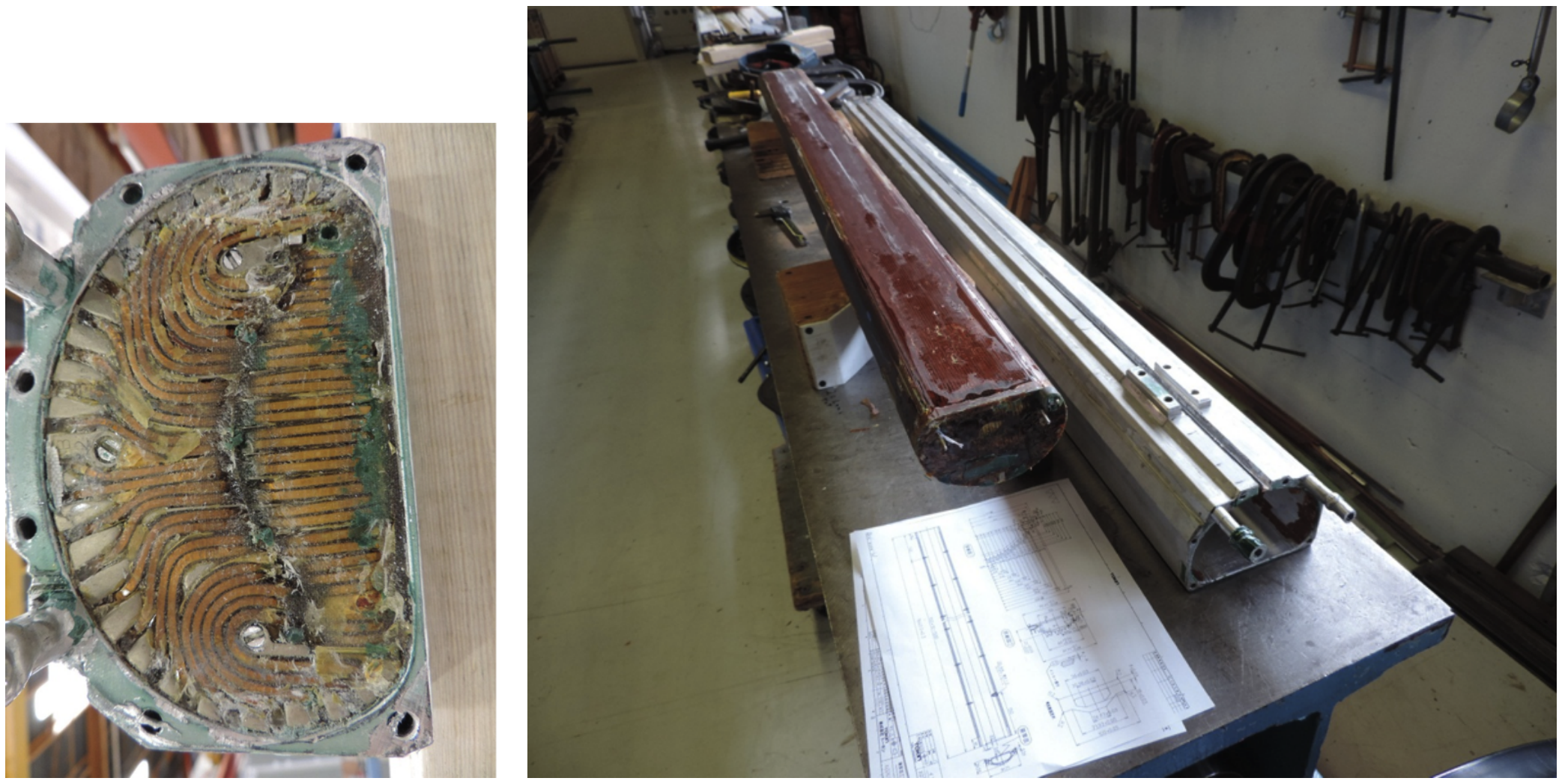

(a) Encased Coil End (b) Coil Removed

Figure 10.24: (a) View of the magnet end. The coil block was epoxy impregnated inside the aluminum case. (b)Coil block removed from the aluminum case.

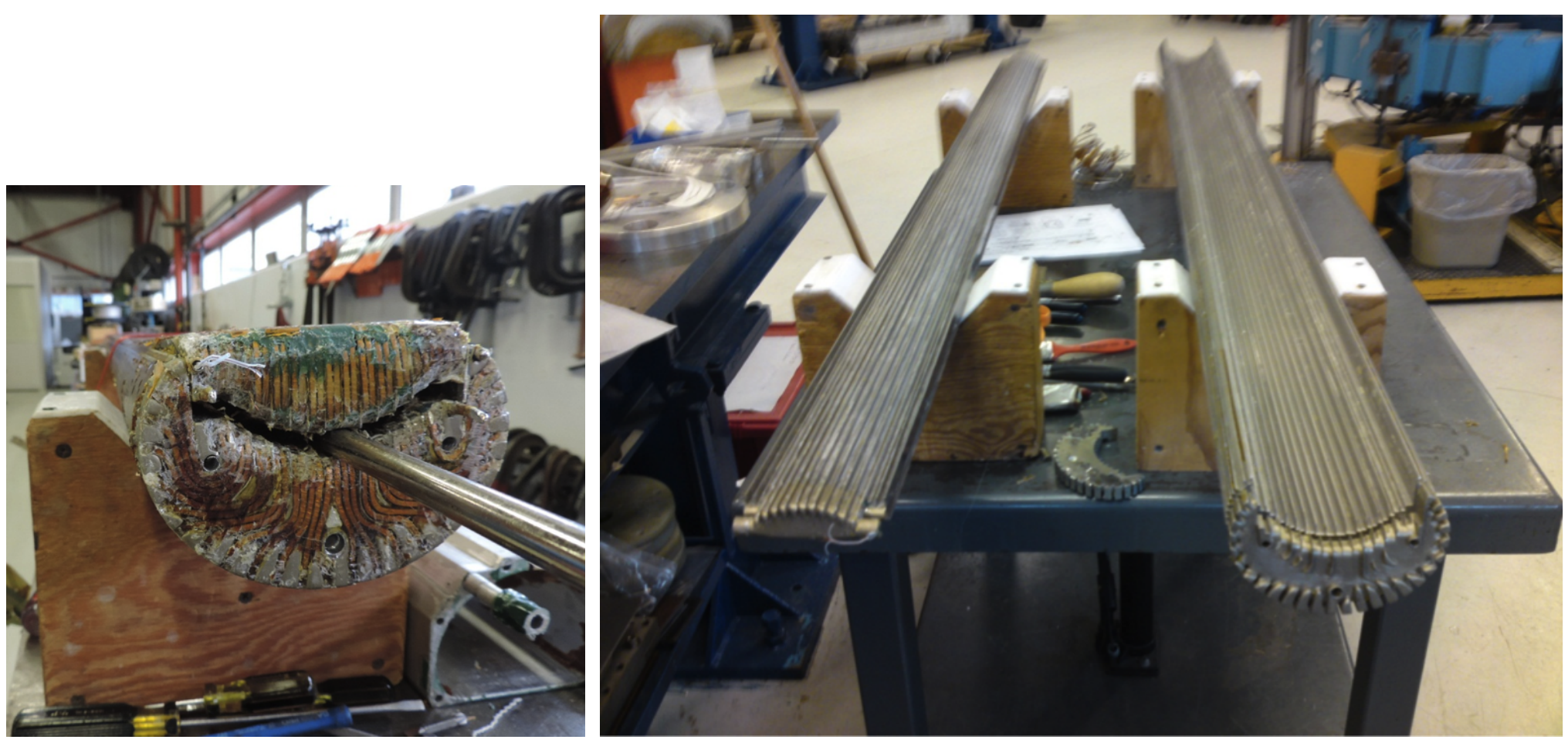

(a) Separating the Coils (b) The mandrels

Figure 10.25: (a)Separating the inner and outer coils (b)Inner (left) and outer (right) coils aluminum mandrels after removing the superconductor from slots.

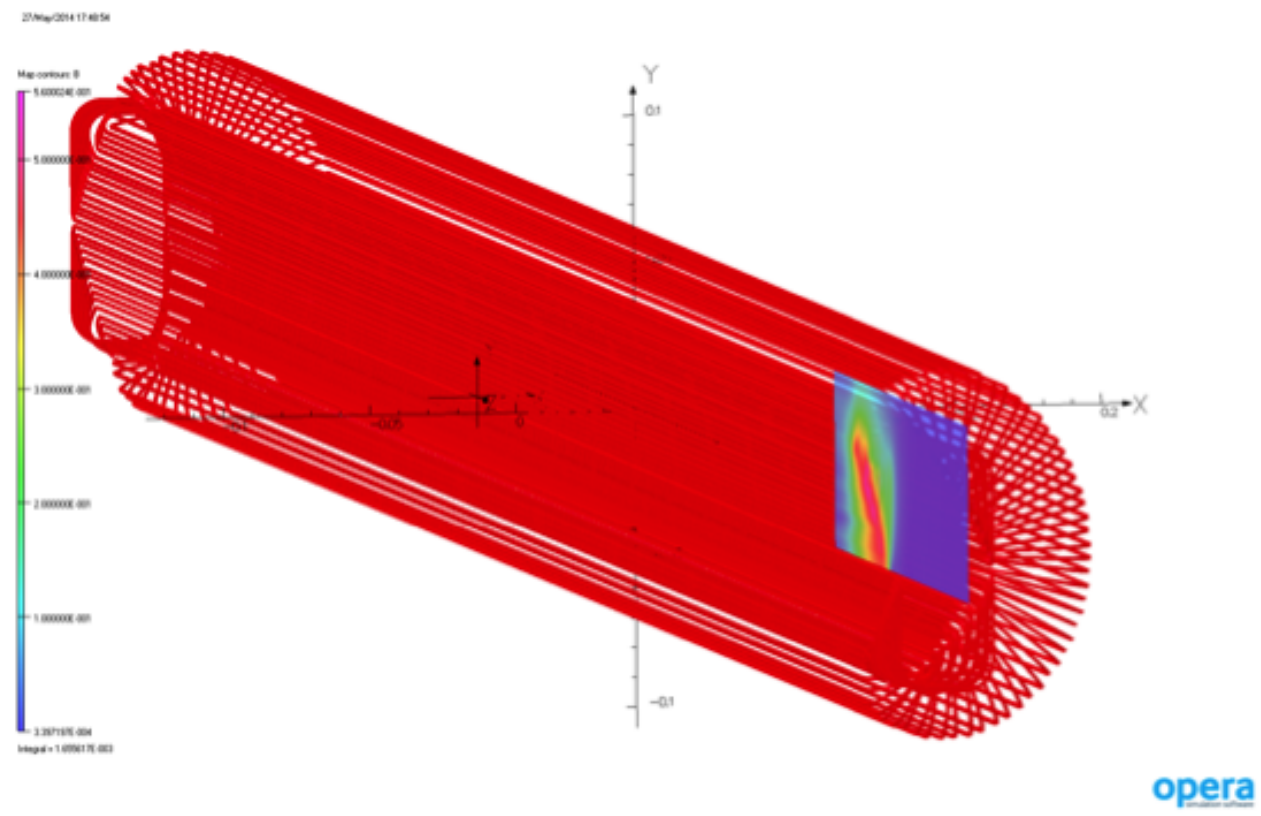

Figure 10.26: The inflector magnet with open ends. The shield peak field is 0.56 T.

**Development of Superconductor for a New Inflector**

In the present double cosine theta design, the operating current of the inflector is 2,685 A. The conductor of the original inflector had bare dimensions of 2 mm $\times$ 3 mm. In an optimized design for the new inflector, a Cu stabilized conductor has been developed aiming at a similar conductor size. This was done using the Cabling Facility of FNAL's Technical Division. This facility includes a compact cabling machine with 42 spools and electronic synchronization for lay angle control, a re-spooler, sets of forming fixtures, mandrels and measuring devices, and has been mostly used to develop and fabricate wide Rutherford-type cables for dipole and quadrupole accelerator magnets. To fabricate the small cables required by the new inflector design, the idea was developed of flat-rolling a composite round cable, or 6 around 1, made of seven 0.8 mm wires. This was done to assess the feasibility of a rectangular cable of appropriate size without the need of a mandrel. Cu wires of 0.8 mm diameter were used for practice. The transverse deformation was applied with the turk-head of the cabling machine. The turk-head forming fixture is composed of two vertical rolls $\sim$20 mm wide and two horizontal rolls 1.2 mm thick, both with variable gaps (see Fig 10.27(a)). Because the minimal horizontal gap presently allowed by the machine was larger than the desired cable width, the 6 around 1 cable was free to expand laterally under compression, as shown in the schematic of Fig. 10.27(b).

Using this technique, Cu cable samples were produced as shown in Table 10.5 in order to optimize the various parameters into play. These include cable pitch length (the shorter, the more compacted and rigid is the cable), thickness (the smaller, the more compact and mechanically stable is the cable), width, tension of the central wire and of the peripheral strands (not shown in Table), absence of crossovers (i.e., two strands crossing each other), and amount of residual twist produced within the cable (the lower the better). A picture of a good quality cable is shown in Fig. 10.28.

The next step was that of producing superconducting NbTi cable to measure the effect of plastic deformation on the strand superconducting properties. A 0.804 mm wire with Cu/SC Ratio of 1.34 from Oxford Superconducting Technology was used from TD inventory. The 6 around 1 cable parameters obtained with the Cu were used as starting parameters for the

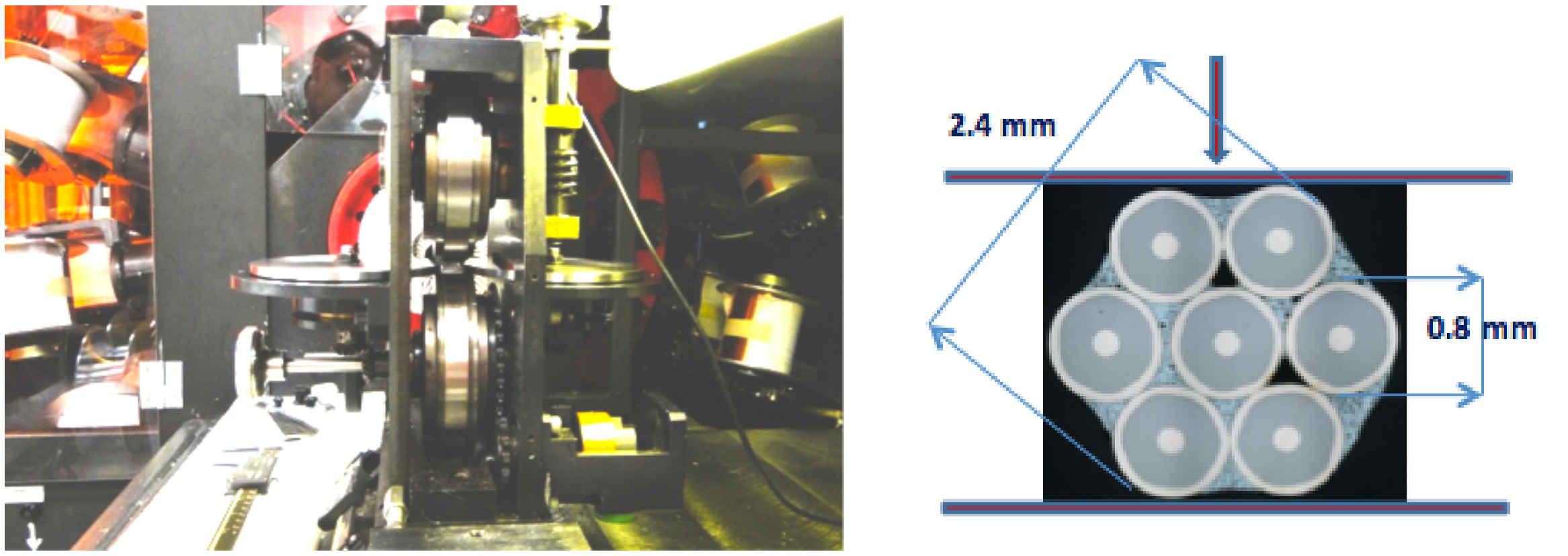

(a) turk-head (b) 6 around 1 configuration

Figure 10.27: (a) Picture of the whole turk-head (b) 6 around 1 cable configuration shown with a schematic of the two compressing vertical rolls of the turk-head magnet.

Table 10.5: Parameters of the 6 around 1 cables produced out of Cu wires of 0.8 mm size.

| Cable ID | Pitch length, mm | Thickness under tension, mm | Width under tension, mm | Crossover | Appearance | Inherent twist | Top roller, mm |
|---|---|---|---|---|---|---|---|
| R&D_CF_01_14 Id 1 | 60.0 | 2.4 | 2.4 | no | good | round | 0.900 |
| R&D_CF_01_14 Id 1A | 60.0 | 1.257 | 4.1 | yes | very compressed | no | -0.006 |
| R&D_CF_01_14 Id 2 | 40.0 | 1.375 | 3.5 | no | very compressed | no | -0.006 |
| R&D_CF_01_14 Id 3 | 51.8 | 1.35 | 3.71 | no | good | no | -0.006 |
| R&D_CF_01_14 Id 4 | 60.0 | 1.36 | 3.73 | yes | center strand slightly exposed | no | -0.006 |
| R&D_CF_01_14 Id 5 | 55.9 | 1.354 | 3.61 | no | center strand slightly exposed | no | -0.006 |
| R&D_CF_01_14 Id 6 | transitional | section | | | | | 0.300 |
| R&D_CF_01_14 Id 7 | 55.9 | 1.72 | 3.75 | no | good | slight | 0.300 |
| R&D_CF_01_14 Id 8 | 55.9 | 1.56 | 3.43 | no | good | no | 0.150 |
| R&D_CF_01_14 Id 9 | 55.9 | 2.5 | 2.5 | no | good | round | 0.900 |
| R&D_CF_01_14 Id 10 | 55.9 | 1.273 | 4.05 | slight | fair | no | -0.006 |
| R&D_CF_01_14 Id 11 | 9.0 | 2.605 | 2.605 | no | poor | round | 0.900 |
| R&D_CF_01_14 Id 12 | 9.0 | 1.375 | 3.6 | no | poor, very compressed | yes | -0.006 |

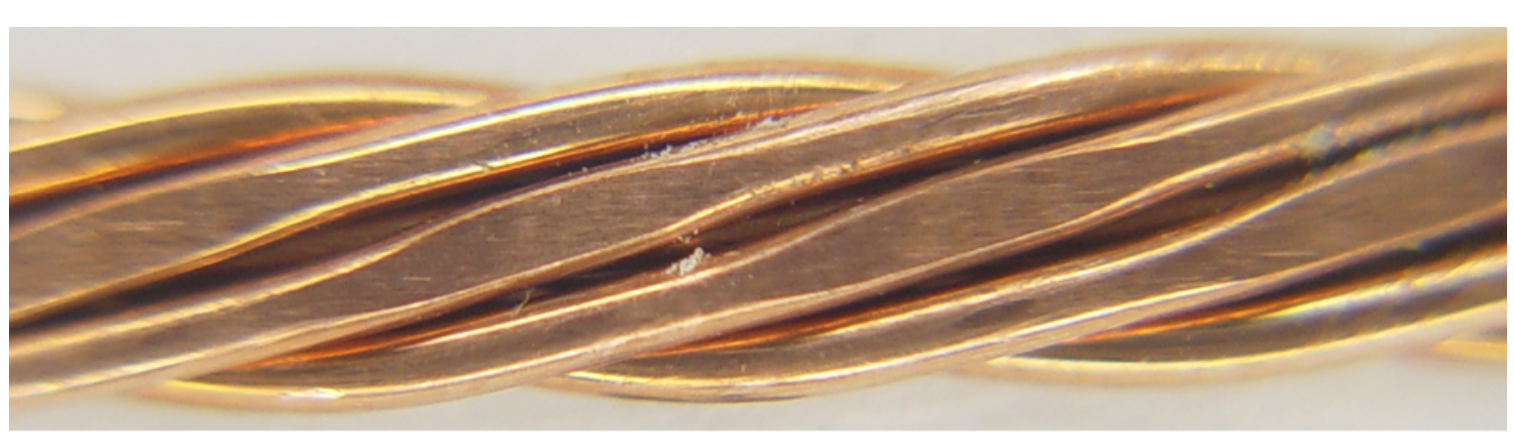

Figure 10.28: Picture of good quality 6 around 1 cable with thickness of 1.72 mm.

NbTi cables, which then required their own optimization since the mechanical properties of multifilamentary NbTi are very different than for the Cu. Superconducting cable samples were produced aiming at various thickness values, as listed in Table 10.6.

Table 10.6: Parameters of the 6 around 1 cables produced with seven NbTi wires.

| Cable ID | Pitch length, mm | Thickness under tension, mm | Width under tension, mm | Thickness, mm | Width, mm | Crossover | Appearance | Inherent twist | Top roller, mm | 0.8 NbTi outer strands | Center strand | To be tested |
|---|---|---|---|---|---|---|---|---|---|---|---|---|
| R&D_CF_02_14 Id 1 | 55.9 | 1.497 | 3.28 | 1.56 | 3.45 | no | good | yes | -0.206 | 6 | NbTi | Y |
| R&D_CF_02_14 Id 2 | 60.6 | 1.771 | 3.14 | 1.875 | 3.27 | no | good | yes | -0.115 | 6 | NbTi | Y |
| R&D_CF_02_14 Id 3 | 60.6 | 2.51 | 2.51 | 2.51 | 2.51 | no | good | round | 0 | 6 | NbTi | |

The critical current $I_c$ of strands extracted from the cables above was tested at 4.2 K and up to 5 T. Some of the results are shown in Fig. 10.29, which also pictures the cross sections of cables Id1 and Id2 from Table 10.7. In these NbTi cables, the central strand suffered much less than the outer strands, which showed instead considerable $I_c$ degradation and very low n-values, which is an indication of superconductor damage as well.

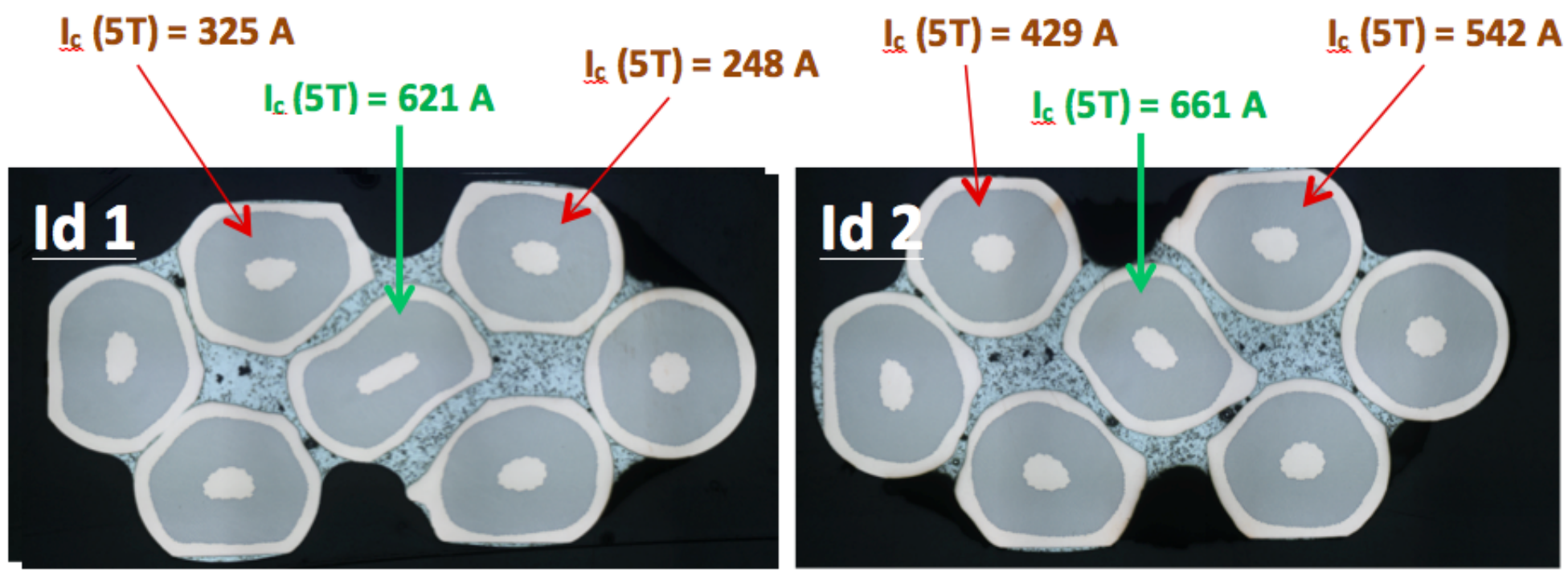

Figure 10.29: Cross sections of NbTi cables made with seven superconducting strands and showing $I_c$ data at 5 T. (Pictures by Marianne Bossert.)

Following such results, the central superconducting strand was replaced with a Cu one, as Cu is a much softer material than NbTi, and the superconducting cable samples, as listed in Table 10.7, were produced again aiming at various thickness values.

Table 10.7: Parameters of the 6 around 1 cables produced with six NbTi wires around a central Cu strand.

| Cable ID | Pitch length, mm | Thickness under tension, mm | Width under tension, mm | Thickness, mm | Width, mm | Crossover | Appearance | Inherent twist | Top roller, mm | 0.8 NbTi outer strands | Center strand | To be tested |
|---|---|---|---|---|---|---|---|---|---|---|---|---|
| R&D_CF_03_14 Id 1 | 60.6 | 2.40 | 2.40 | 2.40 | 2.40 | no | good | round | 0 | 6 | Cu | |
| R&D_CF_03_14 Id 2 | 60.6 | 1.785 | 3.23 | 1.865 | 3.25 | yes | fluffy | yes | -0.112 | 6 | Cu | Y |
| R&D_CF_03_14 Id 3 | 40 | 1.713 | 3.13 | 1.800 | 3.16 | no | good | yes | -0.112 | 6 | Cu | Y |

The critical current $I_c$ of strands extracted from the cables above was tested at 4.2 K and up to 5 T. In these NbTi cables with a central Cu strand, the latter absorbed most of the deformation, therefore better preserving the integrity of the outer NbTi wires. This is shown in Fig. 10.30, which also pictures the cross sections of the cables from Table 10.7. This can also be seen from Fig. 10.31, which demonstrates how the cable with only 6 NbTi strands and 1.8 mm thickness performs better over all of the magnetic field range and offers more Cu stabilizer than the 100% NbTi cable, despite the latter being thicker at 1.875 mm, but suffering more damage and therefore seeing larger critical current degradation.

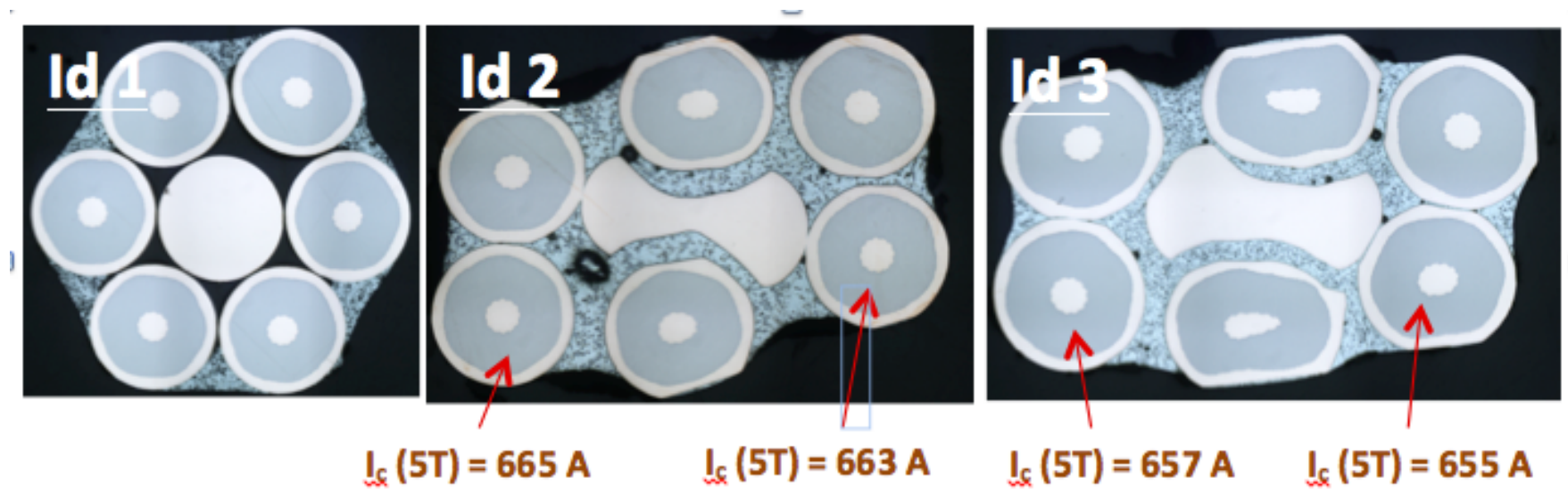

Figure 10.30: Cross sections of NbTi cables made with six superconducting wires around a central Cu strand and showing $I_c$ data at 5 T. (Pictures by Marianne Bossert.)

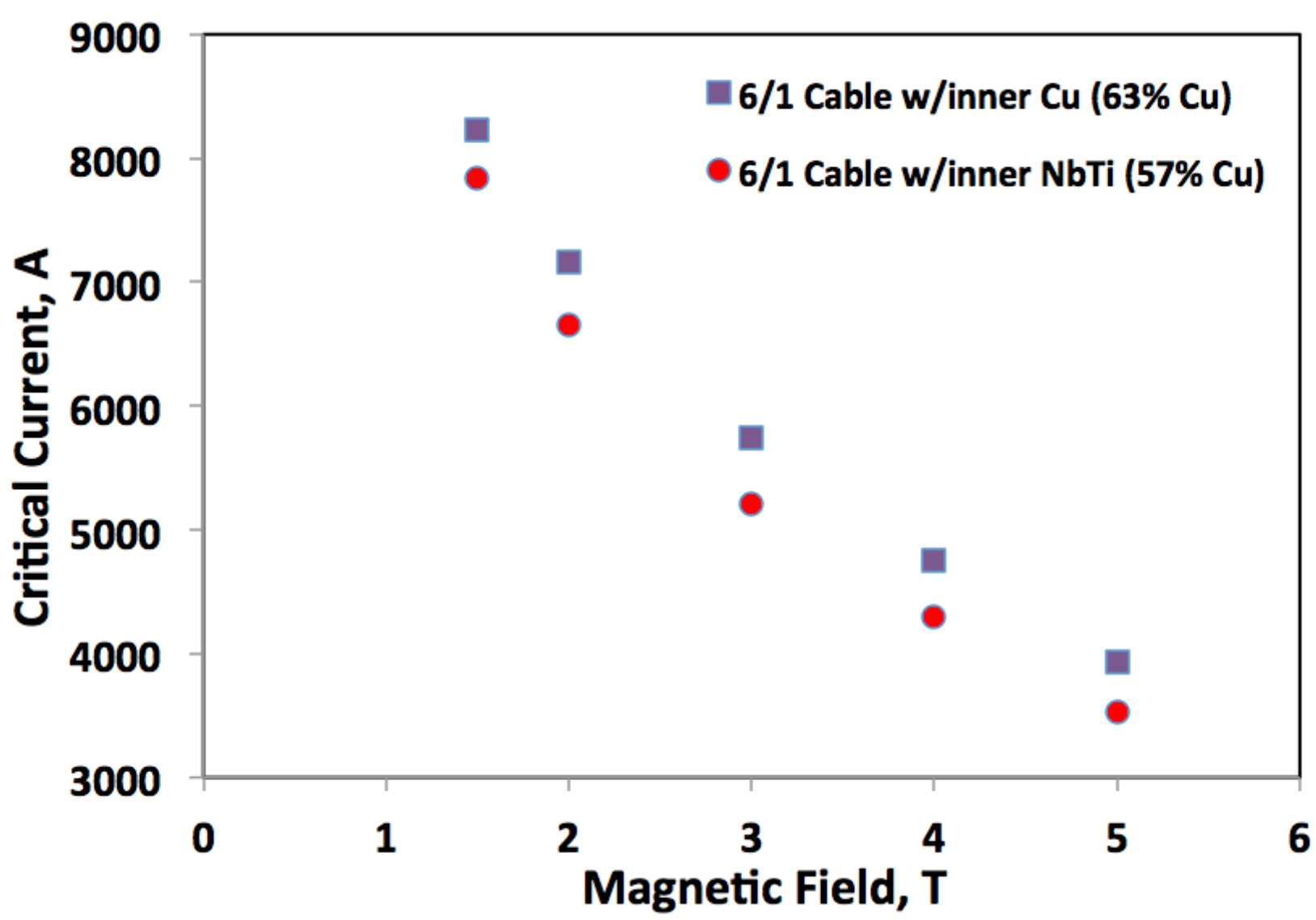

Figure 10.31: $I_c$ as a function of magnetic field $B$ for the 6 around 1 cable made with only NbTi strands and that made with a central Cu wire.

The cable performance at the inflector operation temperature of 4.6 K is obtained by parameterizing and fitting the data acquired at 4.2 K. At 4.6 K and at the maximum field of 3.5 T seen by the inner layer of the magnet under design, the 6 around 1 cable made of six NbTi wires around a central Cu strand is expected to carry 4,800 A, which offers 80% margin with respect to an operation current of 2,685 A. A length of 60 m of such 1.8 mm × 3 mm cable was fabricated for magnet practice winding, and a length of round cable, which is sufficient to wind the whole magnet (i.e. 170 m), was also fabricated. This latter cable will be available to be used as-is if needed, or to be flat-rolled to size, according to the feedback produced by the practice magnet winding. Constraints associated to the winding technology will determine whether further cable R&D will be needed (for instance if cable twist were to be considered excessive for winding, solutions would have to be proposed).

To accommodate instead an inflector magnet design with 10 mm larger aperture, which produces a larger maximum field of 5 T on the outer coil, a similar 6 around 1 study was carried out using 1 mm NbTi and Cu wires. The results on the effect of the central softer Cu wire were very similar for all the produced cables, which again aimed at various thickness values. The superconducting properties of the NbTi peripheral strands were preserved with very good $I_c$ retention with respect to the virgin wire. Based on extracted strands $I_c$ data, a cable of 2.3 mm × 3.85 mm is expected to carry 4,800 A at 5 T and 4.2 K. A sample of the latter cable was produced for magnet practice winding of the outer layer.

### 10.5.2 The Superconducting Passive Shield

For any given magnetic field value $B_1$ to be shielded by a superconducting slab exposed to an external magnetic field $B_{ext}$ (see the schematic in Fig. 10.32), the thickness $d_{min}$ that is requested of the slab is inversely proportional to its critical current density, $J_c(B_{ext})$, according to the following simple formula:

$$d_{min} = \frac{B_1}{\mu_0 J_c} \tag{10.1}$$

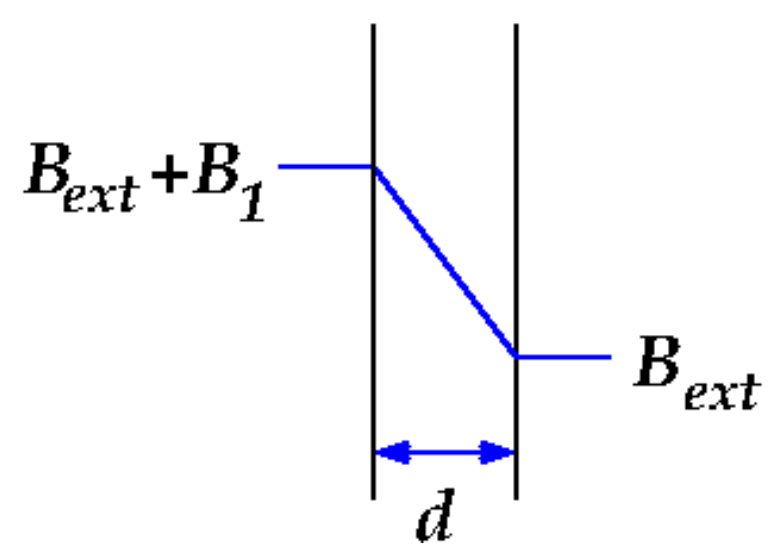

Figure 10.32: $B_1$ is the magnetic field value to be shielded by a slab of thickness $d$ exposed to an external magnetic field $B_{ext}$.

For the shield of the inflector magnet, $B_{ext} \simeq 1.5$ T and $B_1 \geq 0.1$ T. The $J_c$ can be obtained from either testing the critical current $I_c$ of a superconducting sample and normalizing it to the transverse superconducting area of the sample, or from magnetization measurements. The latter are especially important for accuracy at low magnetic fields below 2 T, where self-field effects are non-negligible. This topic is expanded at the end of this section.

In order to test the critical current $I_c$ of NbTi foil samples, an existing experimental setup of the Superconducting R&D lab in FNAL's Technical Division (TD) was modified and re-commissioned. New parts for the sample holder have been procured to replace with pressure contacts the soldered contacts between sample and electrical leads, and between sample and voltage taps. This was done since bulk NbTi cannot be soldered to Cu. The new setup is pictured in Fig. 10.33(a-b). In the middle is the transverse view of the NbTi foil sample mounted in its holder. The sample ends are coated with Indium, inserted within two circular half lugs made of Cu, and clamped to the Cu current carriers. Stainless steel screws are used as voltage taps. They are held on the sample face in its central area by means of G-10 holders, as seen on the left. This setup allows for the sample rotation to perform critical current tests at different field orientations with respect to the sample. The $I_c$ tests on shield samples were performed at 4.2 K from 0 T to 2 T in the parallel and perpendicular field orientations shown in Fig. 10.33(c) The samples were 3 mm wide and 38 mm long.

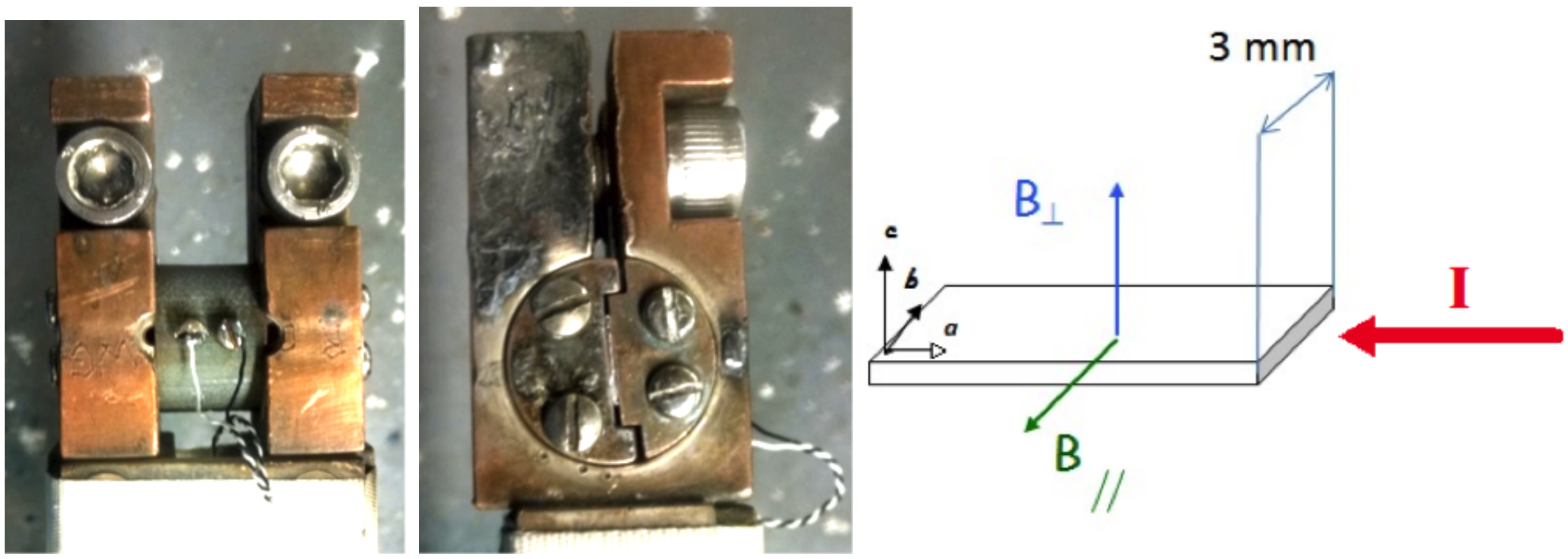

(a) Sample Holder (b) Sample Holder (c) $I$ and $B$ geometry

Figure 10.33: (a-b)Transverse view of a NbTi foil sample (b) mounted in its holder. The sample ends are coated with Indium, inserted within two circular half lugs made of Cu, and clamped to the Cu current carriers. Stainless steel screws, used as voltage taps, are held on the sample face in its central area by means of G-10 holders.(c) Magnetic field orientations used to test the $I_c$(4.2 K) of superconducting shield samples, which were 3 mm wide and 38 mm long.

Another factor to take into account for the characterization of these sheets is their $I_c$ anisotropy. The $I_c$ is typically the largest along the rolling direction, which is shown for instance in Fig. 10.34 for the 25 cm wide and 100 $\mu$m thick Luvata NbTi sheet. In the case of samples cut out of the Nippon Steel Corporation shield used in the first inflector, the rolling direction was unknown. In the following plots the orientation of the latter samples is therefore indicated by their cross section being transverse or parallel to the beam direction with respect to the shield location in the first inflector magnet. Also, because the Japanese shield is made of dozens of thin superconducting layers embedded in Nb and Cu, in Figs. 10.35(a-b) the engineering critical current density $J_e$ was used to compare the performance of samples from the original shield with the Luvata samples. To obtain $J_e$ the $I_c$ is normalized to the total cross section of the sample.

Optical microscopy of samples from the first E821 inflector's shield confirmed that the rolling direction of the shield was that parallel to the beam with respect to its location in the first inflector magnet. This can be seen from Figs. 10.36(a-b), where in the former the superconducting layers in the cross section cut transversely to the beam are flatter and less inhomogeneous than those in the cross section cut parallel to the beam. In these same cross sections one counts 29 Cu layers and 30 intermediate NbTi/Nb layers within an overall slab thickness of 0.22 mm to 0.24 mm. This means that in the first inflector the original 0.75 mm shield developed by Nippon Steel Corporation had been used after further rolling to make it thinner.

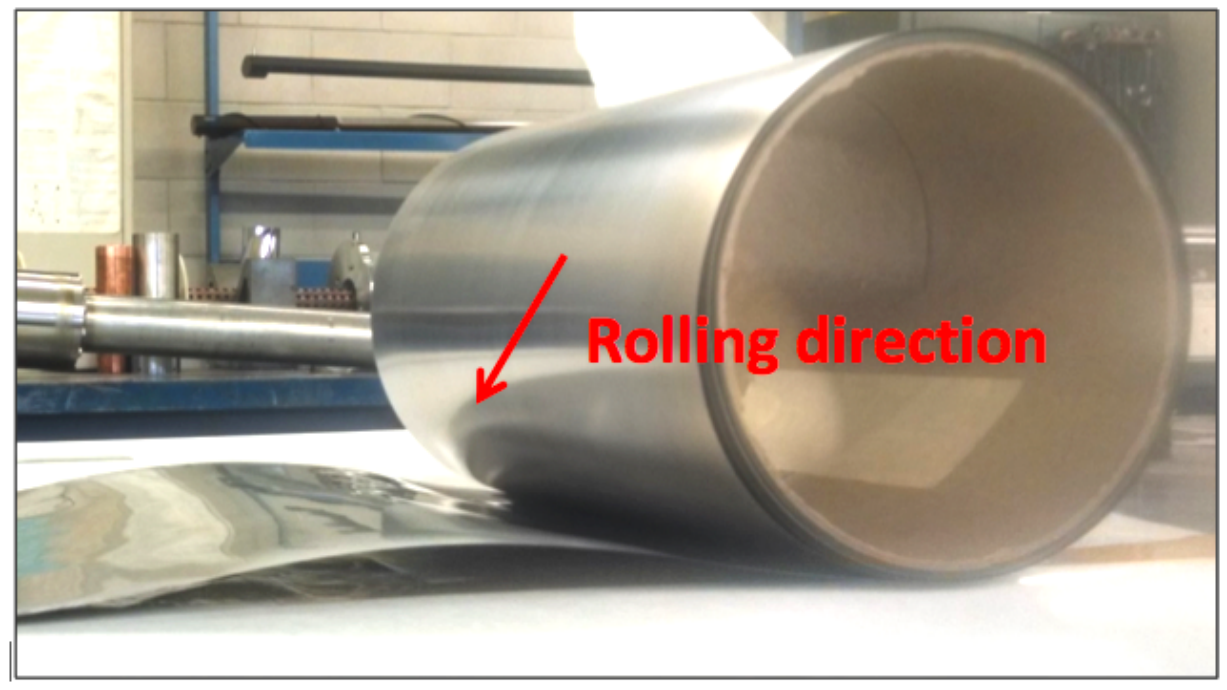

Figure 10.34: Picture of Luvata NbTi sheet showing also its rolling direction.

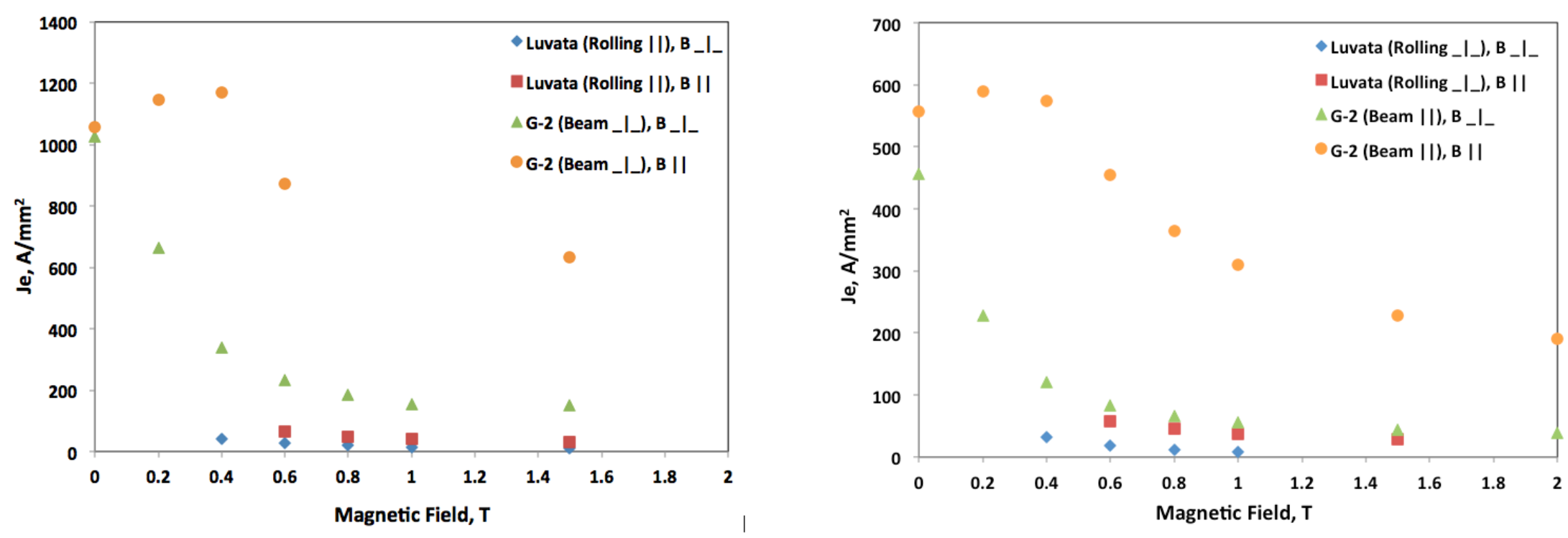

(a) ∥ to rolling (b) ⊥ to rolling

Figure 10.35: (a) Comparison of the engineering critical current density $J_e$ along the rolling direction as function of both parallel and perpendicular magnetic fields between samples from the first inflector shield and the recently acquired Luvata samples. (b) Comparison of the engineering critical current density $J_e$ perpendicularly to the rolling direction as function of both parallel and perpendicular magnetic fields between samples from the first inflector shield and the recently acquired Luvata samples.

The comparison study showed that in parallel field the $J_c$ of the original Japanese shield is more than 10 times larger than the Luvata. It also showed that the original shield is strongly anisotropic, in that the difference in performance between the rolling direction and its orthogonal one is greater than a factor of 3. This can be compared to the bulk NbTi sheet by Luvata where such difference is less than 10%. The difference in performance in parallel and magnetic fields is also larger for the multilayer shield than for the Luvata slab.

As mentioned above, the $J_c$ of a superconductor can also be derived from magnetization measurements, where a slab of thickness $d$ is exposed to an external cycling magnetic field B, which induces in the slab two opposite and equal currents. An example of such hysteresis curve is shown in Fig. 10.37(a). The associated magnetic field profile can be seen in

$$\Delta M = \frac{\mu_0 J_c(B) d}{2} \tag{10.2}$$

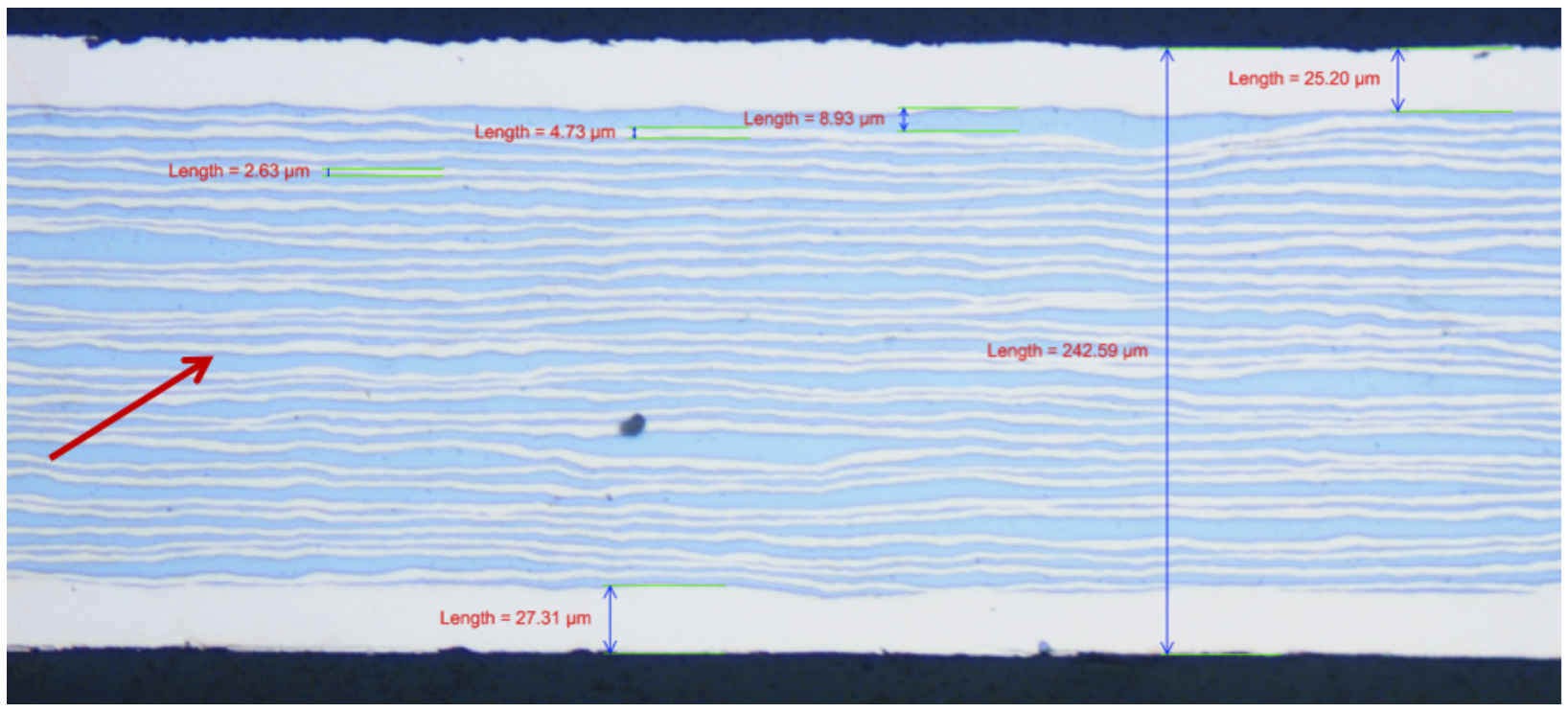

(a) ∥ to beam

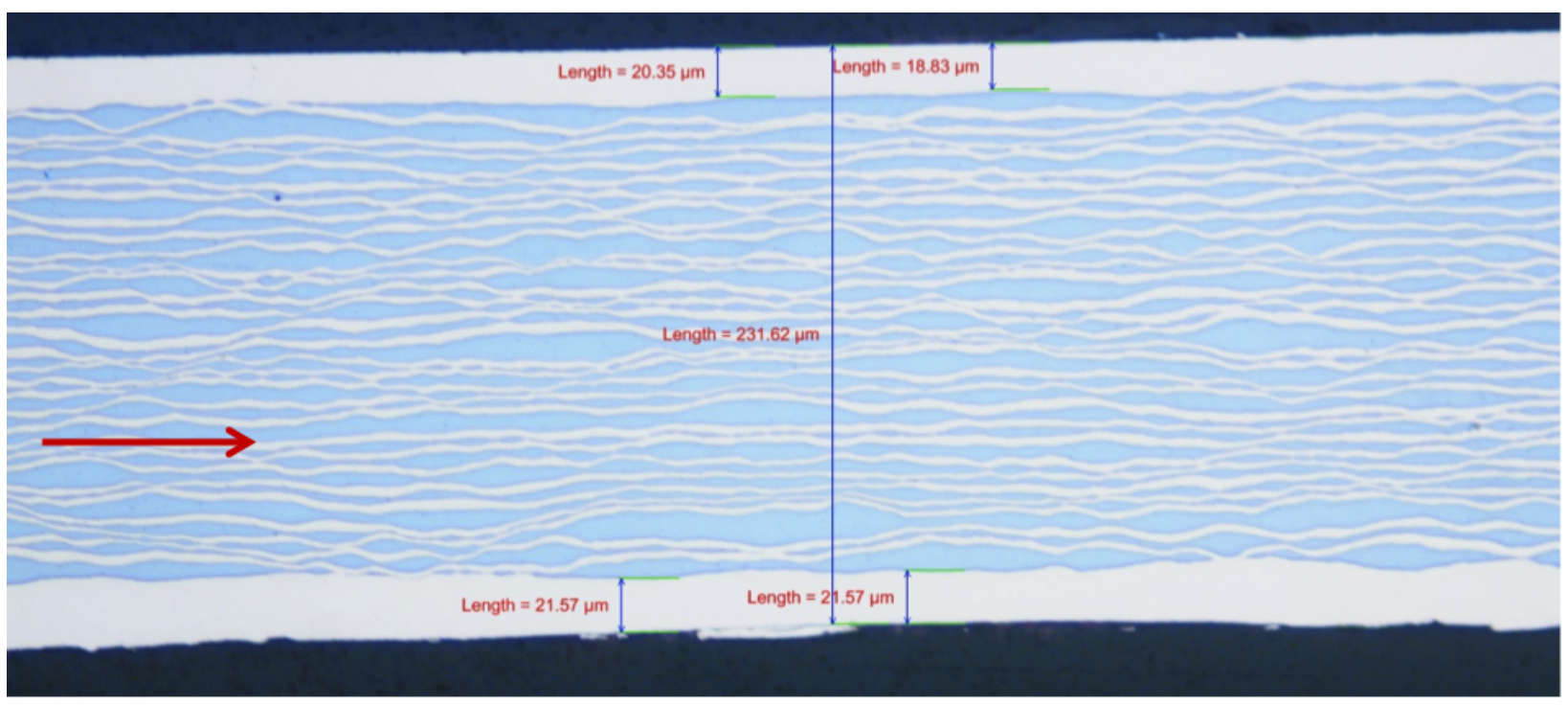

(b) ⊥ to beam

Figure 10.36: (a) Microscopy of original shield sample cut transversely to the beam (shown in Figure as entering the sheet) with respect to its location in the first inflector magnet. (b) Microscopy of original shield sample cut parallel to the beam with respect to its location in the first inflector magnet (Pictures by Allen Rusy, FNAL).

Fig. 10.37(b). The amplitude $\Delta M$ of the magnetization curve follows the following equation:

Magnetization measurements of superconducting samples are performed in TD using a balanced coil magnetometer (Fig. 10.38(a)). The sample is placed inside one of two balanced pick-up coils, and its magnetization is derived from an integrated voltage signal. These measurements are currently performed in parallel magnetic field. Using such setup, samples of the NbTi Luvata foil (Fig. 10.38(b)) of both orientations (i.e. with current parallel and orthogonal to the rolling direction) were tested at 4.2 K up to 3 T, and their $J_c$(4.2 K) were obtained from Eq. 11.2. The $J_c$ results at 4,2 K are shown in the plot of Fig. 10.39, where they are also compared with the $J_c$ data obtained with transport current measurements. Below 2 T the transport current data produce lower $J_c$ values due to self-field, and the data converge in the vicinity of 2 T.

The comparison of Eqs. 10.1 and 10.2 produces the simple following correlation for our

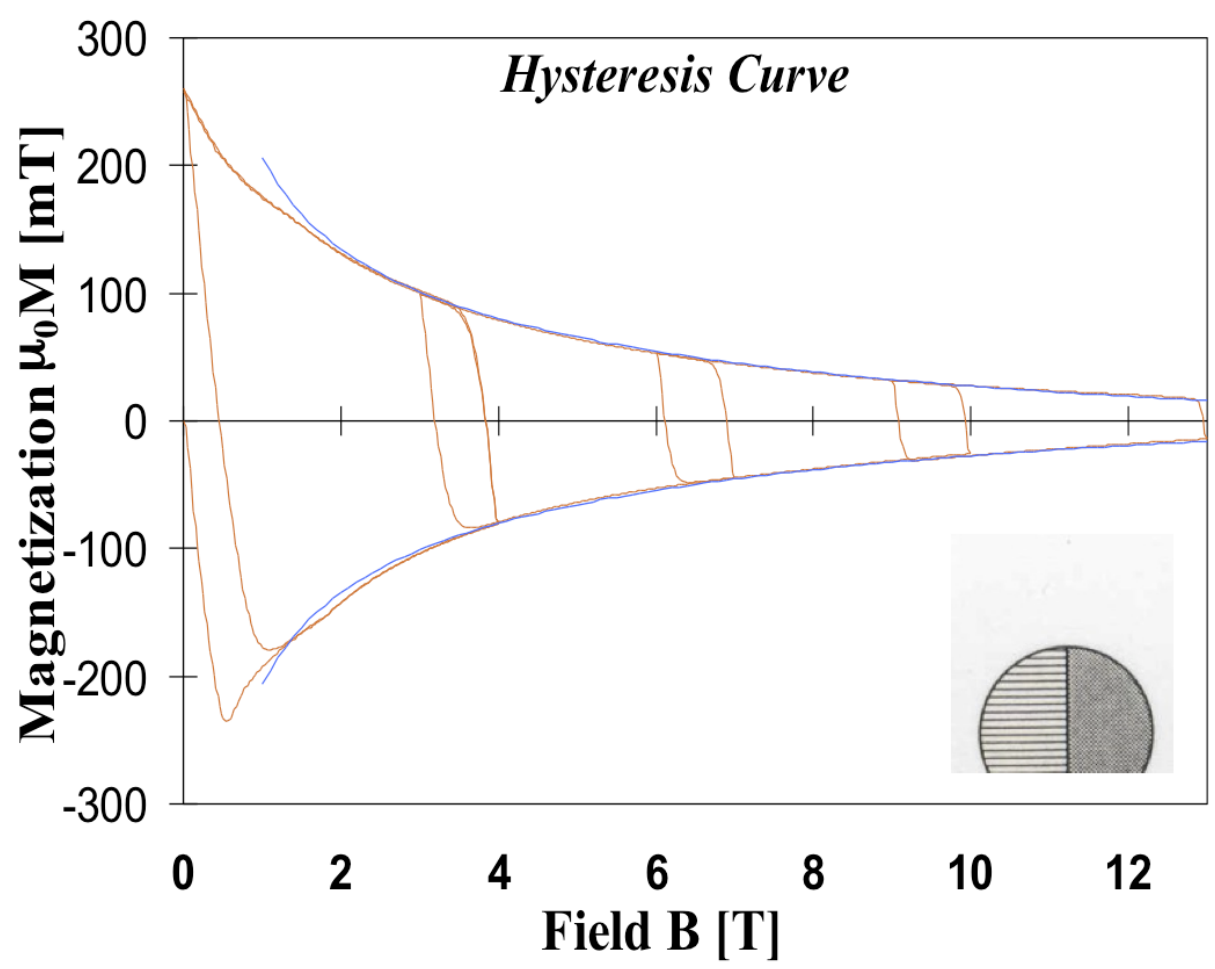

(a) Hysteresis

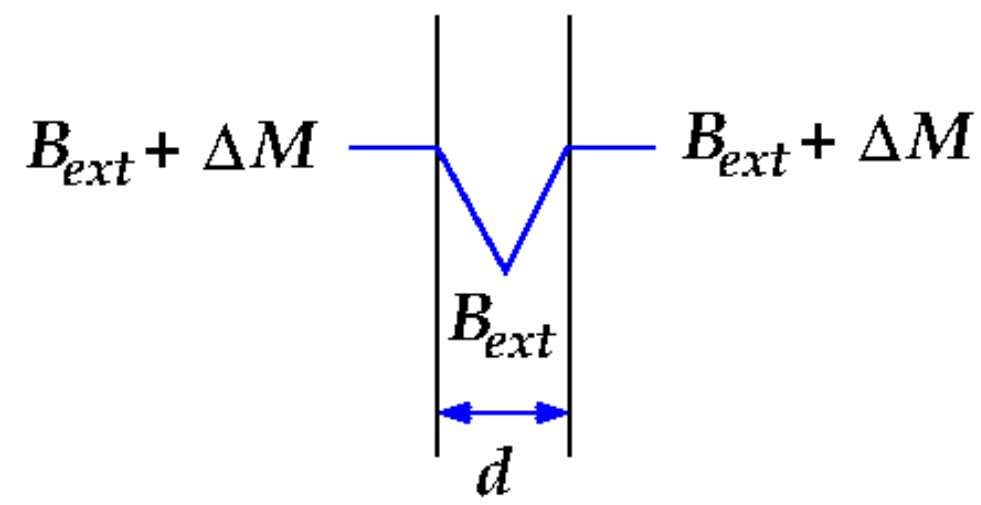

(b) $B$-field Profile

Figure 10.37: (a) Typical magnetization curve of superconducting material. (b) Magnetic field profile of superconducting slab exposed to external magnetic field.

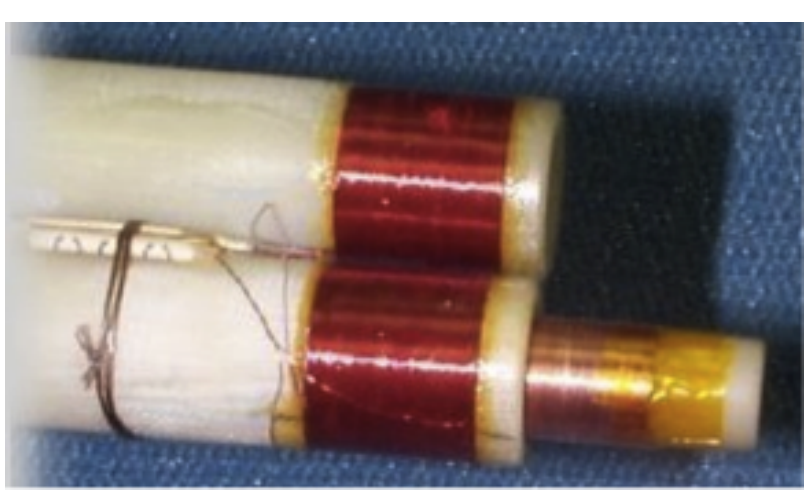

(a) Magnetometer

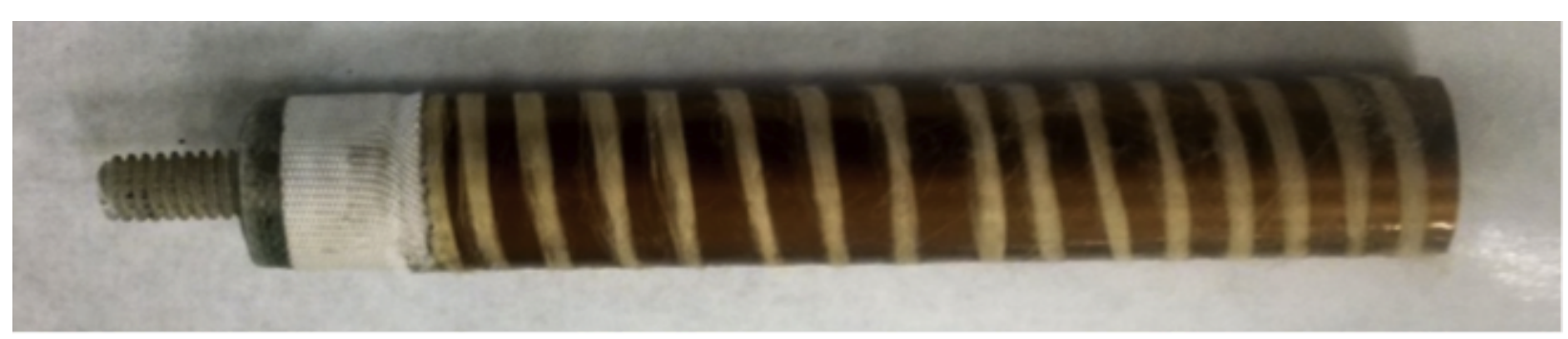

(b) Sample

Figure 10.38: (a) Balanced coil magnetometer used in TD to measure magnetization of superconducting samples. (b) Magnetization sample made by winding a number of turns of the NbTi Luvata foil.

experiment:

$$B_1 = 2\Delta M(B_{ext}) \tag{10.3}$$

The $\Delta M$ measured in parallel field at 1.5 T for the Luvata foil sample in the rolling direction was of 0.07 T, i.e. $J_c$(1.5 T) 110 A/mm$^2$, producing a shielding field $B_1$ of 0.014 T. This is consistent with Eq. 10.1, which for superconducting slabs 100 $\mu$m thick requires a $J_c$(1.5 T) of 800 A/mm$^2$ in order to shield 0.1 T. A shield 250 $\mu$m thick like that used in the first inflector would require a $J_c$(1.5 T) of 350 A/mm$^2$to 400 A/mm$^2$to shield 0.1 T. Transport measurements performed so far in parallel and perpendicular fields respectively showed a $J_c$(1.5 T) between 149 A/mm$^2$ and 633 A/mm$^2$in the rolling direction, and a $J_c$(1.5 T) between 44 A/mm$^2$ and 228 A/mm$^2$ in the orthogonal direction. More accurate values could possibly be obtained with magnetization measurements.

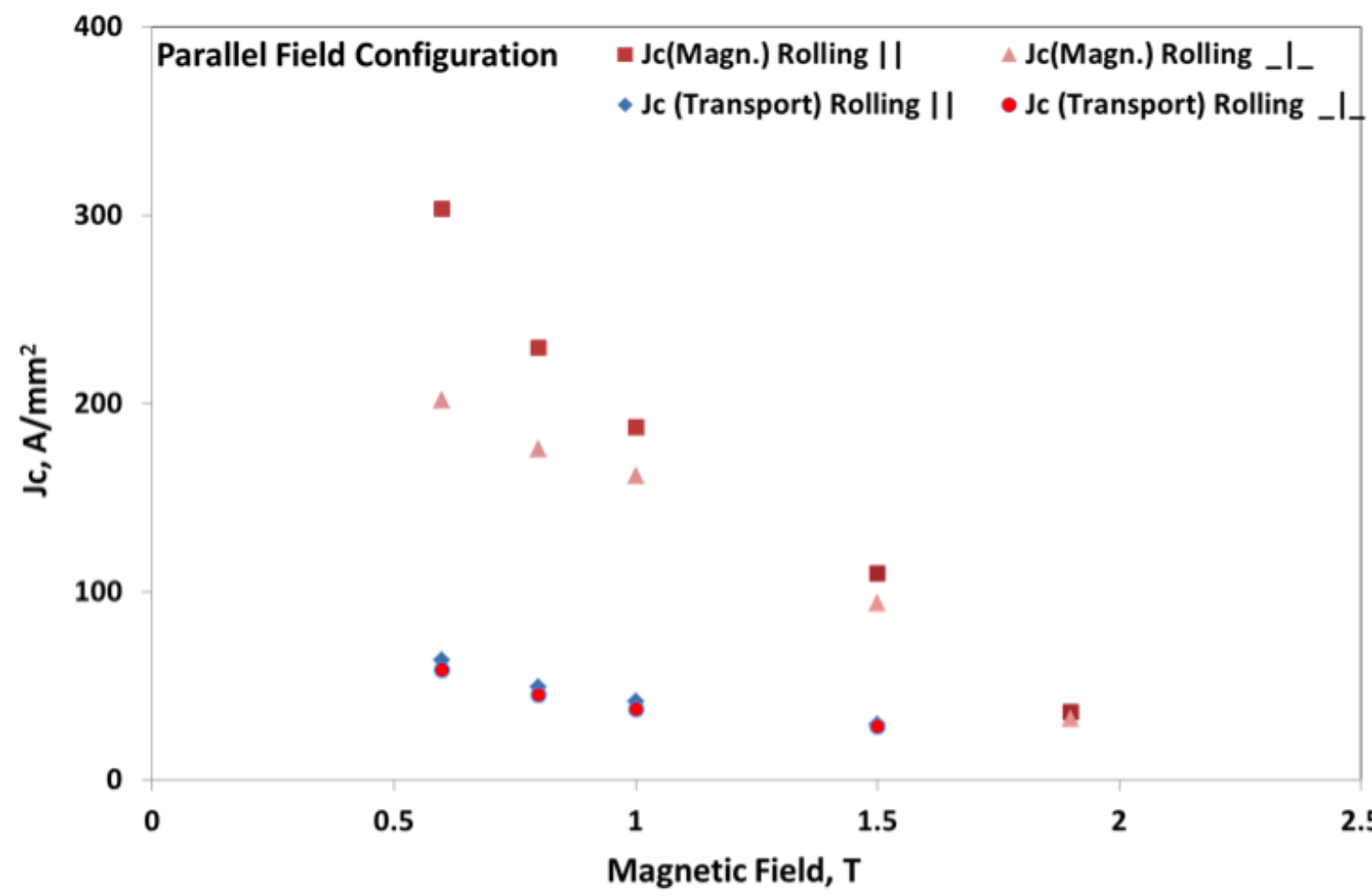

Figure 10.39: Comparison of $J_c$(4.2 K) obtained in parallel magnetic field for samples of the NbTi Luvata foil of both orientations (i.e. with current parallel and orthogonal to the rolling direction) using magnetization and transport current measurements.

### 10.5.3 Double Magnet, Using the Direct Winding Technique

A technique of "direct winding" of superconducting coils has been developed at Brookhaven National Laboratory by Brett Parker. The E989 project has engaged him to come up with a design of a new inflector that uses this technology. The BNL Direct Wind technique is a computer controlled magnet production process where:

- We temporarily bind round superconducting cable to the outer surface of a support structure
- Fill in empty space in the winding pattern with a combination of G10 and epoxy
- After which we apply a compression wrap of s-glass fiber under tension to provide coil pre-stress to counter the Lorentz forces on the conductor during operation.

A complete coil package usually consists of several coil layers. Since we can make small field tuning corrections in successive coil layers based upon magnetic measurements of earlier layers we have been able to satisfy challenging magnetic field quality goals. In this manner we have fabricated rather complex, compact, self-supporting coil structures for a variety of projects[26].

A seemingly natural inflector magnet configuration is one that has nested Direct Wind dipole structures that are adjusted so to have their mutual external fields cancel; however, unless these dipole coils are aligned on a common center (which increases the required kicker strength) this geometry leads to large conductor peak fields due to flux that has to be "channeled" between the inner and outer coil structures. This peak field increase could be mitigated by adding a significant defocusing gradient to the dipole field but this has other unfavorable consequences for the beam optics.

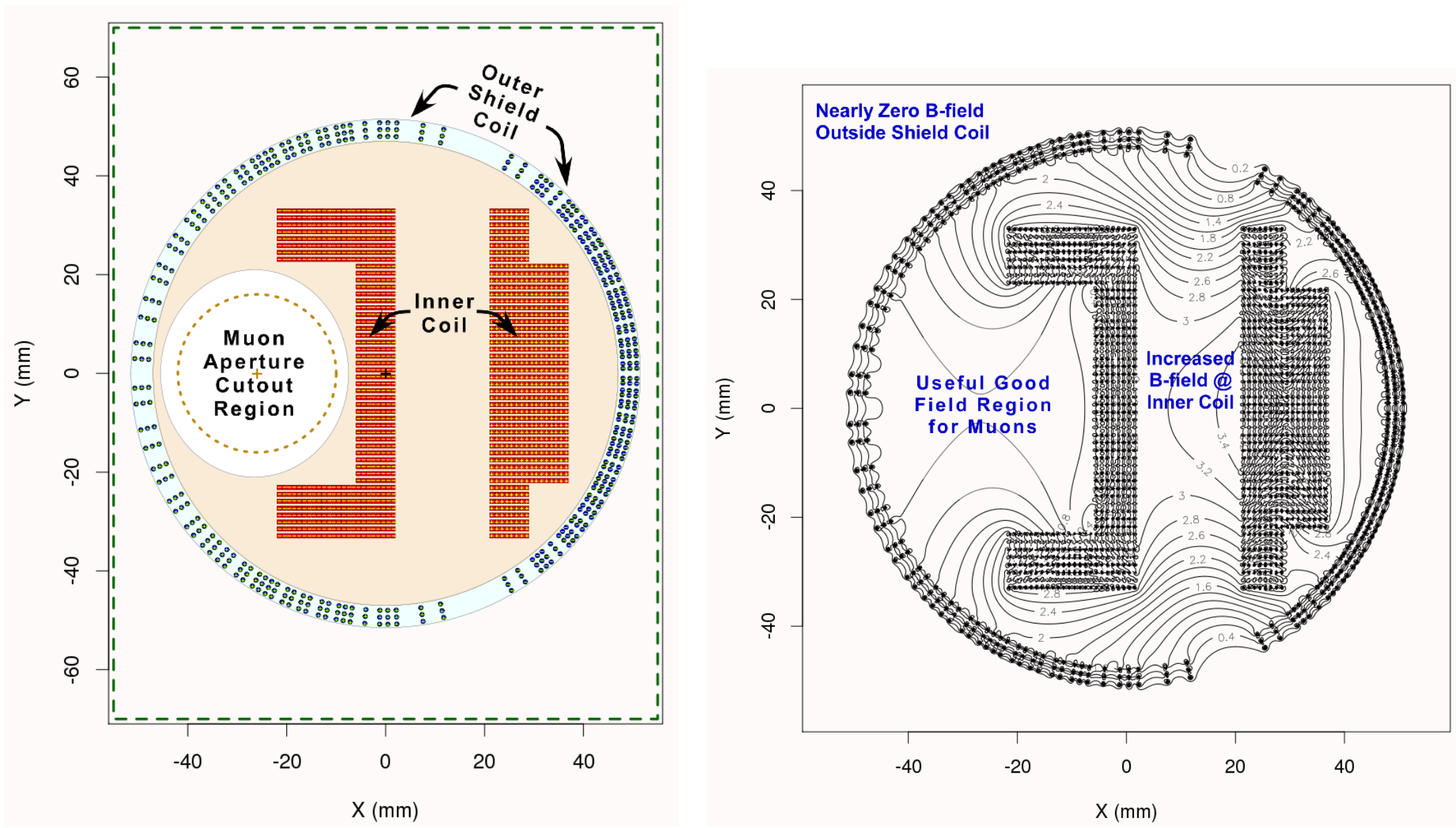

(a) Direct Wind Coil (b) Magnetic Design

Figure 10.40: (a) Conductor Cross Section for an open end coil inflector magnet. The dipole field in the muon aperture results from the combination of the external field due to a nearby coil plus the field generated by a three layer Direct Wind shield coil. (b) Contours of Constant $|B|$ Plotted for the open end coil inflector magnet. Here we see that the $B$-field from the inner coil and outer shield coil is well contained within the boundary of the outer shield coil. Horizontal space is constrained so increasing the space available to the muon beam means reducing the space for flux inside the inner coil which results in increased coil peak field.

Thus in order to maximize the clear aperture for incoming muons while minimizing the increase in required kick angle we have settled upon a hybrid coil scheme shown in Fig. 10.40(a). Motivation for this scheme comes from consideration of the E821 design (see Fig. 10.6), which can be viewed as consisting of two modified dipole windings of opposite polarity placed side-by-side. The flux generated by the coil turns around the main muon aperture is augmented by the external field due to coil turns next to the main aperture. Similarly in Fig. 10.40(a) we have a side coil structure whose external field provides part of the dipole field with the rest coming from the internal field of an outer dipole shield coil; the external field of the inner coil and outer shield cancel each other outside the structure but they add coherently in the main muon aperture as shown in Fig. 10.40(b).

Because the inner coil does not have to pass the muon beam, we choose to wind it as a series of closed end, racetrack coils rather than via Direct Wind. This choice opens the possibility to use a larger and more robust (e.g. added superconductor yet also more room for copper stabilizer) rectangular conductor for winding the inner coil. The inner coil is shaped to both fit inside the shield coil and to provide improved field uniformity on the muon aperture. Reducing the space inside the inner coil gives more space for the muons on the outside at the cost of raising the peak field at the inner coil; however, with its inner conductor having increased operating margin, this is tolerable especially as the inner coil closed ends helps to keep field contained inside.

The inner coil is wound on a solid structure with integrated cooling channels so that the conductors are well supported and conduction cooled. When assembled with some additional parts we come to a smooth outer surface around the inner coil structure upon which we can lay down a three layer Direct Wind coil structure using round cable. Before winding each of the shield layers we make magnetic measurements of the field generated by what has been wound so far and compare to our calculated expectations. In this manner we can derive small corrections before winding each successive shield coil layer to ensure that when the inner coil and external shield coil are powered in series that the external field is canceled as well as possible.

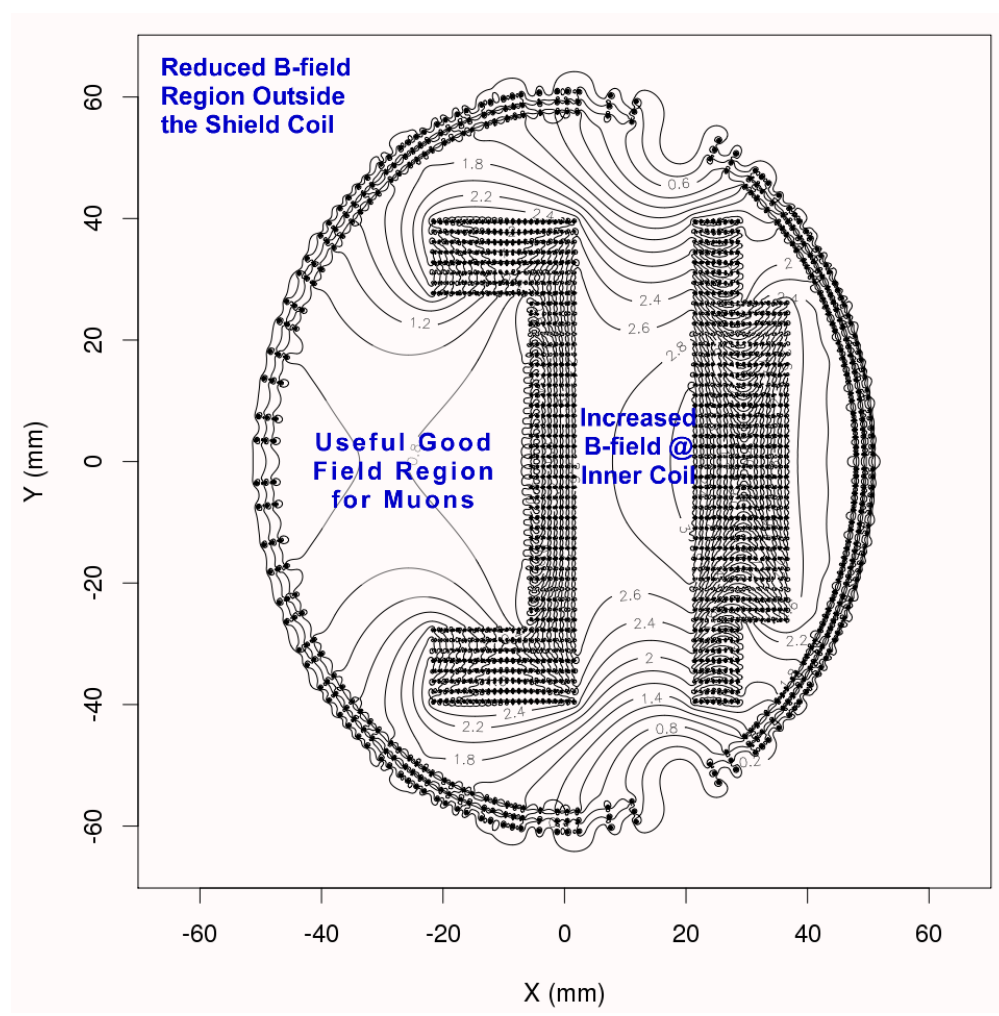

Figure 10.41: Contours of constant $|B|$ plotted for vertically expanded inflector magnet. Here we have the same basic coil structure as for Fig. 10.40 except that all vertical dimensions have been expanded by 20% resulting in 20% greater vertical aperture for muons. The field extending beyond the shield coil is small and it is anticipated that it could be reduced further with minor re-optimization of conductor positions.

Preliminary estimations based upon field calculations of the middle (effectively 2d) section of the inflector magnet indicate that it is possible to keep the external field in the middle well below levels acceptable for the passive superconducting shield. Optimization of the coil "ends" is currently in progress and we do not yet have results available to what level the inflector end field components can be suppressed before reaching the superconducting shield. However we anticipate that a combination of tailoring the end field of the shield coil, as has been successfully demonstrated for the SuperKEKB cancel coils [27], will be necessary along with moving some fraction of the inner coil turns out of their racetrack planes (in a manner similar to making "bedstead turns") in order to truncate the coil end field to an acceptable level. This is the main remaining design optimization task upon which work will continue.

We will also investigate manufacturing a non-circular shield coil as shown in Fig. 10.41. The configuration shown in Fig. 10.41 comes from stretching the previous round configuration by 20% vertically. The benefit of the stretched configuration is gaining 20% in vertical aperture to be able to transport more muons. However, in both theory and practice optimizing the coil configuration for non-circular coils and deriving winding pattern corrections from magnetic measurements is much more complex, so that there is increased risk that the production magnet would have significant field errors. At this point in time completing the end field optimization for circular coils is of highest priority but once completed it will be natural to extend the optimization to include non-circular coils.

## 10.6 Muon Storage Simulations Using a New Inflector

Several aspects of a new superconducting inflector magnet are simulated[1] to study their impact on the fraction of muons transmitted into the storage region. There is significant ongoing work on simulations, and this report summarizes progress to date. Initially studies focused on the E821 inflector, and the beam that could be stored using it at Fermilab. As the end-to-end beam has been developed, these calculations have become more sophisticated. The information presented below represents our latest, but not final results. While promising, it is very much a work in progress. Simulations for E821 predicted that an inflector with open ends (and 18 mm horizontal aperture) would store 1.75 times as many muons as one with both ends closed.

The latest simulation begins with protons hitting the production target, the creation and decay of pions, and propagation of the muon beam around the delivery ring and into the muon storage ring. The temporal distribution of the muons is assumed to be the same as that of the protons on target, with base width of 120ns. There are 10,000 muons in the distribution. The kicker field profile is for the E989 plate geometry, and the kicker pulse is assumed to have 20 ns rise and fall times with 80 ns flat top. The distribution of muons is tracked through the hole in the backleg iron, the storage-ring fringe field and trough the inflector. The simulation includes scattering in the upstream cryostat window, and the coils on both ends of the inflector. The downstream cryostat window is not included. There is scattering in the quadrupole plates as the muons enter the storage ring. The quadrupole plates are assumed to be in the E989 geometry, with the short plates in Q1 at the nominal 5 cm radius from the beam center, and the longer plates at 7 cm. The results from the most recent simulation are shown in Fig. 10.42.

The red curve shows the capture efficiency as a function of kicker field for the baseline inflector configuration. The error bars are statistical. $\beta_x, \beta_y$, and $\eta_x$ have all been optimized for capture efficiency using similar scans. For each set of Twiss parameters the distribution is "refocused" upstream of the magnet iron so that it will have the desired phase space parameters in the inflector. That is, the $\beta$-functions in the inflector are optimized for capture efficiency.

The green curve corresponds to the case where the overlapping inflector end coils are eliminated but with no change in aperture. To estimate the benefit of a larger inflector bore we increase the aperture of the inflector to $\pm 18$ mm from $\pm 9$ mm (blue curve) (and eliminate scatter in the end coils). This is a bit artificial since it assumes no change in the radial position of the inflector axis (see Fig. 9.6). The black curve assumes that the inflector axis is displaced radially by $77 + 9 = 86$ mm and we find that the optimum kick field increases. To see if the reason that with the 86 mm offset (black) was worse than the 77 mm offset(blue) was caused by scattering in the quadrupole plates, the quadrupole scattering was turned off, which gave the dashed curve; a small effect. At present it is not clear why the efficiency is worse for the case with the 86 mm offset compared to the 77 mm offset. We are continuing to study this effect to determine its source.

[1]This material complements the discussion in Section 8.3.2

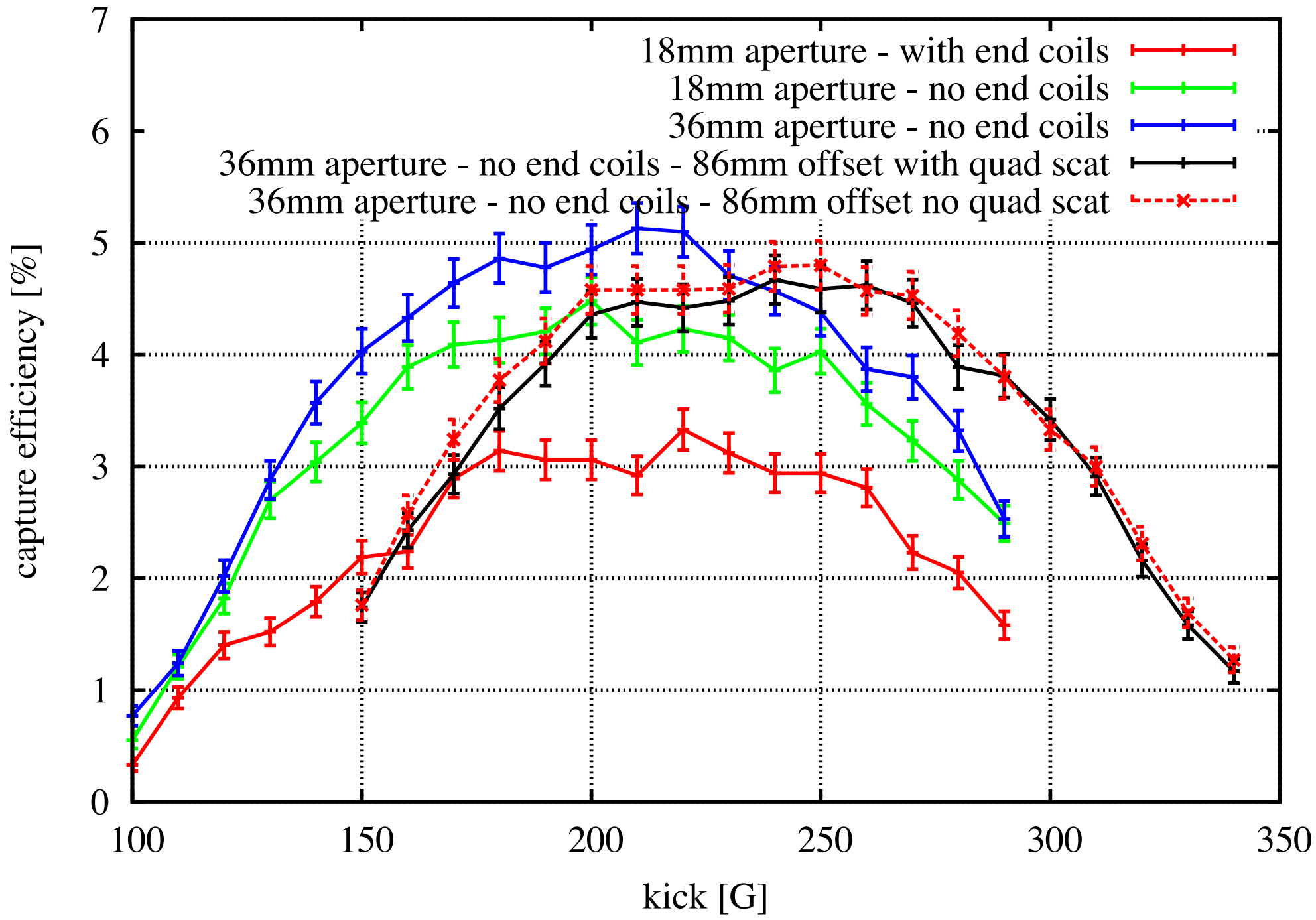

Figure 10.42: Capture efficiency vs kicker magnetic field. The red line (18 mm aperture with end coils) represents the baseline design using the E821 inflector.

### 10.6.1 E821 Inflector Simulation

I this section we report on the earlier studies. We are actively comparing these results to the recent one discussed above. The options studied were the following with the E821 setting shown in parentheses: *a*) open-end vs closed-end (E821) geometry, *b*) 40 mm vs 18 mm (E821) horizontal aperture, *c*) sensitivity to beam phase-space matching. Results of the simulation are presented as improvement factors defined as the fraction of stored muons with the new inflector divided by the baseline E821 inflector. The baseline E821 storage rate is also presented. Assuming all improvements add coherently, a new open-ended inflector with a 40 mm horizontal aperture is expected to increase the fraction of stored muons by a factor of 3.8 compared to the E821 inflector.

The E821 inflector magnet is simulated using a `GEANT`-based software, which allows particle tracking beginning at the upstream end of the inflector. Within this framework, the closed ends of the inflector are constructed using distinct volumes of aluminum (1.58 mm), copper (0.39 mm), and niobium-titanium (0.43 mm). An additional 4 mm of aluminum is added to each end to model the window, flange, and cryostat. Between the end-caps, a "D"-shaped vacuum beam channel is constructed to approximate the double cosine theta geometry. The magnetic field within the beam channel is the vector sum of the main magnet fringe field and the 1.45 T field ($\int \vec{B} \cdot d\vec{\ell} = 2.55$ Tm) produced by the inflector magnet.

The E821 muon beam is simulated by uniformly populating a $40\pi$ phase space ellipse. The phase space axes are determined by the beam Twiss parameters, $\alpha$ and $\beta$ in both horizontal ($x$) and vertical ($y$) directions. The nominal Twiss parameters are determined by maximizing the transmission rate through the inflector and shown in Table 10.8 when the beam is localized at the "downstream"-end of the inflector (*i.e.* nearest to the ring). The beam momentum, $|P|$ , is generated by sampling a Gaussian distribution with mean equal to the magic momentum $P_{\rm magic}$ and width $\delta P/P = 0.5\%$. The longitudinal width of the beam, or equivalently the width in time, is 25 ns.

Table 10.8: Nominal muon beam Twiss parameters.

| Direction | Emittance ($\varepsilon$) | $\alpha$ | $\beta$ |
|---|---|---|---|
| Horizontal (x) | 40 | -0.544 | 2.03 |
| Vertical (y) | 40 | -0.0434 | 19.6 |

All muons passing into the storage region are given a "perfect kick" to place them onto a stable orbit. This kick is modeled by applying a 220 Gauss magnetic field throughout the kicker volume for the first revolution. Finally, the storage rate is defined as the fraction of muons surviving 100 revolutions around the storage ring. No muons are allowed to decay in this simulation.

### 10.6.2 Open-ended vs. Closed-ended Inflector Geometry

The E821 inflector magnet was constructed with a closed end (*i.e.* the superconducting coils wrapped around the end of the magnet) because this greatly reduced magnetic flux leakage into the muon storage region. The impact of the closed end on the horizontal and vertical emittance was studied analytically and with the `GEANT` tracking software. In the analytic approach, the fraction of muons traversing the inflector ends is studied by comparing the horizontal and vertical beam widths ($\sigma_x, \sigma_y$) after multiple scattering in the material. In this study, a beam filling the horizontal aperture of 18 mm grows to a size of $\approx 35$ mm, suggesting that approximately half ($18/35 = 51\%$) of the beam will fail to exit the inflector aperture. With two closed ends the net effect is to lose between $50 - 75\%$ of the incoming beam.

The tracking simulation approach removes the end coils, flange, and window from the GEANT inflector material without altering the magnetic fields. Table 10.9 summarizes the muon storage rates assuming an open and a closed inflector magnet. The beam parameters and inflector aperture are identical in both simulations. Values in parentheses show the results of an incoming beam with a momentum spread of 2% instead of the nominal 0.5%.

Table 10.9: Summary of E821 Inflector Simulations.

| Inflector Geometry (Upstream-Downstream) | Muons Generated | Muons Surviving | Storage Fraction |
|---|---|---|---|
| Open-Open | 5000 (20000) | 664 (691) | 13.2±0.3 (3.4±0.1) |
| Closed-Open | 5000 (20000) | 522 (593) | 10.4±0.3 (2.8±0.1) |
| Closed-Closed | 5000 (20000) | 323 (395) | 6.5±0.3 (1.9±0.1) |
| Improvement Factor ≡ Open-Open/Closed-Closed | | | |
| | 5000 (20000) | - | 2.1× (1.7×) |
| Improvement Factor ≡ Closed-Open/Closed-Closed | | | |
| | 5000 (20000) | - | 1.6× (1.5×) |

### 10.6.3 Sensitivity to Beam Phase-space Matching

A consequence of the limited inflector aperture is gross phase space mismatching into the storage region. This is seen by studying the amplitude of the muon beam ($A$), which is defined as $A = \sqrt{\beta\varepsilon}$. The maximum horizontal size of a beam clearing the inflector is ±9 mm, thus, a beam with $\varepsilon = 40$ mm-mrad must have $\beta_x < 2.5$ m and $\beta_y < 19.6$ m. As this beam propagates into the storage region the horizontal $\beta$-function subsequently undergoes large oscillations with $\beta^{\max} = 28$ m and $\beta^{\min} = 2.5$ m. This corresponds to a modulation of the horizontal beam amplitude ($A$) of $r = \sqrt{\frac{\beta^{\max}}{\beta^{\min}}} = 3.4$. This oscillation causes significant beam to be lost on the collimators in the ring.

An alternative to these large oscillations is to perfectly match the $\beta$-functions into the storage ring. Assuming a drift space within the inflector ($\vec{B} = 0$), then the $\beta$-function at the inflector is defined as $\beta^{\text{inf}} = \beta^{\text{ring}} + s^2/\beta^{\text{ring}}$. The resulting $\beta$-functions ($\beta_x^{\text{inf}} = 7.6$ m and $\beta_y^{\text{inf}} = 19.2$ m) requires the incoming beam to be 2.38 times larger than the inflector aperture. Thus, only $1/2.38 = 42\%$ of the beam will clear the inflector. This conclusion follows the GEANT-based tracking result, which shows 53% of the beam clearing the inflector aperture.

### 10.6.4 Increased Horizontal Aperture

The E821 inflector was constructed with a ±9 mm horizontal aperture in part due to the double cosine theta magnet geometry and the limited space between the outer main magnet cryostat and the muon storage region. The horizontal aperture also constricts the available phase space in the muon storage region, whose aperture is ±45 mm.

An augmented inflector "D"-shaped aperture of $\pm 20 \times \pm 28$ mm$^2$ is modeled in the `GEANT` tracking software. In this study, the main magnet fringe field is assumed to be identically canceled within the inflector beam channel for simplicity. The horizontal beam size is increased allowing for ideal matching to the storage ring $\beta$-function, corresponding to $\beta_x = 7.6$ m. The horizontal and vertical $\alpha$ Twiss parameters are set to zero in this scenario.

Table 10.10 summarizes the muon storage rates for the two apertures (18 vs 40 mm) and the two end coil inflector geometries (open vs closed) [2].

Table 10.10: Summary of E821 Inflector Simulations. The "D"-shaped aperture shown in Fig. 10.6(a) was used. The vertical aperture was 56 mm, the horizontal (radial) aperture was 18 mm, or 40 mm.

| Inflector Aperture (Open or Closed ends) | Muons Generated | Muons Surviving | Storage Rate |
|---|---|---|---|
| 18 mm Aperture ($A_{\pm 9}$) | | | |
| (open ends) | 120000 | 11444 | 9.5±0.1 |
| (closed ends) | 120000 | 5117 | 4.2±0.1 |
| 40 mm Aperture ($A_{\pm 20}$) | | | |
| (open ends) | 120000 | 19161 | 15.9±0.1 |
| (closed ends) | 120000 | 8706 | 7.2±0.1 |
| Improvement Factor $\equiv A_{\pm 20}/A_{\pm 9}$ | | | |
| (open ends) | - | - | 1.7× |
| (closed ends) | - | - | 1.7× |
| Improvement Factor $\equiv A^{\text{Open}}/A^{\text{Closed}}$ | | | |
| (18 mm Aperture) | - | - | 2.2× |
| (40 mm Aperture) | - | - | 2.2× |
| Improvement Factor $\equiv A^{\text{Open}}_{\pm 20}/A^{\text{Closed}}_{\pm 9}$ | | | |
| | - | - | 3.8× |

### 10.6.5 Summary and Future Simulations

The most recent, but preliminary, simulation that includes

It is clear that we need a full tracking simulation to replace the phase-space models used thus far. The incoming beam to the inflector needs to be optimized for the 18 mm inflector opening, and for 40 mm new inflector. The inflector team will work with the beamline experts to make sure that these essential calculations are high priority.

[2]Note that these storage rates are computed with a different muon beam and therefore can not be compared directly to the rates in the previous sections.

## 10.7 ES&H

The superconducting inflector is in a cryostat that includes one section of muon beam tube. The cryostat vacuum is separate from the beam vacuum chamber, so that the inflector can be operated independently of whether the muon beam chamber is evacuated. The cryogenic system, and its operation will follow all Fermilab safety standards for cryogenic and vacuum system operations. This includes, but is not limited to Extreme Cold Hazard, Oxygen Deficiency Hazards. The cryogens involved are liquid helium and liquid nitrogen. No flammable liquids or gases will be employed. The existing E821 inflector was operated at Brookhaven National Laboratory where similar safety requirements were in place.

## 10.8 Risks

### 10.8.1 Relocation Risk

The relocation risk was minimized by careful disassembly and shipping. The E821 inflector is on-site a Fermilab, and as soon as cryogenic capability is available in MC-1, we will set up a test stand in the experimental area outside of the ring to cool and power the inflector.

### 10.8.2 Other Risks

There is the possibility that some mechanical aspect of the E821 inflector has deteriorated in the 12 years since it was operational, causing the magnet to quench repeatedly before reaching full current. This risk is probably small, since it was tested at KEK, shipped to BNL, installed, and was brought to full current with only a few training quenches. It was very robust in subsequent operation at BNL. The plan to test it as soon as possible at Fermilab will clarify this risk.

At Brookhaven a helium leak in the valve box or lead-pot developed. After examining the situation after transport, it has been decided to re-build this part of the inflector system, rather than repair the existing one. There is a small risk that the leak was in the magnet itself, but this is viewed as extremely unlikely by Akira Yamamoto, who supervised the engineering design and construction, and Wuzheng Meng, who did the magnetic design and was responsible for its operation at BNL.

The most sensitive part of the re-installation is reconnecting the inflector leads. Our technician Kelly Hardin was involved in the disassembly at BNL, and understands the issues involved in the reconnection very well.

## 10.9 Quality Assurance

Proper quality assurance is essential in the transport and reassembly of the inflector magnet. The mechanical aspects, heat shield, etc. will be carefully examined for issues, once the inflector arrive at Fermilab. It will be determined as quickly as possible whether the inflector meets the Muon g-2 requirements for performance and reliable operation. Quality Assurance will be integrated into all phases of the transport and reassembly work. including design, procurement, fabrication, and installation.

## 10.10 Value Engineering

The baseline is to begin the experiment by re-using the existing E821 Inflector. A new inflector with a much larger horizontal aperture could permit two to three times as many muons to be stored. A gain of this factor would significantly improve the statistical reach of the experiment, and permit more beam time to be used for systematic studies.

Thus a new inflector would present a significant opportunity to improve the experiment and to use running time more effectively. At present there is insufficient funding in the project to develop and produce a new inflector. Nevertheless, given the potential benefit, we will continue to support the preliminary design of a new inflector, in order to understand the cost of such a device. Should sufficient contingency be earned back from other areas of the project, we will proceed with final design and implementation.

# Chapter 11

# Beam Vacuum Chambers

The muon storage volume, which lies within the 1.45 T magnetic field, is evacuated in order to minimize multiple scattering of muons and positrons. This is accomplished by a set of aluminum vacuum chambers, which also provides mechanical support for:

- the beam manipulation systems: the electrostatic plates of the quadrupole system, the collimators, and plates of the magnetic kicker system.
- the positron detection systems: the trace-back straw trackers and auxiliary detectors such as the fiber harp.
- the magnetic field measurement systems: $\sim$ 400 fixed NMR probes surrounding the storage volume, a set of rails for the trolley NMR system, and the plunge probe system.

The chambers from BNL E821 will be reused for E989, and we will make changes as described in the section below. The chamber design is detailed in the BNL E821 design report[1], and so only a brief discussion is given here. Figure 11.1 shows the layout. The system comprises mainly 12 large vacuum chambers, separated by 12 short bellows adapter sections.

A simplified FEA model of a large vacuum chamber is shown in Figure 11.2, depicting the top plate and the contains 15 grooves for mounting the NMR probes. The 15 grooves on the bottom and flange ports are not shown. The FEA model predicts that the top and bottom surfaces deflect by 0.453 mm under vacuum load[2]. This is in agreement with the measurement of 0.45 mm[7]. The FEA model reconfirms that the chamber has a safety factor of 2.9, and the wall stresses are below 12000 psi, as required by the ASME Pressure Vessel Code for pressure vessels for Aluminum 6061-T6.

The 12 vacuum chambers and 12 bellow adapter sections are bolted together and placed in between the upper and lower pole pieces. The average radius of this structure is mechanically fixed and cannot be adjusted. In the E821 design, the vacuum chambers are electrically connected to one another via bolting hardware, except at a single location where there is a thin dielectric sheet. This prevents eddy currents, such as arising from energy extraction of the magnet, from traveling completely around the ring. For E989, we will modify the electrical connections slightly as described below. Finally, all chamber materials including bolting hardware are non-magnetic.

Figures 11.3 and 11.4 show the cage system and how it resides inside a vacuum chamber. The cage system holds the quadrupole plates, kicker plates, and the rails used by the trolley.

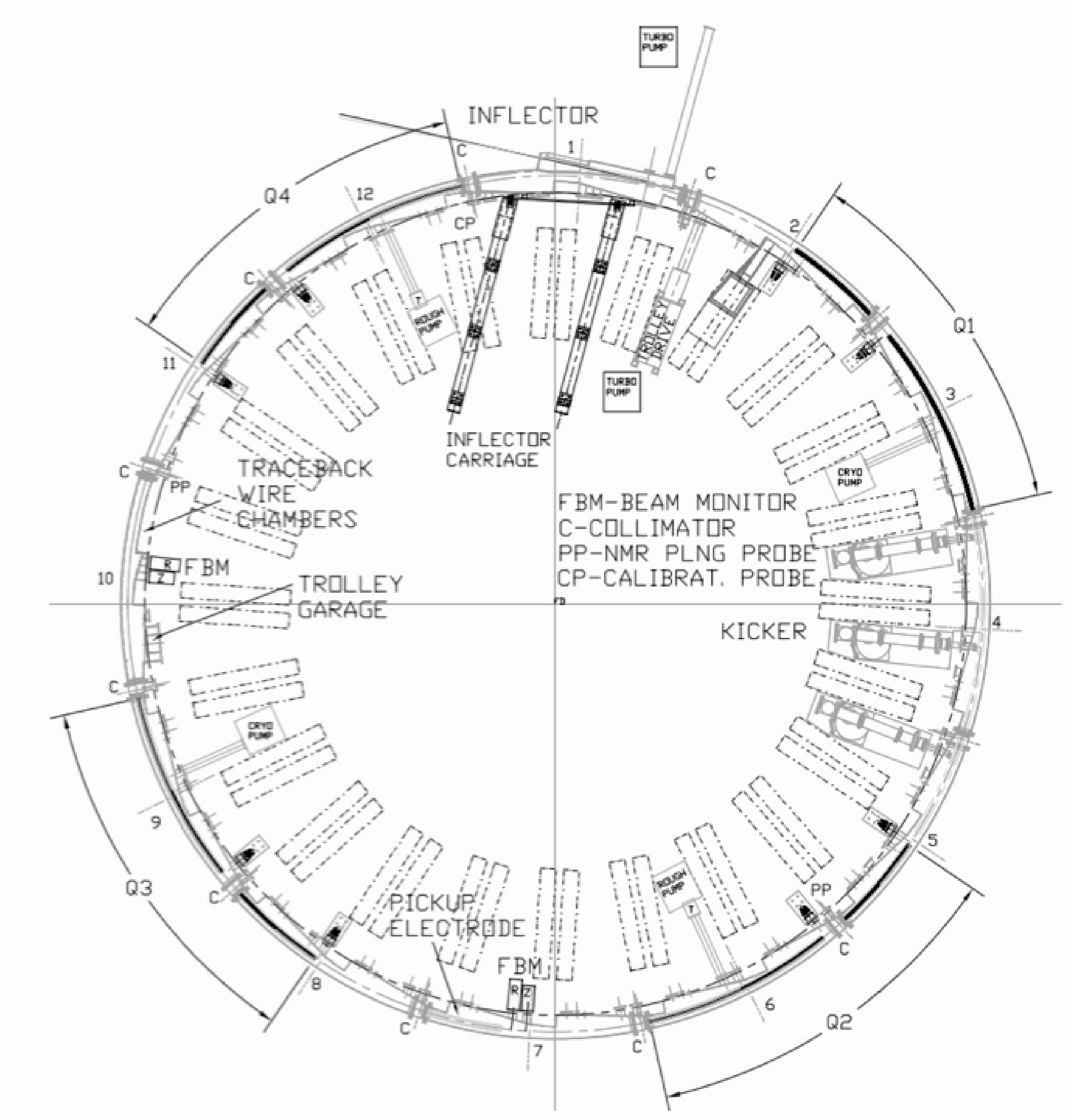

Figure 11.1: Layout of the BNL E821 beam vacuum chamber system.

Screws allow for adjusting the position of the cage within the vacuum chamber system. The position of the cage system plays an important role, and has the following requirements. (1) The rail system from neighboring vacuum sections must line up to allow smooth motion of the trolley as it travels between sections. And (2), since the quadrupole plates and kicker plates positions define the beam storage region, these devices should place the beam in the most uniform portion of the magnetic field. The beam center should be at the geometrical center between upper and lower pole faces. The E989 method to survey and align these cages are described in the following subsection.

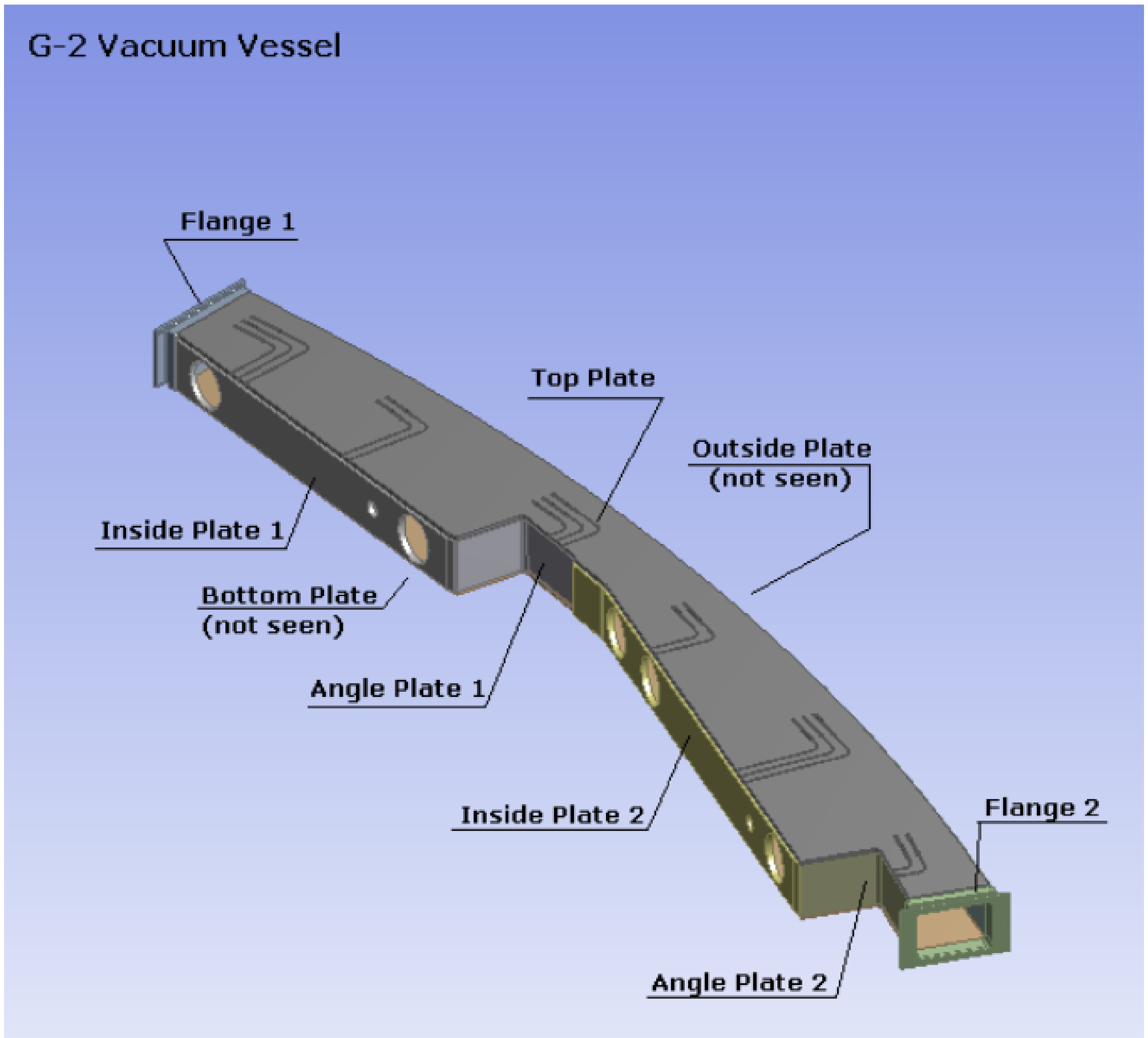

Figure 11.2: Simplified mechanical model for stress and strain analysis.

## 11.1 Changes to the E821 Design

For E989, we are proposing to make the following changes. For E821, the magnetic field gradient near the azimuthal boundary between two yokes was found to be sufficiently high, causing the fixed probes in that region to have readout difficulties. The reason for the higher gradients are discussed in chapter 9. For E989, while we expect to minimize the gradient with improved shimming, we will also lengthen or cut new grooves to optimize the probe placement. The maximum number of affected grooves are shown in reference [3].

In E821, the trace back system operated in air and was located in vacuum chamber sector 10, which was modified to be without a 'scallop'. For E989, a straw station will be in vacuum chamber 11, and vacuum chamber sector 10 will have its scallop shape reinstalled. Straw stations will also be placed in sectors 2 and 8. The inner radius vertical side walls of sectors 2, 8 and 11 will be modified to accept the straw chamber flange. Figure 19.5 shows the locations of the proposed changes.

Finally, the vertical inner radius surface of the vacuum chamber will be lined with insulation. This will improve the thermal stability of the magnet iron, which is critical for the field uniformity as discussed in Section 9.4.

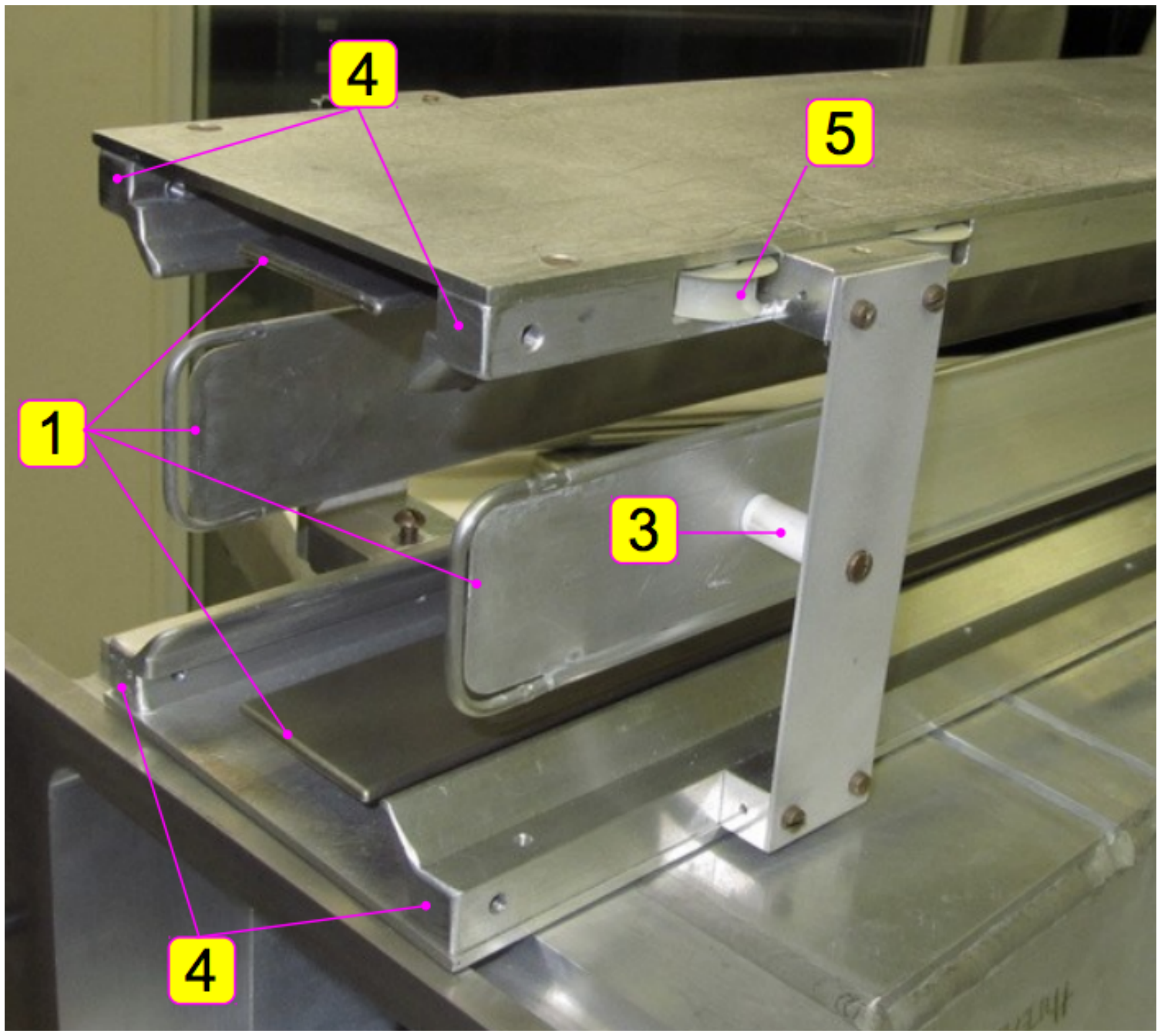

Figure 11.3: Picture of a cage system showing the (1) quadrupole plates, (3) macor (insulator) supports, (4) trolley rails, and (5) a wheel for guiding the cable that pulls the trolley

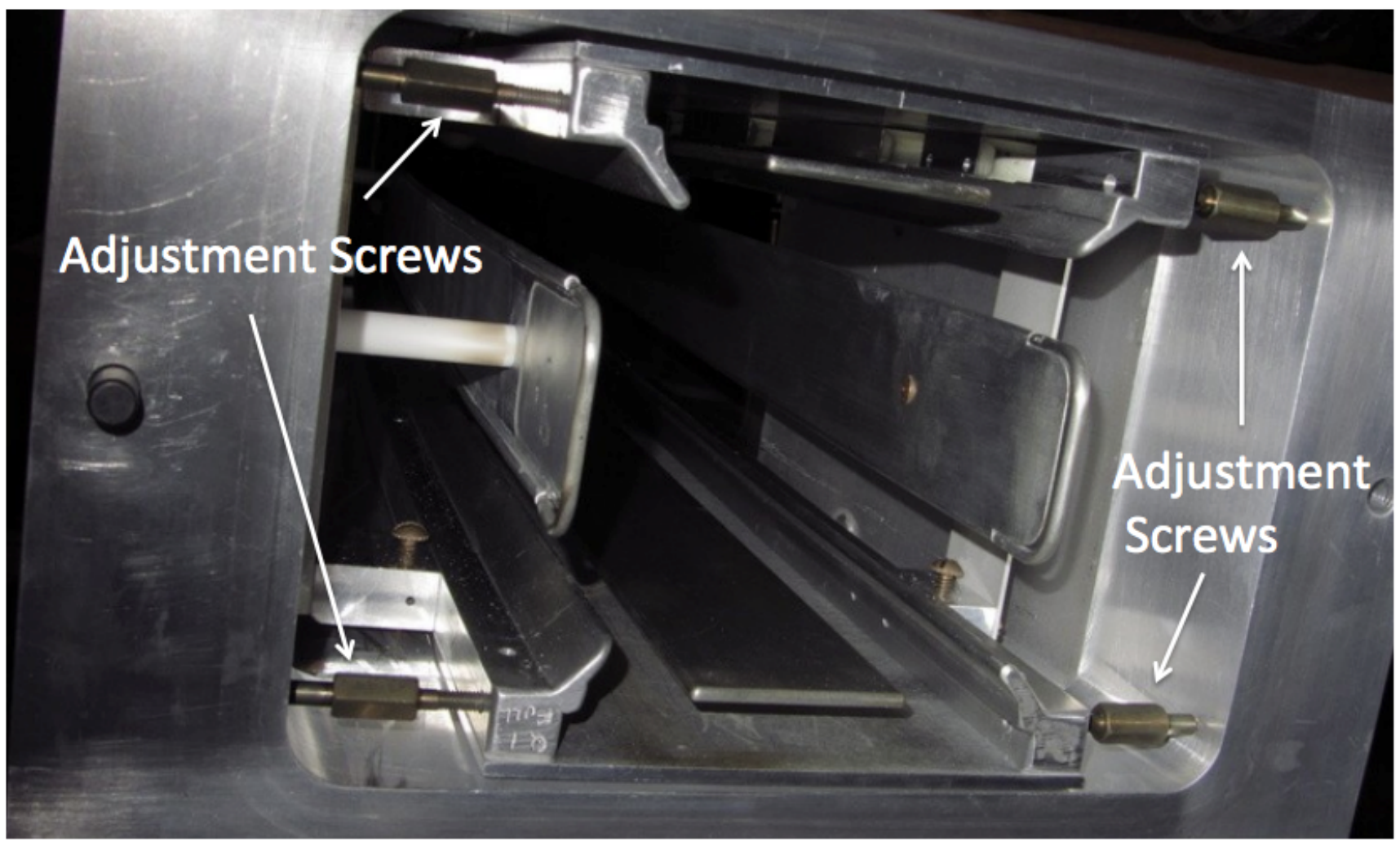

Figure 11.4: Picture of a cage system inside a vacuum chamber showing the adjustment screws to center the quadrupole plates with the geometrical center of the pole pieces.

## 11.2 Vacuum Chambers

This WBS refers to the actual chambers, the small bellows, the piping to the pumps, and the bolting hardware. We will be making major modifications to sectors 2, 8, 10, and 11.

This WBS also covers the reassembly labor effort.

Chamber sectors 2, 8 and 11 would be re-machined to accept the new in-vacuum straw trace back chambers. For sector 10, the 'scallop' portion must be reinstalled. The grooves housing the fixed probes in near the boundary between two yokes will be modified, allowing the affected probes to operate in a lower gradient region of the field.

### 11.2.1 Chamber Electrical Grounding

As discussed in chapter 9, each large metallic component will have a single low impedance connection to the star-shaped structure that defines the ring ground ("the star ground"). Since the bolts tying the chambers together present a high impedance at high frequencies, they cannot serve as a grounding path for the chambers to one another. Therefore, as shown in figure 11.5, each of the 12 large chambers will have a single dedicated low impendance connection to the star ground. In addition, we wish to cut the path of eddy currents circulating through all 12 chambers, which would arise from a magnet quench or energy extraction. Dielectric sheets will be inserted in between chamber flanges to increase the impedance for these eddy currents, which are at low frequencies.

Devices attached to the chambers, such as the NMR electronics hardware, straw chambers, kickers, and quads will typically be electrically isolated from the chambers. In particular, the vacuum pump ports will have ceramic breaks. Finally, there are G10 sheets on top and bottom of the chambers, isolating them from the poles.

### 11.2.2 Chamber Alignment

At Fermilab, the positions of all objects are given in terms of the DUSAF coordinate system. While the chamber positions do not need to be known to good accuracy, the trolley rails, which are mounted within the cages, need to be known to approximately 0.5 mm. Deviations from perfect rail alignment may couple with gradients in the magnetic field to change the field integral seen by the muons compared to that measured by the trolley. Assuming the goals of the shimming effort (Section 15.8) are met (in particular the azimuthal field uniformity) the corresponding error should be negligible for E989 [4]. Fiducial cups will be aluminum-welded to the inner radius of each chamber, and spherical reflectors will be periodically inserted into these cups. In combination with a laser tracker, the spherical reflectors allow for periodic monitoring of the position of each chamber in the DUSAF coordinate system. The position of the cages will be referenced to these fiducial cups. Therefore, the cage positions will be known in the DUSAF coordinate system.

A second and important in-vacuum cross-check will be made by the introduction of known, precise gradients in the magnetic field with the use of the surface correction coils. By tracking the NMR readings taken by the trolley probes throughout the ring in this dedicated run configuration, measurements of gradients different from that introduced by the surface coils would be indicative of cage misalignments.

## 11.3 Vacuum Pumps

The vacuum level must be less than $10^{-6}$ Torr in the region of the quadrupoles. This is to minimize the trapping of ionized electrons due to the residual gas. However, there is a vacuum load of $\sim 2.1 \times 10^{-5}$ Torr $l/s$ from each of the two straw tracker trace back system [5]. From this requirement alone, the minimum pumping speed is 21 liters/sec at $10^{-6}$ Torr. However, each pump is attached to the vacuum chamber through a large pipe. As the pumps will likely contain ferromagnetic material and generate transients that would affect the magnetic field uniformity, they must remain sufficiently far from the vacuum volume. For E821, this distance was 1-2 meters. Therefore, extra piping will increase slightly the pumping speed requirement. Finally, the vacuum chamber system should remain clean, as the quadrupole and kicker plates carry high voltage and the high current, respectively. We will ensure this by utilizing dry (oil-free) roughing and turbo molecular pumps, and also cryogenic pumps.

We will utilize 5 Turbo Molecular Pumps (TMPs) at stations 6, 12, 10 (trolley drive), 8 (straw chamber), and 10 (near straw chamber) respectively. Each of these TMPs would be backed by a roughing pump. Cryogenic pumps will used on chambers 3 and 9. Detailed description of the pumps are given in the reference[6].

## 11.4 Mechanical Interface

As mentioned above, the vacuum chambers must provide the mechanical interface for several systems. This WBS covers the following activities needed for the NMR system:

- Modifications to the upper and lower grooves to improve the S/N of fixed probes near the boundary between yoke pieces.

- Calibration of the trolley position in DUSAF coordinates: for a given motor or position encoder reading, what is the actual position of the fiducial marks on the trolley in DUSAF coordinates.

- Calibration and operation of the plunge probe motors in chambers 10 and 1. Calibration refers to converting a given motor encoder reading to an actual position of the probe head.

There are two plunge probes, one integrated into chamber 1 and the other housed on top of a vacuum bellow that connects chambers 10 and 11. The mechanism to move the vacuum bellow plunge probe is shown in Figure 11.6. The probe itself is in air. The probe is moved radially by piezo electric motors. No changes are needed for the plunge probe mechanism, other than to upgrade the motor controller and computer.

The chamber 1 plunge probe mechanism moves radially as well as vertically. Therefore, the titanium vacuum bellow is significantly larger than the other (figure 11.6), and caused a noticeable shift in the absolute calibration. For E989, we plan to house the piezo motors and probe in vacuum, therefore bypassing the use of the large titanium below.

## 11.5 ES&H, Quality Assurance, Value Management, Risk

The vacuum chamber system should pose no health hazard, since the chambers, when evacuated, have a 2.9 safety factor for stress before yield. Quality assurance and value management concerns are minimal since we are reusing or modifying a few E821 chambers.

There is a risk that the new in-vacuum straw traceback chambers provide too much gas load to the vacuum. In that event, a corrective action could be to add additional pumping. This risk is addressed in the straw trace back discussion of this document.

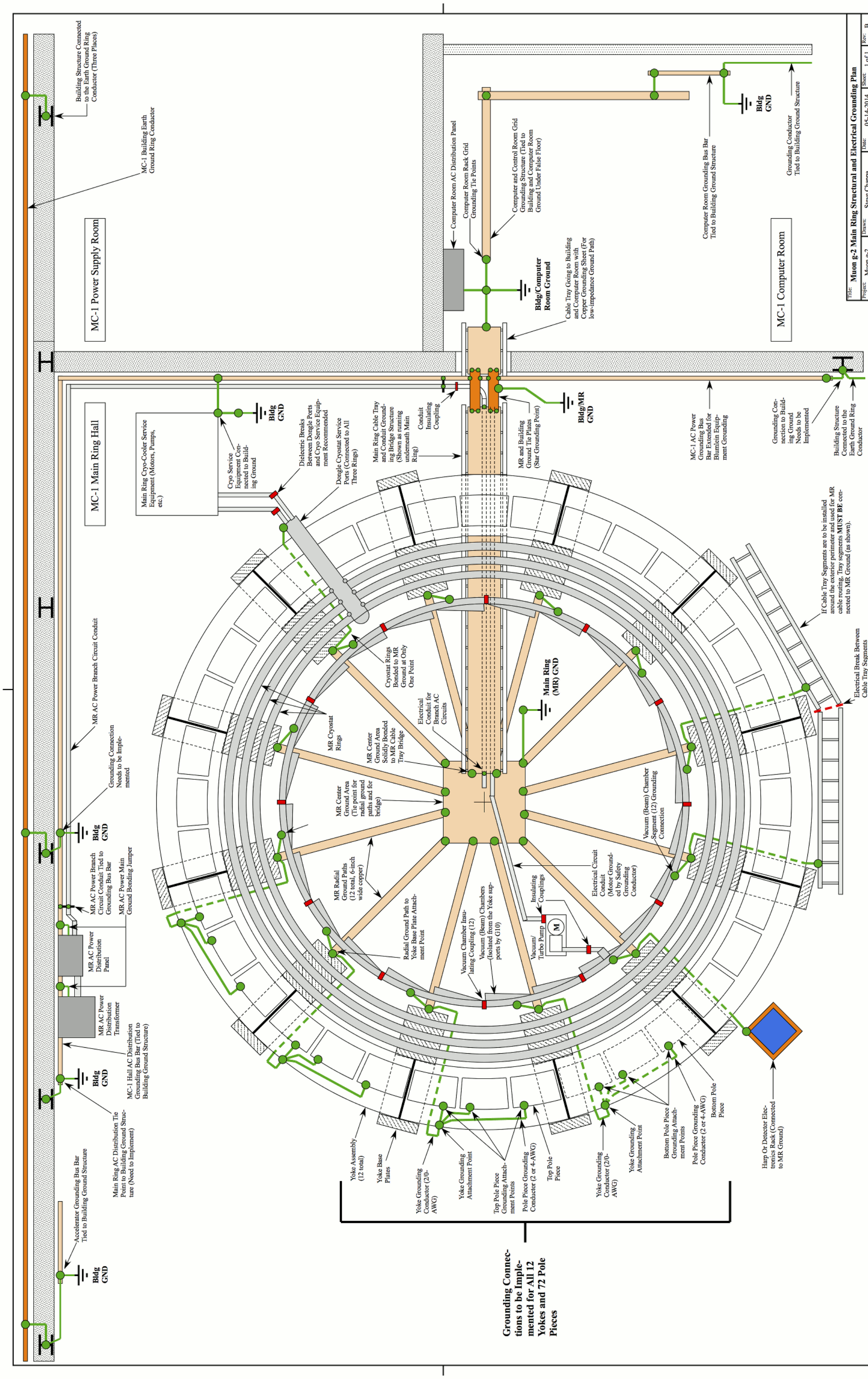

Figure 11.5: The grounding of g-2 metallic structures.

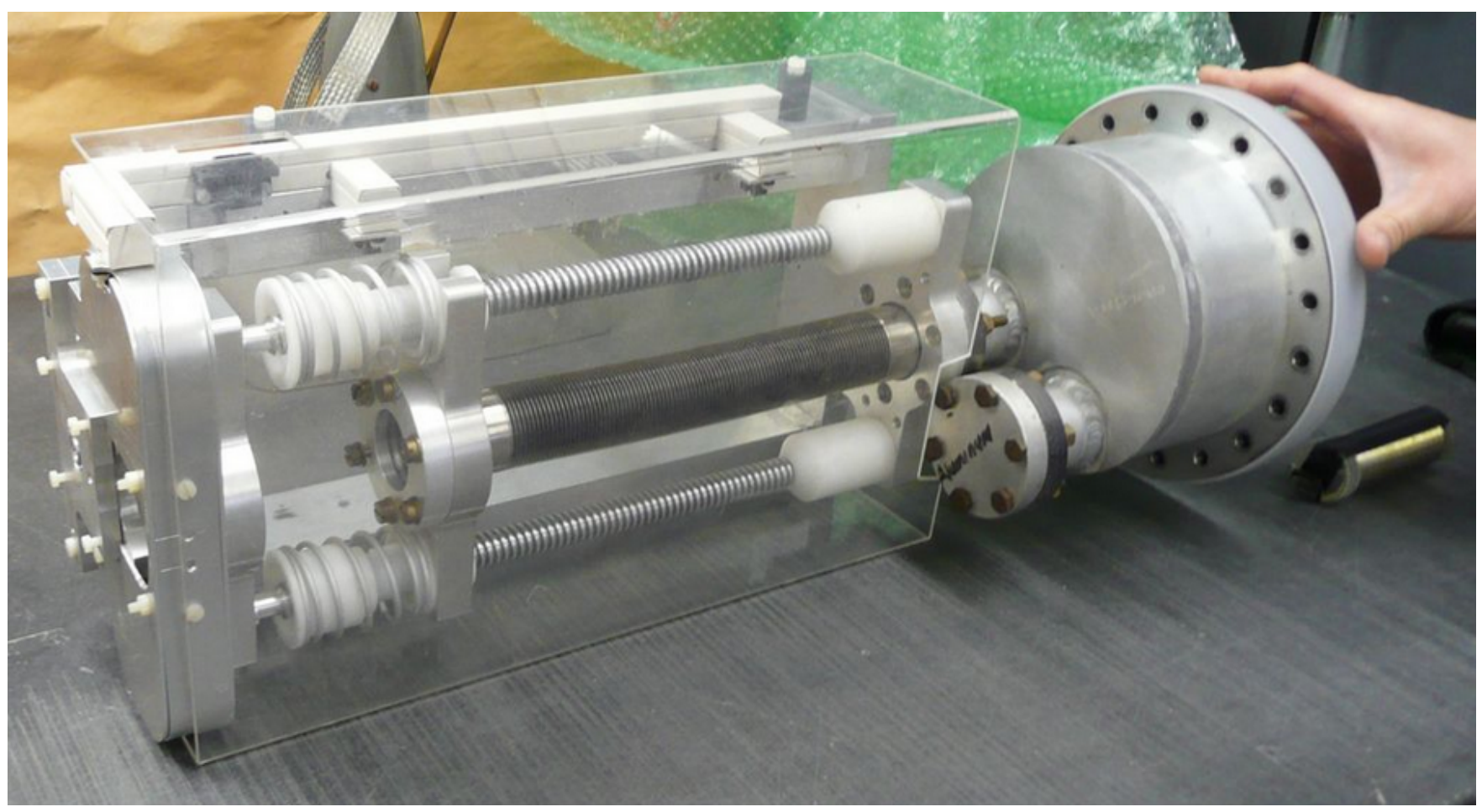

Figure 11.6: The plunge probe mechanism. The probe lies inside the titanium vacuum bellow.

# Chapter 12

# The Fast Muon Kicker

Injected muons exit the downstream end of the inflector magnet, and enter the good field region of the main dipole. The trajectory of the muons exiting the inflector is a circle displaced 77 mm radially from the closed orbit of the storage ring. The muons emerge from the inflector into the full 1.41T field of the dipole but with trajectories that are on average, tangential to the displaced circle. The muons cross the closed orbit of the storage ring, that is the orbit at the magic radius, about 90° azimuthally around the ring from the end of the inflector. As shown in Fig. 12.1, The muons intersect the closed (ideal) orbit at an angle of $\theta_0 = 10.8$mrad. The minimum crossing angle, namely $\theta_0 = 10.8$mrad obtains for trajectories that are tangential at the inflector exit. Any angle, positive or negative, with respect to the tangent line results in crossing angle greater than the minimum.

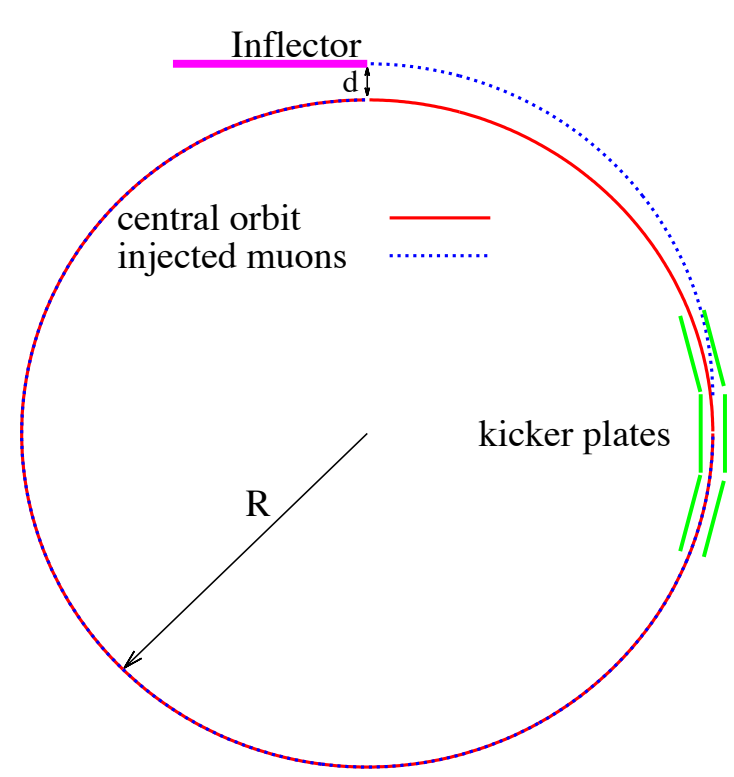

Figure 12.1: Muons exit the inflector displaced $d = 77$mm from the central orbit and cross the central orbit at an angle of 10.8 mrad 90 degrees around the ring. The 10.8mrad kick directs muons onto the central orbit.

Simulation of the injection of a distribution of muons that includes scattering in the inflector coil ends, and scattering in the quadrupole plate indicates that the total kick angle that achieves maximum storage efficiency is 10.8± 0.4 mrad. Our target for the kicker system is a 12.8 mrad kick, so that there is sufficient operating margin, and in anticipation of a new larger aperture inflector that would displace the center of the inflector aperture further from the magic radius.

The fast kicker is a pulsed magnet with vertical field that directs the muons onto the ideal orbit, by compensating the crossing angle. Ideally, the centroid of the injected bunch, on exiting the field of the kicker plates, will coincide with the closed orbit of the storage ring, thus eliminating residual coherent betatron oscillation. The 10.8 - 12.8 mrad kick corresponds to an integrated vertical field of 1.1 - 1.3 kG-m.

The E989 kicker will be comprised of three independent 1.27m long magnets, each with a dedicated pulse forming network as current generator. Muons are delivered to the storage ring in pulses with 95% transverse emittance near $40\pi$ mm-mrad, pulse length of about 120ns and at a peak repitition rate 100Hz and an average rate of 12Hz. The ideal kicker field maintains a flat top at the required 285 - 336 Gauss, for the full 120ns, and then returns to zero before the lead muons complete a single revolution and re-enter the kicker aperture 149ns later. (The revolution period of the ring is 149 ns). The rise and fall time of the kicker pulse is ultimately determined by the finite length, and transit time of the pulse in each of the three strip line magnets. Faster rise and fall obtains with shorter magnets.

The injection of muons into the storage ring is complicated by several requirements:

1. Since the magnet is continuous, any kicker device has to be inside of the precision magnetic field region.

2. The kicker hardware cannot contain magnetic elements such as ferrites, since they will spoil the uniform magnetic field.

3. Any eddy currents produced in the vacuum chamber, or in the kicker electrodes, must be negligible by 10 to 20 $\mu$s after injection, or must be well known and corrected for in the measurement.

4. Any kicker magnet hardware must fit within the real estate occupied by the E821 kicker, which employed three 1.7 m long devices.

5. The kicker pulse should be shorter than the cyclotron period of 149 ns

## 12.1 Requirements for the E989 Kicker

The need for a fast muon kicker was introduced in Section 3.2. Direct muon injection was the key factor that enabled E821 to accumulate 200 times the data as the preceeding CERN experiment. Since E989 needs more than twenty times as much data as E821, it is critical that the limitations of the E821 kicker be mitigated. The layout of the E821 storage ring is shown in Fig 12.2. The three kicker magnets are located approximately 1/4 of a betatron wavelength around from the inflector exit.

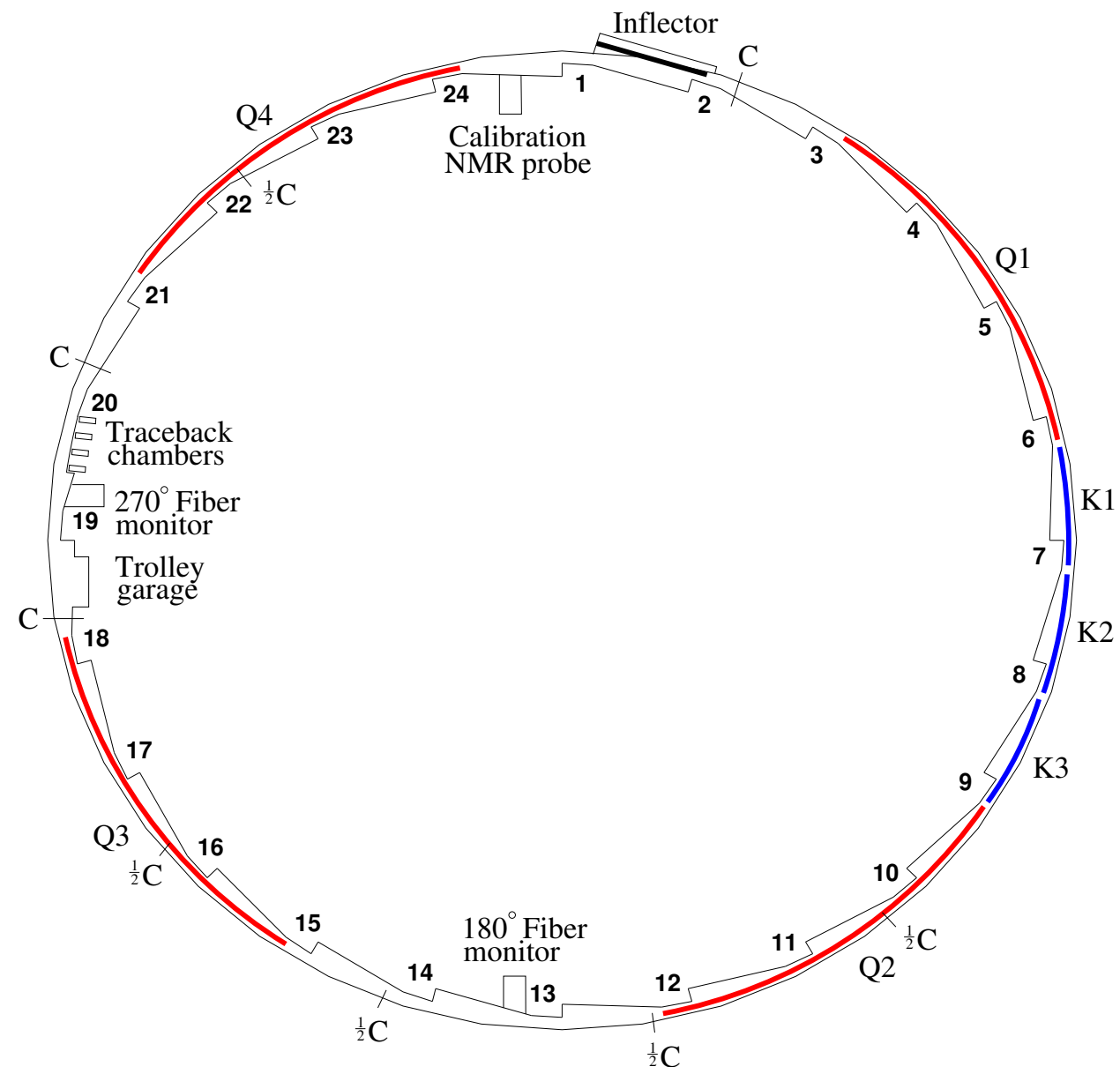

Figure 12.2: The layout of the storage ring, as seen from above, showing the location of the inflector, the kicker sections (labeled K1-K3), and the quadrupoles (labeled Q1-Q4).

### 12.1.1 The E821 Kicker and its Limitations

The E821 kicker [1] consisted of three identical sectors with 1.7 m long parallel plates carrying current in opposite directions, located as shown in Fig. 12.2. Each section was powered by a pulse forming network where a HV capacitor was resonantly charged to $\simeq$ 95 kV, and then shorted to ground by a deuterium thyratron, giving a characteristic damped LCR oscillating current and magnetic field. The resulting LCR pulse is shown in Fig. 12.3(Left). The LCR pulse was much wider than the muon pulse length, in fact significantly longer than the cyclotron period of 149 ns. This is emphasized by the series of red gaussians which are separated by the 149 ns revolution period. Thus the beam is kicked several times before the LCR pulse dies away. Simulations show that the peak field achieved with the LCR pulser was only 120G and yielded a kick angle of barely 6 mrad. Thanks to the multi-turn width of the pulse, the cumulative kick was sufficient to store muons.

The kicker units began sparking around 95 kV, limiting the pulse current. The number of muons stored vs. kicker high voltage in the E821 experiment is shown in Fig. 12.3(Right). At the maximum accessible voltage, the number of stored muons had not yet plateaued. It is not entirely clear how many muons might have been stored if it had been possible to increase the voltage until the maximum number of stored muons was reached. Simulations suggest that the number stored would have increased by 30%. But even had that higher capture efficiency been reached, there would have remained a significant coherent betatron oscillation do to the temporal nonuniformity of the kick over the length of the muon pulse.

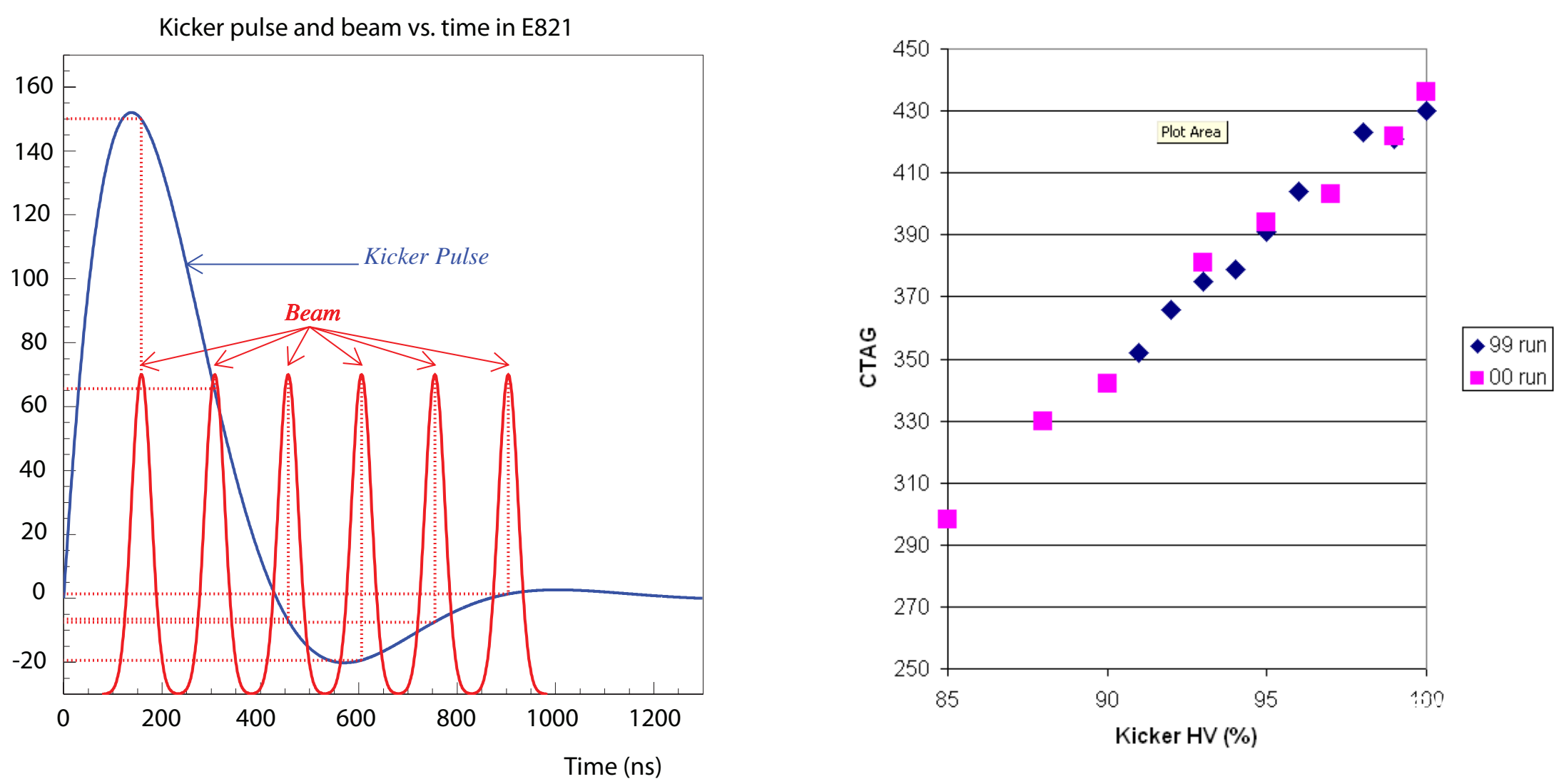

Figure 12.3: (Left)The E821 kicker LCR waveform (blue). The red pulses represent the injected beam, which has a cyclotron period of 149 ns. (Right) The number of stored muons versus kicker high voltage (arbitrary units).

## 12.2 New Kicker Design

The design of the kicker for E989 attempts to address the shortcomings of the E821 kicker, specifically with respect to the pulse shape and pulse amplitude. The E989 pulse forming network is based on a Blumlein triaxial transmission line as an alternative to the E821 LCR PFN. The kicker plates are redesigned with somewhat higher efficiency, that is, higher magnetic field between the plates per unit current through the plates. Fig. 12.4 is a rendering of the 1.27 m long plates of the new kicker magnets along with hardware for mounting the plates in the vacuum chamber and the tracks for the NMR trolley.

The E989 current generator is a variation on a transmission line PFN. Consider each 1.27 m long pair of kicker plates a transmission line with impedance $Z_L$, and imagine, at least conceptually, that each is terminated with a resistive load $R = Z_L$. If the impedance of the PFN transmission line matches the impedance of kicker plates and load, then the current pulse will have width $\tau = 2L/c$ where $L$ and $c$ are the length and group velocity of the PFN, and peak current $I = V/Z$, where $V$ and $Z$ are the peak charging voltage and the impedance of the line respectively. Then the rise time of the pulse is limited only by the switching time of the thyratron of about 20-30ns.

The configuration with transmission line, kicker magnet, and load all in series, is however impractical in this application, because of the relatively high impedance of the kicker magnet plates. The estimated impedance of the kicker plates is nearly 600 Ω. As will be shown below, the current required to achieve the requisite $\sim 280$ G field between the plates is about 4.7 kA. If the impedance of the PFN is matched to the impedance of the kicker plates, a charging voltage of nearly 2.8 MV would be required to push 4.7kA through the load. Such an arrangement would also require terminating the kicker plates with a resistor with result that the plates would float to high voltage with each pulse.

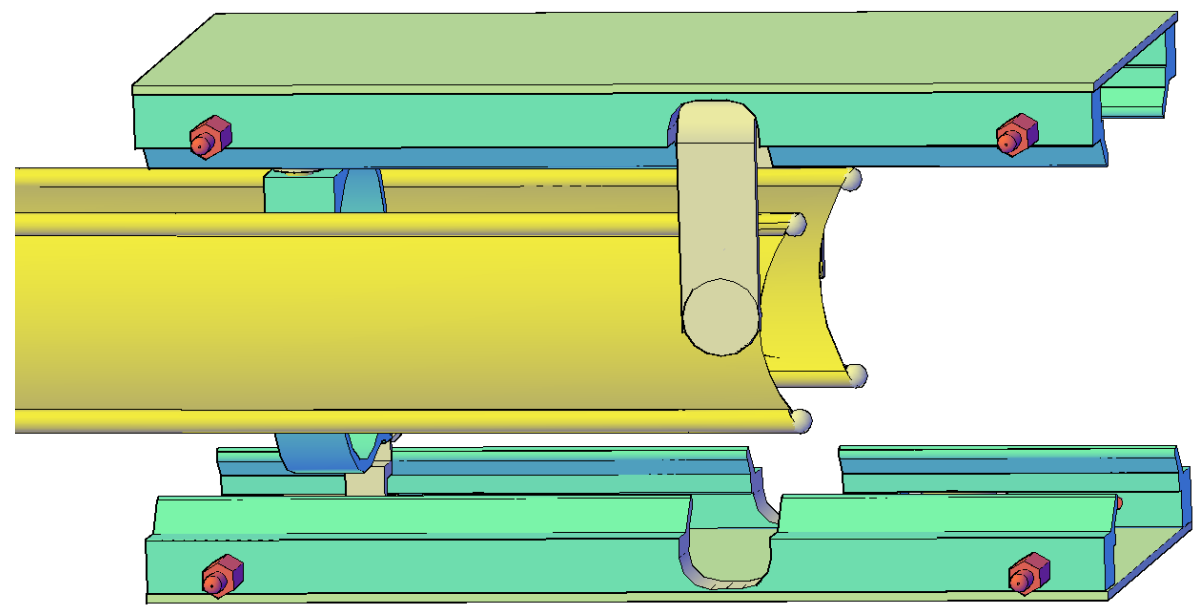

Figure 12.4: The kicker plates for E989. The current pulse is fed to the 1.27m long plates at the far end to the left of the plot. A jumper connecting the plates is placed at the near end that closes the circuit as shown. The plates are suspended from the top of the vacuum chamber. The NMR trolley rails are shown in green. The trolley will roll between the plates.

### 12.2.1 Current source

Instead of terminating the kickers with a resistive load, the load resistor is placed between the pulse forming network and the kicker plates. The impedance of the PFN transmission line is simply matched to the load resistor. The load resistance, $R = 12.5\Omega$, is chosen so that the requisite charging voltage (V=IR=54kV) is well within reach of an available thyratron switch. The reflections that will inevitably arise from the imperfect match at the junction of load resistor and kicker plates, will be confined to the plates, and dissipated on the timescale of the plate round trip transit time of about 10ns. The triaxial line and matched terminating resistor act as a current source. At the transition through the load resistor to the kicker, the transmission line is tapered to near zero impedance. The configuration is then equivalent to a current source as there will be zero voltage at the end of the taper. A schematic of the power supply, charging circuit, Blumlein current generator (see next section) and kicker is shown in Fig. 12.5.

## 12.3 Blumlein Pulse Forming Network

The pulse forming network developed for the kicker is a Blumlein triaxial transmission line. The Blumlein is shown schematically in Fig. 12.6. The LCR circuit used in E821, and a coaxial transmission line are included in the Fig. 12.6 for comparison. The equivalent circuit for a Blumlein is a pair of series bi-axial lines with a shared conductor and it is so rendered in Fig. 12.7. The width of the pulse

$$\tau = \frac{2L}{v} = 2L\frac{\sqrt{\mu\epsilon}}{c}.$$

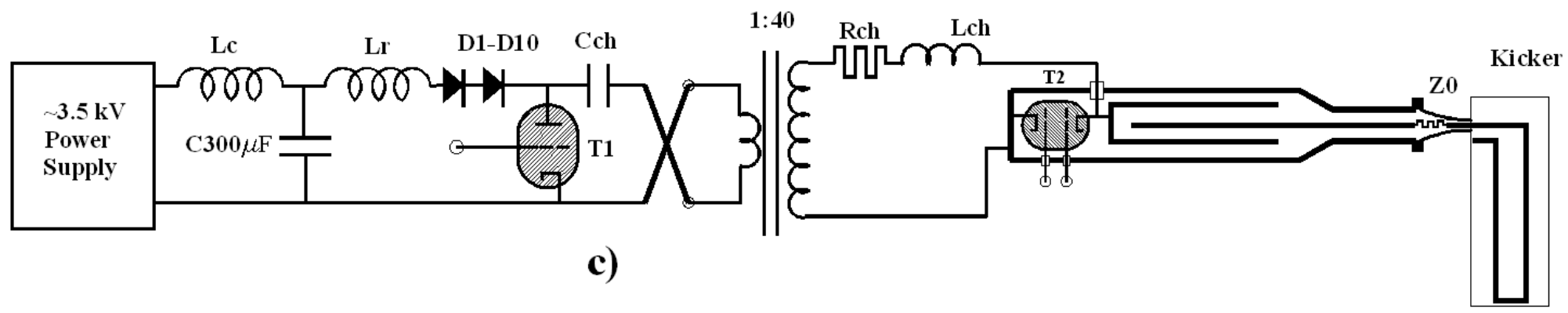

Figure 12.5: Schematic of the kicker, pulse forming network and charging circuit. The Blumlein, resistive load ($Z_0 = 12.5\Omega$) and kicker are in series to the right of the figure. In the final installation the load resistor is mounted near the vacuum chamber coupling directly to the kicker plates. The Blumlein connects to the resistor via high voltage coax.

For the bi-axial line the voltage at a matched load is half the charging voltage. For the Blumlein, output voltage and charging voltage are one and the same[2]. Another advantage of the Blumlein as compared to a bi-axial transmission line is that the base of the thyratron can be fixed at ground potential. A bi-axial pulse forming network would require that the base of the tube float to high voltage when the thyratron is switched.

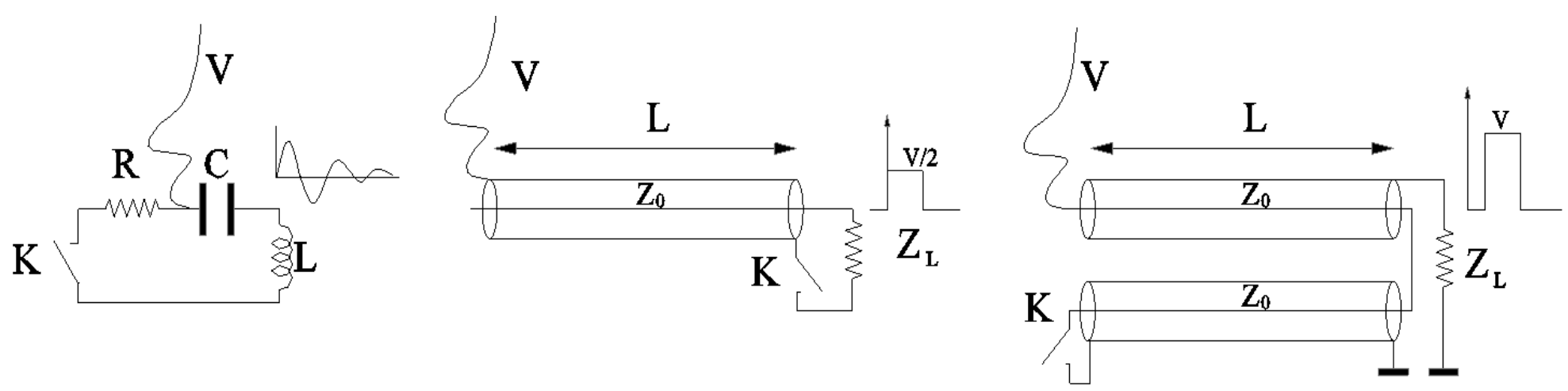

Figure 12.6: The overdamped LCR circuit at left was used in E821. The line labeled "V" indicates the charging voltage and $K$ the thyratron switch. At center is a coaxial transmission line PFN. The Blumlein equivalent circuit is at the right. The corresponding pulse shape is shown for each of the configurations. Note that for both coaxial and triaxial lines, pulse width is proportional to twice the line length. Voltage across a matched load for the Blumlein is twice that of the coax. The Blumlein pulse is delayed by half of the pulse width.

The schematic of the $12.5\Omega$ Blumlein prototype is shown in cross section in Fig. 12.8. The middle conductor is connected through a large resistance and inductance to the high voltage power supply. Current flows through the load $Z_L$ from the central conductor during the charging cycle. The thyratron (T) shorts the middle conductor to the outer conductor and after a delay of $T/2$, where $T$ is the width of the current pulse generated by the line, the current flows through the resistive load and into the kicker.

Some details of the Blumlein are shown in Fig. 12.9. The basic electrical properties of the pulser with an equivalent circuit are computed with SPICE. Each of the two series coaxial lines are modeled with discrete elements as shown in Fig. 12.10(Top). The kicker load is represented with characteristic capacitance and inductance. The current pulse through the kicker when the switch is closed, as computed with SPICE, is shown in Fig. 12.10(Bottom).

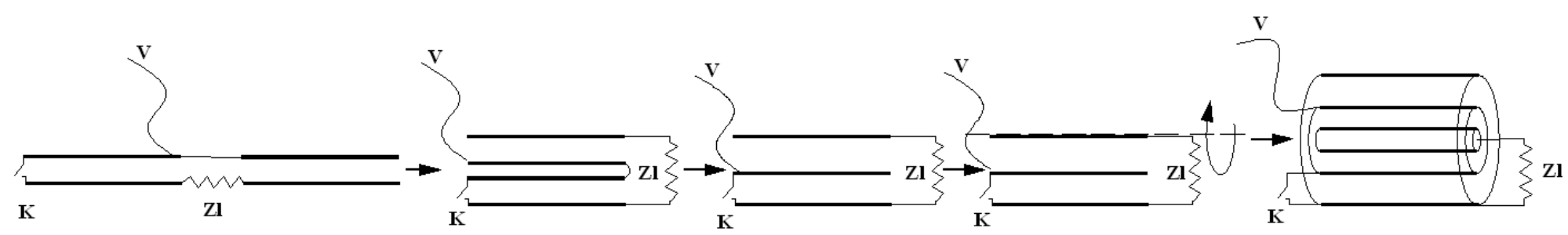

Figure 12.7: Topological modification of series coaxial lines into a tri-axial Blumlein transmission line. (One can choose the rotation axis coinciding with the lower plate).

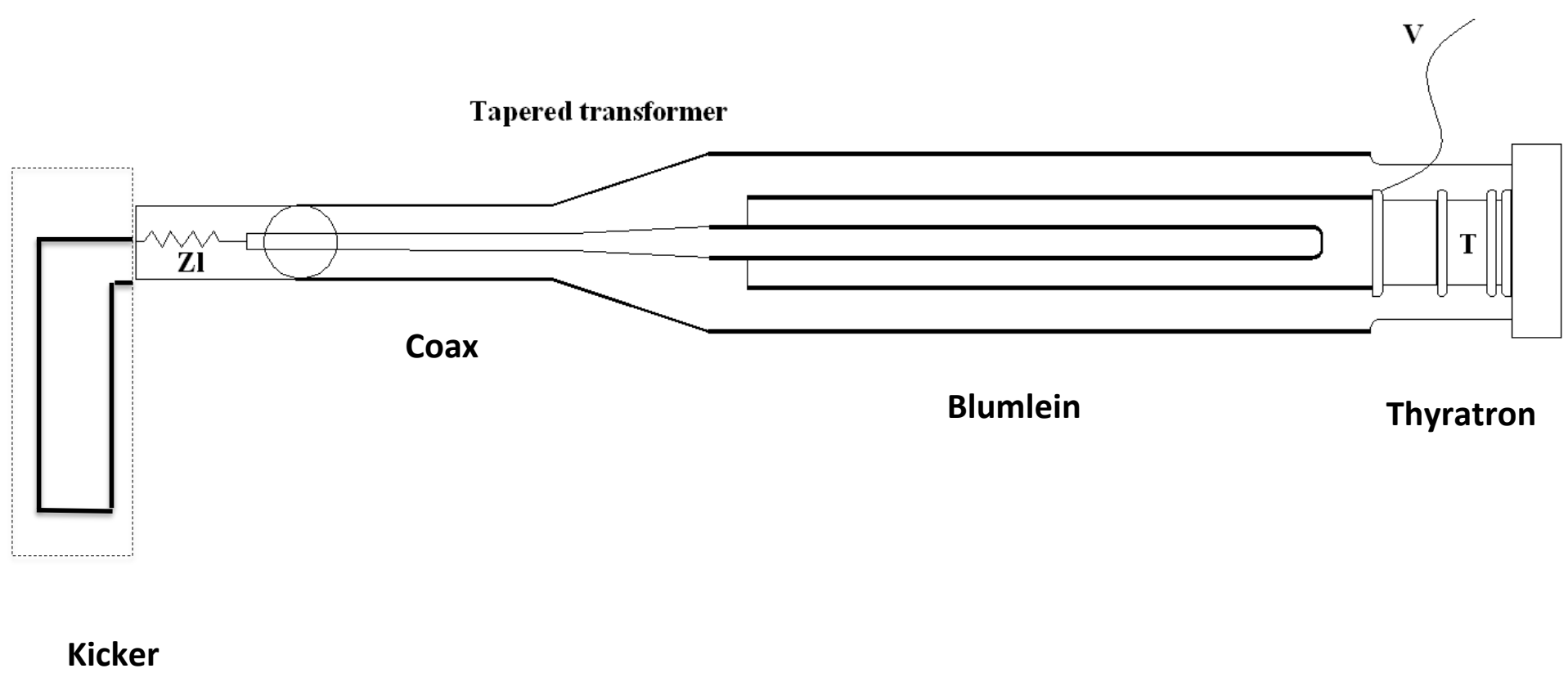

Figure 12.8: Middle conductor is charged to high voltage via the line labeled "V". The centermost conductor is coupled via a high voltage coaxial cable to the resistive load that is mounted directly to the input of the kicker. The thyratron (T) shorts the middle conductor to ground. The volumes between conductors, around the thyratron and load are all filled with transformer oil.

The impedance of the triaxial line is equivalent to the sum of the impedances of the series bi-axial lines. The middle conductor in Fig. 12.8 that is charged to high voltage serves as the inner conductor for one bi-axial line and the outer conductor for the other. The impedance of each of these bi-axial components is 6.25Ω. The output of the PFN is coupled to the load with four parallel 50Ω high voltage coaxial cables, with combined impedance of 12.5Ω. The blumlein and transition hardware is shown in Fig. 12.11.

A schematic of the Blumlein pulser connected via high voltage coax to the kicker inside the muon ring vacuum chamber is shown in Fig. 12.12. Also shown is the electronics rack with 1500 V power supply, thyratron driver, and thyratron trigger pulser. The cylindrical container sitting on the floor beside the rack is the oil tank with high voltage transformer. While the single high voltage transformer can in principle charge all three Blumleins for each of the three kickers, independent supplies will be deployed so that failures in the electronics of one PFN and kicker are not propagated to the others.

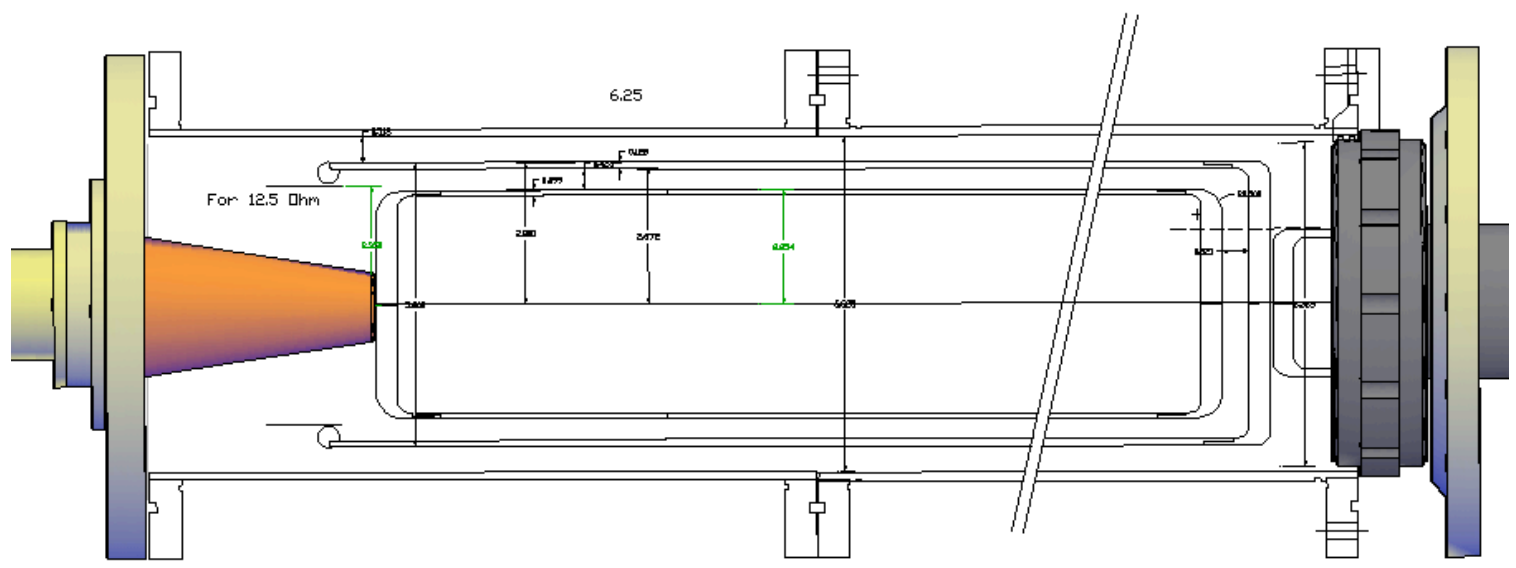

Figure 12.9: Cross section of 12.5Ω. Blumlein. The central conductor couples through the orange transition at the left of the figure to the load resistor and kicker. The middle conductor connects through the U bracket near the right to the thyratron (not shown). The penetration of the high voltage charging line through the outer conductor to the middle conductor is not shown.

### 12.3.1 Thyratron switch

A thyratron switch shorts the middle conductor of the Blumlein to ground, initiating the current pulse that drives the kicker. The peak current required to achieve the approximately 1.1 kG-m field depends on the cross sectional shape of the kicker plates (determining relationship between current and B-field), and plate length (limited to less than 5.1m). As will be shown in the next sections, a peak current in excess of 4kA is required. Table 12.1 summarizes the parameters of the kicker thyratron used in the E821 experiment and the high current two grid tube (thyratron model CX1725X[4]) that is used for the new kicker current generator.

| Thyratron | CX1699 (E-821) | CX1725X |
|---|---|---|
| Peak Voltage [kV] | 130 | 70 |
| Peak Current [kA] | 3 | 15 |
| dI/dt [kA/$\mu$s] | 10 | <300 |
| Repetition rate [Hz] | 400 | 2000 |
| Grids | 4 | 2 |

Table 12.1: Evidently the E821 tube can not switch the requisite (>4kA) current. The CX1725X tube proposed for the new kicker pulse generator is faster (dI/dt), in addition to the higher current rating, and can more easily manage the repetition rate.

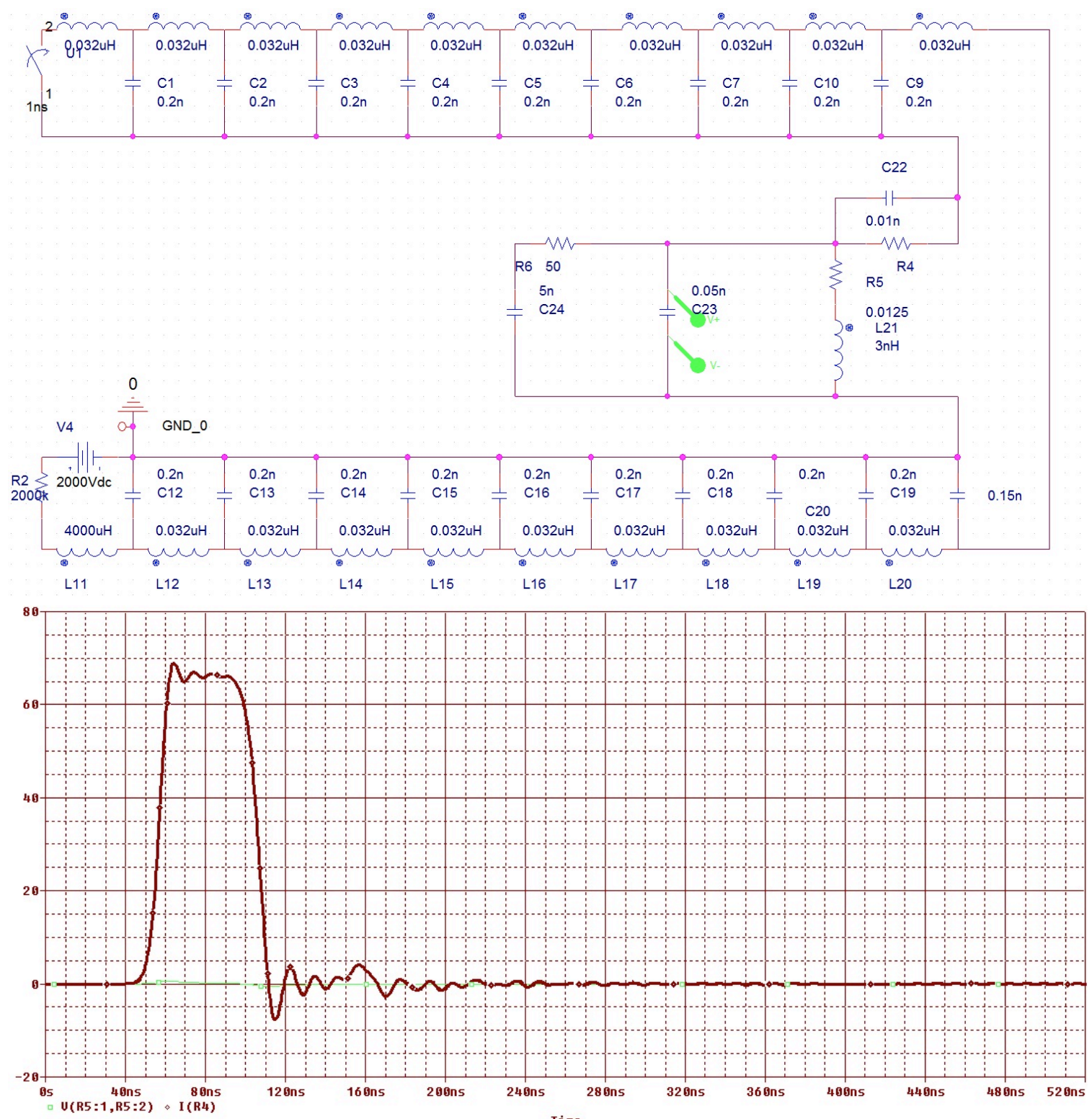

Figure 12.10: Discrete circuit element of the Blumlein that is shown as the right hand schematic in Fig. 12.6. Each of the two transmission lines is assembled with lumped inductance and capacitance. The kicker load is represented as inductance with small capacitance.

Note that in the Blumlein configuration, the thyratron current is twice the load current. The CX1725X can deliver up to 7500A to the load, well in excess of the anticipated requirement.

## 12.4 Kicker Plate Design

The geometry of the kicker plates is optimized for higher efficiency than the E821 design, that is higher midplane magnetic field for a given current through the plates. The field profile for the proposed plate geometry is shown in Fig. 12.13(Right) as compared to the E821 geometry in Fig. 12.13(Left). The calculation assumes a pulse width of 100ns and rise time of 20ns. As can be seen in Fig. 12.13, the E-821 plate geometry yields uniform vertical field over a larger region than the E-989 geometry, at the cost of significantly lower fields near the top and bottom edges of the plates. The advantage of the proposed E-989 geometry is that the field in the midplane is 65 Gauss for a plate current of 1000 A, as compared to only 35 Gauss/kA for the E821 kicker magnet. The relative efficiencies of the two geometries corresponds to the ratio of the perimeters of the plates.

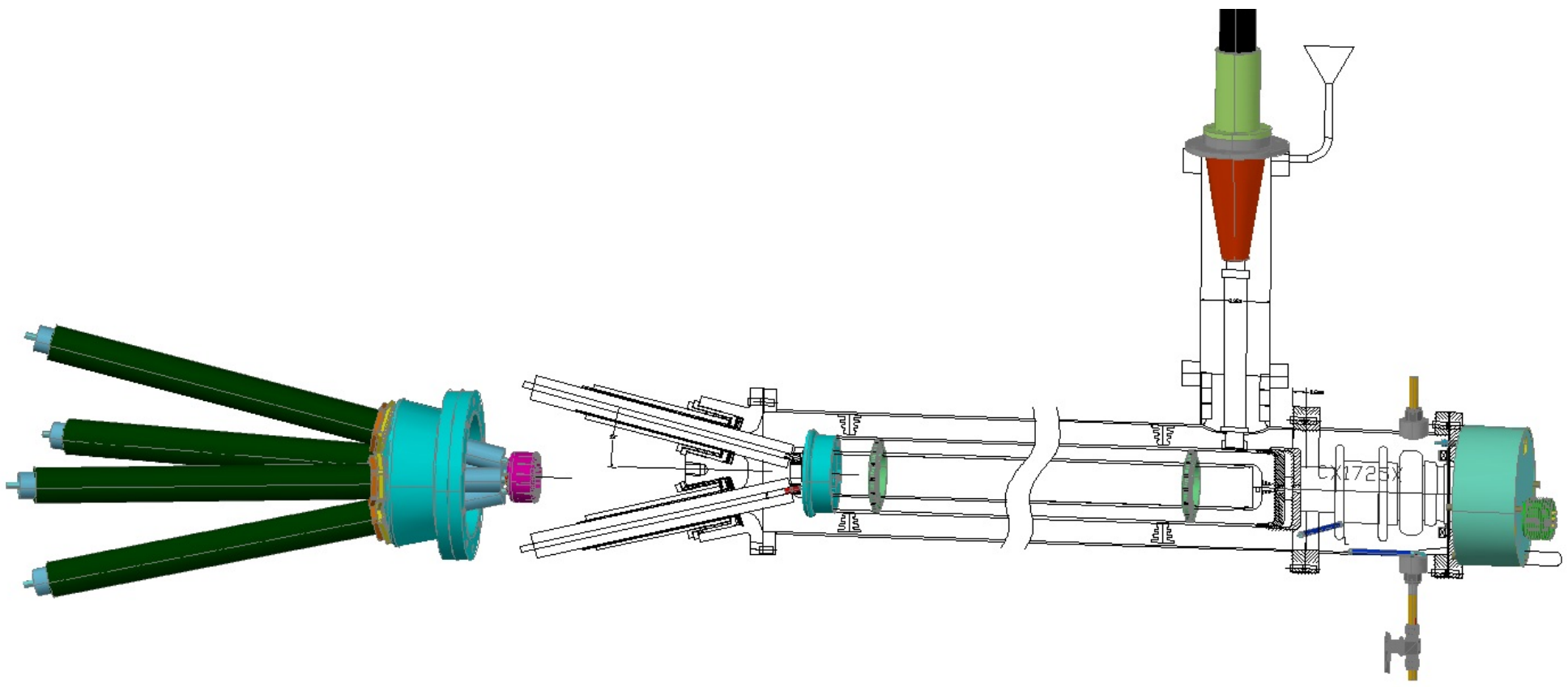

Figure 12.11: Transition from Blumlein (far right) to 4-50Ω cables. The 4 cables are coupled to the centermost conductor of the Blumlein. The high voltage charging transformer couples to the middle conductor via a feedthrough at the right. The thyratron switches the middle conductor to ground (outer conductor)

A full tracking simulation of the injection process shows that the somewhat weaker field away from the midplane in the E-989 geometry degrades capture efficiency by only a few percent. Also, by distancing the kicker plates from the vacuum chamber, long lived eddy currents induced in the aluminum chamber that might effect the local magnetic field are reduced.

### 12.4.1 Kicker Plate Length

There are 5.1m available in the storage ring for kicker plates. The required 10.8 to 12.8mrad kick corresponds to an integrated field of $1.11\mathrm{kG-m} < \int \mathbf{B} \cdot d\mathbf{dl} < 1.31\mathrm{kG-m}$. Given the relation between B-field and peak current for the proposed plate geometry of 65G/kA the current requirement can be written $17\mathrm{kA-m} < I_{peak}L < 20\mathrm{kA-m}$. The rise and fall time of the current pulse depends on the length of the plates, and in particular increases with plate length. The peak current through the 12.5Ω resistive load is limited by the tube voltage maximum of 70kV to 5.6kA and therefore according to the inequality above the length $L$ of the plates must be at least 3.6 m. In E989 there will be three, 1.27m long kickers corresponding to a total length of 3.8m, thus leaving a margin of about 10%. If the system is limited for any reason to $I < 5.6$kA, there is space available in the storage ring for a fourth 1.27m long kicker. Note that in any event, the new kicker system will operate at voltage well below the $\sim$ 95kV breakdown voltage of the E821 kickers.

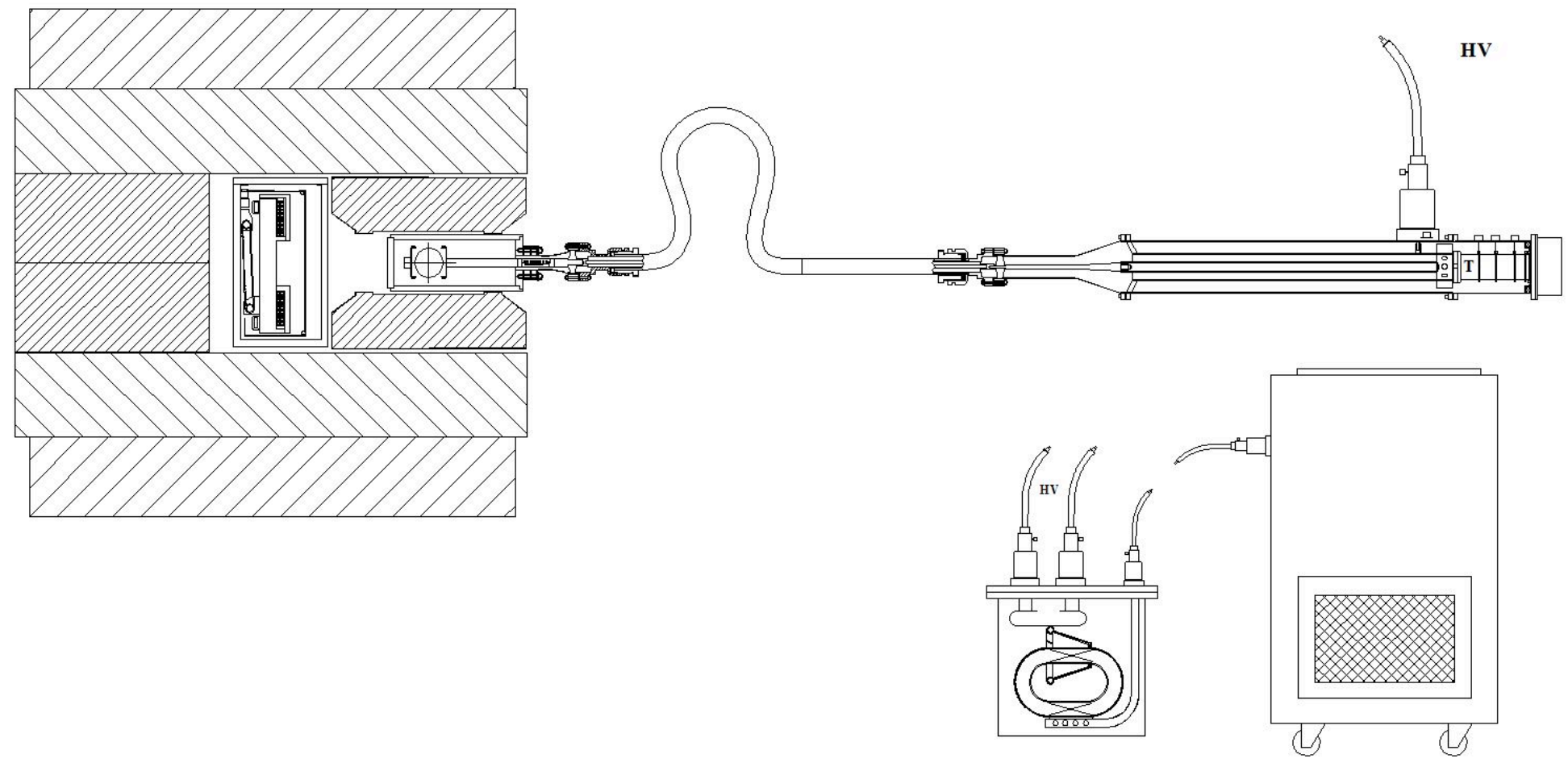

Figure 12.12: Blumlein is coupled through four parallel 50Ω high voltage coaxial cables to the kicker magnet inside the ring vacuum chamber. (Only a single coax is shown). High voltage power supply and thyratron driver are in the electronics rack at right. The 1:85 high voltage transformer that provides charging voltage to the PFN is in the oil tank to the left of the electronics rack.

### 12.4.2 Trolley rails

Whereas in the E821 configuration, the kicker plates served as rails for the NMR trolley, those functions will be separated in the new implementation as can be seen in Fig. 12.14. The NMR trolley will roll between the plates. The kicker plates will be suspended within a cage as shown in Fig. 12.15. Care will be taken to ensure the stability of the plates with respect to the time dependent forces associated with the current pulse. At the same time it is desireable to minimize the thickness of the plates so as to minimize scattering of decay electrons.

## 12.5 Kicker R&D at Cornell

A laboratory has been outfitted at Cornell to build and test a prototype Blumlein pulse forming network and fast kicker magnet. The electronics that has been recovered from the E821 experiment and re-assembled includes: high voltage power supply, high voltage charging transformer, and trigger pulser. Thyratron and driver were procured to match requirements of the new current generator and kicker plate geometry. The 9 meter long prototype 12.5Ω Blumlein is shown in Fig. 12.16,12.17 and 12.18. In order to maximize the pulse width the space between the concentric cylinders of the Blumlein is filled with castor oil with permittivity $\epsilon = 4.7$ (rather than the lower permittivity silicon oil). The vacuum chamber with prototype kicker plates is shown in the top left of Fig. 12.19(Left) and Fig. 12.20.

In order to better understand the dependence of pulse shape on kicker length (and kicker inductance), measurements were performed with kicker plates with different effective lengths.

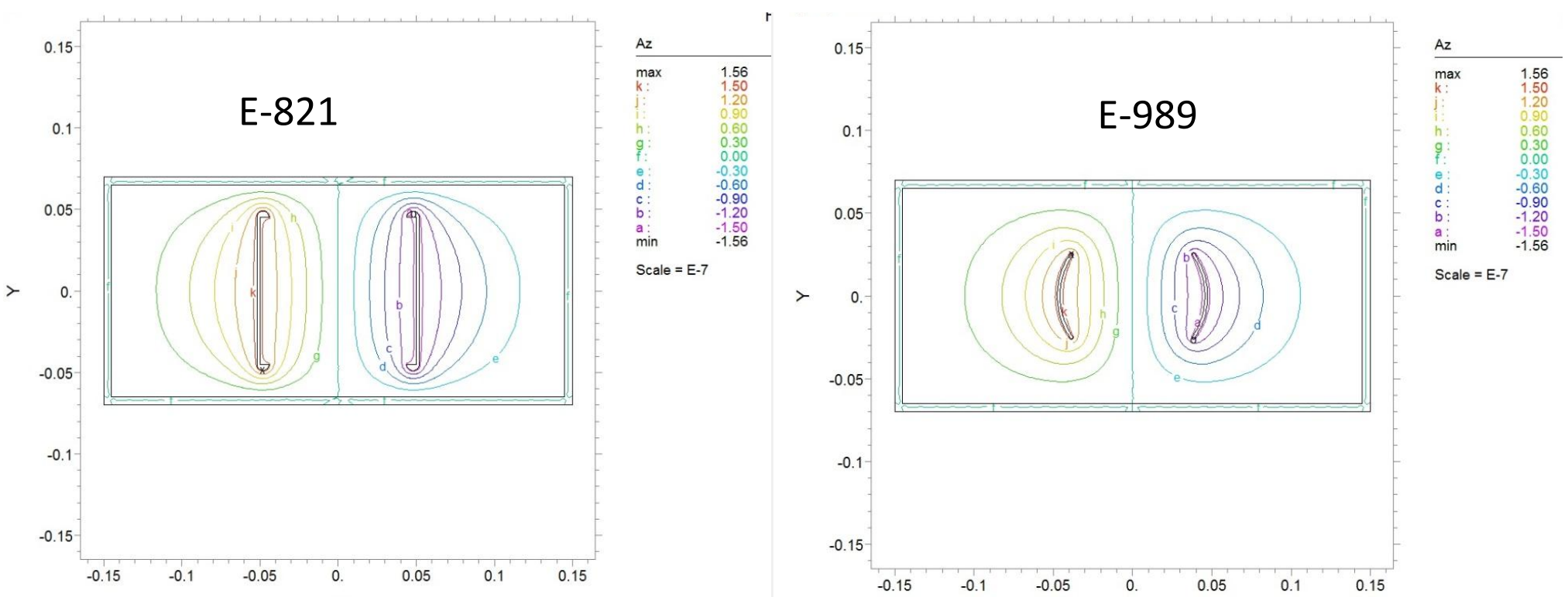

Figure 12.13: (Left) E821 kicker plate geometry and field lines computed with FLEX PDE. The boundary condition at the vacuum chamber surface includes the effect of induced currents due to fast rise time. Note the high density of field lines at the edge of the plate that also serves as the trolley rail. (Right) Proposed kicker plate geometry and magnetic field lines.

A photograph of the shorting bar used to change the effective plate length, and some of the measurements are shown in Fig. 12.19. The rise time clearly increases with the length of the kicker plates, indicating that the pulse shape is dominated by the properties of the load, rather than the thyratron switch. The length of the kicker plates is chosen so as to optimize the pulse shape consistent with the integrated field requirement and the peak current available from the pulse forming network. The optimimum length is 1.27 m (that is as short as is compatible with the available peak current). The Blumlein is coupled to the kicker using four 50Ω high voltage coaxial cables in parallel.

Three 12.5Ω Blumleins, each outfitted with the 2-gap CX1725X thyratron will drive three 127cm long kickers. If necessary, the rise and fall time of the current pulse can be made shorter by reducing the kicker length and the integrated field recovered by adding a fourth kicker (and Blumlein). With four kicker magnets, the 1.31kG-m integrated field could be achieved with magnet length of 90cm operating at 5146A.

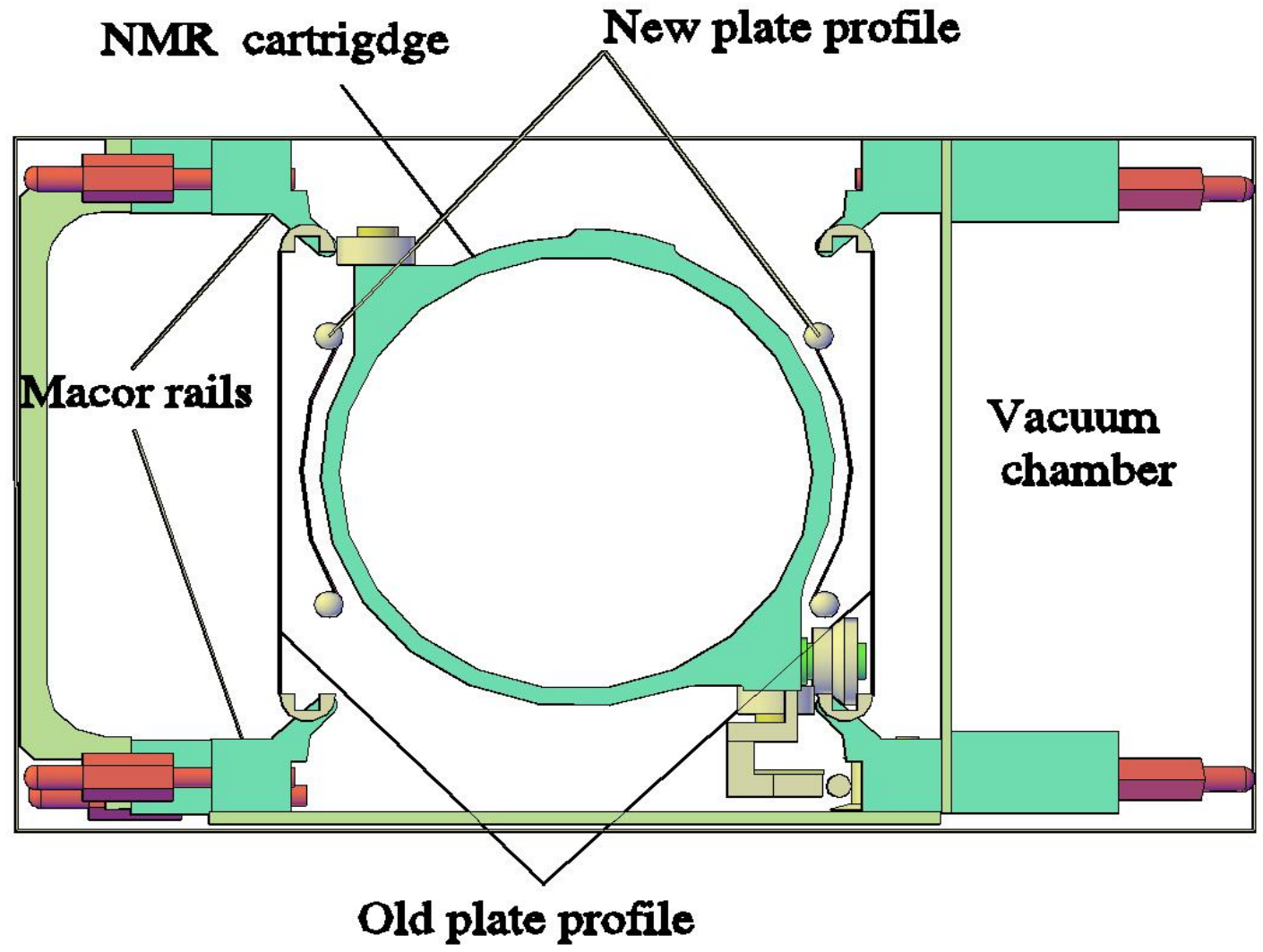

Figure 12.14: (Left)E821 kicker plates, new plates, NMR trolley, and new rails are all superimposed.

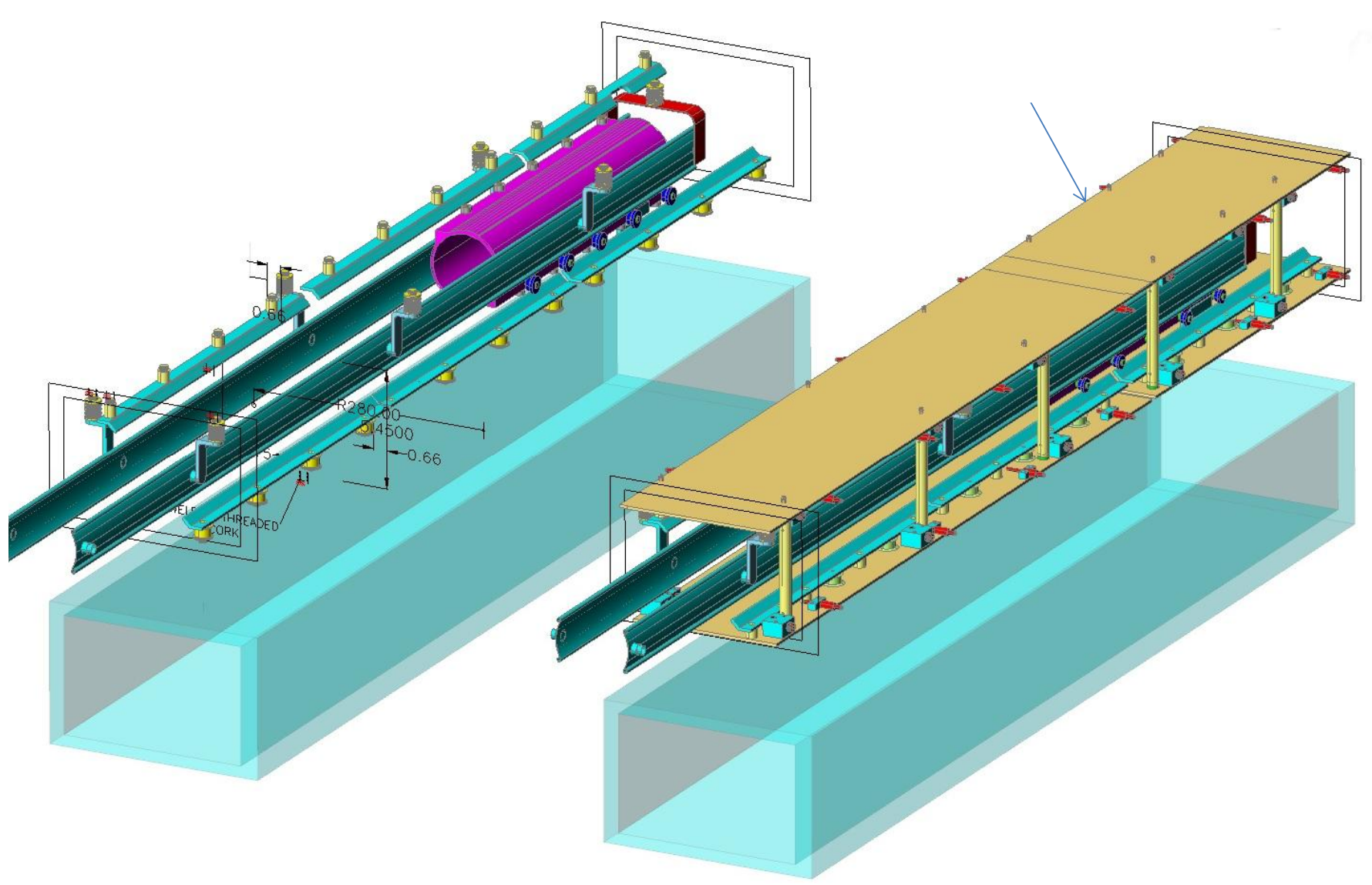

Figure 12.15: (Left)Plates suspended in cages with top and bottom of cage removed. (Right) Plates inside cages and outside of vacuum chamber below.

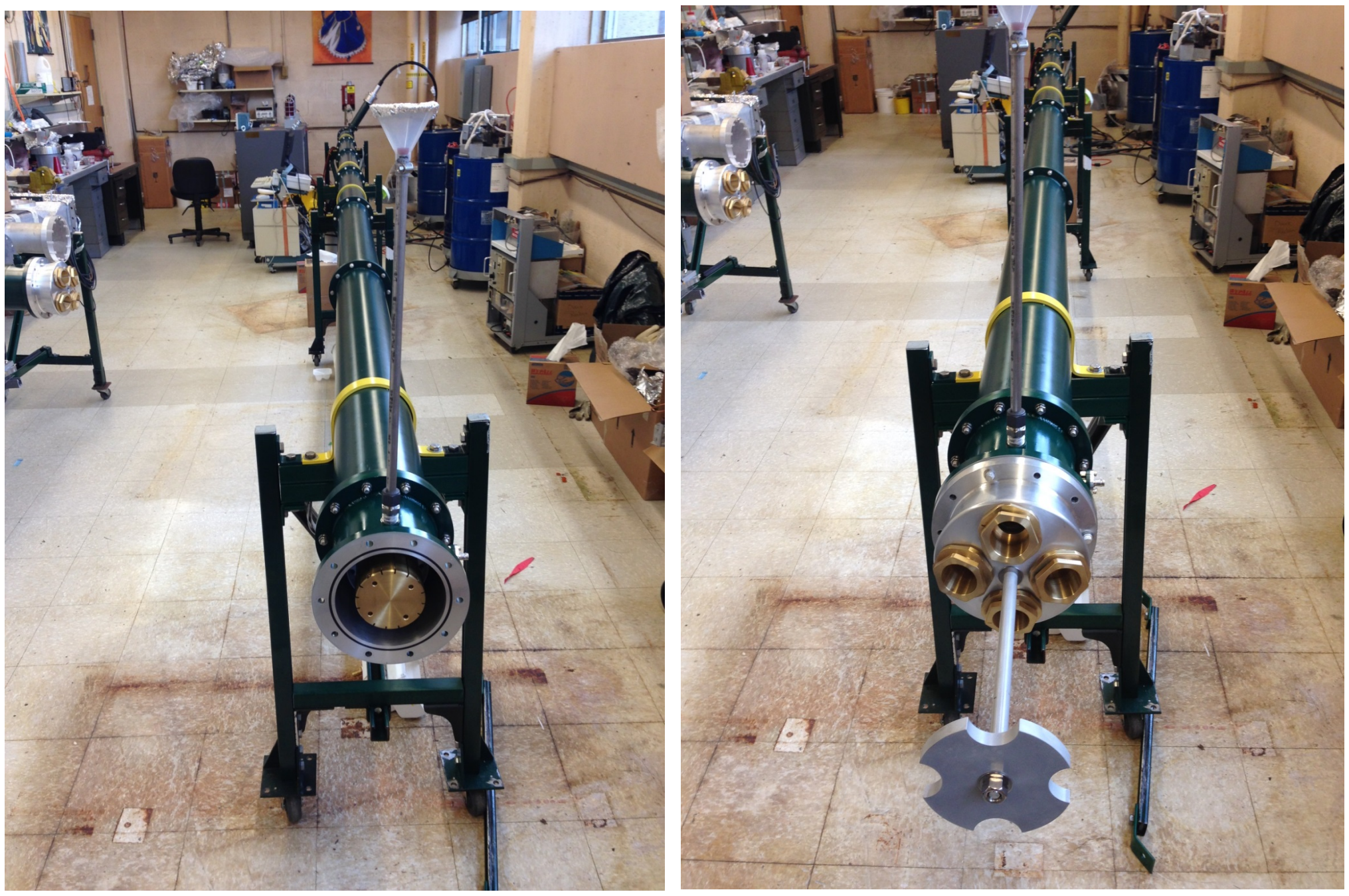

Figure 12.16: (Left) Blumlein. The black cable visible in the background couples the charging transformer to the middler conductor of the Blumlein. The thyratron is mounted horizontally inside the green cylinder at the far end. The space between the conductors of the Blumlein and the volume around the thyratron is filled with transformer oil. The central conductor of the Blumlein, that couples to the four coaxial cables is seen in the foreground. (Right) The fixture that will capture the four coaxial cables is mounted on the near end of the Blumlein.

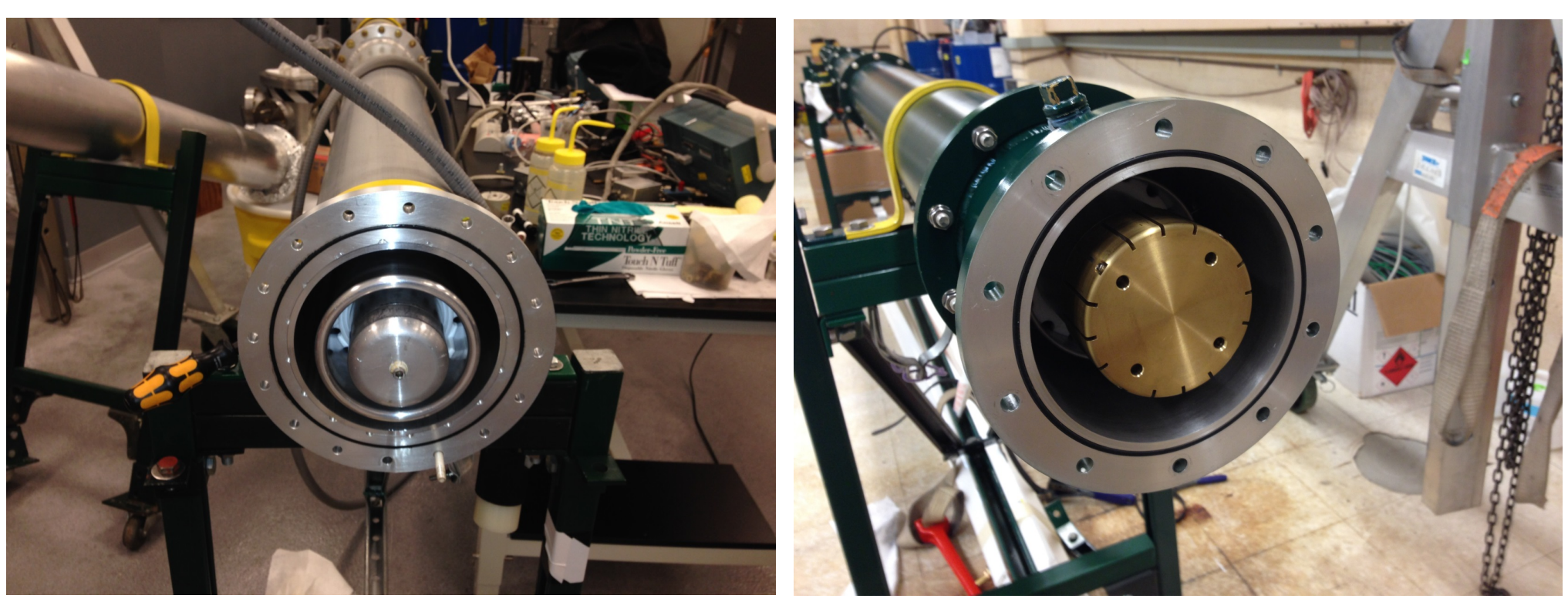

Figure 12.17: (Left) The three concentric cylinders of the Blumlein. The middle cylinder is charge to high voltage. The thyratron switch shorts the middle conductor to the outer cylinder. The current pulse propagates through the central conductor to the load. (Right) The central conductor with cap that couples to cables. The middle conductor is hidden from view.

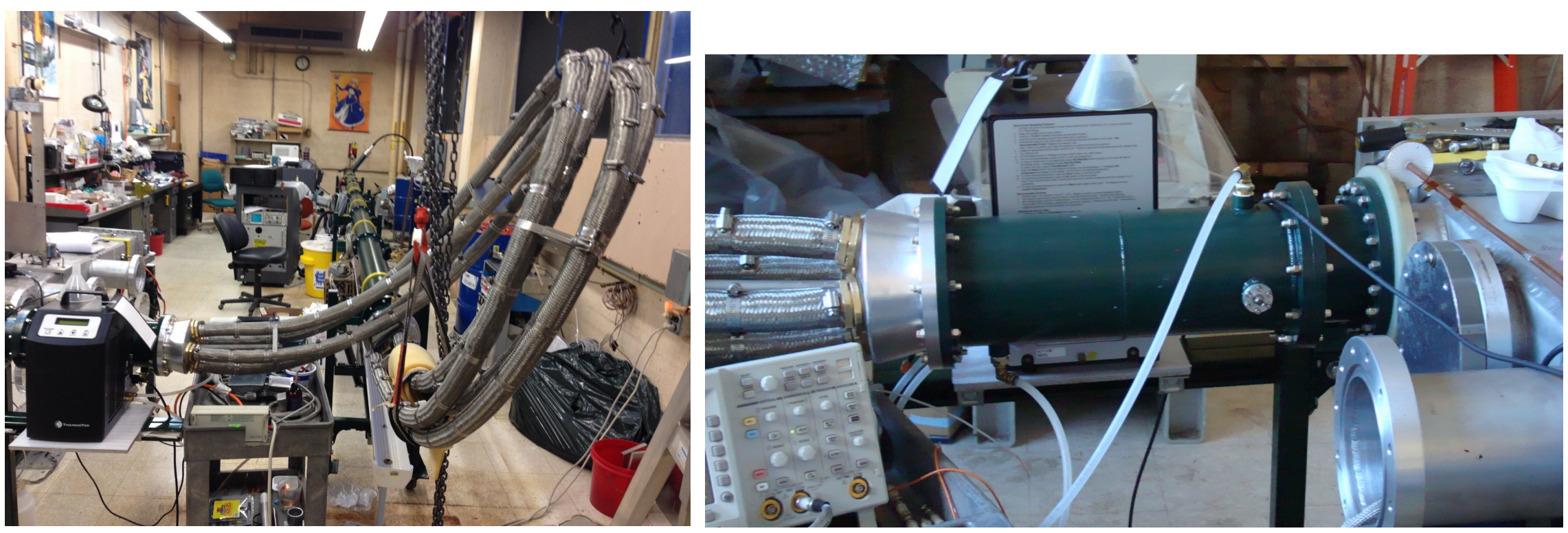

Figure 12.18: (Left) Blumlein with thyratron end in background and transition to coax in foreground. Four parallel coaxial cables connect Blumlein to load resistor (shown in photograph at right) at vacuum chamber at left of photograph. (Right) The four coaxial cables from the Blumlein are coupled to four 50Ω resistors inside the cylindrical chamber. The downstream end of the resistors are coupled together and then to the kicker magnet inside the vacuum chamber at far right of photograph.

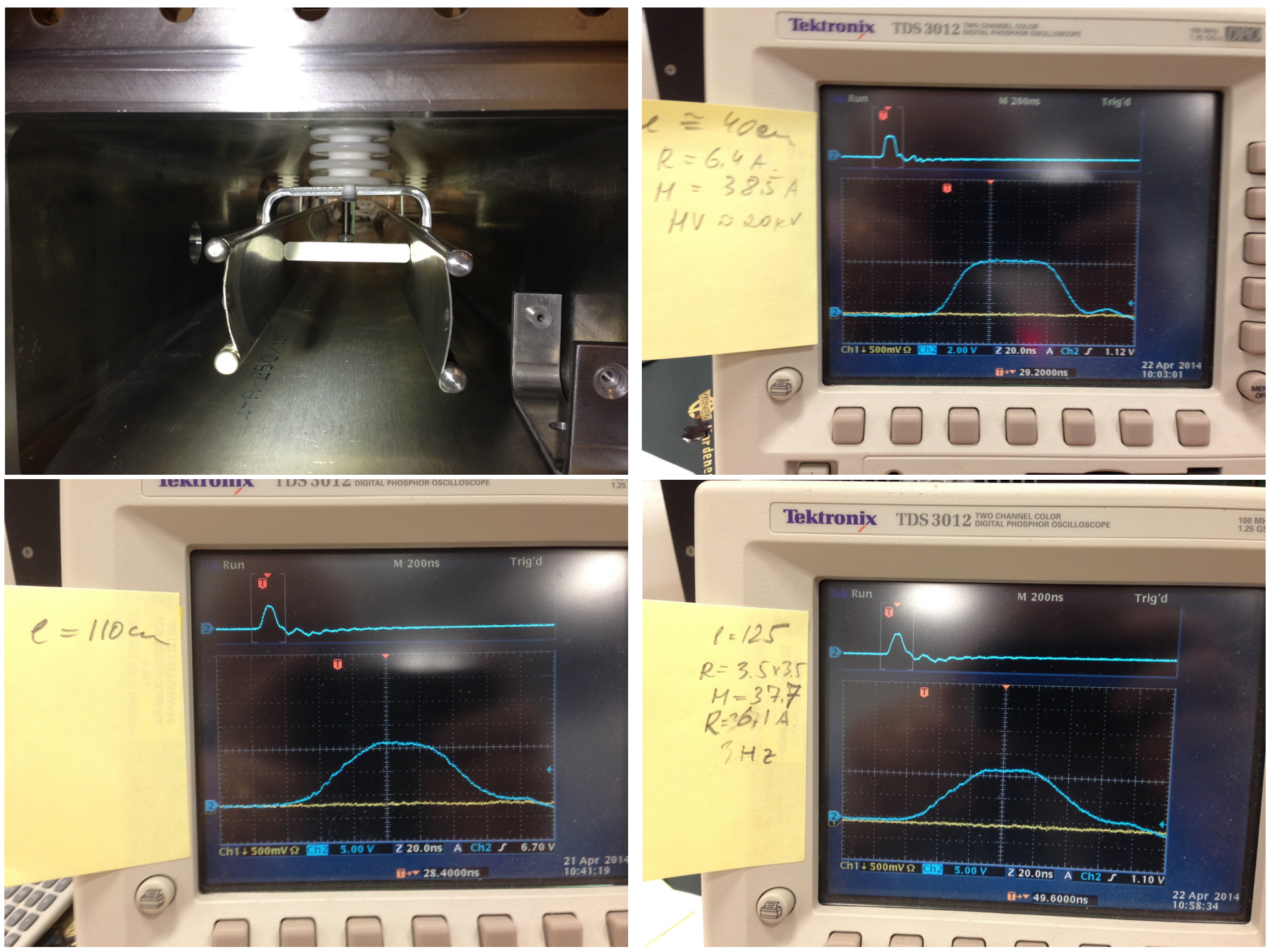

Figure 12.19: (Top left) Prototyper kicker plates with shorting bar. (Top right). Kicker magnetic field pulse measured with loop pickup with effective kicker length of 40cm. The time scale on the oscilloscope is 20ns/division. The rise and fall time of the pulse is 20ns, with a 50ns flat top. The total pulse length is consistent with the length of the Blumlein. (Bottom left) Magnetic field for 110cm kicker. (Bottom right) Magnetic field for 125cm kicker. The rise and fall time increase with the length of the kicker. The pulse shape is also observed to be sensitive to the thyratron reservoir current.)

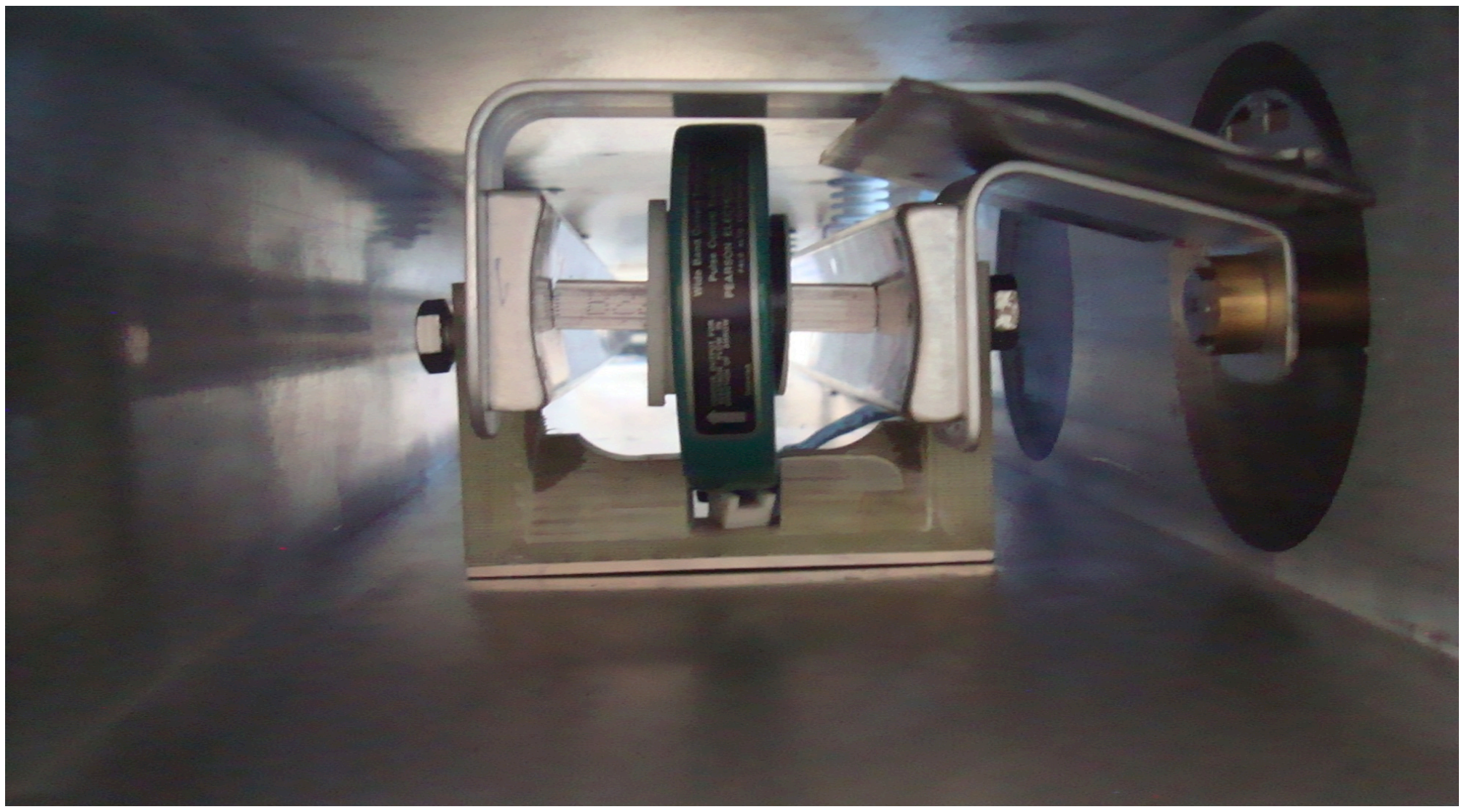

Figure 12.20: Kicker plates mounted in the vacuum chamber with couplings through chamber wall to 12.5Ω termination resistor. Here a current transformer is mounted on a shorting bar. The shorting bar can be moved along the length of the plates and the current measured as a function of plate length.

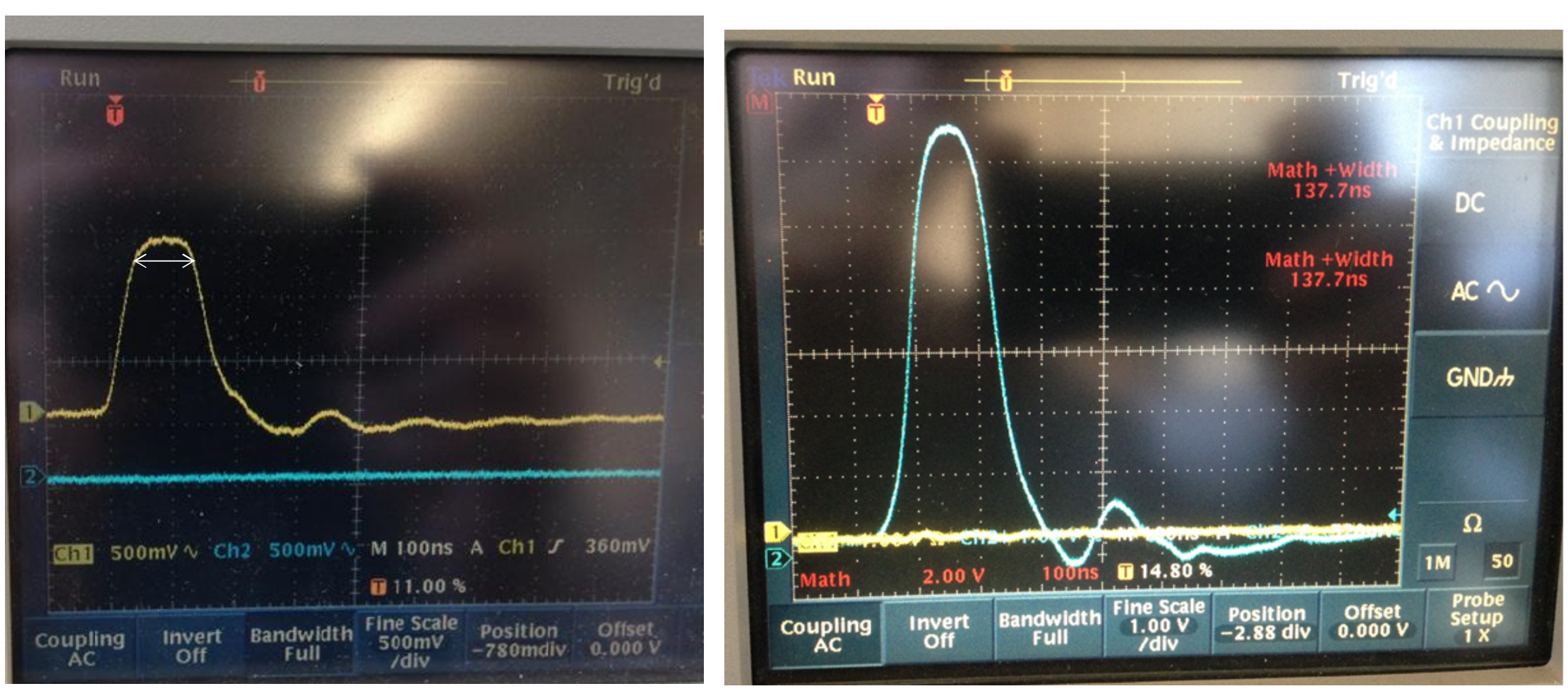

Figure 12.21: (Left)Current pulse with 35 cm plates. (Right) Current pulse with 127cm plates. Note that the rise and fall times are longer for the pulse at right and the flat top narrower. The tail that extends beyond 149 ns has negligible dependence on the length of the plates and is apparently due to mismatch of Blumlein, coax cable and load resistance.

## 12.6 Kicker Field Measurement

Measurement of the time dependent field of the kicker will ultimately determine the effectiveness of the design choices. Furthermore, it is essential to measure, and ideally eliminate, fields due to the eddy currents in the vacuum chamber and kicker plates, that are generated by the kicker pulse. If the eddy currents have a long decay time, any persistent field will introduce a systematic shift in $a_\mu$.

A Faraday magnetometer has been designed and tested to measure the time dependence of the kicker field and associated persistent fields modeled on the device used in E821. The cylindrical TTG birefringent crystal is mounted in a G10 tube that passes through the vacuum chamber and between the kicker plates. Polarized light from the green laser mounted above the vacuum chamber, passes along the axis of the tube, through the crystal and then into a Wollaston prism that is mounted below the vacuum chamber as shown in Figure 12.22 and 12.23. The polarization direction of the light emerging from the crystal is rotated by an angle proportional to the magnetic field. The prism directs orthogonal polarizations into the two arms. The difference of the amplitude of the orthogonal components is a measure of the change in field. Ports machined in the prototype vacuum chamber transmit polarized laser light into the chamber and then through a birefringent crystal that is mounted between the plates in the laboratory as shown in Figure 12.24.

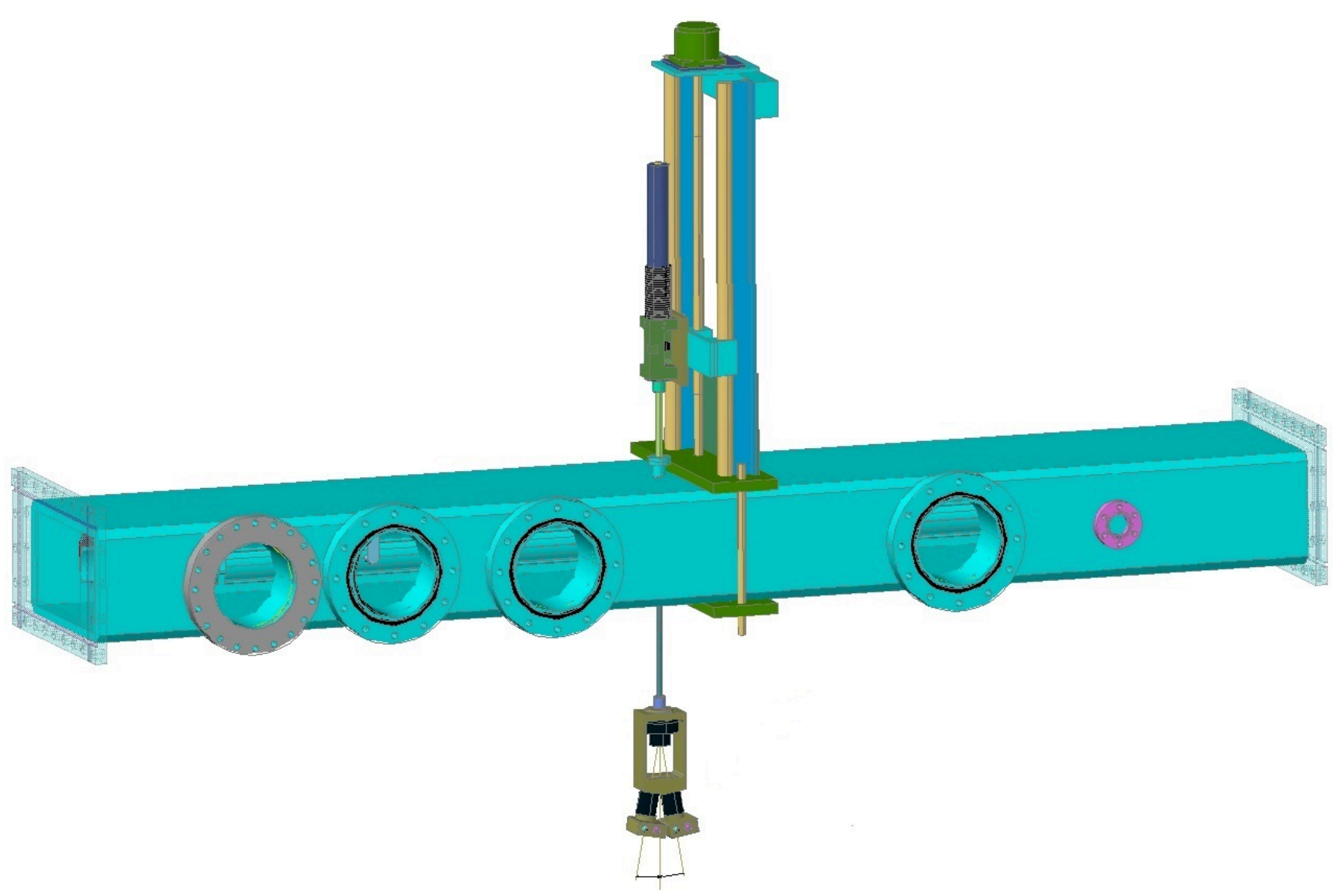

Figure 12.22: The laser is mounted above the vacuum chamber on a stage so that the laser, G10 tube and crystal can be used to measure the B-field along the vertical axis. The light from the two arms of the prism is detected with photo-diodes that are not shown.

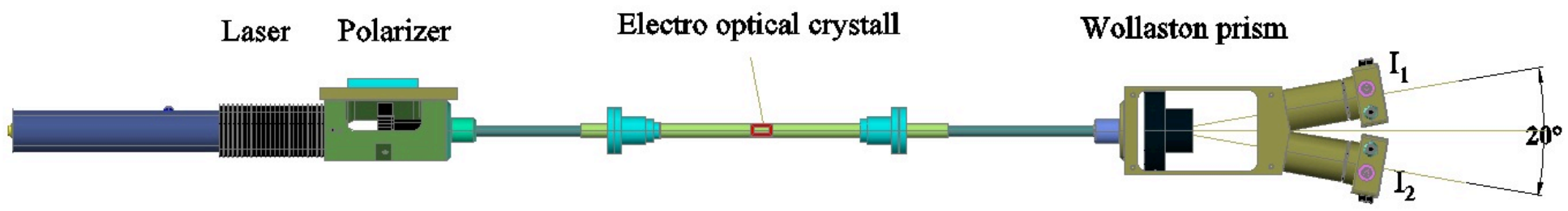

Figure 12.23: Magnetometer showing laser, polarizer, electro-optical crystall (TGG) and prism.

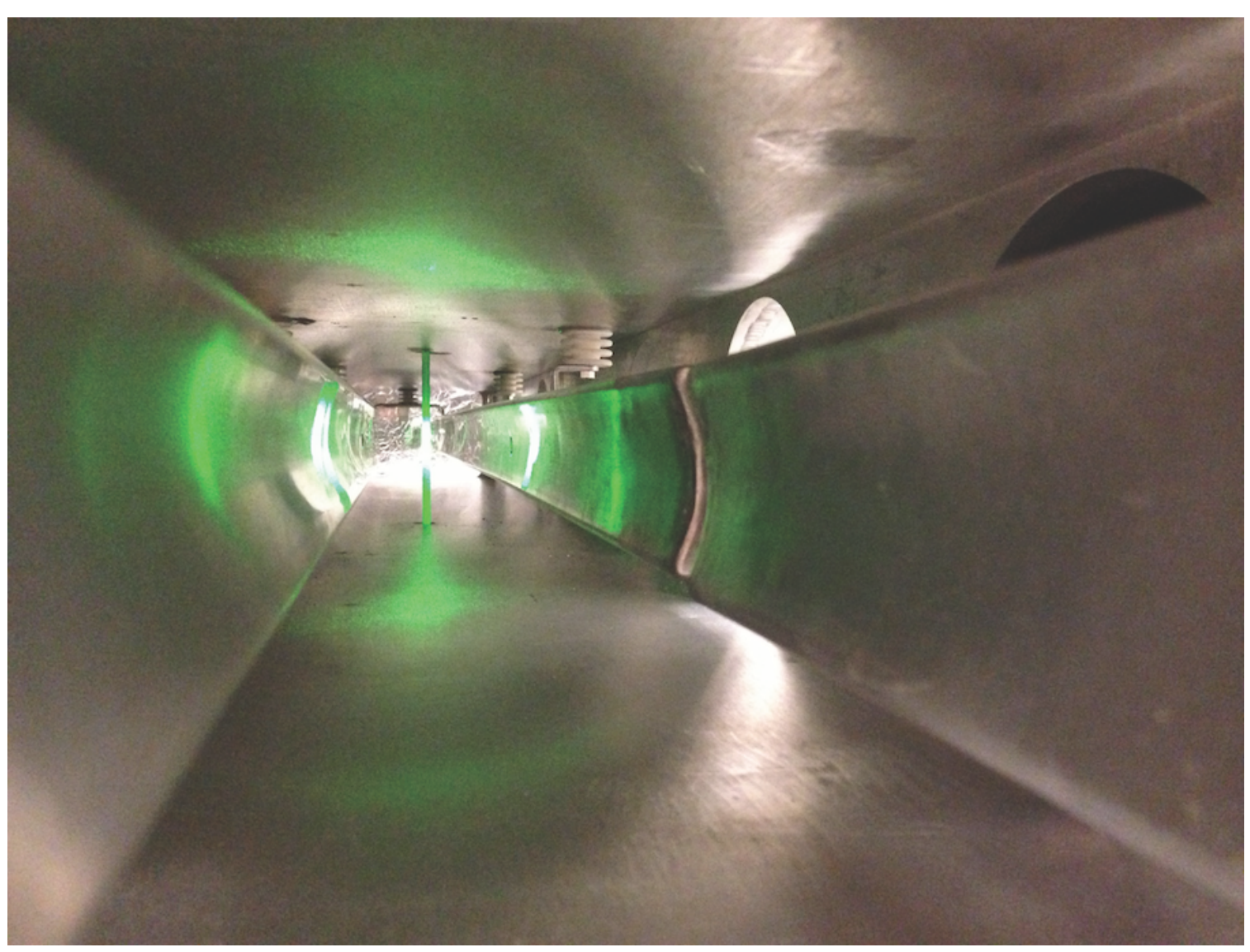

Figure 12.24: Green laser light passes through the crystal that is mounted between the kicker plates.

# 12.7 Risks

## 12.7.1 Performance Risk

The kicker system is designed to provide an integrated field of 1310 G-m for the duration of the length of the injected muon pulse ($\sim$ 120ns), and then drop to zero field, 149ns after the first muons entered the ring. Failure to achieve the specified field value will result in reduced muon capture efficiency and increased coherent betatron oscillation of the muons that are captured. Failure to turn off after 149 ns will likewise compromise capture efficiency and contribute to coherent betatron motion. The risk of less than optimal system performance are increase in statisical error (fewer muons) and additional systematic error (increased coherent betatron motion). The measured kicker current pulse is shown above 12.21(Right) and a numerical representation in Figure 12.25(Left). A tracking study quantifies the performance penalty due to the imperfect nature of the pulse shape (finite rise and fall time), and the nonuniformity of the kicker field resulting from the optimization of the plate geometry for efficiency. The study is summarized in Figure 12.25(Right). There is a negligibly small capture efficiency penalty due to the spatial non-uniformity of the kicker pulse. The effect of the finite rise and fall time, and the tail that persists beyond 149 ns, is a loss of capture efficiency of about 17%. The effort to eliminate the impedance mismatch responsible for the tail is continuing. The source of the mismatch may be deterioration of the Blumlein transformer oil, nonstandard impedance of the cables, and/or dimensional errors in the Blumlein. An impedance transformer that includes tunable resistance and capacitance to match blumlein to cables is being developed.

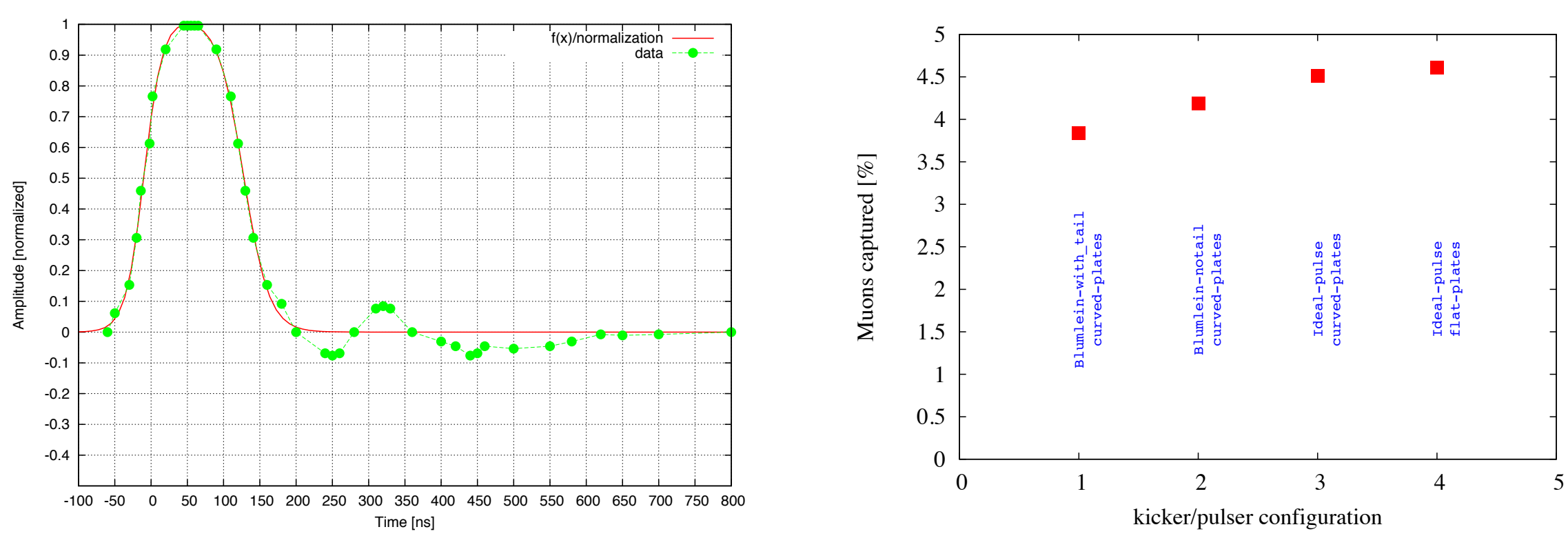

Figure 12.25: The capture efficiency for each of four kicker configurations is shown in the plot. Configuration 4-Perfect rectangular pulse shape and perfectly uniform magnetic field; 3-Perfect rectangular pulse with field profile corresponding to the curved plates; 2-Pulse shape as shown by the red curve at left; 1-Pulse shape as shown by green points and dashed green curve at left.

As noted above, the pulse rise and fall times can be reduced, and the flat top increased, by reducing the length of the kicker plates. If necessary, a fourth kicker and Blumlein driver could be added to increase the integrated field.

A failure of the thyratron tube, a breakdown internal to the Blumlein-PFN, or a breakdown of the plates inside the vacuum chamber would have more catastrophic consequences, as very few muons will store without an operational kicker. The risk of system failure will be mitigated by operating the kicker system continously for a month at the design repitition rate and at voltage 10% above design to demonstrate integrity prior to installation in the ring. The current generator is designed to operate at less than 70kV, well below the level at which the E821 system was limited by breakdown.

There is some risk that the kicker will excite a long lived eddy current in the vacuum chamber that will in turn generate a lingering magnetic field that will alter the muon precession frequency. The characteristics of persistant field will be calculated and measured.

The vacuum feethrough will be cooled with Fluorinert, rather than transformer oil, so that in advent of a leak to vacuum, the storage ring vacuum system is not contaminated with oil.

The high speed switching of high voltages and currents through the stripline plates can be a significant source of electro-magnetic noise. A series of tests to determine the effect of the noise on the detector electronics will be performed and grounding scenarios to minimize its impact on the experiment identified.

## 12.8 Quality Assurance

The quality of the kicker system will be assured by extensive testing in advance of installation into the ring. While a single power supply and high voltage transformer has the capacity to power all three kicker/PFN system, we are planning to use independent supplies so that in the event of a high voltage problem with one of the systems, the others are unaffected.

## 12.9 ES& H

The kicker system will operate at high voltage $\sim$ 70kV, however there will be no exposed high voltage electrodes. All external surfaces of the Blumlein will be fixed at ground potential. As there are no diodes in the charging circuit, the time constant for dissipation of stored charge is a few seconds. A procedure for de-energizing, in order to dissipate stored energy, in the event that repair or dissassembly is required will be established. Each of the three Blumlein tri-axial lines will be filled with castor oil and an oil containment system implemented. While there is the danger of a spill, (75 liters/line), the oil itself is not hazardous.

## 12.10 Value management

We are reusing as much as possible, components from E821 including charging power supplies and transformers. The E821 thyratrons are not suitable and will be replaced with the higher current, lower voltage, 2 gap CX1725X thyratron.

# Chapter 13

# The Electrostatic Quadrupoles (ESQ) and Beam Collimators

## 13.1 Introduction

The Electrostatic Focusing Quadrupoles (ESQ) in the $(g-2)$ storage ring are used to confine muons vertically. The ESQ were first used for beam storage in the final muon $(g-2)$ experiment at CERN [1], and in the muon $(g-2)$ experiment E821 at BNL [2]. Our baseline plan for E989 is to reuse the existing E821 ESQ after careful refurbishment and necessary upgrades as will be discussed in the following sections. This decision was made after we have carefully considered alternatives, e.g., weak magnetic focusing, and alternating skew electrostatic quad focusing.

## 13.2 Requirements for ESQ and Beam Collimators

The E989 ESQ must

1. Produce vertically focusing electrostatic quadrupole field in the muon storage region.
2. Have operating point in resonance-free region around $n = 0.185$.
3. Have stable operation during extended periods of time in a vacuum of $10^{-6}$ Torr or better.
4. Have reliable operation in pulsed mode with time structure of the beam as described in section 7.2 (12 Hz average rate of muon spills that comprises sequences of four consecutive spills with 10 ms spill-separations).
5. Be optimized for muon storage efficiency.
6. Have minimum possible amount of material at places where the trajectories of incoming muons and decay positrons intercept the parts of ESQ.

7. The quality as well as long- and short-term stability of the electrostatic quadrupole field must be sufficient to keep the beam-dynamics systematic uncertainties well below the E989 goal.

The E989 ESQ and Beam Collimators must

1. Provide effective scraping of the injected beam to remove muons outside the storage region.
2. Be made from non-magnetic materials to not deteriorate the quality of the dipole storage magnetic field.
3. The thickness and the shape of collimators must be optimized to satisfy the two conflicting requirements *i)* efficiently remove muons outside the storage region and *ii)* has little effect (e.g. multiple scattering, showering, etc.) to the decay positrons.

## 13.3 Design of ESQ

Since we are planning to reuse the E821 quadrupoles for E989, the basic features of the mechanical and electrical design of ESQ in E989 are the same as in E821 as described in [3]. In the present document we describe them and include the main points that aim to improve the muon ring acceptance and reduce muon losses as well as certain systematic errors associated with the coherent betatron oscillation frequencies.

### Mechanical Design

Fig. 13.1 shows a schematic top view of the muon $(g-2)$ storage ring indicating the location of four ESQ Q1-Q4 inside the scalloped vacuum chambers. Ideally, the ESQ plates should fill as much of the azimuth as possible, but space is required for the inflector and kicker magnets. For symmetry reasons, the quadrupoles are divided into four distinct regions Q1-Q4. Gaps at 0° and 90° for the inflector and kicker magnets, along with empty gaps at 180° and 270° provide a four-fold symmetry. Overall, the electrodes occupy 43% of the total circumference. The four-fold symmetry keeps the variation in the beta function small, $\sqrt{\beta_{\rm max}/\beta_{\rm min}} = 1.04$, which minimizes beam "breathing" and improves the muon orbit stability.

Each quad segment consists of a "short" quad of 13° and a "long" quad of 26°, (see Fig. 13.2), for two reasons: 1) to make every quadrupole chamber independent of others, facilitating their development, testing, etc., and 2) to reduce the extent of low energy electron trapping. Therefore there are two high voltage vacuum-to-air interfaces for each quadrupole segment.

A schematic representation of a cross-section of the electrostatic quadrupoles is shown in Fig. 13.3 with the various dimensions indicated. Shown are the four flat aluminum plates ("electrodes") symmetrically placed around the 90-mm-diameter muon storage region. The four NMR trolley rails are at ground potential. Fig. 13.4 shows a picture of the quadrupoles at the downstream end of one chamber. Fig. 13.5 shows one segment of the muon $(g-2)$ electrostatic quadrupoles at BNL outside its vacuum chamber.

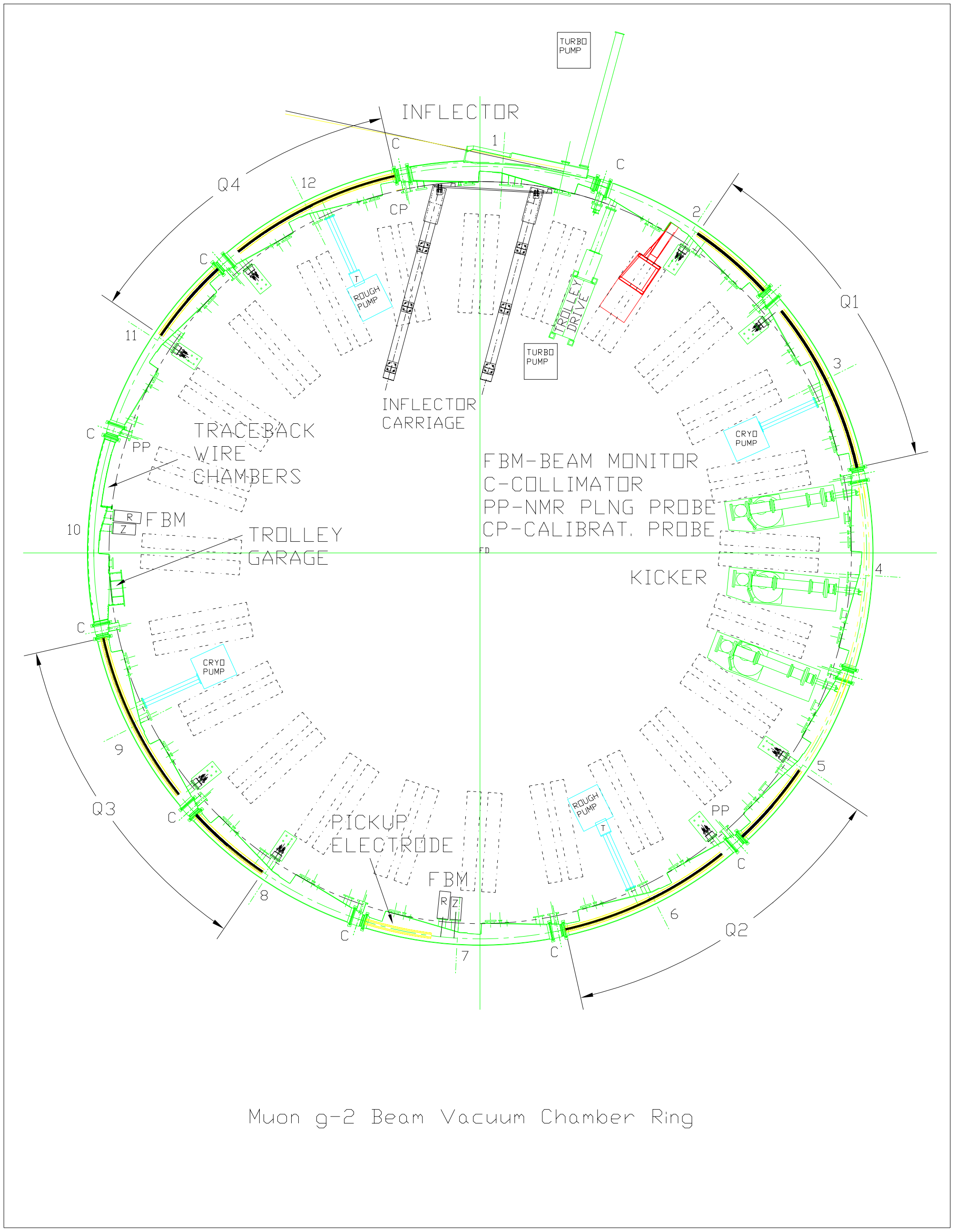

Figure 13.1: A schematic view of the muon $(g-2)$ ring as well as the location of Q1, Q2, Q3, and Q4, the four-fold symmetric electrostatic focusing system.

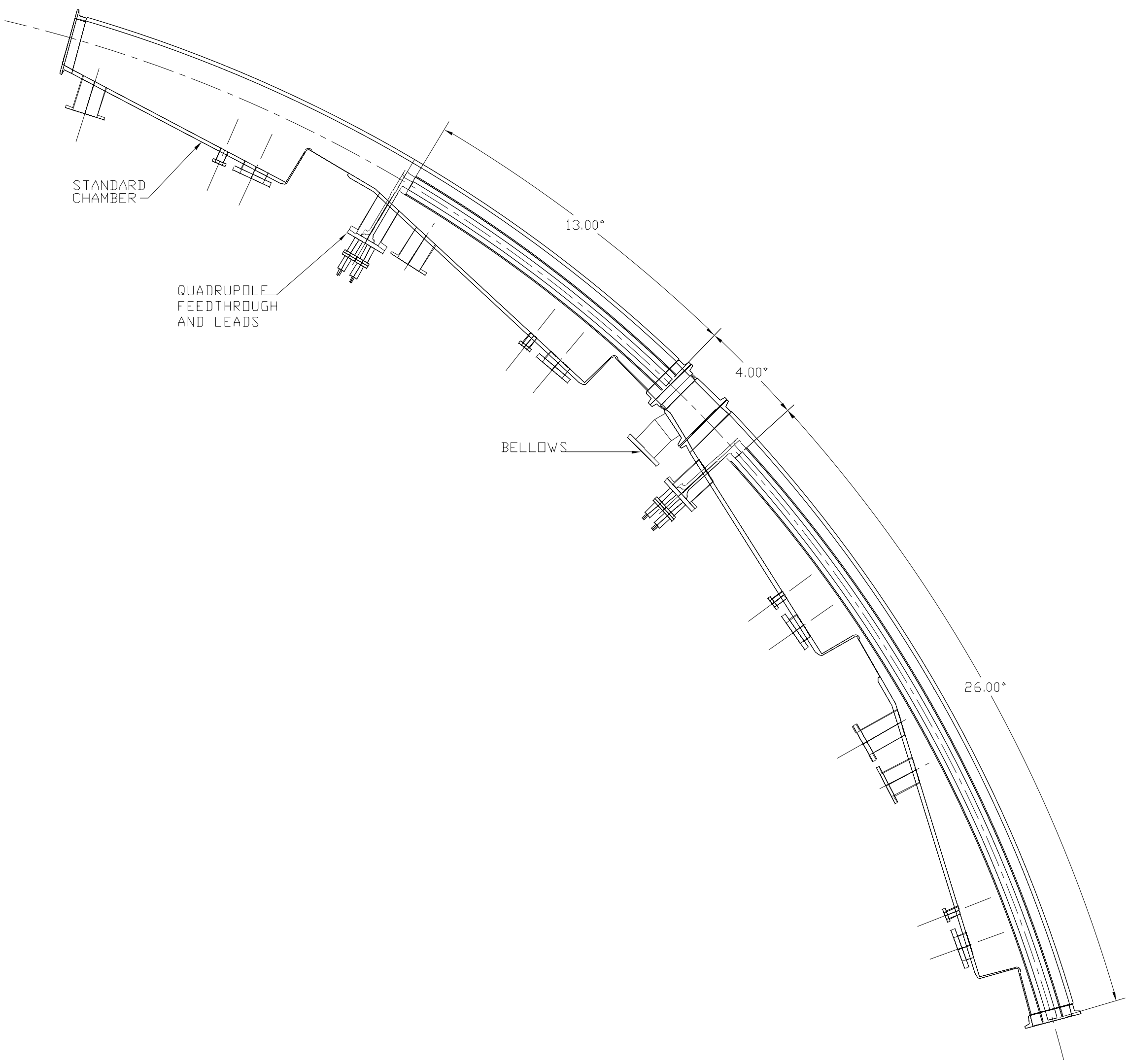

Figure 13.2: A schematic view of a short quad of 13°, and the adjacent long quad of 26°. The high voltage feeding leads break the quad symmetry at the upstream end of the plates to quench the low energy electron trapping and guide them outside the magnetic field region, where they can be released. Some of the bellows are equipped with collimators where the muon beam is scraped immediately after injection.

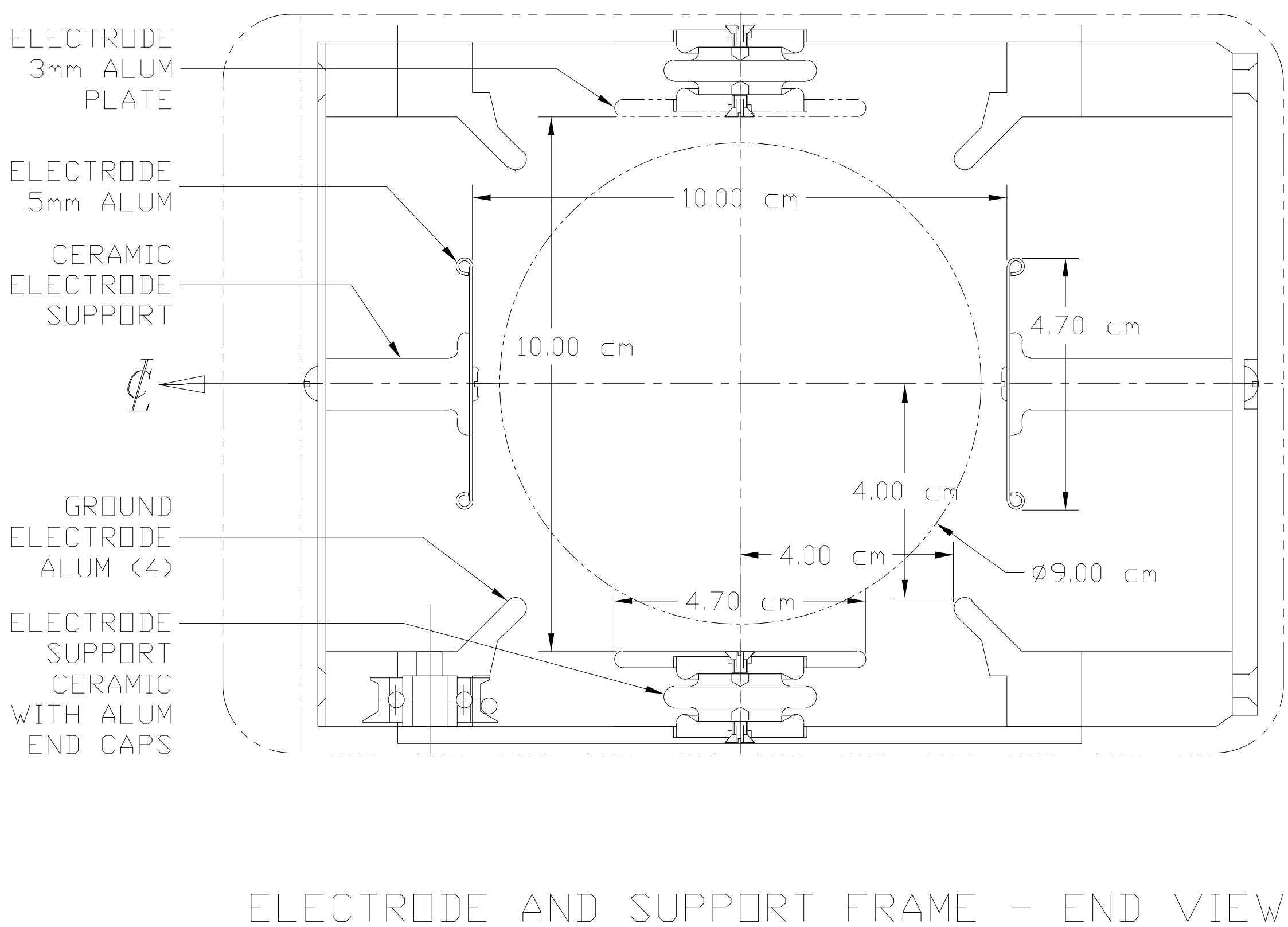

Figure 13.3: A schematic of the quadrupole cross-section. The rails in the corners are kept at ground potential. Most of the side support insulators are replaced with uniform diameter insulators of 0.5 cm.

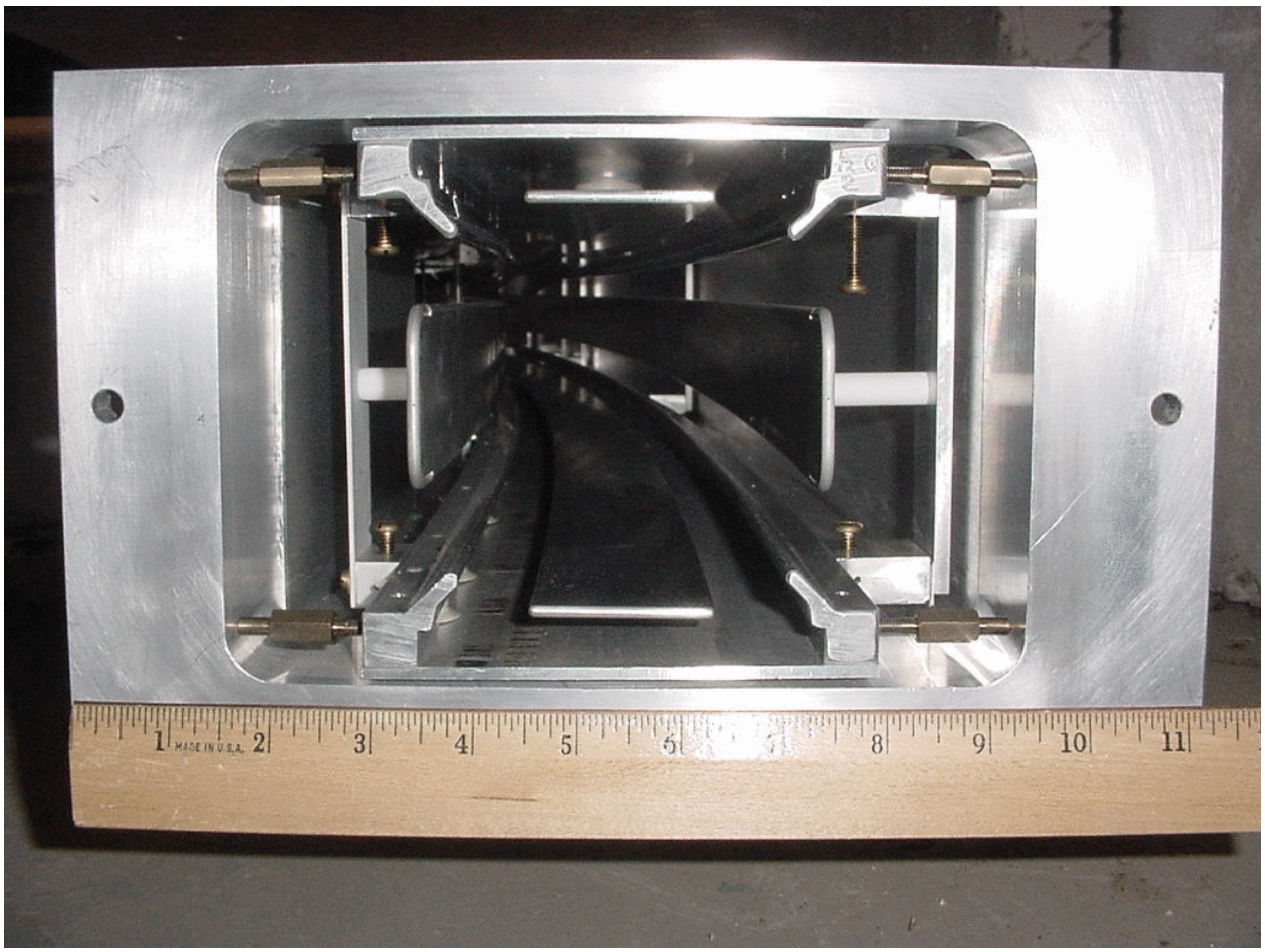

Figure 13.4: A photograph of the downstream end of a vacuum chamber with the cage and quads showing.

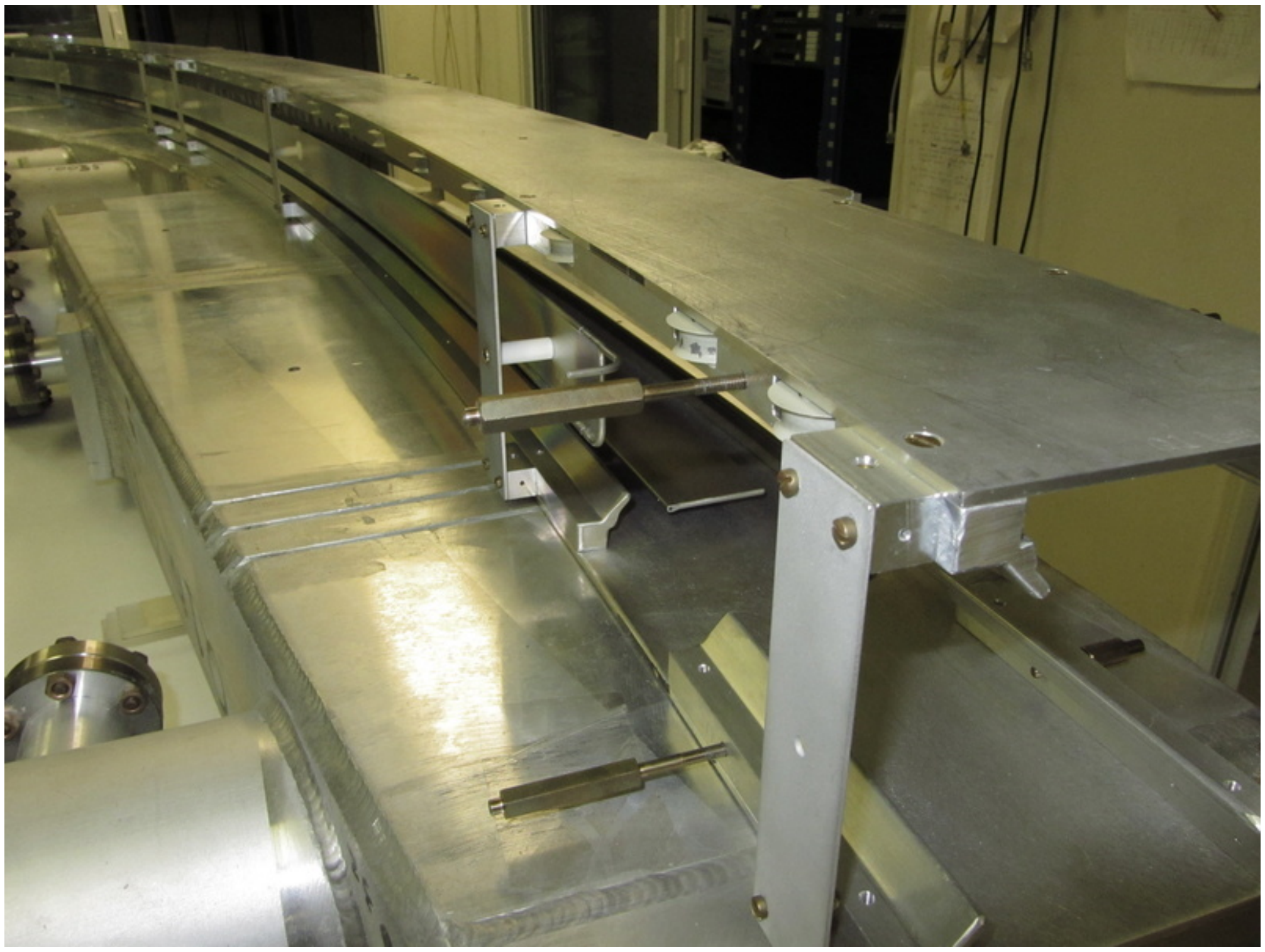

Figure 13.5: One cage (placed here up-side-down) that holds the plates of the electrostatic quadrupoles of the muon $(g-2)$ experiment.

The placement accuracy was 0.5 mm for the horizontal (top/bottom) quad electrodes, and 0.75 mm for the vertical (side) quad electrodes. When measured by the surveyors the electrodes were found to be well within those values.

## Electrical Design

The electrical diagram of the quadrupoles is given in Fig. 13.6. The storage capacitance is 1.5 $\mu$F, rated at 40 kV [4]. The high voltage (HV) switches are deuterium thyratrons, models CX1585A and CX1591, made by EEV [5], rated at the minimum to 40 kV; and 5 kA maximum current. The baseline plan is to use HV power supplies PS/LKO4OPO75-22 (positive polarity) and PS/LKO4ONO75-22 (negative polarity) from Glassman. They are capable of delivering up to 40 kV at 75 mA maximum [6]. Their voltage regulation is better than 0.005% whereas the ripple is better than 0.025% RMS of rated voltage at full load.

The capacitance of the distribution cables (RG35B) is between 1 and 3 nF depending on how many quadrupole plates are fed, deployed in the star configuration, by the same HV pulser unit. There are four pulser units, two for applying standard high voltage (i.e. a single voltage value per pulse), one for each polarity, and two for scraping high voltage (i.e. two different voltage values per pulse), again one for each polarity. Following the E821 scraping scheme [3], two of the quadrupoles will be used to scrape the injected beam horizontally, by moving the beam sideways, while all the quadrupoles will be used to scrape the beam vertically.

HV monitors will be used to record traces of voltages on the quadrupole plates. Their location is indicated in Fig. 13.6. Fig. 13.7 shows the (home-made) HV monitors output waveforms as recorded by an oscilloscope. Voltage traces from each pair of quadrupole plates will be continuously digitized during each fill at 25 MHz sampling rate using 8-bit 500 MHz waveform digitizers originally built by the Boston University for the MuLan experiment. The sampling frequency of digitizers will be downgraded for E989.

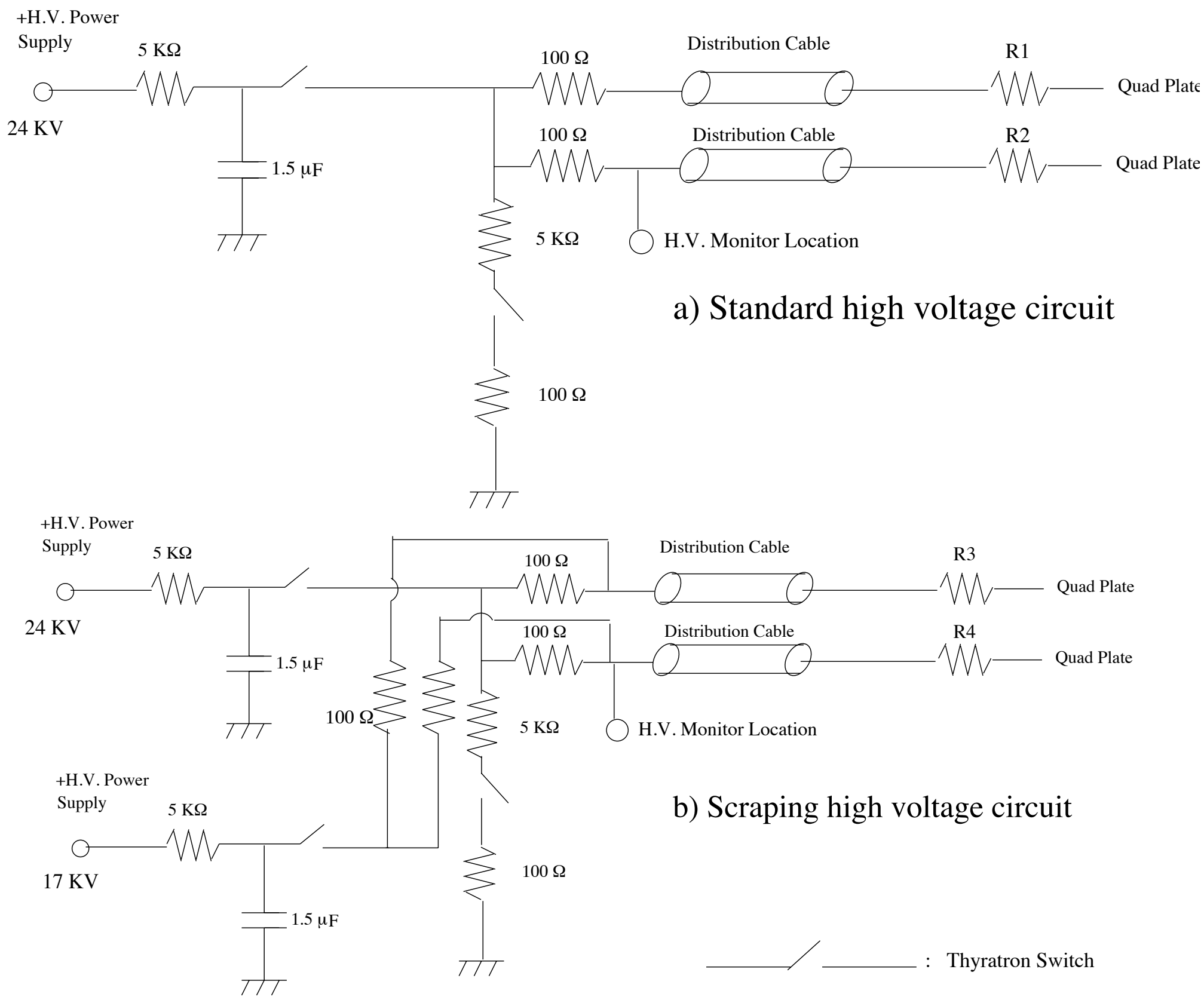

Figure 13.6: A schematic of the scraping and standard HV pulsing systems. The Thyratron switch model used in E821 was the CX1585A produced by English Electric Valve, good to 40 kV.

## 13.4 Design of Beam Collimators

Beam collimators in the $(g-2)$ storage ring are used to remove muons outside the 9-cm-diameter storage region. In E821 the collimators were 3-mm-thick copper rings having inner and outer radii of 4.5 and 5.5 cm. Eight collimators were installed around the storage ring, the locations are indicated in Fig. 13.8. As will be discussed below, to reduce systematic uncertainties, the design of collimators will be improved in E989, new collimators will be manufactured. In E821, the inner half circles of some collimators were removed to avoid scattering of those low-momentum muons which, because of limitations of the kicker, would otherwise strike the collimator on the first turn and be lost. A photograph of one of the half-aperture collimators is shown in Fig. 13.9. In E989 the kicker will have enough strength, therefore all half-collimators will be replaced by full-collimators. To improve the efficiency of collimators, the thickness of collimators in E989 will be increased [7]. The optimal thickness will be found as a compromise between conflicting requirements to the collimators, provide efficient scraping of the beam muons and have low distortion of the magnetic field and low scattering of the decay positrons. The studies will be completed by the Final Design.

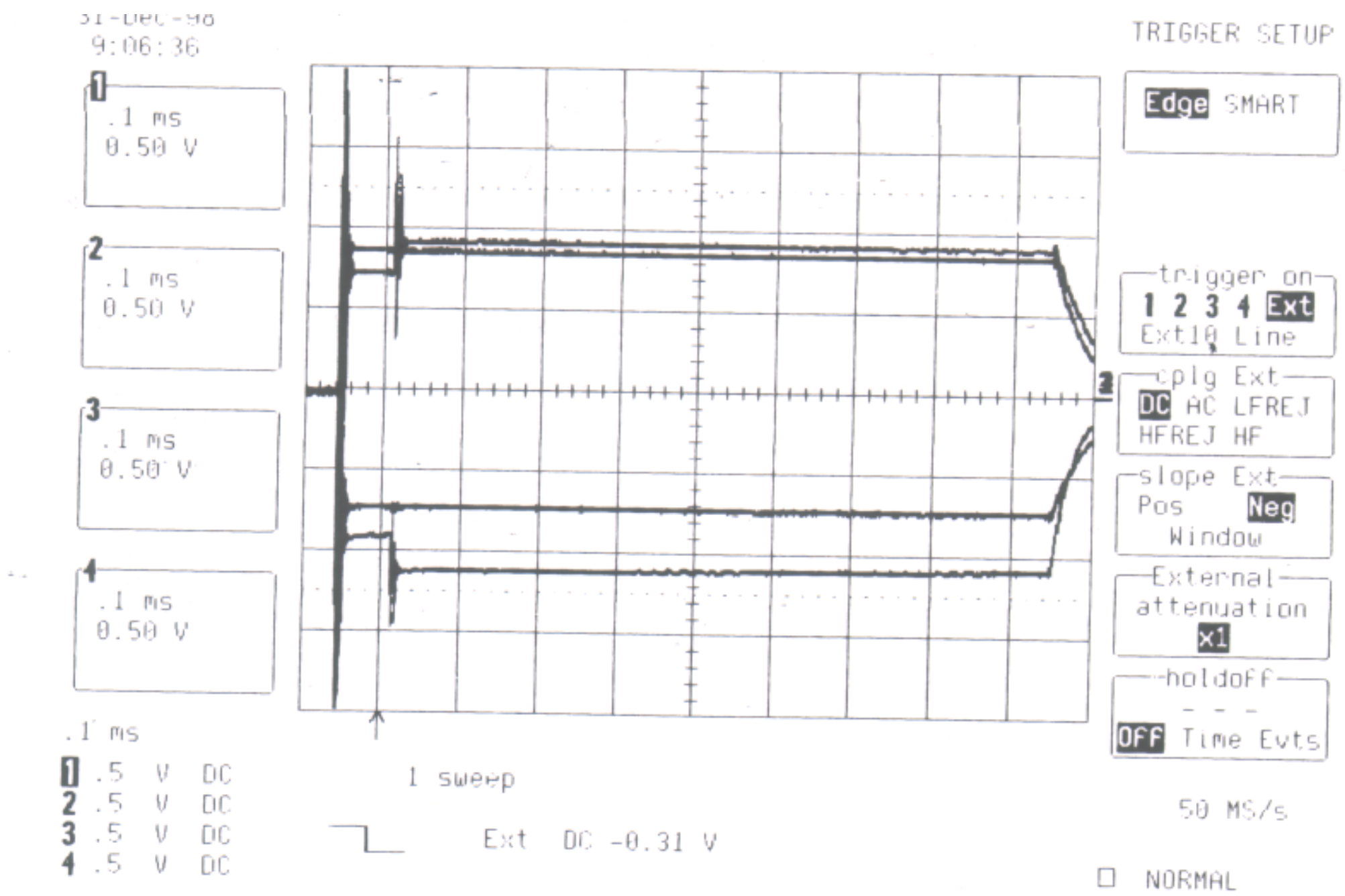

Figure 13.7: The output of the HV monitors as recorded by the oscilloscope.

Muons outside the storage region hit the collimators, lose energy and are lost after several turns. In a storage ring with perfectly uniform dipole magnetic field and ideal quadrupole focusing, no further beam losses occur. In reality, the higher field multipoles provide a perturbative kick which causes some muons to eventually be lost during the measurement period. Muon losses distort the shape of the decay time histogram and can bias the $\omega_a$ determination. To reduce such losses the beam is *scraped* to create about 2-mm-wide buffer zone between the beam and collimators. During scraping the quadrupole plates are charged asymmetrically to shift the beam vertically and horizontally and move the edges of the beam into the collimators. After scraping the plate voltages are symmetrized to enable long-term muon storage. We are also studying an alternative scraping method based on excitation of Coherent Betatron Oscillations (CBO). Muons with trajectories close to the beam collimators can be removed from the storage ring by amplifying the amplitude of their betatron oscillations and making them to hit the collimators. The excited CBO will then be damped by applying counter-perturbation. CBO excitation and damping can be accomplished by pulsing a pair of dedicated plates (quadrupole, kicker or custom-built plates) at the CBO frequency.

The collimators can be manually rotated into the "beam position" for data taking, or into the "out-of-beam position" to run the NMR trolley around the storage ring for mapping the storage magnetic field. For E989 we are planning either to fully automate the rotation of collimators or to install sensors of collimator rotation status. The information on the status of collimators will be important in preventing human mistakes by inhibiting trolley operations if one of the collimators is in a wrong position.

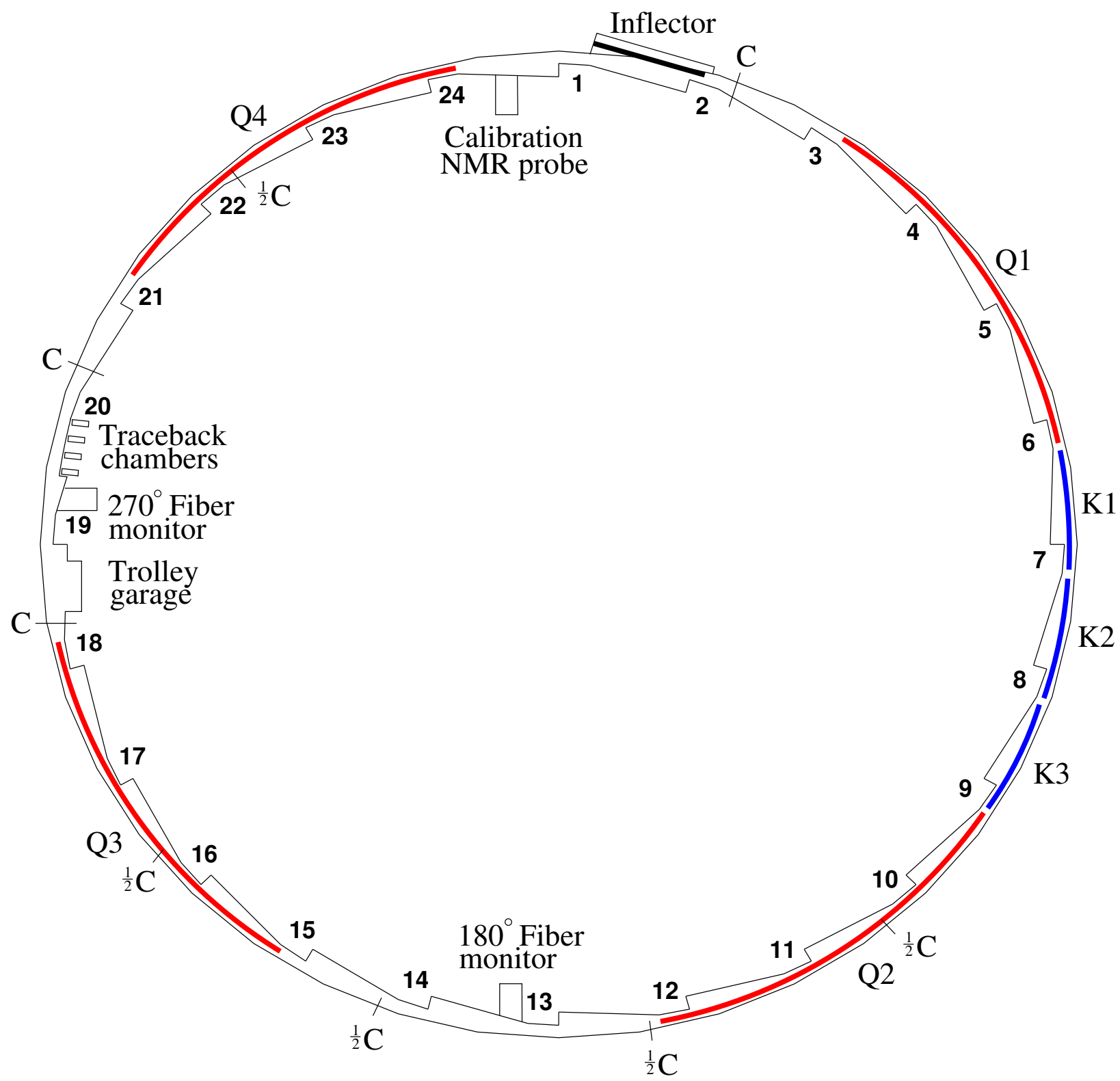

Figure 13.8: Schematic diagram of the E821 ring showing the location of the half and full collimators. For the FNAL experiment, there will only be full collimators.

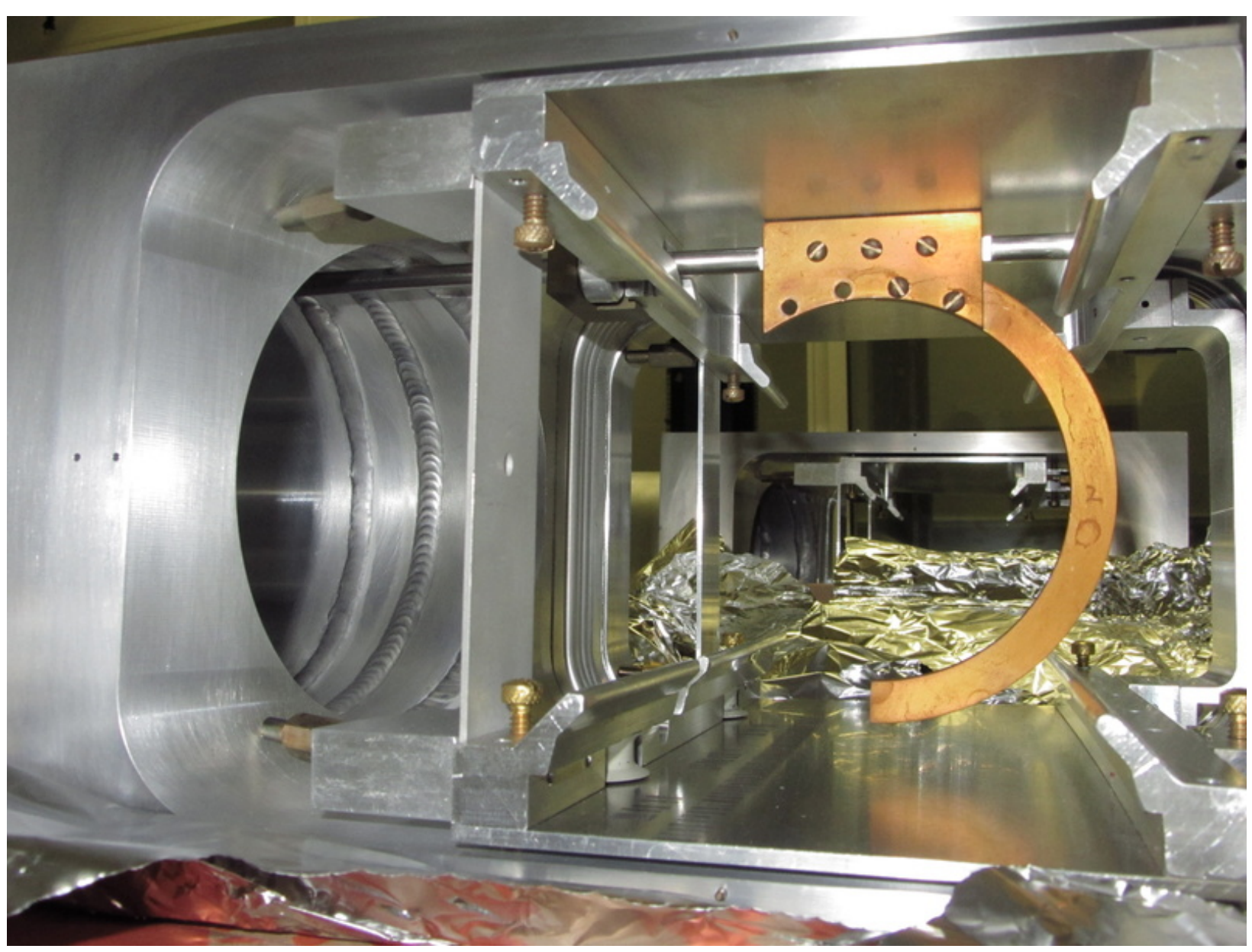

Figure 13.9: Photograph of a half-aperture beam collimator from E821.

## 13.5 Beam Polarity and ESQ

The great success of the quadrupole system is based on the fact that it allowed the storage of positive and negative muons for more than 0.75 ms in the storage ring, even though the azimuthal quad coverage was almost half that of the last muon $(g-2)$ experiment at CERN. The vacuum requirements were in the low $10^{-6}$ Torr for the positive muons and low $10^{-7}$ Torr for the negative muons. Higher vacuum pressures were tolerated for limited operation periods. Those requirements allowed a speedy recovery after any unavoidable opening up of the vacuum chambers during the initial stages of the runs, related mostly to issues other than the quadrupole operations.

For E989 we focus on positive muon storage only, due to the following advantages:

- It allows us to improve the $E$-field quality by restoring the normal quadrupole field in the lead region. The plan is now to connect the leads at the center of the plates, expose the $E$-field from the leads for a couple of centimeters, and, if practical[1], hide them behind a ground shield. The aim is to shield the muon storage region from the $E$-field generated by the leads.
- Due to the relaxed vacuum requirements associated with $\mu^+$ running, we will be able to raise the high voltage and keep it there for longer times. This may have an impact on the muon lifetime measurement or other systematic error measurements.
- For E821, the quadrupoles required a lengthy conditioning period (a couple of hours, depending on pressure) after every trolley run. For positive muon storage, plus an automated conditioning system, we expect to minimize this recovery time by a factor of two to three. Quadrupole conditioning is much more straight forward in the positive polarity than in the negative polarity. The main reason is that in the negative polarity the support insulators are intercepting the low energy trapped electrons, which, depending on the trapping rate, could cause sparking. The conditioning process in the negative polarity was very delicate and lengthy. One of the possible models for why it worked was that the slow conditioning creates a thin conducting layer on the insulator surface, allowing them to slowly move and thus avoid accumulating a critical level. For the positive polarity there are no insulators in the way of the trapped electrons. We will write a computer software program that will be able to condition the quadrupoles taking into account the vacuum pressures and sparking history.
- For positive muon storage we expect the voltage on the plates to be more stable as a function of time and from pulse to pulse.

[1] Preliminary Design studies indicate that unshielded HV leads have little effect on beam dynamics. Since the space is very tight, we will reevaluate the need and feasibility of shielding during the Final Design.

## 13.6 ESQ Improvements for E989

Quad upgrade and testing aims to produce an ESQ focusing system that maximizes muon statistics and minimizes potential systematic errors. A large number of improvements will be implemented to ESQ system based on the experience we accumulated during E821 operations. For E989 we require improvements in a number of areas:

1. Operate the quadrupoles at a higher $n$-value to primarily change the horizontal coherent betatron oscillations (CBO) frequency away from near twice the muon $(g-2)$ frequency. The CBO frequency, being very close to twice the $(g-2)$ frequency, see Table 13.1, pulled the $(g-2)$ phase and was a significant systematic error that required special attention during data analysis. We aim to operate at $n \approx 0.18$ to reduce it by more than a factor of three. Other improvements, e.g., properly matching the beamline to the storage ring (requiring a proper inflector channel) is expected to reduce it by at least another factor of three. Overall the CBO systematic error can be reduced to the level required by E989.

2. Reduce the muon losses by more than an order of magnitude to reduce the Lost Muon systematic error. We will achieve this goal by moving the operating point to $n \approx 0.18$, beam scraping by 2.6 mm in the horizontal and vertical directions after injection, and by keeping the radial $B$-field below 50 ppm (this level of radial $B$-field displaces the average vertical position by about 2 mm). The region around $n = 0.18$ is more resonance free than the previous $n$-values we ran with, see Fig. 13.10. We will refine the quadrupole operating mode by running precision beam dynamics tracking simulations to more accurately predict the muon population phase-space after scraping.

3. Shield the muon storage region from the modified quadrupole field due to the HV feeding lead geometry. This region is less than 5% of the good quad coverage around the ring, but it can still influence the muon loss rate. Preliminary Design studies indicate [8] that the effect on beam dynamics is small, therefore, this is a low priority task.

4. The quadrupole voltage monitors were home-made with limited success in achieving an adequate frequency compensation. We now plan to equip every quad plate (32 in total) with a commercially available frequency compensated HV monitor. This will improve the voltage stability readout by an order of magnitude. In addition, we will cross-calibrate the frequency compensation of each monitor with the electric field in the quad region measured using the Kerr effect.

5. Improve the reliability of the HV-vacuum interface regions with a goal of reducing sparking by at least an order of magnitude. The base design is to cover the interface with dielectric shielding capable of holding high electric fields. Alternative design calls for increasing the spacing between positive and negative leads in the air side of the interface or switching to oil-filled air-vacuum interfaces.

6. The outer Q1 plate and support insulators are estimated to have reduced the stored muon population by about 40%. We now plan to address the muon loss issue by a number of alternative modifications which will be discussed below.

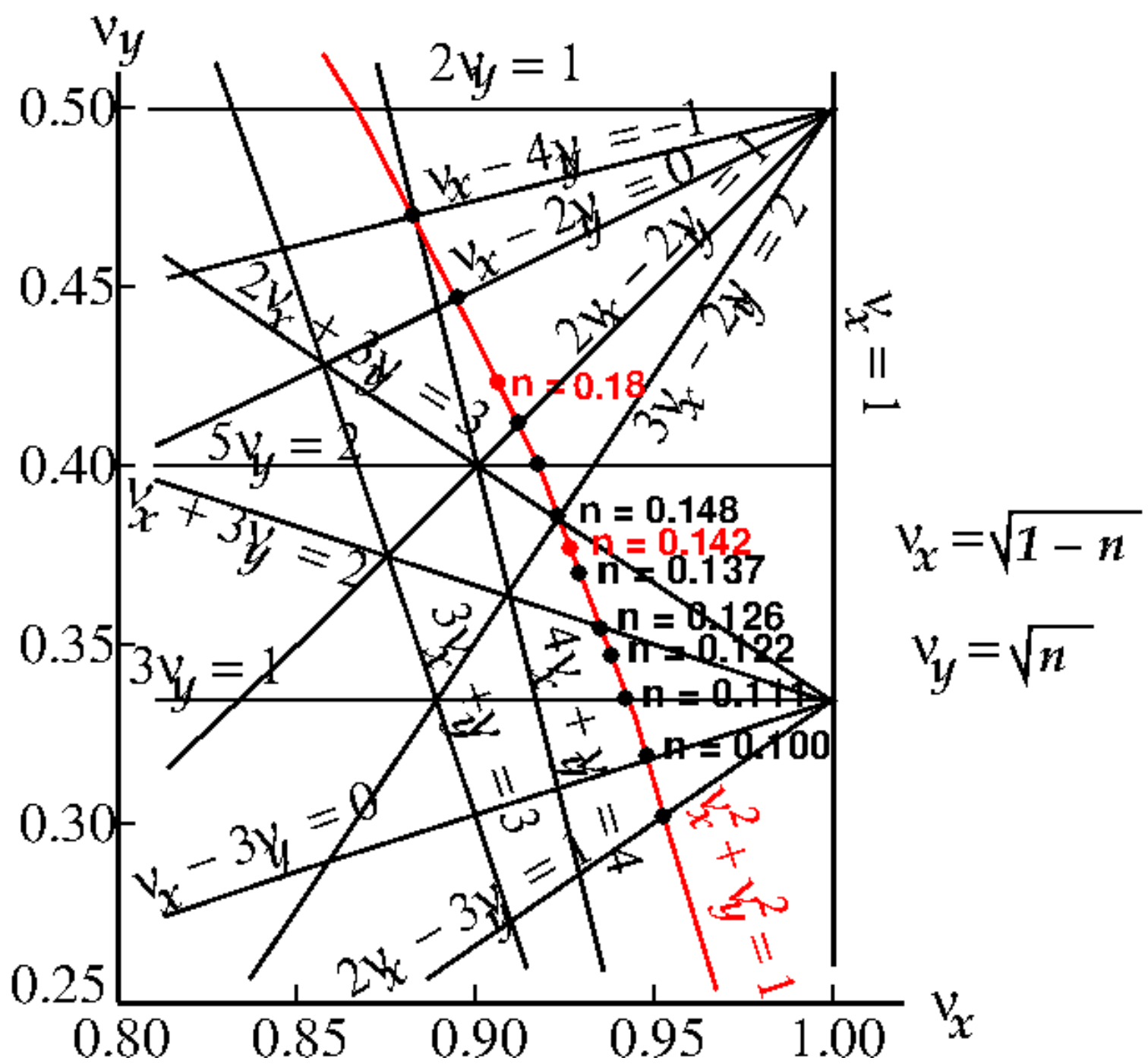

Figure 13.10: The vertical vs. horizontal tune plane together with a number of potential resonance points. The $n-$values $n = 0.142\,; 0.18$ are indicated in red. $n = 0.18$ lies between the resonance lines $\nu_x - 2\nu_y = 0$ and $2\nu_x - 2\nu_y = 1$.

7. Measure the plate vibration during pulsing and stiffen the plate support as needed.

8. Mechanical improvements of ESQ system include

   - increasing the rigidity of quadrupole cages;
   - modification of bolted connections between the parts of quadrupole cages to simplify alignment;
   - improvement of alignment screws (see Figs. 13.4, 13.5);
   - grinding the inner sides of bellow sections to flat to make them more suitable for alignment purposes (see adjustment screws in Fig. 13.9);
   - improving the rigidity of top and bottom quadrupole plates (see Fig. 13.4) against vibrations due to pulsed HV. Preliminary Design studies indicate [9] this is a non-negligible effect, therefore we will continue to investigate this problem and develop a solution. The solution can be as simple as relocating the HV standoffs by 10-20 cm towards the ends of the plates.

All labor-intensive mechanical modifications of the ESQ system, alignment and survey will be performed at Fermilab. Design, R&D studies and prototyping will be pursued at BNL.

Table 13.1: Comparison of high-$n$ and very high-$n$ values.

| **Parameter** | **n=0.142** | **n=0.18** |
|---|---|---|
| horizontal tune, $\nu_x$ | 0.926 | 0.906 |
| vertical tune, $\nu_y$ | 0.377 | 0.425 |
| $f_{\rm CBO}$ | 495 kHz | 634 kHz |
| $f_{\rm CBO}/f_a$ | 2.15 | 2.76 |
| $1/(f_{\rm CBO} - 2f_a)$ | 27$\mu$s | 5.7$\mu$s |
| HV | 25 kV | 32 kV |

## 13.7 Upgrade to Higher $n$-Value Operation

The maximum voltage we used during the muon runs on the ESQ of E821 was 25.4 kV, resulting in a field focusing index of 0.144. We now plan to raise the maximum voltage to 32 kV for a field focusing index of $n \approx 0.18$. We expect that the higher $n$-value will

1) increase the ring admittance and most likely the muon storage efficiency;
2) reduce the muon losses during storage;
3) reduce the coherent betatron oscillation (CBO) systematic error.

We will test the quads up to 35 kV, about 10% higher voltage than the anticipated nominal voltage level.

The main issue in E821 was to be able to hold the high voltage without sparking for about 1 ms. This is a very demanding task, especially for storing negative polarity muons, due to low energy electron trapping in the quad region. We were able to achieve this task by designing the HV feeding leads in a way to quench the low energy electron trapping, see Figs. 13.11, 13.12, 13.13.

Vladimir Tishchenko is the L3 manager for the ESQ system and Yannis Semertzidis was the former L3 manager for the same system. The ESQ system currently consists of 8-chambers, 4-pulser systems, 6-HV-power supplies, and a HV monitoring system. To upgrade the ESQ system to higher operating voltage we will:

- Refurbish the HV pulsers to operate at a maximum voltage of ±35 kV, from the present ±25 kV used in E821.
- The side insulators are all varnished due to the negative muon operation at BNL, see Fig. 13.14. The insulators will be either cleaned or will be replaced by new ones.
- Optimize the ESQ for positive polarity muon storage. The leads will be re-configured to quench the low energy electron trapping more efficiently aiming to achieve higher electric field gradient by 30% compared to E821. Achieving this goal will help eliminate the CBO systematic error as well as substantially reduce muon losses.
- Expand the HV vacuum chamber/air interface tube aiming to significantly reduce the sparking in the vacuum side of the leads.
- Modify the geometry of the HV–vacuum interface to reduce sparking in the air side of the HV lead system or immerse it in oil that can withstand the $E$-field strength.

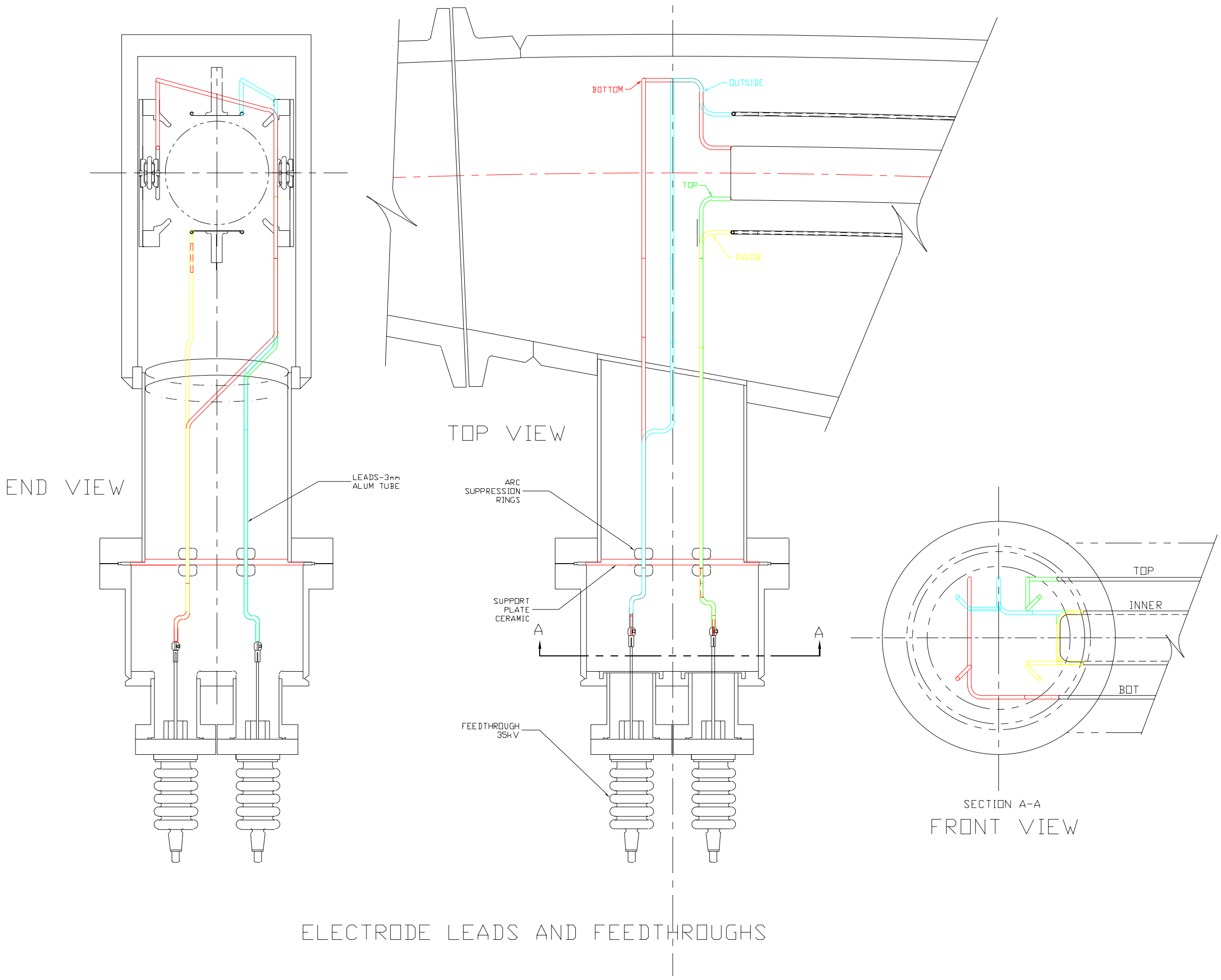

Figure 13.11: Various aspects of the quadrupole high voltage feeding lead geometry, designed to minimize low energy electron trapping.

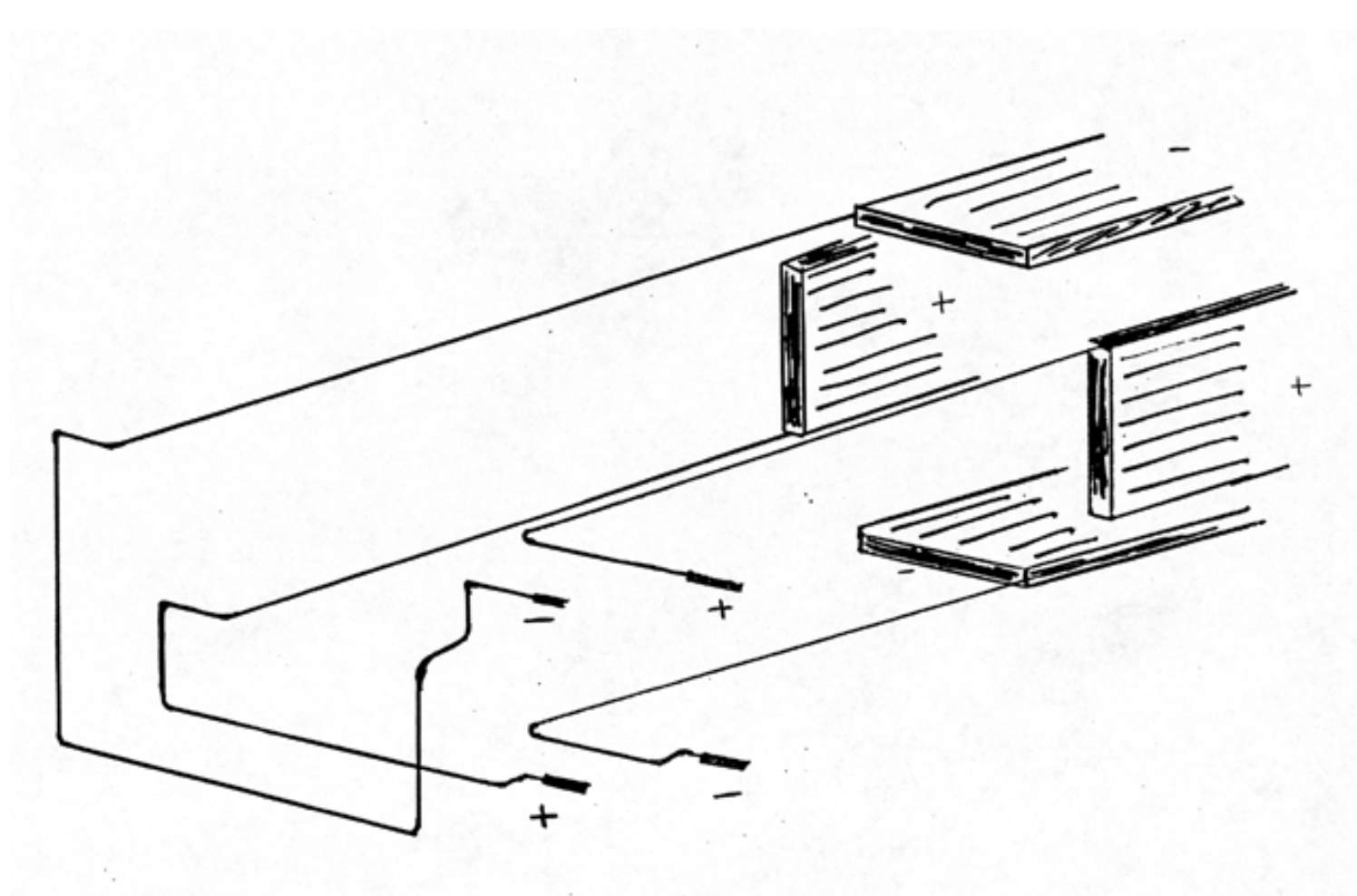

Figure 13.12: Early stages of a hand drawing indicating the high voltage feeding lead geometry.

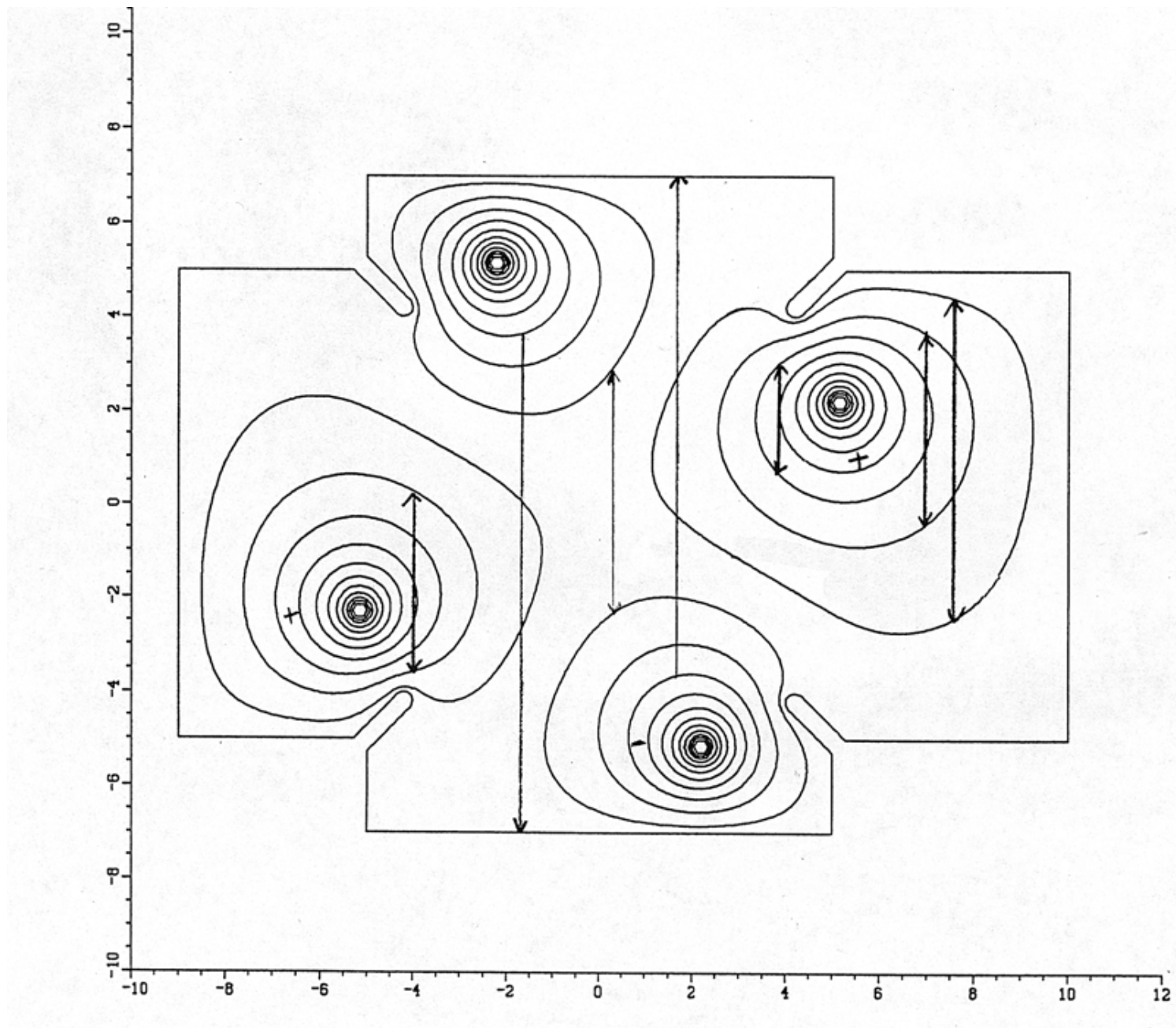

Figure 13.13: A cross section of the lead geometry (vertical [cm] vs. horizontal [cm]). The schematic shows the equipotential lines from an OPERA calculation as well as the low energy electron trapping regions derived from energy conservation. The lead-geometry was designed to optimize the quenching of the electron trapping for the negative muon storage polarity.

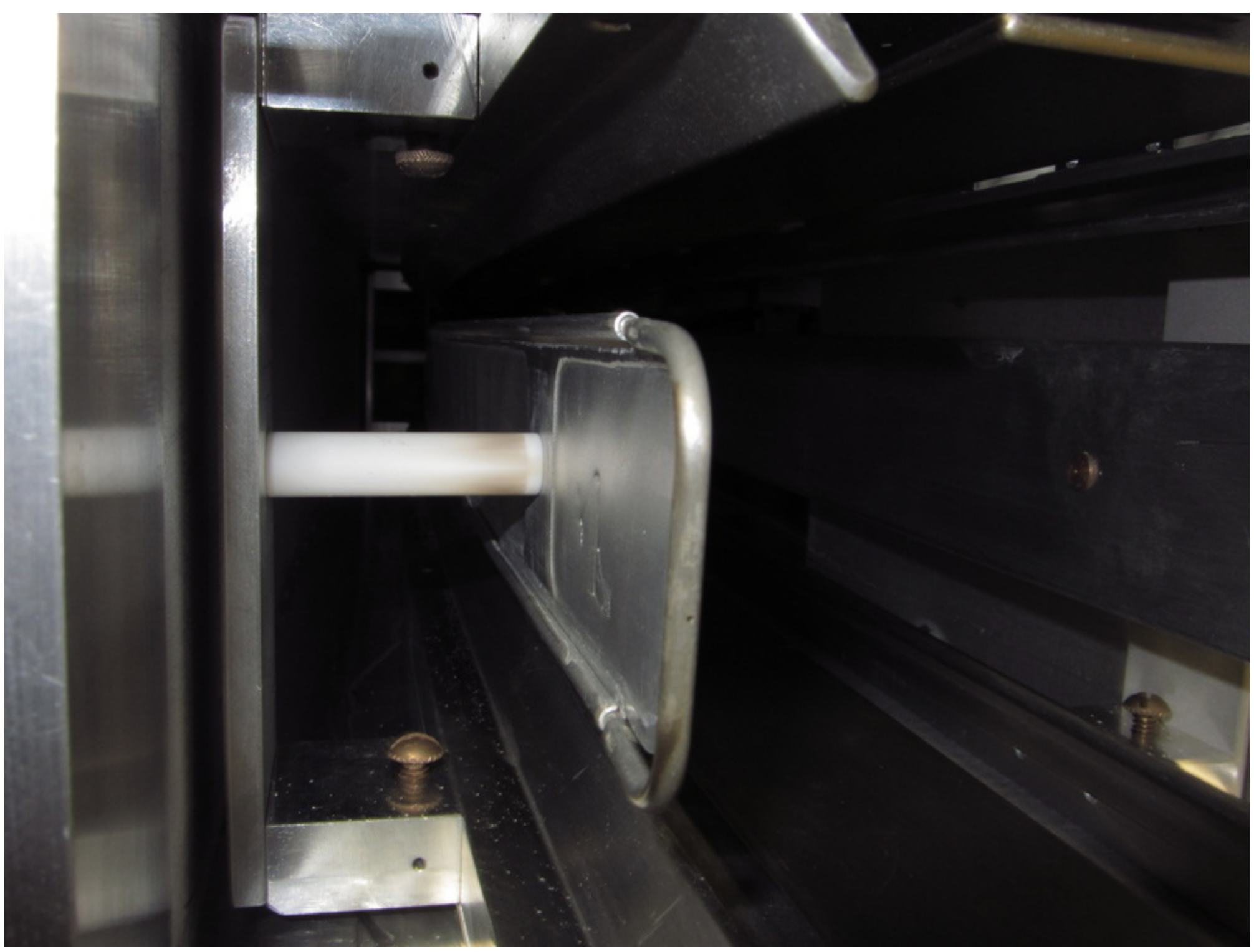

Figure 13.14: The side support insulators are varnished due to trapped electron obstruction during negative muon operations at BNL (darkened appearance close to the plate).

The later is applied routinely in HV applications but it is harder to gain access to it. The sparking rate in the positive polarity in E821 was dominated by sparks at those locations (approximately one spark per 0.5-1 million pulses).

- Shield the electric field generated by the leads from the muon storage region.
- Measure or place strict limits on the magnetic field generated by the trapped electrons.
- Calibrate the pulse shape output of the commercial HV monitors by measuring the electric field generated by the plates using the Kerr effect. The bandwidth (BW) of the Kerr effect measurement is in the GHz range and therefore it is not limited by the level of frequency compensation due to the large capacitance of the components involved.
- Measure the vibration parameters of the quadrupole plates when pulsed using a laser light and a split diode detector. The quad plates can flex under the electromagnetic forces when pulsed. This flexing is (crudely) estimated that it can be of order 10 mm if the pulse duration is of order 1 s. However, for 1 ms the plates can only move by about 10 $\mu$m, much below our specs. We will setup a laser system to measure the plate motion due to the impulse of the electrostatic pulse.

During Preliminary Design the beam dynamics studies were performed to ensure that the new higher-$n$ operating point is located far from major betatron and spin resonances [10, 11]. Ref. [10] suggested $n = 0.185$ for E989 as being safely below the $N = 3$ CBO observational resonance at $n = 0.196$. Analysis of betatron resonances around $n = 0.185$ indicated that in the region between $n = 0.17$ and $n = 0.198$ only $6^{\text{th}}$-order or greater resonances exist which are weak [11]. During adiabatic transitions from scraping to production regimes in E821 the $(g-2)$ storage ring had to cross $\nu_x + 3\nu_y = 2$ resonance (see Fig. 13.10). Resonance crossing lead to partial population of space cleared during scraping. For E989 we will try various scraping voltages in order to find the optimal figure of merit.

In the region $n \approx 0.18$ there is a $k = 0, i = 0, j = -2$ spin resonance at $n = 0.1865$ that would be driven by a radial or longitudinal magnetic field which goes as $x^i y^j \cos(ks/R)$. The analysis showed [11] that field focusing index $n = 0.185$ provides safe operation point.

## 13.8 Upgrade of ESQ Q1

According to `Geant4` simulations [12, 13], the outer plate of quadrupole Q1 and support insulators reduce the fraction of stored muons by about 40%. The baseline plan of addressing the muon losses is to relocate the outer plate of Q1 from $x = -5$ cm to $x = -7$ cm to allow for the uninhibited injection of the muon beam. Fig. 13.15 shows the `OPERA` model of the quadrupole plates in a quadrupole cage. The plate width is adjusted so that only the normal quadrupole field is dominant, and the 20-pole is kept at the 2% level. Every other multipole is below 0.1%, including the sextupole, octupole, etc. Fig. 13.16 shows the current plan for providing a “massless” outer Q1 plate, by placing it outside the muon path. In order to restore an acceptable field quality, the plate voltage also needs to be raised by about a factor of two, see Fig. 13.17. Another parameter we can use to improve the field quality is to work with the plate geometry (width, shape, etc.). The requirement of increasing the voltage by a factor of about two we believe we can achieve in the positive muon polarity and we will test it with the test setup at BNL, and with a magnetic field later on at Fermilab.

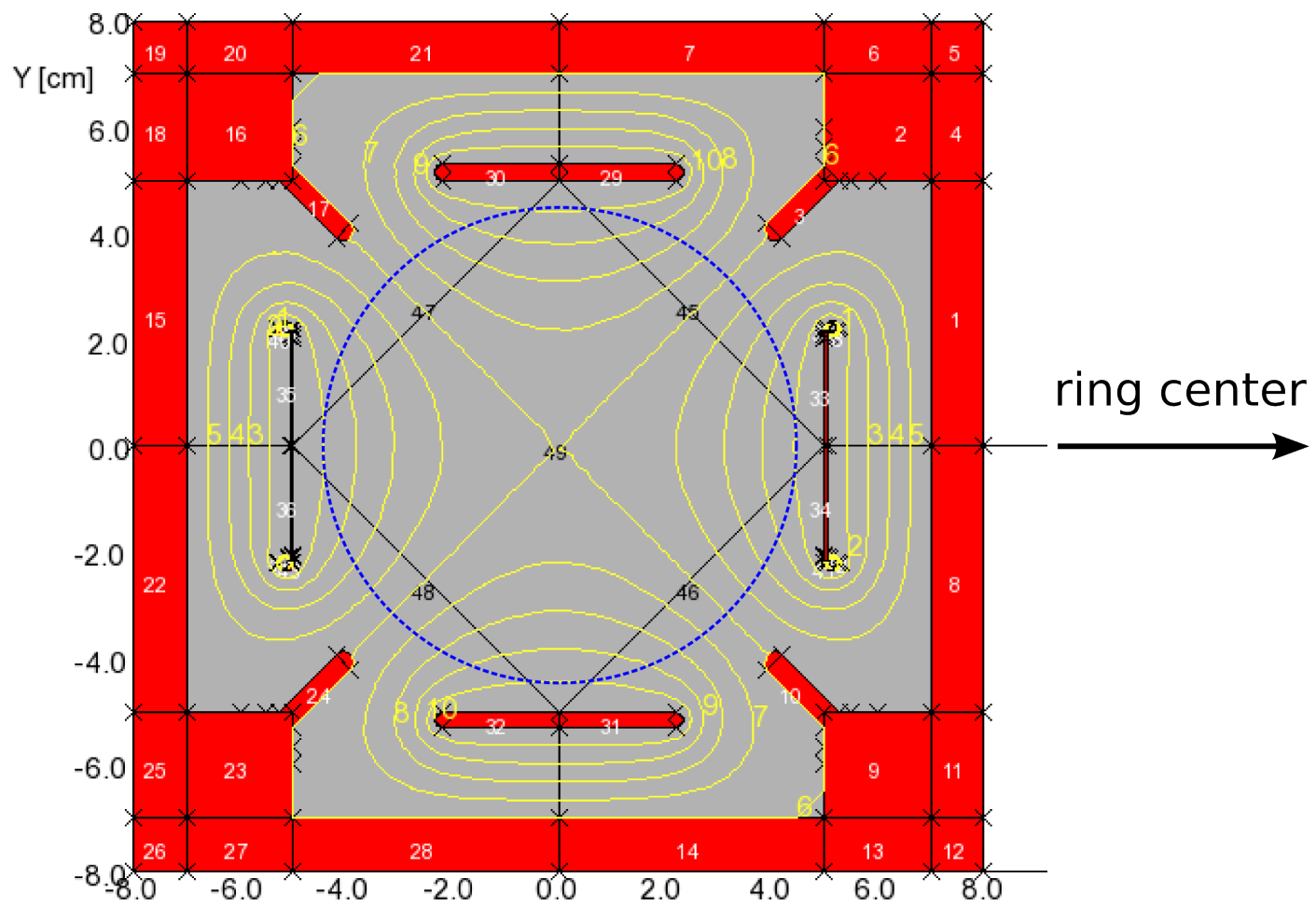

Figure 13.15: An `OPERA` model of the (normal: Q2, Q3, and Q4) electrostatic quadrupole plates. The top/bottom plates are at a positive voltage and the side electrodes are at (the same) negative voltage. The yellow curves represent the equipotential lines. The 90-mm-diameter muon storage region is indicated by the blue dashed circle.

Since a very high voltage (up to 70 kV) is needed to maintain acceptable quality of the quadrupole field, we are planning to operate the Q1 outer plate at a DC voltage. The horizontal scraping will be accomplished by quadrupoles Q2 and Q3.

## 13.9 Coherent Betatron Oscillations Systematic Error

The average position and width of the stored beam can vary as a function of time as the beam alternately focuses and defocuses in the ring. This is the result of a mismatched injection from the beam-line into the $(g-2)$ ring via a narrow line, the so-called inflector magnet. This imposes an additional time structure on the decay time spectrum because the acceptance of the detectors and the $(g-2)$ oscillation phase depends on the position and width of the stored muon ensemble.

The CBO frequency in E821 was close to the second harmonic of $\omega_a$, so the difference frequency $\omega_{\rm CBO}-\omega_a$ was quite close to $\omega_a$, causing interference with the data fitting procedure and thereby causing a significant systematic error (see Chapter 4). This was recognized in analyzing the E821 data set from 2000. In the 2001 running period the electrostatic focusing field index, $n$, was adjusted to minimize this problem. This greatly reduced the CBO systematic uncertainty. We will follow this strategy again but this time we will increase the quad voltage by another 30% to decrease the CBO systematic error by more than a factor of three, see Fig. 13.18.

In addition, the anticipated new kicker pulse shape will better center the beam on orbit. On the detector side, we plan to increase the vertical size of the detectors compared to E821 (from 14 to 15 cm). This reduces the fraction of lost electrons passing above or below the detector, and therefore the sensitivity of the detector acceptance to beam position and width.

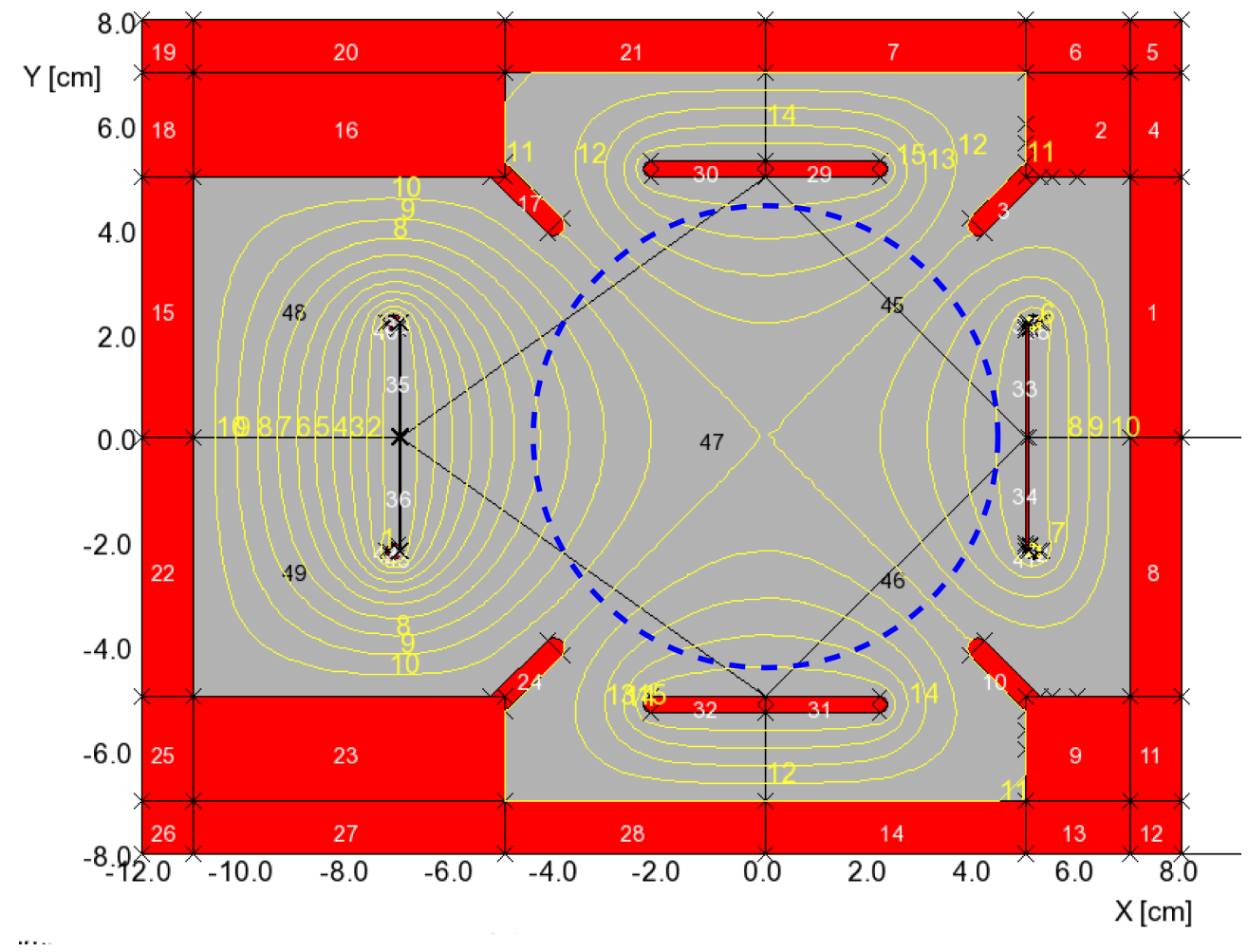

Figure 13.16: An OPERA model of the electrostatic quadrupole plates for Q1. The left plate is displaced to the outside by 2 cm to allow the muons to enter the storage region without having to cross the plates or the support insulators. In order to restore a good field quality (indicated by the symmetric equipotential lines in the center region), the voltage on the left plate is about twice that on the right plate. The 90-mm-diameter muon storage region is indicated by the blue dashed circle.

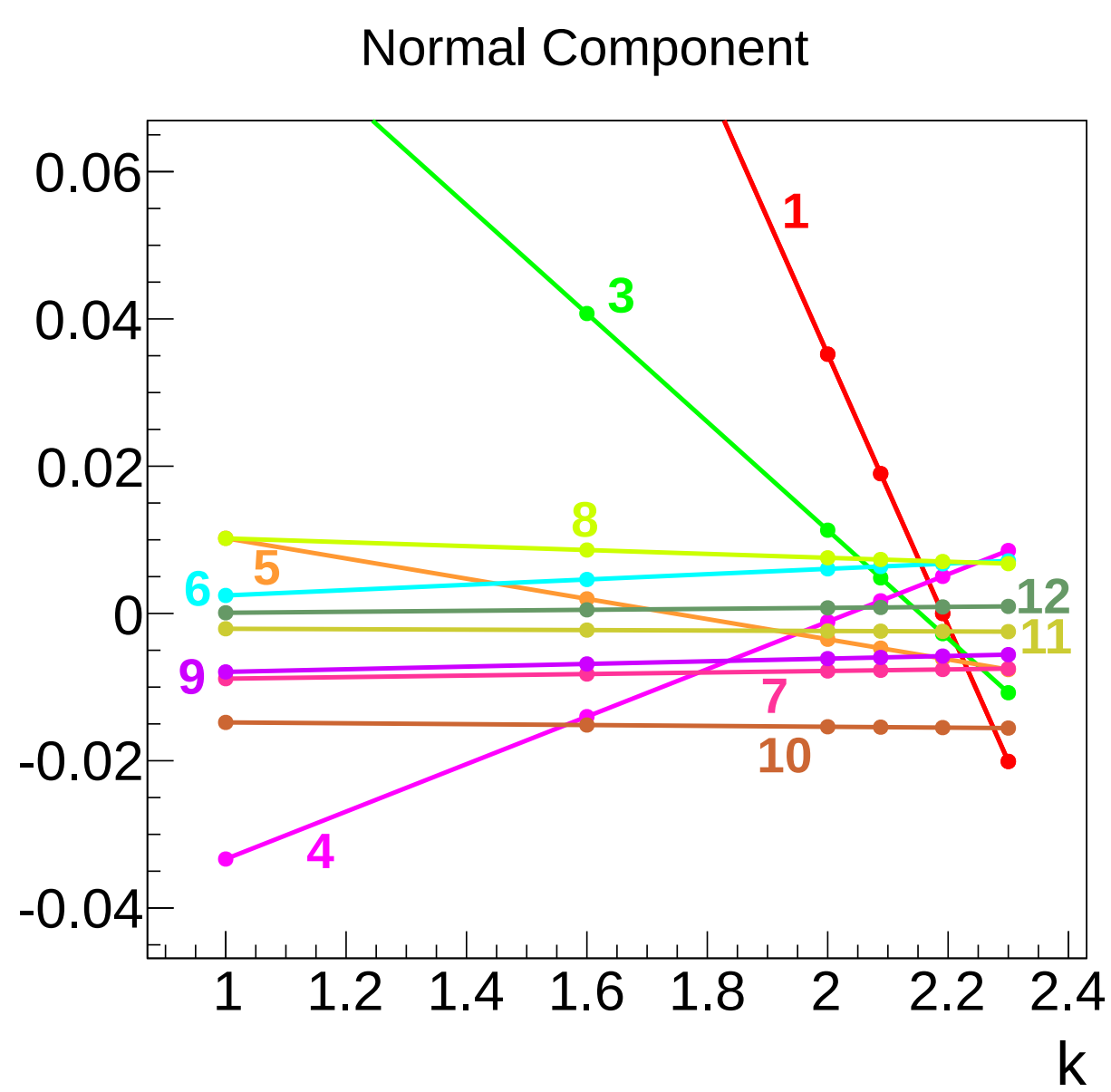

Figure 13.17: Results from OPERA as a function of the voltage multiplication factor for the displaced (outer Q1) plate. Most of the multipoles are below 1% but not all.

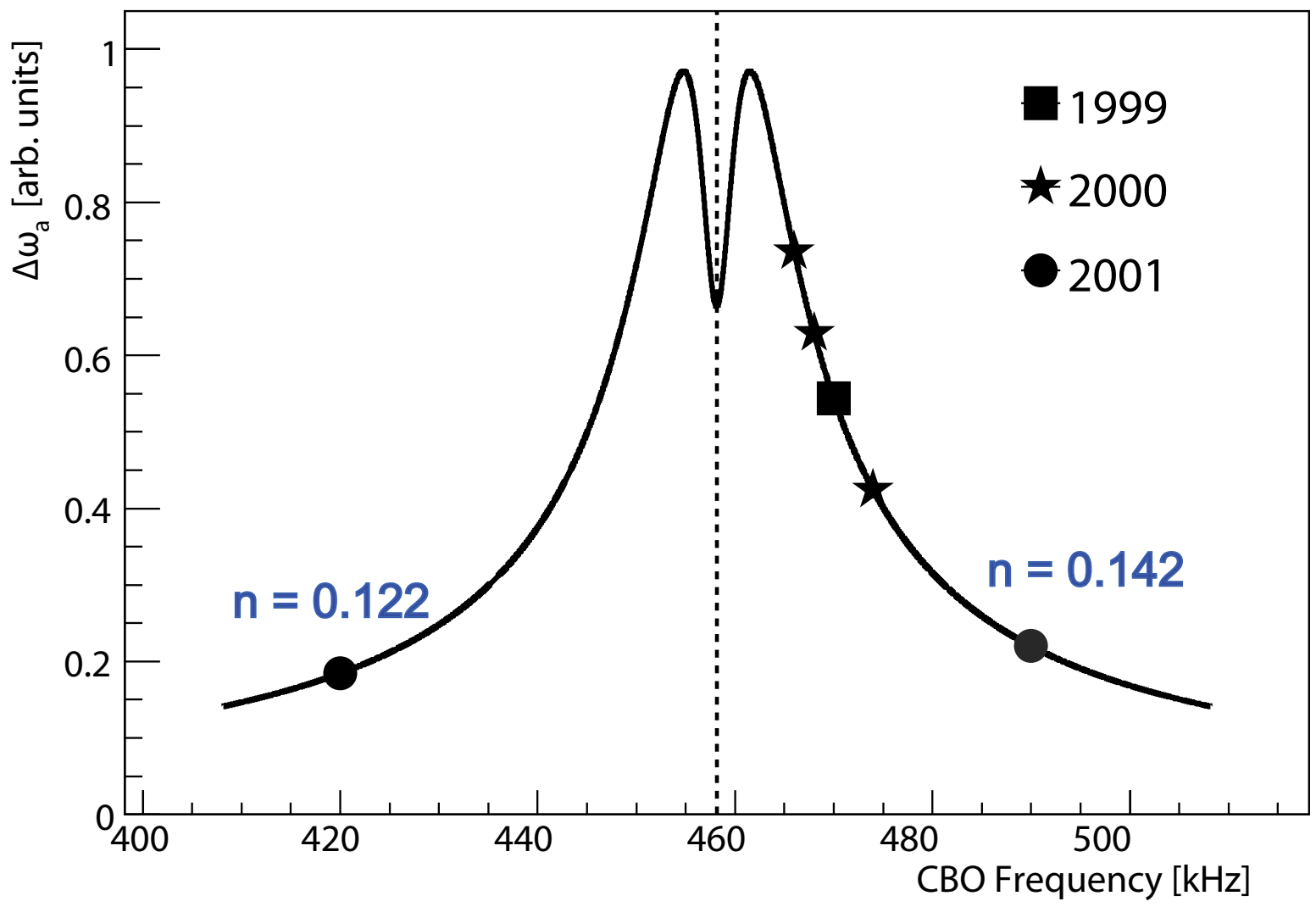

Figure 13.18: The CBO systematic raw error (arbitrary units) as a function of CBO frequency. The year notation indicates the frequencies ran with in E821. For E989 we plan to use much higher field focusing index (see quad section) with a projected CBO frequency of 634 kHz. This frequency will significantly reduce the CBO systematic error.

In an ideal world, where the detector resolution is uniform around the ring, the CBO systematic error averages to zero when all the detected positron pulses are summed up. However, for E821 the kicker plate geometry broke significantly the detector resolution symmetry around the ring resulting to a non-zero average. With the new design we expect to significantly restore this symmetry.

The combined efforts should reduce the CBO uncertainty by at least a factor of four to well below 0.02 ppm. If a new inflector with wider horizontal aperture is used, then it is feasible to eliminate the CBO systematic error to well below our sensitivity level.

## 13.10 Collimators and Lost Muon Systematic Error

The E821 lost muon systematic error was 0.09 ppm. In this section we discuss how we will decrease the lost muon rate with an improved storage ring/collimator system. The distortions of the vertical ($y$) and horizontal ($x$) closed orbits (CO) due to radial ($B_r$) and vertical $B_y$ multipole magnetic field distortions are:

$$\Delta y_{\mathrm{CO}} = \sum_{N=0}^{\infty} \frac{R_0}{B_0} \frac{B_{rN} \cos(N\Theta + \phi_{yN})}{N^2 - \nu_y^2} \tag{13.1}$$

$$\Delta x_{\mathrm{CO}} = \sum_{N=0}^{\infty} \frac{R_0}{B_0} \frac{B_{yN} \cos(N\Theta + \phi_{xN})}{N^2 - \nu_x^2}\,, \tag{13.2}$$

Table 13.2: Distortion of the closed orbits for E821 (FNAL) tune values and $B_{rN}/B_0$ and $B_{yN}/B_0 = 10$ ppm.

| $N$ | $y_{CO}$ (mm) | $x_{CO}$ (mm) |
|---|---|---|
| 0 | 0.53 (0.40) | 0.08 (0.09) |
| 1 | 0.08 (0.09) | 0.53 (0.40) |
| 2 | 0.02 (0.02) | 0.02 (0.02) |
| 3 | 0.01 (0.01) | 0.01 (0.01) |

where $N$ is multipole component, $R_0 = 7112$ mm is equilibrium radius, $B_0 = 1.45$ T is central value of the dipole magnetic field, $\nu_x$ and $\nu_y$ are horizontal and vertical tunes, respectively.

For E821, the average radial magnetic field $B_{r0}$ drifted by typically 40 ppm per month, which was correlated with temperature changes. About once a month $B_{r0}$ was adjusted with the current shims to maximize the number of stored muons, i.e., centering the beam vertically in the collimators. From equ. (13.1) $B_{r0}/B_0 = 40$ ppm changes the vertical closed orbit by 2 mm. At FNAL we plan much better temperature control compared to E821. $B_{y1}/B_0$ was shimmed to $< 20$ ppm, which distorted the horizontal closed orbit by $< 1$ mm. For the FNAL experiment, we want both of these components $< 10$ ppm. Other components are less important since $\nu_y^2 \approx 0.18$ and $\nu_x^2 \approx 0.82$ are closest to the integers 0 and 1, respectively (see Table 13.2). For E821 we used $\nu_y^2 \approx 0.13$ and $\nu_x^2 \approx 0.87$.

The E821 collimators were circular with radius 45 mm. The E821 beta functions vs. ring azimuth are shown in Fig. 13.19. The FNAL experiment collimators will be oval with the $x$ and $y$ axes modulated by the square root of the beta functions, i.e., $\pm 0.8$ mm in $x$ and $\pm 0.7$ mm in $y$.

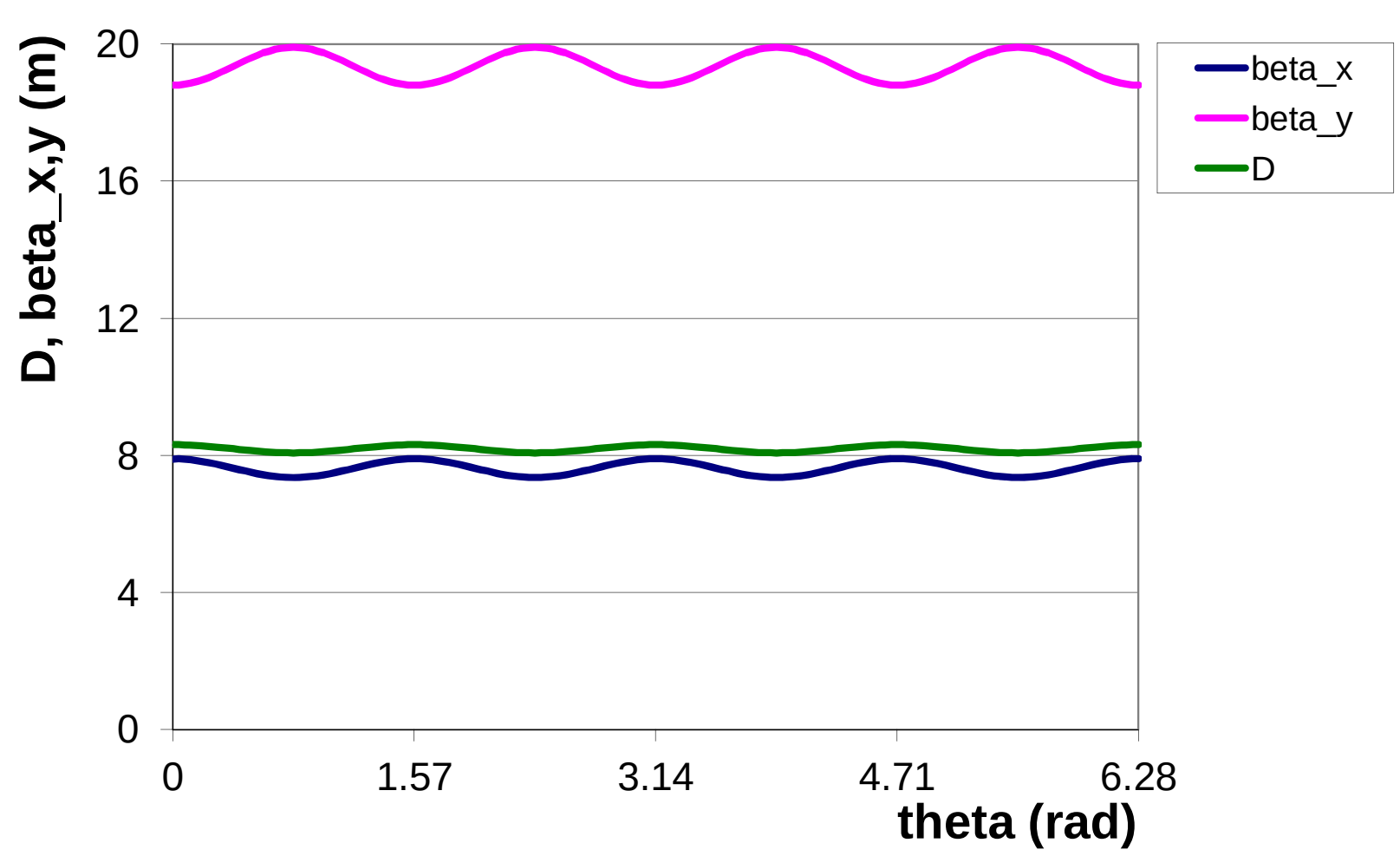

Figure 13.19: E821 horizontal and vertical beta functions.

Fig. 13.8 shows the E821 collimator ring placement. Since the E821 kick extended over many turns, we needed "half" collimators just after the kicker and at $\pi$ radial betatron phase advance, so that the muons would survive enough turns to get the full kick. The FNAL kicker is being designed to give the full kick on the first turn. Thus we can go from 3 full collimators and 5 half-collimators to eight full collimators.

We purposely distorted the vertical and horizontal closed orbits by 2.6 mm during scraping for the first $10^2$ turns, but 2.6 mm was not large compared to the above effects. Indeed, when there were large temperature variations, we sometimes observed that the lost muon rate went *up* after scraping ended(!), which can happen in the presence of a significant radial $B$-field with proper polarity. With better control over the horizontal and vertical orbit distortions due to $B_{r0}$ and $B_{y1}$, oval collimators to match the ring beta functions, and eight full collimators, we anticipate a lost muon rate at FNAL which will be about ten times lower than E821. The exact lost muon rates will be calculated with tracking simulation. The collimator positions should be surveyed to better than 0.3 mm. The coefficient of expansion of steel is $1.3 \times 10^{-5}$/C; multiplying times the radius of 7.1 m gives 0.1 mm/C.

The collimators are able to be put into the "beam position", or into the "trolley position". The latter is required to run the NMR trolley. We will put one collimator into the beam position and record the lost muon rate with the lost muon detector. This takes about ten minutes of data collection. Then we put a second collimator into the beam position. We will have from simulation how much the lost muon rate should decrease with two collimators perfectly aligned with respect to the closed orbit. If we don't observe this decrease, we will remotely position the second collimator in $x$ and $y$ until we achieve the desired result. Then we put in the third collimator, etc.

## 13.11 Alignment and Survey

As it was discussed in the previous section, the collimator positions should be surveyed to better than 0.3 mm. The placement accuracy requirements of ESQ plates in E821 was 0.5 mm for the horizontal (top/bottom) quad electrodes, and 0.75 mm for the vertical (side) quad electrodes. When measured by the surveyors the electrodes were found to be well within those values. To improve the quality of quadrupole focusing field and reduce systematic uncertainties we require that the ESQ plates must be positioned and surveyed to 0.3 mm or better. Based on our experience in E821 we believe that this accuracy level is within the reach of modern survey technologies. Mechanical modifications of quadrupole cages summarized in section 13.6 is an important part of improvements that lead to better alignment accuracy.

## 13.12 Quality of Quadrupole Focusing Field

Due to discrete structure of ESQ, flat geometry of ESQ plates, scalloped vacuum chambers, imperfect placement of ESQ plates discussed in the previous section, etc., the real focusing field of ESQ contains non-quadrupole multipoles. Important considerations of the multipoles of the focusing field include

1. Distortions of the closed orbit which affect

    - $\omega_p$ systematic error;

- Lost Muon systematic error;
- Pitch correction systematic error;
- $E$-field correction systematic error;

2. Betatron resonance excitation which affects the Lost Muon systematic error.

3. Coherent betatron oscillation (CBO) de-coherence, which affects the CBO systematic error.

In order to have acceptable beam dynamics systematic errors, non-quadrupole field multipoles must be kept below appropriate levels. Preliminary specs on ESQ multipoles have been defined by the Preliminary Design studies [14, 15]. Below we give a brief summary of the results.

- **CBO** ESQ field multipoles (presumably the 20-pole) lead to faster CBO de-coherence, which reduces the CBO systematic uncertainty. Potentially one can use this fact to optimize the ESQ field to reduce the CBO systematic error in E989, but more rigorous studies to better understand the effect are under way (see section 4.5.2).

- **Betatron resonance excitation** The closest 20-pole resonance is relatively narrow ($\delta n = \pm 10^{-3}$) and far ($\Delta n \approx 0.02$) from the $n = 0.185$ operating point in E989. Based on systematics measurements in E821, we concluded that this should not be an issue for E989.

- **$E$-field correction** A preliminary systematics study shows a significant (up to $\Delta\omega_a \approx$ 30 ppb) effect only for non-magic muons (with $\Delta p/p_{\rm magic} \approx 0.25\%$) with large betatron oscillations ($A_x = 40$ mm). To estimate the effect in the entire E989 data set a realistic data sample of stored muons will be simulated (see section 4.5.2).

- **Pitch correction** The effect is sizable ($\Delta\omega_a \approx 9$ ppb) only for muons with large betatron oscillations $A_y \sim 40$ mm. Like in the previous case, a better estimate will be made when a sample of stored muons will become available.

- $\boldsymbol{\omega_p}$ The expected distortion of the closed orbit due to quad misalignment is $\Delta x <$ 0.2 mm. The "Muon Distribution" component of the total $\omega_p$ systematic error will be addressed by improving the uniformity of the storage magnetic field (see section 15.3 for more details).

Summarizing, Preliminary Design studies show that the chosen baseline design of this WBS element meets the systematics goals of the experiment. We will continue our studies during the Final Design to finalize the specs on ESQ plate placement accuracy and magnitudes of field multipoles.

## 13.13 ES&H

Potential hazards of the ESQ system are power system and $X$-rays.

The system contains both low voltage, high voltage (up to 75 kV) and high current circuits. There are no exposed electrical terminals. All electrical connections are bolted and enclosed. Cables will either run along the floor in a cable tray or in a double-grounded conduit. The power supplies and the thyratrons are fused. We will use lock out/tag out when servicing the unit. When the power supplies are disabled, the storage capacitors will also be shorted to ground with a safety relay. We do not anticipate that we will need to work on the unit hot. There are no requirements for emergency power. There will be a remote control unit in the control room. The operation of ESQ will be limited to system experts and trained personnel.

Soft $X$-rays can be produced in the system in spark discharges. Even though the ESQ system is designed to have no sparks during normal running conditions, sparks are most likely to occur during conditioning of the system. Aluminum vacuum chambers with 1-cm-thick walls provided adequate shielding against X-rays in E821. Due to higher operating voltage in E989 the shielding by vacuum chambers may not be sufficient. We are planning to develop an integrated X-ray safety plan together with the kicker group.

One of the alternative designs of the outer plate of Q1 quadrupole includes beryllium foil. Beryllium is ideal material for such purpose due to its mechanical, electrical and magnetic characteristics. Most importantly, muon scattering in beryllium will be significantly reduced in comparison with aluminum plate due to lower $Z$ of beryllium. However, beryllium is a well-known health hazard. We are not planning to machine beryllium in the Lab. The foil will be produced and, presumably, assembled by a certified commercial company. During running the beryllium plate will be enclosed in vacuum chamber inaccessible to regular personnel. Only certified personnel will be allowed to perform work on a modified ESQ Q1.

The ESQ design will be reviewed by the (PPD or AD) electrical safety committee. Proper Operational Readiness Clearance will be obtained before unattended operation of the systems. Job Hazard Analyses will be performed for any work tasks that involve working on the high voltage systems.

## 13.14 Risks

The baseline design is to displace the Q1 outer plate by about 2 cm (the needed displacement will be determined more accurately by R&D studies). If the baseline design cannot be achieved for various reasons, we will consider the following alternatives for Q1 outer plate, *i)* a plate made from a thin beryllium foil, *ii)* a plate made from a thin wire mesh, *iii)* a plate made from a thinner aluminum foil, *iv)* a plate from other alternative materials (e.g. fiber carbon). This will lead to the following consequences to the Project

- More effort will be needed for R&D studies of alternatives.
- Muon scattering in any material will reduce the fraction of stored muons and hence increase the time required to reach the statistical goal of the experiment. The preferable material is beryllium.
- Beryllium foil will increase the cost of the Project. The cost of the beryllium material for the plate is about $16 k per meter. Thus, to cover a 5-m-long quadrupole plate at

least \$90 k in addition will be required not including the manufacturing and assembling expenses.

- Beryllium is a hazardous material. Special handling requirements will complicate ESQ plate installation and adjustment procedure.

The baseline design is to increase the operating voltage of ESQ to $\pm 35$ kV. The CBO systematic error will be more challenging to address if this goal is not reached. This will also increase muon losses.

The ESQ system requires good vacuum to operate properly ($10^{-6}$ Torr or better). Bad vacuum conditions may lead to inability of ESQ to operate at nominal voltage. One potential source of vacuum leak is the tracker system. If the leak is too large, additional vacuum pumps may be needed to pump the vacuum chambers equipped by the tracker system. If high vacuum conditions are not met with installation of additional vacuum pumps, we may consider taking production data without tracker system and taking special runs with the tracker system to measure the distribution of muons in the storage ring. The disadvantage of such a mode of operation is that the tracker runs will be excluded from the production dataset.

## 13.15 Quality Assurance

Reliable operation of the quadrupole system is necessary to achieve the experiment's goals. We have planned a testing program that includes computer simulations and extensive hardware testing of the ESQ system in advance to installation into the experiment to insure reliability, and this is accounted for in the cost and schedule estimation.

BNL has established a test stand to assess performance of the ESQ system. The test stand will include vacuum system, high voltage electrical system, high voltage monitors, electrooptic high-voltage system and the data acquisition system. It has already been or will be used to

1. Study the stability of the high voltage with and without magnetic field by pulsing the plates 10% above nominal voltage.

2. Study the mechanical stability of the quadrupole plates under high voltage stress.

3. Perform R&D studies of the Q1 outer plate.

4. Perform R&D studies of high voltage leads.

5. Test the procedure of conditioning the ESQ system.

6. Measure the X-rays exposure level due to sparking.

7. Develop and test the data acquisition system.

We are planning to install the ESQ system a year in advance of the start of the experiment. This will allow us to test the system in real experimental environment and will give us sufficient time to make alternation if necessary without delaying the schedule of the experiment.

To assure the quality of the future experimental data and to identify potential problems we will continue doing precision computer simulations of two types, `OPERA` simulations of the electric field produced by both quadrupole plates and high voltage leads, and tracking simulations of muons in the electric and magnetic field using `Geant4` and/or independent dedicated tracking program developed by Y. Semertzidis for E821. The computer simulations are backed up by analytic calculations where possible.

## 13.16 Value Management

The reference design is lower cost than other alternatives we have considered (see discussion above) and this is the design we will use, provided it meets the requirements. The design process has benefitted from the experience gained in E821.

The baseline design is to re-use the existing E821 electrostatic focusing quadrupoles. Some components require cleaning and refurbishing. To meet the statistics goal of the E989 experiment and maximize the number of stored muons we are planning to upgrade the outer plate of the quadrupole Q1. To meet the systematics goals of E989 we are planning to modify some components of the ESQ system (improve rigidity of ESQ cages to meet new requirements on alignment precision, redesign high voltage leads to provide the electrostatic field of better quality, upgrade some components of the high voltage power system to enable operation at higher field focusing index, etc.). Where possible, the upgrade will reuse the existing components from E821.

We are planning to re-use the existing Boston waveform digitizer electronics used in muon lifetime measurements by the MuLan collaboration [16]. The digitizers will be used to record HV traces from each quadruple plate (32 channels total).

## 13.17 R&D

Work is well underway on R&D studies of quadrupole Q1. The `Geant4` simulations conducted independently by N.S. Froemming [12] and T. Gadfort [13] were important in guiding the choice of material for the outer plate of Q1. The preliminary `OPERA` simulations were important in making the choice of the baseline design of quadrupole Q1 (Fig. 13.16). The tracking simulations were important in understanding the muon beam dynamics in $(g-2)$ storage ring with skew and upright quadrupoles [17, 18]. More precision computer simulations will be conducted to finalize required tolerances and to quantify systematic uncertainties related to ESQ system. Extensive tests of a prototype of quadrupole Q1 will be conducted in a test stand at BNL.

# Chapter 14

# Ring Instrumentation and Controls

This chapter describes the technical design of the g-2 cryogenic and vacuum control system and other process systems supporting the experiment. This control system will be a copy of the typical Siemens S7-400 PLC (Programmable Logic Controller) control system as deployed by the Fermilab PPD/Mechanical Department. The g-2 cryogenic and vacuum system will be located on the Muon campus in the MC1 building and Muon g-2 experimental hall. This area is classified as ODH Class 0 area and has several large cryogenic and gas components. Cryogens include Liquid Helium and Liquid Nitrogen. This cryogenic system has approximately 600 electronic input sensing devices and 100 output devices. Input devices include temperature sensors, pressure transmitters, vacuum gages, level probes, and strain gages. Output devices include solenoid valves, control valves, and vacuum valves and pumps. All electronic and electrical control system equipment is air cooled and does not require any forced air cooling or water cooling. Cabinet air vents are provided for certain devices where appropriate. The control system equipment components are all commercially available products which are UL listed. The cryogenic control system has been designed and will be built following all the required rules and standards such as the NEC and NFPA 70E. All premises wiring is to be installed by Fermi Electrical contractors and licensed electricians.

## 14.1 Cryogenic/Vacuum Control System

The G-2 process controls also known as the slow controls will have a Siemens S7-400 PLC with S7-300 associated I/O modules as the master control system. There will be sub systems controlled by other PLCs that will report to the Master PLC. The Master PLC will also provide system data and interlocks to these subordinate PLC systems. The DAQ control system known as MIDAS will also be provided data and interlocks from this S7-400 master PLC. See the Control System Architecture drawing shown in figure 14.1.

### 14.1.1 Piping and Instrumentation Diagram

The piping and instrumentation diagram is shown in figure 14.2.

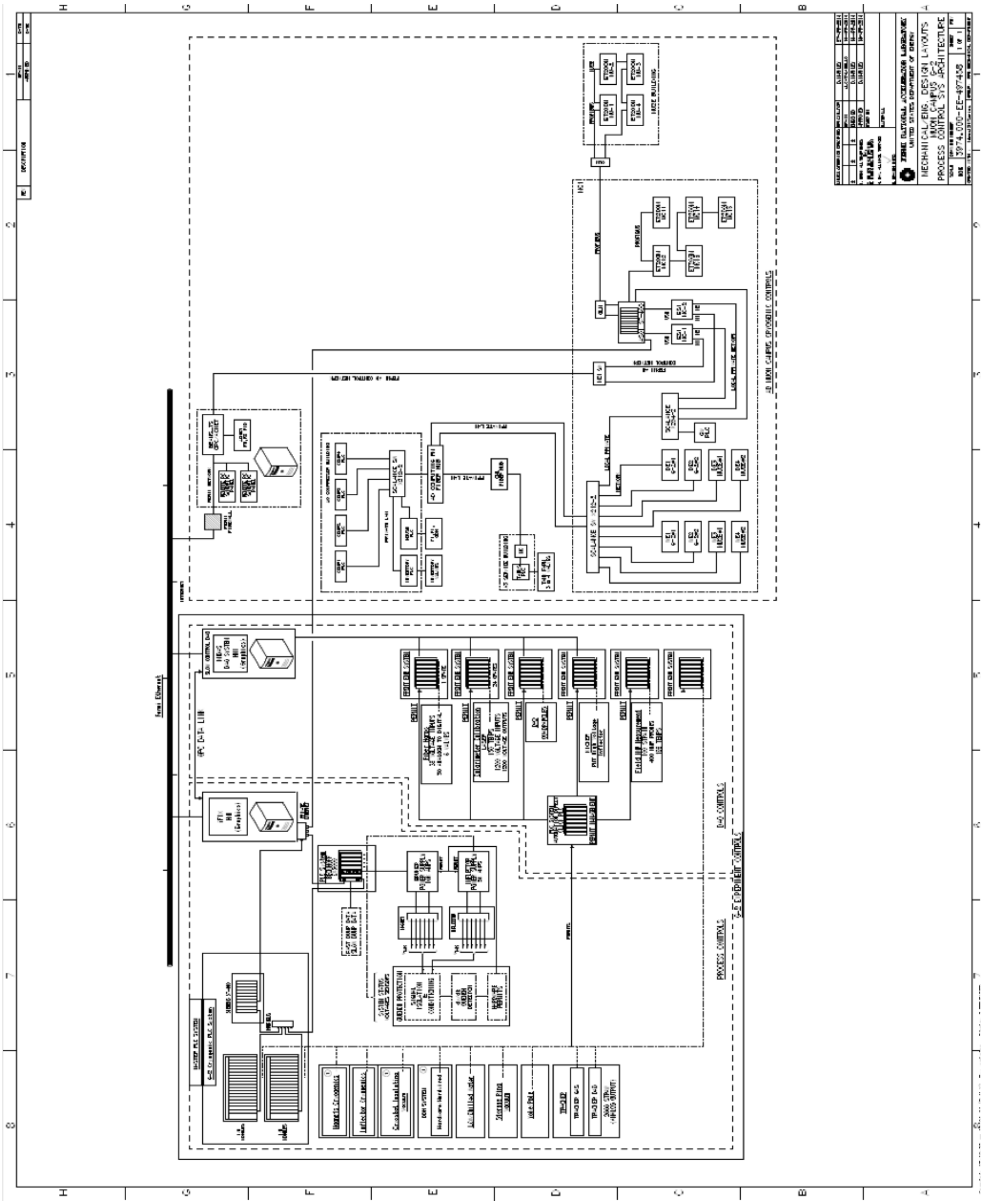

Figure 14.1: G-2 Controls System Architecture.

### 14.1.2 Programmable Logic Controller

The g-2 cryogenic/vacuum system will be controlled by a Siemens S7-400 PLC with S7-300 associated I/O modules (or equivalent industrial controls system) networked on a Profibus network. This PLC system will be programmed using the Siemens S7 engineering programming software (or equivalent software meeting IEC 61131-3 standard). Siemens S7-400 PLC systems are currently in use at several Fermilab projects: LAPD, LBNE 35 Ton, Super CDMS, Microboone, and NML/CMTF.

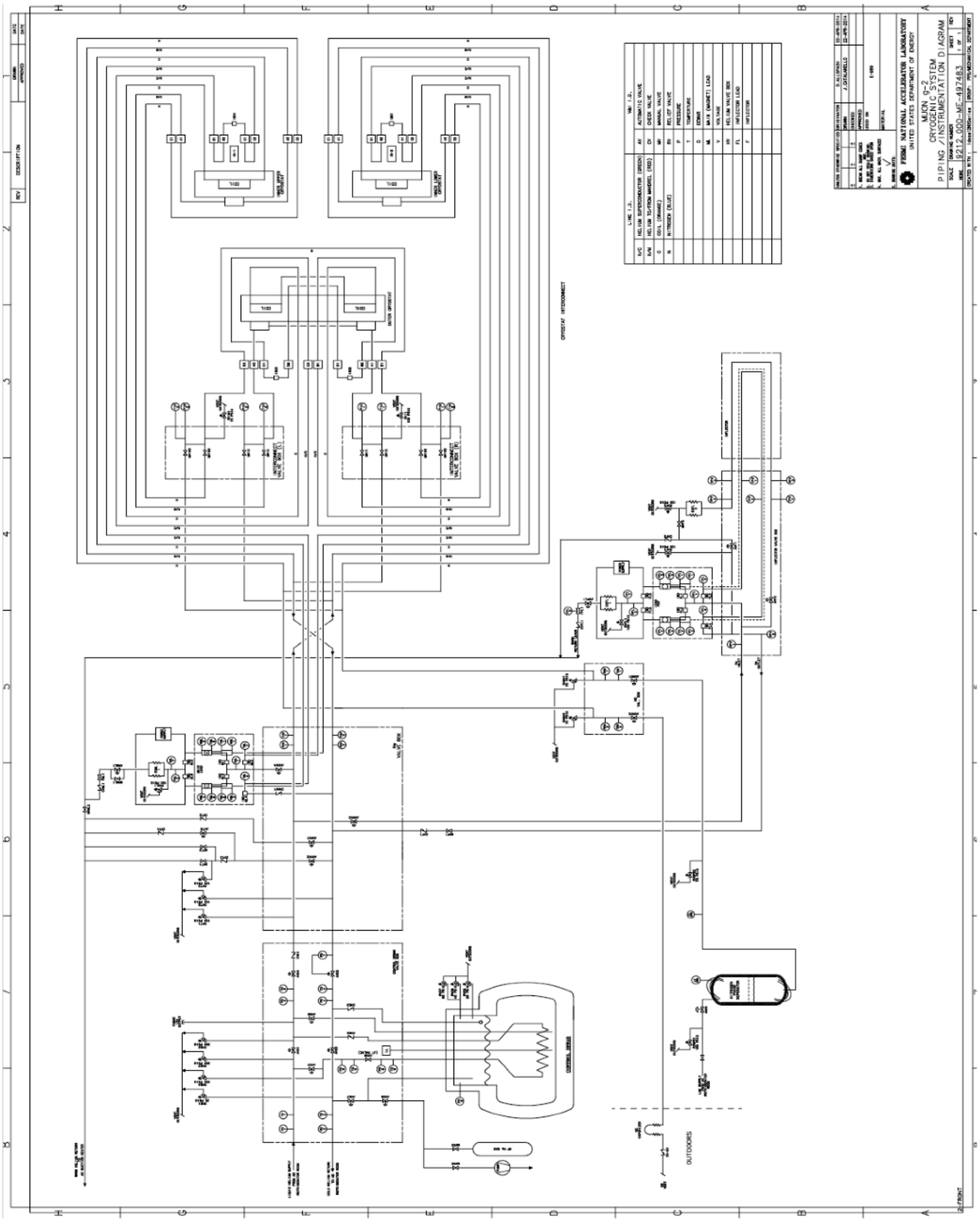

Figure 14.2: The G-2 Piping and Instrumentation Diagram.

### 14.1.3 Instrumentation

The g-2 cryogenic/vacuum system instrumentation will consist of commercial transducers, transmitters, valves, positioners, strain gauges, and thermometers. These commercial devices will have conventional signals that conform to the normal PLC module input and output signal ranges. The locations of these devices are given in reference [1]. PSIG, PSIA, and differential pressures will be measured by commercial transmitters that provide a 4-20 mA signal when available. A voltage range of 0-10 VDC will be an alternate range. Temperature will be measured by a Platinum 100 ohm RTD when possible, because the PLC system can read them directly. There are a number of Silicon diodes in the detector which will be measured by a Lakeshore temperature transmitter that provides a 4-20mA output to the PLC system. Any Cernox RTDs will also be measured with a Lakeshore temperature transmitter that provides a 4-20 mA signal to the PLC.

### 14.1.4 Strain Gauges

G-2 has strain gauges (full bridge style) mounted on the straps and radial stops, which are inside the cryostats and so are at cryogenic temperatures. Room temperature strain gauges are also mounted on the outer cryostat pushrods. Strain gauges will be used during commissioning to make sure that the forces agree with expectation. It is not necessary to read out all these strain gauges during regular experiment operation.

All strain gauges (4 wires per gauge) will be connected to a patch panel in the control room. A select subset of strain gauges will be read out with 2 Vishay D4 units, which will be connected to a Windows PC via USB.[1] Each Vishay D4 unit can read out 4 full-bridge strain gauges. A full description can be found in reference [2].

### 14.1.5 Vacuum Pump Control

Vacuum Pump stations are used to maintain various vacuum systems. These stations can be controlled by the Master PLC system using discrete I/O. Each pumping station also has isolation control valves. The vacuum pumps for the ring cryostats and beam vacuum chambers are described in tables 14.1.5 and 14.1.5.

The pump station remote control interface is shown in figure 14.3.

### 14.1.6 Programmable Logic Controller Input/Output

The g-2 cryogenic/vacuum system will be controlled by a Siemens S7-400 PLC with S7-300 associated I/O modules (or equivalent industrial controls system) networked on a Profibus network. The I/O modules convert digital PLC values to the field instrumentation and visa-versa. Table 14.1.6 summarizes the input/outputs.

[1] Any strain gauge can be read out by manually plugging it into one of the Vishay D4 units.

Table 14.1: Summary of vacuum pumps for the ring cryostats. The Turbo Molecular Pumps (TMPs) pump helium and air to maintain the cryostat insulating vacuum pressure at less than 1E-4 Torr.

| Pump Type | Location |
|---|---|
| Roots Blower | Lead Pot 1 |
| Backup to Roots Blower | Upper 11 Position |
| TMP | Upper Lead Pot 2 |
| TMP | Upper 7 |
| TMP | Lower 7 |
| TMP | Lower 11 |
| TMP | Outer 7 |
| TMP | Outer 11 |
| TMP | Lower 3 |

Table 14.2: Summary of vacuum pumps for the ring beam vacuum chambers. The Turbo Molecular Pumps (TMPs) pump helium and air to maintain the cryostat insulating vacuum pressure at less than 1E-7 Torr.

| Pump Type | Location |
|---|---|
| TMP | Chamber 6 |
| TMP | Chamber 12 |
| TMP | Chamber 8 (contains straw tracker) |
| TMP | Chamber 10 (contains straw tracker) |
| TMP | Chamber 2 (contains trolley drive) |
| Cryo Pump | Chamber 3 |
| Cryo Pump | Chamber 9 |

### 14.1.7 Human Machine Interface

Human Machine Interface (HMI) controls will be provided through GEFANUC's iFIX software. iFIX connects to the S7-400 through Private Ethernet using an Industrial Gateway Server (IGS) driver included with the iFix software. iFIX will handle all operator security, computer alarming, and remote operator controls via the PPD-iFIX server. iFIX will also provide historical data through the PPD-iFIX historian. This historical data will be viewable in iFIX picture displays or on the web through the iFIX Proficy portal server. An example of an HMI for the LAPD experiment is shown in figure 14.4.

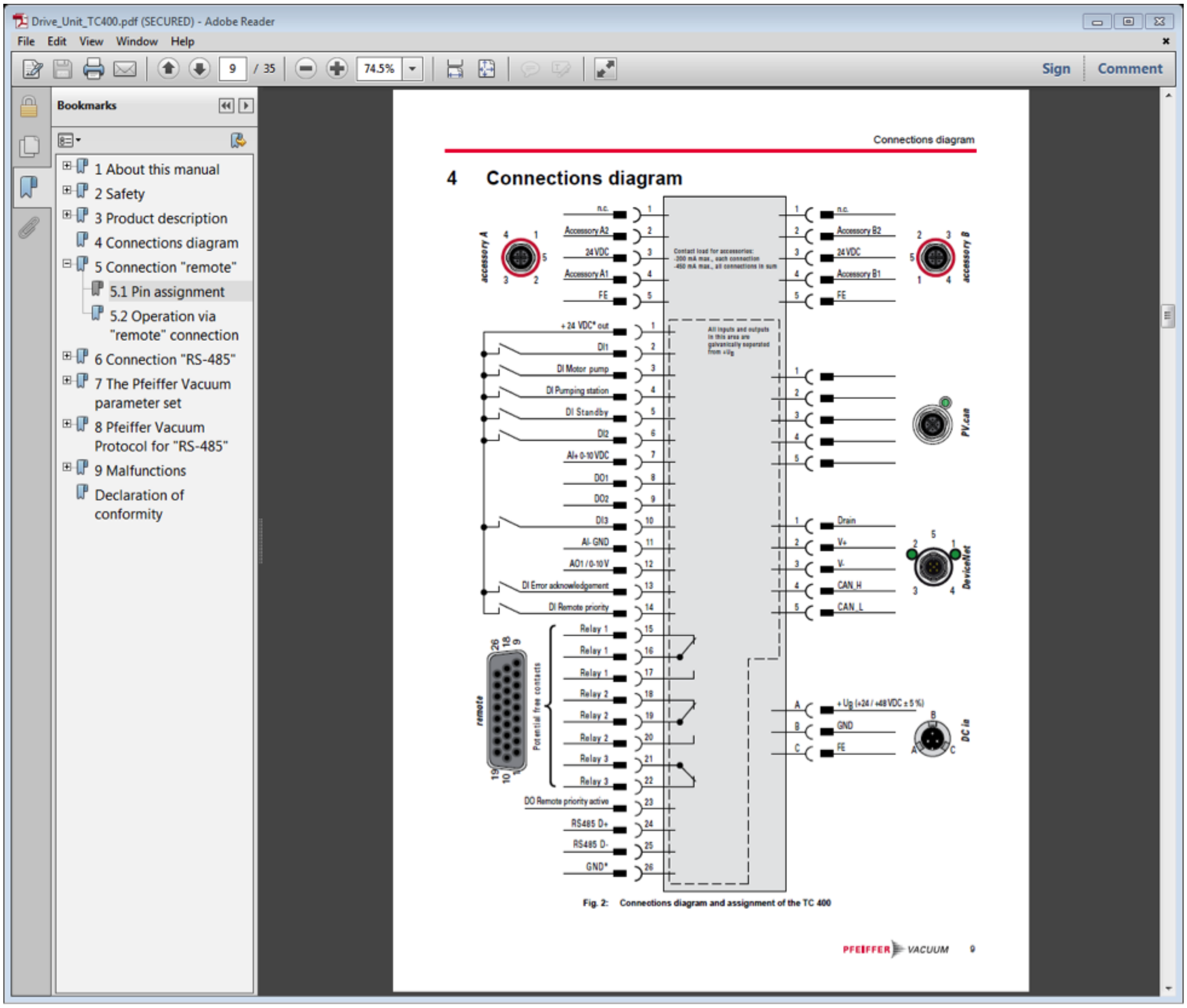

Figure 14.3: The Pump Station Remote Control Interface.

### 14.1.8 Helium Refrigerator Controls

The Helium liquefier system controls were designed and built by the Accelerator Division cryogenic department. This control system also consists of a Siemens S7-400 PLC as the core controller. This system will easily interface with the G-2 experiment master Siemens S7-400. The AD cryo S7-400 and the G-2 S7-400 PLC will share a private network. This private network will allow PLC variables to be shared in either direction and also allow the iFIX HMI to control the Helium refrigerator components.

### 14.1.9 System Communication and Data Sharing

Data will need to be shared between multiple systems outside of the Siemens master PLC system. There are a number of methods and protocols that allow data sharing such as OPC, MODBUS TCP/IP, SQL, and others. The Helium refrigerator communication will be through the private network that links the G-2 Master S7-400 PLC and the AD cryogenic control system S7-400 PLC. The DAQ system HMI is known as MIDAS. The most likely communication path between the G-2 Master Siemens S7-400 PLC and MIDAS will be OPC over Ethernet, but SQL is also possible. The magnet DC control system will be run by a Beckhoff CX5000 PLC. The Beckhoff CX 5000 PLC will be linked to the G-2 Master Siemens S7-400 PLC through PROFINET which is an industrial protocol supported by both systems

Table 14.3: Summary of types of input/outputs connections to the PLC

| Subsystem | 4-20mA 0-10 VDC Analog inputs | Silicon Diode inputs | RTD (PT100's) inputs | Discrete (digital) 24 VDC inputs | Discrete (digital) 24VDC outputs | 24 VDC Analog outputs |
|---|---|---|---|---|---|---|
| Magnet Cryo | 23 | 68 | 10 | 20 | 8 | 7 |
| Inflector Cryo | 10 | 10 | 10 | 10 | 10 | 4 |
| Cryostat Vacuum | 10 | 0 | 10 | 40 | 40 | 2 |
| ODH | 10 | 0 | 2 | 10 | 6 | 2 |
| LCW/Chilled Water | 10 | 0 | 10 | 10 | 10 | 4 |
| Storage Ring Vacuum | 7 | 0 | 7 | 30 | 30 | 4 |
| Yoke/Pole | 10 | 0 | 10 | 10 | 10 | 10 |
| Tracker | 10 | 0 | 10 | 10 | 10 | 10 |
| DC System | 10 | 0 | 10 | 10 | 10 | 4 |
| Subtotal | 100 | 78 | 79 | 150 | 134 | 47 |
| Total | 588 | | | | | |

## 14.2 Life Safety and System Reliability

### ODH System

The ODH system will utilize six MSA O2 heads. Two O2 heads will be located near the ceiling of the g-2 experimental hall, with another four O2 sensors located near the floor of the hall. There will be an ODH warning horn and strobe lamp. These will be centrally located in the hall. There will be two ventilation fans used to maintain the ODH risk class zero status in the g-2 hall. One fan will exhaust air out of the g-2 hall at the ceiling venting it outside. The second fan will supply fresh air to the building near the floor outside of the rings. These fans are controlled by the S7 PLC and can also be run locally using a switch mounted at the fan controls. The ODH system is hardwired to both fans such that during an ODH alarm both fans run. The O2 Sensors are MSA model A-UltimaX-PL-A-14-03D2-0000-100 and have a span of 0-25%. Each O2 sensor is to be wired to an MSA electronic controller which provides an analog output signal wired to the S7 PLC. This MSA electronic unit also provides relays which have three O2 level alarms thresholds, 18.5%, 18%, and 17.5%. The relay output that is set at 18.5% is wired directly to the ODH warning horns and strobe lamps located in MC1 and FIRUS. The MSA electronic unit also provides a trouble relay output which is also wired to the PLC and FIRUS. The trouble output is wired in a failsafe manner, such that loss of power or blown fuse to the ODH controls will generate a trouble alarm. The MSA equipment is wired directly to its own self-contained control circuitry in its own enclosure. This self-contained enclosure has its own power supply which is independent of the PLC control system, allowing the ODH system to function independently of the PLC control system. The power for this ODH system comes from a U.P.S. (Uninterruptible Power Supply).

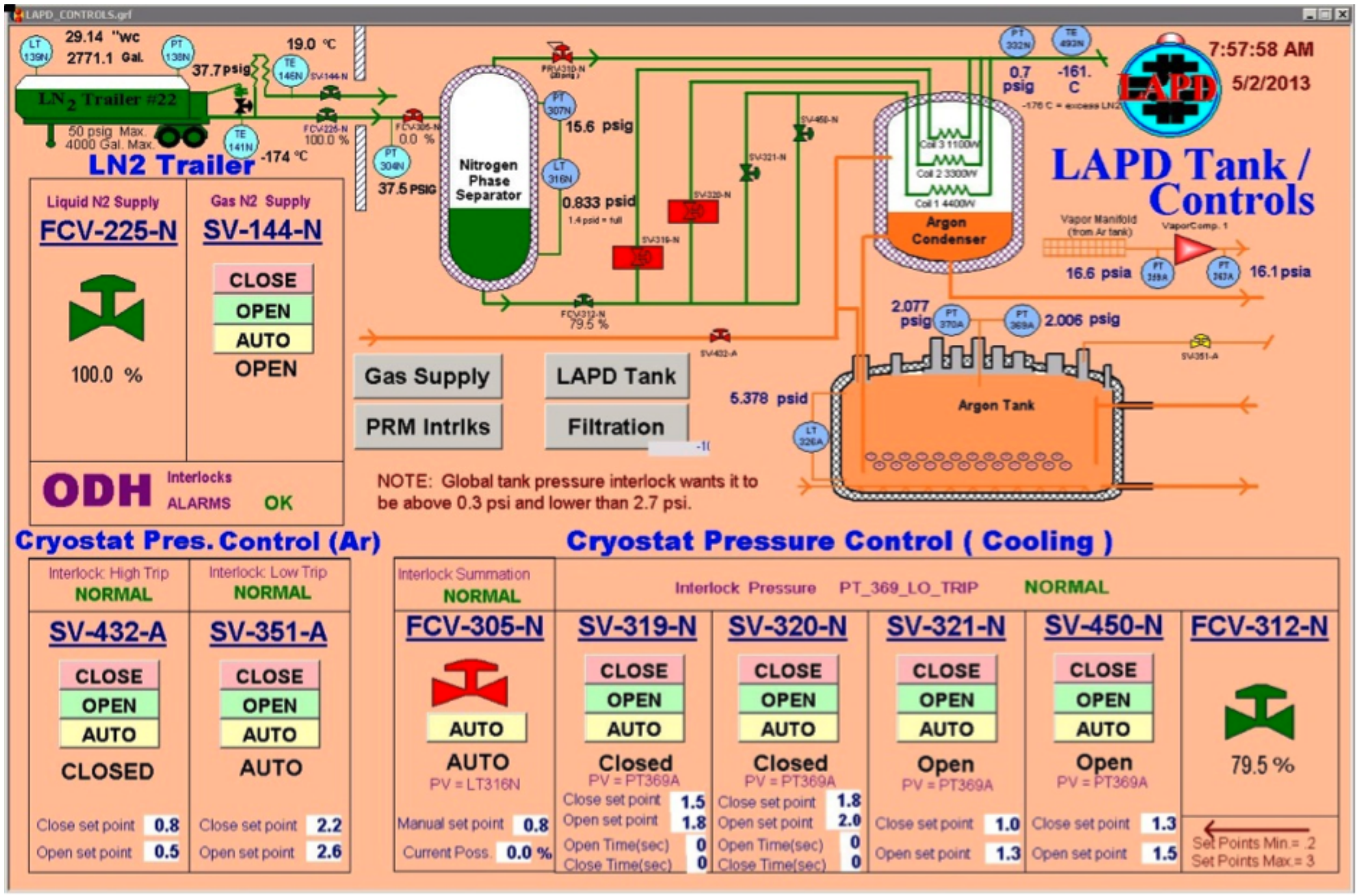

Figure 14.4: Human Machine Interface for the LAPD experiment.

### 14.2.1 Uninterruptible Power Supply (U.P.S.)

The control system U.P.S. is to be a commercial unit such as those manufactured by Powerware. The U.P.S. input power is fed from a premises powered outlet using the U.P.S. input line cord. This U.P.S. system will be natural gas generator backed. The diesel generator will be auto start with auto switchover on commercial power loss. There will be other loads on this generator as well. The U.P.S. has standard outlets located on the rear of the cabinet. An APC surge protector is located on the U.P.S. and its input power cord is plugged into the U.P.S output outlets. All relevant control system loads are plugged into the APC surge protector output outlets.

### 14.2.2 PLC Reliability and Redundancy

Siemens SIMATIC (S7 PLC and ET200M I/O modules) components meet all relevant international standards and are certified accordingly. Temperature and shock resistance are defined in the SIMATIC quality guidelines, as are vibration resistance or electromagnetic compatibility. The Siemens S7 PLC system equipment can be redundant at many different levels, from the PLC CPU (Hot Backup) to the module and instrument level. We expect to have the redundancy at the PLC level.

# Chapter 15

# The Precision Magnetic Field: $\omega_p$

In this chapter the requirements and design for the precision magnetic field measurement system are presented, followed by the requirements and procedures for shimming the storage ring magnetic field.

## 15.1 Precision Magnetic Field Measurement

### 15.1.1 Relation between $a_\mu$ and $\omega_p$

In an idealized experiment, the anomaly $a_\mu$ could be extracted by measuring the difference frequency $\omega_a$ between the muon spin $\omega_s$ and cyclotron frequencies $\omega_c$ in a storage ring with a perfectly homogeneous magnetic field $B$ without focusing fields (see Eqn. 3.6):

$$\omega_a = -\frac{Qe}{m_\mu} a_\mu B, \tag{15.1}$$

where $e > 0$ and $Q = \pm 1$. The lab-frame magnetic field $B$ is measured using proton nuclear magnetic resonance (NMR) and expressed in terms of the free proton angular precession frequency $\omega_p$, via $\hbar\omega_p = 2\mu_p|\vec{B}|$. The proton gyromagnetic ratio $\gamma_p \equiv 2\mu_p/\hbar = 2\pi \times 42.577\,4806(10)$ MHz/T [1], so $\omega_p \approx 2\pi \times 61.79$ MHz in the 1.45 T field. Expressing $B$ in terms of $\omega_p$, the muon anomaly $a_\mu$ is:

$$a_\mu = \frac{\omega_a}{\omega_p}\frac{2\mu_p}{\hbar}\frac{m_\mu}{e} = \frac{\omega_a}{\omega_p}\frac{\mu_p}{\mu_e}\frac{m_\mu}{m_e}\frac{g_e}{2}, \tag{15.2}$$

using $\mu_e = g_e e\hbar/4m_e$. Experiment E989 will measure $\omega_a/\omega_p$, the additional ratios appearing in Eqn. 15.2 are well known from other experiments: $\mu_e/\mu_p = -658.210\,6848(54)$ (8.1 ppb) [1], $g_e/2 = 1.001\,159\,652\,180\,73(28)$ (0.28 ppt) [2], and $m_\mu/m_e = 206.768\,2843(52)$ (25 ppb) [1] [1]. The latter ratio is extracted using Standard Model theory from the E1054 LAMPF measurement of the ground state hyperfine interval in muonium $\Delta\nu_{\mathrm{Mu}}(\mathrm{E1054}) =$

[1]Recently the magnetic moment of the proton was measured with a factor of 3 greater precision [4], which may further reduce the uncertainties in Eqn. 15.2.

4 463 302 765(53) Hz (12 ppb) [3]. The measurement is compared with the SM theoretical prediction to extract $m_e/m_\mu$, which appears as a parameter in the prediction [1] :

$$\Delta\nu_{\rm Mu}({\rm Th}) = \frac{16}{3} c R_\infty \alpha^2 \frac{m_e}{m_\mu}\left(1+\frac{m_e}{m_\mu}\right)^{-3} + \text{higher order terms.} \qquad (15.3)$$

The theory uncertainty has a 101 Hz (23 ppb) contribution from uncertainty/incompleteness in the theory calculation, but is dominated by the uncertainty in the mass ratio $m_e/m_\mu$. The uncertainties on the Rydberg $R_\infty$ and fine structure constant $\alpha$ (extracted using QED theory from a measurement of $a_e$ [1]) are negligible in comparison. The hyperfine interval is dominated by QED contributions, but there are higher-order corrections including a weak contribution $\Delta\nu_{\rm Weak} = -65$ Hz from $Z^0$ exchange, a hadronic contribution $\Delta\nu_{\rm Had} = 236(4)$ Hz, and a hadronic light-by-light contribution of 0.0065 Hz (see references in [1]). Setting $\Delta\nu_{\rm Mu}({\rm E1054}) = \Delta\nu_{\rm Mu}({\rm Th})$ determines the mass ratio $(m_\mu/m_e)$ to 25 ppb, but the result is the theory-dependent.

Arguably, the line of arguments above could be broken by new physics that may contribute to $a_\mu$. This new physics might also contribute to $\Delta\nu_{\rm Mu}$ and must be included for consistency when extracting $m_\mu/m_e$. For instance, given that the difference between $a_\mu^{\rm E821}$ and $a_\mu^{\rm SM}$ is roughly twice the weak contribution to $a_\mu$, a comparable new physics contribution to $\Delta\nu_{\rm Mu}$ would perturb the extracted value of $m_\mu/m_e$ by almost 30 ppb.

To extract a value of $a_\mu$ less dependent on assumptions, note the ratio of muon to electron masses can be expressed as $m_\mu/m_e = (g_\mu/g_e) \times (\mu_e/\mu_\mu)$, so Eqn. 15.2 becomes:

$$a_\mu = \frac{\omega_a}{\omega_p}\,\frac{\mu_p}{\mu_\mu}\,(1+a_\mu) = \frac{\omega_a/\omega_p}{\mu_\mu/\mu_p - \omega_a/\omega_p}. \qquad (15.4)$$

The advantage is that the quantities on the right-hand side are determined experimentally with a minimum of theory. The magnetic moment ratio $\mu_{\mu^+}/\mu_p = 3.183\ 345\ 24(37)$ was determined to 120 ppb essentially directly by E1054 from muonium Zeeman ground state hyperfine transitions measured in a 1.7 T field [1, 3]. The result is based solely on measured quantities, the validity of the Breit-Rabi Hamiltonian to describe the experiment, and a small (17.6 ppm) bound-state QED correction to the $g$-factor for a muon bound in muonium (where the uncertainty on the correction is sub-ppb). In this approach, $a_\mu$ is extracted solely from measured quantities, but limited by the experimental precision on $\mu_\mu/\mu_p$ of 120 ppb. A more precise, independent measurement of $\mu_\mu/\mu_p$, planned at J-PARC, would be very helpful. Note however, that even in the absence of a new measurement, any BSM theory can be tested against E989 at the 140 ppb level as long the potential BSM contributions to $\Delta\nu_{\rm Mu}$ are considered simultaneously.

### 15.1.2 Physics Requirement on $\tilde{\omega}_p$

Based on the above approach for $a_\mu$, our goal for the total uncertainty on $\tilde{\omega}_p$ in E989 is $\delta\tilde{\omega}_p \leq 70$ ppb, roughly a factor of 2.5 times smaller than was achieved in E821. Here $\tilde{\omega}_p$ refers to the free proton precession frequency $\omega_p$ weighted by the muon distribution in the storage ring.

### 15.1.3 Design of the E989 Field Measurement System

E989 will largely use the principles and field measurement hardware originally developed at the University of Heidelberg and Yale which were employed successfully in E821 at BNL [7] and E1054 at LANL [3]. The E821 field measurement electronics and the underlying physics are described in [5]. The calibration of the field measurements in terms of the equivalent free proton precession frequency using an absolute calibration probe is described in [6]. Details of the E821 field analysis, systematics, and of the hardware are described in the final E821 paper [7], and in several theses and notes [8, 9, 10, 54].

While E821 achieved an uncertainty $\delta\omega_p \approx 170$ ppb, E989 will have to implement specific changes to the hardware and techniques to reduce the systematic errors to the final goal of $\delta\omega_p \approx 70$ ppb. The E989 field measurement hardware, techniques, and changes from E821 will be discussed in the rest of this chapter.

### 15.1.4 Error budget for the $\omega_p$ measurement

The systematic errors on the field measurement from E821 are listed below in Table 15.1. The last two columns list the uncertainties anticipated for E989 and the sections in this chapter where these uncertainties are discussed in detail.

| Source of uncertainty | R99 [ppb] | R00 [ppb] | R01 [ppb] | E989 [ppb] | Section |
|---|---|---|---|---|---|
| Absolute calibration of standard probe | 50 | 50 | 50 | 35 | 15.4.1 |
| Calibration of trolley probes | 200 | 150 | 90 | 30 | 15.4.1 |
| Trolley measurements of $B_0$ | 100 | 100 | 50 | 30 | 15.3.1 |
| Interpolation with fixed probes | 150 | 100 | 70 | 30 | 15.3 |
| Uncertainty from muon distribution | 120 | 30 | 30 | 10 | 15.3 |
| Inflector fringe field uncertainty | 200 | – | – | – | – |
| Time dependent external $B$ fields | – | – | – | 5 | 15.6 |
| Others † | 150 | 100 | 100 | 30 | 15.7 |
| Total systematic error on $\omega_p$ | 400 | 240 | 170 | 70 | – |
| Muon-averaged field [Hz]: $\tilde{\omega}_p/2\pi$ | 61 791 256 | 61 791 595 | 61 791 400 | – | – |

Table 15.1: Systematic errors for the magnetic field for the different run periods in E821. R99 refers to data taken in 1999, R00 to 2000, R01 to 2001. The last two columns refer to anticipated uncertainties for E989, and the section in this chapter where the uncertainty is discussed in detail. †Higher multipoles, trolley temperature and its power supply voltage response, and eddy currents from the kicker.

It is important to note the steady reduction in uncertainties achieved in E821. The goal of 70 ppb uncertainty on $\omega_p$ for E989 in Table 15.1 reflects the current estimates of what can be achieved based on the experience in E821.

### 15.1.5 Overview of Precision Magnetic Field Measurement

Pulsed Nuclear Magnetic Resonance (NMR) is at the heart of the magnetic field shimming, measurement and control systems, since it can measure magnetic fields to absolute accuracies of tens of parts per billion (ppb).

The pulsed NMR hardware developed for E821, which produced and detected the free induction decay (FID) signals from protons in water or in petrolatum, has already demonstrated single shot precision at the level of 20 ppb [5], and absolute calibration in terms of the free proton precession frequency with an accuracy of of 35 ppb [6]. The challenge of the field measurement is to effectively transfer this absolute calibration to the many NMR probes required to monitor the field in the large volume and over the long periods of time during which muons are stored.

There are four major tasks required from the NMR system:
(1) Monitoring the field when muon data are being collected using fixed probes [15.2.1];
(2) Mapping the storage ring field when the beam is off using the NMR trolley [15.3];
(3) Providing an absolute calibration chain relating field measurements to the Larmor frequency of a free proton [15.4.1];
(4) Providing feedback to the storage ring power supply when muon data are collected [15.5].

We start with a brief description of NMR, explain the field measurement principles in more detail, then describe the required hardware.

#### Field Measurement with Pulsed NMR

Precision measurements of the magnetic field are made by detecting the free induction decay (FID) signal of protons in materials containing hydrogen such as water or petrolatum[2] using pulsed NMR [11, 12, 5]. The probes used in E821 for these purposes are shown in Figs. 15.1 and 15.2. The material samples are located in small volumes (typically <1 cm$^3$) surrounded by a coil $L_s$ and the rest of the body of an NMR probe. Several hundred fixed probes are located around the azimuth of the ring, just above and below the muon storage volume. Other sets of probes are pulled through the storage volume in a trolley used to determine the field seen by the muons. A final set of probes is used for calibration.

In a typical measurement, an RF pulse at $\omega_{\rm ref} = 2\pi \times 61.74$ MHz, is used to produce a linearly polarized rf magnetic field in the coil $L_s$, orthogonal to the storage ring dipole field. This rotates the magnetization of the protons in the sample so that it is perpendicular to the main field of 1.45 T. After the $\pi/2$ pulse, the proton spins precess coherently in the external field at the proton magnetic resonance (Larmor) frequency $f_{\rm NMR} \approx 61.79$ MHz. The magnetic field from the rotating magnetization induces an EMF in the coil $L_s$ which is called the NMR signal. This signal typically decays exponentially with a time constant $T_2^*$ of several ms. The coil (which is used for both excitation and detection) is part of a circuit with a capacitor $C_s$ in series to form a resonant circuit at the NMR frequency $f_{\rm NMR}$ with a quality factor $Q$ typically between 30 and 100. An additional parallel coil, $L_p$ or series capacitor is used to match the impedance of the probe at $f_{\rm NMR}$ to the 50 $\Omega$ impedance of the cable. The NMR signal propagates through a cable to a duplexer which directs the signal to a low-noise preamplifier. The amplified signal is mixed with the synthesizer frequency $\omega_{\rm ref}$, and the difference frequency $f_{\rm NMR} - f_{\rm ref} \equiv f_{\rm FID}$ goes through a low pass filter and is amplified and digitized. This signal is referred to as the FID (free induction decay). The frequency of the FID, $f_{\rm FID}$, is sensitive to the local field value and is of order 50$\pm$5 kHz. The exact frequency can be determined from the digitized signal by identifying and counting zero crossings of the FID and counts of the digitizer clock until the signal has decayed to about $1/e$ of its peak value, which typically takes several ms [3]. Other analysis techniques are under development. The local magnetic field is then characterized by the frequency $f_{\rm NMR} = f_{\rm ref} + f_{\rm FID}$ with resolution approaching 20 ppb.

[2] The default is to use petrolatum as the NMR sample in the E989 fixed probes. We will refine and complete the studies of the temperature dependence of the susceptibility and shielding in petrolatum before the end of 2014.

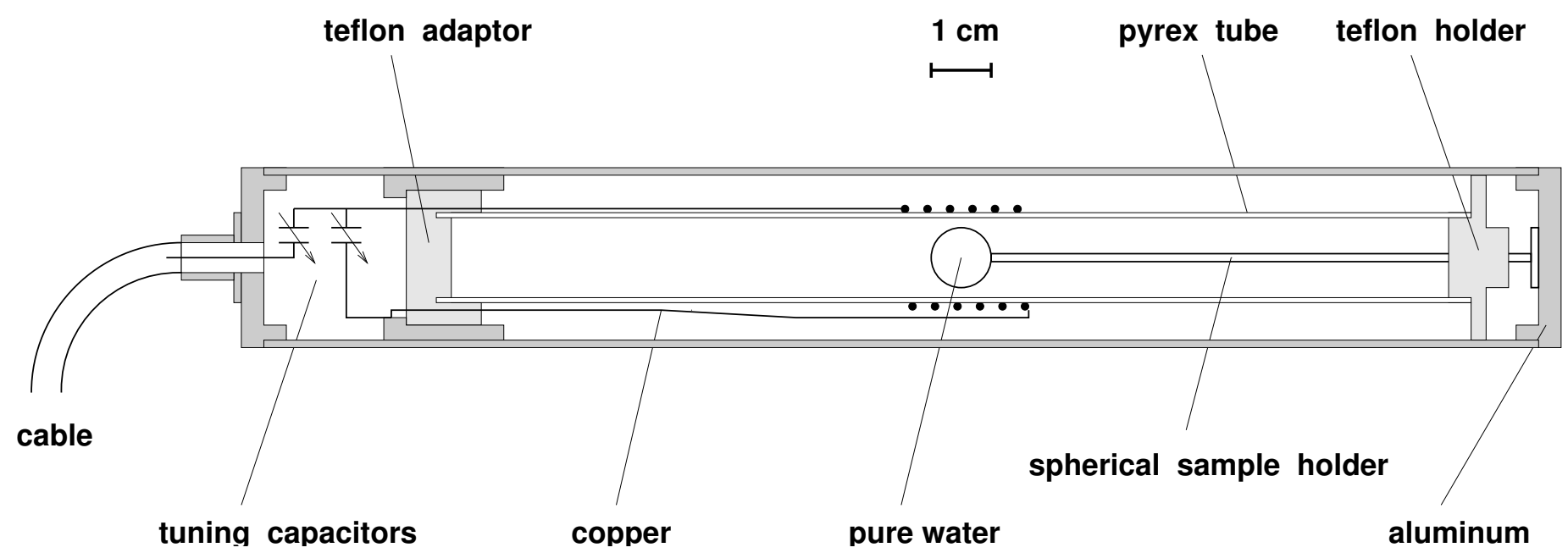

(a) Absolute calibration probe

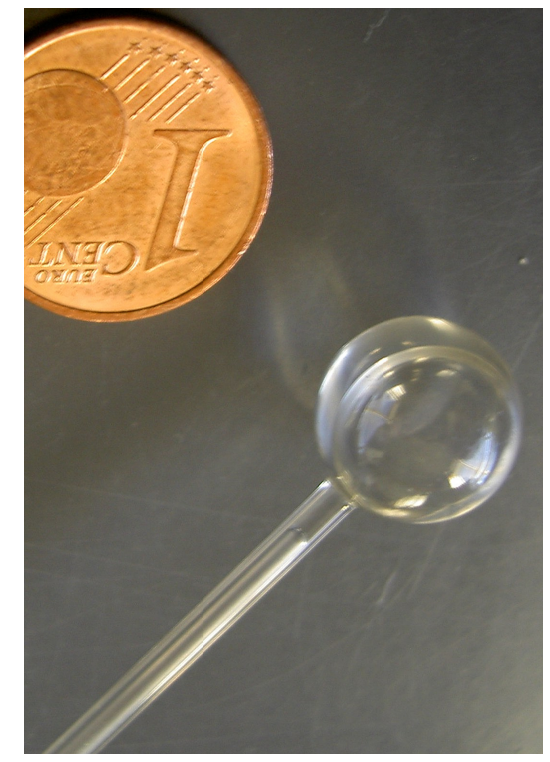

(b) Spherical Pyrex container

Figure 15.1: (a) Absolute calibration probe featuring a spherical sample of water. This probe and its driving and readout electronics are the very same devices employed in reference [3] to determine $\lambda$, the muon-to-proton magnetic-moment ratio. (b) The spherical Pyrex container for the absolute probe.

As discussed below, the reference frequency $f_{\rm ref} = 61.74$ MHz is chosen such that $f_{\rm ref} < f_{\rm NMR}$, and is obtained from a frequency synthesizer phase locked to a Rb frequency standard stabilized by GPS (see section 15.2.4). The same Rb standard will provide the time base for the $\omega_a$ measurement.

## 15.2 The Fixed Probe NMR system

The purpose of the fixed probe system is to accomplish the first major field measurement task of Sec. 15.1.5; monitoring the field continuously while muon data are being collected. The fixed probe system consists of the fixed NMR probes and the accompanying VME system, DAQ, pulser, mixer, multiplexers, and digitizers. A block diagram of the recommended system is shown in Fig. 15.3.

[3]Corrections are applied to handle lineshapes for which the instantaneous FID frequency varies with time [13].

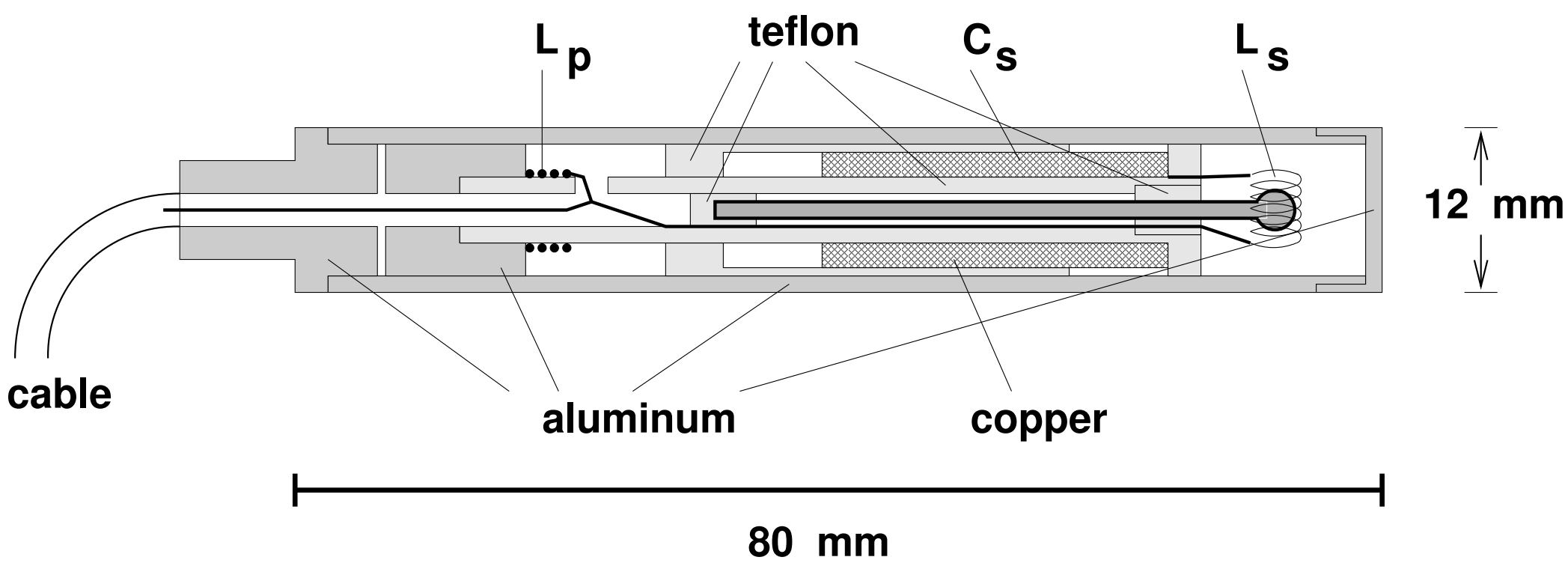

(a) Plunging probe

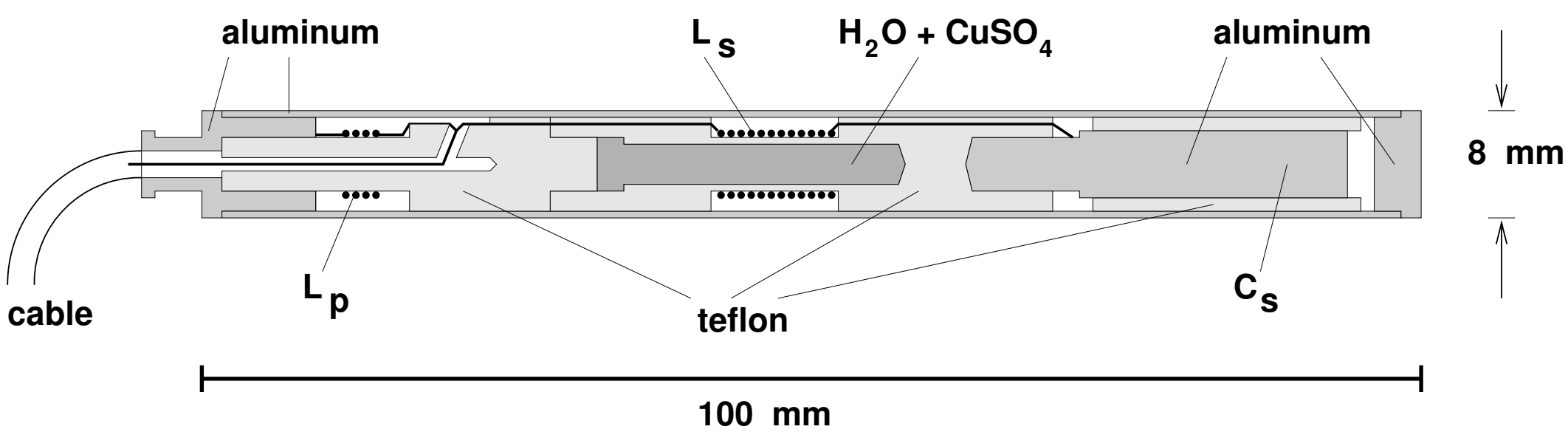

(b) Trolley and fixed probe

Figure 15.2: (a) Plunging probe, which can be inserted into the vacuum at a specially shimmed region of the storage ring to transfer the calibration to the trolley probes. (b) Schematic of the probes used in the trolley and as fixed probes in E821. The resonant circuit is formed by the two coils with inductances $L_s$ and $L_p$ and a capacitance $C_s$ made by the aluminum housing and a metal electrode. The active sample volume has a diameter of 2.5 mm and a length determined by the coil $L_s$ of 15 mm.

The design consists of a set of 378 NMR probes at 72 locations in azimuth around the ring. The number of probes at each azimuthal position alternates between two probes at radii of 7112 and 7142 mm, or three probes at radial positions of 7082, 7112, and 7142 mm, where the probes are placed in matching grooves on the upper and lower surfaces of the storage ring vacuum chambers. From this geometry the fixed probes provide a good monitor of the dipole field around the ring, with some sensitivity to changes in the skew and normal quadrupole components.

Groups of 20 probes are connected to a single analog multiplexer. Twenty multiplexers are required to handle all of the fixed probes, the plunging probe and the absolute calibration probe (see below). In a typical BNL E821 measurement sequence, one probe from each multiplexer was excited and its FID passed to a frequency counting module (there were 20 of these). Roughly 0.2 s later, a second probe from each multiplexer was selected, excited, and read out. Given there were 5 probes read out per second per multiplexer and 20 multiplexers,

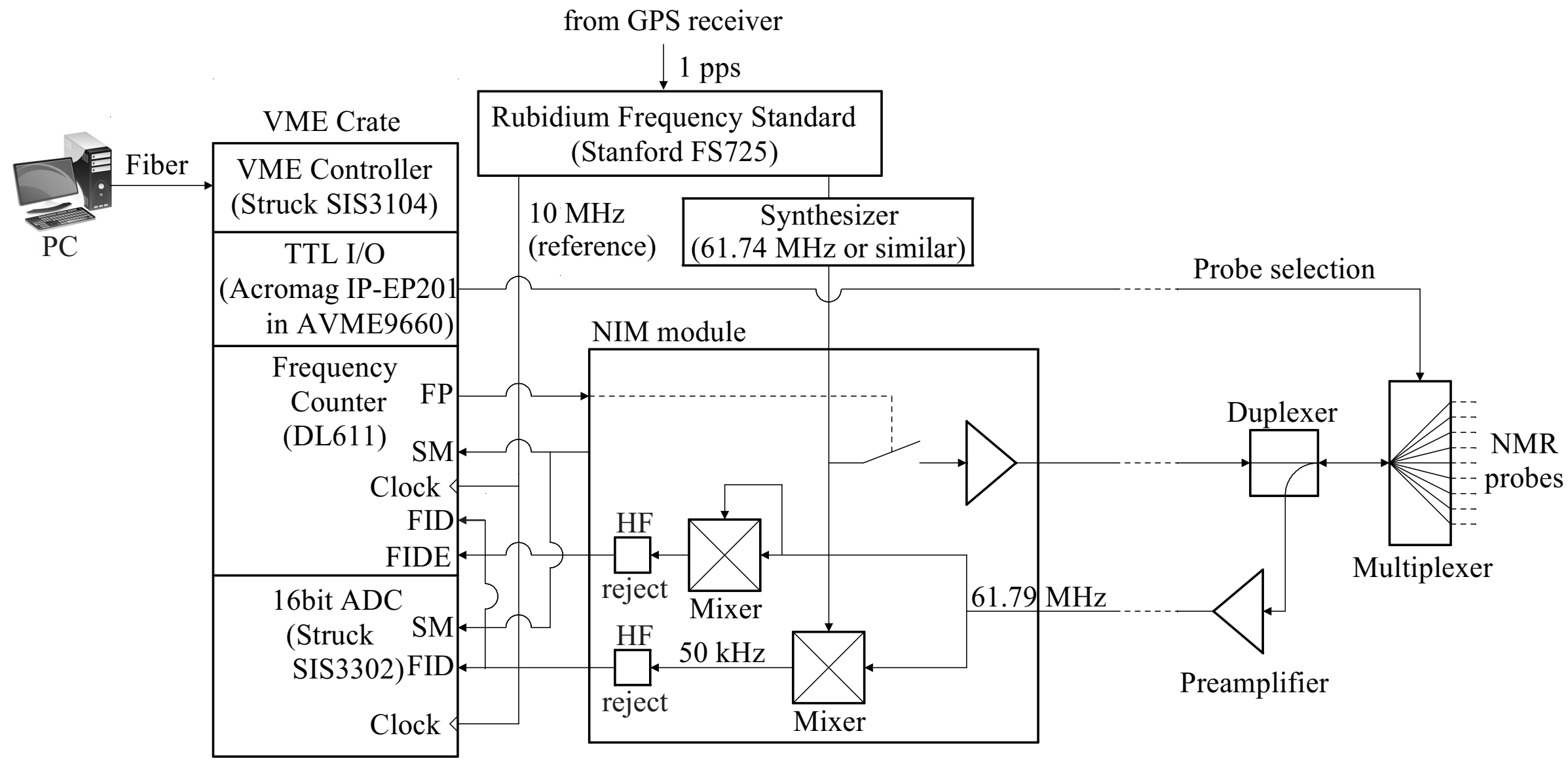

Figure 15.3: A schematic of the fixed probe system. The system consists of a set of NMR probes close to the muon storage volume connected to multiplexer boxes that sit on the storage ring magnet. The multiplexers, which contain the duplexers and preamplifiers, are connected to NIM modules and a VME system and DAQ that sit in the counting house. The functions of each element are described in the text.

100 probes were read out per second. The complete set of fixed probes was read out every 4 seconds during data-taking periods when muons are stored. For E989, with current PCs and faster VME64 hardware, all probes can now be digitized and read out within the 1.33 second beam supercycle.

In E821, roughly half of these fixed probes were used to monitor the storage ring field. Of the remaining probes, some were noisy, and a significant number were located in regions near the pole-piece boundaries, where local magnetic field gradients were sufficiently large that they reduced the free induction decay time $T_2^*$ in the probes to a millisecond or less.

In E989 several steps will be taken to increase the number of useful probes. First, finite element analysis of the vacuum chambers indicates that the grooves containing the probes can be extended without significantly increasing the deflection of the chambers under vacuum. This will allow probes to be moved farther from pole boundaries, increasing the number of useful probes. Second, in E821, at least initially, many of the fixed probes used water in their sample volumes and over the course of the experiment experienced loss of the water. In E989, the water samples will be replaced with petrolatum (CAS 8009-03-8). Petroleum jelly was observed in E821 to have low evaporation while providing a proton NMR signal comparable in magnitude and frequency to water. It has favorable relaxation times; measurements at U. Michigan indicate $T_2$ of order 40 ms. A third factor contributing to a more robust fixed probe measurement comes from the refurbishment of the E821 probes (see below), in which defective electrical contacts which led to poor signal quality will be improved. Finally, the NMR electronics of E821, which extracted precession frequencies by counting zero-crossings of FIDs mixed down to the 50 kHz range, will be replaced with a high-performance set of digitizers. The 16 bit (13 bits effective), 10 MS/sec digitizers (Struck SIS 3302) will allow useful information to be extracted from probes with short signals. In addition, as explained below, higher resolution is made possible by extracting frequency information from the entire FID rather than just the zero crossings.

### 15.2.1 Fixed Probes for E989

The same basic probe design from E821 will be used for E989 (see Fig.15.2(b)). Materials used to construct the probes, mostly aluminum and teflon (PTFE), have low susceptibility and the coax cable has copper conductors instead of the more common copper-plated steel. Each probe's outer aluminum shell has a diameter of 8 mm, which fits in grooves machined into the outside surface of the top and bottom plates of the vacuum chamber. The probe's outer shell and the inner body form the capacitor $C_s$, which is in series with $L_s$ and makes a resonant circuit. $C_s$ is adjusted by moving a PTFE sleeve in and out to tune the circuit to the frequency $\omega_p/2\pi$. A circuit quality factor $Q \approx 30$ corresponds to a bandwidth of 3% which is the range over which the magnetic field can be measured without retuning the probes. A parallel coil $L_p$, allows for matching the impedance of the probe to 50 $\Omega$ for optimal transmission.

**Refurbishing existing probes:**

Resurrecting the existing E821 measurement system requires a complete working set of probes provided either by refurbishing existing ones or constructing new ones. In E821 the sample volume was filled with water doped with $CuSO_4$. This dopant is paramagnetic and is used to shorten the magnetization recovery time $T_1$ so measurements can be made more frequently. For pure $H_2O$, $T_1 \approx 3.5$s, so consistent free-induction-decay NMR measurements could only be taken every 30 seconds. However, paramagnetic impurities also shift the measured frequency, and changes in the $CuSO_4$ concentration, for example due to slow evaporation of the water, will affect the stability of the measurement. An examination of probes from E821 indicated that in some probes water had leaked and corroded part of the probe. These probes will be rebuilt. To prevent similar difficulties in E989, we will use petroleum jelly in place of the $CuSO_4$-doped water. This idea was implemented in some fixed probes of E821 by R. Prigl. We have recently measured the NMR relaxation times of a sample of petrolatum at 0.4 T using saturation-recovery (for $T_1$) and a standard spin-echo sequence (for $T_2$) at Michigan. We find $T_2 \approx T_1 \approx 40$ ms, which is long enough for very high resolution frequency determination, and will not restrict $T_2^*$ which is typically limited to a few milliseconds by local field gradients. The temperature dependence of the diamagnetic shielding is also crucial. Preliminary work in E821 indicated the temperature coefficient of the chemical shift $d\sigma/dT$ is smaller than the shift in water, providing greater stability. Recently at Washington, $d\sigma/dT$ for petrolatum was measured to be $\approx +1$ ppb/K, roughly a factor of 10 smaller than in water. The volume magnetic susceptibility $\chi_V$ of petrolatum was measured to be roughly -8$\times 10^{-6}$ in SI units, with temperature dependence $d\chi_V/dT$ roughly $-1 \times 10^{-9}$. The overall temperature dependence of petrolatum probes should be more than a factor 4 smaller than the fixed probes with water samples of E821.

The scale of the refurbishing effort was estimated from tests on about 40 fixed probes from E821. The results indicated that about 1/4 of the probes had water leaks and corrosion, and will need to be rebuilt. Another area of some concern occurs where the coil wire is bonded to the aluminum parts with low temperature solder. This connection requires inspections and probes with broken connections will require re-soldering. All probes will be filled with petroleum jelly from a single source, tuned, and checked for correct functioning.

An existing dipole magnet at University of Washington (see figure 15.4) has been repurposed to provide the 1.45 Tesla field required for testing the NMR probes. The nominal field uniformity is 100 ppm/cm over the entire pole face but there are regions with better than 10 ppm uniformity over the active volume of the probe. Active shimming with opposing Helmholtz coils further cancels linear gradients and FIDs lasting 4 to 5 ms can be obtained. A subset of the E821 electronics initiates and reads out the free induction decays. The magnet permits testing the entire fixed probe system over a wide range of field gradients.

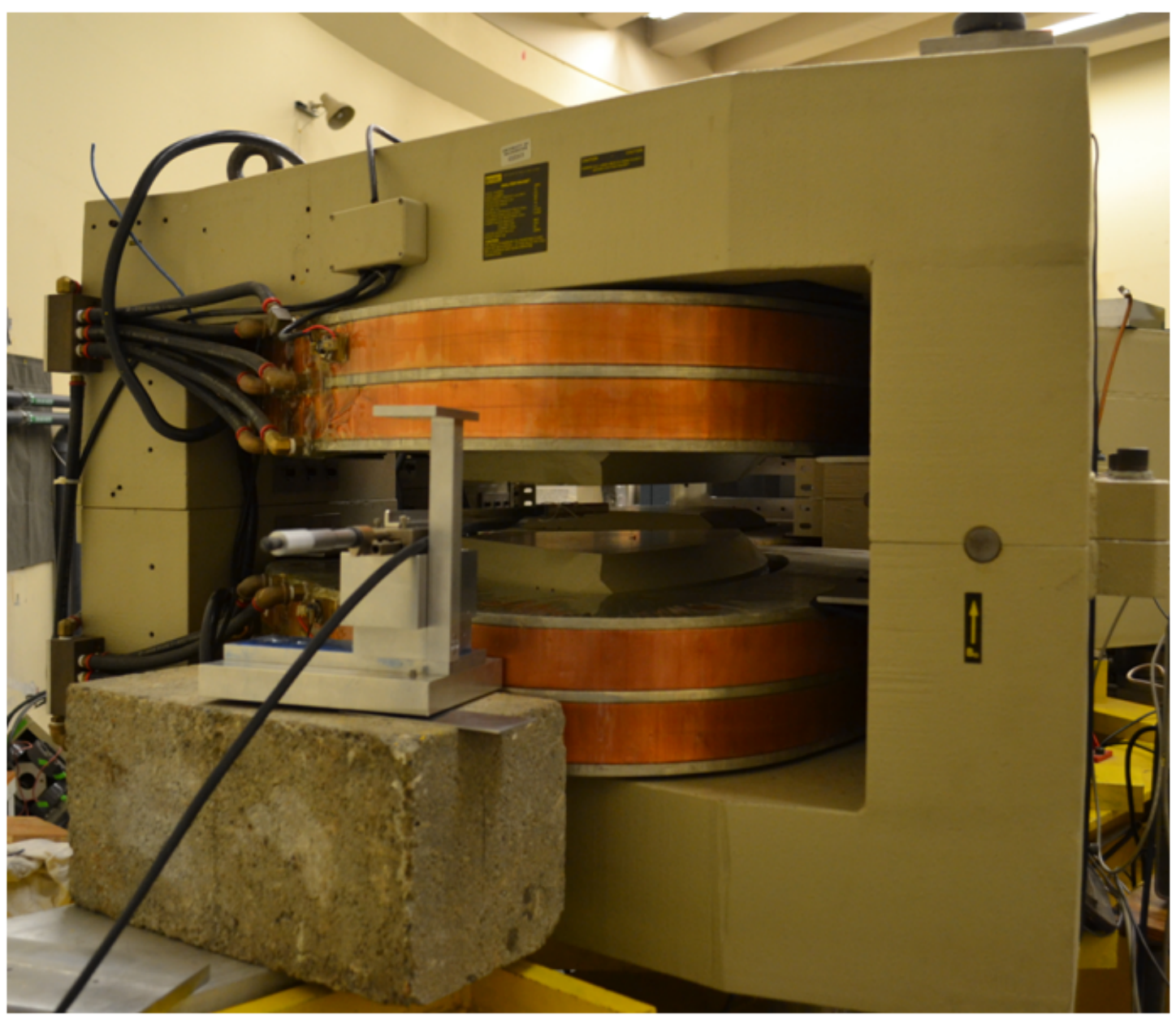

Figure 15.4: Dipole magnet at the University of Washington shown with field mapping in progress. It provides a 1.45 Tesla field sufficiently uniform for testing probes and electronics for the E989 fixed probe system.

**Constructing new probes:**

Any new probes constructed are required to fit in grooves machined in the vacuum chambers. The E821 probes were reverse engineered and CAD drawings produced. In consideration of the quantity, new parts will be produced using computer numerically controlled mills and lathes at the UW machine shop. Metric sized aluminum tubing 8 mm $\times$ 0.5 mm and 7 mm

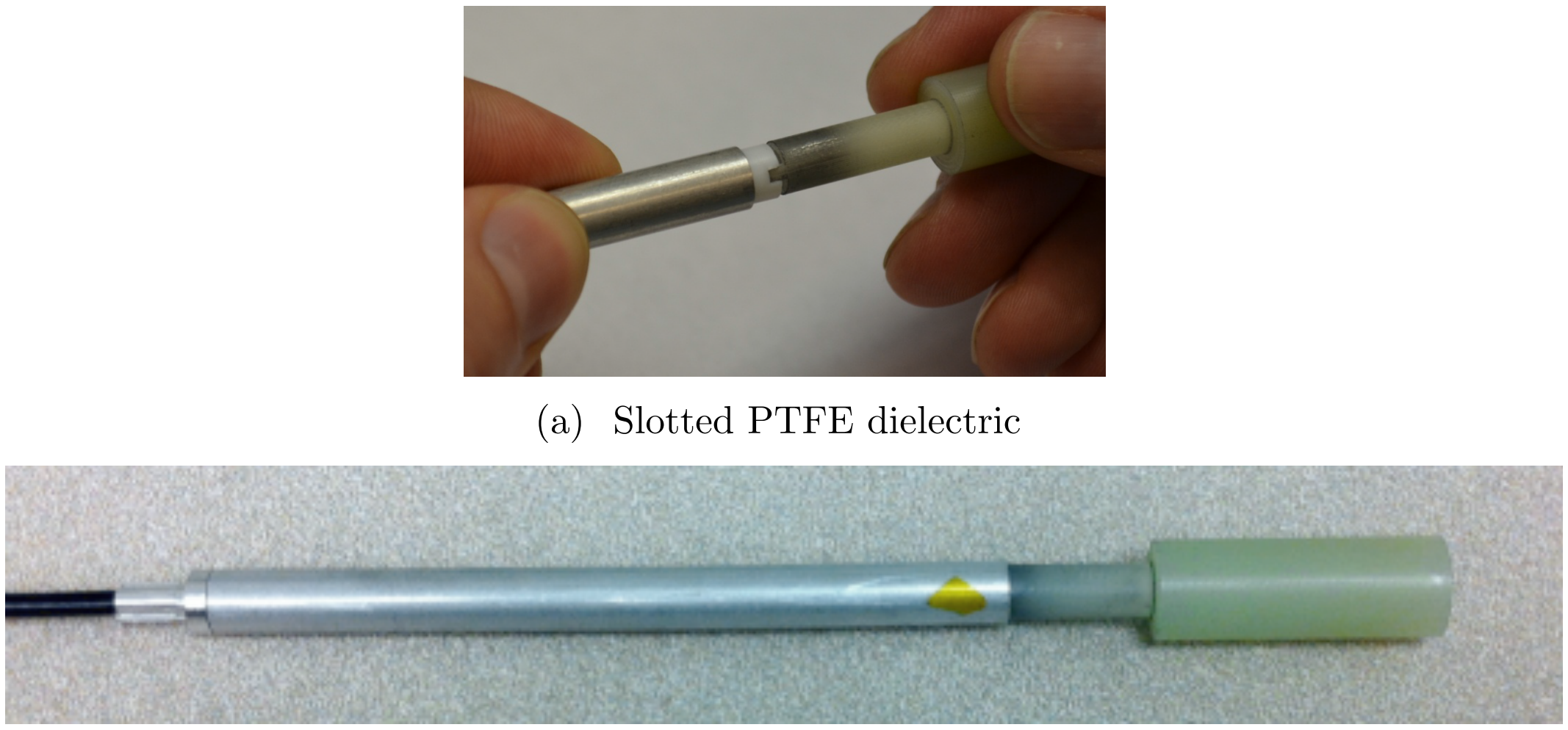

(a) Slotted PTFE dielectric

(b) pNMR probe with crimp connection

Figure 15.5: (a) The position of the PTFE dielectric can be adjusted within the assembled probe by using a hollow core screwdriver. (b) An assembled NMR probe during frequency tuning with the new crimp type connection for the coax cable.

PTFE rods are available in the European market. The new design concentrates on three parts of the probe. Firstly, as in the E821 design, the resonant frequency of the LC circuit is tuned by adjusting the position of a PTFE dialectric cylinder on the threaded inner conductor of the capacitor $C_s$. In the new design, only the end cap shown in figure 15.2(b) needs to be removed. Milling a slot in the end face of the PTFE capacitor dialectric allows it to be adjusted with the hollow core screw driver shown in figure 15.5(a). The resonant frequency is tuned while observing the reflection coefficient on a vector network analyzer.

Secondly, we use an ultrasonic flux free soldering iron to connect $L_s$ and $c_S$. Ultrasonic sound replaces the flux normally used to clean surfaces while soldering. The third improvement is the way the electrical connections a re made between the coax braid, probe end cap and coil $L_p$. We have successfully demonstrated mechanical crimp connections that will replace the conducting epoxy used in E821 probes.There are major advantages to crimp type connections. The electrical contact between the cable braid and the probe end cap are significantly improved. During the process of crimping the insulating aluminum oxide layer is mechanically broken which results in a low ohmic joint. In addition, the crimped connection is protected from corrosion since it is air tight and thus keeps oxygen away from the contact areas. It is mechanically strong. The crimp includes the coax outer jacket in the same manner as the commercial SMA connector on the other end of the cable and provides strain relief (see figure 15.5(b)). In the E821 design, strain relief was provided by a shrink tube which was sometime not sufficient to prevent the epoxy joint from breaking. Where possible we intend to reuse the proton sample holders from the old probes to reduce the number of these complex parts that have to be made. After new parts are completed coils will be wound and the same procedure as for refurbishing will be followed.

**Testing the probes:**

For each probe, the resonant circuits are tuned to 61.79 MHz and 50 Ω impedance using a vector impedance meter. The resonance is then excited in the probe and if the free induction decay (FID) is observed with sufficient signal to noise ratio SNR, the probe passes the test. In the E821 system the SNR at the beginning of the FID was ≈ 300:1.

### 15.2.2 Multiplexers

Groups of up to 20 fixed probes are connected to a single multiplexer located a few meters away (probes have 5 meter cable lengths), which sits on top of the magnet. The multiplexer design is described in [5, 8]. It is a self-contained electronic module that selects one of 20 NMR probes based on a 5 bit address set by the NMR DAQ. A cable terminated with DB-9 connectors connects the multiplexer to TTL level digital output lines from the VME subsystem. The TTL signals control PIN diodes in the multiplexer that direct a $\pi/2$ RF pulse from the NIM pulser/mixer module to tip the spins in the selected NMR probe. When the circuit is closed, there is a measured 2.5 dB insertion loss. To improve the isolation when the circuit is open, effective-$\lambda/4$ lines are incorporated into the switches. A duplexer (basically a transmit/receive switch) in the multiplexer directs the high power RF pulse towards the selected NMR probe, while isolating the sensitive low-noise preamplifier. After the RF pulse, the duplexer directs the returning low-level FID precession signal to a low-noise preamplifier with a gain of approximately 60 dB. The switches, duplexer, and TTL integrated circuits are fabricated with discrete surface mount components, and replacements are readily available.

The preamplifier consists of two RF amplifier modules, a UTO-101 and GPD-201 connected in cascade. These were made by Avantek at the time the multiplexers were developed and are currently sourced by Teledyne-Cougar. The first stage has a noise figure of 1.7 dB with a typical gain of 27 dB. The wireless communication industry has advanced the state of the art in RF amplifiers since the days when the multiplexers were designed providing many more options. For example, the Minicircuits ultra low noise MMIC amplifier PSA4-5043+ has an equivalent gain and an improved noise figure of 0.7 dB and also recovers rapidly from an overload condition such as is experienced when probes are excited. The amplifier is available both as a surface mount chip and as an evaluation module (TB-653+) with connectors that could be used in the multiplexer with minor circuit board modifications. This preamplifier has the dual advantage of better noise performance which leads to longer useful FID times and no longer having to rely on obsolete parts. We plan to evaluate the unit in the near future.

### 15.2.3 Pulser and Mixer

The Pulser and Mixer is a single width NIM module located adjacent to the VME crate and connected by cables to a single multiplexer. This functions as the receiver and transmitter (pulse generator) for the NMR fixed probes. One shot multivibrators triggered by the fire pulse (FP) from the VME subsystem generate a logic pulse with a width adjusted to the desired length for the $\pi/2$ pulse. A jumper on the circuit board determines whether this generated logic pulse or the input FP pulse determines the duration of the $\pi/2$ pulse which then gates on the $f_{\rm ref} = 61.74$ MHz reference and sends it through a custom made class-C amplifier to the multiplexer. The measured RMS power in a 50 ohm load is about 0.8 Watts and length of the $\pi/2$ pulse is 7 $\mu s$. In the receiver part of the module, the amplified probe signal from the multiplexer is mixed with the reference frequency $f_{\rm ref}$ and the difference frequency (about 50 kHz) is the free induction decay signal (FID). In a second branch of the receiver, the envelope of the probe signal (FIDE) is obtained by mixing the preamplifier output with itself and filtering out the $2\omega_L$ component. The $\pi/2$ pulse causes some saturation in the preamplifier and a TTL signal (SM) is provided to delay processing the FID for roughly 10 $\mu$s after the end of the RF pulse. In E821 this pulse was used by the custom Heidelberg DL611 frequency counter module to enable the start of the zero crossing counter. The delay accommodates both the dead time in the preamplifier and transients such as those from ringdown of the $L_s$ coil and decay of the transient response in the low-pass filter after the mixer.

The inputs and outputs of the NIM modules are shown in Fig. 15.3. The inputs are the fire pulse (FP) which initiates the free induction decay, the synthesizer reference $f_{\rm ref}$ and the returning FID signal from the multiplexer preamplifier. The outputs are the high power $\pi/2$ pulse, the SM signal, the envelope function (FIDE) and two channels of FIDs, one for the DL611 zero-crossing counter and one for the waveform digitizer. This module contains mostly surface mount and integrated circuit components and replacements are readily available. RF amplifiers GPD-201, GPD-202, and 3 GPD-462 were made by Avantek when this board was designed and they continue to be sourced by Teledyne-Cougar. The most likely point of failure in the class C amplifier is a DU2820S MOSFET, which is still available and could be replaced if necessary with little difficulty. Replacement amplifiers exist but they tend to be wide band class A amplifiers, which are less efficient. In constructing new pulser/mixer modules, using class-A or class-AB amplifiers would require mounting external to the NIM module to dissipate the additional heat. Modular high performance commercial RF amplifiers are available from Amplifier Research and other vendors.

### 15.2.4 Frequency Reference

The NMR system requires a frequency synthesizer at 61.74 MHz, which is amplified and pulsed for manipulating the proton spins in the NMR probes and a 10 MHz clock for the waveform digitizers. Since the NMR clocks and the clock used by the waveform digitizers to determine the muon spin precession frequency are phase-locked to the same master clock, the variations in the master clock frequency drop out to first order, and the ppb-level accuracy requirement is reduced to be less than about $10^{-5}$. Nevertheless, to simplify the study of systematic uncertainties in the magnetic field, our goal is to have relative uncertainties in the reference clock frequencies be less than 1 ppb. This uncertainty must be achieved for short times, that is a few milliseconds for a single NMR measurement where phase noise is the limiting factor and long times over the course of the experiment.

Figure 15.6 shows how the reference frequencies are generated. The master clock, an EndRun Technologies Meridian Precision GPS Frequency Standard, is an atomic rubidium oscillator disciplined by GPS for better long-term stability. It generates a master frequency that is a sine wave at 10 MHz and an IRIG time code signal used by DAQ computers for their time stamps. The master frequency is distributed to the data acquisition system for $\omega_a$, where frequency synthesizers are used to transfer the master clock frequency to the clock frequency needed by the waveform digitizers. It is also used as input for the frequency synthesizer responsible for the NMR reference frequency $\omega_{ref}$. Frequency synthesizers in both systems should have resolutions of about 1 ppb, to allow for some variability in the frequency that can be used for blinding schemes and for tests of systematics. It is the synthesized frequencies that must fulfill the accuracy requirement discussed above. Synthesizers that fulfill the requirements given above are available from vendors including Precision Test Systems, Symmetricon, Agilent, and Stanford Research Systems. Distribution amplifiers from vendors including Stanford Research Systems and Mini-Circuits can distribute the master clock and the needed reference frequencies to the places where they are needed. There is experience in the collaboration in the specification and use of these systems from E821 and, more recently, from the clock system used for the MuLan experiment at PSI in Switzerland.

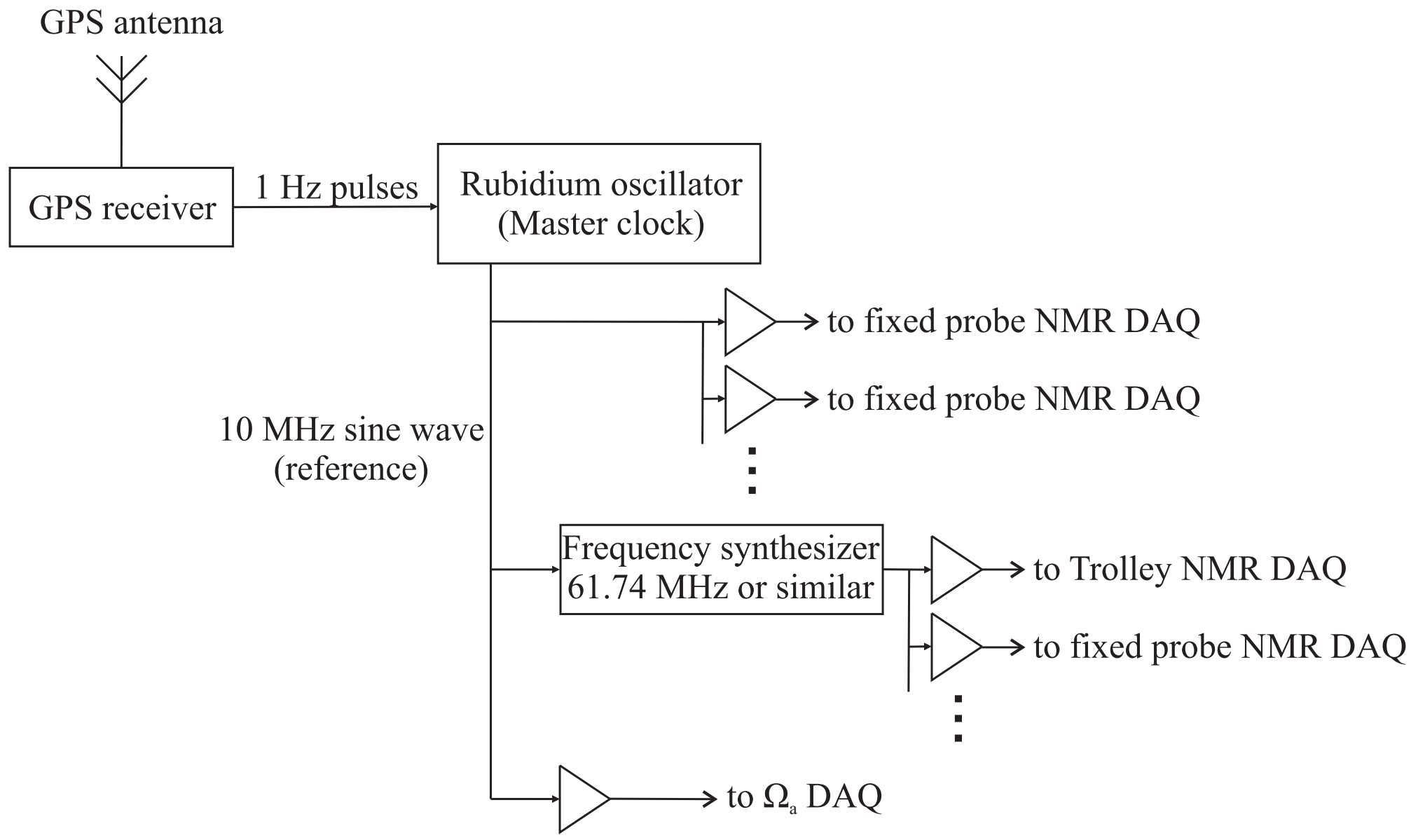

Figure 15.6: Scheme for common master-clock.

### 15.2.5 NMR VME System and DAQ

The fixed probe NMR system will be controlled from a VME crate performing similar functions as in E821 but with different hardware.

The NMR DAQ system (see Fig. 15.3) performs the following tasks: (1) Selects probes in the multiplexers, (2) Initiates Free Induction Decays (FID) in selected probes, (3) Digitizes the mixed down FID waveforms from the NIM modules, (4) Extracts frequencies $\omega_p$ from the digitized waveforms for displaying and controlling the field and (5) Records frequencies and selected digitized waveforms along with a timestamp.

Hardware for the new system includes a 6U VME 64x crate and Struck SIS1100 VME crate controller which connects via a fiber optic cable to a Struck SIS3104 PCIe card on a PC running Scientific Linux. This allows the VME crate to be placed in the experimental hall and controlled by the PC in the control room. The new controller is compatible with the VMEbus IEEE-1014 standard E821 crate, and with the faster 2eVME bus cycles of VME64x system of E989. The 5 bit multiplexer addresses are provided by an Acromag AVME9660 VME carrier card with four 48 line digital I/O daughter boards. One of these, an IP-EP201 includes a programmable Altera Cyclone II FPGA controller and clock. Acquisition code executes in the PC under LINUX, which is not a real time operating system. Critical timing requirements are achieved with the FPGA controller which will be programmed to sequence through probe addresses and trigger the waveform digitizers. The controller can also generate the appropriate length $\frac{\pi}{2}$ pulse to initiate free induction decays in the probes which can be varied to maximize the amplitude of the FID. A jumper selectable option will be added to the NIM modules to allow the width of this pulse to be controlled directly or from the default internal logic. Programming the FPGA controller requires some familarity with VHDL, its native language. Example software exists to map VME registers directly to the 48 lines of digital I/O so they may be controlled from code in the PC. FPGA sequencer code can be developed and debugged in C++ and then translated into native VHDL code for compiling and down loading to the FPGA. A custom breakout panel maps the 50 pin ribbon cable connectors on the VME module to independent DB-9 connectors for cabling to the multiplexer units. The mixed down FID waveform signals are read by 3 Struck SIS3302 8 channel 16 bit waveform digitizers to complete the complement of VME modules. These digitizers are described in section 15.2.6. A Meinberg TCR170PEX IRIG Time Code Receiver provides a timebase for the PC disciplined by the Meridian master GPS receiver. This PCIe board has a kernel driver which allows it to be used as a reference source for the NTP daemon and LINUX file system timestamp. A timestamp with accuracy better than a millisecond will be recorded with each set of NMR measurements.

Data acquisition software will be based on MIDAS, a system developed at TRIUMF and PSI that supports VME, RS232, GPIB, ethernet and USB hardware. It is used by many groups in the collaboration, integrates with ROOT for analysis and is officially supported by Fermilab under Scientific Linux. The MIDAS architecture supports multiple Front End modules that interface with the hardware and allow multiple data streams to be associated with a single event. While NMR measurements are in principle asynchronous with respect to accelerator operations, they can be synchronized to beam delivery events using available timing signals from the accelerator. The FPGA sequencer can vary the start delay so measurements can be made in quiet times as well as during beam spills and kicker events. Field measurements average the field over 5 ms. Measurement start times can be varied relative to the 17 millisecond period of ambient 60 Hz fields for a high precision determination of their contribution to the ring field.

In E821, the DL611 modules, which are described in [5], counted the number of zero crossings, $k$, of the FID until its envelope represented by the FIDE signal fell below an adjustable threshold or until a maximum time had elapsed. Over the same interval the ticks, $N$, of an external stabilized clock at $\nu_{\rm clock}$=20 MHz were counted. For both counters the start and stop coincided with a positive zero crossing of the FID. The frequency of the FID is $\nu_{\rm FID} = k \times \nu_{\rm clock}/N$. For FID signals exceeding 1 ms, the uncertainty on the frequency due to the discretization of the clock counts is roughly 8 ppb or less. The uncertainty on the frequency due to the finite signal-to-noise of the FID (typically $S/N \approx 100$ when FIDE falls below threshold) is given roughly by $\sigma_s/(S/N)$ where $\sigma_s$ is the signal line width, and $S/N$ is the signal-to-noise ratio. This is typically of order 25 ppb on a single NMR FID above threshold for a millisecond or more [5]. This motivated the decision to replace the DL611 hardware with waveform digitizers in E989.

A second MIDAS DAQ system will be assembled around E821's VME crate and DL611 zero crossing counters. The crate and some modules are known to work and code to access them has already been written. Given the modular nature of MIDAS an additional Front End module developed from the existing code base is all that is required to add the DL611s to the event stream. With this system in place, frequency readouts using both waveform digitizers and DL611 hardware can be compared.

### 15.2.6 Digitizer

With the DL611 hardware, the precession frequency is obtained directly from internal registers in the module. A more precise frequency determination can be achieved by first acquiring FID waveforms with fast, low-noise digitizers, such as the Struck SIS3302. These VME modules can simultaneously digitize 8 independent channels (8 multiplexers) with 16-bit ADCs (13 effective bits). The digitizers get their signals from the FID outputs on the NIM Pulser and Mixer modules, and sample at a frequency of 10 MHz for roughly 5 ms. Digitized waveforms require additional processing for frequency determination and various algorithms were explored for frequency extraction. One of these is a software version of the DL611 hardware where the digitizer's sample rate replaces the external clock, but with enhanced zero-crossing resolution, below the clock period, by interpolating between points on the waveform. Other methods use an FFT algorithm and extract frequencies from peaks in the power spectra. The most robust method found performs a 90 degree rotation in the complex frequency domain followed by an inverse FFT to transform back to the time domain which gives a function orthogonal to the original FID. The arctangent of this orthogonal pair gives the instantaneous phase advance of the precessing protons. With homogeneous fields, there is a linear relationship between phase and time in the FID, and frequency is just the slope of this line (from a least squares fit).

Figure 15.7 Compares the frequencies extracted from simulated FID waveforms by counting zero crossings with the method using continuous phase advance. Simulated waveforms have identical frequencies and differ only in Gaussian noise. The $x$ scales extend from -50 ppb to +50 ppb on both plots. The widths of the histograms represent the precision by which each method can extract the frequency in the presence of noise. The width from counting zero crossings (left) is about an order of magnitude larger and like the actual DL611 hardware, is dominated by uncertainties in the times of the last zero crossings where the signal to noise ratio is smallest. These plots show the advantage of using the entire digitized FID waveform in determining the frequency.

Linear gradients in the magnetic field over the probe volume reduce the precision of frequency measurements by decreasing the length of time the protons precess coherently ($T_2^*$). With non-linear gradients the lineshape is asymmetric and the relationship of phase with time is no longer linear so the instantaneous FID frequency varies with time. The average field, which is the field seen by the muons, corresponds to the centroid of the line. This can be determined by extrapolating the instantaneous FID frequency to its value at t=0 [13]. The FID frequency variation can be of order 15 ppb [13], and is relevant primarily for the absolute calibration. Digitization of FIDs from the abolute calibration probe is essential to reduce possible systematic uncertainties on the $B$ field measurement from lineshape asymmetries and field inhomogeneities.

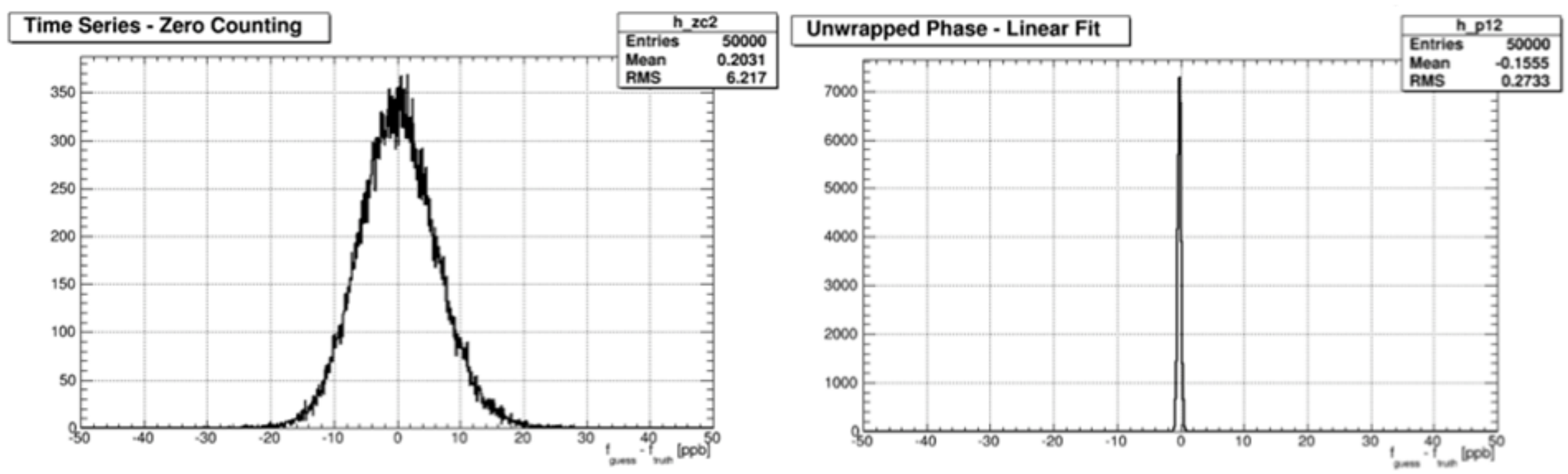

Figure 15.7: The figure shows histograms of frequencies extracted from simulated free induction decays with added Gaussian noise. In the left plot frequencies are determined from counting zero crossings (a software implementation of the DL611 hardware counters used in E821). In the plot on the right frequencies are determined from the phase advance of precessing protons vs. time in the FID waveform.

The digitizers will only provide these benefits if the digitization noise is comparable or less than the noise on the signal. For E821, FID signals were of the order of a volt. When the FID envelope (FIDE) dropped below threshold at $\approx$ 100 mV, the $S/N$ of 100 implies signal noise was of order 1 mV. For the SIS3302, with a -1V to 1V full scale range, and minimum 13 effective bits, the digitization noise is predicted to be 0.24 mV, safely below the FID noise.

With a sampling frequency of $10 \times 10^6$ samples/s, 16-bit resolution, and a period of $5 \times 10^{-3}$ s, the waveform from a single probe produces $\approx$ 100 kB of data. Muons are injected into the storage ring in the first 0.5 seconds of the 1.33 second beam supercycle. This leaves a quiet period of 0.82 seconds without field perturbations from the kicker field. Reading 400 probes once in half of the beam cycle would require a data rate of about $2 \times 30$ MBytes/s (40 MBytes/1.33s) which is well within the 80 MBytes/s bandwidth that can be achieved with VME64x crates. Varying the start time in the cycle allows the probes to be read in both quiet and kicker periods to study residual effects of the kicker fields. In addition, the T1 relaxation time for Petrolatum is about 40 ms permitting probes to be read out every 0.4 seconds. Acquiring waveforms continuously for a year at this rate would consume about a petabyte of storage. There would be no need to store all digitized waveforms. After the frequency information has been extracted, a smaller subset of the digitized FIDs would be retained for off-line analysis. Recording waveforms once per minute, for example, would reduce the required yearly storage to 17 terabytes, an amount easily obtained on todays modern servers. The actual rate would be determined by the needs of the analysis.

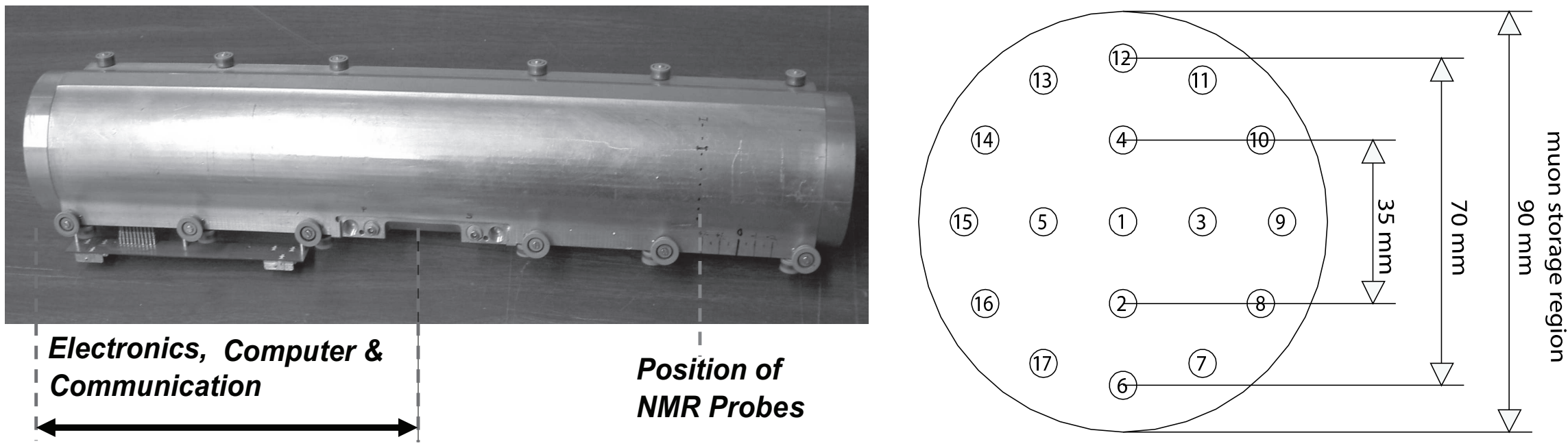

(a) NMR Trolley (b) Distribution of NMR probes

Figure 15.8: (a)Photograph of the $\sim$ 50cm-long NMR trolley, which measures the magnetic field in the storage ring. The array of 17 NMR probes, which are located inside the trolley housing, are 82(1) mm behind the front of the trolley. Electronics occupies one end of the device. At the other end where the probes are located, the field perturbation from the electronics is less than 2 ppm and is accounted for as part of the calibration procedure. (b) The probe numbers and placement are given by the schematic.

We also note that a high performance digitizer is also useful in analyzing the lineshapes of the calibration probes. These lineshapes contain information about perturbations due to probe materials, the amount of water in the thin neck of the spherical water sample and other effects. This information can be used to reduce the uncertainties in the absolute calibration.

## 15.3 Trolley

The trolley performs the second major task of the field measurement system in Sec. 15.1.5: it determines the magnetic field distribution over the muon storage volume around the ring intermittently when the beam is off. It uses an in-vacuum NMR trolley system developed for E821 at the University of Heidelberg [9], shown in Fig. 15.8(a). This trolley contains 17 NMR probes arranged in concentric circles as shown in Fig. 15.8(b). Each probe measures the field at several thousand points around the ring. The trolley was designed and built to minimize distortions of the magnetic field in the ring. On board is a fully functional CPU which controls the FID excitation and FID zero crossing counter. Additional sensors connected to the microcontroller are used to measure position, pressure, and temperature. The trolley is pulled around the storage ring by two cables, one from each direction circling the ring. One of these cables is a thin co-axial cable with only copper conductors and Teflon dielectric and outside protective coating (Suhner 2232-08). It carries simultaneously the dc supply voltage, the reference frequency $f_{\mathrm{ref}}$ and two-way communication with the spectrometer via RS232 standard. The other cable is non-conducting nylon (fishing line) to eliminate pickup from the pulsed high voltage on the kicker electrodes.

**Magnetic Field Maps**

From the trolley field measurements, the multipole composition of the field averaged over the ring azimuth is extracted, then folded with the multipole expansion of the measured stored muon beam profile. Because muons cannot be stored while the trolley is in the storage volume, trolley field maps must alternate with periods in which muon spin precession data are taken. Over the duration of E989, trolley runs will be executed on intervals of non-integer diurnal cycles such that the overall distribution of trolley runs versus time is uniform. This reduces correlations between trolley runs and possible biases from day/night or other potential periodic changes in the storage ring field.

During mapping, the trolley is moved into the storage ring and pulled continuously clockwise or counterclockwise through the entire storage volume over the course of roughly 2 hours. The field is sampled at 6000 locations in azimuth by each of the 17 probes (which are cycled through continuously) for a total of 100,000 field points. During these runs, a cross-calibration of the field observed by the fixed probes and the field measured by the trolley probes is performed. This is required to determine the magnetic field encountered by the stored muons in the storage volume from the measurements of the surrounding fixed probes taken at the same time. Because this cross-calibration between the trolley measurements and fixed probes will be slightly different each time the magnet is powered up or down (because the field shape changes slightly), trolley runs must be taken every time the magnet current is changed. This cross-calibration is sensitive to magnet temperature and current, both of which also change the field shape slightly. Improved insulation of the magnet and experimental hall floor, as well as the more uniform thermal environment in the E989 experimental hall versus E821 should keep the magnet temperature more stable and uniform. This will also reduce the magnitude of changes to the current required to stabilize the field. Reductions in both factors will reduce changes in the cross-calibration and allow better tracking of the storage ring field between trolley runs. Note that these changes in cross-calibration can be determined in advance of data-taking by measuring the difference in fixed probe and trolley probe measurements as the current in the magnet is changed deliberately, and as a function of the experimental hall/magnet temperature. Finally, the frequency of trolley runs will be adjusted and additional insulation can be used to ensure that the uncertainty on the field tracking goals are met.

The performance of the system can be gauged from E821. The magnitude of the field measured by the central trolley probe is shown as a function of azimuth in Fig. 15.9 for one of the trolley runs in E821. The inset shows that the fluctuations in this map that appear quite sharp are in fact quite smooth, and that the noise is small in comparison.

Since the NMR frequency is only sensitive to the magnitude of $B$ and not to its direction, in addition to the main vertical field $B_y$, in principle there could be significant radial fields and fields in the azimuthal direction. In practice, the magnitude of radial field will be measured (see section 15.8.2), pole tilts will be measured with a precision electrolytic tilt sensor and adjusted, and surface correction coils will be used to ensure the local radial field is less than 50 ppm. In E821 the average radial field achieved was $20 \pm 10$ ppm [52]. Longitudinal components of the field will be similarly restricted, especially by improved shimming. Hall probe magnetometers will measure the radial and longitudinal field components prior to vacuum chamber installation, and the radial field will still be accessible via vacuum ports subsequent to chamber installation. This iterative sequence of measurements and corrective actions should ensure the difference between the azimuthally-averaged $|\vec{B}|$ and $|B_y|$ is less than 10 ppb.

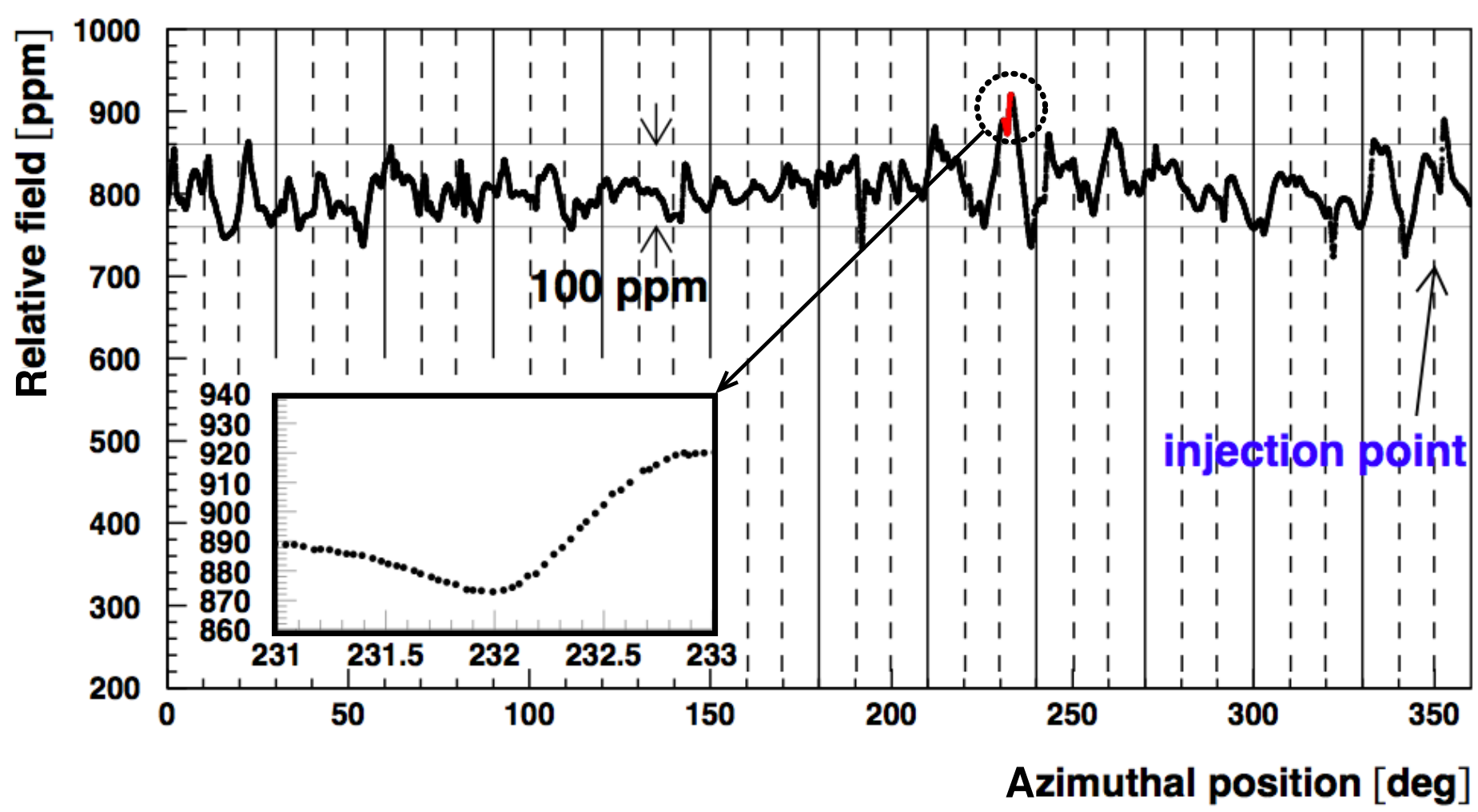

Figure 15.9: The trolley measurement of the magnetic field at the storage of the storage region vs. azimuthal position. Note that while the sharp fluctuations appear to be noisy, when the scale is expanded the variations are quite smooth and represent true variations in the field.

Since the storage ring has weak focusing, the average over azimuth is the important quantity in the analysis. This is achieved using the trolley field maps by averaging the trolley probe measurements over azimuth. A contour plot for an azimuthally-averaged field map from E821 is shown in Fig. 15.10(b). Such azimuthally-averaged field maps are then expressed in a two-dimensional multipole distribution over the radial and vertical directions, $x$ and $y$. It is more natural to use cylindrical coordinates $(r, \theta)$ where $r = 0$ is the center of the storage region and $\theta = 0$ points radially outward from the center of the ring. This yields the expansion for the field components:

$$B_y(r,\theta) = \sum_{n=0}^{n=\infty} \left(\frac{r}{r_0}\right)^n [\quad a_n \cos n\theta + b_n \sin n\theta] \tag{15.5}$$

$$B_x(r,\theta) = \sum_{n=0}^{n=\infty} \left(\frac{r}{r_0}\right)^n [-b_n \cos n\theta + a_n \sin n\theta]\,, \tag{15.6}$$

where the normal multipoles, $a_i$, and skew multipoles, $b_i$ are normalized at $r_0 = 45$ mm. In practice the $B_y$ series is dominated by the dipole term $a_0$ and the expansion is limited to $n \leq 4$. The radial field distribution $B_x$ is dominated by the radial dipole term $b_0$ which is measured separately with a Hall probe and is of order 20 ppm. NMR is nominally sensitive to the field magnitude, but effectively measures just the magnitude of the vertical field $|B_y(\vec{r})|$ since the orthogonal field components, which are of order 50 ppm or less, perturb the field magnitude by less than a few ppb.

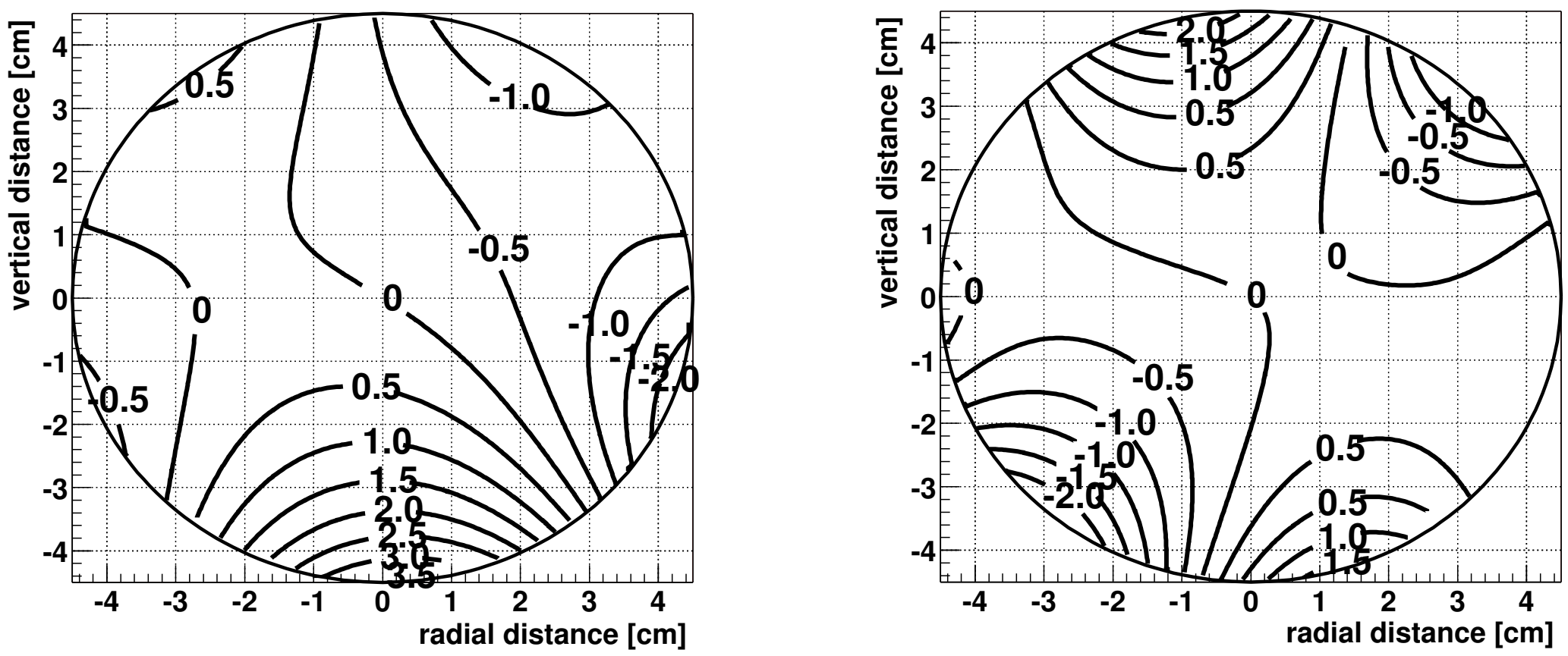

(a) Calibration position (b) Azimuthal average

Figure 15.10: Homogeneity of the field (a) at the calibration position and (b) for the azimuthal average for one trolley run during the 2000 period. In both figures, the contour spacing is 500 ppb.

An example of the multipole decomposition from BNL E821 is shown in Table. 15.2.

Table 15.2: Azimuthally-averaged multipoles at the outer edge of the storage volume (radius = 4.5 cm) obtained in E821. The new experiment aims for a factor of two or better reduction in the average multipole amplitudes and also better azimuthal uniformity.

| Multipole [ppm] | Azimuthal Average Normal | Skew |
|---|---|---|
| Quadrupole | 0.24 | 0.29 |
| Sextupole | -0.53 | -1.06 |
| Octupole | -0.10 | -0.15 |
| Decupole | 0.82 | 0.54 |

**Fixed probe tracking of the magnetic field**

During data-collection periods the field will be monitored with the fixed probes. To determine how well the fixed probes allow monitoring the field in the storage ring acting on the muons, the field average determined by the trolley, and that predicted by the fixed probes will be compared for each trolley run. The results of this analysis for the E821 2001 running period is shown in Fig. 15.11. Disregarding hysteresis shifts identified in the figure with vertical lines, the RMS distribution of these differences is ∼100 ppb.

The uncertainty in interpolating the storage ring field between trolley runs, by using the fixed probes, will be reduced from 70 ppb achieved in E821 to 30 ppb in E989 (see Table 15.1). The improvements are the result of several changes. First, the E989 experimental hall floor will be monolithic and much more mechanically stable than the E821 floor which was composed of three concrete sections. Second, the E821 experimental hall had very poor temperature control (day-night changes of nearly 4°C were common), poor insulation, and large changing temperature gradients across the magnet. In E989, temperature stability and uniformity in the hall is a high priority. The HVAC system will hold the hall stable and uniform to $\pm 1^\circ$ C during data collection periods. This is at least a factor two more stable than E821. Also, the thermal insulation around the magnet will be improved, and heat flow from gaps around the magnet coil cryostats will be eliminated. These major improvements reduce the temperature-induced changes in the field by a factor of two.

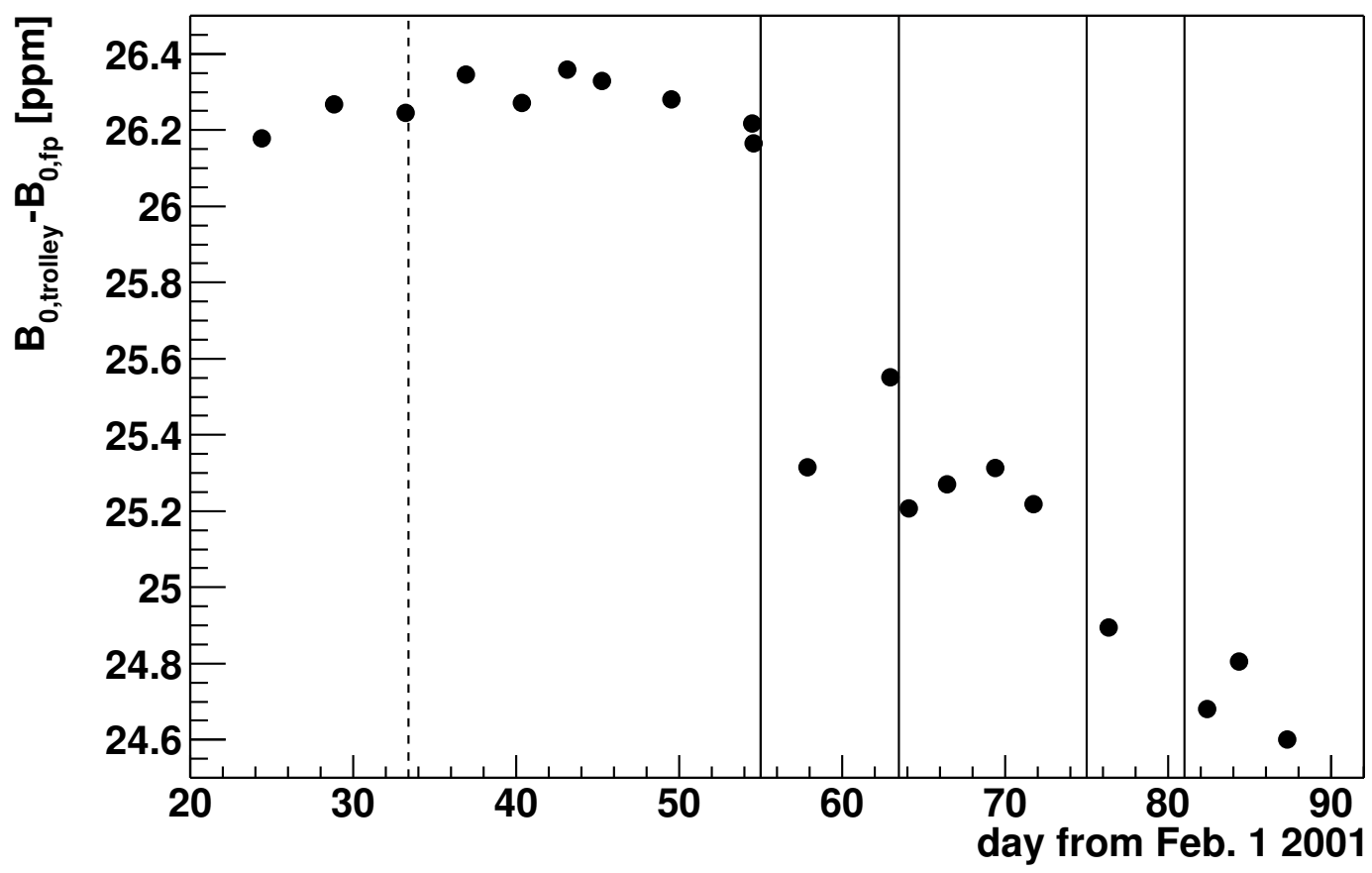

Figure 15.11: The difference between the average magnetic field measured by the trolley and that inferred from tracking the magnetic field with the fixed probes between trolley runs. The vertical lines show when the magnet was powered down and then back up. After each powering of the magnet, the field does not come back exactly to its previous value due to hysteresis, so that only trolley runs taken between magnet powerings can be compared directly.

The RMS difference in field average predicted from the fixed probes and that measured by the trolley was roughly 100 ppb in E821, where trolley runs were taken approximately every 3 days. For the same time interval between trolley runs in E989, this RMS difference should be reduced by a factor 2 given the improvements in the hall floor, HVAC, and insulation. Assuming the improvements only reduce the RMS difference to 70 ppb, the goal of 30 ppb on field interpolation with the fixed probes can be achieved by more frequent trolley runs. In E821 the interval between trolley runs was roughly 3 days, and did not need to be more frequent since the experiment was statistics limited. For E989, this interpolation uncertainty must be reduced significantly compared to E821. Trolley runs which are three times more frequent will reduce the interpolation uncertainty by approximately another factor of $\sqrt{3}$. Improvements in the fixed NMR probes as well as the trolley NMR probes (see Section 15.3.6) should render them less sensitive to temperature variations. OPERA studies of the magnet indicate a small sensitivity of the difference between the fixed probes and trolley probes to small changes in the magnet current. This is due to nonlinearity in the $B-H$ curves of the yoke and pole pieces, leading to slight changes in the field shape. In E821, the current in the magnet was adjusted based on feedback from the NMR. In E989, due to improved climate control the magnetic field should be much more stable passively, and the feedback corrections (changes to the current) will be smaller. The effect of slight changes in the magnet current on the difference between trolley probes and fixed probes can be measured and compared with the OPERA model. The results of these studies and changes to the field due to temperature changes should allow corrections to be made to the fixed probe readings that incorporate changes to both the magnet temperature and the magnet coil current. The ensemble of these efforts should combine to further reduce the 100 ppb RMS uncertainty in E821 to 30 ppb in E989.

**Uncertainty from the muon distribution**

In the simplest approach, the value of $\omega_p$ entering into the determination of $a_\mu$ is the field profile weighted by the muon distribution. The multipoles of the field, Eq. (15.6), are folded with the muon distribution,

$$M(r,\theta) = \sum[\gamma_m(r)\cos m\theta + \sigma_m(r)\sin m\theta], \tag{15.7}$$

to produce the average field,

$$\langle B\rangle_{\mu-\mathrm{dist}} = \int M(r,\theta)B_y(r,\theta)rdrd\theta, \tag{15.8}$$

where the moments in the muon distribution couple moment-by-moment to the multipoles of $B_y$ (expressed in terms of the free proton precession frequency). The determination of $\langle B\rangle_{\mu-\mathrm{dist}}$ is more accurate if the field is quite uniform (with small higher multipoles) so the number of terms is limited, and the muons are stored in a circular aperture, thus reducing the higher moments of $M(r,\theta)$. This simple approach to determining a weighted average requires correction because closed-orbit distortions perturb the radius of the stored muon beam around the storage ring azimuth.

In E821 the weighted average was determined using two techniques. One used a muon tracking calculation and a field map to determine the field seen by each muon. The second determined the average field from the dipole and quadrupole components of the magnetic field coupled with the beam center determined from a fast-rotation analysis. These two agreed extremely well, validating the choice of a circular aperture and the $\pm 1$ ppm specification on the field uniformity that were set in the design stage of the experiment. Part of the E821 muon distribution uncertainty came from radial and vertical offsets of the beam. These offsets couple with normal and skew quadrupole moments of the field. For E821, these corrections were of order 12 ppb for the skew quadrupole (limited by the RMS scatter of the skew quadrupole), and 22 ppb for the normal quadrupole. The latter was limited primarily by a lack of knowledge of the muon beam radial position on a run by run basis. Sextupole and higher skew multipoles of the beam were less than $10^{-3}$ in E821 and did not require correction. No correction was made for the 11 ppb normal sextupole field contribution in E821. These techniques worked quite well in E821, and the uncertainty on $\langle B \rangle$ weighted by the muon distribution was conservatively estimated as $\pm$ 30 ppb [7].

In E989 we anticipate several improvements. First, the muon beam distribution will be monitored every fill with higher precision by the fast rotation analysis and new muon trackers. Second, improved field uniformity and improved magnet stability (so the multipole decomposition is more stable) will couple with the improved knowledge of the beam. The combined improvements will reduce the normal and skew quadrupule uncertainties, which were 12 ppb and 22 ppb in E821, to below 10 ppb for E989. By measuring and correcting the normal sextupole contribution to 20% or better of its value (which was left as a 11 ppb uncertainty for E821), that contribution can be made negligible. The effects due to these higher multipoles should also be reduced due to higher field uniformity and stability, and by moving the outer trolley probes to a slightly larger radius. Measurements of the higher multipoles will also be made before the vacuum chambers are installed with the shimming trolley (discussed below), and the influence of the vacuum chambers on the field will be determined. The increased knowledge of the field and muon positions around the ring afforded by these system upgrades will be fully exploited with full tracking calculations in the analysis.

To summarize, improved field homogeneity and stability, as well as improvements in muon tracking techniques and efficiency afforded by considerable advancements in computing power since E821 should allow E989 to achieve 10 ppb uncertainty on the muon distribution contribution to the determination of $\omega_p$.

## 15.3.1 General trolley requirements

In the following sections, we will first specify the requirements on the trolley for E989 based on the field mapping tasks outlined above, then discuss the conceptual design of future upgrades and efforts related to the trolley system in detail.

As can be seen from Table 15.1, trolley related systematic errors in the BNL E821 experiment were significant and require improvement to meet the physics goals in E989. The two main sources of uncertainty stem from the calibration procedure of the trolley probes to the plunging probe (90 ppb) and errors related to position uncertainties during the actual trolley runs (50 ppb). Additional smaller effects (like temperature or voltage drifts) were grouped into one systematic error (Others) together with non-trolley related systematics in the field measurement. For the new E989 experiment, the trolley system will be used in a very similar fashion to E821. Given the required improvements in the overall systematic errors for the field measurement, we will need some changes for the new system.

#### Trolley measurement of $B_0$

In E821, the uncertainty on the trolley measurement of $B_0$ was 50 ppb (see Table. 15.1), due primarily to nonlinearities in the trolley position readout. This 50 ppb uncertainty will be reduced to 30 ppb in E989 by mitigating these nonlinearities with use of a barcode reader system. The uncertainty of this improved system will be less than 5mm, a considerable improvement over E821. Second, uncertainties due to trolley rail irregularities will be reduced, as discussed below. Third, a more sophisticated feedback algorithm to stabilize the magnet during trolley runs will be developed. Fourth, the uncertainty on the field integral in azimuth depends on position uncertainty coupled with field gradients. In E989 we aim to reduce the field gradients in azimuth by a factor of two compared to E821. These steps should reduce the uncertainty on the trolley measurement of $B_0$ to 30 ppb in E989, as discussed in detail below.

The requirements for the actual measurement during a single trolley run remain the same as in E821. An individual NMR frequency measurement will have a precision at least as good as 20 ppb as was achieved in E821. The field will be measured at 6000 points around the ring for each probe. A single trolley run should be accomplished in at most two hours. While about 1 hour is required for the mapping of the 6000 data points for each probe, the return trip can be sped up to reduce the interruption of the spin frequency measurement. Trolley runs should be repeated more frequently than in E821 where an interval of 2-3 days was typical. Increased frequency of trolley measurements will reduce the error associated with the fixed probe interpolation and reduce uncertainties associated with temperature changes in the storage ring.

As mentioned above, uncertainties in the transverse position of the trolley and longitudinal nonlinearities coupled with field gradients are a major systematic error category in the measurement. While we plan on having improved overall shimming in E989 and hence reduced gradients, we will also put effort into reducing the position uncertainties. In E821 the longitudinal position of the trolley during field mapping was inferred from two sources. First, the unwinding of the trolley cable was measured by optical rotary encoders in the drums (see Fig. 15.13). Second, the observed change in the NMR frequency in the fixed probes due to the small but measurable changes in the magnetic field induced by the trolley (maximally in the vicinity of the onboard electronics). The optical encoder read out diverged from perfectly linear behavior maximally by a few cm, and this contributed a 50 ppb systematic uncertainty on $\omega_p$. Together with improved azimuthal field homogeneity (see section 15.8), a more precise longitudinal position resolution in E989 will reduce this uncertainty to negligible levels compared to the overall error on $\omega_p$.

During its movement, the trolley rides on two rails, which determine its transverse position with respect to the center of the muon distribution. The rails were not continuous at the junctions between adjacent vacuum chambers. Slight misalignment of neighboring rails at these gaps led to possible tranverse deviations of the trolley's center introducing an estimated systematic error of 10 ppb. In E989 we will reduce these misalignments and precisely verify the deviation of the rails from their nominal position around the full azimuth. Here, we will employ two methods, namely (1) optical survey with precision laser tracking and (2) the introduction of known (measured), transverse gradients by means of the pole surface coils during dedicated special trolley runs. Further considerations related to uncertainties due to the knowledge of the trolley's transverse position are discussed in Section 15.3.5.

During the calibration, relative position uncertainties between the plunging and trolley probes contribute to the overall systematic error. In E821, the transverse reproducibility was estimated to be 1 mm, whereas the relative position uncertainty in azimuth was determined to 3 mm. Together with field gradients, this resulted in systematic error contributions of 20 ppb and 30 ppb, respectively. For E989, we aim to cut these contributions in half (at least in the azimuthal direction) by more precise alignment of the probes' active volumes in a repeatable way. For this, we plan to allow positioning of the plunging probe in all three directions and perform careful testing of the calibration transfer from the absolute calibration probe to the trolley in a homogeneous solenoid test magnet (see Section 15.3.9) before data taking. Additionally, precisely known gradients at the calibration station will be induced with the surface correction coils to allow systematic studies of the probes' active volumes.

Other effects contributed to the E821 measurement of $B_0$ with a summed contribution of 50 ppb. These included the temperature and trolley power supply voltage dependence of the NMR measurement and an estimate of the influence of higher multipoles. The dependence of the FID frequency measurement on the supply voltage of the trolley was measured to be 270 ppb/V and a voltage stability of 50 mV was achieved. In the new experiment, a modern power supply will significantly reduce the voltage drift and make this contribution totally negligible. We will also reduce the effects of external temperature changes on the extracted NMR frequency. The new NMR probes will be less sensitive to temperature variations. In addition, an increase in the heat dissipation of the trolley would reduce the change of its temperature during the measurement. Overall, temperature-related effects should be much smaller in E989, and we can carefully study them in the solenoid test magnet under controlled temperature conditions.

It should be noted that any additional modifications must not disturb the field observed by the trolley NMR probes at an appreciable level. The maximal distortion of the field caused by the trolley electronics in E821 was about 2 ppm and future changes cannot introduce any major additional magnetic contribution compared to this level.

In the following sections, we will describe the baseline design for future upgrades and activities that are aimed to meet the above outlined requirements for the E989 experiment.

### 15.3.2 Garage

The trolley garage shown in Fig. 15.12 is attached to one of the vacuum chambers and serves the purpose of storing the trolley inside the vacuum outside the muon storage aperture during the main periods of muon spin precession measurement. A set of threaded rods driven by a non-magnetic piezo motor provides the mechanism to move cut-outs of the rails into the muon storage region and retract them.

The system has been tested recently at Argonne National Laboratory and preliminary findings indicate the mechanical system is in good health and will not require repairs or major changes. More rigorous testing will be performed to understand if this level of mechanical integrity can be relied upon for the lifetime of the experiment. The lack of electronic motion stops at both end positions of the rails has possibly put stress on the mechanics over the course of the E821 operation. Adding position indicators at these two extreme positions of the garage with feedback into the new motion controller system (see Section 15.3.4) will allow for smooth stopping of the rails following translation from or to the parked position. We will use slotted optical sensors as position stops. Since this equipment will be housed near the storage volume, the elected option must comply with the non-magnetic and low-current requirements so that the magnetic field homogeneity is not disturbed. This will be verified with a combination of studies in the test magnet (see section 15.3.9) and calculations. In addition, the piezo-electric motor will be replaced with its successor model from the same company. Spare motors for both the garage and drive mechanism will provide redundancy over the life of the experiment.

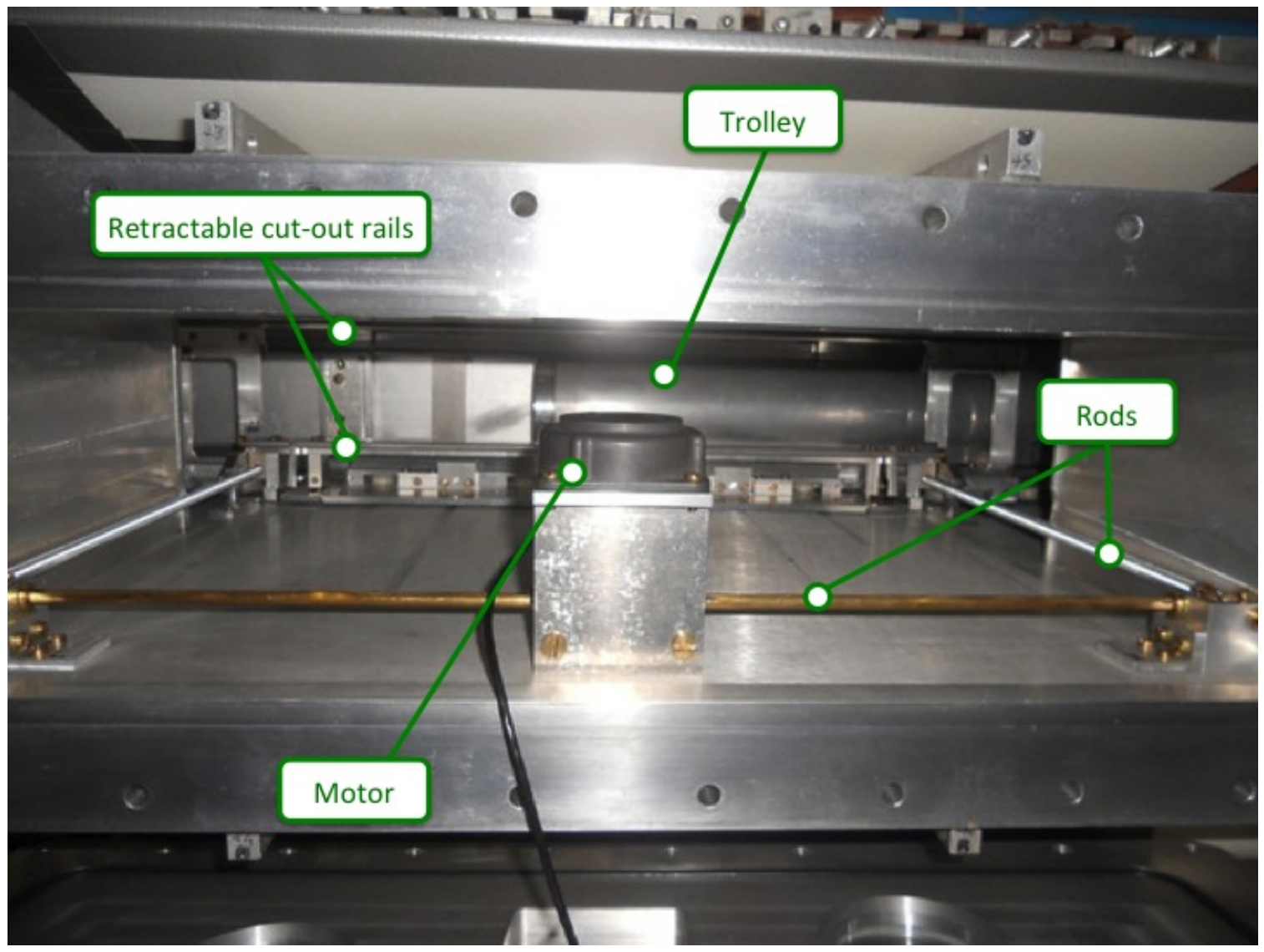

Figure 15.12: Trolley garage with the piezo motor, the driving rods, the rails and the trolley partially in the parking position where it can be retracted from the storage region.

As can be seen in Figure 3.5, the trolley garage is attached to one of the twelve vacuum chambers. Since the upgrades to other systems (such as the alignment of the trolley rails) will require work on the vacuum chambers, upgrades to the garage will be coordinated with these other activities over the next two years.

### 15.3.3 Drive

The trolley drive mechanism shown in Figure 15.13 is located about 120° away from the garage. It is connected to one of the vacuum chambers via a long port and sits on the inner side of the storage ring. It will be moved from its previous 1 o'clock location to a new 9 o'clock (see Fig. 11.1) location to accommodate the incoming muon tracker station. The drive contains two cable drums that are each driven by a piezo-electric motor. Two cables are required to pull the trolley a full 360 degrees in both directions during the NMR measurement of the storage field and the return trip. Since the cables remain attached to the trolley during its storage in the garage, one of the two cables passes through the kicker region. To prevent any damage to the onboard trolley electronics from electronic pickup on the cable from the kicker pulses, this cable is a non-conducting fishing line. The other cable is an all-copper double-shielded cable with an outer coating suitable for in-vacuum operation. This cable provides the power, reference frequency signal and the communication with the trolley microcontroller. The drive mechanism also incorporates one optical rotary encoder for each drum to monitor the angular motion and tension sensors measuring the pulling force on the cables.

All relevant mechanical and electrical components of the trolley drive have been tested individually and are in good health. The electrical components, their functions and planned

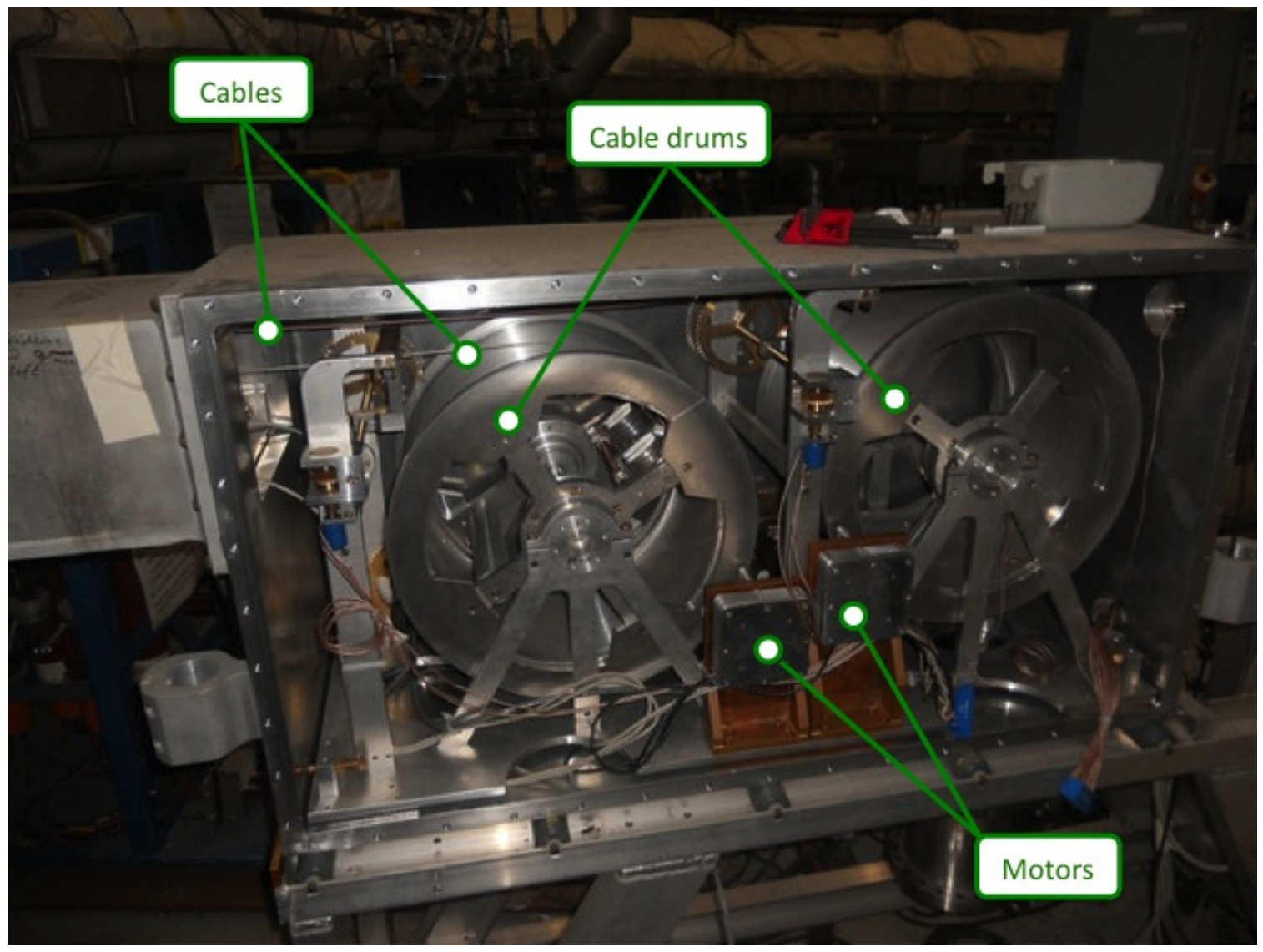

Figure 15.13: Trolley drive with the cable drums, motors and cables.

upgrades are described below:

- **Two piezo motors and drivers**: One motor controls the motion of each drum that houses either the coaxial cable or the non-conducting fishing line. These motors and their drivers will be upgraded to the most recent and supported commercial model. This upgrade will allow for a quicker return trip of the trolley and therefore an improved duty cycle for the muon spin precession measurement.

- **Two optical encoders**: These sensors record the rotation of each large drum. As they only communicate via RS232, we plan to replace them with units that can readily communicate with the new motion controller (see section 15.3.4) via the modern Synchronous Serial Interface (SSI) protocol. Due to the large size of the drum, the encoder position readings are not perfectly linearly related to the motion of the trolley. However, this system will only serve as a cross-check to the trolley's primary longitudinal position determination with a barcode reader (see section 15.3.5) and to verify the trolley's movement is as expected.

- **Two tension sensors**: These units monitor the force on a guiding wheel imparted by tension on the trolley cables. This will be used as-is in a feedback system with the motion controller to reduce the motor speed when the cable tension exceeds a defined threshold. The threshold will be set far below the point where any of the two cables could disconnect from the trolley which would require access to the vacuum chambers and cause a longer downtime during experimental data taking.

The above components have all been successfully tested independently, and in E989 the full system will be operated and moderated by the 8-axis commercial motion controller unit DMC-4183 from Galil as described next.

### 15.3.4 The new motion control system

The control of the piezo-electric motors and readback of associated sensors for the trolley garage, drive and the plunging probe was based on the Siemens SAB80C535 microcontroller in E821. As this microcontroller is obsolete today, we will replace the control system with one single 8-axis Galil DMC-4183 motion controller[4]. This new central unit is schematically depicted in Figure 15.14 and will interface with all five motors from the drive (2 motors), garage (1 motor), and the plunging probe (2 motors). Feedback from the additional sensors will allow to program appropriate operation for each system. Two additional axes are spares for future extension or as replacement in case of failure of one of the five motor channels.

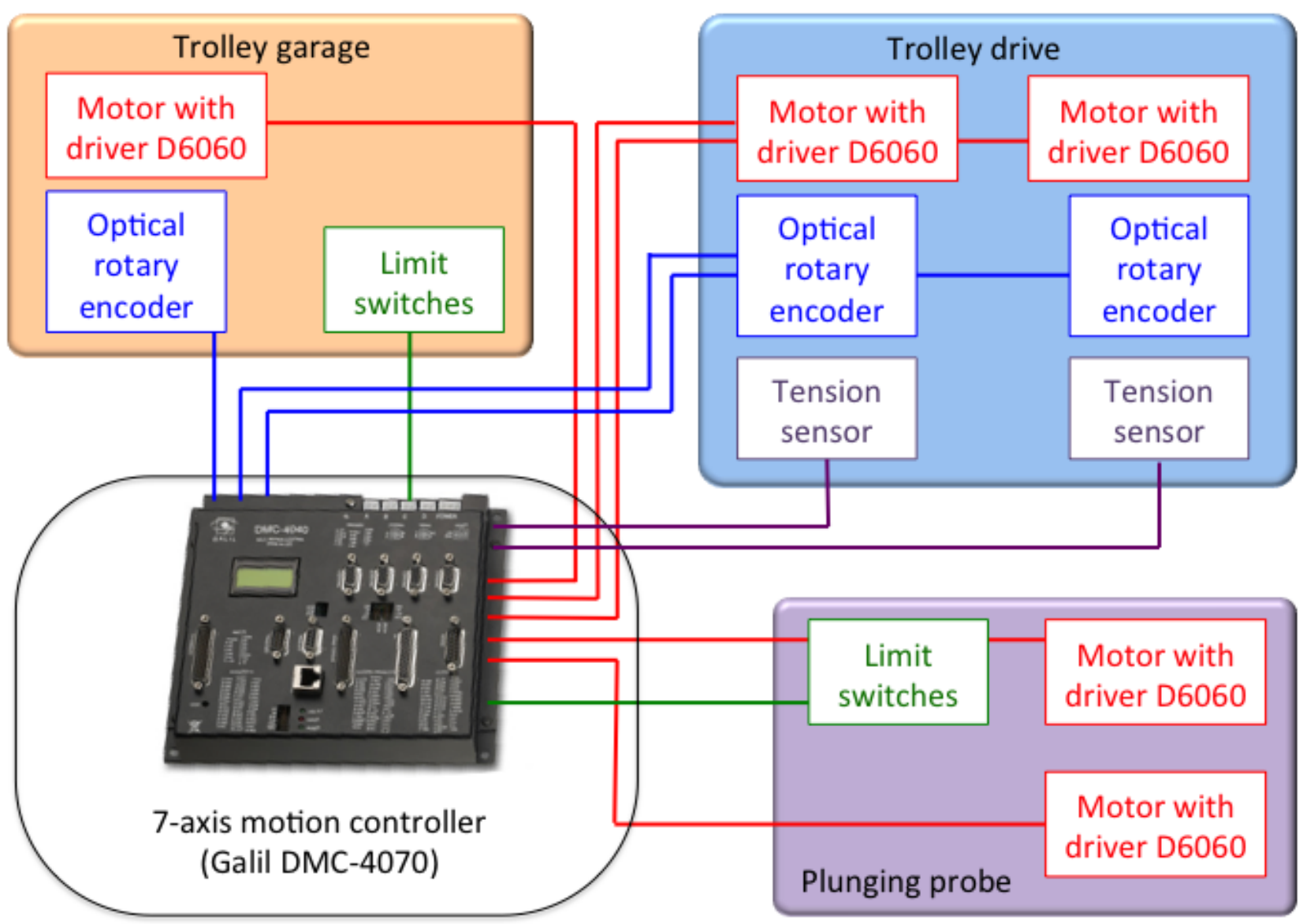

Figure 15.14: Schematics of the new integrated motion control system for the trolley garage, drive and the plunging probe. Centeral unit for the control is a 8-axis Galil DMC-4183 motion controller which also interfaces to the various feedback sensors.

The Galil motion controller was chosen based on its wide range of applications with motors and encoders and in-house experience with this specific device. The employed Shinsei non-magnetic piezo-electric motors (USR60) require a DC voltage of 12V (max. 4A current) which will be provided by separate power supplies. The control of the motor's spinning direction requires two TTL signals (clock- and counterclock-wise) and a variable DC voltage (0 - 3.2 V) sets its speed. Each axis on the Galil provides such signals in form of a $\pm 10$ V signal which can be split by a small adapter board built in-house to generate the two required signals for the driver. Replacements of the current 24-bit absolute rotary encoders will be interfaced by the motion controller by ordering its interconnect module with an available SSI option. The possibly added incremental encoder on the garage motor comes with an

[4]http://www.galilmc.com/products/dmc-41x3.php

ABZ-phase output which can be processed by the Galil DMC-4183 by default. Each axis of the motion controller has two limit switch inputs for the forward and reverse direction of the associatied motor. Their activation will immediately inhibit the current motion of the motor. This feature will be used for the trolley garage where we add new limit switches to allow for smooth stopping at the end positions of the rails and the plunging probe mechanism which is already equipped with such limit switches.

Once the motion controller has been purchased, we will start the implementation of the programming in conjunction with the integration of the above mentioned hardware. While the plunging probe is a system by itself, the garage and drive mechanisms can be interlinked in this new control scheme. Parking and restoring the trolley into the storage area requires the simultaneous movement of the garage motor as well as appropriate (un-)winding of the cables. In the implementation phase, we will program such integrated motion control for these systems and test their functionality with a mock setup of one vacuum chamber with the garage, the trolley drive and a dummy trolley. In the initial phase, programming of the Galil system will be facilitated by using the free GalilSuite software from the vendor. In-house experience at Argonne will help to develop a first version of a basic motion control quickly. In the long-term, integration into the MIDAS data acquisition will be achieved by using the C++ libraries for Linux available from Galil.

### 15.3.5 Position Measurement

The measurements of the trolley's position in both the longitudinal and transverse directions relative to its motion plays an important role in the evaluation of several systematic error sources. Uncertainties in the trolley's position convoluted with the local field gradients give rise to uncertainty in determining the average $B$ field over the ring. The same effects introduce uncertainty in the cross-calibration with the plunging probe.

As stated in the requirements section 15.3.1 above, some improvements in the determination of the trolley's position compared to the E821 experiment are necessary. Together with the better shimming of the magnet (see section 15.8) and hence reduced field gradients, this will significantly reduce position related systematic errors for E989.

During the calibration procedure of the trolley probes in a specially shimmed region in the ring, the plunging probe and the trolley probes (more precisely their active volumes) must be positioned repeatedly at the same position. The uncertainty in E821 for the relative azimuthal alignment was estimated to be 3 mm. The NMR probe active volumes relative to their trolley housing will be calibrated with deployment of precisely known gradients with active surface coils. The trolley itself was positioned by eye in E821, and so we foresee improvements here by means of a well-defined stop mechanism or an external laser survey system viewing a fine positioning grid through the viewing port. The plunging probe from E821 had motion limited to the vertical and radial directions. Adding motion in the azimuthal direction inside the vacuum chamber will also help in reducing position uncertainties in calibration (see Section 15.4.1).

While the trolley moves on the rails around the ring, the transverse position of the 17 NMR probes relative to the central muon orbit is mainly defined by the precision alignment of the rails. In a recent analysis with E821 trolley data, we have studied the effects of the azimuthal variations of the first higher multipoles of the field. Decomposition of these multipole distributions into a finite series of orthogonal functions allowed an analytical study of the coupling with possible rail displacements. The worst-case scenario of in-phase coupling of rail distortions with the largest quadrupole moment while maintaining consistency with the E821 spec of less than 0.5 mm rail deviation gives a maximal systematic shift of 50 ppb. For E989, reduction of the azimuthal variation of the field by improved shimming will reduce this by a factor of 2 or more. Special focus during the shimming needs to be placed on the dominant quadrupole variation to significantly reduce this systematic uncertainty.

Since the rails are a mechanical system tied to the vacuum chambers, special focus in the chamber alignment will be needed to meet the precise rail alignment of $\pm 0.5$ mm or better. For that purpose, we have developed an alignment strategy for the rails together with the metrology group at Fermilab. After mechanical improvements of the rail fixture, curvature, and positioning inside the vacuum chambers are performed in conjunction with other work on the cages, these stringent alignment requirements must be verified. We anticipate a combination of two measurements to have a consistent cross-check of the trolley's transverse movement. The first verification is based on precision survey of the trolley on the rails by means of laser tracking. After mapping the shape of all individual rail sections with this method, the rail location inside the chambers are referenced to external markers mounted on the chamber walls. Once all chambers are mounted in the storage ring, these external markers will allow to determine the final position of each rail since FEA modeling showed that the deflection of the chamber walls is small when evacuated. We are currently developing the specific implementation of all required steps together with the metrology group at Fermilab with a test setup. A second technique to measure the trolley's transverse position involves imposing radial and vertical gradients using the surface coils to observe the changes in the trolley NMR probe readings around the ring.

In E821 the longitudinal position measurement of the trolley was achieved with a combination of optical rotary encoders and potentiometers monitoring the cable unwinding as well as the response spikes in the NMR frequency of the fixed probes due to the passing electronics of the trolley. The overall estimate of the longitudinal uncertainty was about one centimeter. We aim to reduce this uncertainty to 5 mm or better by refurbishing an onboard barcode reader that was mostly unused in E821 due to unreliable operation. As can be seen in Figure 15.15(a), the vacuum chambers are equipped with marks around the ring. The continuously spaced marks have a spacing of 2.5 mm while the larger spaced, unique patterns serve as absolute reference marks.

Figure 15.15(b) shows the basic principle of the barcode reader. A pair of an LED and a photo-sensor (CLOCK) is placed over the regular marks and is constantly sensing the photocurrent. At the detection of a transition edge between marks, the onboard microcontroller in the trolley (see section 15.3.7) sees an interrupt and pulses two additional LEDs. Both of these are offset in the direction of motion from the CLOCK LED. One of the LEDs and its photo-sensor (DIR) are also above the regular pattern marks. In combination with the CLOCK sensor, this determines the relative trolley position and its direction of movement. The readout of a third LED / sensor pair (CODE) above the unique marks allows for determining the absolute trolley position at the occurence of those patterns.

For E989, we will build a new barcode reader board to achieve the required longitudinal position measurement. In a first step, we are designing a stand-alone new barcode reader centered around a Microsemi SmartFusion evaluation kit. This will be mounted on a small carriage that can be pulled through individual vacuum chambers by hand or with the trolley drive. With available ADC channels on-board, full digitization of the entire barcode marks and their contrast level will be performed. Based on the results, we can empirically determine whether the contrast level is high enough to have a fixed threshold for detecting a pattern transition. This is important as the current trolley electronics is based on such a fixed threshold. We will also optimize the optical arrangement of the LED and sensors using the proto-type barcode reader with full digitization. Once a reliable operation with the existing barcode marks has been proven, the optimized prototype will be adapted to match the trolley electronics interface and meet the non-magnetic requirements. As we have the electronics schematics from E821, this is a straight-forward task. Then, the system will be tested for reliability and that it works under vacuum without overheating.

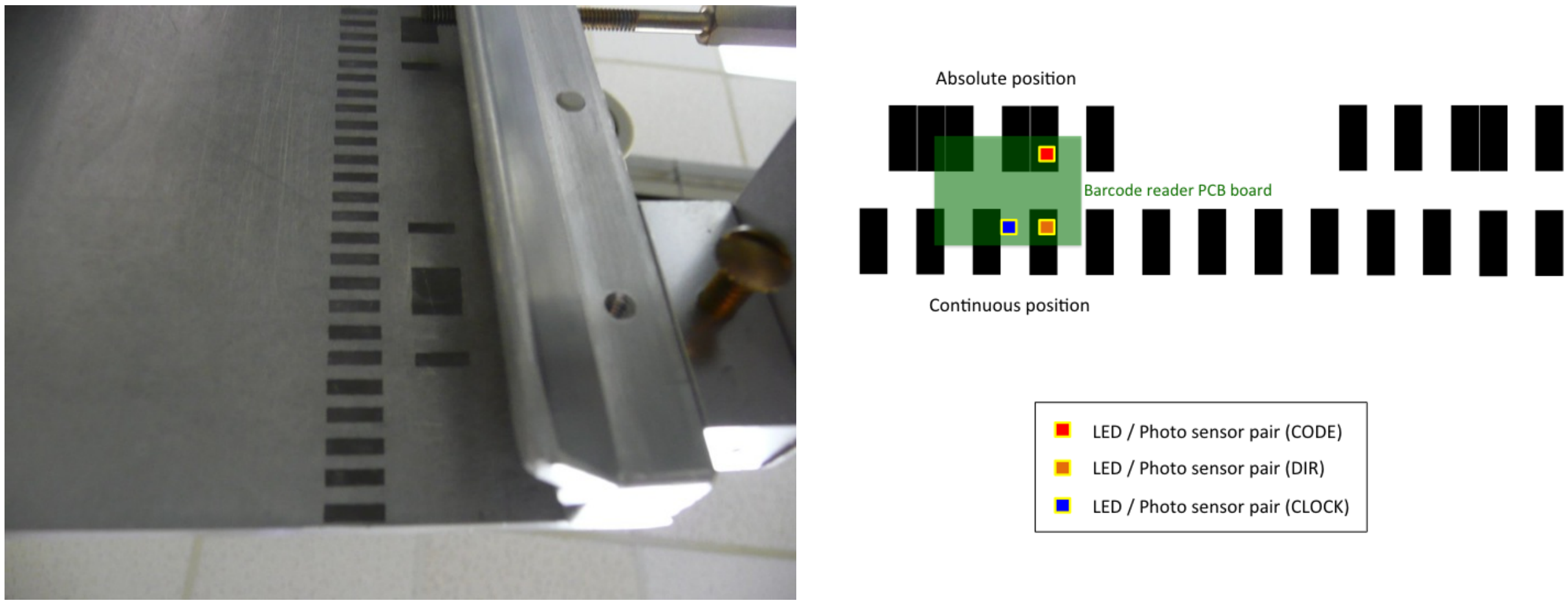

(a) (b)

Figure 15.15: (a) Trolley bar code marks on vacuum chamber plate. (b)

### 15.3.6 Probes

The 17 trolley NMR probes (see Figure 15.2(b)) are identical to the fixed probes. No major work should be necessary for future use in E989 except for the standard fabrication procedure performed by collaborators from the University of Washington. These activities (see section 15.2.1) will include filling with petroleum jelly, tuning for operation at 61.79 MHz, and impedance matching to the 50 Ω cable as well as testing of a normal NMR response at 1.45 T. To reduce the uncertainties in the absolute calibration of these probes, minor changes to the coil windings and sample may be made so that the active volume of the probes is better defined. This can be achieved by restricting the sample to the most homogeneous part of the field produced by the coil $L_s$, and by ensuring the number of windings and their positions in the 17 probes are identical.

Alternate distributions of the NMR probes within the trolley (see Figure 15.8(b)) were studied and the configuration used by E821 already heavily overconstrains the determination of the important multipole amplitudes present inside the homogeneous field. Therefore, no change to the probe configuration inside the trolley is necessary.

### 15.3.7 Frequency Measurement

The NMR frequency measurement for the 17 NMR trolley probes is all integrated into the onboard electronics which were developed for the Brookhaven E821 experiment. At its heart sits the Motorola 68332$\mu$C microcontroller with a multitude of functionality. Power, RS232 communication, and the NMR reference frequency are multiplexed over a single double-shielded co-axial cable. The remaining NMR components (RF pulse amplifier, multi- and duplexer, signal preamplifier and frequency counter) are all integrated into the trolley housing together with temperature and pressure sensors and the barcode reader. The development of this minimally magnetic, low-power and low-noise system was a major effort in E821. Preliminary tests in June 2014 of individual components of the electronics have not revealed any obvious damaged components.

As the central microcontroller in the existing E821 electronics is outdated, it will be replaced with a modern SmartFusion2 chip from Microsemi which comprises both an ARM processor and an FPGA. In addition, full digitization of the NMR signal will replace the zero-crossing counting implemented in hardware in the old electronics. This will make the trolley readout compatible with the rest of the NMR probes used in E989. Depending on the maximum bandwidth of the current communication interface over the coaxial line, an upgrade of the communication protocol and interface might become necessary to facilitate offloading the increased data rate to the DAQ computer.

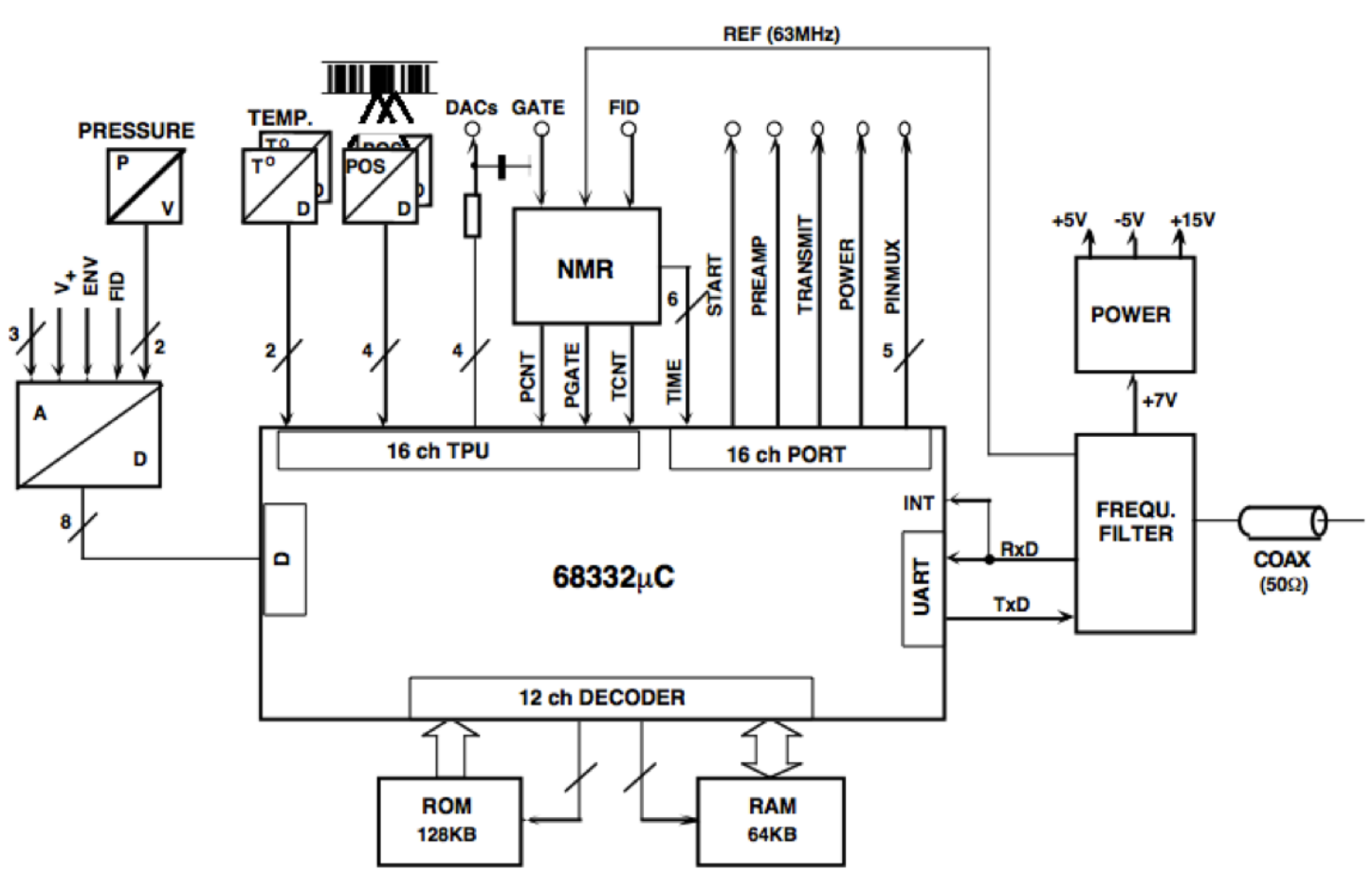

Figure 15.16: Schematics of central electronics on board of the trolley.

Even the low electric power of less than 1 W leads to changes in the temperature of the trolley electronics and the probes of a few °C over the course of a trolley run. As the measured NMR frequency is temperature dependent, minimization of the temperature changes will help to reduce the associated systematic error. The influence of the temperature on the NMR measurement will be carefully studied in a test solenoid (see section 15.3.9). We also started investigating whether more heat can be dissipated via radiation to the vacuum chamber walls by increasing the surface emissivity of the trolley and by adding additional heat sinks to optimize the heat flow. Finally, reverse-direction trolley runs will provide a powerful *in situ* test temperature-dependent effects.

### 15.3.8 Trolley DAQ

The DAQ computer communicates with the onboard microcontroller using the RS232 protocol over the single co-axial cable connected to the trolley. A new DAQ computer will perform this function in E989 and will provide all necessary user interfaces to execute commands on the trolley microcontroller. The same DAQ infrastructure will be used to communicate with the trolley drive, garage and the plunging probe mechanism via the new motion controller as described in Sec. 15.3.4.

### 15.3.9 The 4T test solenoid

The homogeneity of the magnetic field must be known to the 70 ppb level for the new E989 experiment. Achieving this precision goal requires economical testing of the NMR equipment over prolonged periods at a precision of greater than 20 ppb. This requires a magnetic field of 1.45T with high stability and homogeneity. The main purpose is the testing of the in-vacuum NMR trolley system and a precise reevaluation of the absolute calibration probe (see Section 15.4.1).

The most stringent requirement for the magnet comes from the plan to repeat a careful evaluation of the absolute calibration probe at the nominal B field of 1.45T for $g-2$. As the overall aim is a reduction of 50 ppb to 34 ppb as was achieved in a 1.7T field with the same probe [6], the gradients and drift should be significantly smaller than this goal. The envisoned tests will include both systematic studies of the NMR measurements with the same probe under changing external factors (e.g. temperature, power supply voltages, ...) and the comparison of different probes at the same location within the magnet. The latter requires a stability of the field when probes are swapped at the level of 10 ppb in order to be negligible with respect to the precision goal of these studies. Assuming that probes can be placed into the center of the test magnet at time intervals of 3 minutes, the 10 ppb requirement translates into a maximal allowable drift of 200 ppb per hour.

The field gradients coupled with uncertainties in the positioning of a probe also limit the reproducibility between exchanges of probes. Individual probes can be positioned with an accuracy of $\pm$ 0.5 mm on a well designed platform. Together with the 10 ppb specification above, we therefore can tolerate gradients of no greater than 200 ppb per centimeter in a typically few-centimeters large calibration region at the center of the test magnet.

In order to test the trolley system, the bore of the magnet must be large enough to conveniently fit the device. The most stable and homogeneous magnets available are persistent MRI magnets which have a solenoid field. Due to the orientation of the NMR probes inside the trolley, the current 50 cm long trolley would need insertion perpendicular to the axis of the magnet. At minimum a bore size of 60 cm is required to accomplish that. The active volumes of the NMR probes are located 8 cm away from one end-cap. Hence, a bore of about 90 cm would be ideal to operate the probes in the most homogeneous center of the magnet.

Table 15.3 summarizes these requirements for the planned E989 systematic studies being performed in this magnet. The last column shows the specifications of an available Oxford OR66 magnet which is shown in Figure 15.17.

| | Requirement for E989 | Oxford OR66 magnet |
|---|---|---|
| Magnetic field | 1.45 T | $\leq$ 4 T |
| Stability | $<$ 200 ppb/hr | 90 ppb/hr |
| Homogeneity at center | $<$ 200 ppb/cm | $\sim$10 ppb/cm |
| Bore diameter | $>$ 60 cm | 68 cm (with gradient coils)<br>90 cm (w/o gradient coils) |

Table 15.3: Requirements for the E989 and specifications for the Oxford OR66 magnet at full field of 4 T. The quoted homogeneity is achieved with passive and active shimming which requires the gradient coils being installed as they contain the active shim coils.

This magnet has been shipped cold from its former medical research application in a VA hospital in San Francisco to Argonne. An outside company will ramp the magnet up to 1.45 T and will shim the shape to reach its optimal homogeneity. We are currently designing a suitable platform to insert the trolley and other stability monitoring NMR probes to support the calibration procedure. This configuration will also allow studyies of various systematic uncertainties and characterizations of ferrous properties of materials being considered for use in the experiment. A temperature controlled enclosure will help to understand the temperature dependence of the NMR frequency extraction.

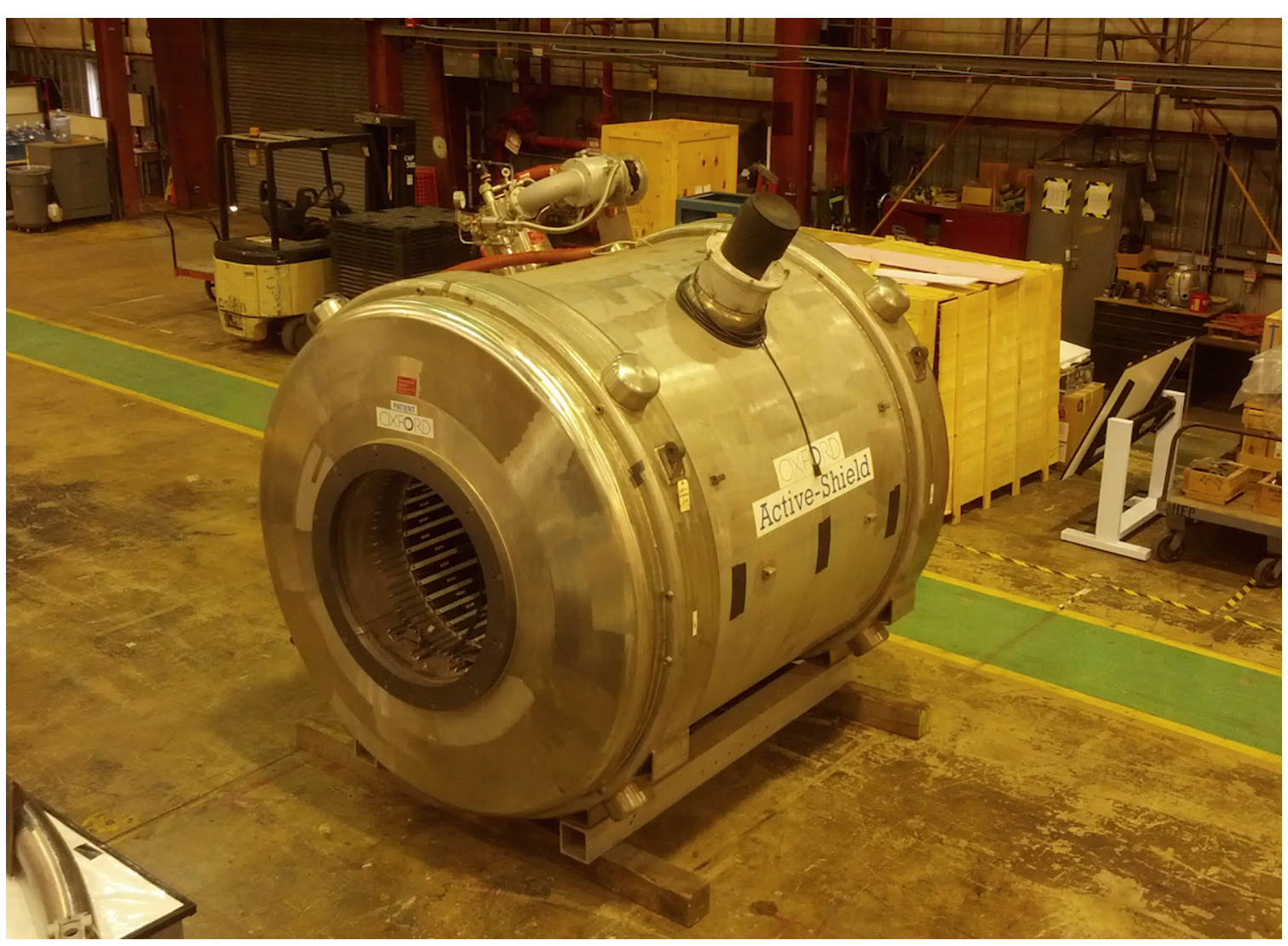

Figure 15.17: The 4T, test solenoid OR66 from Oxford Instruments inside building 366 at Argonne National Laboratory waiting for installation.

The test magnet will be available for NMR related studies during the entirety of the $g-2$ experiment's lifetime. Especially in the early phase during the installation and shimming of the $g-2$ storage ring, it will be solely dedicated to the extended NMR tests described above. Once the $g-2$ usage reduces, we have the freedom to make this test facility available to other users. Since the magnet power supply came with the magnet, we can infrequently ramp the field to other desired values. That offers the flexibility for the $g-2$ collaboration to come back with dedicated studies at 1.45 T even during data taking periods. For example, a careful repetition of the calibration probe's evaluation at the end of data taking will be an important and desired cross-check to complete the analysis.

# 15.4 Absolute calibration of the magnetic field

## 15.4.1 Necessity for calibration

The magnetic field map of the storage volume is made by the trolley, using pulsed NMR on protons (see section 15.3). However, these protons are not free - they are in hydrocarbon molecules in a macroscopic sample of petrolatum, which in turn is surrounded by probe materials - teflon, copper wire, and aluminum, all of which perturb the magnetic field at the proton. For each trolley probe $i$ at location $\boldsymbol{r}_i$, these perturbations must be measured and a correction $\delta_i$ applied to the raw Larmor precession frequency $\omega^i_{\rm raw}(\boldsymbol{r}_i)$. Then the free proton precession frequency, $\omega_p(\boldsymbol{r}_i)$, can be extracted from the raw measurements, $\omega^i_{\rm raw}(\boldsymbol{r}_i) = (1-\delta_i)\omega_p(\boldsymbol{r}_i)$.

The procedure that determines the individual corrections $\delta_i$ for each trolley probe is called absolute calibration. This is the third major field measurement task outlined in Sec. 15.1.5.

This absolute calibration of the trolley probes is performed with two special NMR probes; the plunging probe and absolute calibration probe (see Figs. 15.2(a) and 15.1(a)). The probes used for BNL E821 absolute calibration (see [6]) can be used for E989, though improved versions are being developed for comparison and to reduce some systematic uncertainties. The details are described below.

### Absolute calibration probe

The Larmor angular precession frequency of a proton in a pure water sample in the absolute calibration probe $\omega_{\rm probe}$ is related to the Larmor precession frequency of a free proton at the same location, $\omega_p$, by a total correction factor $\delta_t$,

$$\omega_{\rm probe} = (1-\delta_t)\omega_p, \text{ where} \tag{15.9}$$

$$\delta_t = \sigma(\mathrm{H_2O}, \mathrm{T}) + \delta_b + \delta_p + \delta_s. \tag{15.10}$$

Contributions to $\delta_t$ come from the diamagnetic shielding of protons in a water molecule $\sigma(\mathrm{H_2O,T})$, the shape- and temperature-dependent bulk susceptibility of the water sample $\delta_b(T)$, paramagnetic impurities in the water sample $\delta_p$, and the magnetic field from the magnetization of paramagnetic and diamagnetic materials in the probe structure, $\delta_s$ [6]. These correction are discussed below.

The absolute calibration probe has a near-spherical water sample ($\delta r/r < 1\%$, see Figs. 15.1(b)) and is described in detail in [6]. The Larmor angular frequency observed of a proton in a spherical water sample is related to that of the free proton through [11, 1]

$$\omega_{\rm L}({\rm sph}-{\rm H_2O},T) = [1-\sigma({\rm H_2O},T)]\,\omega_{\rm L}({\rm free}). \tag{15.11}$$

Measurements of $\sigma(\mathrm{H_2O},T)$, the temperature-dependent diamagnetic shielding of the proton in a water molecule, are described in [15, 23, 14] with the result:

$$\sigma({\rm H_2O},T) \;=\; [25.680(2.5) + 0.01036(30)\times(T-25.0°{\rm C})]\times 10^{-6}. \tag{15.12}$$

The bulk correction in Eqn. 15.10 is given by $\delta_b(T) = (\epsilon - 4\pi/3)\,\chi(T)$ in cgs units, where $\epsilon$ is a sample shape dependent factor and $\chi$ is the susceptibility of water $\chi(\mathrm{H_20})=-0.720(2)\times 10^{-6}$ [24]. The temperature dependence of $\chi(\mathrm{H_2O})$ has been measured [25]:

$$\chi({\rm H_2O},T) = \chi({\rm H_2O},20°C)\times\Big(1 + a_1(T-20) + a_2(T-20)^2 + a_3(T-20)^3\Big), \tag{15.13}$$

where $a_1 = 1.38810\times 10^{-4}$, $a_2 = -1.2685\times 10^{-7}$, and $a_3 = 8.09\times 10^{-10}$ for $T$ in Celsius. A spherical sample shape is used in the absolute calibration probe because the shape-factor for a sphere $\epsilon = 4\pi/3$, which eliminates the bulk magnetic susceptibility correction, $\delta_b = 0$. (For an infinitely long cylinder $\epsilon = 2\pi$ when the axis is perpendicular to $\vec{B}$. The precession frequency observed using a cylindrical water sample should be 1.508(4) ppm higher than that of a spherical sample in the same field.) The sphericity of the glass bulb containing the water sample was measured for BNL E821 with an optical comparator, and corrections were made to $\delta_b$ [6]. Corrections were also made for bubbles or excess water in the neck of the sample, and other imperfections. Using pure, deionized, degassed, multiply-distilled water in the absolute calibration probe reduces $\delta_p$.

Several techniques can be applied to reduce the magnitude and uncertainty on the remaining corrections. The presence of impurities can be monitored by measuring the magnetization time-constants $T_1$ and $T_2$ using appropriate pulse sequences. The perturbation due to probe materials $\delta_s$ can be minimized by using material with low susceptibility and testing them for magnetic impurities. The residual influences must be measured so that $\delta_s$ is known at the 10 ppb level or better. Tight mechanical tolerances ensure symmetry about the cylindrical axis of the probe and reduce the sensitivity to the orientation of the probe [26]. Symmetry about the midplane helps make the lineshape symmetric (reducing time-dependence in the FID zero-crossing rate).

These techniques were applied and the properties of the absolute calibration probe developed for LANL E1054 and BNL E821 were measured carefully at LANL in a very stable MRI solenoid at 1.7 T so the absolute field in terms of the free proton precession frequency was determined by the probe to an accuracy of 34 ppb [6]. Measurements made by the absolute calibration probe were expressed in terms of the equivalent free proton frequency at 1.45 T to a reduced precision of 50 ppb in BNL E821. The main reason for the reduced precision was uncertainty in the probe temperature during the absolute calibration procedure. The precision will be improved for FNAL E989 to the 35 ppb level proven for LANL E1054 as discussed below.

First, the properties of the absolute calibration probe which were determined in the E1054 solenoid at 1.7 T to an accuracy of 34 ppb [6] will be repeated through extensive testing in the MRI solenoid at Argonne at 1.45 T. By constructing a thermal enclosure for use in the solenoid with 0.1 °C stability, and with a higher performance ADC for the FID, the properties of the absolute probe should be determined at least as well as E1054, but now at the correct field for E989 and with greater control over the temperature. Another improvement is that construction of a new calibration platform will allow a few additional probes to be used to track the changes of the solenoid field during these calibration studies. The new platform will exchange probes with sub-mm position uncertainty, with the perturbation to the local field minimized and measured. These procedures should reduce the uncertainties on the absolute calibration probe from 50 ppb in E821 to 35 ppb in E989.

The absolute calibration procedure is performed with the ring up to air. It involves measuring the field at a fixed location in the ring with the center trolley probe, moving the trolley away, then measuring the field at the same location with the absolute calibration probe (which is mounted on a G10 support attached to a vacuum chamber flange). This is performed at times of the the day when the experimental hall temperature and magnet are most stable. The probe temperature will be monitored and determined with less than 0.5 °C uncertainty in the absolute temperature so the associated uncertainty in the diamagnetic shielding is less than 5 ppb. Calibrations performed at different temperatures should be consistent once the known temperature-dependent changes in shielding are accounted for.

An additional shift of the calibration in E989 versus those in the solenoid at LANL arises due to the presence of magnetic images of the magnetized probe materials and water sample of the absolute calibration probe which appear in the high-$\mu$ iron of the pole pieces. These effects are at the level of 40 ppb and were insignificant for BNL E821, but they will be determined as part of the calibration procedure in E989. This will be done by inserting a small diameter NMR probe (like one of the trolley/fixed probes) between the g-2 storage ring pole pieces. The shell of the absolute calibration probe (after removing the water sample) will be placed around the small probe, the field will be measured, then the shell removed and the field measured again by the small probe. The shift in field measured by the small probe yields $\delta_s$ in Eqn. 15.10. Measuring $\delta_s$ a function of the height above the pole pieces accounts for most of the magnetic image effects. These shielding effects will be compared between measurements made in the Argonne solenoid and the storage ring and cross-checked with predictions.

It is important to note the same absolute calibration probe intended for use in E989 was used in LANL E1054 to determine $\mu_\mu/\mu_p$ [3]. By using the same absolute calibration probe as E1054, there is a direct robust link of our E989 magnetic field to the muon magneton; proton NMR has only the role of a intermediate fly wheel. This link is independent of possible future changes in fundamental constants in the regular adjustment procedures [1], unless the muon magneton will be remeasured better experimentally or the theory-dependent value of $\mu_\mu/\mu_p$ is used to extract $a_\mu$. In these case $\mu_\mu/\mu_p$ is largely independent of the properties of the probe. This link is important if the value of $\mu_\mu/\mu_p$ directly extracted from the muonium experiment is used to extract $a_\mu$. The same probe will also be used as part of the calibration of the new J-PARC muonium experiment, which should reduce the experimental uncertainty on $\mu_\mu/\mu_p$.

Finally, as an important cross-check, we are developing a second absolute calibration probe using a spherical water sample but with several improvements. First, the water sample will be contained in a spherical shell made from machined Macor hemispheres. Prototype shells were constructed, the interiors were filled with a casting material and the casts ex-amined on a Suburban Tool MV-14 10× optical comparator, with a Mitutoyo micrometer, and from analysis of images taken with a CMOS camera. The interiors were spherical to $\delta r/r \leq 0.5\%$, where an asphericity of 0.5% leads to a fractional uncertainty $\delta B/B \approx 15$ ppb [17]. The machined shell incorporates a sealable plug, eliminating the long glass stem that lead to line broadening in E1054/E821. Other improvements: to reduce $\delta_s$, the probe shell will be made of a zero-susceptibility combination of thin aluminum (paramagnetic) and copper (diamagnetic) cylinders, or will be omitted completely, reducing the pole-piece image effects. Smaller, non-magnetic tuning capacitors than those of the E1054/E821 probe will be used, and moved farther from the sample, reducing $\delta_s$. Near zero-susceptibility wire (from Doty Scientific) will be used for the coil. The coil itself will be wound with a slight non-uniformity to achieve higher RF uniformity over the water sample, and by the reciprocity principle, give a more uniform $B$ field measurement. The sample will be held and the coil wound on thin-walled cylinders of low-susceptibility material (Macor, Pyrex or quartz). This probe will be compared with the E1054/E821 probe and will provide some redundancy and reduce risks in case of loss or damage. The reduced susceptibility and greater sphericity should lead to smaller systematic uncertainties. Having two probes to compare many times will provide an important check on the stability of the absolute calibration. The use of a precision cylindrical absolute calibration probe is also being considered.

### Cross-calibration of the trolley with the plunging probe

The calibration of the trolley involves shimming a section of the storage magnetic field to be as uniform as possible. This is important because absolute calibration involves measuring the field at the same location with a trolley probe, then a calibration probe. If the field is not uniform, then a difference in the trolley and calibration probe positions leads to the probes measuring different fields. This effect can not be separated from the calibration offset $\delta_i$ which is sought. Improving the field homogeneity reduces the requirements on probe positioning accuracy.

Good local field homogeneity is achieved by tailoring the surface correction coil currents temporarily to reduce the multipoles locally. Additional sets of coils (on the other side of the surface coil PCBs and/or on coils between the pole pieces and yoke) will be used to remove gradients in the azimuthal direction. This shimming needs to be done only over a restricted volume that encompasses the active volumes of the trolley probes.

Ideally, the field in the homogeneous field region is measured by the trolley probes; then the trolley is moved away and the absolute probe determines the field at the locations just measured by the active volumes of the 17 trolley probes. The magnet is stabilized during this procedure, which is repeated many times. Since the absolute probe is too large to reach all of the locations measured by the trolley probes, an intermediate probe called the plunging probe is also used (see Fig. 15.2). The plunging probe must be calibrated by the absolute calibration probe.

To cross calibrate the trolley probes with the plunging probe during E821, the field was measured by the 17 trolley probes at the calibration location. The trolley was moved away, and the plunging probe, which sits on the end of a 1 meter long, hollow G10 arm, was moved in the radial and vertical directions to measure the field at the locations formerly occupied by the trolley probes. The difference between the measurements calibrates each trolley probe with respect to the plunging probe (after correcting for drifts in the field at the level of 100 ppb, which were measured by fixed probes $\approx 10^\circ$ away from the calibration location).

Errors in the E821 calibration arose from uncertainties in the azimuthal positions of the probes, uncertainties in the transverse positions of the probes, and uncertainties in the comparison of the trolley probes with the central trolley probe which is calibrated directly by the absolute calibration probe.

In the azimuthal direction, errors in the E821 calibration procedure arose both from uncertainties on the positions of the centers of the active volumes inside the trolley along its axis (unknown at the level of a few mm) and of the trolley itself. The position uncertainty on the location of the active volumes of the calibration probes was also at the 1 mm level. Coupled with field gradients in the azimuthal direction gave a 30 ppb contribution to the uncertainty in the relative calibration.

In the transverse plane, the trolley probes are fixed with respect to the frame inside the trolley that holds them, and variation in the trolley transverse position is restricted below a mm by the rail system on which it rides. The vertical and radial positions of the trolley probes with respect to the plunging probe are determined by applying a sextupole field with the surface coils and comparing the change of field measured by the two probes. Magnetic field inhomogeneities of order 20 ppb/mm in the calibration region used in E821 (see Fig. 15.10(b) and Table 15.2) coupled with radial and vertical position uncertainties of order 1 mm to give a 20 ppb uncertainty. Field drifts and uncertainty in water sample temperature also contribute to the uncertainty.

Each trolley probes was calibrated against the plunging probe approximately 6 times in the 2001 run of E821. The RMS scatter of these 6 measurements of the relative calibrations of each probe versus the center trolley probe, averaged about 140 ppb. The resulting uncertainty on the relative calibration was estimated as 70 ppb, and is consistent with the uncertainty expected from the positional uncertainties described above.

The calibration of the center trolley probe with respect to the absolute calibration probe had an uncertainty of 50 ppb, so the total trolley probe calibration uncertainty was 90 ppb (the sum of 70 and 50 ppb in quadrature).

Changes to this procedure can reduce the relative calibration uncertainties to 30 ppb for E989. For E821, the plunging probe and G10 rod were inside a close-fitting metal tube with an enclosed end, and were operated in an air atmosphere. The tube was attached to a titanium bellows that allowed the probe to be translated in the radial and vertical directions in the calibration region, while the storage ring remained under vacuum. However, the probe could not be translated in azimuth. Since the trolley could not be positioned with high precision in the azimuthal direction, there were uncertainties at the few mm level between the plunging probe azimuthal position and the those of the active volumes of the trolley probes (which were not visible from the exterior).

Another limitation of the E821 trolley calibration with the plunging probe was due to the metal bellows and other parts of the plunging probe translation mechanism. These would move when the plunging probe position changed, perturbing the field at the location of the plunging probe at the 10 ppb level. To reduce this perturbation for E989, the bellows and metal sheath will be removed and the plunging probe operated in vacuum. This reduces the total mass of moving material significantly, reducing the local perturbation to the field, including those from magnetic images, during calibration. The residual perturbation will be measured.

Another improvement will be to provide 3-dimensional positioning of the plunging probe with sub-mm accuracy. This will be done with three high-vacuum compatible, non-magnetic linear translation stages. The stages will provide simultaneous radial motion of 25 cm, vertical motion of 10 cm, and azimuthal motion of 2.5 cm, and will be driven by the same non-magnetic Shinsei USR60-E3N ultrasonic motors with encoders, Shinsei D6060S drivers and Galil controller as the trolley drive and garage.

In addition, the trolley position during calibration will be controlled and measured more precisely than E821 by using physical stops and/or optical survey or other high accuracy position readout of the trolley and plunging probe positions. A system under study consists of several CMOS cameras which observe the plunging probe from roughly 50 cm away with $\geq 15°$ field of view. Translation of the probe with respect to fiducial markers in the calibration region can be detected with the imaging system, allowing the probe position in 3 dimensions to be determined by triangulation. With a $2592\times$ 1944 pixel array on a non-magnetic board camera (such as the Leopard Imaging LI-OV5640-USB-72) and low-distortion optics, sub-mm position resolution is possible. The trolley position can also be determined with the same camera system, enabling sub-mm relative position accuracy.

Another improvement is that the locations of the active volumes of all the trolley probes will be made more uniform by adjustments to the coil windings and sample position. The active volumes will be determined beforehand in the Argonne MRI solenoid to sub-mm accuracy [5], and precision fiducial marks on the exterior of the trolley will be added. This will allow positioning of the trolley and plunging probe active volumes with mm accuracy or better. Also, by aligning the active volumes of the trolley probes in azimuth to within a mm, the volume which needs to be shimmed for calibration is reduced.

Finally, we anticipate automating these procedures with the closed-loop positioning system outlined above so each trolley probe can be calibrated dozens or even hundreds of times. By reducing the position uncertainties on the trolley probe active volumes to a millimeter or less in azimuth, vertical and radial, and by positioning the trolley and plunging probe with millimeter or better accuracy during calibration, the position uncertainties will be reduced by a factor two or more compared to E821. The RMS scatter of relative calibration measurements should be reduced from 140 ppb achieved in E821 to less than 100 ppb assuming no improvement in the field shimming. Using petroleum jelly in the trolley probes, which has greater immunity to evaporation than water and a susceptibility with a smaller temperature dependence, might reduce this RMS scatter even further, and improved hall temperature stability will help. In addition to these steps, significant attention will be paid to monitoring the probe temperatures, and additional coils might be used to further reduce local field gradients and changes during calibration. By automating the trolley probe calibration and taking dozens of measurements, field drifts can be averaged out and the error on the mean can be brought down to 30 ppb.

In a similar way, the RMS scatter of 50 ppb in E821 on the absolute calibration of the center trolley probe should be reduced. By automating and repeating the calibration many times, the goal of calibrating the central trolley probes with respect to the absolute calibration probe to an accuracy better than 20 ppb should be achievable. This provides a redundant check on the calibration of several of the trolley probes.

[5] By superimposing a field gradient in azimuth, the location where the line center does not shift determines the effective center. The linewidth will be broadened and give some information regarding the size of the active volume.

We note that the calibration with the plunging probe can be done with the ring under vacuum as many times as is necessary during E989. If necessary, calibration with the absolute calibration probe can be done with the ring backfilled with nitrogen to reduce effects due to the paramagnetism of $O_2$ which appear at the level of 30 ppb (which can be measured as in E821). Calibration with the absolute probe will be done before and after muon data-taking, and any time the ring is let up to air. More frequent absolute calibration will be performed if necessary to achieve a total trolley calibration uncertainty of 30 ppb.

### 15.4.2 $^3$He Magnetometry

We plan to develop a second absolute calibration probe using the NMR signal from polarized $^3$He to cross-check and possibly replace the water-based absolute calibration probe.

There are several potential advantages to using hyperpolarized $^3$He in place of distilled water in an absolute calibration probe, which will lead to reduced systematic uncertainties. The diamagnetic shielding factor (see equation 15.10) for $^3$He has been calculated to be $\sigma_{^3\text{He}} = 59.967\ 43(10) \times 10^{-6}$ [1]. While the shielding correction is larger than for $H_2$ and $H_2O$, the uncertainty is much smaller, and the temperature coefficient is about 100 times smaller [24, 40]. The dependence on the shape of the $^3$He volume is much smaller, unlike the water sample which has to be spherical (the precise impact of cell-shape effects will be investigated). With a gaseous sample, motional narrowing eliminates line-shape distortion and the FID produces a Lorentzian line shape whose center is well defined. For $H_2O$ the line shape must be analyzed in the same way as reference [14] in order to accurately transfer the calibration.

Our experience with hyperpolarized xenon suggests that that signal-to-noise is comparable to the E821 $H_2O$ calibration sample. NMR with hyperpolarized $^3$He produced by laser optical pumping is also practical because the NMR signal per atom is of order $10^5$ times larger than that from protons at 1.5 T, compensating for the much lower concentration in the gas phase. Another advantage is that samples can be made smaller, e.g. 5 mm diameter and thus average the field over a smaller volume. Finally, the magnetic images of a $^3$He probe in the E989 pole pieces would be smaller than for a water probe because of the lower electron density, leading to smaller systematic shifts.

Given the advantages above, we expect the field can determined in terms of the $^3$He precession frequency with smaller systematic uncertainties than water. To get an absolute calibration in terms of $\omega_p$ would require knowledge of $\mu_{^3\text{He}}/\mu_p$, the ratio of magnetic moments of a free $^3$He to a free proton. However, the most precisely measured related quantity is $\mu'_{^3\text{He}}/\mu'_p = -0.761\ 786\ 1313(33)$ (4.3 ppb), the magnetic moment ratio of a bound $^3$He to that of proton in a spherical water sample [18, 1]. Still, since this ratio is known to 4.3 ppb, measurements with the water-based absolute calibration probe should agree with a $^3$He based probe to 4.3 ppb. Performing this cross-check would be quite valuable. Therefore we plan to develop an independent absolute calibration probe based on a $^3$He sample. This will be carried out at Oxford, working with Michigan and other collaborators. The aim is to use this to verify the calibration of the water-based absolute calibration probe using the test magnet described in section 15.3.9. This will lay the ground for the possible use of $^3$He as a magnetometry standard in the future. Further work to measure $\mu_{^3\text{He}}/\mu_p$ or $\mu_{^3\text{He}}/\mu_B$, would also be very valuable.

Hyperpolarized $^3$He can be provided from a spin-exchange (SEOP)[42, 41] or metastability-exchange (MEOP)[43, 44] set up. The Michigan group have extensive experience with SEOP and have worked closely with MEOP systems and will cooperate with expert groups at Ecole Normal Superieur and Simon Fraser, and Oxford, to develop an effective system. A decision of which pumping technique to use will be taken in 2014.

The MEOP approach has advantages for the calibration because it can be applied at room temperature and therefore in-situ in the probe. In the MEOP scenario, a turnkey 1083 nm laser light from a fiber laser (*e.g.* Keopsys CYFL-GIGA series) is distributed to each of the probe cells by a fiber. A discharge is struck in the cell to produce the excited state. Polarization of 10% or more is expected in a few minutes. A second option is hyperpolarized $^3$He produced in a separate cell and transferred to the calibration cells through PFE tubing similar to the polarizers we have used for medical imaging work [45]. In either scenario, the dominant corrections, which are small, will arise due to cell-shape effects ($\delta_b$ in equation 15.10), i.e non-sphericity of the cell, but more importantly the polarized gas residing in any tubing or pull off left over from the cell filling procedure.

High-field hyperpolarization magnetometry using the MEOP technique will be developed at Michigan working with long-term collaborators on a search for the electric dipole moment of Xe atoms [46] and medical-imaging work [47]. One of the challenges is producing high magnetization at high field (signal-to-noise is proportional to magnetization, i.e. the product of polarization and gas density). High-field MEOP polarization of $^3$He has been recently studied by the ENS group [48, 49], who show that, due to higher polarization rates, nuclear polarization, of 80% at 1.33 mbar and 25% at 67 mbar, have been achieved. The magnetization at 67 mbar is essentially identical to protons in $H_2O$ at 1.5 T, though the signals may be slightly smaller due to the difference of gyromagnetic ratios (32.4 MHz/T for $^3$He and 42.6 MHz/T for protons).

The development work at Michigan will make use of our 1.45 T magnet and will enable probe development, polarization, and studies of temperature dependence similar to those planned for petroleum jelly. SEOP polarized samples are also under consideration, and we have significant experience in these techniques and equipment.

## 15.5 Feedback to the magnet power supply

A feedback system similar to E821 will be used to stabilize the storage ring magnetic field at the 1 ppm level or below. This is the fourth major field measurement task outlined in Sec. 15.1.5.

Without feedback, the field in the storage volume will change due to changes in the magnet current, mechanical distortions of the magnet (floor shifting, mechanical creep of magnet yoke, poles, or cryostat), and from changes in the magnet gap from thermal expansion/contraction of the yoke and poles. Both mechanical and thermal distortions should be reduced compared to BNL E821 due to the greater mechanical stability of the thick, monolithic concrete floor, and much improved experimental hall temperature stability ±2°F, and improved magnet insulation.

The storage ring magnet power supply can be set with 17 bit resolution (7.6 ppm). The current is regulated by a Bruker DCCT in the power supply, providing long term current stability of 0.5 ppm (8 hours) and ± 0.3 ppm or less over several minutes to several hours. Fine regulation of the current (much finer than the 7.6 ppm resolution of power supply DAC) is achieved by sending a current through the additional "Trim" input winding around the DCCT core, which effectively changes the current set point.

The magnitude of the required trim current is determined using the fixed probe NMR system. Field measurements from a subset of fixed probes are averaged, then compared with a desired field set point. This is done in real time in software on the NMR DAQ PC. Since the mechanical and thermal time constants of the magnet are very long (hours), and since the Bruker power supply has excellent short term stability, this error signal does not need to be determined faster than 1 Hz. In BNL E821, the error signal determined the setting of a 12-bit Jorway Model 31 (voltage output) DAC residing in a CAMAC crate, that was sent to the DCCT trim input. This scheme achieved $\mathcal{O}(1)$ ppm level of stability. In E989, the CAMAC crate and Jorway DAC will be replaced with a stable 16-bit current output DAC board in the NMR VME crate or a separate PCIe or USB board on the NMR PC, or a Keithley 6220 current source. The latter has the advantage that it can be positioned near the power supply DCCT, and controlled remotely by GPIB or RS-232. This avoids long cable runs of the analog feedback signal which would be susceptible to noise pickup.

The gain of the DCCT trim input winding will be determined in bench tests before the end of 2014. The optimal feedback loop scheme and loop coefficients will be determined once the magnet is cold and powered, before data-taking, through Zeigler-Nichols or other established tuning schemes. Field perturbations for tuning schemes can be introduced through a single coil winding on the inner radius of the vacuum chamber of the outer cryostat. A coil in this location produces a field similar to that from the outer coils, and ppm scale perturbations can be introduced with currents of a few hundred mA in a single winding.

Feedback will be used during muon data-taking, trolley runs, and probe calibration. With a more stable thermal and mechanical environment for the magnet, and an optimized feedback loop, we anticipate better magnetic field stability in E989 compared to BNL E821, and sub-ppm stability should be achievable.

## 15.6 Time-dependent perturbations to the $B$ field

In BNL E821, ramping of the AGS magnets changed the field in the experimental hall by ≈ 0.5 mG (34 ppb) over each cycle. The proton beam was extracted and muons were stored during the flat-top of the cycle, with no effect on the g-2 measurement. However, during the ramp up or down, the field in the storage volume changed by roughly 10-15 ppb as measured by the fixed probe NMR system [10]. The main effect is thought to be due to fields produced by currents induced in the outer mandrel [32]. Calculations which treat the magnet as coupled transformers formed by the mandrels and coils, but ignore any other effects (such as skin effects), predict field perturbations of order 25 ppb.

For E989, time-dependent magnetic fields from sources such as the Booster accelerator, and power lines could perturb the field in the storage ring at the ppb level. The error budget for these fields is 5 ppb, and we have carried out extensive studies of the 15 Hz Booster field, transient magnetic fields in MC1, and induced fields from cars and trucks [33]. These measurements began before building construction, and have continued through the re-assembly of the storage-ring magnet. A complete discussion of the Results and Opera 3d simulations are reported in g-2 note #66 [36], and are summarized below.

The principal difference between the BNL magnetic environment and that at Fermilab is the fact that at Fermilab the 15 Hz of the Booster accelerator is derived from the 60 Hz powerlines. In BNL E821, muons were injected into the $(g-2)$ ring storage region asynchronous to the 60 Hz AC power, since the AGS operation is powered by a motor-generator, rather than directly off of the LILCO power lines. This feature averaged out any transient magnetic fields induced in the muon storage region by the 60 Hz power lines.

### 15.6.1 Measurements of Transient Fields

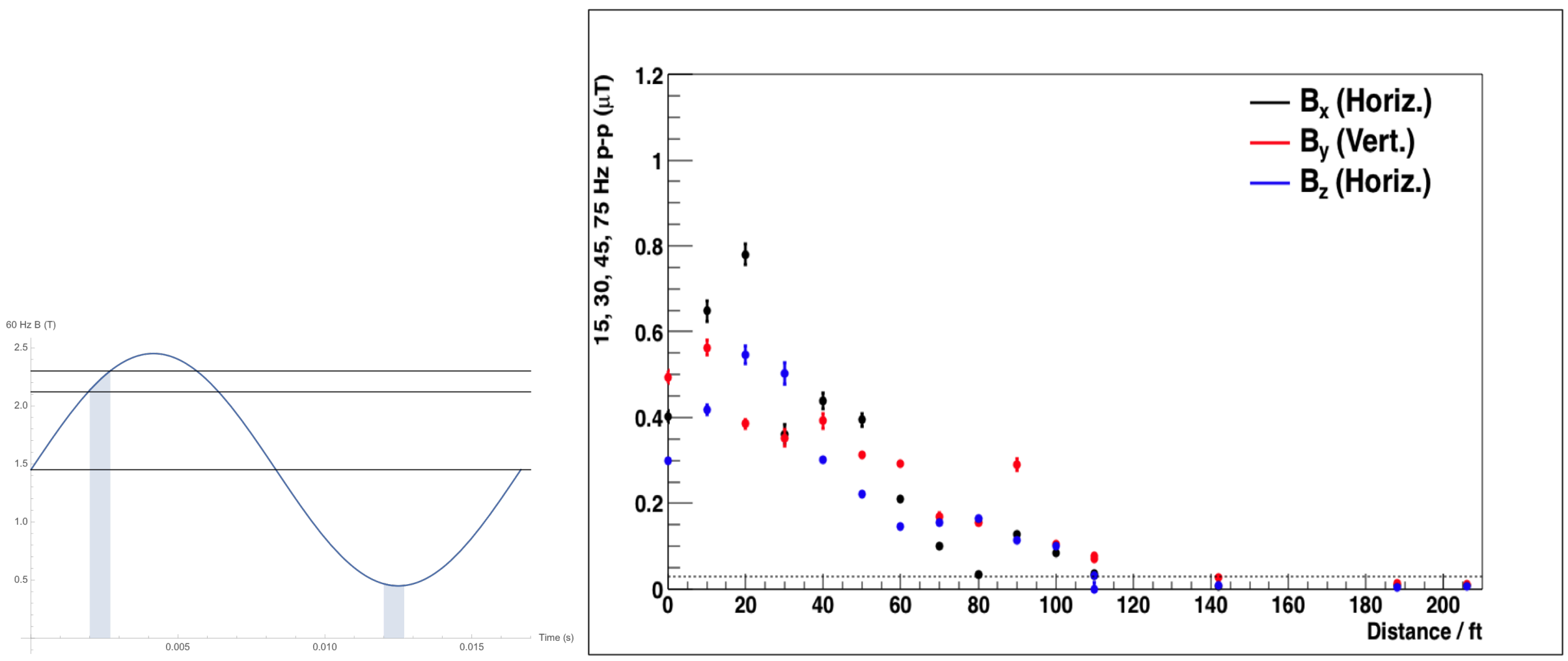

(a) (b)

Figure 15.18: (a) An exaggerated 60 Hz magnetic field on top of the 1.45 T storage region field. The middle line shows what the NMR probe measurements average out to, while the shaded regions show the time covered by the first two muon fills from a single Booster batch. The width of the line corresponds to the 700 $\mu$s ($\simeq 10\gamma\tau_\mu$) measurement time. There is a 10 ms time gap between fills of the storage ring. (b) The booster harmonics (excluding the 60 Hz), as a function of distance from the circumference of the Booster. The dashed line shows the 10 ppb level of the 1.45 T magnetic field. On this graph, some points are inside of the booster circumference, and the closest MC-1 wall is approximately at 200 ft.

At Fermilab, the bunch of muons to be stored will then be injected at the same phase of any 15/60 Hz magnetic field. Due to the asynchronous NMR measurement, the NMR probes will average out this time-varying field as the muons travel around the storage ring. Fig. 15.18(a) gives a pictorial representation of the problem. The fills of the ring will be separated by 10 ms, and the measurement time is $\simeq 0.7$ ms (just over ten muon lifetimes), where ($\gamma\tau_\mu = 64.4$ $\mu$s, and $\tau_{60\text{ Hz}} = 16.7$ ms). The phase of each 0.7 ms measurement period relative to the 60 Hz will be unmeasured, and if unaccounted for, could lead to an error on the magnetic field used in the determination of $a_\mu$.

Because of the close proximity of the Booster Accelerator to the MC-1 building, we measured the 15 Hz Booster field as a function of distance from the Booster toward the MC-1 building. The Booster field vs. distance is shown in Fig. 15.18. It became immediately obvious that the time-varying magnetic field near or inside of MC-1 was dominated by the 60 Hz from the powerlines, and the 15 Hz is negligible at the MC-1 building. We have studied whether this 60 Hz background is large enough to change the central storage-ring field by more than the systematic error budget of 5 ppb, which is 7.25 nT of the main 1.45 T magnetic field[6].

Fourier transforms of thee time varying fields near the booster and in the empty MC-1 Building show that the 15 Hz is almost absent inside of the building, but there is significant 60 Hz, as shown in Fig. 15.19. Analysis of data shows that the booster harmonics are well below the 5 ppb level inside of the empty building, but significant 60 Hz exists there, with the vertical component being roughly 20 ppb [36].

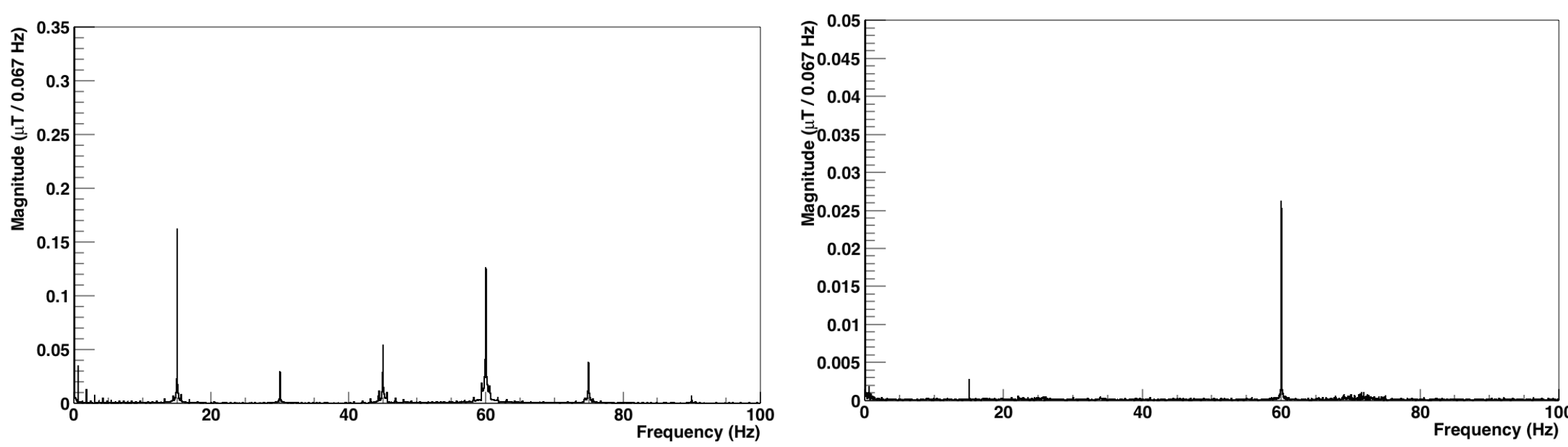

(a) (b)

Figure 15.19: (a)An FFT of the magnetic field at a point close to the booster, approximately 160 ft from the edge of the MC-1 building wall. (b) An FFT of the magnetic field at a point inside the empty MC-1 building.

After the storage ring magnet yokes and cryostats were assembled, but before the pole pieces were installed, additional measurements were carried out. The AC fields were measured at the azimuthal center of each yoke sector at the storage region radius $r_0 = 7.112$ m after the iron yokes and cryostats were installed (see Fig. 15.20(a)). The inner yoke radius is $r_{Yi} = 6.832$ m. Measurements were also taken at the same azimuthal locations at the radii represented by the three red circles: $r_1 = 5.61$ m; $r_2 = 4.36$ m; $r_3 = 0.6$ m. Only four points were measured at $r_3$. The results are shown in Fig. 15.20(b) for the center of the Yoke pieces are $r_0$. The final running period at Brookhaven was for $\mu^-$, and the remnant field between the yoke pieces at the muon storage radius of 7.112 m has a residual vertical field that saturated the fluxgate magnetometer (see Fig. 15.20(b)). At radius $r_1$, all components of the transient field were less than the 5 ppb level, except points 8 and 10, which were near operating vacuum pumps.

[6]At the time of this writing, there has been preliminary work investigating whether the modification of some NMR probe readout times can help measure this phase difference.

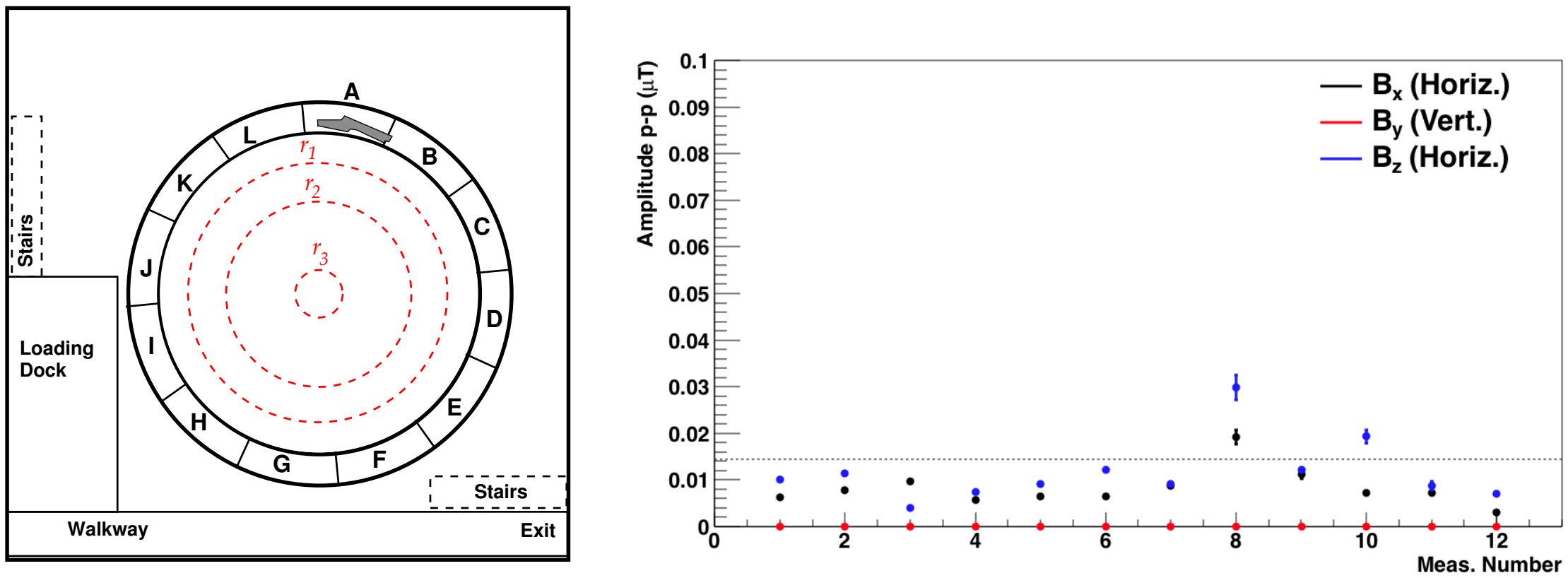

(a) (b)

Figure 15.20: (a)Measurement locations for field measurements. The location where the inflector magnet will be installed in yoke sector A is also shown in the sketch. (b)Peak to peak field magnitudes in 3 dimensions for 15 Hz harmonics (including 60 Hz) at the storage radius $r_0 = 7.112$ m.

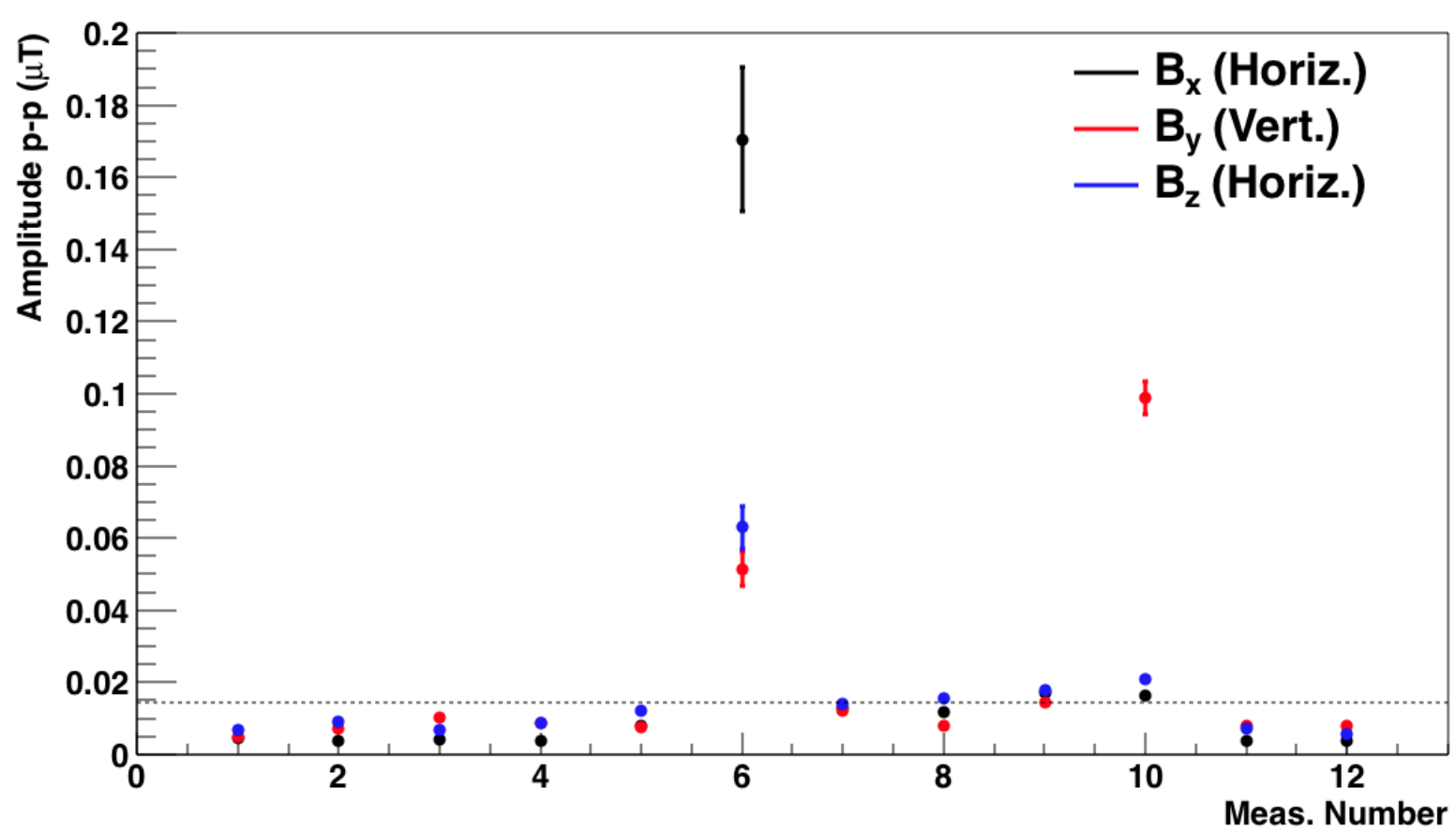

Figure 15.21: Peak to peak field magnitudes in 3 dimensions for 15 Hz harmonics (including 60 Hz) at $r_1$ (4 ft in from $r_{Yi}$). The azimuthal positions corresponded to the center of the yoke pieces (1=A, 2=B, etc.). Points 6 and 10 were located near running vacuum pumps, leading to abnormally large fields. The dashed line represents the 5 ppb acceptable error level. The points far enough from the vacuum pumps lie at, or below, this threshold.

### 15.6.2 Opera Simulations of Transient Fields

An Opera-3d model of the storage ring that included the coils, cryostats, all the iron and the muon vacuum chamber was built. The reader is referred to Ref. [36] for full details. Simulations of vertical transient fields were carried out, with and without the beam vacuum chamber present. A 3° sector of the ring was simulated with axial symmetry. The Opera model is shown in Fig. 15.6.2(a) and the magnetic field is shown in Fig. 15.6.2(b). Once this model was built in Opera, it was used to calculate the effect of a sinusoidal driving field, to determine the shielding factors of the storage ring, with and without the muon beam vacuum chambers installed. The calculated magnetic field at the beam-center height, is shown as a function of radius in Fig. 15.6.2.

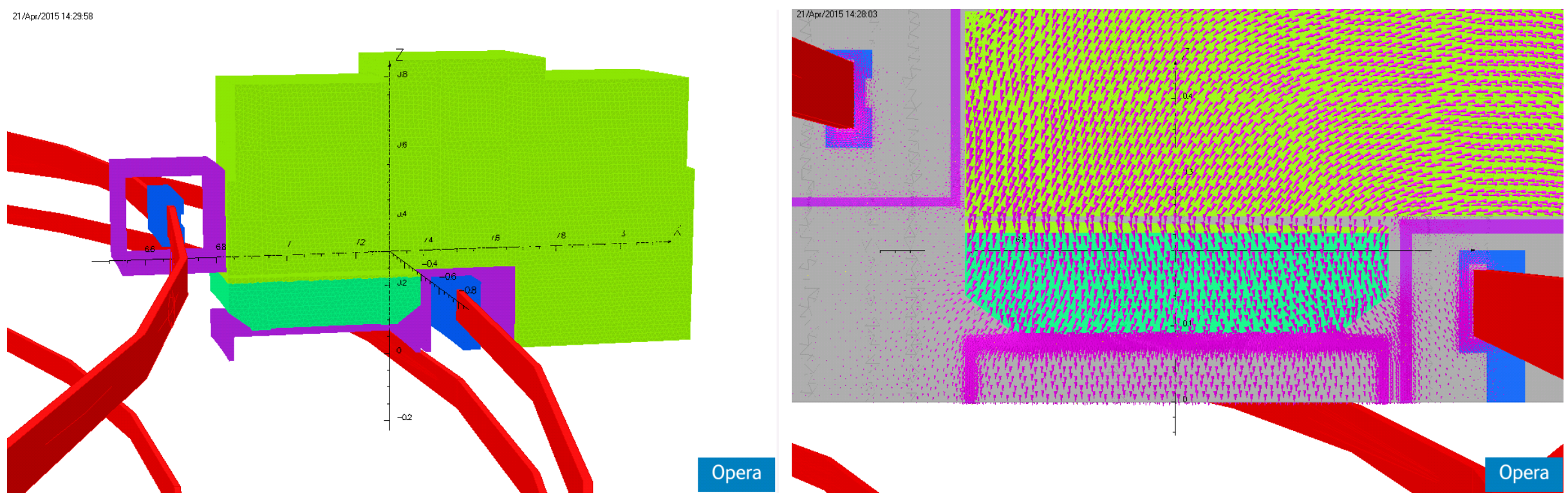

(a) (b)

Figure 15.22: (a) A 3 degree section of the ring. In light and dark green are the yoke and pole pieces respectively, with proper B-H curves applied. In red are the superconducting coils, modeled as perfect Biot-Savart conductors driven at 5200 amps. In blue are the cold aluminum mandrels, modeled with the appropriate conductivity [35]. In purple are the aluminum cryostat boxes housing the superconducting coils, as well as a wide section of the beam vacuum chamber. The apparent blockiness of the cylindrical superconducting coils is a function of the OPERA program visualization, and not a representation of how the program performed the calculations. (b)Close up view of of the storage region, with field lines superimposed upon the geometry. One can easily see that the field lines follow the proper path, and there is a nice uniformity across the storage region. The field magnitude at the center of the storage region was calculated to be within less than 1% of the target 1.45 T.

There is a single dielectric stop for the beam vacuum chamber, around which fields will penetrate to a greater degree as induced current flow is restricted. An OPERA model with a 3 mm slice cut out of the the 3 degree section of the vacuum chamber was constructed, and analyzed to study the effect of this discontinuity. The results are shown in Fig. 15.25, which indicates some magnetic field penetration around the slit. We will study this issue further to make sure it will not cause a problem.

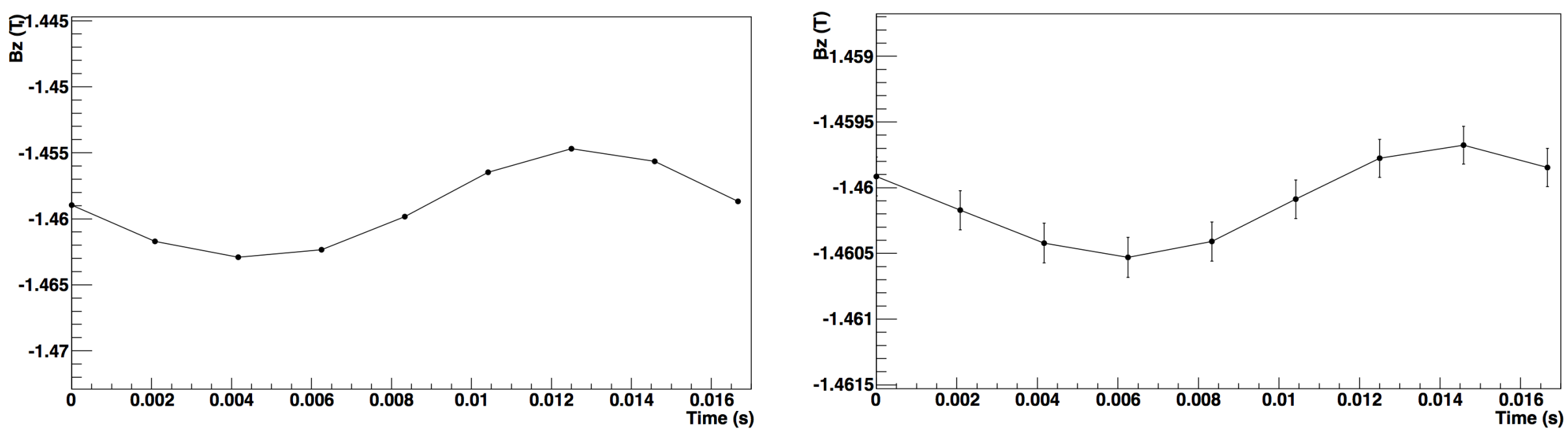

(a) (b)

Figure 15.23: (a)The field magnitude in the vertical direction at a single point at the center of the storage volume for a .1 T external driving sine wave field. This result was produced within the OPERA model where the vacuum chamber was absent. Note the reduced magnitude and opposite direction to the free field. (b)The field magnitude in the vertical direction at a single point at the center of the storage volume for a .1 T external driving sine wave field. This result was produced within the OPERA model where the vacuum chamber was present. Compared to the non-vacuum chamber model, the external field is even further reduced, to the point where the solution accuracy is approaching its lower limits. There is also an unimportant phase shift here due to the larger resistance of the vacuum chamber.

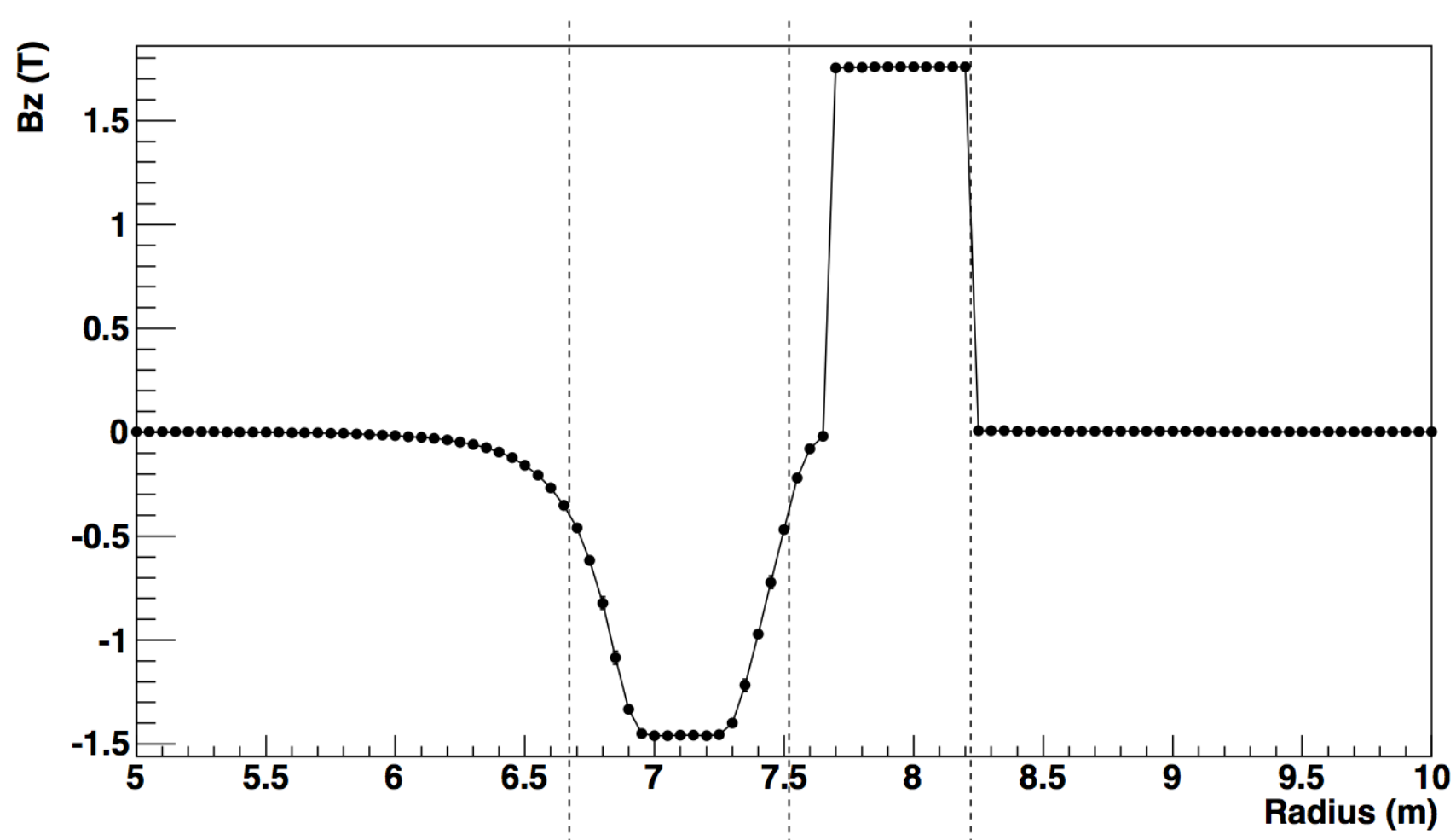

Figure 15.24: Field values for points radially outward from the center of the ring. The first two dashed lines from the left represent the placement of the superconducting coils while the last dashed line shows the location of the outer edge of the iron yoke. The muon storage region is centered at $r_0 = 7.112$ m. Error bars are included in the plot.

### 15.6.3 Summary of Measured and Calculated Fields

The ambient fields in the empty hall have been measured and shown to be of order 20 ppb or less, and calculations show that they are 5 ppb or less with the presence of the ring and vacuum chambers. With the calculated shielding factors, we are confident in the ability of the ring to passively self shield ambient fields from the booster and powerlines to below the 5 ppb acceptable error level. Even if the ambient fields were a factor of four higher, they will still be shielded out. Still of concern are fields from nearby electronics, which have yet to be measured. If their fields are less than 1 mG at the edge of the vacuum chambers, then the field that penetrates the storage region will be below the 5 ppb level. Those which produce fields greater than this will have to be further shielded in some fashion. The next step is to measure these fields as different electronics are produced in the coming months.

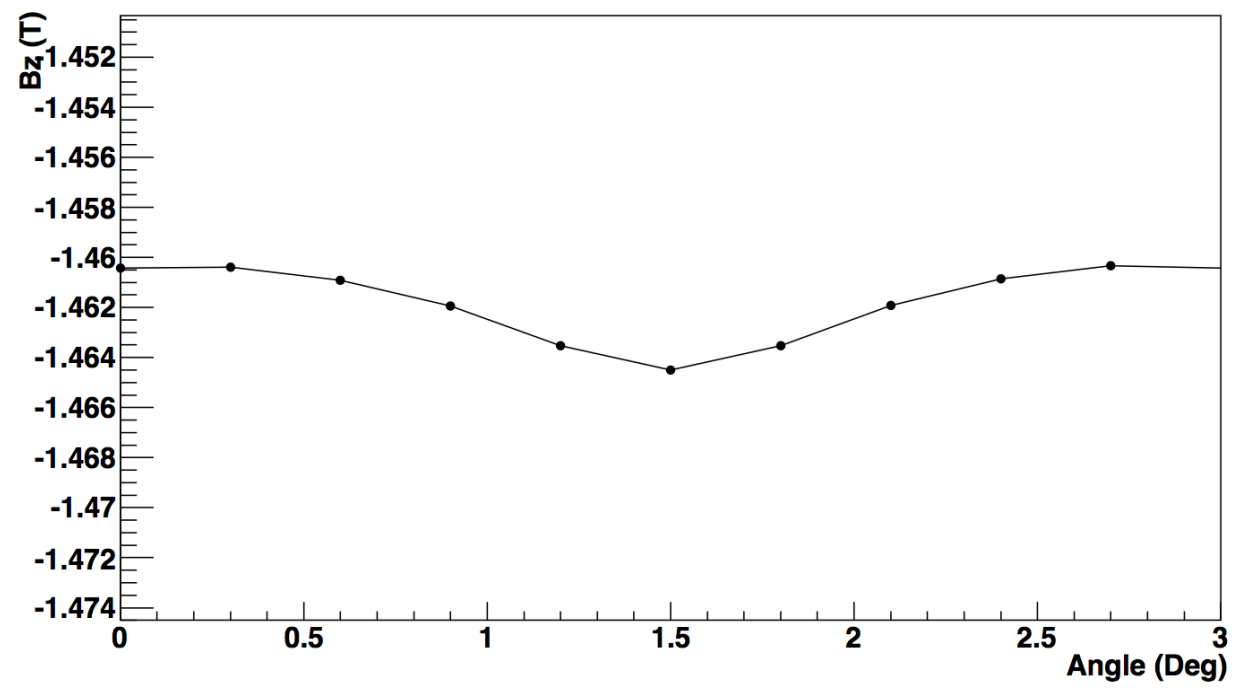

Figure 15.25: Field values for points axially around the 3 degree section of the ring for an OPERA model with a vacuum chamber that has had a slice cut out of it. The vertical field penetrates to a larger degree than the model of a vacuum chamber with no cut. This is for a single time point when the external field is at its maximum. Error bars are included within the plot.

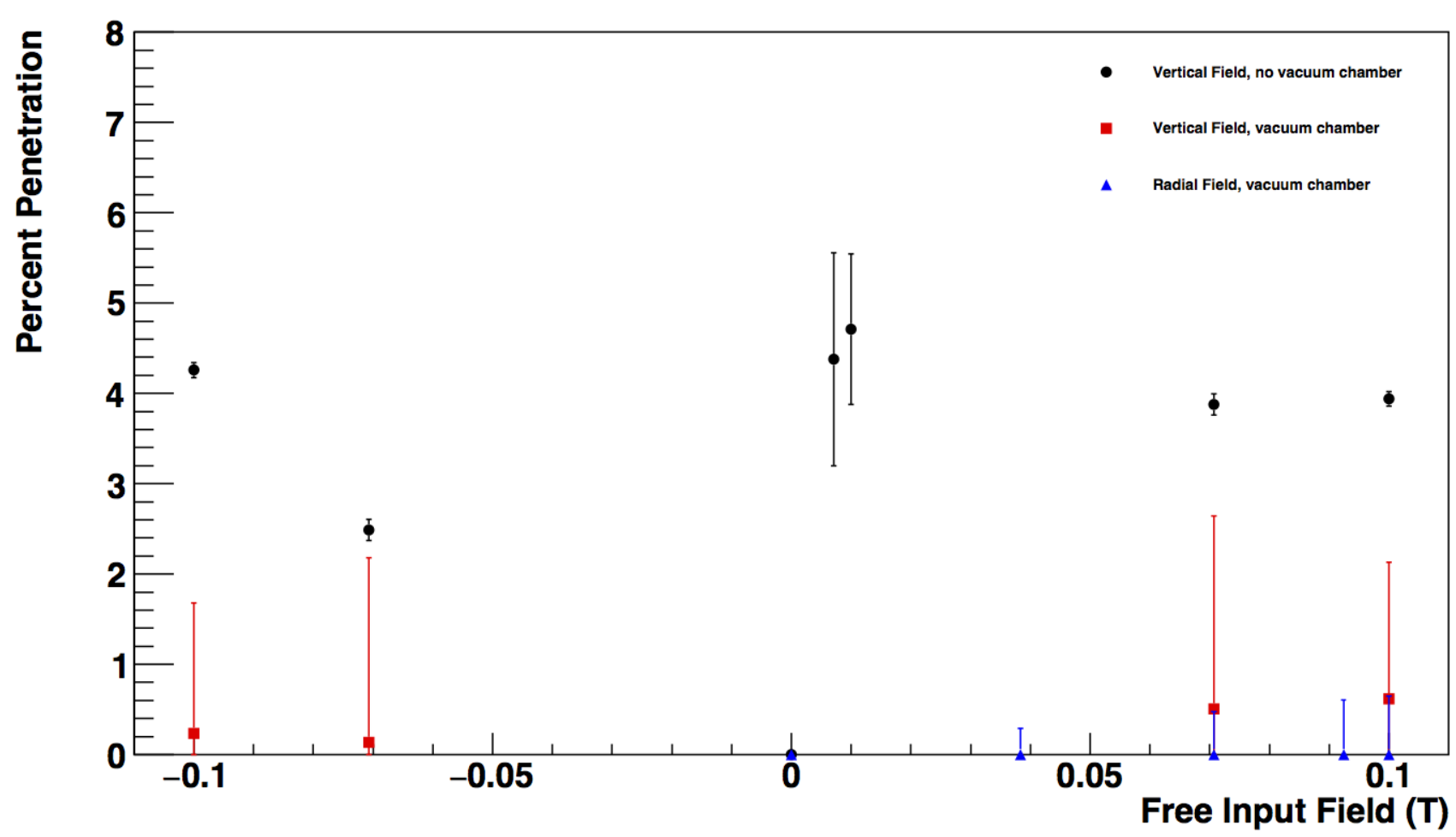

Figure 15.26: Percentages of varying free field values at the center of the storage region for different OPERA 3D models. The shielding of the ring system is shown to passively be quite effective.

The effect of these external fields in E989 can be monitored in several ways. While readings of the fixed probes are asynchronous with accelerator operations and the 60 Hz power grid, they will be time-stamped with an absolute accuracy better than 1 ms. The FIDs have time constants of just a few ms, so fixed probe readings are in principle sensitive to field perturbations at frequencies up to a few hundred Hz. However, the probes are only read out at about 5 Hz, so perturbations at 15, 30, 60 Hz etc will get aliased to additional image frequencies when probe readings are Fourier analyzed.

A scheme of monitoring the transients in the hall with the use of several fluxgate magnetometers has been proposed. We will study that, and other suggestions with additional Opera-3d simulations to determine the most effective method of measuring these transients. We will also have ample time during the shimming process to study these effects, including carrying out shielding studies with fluxgates inside and outside of the vacuum chamber.

We are also concerned about possible effects from the calorimeter and tracker electronics, and we are beginning an active program to study the magnetic fields produced by the electronics as a function of rate.

## 15.7 Other systematic effects on the field determination

Other contributions to the E821 systematic uncertainty on $\omega_p$ are listed in Table 15.4, along with a brief comment on the changes required to meet the E989 goal. The total of these effects was estimated to be 100 ppb in E821 (see [7, 54] and Table 15.1). Details on the sources of uncertainty and how they will be reduced to 30 ppb for E989 are discussed below.

**Kicker Transients:**

One source of field uncertainty in BNL E821 was from transient magnetic fields from eddy currents in the vacuum chambers induced by the pulsed kicker magnet. These field transients were measured in the E821 g-2 storage ring volume, before data-taking, using a magnetometer based on Faraday rotation in a crystal of TGG [53]. The transients affect the kicker region (roughly 5 m out of the 45 m circumference), and were of order 13 mG 20 $\mu$s after the kick, corresponding to a local change of 1.1 ppm [53]. However, the transients decay rapidly so by 30 $\mu$s when data-taking started, and when averaged over the ring circumference, the effect was 20 ppb [54].

For E989, an effort is being made to reduce these transients as a consideration in the new kicker design. Second, rather than being left as an uncertainty, measurements of the transients will be incorporated into the final field result. This can be done with the E821 Faraday magnetometer, which reached noise levels equivalent to a 20 ppb perturbation after some averaging. With more averaging and no other changes, a measurement of the kicker transients can be made with an uncertainty of 10 ppb. This will be done before data-taking with the TGG crystal located in the storage volume.

In E989 additional steps are possible. The Faraday magnetometer performance can be improved by replacing the E821 10 mW $Ar^+$ laser at $\lambda = 514.5$ nm, with an off-the-shelf

| Source of Uncertainty | E821 Magnitude [ppb] | Improvements for E989 | E989 Goal [ppb] |
|---|---|---|---|
| Kicker Eddy Currents | 20 | Average more measurements, reduced transients, improved magnetometer | 10 |
| Higher Multipoles | 30 | Move trolley probes to larger radius, measure higher multipoles | 15 |
| Trolley temperature changes | 45 | Calibration of temperature effects, reduced temperature range | 15 |
| Trolley power supply voltage fluctuations | 20 | Use better power supply, calibration of effects due to voltage fluctuations | 5 |
| Oxygen effect | 40 | Improved measurement of oxygen effects | 10 |
| Image effects | 40 | Reduced susceptibility probes, measurement of image effects | 15 |
| Frequency Reference | 3 | Calibration of effects due to change in reference frequency | 3 |
| Total | $\leq$100 | | 30 |

Table 15.4: Other sources of systematic uncertainty from BNL E821, with changes in preparation to meet the requirements for E989.

405 nm diode laser with 80 mW power (Edmund Optics #64-823 for instance). The Verdet constant at 405 nm (and hence signal) of TGG is more than 50% larger than at 514.5 nm. The higher laser power also improves the ratio of signal to photon shot noise. This allows the transient signal to be measured with a lower noise floor in less time. Measuring the second beam from the polarizing beam cube splitter (analyzer) of E821 doubles the signal to noise and cancels noise from laser intensity changes.

Other improvements being considered are to record the kicker pulse waveforms shot-by-shot so the field transients can be estimated more precisely if the waveforms change. As a redundant check, the Faraday measurements taken inside the storage volume before data-taking could calibrate a second magnetometer positioned slightly outside the storage volume and left in place during data-taking. This would continuously monitor and provide a correction to the field average from kicker-induced transients. The second magnetometer would be build specifically for the transient field measurement, to be sensitive to few mG fields with a bandwidth approaching 1 MHz. This magnetometer could not use the high susceptibility, field-perturbing TGG, but a longer crystal (3-4 cm) of diamagnetic SF59 for comparable performance Using a 405 nm diode laser of a few hundred mW, balanced photodetectors with photocurrents subtracted, AC-coupled, then amplified, can approach mG sensitivity on time scales of a minute, and reduce the uncertainty on the field average from kicker transients well below 10 ppb. This would be developed and implemented if errors from kicker transients must be reduced below 10 ppb to compensate larger uncertainties in other parts of the field measurement.

**Higher Multipoles:**

In E821 the trolley measured the magnetic field in the muon storage region to a radius of 35 mm. The field beyond 35 mm was extrapolated using the measured multipoles up to and including the normal and skew decupoles. Early in E821 before the vacuum chambers were installed (and well before the final field homogeneity was achieved), the field was measured to 45 mm radius using a special shimming trolley. The multipoles higher than decupoles were measured to be less than 8 ppm at 45 mm, and when convoluted with the falling muon distribution at large radius, gave a maximum uncertainty of 30 ppb from higher multipoles.

In E989, this uncertainty will be reduced to 10 ppb. First, with higher field uniformity and stability, the higher order multipoles should be significantly smaller than the worst case estimate given above. The higher order multipoles will be measured with the shimming trolley to 45 mm so we have a map of the field over the full muon aperture before the vacuum chambers are installed. The influence of the vacuum chambers will measured using a modified shimming trolley that is inserted in regions of the vacuum chambers where the probes can be extended to the full storage aperture (with the chambers at atmosphere). Second, if possible, the beam-tube trolley probes will be moved to slightly larger radius to map a larger fraction of the storage volume, which will also reduce these uncertainties. If necessary, when the ring is up to air, we will make consecutive measurements with the trolley probes rotated around the trolley axis. Combining these field maps allows a direct measurement of the higher order multipoles in the presence of the vacuum chambers. These steps will reduce the uncertainty from 30 to 10 ppb.

On a related note, better tracking and monitoring of the muon distribution will reduce the uncertainties on the convolution of the muon distribution with the field. Also, unlike E821, in E989 the multipoles beyond the quadrupole in the field convolution with the muon distribution, such as the normal sextupole contribution, will be corrected for by including the measured values in the convolution with the muon distribution, rather than leaving these contributions as an uncertainty. Given the expected significant improvements in hall temperature stability, these multipoles should be more stable than in E821.

**Trolley Temperature:**

Changes in the trolley temperature in BNL E821 lead to observed fluctuations in the Larmor frequency extracted from the trolley probes of order 30 ppb/°C. This sensitivity can come from the electronics and from actual temperature dependent changes in the magnetic environment in the trolley and probes. Trolley measurements during E821 were made over a temperature range of 23°C to 30°C, with an RMS spread of 1.5 degrees, for an estimated 45 ppb uncertainty. Within a trolley run, the electronics rose in temperaure by about 4 degrees, the probes by about 1.5 degrees [9].

These temperature effects will be studied extensively in a thermal enclosure in the MRI solenoid at Argonne, so temperature effects can be corrected rather than left as an uncertainty. The temperature dependence of the trolley electronics will be measured. The perturbations to the trolley probe readings from the temperature dependence of the susceptibility of the trolley aluminum shell will be measured (a fractional change in the perturbation of $10^{-3}$ for every 2°C change in temperature is expected). The temperature dependence of the probe response due to temperature dependent changes in the bulk susceptibility and chemical shift of the NMR sample will also be measured. The temperature sensors in the trolley will be calibrated to 0.1°C. Also, the possibility of using petrolatum in the trolley probes will be investigated since its temperature dependence (at least of the bulk susceptibility) might be lower than in water. Changing the trolley probe sample shape to spheres might also reduce the temperature sensitivity (since it eliminates the temperature dependence of the bulk susceptibility of the sample). This will tested in the MRI magnet.

Finally, applying a high vacuum compatible, non-magnetic, high-emissivity coating to the trolley (Cerablak HTP with $\epsilon > 0.9$, made of carbon encapsulated in an aluminum phosphate glass) should lead to a reduction in the temperature rise of the trolley during trolley runs, roughly by the ratio of emissivities, $\epsilon_{\rm E989}/\epsilon_{\rm E821} \geq 0.8/0.2 \approx 4$, a minimum of several degrees. This will reduce the magnitude of all temperature dependent changes in the trolley.

Perhaps most important, the much more stable experimental hall temperature in E989 compared to BNL, will reduce the range of temperatures of trolley operation from 7°C in E821 to 2°C in E989. This alone should reduce temperature-related uncertainties by a factor of 3. Between careful calibration of temperature effects, and efforts to reduce the temperature range of the trolley, a systematic uncertainty of 15 ppb should be achievable.

**Trolley Power Supply:**

The E821 trolley power supply operated around 9.9 V and drifted around 50 mV. These fluctuations in the trolley supply seemed to lead to FID baseline shifts and uncertainties in counting of zero-crossings, leading to a voltage sensitivity of roughly 400 ppb/V observed in the probe readings, for an uncertainty of 20 ppb. This will be replaced in E989 with a modern stable linear supply. With and stability and monitoring at the level of 10 mV, this uncertainty should be below 5 ppb in E989.

**Oxygen Effect:**

The oxygen effect is described in [7], and is due to the paramagnetism of $O_2$ which is inside the trolley and perturbs the field seen by the probes. It also affects the plunging probe. The problem arises when the absolute calibration is performed since the ring is then filled with air, creating a different field environment. The resulting correction (which was measured during E821 in the storage ring magnet) is about 40 ppb (with roughly a 20 ppb uncertainty). The correction will be measured again carefully in E989 in the ring and the MRI solenoid for the trolley, plunging probe, and absolute calibration probe. Given the much greater stability of the solenoid, it should be possible to reduce the uncertainty to 10 ppb.

**Image Effect:**

When the plunging probe and absolute calibration probe are inserted in the gap between magnet pole pieces, the field at the NMR sample is perturbed by the magnetic images of the probe materials in the pole pieces. When the probe moves vertically in the storage volume to calibrate the trolley probes, the magnitude of the images change, hence the effective shielding properties of the probe are position dependent. The magnitude of these effects is estimated as 40 ppb from knowledge of the probe materials and their susceptibilities. In E989, the position dependence of the shielding will be reduced by using combinations of materials with near zero susceptibility (see section 15.4.1), and the residual effects will be measured in the ring to reduce the uncertainty to 15 ppb.

**Frequency Reference:**

The frequency extracted from an FID had a small 3 ppb dependence on the synthesizer reference frequency. This will be studied again in E989 to confirm that the effect is small, but not serious enough to warrant any changes. The effect should actually disappear when the FIDs are digitized in E989 and zero-crossings can be interpolated to give fractional clock counts.

In combination, all of the efforts detailed above should reduce the systematics observed in E821 from 100 ppb to 30 ppb in E989.

## 15.8 Magnet Shimming

### 15.8.1 Overview

The main technical objective of the $g-2$ storage ring shimming is to produce the most uniform field possible. Both the muon distribution and the average field can be described by multipole expansions (see Sec 15.3). The field uniformity requirement is that the field varies less than 1 ppm over the 4.5 cm storage radius. Specifically, each multipole contribution should vary by less than 1 ppm when averaged over azimuth. In order to enhance additional systems within the field measurement, we will aim to reduce the variation of these individual multipole contributions to 100 ppb. Improved field uniformity at any local azimuthal position is also desirable since the performance of the trolley NMR probes relies on keeping the field gradients as small as possible in order to optimize the measurement of the free induction decay (FID) signal (see Sec 15.2.1). Furthermore, small field gradients reduce the uncertainty contribution from the position uncertainty of the probes. Since the probes sample the field over a non-negligible volume, the requirements on knowledge of the position are relaxed if the field gradients are minimized.

The $g-2$ magnet was designed to produce a field uniformity in the muon storage region of better than a few parts in $10^4$ prior to shimming. This design spec was achieved by using high-quality steel for the magnet yoke, and ultra-low carbon steel (ULCS) for the pole pieces. Upon assembly, the field uniformity was improved by more than two orders of magnitude through a shimming sequence. The general shimming strategy implemented in E821 was two-fold: passive shimming via precision alignment of ferromagnetic materials and active shimming utilizing current distributions. We will base our general shimming procedure on that of E821 and the experience gained therein, supplemented by modeling results.

The $g-2$ superconducting coils, yoke, pole pieces, and shims have been simulated with `OPERA-2D` as well as `OPERA-3D` [38]. The results of these simulations are compared both with `POISSON` simulations and results obtained during the development of E821 [27, 28, 29, 30, 31, 37, 39]. One critical aspect of the simulation is the use of realistic B-H magnetization curves. Although the steel is not fully saturated at 1.45 T, the response is not perfectly linear. This non-linearity is partially responsible for generating higher-order multipole moments in the simulations, and must be recognized during the actual shimming procedure. Our `OPERA` simulations will allow for a sophisticated shimming plan that improves the overall uniformity of the field in E989 compared to E821.

The shimming consists of the following elements: (1) Passive or mechanical shimming using precise positioning of materials, and (2) Active or current-based shimming using surface correction coils. These are discussed in the following sections.

### 15.8.2 Passive Shims

#### Shimming Procedure

Passive shimming refers to the set of mechanical adjustments that are performed during the assembly of the ring and remain fixed during the running period. The strategy begins with adjustments far from the muon storage region and work inwards with finer and finer adjustments. The principal passive shimming elements consist of the following:

1. Iron pieces on the yoke
2. Alignment of the pole faces
3. Wedge shims in the air gap between the pole piece and yoke
4. Edge shims in the gap between upper and lower pole faces
5. Iron pieces on the pole surface near the azimuthal gaps between adjacent pole faces

Throughout the physics measurement, the NMR trolley described in Section 15.3 will travel around the ring to map out the magnetic field. Due to the limited space inside the vacuum chambers, the NMR probes inside the trolley only extend to $r$=3.5 cm, which is smaller than the extent of the muon beam ($r$=4.5 cm). However, prior to the installation of the vacuum chambers, more space is available between the pole pieces and a larger trolley will be used. This shimming trolley (see Figure 15.27) consists of 25 NMR probes; one is at the center of the muon storage region, eight are at a radius of 2.25 cm, and 16 are at a radius of 4.5 cm. This allows a mapping that extends to the outer radius of the muon storage region. The shimming trolley probes use the same electronics as the fixed probes.

In E821 the shimming trolley was positioned on the end of a $\approx 7$ meter turntable arm positioned about the center of the storage ring. Rotating the turntable allowed the shimming trolley to map the field at various azimuthal positions. In E989, the trolley will be moved manually or with a small arm and motor.

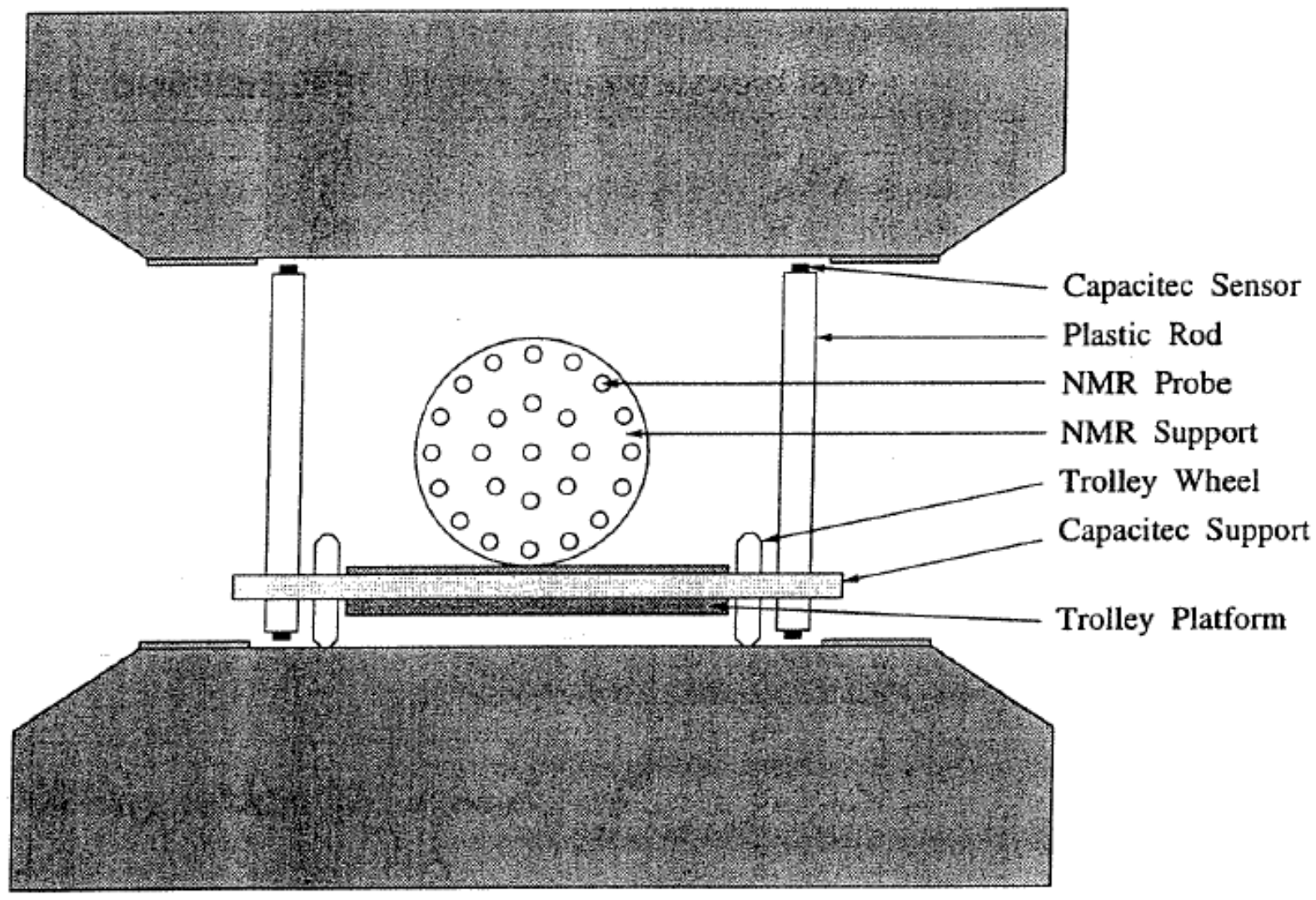

Figure 15.27: Schematic depiction of the NMR shimming trolley situated on a platform. Capacitive sensors on plastic rods help measure the pole piece alignment.

The platform on which the shimming trolley rests also allows for a precision measurement of the vertical gap between the upper and lower pole faces. This information can then be used to re-align the pole pieces ultimately producing a more uniform field. In E821, plastic rods with capacitive sensors on each end allowed for a determination of the relative parallelism between the poles [39]. For E989 the plastic rods might be replaced by quartz rods or other materials with very low coefficients of thermal expansion and low susceptibility. Furthemore, we are upgrading the capacitive sensors to a series whose readings are less sensitive to temperature and contain fewer field-perturbing elements. Careful attention will be paid to systematic effects due to changes in the rod angles and differences in length.

The data from the shimming trolley is analyzed and used to inform the next iteration of mechanical adjustments necessary to proceed to the next stage of the shimming procedure. A two-dimensional slice of the $g-2$ magnet used for `OPERA-2D` simulations is shown in Figure 15.28 for reference.

**Yoke Iron** The yoke is subdivided into twelve 30° sectors, as described in Section 9.2. Long wavelength azimuthal variations in the field uniformity are addressed by adjusting the positioning of pieces of iron on the outer surface of the yoke. In particular, an increase in the air gap between the top piece of steel and the upper yoke plate (see the label "Air gap" in Figure 15.28) leads to an increase in the overall reluctance of the magnetic circuit. In this manner, rough adjustments to the dipole field can be achieved on a sector-by-sector basis. In other regions of the ring, steel shims will be added to the outside of the yoke in order to compensate for the various holes and penetrations that are required for items like vacuum feedthroughs, the inflector, etc.

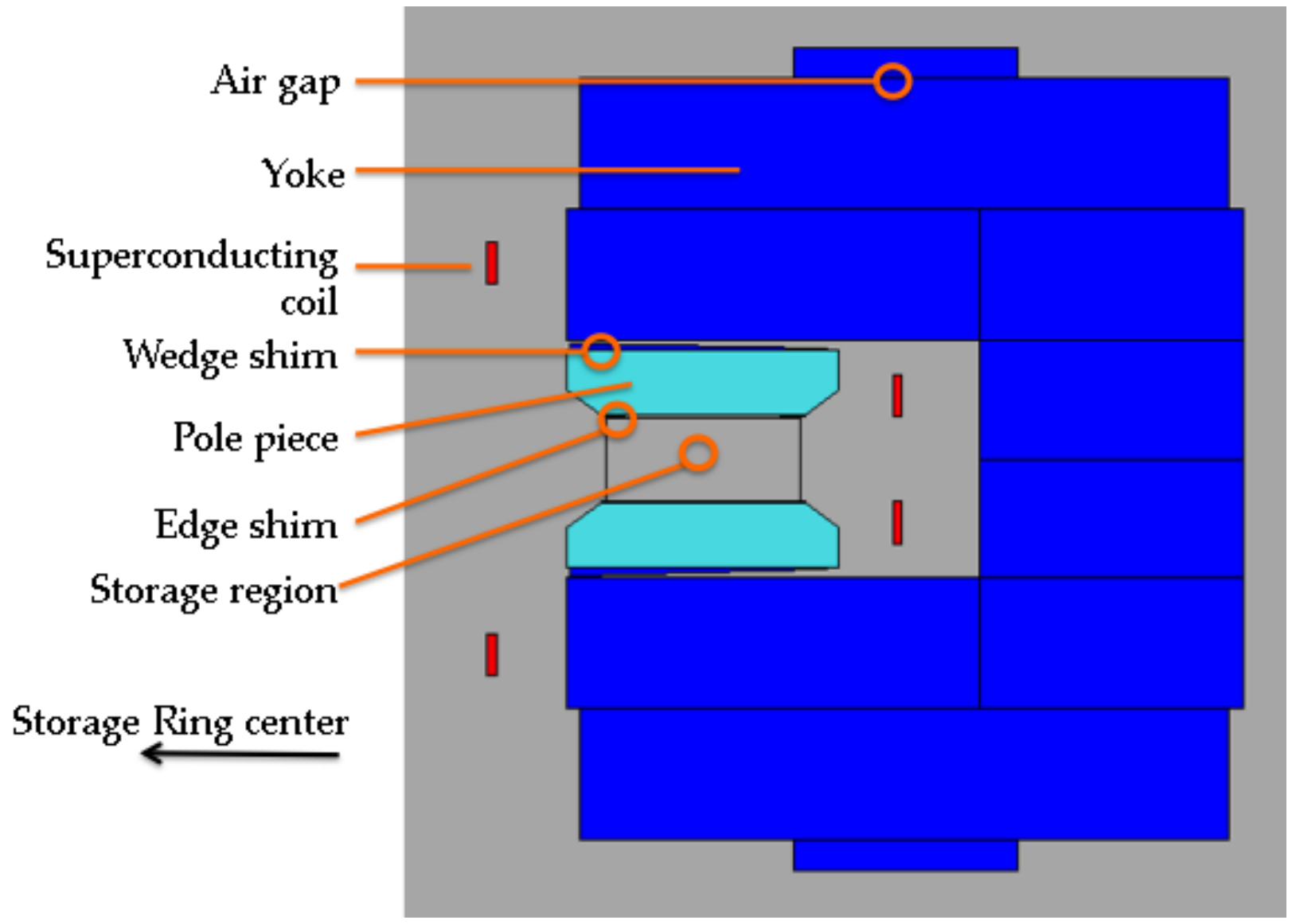

Figure 15.28: OPERA-2D model of the $g-2$ magnet. The yoke and wedge shims are steel and shown in blue. The pole pieces (cyan) are made from ultra-low carbon steel (ULCS). The current in the superconductor coils is into the page for the inner coils and out of the page for outer coils.

**Pole Piece Alignment** The Capacitec sensors mounted to the shimming trolley apparatus will measure the gap between the upper and lower pole pieces to a precision of $< 1$ $\mu$m. OPERA-3D simulations show that an increase in this gap size of 25 $\mu$m corresponds to a 250 ppm decrease of the dipole field [55]. A 50 $\mu$m change in the gap over the 56-cm radial extent of the pole corresponds to a change in the quadrupole moment of 120 ppm.

In E821 the gap height around the ring varied by $\pm 23$ $\mu$m rms, with a full range of 130 $\mu$m. Adjacent poles were matched to $\pm 10$ $\mu$m to reduce field distortions caused by discrete steps in the pole surfaces. Poles were leveled to $\pm$ 50 $\mu$rad using measurements from a precision electrolytic tilt sensor (see section on the radial field).

In E989 we will build a shimming trolley that is similar to the one used in E821 with a couple of upgrades. We plan to use several new copper-kapton thin-wand capacitive probes (Model GPS-7G-A-200-FX-5509-6108) instead of the button probes used in E821 which contained stainless steel. New quartz rods will be necessary to mount the probes and help provide the proper cable strain relief. A new set of 200Hz-BNC Dual Channel Linearized Capteura Amplifier Cards from Capacitec, Inc. will be used to read out these probes. The combination of these probes and amplifiers produce a linear output voltage in the range of 0 to 10V when the probe is a distance of 0 to 1.27 mm from the surface of interest. The quoted precision of this equipment allows a gap measurement precision of 0.07 microns. The overall temperature stability of these probes is 0.1 micron/degree Celsius, which is also improved with respect to the ones used in E821. The significantly improved temperature stability and uniformity of the experimental hall in E989 over E821 will help by reducing the size of thermal distortions, where changes of a few degrees Celsius change the gap by 10 $\mu$m or more due to thermal expansion of the steel. Finally, we intend to mount a vertical plate to the shimming trolley at the inner radius in order to enable monitoring by a laser tracking system. This system will hold non-magnetic reflectors on the plate and use a series of optical interferometers that are standard instruments used by the metrology department. This system will provide the location of the shimming trolley in the coordinates of the storage ring, and will allow for a more automated feedback procedure when analyzing the shimming trolley NMR data. Recommendations can then be made for each sector as we iterate on the mechancial adjustments necessary to properly align the pole pieces. The combination of these various system upgrades will allow E989 to exceed the benchmarks achieved by E821.

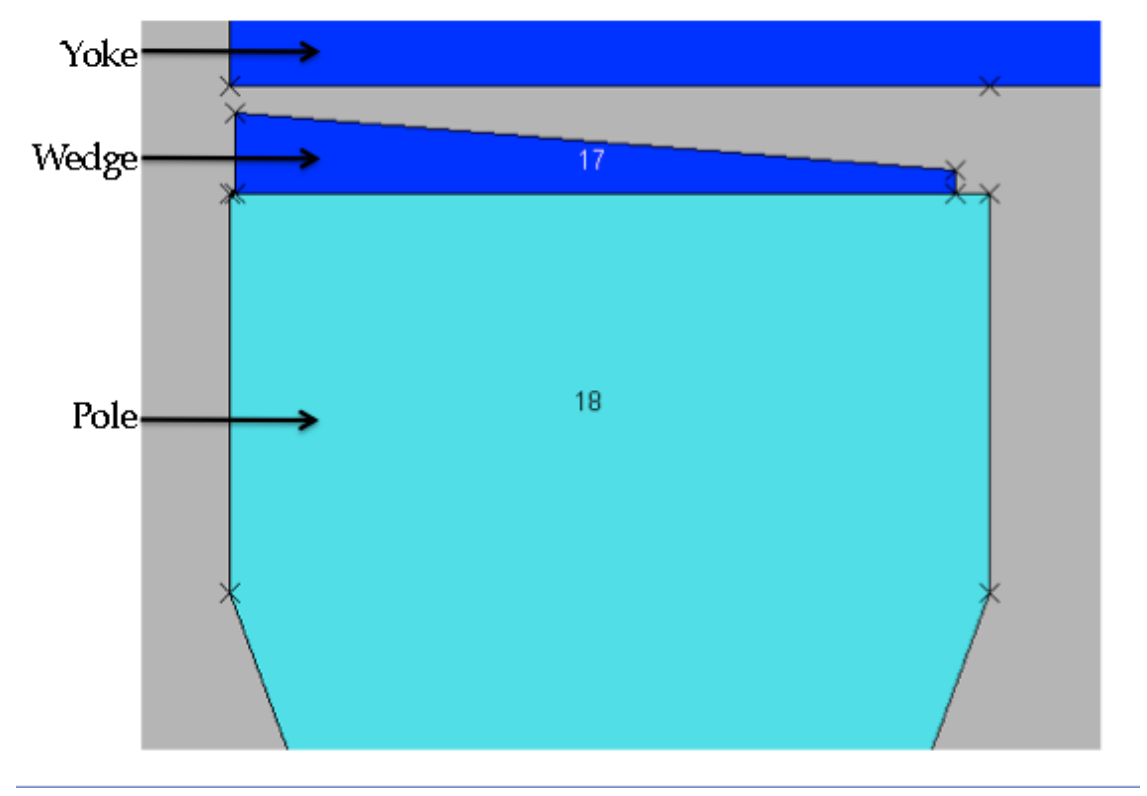

Figure 15.29: OPERA-2D model of the $g-2$ magnet, zoomed in on the air gap between the yoke and pole pieces. The wedge shims are radially adjustable.

**Wedge Shims** Wedge shims are inserted into the 2-cm air gap between the pole piece and the yoke, as depicted in Figure 15.29. The gap is designed to isolate the high-quality precision pole pieces from the yoke steel, which contains some magnetic inhomogeneities. Each 30° sector contains 72 wedge shims, which are 9.86 cm wide (azimuthally) and 53 cm-long (radially)[50]. This is shorter than the 56 cm-long pole pieces (radially), to accommodate radial adjustments. At the inner radius, the wedge shims are 1.65 cm thick, while at the outer radius, they are 0.5 cm thick. Viewed from above, each wedge shim is rectangular. Thus the space between adjacent wedge shims increases as the radial coordinate increases.

The angle of the wedge shims was calculated to be 20 milliradian in order compensate for the intrinsic quadrupole moment produced by the C-shaped magnet. Due to the asymmetry in the C-magnet, the field lines tend to concentrate in the gap near the return yoke. The dipole field is determined by the average thickness in the air gap above the storage region. The average wedge thickness is adjusted by translating the radial position of the wedge shims. Because of the shallow angle of 20 mrad, a radial movement by 50 $\mu$m changes the gap by 1 $\mu$m, allowing fine control for the dipole field. OPERA-2D simulations show that inserting the wedges into the air gap (towards the return yoke) radially by 50 $\mu$m produces a 5.4 ppm increase in the dipole field. The quadrupole and higher-order multipoles are each affected by less than 0.1 ppm for this adjustment. We will reuse the wedge-shims that were constructed for the E821 experiment. The angle correctly compensates for the intrinsic quadrupole of the C-shaped magnet, so we will just clean and refurbish these components.

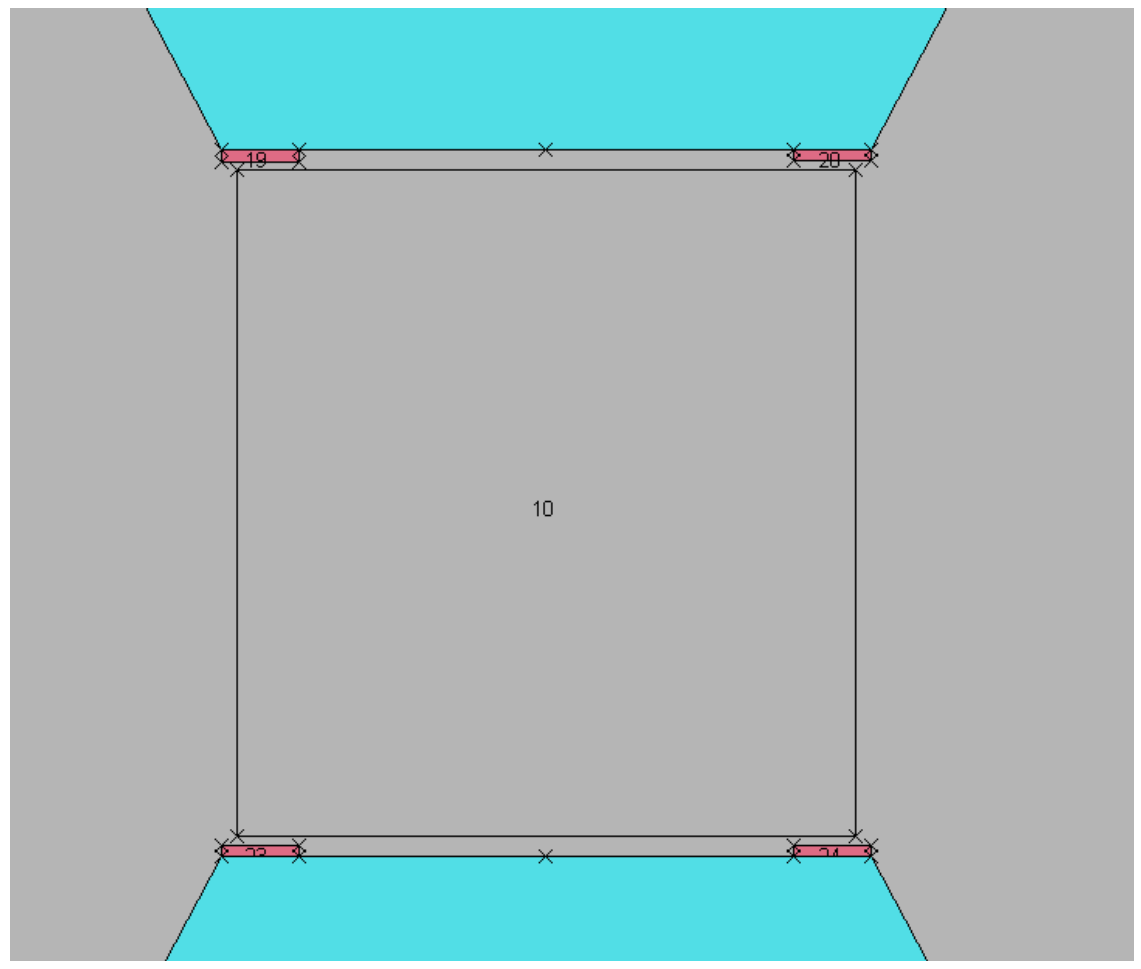

Figure 15.30: `OPERA-2D` model of the $g-2$ magnet, zoomed in on the storage region. Edge shims are mounted on the pole pieces. "Inner" refers the shim at smaller radius (closest to the center of the ring), while "outer" refers to the shim at the larger radius (closest to the return yoke).

**Edge Shims** Each of the 36 pole pieces has four removable edge shims mounted on the surface closest to the muon storage region, as shown in Figure 15.30. Each shim is 5 cm wide (radially), spans one pole piece (10° azimuthally), and is positioned at either the inner or outer edge of the pole faces. Variation of the thickness of the edge shims can produce predictable multipole corrections.

In E821, the shims were ordered oversized (3.2 mm for the outer shims and 4.4 mm for the inner shims) and then ground down to tune the quadrupole through octupole moments. A first pass was performed to uniformly grind the shims as a function of azimuth. In E821 a final pass was planned to optimize the thickness of the edge shims pole-piece-by-pole-piece, but this did not occur due to time constraints.

We have studied the effect of systematic shim thickness variations in `OPERA-2D`. Since the 2D model assumes vertical symmetry, the upper and lower edge shims are always adjusted simultaneously. Symmetrically increasing the thickness of both the inner and outer edge shims primarily affects the sextupole moment. `OPERA-2D` simulations helped us determine that a 100 $\mu$m increase in the edge shim thickness in all four corners increases the sextupole moment by 10.8 ppm. Asymmetric thickness adjustment leaves the sextupole moment unchanged and allows fine tuning of the quadrupole and octupole moments. Increasing the outer edge shim thickness by 100 $\mu$m while decreasing the inner edge shim thickness by the same amount increases the quadrupole and octupole moments by 13.2 ppm and 5.6 ppm, respectively. Although the simulation utilized vertical symmetry, this model can be extended to up-down and diagonal (skew) asymmetries.

We plan to use a similar shimming strategy in E989. An Ultra-Low-Carbon-Steel (ULCS) material called ARMCO Pure Iron is manufactured by AK Steel, and we have obtained a sample with dimensions 297 mm $\times$ 210 mm $\times$ 3 mm. It has excellent magnetic properties that are appropriate for the edge shim application: high permeability, uniformity, and machinability. We cut six small bars with volume 100 mm $\times$ 20 mm $\times$ 3 mm. Prior to sending them to KJS Associates for magnetic classification, we practiced introducing several types of machining stress: we ground off 200 micron from several samples, added countersunk holes, and additionally performed stress tests by hammering on the samples. Figure 15.31 show a sample B-H curve that has been characterized (red), as well as some black data points that come from the AK Steel data sheet. Ultimately, we found that the samples were relatively resilient against usual stresses. All samples were measured to have magnetization curves within the margin of error of the data sheet and measurement uncertainty [56]. Thus we will order oversized edge shims, map the field, calculate the appropriate thickness required, grind the shims, and iterate. Based on the experience of E821 and the extensive `OPERA` simulations, we believe this phase of the shimming will require only two iterations.

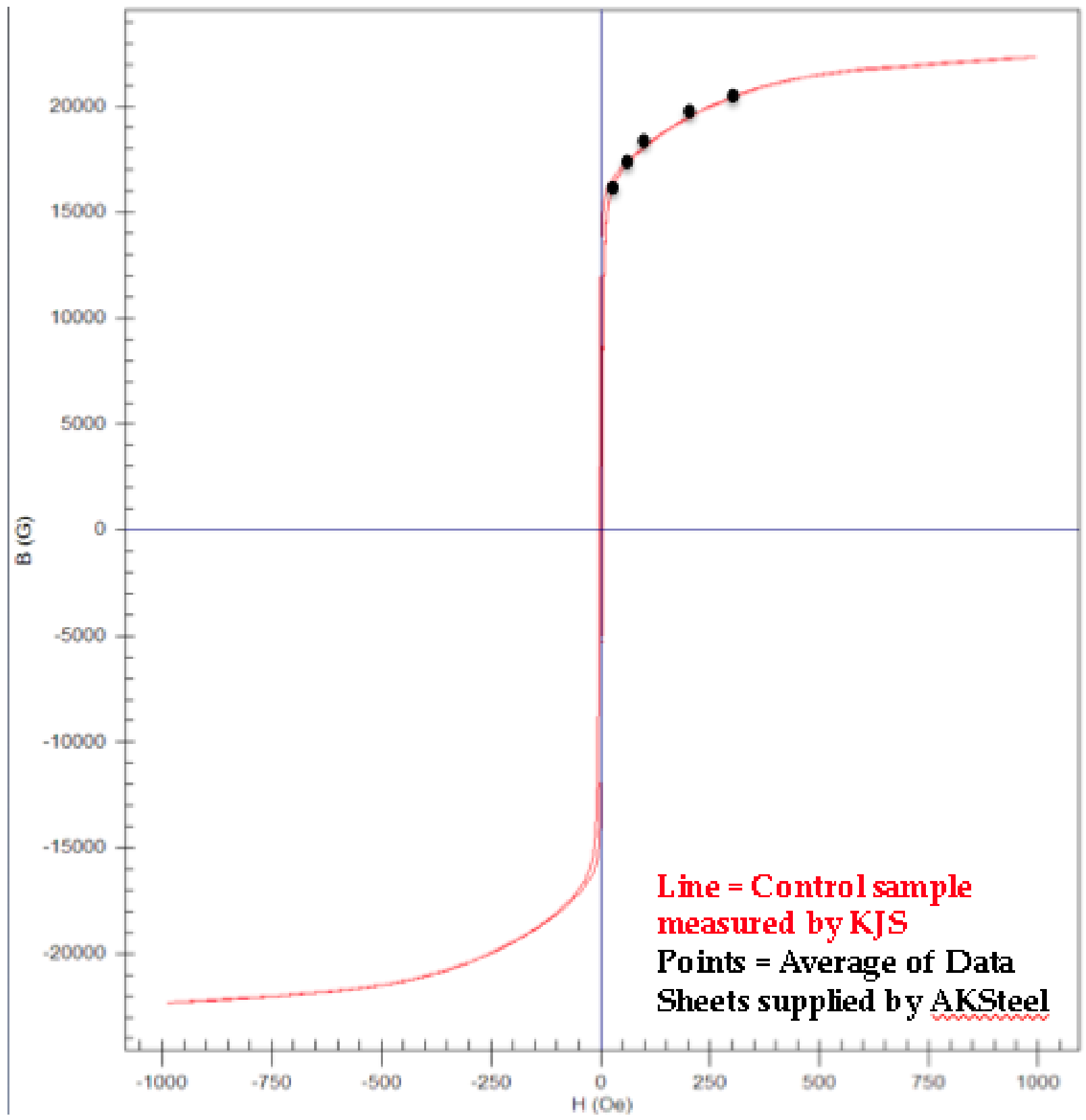

Figure 15.31: Test results of a control sample of ARMCO Pure Iron steel, manufactured by AK Steel and tested by KJS Associates. The measured B-H curve is shown in red and contains an uncertainty of 2%. The black data points correspond to the AK Steel data sheet for these samples, with an uncertainty of 1%. Additional tests of other samples showed expected reproducability.

**Gap shims** Significant variations in the magnetic field occur at the azimuthal boundaries between adjacent pole pieces, as shown in figure 15.32. The effect is even more pronounced at the pole piece surface than in the storage region, jeopardizing the effectiveness of the fixed probes located near the inter pole piece gaps. In E989, we plan to reduce the azimuthal variations in the field by shimming the gaps with thin iron plates. The basic concept is to span the surface of adjacent pole pieces with high quality steel plates varying from 10 to 100 $\mu$m. Simulations show that a local change of 10 $\mu$m in the air gap between the pole pieces results in a 40 ppm shift in the dipole field. This should be a fairly short range effect (azimuthally) that will reduce the local field gradients and improve the performance of the fixed NMR probes mounted in the vacuum chambers.

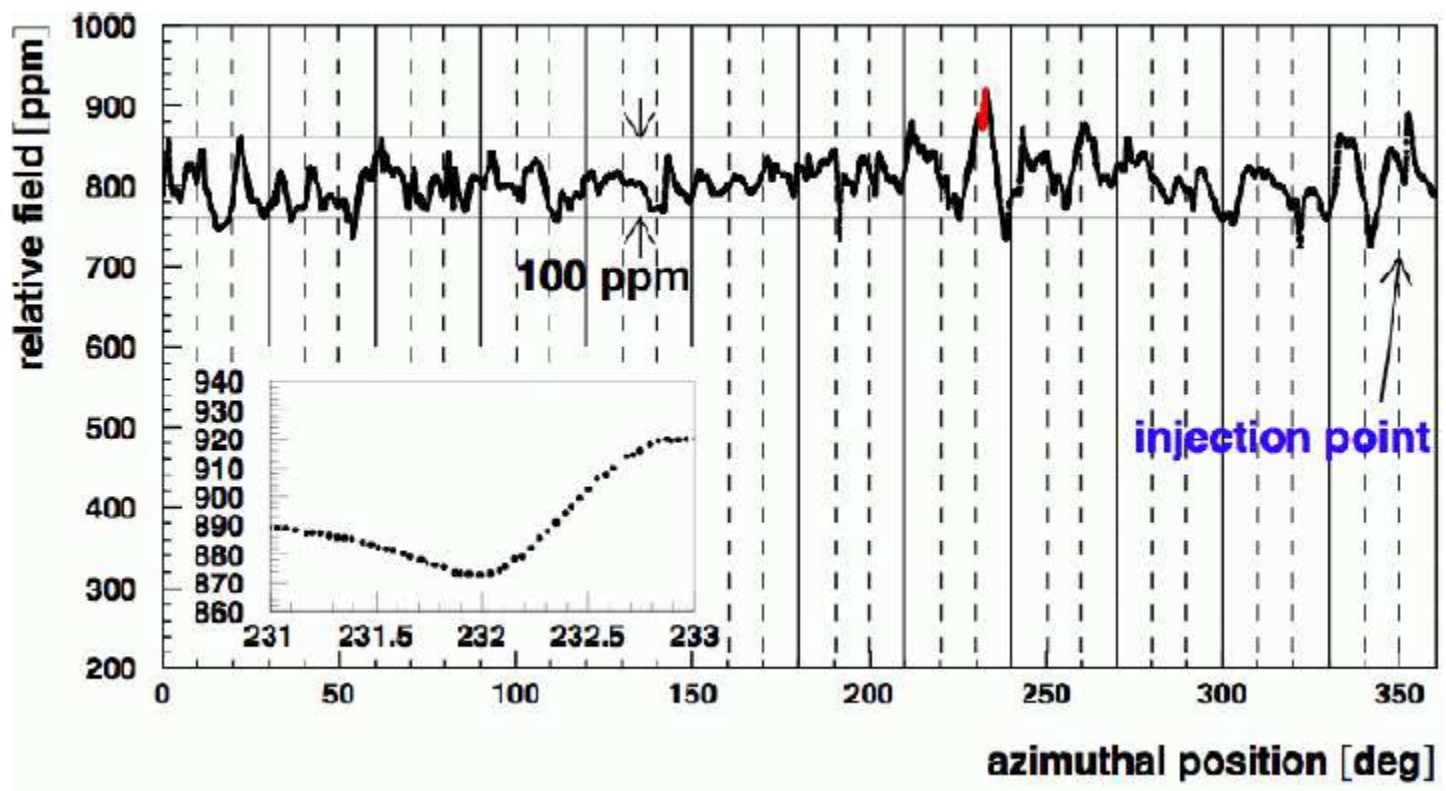

Figure 15.32: The magnetic field determined by the center NMR trolley probe versus azimuthal position in the storage ring during one trolley pass (reproduced from Ref [7]). The solid vertical lines denote boundaries between the 12 yoke sectors. The dashed vertical lines denote the pole piece boundaries.

### Radial Field

For E821, an auxiliary measurement of the radial component of the magnetic field was performed during the passive shimming phase prior to the installation of the vacuum chambers. In the storage region, the direction of magnetic field is principally vertical. The presence of a radial field component has a significant impact on the muon storage beam dynamics, affecting both the mean vertical position and the vertical betatron oscillations. Quantitatively, the radial field component needs to be measured to $\approx$ 10 ppm of the total vertical field. However, the NMR probes only measure the total magnitude of the magnetic field without providing information about the separate vertical and radial components. In E821 an auxiliary measurement using Hall probes was implemented to quantify the radial component of the field and we plan to repeat this procedure.

Figure 15.33 shows a schematic representation of the E821 setup used to measure the radial field [39]. Two Hall probes (BH-206, F.W. Bell) were vertically aligned to measure the radial magnetic field, with the Hall currents running in the $z$ and $y$ directions. To ensure alignment of the setup with respect to the gravitational vertical direction, electrolytic tilt sensors (RG33A, Spectron Systems Technology, Inc.) were mounted to the support structure. Finally, to account for potential misalignment of the Hall probes with respect to the support structure, the measurements were repeated after rotating the entire structure by 180° about the vertical axis and taking the difference of the Hall voltages.

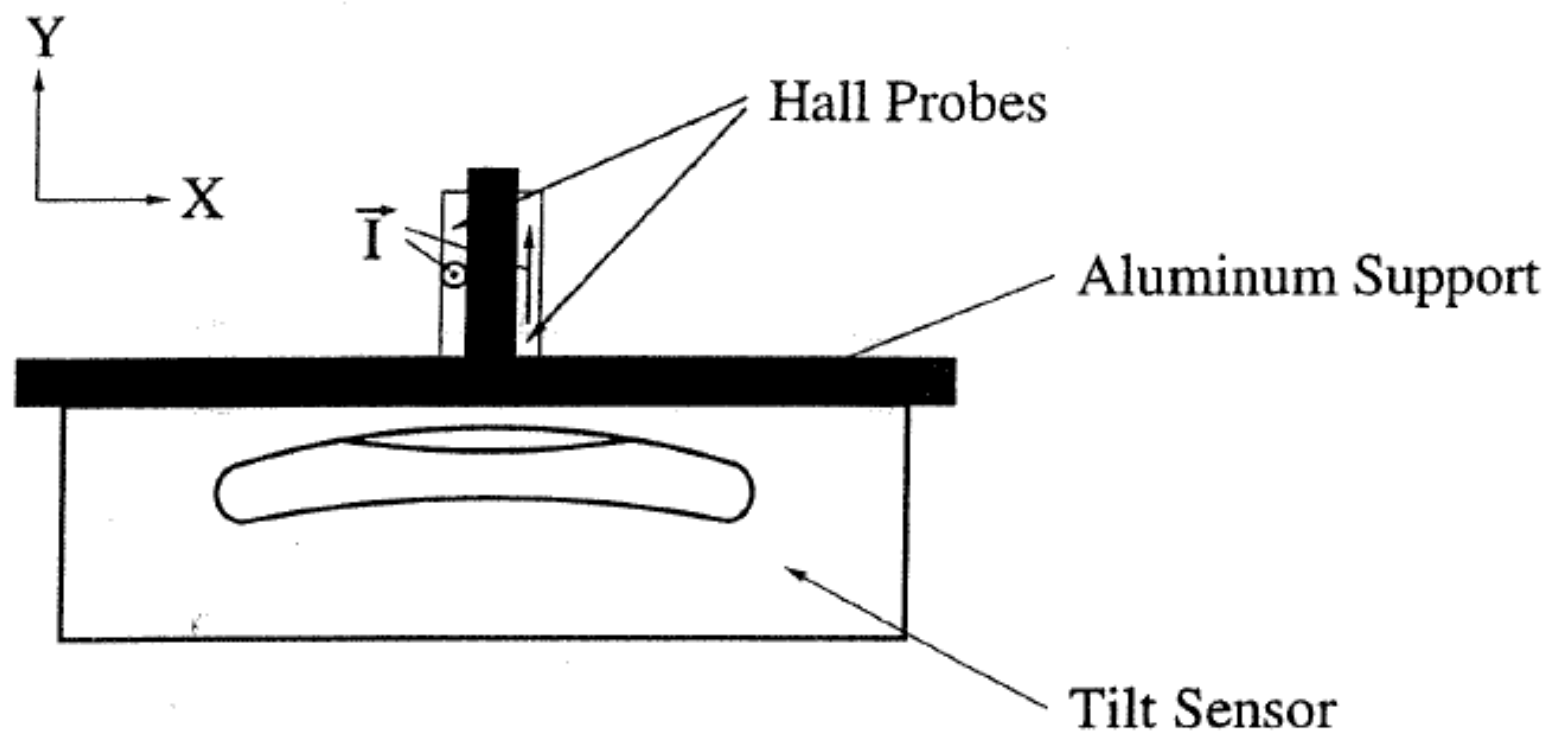

Figure 15.33: A schematic representation of the radial field measurement configuration in E821 [39]. Two Hall probes are mounted to measure the radial component of the field ($x$) with Hall currents oriented in the $z$ and $y$ directions. The rigid configuration is equipped with a tilt sensor. Rotating the entire setup 180° about the $y$-axis isolates the radial component.

Figure 15.34 shows the data from the Hall probes overlaid on the expected radial field determined from the multipole expansion of a field measurement. The overall precision of the radial field measurement was significantly better than the required 10 ppm. Figure 15.35 shows the radial field measurement (dots) from the Hall probe as a function of the azimuthal position around the ring. The line in this plot represents the measured pole tilt derived from the capacitive sensor data described above. The tracking of these two curves demonstrates the dependence of the radial field on the pole alignment.

### 15.8.3 Active Shims

Active shimming refers to the adjustment of current distributions to minimize any residual field non-uniformities that remain after the passive shimming is complete.

The principal active shimming controls consist of the following:

1. Control of the main superconductor current
2. Surface correction coils on printed circuit boards mounted between the pole face and the vacuum chamber
3. Dipole correction loops placed in the gap between the wedge shim and the yoke
4. Gap correction loops located in the azimuthal gaps between adjacent pole faces

#### Main Current

The central value of the dipole field is determined primarily by the current in the main superconducting coils. The nominal current is 5200 Amp per turn. `OPERA` simulations show that an increase of 1 A increases the field in the storage region by about 2 Gauss (140 ppm). During beam-on data collection periods, the field is monitored by the fixed NMR probes. The shape of the magnet gets distorted due to temperature variations which leads to a change in the magnetic field, so a feedback loop is utilized to stabilize the dipole field accordingly.

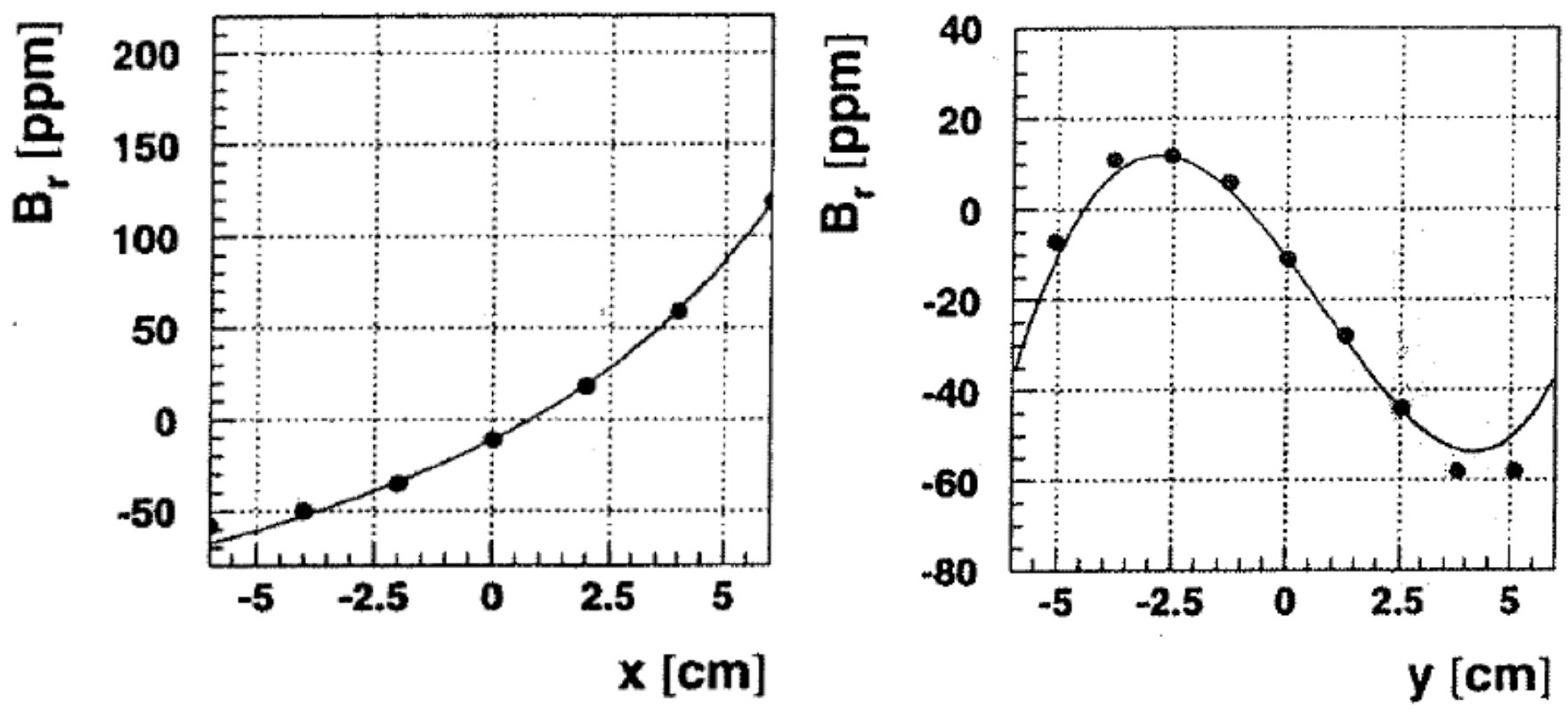

(a) (b)

Figure 15.34: Radial component of the magnetic field in ppm as a function of (a) radial position $x$ and (b) vertical position $y$. The dots show the data from the Hall probe, while the solid lines represent the field variation expected from the multipole coefficients calculated from the absolute field measurement.

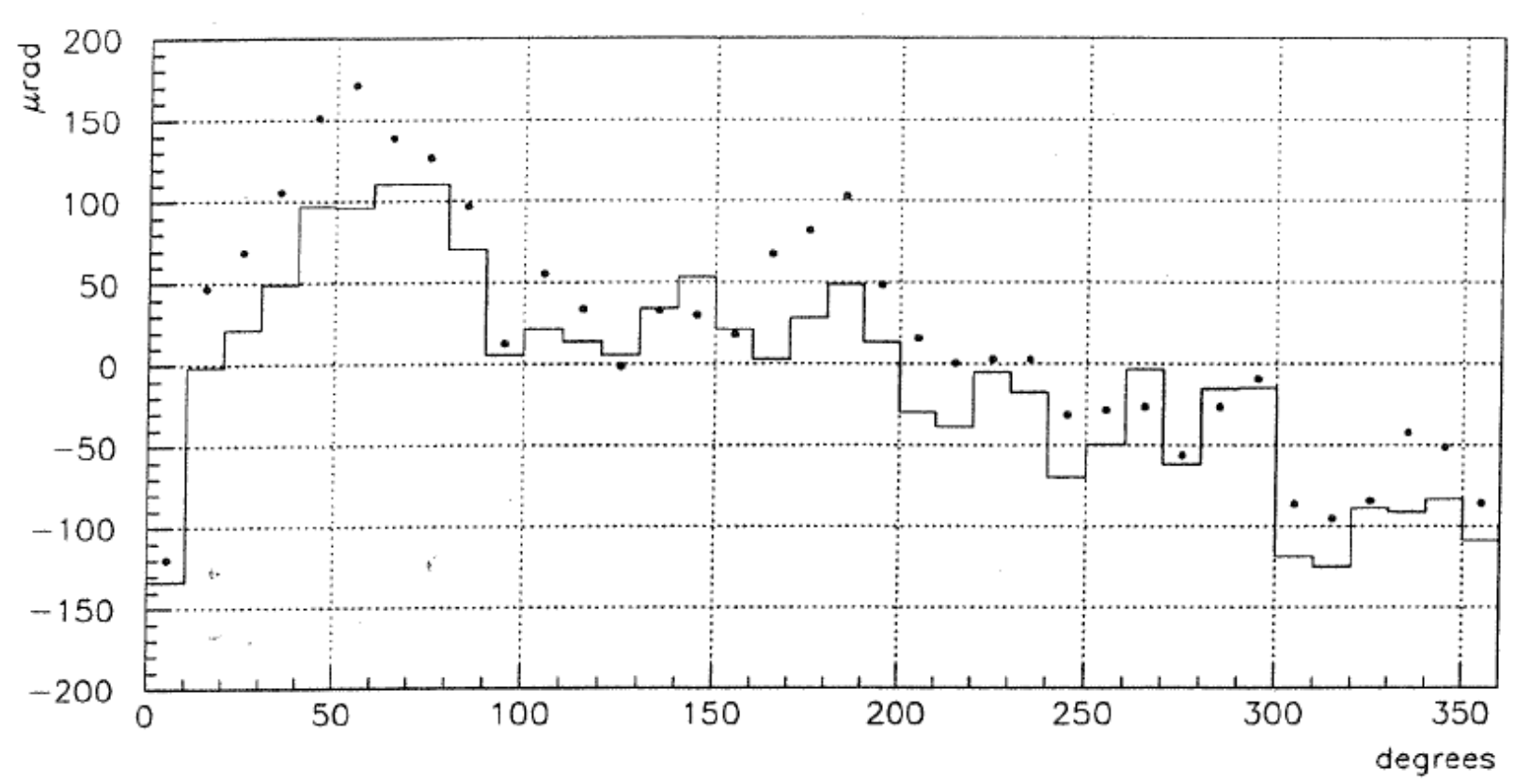

Figure 15.35: The radial field measurement (dots) from the Hall probe and the average pole tilt (line) from the tilt sensor are shown as a function of the azimuthal position around the ring.

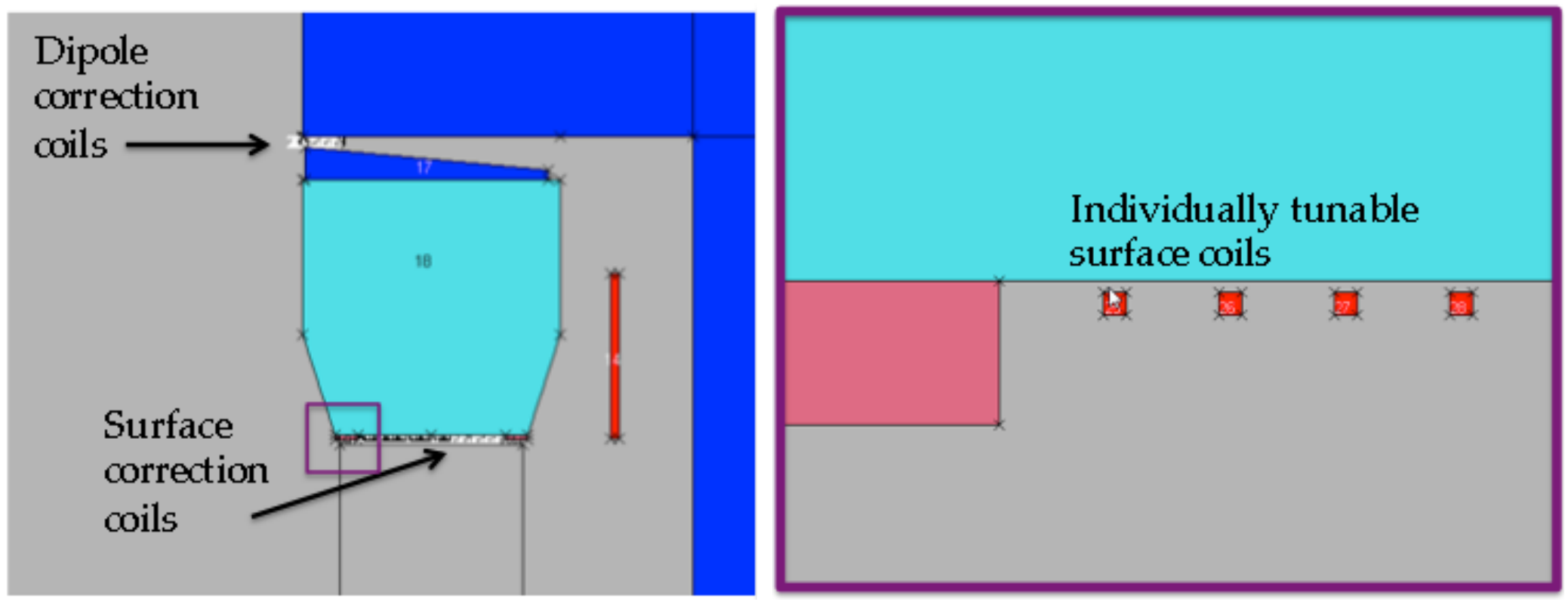

(a) (b)

Figure 15.36: OPERA-2D depiction of the dipole correction coils and the surface correction coils. (a) Schematic overview showing the positions of the current traces on the printed circuit boards. The purple box is zoomed in and displayed in panel (b). Individual traces are adjusted to tune various multipole contributions.

**Surface Correction coils**

Correction coils on the surface of the poles permit fine control of static, and slowly varying errors. The surface coils can be used to correct the lowest multipoles and adjust the field by up to tens of ppm, thus providing significant overlap between the iron shimming and the dynamic shimming. These coils have been constructed to generate moments over the entire 360° azimuth. The coils were designed with printed circuit boards, with 120 wires running azimuthally around the ring on the top and bottom pole surfaces facing the storage ring gap, and spaced radially 2.5 mm apart. The boards must be thin enough to fit between the pole faces and the vacuum chamber. With the pole-to-pole spacing of 180 mm and a vacuum chamber height of 165 mm, this allows up to 7.5 mm for each board and its corresponding insulation and any epoxy or glue that is necessary to affix the boards to the surface of the poles. We have studied the surface correction coils in OPERA-2D (see Figure 15.36) and verified that the expected residual multipole contributions can be compensated with the appropriately applied current distributions. One additional benefit of this system is that we can induce known multipole distributions in the NMR probe region and use the surface correction coils to help determine our position resolution of the trolley probes. A summary of the principal current distributions is shown in Table 15.8.3. E821 used these coils successfully to shim out the final few ppm for the higher order multipoles.

For E989, we plan to fabricate new printed circuit boards at Fermilab that extend over each of the 72 pole pieces. Each board should contain 100 azimuthally directed traces with a radial spacing of 2.5 mm, and a length of one pole piece ($\approx$ 1.25 m). The conducting elements should be positioned accurately to better than 50 $\mu$m in the radial direction. Since these boards will placed in the gap between the poles and the vacuum chamber, resistive heat dissipation is an important consideration. We wish to keep the total resistance per channel below 4 Ω. At a maximum anticipated current of $\pm 2$ Amp/channel, this amounts to 16 W/channel, or 3.2 kW total power if operating all traces at the maximum current simultaneously. This would be an unusual, upper-limit implementation mode for the surface correction coils, as Table 15.8.3 indicates the radial dependence of the current necessary to tune individual moments.

Table 15.5: Leading order current distributions as a function of the radial coordinate ($x$) that needed to correct for various multipole components, and the maximum range that can be corrected with less than 1 Amp. The currents are calculated at the fixed vertical position of the boards ($y = a = 9$ cm). Geometric and finite-size-effect corrections to the currents are necessary to compensate for residual higher order moments.

| Multipole | $K(x)(y = a)$ | Maximum range (ppm) |
|---|---|---|
| Quadrupole | $a$ | 20 |
| Sextupole | $2ax$ | 10 |
| Octupole | $3ax^2 - a^3$ | 8 |
| Decupole | $4ax(x^2 - a^2)$ | 6 |

Because these coils extend azimuthally around the entire ring, interconnects between adjacent boards must be designed. Low resistance (10 mΩ), high quality blade connectors can be introduced at the union of adjacent boards. With 36 connectors, this only contributes 0.36 Ω to the total loop resistance per conductor. This additional resistance is taken into account when determining the design resistance for each trace. If necessary, multiple conducting layers at each radial position can be utilized in order to reduce the resistance to the appropriately low values. We will finish each board with a protective cladding covering the traces to provide electric isolation from the poles and vacuum chamber.

There are 36 upper poles and 36 lower poles. Overall, 70 of the surface coil boards have identical designs, and two boards (one upper and one lower) will be specialized in order to accommodate leads to the power supply boards. The external leads will be routed in a manner that minimizes the production of unwanted stray fields.

The baseline design for the power supplies for the surface correction coils derives from the design for the DECAM Heater Controller Crate used in the Dark Energy Survey [57]. The circuit specifications for that application are very similar, and allow us to build upon previous developments at Fermilab. The power supplies must be able to provide bi-directional, continuous DC currents of up to 2 Ampere for 200 total channels, with individual control of the current on each channel. The types of ICs utilized in the DECAM Heater board will be able to provide these currents without overheating, assuming the implementation of reasonably standard crate cooling. The SCC resistance will be $< 4\Omega$. Each trace on the surface correction coil amounts to a 7m-diameter circular loop, so our design calls for filters to help mitigate noise effects. In order to cancel higher-order magnetic field multipoles, fine tuning of the individual currents must be provided. To achieve 100 ppb resolution for those moments, we need at least 5 mA resolution per channel. Over the dynamic range that the SCCs will need to shim, this requires a minimum of 10 bits for the DACs. We plan to use a 12-bit DAC to allow for finer resolution of the currents and better control of the magnetic field. A schematic of the DES Heater Controller Driver is included for reference in Figure 15.37. The specific components relevant to the $g-2$ SCC are shown in the simplified block diagram in Figure 15.38.

We plan to use three standard 6U VME crates, each with 17 SCC driver cards, with 4 channels per board. Accounting for adequate cooling, each crate would require about 9U of rack space. Our baseline plan is to reuse available VME crates at Fermilab.

Table 15.6: A summary of design requirements for the Surface Correction Coils and associated Power Supplies

| Category | Specificiation | Comments |
|---|---|---|
| Number of Boards | 72 | 1 per pole piece; 2 specialty boards |
| Channel per Board | 100 | 200 total (100 upper, 100 lower) |
| Resistance per trace | $< 4\ \Omega$ | Includes 36 blade connectors per channel |
| Channel spacing | 2.5 mm | Center channel at r=711.2cm |
| Current range | -2 A<I<+2 A | D C output, Each channel tuned separately |
| Current resolution | 5 mA | Controlled by DAC chosen |

As an alternative for the surface coils, we also investigated the possibility of routing channels in a G10 board and inserting 12-14 AWG wires for each of the traces. This alternative has the benefit of offering a very low resistance ($< 1\ \Omega$ per conductor), but will limit the precision of the placement of the coils. The mechanical construction of the boards would also be more time consuming. Our baseline alternative is to route traces on a circuit board.

Finally, we will explore the option of using the reverse side of the boards for active current shims in the azimuthal gaps between adjacent pole pieces. The baseline design calls for using passive mechanical shims in these gaps instead, but the continued value engineering efforts will revisit this issue.

### Dipole correction loops

The “continuous” ring was built with 10° pole sections, 36 of which form an almost continuous ring. Dipole correction coils were located in the air gaps of each 10° pole of the E821 storage ring, as depicted in Figure 15.36 (a). These coils consisted of 50 turns of copper wire wound in a rectangular shape. The dipole correction coils were capable of tuning each pole section independently, but were not used. Instead E821 used an active NMR feedback loop to stabilize the overall field by adjusting the main superconductor current. It is possible that the field in E989 could be stabilized in each pole piece separately by using the local NMR feedback to adjust the dipole correction currents. This would be particularly useful if temperature gradients over the 14 m diameter ring lead to different field distortions in different locations. However, given the much improved thermal stability and uniformity in MC-1 over E821, it is unlikely the dipole correction loops will be required.

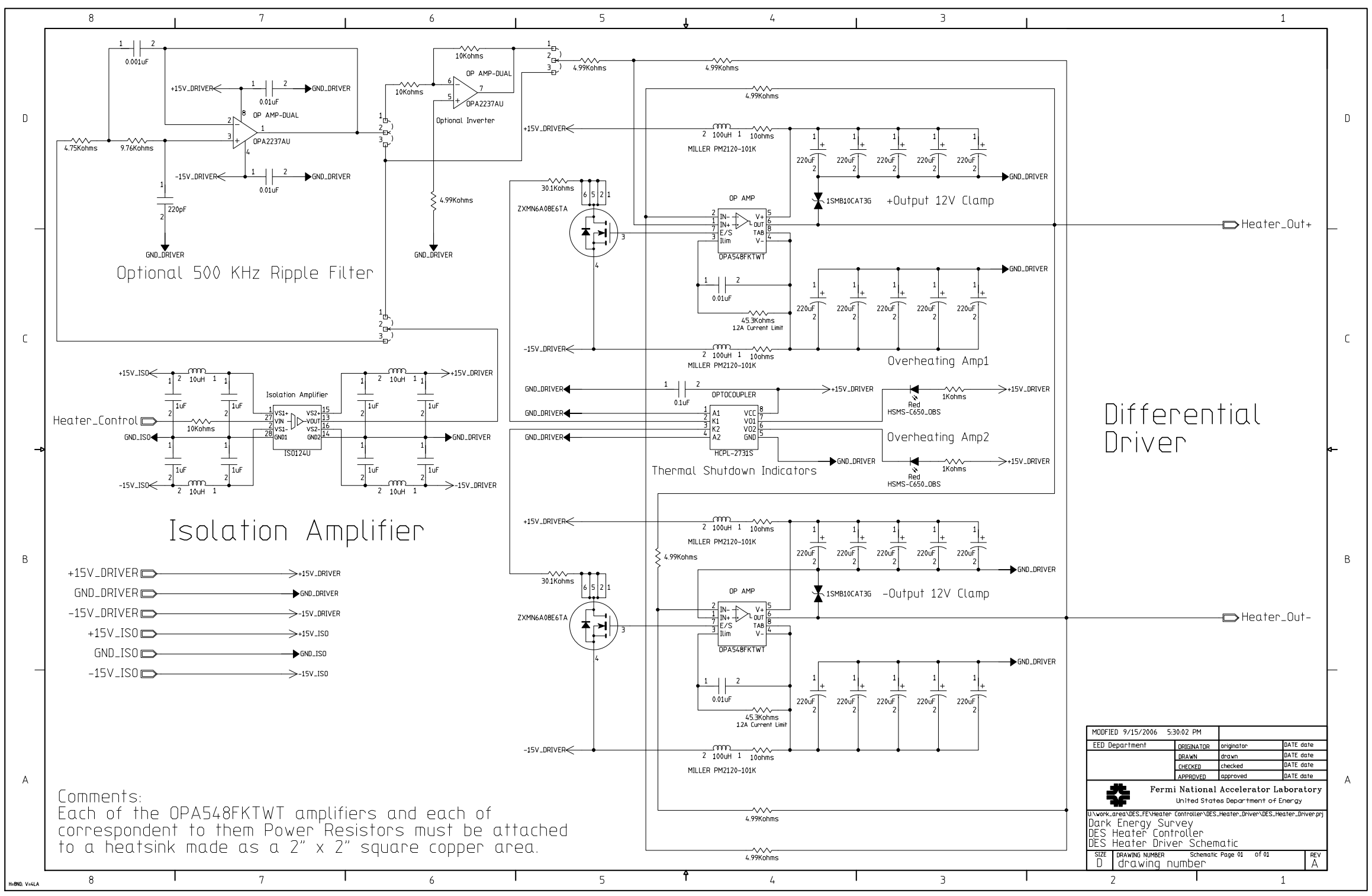

Figure 15.37: A schematic of the circuit used for the DES Heater Driver Board.

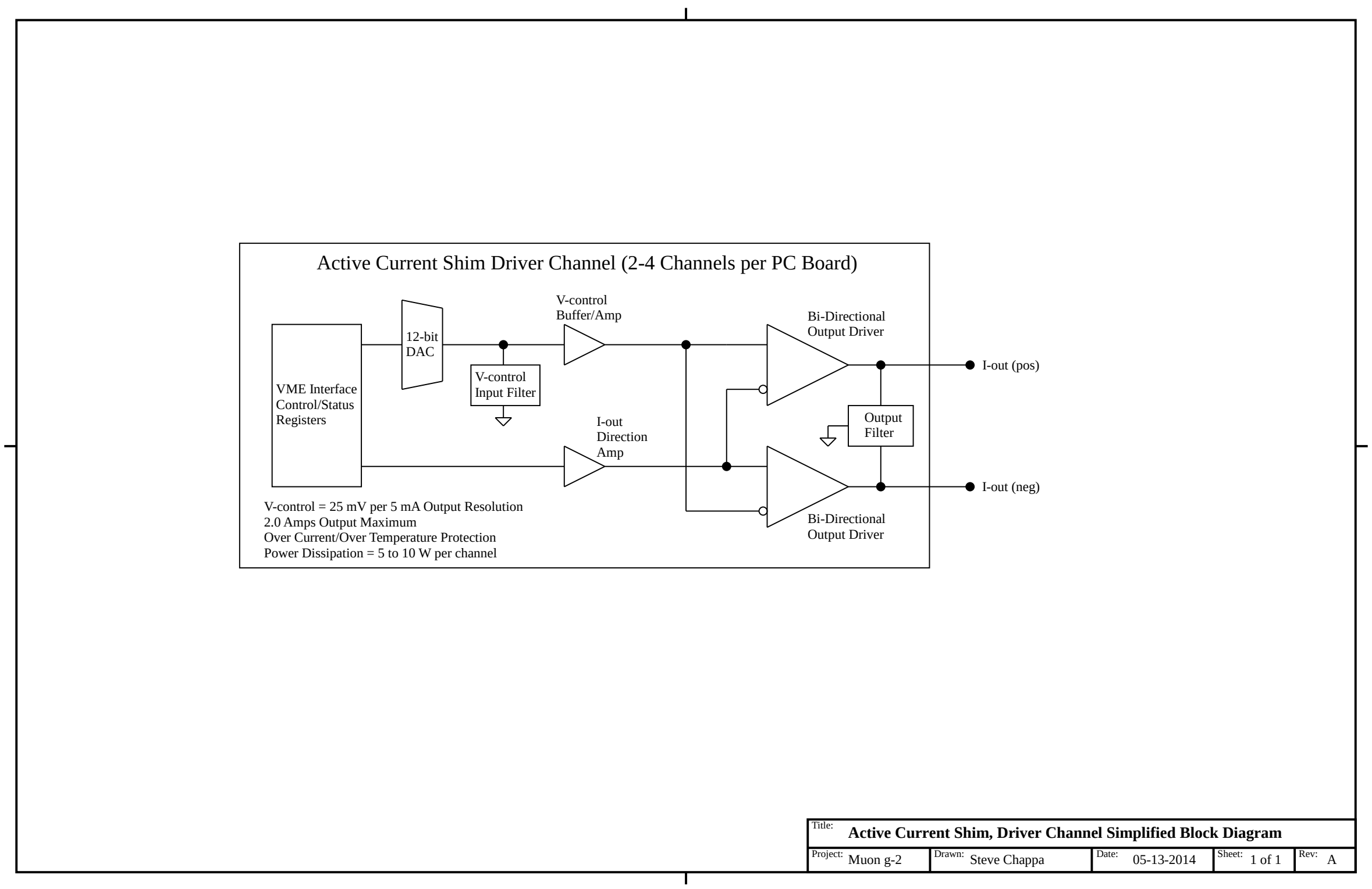

Figure 15.38: A simplified block diagram used to lay out the requirements of the Surface Correction Coils.

**Gap correction loops**

We want to minimize the azimuthal variation of the magnetic field, as explained in Section 15.8.2. Thus, E989 will use `OPERA-3D` to study the possibility of adding small loops to the bottom of the surface correction coil boards at the azimuthal positions between adjacent poles. We would primarily have control over the dipole moment, with limited ability to modify the higher order multipoles.

### 15.8.4 E821 results

E821 successfully implemented many of the passive and active shimming techniques described above. Table 15.8.4 shows the historical progression of the uniformity of the field as a function of time during the commissioning phase of the experiment. As they adjusted shims closer to the storage region, the higher order multipoles became more controlled. The final column shows the principal changes that were implemented at that step. We plan to use this experience to compress the shimming schedule for E989.

Table 15.7: Quadrupole (Q), Sextupole (S), Octupole (O), and Decupole (D) multipoles, broken down into normal(n) and skew(s) components, in ppm, evaluated at the storage radius ($r = 4.5$ cm).

| Date | $Q_{\rm n}$ | $S_{\rm n}$ | $O_{\rm n}$ | $D_{\rm n}$ | $Q_{\rm s}$ | $S_{\rm s}$ | $O_{\rm s}$ | $D_{\rm s}$ | Action |
|---|---|---|---|---|---|---|---|---|---|
| Jun 1996 | -169.12 | 112.03 | -34.16 | 23.71 | 27.06 | 5.82 | 3.12 | 0.46 | Initial configuration |
| Nov 1996 | 5.52 | 3.19 | -1.11 | 1.95 | 9.13 | 5.32 | 0.85 | 0.45 | Edge shims ground uniformly |
| Jul 1997 | 5.26 | 2.94 | -1.03 | 1.45 | 12.26 | 2.78 | 0.36 | 0.25 | Edge shims ground in each pole piece+ pole alignment |
| Aug 1998 | 7.73 | -5.29 | -2.79 | 0.38 | -2.07 | -0.02 | -0.25 | 0.71 | Final passive shimming |
| Sep 1998 | -2.54 | -1.25 | -2.70 | 0.34 | -2.39 | -0.18 | -0.28 | 0.42 | Active shimming commissioned |
| PRD | 0.24 | -0.53 | -0.10 | 0.82 | 0.29 | -1.06 | -0.15 | 0.54 | Publication |

## 15.9 Value Management and Alternatives

We are realizing significant savings in the magnetic field measurement system by refurbishing as much of the E821 hardware as possible, rather than building a completely new system. At the same time, we are making improvements (such as the change to the NMR sample material, improved shimming and magnet temperature control, improved calibration apparatus, ...) that improve operational characteristics and reliability, and which are necessary to reach our goal of an uncertainty on $\omega_p \leq 70$ ppb. Some examples of the various alternatives that are considered and the value management principles used to make down-selections are highlighted here.

### Probes

The determination $a_\mu$ in terms of $\omega_a/\omega_p$ and $\mu_\mu/\mu_P$ requires the storage ring magnetic field be measured in terms of the muon distribution weighted free proton precession frequency, $\omega_P$. In principle this could be done by injecting a 3.1 GeV/$c$ polarized proton beam into the storage ring with a proton spin analyzer (polarimeter) in the ring based on the spin dependence of $p$-carbon elastic scattering. The anomalous precession frequency would be close to 40 MHz, and the phase space of the protons and muons would have to be matched or measured to mm precision. Proton beam measurements of the field distribution would have to alternate with muon injection. Developing a polarized proton source, 3.1 GeV/$c$ accelerator, polarimeter, proton beam position monitoring hardware etc. would add significant cost and technical challenge to the experiment, and it is not clear that the field determination could be made to 70 ppb. Further, during periods of muon injection, an NMR-based system of fixed probes outside the storage volume would still be required to monitor the field and provide feedback to the power supply to stabilize the field. A similar NMR system would be required to shim the magnet.

Other field measurement technologies such as Hall probes have been considered. One advantage is that a 3-axis device could make separate measurements of $B_x,\ B_y,\ B_z$. However, Hall probes have a significant temperature dependence ($\sim$ 10 ppm/°C), resolutions at the 1 ppm level (versus 20 ppb for NMR), and fluctuating offsets. Commercial technology is currently inadequate for the level of accuracy sought in E989. Further, they would require frequent calibration in terms of an equivalent free proton precession frequency, so an NMR-based absolute calibration probe would still have to be developed and tested.

Finally we note that pulsed NMR is preferred over CW techniques since the latter typically requires a small field modulation coil that perturbs the local field, introduces image fields in the iron poles and yokes, and is potentially difficult to calibrate at the precision sought in E989. In addition, the lineshape analysis required in CW techniques to achieve ppb levels of precision is substantially more difficult than the analysis methods required for analyzing pulsed NMR FIDs. Newer approaches replace the field modulation with frequency modulation, but still must deal with the time constants of the probes and of the samples, and still need a sophisticated signal analysis to determine the resonance frequency precisely from the time domain response.

### Trolley

We are investigating improvements with respect to systematic uncertainties by changing the size and shape of the trolley probes. Smaller probes could allow positioning of the probes to slightly larger radii and hence improve the determination of higher multipole moments of the magnetic field. However, the expected gains need to be understood to verify such change as the current probes are certainly already close to optimal.

A major alternative could be the design of a new trolley system with much reduced onboard functionality. Given the crucial role of this system in the g-2 experiment, a replacement of the existing trolley would eliminate such single point of failure. In this alternative version, the onboard electronics would ideally be reduced to only host the multi- and duplexer, a preamplifier, and a small control unit including some of the sensor functionality (temperature, pressure, position measurement). This scheme requires that the NMR analog signal can propagate with minimal distortion over the 45-m-long co-axial cable to the outside of the vacuum where electronics for frequency determination would be located. Similarly, the RF pulse would be sent to the trolley from the outside eliminating the need of the RF amplifier onboard. The processing of the NMR signal could then happen analogously to the fixed probes in the proposed new readout system with full digitization. However, such an alternative development would require major cost and labor resources and the actual implementation has to be based on a more detailed cost-benefit-analyses. Thus our baseline calls for refurbishing the existing E821 trolley and a new trolley system remains a risk/alternative.

### Absolute Calibration

The baseline plan for the absolute calibration involves using reusing the spherical water sample that has been used in the Brookhaven $g-2$ experiment, as well as the E1054 experiment that determined the ratio $\mu_\mu/\mu_p$. An alternative going forward is to develop a polarized $^3$He sample instead of water. This alternative benefits from lower uncertainty in the diamagnetic shielding factor, reduced sensitivity to some geometric effects, and a much smaller temperature coefficient. The addition of the $^3$He calibration system would provide a valuable cross-check of the spherical water probe, and thus remains a viable alternative going forward, as described in section 15.4.2.

### Shimming

The magnet shimming team has decided to refurbish equipment wherever possible. This includes reusing most of the shim kits from the E821, including the iron pieces on the yoke, the poles, and the wedge shims. New edge shims are being ordered to provide the flexibility required to tune the residual sextupole and octupole moments that remain after the rough shimming procedure.

One alternative that remains involves the possibility of reusing the existing edge shims. A significant amount of rust has formed in a non-uniform manner across the exposed face of some edge shims, so these would likely need to be resurfaced. The inevitable reduction in the total magnetic thickness of these shims would require the addition of very thin steel shims on top of the existing edge shims. We have identified a variety of vendors that have 1-mil thick AISI 1008 shim stock that could be appropriate for this application, at a cost as low as $2.5 per sq-ft. We need to investigate the magnetic uniformity of the candidate material and ensure that we can securely fasten these foils to the existing edge shims. In the case that we need thinner edge shims to correct a higher order multipole, we have also simulated the possibility of adding non-magnetic (or low-magnetic) spacers between the pole pieces and edge shims. We have found that this has an effect that is proportional to the reduction of the

thickness of the edge shims. One concern with this method is that the steel screws that fasten the edge shims to the poles would create magnetic shorts that could produce non-uniform effects, so the ability of this method to meet the technical uniformity requirements has not been verified. Further value engineering will be applied to see if we can reduce the cost of the edge shim system without jeopardizing the technical requirements by employing these low-costs methods. Our current baseline calls for repeating the E821 method of ordering oversized edge shims and then grinding them to the appropriate thickness.

A final example of value management in the shimming system comes from the Surface Correction Coils. The driver boards for these coils are based upon the design for the Dark Energy Survey's DECAM Heater Controller Board. Our engineers identified many similarities between our requirements and these existing boards, and plan to use that as a base design for our system.

**Test Magnet**

The HEP division at Argonne has acquired a 4T solenoid magnet (see Section 15.3.9). This solenoid has a measured stability of better than 0.1 ppm/hour, a shim set for ppm-level homogeneity over a 10 cm diameter volume, and a 680 mm bore diameter, essential for extensive performance tests of the NMR trolley in advance of installation in the $g-2$ storage ring. We will also perform systematic studies of the absolute calibration probe in this magnet. They will also be essential for determining the magnetic influence of the calorimeter and tracker hardware on the storage ring field. The acquired magnet has replaced the original option of bringing a 1.5 T solenoid from Los Alamos. The decision basis for this 4 T option were reduced costs for the transfer and installation, and expected lower operational costs. The system was also readily available leading to a shorter time until it is fully operational for the essential systematic tests of the NMR equipment.

## 15.10 ES&H

The trolley garage, which is part of the field monitoring system, is a vacuum vessel. Lasers are used during survey/alignment activities and calibration activities. The storage ring magnetic field is at 1.45 T and has a strong fringe field in the interior of the ring. The hazards encountered in the field monitoring operation are therefore Laser Hazards, Vacuum Vessel Hazards, and High Magnetic Field Hazards. Engineering review will determine the necessary requirements on the vacuum vessels. Job Hazard Analysis will be performed for any testing, installation, or operational task that involves personnel working in the high field environment or using lasers.

In addition there will be three or four 19" racks of field measurement electronics. These will typically draw a few kW of power each, and do not produce high voltages or large currents. All personnel working on this equipment will be trained to ensure safe operation.

## 15.11 Risks

### Fixed Probe System

The fixed probe system is essential for field monitoring during data taking. The technology required to monitor the field at the required level already exists so the risks are primarily in two other categories; (1) the risk that the refurbishment of the fixed probe system can not be completed on time, (2) the magnet stability is worse than anticipated.

The refurbishment of the fixed probe electronics requires that new preamplifiers for the NMR signal and a new RF pulse amplifier must be found since the vendors for the E821 components are not in business. These risks are minor or non-existent as new candidate components (with higher performance in some cases) have been identified and will be tested. The risks to the schedule come primarily from the time and effort required to refurbish the roughly 400 fixed probes and the NIM crate and multiplexer electronics. The mechanical work on the fixed probes can be distributed to additional university or laboratory machine shops so it may be done in parallel. Filling the samples and tuning the probes can also be done in parallel by any group with a vector impedance meter or network analyzer. In E821 such work was done by undergraduates. If the electronics work falls behind schedule, it can also be done in parallel at the Electronic Design Facility at Boston University, at Argonne, or Fermilab once new boards have been designed.

The second risk regarding magnet stability is more serious but unlikely as the new building is designed specifically for magnet stability. If the anticipated gains in magnet stability do not materialize, additional insulation can be applied around the magnet and on the experimental hall floor. If necessary, an inexpensive, easily installed and removed thermal enclosure from aluminum framing and foam board insulation can be constructed.

### Absolute Calibration System

The calibration of the trolley probes requires that the absolute calibration probe and plunging probe and their positioning systems have been extensively tested. The risks here are that the probes are damaged, and that the system does not perform to specifications.

If the calibration probes are damaged, they can be be remade. The most delicate part is the highly-spherical glass bulb used in the absolute calibration probe. This part has survived nearly 20 years already, and the vendor, Wilmad LabGlass has already been contacted regarding making replacements. Spare absolute calibration probes will be prepared for E989, and ready for use long before data-taking.

The risk that the new calibration system does not perform to specifications is small. The changes from E821 are relatively minor but should be effective (primarily adding motion in the azimuthal direction for the plunging probe, adding a closed loop positioning system, and better determination of the precise location of trolley probe active volumes). These systems involve little technical risk and their performance can be tested thoroughly before the experiment takes data. The worst schedule risk is losing roughly 2-3 weeks of muon data. If the calibration system is not ready when the experiment starts, muon data can still be taken while the calibration system is being prepared. When the system is ready, roughly 2-3 weeks of muon data would be lost for installation as it would involve letting the vacuum chambers up to air, installing the system, testing it, and pumping back down.

**Trolley**

The trolley and its associated mechanics are a central piece in the measurement of the magnetic field of the storage ring. A major risk would be the partial or complete failure of the onboard electronics. Depending on the severity of the failure mode, the consequences could range from a replacement of the broken component, a redesign of parts of the electronics up to the need of a complete redesign of the trolley electronics. While the probability is low for this to happen, a realization of this risk could have both significant cost and schedule impact. Mitigation of the risk is hence important. It involves careful refurbishment of the system with guidance from former experts as well as the refurbishment of both existing trolleys so that both are fully operational. For that purpose, we bring together a group of current E989 and former E821 collabortion members and the electronics engineer from Heidelberg in June at Argonne National Laboratory. During the 9 days of scheduled work together, we will carefully learn how to communicate with the central microcontroller and transfer the knowledge to the new collaborators that are in charge of this system.

Damage of the trolley garage or drive mechanism during the shipping was prevented by careful packing. Immediate mechanical inspection and testing of both devices at Argonne have shown no indication of any mechanical shortcomings so that we were able to retire these risks. Of course, final shipment to Fermilab for installation at the experimental hall will be handled in a similar careful way.

Another risk is associated with the position measurement upgrades to determine the longitudinal position of the trolley during its data taking. There is a small possibility that the anticipated upgrade of the barcode reader does not succeed because of remaining overheating or other unforeseen issues. In the case of this event, alternative solutions must be sought to meet the requirements. To mitigate this risk, we are currently building a stand-alone prototype barcode reader. With it, we can test early that the replacement barcode reader will work with the barcode marks and in vacuum.

**Shimming**

The shimming procedure used in E821 has been examined and provides the basis for shimming the field in E989. Careful review of past safety procedures will be necessary to ensure the successful, safe shimming of the field. An examination of associated risks reveals two main categories of risks associated with the shimming procedure: damage to equipment and delay of the experiment. To mitigate each category, we will begin as early as possible with a well-formulated plan.

Damage during shipping is a risk for the yoke, poles, wedge shims, edge shims and dipole correction coils. To address this risk, we have shipped most of the steel well in advance of the installation in the experimental hall at Fermilab. If there were any unexpected accidents in transit or during the installation of the steel in the hall, this would allow us the necessary time to order replacements for these parts. Additionally, there are ongoing risks to the materials during both the shimming procedure and the subsequent running due to the enormous energy stored by the $g-2$ magnet. The stray fields are significant enough to attract loose ferromagnetic materials in the experimental hall towards the steel. These pieces could potentially impact and damage the precisely manufactured surfaces of the shims and poles, causing major distortions to the field but more seriously, posing a significant risk to human safety. To address these issues, we will follow safety procedures to ensure that no loose magnetic materials are left in the hall when the ring is powered. We will check that the shims are securely fastened to the pole and yoke pieces. We will continue to examine the forces on the various screws and bolts in simulations to ensure that sufficient safety factors are utilized. In all cases, following disciplined safety procedures will prevent potentially damaging incidents with both people and equipment.

The shimming procedure calls for ordering oversized edge shims and then grinding them down to the appropriate thickness. There is a schedule risk associated with grinding off too much material. We would then need to reorder the shims and recommence the grinding step of the shimming, which would delay meeting our shimming goals. To mitigate this we will continue to compare our simulation results with past experience to get a solid understanding of the dependence of the multipole moments on the shim thickness. We will proceed with a conservative plan to grind in a couple of iterations, so as to prevent "overshooting" the required thickness. A technique of stacking thin steel foils on the edge shims is being investigated as both an alternative to ordering new shims and as a potential mitigation technique for the possibility of overgrinding.

A schedule risk would be realized if the shimming procedure fails to achieve the required uniformity. This could occur in a variety of ways, for example if detector systems introduce large, non-symmetric distortions to the field or if the finite-size effects of surface coils limit our fine tuning ability. If we do not achieve our uniformity goals, we would have to make improvements in other areas - namely, better knowledge of the probe positioning, better absolute calibration and better temperature control. To address these issues, we will continue to study the magnet in `OPERA` and advance the simulation plans. We will get an early start on the fabrication of the printed circuit boards, and understand their requirements and technical capabilities. We will also remain involved with the other teams to ensure their systems do not introduce unmanageable distortions to the field, due to either materials or currents. A test stand with an $\approx 1.45$ T field is under development at ANL, and will be available to test proposed systems in advance. These steps will help ensure that the uniformity goals are achieved.

**Other Performance Risks**

Many of the risks that would prevent us from achieving the uncertainties outlined in Table 15.1 can be mitigated by spending enough time on trolley runs and trolley calibration so the goals are met. These activities will often conflict with data-taking which reduces the statistical uncertainties on $\omega_a$. A balance between these activities will be established that brings the uncertainties down on $a_\mu$ most efficiently. This will depend on actual event rates and magnet stability. If the storage ring dipole and quadrupole fields are at least a factor of two more stable than E821, and if the field measurement hardware and shimming performance goals outlined above are met, the target for the precision of $\omega_p$ should be achievable without major risks to the schedule.

## 15.12 Quality Assurance

It is necessary to test the NMR hardware before shimming and installation in the g-2 storage ring. This requires the development of independent test-stands that include a set of NMR probes, NMR electronics, DAQ, and a magnet. We have located at least 4 magnets suitable for these purposes, where the requirements on the magnet depend on the hardware component being tested.

To test the fixed probes requires a vector impedance meter or network analyzer and a magnet at 1.45 T with field gradients less than 20 ppm/cm. The latter is sufficient to ensure an FID of millisecond duration, sufficient to confirm the probe works. An electromagnet suitable for testing fixed probes and basic functioning of the NMR hardware has been prepared during 2013 at the University of Washington.

Precision tests of the NMR hardware - such as single shot frequency resolution, temperature dependence of NMR signals, reference frequency dependence of the electronics, aging effects, measurements of $T_2$, etc. require magnets with stability of 10-100 ppb per hour and field gradients of <200 ppb/cm. Such a magnet is available to the group at University of Michigan (with access to a large bore persistent mode MRI magnet). The University of Massachusetts group has unrestricted access for several years to a small bore (89 mm) persistent mode superconducting magnet from Cryomagnetics, with better than 0.01 ppm/hr stability and a shim set to achieve sub-ppm/cm homogeneity over an 8 $cm^3$ volume. The small bore cannot accommodate the E989 NMR probes (it can only accommodate the plunging probe), but many sensitive tests of the NMR electronics and behavior of the NMR probe samples (temperature effects etc.) can be measured at the 10 ppb level using custom probes.

A critical test magnet system has been shipped to ANL and is in the process of being commissioned. This persistant-mode, large bore superconducting MRI magnet will facilitate the testing of equipment directly related to the field measurement, as well as equipment from other groups that runs the risk of adversely affecting the magnetic field. The proximity of this test facility near Fermilab is an important step that will allow quick feedback on candidate prototypes and will see regular use during the buildup of the experiment.

Extensive early testing of the NMR hardware will allow the identification of problems and the implementation of solutions in advance of installation in the g-2 ring. By having several absolute calibration probes, repeated calibration of the trolley probes, and extensive investigations of potential systematics, we intend to produce a robust result on $\omega_p$.

# Chapter 16

# The $\omega_a$ Measurement

The anomalous spin precession frequency $\omega_a$ is one of the two observables required to obtain the muon anomalous magnetic moment, $a_\mu$. In order to ensure that the experiment's proposed goal of 140 ppb precision in $a_\mu$ is achieved, the error budget allows for a 100 ppb statistical uncertainty combined with equal 70 ppb systematic uncertainties from each of the $\omega_a$ and $\omega_p$ analyses. This chapter summarizes the procedure for the $\omega_a$ measurement, with Chapters 17-22 elaborating upon the design of each subsystem. First the decay kinematics are reviewed and the encoding of the muon spin information into the data set is explained (Section 16.1). From that basis several complementary analysis methods are described (16.1.2). Then a review of uncertainties is presented, first the statistical (Section 16.2) and then the systematic uncertainties intrinsic to the detectors (Section 16.3). Then the detector system organization is outlined and broken down into subsystems (Section 16.4). These subsystems include stored muon monitoring, decay positron tracking, electromagnetic calorimeter, signal digitization, data acquisition, and slow control systems.

## 16.1 Measurement Overview

In this experiment the polarized positive muons are stored in a magnetic ring. Their spins precess at a different rate than their momenta. The anomalous precession frequency, $\omega_a$, is the difference between the ensemble-averaged muon spin precession and cyclotron frequencies. Direct measurement of the muon spin is not practicable so an indirect measurement is made via the decay positrons. Muon decay proceeds through the weak force and therefore is parity violating. The consequence of this behavior is that the emitted positron momentum is correlated with the muon spin direction. Therefore by measuring the decay positrons and analyzing their energies, a measurement of the muon spin is possible. We have elected to use calorimeters to measure the positron energy and time. The following sections describe the details of the muon decay pertinent to understanding the design of the calorimeters as well as the possible methods of analysis.

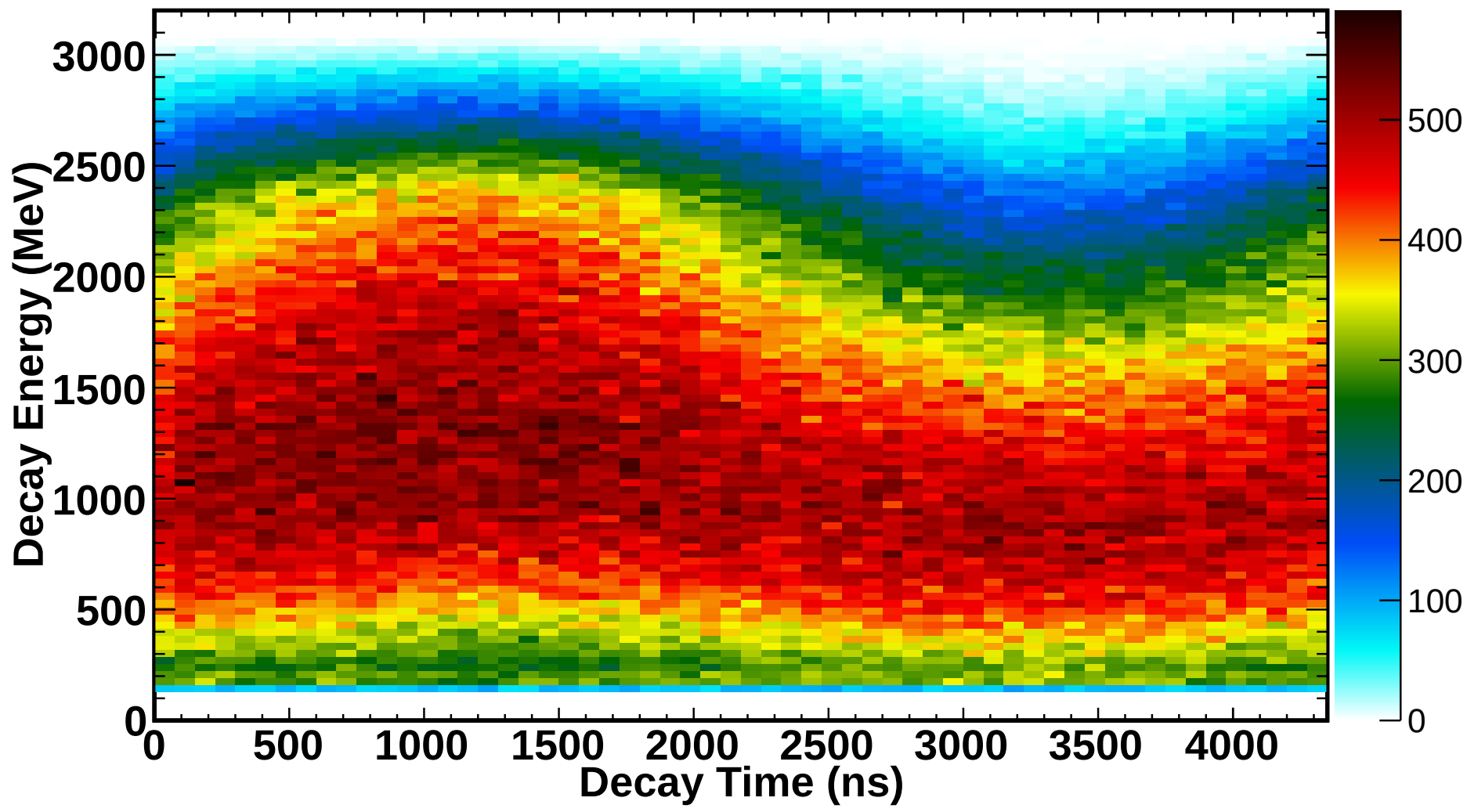

Figure 16.1: The decay positron energy spectrum as a function of time, modulo a complete 4362-ns$(g-2)$ period. The muon spin and momentum are aligned at $\pi/2$ and anti-aligned at $3\pi/2$ in this figure, corresponding to about 1090 and 3271 ns, respectively.

### 16.1.1 Muon Decay and Boost Kinematics

In this section we summarize the most important aspects of muon decay with respect to the detector design. For a comprehensive discussion of the kinematics of muon decay, see Section 3.5. Starting in the muon rest frame, the angular distribution of emitted positrons from an ensemble of polarized muons is

$$dn/d\Omega = 1 + a(E)\left(\hat{S}_\mu \cdot \hat{P}_e\right), \tag{16.1}$$

where $\hat{S}_\mu$ is the muon spin direction and $\hat{P}_e$ is the positron momentum direction. The asymmetry $a$ depends on positron energy $(E)$ and is such that the higher-energy positrons are emitted parallel to the muon spin. To boost to the laboratory frame we define $\theta^*$ as the angle between the positron momentum and the Lorentz boost,

$$E_{e,lab} = \gamma(E_e^* + \beta P_e^* \cos\theta^*) \approx \gamma E_e^* \left(1 + \cos\theta^*\right). \tag{16.2}$$

The starred quantities indicate the CM frame. The magic momentum requirement fixes $\gamma$ at 29.3. Due to the correlation between the muon spin and the positron momentum direction the angle between the positron momentum and the boost direction from the muon center-of-mass frame (CM) to the lab frame acts as an analyzer of the muon spin. The maximum positron energy in the lab frame occurs when the positron decay energy $(E_e^*)$ is the maximum and the positron momentum is aligned with the boost direction $(\cos\theta^* = 1)$. Figure 16.1 shows the decay positron energy spectrum as a function of time for one $(g-2)$ period.

The modulation of the decay energy spectrum occurs at the frequency $\omega_a$. Therefore by measuring this modulation a precision measurement of $\omega_a$ is possible.

### 16.1.2 Analysis Methods Summary

The standard analysis procedure is to identify individual decay positrons and plot the rate of their arrival versus time using only events having a measured energy above a threshold. The top panel of Figure 16.2 shows the result of this analysis method applied to simulation data. This method is named the $T$ (time) method; it was the dominant analysis technique used in the Brookhaven experiment and it is well tested against systematic errors.

The rate of detected positrons above a single energy threshold $E_{th}$ is

$$\frac{dN(t;E_{th})}{dt} = N_0 e^{-t/\gamma\tau_\mu}\left[1 + A\cos(\omega_a t + \phi)\right]. \tag{16.3}$$

Here the normalization, $N_0$, average asymmetry, $A$, and initial phase, $\phi$, are all dependent on the threshold energy. A parameterization of this function is used to fit the results from the $T$ Method analysis and extract $\omega_a$. The $T$ method is sufficient to reach the experimental goal. The method is well understood and proven to work due to its implementation by the E821 collaboration.

However it is possible to extract more statistical precision from the data set. The information of the muon spin is encoded via the positron momentum. By weighting events in proportion to their energy, or the asymmetry associated with their energy, the statistical precision is improved. As in the $T$ method, the data stream from the calorimeters must be first deconstructed into individual events and then binned in time before being fit to extract $\omega_a$.

An alternative approach is to not identify individual decay positrons but rather to digitize the detector current vs. time, which is proportional to the energy deposited in the calorimeter vs. time from the decay positrons. From the relationship depicted in Figure 16.1, the energy in the detector will oscillate at the $g-2$ frequency. That oscillation can then be fit to determine $\omega_a$. This approach is called the $Q$ (charge) method and it is inherently energy weighted, with a near-zero threshold. It was not used in E821 due to memory and readout limitations, but will be implemented in E989. This method has the attractive feature of being immune to the systematic effect of pileup (described in Section 16.3.2).

An example of this the spectrum that would result from a $T$ or $Q$ method analysis using the same simulated data set is shown in Figure 16.2. We summarize and compare the important features of the three $T$-based and the $Q$-based methods:

- ***T* Method:** Events in the calorimeter are individually identified, sorted and fit to obtain time and energy. The events vs. time-in-fill histogram is built from all events with reconstructed energy above a threshold. All events in the histogram are given equal weight. The figure-of-merit (FOM) is maximized for a positron energy threshold of 1.86 GeV, as discussed below. The quantity $\omega_a$ is obtained from a fit to a pileup-subtracted histogram. This is the standard method used in E821 and the benchmark for determining the statistical and systematic requirements for the E989 experiment.

- ***E*-Weighted Method:** Identical to the threshold $T$ method except that the histogram is built by incrementing a time bin with a weight equal to the energy of an event, therefore producing the Energy vs. Time-in-Fill histogram.

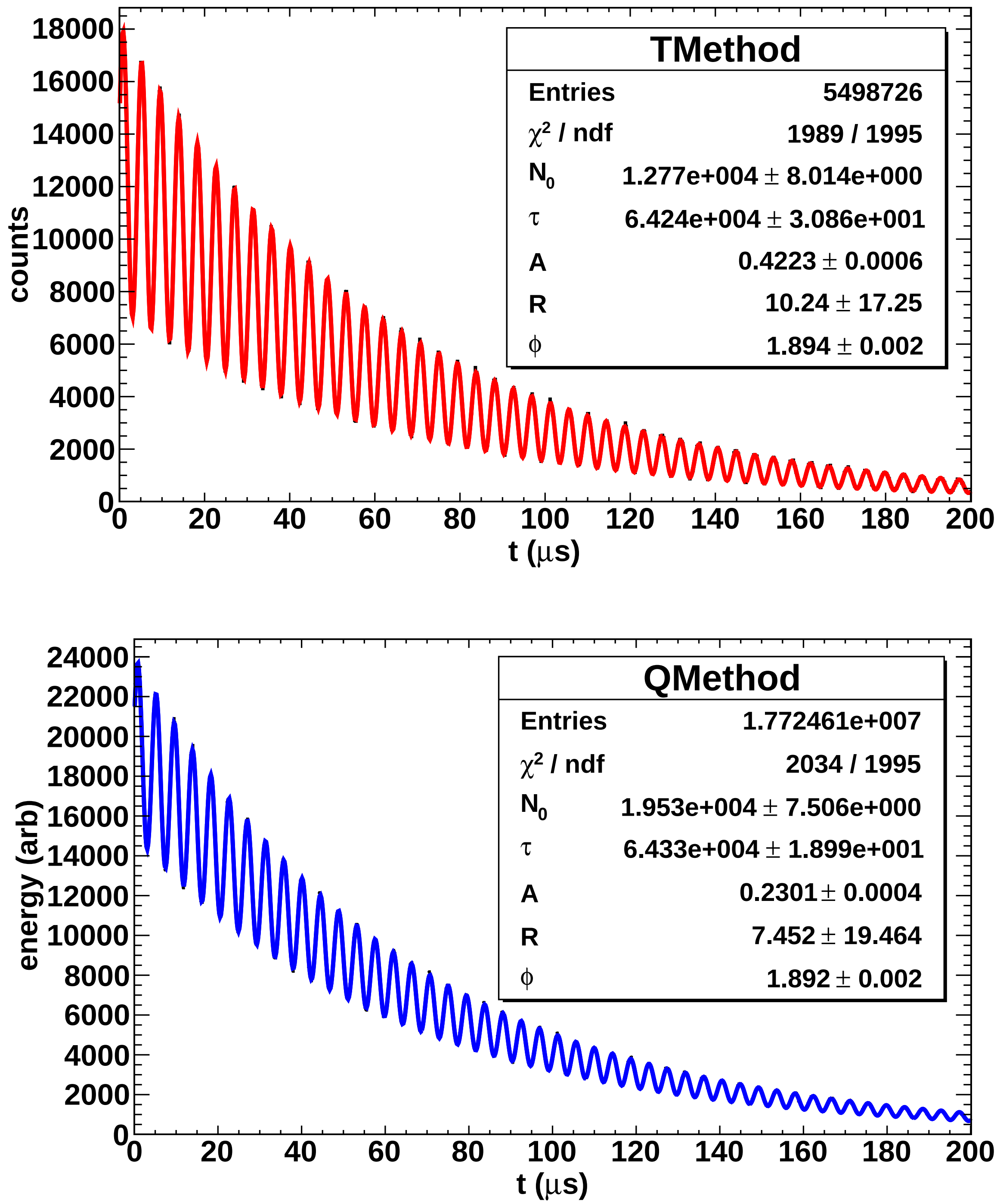

Figure 16.2: Top: Monte Carlo data analyzed using the $T$ method with a threshold cut at $y = 0.6$. Bottom: Same data analyzed using the $Q$ method. Detector acceptance is included. The asymmetry $A$ is much higher for the $T$ method; however, the $Q$ method has many more events $(N)$. The $\omega_a$ Monte-Carlo truth is $R = 0$ and the uncertainty in $R$ is a measure of the precision, in ppm. Both methods give a similar statistical uncertainty and acceptable fit central values.

- ***A*-Weighted Method:** Identical to the threshold $T$ method except that the histogram is built by incrementing a time bin with the average value of the asymmetry corresponding to that positron energy. This technique yields the maximum possible statistical power for a given threshold using the event identification technique.

- ***Q* Method:** An alternative approach whereby the detector current—a proxy for the deposited energy—is digitized and plotted as a function of time. Individual positron events *are not* identified. This procedure leads to a histogram of energy vs. time-in-fill. No attempt to correct for pileup is necessary here and a very low threshold is desired. The statistical power of this approach almost reaches that of the threshold $T$ method.

All of the mentioned analysis techniques will be applied to the data set by different research groups throughout the collaboration. The analysis will be conducted in a blind fashion. Next we provide an overview of the statistical and systematic uncertainties. We conclude the chapter by giving an overview of the detector systems (full details are given in subsequent chapters).

## 16.2 Statistical Uncertainty

The $T$ and $Q$ methods lead to similar histograms (see Figure 16.2) with different bin weights and asymmetries. A fit is performed using Equation 16.3 and the relevant parameter, $\omega_a$, is obtained. The optimization of the experimental system follows from minimizing the uncertainty on that parameter, namely $\delta\omega_a$. A detailed study [1] of the statistical methods gives guidance to the statistical power of any data set built using various weighting methods. The uncertainty on $\omega_a$ can be parameterized as

$$\delta\omega_a = \sqrt{\frac{2}{N(\gamma\tau_\mu)^2} \cdot \frac{\langle p^2 \rangle_y}{\langle pA \rangle_y^2}}, \tag{16.4}$$

where $N$ is the integrated number of decay positrons in the analysis, $p$ is the weight function and therefore is method dependent, $A$ is the asymmetry, and $\langle f \rangle_y$ is the value of $f$ averaged over all detected positron energies above threshold. The parameter $y$ is the fractional decay positron energy with respect to a maximum value; therefore $y$ ranges from 0 to 1, with $y = 1$ corresponding to approximately 3.1 GeV.

Figure 16.3 shows *differential* plots of $N$, $A$, and $NA^2$ vs. energy for a uniform acceptance detector. This plot illustrates the importance of the higher-energy positrons (those with the greatest asymmetry). The asymmetry is negative for lower-energy positrons; thus, a single low threshold can be expected to dilute the average asymmetry. The modification of the ideal curves due to the detector acceptance is significant, as the detector placement has been designed to greatly favor the higher-energy events. low energy positrons are more likely to curl between detectors and be missed. The acceptance impacts the values of $N$ and $A$, which are functions of the energy-dependent detector acceptance. In the $T$ method, each event carries the same weight ($p = 1$) and the uncertainty $\delta\omega_a$ (Eq. 16.4) reduces to

$$\delta\omega_a = \frac{1}{\gamma\tau_\mu}\sqrt{\frac{2}{NA^2}}. \tag{16.5}$$

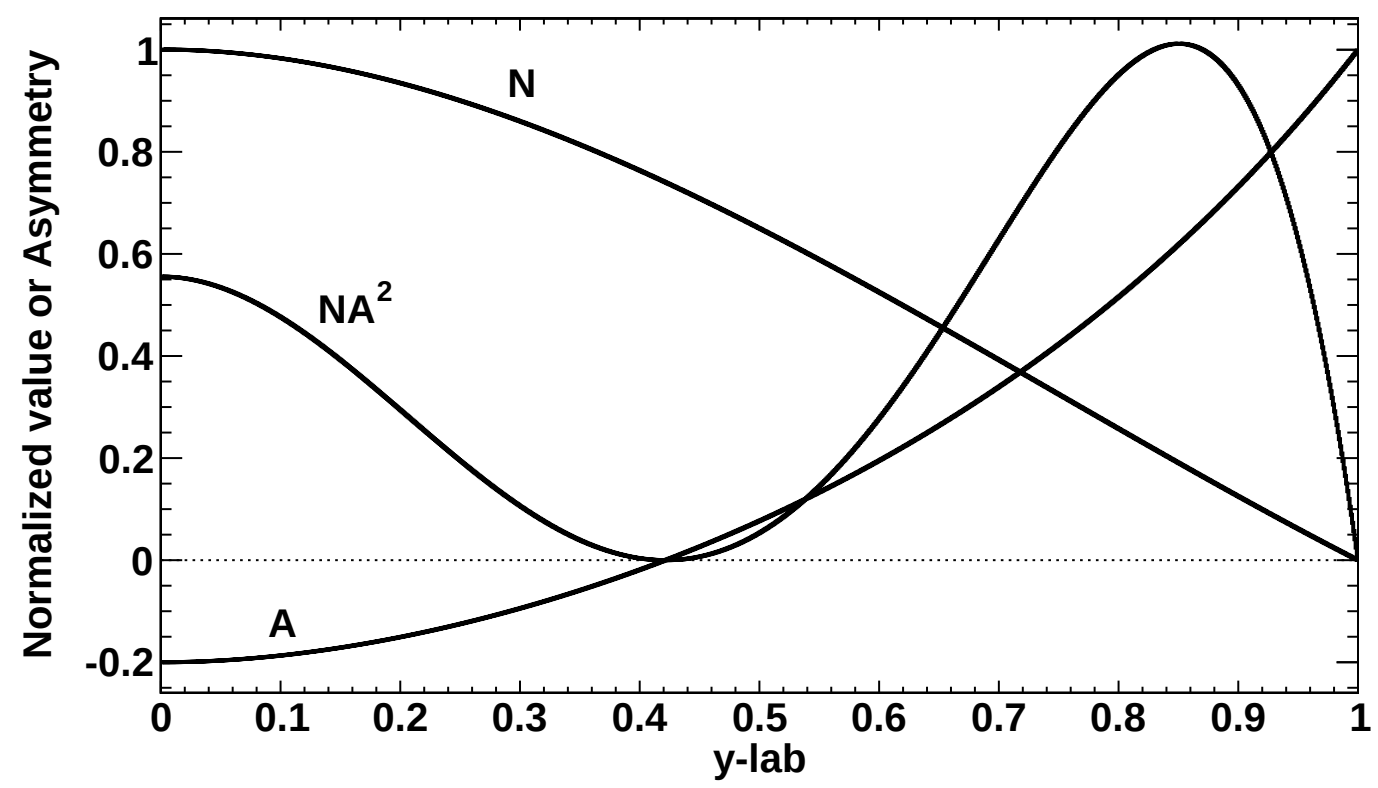

Figure 16.3: The differential distributions: normalized number of events ($N$), asymmetry ($A$), and the figure of merit ($NA^2$). Note, this plot assumes uniform detector acceptance across the full energy spectrum.

The figure of merit (FOM) that should be maximized to minimize $(\delta\omega_a)^2$ is $NA^2$. The value of the threshold that maximizes the FOM corresponds to $A \approx 0.4$ and an energy of 1.86 GeV. Therefore the relative uncertainty in $\omega_a$ is

$$\frac{\delta\omega_a}{\omega_a} = \frac{1}{\omega_a} \cdot \frac{\sqrt{2}}{\gamma\tau_\mu AP} \cdot \frac{1}{\sqrt{N}} \approx \frac{0.0398}{\sqrt{N}}. \tag{16.6}$$

The variable $P$ represents the average polarization of the muons. The end-to-end beamline transport simulations project a value of 0.95 for the stored muon polarization. For a statistical uncertainty on $\delta\omega_a/\omega_a$ of 100 ppb, $N = 1.6 \cdot 10^{11}$ fitted events will be required. The $T$ Method is sufficient for reaching the goal of E989 and all benchmarks and estimates are based solely on this method. However there is an opportunity for additional precision by incorporating other analysis techniques.

The $Q$ method is an energy-weighted ($p = y$) analysis with a single very low threshold, since events do not need to be individually identified. The computation of $\langle p^2\rangle_y/\langle pA\rangle_y^2$ in Eq. 16.4 is non-trivial. We conducted a simulation that included the detector acceptance to determine a realistic FOM for the $Q-$ and the $A-$ and $E$-weighted methods. Figure 16.4 shows the results vs. threshold energy. The meaningful FOM for the $Q$ method corresponds to the near-zero threshold end of the curve, which is circled. It should be compared to the $T$ method at its peak.

The $Q$ method was not employed in E821 owing to the high energy threshold, lack of sufficient memory in the digitizers, and limitations on the transfer speed to the DAQ. These technical limitations are easily overcome with today's large memories in such devices and faster bus speeds. Note that the data sets in the $T$ and $Q$ method are not identical, but substantial overlap exists. For example, in the $T$ method, all events below $\sim 1.86$ GeV do not contribute and all events above are weighted with $p = 1$. The $Q$ method includes all events that strike the detector and weights each by its energy, $p = y$. Therefore, a combination of the results of the two methods will enable a small reduction in the final uncertainty of $\omega_a$; more importantly, the two methods will also serve as an important cross checks on the final result because they have very different sensitivities to various systematic uncertainties.

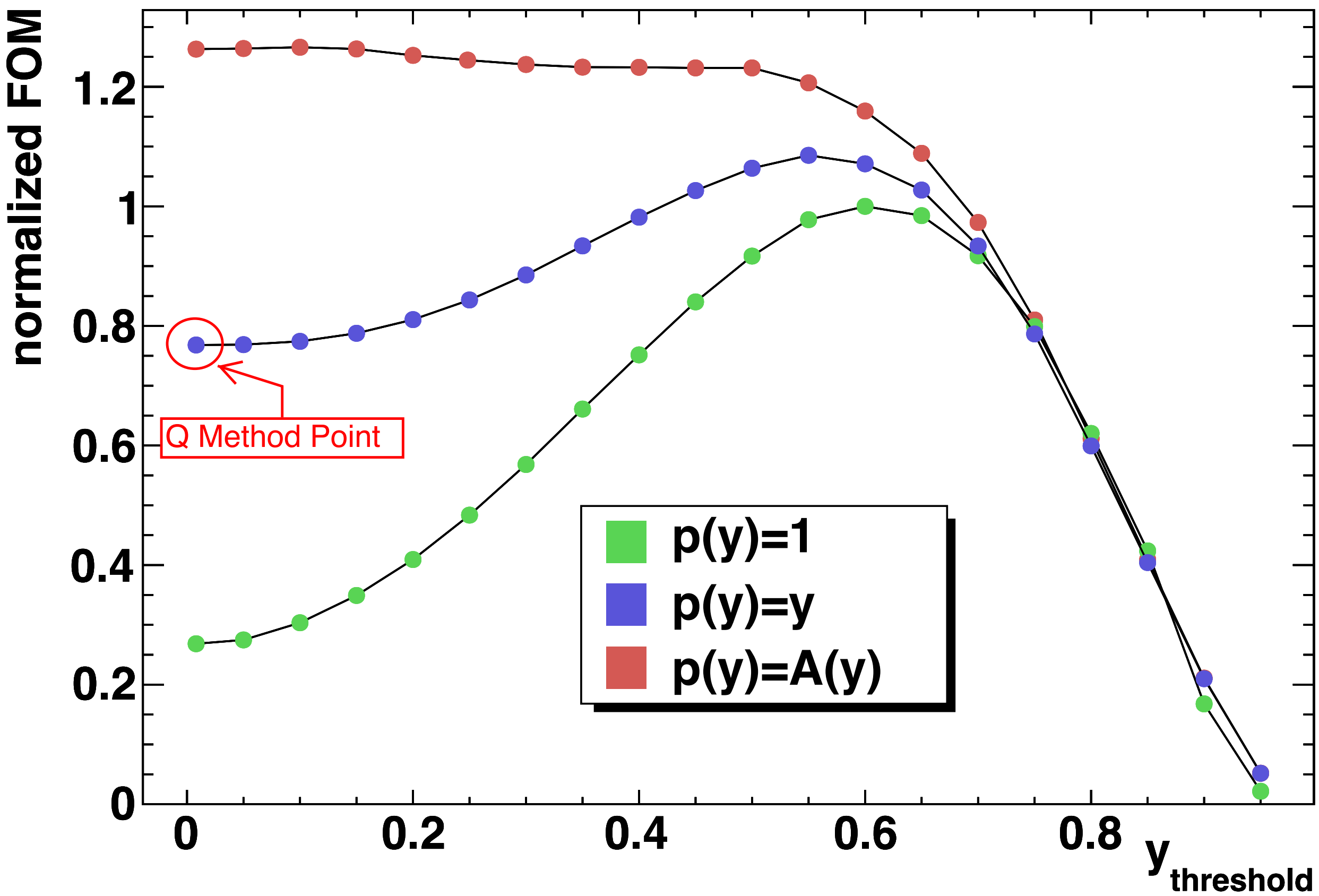

Figure 16.4: The figure-of-merit for the $T$ ($p(y) = 1$), energy-weighted ($p(y) = y$), and asymmetry weighted ($p(y) = A(y)$) methods are plotted versus threshold energy. They were calculated from the output of a simulated data set including detector acceptance. Two points are of particular interest. The first is the maximum of the unweighted distribution, which occurs at y=0.6 (1.86 GeV), and is therefore used for the $T$ method threshold. The second is the low-threshold limit for the energy-weighted distribution, which is circled in the figure. It corresponds to the $Q$ method analysis.

Table 16.1: Detector-specific systematic uncertainties in E821 and proposed upgrade actions and projected future uncertainties for E989.

| E821 Error | Size [ppb] | Plan for the New $g-2$ Experiment | Goal [ppb] |
|---|---|---|---|
| Gain changes | 120 | Better laser calibration; low energy threshold; temperature stability; segmentation to lower rates | 20 |
| Pileup | 80 | low energy samples recorded; calorimeter segmentation; Fast Cherenkov light; improved analysis techniques | 40 |

## 16.3 Detector-related Systematic Uncertainties

In this section we discuss the primary systematic error issues related to the Detectors, Electronics, DAQ, and the Offline Analysis. Table 16.1 lists the Gain and Pileup uncertainties and projections for improvements in the new $g-2$ experiment. The traditional $T$ method analysis is assumed because uncertainties can be reliably projected based on our considerable experience in these analysis efforts. Since the $Q$ method is new, we have not included its positive and partially independent impact on the final statistical result, nor are we presently able to fully project associated systematics. This topic is an active study in the collaboration. One key attractive feature of the $Q$ method is pileup immunity; there is no correction necessary, so that systematic uncertainty is absent. Our preliminary studies also indicate that the sensitivity to a systematic gain variation vs. time is also reduced. However, there are unexplored issues with acquiring a large body of fully integrated waveforms that remain to be studied.

### 16.3.1 Gain Changes and Energy-Scale Stability

The error budget for E989 assigns a 20 parts per billion limit to the gain systematic error. To connect this limit to hardware specifications, we conducted a simulation and analysis where gain perturbations could be applied. In general, the real data analysis plan is to correct any systematic hardware gain drifts over the short term time scale of a fill, event by event for each calorimeter station and crystal. The gain correction function must be prepared by detector response using both the known laser calibration pulses and evaluating the time stability of the pileup-corrected overall energy spectrum. Since the event rate changes by more than four orders of magnitude over a 700 $\mu$s fill, one might expect a hardware gain instability to vary with the rate. Accordingly, our simulation assumed a fill-scale gain perturbation of the form $G(t) = 1 + \epsilon \exp(-t/\tau_\mu \gamma)$, where $\tau_\mu \gamma$ is the 64.4 $\mu$s time-dilated muon lifetime, and $\epsilon$ is the magnitude of the unknown perturbation. In the above description, $G(t)$ represents the difference between the true gain vs. time behavior of the detector and electronics systems and the corrected one; that is, where $G(t)$ is the error in the gain correction, not its actual magnitude.

Monte Carlo techniques were used to build the standard $g-2$ decaying oscillation spectra with the expected E989 statistics, realistic detector effects, and built-in gain variations according to the above function. For each oscillation spectrum, a complementary laser-calibration data set was generated; the calibration system will provide an independent gain measurement every 5 $\mu s$ with relative uncertainty no worse than $4 \times 10^{-4}$. For the purpose of this study, a statistical uncertainty of $4 \times 10^{-4}$ was assumed. These laser data are fit and the resulting function used to correct the $g-2$ spectrum. The results for $\epsilon = 5 \times 10^{-3}$ are shown in Figure 16.5 for the $T-$based analysis method. The figure shows that correcting the gain with laser calibration data brings the $\omega_a$ perturbation well below the limit of 20 ppb. The target gain stability for E989 is $\delta G/G < 10^{-3}$ over a 700 $\mu s$ fill, which this study shows is sufficient to achieve the gain systematic error goal while leaving breathing room for additional effects introduced by gain oscillations at $\omega_a$. Such effects are the subject of ongoing studies.

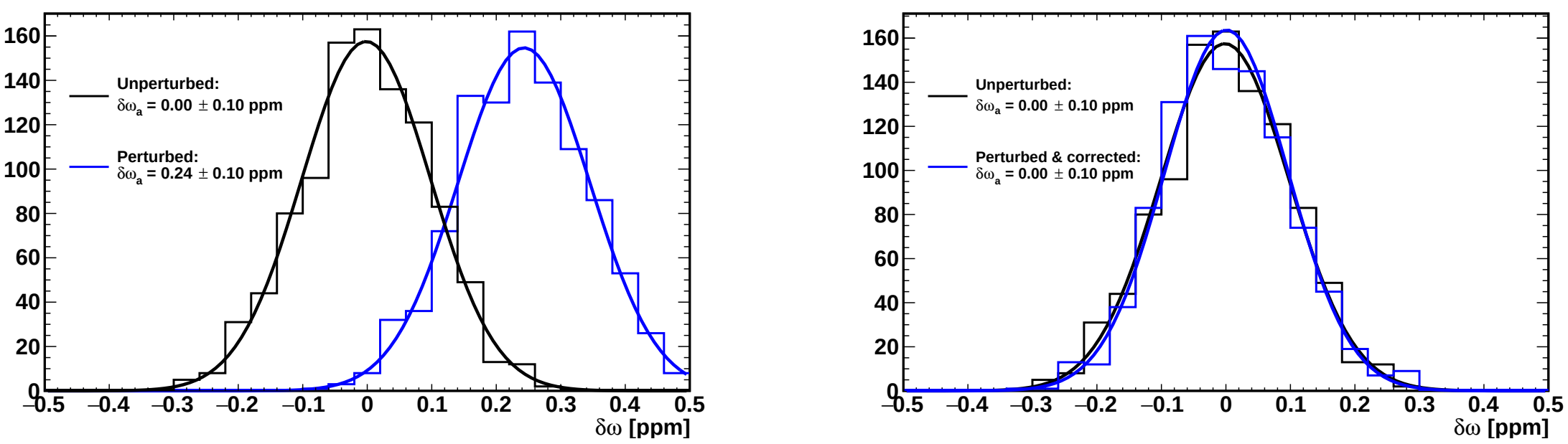

Figure 16.5: Histogram of extracted precession frequencies from 1000 simulated $T-$method histograms with and without an exponential gain perturbation of the form $5 \times 10^{-3} \exp(-t/\tau_\mu \gamma)$. The 100 ppb statistical uncertainty matches that expected in E989. The left plot does not include a correction from the simulated laser calibration data while the right plot does. After including the correction, the mean and width of the extracted $\omega_a$ distribution is restored to that of the unperturbed spectrum.

The hardware gains of the E821 detectors [2] were determined to be stable to ≈0.15% from early-to-late times within a storage ring fill. This limit was established by plotting the average energy for each $(g-2)$ period versus time after the PMTs were switched on. The gating circuitry in the tube base that allowed the PMTs to be turned off to avoid the initial burst of pions entering the ring also resulted in a small variation in the gain. For gain variations like this one, where the time constant is long compared to the $(g-2)$ oscillation period, the coupling to the $\omega_a$ frequency is small and, after correction, the residual systematic error was less than 20 ppb.

Several aspects of the current plan will be different. The first is that we will use silicon photo-multipliers (SiPMs), which can be saturated from a light burst and then recover with the same time constant as a low-light pulse. Each pixel recovers with a common time constant. Importantly, we do not intend to switch off these devices during injection because the anticipated hadronic-based flash will be (largely) absent. The initial pion flux at the target location will be reduced by the factor of $10^5$ owing to the long beamline. Beam protons will be eliminated in the Delivery Ring using a kicker system timed to fire when the muon and proton bunches are well separated in space.

If the gain oscillates at a frequency $\omega_a$, with an amplitude that varies in time, and with a phase that differs from that of the actual $\omega_a$ oscillation, then a direct error on the measured value of the anomalous precession frequency is produced. The average rate at which energy is deposited into the calorimeters oscillates with frequency $\omega_a$, and therefore any rate dependence in the gain of the detectors produces gain oscillations. In E821, we were able to demonstrate that the gain dependence on rate was small enough that its effect on $\omega_a$ was typically less than 20 ppb. In the new experiment, the increased beam rates will be partially offset by increased detector segmentation and our proposed monitoring system will be greatly improved compared to that employed in the past.

In E821, a UV-laser system was used to periodically pulse the scintillator in the detectors and thus monitor the complete gain and reconstruction chain during data collection against an out-of-beam reference counter. Unfortunately, the light distribution system included too many branches and not enough sub-branch reference detectors. Additionally, the laser intensity varied significantly on a shot to shot basis, making the per shot corrections large. Small fluctuations cascaded so that gain stability could be monitored to no better than a few tenths of a percent, which was not quite good enough to build a sensitive gain correction function. The system being designed for E989 will use cascaded distribution systems having multiple monitors at each stage. This is described in Chapter 17, Section 17.4.3.

The largest contribution to the gain systematic error in E821 came from analysis reconstruction induced gain oscillations at the $\omega_a$ frequency. The interpretation of the energy of a pulse from the fit to the waveform had a small bias. When a hardware signal rose above the waveform digitizer (WFD) trigger threshold, a pre-set minimum number of sequential samples was recorded. These data were fit offline to determine the peak height, time and the linear background under the pulse. However, if a trigger pulse was followed or preceded closely by another pulse, both pulses were fit together with a common background term, and the fitting region became longer compared to what is routinely used for a single pulse. In these pulses, the fitted energy was found to depend on the length of the fitting region, which was varying because of the hardware limitation. Because the data rate oscillates at frequency $\omega_a$, and is higher at early than at late decay times, it follows that the fitting region length oscillated at frequency equal to $\omega_a$ and was, on average, longer at early times compared to late times. This produced a small, effective gain oscillation having an amplitude that decreases with time. A systematic error on $\omega_a$ results.

Given the current capabilities in data throughput, the new electronics will record all samples rather than triggered, fixed-length isolated islands. This avoids the intrinsic bias in the recorded data and allows reconstruction routines to compensate for the waveform islands that have more than one pulse. In addition, we will have one other new tool that will provide powerful information related to energy scale and gain. As discussed in Chapter 19, a large-acceptance tracker system will be built just upstream of at least two calorimeter stations. This system will reside inside a modified vacuum chamber. It will be capable of providing high-precision tracking with good momentum definition these calorimeter stations, which will provide an absolute energy scale. The position information obtained will also inform the calorimeter cluster algorithm development. The energy scale obtained from the directly calibrated stations can be bootstrapped to other calorimeters by comparing the average energy distributions from decay positrons, which are expected to be similar. In summary, we expect that the largest of the gain systematic errors from E821 will be eliminated by the design of the electronics and data acquisition systems, combined with the verification from the tracker. The smaller contribution will be reduced by a more precise hardware gain monitoring system.

### 16.3.2 Pileup

The term "Pileup" refers to the overlap of events in the calorimeter that originate from separate muon decays, too close to each other in time and space to be resolved into individual pulses. When two pulses overlap, the result is that the two individual events are lost, and one event with the sum of their energies is gained. In general, a finite time offset between the two pulses exists and the recorded pulse shape is widened so that the combined amplitude is somewhat less than the sum of the individual amplitudes. Because the fraction of pileup events increases with rate, a component is introduced into the time spectrum that decays with half the nominal lifetime, or 32 $\mu$s, preventing a precise fit to the five-parameter function in Equation 16.3.

A more serious issue is that the muon spin precession phase varies with the energy of the pulse. A high energy positron has a larger radius of curvature and therefore a longer time-of-flight to the calorimeter, so it carries the phase of a muon that decayed earlier than one that produced a low energy positron. When two low energy pulses are lost and an apparent high energy pulse is gained, the high energy pulse still has the phase of the low energy pulses. The varying fraction of pileup over the fill causes an average early-to-late phase shift that directly distorts the fitted $\omega_a$. Consequently, the $(E, t)$ distribution of pileup pulses must be constructed and subtracted from the spectrum before it is fit.

The pileup distribution can be constructed based on the assumption that the probability of a pulse at time $t$ is, to a good approximation, the average of the probabilities that it is found at times $t + \delta t$ and $t - \delta t$, provided that $\delta t$ is small compared to the precession period. Consequently, secondary pulses from the "shadow" just before or after a primary pulse can be added to the primary pulse to form a constructed pileup event. To the extent that all fills have equal initial intensities of stored muons, the probability of a pulse at time $t$ in one fill is also nearly the same as the probability of a pulse at the same time $t$ in a different fill; this can provide another independent source of "shadow" events.

In E821, the construction of these distributions was complicated by the fact that only short islands around each pulse were stored, with a threshold of nearly 1 GeV required to store an island. The pileup distribution could only be fully reconstructed in a straightforward way at energies greater than twice this threshold. As described in the following chapters on electronics and data acquisition, in E989, the full waveform for the entire fill will be available. It will therefore be possible to construct pileup down to a very low energy threshold.

The unresolved pileup fraction scales quadratically with rate in each segment of the detectors. The effective size of the segment depends on the geometric extent of the shower. Our simulations demonstrate that an array of $PbF_2$ crystals, having 54 independent segments (see Chapter 17), and a smaller Molière radius compared to the Pb/SciFi used in E821, will provide an effective three-fold reduction in the intrinsic pileup based on the implementation of a very simple and robust shower separation routine and a 9-element cluster algorithm. The simulation includes a representative stored muon ensemble in the ring and correct spin physics in precession and decay.

Further improvements will also be associated with the correction for unresolved pileup. After following the pileup subtraction procedure described above, we will be able to check the result using an applied artificial deadtime (ADT). The ADT is the time established in the analysis software below which two pulses are not resolved (even if they can be). The analysis proceeds by sorting data using a series of ADT values beginning with the intrinsic, device-specific constraints, and artificially extending to much larger values that exaggerate the pileup. The extraction of $\omega_a$ is then done for each data set, and $\omega_a$ will be plotted as a function of ADT. In principle, $\omega_a$ should not depend on ADT, but a small correction could be included by taking the deadtime-free value that occurs at the zero-ADT extrapolation point. We have spent considerable laboratory bench time and offline pulse-reconstruction efforts to determine and optimize the minimum hardware ADT that our detectors will permit. Our laboratory tests demonstrate that pulses separated by 5 ns or more can be resolved easily for most pulse-amplitude ratios expected.

In addition to the work that was done for E821, we have also carried out a precision muon lifetime analysis with a pileup correction algorithm based on this pileup construction and ADT extrapolation concept. The work is well documented [4, 3]. The algorithms will be tested using the one station that has a high-resolution tracker (see Chapter 19) that can resolve pileup events at the few mm level and provide the corresponding momentum of each. Comparing identified pileup events from the tracker to the interpretation of the same events by the calorimeter will give a great degree of confidence in the methods.

For reference we comment on what had been achieved in the past. The pileup systematic error of 80 ppb in the E821 experiment was obtained from three components listed below. The first two were correlated and add linearly. The third is not correlated so it was added in quadrature to the other two.

1. Pileup efficiency, 36 ppb. This is due to an estimated 8% uncertainty in the amplitude of the constructed pileup spectrum.

2. Pileup phase, 38 ppb. This is the error due to the uncertainty in the phase of the constructed pileup spectrum.

3. Unseen pileup, 26 ppb. This is the error due to pulses so small that they cannot be reconstructed and therefore they are not included in the pileup construction. In general, the energy from these pulses cancels out, because they occur as often in the pedestal, where they lower the fitted pulse energy, as under the pulse, where they raise it. This error accounts for the potentially incomplete cancellation.

We expect that the segmented detectors, better laser calibration, more complete waveform record storage, verification of methods by using the tracker, and the use of our more modern extrapolation algorithms will lead to a comprehensive pileup correction with minimal uncertainty. We assign up to 40 ppb here to account for any difficulties in the anticipated analysis. As mentioned earlier, the $Q$ method is complementary to the traditional $T$ method and has different sources of systematic errors. The most significant difference is the effect of pileup—it will be greatly reduced for the $Q$ method.

## 16.4 Detector System Oveview

Figure 16.6 shows the locations of the calorimeters and trackers with respect to one of the 12 vacuum chamber segments. Two decay positron trajectories are indicated in the figure corresponding to high- and low energy events. The decay positrons have momenta below the muon storage momentum and therefore curl to the inside of the ring through the opening in the $C$-shaped magnet. Electromagnetic calorimeters are used to intercept the positrons and provide a measurement of energy and time of detection.

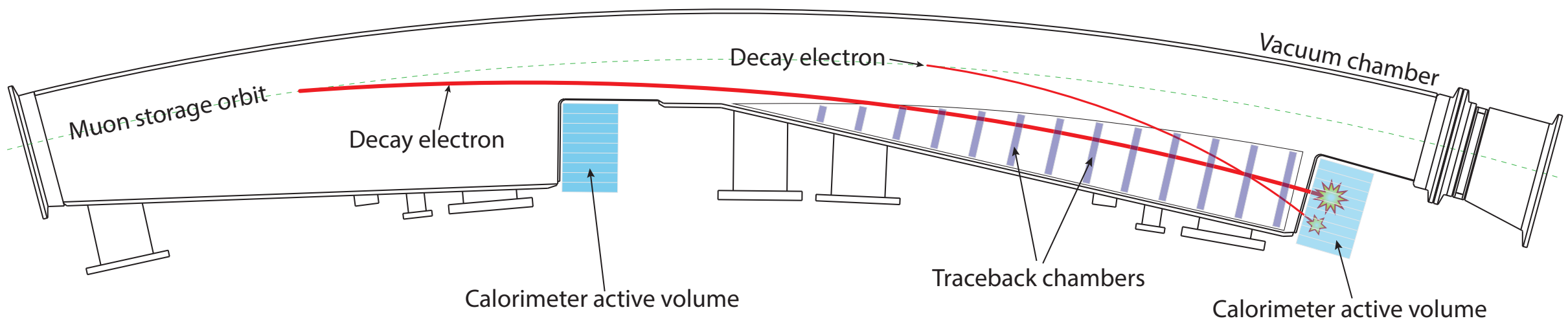

Figure 16.6: Scalloped vacuum chamber with positions of calorimeters indicated. A high- (low-) energy decay electron (or positron for the preferred positive muon storage) trajectory is shown by the thick (thin) red line, which impinges on the front face of the calorimeter array.

To operate the calorimeters and record the data several subsystems are required and shown schematically in Figure 16.7. First a segmented array of $PbF_2$ crystals absorbs the energy of the positrons and converts it to Cherenkov light. There are 24 of these arrays spaced evenly around the ring. Each station contains 54 crystals for a total of 1,296 individual $PbF_2$ crystals. Then the light is detected by SiPMs. These devices produce a pulse that is digitized separately for each block by custom digitizers running at 800 MSPS with 12 bit depth. The digitized signals are passed to a farm of graphics processing units (GPUs) where they are reduced to a form suitable for storage. In the case of the event reconstruction style analysis ($T$ Method) regions of interest or “islands” are selected for recording. In the case of the current readout analysis ($Q$ Method) the waveforms are summed across several digitizer samples as well as within individual stations to reduce the size of the recorded data.

The laser calibration subsystem provides the means to monitor the detector gain stability. The proposed SiPM readout devices are particularly sensitive to bias and temperature stability. We have dedicated subsystems to provide a stable bias supply to each device and to monitor temperature inside the calorimeter enclosures. Two tracker stations will gather data that contains a large number of resolved two-track events that might appear as unresolved pileup in the calorimeters. These data will be crucial to develop our calorimeter cluster and pileup-subtraction routines. Single track events will be used to determine the absolute energy calibration. The fiber harp system and the trackers are both needed to determine the stored muon beam distribution, which must be known to make the electric field and pitch corrections. The slow control data from the entire experiment will be gathered by a dedicated subsystem, which will monitor the performance and health of all the subsystems described in this TDR. The chapters which follow separately describe the requirements and design of these subsystems.

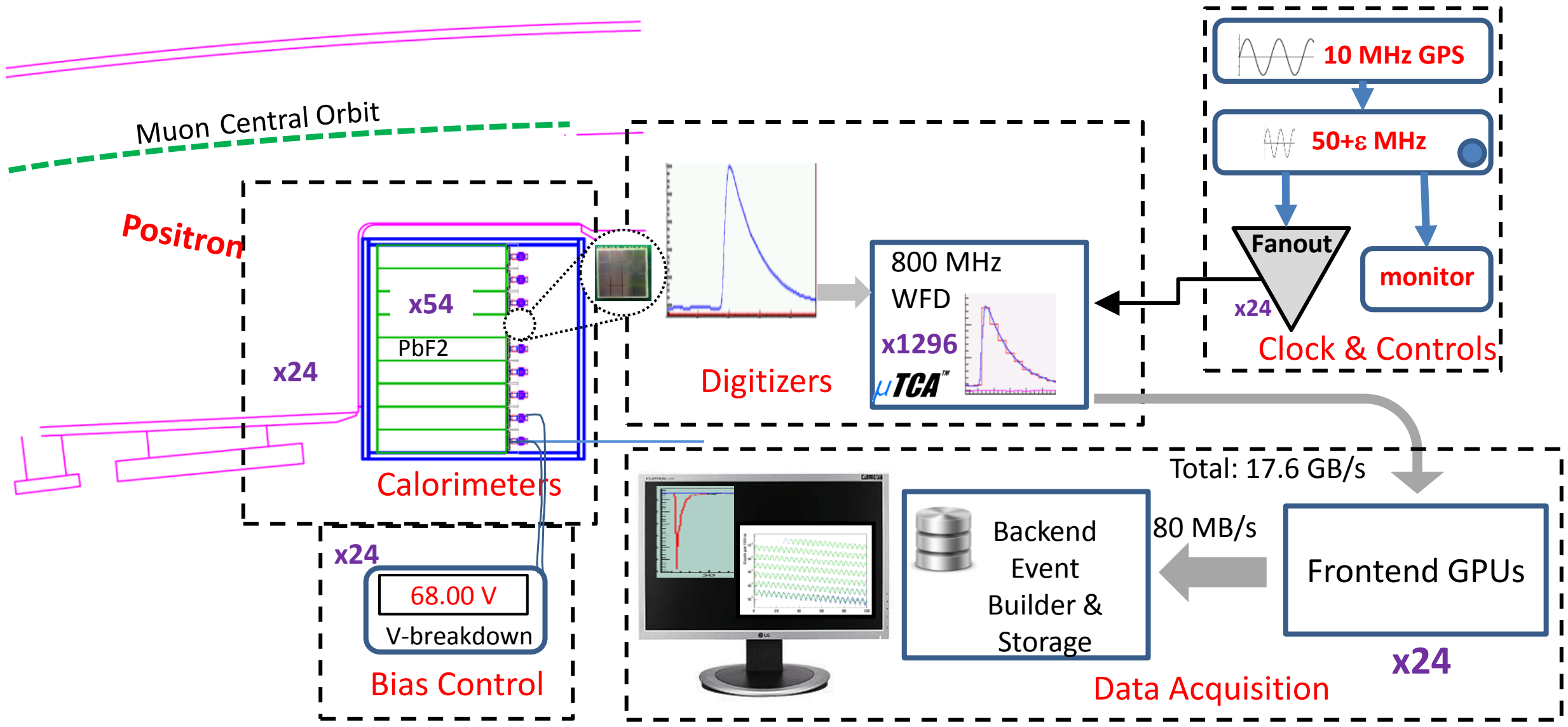

Figure 16.7: Schematic of the relationship of a number of the $\omega_a$ subsystems. The dashed boxes represent distinct responsibilities of different groups within the collaboration.

### 16.4.1 Calorimeter Subsystem Considerations

The calorimeters will be placed adjacent to the storage ring vacuum chambers, and located at 15 degree intervals around the ring. The 24 stations and their locations are constrained by the reuse of the E821 vacuum chambers, see Figure 16.6. These parameters were optimized in a study preceding E821 construction and the conclusions remain valid for E989. The number of emitted decay positrons vs. decay energy is shown in Figure 16.8. The geometry is designed to favor the high energy positrons because they are most correlated with the muon spin. low energy positrons have a higher chance to curl in between calorimeter stations and be lost.

The design of the new calorimeters is constrained by the unusual experimental demands. It is important to emphasize that the relevant time scale for most systematic uncertainties is one 700 $\mu$s long measuring period. The initial instantaneous event rate of several MHz drops by almost five orders of magnitude during the 700 $\mu$s measuring period; thus, any rate-dependent detector or readout response changes must be accurately known. The overall measurement system must be extraordinarily stable for each short-term storage ring fill; however, long time scale drifts can generally be tolerated. As discussed in Section 16.3.1, if the gain vs. time-in-fill, $G(t)$, is not constant, then $\omega_a$ might be incorrectly determined. Similarly, a time shift $\Delta t$ owing to the clocking system or other influence can also change the fitted frequency. From our simulations the stability conditions that ensure less than a 50 ppb shift to $\omega_a$ require that $\Delta G <$ 0.1% and $\Delta t <$ 10 ps over a 200 $\mu$s interval.

As described in Section 16.3.2, incorrectly treated pileup can lead to a large systematic uncertainty. As illustrated in Figure 16.6 low energy positrons have a shorter flight path to the detector compared to higher-energy positrons; thus they correspond to muons having a different average spin at the time of the decay. The rate of fake high energy positrons coming from coincident low energy positrons has a $\sim e^{-2t/\gamma\tau}$ time dependence. This means the pileup rate falls twice as fast as the muon population decays. To minimize pileup, the calorimeter response must be fast (few ns) and the readout system must record information to enable the distinction between closely occurring pulse pairs, which strike the same detector elements. This information should also provide a mechanism to correct the data, on average, by removing the pileup events. Furthermore, if the detector segmentation is optimized, many simultaneous lower-energy positrons will be recorded in independent area of a calorimeter station and thus will not be interpreted as a pileup event. The goal in the detector design is to reliably resolve same-element pulses separated by 5 ns or more, to segment the detector to minimize pileup, and to accurately subtract unresolved pileup.

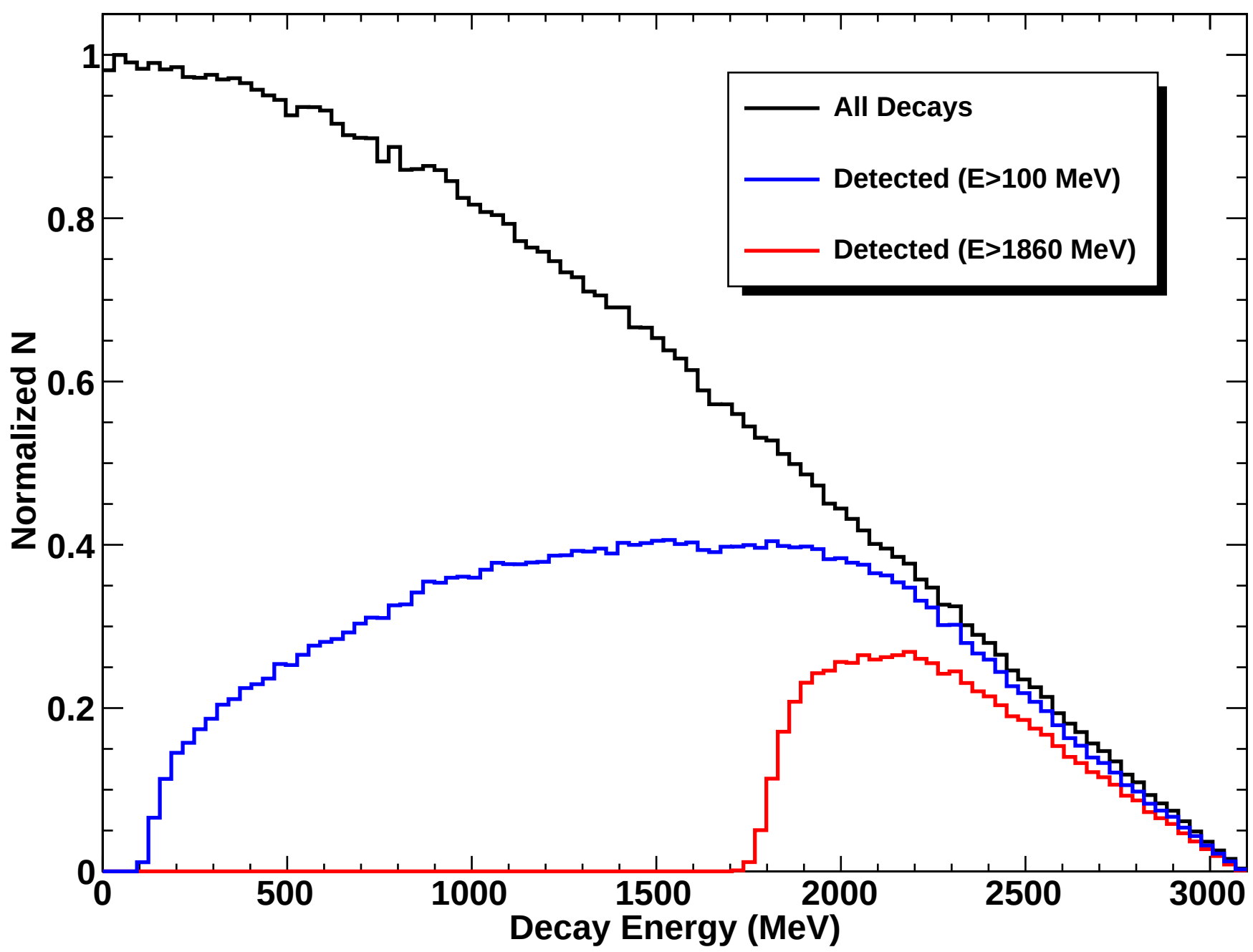

Figure 16.8: Number of decay events vs. decay energy (black), number of events that deposit detectable energy over 100 MeV (blue), and number of events that deposit detectable energy over 1860 MeV (red). This plot is generated from a simulation using full geometry, including pre-showering effects, and tracking the particles from muon decay through showering and energy deposition in the calorimeter.

The calorimeter resolution must be moderately good near 1.8 GeV (better than 10%) to provide adequate energy discrimination. However, it also must be compact to avoid a preponderance of positrons that strike the inside face of the detector. Higher density materials enable a more compact detector but yield inferior energy resolution. For E989, we are aiming for a factor of 2 improvement in resolution compared to E821. The material $PbF_2$ meets both criteria of compactness and resolution. The energy resolution of the $PbF_2$ calorimeter system was tested at SLAC and shown to be 2.8% at 3.5 GeV.

### 16.4.2 Electronics and Data Acquisition Subsystems Considerations

To guarantee deadtime-free calorimetry readout, the signal from each of the 1,296 active calorimeter channels are continuously digitized for every 700 $\mu$s muon fill. Those waveforms are then transferred to the DAQ system for data reduction – isolation of the time windows containing electromagnetic showers – and storage. The DAQ must apply an energy threshold to identify showers within a station, so the 54 waveforms from that station must be summed to keep the threshold independent of the incident positron position in a crystal. All 54 WFD waveforms must therefore be transferred to the same frontend DAQ system, which will use the waveform sum to perform data reduction on the digitization stream: identification of time islands with activity ($T$ method) and time rebinning of the waveform ($Q$ method).

The energy range of interest for an individual calorimeter element is 25 to 3100 MeV for single events. Allowing pileup, suggests pushing the upper limit close to 5000 MeV. A digitizer with 12-bit depth (4096 channels) is ideal. It will allow good pulse definition, important for the energy resolution requirements, and it will not saturate for the highest energy events. As discussed in Chapter 17, the pulse shape rise time is approximately 2-4 ns. The laboratory tests have led to selection of a sampling rate of 800 MSPS. Therefore, each digitized waveform corresponds to 560K 12-bit words for each muon fill.

A precision oscillator ("clock") will provide the time base from which the $\omega_a$ frequency is measured. It must be controlled to provide negligible error compared to the anticipated 100 ppb uncertainty on $\omega_a$. In order to achieve this, the clock must have jitter that is significantly less than the 1.25 ns sampling period of the waveform digitizers. It must also have very low ($< 10$ ps) systematic shift across the time of a single fill. This latter requirement is important because of the large variation in event rate within a fill. A systematic time-slew that is correlated with muon or positron intensity would bias the result. The clock system must also enable a convenient blinding scheme such that the actual precise clock frequency cannot be known to the data analyzers.

The data acquisition system must provide a deadtime-free readout of calorimeter segments using the waveform digitizers. Onboard memories in the digitizers will buffer the raw data and allow its asynchronous readout, thus decoupling the data acquisition cycles from storage ring fills. A frontend layer of multicore CPUs/GPUs will process the digitized records of each fill from every calorimeter segment into $T$-method, $Q$-method and other derived datasets. A backend layer of multicore CPUs/GPUs will handle the assembly of event fragments from the frontend layer and transfer of assembled events to the mass storage. Each stored event will represent a complete deadtime-free history of the entire activity in the detector system for every fill cycle.

# Chapter 17

# Calorimeter

This chapter illustrates the design concept for the 24 electromagnetic calorimeters. The primary purpose of the electromagnetic calorimeter is to measure the energy and time of arrival of the daughter positrons from stored muon decay. The physics goals and subsequent requirements are reviewed. The recommended design for each subsystem — Absorber, Photodetection (SiPM), Bias Control, Laser Calibration, and Mechanical — is then presented. Finally, alternative designs, ES&H, risks, quality assurance, and value management are discussed.

## 17.1 Physics Goals and Requirements

After a muon decays into a positron and a neutrino, the positron doesn't end up with sufficient energy to fly along the magic orbit in the ring. It curls inward where it hits a segmented *lead fluoride calorimeter* readout by *silicon photo-multipliers* (SiPM). The primary physics goal of the calorimeter is to measure energy and hit time of daughter positrons.

The requirements on the energy and time measurements are:

- Relative *energy resolution* of the reconstructed positron energy summed accross calorimeter segments must be better than 5 % at 2 GeV. The modest specification on energy resolution is motivated by the purpose of the energy measurement which is to select events. Deposited energy is not a direct observable in the experiment.

- *Timing resolution* of the hit time extracted from the fit of the SiPM current pulse must be better than 100 ps for positrons with kinetic energy greater than 100 MeV in any combination of temporal and spatial pileups.

- The calorimeter must be able to *resolve* two *showers* by temporal or spatial separation. The calorimeters must provide 100% efficiency in the discrimination of two showers with time separations greater than 5 ns. Showers that occur closer in time than 5 ns must be further resolved spatially in more than 66% of occurrences. These requirements correspond to the systematic uncertainty on unidentified pileup events of 40 ppb.

- The *gain (G) stability* requires a maximally allowed gain change of $\frac{\delta G}{G} < 0.1\,\%$ within a 200 $\mu$s time period in a fill. In addition, the arrival of a pulse should not affect the gain for a second pulse arriving a few nanoseconds later on the same channel, unless that change is understood and can be applied to the interpretation of a following pulse in a reliable manner. The long term gain stability (intervals of multiple seconds) is more relaxed and must be $\frac{\delta G}{G} < 1\,\%$. To verify the overall gain stability, each of the 24 stations must be equipped with a calibration system that must monitor the gain continually during the muon spills with a precision of $\frac{\delta G}{G} \sim 0.04\,\%$. These requirements correspond to the systematic uncertainty on rate dependent gain effects of 20 ppb.

- Efforts must to be made to preserve *fidelity* of the Cerenkov light *pulse shape* through the analog and digital signal chain.

## 17.2 Evaluation Methodologies

The baseline calorimeter design builds upon extensive testing and simulation efforts. Prototype detectors were built. Various silicon photomultiplier (SiPM) and photomultiplier tube (PMT) candidates were tested. Several iterations of the electronics boards needed to operate the SiPMs were built and tested. Both laboratory and test beam studies were performed. We employed simulations to study detector performance, sensitivity to $\omega_a$ and pulse-shape fitting.

### 17.2.1 Test beams

Laboratory tests with beams and lasers provided crucial results on pulse shape and energy resolution for different component choices, and also validated and informed our simulation efforts. The Fermilab Test Beam Facility (FTBF) was used several times to evaluate prototype calorimeters. In particular, our first effort in which 0.5-mm pitch tungsten plates alternated with 0.5-mm layers of scintillating fiber resulted in a publication [2]. A larger prototype was then built and tested, see Fig. 17.1.

The recommended design based on $PbF_2$ crystals was tested using the 7-crystal array shown in Fig. 17.1. These crystals were compared directly to the W/SciFi detector and to a custom $PbWO_4$ crystal during the April 2012 FTBF period. The right panel of Fig. 17.1 shows the arrangement of $PbF_2$ crystals during assembly and displays a front view of the full test setup. Various readout methods, wrappings and couplings were employed.

Two test beam evaluations of advanced $PbF_2$ calorimeter prototypes were completed at SLAC National Accelerator Laboratory's End Station Test Beam facility. The SLAC prototypes included SiPM photodetectors with custom amplifier boards designed by electrical engineers at the University of Washington. In both tests, SLAC provided an electron beam with energies ranging from 2.5 to 4 GeV, similar to the energies we expect from decay electrons in the final experiment. Scanning over this range of electron energies enabled evaluation the calorimeter's energy resolution and linearity.

The first test occurred in November of 2013 and featured a $3 \times 3$ array of crystals; the second occurred in July of 2014 and featured a larger $4 \times 7$ array. Both of these tests were highly successful in confirming the baseline calorimeter's ability to meet its design requirements. In order to investigate the effects of different wrapping types on the detector response, the $4 \times 7$ array used in 2014 was prepared as a $4 \times 4$ array of white, reflective paper wrapped crystals adjacent to a $4 \times 3$ array of black, absorptive paper wrapped crystals. Results from the 2014 test beam experiment have been published [3]. This document will focus on results from the 2014 test because the design and techniques applied there incorporated lessons learned from the 2013 test.

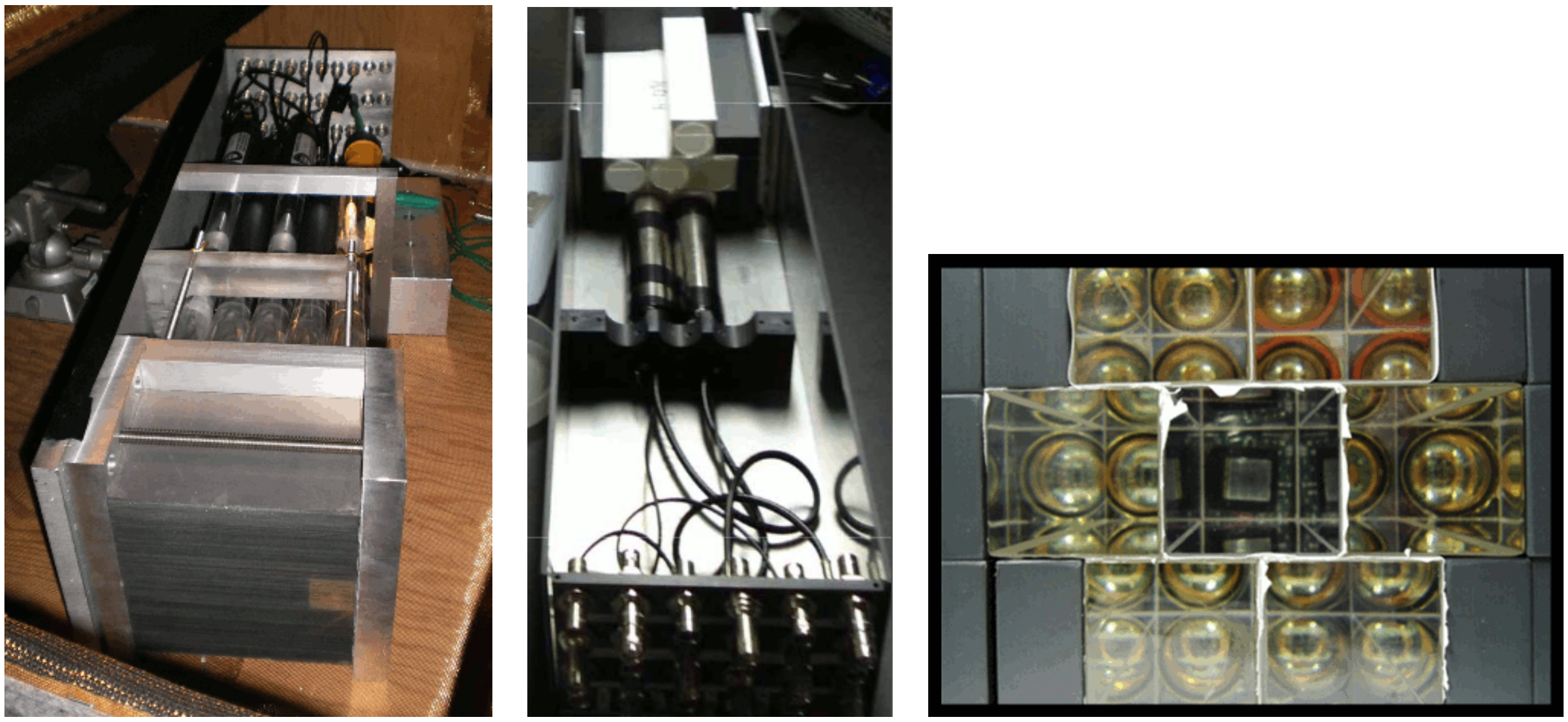

Figure 17.1: Left: Monolithic block of W/SciFi having 0.5 mm thick pure tungsten plates alternated with 0.5 mm diameter ribbons of blue scintillating fiber. The readout side is divided into 25 individual elements. Tapered light-guides direct the light from a $3 \times 3$ cm$^2$ area to a PMT. Center: Crystals being prepared for test beam. Here, PMTs are used for the outer elements and a SiPM will be placed on the center crystal and alternatively a very fast Hamamatsu R9800 PMT for comparison. Right: Front picture of the 7-crystal test array used in the FTBF. In this configuration, a SiPM is visible on the center channel, while PMTs are used on the remaining elements. These crystals were wrapped in white Millipore paper.

A light-tight housing for the tested prototype was constructed by CENPA engineers and featured a cold-air temperature stabilization system that was required to maintain acceptable SiPM gain stability. A cooling concept based on pressurized chilled and dried air blowing directly on SiPM boards was commissioned during this test run. While it successfully provided stability, the achieved temperature uniformity was not great and the system would be difficult to implement in the final design. A new water-cooling design is under preparation for the final calorimeter housing. Signals from the SiPMs were fed into a pair of Struck's 4-channel SIS3350 500MS/s 12-bit digitizers and a CAEN DT5742 16-channel 12-bit digitizer running at 1 GS/s.

The SLAC test provided an invaluable opportunity to test the calorimeter's laser calibration system. At the center of the calibration system is a PicoQuant pulsed diode laser, distributed through fiber optics to each crystal, and ultimately each SiPM, individually. A remote controllable neutral density filter wheel is placed in the path of the laser beam and used to vary the pulse energy seen by the calorimeter. By observing how the statistical width of the reconstructed energy distribution in each SiPM varies with the mean, one can obtain relative calibration constants for each of the calorimeter segments. In principle, this technique does not rely on each fiber delivering the same amount of light, as long as the light level is low enough that the SiPM responses can be treated as linear. Once relative calibration constants are established, long term stability is ensured by periodically firing a fixed-energy laser pulse into every segment. Undesired fluctuations in laser pulse energy are corrected for by a suite of dedicated laser monitors. These techniques produced a reconstructed energy that was independent of both time and impact position and revealed that our calorimeter yields $(1.45 \pm 0.05)$ pe/MeV when the crystals are wrapped in white, reflective paper and $(0.76 \pm 0.04)$ pe/MeV when the crystals are wrapped in black, absorptive paper (fig. 17.3).

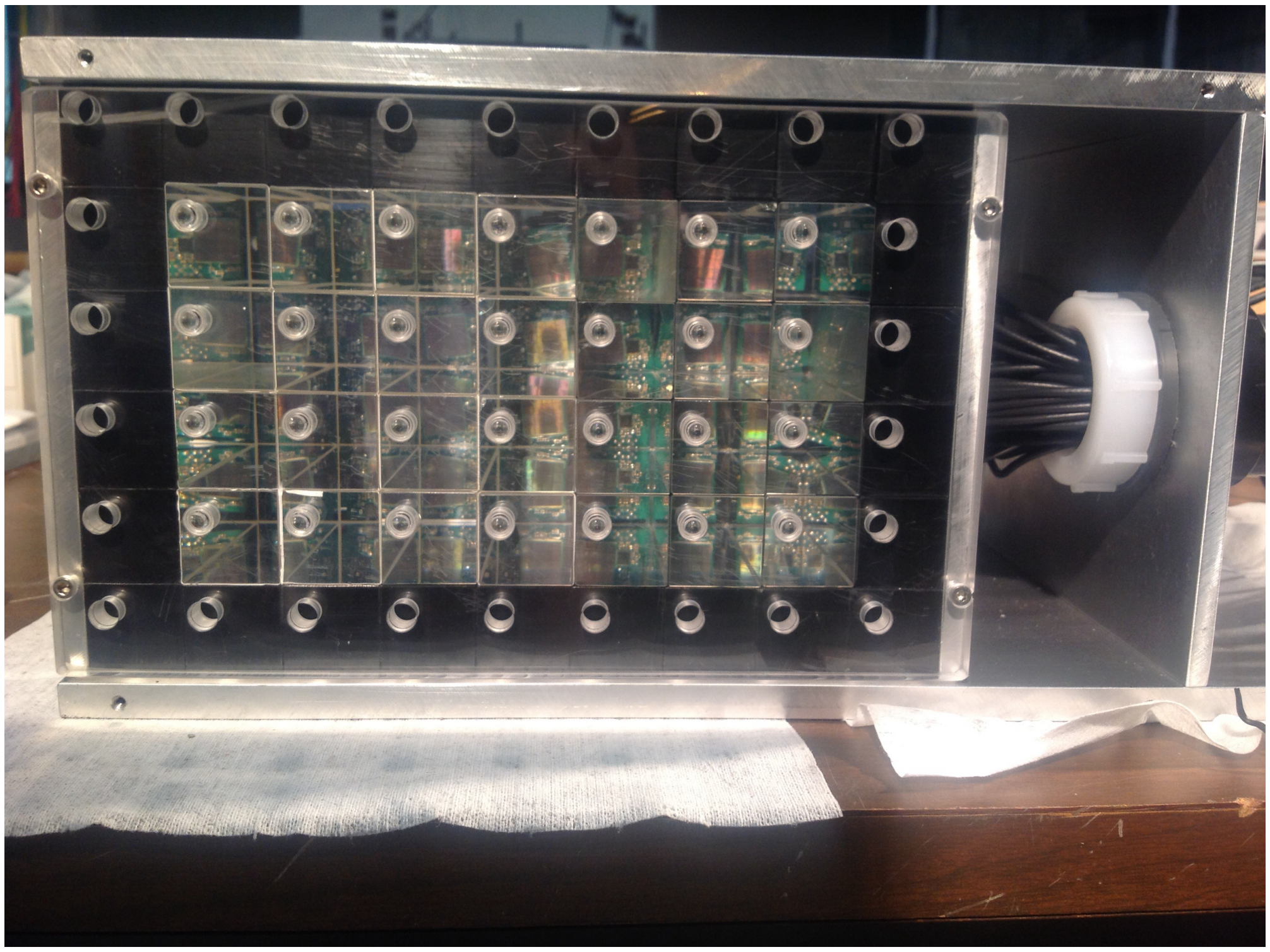

Figure 17.2: A calorimeter test-box was used at SLAC to measure the light yield of lead fluoride, better understand SiPM photo-efficiency, and characterize energy resolution. Analysis of data obtained using this test-box proved that our design meets $g-2$ requirements.

In order to achieve accurate energy measurements and the best resolution, the energy seen by each detector must be summed together on an event by event basis. For linearity and resolution tests, sums were taken over $3 \times 3$ sub-arrays. The sums were executed using calibration constants obtained with the aforementioned laser system.

After conducting energy sums across our array, we were able to demonstrate that the *energy resolution* of our calorimeter exceeds the requirement of 5 % at 2 GeV for both white and black wrapped crystals, see fig. 17.4. An energy scan in the 2.5 – 4.0 GeV range, depicted in fig. 17.3, proved an excellent linearity of the $PbF_2$ calorimeter.

The SLAC facility featured a remote controlled x–y table that allowed us to scan the beam over the face of our detector. Our reconstructed energy was independent of the beam position and incident angle (we tested up to 20°). By examining the distribution of energy between crystals we were able to locate the beam position. This segmented nature of the detector contributes significantly to its pileup resolution capabilities. Additionally, the test beam was used to investigate the calorimeter system's long term gain stability. Using the techniques described above, we found that gain stability is correctable to better than $10^{-4}$/hour (fig 17.5).

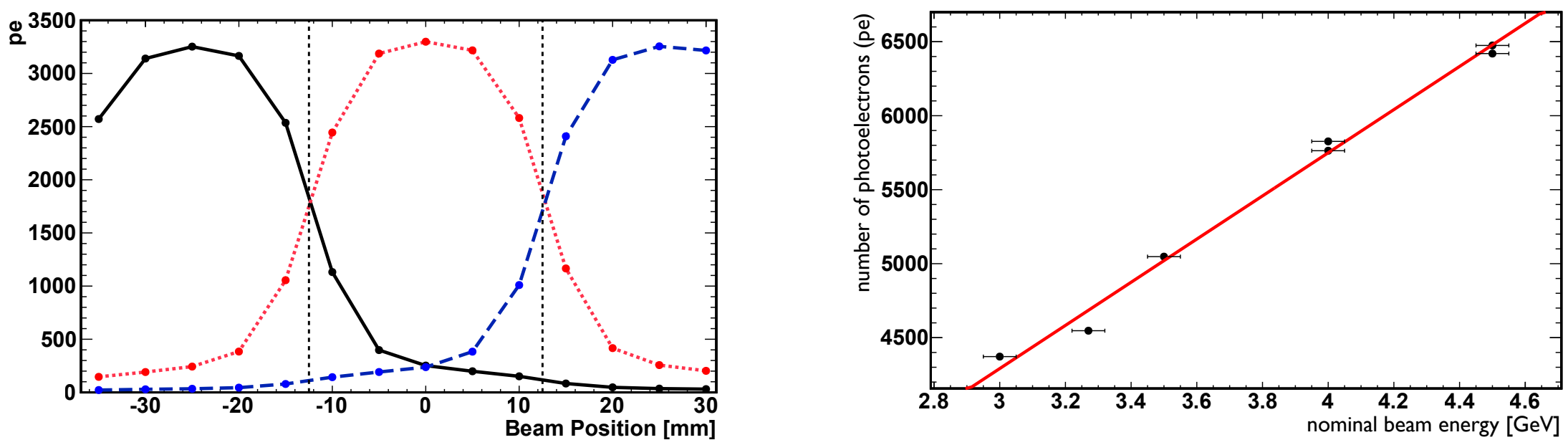

Figure 17.3: Left: Reconstructed energy in three adjacent crystals as the electron beam is scanned across. The calibration constants obtained from the technique described in the text bring their peaks to the same height. pe stands for photelectrons. Right: Measured pe in a $3 \times 3$ cluster of crystals wrapped in white, reflective paper as a function of nominal beam energy. The uncertainties on the nominal beam energies are set to 50 MeV, the stated confidence of the beam operators. Absolute scale is obtained from the laser calibration and allows an extraction of pe/MeV.

## 17.3 Lab tests

*Timing response* of SiPMs was studied using a 407 nm pulsed diode laser. A laser shot was split into two channels, one of them was optionally delayed and fed through crystals into two different SiPMs. The hit times were extracted from pulse fits. An experiment was repeated many times, and a histogram of these time differences was created. The histogram was well fit with a Gaussian distribution, with mean value identical to the known delay between the two laser shots, and with the sigma of 40 ps. The sigma was independent of pulse heights. The same study was repeated for the pileup events: both the prompt and delay pulses were fed into a single crystal read by a SiPM. Also in this case, the histogram of hit time differences was statistically compatible with the Gaussian distribution, its mean value was not biased, and the sigma turned was 70 ps. Both the studies were performed with 500 MHz digitizers. This timing response meets the experimental requirements. Upgrading from 500 MHz to 800 MHz in the final digitizer design will likely further improve the timing response.

SiPM *bulk recovery time* (temporary bias voltage drop) was understood in a dedicated systematics study. An Agilent 33521A function generator was used to generate LED pulses at a rate of $1 \cdot \exp(-t/\tau_\mu)$ MHz, where $\tau_\mu$ is the boosted muon lifetime of about 64.4 $\mu$s. This setup simulates a $(g-2)$ muon fill. At an adjustable time during this fill, the SiPM was illuminated with a fixed-energy 407 nm laser shot and the resulting pulse was digitized and recorded. By observing the SiPM response to a fixed-energy laser shot at varying times during and outside of a simulated fill, the SiPM average gain function, $G(t)$, can be extracted. The energy of the LED pulses was also varied to investigate how $G(t)$ changes with average current; it was determined that $G(t)$ scales linearly with average current in the relevant regime.

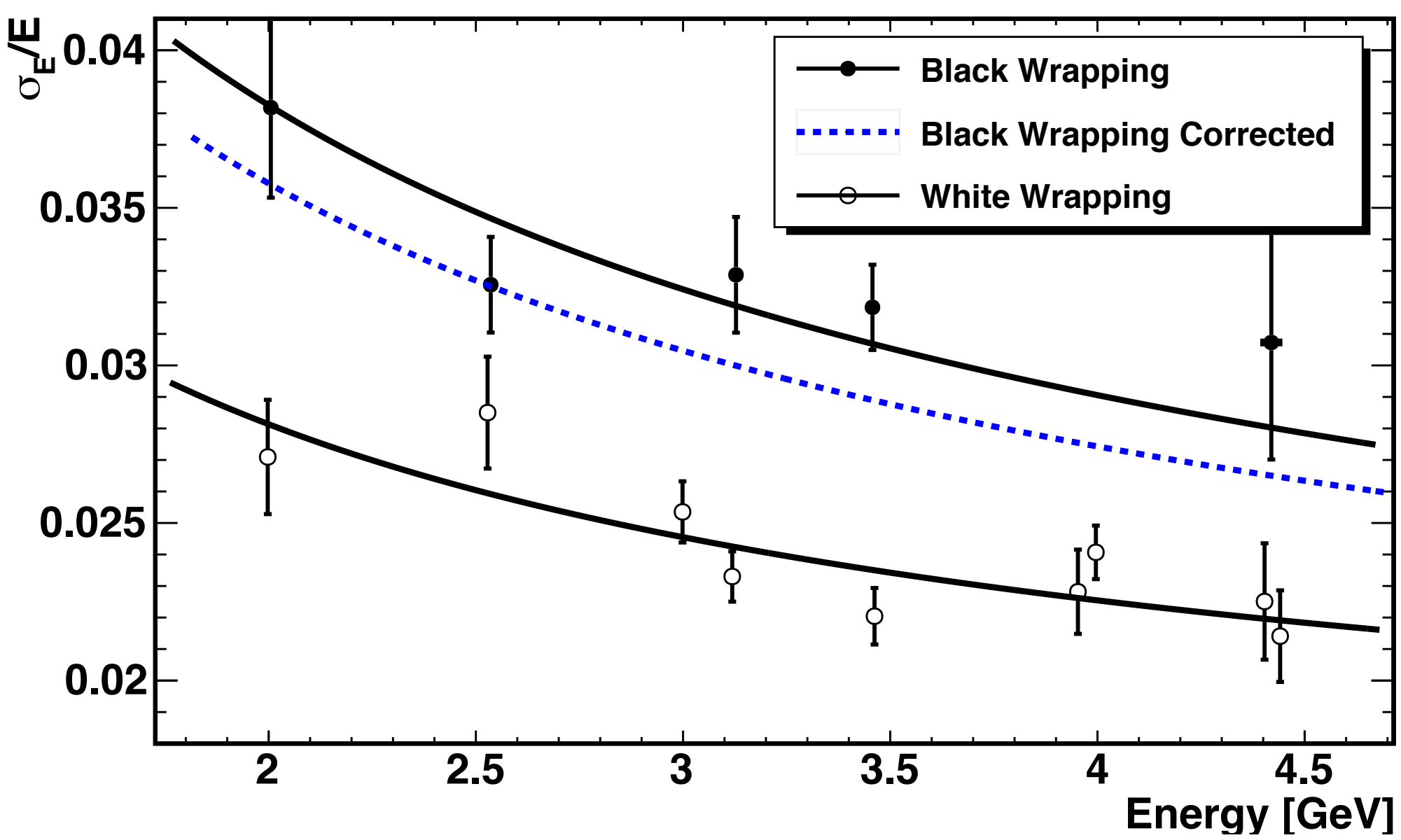

Figure 17.4: Energy resolutions of $3 \times 3$ arrays of $PbF_2$ crystals with black and white wrappings as a function of energy. Fit functions are of the form $\sigma_E^2/E^2 = (1.5\,\%)^2 + a^2/E$. The blue dashed line is the result of correcting the black-wrapped curve for dead channels SiPM channels discovered after the fact.

It was shown in section 16.3.1 that the target value for gain perturbations during the fill is $< 10^{-3}$. The result of this test, shown in fig. 17.6, is that the current SiPM boards meet the experimental systematic error requirement of gain stability better than $10^{-3}$ during the fill.

### 17.3.1 Calorimeter rate simulation

It is instructive to estimate the event rate for a single calorimeter station and for the hotest element of the designed array of 54 crystals. The average number of positrons above an energy threshold $E$ incident on a calorimeter at a fixed time after injection follows the form given in Equation 17.1. Here $N_\mu(E)$ is the number of muons that decay to positrons above energy $E$ and $\epsilon(E)$ is the calorimeter acceptance. Using the expected number of stored muons per fill of $\sim$ 16,000 (see Chapter 5) and the acceptance derived from our full GEANT-based simulation [6], we find the average rate at the fit start time of 31 $\mu$s to be nearly 3 MHz for a full calorimeter, as shown by the solid black line in Fig. 17.7. What the detector experiences is the local instantaneous rate, which is modulated by two frequencies. The first is the normal $g-2$ frequency having period $\sim 4360$ ns. The modulation depends on the asymmetry, $A(E)$ as parameterized in Eq. 17.2. For a threshold as low as 100 MeV, the asymmetry is small and the resulting rate is shown by the solid blue line in Fig. 17.7. The "in-phase" peak rate rises compared to the average value by a small factor.

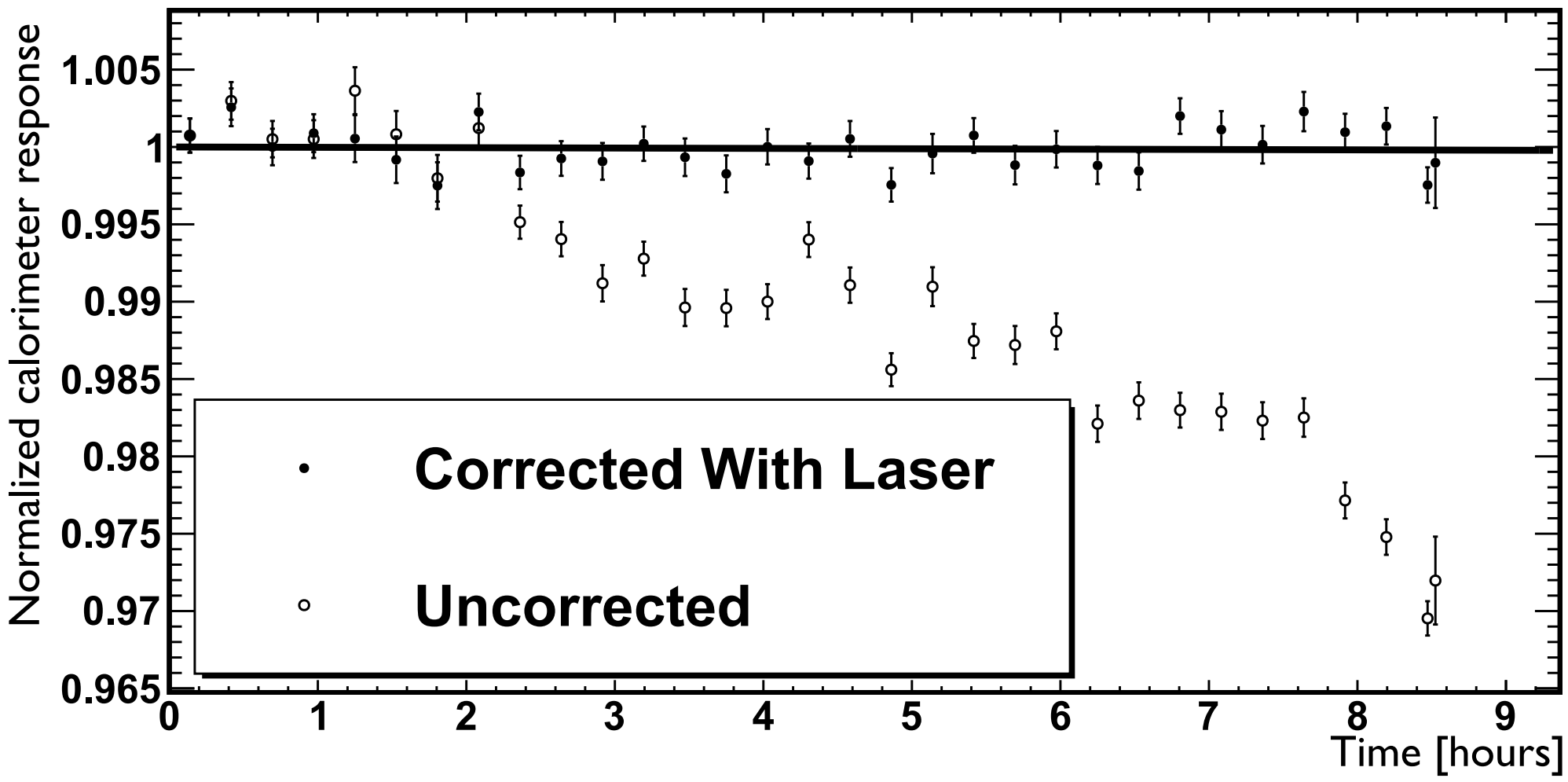

Figure 17.5: Reconstructed energy, normalized to 1, from a 3 GeV electron beam firing continually over a 9 hour period. After correction with the laser system, the reconstructed energy is stable at the level of $10^{-4}$/hour.

$$R_{\rm exp}(E,t) = N_\mu(E)\cdot\epsilon(E)\cdot(\gamma\tau_\mu)^{-1}\exp\left(-t/\gamma\tau_\mu\right) \quad (17.1)$$

$$R_{g-2}(E,t) = R_{\rm exp}\cdot\left[1 + A(E)\cos\left(\omega_a t\right)\right] \quad (17.2)$$

$$R_{\rm FR}(E,t) = R_{g-2}\cdot\left[1 + A_{\rm c}\exp\left(-t/\tau_{\rm FR}\right)\cos\left(\omega_{\rm c} t\right)\right] \quad (17.3)$$

More importantly is the effect of the time-structured beam profile at injection. The $\sim$ 80 ns flat-top of the kicker will efficiently store the central portion of the $\sim$ 120-ns-long entering beam bunch. As this bunch rotates at the cyclotron frequency, $\omega_c = 149$ ns, a large intensity oscillation will be imprinted on the distribution. Using a simplified model with a cosine dependence, this "fast-rotation" effect can be described by an amplitude, $A_{\rm FR}$, a damping time constant $\tau_{\rm FR}$, and the cyclotron frequency $\omega_{\rm c}$. The damping occurs as the beam spreads out owing to the momentum dispersion. An effective damping factor is used based on the E821 experiment. The final rate is described by Eq. 17.3 and is depicted by the solid red envelope in Fig. 17.7 Note that $\omega_c \gg \omega_a$. The peak rate rises at fit start to about 4 MHz.

Unlike E821, detectors in E989 will not be gated off during injection. Therefore both the average and instantaneous rates at $t = 0$ in Fig. 17.7 are useful guides for planning purposes.

Table 17.1: Projected Rates in Calorimeters

| Region | $\epsilon$ | Instantaneous Rate | $\sim N_{\rm stored}$ |
|---|---|---|---|
| Full Calorimeter ($E > 100$ MeV) | 0.0191 | 2.9 MHz | linear |
| Hot Crystal ($E > 25$ MeV) | 0.0031 | 0.48 MHz | linear |
| Hot Crystal Pileup Fraction | – | 0.24% | quadratic |

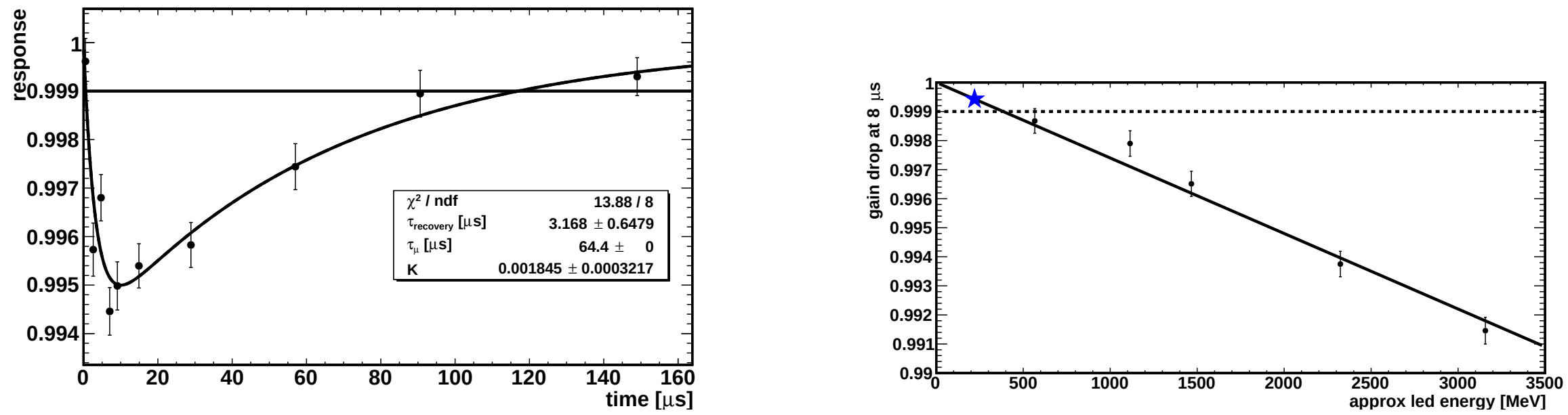

Figure 17.6: Left: Results from lab test measuring SiPM $G(t)$. The vertical axis shows the ratio of SiPM pulse area during an LED fill to pulse area outside of an LED fill, where an LED fill is described in the text. The minimum of this function occurs at $\sim 8\,\mu s$. Right: $G(8\,\mu s)$ scaling as energy of LED shots during the LED fill is varied. The scaling is linear. The blue star denotes the average pulse energy expected in $(g-2)$, the dashed line denotes the experimental goal. Extrapolating LED energy to the expected $(g-2)$ energy indicates that our system exceeds the design requirement.

The instantaneous rates are summarized in Table 17.1. The term "Full Calorimeter" is the rate as if the entire calorimeter were readout as one monolith. The crystal where the maximum rate is expected is called the "Hot Crystal." A crystal is assumed to be hit if it absorbs more than 25 MeV in any event, no matter where the incoming positron strikes. Figure 17.8 shows the instantaneous rate of hits over 25 MeV for each individual calorimeter crystal in the array.

Finally the fractional rate of pileup is also calculated for the hot crystal. The pileup rate is given by

$$\langle R_{\text{pileup}} \rangle = \langle R \rangle^2 \cdot \Delta t \tag{17.4}$$

where $\Delta t$ is the resolving time of the calorimeter. Working with the rate in the hot crystal and a resolving time of 5 ns we calculate the maximum pileup fraction to be 0.24%, or a rate of about 1 kHz.

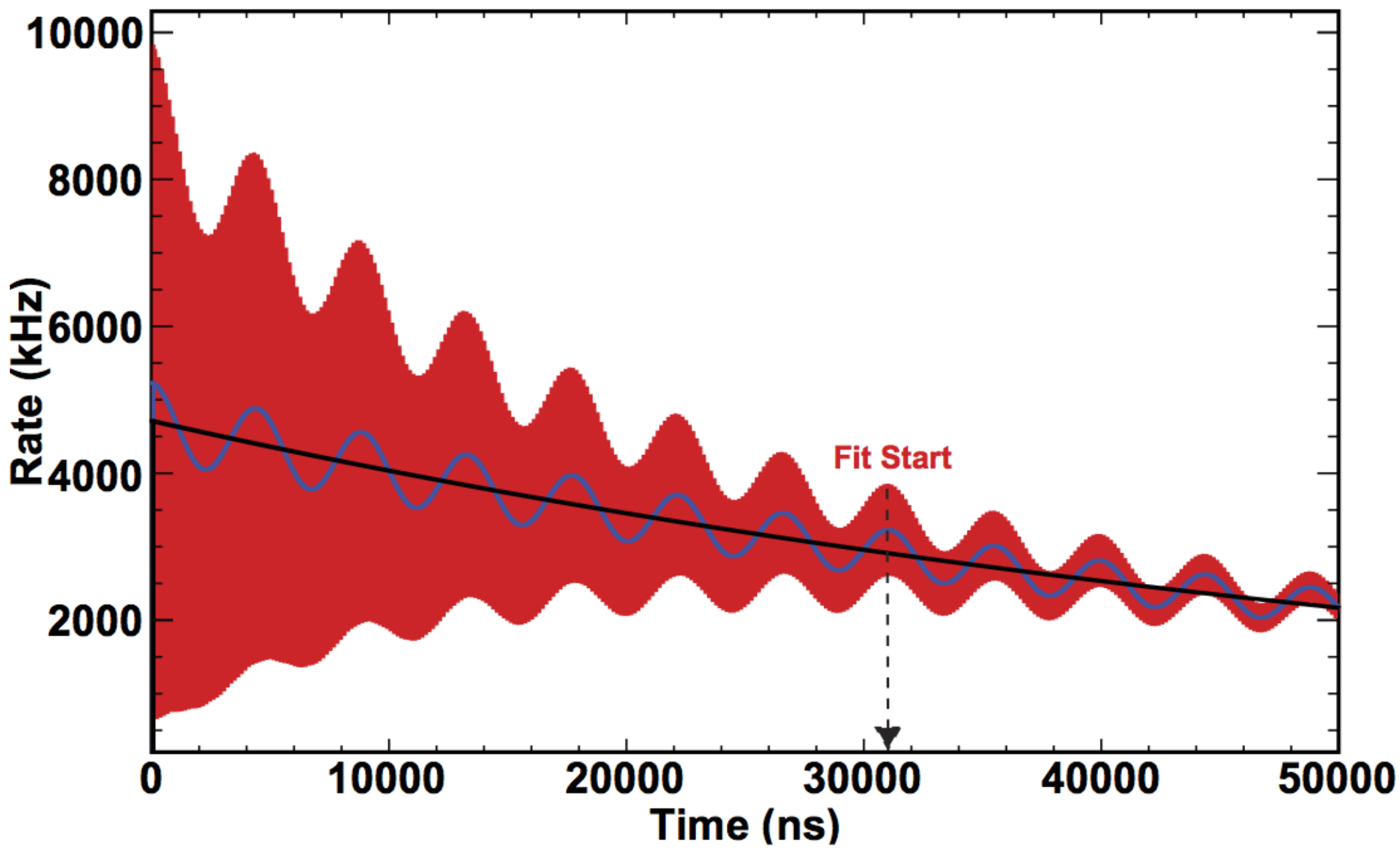

Figure 17.7: The three curves represent the instantaneous rate for: decaying muons (black), with $\omega_a$ oscillation (blue), and with fast rotation (red). This data was simulated using the standard run conditions outlined in the TDR and a low-energy threshold of 100 MeV. This plot is for one calorimeter station.

## 17.4 Baseline calorimeter design

The calorimeter system includes the following subsystems: absorber, photodetection, bias control, calibration, and mechanical. Over the past several years, the calorimeter design has gone through an extensive down-select process for absorber and readout technologies.

- *Absorber:* Each of the 24 calorimeter stations consists of a $6 \times 9$ array of lead fluoride ($PbF_2$) Čerenkov crystals. The crystal is $25 \times 25 \times 140\,\mathrm{mm}^3$, and wrapped in a single layer of highly reflective Millipore paper.

- *Photodetector:* Each crystal is read out by a monolithic 16-channel Hamamatsu MPPC (S12642-0404PA-50). (Multi-Pixel Photon Counter, also called silicon photomultipliers or SiPM's). This SiPM has an active area of $12 \times 12\,\mathrm{mm}^2$, and 57 344 50-$\mu$m pixels. The current pulse output by SiPM is amplified and converted into a voltage signal in a custom made amplifier board that the SiPM is soldered to. The output stage of the amplifier board features a digitally controlled variable gain amplifier which is AC coupled to a differential pair of coaxial cables to avoid ground loops, and maximize signal to noise ratio. The recommended design does not include pole zero correction, or any other tool to shape the output pulse. Preserving the intrinsic Čerenkov light pulse shape is among our goals. The light yield of the crystals is MC-predicted and measured to be 1.0 registered photo-electrons per 1 MeV of deposited energy.

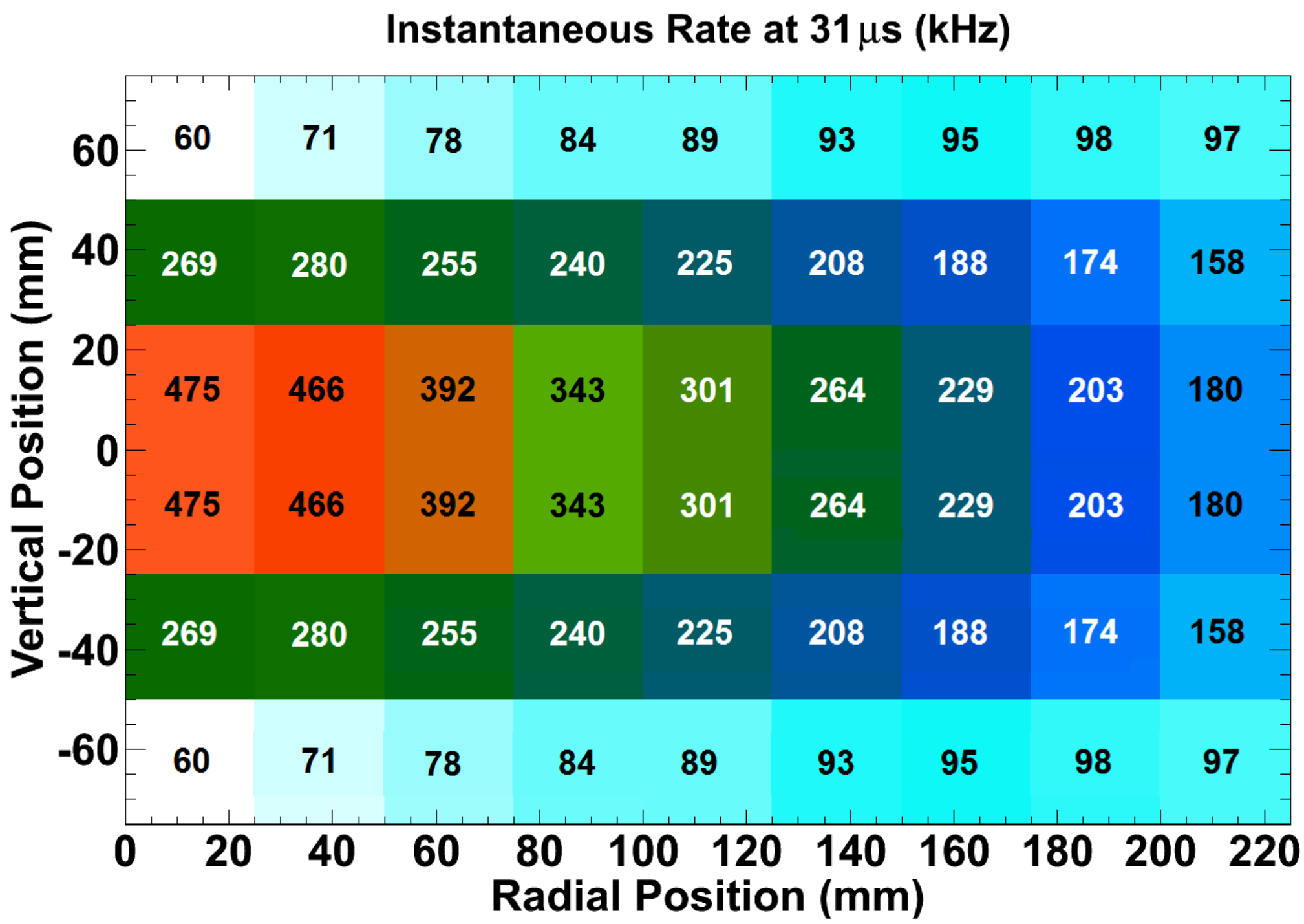

Figure 17.8: Projected instantaneous rate in each calorimeter crystal 31 $\mu$s after muon injection. Individual crystal acceptance was calculated from a GEANT simulation with showering and energy depostion in the crystals. A threshold of 25 MeV was applied in analysis to act as a hardware threshold. The acceptances were then combined with the overall rate scale outlined in the TDR to determine the rate in each crystal. The smaller radial positions are closer to the muon storage region, as seen by the enhanced rate.

- *HV bias control:* Reverse bias voltage applied to a SiPM is provided by a commercial HV bias power supply that maintains better than 1 mV stability over the critical 700 $\mu$s time window. The voltage output ranges from 60 V to 80 V, and can be set digitally. Each output channel will serve about 12 SiPMs, allowing four or five individually tunable bias values per calorimeter.
- *Laser calibration:* The calorimeter gain is calibrated and continuously monitored by a state-of-the-art high-performance laser-based distributed system. The unique system is critical to keep systematic contributions from any energy-scale instability well below the statistical precision of the measured $\omega_a$ frequency. Additionally the system will be used to initially tune and set gains for all crystals.
- *Mechanical:* A calorimeter housing is a moveable light-tight enclosure that provides sufficient cooling power to temperature stabilize the crystals, SiPMs, and amplifiers. The platform, which rides on a set of rails, allows easy insertion into or out of the ring in the radial direction. The absolute position of the calorimeter is of lesser importance than reproducibility of the position.

Several factors that influenced the technology choice are:

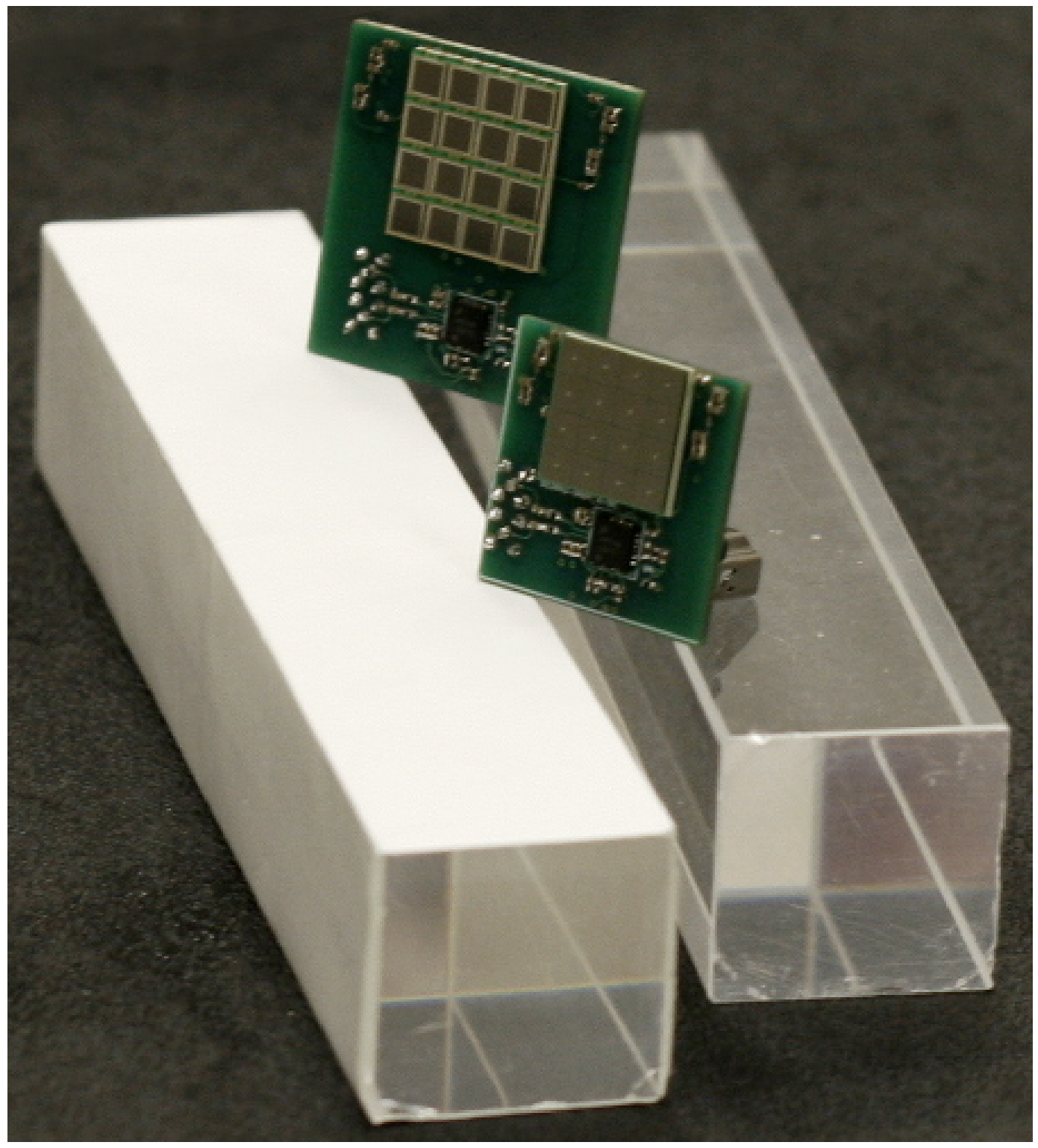

Figure 17.9: Sample $25 \times 25 \times 140\,\mathrm{mm}^3$ $PbF_2$ crystals (bare and wrapped in Millipore paper) are pictured together with a 16-channel monolithic Hamamatsu SiPM mounted to our transimpedance amplifier board (front). Behind it, one of alternative SiPM designs manually assembled from 16 individual SiPMs is shown. A Millipore wrapped crystal read out by a monolithic 16-channel SiPM is the core of our recommended design.

- Each of the 24 calorimeter stations will be located in the fringe field of the central storage ring, directly adjacent to the muon storage volume in a cutout of a scalloped vacuum chamber (see Fig. 16.6). The space is highly constrained vertically (17 cm) and longitudinally (owing to vacuum interconnects and flanges). Strict limits exist on the allowed magnetic field perturbation from the absorbers, electronics and mechanical housings. All the components and materials used for calorimeter construction are carefully tested using the University of Washington test magnet.

- The absorber must be dense to minimize the Molière radius and radiation length. A short radiation length is critical to minimize the number of positrons entering the side of the calorimeter while maintaining longitudinal shower containment.

- The intrinsic signal speed must be very fast with no residuals on either leading or trailing edge since the leading edge reports on hit time, and the quality of the trailing edge is essential for reliable pileup correction.

- The energy resolution should be good—it is used to select events—but it need not be "excellent." A resolution of $\sim 5\,\%$ at 2 GeV is considered sufficient and improves upon the E821 calorimeter system by a factor of 2.

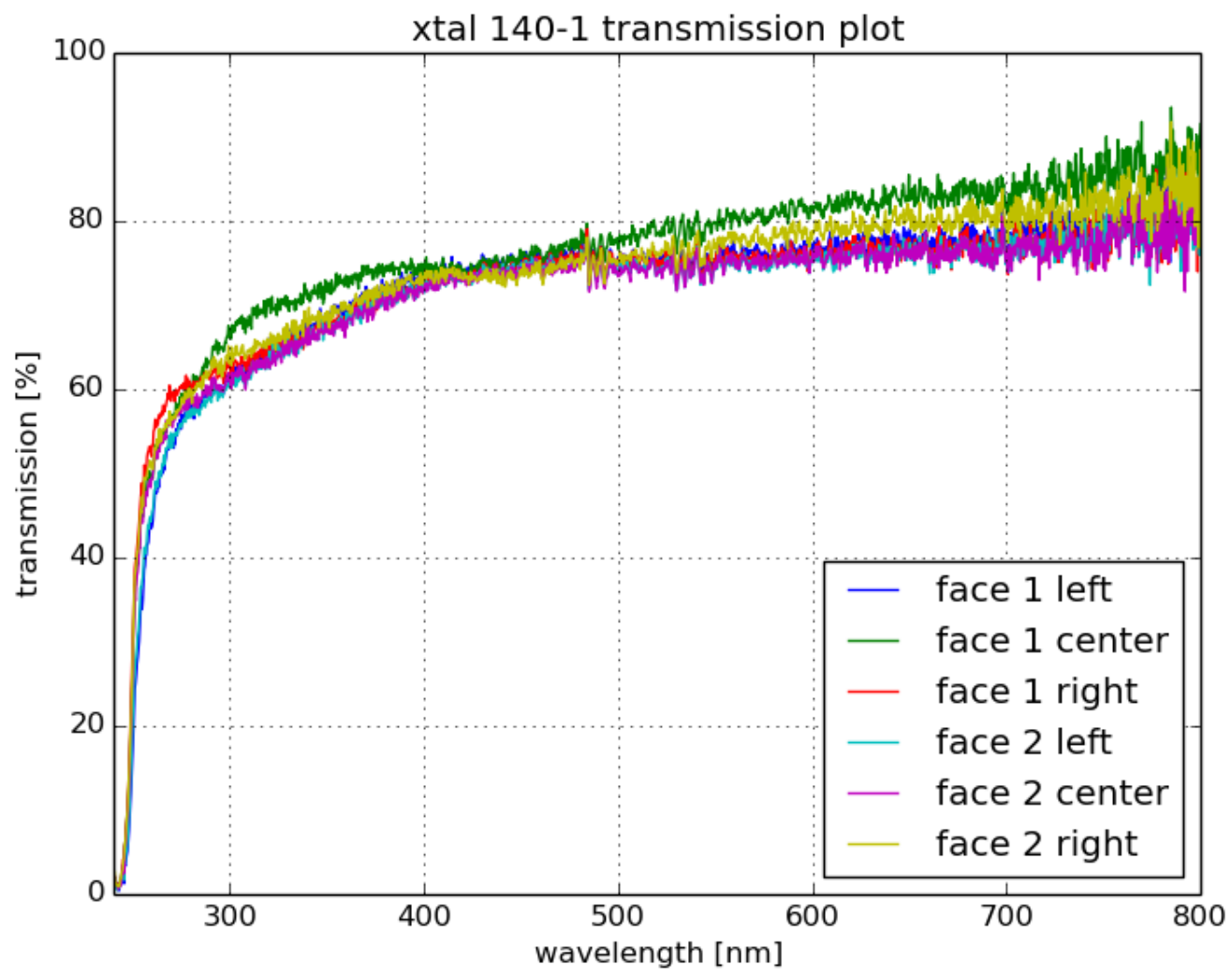

Figure 17.10: Transverse transmission efficiency vs. wavelength through a $PbF_2$ crystal is measured in the transverse direction. The curves correspond to transmission measurements taken with an Ocean Optics spectrometer at different locations on the crystal's faces.

### 17.4.1 Absorber Subsystem

The default material choice following an extensive comparative evaluation program (see Sec. 17.5) is lead-fluoride crystal ($PbF_2$). This crystal's combination of good energy resolution and a very fast Cherenkov signal response outperformed the other absorber options that we considered (see Sec. 18.4). It has very low magnetic susceptibility, a radiation length of $X_0 = 0.93$ cm, and a Molière radius of 1.8 cm. Our extensive test-beam program, as described earlier, verified the resolution and light yield. Table 17.2 presents a summary of the properties of the crystals.

The Shanghai Institute of Ceramics (SICCAS) provided the prototype crystals and a competitive quote for the 1325 elements required for the full system (plus spares). We own and have used instrumentation to measure the spectral response of the crystals over the range 230 nm to 800 nm, see Fig. 17.10. We have also made atomic force microscopy (AFM) measurements on a crystal to determine the surface quality so that we might properly represent it in our light-propagation simulations. The crystal procurement plan involves local oversight by the Shanghai University members of the collaboration, who are local to the vendor. They worked closely with SICCAS to prepare the requisition, which has now been enacted. The crystals will be sent to the University of Washington team for wrapping and assembly. $PbF_2$ crystals are relatively easy to handle; they are only slightly hydroscopic.

Table 17.2: Properties of lead-fluoride crystals

| | |
|---|---|
| Crystal cross section | $2.5 \times 2.5$ cm$^2$ |
| Crystal length | 14 cm ($> 15X_0$) |
| Array configuration | 6 rows, 9 columns |
| Density of material | 7.77 g/cm$^3$ |
| Magnetic susceptibility | $-58.1 \times 10^{-6}$ cm$^3$/mol |
| Radiation length | 0.93 cm |
| Molière radius $R_M$ | 2.2 cm |
| Molière $R_M$ (Cherenkov only) | 1.8 cm |
| $KE_{threshold}$ for Cherenkov light | 102 keV |

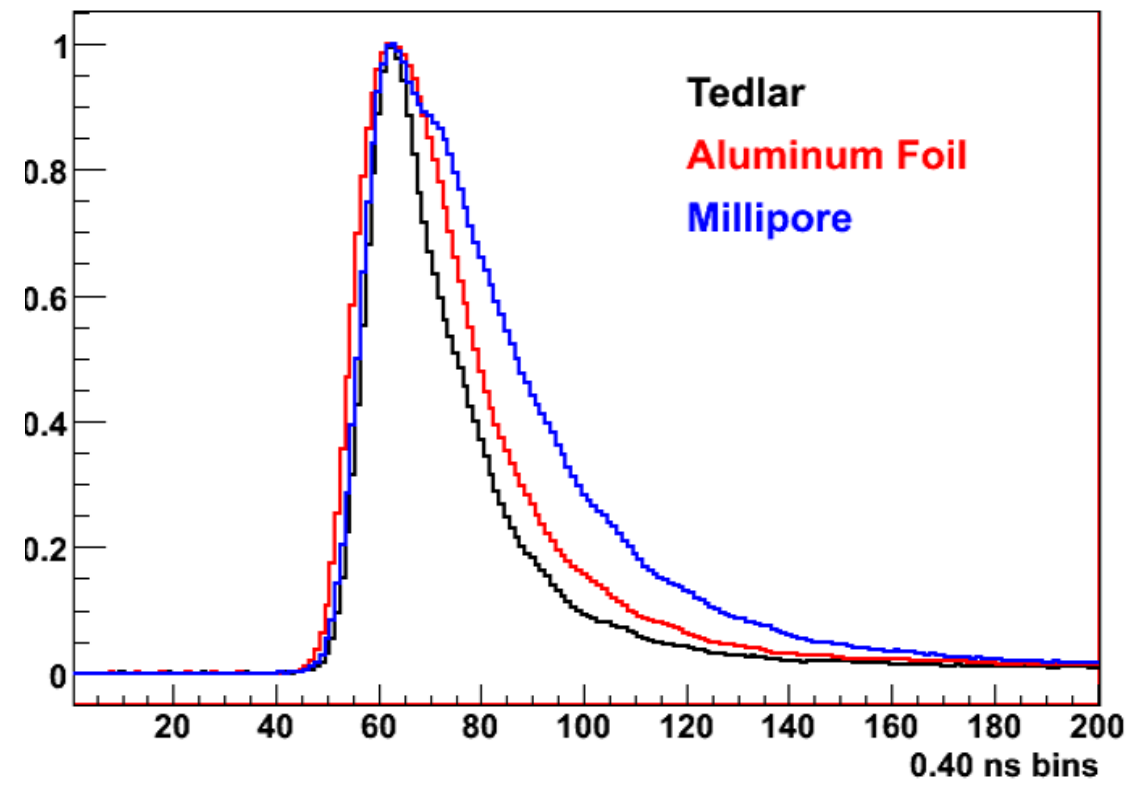

Figure 17.11: Left: Normalized response of $PbF_2$ crystal wrapped in absorptive black tedlar, reflective aluminum foil, and white Millipore paper. A 29-mm Photonis PMT was used to make these comparisons.

Detailed Geant4 ray-trace simulations and direct laboratory measurements have been used to study the light collection efficiency of the crystals subject to various wrapping schemes and couplings to the photo-sensitive readout. We have focussed on two extremes, namely an all-black Tedlar absorptive wrapping, and a diffuse reflective white Millipore paper wrapping. The black wrapping largely transmits only the direct Cherenkov light cone, while the white wrapping allows light to bounce multiple times within the crystals, eventually leading to a higher overall photon yield. Both wrappings have advantages, and we've selected the white Millipore paper as our baseline choice. For the shortest pulse occupancy time, the black wrapping excels. For the greatest light yield, the white wrapping is better. Shorter-duration pulses improve pileup rejection; higher light yield improves energy resolution. We have evaluated both wrappings and an aluminum foil wrapping in a test-beam using a standard 29-mm Photonis PMT for readout. Figure 17.11 shows the results with the amplitudes normalized. The black wrapping (labeled Tedlar in the figure) response shape reached the limit of the PMT response time.

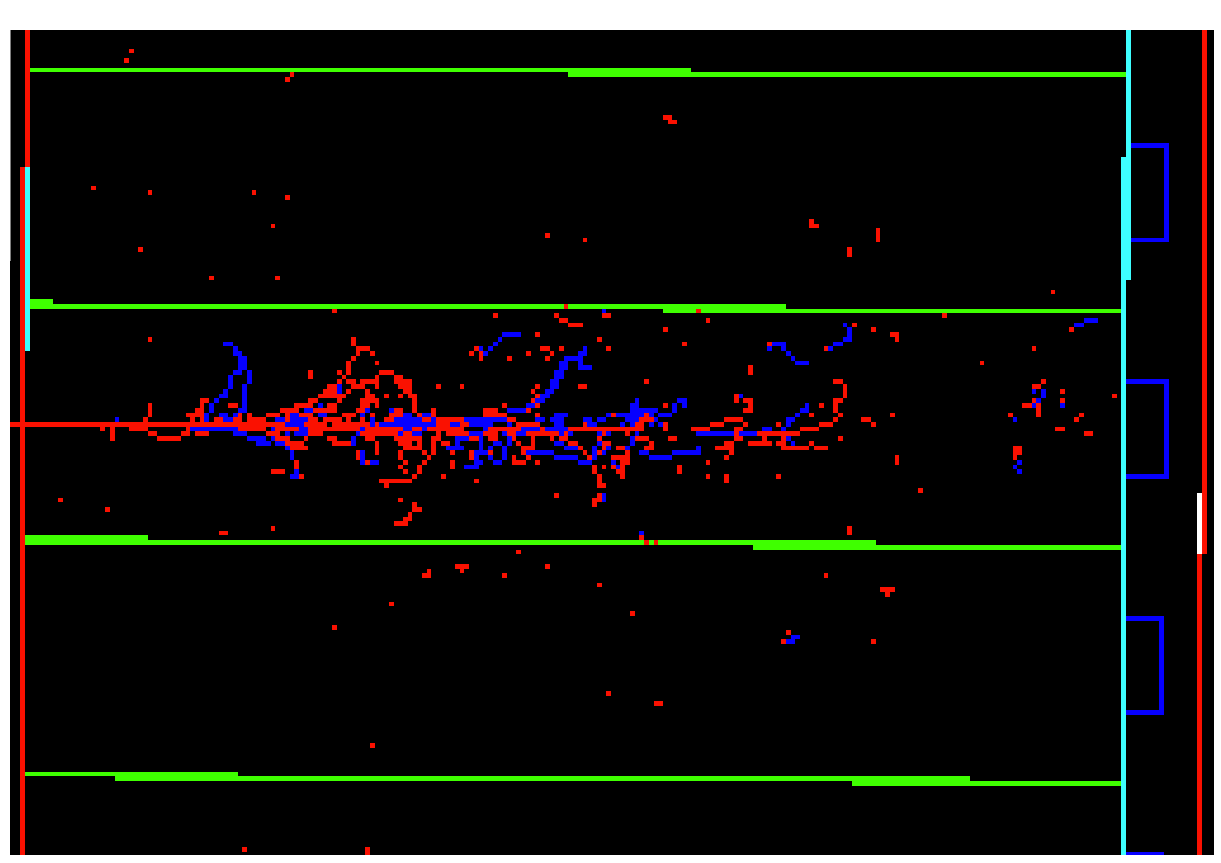

Figure 17.12: A single shower showing secondary positrons (blue) and electrons (red) in a $2.5\,\text{cm}\times2.5\ \text{cm}\times15X_0$ deep $PbF_2$ crystal, subject to a 2 GeV positron incident from the left. Photons have been removed for clarity.

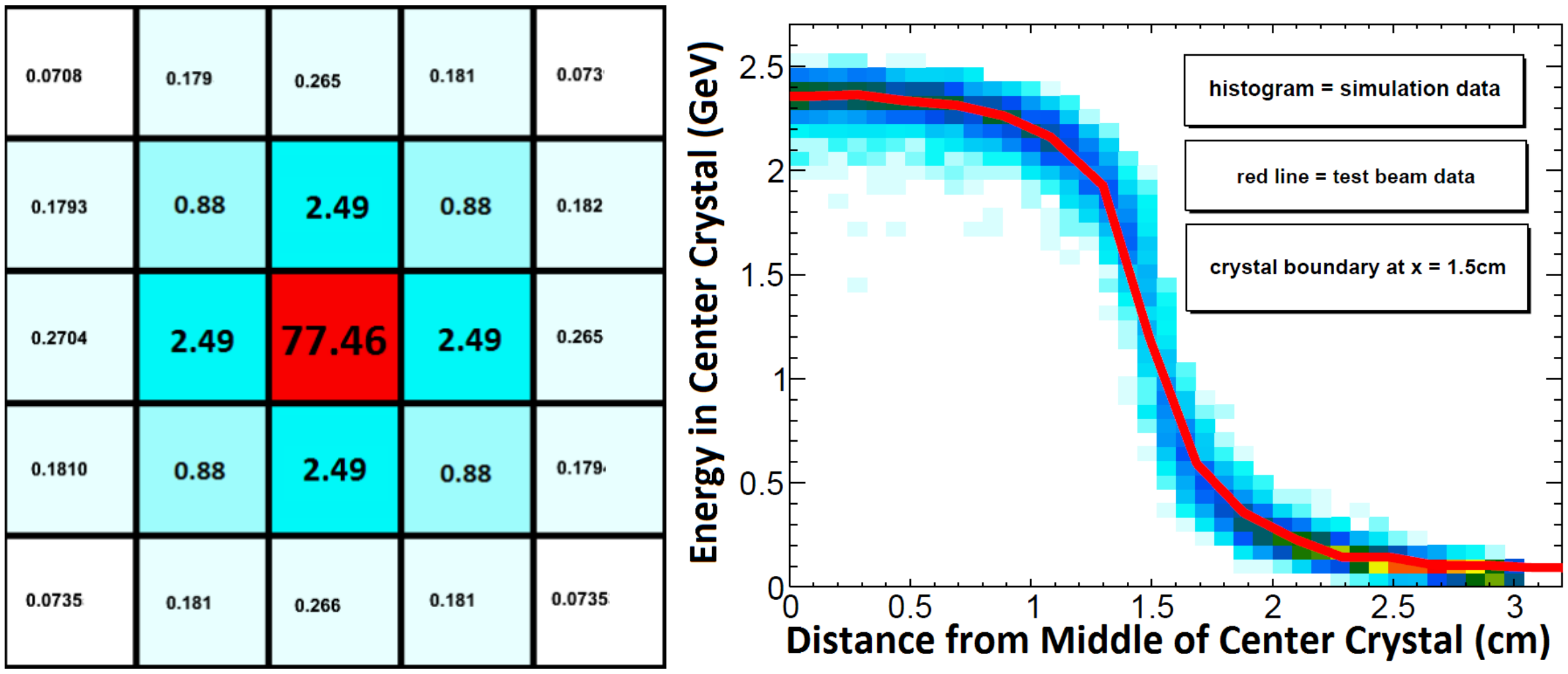

Figure 17.13: Left: Schematic representation of energy deposition in a section of the segmented electromagnetic calorimeter. Each cell is one crystal with dimensions (2.5 x 2.5 x 14 $\text{cm}^3$). The numbers represent the percentage of the kinetic energy deposited in each crystal. This data was produced from a GEANT-4 simulation with a positron incident on the center of the central crystal. The results do not change for positrons in the range of 0.5 GeV to 3 GeV. Right: A comparison of test beam data and simulation data. This plot shows the shower leakage into neighboring crystals as a function of beam incidence position.

GEANT-4 simulations were used to optimize the individual crystal size and the array matrix configuration. A visualization of a typical 2 GeV positron shower is shown in Fig. 17.12. A driving specification for an array of crystals is the reduction in pileup to be realized by spatial separation. Candidate arrays of $5 \times 7$ and $6 \times 9$ (height by width) segmentation using $3 \times 3$ cm$^2$ or $2.5 \times 2.5$ cm$^2$, $15X_0$-deep crystals, respectively, will fit the space constraints. A simulation with full showering and cluster reconstruction using a simple and robust two-shower separation algorithm was used to choose the best arrangement. Not surprisingly, the higher-granularity array is best. We find that it will provide at least a 3-fold reduction in pileup compared to a monolithic design. These conclusions were arrived at from a combination of simulation and direct measurement at the FTBF.

Energy sharing among neighbor crystals is shown in Fig. 17.13 for a shower that strikes the center of the middle crystal. The simulation is calibrated against the test-beam measurements in which an electron beam was directed into a crystal at various known positions and the ratio of neighboring crystal responses was recorded. Fig. 17.13 also shows a histogram of data vs. the simulation prediction. The agreement is excellent and verifies the model used

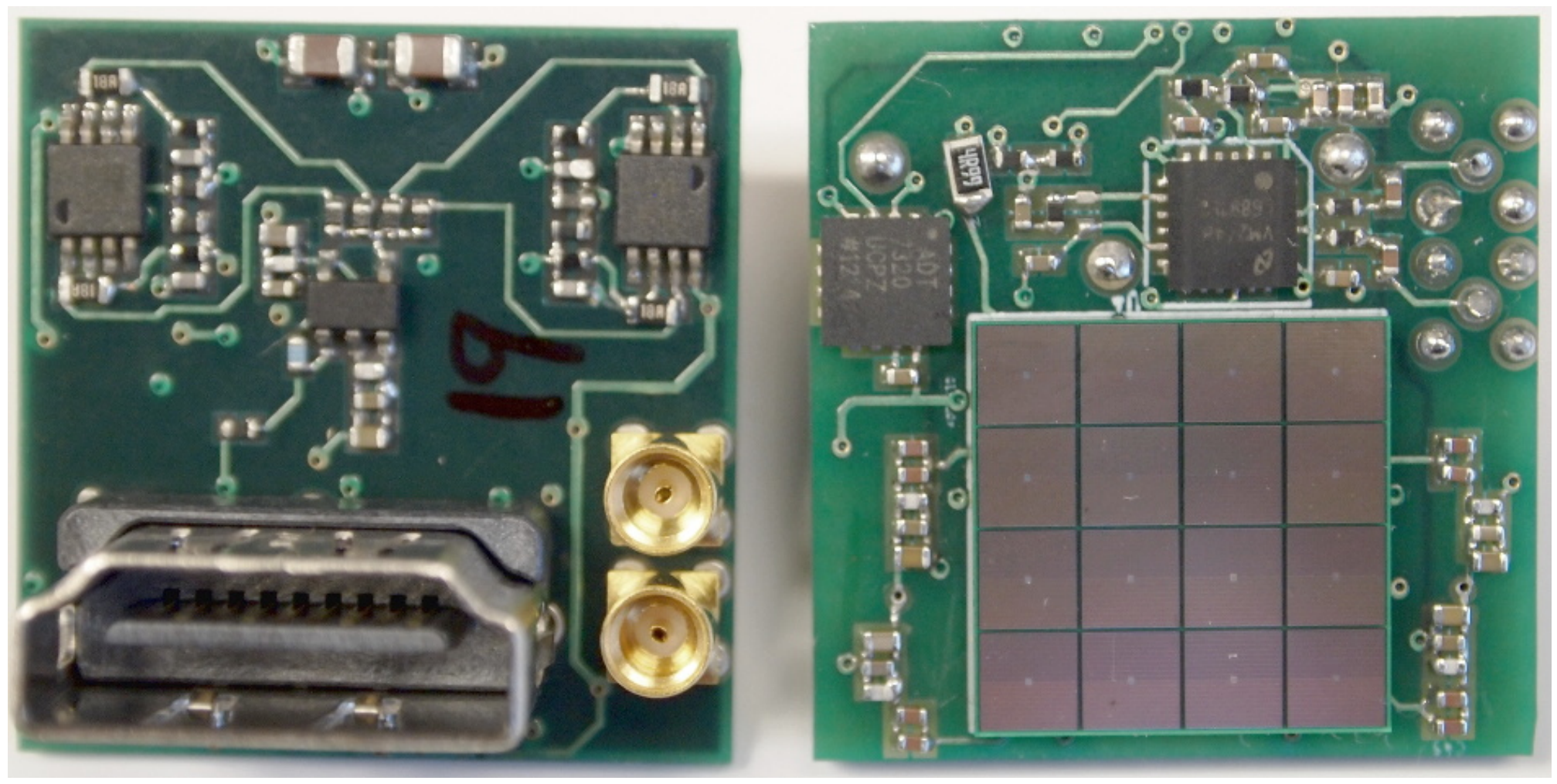

Figure 17.14: A baseline prototype of surface-mount 16-channel SiPM soldered on the amplifier board. The two MMCX connectors represent the AC coupled differential voltage signal out. The common bias voltage in, the board low voltage, and SPI lines to regulate gain are supplied through the HDMI connector.

to optimize array size and to evaluate pileup by shower separation.

### 17.4.2 Photodetection Subsystem – SiPM

In the baseline design, silicon photomultipliers (SiPMs) read out the crystals. While challenging and relatively new devices, they are increasingly preferred over traditional PMTs in many nuclear and particle physics applications. As such, the body of experience in their use is growing rapidly and the variety of SiPM devices from many manufacturers is increasing. They work as pixelated Geiger-mode counters. The default SiPM we are considering has 57,600 50-$\mu$m-pitch pixels on a $1.2 \times 1.2\,\mathrm{cm}^2$ device. When a photon strikes a pixel, it can cause an avalanche that is summed together with the other struck pixels in a linear fashion to produce the overall response. Quenching resistors are intrinsic to the device to arrest the avalanche and allow the device to recover. The recovery time constant of a fired pixel is typically 10's of ns. Those pixels that are not struck, meanwhile, remain ready for a next pulse. In general, the concept is to have a pixel count that greatly exceeds the highest photon count that would strike the device. For example, for our crystals, a working number is 1.5 PE/MeV (for SiPM devices PE represents a converted photon). With a range of up to 3.1 GeV for single events, the occupancy fraction remains no more than about 5 %, which is in a near-linear regime and allows for a good measurement of any closely trailing second pulse.

The selection of SiPMs over PMTs is pragmatic. SiPMs can be placed inside the storage ring fringe field, thus avoiding the awkward, long lightguides that would be needed for remote PMTs. They have high photo-detection efficiency, they will not perturb the storage ring field, and they can be mounted directly on the rear face of the $PbF_2$ crystals. Large-area SiPM arrays are cheaper than same-size PMTs, their cost is falling, and their performance characteristics continue to improve. We have spent the last 2 years developing lab tests to evaluate these devices. The collaboration has designed and built a series of custom pre-amplifier and summing amplifier boards. The most recent version features a multi-staged transimpedance amplifier with a remotely controllable, variable gain output state amplifier (LMH6881).

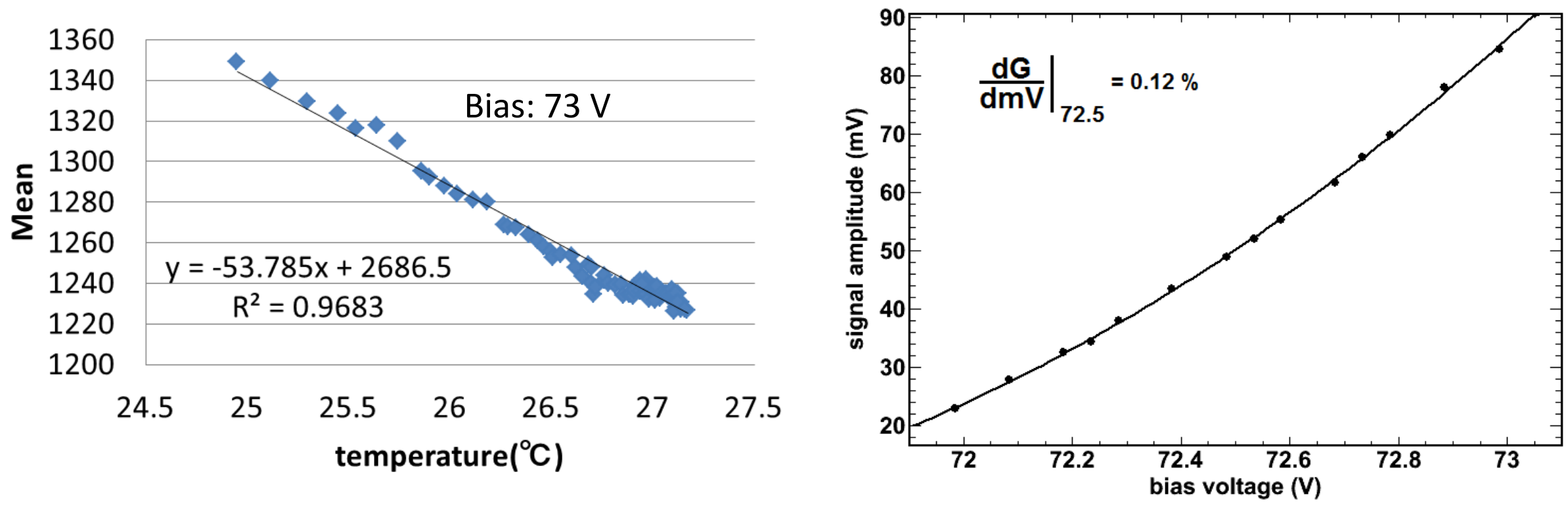

Figure 17.15: Left: Gain vs. temperature dependence of the previous generation 16-channel SiPM by Hamamatsu. Right: Gain vs. bias voltage. The measured change in gain is 0.12 % per mV.

Large-area SiPM devices are packaged as arrays of smaller individual channels. While the market is constantly evolving, we are presently using a Hamamatsu surface-mount 16-channel SiPM having 57,600 50-$\mu$m pixels in a $1.2 \times 1.2\,\mathrm{cm}^2$ active area. It is reasonably well-matched to the surface $2.5 \times 2.5\,\mathrm{cm}^2$ crystal face. Model number S12642-0404PA-50 forms our baseline design. It features Ni-based quenching resistor, through-silicon vias to avoid parasitic inductances in series with output current pulse, optical trenches to minimize cross-talk, and it is made of high-purity Si wafers to minimize dark current, after-pulsing, and to increase charge delivered by a fired pixel. Figure 17.14 shows this SiPM mounted on a prototype amplifier board.

One of the challenges of using SiPMs is their particular sensitivity to temperatures. Figure 17.15 (left panel) shows the gain change of our SiPM device vs. temperature. The gain change is 4 % per °C. This measurement was performed with an older product built with a poly-crystalline Si quenching resistor. Our recommended SiPM uses a Ni-based resistor and its temperature dependance is reduced[1]. The exact temperature dependence of the baseline SiPM will be evaluated during the next test run at SLAC. The calorimeter design is prepared to handle the temperature dependance currently observed. While short-term shifts are unexpected, the overall SiPM environment must be maintained at a fairly constant temperature in order to simplify the global calibration of gain during the running period.

The response of a SiPM is also quite sensitive to the bias voltage stability above Geiger-mode breakdown threshold. The right panel of Fig. 17.15 shows a lab measurement of a SiPM voltage amplifier board output vs. bias voltage. The slope is gain is 0.12 % per mV near the working bias of 72.5 V, leading to the need to have a separate bias control subsystem, which we describe in Subsection 17.4.2.

[1]The Ni fraction is tiny; we evaluated the magnetic perturbation of this SiPM in our test magnet and the effect of Ni is negligible.

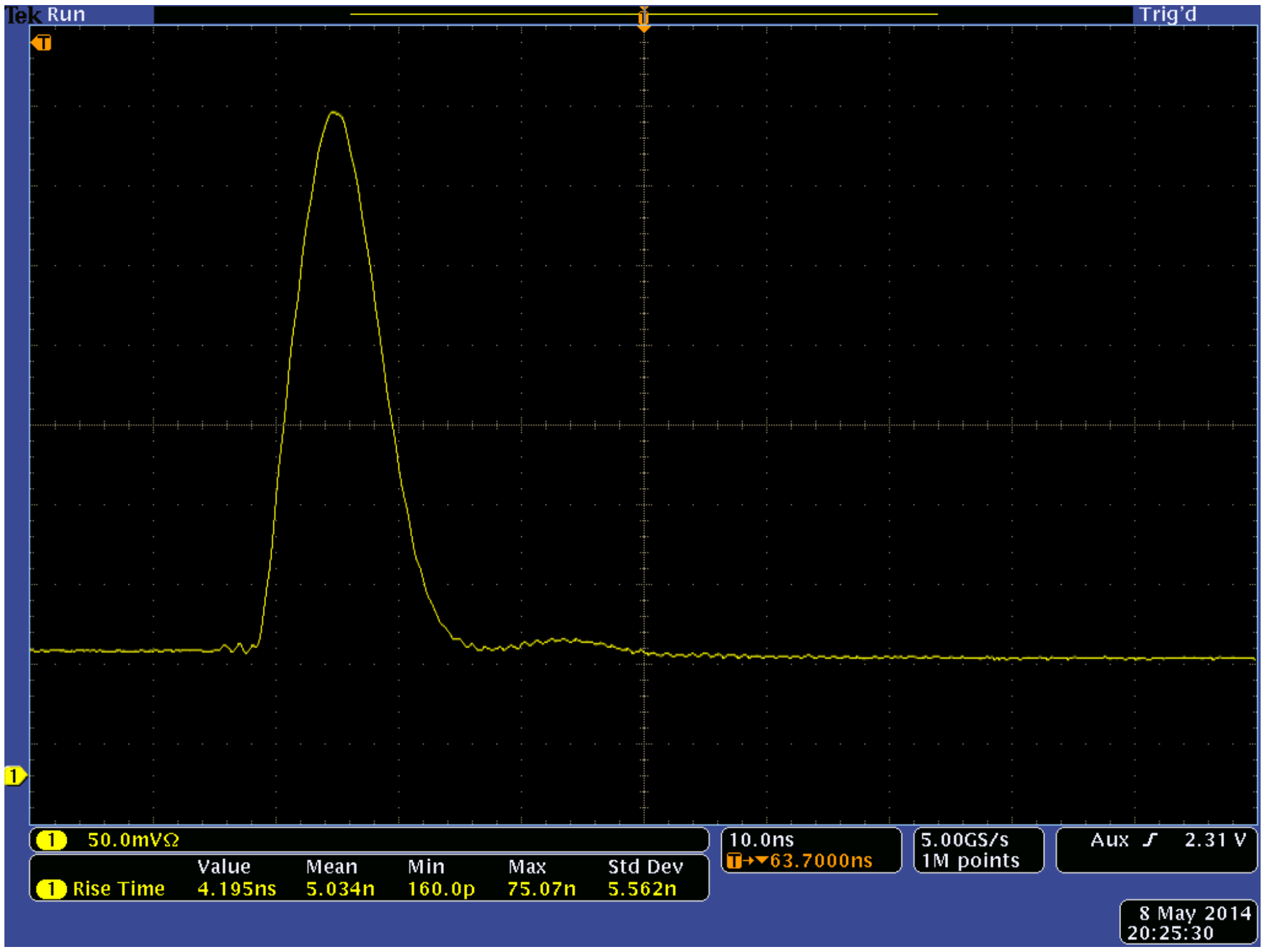

Figure 17.16: A scope trace of the SiPM response to a 2.5 ns wide LED pulse. For the purpose of the measurement, a balun transformer was used to convert the differential output into a single ended one.

The most recent amplifier board (Fig. 17.14) is based on a concept of a multi-staged, transimpedance op-amp. In the first step, current pulses from 4 SiPM channels are added together and converted into voltage pulses in a fixed-gain transimpedance amplifier which is operated at a constant gain of 600 Ω. This first stage is designed around a THS32302 op-amp, which is operated in current mode. Avoiding any shunt resistors, extremely low input impedance of the op-amp, and 2 GHz bandwidth are keys to short pulse widths. In the next step, the four partial sums are added together using a THS3201 op-amp operated at unity gain in voltage mode. The final stage drives an AC coupled differential pair of coax cables, and is designed around a LMH6881 digitally controlled variable gain amplifier. A 2 Vpp amplitude at the output of the amplifier board corresponds to energy deposited in the crystal by a 4 GeV positron. A representative pulse shape response to a 2.5 ns-wide LED pulse is shown in Fig. 17.16. The pulse FWHM here is $\sim 8$ ns.

Given this working electronics board, measurements were carried using our fast laser system (pulse width $< 1$ ns) to determine the two-pulse separation and the sensitivity to high rate. A scope shot of two such pulses, separated by 8 nsec is given in the right panel of Fig. 17.17. These pulses can be easily resolved in a fit. The left panel demonstrates that the SiPM board is capable of operating well at a rate of 10 MHz. These and other test to date show that the current design meets the demands of the experiment.

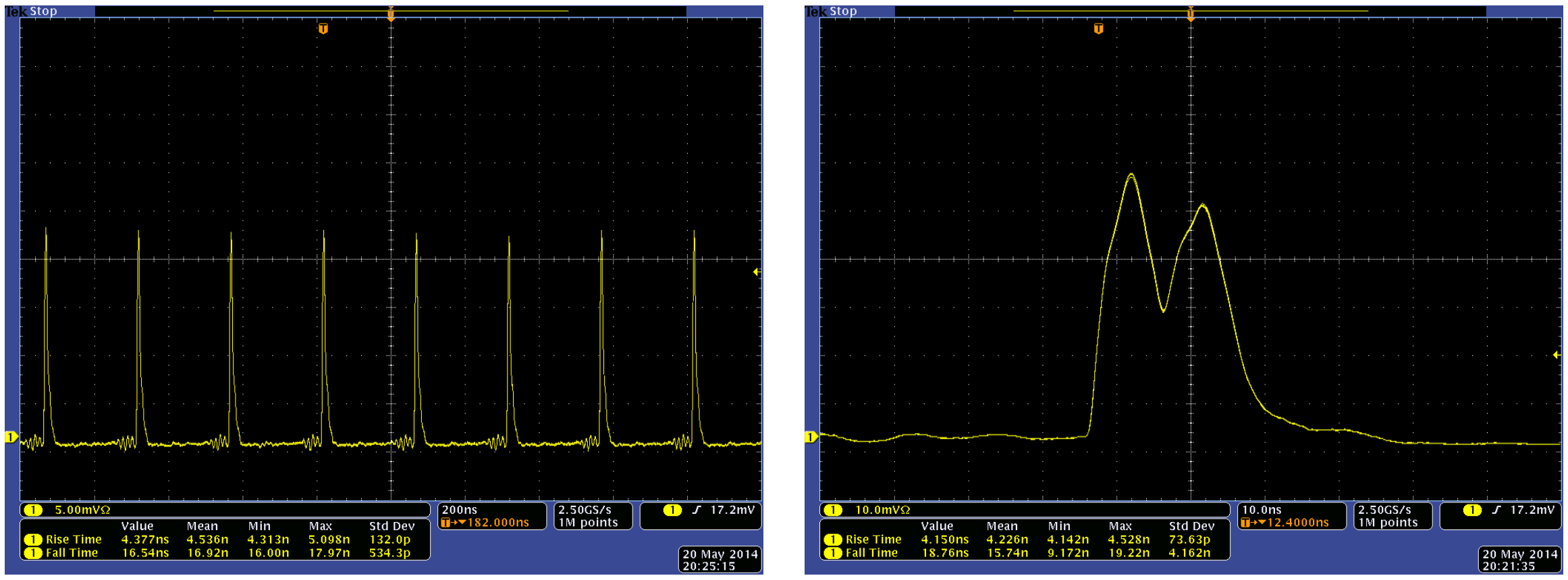

Figure 17.17: Left: A baseline SiPM prototype is stable running under 10 MHz laser shots. Right: An example what a SiPM response to a 8 ns pileup looks like.

### Bias voltage system for the SiPMs

Several possible approaches and technical implementations have been considered for the SiPM Bias Voltage System (BVS). In the initial stages of the E989 proposal and instrumentation development, the only viable option was a dedicated design and fabrication of modules specific to this experiment. None of the commercially available units at the time was able to produce the combination of specifications thought to be required for the E989 Calorimeter SiPMs: 1 mV or better precision of setting and stability of maintaining bias voltages in the 60 − 80 V region, while maintaining a time-average current of $\sim 50\,\mu$A for each channel, with significantly higher instantaneous peak currents associated with storage ring filling times. Each channel was expected to require its own distinct value of bias voltage, accurately maintained over time.

Two prototype designs were developed at the University of Virginia, and four 16-channel BVS boards were built and instrumented in a dedicated chassis jointly with the group from James Madison University. The prototype systems were operational in June 2014, and were tested in-beam during a test beam time at SLAC in July 2014. Sadly, both designs turned out to suffer from instabilities and bias voltage variation with load far outside the specified limits, requiring a thorough redesign.

At the same time, evolution in the design and production of the Geiger-mode avalanche photodiodes have resulted in significant changes in requirements for E989 Calorimeter BVS. The main result of these developments is that more recently produced batches of SiPMs cluster far more narrowly in gain and bias voltage requirements than the older units did.

On the other hand, in-beam as well as laboratory laser tests and evaluations of recent generation SiPMs have yielded wider plateaus of acceptable gain vs. bias voltage, relaxing the narrow requirement for bias voltage compared to the initial specifications.

Thirdly, in 2015 manufacturers iseg/Wiener and CAEN introduced several new high precision high voltage modules with increased stability, as well as high delivered current, 10 mA or more per channel. These modules make it feasible to design a central calorimeter bias voltage system in which typically some 12 channels share the same bias voltage. This drives down channel multiplicity and cost by an order of magnitude. The location of the system in the center of the storage ring further reduces the risk that the BVS would affect the storage ring magnetic field, thus removing the need for extreme control of magnetic properties of each component in the system.

At the time of this writing we are testing three modules: iseg EHS 8401p F, Wiener MPV8120LI, and CAEN A1539 in their respective MPOD and CAEN chassis enclosures. They were delivered to the University of Virginia in the last week of May (iseg/Wiener) and April (CAEN). All are capable of delivering 10 mA per channel, or more, and thus provide a large current reserve even if each channel is split among as many as 12 Calorimeter modules. Following basic bench-top tests, the units will be tested with an array of calorimeter crystals and SiPMs, along with the E989 laser pulsing system.

Initial tests indicate that all units maintain stability at the sub-mV level, which provides a solid foundation for a BVS system that meets all E989 requirements. The iseg/Wiener units have the advantage of having individual isolated grounds for each channel, an essential feature to prevent the occurrence of ground loop noise. The realistic tests with calorimeter module array(s) will determine what modifications, if any, are needed on the breakout boards in order to ensure that they provide the charge locally to each channel on the $\mu$s time scale, to accommodate the transient current draw associated with storage ring filling times.

### 17.4.3 Laser Calibration System

A high performance calibration system is required for the on-line monitoring of the output stability of each individual tower in all calorimeter stations. It is estimated that the detector response must be calibrated with relative accuracy at sub-per mil level to achieve the goal of the E989 experiment to keep systematics contributions to the accuracy on the measured observables at 0.02 ppm level. This is a challenge for the design of the calibration system because the desired accuracy is at least one order of magnitude higher than that of all other existing, or adopted in the past, calibration systems for calorimetry in particle physics.

Almost 1,300 channels must be calibrated during data taking; the proposed solution is based on the method of sending simultaneous light calibration pulses onto the readout photo-detector through the active sections (crystals) of the calorimeter. Light pulses must be stable in intensity and timing to correct for systematic effects due to drifts in the response of the crystal readout devices. A suitable photo-detector system must be included in the calibration architecture to monitor any fluctuation in time of the light source intensity and beam pointing as well as any fluctuation of the transmitted light along the optical path of the light distribution system, which could occur due to mechanical vibrations or optics aging.

Some guidelines are defined to select the light source(s) and to design the geometry of the light distribution and monitoring; the following criteria are adopted to select the light source type:

- light wavelength must be in the spectral range accepted by the detector and determined by the convolution of the spectral density of the Cherenkov signal produced by electrons in $PbF_2$ crystals with the spectral transmission of the crystals, and with the spectral Q.E. of the photo-detector; Q.E. is peaked around 420 nm for SiPMs.

- the luminous energy of the calibration pulses must be in the range of the electron deposit in the crystals, typically 1-2 GeV; this corresponds to a luminous energy on each tower of a calorimeter station of about 0.01 pJ, or to about 13 pJ for simultaneous excitation of all calorimeter readout channels (1300). The numbers quoted above merely indicative of the order of magnitude and they are derived by assuming that the readout of each crystal will produce about 1.5 photo-electrons per MeV with 30% P.D.E. (Particle Detection Efficiency) for SiPMs and with 23% coverage of the crystal readout face.

- the pulse shape and time width must be suitable to infer on the readout capability in pile-up event discrimination; pulse rise/trailing time must be of the order of some hundred of picoseconds, the total pulse width should not exceed 1 ns. This implies a peak power per pulse at the source of some Watts (some nJ in a 1 ns wide pulse), assuming the conservative value $0.001 < T < 0.01$ for the total intensity transmission factor of the calibration system.

- the pulse repetition rate must be of the order of 10 KHz; this value will be tuned to obtain the best compromise between the need of having enough calibration statistics in the time interval (some tens of microseconds after the muon injection in the ring) when the maximum rate is achieved in the readout devices and the need to avoid saturation of the DAQ bandwidth.

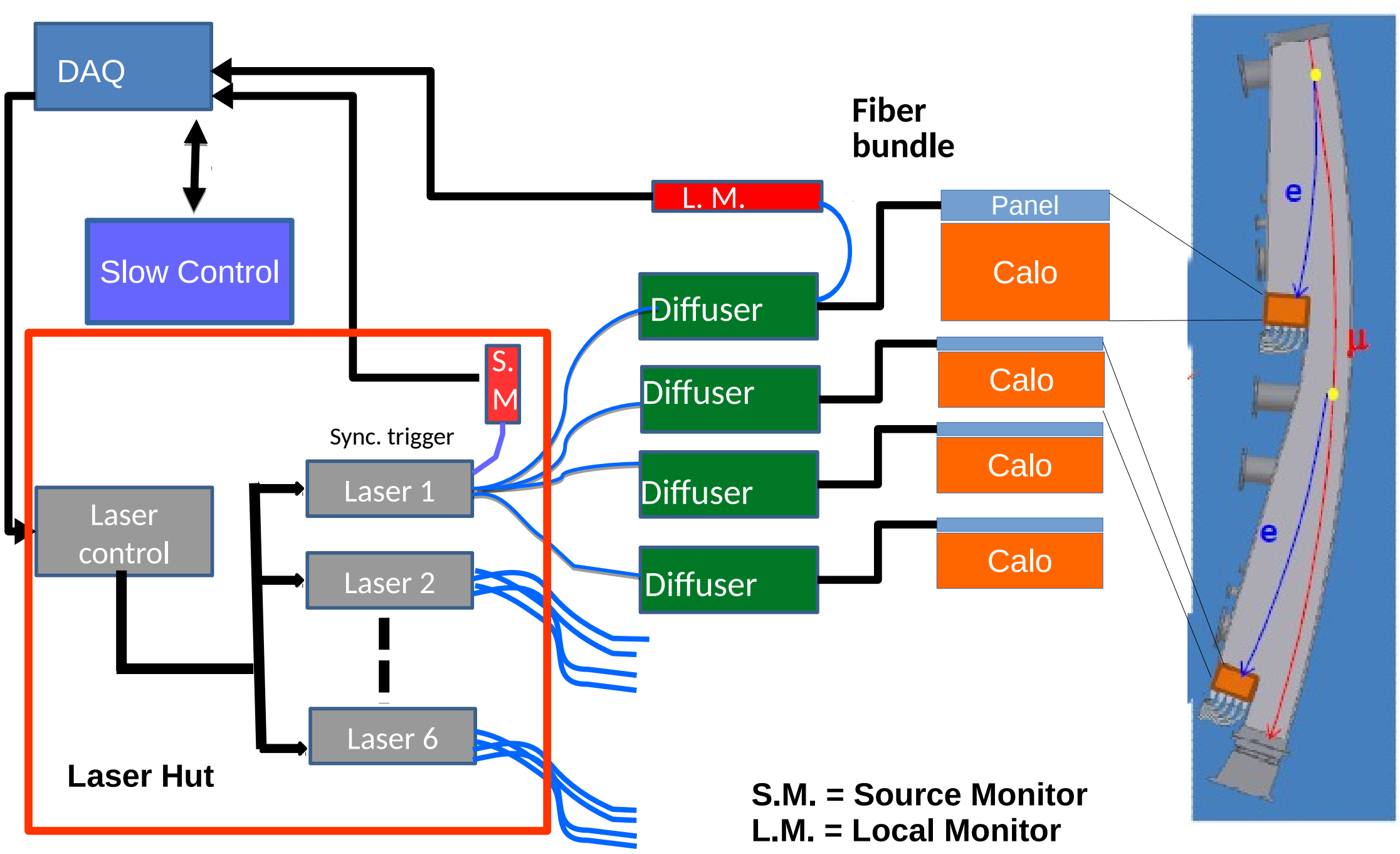

Figure 17.18: Sketch of the laser calibration system of the calorimeter.

A number of commercial diode lasers, for example LDH series from Picoquant, bare diodes from Roithner LaserTechnik, Picopower-LD series by Alphalas or others, cope with the criteria listed above and have been considered as a source for the calibration pulses. The final choice will be made after the completion of all tests required to qualify, in terms of light transmission and time stability, all other optical elements of the calibration system. Guidelines for designing the light distribution chain are listed below:

- High sensitivity monitors of the transmitted light at the end-point of each individual section of the distribution chain must be used to ensure online control of the system stability and to have information for applying feed-back correction to the source operation parameters, if needed.

- The optical path must be minimized in order to limit the light loss due to self-absorption in the optical fibers. The number of cascade distribution points must be also minimized to reduce the unavoidable light loss in the couplers between different sections.

- The laser source and its control electronics should be located outside the muon ring in order to avoid e.m. perturbations of the local field induced by the current flow used to excite the laser. Consequently, a suitable geometry should include: 1) a primary light distribution point outside the ring, 2) a number of silica launching fibers about 20 meter long, used to feed light from the primary distribution point to the secondary ones, and 3) secondary distribution points located close to each calorimeter station. From the secondary distribution points, short bundles of plastic (PMMA) fibers, about 1 meter long, feed the light through a panel to each individual tower. of the calorimeter station.

- Optical fiber selection: multimode silica fibers (20 dB/Km attenuation at 400 nm) are the best solution for long path light transmission and in terms of robustness against solarization or other aging affects due to large values of transmitted light intensity. For the shorter fiber bundles, where the transmitted intensity is at least one order of magnitude lower and the distance to be covered very short, PMMA fibers (200 dB/Km attenuation at 400 nm) are considered, also for budget reasons.

A possible geometry fulfilling all the requirements set by the guidelines listed above is shown in Figure 17.18. The light generated by the laser source (a pulsed diode laser) is divided in 4 or more parts, coupled into the launching fibers and sent to the secondary distribution points located inside the ring and near each calorimeter station. A battery of 6 synchronized lasers will be necessary to provide the required light power. They will be located, together with the optics and the fiber couplers, onto an optical table inside a laser hut with temperature and dust controls. The quartz launching fibers (about 20 meter long, one per calorimeter station) route the light to secondary distribution devices located near the calorimeter stations, each distributor serving one station. A small fraction of the light exiting each laser source and each light distributors is routed to source and local monitors respectively. Their analog signal is returned to the DAQ system for both on-line checking of the system stability and futher offline monitoring of the calibration signal. Interface with DAQ is also required for slow control signal recording and communication with the timing signal controls is used to trigger the electronics of the laser driver. The monitors are designed to factor out instabilities in the laser system: assuming the laser-induced signals from all calorimeter elements ($Sci$) and the monitor ($SM$) are subject to the same laser fluctuations, the fluctuations should be eliminated in the ratio $Sci/SM$ . These ratios should then reveal fluctuations in the calorimeter response, provided other fluctuations in $SM$ are stable to the degree required.

The following approach is considered to optimize the monitor stability:

- We use zero gain PIN diodes which are much more stable than SiPMs to variations in bias and temperature;
- We expose them to much higher light levels to minimize photostatistical fluctuations;
- We equip the monitor with electronics specifically designed to get high stability. This electronic board is currently under construction;
- We use a redundant system, with two photodetector for each monitor;
- We minimize pointing fluctuations by incorporating diffusion and mixing elements;
- We incorporate a radioactive source, for absolute calibration.

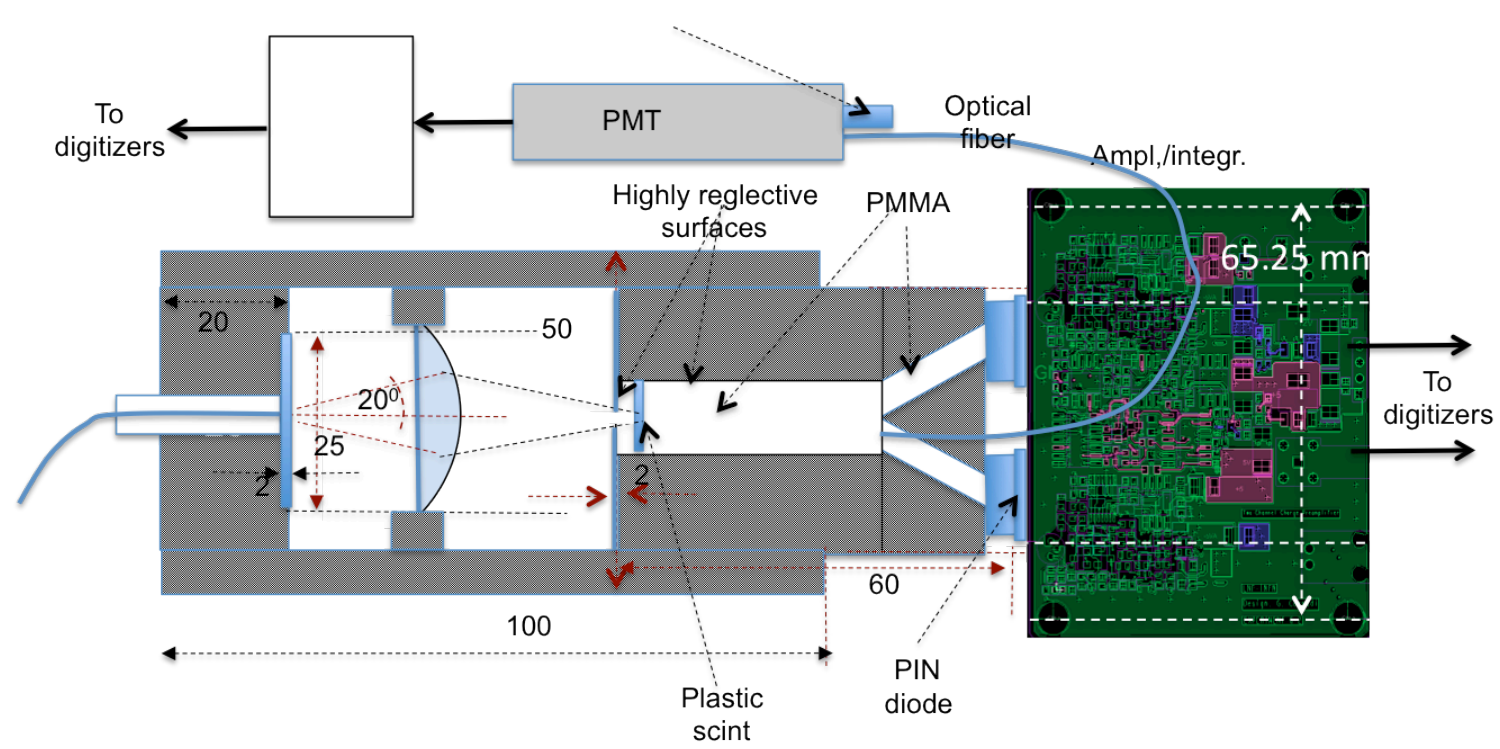

Figure 17.19: Schematic illustration of the monitor prototype.

Based on these consideration, a monitor prototype (whose geometry is sketched in Figure 17.19), has been developed.

The laser light is injected by a fiber from the left and is incident on a diffuser engineered to emit in a $20^o$ angle. The diffused beam is focused onto fast plastic scintillator disk located near the entrance to a diffusing chamber and faced up against a PMMA mixer. The fast scintillator acts as an ideal diffuser because it emits isotropically. The cavity filled with PMMA acts as a mixing chamber. This combination should be effective in minimizing beam-pointing fluctuations which could be produced by fluctuations in the emittance of the laser. The photodetectors include two large-area PIN diodes for redundancy. The PIN diodes are relatively slow ($\sim$ 10 ns pulse width) compared to the laser pulses but fast enough given that they will be integrated by their electronics for stability. The third photodetector is a PM with a 8 mm diameter cathode and a Cockroft-Walton base which generates its own HV and operates at low voltage. Laser fluctuations are eliminated in the ratios of these three signals which are used monitor other eventual instabilities in the individual signals. PM stability is monitored independently by exposure to the signal from a NaI crystal on which a very small quantity ($\sim$ 10 counts/sec) of $^{241}$Am has been deposited. Since the PM is also sensitive to the laser signals it serves as an absolute reference. The prototype of the preamplifier/integrator card for the PIN diode signals, currently under development, is also shown in Figure 17.19.

Qualification tests of the individual components, including comparison measurements of different options, have been extensively performed:

- laser source: as an alternative to a single, powerful laser light source, the possibility to use four or six synchronized laser heads with lower power (diode lasers), controlled by the same driver has been chosen. This solution has the advantage that, in case of one laser failure, no calibration stop will occur during data acquisition. Moreover, pulsed diode lasers have excellent stability characteristics and versatility in driving rates.

- light distributors: baseline solution uses simple beam splitters and coupling into multimode launching fibers. This solution should guarantee intensity stability of the distributed light against geometrical effects due to beam-pointing instabilities. Beam-splitters made with the linear circuit technology could also be considered as primary distributors if commercial devices, nowadays widely used only in the IR range for telecommunication, will be produced for the near-UV/visible range. A prototype will be tested soon. Concerning secondary distributors, extensive tests have been performed to compare an integration sphere versus an engineered diffuser. The former shows a higher degree of output uniformity (up to 99.3 percent for a 1 mm diameter fiber connected to one sphere port) at the price of a higher factor in intensity loss. The latter shows a satisfactory degree of uniformity (see Figure 17.20) with a much higher transmission factor and has been retained for the baseline solution of the light distribution system. The comparison between the integrating sphere and the engineered diffuser has been published in Ref. [7].

The complete chain of the baseline solution of the distribution system, from the laser head to the single calorimeter tower, has been tested experimentally by sending laser signals to the rear faces of the $PbF_2$ crystals, where the SiPM photodetectors are placed. Here, one important requirement for the laser calibration pulses, beyond stability, is uniformity. In our baseline system, the light from the 54 plastic fibers of the bundle is reflected inside a panel placed in front of the calorimeter (see Fig. 17.18) and sent into each tower. The uniformity of the illumination of the rear facet of the $PbF_2$ crystal is shown in Fig. 17.21, where a continuous wave laser at 532 nm has been used. The obtained uniformity is enough to ensure a comparable illumination of all SiPMs in the calorimeter station, which have a surface covering only 23% of the crystal faces and, due to mounting constraints, are not placed in the same position with respect to the crystal axis.

### 17.4.4 Mechanical Subsystem

Each calorimeter station will comprise a number of individual crystals that are made out of dense, usually brittle, material. In addition, the detector including the photo-sensitive SiPM must be in a light tight encapsulation. A set of crystals and photo-detectors forming one calorimeter station weighs $\sim$ 40 kg. The housing must provide the light tightness, proper stability to carry the weight, feedthrough for bias and low voltage, and a mechanism for easy lifting of the entire box and insertion into or out of the ring in the radial direction.

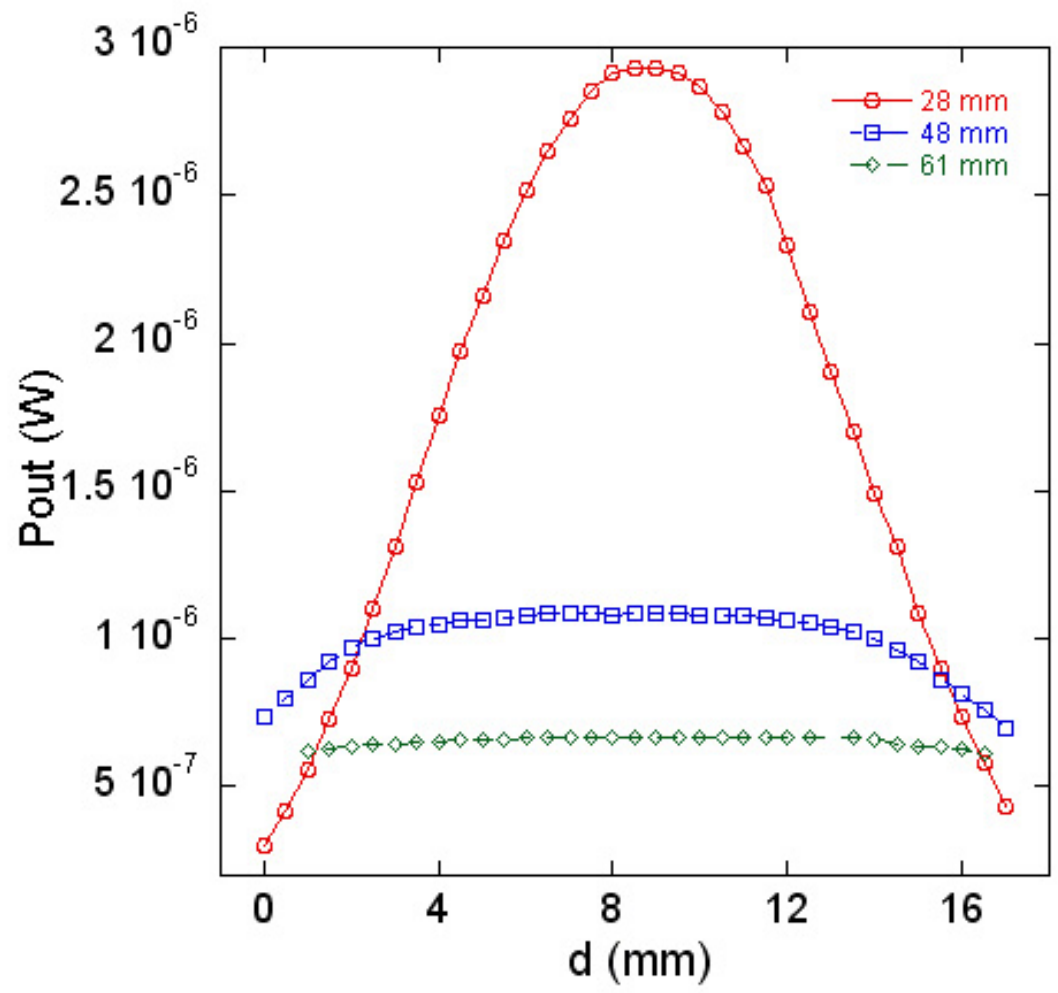

Figure 17.20: Mapping of the illumination uniformity after the 20 degrees diffuser. The light is collected with a 1 mm diameter, 0.49 N.A. optical fiber, at different distances from the diffuser. The input power on the diffuser is 1.3 mW.

The locations of the 24 calorimeters in the experiment are fixed by the design of the scallop-shaped vacuum chambers. Several vacuum ports, bellows and the magnet's pole gap impose spatial limitations on the design. Specifically, the length of the calorimeter station along the positron's trajectory cannot exceed 38 cm. The pole gap limits each station to less than 17 cm. There are no tight limiting factors in the radial dimension that we can see would constrain the calorimeter station.

For installation, maintenance and access to the vacuum chamber or the magnet, each calorimeter station must be easily removable. We will determine later the degree of alignment necessary, but a system of pins should allow for a reproducible position. The absolute position is less critical. The calorimeter housing and retractable platform must allow for routing of a variety of cables (detector signals, bias voltage, control signals, monitoring signals) and service lines (e.g. cooled air pipes). The mechanical design must incorporate space for the readout electronics crates to be placed on-board to be compatible with the moving mechanism.

Heat dissipated by a SiPM and its amplifier board is removed by dried cool air. The maximum heat power under a steady very high rate hypothesis (very long high rate laser run) could reach up to 1.2 W per SiPM board. The cooling system is design with a power limit of 2 W in mind, and was successfully commissioned during a SLAC run. The system consists of a small cooler serving as a source of cooling power for pressurized air dried in a large desiccant containter. A distribution system delivering a steady flow of cool air directly to the center of a SiPM board forms the back wall of the calorimeter box.

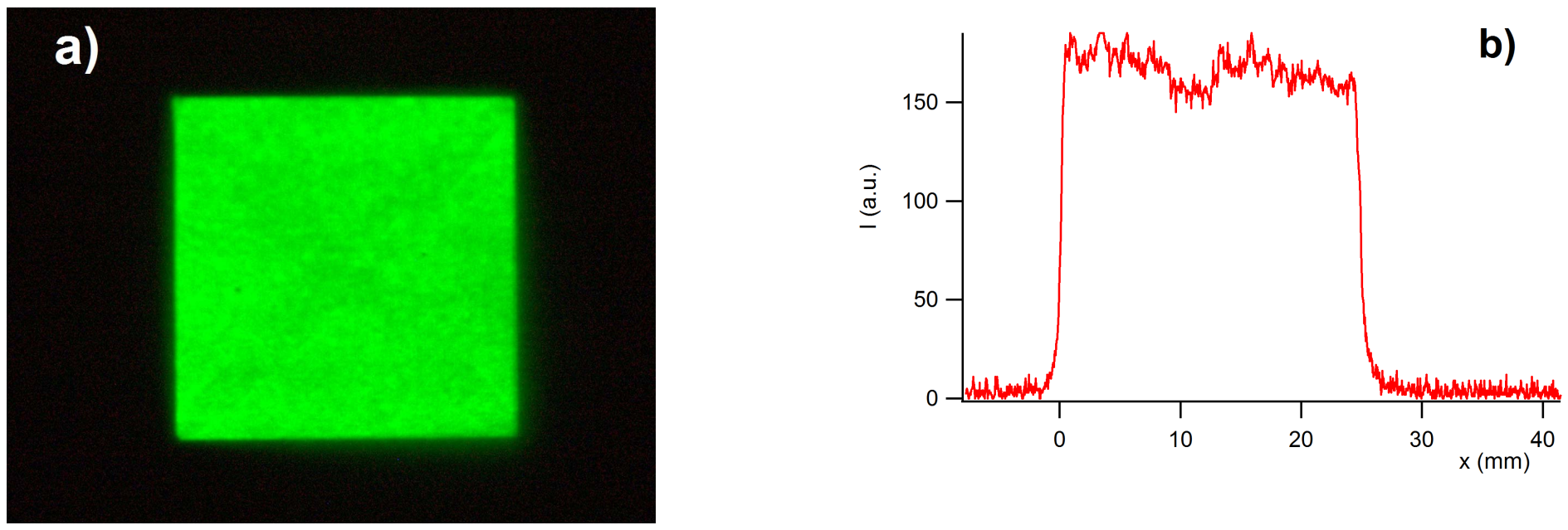

Figure 17.21: In a) image of the laser light exiting the rear facet of the crystal, diffused by a paper sheet. In b) Horizontal cut of the image in a).

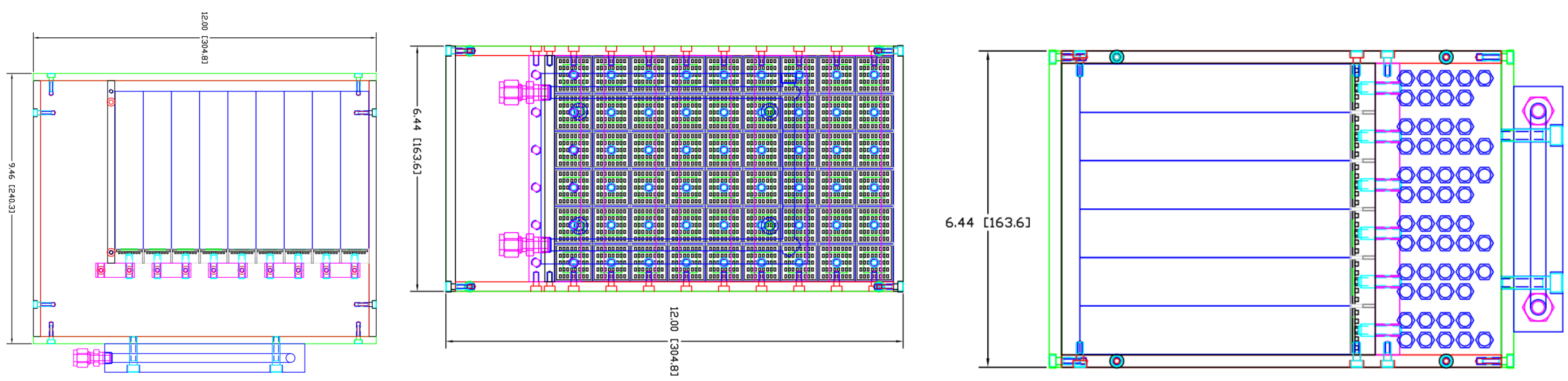

Figure 17.22: Proposed calorimeter light-tight housing. Left: Top view of 9 crystal columns. The storage ring is on the right. The extra space on the left side is used for panels, cooling, servicing. The rear part of the box includes a cooling channel. The front accommodates the calibration plate (not shown here). Middle: Side view of 6 rows of crystals. The conceptual plan for the chicanes where signal cables pass through is indicated. Right: Rear view showing the 54 SiPM amplifier summing boards.

A mechanical housing system was built for the test beam and a full-scale version has been designed by CENPA engineers and costed for the experiment. It ensures a light-tight environment, provides cooling as necessary for stable SiPM board operation, includes patch panels and chicanes for cable runs and has a front-end that will mate to the calibration interface plate described in fig. 17.23. The housing has serviceable doors that will allow easy access to the crystals and readout devices. Figure 17.22 shows three engineering drawings of the proposed system. The front panel of the light-tight housing holds the calibration plate. Figure 17.23 shows our current design in perspective.

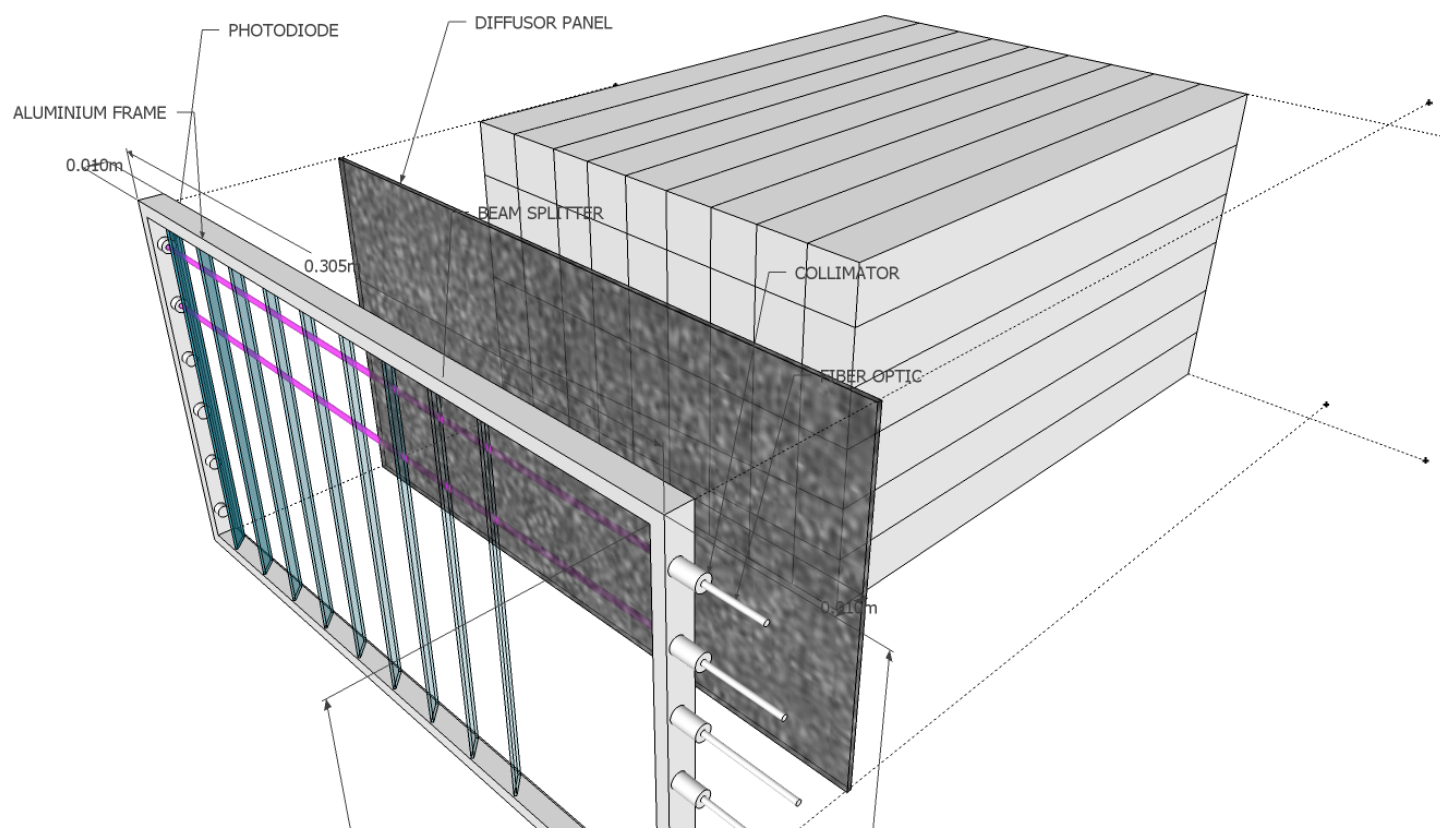

Figure 17.23: Proposed front calibration plate. Six optical fibers penetrate from the inner radial direction. The light is split using a series of beam splitter plates. An alternative desing based on a diffusor panel that spreads the light across the crystal front faces is obsolete now.

## 17.5 Alternative Design Considerations

Two alternative calorimeter material options and one alternative readout option were tested. These included a home-built tungsten-scintillating fiber sampling calorimeter, which is dense ($X_0 = 0.7$ cm), and has a fast-scintillator signal response [2]. Unfortunately, it did not exhibit acceptable resolution in the as-built W:SciFi 50:50 ratio and necessary modifications would reduce the density. Next we tested a custom undoped lead tungstate ($PbWO_4$) crystal. The idea was to reap the benefits of the higher light yield of $PbWO_4$, but to avoid the slow scintillator light component that is prohibitively long for our application. Although its resolution was excellent, the intrinsic pulse FWHM of 15 ns greatly exceeds the 4 ns width measured for $PbF_2$. There were no benefits of this crystal from cost or other perspectives. The choice of absorber is fixed. Owing the long lead time and proven design, we have ordered the crystals for $g - 2$. The decision has been taken.

We also evaluated fast photo-multiplier tubes as alternatives to SiPMs. The Hamamatsu 9800 is an excellent PMT, having a fairly compact footprint and intrinsic fast response. We are using it regularly to benchmark the intrinsic light output time distribution from our crystals. Unfortunately, it is not a good choice for full implementation in the experiment owing to the need to place these PMTs at least 1.5 m from the calorimeter arrays. Because of the rear-face readout from the geometry, the guides would require a rapid 90-degree bend toward the radial direction and then a second bend to put the PMTs out of plane. The high cost of the PMTs (about 5 times higher than the SiPMs) and the awkward light-guide constraint were deemed to be major issues compared to the development of SiPMs that can be located onboard the crystals. Turning now to a PMT-based solution would increase costs significantly and introduce delays, as the design and testing program would have to start anew for this option. We have committed to using on-board, non-magnetic, fast SiPM readout.

One option to increase the fractional readout area on the rear face of the crystals (presently $144\,\text{mm}^2$ / $625\,\text{mm}^2$ for the 16-ch MPPC) is to use larger area arrays of tile-able SMT packaged SiPMs on custom-designed PCBs. For example, an ideal $5 \times 5$ array of $9\,\mathrm{mm}^2$ can be made using single channel Hammamatsu devices. Larger area coverage would allow the use of smaller, thus in principle faster pixels, while maintaining sufficient overall photon detection efficiency. We built such a 16-channel segmented SiPM board from $25\,\mu$m products and evaluated against a monolithic $50\,\mu$m pixel board. They turned up having exactly the same pulse shape width. The sub-optimal leads that the small product comes with suffer with high parasitic capacitances that deteriorates pulse shape. Also manually assembled SiPM board is significantly more expensive and the cost per unit area of the SiPMs packaged in these smaller units is nearly a factor of two higher.

## 17.6 ES&H

The 1300 SiPMs all receive a low-current (limited to $150\,\mu$A) $\sim$ 70 V bias voltage, which is delivered to the enclosed housing through ribbon cables from a custom-built bias control system.

A laser system will be used to distribute calibration pulses. Apart from the laser hut, the light will be entirely contained in optical fibers, with no possibility of escaping under normal use. The laser hut will be located in the electronics control room and will have appropriate FNAL-supported safety and security requirements included and proper training for any operators.

The mechanical weight of the calorimeters is only 40 kg each and they will be supported on railed housings. There is no vacuum insertion, but these detectors will be placed near the storage ring magnetic field and, as such, care must be taken when servicing them to ensure that no magnetic tools are used (a general requirement for any access to the storage ring area).

## 17.7 Risks

### 17.7.1 Performance Risks

In the very unlikely case that the gain stability of the system in actual use fails to meet the specifications, other analysis techniques will have to compensate for the energy scale information loss. This situation happened in E821, where the laser system did not meet the performance goals. Instead, E821 was able to determine the stability of the gain from the data itself. It is not ideal, but did mitigate the risk.

### 17.7.2 Schedule Risks

An NSF Major Research Instrumentation proposal was granted in 2013 to a consortium of six domestic and one international partners within the full collaboration. A substantial matching component was arranged from the domestic and international universities involved. The MRI funds provide already the full necessary capital to fund the detectors, electronics and DAQ systems related to the measurement of the muon precession frequency, $\omega_a$. This funding supports the purchase of all $PbF_2$ crystals, the SiPM readouts, the board fabrication and design, and the mechanical housing, as described in this Chapter.

In early 2014, the contract for the purchase of 1325 $PbF_2$ crystals from the Shanghai SICCAS High Technology Corp was completed. The first batch of 25 crystals was received in May, 2014.

Separately, the Italian groups were granted partial funds, and are awaiting approval for the rest of resources from INFN for their production of the calibration subsystem.

Since the MRI funds were secured, all associated funding-based schedule risks are extinct. The demanding performance parameters applied to the bias control and laser gain monitoring system could require revision cycles that impact the schedule. We are mitigating the risk with an aggressive R&D program schedule that will accommodate several design iterations.

## 17.8 Quality Assurance

Our local Shanghai University collaborators are able to monitor the production of the crystals at SICCAS and intervene if any problems arise. The crystals will be shipped to the University of Washington where they will be measured, and tested for transmission efficiency, then wrapped. A traveler document system has been written to keep track of crystals and measurements. The SiPM boards will be built and then tested using a custom light scanner that can calibrate each device. Finally, the individual crystal-SiPM packages will be assembled into an array and tested with a calibrated laser front panel plate. We have a SiPM test laboratory at UW to evaluate the production SiPM boards and will prepare a program using undergraduate students to evaluate each piece in the assembly line.

## 17.9 Value Management

Competitive quotes had been obtained in order to prepare the MRI Proposal. Local fabrication at universities with largely overhead-free labor will keep costs in check. We have finished an aggressive program of SiPM vendor evaluations and board designs to maximize readout coverage at competitive cost.

## 17.10 R&D

We have used test beam opportunities at Fermilab and SLAC repeatedly as necessary. We intend to return to SLAC for another electron test beam to demonstrate calorimeter performance and long gain stability with prototypes of the final HV power supplies and digitizers.

We continue to use our local laboratory tools to evaluate SiPM performance and have several student projects ongoing to map out gain functions and other performance characteristics.

# Chapter 18

# Calorimeter Backend Electronics

## 18.1 Backend Electronics

The calorimeter backend electronics for E989 encompass the systems for the distribution of the clock and synchronization signals to the experiment and for the digitization of the waveforms from each channel of electromagnetic calorimetry. The backend electronics for the tracker and the auxiliary detectors are discussed within those sections. The calorimeter and tracker backend electronics will both be implemented as $\mu$TCA Advanced Mezzanine Cards (AMCs).

The backend electronics will also be used to capture the signals from the entrance counters (see Ch. 20), the fiber beam monitors (see Ch. 20), the electrostatic quadrupoles (see Ch. 13), the fast muon kicker (see Ch. 12), and the laser calibration system (see Ch. 17). The channel counts for these systems are

| | |
|---|---|
| Entrance counter: | 2, |
| Fiber harp: | 28, |
| Quadrupoles: | 4, |
| Kicker: | 6-9, |
| Laser: | 42. |

### 18.1.1 Physics Goals

The clock system must provide a frequency-stabilized and blinded clock signal that provides the time basis for the determination of $\omega_a$ and a second frequency-stabilized clock, tied to the same master clock, for the precision magnetic field measurement.

As discussed in Chapter 16, the calorimeter backend electronics contribute to three fundamental areas in the determination of $\omega_a$: the determination of the positron arrival time at the calorimeter, the determination of the positron energy, and the separation of multiple positrons proximate in time (pileup). Given the continuous distribution that the muons reach in the storage ring and the random decay times, waveform digitizers (WFDs) best fulfill these roles. The SiPM output response is deterministic, so a fit to the digitized waveform can determine the arrival time and the energy of each positron and can resolve overlapping signals. The WFDs will digitize each muon fill in its entirety, and the frontend DAQ computers will derive the persistent $T$ method and $Q$ method datasets from these continuous waveforms. This scheme eliminates hardware-level deadtime and associated efficiencies that would couple to muon intensity and introduce systematic biases that are difficult to control at the sub-part per million level.

Table 18.1: Summary of the minimum clock and digitization requirements for the calorimeter backend electronics, which are discussed in more detail in Section 18.1.2.

| Feature | Driving Consideration | Requirement |
|---:|---|---|
| Digitization rate | Pileup identification | $\geq$ 500 MSPS |
| Bandwidth | Pileup identification | $\geq$ 200 MHz |
| Bit depth | Energy resolution | 12 bits |
| Station readout rate | Fill length and repetition rate | $\geq$ 3 Gbit/s (avg.) |
| Clock stability over fill | Negligible $\omega_a$ systematic contribution | $<$ 10 ps over 700 $\mu$s |
| Frequency upconversion | Time base for frequency $\omega_a$ | $<$ 1 Hz |
| Clock jitter | Signal fidelity for time extraction | $\leq$ 200 ps |

The WFD must convert the waveforms from analog to digital while retaining the signal fidelity necessary to meet the calorimetry requirements on energy resolution and pileup differentiation. The system must convert the distributed master clock frequency to the required sampling frequency range while maintaining the timing requirements, without allowing circumvention of the experimental frequency blinding. The digitized waveforms must be transferred without loss to the DAQ frontends for data reduction. The system must also provide the support and infrastructure to capture samples for pedestal determination, gain monitoring and correction, and stability cross checks of the gain monitoring system.

These considerations lead to the minimum requirements summarized in Table 18.1.

### 18.1.2 Requirements

#### Clock and synchronization distribution

To avoid systematic biasing of $\omega_a$, the distributed master clock must be held stable against systematic phase shifts or timing drifts to under 10 ps over the 700 $\mu$s fill [1]. To help maintain signal fidelity and the subsequent extraction of the positron arrival time from fits to the digitized waveform, the random timing jitter should be smaller than the ADC signal sampling window (the ADC's aperture delay), which is of order $100 - 200$ ps for the required digitization rates. The frequency upconversion within the WFDs must maintain these requirements, and the upconversion factor must be determined to better than a part per billion.

Synchronization signals such as begin-of-fill must be distributed to each backend electronics channel to signal the time to capture the data. The synchronization signals will flag the specific master clock cycle on which to begin data acquisition for each muon fill. These signals will arrive with the granularity of the master clock. The muon beam entrance-counter signals will be digitized by this WFD system for one of the calorimeter stations and will provide a precise start time measured with the sampling frequency (800 MHz), not master clock (40 MHz) precision. Additionally, the laser system will be pulsed in advance of every fill to dead-reckon the time-alignment of all calorimetry channels.

### Waveform digitization

**Signal requirements:** The energy resolution budget (5% near the 1.5 GeV threshold for fitting) determines the WFD minimum bit depth. Assuming a typical $3 \times 3$ array of crystals summed to determine the energy, having 8 bits at 1.5 GeV would already contribute 1.2% to the energy resolution. This energy is about half the maximum energy range, and the system should have the overhead for complete study of the pileup energy distribution, which requires 10 effective bits. The effective number of bits is typically between 1 and 2 bits lower than the physical ADC bits. A digitization depth of 12 bits provides the appropriate resolution.

The signal separation characteristics will be determined by a combination of the crystal wrapping[1], SiPM and amplifier response (see Figure 17.16), total cable, and WFD bandwidth. The WFD bandwidth must be large enough to avoid significant stretching of the pulse shapes, with the rise time remaining under 2 ns (if the final wrapping choice allows). The overall pileup requirement for the experiment (see Section 17.1: the system must be able to distinguish pulses separated by 5 ns) drives the specification for the digitization sampling rate. Laboratory tests at 800 megasamples per second (MSPS) show clear separation of pulses with this separation (see Figure 17.17). Figure 18.1 shows the fits for two pulses measured in the lab with 5 and 10 ns separations at 500 MSPS sampling rate. We can clearly resolve even the 5 ns separation at 500 MSPS, but we will lose fidelity at lower sampling rates. A higher digitization rate will give more headroom to cope with higher intensity muon beams, and there are 800 MSPS ADCs available that afford a good balance between performance and cost. Furthermore, with the improved SiPMs now in hand, their inherent signal characteristics are much faster, with rise times of order 2 ns and under 4 ns fall times. These times also push us towards the higher sampling rate. Simulations based on the Čerenkov photons show that, at 800 MSPS, the presence of a second soft cluster can be identified 100% of the time down to a time separation of 3.5 ns. With the additional factor of 3.0 that comes from the $PbF_2$ segmentation, this performance can accommodate beam intensities of a factor of 2.5 over the proposed intensity (10.0 over the Brookhaven E821 intensity). Our baseline implementation will be based on the 800 MSPS rate.

**Physical requirements:** The WFD crates will be located just over 1 m from the dipole field of the storage ring, where the fringe field ranges from 30 to 60 Gauss. Ideally, perturbation of the storage ring field by the resulting magnetization of the materials in each WFD station would be limited to well under a part per million. We can roughly limit the acceptable level of a magnetic material by assuming that a magnetized sphere of that material in a uniform fringe field should at most create a static perturbation of 0.1 part per million. A predominantly Aluminum chassis will be no problem: 15 kg would result in a perturbation under 0.1% of this limit. For ferromagnetic materials, however, the total mass must kept under about 200 g, which may require the power supplies to be located farther away. An Aluminium $\mu$TCA crate was shipped to Fermilab during summer 2014 for characterization using a test magnet. The results of this test revealed the need to change two pieces of the crate that have been addressed to VadaTech. The first piece is the metal connector between the crate power output and the power module that delivers power to the modules in the crate. This connector will from now on be made out of plastic. The second piece is the steel card cage within the $\mu$TCA crate that will be made of Aluminium. Final assessment will be made after the final version of the $\mu$TCA crate from VadaTech arrives at Fermilab during summer 2015.

[1] The GEANT4 Cherenkov light simulation (Section 17.4.1) shows that the wrapping material creates rise time contributions that vary from 1 to 3 ns depending on the material used.

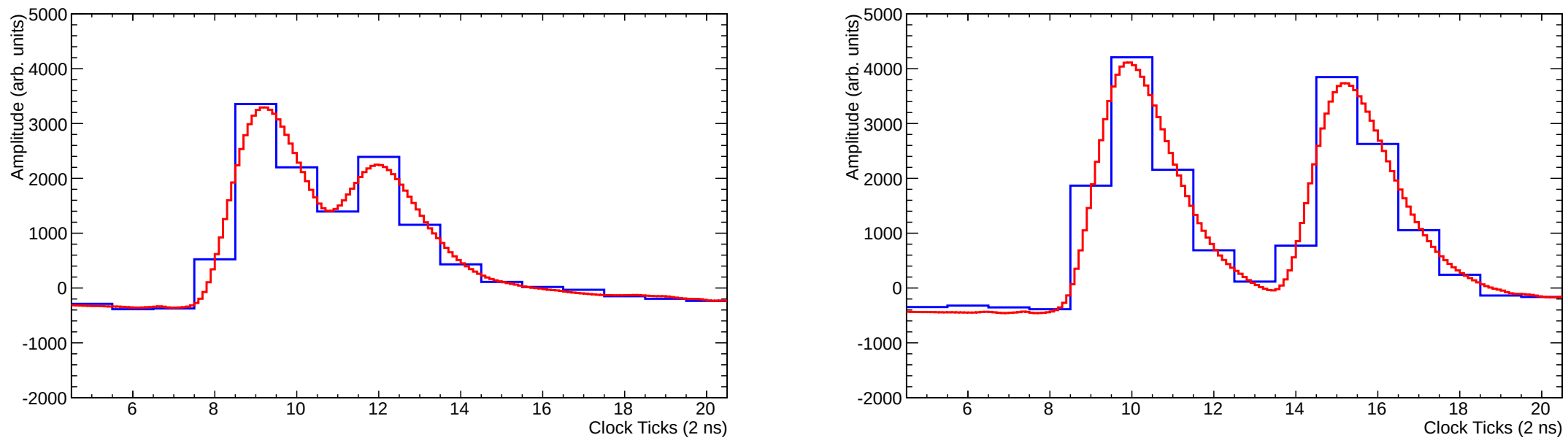

Figure 18.1: Fits to test data (see Section 17.4.1) with two pulses separated by 5 ns (left) and 10 ns (right) with a 500 MSPS sampling rate. The fits (red) clearly resolve the two peaks in the data (blue), even for the 5 ns separation.

**DAQ requirements:** During experimental running, muons will be stored in the storage ring for 700 $\mu$s fills. The basic fill structure will be two groups of eight fills, with the fills within a group occurring at 10 ms intervals and with the groups of eight occurring at 197 ms intervals (see Figure 7.2). This basic structure repeats every 1.4 s for an average fill rate of 12 Hz.

To eliminate dead-time, up to a 700 $\mu$s waveform for each calorimeter channel will be digitized and transferred to the DAQ frontend system for data reduction. Each WFD station must provide adequate buffering and throughput to support the average data rate. Assuming an 800 MSPS digitization rate, the average rate is 5.9 Gbit/s for 12-bit samples being transmitted as 16-bit words and 4.4 Gbit/s if bit-packed, i.e., 12-bit samples being transmitted as 12-bit words. The experiment may run with a shorter 600 $\mu$s sampling time per fill, which would reduce the data throughput to 5.1 Gbit/s (not bit-packed) and 3.8 Gbit/s (bit-packed) for each station.

### 18.1.3 Baseline Design

#### Clock distribution

The clock system will distribute a high-precision clock and synchronization signals to each backend DAQ crate. It will provide an external time reference that is fully independent of accelerator timing to ensure that the $\omega_a$ measurement is not biased by synchronous accelerator or ring events.

The clock system will utilize a series of off-the-shelf components in conjunction with a fiber optic distribution and encoding system that was developed for the CMS experiment [5]. An optical clock distribution system minimizes concerns due to pickup and ground loops. Overall design and consideration of the clock system follows from experience gained in Brookhaven E821 [2] and MuLan [3].

A block diagram of the clock system is shown in Figure 18.2. The principal clock source for both the $\omega_a$ and $\omega_p$ measurement will be produced by a Meridian Precision GPS TimeBase, a GPS-disciplined oscillator. The utilization of the clock in the $\omega_p$ measurement is described in Section 15.2.4. The Meridian module is supplemented by a "low-phase-noise" output module to minimize jitter. The GPS system additionally provides time-stamps. The GPS clock produces a 10 MHz output signal which is fed to a Stanford Research Systems SG380 series RF signal generator providing a shifted $\omega_a$ clock of 40 MHz plus a small offset ($\epsilon$) that will be blinded.

The 40 + $\epsilon$ MHz clock is then carried on double-shielded RG-142 cable from the signal generator to the encoding and fanout system, which resides in a single $\mu$TCA "clock crate." The analog clock signal is injected into an FC7 board [5] equipped with a standard FCM DIO 5CH TTL A module [6]. The FMC card will additionally receive input signals from the Fermilab accelerator and other Muon $g-2$ subsystems. To define the "begin of fill" signal, we have identified the "recycler beam sync extraction event to muon" signal. This signal will be distributed to all backend crates and, after time-alignment, will synchronize and initiate data acquisition (i.e., provide a common start.) The other control signals received by the FC7 will be utilize to define run mode and synchronize the DAQ across systems.

With the 40 MHz clock and control signals in the FC7, the clock frequency is multiplied by four and the control signals are encoded via the TTC protocol developed for the Large Hadron Collider (LHC) experiments [8]. The encoded clock and controls will then be sent via optical fiber to an AMC13 card working in dual-star mode within the clock crate. The AMC13 will distribute the recovered clock and controls along the $\mu$TCA backplane to two fanout FC7 boards which, in turn, will each produce 16 copies of the encoded clock and control signals. These outputs will carry signals to the backend $\mu$TCA crates.

A single fiber carrying the clock and control signals is distributed to the AMC13 board on each backend $\mu$TCA crate. The AMC13 receives the signal, recovers the base 40 MHz clock, and decodes the control signals onto a parallel stream. The clock and control signals are then sent along the backplane to each of the WFDs. On the WFD, the clock frequency is upconverted to the 800 MHz base sampling rate.

The clock crate will drive a single clock/control fiber to each of the 24 calorimeter backend crates, 1 tracker crate, 1 auxiliary detector, 1 laser calibration system crate, and the DAQ system.

Monitoring of the clock system will occur at several stages. A second Rubidium oscillator will reside in the same crate to check the master clock stability. The stability of the difference between master and reference clock frequencies plus a blinded offset will be monitored directly in the counting room, using a data login featured oscilloscope. At the receiving end, the AMC13 will verify clock functionality with an internal counter compared to a local oscillator. Further, direct tests on time slewing and other systematic effects will be performed using the clock signals as seen by the WFDs.

To date, the individual components (GPS-disciplined clock source, frequency synthesizer, AMC13, backplane distribution) have all been tested. In the near future, we will assemble a complete system in order to begin integration testing.

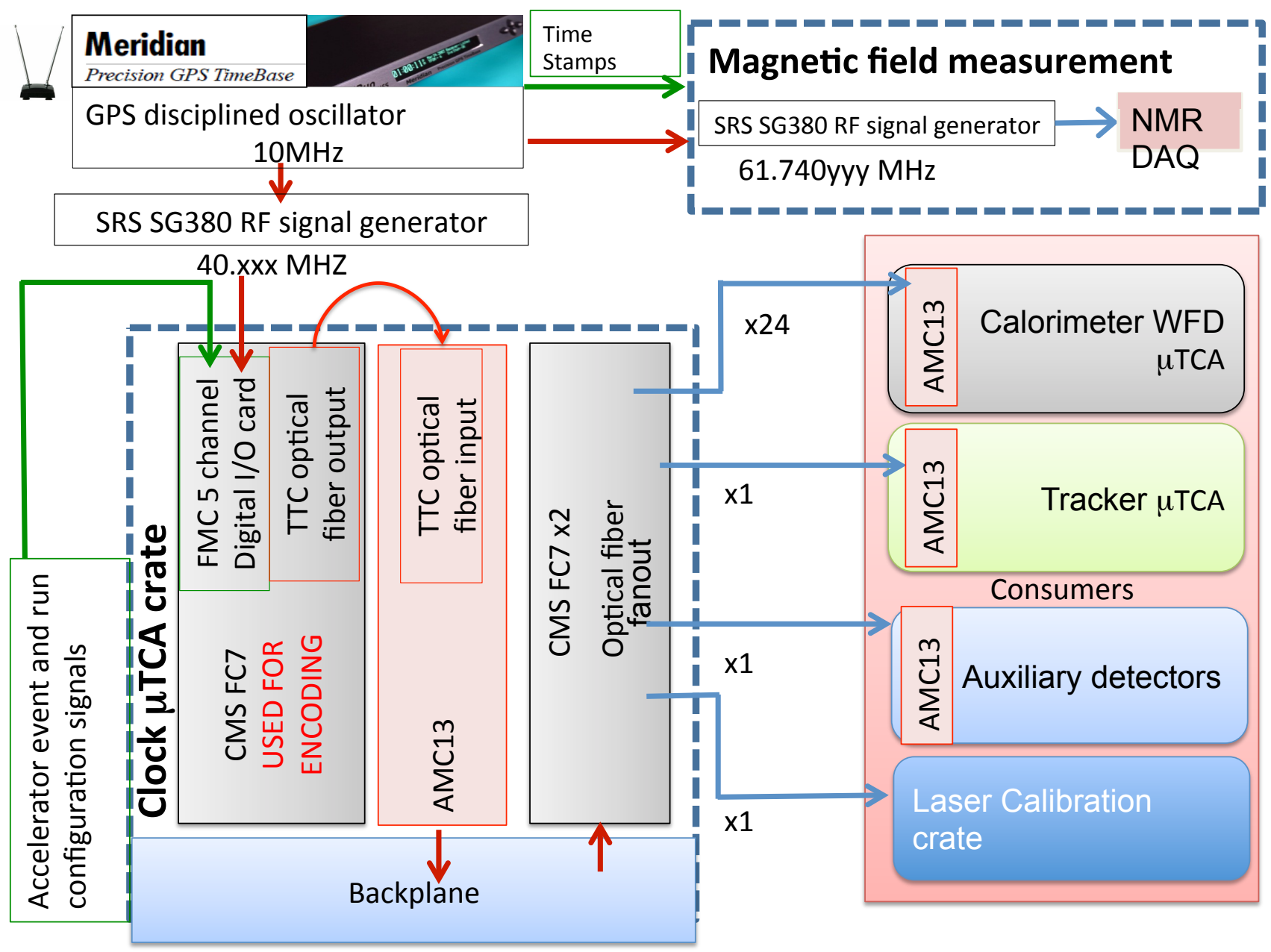

Figure 18.2: Block diagram of the clock distribution system.

### Clock and Commands Center (CCC)

As described above, three FC7 boards will be utilized in the clock crate: one will encode the clock and command signals using the TTC protocol (the Clock and Commands Center, CCC) and two will be used to fan the TTC signals out to backend crates via optical fiber. The CCC FC7 board will receive the clock and relevant timing signals from other systems. Logic on the CCC will then translate the signals into a valid TTC transmission, where the proper command bits are encoded along with the distrubted clock. For the specific example of a normal run mode, where muons will be injected into the Muon $g-2$ storage ring, Figure 18.3 shows the approximate timing for the command logic.

The CCC will receive the begin-of-fill (BOF) signal which will be utilized to initiate data acquisition operations. The BOF signal will arrive at the CCC about 3 $\mu$s prior to beam entering the storage ring. This time is sufficient for logic, handshaking, and distribution of the relevant signals. Within 1 $\mu$s after the BOF is received, the CCC will acquire the "DAQ ready" signal and Laser Calibration System inputs. To initiate data acquisition in the backend crates, we will require both BOF along with the DAQ-ready signal. If the DAQ is not ready to acquire data, the backend electronics will not acquire data. Since our DAQ system is designed to keep pace with data acquisition at all times, a DAQ not-ready state at begin-of-fill time is likely to indicate a problem that needs to be address. The logic "AND" between DAQ ready and begin-of-fill signal will set the trigger of the experiment (A CHANNEL bit of TTC signal will be set to 1). This operation will take about 100 ns after all the inputs have been received and checked. Additional information can be encoded and sent to the backend through the B CHANNEL TTC signal. At this stage the A CHANNEL and B CHANNEL info will be encoded and distributed. The final TTC signal will be then received by all of the consumers after about 2 $\mu$s from the real BOF arrival time. This signal will be delayed and time-aligned as needed so that all backend consumers receive the information at a well-defined time prior to the arrival of beam.

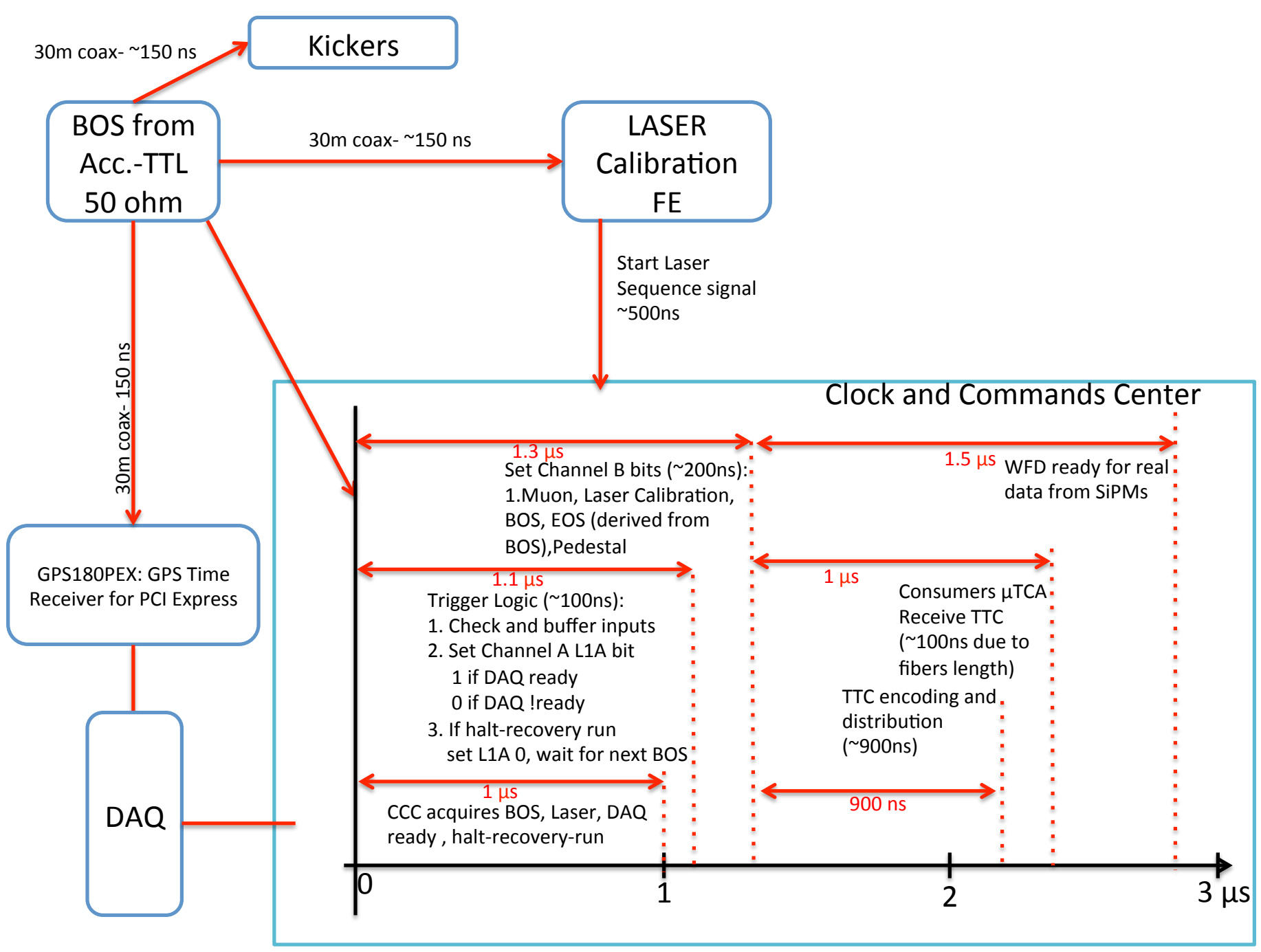

Figure 18.3: A rough estimate of the trigger and commands center time scale

The system outlined here provides a "common start" to the DAQ system, but is not the source of the precision arrival time of the beam. Precise begin-of-fill timing will be provided by the entrance counter signal.

Examples of other run types include laser and pedestal runs. These run types could be initiated synchronously with beam or between muon fills. The CCC system will also be able to initiate a global reset signal to all backend crates which can be utilize to resynchronize the data acquisition process.

### Waveform digitization

The proposed system draws heavily on the hadronic calorimeter and DAQ upgrade [4] underway for the CMS experiment at the LHC, which uses $\mu$TCA technology. The WFDs for each calorimeter station will reside in a single Aluminum VadaTech VT892 $\mu$TCA crate as a set of 12 five-channel AMCs. The first five Aluminum VT892 crates have been delivered in May 2014. The first 10 modified $\mu$TCA crates, which resulted from the magnetic field testing at Fermilab, should be delivered by summer 2015, and the currently owned crates will be sent back for upgrade. The remainder of the crates will be acquired during the project implementation phase.

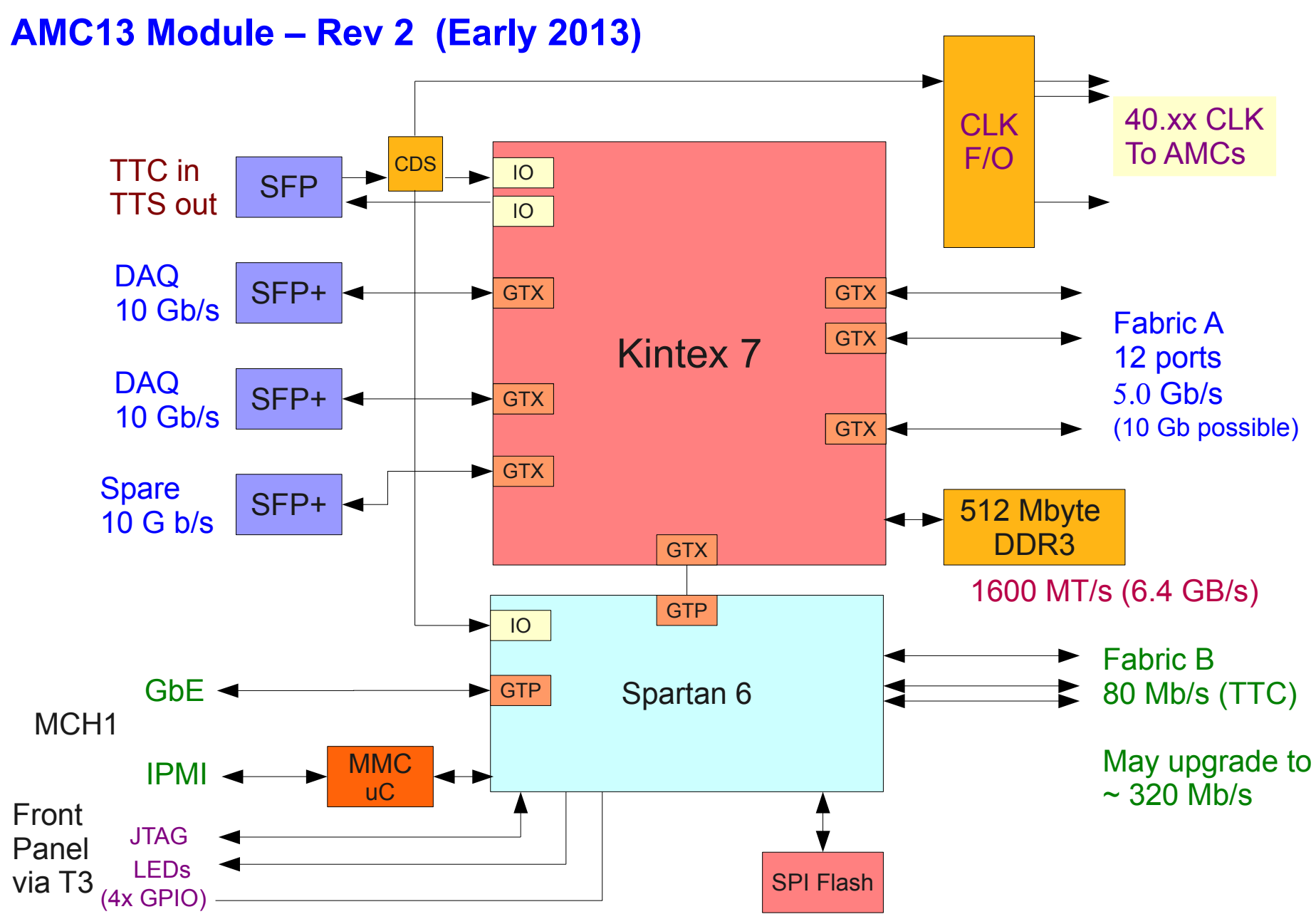

Figure 18.4: Block diagram of the CMS-designed AMC13 $\mu$TCA card that will control the $g-2$ WFD readout.

A VT892 crate accommodates 12 full-height AMC cards. Eleven AMCs will instrument the 54 calorimeter channels for one calorimeter station. The 55$^{\text{th}}$ channel will be used for the laser monitoring source. The 12$^{\text{th}}$ WFD AMC will reside in the crate, which leaves five calorimeter hot spares per station. The $\mu$TCA choice brings a robust system designed for remote operation and monitoring with cooling, power distribution, and clock distribution capabilities already engineered. There will be a separate VT892 crate for each of the 24 calorimeter stations.

**$\mu$TCA infrastructure** The heart of the CMS $\mu$TCA data acquisition system is the Boston U. (BU)-designed AMC card, the AMC13 [7], which replaces a second (redundant) $\mu$TCA Carrier Hub (MCH) in the $\mu$TCA crate. The AMC13 is shown in block diagram form in Figure 18.4. The AMC13 has the responsibility for distribution of external clock and synchronous control signals to the AMC cards, for readout and aggregation of the data for each fill from the AMC cards, and for transfer of the data to the DAQ frontend computers over one to three 10 Gbit/s optical ethernet links. The AMC13 has gone through an extensive prototyping and testing cycle in CMS, and BU has recently overseen the first large production run of AMC13 modules. CMS has demonstrated that the AMC13 is compatible with the MCH from both VadaTech and N.A.T. Based on feedback from CMS related to reliability and support, we will deploy the VadaTech MCH in our $\mu$TCA crates. Figure 18.5 shows the Cornell U. test stand populated with the AMC13, a VadaTech MCH, and one of our Revision 0 five-channel WFD prototypes.

The $\mu$TCA backplane connects each of the 12 WFD AMC cards to the AMC13 and MCH

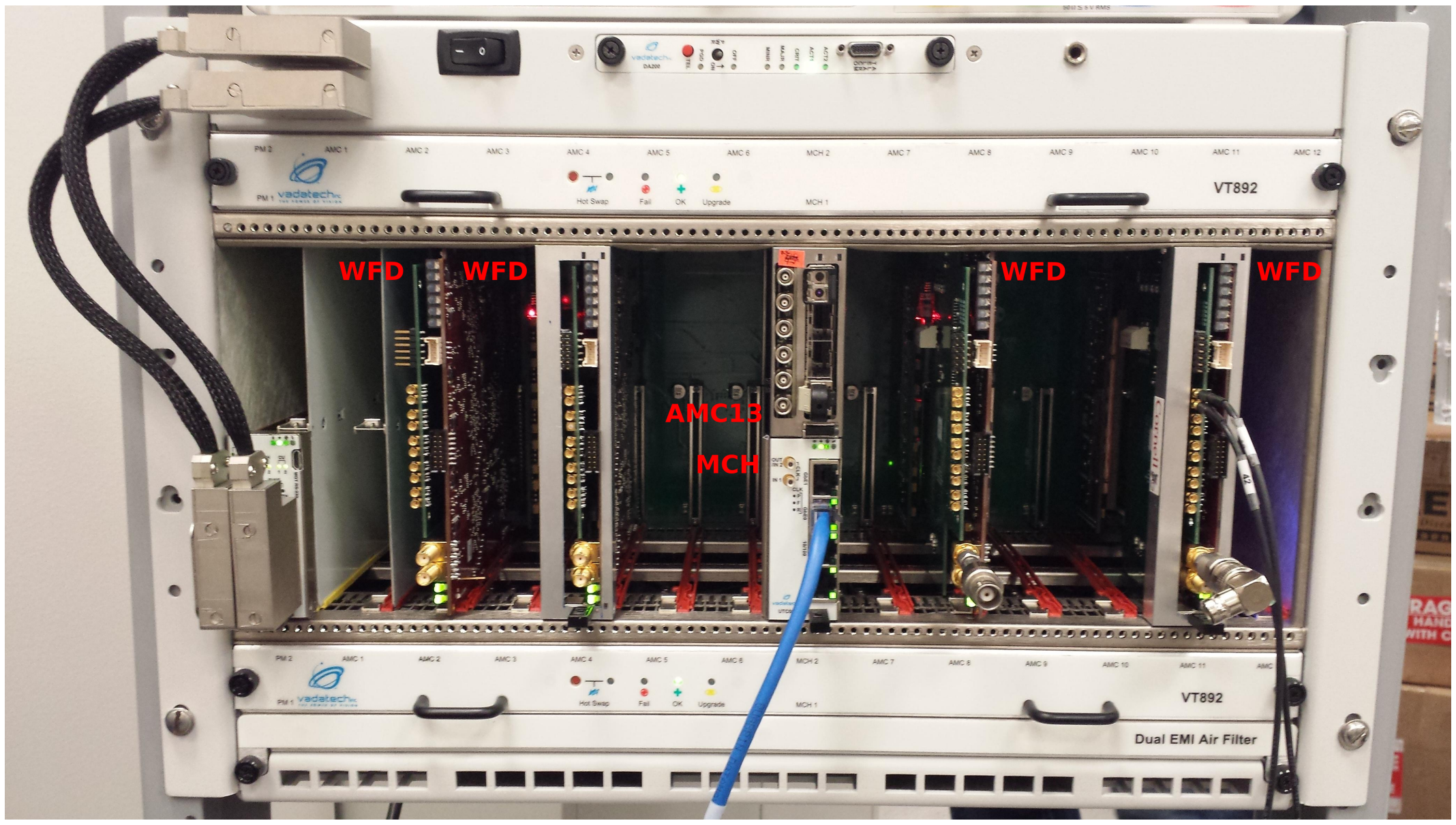

Figure 18.5: The Cornell U. $\mu$TCA test stand based on the VadaTech VT892, populated with a CMS AMC13 (upper of center pair of modules), a VadaTech UTC002 MCH (lower of the central pair), and one of the Revision 0 five-channel WFD prototypes (at right) under testing.

in a star topology (Figure 18.6). High-speed readout occurs over Fabric A and is managed by the AMC13's Kintex-7 FPGA. The star topology and AMC13 implementation allow for parallel readout of the 12 WFD cards at rates up to 5 Gbit/s per link (60 Gbit/s aggregate). For continuous readout at the average 12 Hz muon fill rate, only a 0.6 Gbit/s link to each card is necessary for an 800 MSPS sampling rate with the 12-bit samples being transmitted as 16-bit words. There is plenty of overhead in the backplane transmission rate to support the average readout rate. The system can also accommodate readout in the smallest inter-fill time (10 ms), which requires a link rate of 4.5 Gbit/s assuming transfer of 16-bit words, or 3.4 Gbit/s if bit-packed. The WFD system therefore has the flexibility to support a wide range of beam delivery options.

The AMC13 receives the clock and synchronous signals from the clock system via optical fiber using the TTC protocol [8]. With the 40 MHz master frequency, the clock and synchronous TTC data are distributed as a 160 MHz biphase-encoded signal. The AMC13 recovers the data and clock from the optical signal, routes the TTC data over Fabric B via the Spartan-6 FPGA, and fans out the clock over the CLK1 fabric (see Figure 18.7.) On each WFD AMC card, a Texas Instruments (TI) clock synthesizer chip (LMK04906) will upconvert the master frequency to the 800 MHz frequency needed for digitization.

The AMC13 event builder has been upgraded from its specific CMS HCAL implementation to support both Muon $g-2$ and broader CMS deployment [9]. In particular, the event builder supports much larger event block sizes than it did in its original 4 kB HCAL implementation. A beta version of this new firmware, including the communications firmware block for the AMC side of the 5 Gbit/s link, has been delivered to Cornell U. in spring 2014

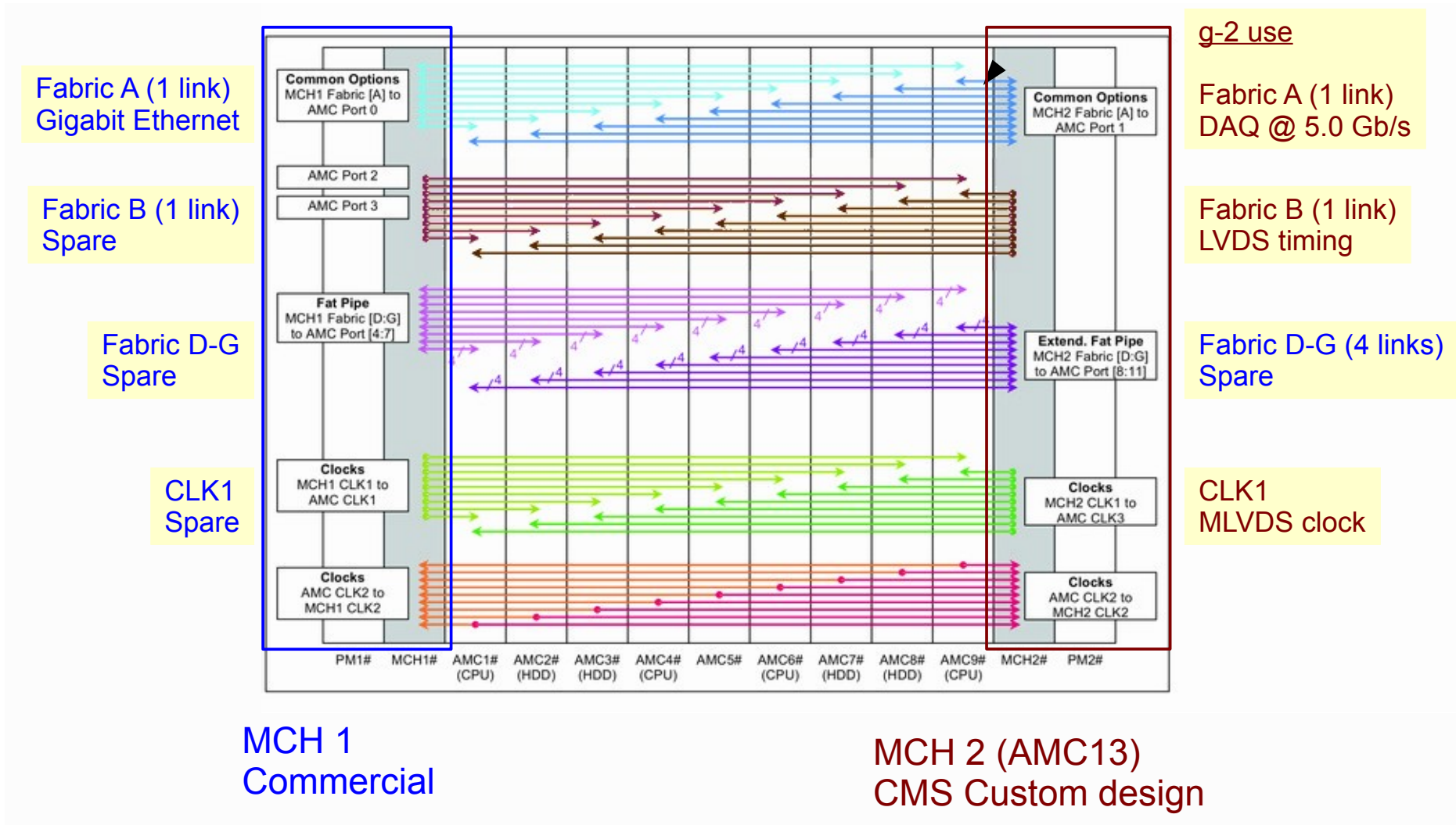

Figure 18.6: Dual-star backplane configuration for use with $g-2$ AMC13. High speed data transfers proceed over Fabric A. Timing and synchronization proceed via Fabric B.

and has been shown to work. Several upgraded versions have been successfully tested since then.

Each AMC13 communicates the event data to its corresponding DAQ frontend computer via one to three 10 Gbit/s TCP/IP optical links. A Muon $g-2$ specific modification of the AMC13 firmware that supports TCP/IP has been released to Cornell U. and U. of Kentucky for testing. The implementation will use the 512 MB memory buffer on board the AMC13, which can accommodate eight fills of data without compression. Significantly more buffering is available on board the WFD AMCs themselves, as discussed below. With the buffering, the AMC13 can communicate asynchronously with the external DAQ system to support the average data rate of about 6 Gbit/s per station. Only one of the three 10 Gbit/s optical links for communicating with the DAQ frontend computers is therefore needed.

The AMC13 has substantial on-board processing capability in the Kintex-7 FPGA that could be used for a variety of tasks. These tasks could include lossless encoding/decoding of the data on-the-fly to allow deeper buffering of the data or to prepare the $T$ or $Q$ method datasets. The latter would substantially reduce the required bandwidth between the AMC13 and the DAQ frontends as well as relieving processing requirements on the DAQ, if needed.

**Waveform digitizer hardware** The baseline WFD design is centered on the TI ADS5401, an 800 MSPS 12-bit ADC with an input bandwidth of over 1.2 GHz. The block diagram for the anticipated final version of the five-channel AMC card is shown in Figure 18.8 and the Revision 0 of the five-channel prototype based on the 800 MSPS ADS5401 is shown in Figure 18.9. The WFD AMC has three boards. The main board is the heart of the system, with the ADCs, buffer memory, channel and fabric FPGAs, clock synthesizer, and $\mu$TCA management controller (MMC). The power supply and regulation, which takes the 12 V payload power distributed by the crate and converts it to the voltages needed on board, is implemented on one of the mezzanine cards. On the one-channel prototype (based on the 500 MSPS ADS5463 from the conceptual design), this block was at the lower right on the main board, but it suffered from power regulator failures that rendered the board irreparable. In addition to switching to a different regulator, we have moved the power supply to a daughter card. If a regulator fails, we can easily replace the daughter card. In addition, this design freed up space on the main board to accommodate all five WFD channels. The third board contains the analog frontend (AFE) and connects to the main board via shielded connectors. This design gives us the ability to fine tune the AFE design without affecting the main board where the cost driver components reside. The design also provides flexibility for future deployment. In the baseline design, the AFE has a differential input impedance of 100 Ω, with a maximum 2 V peak-to-peak differential signal, and will bandwidth limit the signal to 200 MHz.

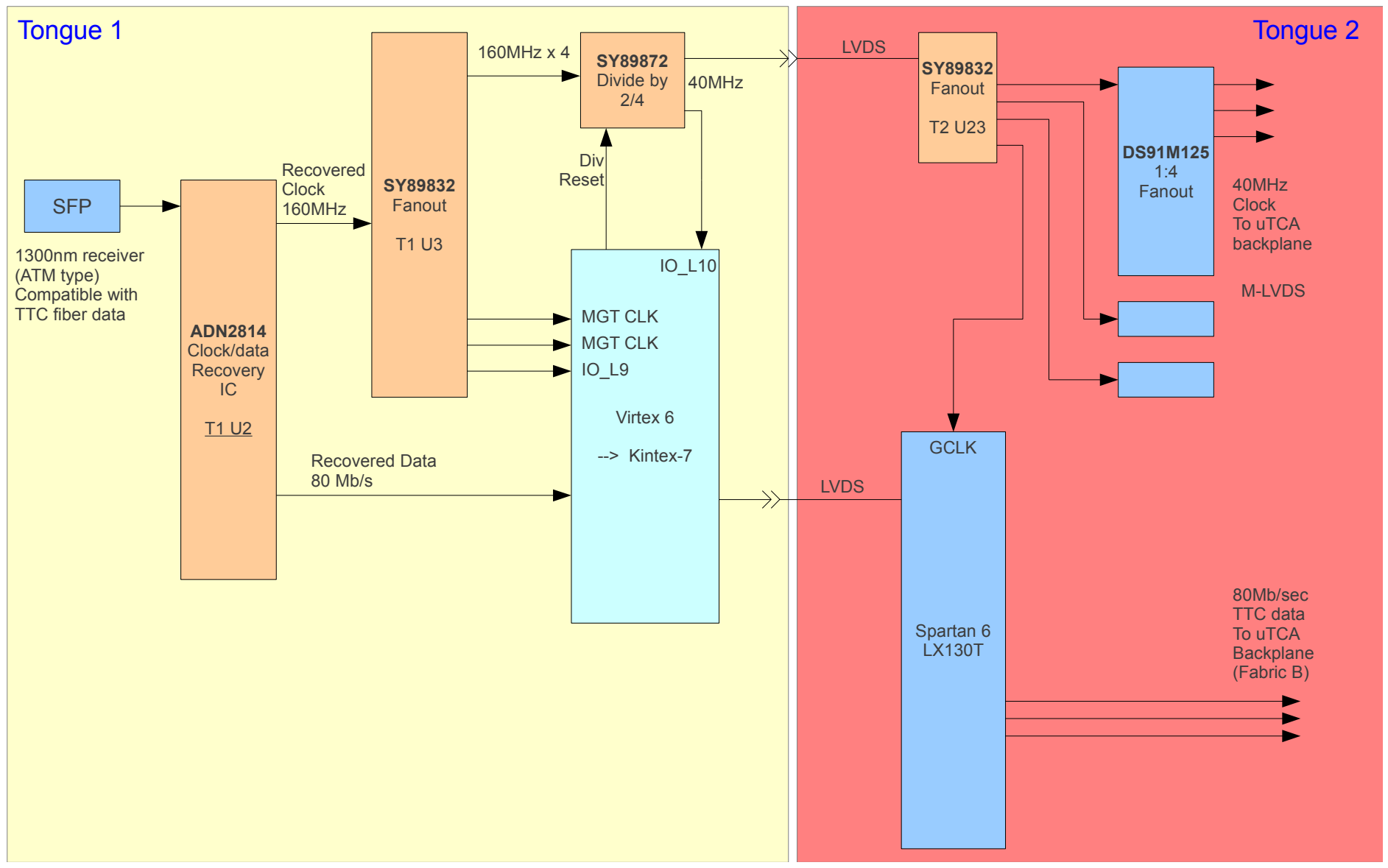

Figure 18.7: Timing paths through the AMC13 for the LHC. The $g-2$ experiment will utilize the FPGA-free LVDS path, which distributes the 40 MHz master clock.

On the main board, each channel has a dedicated Kintex-7 FPGA (XC7K70T-2FBG484C) that controls the data flow out of the ADC to a dedicated 64M×16 bits SDRAM memory buffer as well as the streaming of data over a high-speed serial link to the fabric FPGA. At an average 12 Hz fill rate, the SDRAM can buffer over nine seconds of data with no bit-packing. The fabric FPGA is another Kintex-7 (XC7K160T-1FBG676C), which controls the data streaming out of each channel and over the backplane Fabric A to the AMC13 event builder. Both the channel to fabric FPGA link and the fabric FPGA to AMC13 link operate at 5 Gbit/s. The data will transfer out of the five channels sequentially, so the data from all cards can be transferred to the AMC13 in just under 9 ms for the 800 MSPS 12-bit baseline. The fabric FPGA also supports the 10 Gbit/s ethernet link and IPbus [10], an IP-based protocol for controlling $\mu$TCA hardware devices.

The AMC card must frequency lock on the blinded master $40+\varepsilon$ MHz clock and upconvert it to an $800 + \alpha \cdot \varepsilon$ MHz clock for the ADCs, where $\alpha$ is the upconversion factor between the master and ADC sampling clock. The WFD cards will receive the 40 MHz clock via the

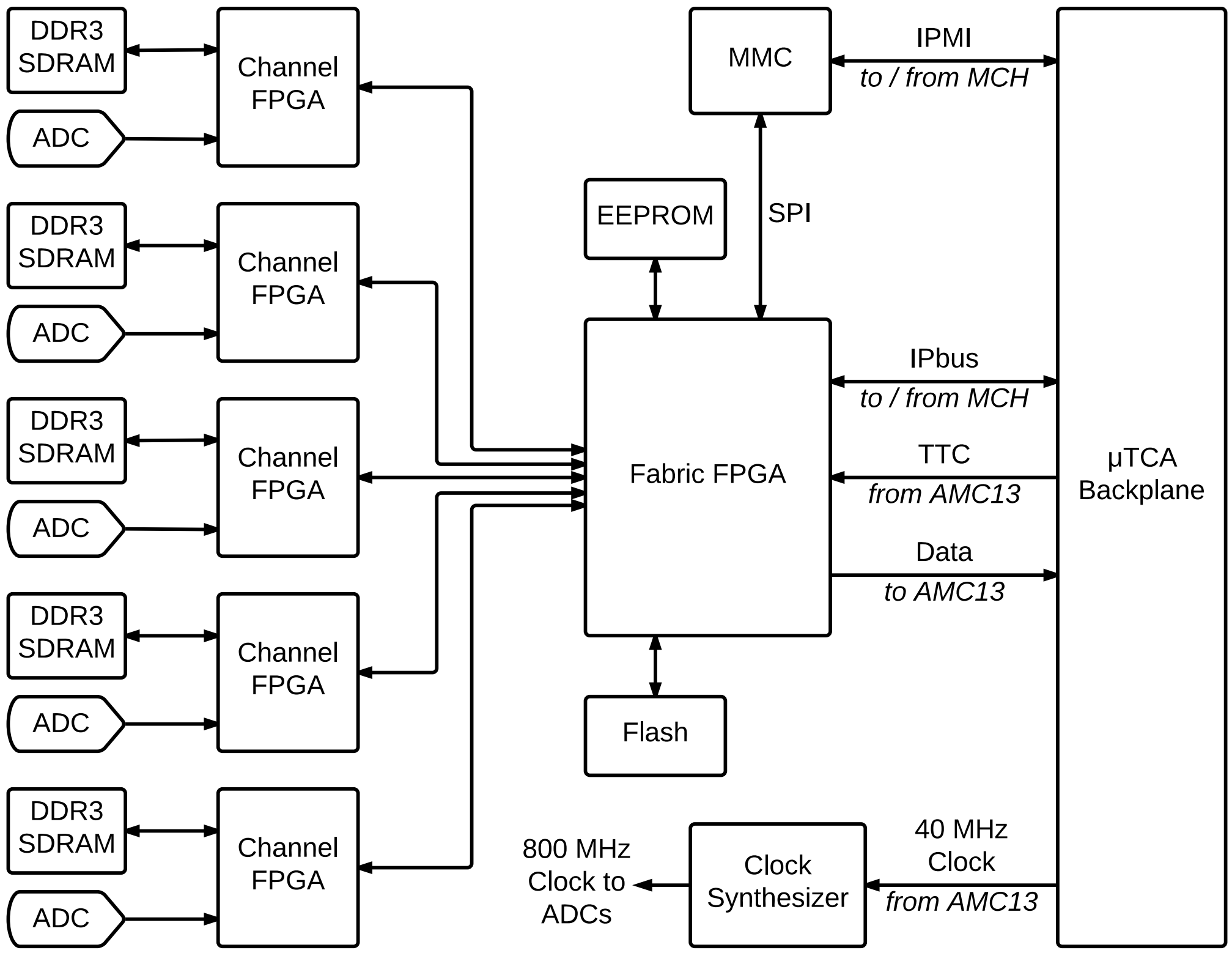

Figure 18.8: Block diagram of the anticipated final version of the five-channel WFD AMC card.

$\mu$TCA backplane. The clock is distributed by the AMC13 via the FPGA-free LVDS clock path shown in Figure 18.7. The full clock path will need significant testing to verify that it will have a highly stable duty cycle, slew and wander within the phase stability specifications over a fill, and no differential nonlinearities. We have, however, consulted closely with the engineer responsible for the clock for operation of the Cornell Electron Storage Ring (CESR), which also has stringent timing requirements. He does not anticipate any intrinsic difficulty in meeting the $g-2$ stability specification. On timescales of several hundred $\mu$sec, a more important issue is typically environmental noise. We must ensure, for example, that the clock supplied to the ADC will be immune to noise sources, especially those correlated with the fill structure such as the firing of the kicker. Because environmental noise is an issue, we use a single package clock management, based on the TI LMK04906, rather than a discrete component solution. This chip can upconvert the 40 MHz input clock to the 800 MHz range and distribute it over five output channels with a programmable delay on each line. The programmable delay will allow correction of channel by channel timing differences in signal path lengths from the photodetectors at the sub-clock-cycle level. The single package device will be far more immune to external noise than a discrete component solution and has much better overall jitter specifications (under 200 fs) than a discrete component solution. The detector-related systematic error budget of 70 parts per billion for the muon spin precession sets the error budget for the ADC sampling clock frequency to be of order 1 part per billion, i.e., 1 Hz. Extensive tests of the upconversion capability and stability of the TI LMK04906

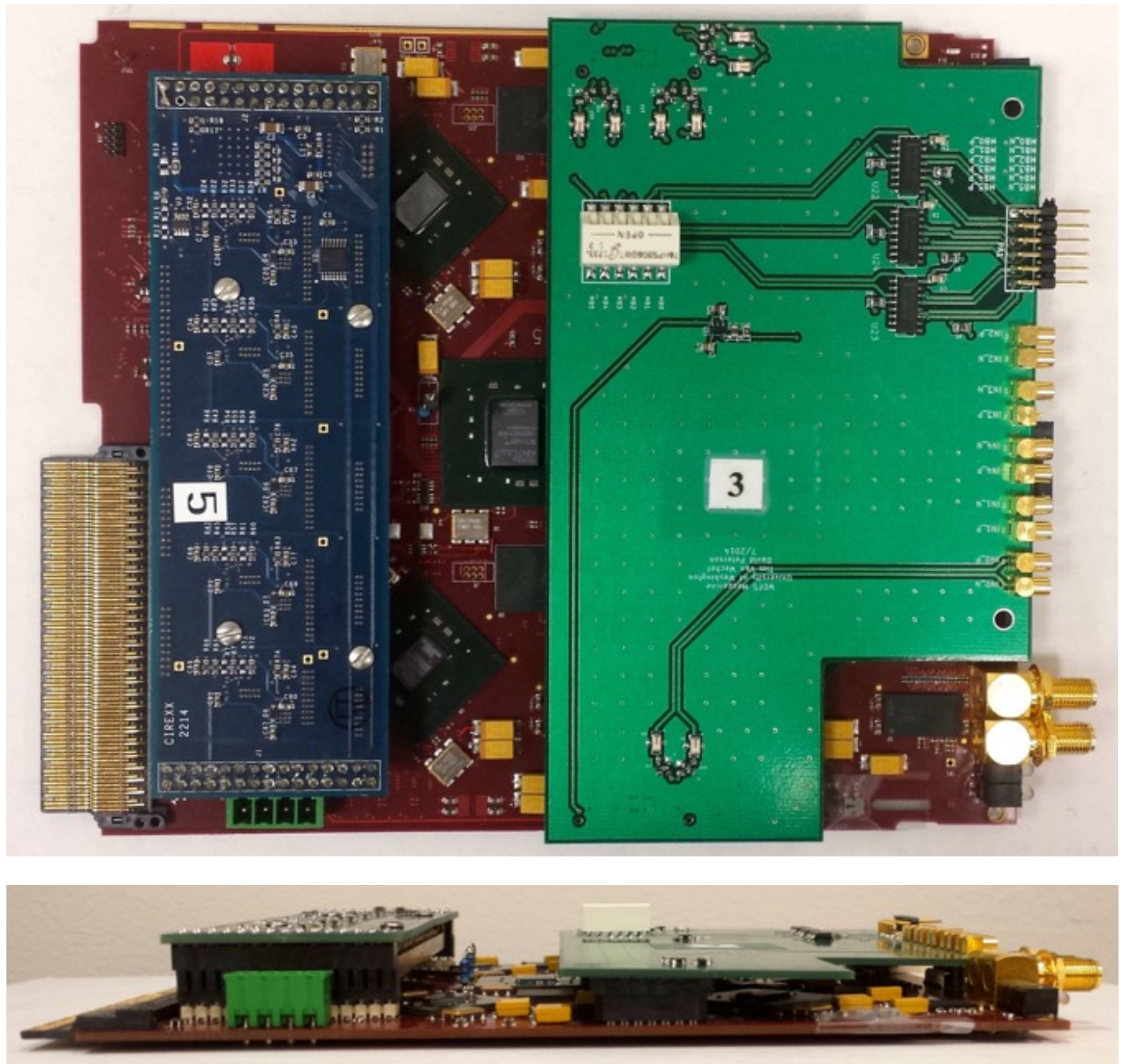

Figure 18.9: The five-channel Revision 0 prototype of the WFD AMC card: the main board is in red, the power supply in blue, and the analog frontend in green.

were performed at Cornell U. and showed that it meets the requirement.

The full clock chain is still undergoing testing to ensure the entire path, from the GPS-stabilized master clock through to the individual ADC channel, satisfies the clock requirements. If not, two backup paths for clock distribution exist that can be used, depending upon the weak link. In one option, the clock could be input to a CERN-developed version of the AMC13 Tongue 3 over copper, independently of the optical TTC signal. This path would bypass the TTC encoding that creates the optical signal as well as the local clock/data recovery on the AMC13. The AMC13 Tongue 2 would still distribute the clock via M-LVDS over the $\mu$TCA backplane. In the other option, the master clock would be fanned out externally and input directly into the front panel of each WFD AMC.

To take full advantage of CMS $\mu$TCA development, we use the same Atmel microcontroller as CMS to implement the AMC MMC. This same microcontroller has been deployed on the CMS HCAL AMC cards as well as on the AMC13 itself.

At the time of writing, the Revision 0 of the five-channel prototye (main board and power board) that succeeded the one-channel prototype has been extensively tested and has met the baseline design goals. The last prototype, the Revision 1 WFD, was designed taking advantage of the Revision 0 testing to implement new capabilities and fix non-critical issues. This Revision 1 is currently under production and assembly and will be tested summer 2015.

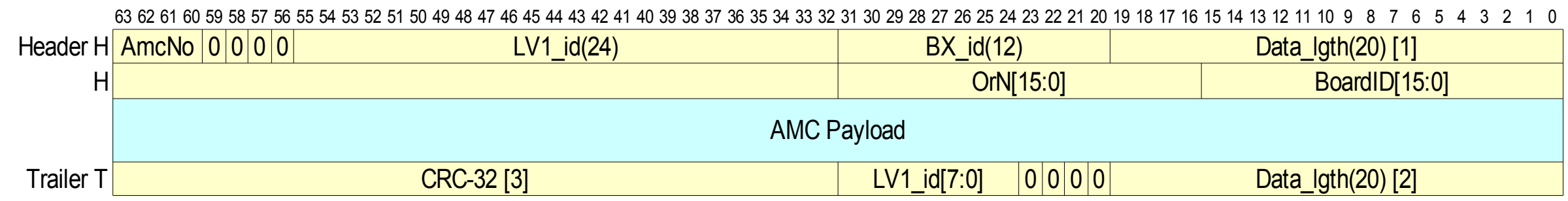

Figure 18.10: The data format for delivering the WFD AMC event payload to the AMC13 event builder.

We will incorporate any modifications (expected to be minor) from the Revision 1 board into the final design, which will go to full production in fall 2015. An AFE board has been provided by U. of Washington for the July 2014 SLAC test beam in order to allow waveform to be sent to the five-channel WFD. This AFE was used to test the WFD Revision 0 at Cornell U. Recently, a more elaborate AFE board was designed at Cornell U. to replace the previous one. This new AFE takes a DC-coupled input signal from the SiPM frontend board via ECDP ribbon cables and offsets this signal using a digital-to-analog converter driven by the main board fabric FPGA. This digital offset will allow utilization of full input range of the ADC. This version was sent for production and assembly.

**Waveform digitizer firmware** The overall operational scheme for the WFD AMC is conceptually straightforward, and makes direct use of the development efforts for the CMS $\mu$TCA environment. IPbus is used to configure the WFD AMC operational mode and its associated registers, and it operates over the gigabit ethernet link managed on the Fabric A connection to the MCH (see Figure 18.6). The TTC protocol provides synchronous configuration and triggering over Fabric B. For example, begin-of-fill signals[2] are delivered synchronously to the WFD AMCs to trigger data collection. Through the TTC broadcast command facility, the WFD can be configured to acquire data for a standard muon fill, a pedestal calibration, or a laser calibration event, which differ in their total sampling times. The sampling times for each data type will be configurable via a corresponding FPGA register.

When the AMC13 receives a begin-of-fill signal over its optical input, it relays it over Fabric B to the AMCs and buffers the begin-of-fill in a FIFO. After building a given event, the AMC13 then reads the next buffered fill from each AMC or waits for that data to become available if it is not yet captured. When a WFD receives the read request, the fabric FPGA will control streaming of the data sequentially from each of the five channel DDR3 memory buffers via the corresponding channel FPGA. The fabric FPGA then packages its event data in the data format expected by the AMC13 event builder, which is shown in Figure 18.10.

Figure 18.11 shows the block diagrams for the fabric and channel firmware designs. The main functionality is now largely implemented and under testing on the five-channel prototype WFD AMCs. The firmware development again takes full advantage of available industry-standard designs and other high-energy and nuclear physics (CMS, in particular) development efforts. BU provides the firmware block for the backplane link to the AMC13. The IPbus firmware is available from its open hardware project. We have implemented our high speed channel FPGA $\leftrightarrow$ fabric FPGA serial link using the Xilinx Aurora link-layer pro-tocol. The remaining control architecture is built on Xilinx's implementation of the AXI4 protocol.

[2]The $g-2$ begin-of-fill signals are the equivalent of the CMS "level-1 accept" signals.

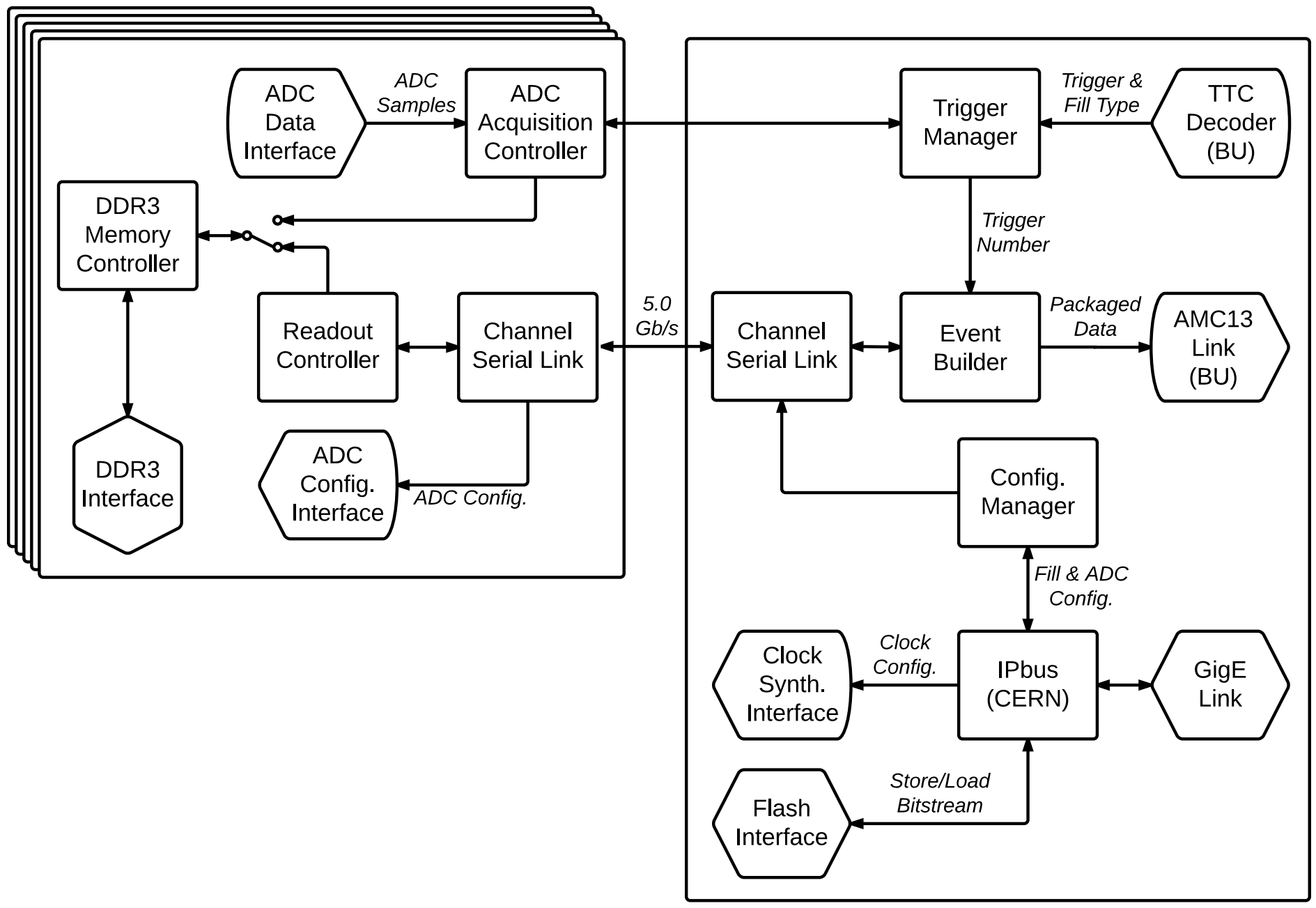

Figure 18.11: Firmware block diagrams for the WFD fabric FPGA (right) and the individual channel FPGA (left) at the time of the Revision 0 five-channel WFD.

The fabric and channel FPGAs have the ability to load their firmware from onboard flash memory at power up. Two "personalities" for the WFD operation will be available for loading. The first one is the standard $g-2$ mode that includes data taking of a 700 $\mu$s fill, laser pulses for calibration and pedestal noise. A TTC command will be issued that causes the WFD to change its acquisition parameters for the three aformentionned data taking type. The second personality is a self- or externally-triggered mode for cosmic ray running. The flash memory is remotely reconfigurable to allow straightforward updating of the WFD installation for bug-fixes, alternate personalities, etc.

**Auxiliary detectors and laser calibration system** Since the previous TDR version in July 2014, the decision was made to use the WFD AMC designed for the calorimeters to capture signals from all the auxiliary detectors (entrance counters, fiber beam monitors, fast muon kicker, quadrupoles) as well as from the laser calibration system. The WFDs for the auxiliary detectors will be gathered in the dedicated "auxiliary detector" $\mu$TCA crate, with the exception of the two WFDs for the entrance counters located in one calorimeter crate. Both the fiber harps and the fast muon kicker should use about a 200 MSPS digitization rate, and the quadrupoles should use about a 20 MSPS digitization rate. The TI LMK04906 clock synthesizer can accomodate those rates, and it was shown at Cornell U. that the WFD can work reliably at those rates. The entrance counter signal will be digitized at 800 MSPS, and the two corresponding WFD channels will be located in one of the calorimeter $\mu$TCA crates. The auxiliary detector $\mu$TCA crate will contain 12 WFDs, i.e., 30% of them being hot spares.

The laser calibration system (see Section 17.4.3) will require one hot spare channel of WFD per $\mu$TCA crate (the 55$^{\text{th}}$ channel) for monitoring the laser pulses at the detector end. A separate $\mu$TCA crate in the laser hut will host additional WFDs for monitoring the source laser pulses. The nominal WFD digitization rate of 800 MSPS will be used.

### 18.1.4 Performance

The overall system is designed to operate to stream at the 5 Gbit/s line rate of the FPGA GTX serial lines. Under nominal running conditions, the required average data transfer rate for a five channel card is just under 0.6 Gbit/s, so the design has almost an order of magnitude of headroom. Cornell U. has established the WFD AMC $\leftrightarrow$ AMC13 backplane link and recorded digitized data coming out of the AMC13, sending an analog waveform to the Revision 0 five-channel WFD AMC prototype. Extensive tests performed recently meet the design requirements. The final WFD prototype was sent for production and assembly and will be tested over summer 2015, leading to the final production of 308 WFD AMCs (15% of hot spare channels).

Simulation and test beam measurements indicate that the proposed baseline meets the basic energy, pulse separation, and random jitter requirements. As noted above, the level of control of systematic timing trends over the 700 $\mu$s fill times with the expected fill structure must still be characterized, but we do not expect a serious problem.

The proposed solution also provides the experiment with significant flexibility. Should the opportunity arise, for example, for a higher average rate of muon fills, there is no intrinsic limitation from the $\mu$TCA-based solution outlined here.

Based on measurements of the power draw of the WFD and the other modules in the crate, we expect each station to consume approximately 420 W of power, which is safely below the maximum power of the 962 W power module available for the VadaTech crate.

The $\mu$TCA solution also provides the platform for the tracker readout boards (see Section 19).

### 18.1.5 Value Engineering and Alternatives

The choice of the $\mu$TCA platform and the CMS AMC13 continue to represent significant value engineering. The $\mu$TCA platform comes with timing, cooling, power, mechanical, and remote monitoring elements all pre-engineered into the system. We found the engineering cost of a custom solution, such as one based around less expensive PCIe technology, would result in a similar total cost but with higher associated risk. The AMC13 choice for data readout with the $\mu$TCA solution has revealed several examples of value engineering. For example, we originally anticipated that $g-2$ would need custom AMC13 event builder firmware. However, with the broader dissemination of the AMC13 throughout the CMS sub detectors, CMS required similar modifications to the original AMC13 event builder firmware. The BU group engineered a common solution that works well for both experiments, halving the development cost for each experiment. Similarly, we have been able to adopt the IPbus control firmware and the Atmel microcontroller MMC firmware, and both have functioned well “out of the box” with only minor modifications for the $g-2$ environment.

We also considered COTS waveform digitizers. When approaching Struck, though, the reply we received was "For an application in the 1500 channel count I tend to assume, that a custom card may be advised to optimize performance and cost to the application." We continued on the path of developing our own.

Given both the increased signal speed in the SiPM response since the CDR and the increased anticipated muon flux, we have moved to a baseline design centered on the 12-bit 800 MSPS ADS5401 ADC from Texas Instruments. This design required a faster speed grade FPGA than the original design based on the TI ADS5463, a 12-bit 500 MSPS.

We have investigated discrete component clock circuitry on the WFD AMC card based on the AD9510 or another similar clock synthesizer. The design included a clock delay line for each channel that can correct for differences in signal path lengths from the photodetectors at the sub-clock-cycle level. Such a design would have significantly more inherent jitter (several tens of picoseconds) and would have greater sensitivity to environmental noise.

For the clock system, we have considered alternate clock sources. The clock for Brookhaven E821 was disciplined by the LORAN-C signal which is now obsolete. Undisciplined Rubidium oscillators would likely deliver the precision necessary; however, the GPS disciplined oscillator provides long term stability as well as additional features like time-stamps, which are of particular use to the field measurement. Options for the frequency synthesizer are limited by the 40 MHz signal necessary for the WFD.

### 18.1.6 ES&H

The $\mu$TCA crate for each calorimeter station will weigh approximately 30 pounds and will be supported by the calorimeter housing (see Section 17.6). Power to each crate will be supplied by a 60 – 70 V supply that connects to an in-crate power module that maintains the stable 48 V on the backplane. When fully populated with the WFDs, the each station will draw approximately 500 W of power. If the magnetic field requirements allow, the power supply will be resident on the crate. If not, the supplies will be located more centrally in the ring, with a few meter cable run and the supply voltage closer to 70 V.

The latter configuration in particular involves high voltage with several amps of current. We will ensure that all our equipment and installation conforms to the Operational Readiness Clearance criteria.

### 18.1.7 Risks

The largest risk in the WFD project arises from the distribution of the optical clock signal through the AMC13 and $\mu$TCA backplane and, in particular, whether that path will meet the frequency and phase drift requirements. To mitigate risk to the project, the WFD AMC cards are designed to allow timing and synchronization inputs via the front panel. This allows for a "brute force" technique virtually identical to that employed in Brookhaven E821 and MuLan, where multiple fanouts were used to distribute analog clock signals directly to each WFD front panel. This alternative would also require a modified clock fan-out and cabling scheme. The total differential cost to the experiment should be under $40k for engineering and production. Biases in the clock translate directly into biasing of $\omega_a$, so the clock must meet its stability requirements. We plan to incorporate in-situ monitoring of the

final upconverted clocks on the AMC modules and will also periodically test the distributed signals at each crate to ensure that they have remained synchronized.

The $\mu$TCA crate will reside about one meter from the storage ring, where the residual magnetic field ranges from 30 to 60 Gauss. The crate, electronics, and power supply can potentially perturb the precision field, both statically and dynamically. The main concern for the static perturbation is the presence of ferromagnetic materials, which must be limited to several hundred grams at the proposed location. VadaTech has previous experience in migrating other $\mu$TCA chassis from their standard steel-based configuration to an Aluminum chassis. They will provide us with a custom Aluminum chassis for the full order and sent us preliminary versions of the chassis for magnetics characterization before filling the full order. They worked with us to identify and control other areas of the crate and modules that contain ferromagnetic materials. Two modifications will be made to the crate to reduce the magnetic field perturbations.

With the final $\mu$TCA crate version in hand, we will assess the magnetic field perturbations from the magnetization of the backend electronics using the fringe fields of the precision magnet that will be established at Fermilab. We will then use our OPERA 3D simulations to extrapolate observed perturbations in the test magnet to the behavior in the fringe field of the storage ring. We will also use pickup coils to assess the dynamical perturbations, and shield the crate as necessary to minimize these. If these mitigations are insufficient, we can always locate the crates somewhat farther from the storage ring, at the expense of modest increases in the signal cable length.

Longer term drifts in the frequency will in principle cancel in the $\omega_a/\omega_p$ ratio since both the $\omega_a$ and the magnetic field measurements are tied to the same master clock. Reliance upon this cancellation would, however, require some care in the procedure that weights the magnetic field measurements with the muon fill statistics. The GPS stabilization of the master clock will minimize this drift and therefore also minimize our need to rely upon strict cancellation of any time dependence in the ratio.

In the CDR, a significant cost risk ($500,000) was associated with a higher digitization rate. Since then, the ADS5401 family of ADCs became available, affording an increase to an 800 MSPS digitization rate. With TI donating the ADCs to Cornell U., the risk decreases significantly and is associated only with the final production-level cost of the higher speed grade FPGAs. The tests performed at Cornell U. with an 800 MSPS digitization rate have shown to be successful, and there is no further cost risk associated with a higher digitization rate.

### 18.1.8 Quality Assurance

Cornell has established a $\mu$TCA test station to assess the performance of the $\mu$TCA platform, the AMC13 modules, and the WFD AMCs themselves. We are entering our third and last stage of prototyping. The initial stage gave us a one-channel design to verify the fundamental design and launch the majority of the firmware development, without facing the board density issue simultaneously. In the second stage, we moved to the full five-channel design with the denser component layout as well as migrating from the ADS5463 (500 MSPS) to the ADS5401 (800 MSPS) ADC. This prototype has undergone significant testing to assure that the baseline requirements for $g-2$ are met. In the current third stage, we keep the baseline five-channel design from the Revision 0.

We will produce enough of the second five-channel prototype to allow us to fully populate the $\mu$TCA crate, as planned for the experiment, so that we can ensure the entire system under full load can meet the specifications and that we do not encounter unanticipated cross talk or clock biasing with the full system.

Several months of burn-in and testing of the production WFD AMC modules will occur after their receipt. After an initial shakedown period to identify infant mortality, each module will be evaluated with an acceptance test suite that incorporates per-module and per-channel criteria. Some of the critical tests will include

- measurement of the phase stability and frequency upconversion factor at the part per billion level using the WFD AMC clock monitoring output,
- cross check of each channels upconversion factor to better than 10 parts per billion using a precision input sine wave,
- measurement of the linearity of each channel, and
- measurement of the noise level of each channel.

All results will be logged in a travelers database referenced by the WFD AMC serial number, which will form the basis of the modules MAC address. With production of the full complement of modules slated for fall 2015, we will have had ample time for testing and shakedown of the modules before installation at Fermilab the following summer.

The U. of Illinois group will test and evaluate each component of the clock system prior to installation. In addition, we will perform detailed in-situ timing for each path in the final experimental configuration. This "timing-in" is necessary to insure that the synchronization signals are delivered to each location simultaneously.

# Chapter 19

# Tracking Detectors

## 19.1 Physics Goals

The primary physics goal of the tracking detectors is to measure the muon beam profile at multiple locations around the ring as a function of time throughout the muon fill. This information will be used to determine several parameters associated with the dynamics of the stored muon beam [1]. This is required for the following reasons:

- Momentum spread and betatron motion of the beam lead to ppm level corrections to the muon precession frequency associated with the fraction of muons that differ from the magic momentum and the fraction of time muons are not perpendicular to the storage ring field.
- Betatron motion of the beam causes acceptance changes in the calorimeters that must be included in the fitting functions used to extract the precession frequency.
- The muon spatial distribution must be convoluted with the measured magnetic field map in the storage region to determine the effective field seen by the muon beam.

The secondary physics goal of the tracking detectors involves understanding systematic uncertainties associated with the muon precession frequency measurement derived from calorimeter data. In particular, the tracking system will isolate time windows that have multiple positrons hitting the calorimeter within a short time period and will provide an independent measurement of the momentum of the incident particle. This will allow an independent validation of techniques used to determine systematic uncertainties associated with calorimeter pileup, calorimeter gain, and muon loss based solely on calorimeter data.

The tertiary physics goal of the tracking detectors is to determine if there is any tilt in the muon precession plane away from the vertical orientation. This would be indicative of a radial or longitudinal component of the storage ring magnetic filed or a permanent electric dipole moment (EDM) of the muon [2]. Any of these effects directly biases the precession frequency measurement. A tilt in the precession plane leads to an up-down asymmetry in the positron angle that can only be measured with the tracking detectors. The goals for the systematic uncertainties that can be directly determined or partially constrained using tracking information are listed in Table 19.1.

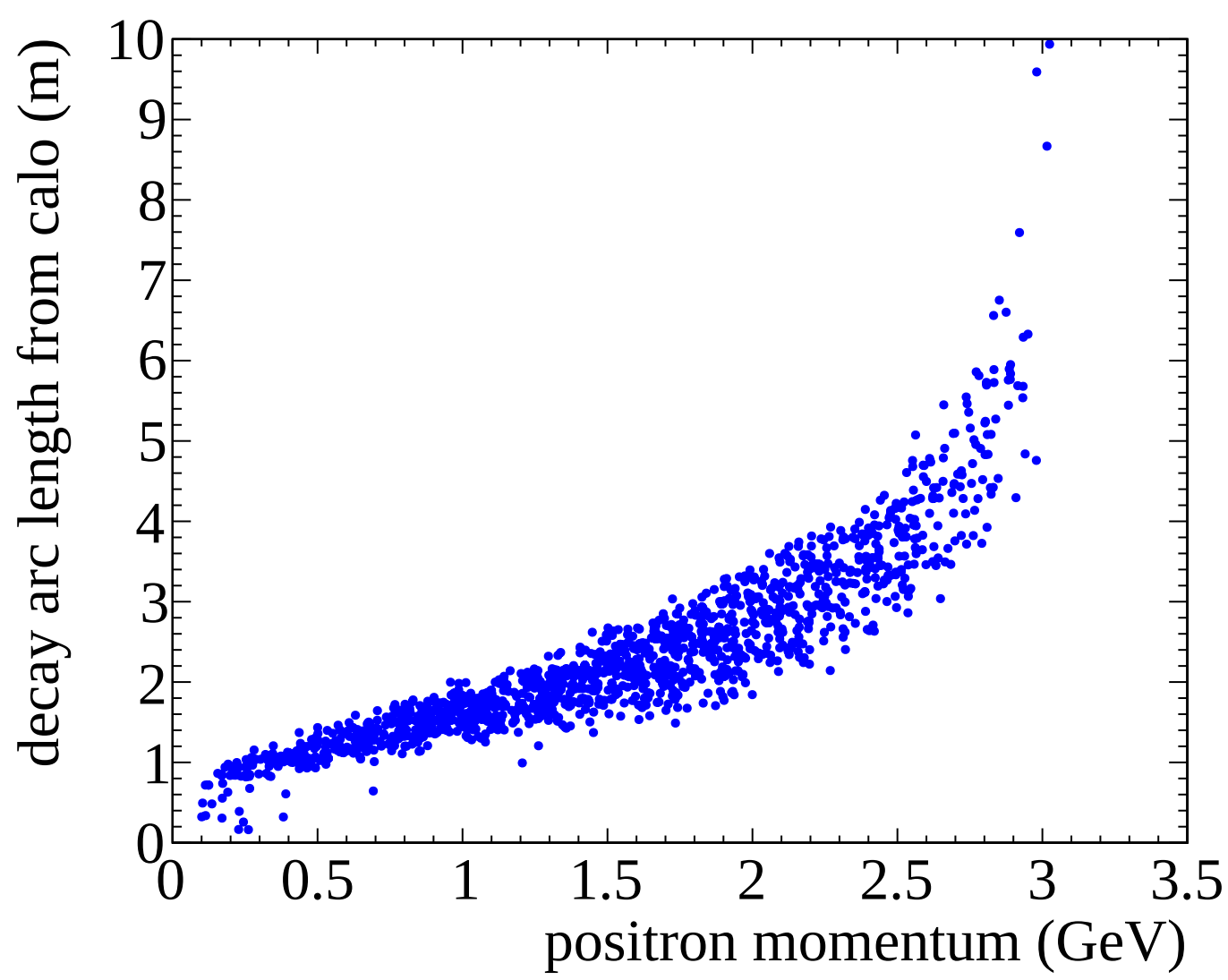

Figure 19.1: Arc length between the calorimeter and the muon decay point as a function of positron momentum.

| Uncertainty | E821 value | E989 goal | Role of tracking |
|---|---|---|---|
| Magnetic field seen by muons | 0.03 ppm | 0.01 ppm | Measure beam profile on a fill by fill basis ensuring proper muon beam alignment |
| Beam dynamics corrections | 0.05 ppm | 0.03 ppm | Measure beam oscillation parameters as a function of time in the fill |
| Pileup correction | 0.08 ppm | 0.04 ppm | Isolate time windows with more than one positron hitting the calorimeter to verify calorimeter based pileup correction |
| Calorimeter gain stability | 0.12 ppm | 0.02 ppm | Measure positron momentum with better resolution than the calorimeter to verify calorimeter based gain measurement |
| Precession plane tilt | 4.4 $\mu$Rad | 0.4 $\mu$Rad | Measure up-down asymmetry in positron decay angle |

Table 19.1: Systematic uncertainty goals for the Muon g-2 experiment. Information from the tracking detectors will be used to constrain these in several ways as indicated in the final column. The first two rows are associated with the tracker's primary physics goal. The second two are associated with the secondary physics goal of the tracker and the main role played by the tracker will be in validating the reductions in the uncertainties provided by the new calorimeters. The final row ia associated with the tertiary physics goal and the improvements are entirely from increased acceptance and statistics in the new experiment.

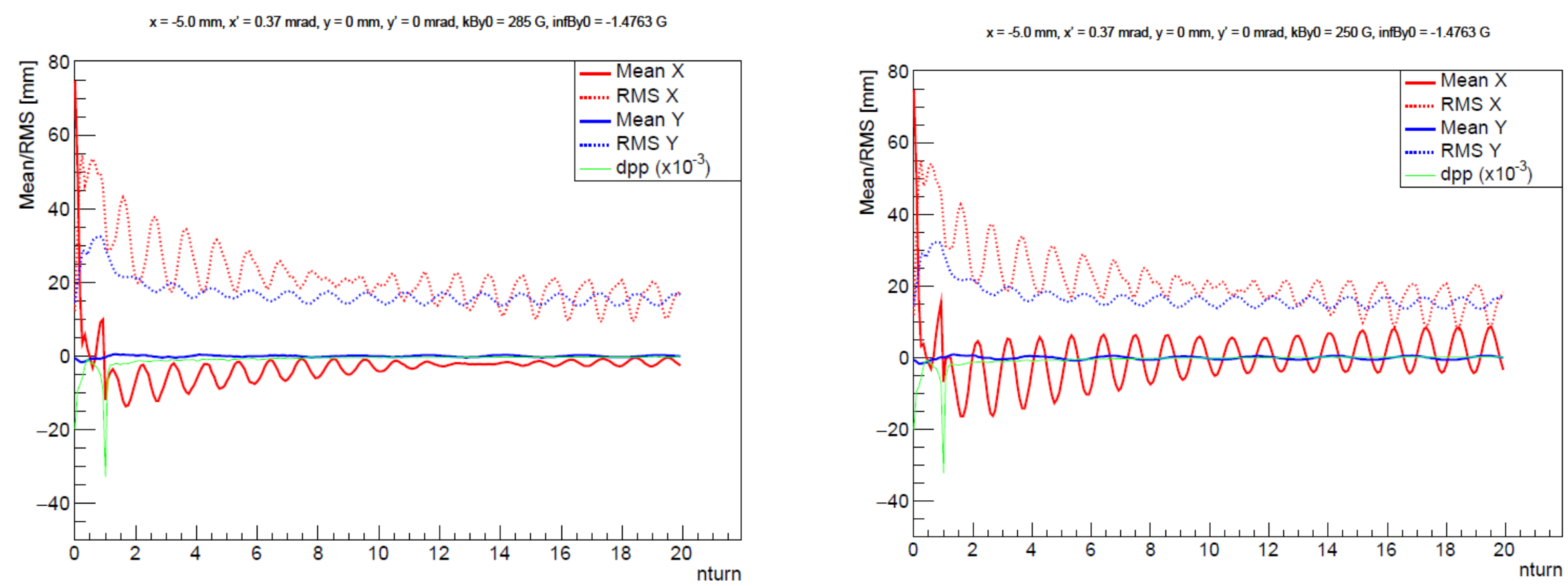

Figure 19.2: The beam position and width for optimal and sub-optimal injection parameters. The solid rd line is the horizontal mean. The dashed red line is the horizontal width. The solid blue line is the vertical mean. The dashed blue line is the vertical width. The green line is the momentum spread of the beam. The distributions demonstrate the need to measure beam positions near millimeter level accuracy.

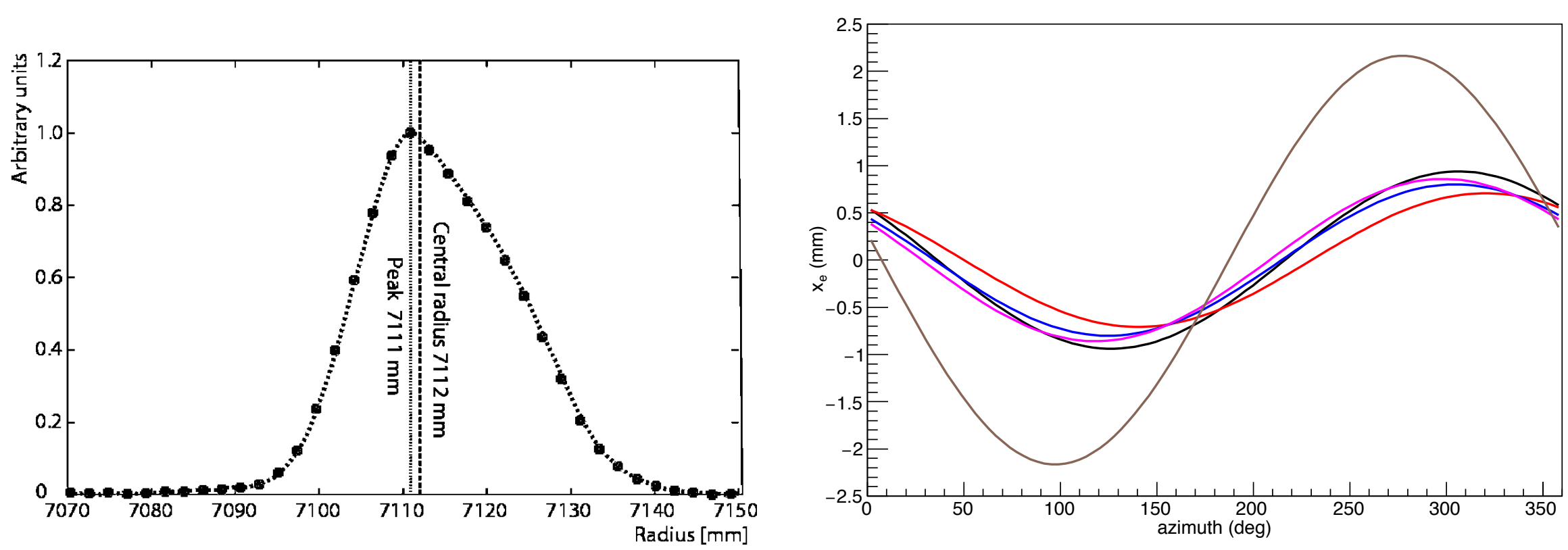

Figure 19.3: Left: The distribution of equilibrium radii measured in the Brookhaven experiment reproduced from Ref. [1]. Right: The equilibrium radius as a function of azimuth around the ring inferred from the non-uniformities in the measured field map for five different trolley runs. The distributions demonstrate the need for millimeter level accuracy and the need for multiple position measurements around the ring.

## 19.2 Requirements

The DC nature of the muon beam requires that the tracker perform well over a large momentum range and for muon decay positions up to 10 meters in front of the first tracking plane. The arc length between the calorimeter and the muon decay point as a function of positron momentum is shown in Fig. 19.1.

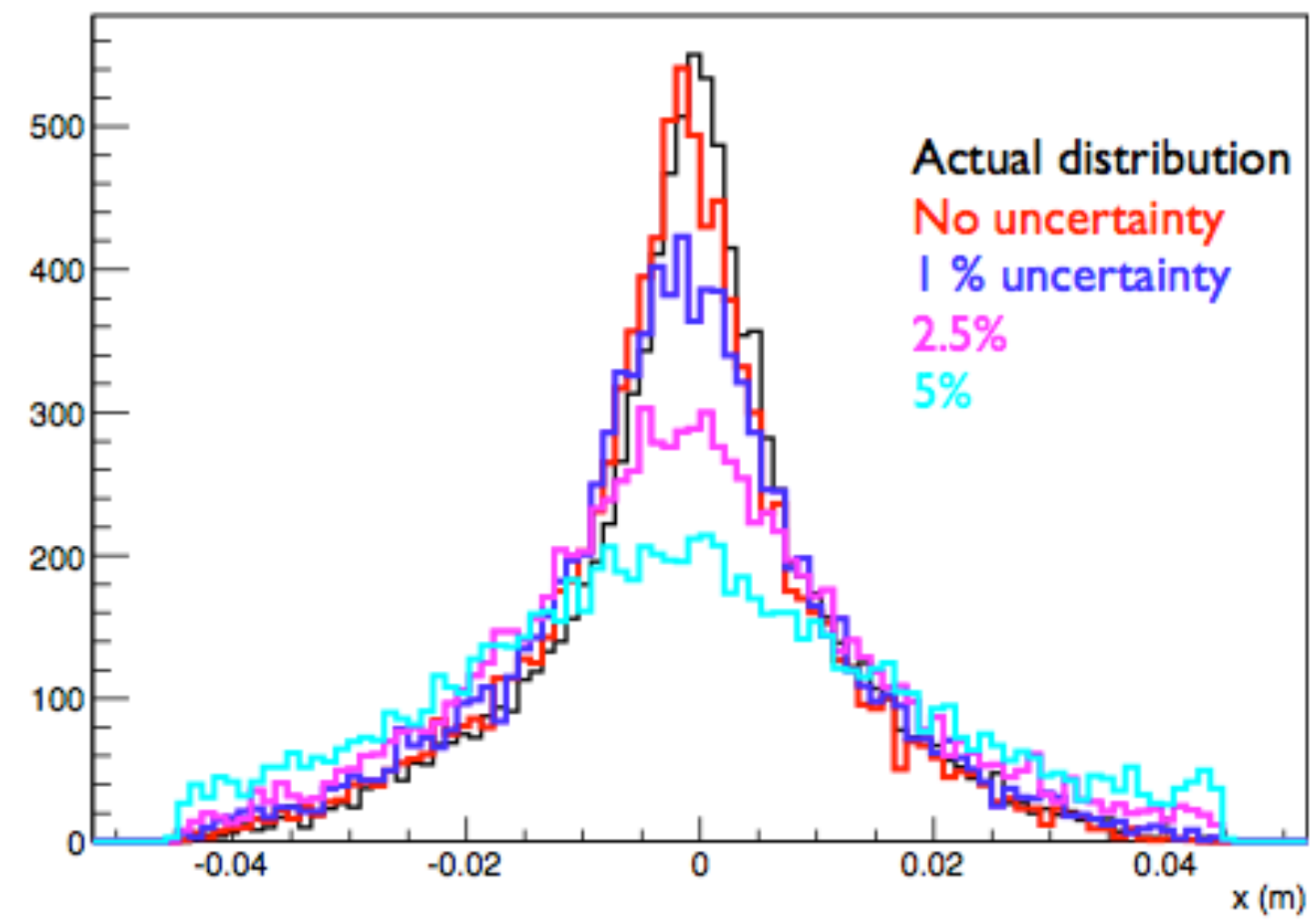

Figure 19.4: The reconstructed beam distribution in the radial dimension for several different assumed uncertainties on the reconstructed curvature of the positron track. A resolution at or below 1% is desirable to reconstruct both the beam mean and width.

The expected change in the radial beam center and width due to betatron oscillations as predicted by the BMAD simulation is shown in Fig. 19.2. The position oscillates with an amplitude of approximately 4 mm and the width oscillates with an amplitude of roughly 2 mm. A measurement with mm level resolution is required to map out this motion. Since the beam has non-integer tune, these features can be seen by a single tracker however, a second measurement position is required for cross checking and redundancy.

The equilibrium radius of the E821 beam is shown in Fig. 19.3. The beam was shifted from the central orbit by roughly 1 mm. Figure 19.3 also shows the equilibrium radii of the E821 beam as a function of position around the ring inferred from the magnetic field measurements. These two distributions also indicate that mm level precision is required. The change in the radii as a function of position around the ring can not be measured by a single tracking station and requires maximizing the number of measurement locations.

Figure 19.4 shows the results of track reconstruction code applied to the BMAD simulation for different values of the momentum resolution of the reconstructed helix. One sees that percent level momentum resolution is desirable. Fast simulations indicate that the position and momentum requirements can be met with a detector with better than 300 $\mu$m resolution per position measurement in the radial dimension in a multi-plane detector with a span of approximatly a meter. Since there is no curvature in the vertical dimension, the resolution requirements are significantly relaxed in that dimension. The long extrapolation from the tracking detector to the muon decay point requires that multiple scattering be minimized and that the material associated with each tracking plane be below 0.5% radiation length.

The trackers are required to reside in vacuum chambers in a vacuum of approximately $10^{-6}$ Torr and have either a vacuum load on the system below $5 \times 10^{-5}$ Torr $l/s$ or include a local increase in pumping speed near the tracker.

The tracker must be located as close to the stored muon beam as possible without inter-fering with the NMR trolley. Any passive material for the tracker should be located outside ±4 cm from the beam center in the vertical dimension to prevent degradation of the positron energy measurement in the down stream calorimeter. Tracking planes should be as close together as possible to maximize acceptance for low momentum positrons while the first and last planes should be as far apart as possible to provide sufficient lever arm for the long extrapolation of high momentum positrons back to the muon decay point.

Any perturbations to the magnetic field due to material or DC currents must be below 10 ppm at the center of the storage region over an azimuthal extent of greater than 2°. Any perturbations due to transient currents on time scales below 1 ms must be below 0.01 ppm since these cannot be detected or monitored with NMR [3]. The requirements are summarized in Table 19.2.

## 19.3 Preliminary Design

The preliminary design is an array of straw tubes with alternating planes oriented 7.5° from the vertical direction. We refer to the plane with negative slope as the U plane and the plane with the positive slope as the V plane with respect to the radial-vertical plane. The DC nature of the beam requires a tracker with multiple planes spread out over as long a lever arm as possible. The required number of planes, along with the need to minimize multiple scattering lead to the choice of a gas based detector. The requirement to place the detectors in the vacuum leads to the choice of straws since the circular geometry can hold the differential pressure with minimal wall thickness.

| Parameter | value | comments |
|---|---|---|
| Impact parameter resolution | 1 mm | Set by RMS of the beam |
| Vertical angular resolution | $\ll 10$ mrad | Set by angular spread in the beam |
| Momentum resolution | $\ll 1\%$ at 1 GeV | Set by calorimeter resolution |
| Vacuum load | $5 \times 10^{-5}$ Torr $l/s$ | assumes $10^{-6}$ Torr vacuum and E821 pumping speed |
| Instantaneous rate | 10 kHz/cm$^2$ | Extrapolated from E821 |
| Ideal coverage | $16 \times 20$ cm | Front face of calorimeter |
| Number of stations | $\geq 2$ | Required to constrain beam parameters |
| Time independent field perturbation | $< 10$ ppm | Extrapolation from E821 |
| Transient ($< 1$ ms) field perturbation | $< 0.01$ ppm | Invisible to NMR |

Table 19.2: Summary of the major requirements and environmental considerations for the tracking detectors.

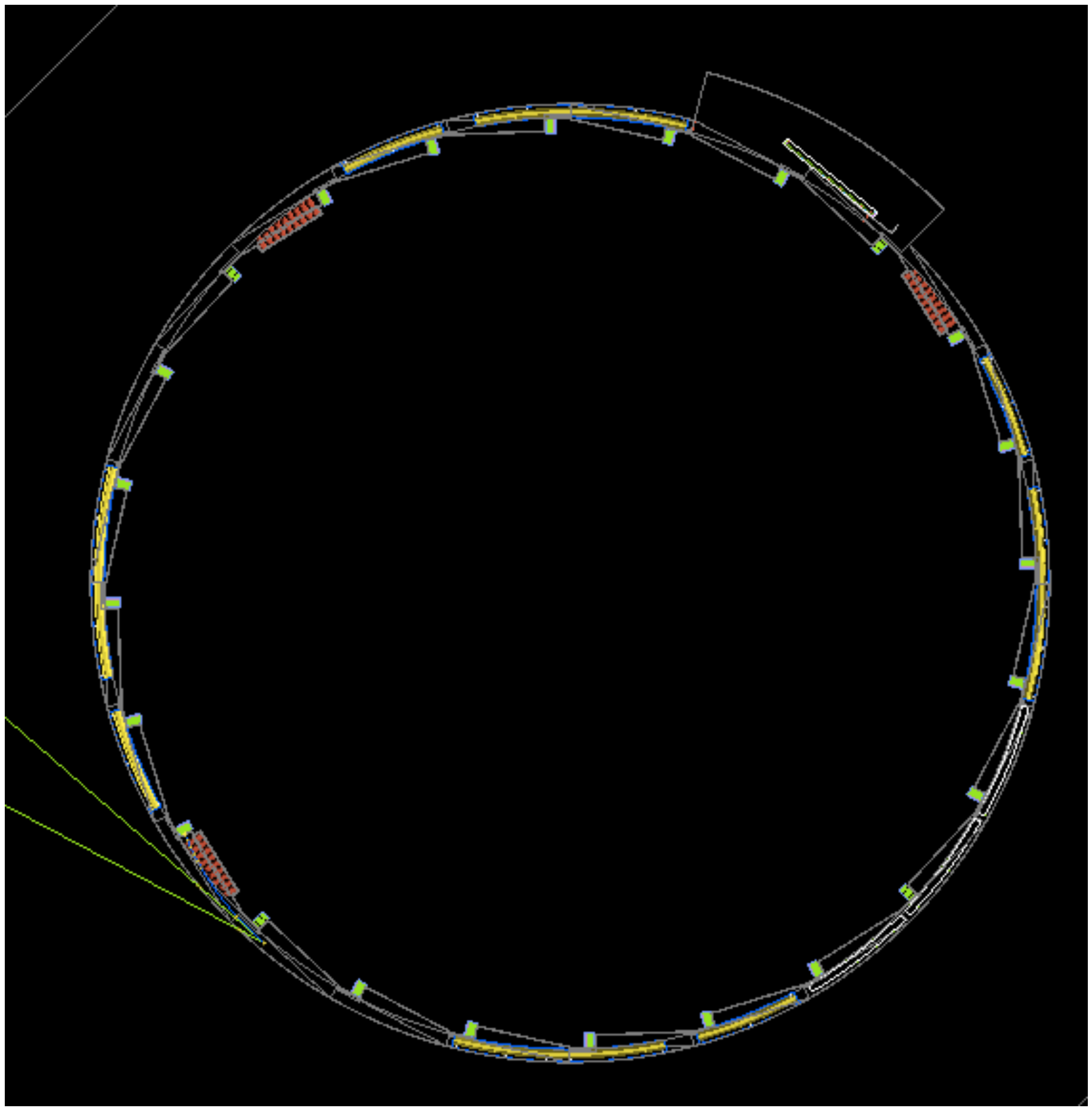

Figure 19.5: Placement of the tracking detectors in the muon storage ring. The detectors can be seen in red in front of calorimeter stations 3, 15, and 21.

### 19.3.1 Mechanical Design

The design is to have three tracking stations placed at approximately 15, 180, and 270 degrees from the injection point respectively as shown in Fig. 19.5. From these locations they have a clear line of sight to the muon beam unobstructed by quadrupoles or collimators. The vacuum chambers in these locations will be modified to provide radial space for the trackers and to contain flanges that allow for installation and servicing of the tracking detectors (see following sub-section).

Each tracking station consists of 8 tracker modules as shown in Fig. 19.6. The modules slot into the 'staircase' walls of the modified vacuum chambers. This design maximises radial coverage whilst avoiding the need to manufacture modules with several different lengths. The tracker modules have four layers of straws arranged as two close-packed doublet planes in a UV configuration oriented $\pm 7.5°$ from the vertical direction.

| | Straws | Modules | Spares | Total |
|---|---|---|---|---|
| Tracking Station | 128 | 8 | 1 | 1152 |
| Total for 3 stations | | | | 3456 |

Table 19.3: Total number of straws in the tracking system including spares.

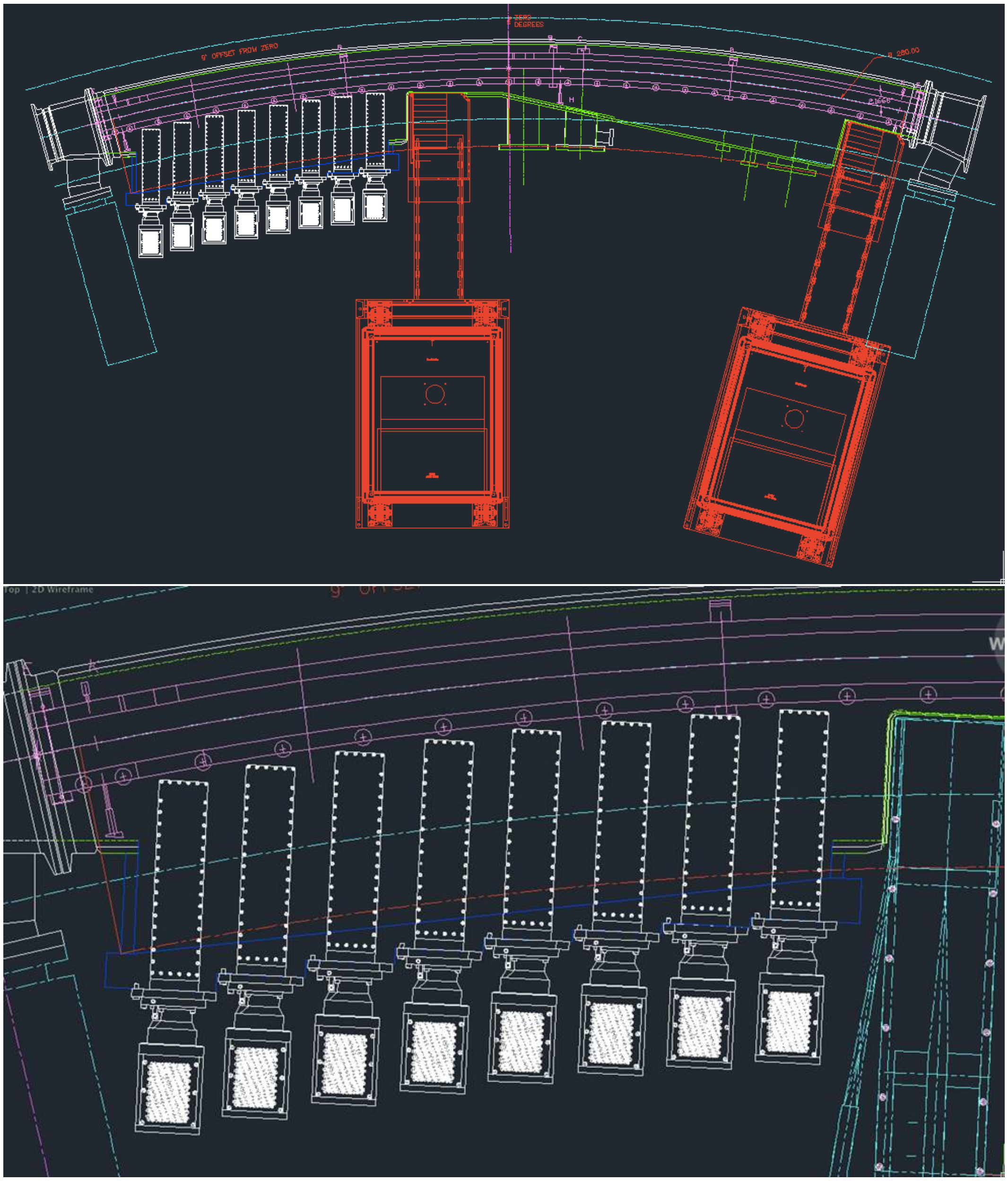

Figure 19.6: Placement of the straw tracking modules in the scallop region of the vacuum chamber. The top figure shows the tracker placement in the upstream section of one of the 12 vacuum chambers and its location with respect to the two calorimeter stations and sleds (red) in the same vacuum chamber. The trolly rail system (purple) is displayed inside the vacuum chamber.

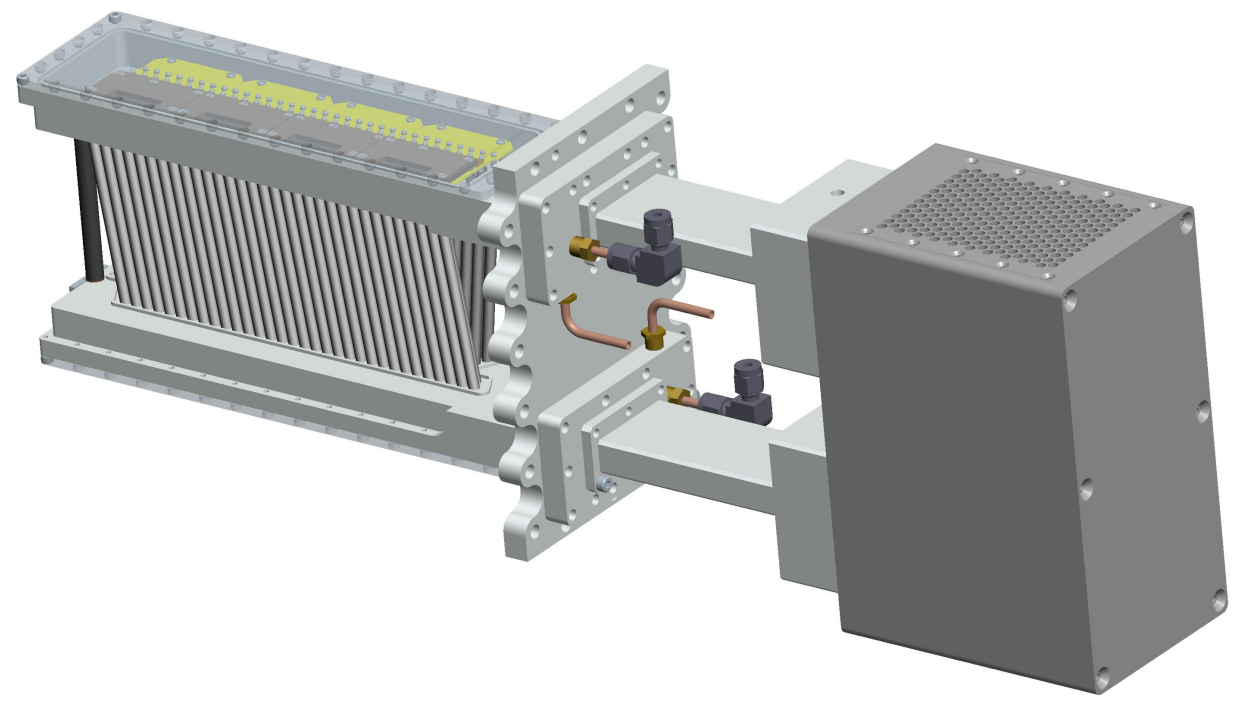

Figure 19.7: Schematic diagram of a tracking module together with the readout electronics attached. The module is 32 straws wide.

| | |
|---|---|
| Straw material | Aluminized Mylar |
| Straw wall thickness | 15 $\mu$m |
| Wire | 25 $\mu$m gold-plated tungsten |
| Straw length | 10 cm |
| Stereo angle | $\pm$ 7.5° from vertical |
| Gas | 50:50 Argon:Ethane |
| Pressure | 1 Atm |
| Operating voltage | 1800 V |

Table 19.4: Summary of the properties of the tracking detectors.

| Material | Thickness | radiation Length (cm) | $X/X_0$ (%) |
|---|---|---|---|
| Gold | 200 Å | 0.3 | $6 \times 10^{-4}$ |
| Aluminum | 500+500 Å | 8.9 | $1 \times 10^{-4}$ |
| Adhesive | 3 $\mu$m | 17.6 | $2 \times 10^{-3}$ |
| Mylar | 6 + 6 $\mu$m | 38.4 | $3 \times 10^{-3}$ |
| Ar:Ethane | 5 cm | $1 \times 10^5$ | $4 \times 10^{-2}$ |
| Total per straw | | | 0.05 |
| Total per station | | | 0.11 |
| Tungsten | 25 $\mu$m | 0.35 | 0.7 |
| Total after hitting 1 wire | | | 0.82 |

Table 19.5: material budget in the active region of a station.

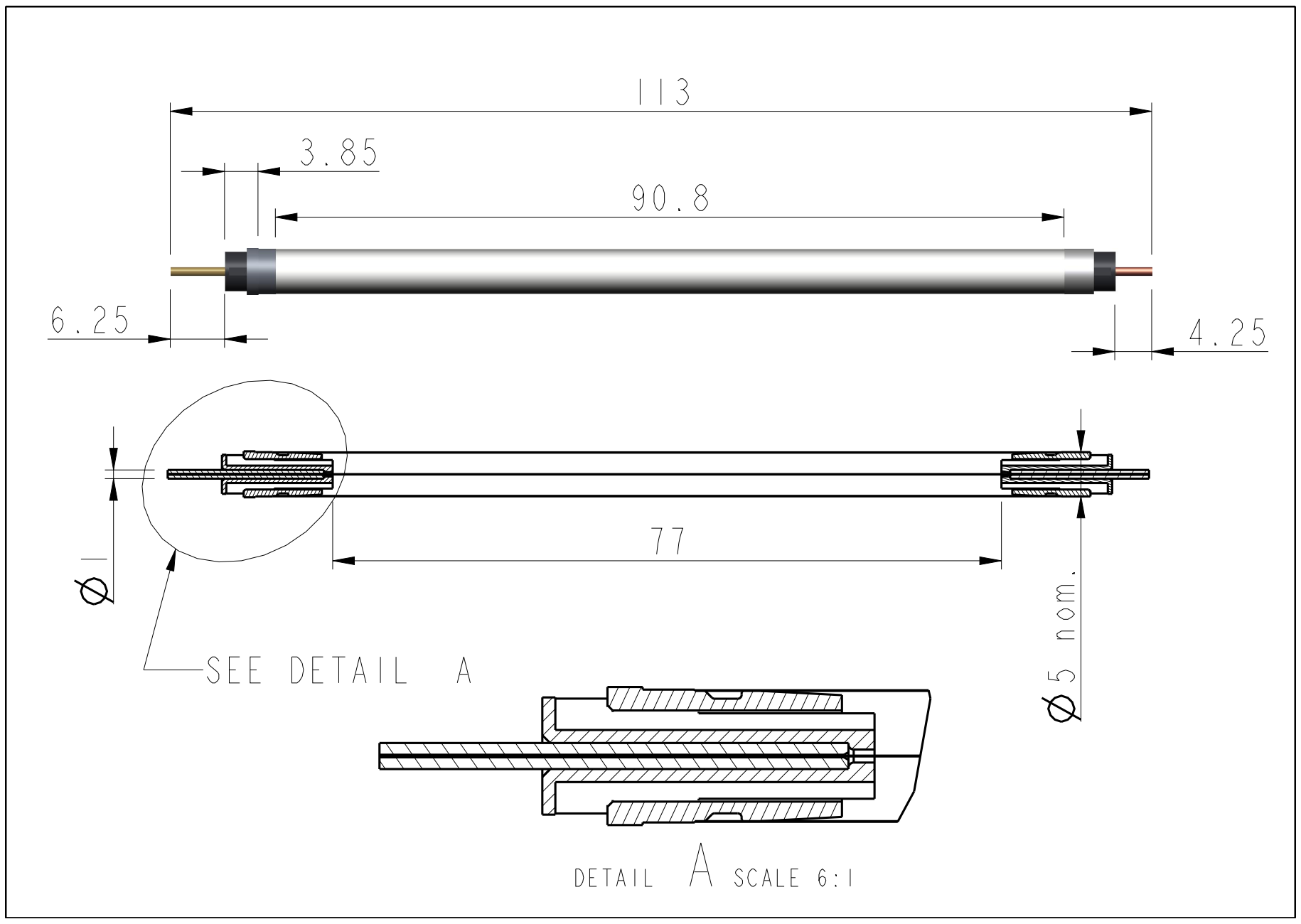

Figure 19.8: Side view and cross-section of assembled straw. The aluminum end-piece on the right is designed to pass through holes in the manifold. The 'top hat' structure of insert on the left will not. Straws are tensioned against this before being epoxied into place. Plastic inserts (shown red) and crimp pins are press-fitted into the aluminum inserts. The pins center the wires and hold them under tension and provide electrical connection to the first-stage electronics.

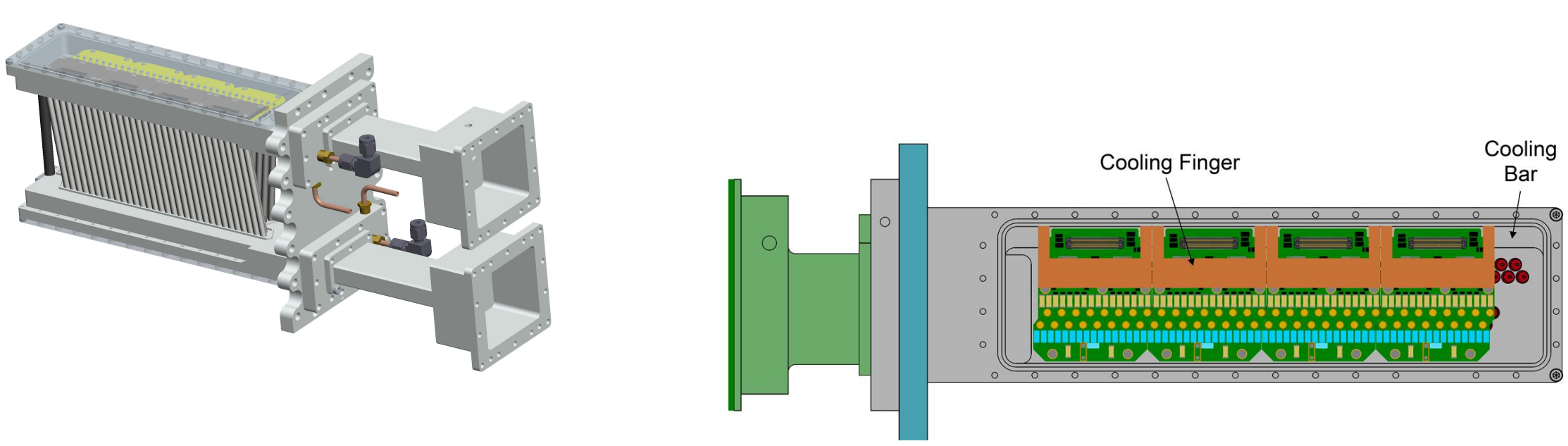

Figure 19.9: Water cooling system for the modules. Heat generating chips are thermally linked connected to a cooling bar machined into the manifold manifold (right). The bar is cooled by water flowing through a longitudinal hole containing a concentric tube for the return flow (left)

The modules are 32 straws wide (i.e. contain 128 channels each). The total channel count including prototypes and spares is listed in Table 19.3.

A schematic diagram of a module is shown in Fig. 19.7. The active height of each station is 7.63 cm. The straws are mounted between aluminum gas manifolds which also house the first stage of the readout electronics.

We have chosen a system based on Mu2e straws [4]. Each straw has a 5 mm diameter and is 10 cm long. The straw wall is made of two layers of 6 $\mu$m Mylar, spiral wound, with a 3 $\mu$m layer of adhesive between layers. The total thickness of the straw wall is 15 $\mu$m. The inner surface has 500 Å of aluminum overlaid with 200 Å of gold as the cathode layer. The outer surface has 500 Å of aluminum to act as additional electrostatic shielding and improves the leak rate. The straws are attached to the manifolds at the ends and tensioned to 65 grams to compensate for expansion under vacuum. The straw parameters are summarized in Table 19.4. The material budget in the active region of each station is given in Table 19.5.

Aluminum endpieces are glued into the straw ends with a combination of silver and structural epoxies. The endpieces locate into holes in the manifold and provide electrical contact for the straws in the aluminum manifolds. The sense wire is 25 $\mu$m gold-plated tungsten centered in the straw. The wire is tensioned to 30-50 grams and held in place by a crimp pin embedded in a polycarbonate insulator inserted into the aluminum end-pieces, as shown in Fig. 19.8. The plastic inserts contain slots to allow gas-flow through the straws.

The wire will be held at a voltage of 1800 V. The drift gas is 50:50 Argon:Ethane. This provides gain $\sim 2 \times 10^6$ and radial position resolutions $\sim 100$ microns.

The top and bottom manifolds of each station extend to the inner wall of the vacuum chamber where they are mounted to a flange. Each manifold is machined from a single piece of aluminum. The pattern of holes for the straws is a compromise between the need to pack the straws as closely as possible to maximize acceptance and the practicalities of machining. Although a prototype with closer packing was successfully machined, the techniques used would be too time consuming for the production trackers. Simulations have verified that the acceptance with slightly wider spaced doublets remains adequate. The box-like structures have aluminum lids with indium seals to close them off. The top and bottom manifolds are bolted to the end-flange again with indium-seals providing the vacuum tightness. This enables the entire station to be inserted into the vacuum as a single piece. The flange contains feed-throughs for the water cooling and is attached to snouts (see Fig. 19.7) which route the electronic signals to the TDC boards. The gas inlet and outlets are also attached to the snouts. This arrangement means that no vacuum feed-throughs are needed. In addition to the support provided by the end flange, the manifolds are held at a fixed separation by carbon fibre posts. This separation must be maintained under forces from tensioning of the straws and sense wires balanced by expansion when the station is in vacuum. A full FEA analysis is needed to determine the distribution of straw and wire tensions in the final design. The eight tracking stations are closely packed in the scalloped region.

### Cooling System

The in-manifold electronics, in particular the 8 ASDQ chips generates 4-5 Watts and require cooling. The trackers are water-cooled. The flow of chamber gas helps to maintain a uniform temperature distribution. The interior space of each manifold has a cooling bar in the form of a 'shelf' machined into one edge. It is drilled longitudinally with a 5mm hole into which a 3mm aluminium tube is inserted. This tube is wrapped with a spiral of wire to keep it concentric with the hole. Chilled water flows in through inner tube and back through the gap between the inner tube and the hole. The ASDQs are thermally connected to the cooled shelf via copper cooling fingers (see Figure Fig. 19.9). Simulations indicate a temperature differential of $\sim 3C$ between input and output water temperature.

### Vacuum Chamber Modifications

There are three design considerations that drive the design of the vacuum chamber modifications. First is the desire to use all possible space (14.8 mm) in the interior of the vacuum chamber for the tracking modules. Second is the desire to be able to replace a single module with a spare with minimal down-time in the event that a straw or wire is broken or a problem arrises in the readout electronics. Third is the desire to attach the tracking modules directly to a flange to simplify insertion and maintain alignment.

These considerations all lead to the need for the flange bolt pattern to be outside the present vacuum chamber height. For the current vacuum chambers, that is impossible since they are constrained to be between the pole pieces of the main magnet. This requires that the vacuum chambers be extended in the radial dimension so that the flange can be added with sufficient clearance from the pole pieces.

The design of the modified chamber is shown in Fig. 19.10. The extension will be welded onto the existing chamber. The vacuum chamber modifications will be done in the Boston University Scientific Instrument Facility (SIF), where the components of the E821 chambers were fabricated. To minimize the number of welds needed, the extension to the chamber will be fabricated of of a single piece of stock, rather than making it from four or more pieces welded together. This design is at the strong urging of Heitor Mourato, the Director of the SIF, who points out the multiple welds will both deform and soften the aluminum, perhaps compromising the integrity of the chambers.

A further design consideration is the desire to add pumping speed locally near the straws. A vacuum port will be added to the bellows in front of the vacuum chamber. A cryo pump will provide 1200 $l/s$ additional pumping speed for Argon. A secondary vacuum wall made of aluminized mylar will also be placed between the tracking station and the muon storage region. This adds material in front of the tracker so the wall will be designed so that it can easily be removed if it is not required.

### Gas Distribution

An 50:50 mixture of Argon:Ethane will be distributed independently to each tracker module with the supply entering the top manifold and the return exiting the bottom manifold. The nominal flow rate is 15cc/min/module. The gas will be stored in a trailer next to the Muon $g-2$ building and the gas purity will be verified for each gas shipment.

The gas system is shown in Fig. 19.11. The entire system can flow either N2 or Ar:Et. Flow rate sensors are placed on the supply and return lines for each module. Gas flow to each module is controlled with a solenoid with shut off capabilities controlled by PLC. Input and output bubblers maintain atmospheric pressure in the system. Check valves are in place upstream of the output bubbler to prevent oil from being drawn back into the system if gas flow is cut off while the straws are under vacuum. The final return is vented to atmosphere on top of the fenced in berm behind the MC-1 building.

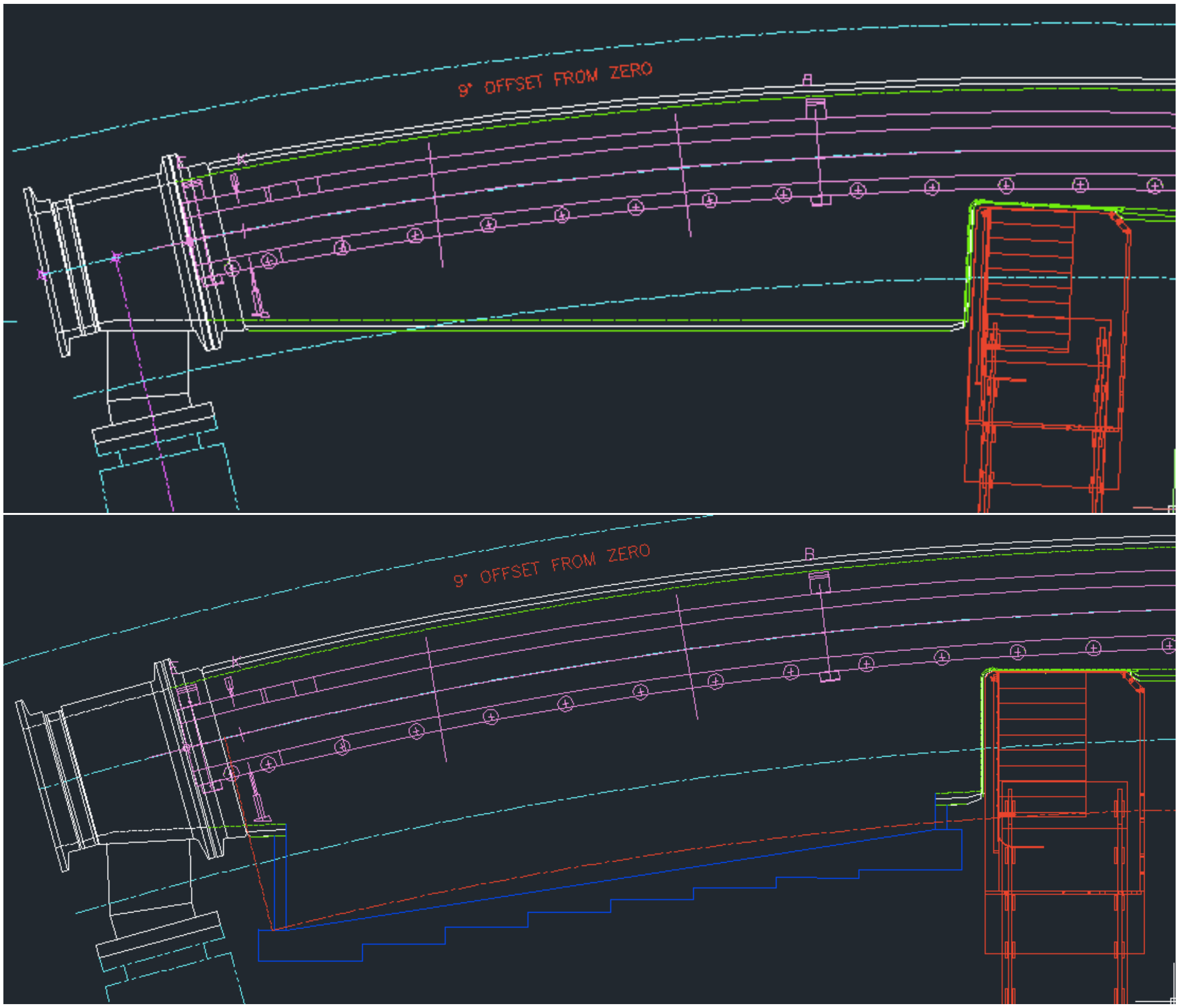

Figure 19.10: Modifications to the vacuum chambers to allow for the tracking stations. The distribution above is an unmodified chamber. The distributions below are a modified chamber. The green lines are the existing vacuum chamber. The blue lines are the extension.

The total charge in the hottest region is expected to be 3 mC/cm. For a pure mixture of 50:50 Ar:Et. we would expect a 20% loss in pulse hight over the lifetime of the experiment based on the literature [6]. The aging process can be eliminated with the proper additive. Based on CDF experience, we plan to introduce part per mil concentration levels of $O_2$ in the gas mixture.

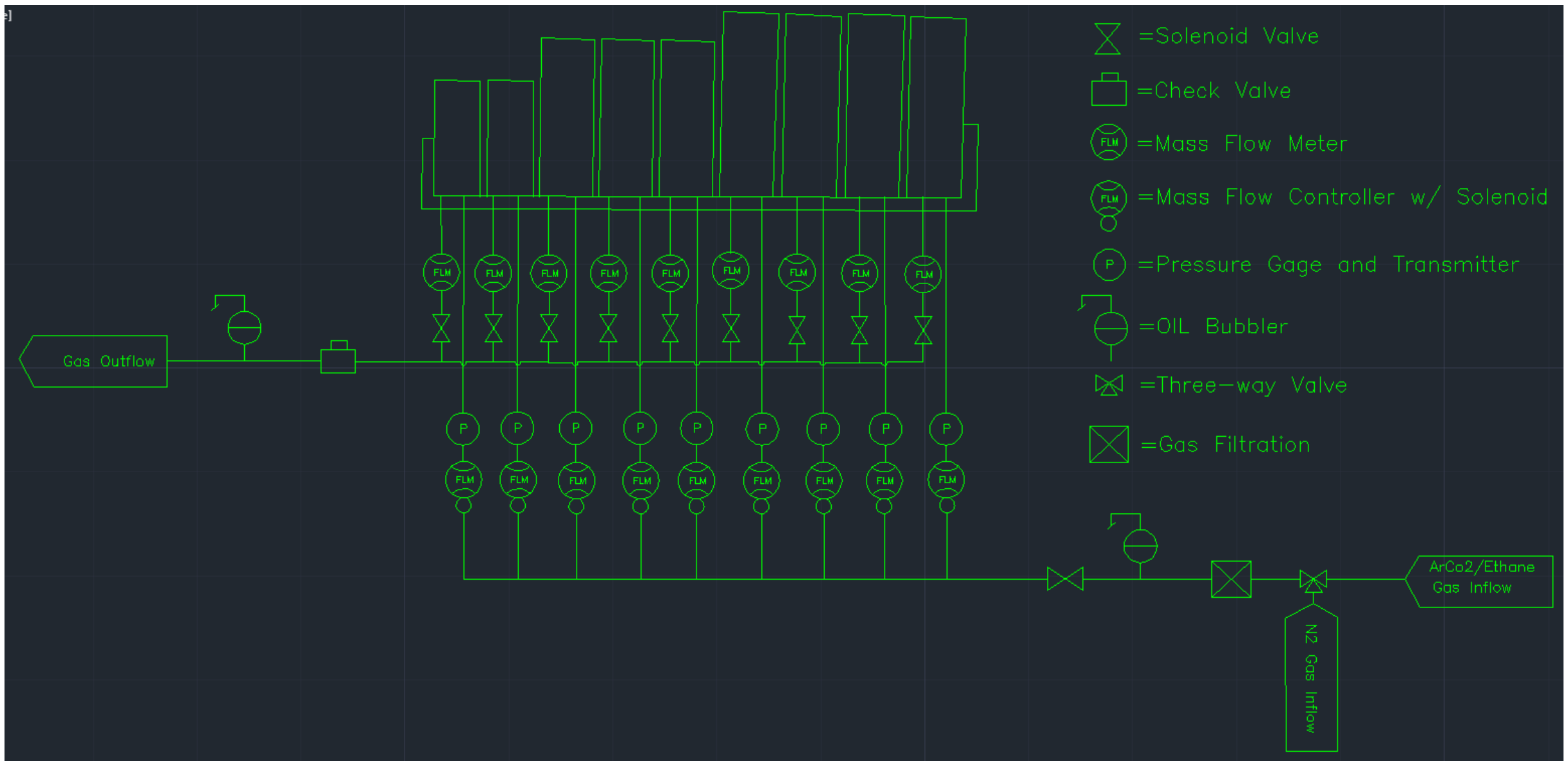

Figure 19.11: Schematic of the gas distribution system

### 19.3.2 Readout Electronics

#### Front-End Readout Electronics

The design of the front-end readout electronics is illustrated in Fig. 19.12. Groups of 16 straws are connected directly to an *ASDQ* board containing two 8-channel ASDQ [7] chips. These ADSQ chips process analog signals and send digital signals over Flexicables to a *TDC* motherboard outside of the gas volume. Each TDC motherboard contains two TDCC boards that serve two ASDQ boards, for a total of 64 straw channels per manifold. A *Logic* board, which serves as an interface between the ASDQ and TDC boards and the back end electronics, and a *High Voltage* card which provides high voltage for the straws, also sit outside the gas volume. Details on each of these modules are presented in the following sections. A CAD view of electronics setup is shown in figure Fig. 19.13.

The orientation of the ASDQ boards inside the gas manifold can be seen in Fig. 19.14 - the straws are mounted directly to the boards. High voltage is supplied by an external line to each of the ASDQ boards separately. The ASDQ board filters the high voltage and supplies this potential to each straw through a 100 kΩ current limit resistor. This HV is blocked from the signal channels by 2 kV 470 pF SMT capacitors. The board is grounded to the aluminum manifold via seven aluminum stand-offs to which the board is screwed.

The connection from the straws to the ASDQs is sketched in Fig. 19.15. A protection circuit consisting of four Shottky diodes in a 2x2 mm DFN package provide bipolar protection for both the primary and inverting ASDQ inputs and a 10 Ω series resistor limits peak current.

The role of the ASDQ ASIC [7] is to provide amplification, shaping, a discriminator and charge measurement for eight straws. Though developed for the CDF Central Outer Tracker, it provides a good match to the Muon $g-2$ tracking detector requirements. It provides fast charge collection (≈ 7ns), good double pulse resolution (≈ 30 ns), low power ($\approx$ 40 mW/ch) and low operational threshold ($\approx$ 2 fC). Baseline restoration and ion tail compensation using a pole-zero cancellation technique are provided. The output of each ASDQ is eight LVDS pairs, with leading edge representing the threshold crossing time and the pulse width proportional to input charge.

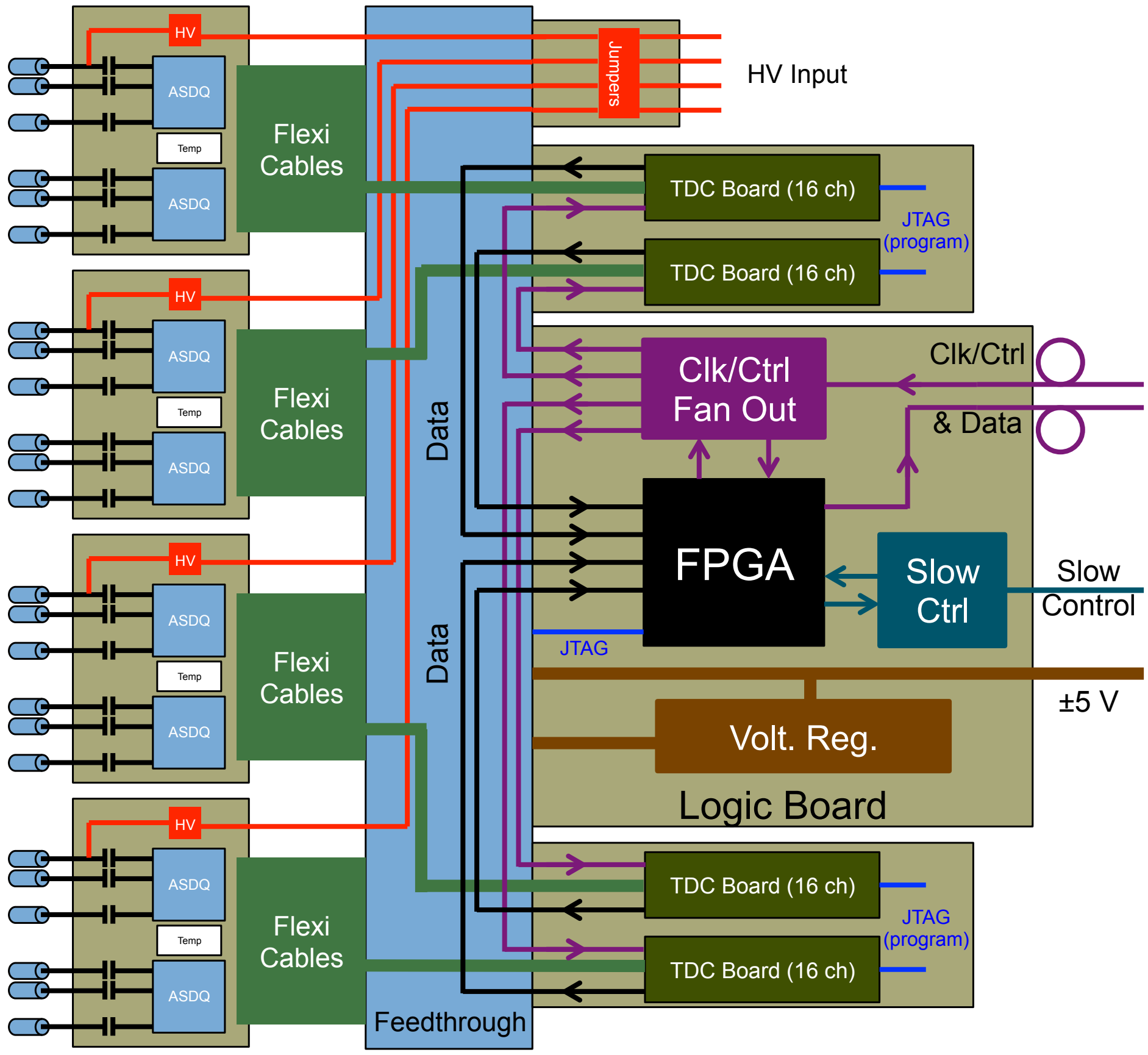

Figure 19.12: Tracker Readout Electronics for 64 straws. 16 straws connect directly to each ASDQ board, which then pass digital signals to a TDC board.

Several tools for control and calibration of the ASDQ ASICs are available. The thresholds for the leading and trailing edge discrimination can be controlled externally. The width of the discriminator pulse can also be adjusted by varying the drain current into the integrating capacitors of the ASIC's discriminator circuit. The ASDQ ASIC is equipped with calibration circuits which, with an external trigger, can feed realistic input pulses into the front end. Using the calibration circuits, the circuit delay time and pulse width-to-input charge relationship can be determined on a channel by channel basis. The ASDQ boards dissipate approximately 1 watt per manifold. Thermal energy is removed from ASICs by a copper cooling finger, which transfers heat to a water cooling-line.

In the previous design, the ASDQ and TDC boards were both located in the gas manifold. Because of cooling problems, the TDC boards have been moved out of the manifold. In this new design, the LVDS pairs produced by the ASDQ are carried through Flexicables to the TDC boards. There are four Flexicables (of different lengths) per manifold, eight in all per station. The Flexicables have a differential impedance of 100 Ω. The Flexicables also carry power (±3 volts, +1.4 volts), control signals, reference voltages and calibration pulses in to the ASDQs, as well as the output of local temperature sensors. The Flexicables are coupled to the ASDQ boards with 80 bin header connectors and Zero Insertion Force (ZIF) connectors on the output (TDC) side. Because of the relatively long path through the Flexicables to the TDC boards, the LVDS signals from the ASDQ are buffered on the TDC boards before being processed by their FPGAs.

The connection between the Flexicables and the TDC boards is shown in Fig. 19.13(b). The TDC motherboards are mounted outside the gas volume in an electronics rack known as the Flobber (Frontend Low voltage Optical Box to BackEnd Readout). Signals which pass between the detector and the Flobber pass through a snout which is mounted to a flange on the manifold. A *Feedthrough* board, adjacent to the Flobber and mounted on the snout, passes signals in and out of the gas volume, serves as a data bus and also provides a seal for the gas volume. The electronics modules within the Flobber dissipate approximately 11 watts per manifold. To keep the modules from overheating, the Flobbers are each equipped with cooling ductwork connected to remote fans.

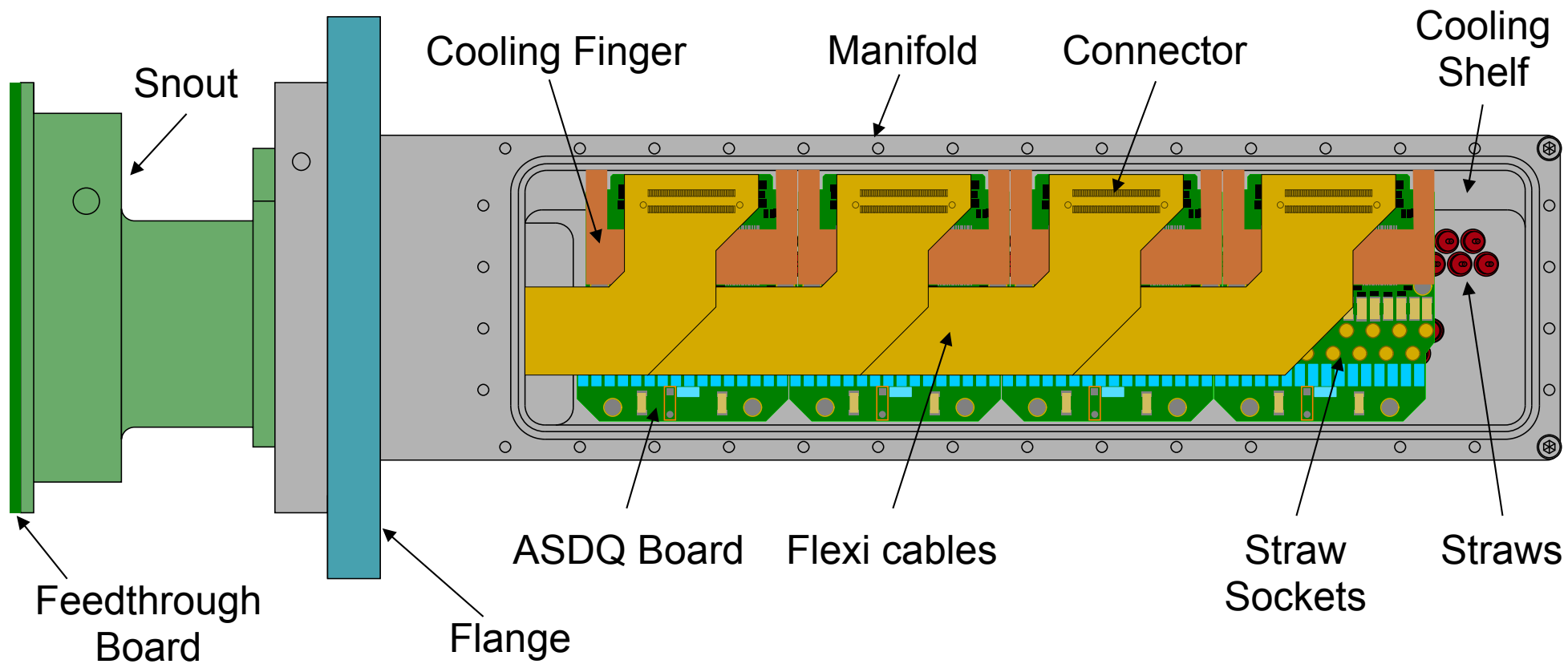

(a) Top View of Manifold Electronics

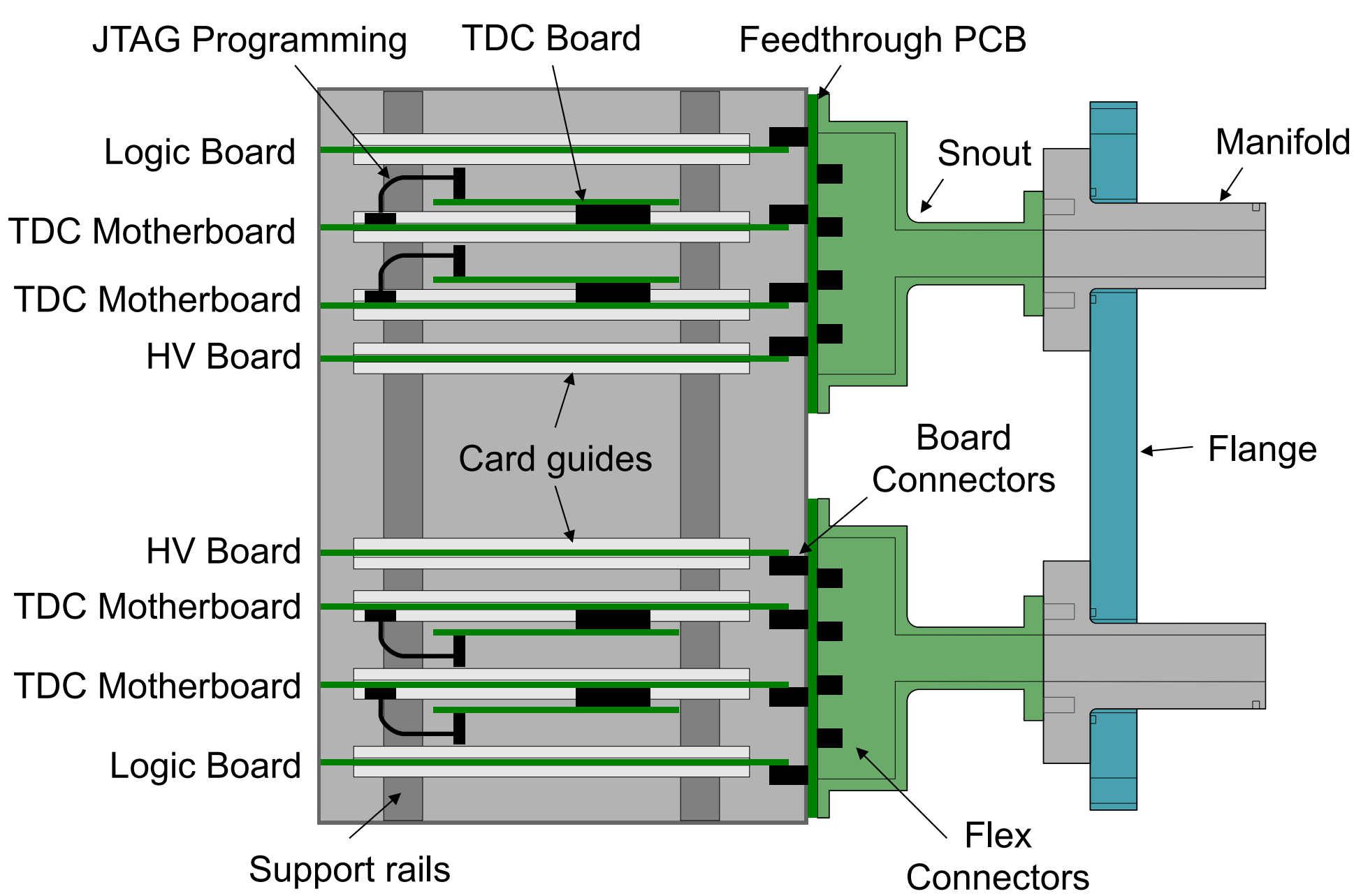

(b) Front View of External Electronics

Figure 19.13: Top down view of electronics and Flexicables within gas volume (Top), side view of external electronics (Bottom).

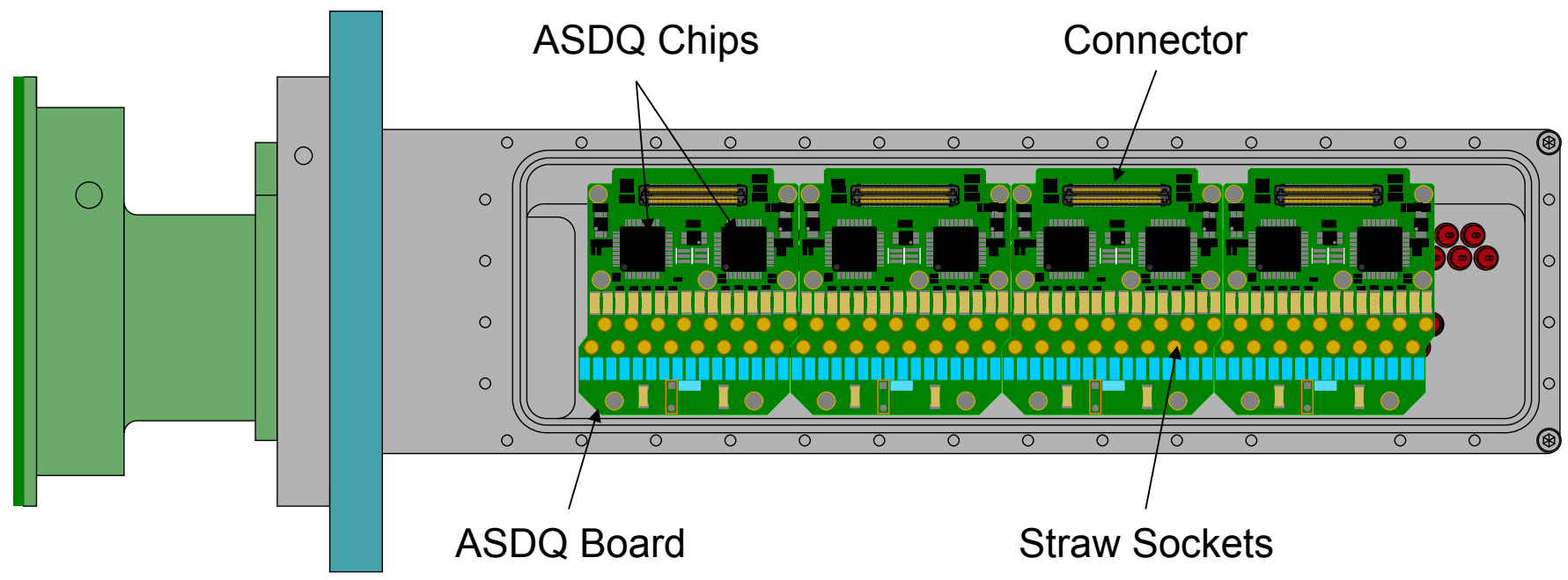

(a) Top View of Four ASDQs boards in Manifold

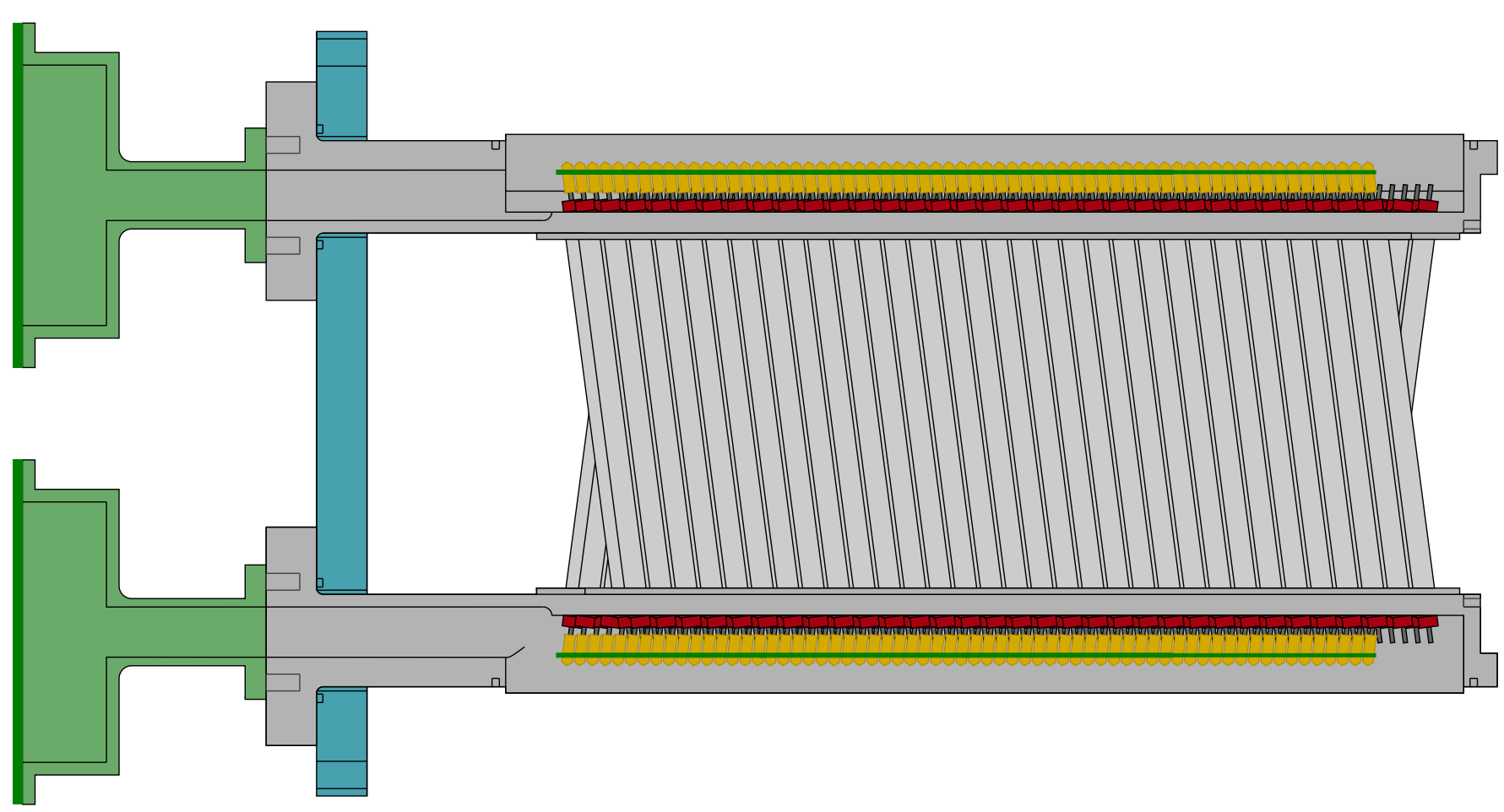

(b) Side View of ASDQ boards and straws

Figure 19.14: Top view of electronics within gas volume (Top), side view of internal electronics and straws(Bottom).

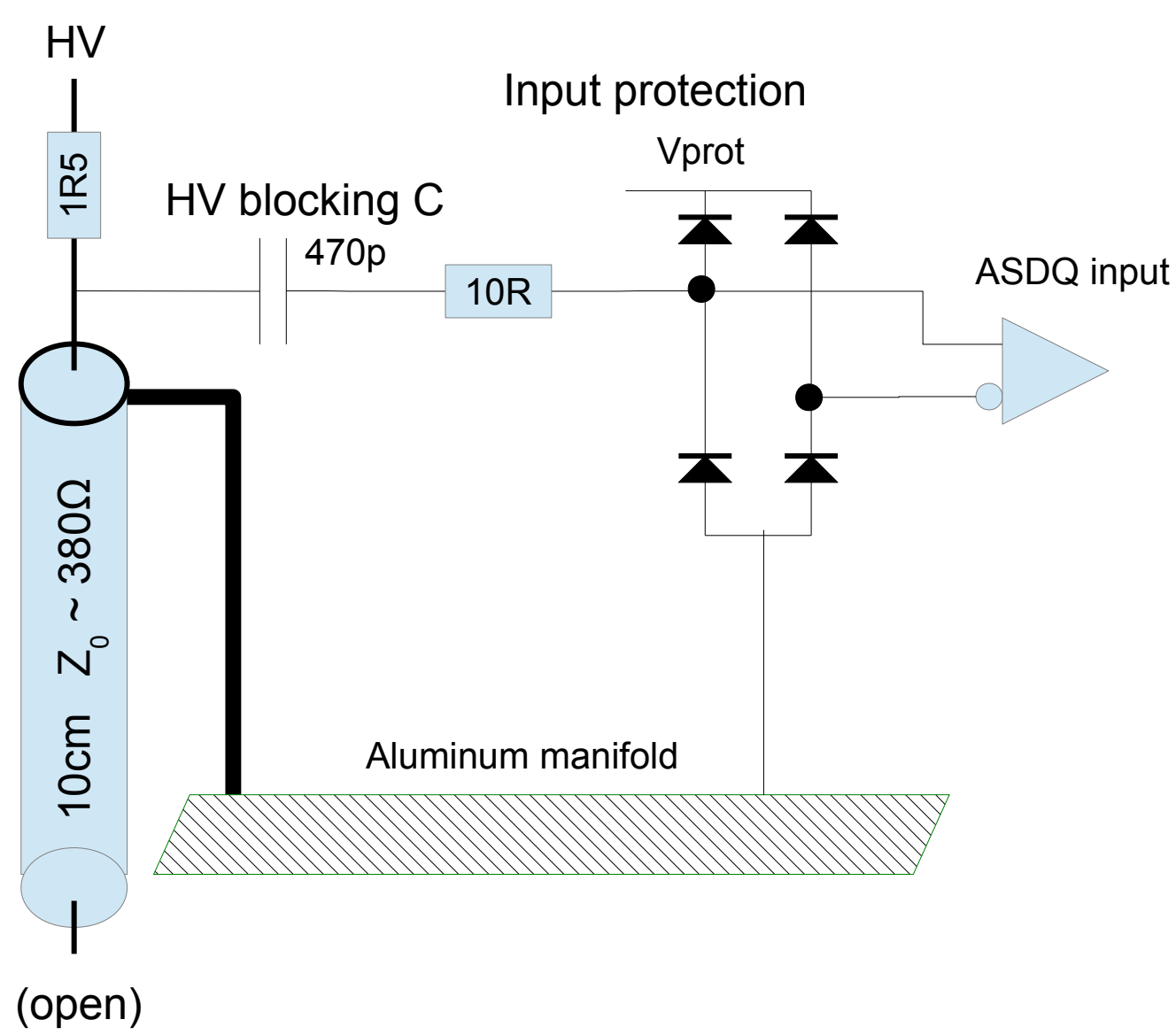

Figure 19.15: Connection between a straw and an ASDQ input.

The TDC modules consist of a TDC motherboard and two daughtercards. Each TDC daughtercard handles two ASDQs (16-channels) and digitizes their output to an accuracy of 625 ps LSB. The TDC circuits are implemented in Altera EP3C5F256C6 FPGAs and the design uses about 50% of the FPGA logic resources. A reference clock of 10 MHz is provided externally on an LVDS signal pair with multiplexed trigger and control signals. The clock is multiplied internally to a four-phase 400 MHz clock for time measurement and internal operation. Up to 2k TDC hits are stored on-chip and read out over a single serial link at 25 MHz. A block diagram of the TDC is shown in Fig. 19.16.

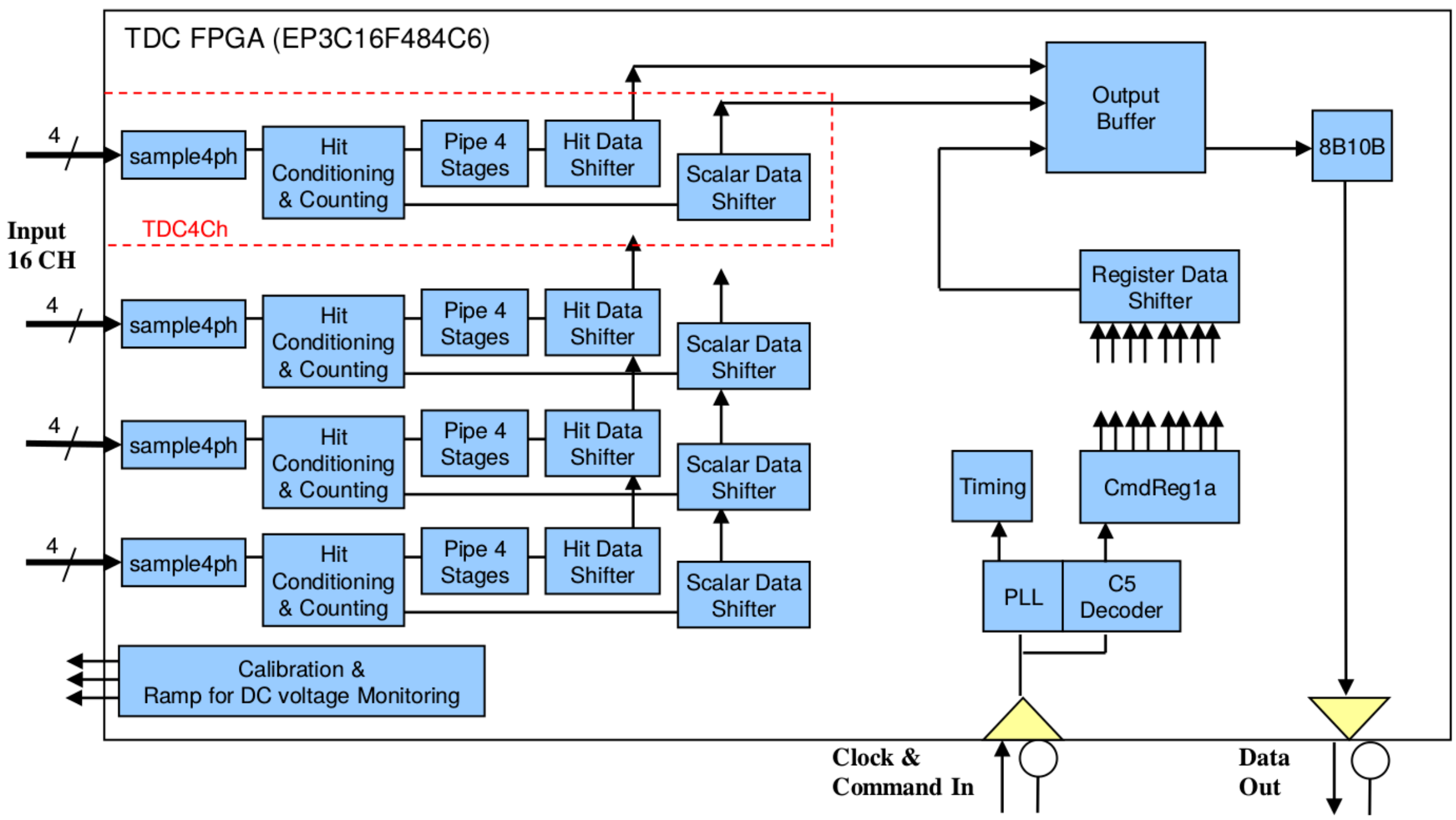

Figure 19.16: TDC Block Diagram

In response to a start-of-spill command, the TDC will commence a capture/readout cycle for one spill. A timing diagram is shown in Fig. 19.17. The start command resets the TDC time stamp to zero and (after a programmable delay) activates the TDC time recording. Up to 2016 hits are recorded during a programmable window, which would typically be 700 $\mu$s to 1 ms. At the end of this window, the TDC can optionally digitize DC voltages or other quantities to be monitored using a Wilkinson ADC scheme with external voltage ramps. At the end of this period the data is transmitted to the Logic Board.

The TDC interface consists of two LVDS pairs carrying serial data. The input clock/command stream is encoded using the C5 (Clock-Command Combined Carrier Coding) protocol at a rate of 10 MHz. The output data is 25 Mbit/sec serial encoded using 8b10b.

The C5 protocol is a DC balanced coding scheme for transmitting messages along with clock in a single channel. Rising edges are equally-spaced at 100 ns and used to recover a 10 MHz clock. Three pulse widths (25%, 50% and 75% of the bit period) are used to encode commands. An idle pattern of 50% pulses is sent when no commands are being transmitted.

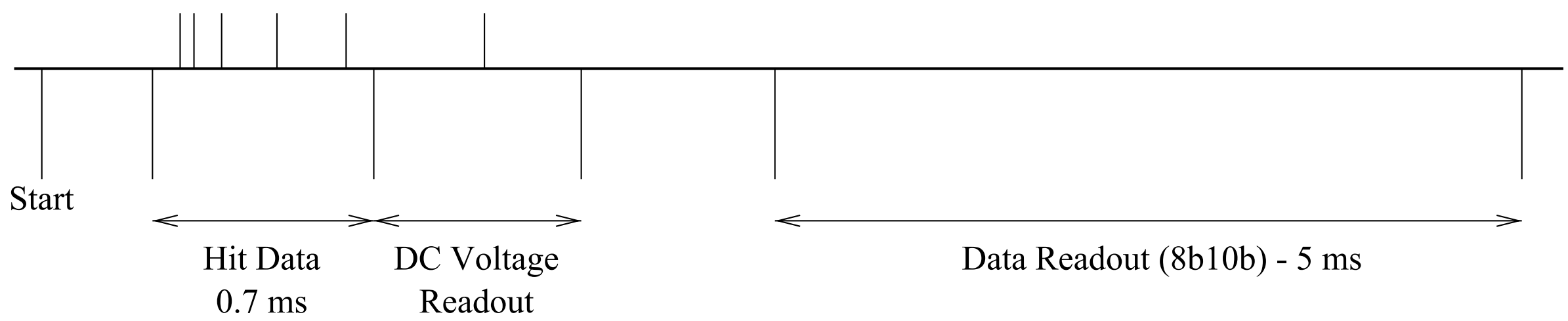

Figure 19.17: TDC Timing Cycle

Various specific patterns are used to trigger actions in the TDC. The C5 data and control codes are shown in Table 19.6 along with the use defined for $g-2$. Several registers control aspects of the TDC operation, as listed in Figure 19.18.

| 4B/5C Codes | Widths of Pulses 0 | 1 | 2 | 3 | 4 | $g-2$ Meaning |
|---|---|---|---|---|---|---|
| Q0 | 0 | 0 | 0 | 0 | 0 | Idle |
| D15 | +1 | -1 | +1 | -1 | 0 | Reset Spill Number |
| D14 | +1 | -1 | +1 | 0 | -1 | Reset Time |
| D13 | +1 | -1 | 0 | +1 | -1 | Set Next Spill Type n |
| D12 | +1 | -1 | 0 | 0 | 0 | Start of Spill |

Table 19.6: C5 Codes for TDC Control. "+1" means wide pulse, "-1" means narrow pulse and "0" means nominal (50%) pulse.

The TDC transmits data using 8b10b coding, a common DC-balanced scheme used in popular networking standards such as Gigabit Ethernet. A detailed description of the coding is beyond the scope of this document, but it is similar in spirit to the C5 scheme. Each 8 bit byte is encoded using a sequence of 10 bits. Codes are chosen to optimize DC balance and to limit the length of runs of successive '0' or '1' values. In addition to the 256 data codes, several control codes are used. A common representation for the symbols is D.x.y for data codes, where x is the decimal representation of the first 5 bits and y is the decimal representation of the final 3 bits. The control codes are represented as K.x.y where x and y are defined as for data. Many codes have two encodings with differing numbers of '1' and '0' bits, which are selected by the encoder logic to maintain DC balance.

When idle, the TDC continuously sends the K.28.5 code, which is a special control code containing a so-called "comma sequence" of binary "0011111" or "1100000" which cannot be found in any bit position in normal codes. This is used by the receiving logic to acquire bit synchronization. After receiving a start command and acquiring hit data, it sends the special K.28.1 code followed by up to 2048 32-bit words, formatted in a data record as shown in Fig. 19.19.

A Logic board is connected to the Feedthrough board and mounted in the Flobber. The Logic board collects data buffers from the TDCs, in 8b10b format, buffers, labels and

<table>
<tr><th>addr</th><th>15</th><th>14</th><th>13</th><th>12</th><th>11</th><th>10</th><th>9</th><th>8</th><th>7</th><th>6</th><th>5</th><th>4</th><th>3</th><th>2</th><th>1</th><th>0</th><th>notes</th></tr>
<tr><td>0</td><td colspan="11">0</td><td colspan="2">$clk_{C5,TDC}$</td><td>$Idle_{sw}$</td><td>En</td><td>reset</td><td></td></tr>
<tr><td>1</td><td colspan="10">0</td><td>Scaler</td><td>Edges</td><td colspan="2">SW Addr</td><td colspan="2">HW Addr</td><td></td></tr>
<tr><td>2</td><td colspan="16">Channel enable mask</td><td></td></tr>
<tr><td>3</td><td colspan="16">TDC start time</td><td></td></tr>
<tr><td>4</td><td colspan="16">TDC selb time</td><td>greater than start</td></tr>
<tr><td>5</td><td colspan="16">TDC end time</td><td>greater than selb</td></tr>
<tr><td>6</td><td colspan="16">TDC readout time</td><td>greater than end</td></tr>
<tr><td>7</td><td colspan="16">Reg 7</td><td>unused</td></tr>
<tr><td>8</td><td colspan="16">Reg 8</td><td>unused</td></tr>
<tr><td>9</td><td colspan="16">Reg 9</td><td>unused</td></tr>
<tr><td>10</td><td colspan="16">Reg 10</td><td>unused</td></tr>
<tr><td>11</td><td colspan="16">Reg 11</td><td>unused</td></tr>
<tr><td>12</td><td colspan="16">Reg 12</td><td>unused</td></tr>
<tr><td>13</td><td colspan="16">spill time 15:0</td><td></td></tr>
<tr><td>14</td><td colspan="16">spill number 15:0</td><td></td></tr>
<tr><td>15</td><td colspan="8">Reg 15</td><td colspan="8">DeMultiPW</td><td></td></tr>
</table>

Figure 19.18: TDC register map

assembles them, then transmits the data packets to the backend electronics. The Logic board has room to buffer up to five spills from each TDC. The data format is outlined in Fig. 19.20. The Logic board reads TDC data from the TDC cards in parallel at 4× 25 Mbits/sec and writes it out at 125 Mbits/sec, in 8b10b format. The Logic board keeps track of the event type and number and compares those values with those of the TDC buffers that it reads out. The logic board notes discrepancies but does not eliminate data.

The Logic Board controls the clock signals used by the TDCs. In normal data-taking, the clock signal for the TDCs is derived from the input C5 data stream. When the backend electronics'C5-clock lands on the Logic board, it is first run through a 1-to-2 fanout IC. The first copy of the C5-clock goes to the Logic board for use in clocking and receiving C5 commands. The second copy goes to a 2-to-4 fan-in/fan-out IC that generates a copy for each of the four TDCs. The fan-in option of the IC allows for the Logic board to generate its own C5-clock signal, from either the externalclock or the on-board oscillator, and fan that out to the TDCs instead. The control over which input should be fanned out is set on the Logic board via the slow control link. **Need chapter reference for slow control here**.

The Logic board has a mezzanine card with low voltage regulators for the ASDQs. The Logic board houses temperature sensors, as well as voltage and current sensors for the ASDQ supplies. If the current monitor senses a current in excess of a safe threshold, it directs the logic board to turn off the regulator. The monitor must then be reset through the slow control system, after which the voltage regulator can be activated again. In addition to power, the Logic board provides external control of the ASDQ ASIC. External voltage references are supplied by the Logic board. Op-amp buffers on the ASDQ board rescale and offset these voltages for use on the ASIC. The Flexicables carry an LVDS pair to trigger the injection of test pulses into the ASIC.

The FPGA on the Logic board reads in control/configuration data from the slow control and passes on that information, with the help of a universal asynchronous receiver/transmitter (UART), to the TDC board. The FPGA can also be used to upload firmware to the TDC

<table>
<tr><td>addr</td><td>31</td><td>30</td><td>29</td><td>28</td><td>27</td><td>26</td><td>25</td><td>24</td><td>23</td><td>22</td><td>21</td><td>20</td><td>19</td><td>18</td><td>17</td><td>16</td><td>15</td><td>14</td><td>13</td><td>12</td><td>11</td><td>10</td><td>9</td><td>8</td><td>7</td><td>6</td><td>5</td><td>4</td><td>3</td><td>2</td><td>1</td><td>0</td></tr>
<tr><td>0</td><td colspan="24">0</td><td colspan="8">0x3C = K28.1</td></tr>
<tr><td>1</td><td colspan="24">0</td><td colspan="8">0x67 = ’g’</td></tr>
<tr><td>2</td><td colspan="24">0</td><td colspan="8">0x2D = ’-’</td></tr>
<tr><td>3</td><td colspan="24">0</td><td colspan="8">0x32 = ’2’</td></tr>
<tr><td>4</td><td colspan="24">0</td><td colspan="8">0x54 = ’T’</td></tr>
<tr><td>5</td><td colspan="24">0</td><td colspan="8">0x44 = ’D’</td></tr>
<tr><td>6</td><td colspan="24">0</td><td colspan="8">0x43 = ’C’</td></tr>
<tr><td>7</td><td colspan="32">Version date: 0x201YMMDD</td></tr>
<tr><td>8</td><td colspan="16">size</td><td colspan="16">Channel Mask</td></tr>
<tr><td>9</td><td colspan="8">0</td><td colspan="24">Spill # [23..0]</td></tr>
<tr><td>10</td><td colspan="32">Time MSB [43..12]</td></tr>
<tr><td>11</td><td colspan="20">0</td><td colspan="12">Time LSB [11..0]</td></tr>
<tr><td>12</td><td colspan="16">Start time (160ns LSB)</td><td colspan="16">End time (160ns LSB)</td></tr>
<tr><td>13</td><td colspan="16">SelB time (160ns LSB)</td><td colspan="16">Readout time (160ns LSB)</td></tr>
<tr><td>14</td><td colspan="32">0</td></tr>
<tr><td>15</td><td colspan="32">1-scalars; 0-both edges;others: 0</td></tr>
<tr><td>16</td><td colspan="32">Scalar Count 15</td></tr>
<tr><td>17</td><td colspan="32">Scalar Count 14</td></tr>
<tr><td>18</td><td colspan="32">⋮</td></tr>
<tr><td>19</td><td colspan="32">Scalar Count 1</td></tr>
<tr><td>20</td><td colspan="32">Scalar Count 0</td></tr>
<tr><td>21</td><td>0</td><td>0</td><td>1</td><td>edge</td><td colspan="4">Ch: 0-15</td><td colspan="22">Coarse time: LSB 5ns, range 0 - 10.485775 ms</td><td colspan="2">Fine</td></tr>
<tr><td>-</td><td colspan="32">⋮</td></tr>
<tr><td>size-1</td><td>0</td><td>0</td><td>1</td><td>edge</td><td colspan="4">Ch: 0-15</td><td colspan="22">Coarse time: LSB 5ns, range 0 - 10.485775 ms</td><td colspan="2">Fine</td></tr>
</table>

Figure 19.19: TDC Detailed Data Format

daughtercards, using a JTAG connection. Although setup and configuration will be handled automatically during actual data-taking, a command line interface is also available for independent test runs.

Two High Voltage boards are also mounted in the Flobber. Each of the High Voltage boards takes four SHV inputs. Each HV line has a bypass capacitor for reducing noise, a current limiting resistor and a jumper for isolating groups of 16 straws, in case of a short. After the current limiting and filtering, the input HV lines are carried to the eight ASDQ boards on separate cables. A return path is also provided.

<table>
<tr><th>-</th><th>31</th><th>30</th><th>29</th><th>28</th><th>27</th><th>26</th><th>25</th><th>24</th><th>23</th><th>22</th><th>21</th><th>20</th><th>19</th><th>18</th><th>17</th><th>16</th><th>15</th><th>14</th><th>13</th><th>12</th><th>11</th><th>10</th><th>9</th><th>8</th><th>7</th><th>6</th><th>5</th><th>4</th><th>3</th><th>2</th><th>1</th><th>0</th></tr>
<tr><td>0</td><td colspan="8">0x3C = K28.1</td><td colspan="24">0</td></tr>
<tr><td>1</td><td colspan="16">0</td><td colspan="16">size</td></tr>
<tr><td>2</td><td colspan="8">0</td><td colspan="24">Spill # [23..0]</td></tr>
<tr><td>3</td><td colspan="32">Time MSB [43..12</td></tr>
<tr><td>4</td><td colspan="7">0</td><td colspan="3">type</td><td>FE</td><td>Er</td><td colspan="4">OOS</td><td colspan="4">$TDC_{EN}$</td><td colspan="12">Time LSB [11..0]</td></tr>
<tr><td>5</td><td colspan="16">TDC0: 31:enabled, 30:OOS, 29: size error</td><td>0</td><td colspan="15">size</td></tr>
<tr><td>6</td><td colspan="16">TDC1: 31:enabled, 30:OOS, 29: size error</td><td>0</td><td colspan="15">size</td></tr>
<tr><td>7</td><td colspan="16">TDC2: 31:enabled, 30:OOS, 29: size error</td><td>0</td><td colspan="15">size</td></tr>
<tr><td>8</td><td colspan="16">TDC3: 31:enabled, 30:OOS, 29: size error</td><td>0</td><td colspan="15">size</td></tr>
<tr><td>9</td><td colspan="24">0 (TDC: 1)</td><td colspan="8">0x3C</td></tr>
<tr><td>-</td><td colspan="24">0</td><td colspan="8">0x67 = 'g'</td></tr>
<tr><td>-</td><td colspan="24">0</td><td colspan="8">0x2D = '-'</td></tr>
<tr><td>-</td><td colspan="24">0</td><td colspan="8">0x32 = '2'</td></tr>
<tr><td>-</td><td colspan="32">⋮</td></tr>
<tr><td>-</td><td colspan="24">0 (TDC: 2)</td><td colspan="8">0x3C</td></tr>
<tr><td>-</td><td colspan="24">0</td><td colspan="8">0x67 = 'g'</td></tr>
<tr><td>-</td><td colspan="24">0</td><td colspan="8">0x2D = '-'</td></tr>
<tr><td>-</td><td colspan="24">0</td><td colspan="8">0x32 = '2'</td></tr>
<tr><td>-</td><td colspan="32">⋮</td></tr>
<tr><td>-</td><td colspan="24">0 (TDC: 3)</td><td colspan="8">0x3C</td></tr>
<tr><td>-</td><td colspan="24">0</td><td colspan="8">0x67 = 'g'</td></tr>
<tr><td>-</td><td colspan="24">0</td><td colspan="8">0x2D = '-'</td></tr>
<tr><td>-</td><td colspan="24">0</td><td colspan="8">0x32 = '2'</td></tr>
<tr><td>-</td><td colspan="32">⋮</td></tr>
<tr><td>-</td><td colspan="24">0 (TDC: 4)</td><td colspan="8">0x3C</td></tr>
<tr><td>-</td><td colspan="24">0</td><td colspan="8">0x67 = 'g'</td></tr>
<tr><td>-</td><td colspan="24">0</td><td colspan="8">0x2D = '-'</td></tr>
<tr><td>-</td><td colspan="24">0</td><td colspan="8">0x32 = '2'</td></tr>
<tr><td>-</td><td colspan="32">⋮</td></tr>
<tr><td>size - 3</td><td colspan="32">size</td></tr>
<tr><td>size - 2</td><td colspan="32">CRC</td></tr>
<tr><td>size - 1</td><td colspan="24">0</td><td colspan="8">0x5C = K28.2</td></tr>
</table>

Figure 19.20: Logic Board Data Format

**Back-End Readout Electronics**

FC7 MicroTCA advanced mezzanine cards developed by CMS will provide the clock and control signals for the logic boards and transfer the hit data from the logic board buffers to the DAQ via a CMS AMC13 card [8]. The FC7 has a Xilinx Kintex-7 FPGA, can accomodate two FMCs and has a connection to the AMC13 via the MicroTCA backplane. A photograph of the board is shown in Fig. 19.21.

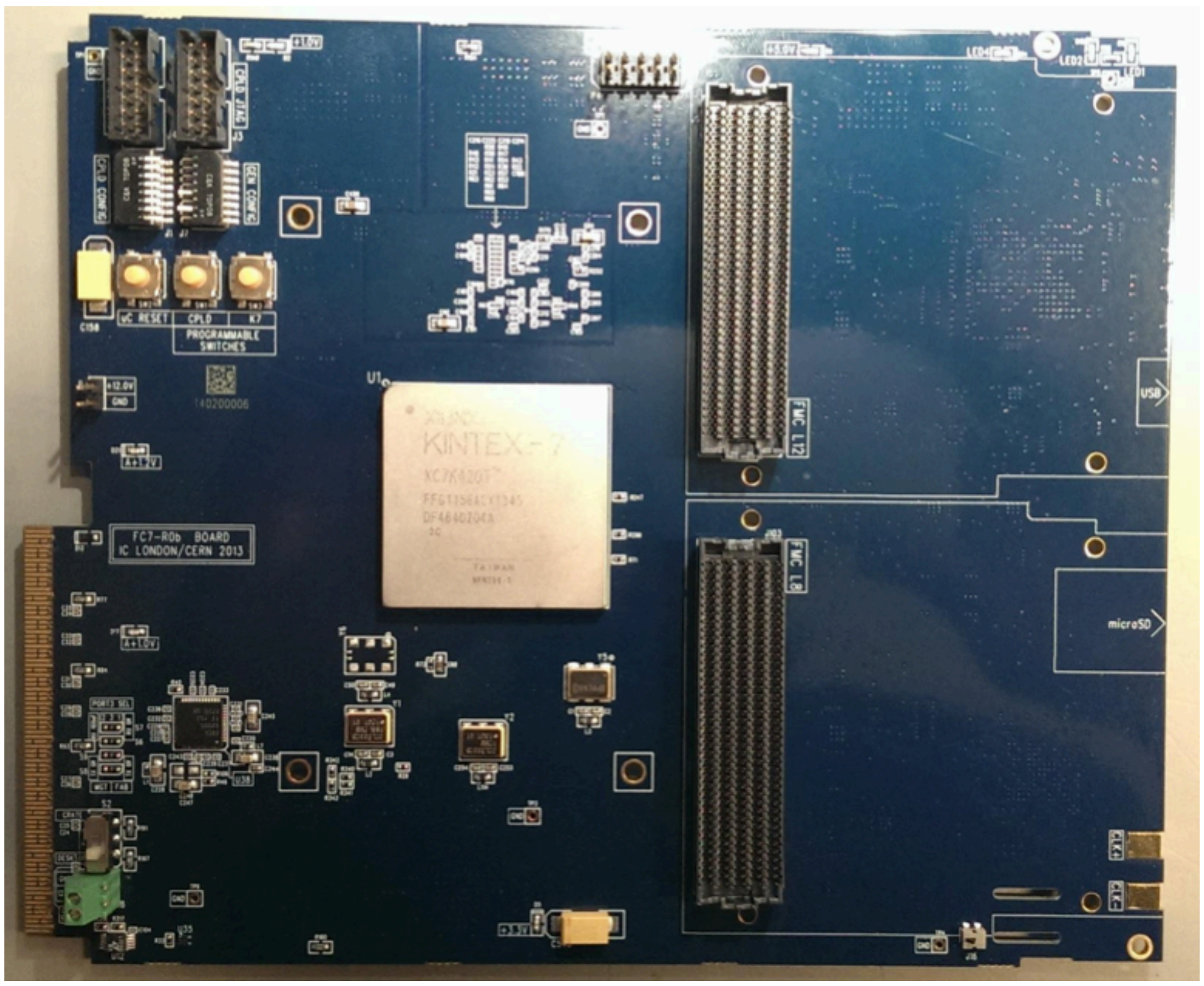

Figure 19.21: A photograph of a prototype FC7. The two FMC connectors are to the right of the Kintex-7 FPGA and allow up to 16 fiber connectors to the front of the board. The backplane connection to the AMC-13/MCH is at the bottom left of the photograph.

It incorporates an MMC providing the IPMI functionality and Gb ethernet via the crate's MCH. The ethernet communication will utilize the IPBus communication tools [9] developed by CMS. The ethernet/MCH interface will be used to upload configuration data and also to send a subset of the data to a local PC to quickly identify issues with data integrity. The FC7 has 30 Mbytes of block ram which is more than an order of magnitude larger than the data volume expected from a single tracker in a single spill. The clock, control and readout data will be transmitted between the FC7s and the logic boards using fiber optic cables. This has the benefit of decoupling the clock/control/data-readout signals from the LV. Potentially this allows the readout crate to be housed in the MC1 counting room away from the storage ring thereby removing the overhead of ensuring that the crate and backend electronics do not perturb the storage ring's magnetic field. One FC7 will be sufficient to readout a single tracker since each logic board will communicate with the FC7 with a single fiber pair (via an SFP) and each of the two FMCs on the FC7 can accommodate 8 fiber pairs. The data from all three trackers can thus be sent to a single MicroTCA crate which has sufficient slots for 12 AMC cards in addition to the AMC13 and MCH. This allows one spare FC7 per tracker in the crate which will enable a quick recovery, with a simple slot re-configuration, should an FC7 fail. The FMCs housing the 16 SFPs and interfacing the optical signals with the FC7 will be the EDA-02707 that is being used for CMS's new Trigger, Control & Distribution System (TCDS) [10]. A schematic of the MicroTCA readout crate is shown in Fig. 19.22.

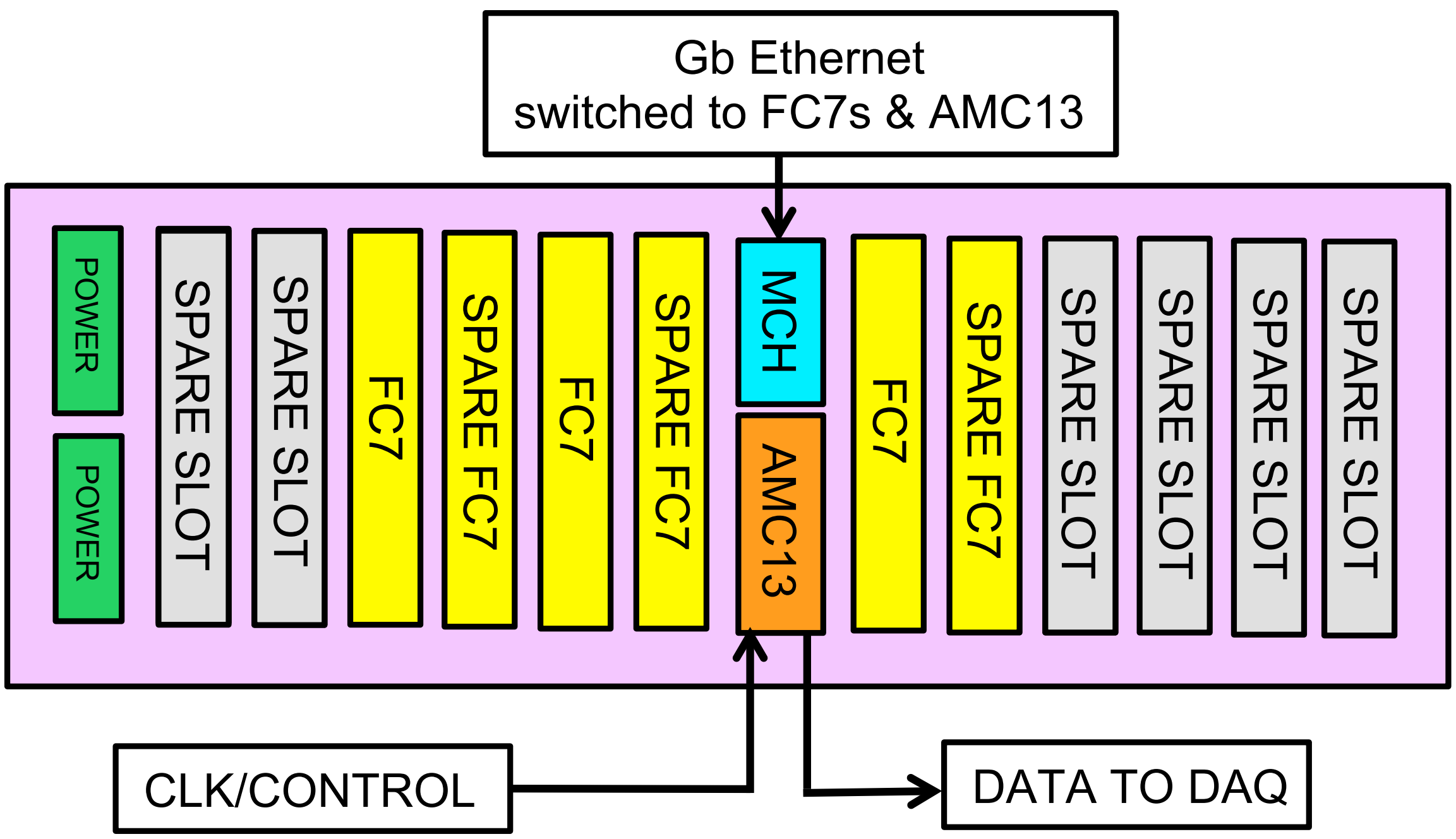

Figure 19.22: A schematic of the MicroTCA tracker readout crate.

The FC7s will receive their clock and control signals from the AMC13 and these will be encoded into the C5 protocol on the FC7 FPGA before being transmitted to the logic boards. The FC7 will receive data in the 8B10B protocol from the logic boards and this will be formatted in the FC7 FPGA to match the data format requirements of the AMC13. The TDCs will begin to accumulate data at a fixed time (O(30) $\mu$s) after the begin of spill signal is received by the AMC13 for a duration of 700 $\mu$s. In order to ensure that the tracker hit times have a common $t_0$ with respect to the calorimeter, sufficient to synchronize the data before track-fitting, the FC7 will multiply the input 40 MHz clock to 800 MHz and will record when the C5 command instructing the TDC to accumulate hits is sent to the logic board with respect to this clock. The C5 commands are sent with a 10 MHz clock and so don't have the necessary granularity to synchronize the data.

A block diagram showing the key features and signal paths of the FC7 is shown in Fig. 19.23.

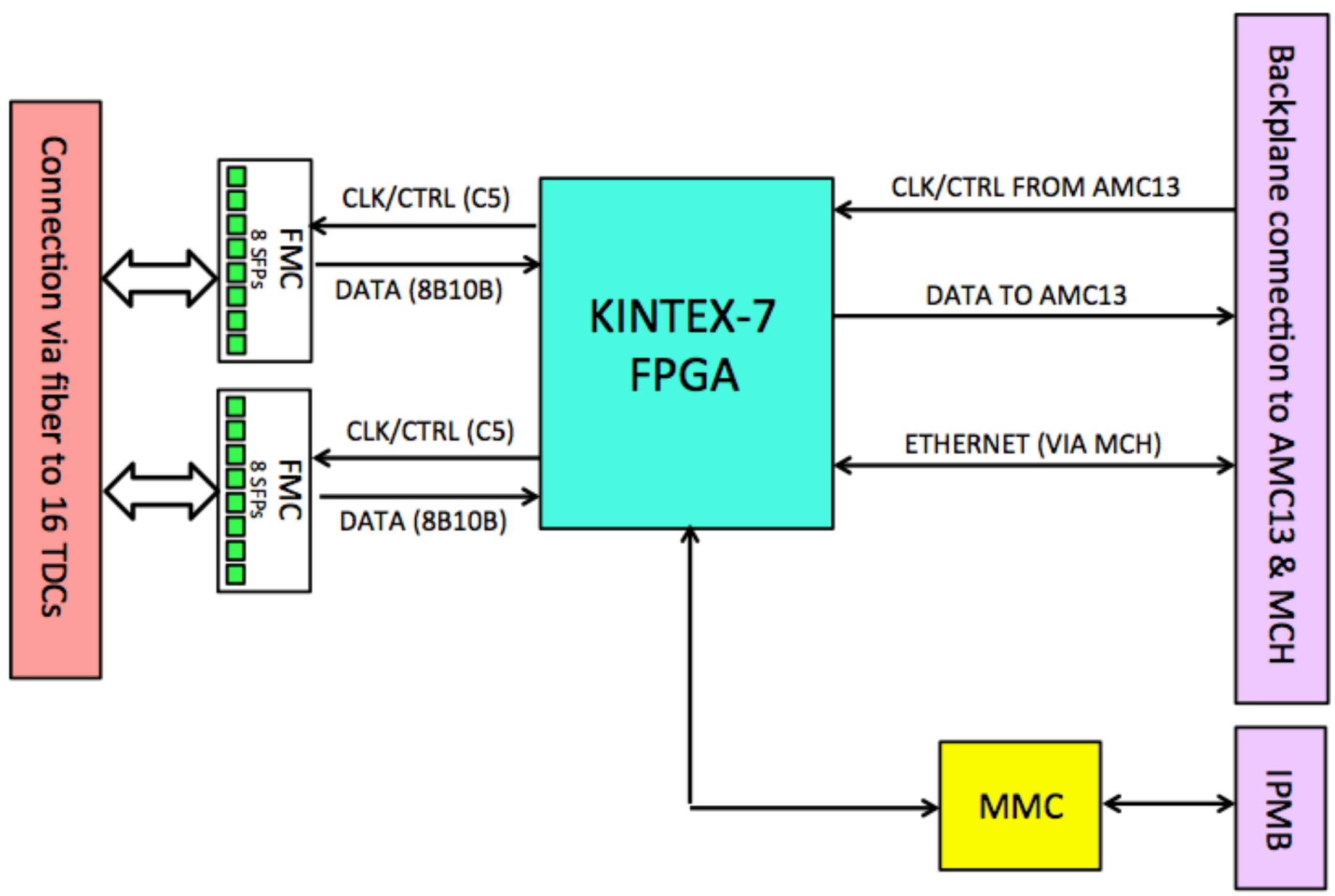

Figure 19.23: A block diagram of the key features and signal paths of the FC7 tracker readout module.

### Low-Voltage Distribution

The low-voltage (LV) system provides ±5V to the logic board board which will be put through regulators to provide the necessary power to the TDCs and ASDQs. The typical operational current is about 2.8A for the +5V supply and 1.4A for the -5V supply. The LV system will provide independent LV to each logic board such that individual logic boards can be powered on and off. The LV boards provide isolated output to avoid ground off-sets between systems which can be a main issue when using long cables.

The LV will be provided by 36W isolated DC-DC converters operating with a 24V DC input supplied by an external commercial AC-DC power supply. The expected power consumption is about 20W so the 36W is chosen to allow for some contingency. One board provides a dual ±5V output sufficient to power a single logic board, so 16 boards are required per tracker. A photograph of the board is shown in Fig. 19.24.

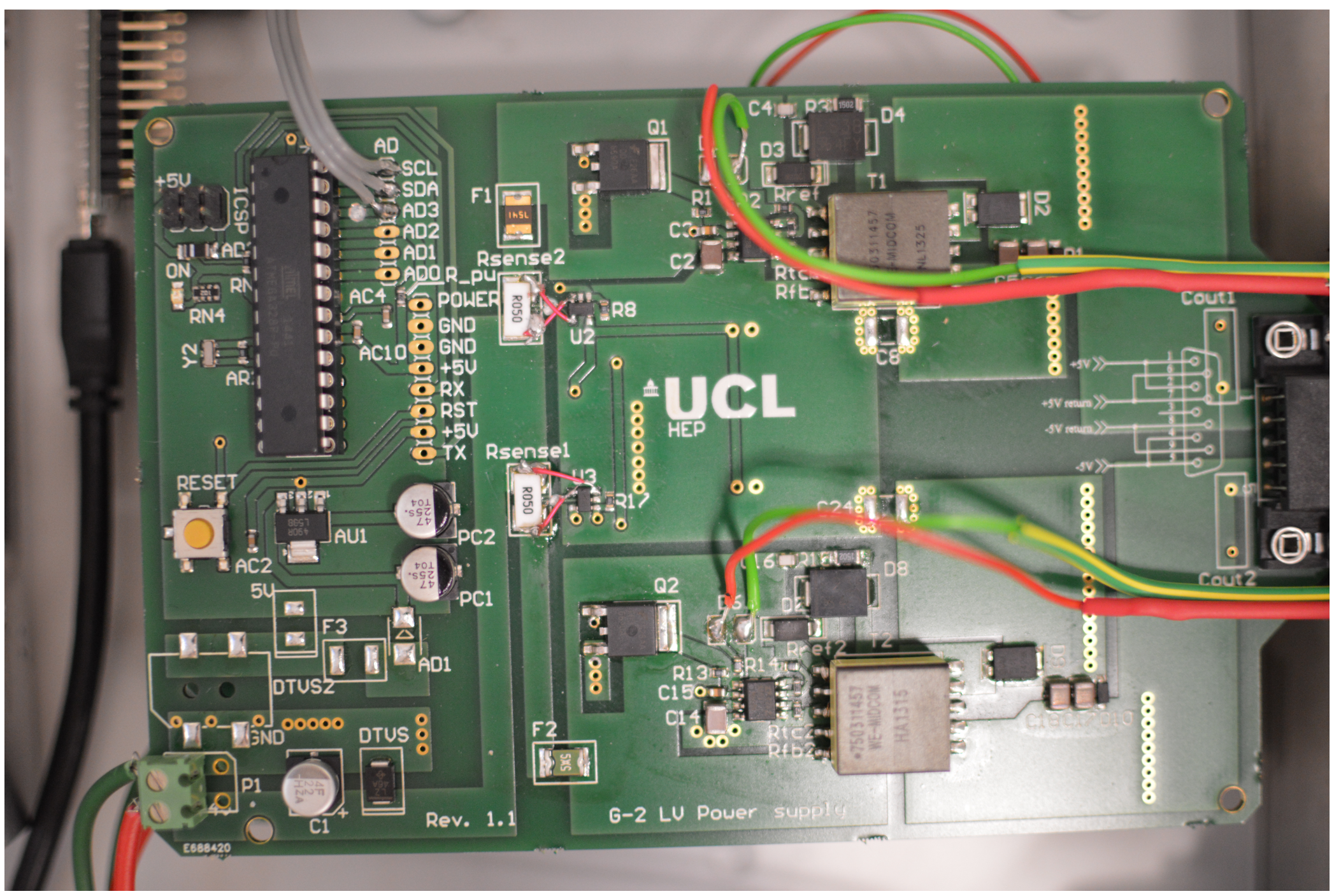

Figure 19.24: A photograph of a prototype low-voltage board.

The boards for one tracker will be housed in a custom 3U crate with a crate controller board which can power individual boards on and off as well as provide monitoring information such as currents and temperatures. The three LV crates will be located in the centre of the ring at a maximum distance of 5m from the logic board so as to minimize signal perturbation and impact on the storage ring field. Furthermore, the LV boards also offer slow control capabilities via direct links to every logic board.

**High-Voltage Distribution**

High voltage will be supplied by a commercial CAEN SY4527LC universal multichannel power supply system containing 12 channel A1535 3.5 kV/3 mA common floating return boards. The nominal HV will be 1800 V. All electronics in the HV chain are rated to 3kV. Two SY4527s will be located in the center of the ring and will distribute HV to the three tracking stations. There is one common HV per 16 straws so each tracking station has 64 HV channels. Trip currents will be set at 10 $\mu$A per HV channel to avoid damaging straw walls if a sense wire breaks.

### 19.3.3 Monitoring and Slow Controls

The details of the experiment-wide Slow Controls system can be found in Chapter 22. The key parts of this system relevant to the straw tracker system are described below. Details of the slow controls data acquisition, storage, and access are not repeated here.

The Slow Controls system will be used to monitor a variety of parameters in the tracking system, as shown in Table 22.1. Voltages and currents for both the LV and HV systems will be monitored using dedicated functionality built into the standard commercial power supplies and interfaced to the experimental slow control system. Ambient pressure, temperature, and humidity measurements (in the regions near the tracking stations) will be monitored using dedicated sensors. These sensors, and their interface to the slow controls system will be identical throughout the experiment and are described in Chapter 22.

A few parameters monitored by the slow controls system will be unique to the tracking system. Since the straw tracker is a gas-based system, both the gas flow and the gas temperature will be monitored. These sensors will integrated into the straw tracker gas system, as described in a previous section. The liquid water cooling system will also require the monitoring of several parameters. The flow and temperature of liquid cooling water will be monitored by the slow controls system at several points in the system. Inside the manifold, several parameters will also be monitored. The temperature of the circuit board will be monitored with a dedicated temperature sensor built into the circuit board. The slow controls interface built into the front-end electronics is described in a previous section. This slow controls interface will be connnected via a USB interface to a computer in the experimental hall. This computer will be running a MIDAS front-end that will be allow this monitored data to be added into the data acquisition system.

The tracking system is critically dependent on both the gas flow and the flow of cooling water. If either of these two systems are comprised, it could result in damage to the components of the system. For this reason, a dedicated interlock system is being constructed (see Sec. 22.2.4). This interlock system will be part of the experiment's main Siemens PLC which is used to ensure the safety of the cryogenic and vacuum systems. It will receive inputs from the flow sensors in the gas and cooling water systems, from the temperature sensors inside the tracker manifolds, and from pressure sensors near the tracker stations. While the logic has not been finalized, the information from these inputs will be used by the PLC to send signals to the LV and HV power supplies in the tracker system to ramp down or turn off the power. The tracker gas system also includes solenoid valves that can be used to quickly turn off gas flow. These signals will also be propagated to the central slow controls system to ensure that data quality can be adequately monitored. Although the exact logic has not yet been finalized, since the interlock system is based around a PLC, enough flexibility is present in the system to meet our needs.

## 19.4 Performance

The straw leak rate has been measured to be $3.5x10^{-6}$ cc/min/per straw for $CO_2$. This corresponds to a vacuum load per tracker station of $7.7x10^{-5}$ Tl/s. Given the additional pumping speed, we expect to be able to maintain a vacuum of $1x10^{-7}$ an order of magnitude below our specification. Furthermore, the permeation of ethane is expected to be significantly lower than $CO_2$. For example, the measured permeation of Methane through Mylar is 40 times less than $CO_2$ providing further head room. A measurement with Ethane is underway.

The expected performance of the tracker technical design is determined by a simulation. The simulation is benchmarked by the performance of a 32 channel prototype module that was tested extensively using a $^{55}$Fe source in the lab and in two beam tests at the Fermilab Test Beam Facility as discussed below. The performance of a single straw is determined using GARFIELD [11]. This program simulates the propagation of electrons and ions in a gas in the presence of electric and magnetic fields based on the measured ion mobility in Argon. Key parameters such as the effective threshold to apply in the simulation are determined using the signal to noise achieved in the prototype and determined in beam. The geometry of the system is determined using a full GEANT4 [12] model of the muon storage ring that includes the proper physics model to simulate muon storage, precession, and decay. The performance of the tracker is determined using a fast tracking software package that takes the positron hit positions from the GEANT4 simulation, applies resolution from GARFIELD and multiple scattering corrections, determines the positron trajectory, and extrapolates back to the point of tangency to determine the muon decay position. The fast simulation currently assumes a uniform magnetic field which is valid for the majority of the tracking volume. A first version of a full tracking code algorithm that uses the GEANT4 hits for input and includes all physics processes for energy loss, multiple scattering, and backgrounds also exists.

The distance versus time $(x - t)$ relation determined from GARFIELD are shown in Fig. 19.25 for the case of zero magnetic field and the 1.5 T nominal magnetic field. The resolution as a function of gain and the resolution as a function of closest approach to the wire at the nominal gain are shown in Fig. 19.26. At a gain of $2x10^6$, the straws are fully efficient for single ionizations assuming the noise levels achieved in the prototype electronics. The average resolution is found to be approximately 100 $\mu$m providing sufficient headroom to the 300 $\mu$m.

The rate in the hottest straw at injection is 200 kHz. At a gain of $2x10^6$, we expect a gain sag of 5% at this rate. Simulations indicate that no noticeable effect is seen since we are fully efficient for single ionizations.

The acceptance to reconstruct at least 5 hits as a function of momentum and as a function of the muon decay distance is shown in Fig. 19.27. There is sufficient coverage at all positron momenta for determination of the muon beam parameters. The loss of acceptance at lower momenta is due to the fact that the lowest momentum positrons originate very close to the calorimeter and the limited available space between the muon decay position and the calorimeter limit the amount of tracker planes the positron can hit. The distance between stations is dominated by the area necessary for the readout PCBs. The azimuthal acceptance of the tracker is shown in Fig. 19.28 indicating we can reconstruct the decay positions of muons up to meters upstream of the tracker. The hit density in the tracker is also displayed in the plot.

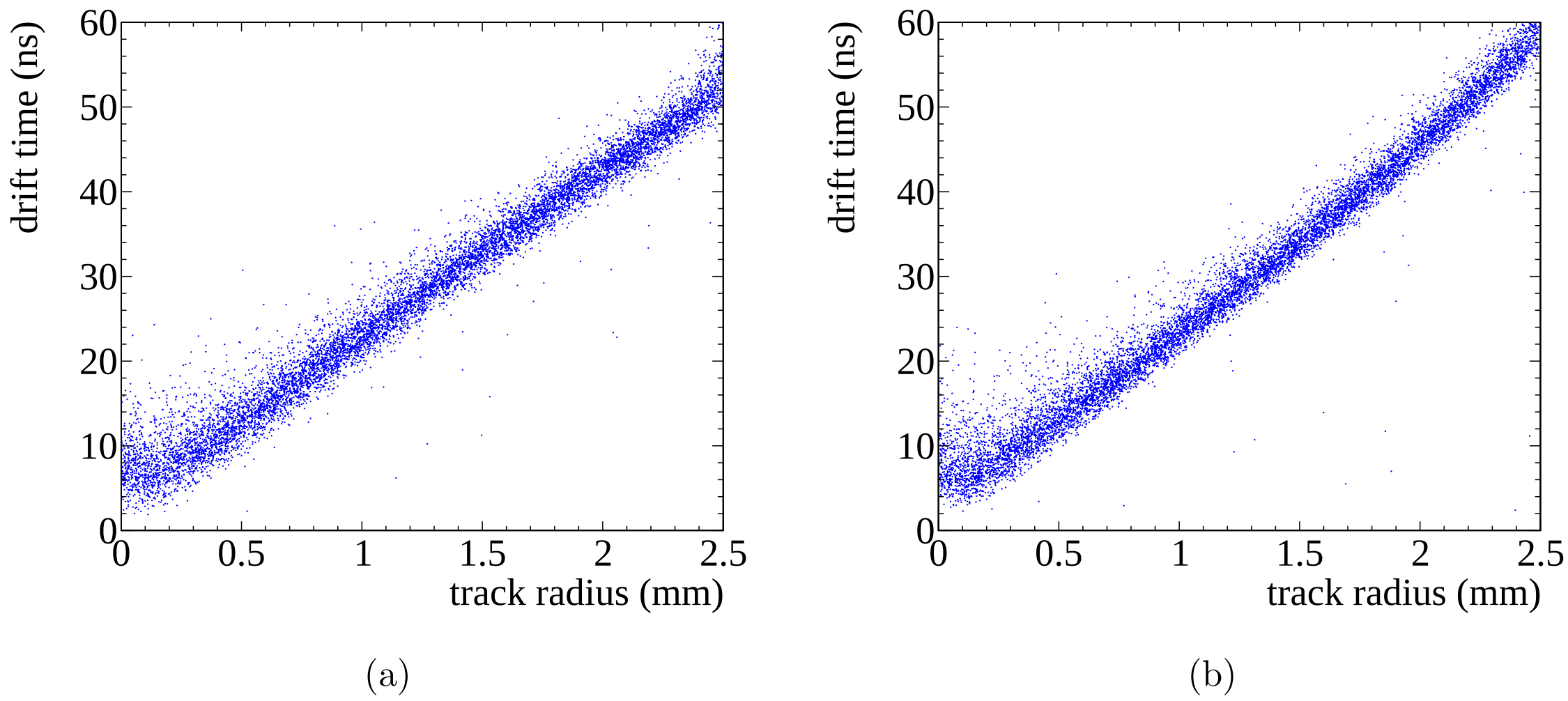

Figure 19.25: Time versus distance relation in a single straw predicted by GARFIELD for 50:50 Argon:Ethane at a gain of 2e6. The distribution on the left is for zero magnetic field. The distribution on the right is for the full 1.5T magnetic field.

The momentum resolution, muon decay position, and positron vertical angle resolution for an Ar:$C0_2$ mixture are shown in Fig. 19.29. At 1.5 GeV, the resolutions are 1 mm on the radial position, 1.5 mm on the vertical position, 0.5% on the curvature, and 1 mrad on the vertical angle. The resolutions on position and momentum rise steeply with momentum but in all cases are below the required values. Both position resolutions become significantly worse above 2.6 GeV. In this region, the muons are decaying between 5 and 10 meters from the first tracking plane and the large lever arm makes a more precise determination impractical. The momentum resolution is worse than for a typical gas based system but is well below the resolution of the calorimeter which satisfies the requirements. The vertical angle resolution is also well below the requirements. The performance is near the specifications for Ar:$CO_2$ and have significant headroom with Ar:Et.

Two large multi-channel prototypes have been constructed. A 32 channel prototype was constructed based on the conceptual design and is shown in Fig. 19.30. The prototype consists of 2 close packed doublet layers, 8 straws wide in U and V configurations. Prototype ADSQ and TDC boards were housed in the gas return manifolds for each view as shown in Fig. 19.31. All components of the prototype were constructed and assembled in Fall and Winter 2013. The prototype was tested using 120 GeV protons at the Fermilab Test Beam Facility for two weeks in January 2014 and April 2014. The January run was conducted with the straws in atmosphere, the April run was conducted with the straws and manifolds entirely contained in a vacuum chamber.

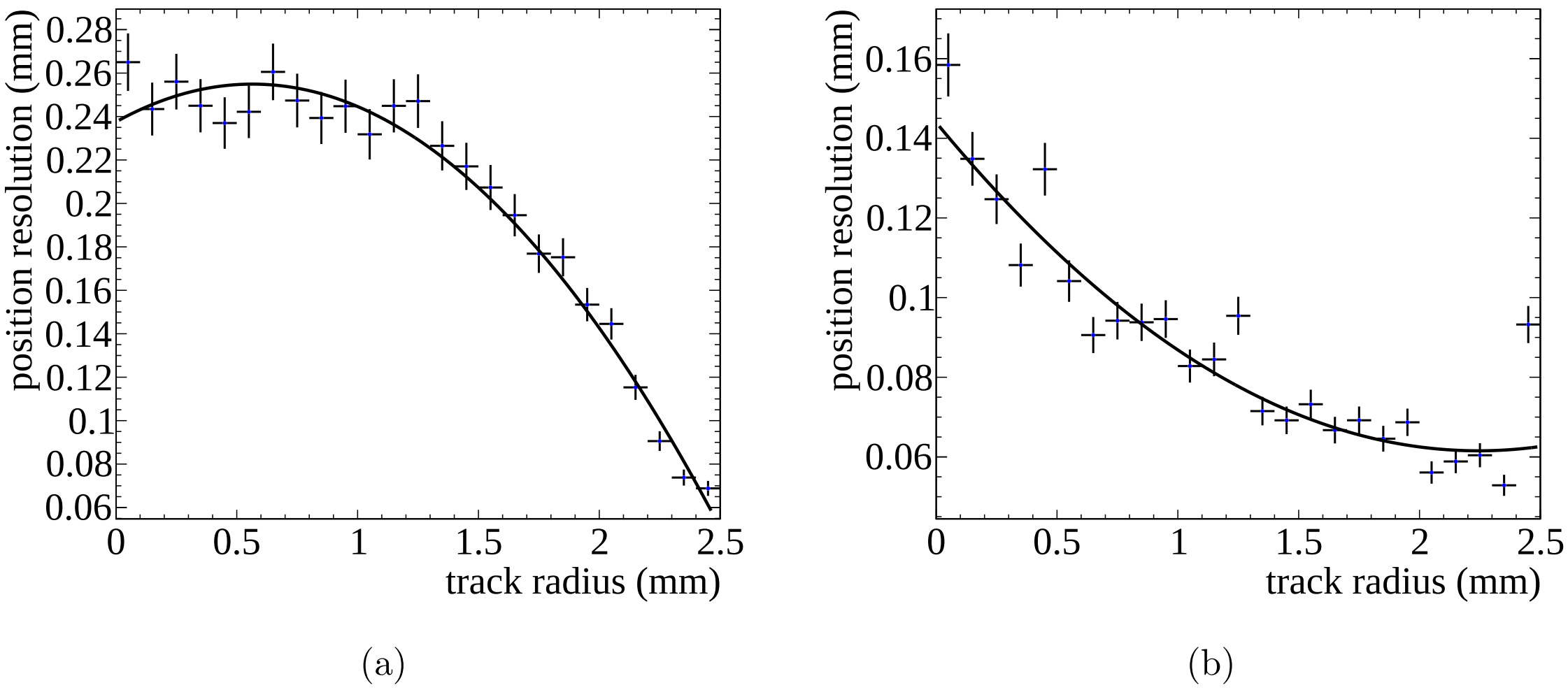

Figure 19.26: (a): The position resolution determined from GARFIELD for a single straw using the derived x-t relation. The distribution on the left is for 80:20 Ar:$CO_2$ achieved in test beam. The distribution on the right is what is expected for 50:50 Ar:Et. The specification is 300 microns.

A 128 channel prototype based on the full preliminary design was constructed in Winter and Spring 2015 and was tested in beam in June 2015 using Ar:$CO_2$, Ar:Et, at atmosphere, and at vacuum. The prototype is shown in Fig. 19.32. Results of the beam test are pending.

The hit count in the prototype is shown for one run in Fig. 19.33. The beam is clearly seen in both the U and V layers. Figure 19.33 also shows the time correlation between hits in the doublet layer. Figure 19.34 compares the time difference between hits in a straw doublet between the prototype and GARFIELD. The agreement in the width of the distributions indicates proper modeling of the drift velocity of the gas. Figure 19.35 shows the anti-correlation in drift times between the straws in a doublet. For the data distribution, the drift times in the V view are shown and the $T_0$ is taken as the average time of the two straws in the U view. The residuals of the correlation are in good agreement indicating proper modeling of the resolutions in GARFIELD once the appropriate threshold is applied to the simulation.

A lessons learned analysis was performed following the prototype construction and the beam test and is documented here [13]. The main conclusions are that first, the expected resolutions reported in the conceptual design report are ambitious. Second, a design iteration was required for the straw end pieces. Third, the resource estimates for the construction used in the basis of estimates were verified. However, the resource type was not. The BOEs at conceptual design relied mainly on student labor for construction. The prototype construction clearly demonstrated that either skilled technicians are required or several more hours would be required. All of the lessons learned are incorporated in the technical design.

The deign of the tracker is driven by requirements for reducing systematic uncertainties on the $g-2$ measurement. However, by measuring the positron vertical angle, the tracker will also be able to limit the size of the muon's permanent electric dipole moment. Considering only the acceptance of the recommended design and requiring at least three stations are hit by the positron, we expect to increase the statistics with respect to the Brookhaven EDM search [2] by approximately a factor of 200 per month and a factor of 3500 for the full run. This gives us enough statistics to improve the limit on the EDM by an order of magnitude very quickly and eventually approach a two order of magnitude improvement assuming systematics can be properly constrained.

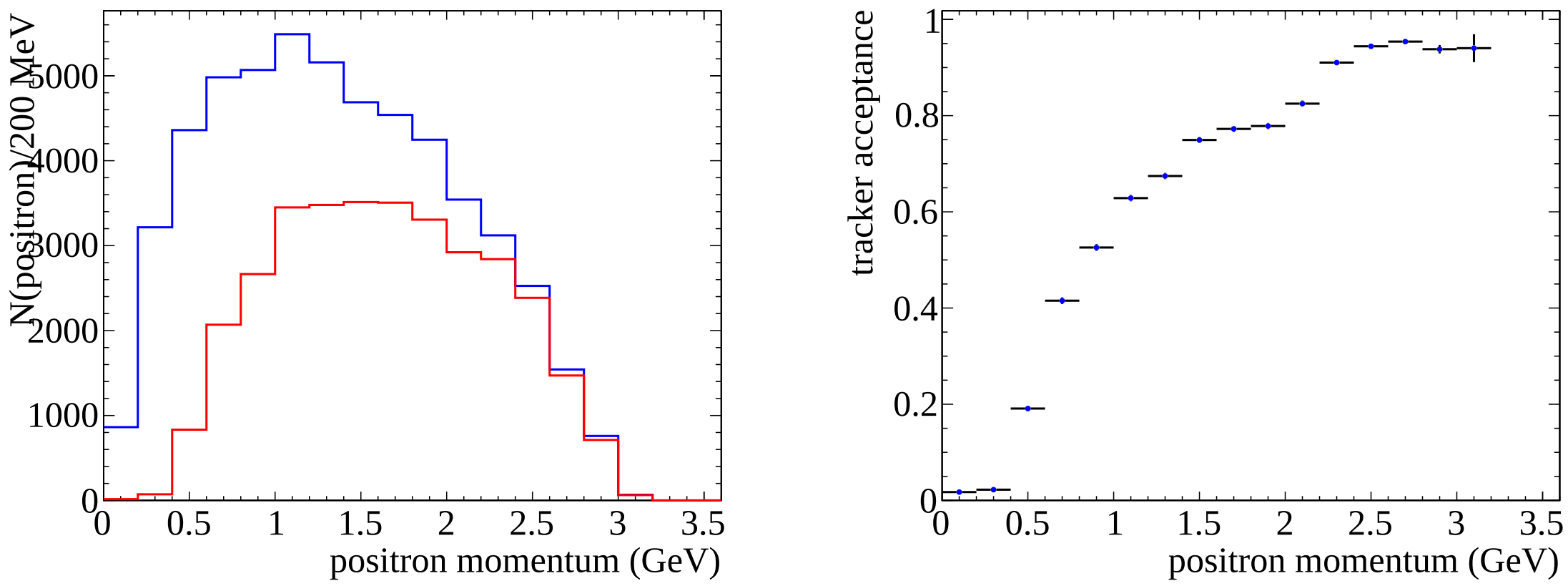

(a) (b)

Figure 19.27: (a): The positron momentum spectrum for positrons incident on the front face of the calorimeter (blue) and also with sufficient hits in the tracking detector to form a track (red). (b): The ratio of the two distributions giving the relative efficiency between the tracker and the calorimeter as a function of positron momentum.

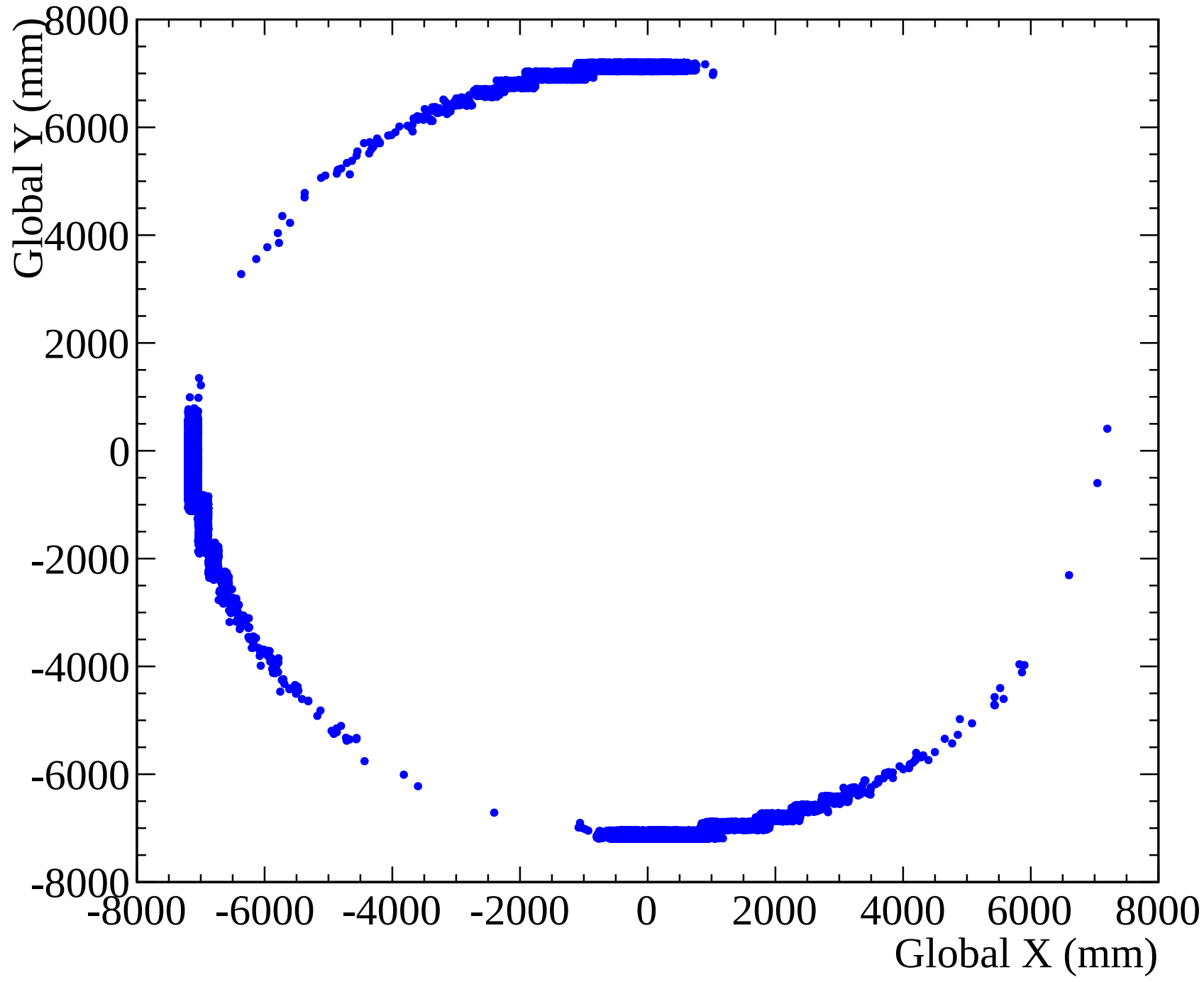

Figure 19.28: Azmithal acceptance of the three trackers.

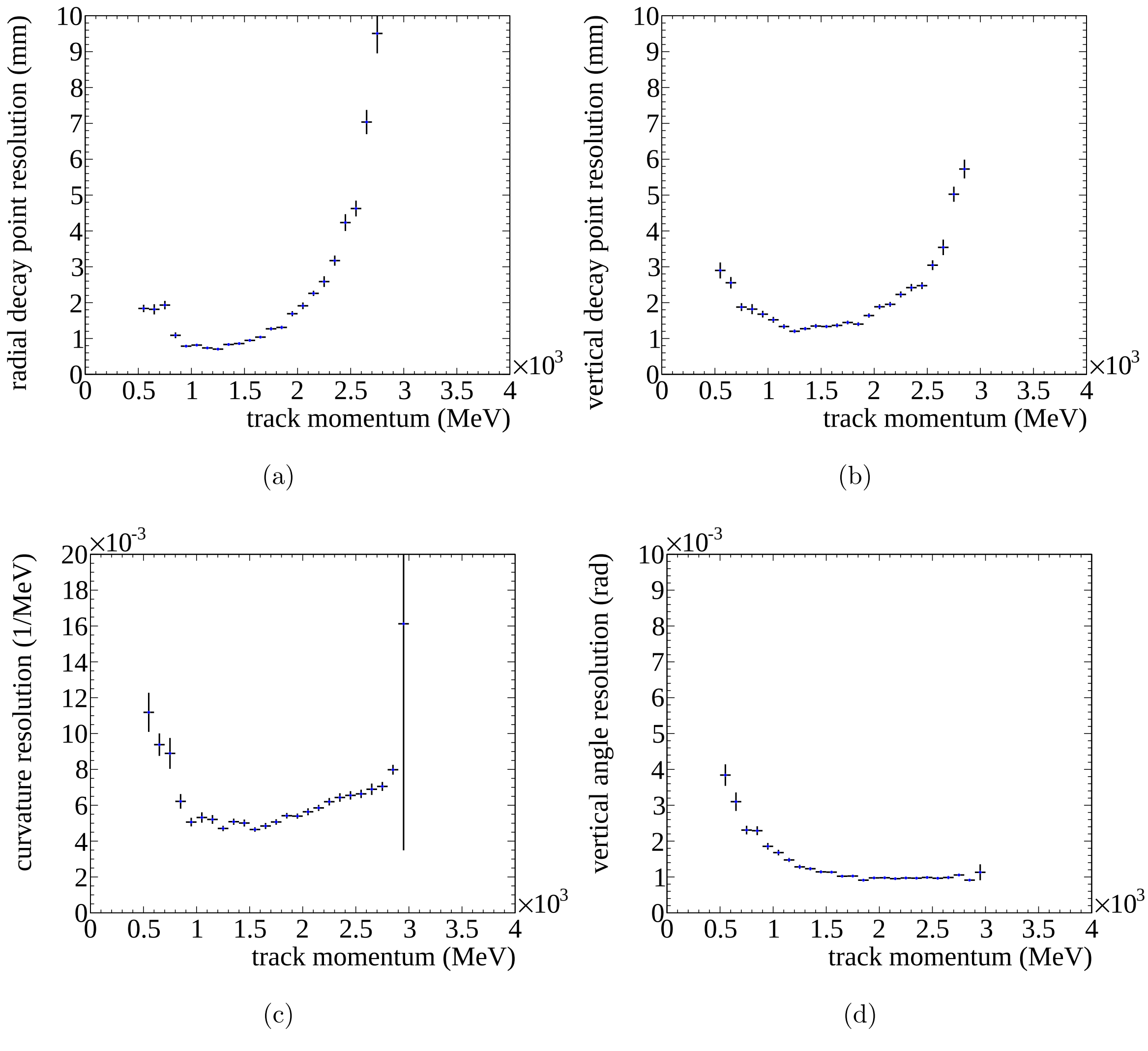

Figure 19.29: Resolutions on the muon and positron parameters that will be measured by the tracker. (a): The radial decay position of the muon. (b): The vertical decay position of the muon. (c): The curvature resolution of the positron. (d): The vertical angle of the positron.

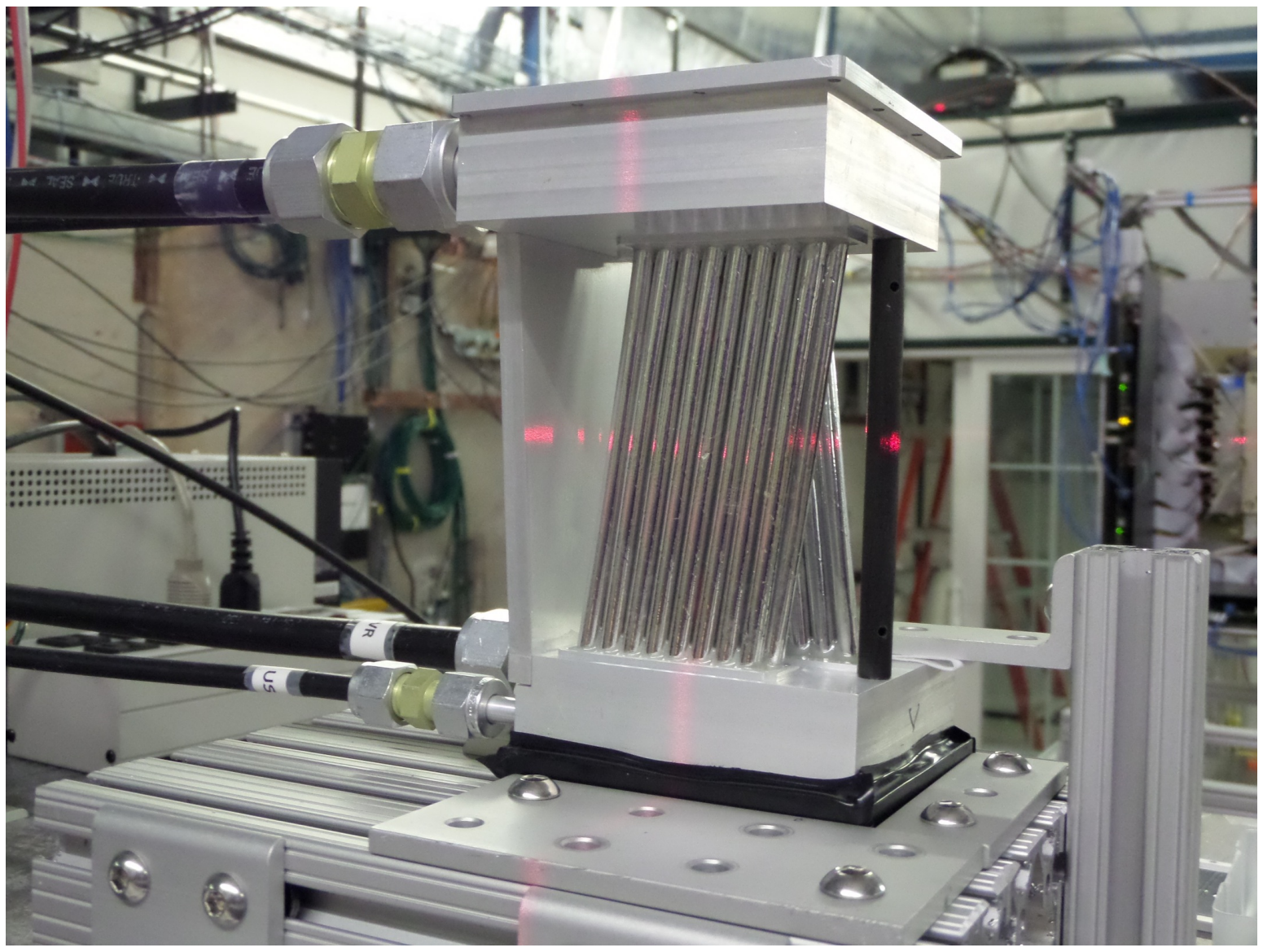

Figure 19.30: 32 channel prototype mounted in the Fermilab Test Beam Facility

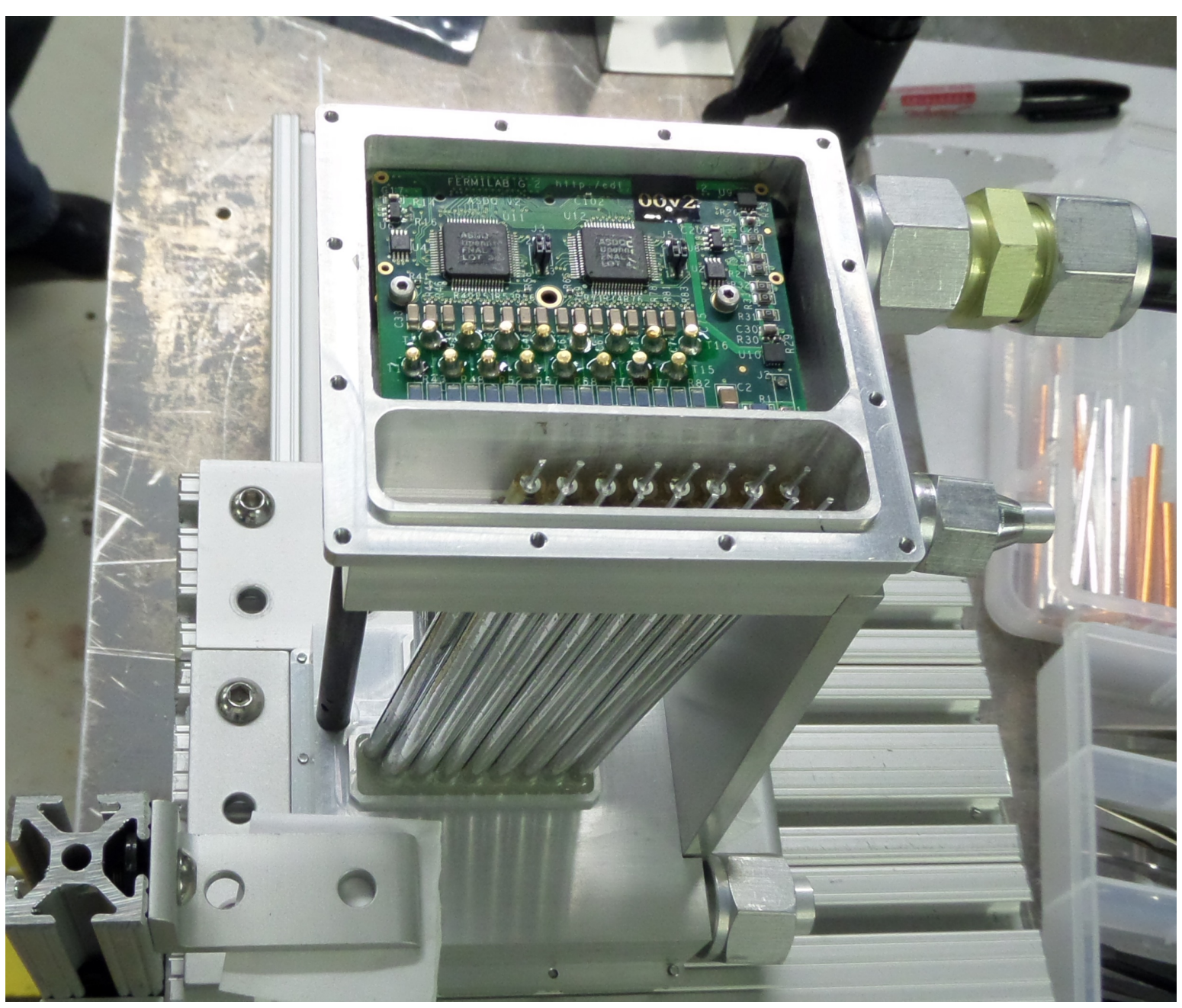

Figure 19.31: ASDQ boards mounted inside the manifold of the 32 channel prototype

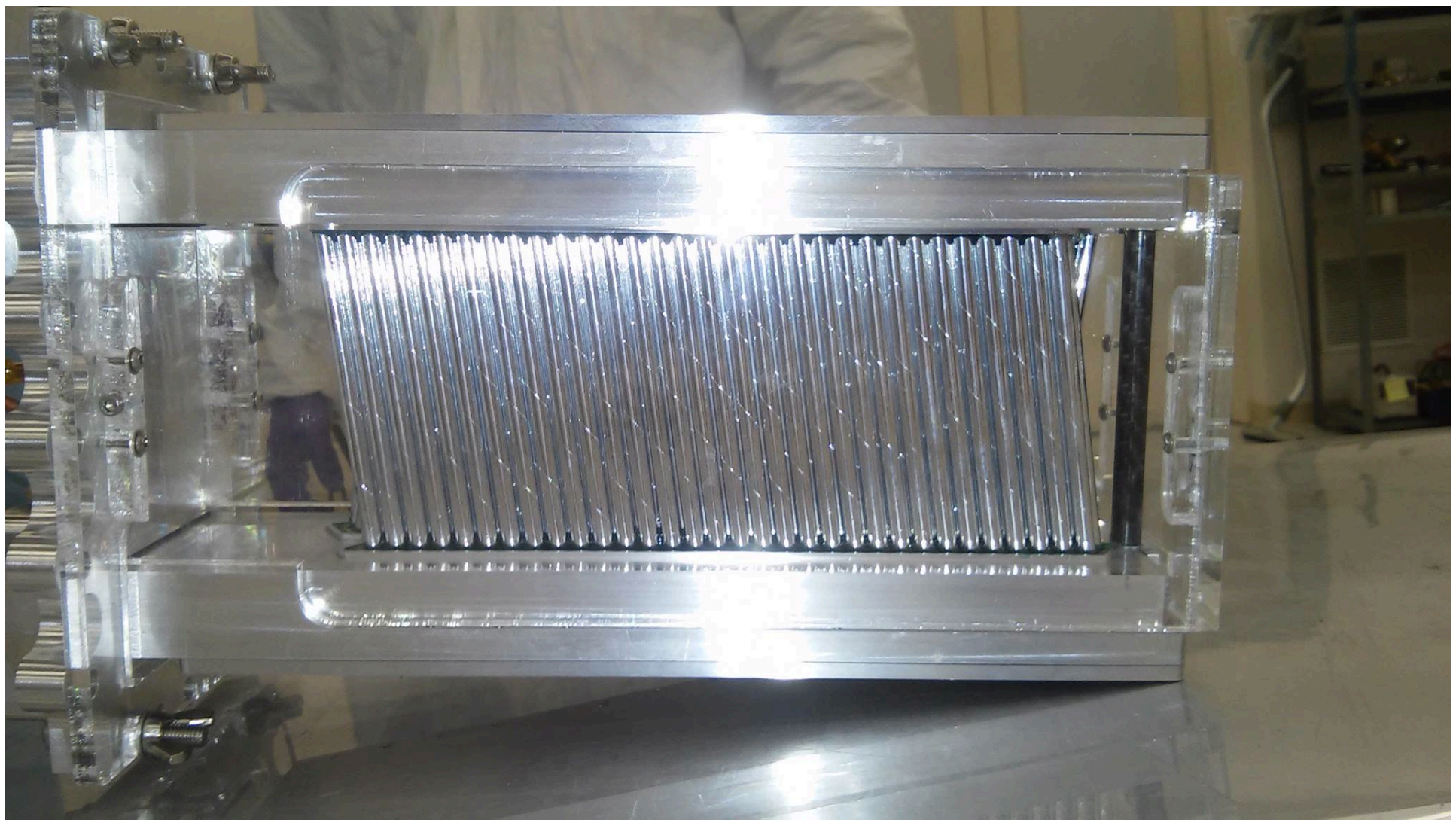

Figure 19.32: 128 channel prototype mounted in the Fermilab Test Beam Facility

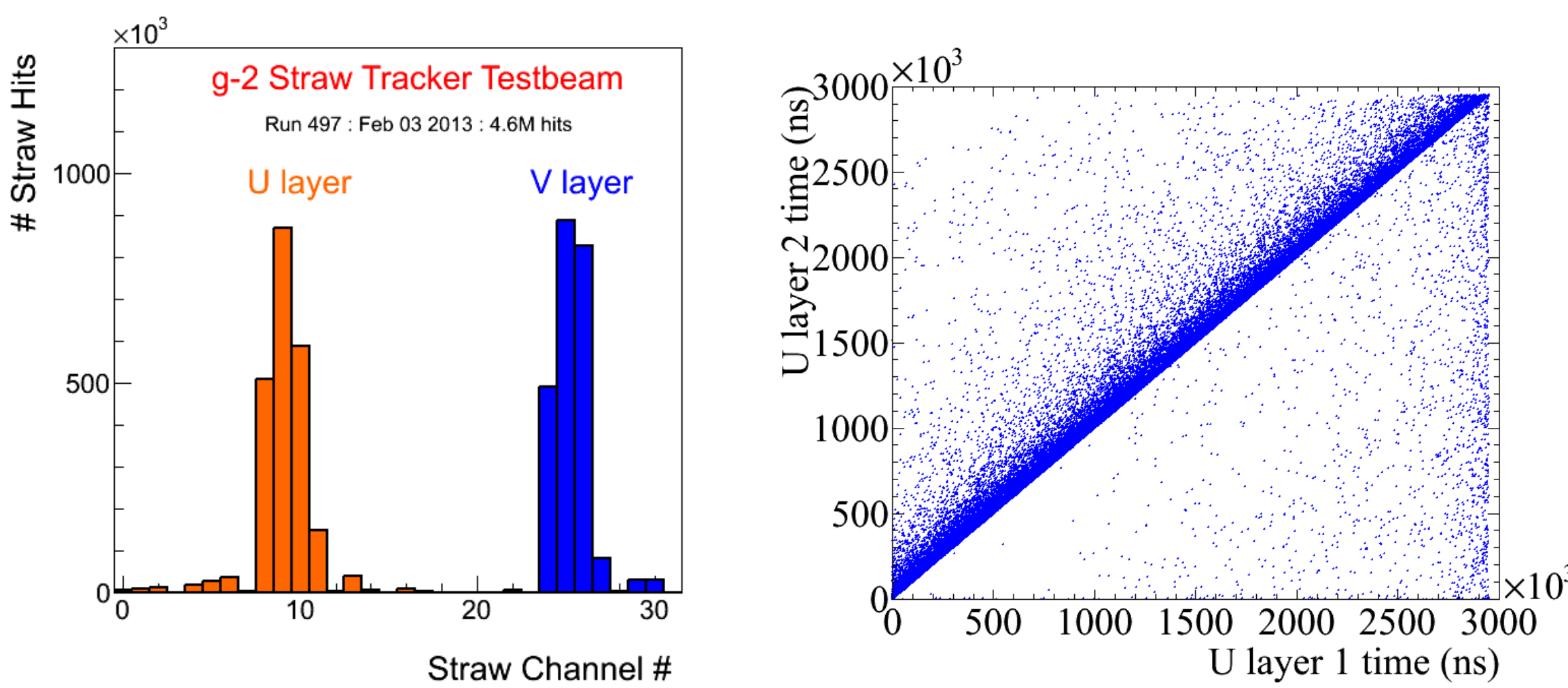

(a)

(b)

Figure 19.33: Data from the 32 channel prototype. (a): The hit distribution in channels showing the beam width and coincidences in the U and V layers. (b): The correlation between times in the two straw layers in the U view.

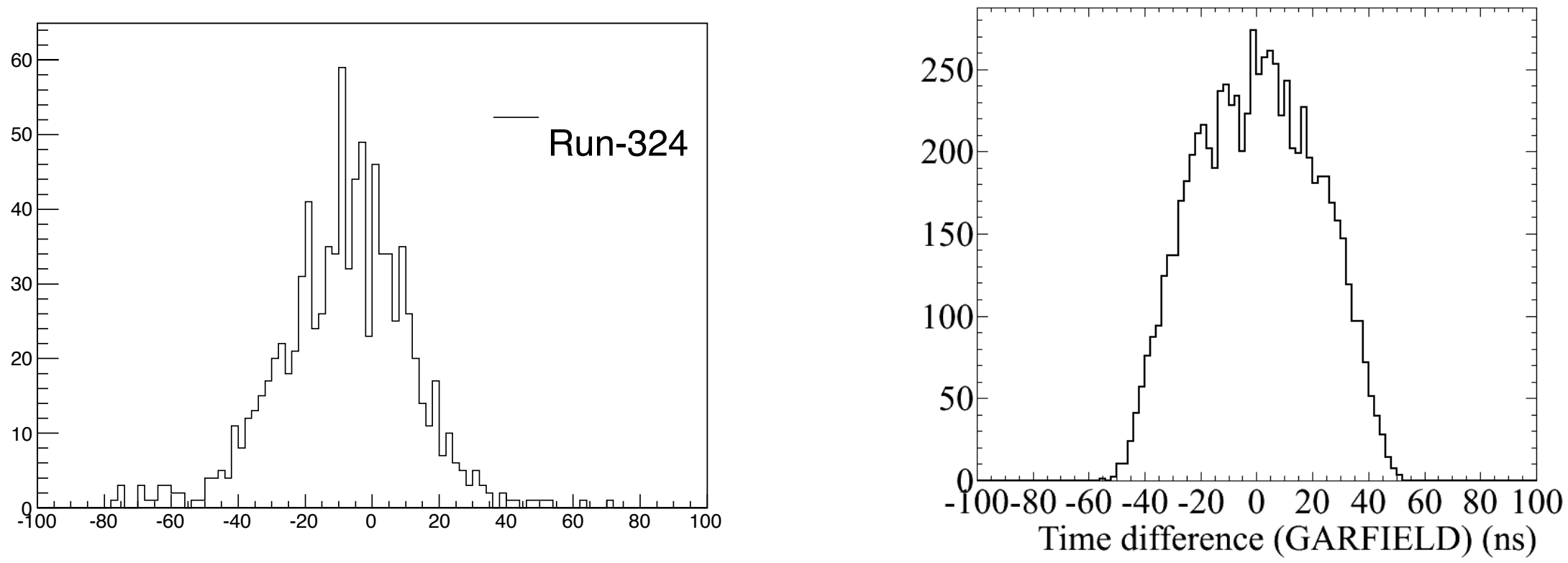

(a) (b)

Figure 19.34: Time difference between hits in a straw doublet. The width of the distribution is determined by the straw diameter, drift velocity, and resolution. The distribution on the left is for data from the beam test. The distribution on the right is from GARFIELD after the effective threshold extracted from data has been applied.

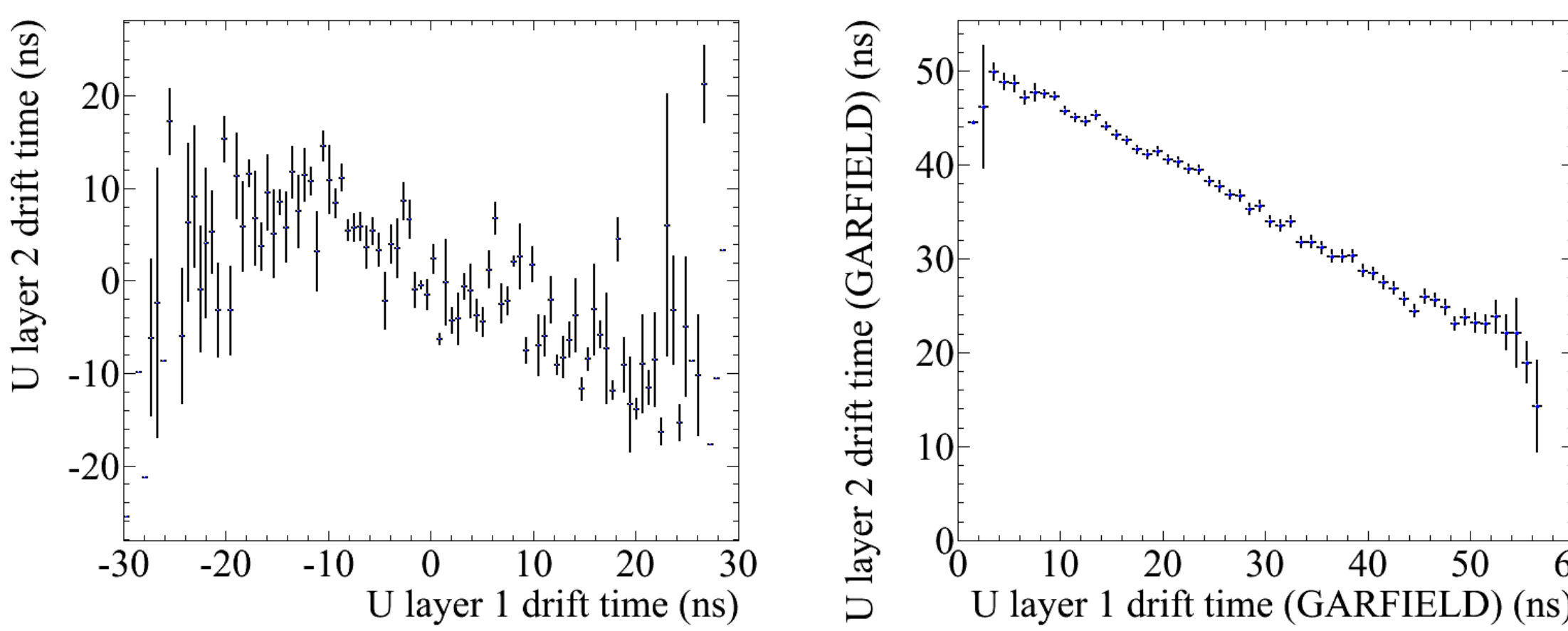

(a) (b)

Figure 19.35: Anti-correlation between the drift times in the straw doublet. The distribution on the left is for data from the beam test. The distribution on the right is from GARFIELD after the effective threshold extracted from data has been applied.

## 19.5 Alternatives

The down selection of the preliminary design has been an evolving process. We are confident that after thoroughly exploring all the alternatives listed below, we have converged on an optimal solution given the requirements and constraints. There are alternatives, mainly regarding the mechanical connection of the straws to the readout electronics, that will be explored and settled prior to construction but we expect the main features to stay intact.

The two leading alternatives to a straw based system for the tracker are both silicon based. The first would use 300 $\mu$m Hamamatsu single sided strip sensors. These sensors were purchased for the DØ Run IIb detector upgrade [14] but never used. Sufficient sensors are in hand to build the $g-2$ tracker. The readout would be based on the FSSRII chip [15] originally designed for BTeV and now being used for instrumentation upgrades for the JLab 12 GeV program. Tracking stations could be made with two sensors at a small stereo angle for a total material budget of 0.5% $X_0$ per station.

The second alternative would use the 50$\mu$m thick Mimosa 26 pixel sensor [16] that has been developed with ILC R&D funding. There is about 25% dead space on the chip which would require a doublet structure to maintain adequate acceptance. Material is also needed in the active region for cooling and for flex cables. A thermal model of the device indicates that heat can be adequately dissipated if the two layers are mounted on blocks of 2.5 mm thick Si foam. After the Si foam and flex cables are added, the material budget is also close to 0.5% $X_0$ per station.

If we had a well defined interaction point and could build something like a 4 layer detector, either of these two alternatives would be preferable to straws. However, the DC nature of the beam requires us to have a multi layer device to sufficiently cover the momentum spectrum of the positrons. Building this out of the silicon options above would add far too much material and the effects of multiple scattering would severely compromise our ability to extrapolate the positron trajectories all the way back to the muon decay position.

For the amplifier, besides the ASDQ chip, we explored using discrete components or building an ASIC. Discrete components were ruled out due to space considerations and also due to power consumption. An ASIC is an expensive alternative particularly since the ASDQ chips are free, but it has the advantage that we could control all material used in the chip to avoid magnetic components such as tin. However we brought the ASDQ and FPGA chips to a 1.5 T test magnet at the Fermilab Technical division and determined that these chips have magnetic properties well within our specifications.

For the TDC, we considered commercially available products such as the 128 channel CAEN 767 or 1190 model multi hit TDCs. This would require bringing all signals out of the vacuum through some sort of feedthrough system. We investigated the feedthroughs being designed for liquid Argon TPC detectors that have the electronics placed inside the cryostat but these would have difficulty operating at the $g-2$ vacuum of $10^{-6}$ Torr. The current design of an FPGA-based TDC is much more simple and cost effective.

For the electronics placement, we have considered designs with both the ASDQ and TDC boards inside the manifold (the conceptual design) as constructed for the 32 channel prototype, a design with both boards outside the manifolds (roughly equivalent to a design produced for Fermilab test beam wire chambers), and a design with the ASDQ boards inside the manifolds and the TDC boards outside the manifolds. The driving constraints are the desire to place the ASDQ as close to the end of the straws as possible without generating excessive heat in the manifold that would require complicated cooling. From the lessons learned from building the 32 channel prototype and operating it in test beam conditions, it is clear that there is not sufficient overhead in signal to noise to allow for the analog signals to be brought outside of the manifold. However, once the signal is digitized there is no constraint on the distance between the ASDQ board and the TDC board. We therefore chose the design with the ASDQ board inside the manifold with modest cooling and the TDC boards outside the manifold. If future versions of the electronics or straw construction yield a higher signal to noise, the gain on the straws can be reduced which will help elevate any issues associated with aging or excessively high rates early in the fill.

For the backend electronics and low voltage distribution we have considered two alternatives. The first is a custom AMC card that acts as a link between the TDC and fronted DAQ that also incorporates the low voltage distribution to the electronics as outlined in the conceptual design. The second is the semi-comercial option of using the FC7 AMC board available from CERN that works readily with our $\mu$TCA architecture and is essentially a large FPGA with configurable inputs and outputs. The low voltage would then be provided by commercial modules. This second choose was chosen since it involves significantly less engineering and provides a large amount of flexibility moving forward.

For the vacuum modifications, we have studied in detail three alternatives. First is a design that makes no chamber modifications and uses the existing ports upstream of the calorimeters. This is the least expensive solution but only allows us to build a tracker with approximately a 0.5 meter lever arm, greatly reducing the performance of the system. Second is a design that adds a flange to the vacuum chambers but maintains the vertical and radial dimensions of the chambers. This is a modest cost and allows for the maximum length for the tracker lever arm and was chosen as the conceptual design. However, the installation, plumbing, alignment, and reproducibility of placement all became serious issues with this design because it required the modules to be placed in the chamber, connections made between the modules and the flange, and finally mounting of the flange. The third design calls for radial extensions to the vacuum chamber that allow for an increase in the vertical size of the chamber. While the most expensive option, this allows modules to be mounted directly to the flange solving several engineering issues that arrived in the conceptual design and greatly reducing risks associated with alignment.

For the straw geometry, we have studied the performance as a function of straw pitch and naturally find the best performance for a pitch equal to the straw diameter. For the conceptual design, we chose a pitch of 5.5 mm. Finite element analysis indicated that this was the minimum pitch that could hold off the vacuum differential across the manifold wall with sufficient safety factor and without having an excessive manifold wall thickness. The 32 channel prototype was built with this pitch. It was found that with this pitch, a significant amount of post processing was required after the manifolds were removed from the CNC machine. The main issue is the relative softness of aluminum and the inability to hold proper tolerances between neighboring holes. We have therefore moved to a design that has a 6 mm pitch. This has been shown to make a major difference in terms of holding tolerances on the machined pieces greatly reducing the labor associated with construction while simulation indicates it is a marginal reduction in performance. Given the increased pitch, we also revisited the choice of doublet versus triplet straw configurations. While the triplet has superior performance in terms of the information gained from a single module particularly if the track goes through a straw gap, the number of modules hit for each track is large enough to counter inefficiencies in a single module. The relatively small increase in performance contrasted with the 50% increase in both channel count and vacuum load lead us to retain the doublet configuration.

For the straw gas, we have studied Ar:$CO_2$ and Ar:Et. Using $CO_2$ as a quencher has the advantage of being non-fammable but has the disadvantage of only being able to operate at low gain. $CO_2$ also has the disadvantage of having a high permeation rate through Mylar. Ethane has excellent quenching properties and a 50:50 mixture of Ar:Et has excellent linearity properties. The two disadvantages are the mixture is flammable and a pure mixture has poor radiation tolerance leading to polymerization of the sense wire. We have now been able to construct a prototype system based on Fermilab ES&H document 6020.3 "Storage and Usage of Flammable Gas" and the prototype has passed an operational readiness review which gives us confidence that we can design a safe device. Since Ar:Et is a standard gas choice, particularly for straws, there are several well known additives that counter act the polymerization process and we have chosen per mil concentration of oxygen based on experience operating the CDF tracking detector. While $CO_2$ meets the requirement specifications, there is no headroom. Ethane as a quencher allows higher gain operation providing significant headroom. We have those chosen Ethane as a quencher.

## 19.6 ES&H

The $g-2$ tracker is similar to other gas-based detectors that are commonly used at Fermilab and the $g-2$ tracker is identical in many cases to the Mu2e system. The main hazard is flammable gas. The area around the tracker will be a class zero flammable area and the operating procedures in the Fermilab ES&H manual for storing and operating a flammable gas system will be followed. Other potential hazards include power systems and compressed gas. The gas will permeate at a small level inside the $g-2$ vacuum and come in contact with the quadrupole high voltage. Any gas leak in the experimental hall will also bring the gas in contact with the high voltage stand-offs and feed-throughs of the kicker and quadrupoles. Because of this, and because using non-flammable gas appears to satisfy the performance requirements, we are precluding the use of flammable gas. These and all other hazards have been identified and documented in the Muon $g-2$ Hazard Analysis [17].

The detector requires power systems with both low voltages with high currents and high voltages. During normal operation, the tracker will be inaccessible inside the storage ring. Power will be distributed to the tracker through shielded cables and connectors that comply with Fermilab policies. Fermilab will review the installation prior to operation.

Gas that will be used for the tracker will be kept in DOT compliant cylinders in quantities limited to the minimum required for efficient operation. The cylinders will be stored in a dedicated location appropriate to the type of gas being used. The storage area will be equipped with fire detection and suppression systems. The installation, including all associated piping and valves, will be documented and reviewed by the Fermilab Mechanical Safety Subcommittee.

The detector itself does not have any radioactive sources. However, $Fe^{55}$ sources will be used to measure the gain of the straws before installation. Usage of radioactive sources will be reviewed to ensure adherence to Fermilab safety policy. In particular, the sources will be properly inventoried and stored and we see no opportunity for producing mixed waste.

Solvents such as ethanol will be used to clean components before assembly and epoxy resins will be used in the assembly process. All chemicals will be clearly labeled and stored in approved, locked storage cabinets and will adhere to the Fermilab safety policy.

# 19.7 Risks

## 19.7.1 Performance Risk

The tracker has been designed assuming a maximum instantaneous rate of 10 kHz/cm$^2$. This value is extrapolated from measurements at the Brookhaven experiment. The Brookhaven experiment had significant contamination from pions that led to a large hadronic flash at the beginning of the fill. This pion contamination has been removed from the Fermilab experiment but there is still a possibility that there will be some unaccounted for background that leads to unacceptable rates. The straws have been designed to operate with $CF_4$ so a faster gas could be used to deal with this. We are also investigating using a circuit to reduce the gain of the straws during injection. This is complicated and would require electrical engineering resources to design if we are required implement this.

The system of collimators used to scrape the muon beam after injection is partially in the line of sight of the tracking detectors. This would limit the acceptance of the tracker and potentially cause high backgrounds early in the fill. To mitigate this risk, we are performing studies to determine alternative locations for the collimators and working closely with the groups associated with the collimator system.

## 19.7.2 Technical and Operational Risk

The greatest technical risk is that the tracking system will in some way affect the precision magnetic field of the storage ring. This risk is being mitigated in several ways:

- All scientists, engineers, technicians, students, and vendors involved in the design and construction of the system are educated on the importance of the magnetic properties of the system.
- The specifications are clearly stated in terms of the static and dynamic effects on the field. These have been documented and agreed on by the collaboration.
- Individual components are taken to an existing 1.5 T test magnet and their static magnetic properties are verified to be within specifications.
- Full magnetic simulation of the detector using OPERA [18] will be added to the existing storage ring OPERA simulation to verify that any static effects can be shimmed out of the field using the existing shimming kit.
- The full detector will be tested in a test solenoid that is being shipped from LANL to Fermilab specifically for this purpose.

- A fast coil will be designed to measure the size and time structure of any transient magnetic fields being produced by the electronics.

As discussed above, all prototype electronics constructed so far have been within the magnetic field specifications.

The vacuum specifications for the g-2 storage ring are set by the electrostatic quadrupoles inside the storage ring. The combination of the electric field from the quads and the magnetic field from the g-2 magnet leads to regions where photoelectrons can be captured in Penning traps. These electrons can eventually interact with residual gas molecules, leading to avalanche and sparking. This is the primary factor influencing the lifetime of the quadrupole plates.

For $\mu^+$ operation, a vacuum of $10^{-6}$ Torr is required. If $\mu^-$ running is required or if the quadrupoles are operated at a greater HV to move to a different tune point, the vacuum may need to be improved to $10^{-7}$ Torr. The leak rate of the straws has been measured by Mu2e and indicates that $10^{-6}$ Torr can be achieved. To mitigate the risk of needing to operate at a higher vacuum we are designing the ability to add higher capacity to the pumping speed near the tracking detectors. We have also added a 25 $\mu$m secondary containment barrier using aluminized Mylar. This greatly increases the efficiency of any local pumping but adds material in front of the detector.

Contaminated gas is a serious risk for any drift chamber. This risk is mitigated in several ways. We will perform a detailed analysis on each batch of gas before it is incorporated into the system. Finally, spare chambers in test stands will use the same gas and will be illuminated with radioactive sources to monitor gain and give early warning of problems.

A broken wire will cause an entire plane of a module to be inoperable. A broken straw will cause an entire module to be inoperable. To mitigate this risk, the system is being designed in a way so that a damaged module can be easily removed and replaced with a spare with approximately 1 day lost to reestablishing the vacuum. We anticipate breaking vacuum at least once every several months to service the NMR trolley so as long as the frequency of problems is much less than this, there is no risk to the run schedule.

## 19.8 Quality Assurance

Proper quality assurance is essential to construct a tracking detector that meets the Muon $g-2$ requirements for performance and reliable operation. Quality assurance will be integrated into all phases of the tracker work including design, procurement, fabrication, and installation.

Individual straws must be leak tight, straight, and be held at the proper wire tension. As the straw modules will be placed in vacuum, leak testing is essential to ensure that the vacuum in the region of the tracking stations is not unduly degraded. The straws will be leak tested before being installed. The straws will be connected to a clean gas system and over-pressured. The leak rate will be measured over an appropriate time interval by measuring the pressure drop. A leak/burst testing apparatus for a single straw has been constructed, as shown in Fig. 19.36. Based on experience with this testing apparatus, a new leak testing apparatus is being designed (see Fig. 19.36 for a design image) which can test multiple straws in parallel. After the assembly of a station, the entire station will be leak tested again. This will be accomplished by placing the straw tube module into a vacuum chamber and flowing clean gas into the module while pumping down the chamber. The module leak rate will be measured by monitoring the pressure in the vacuum chamber versus time.

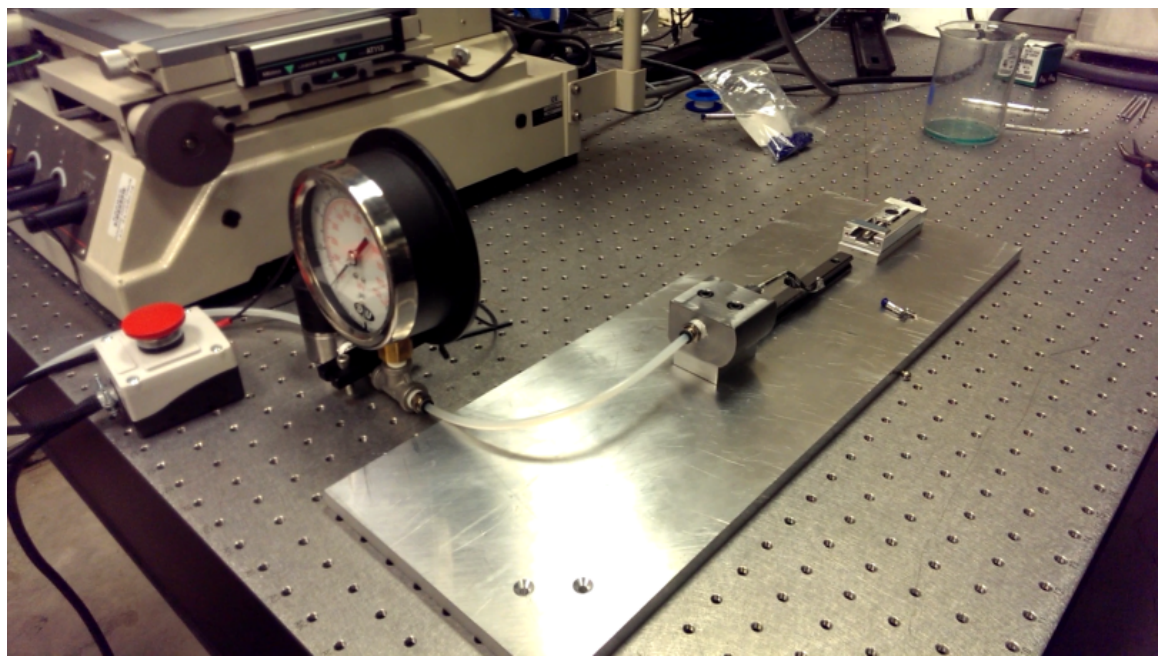
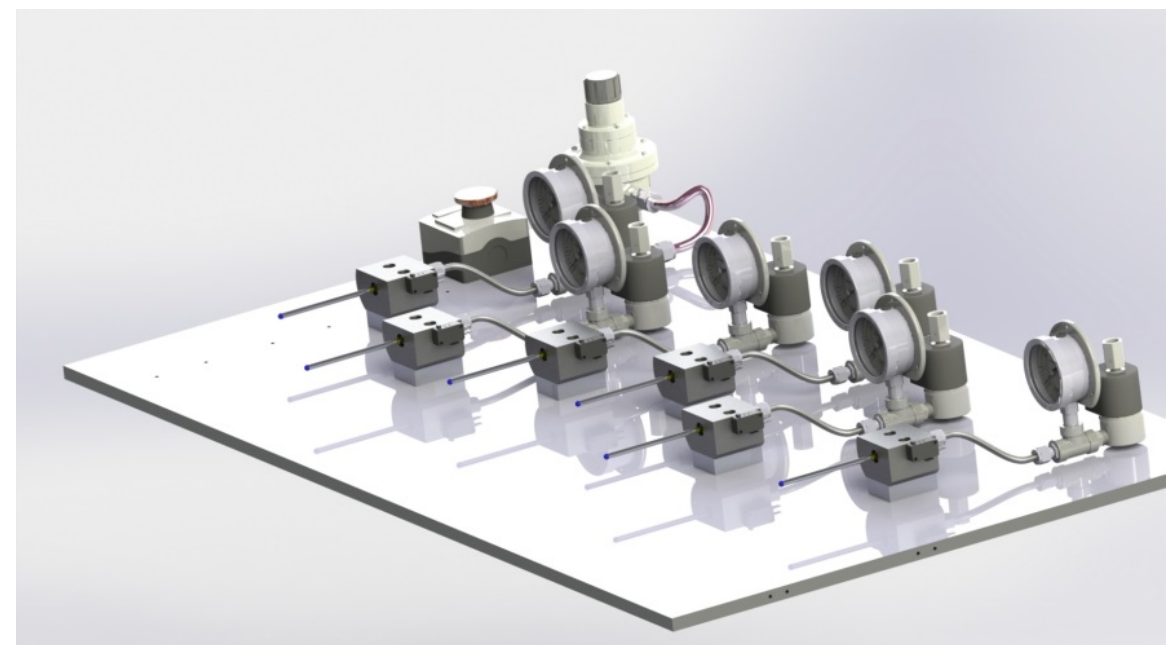

Figure 19.36: Photo of a single-straw leak testing/bursting test apparatus (left). Design drawing of a multiple straw leak testing apparatus (right).

The straws must maintain their shape and be mounted at the proper stereo angle to operate efficiently and to maintain an appropriate distance between the wire and the grounded Mylar surface to avoid breakdown. Straws will first be visually inspected for roundness and straightness before assembly. Flawed straws that escape detection during visual inspection can be identified by non-uniform gas gain and resolution. This will be done as part of the wire position measurement. In addition to visual inspection, the resistance of the straws will also be measured throughout the assembly process. The Mylar straws conduct through the few-hundred Angstrom aluminum and gold coatings. Our experience with these straws has shown that the resistance of the straws is very sensitive to physical damage to straws, likely through damage to the thin conductive coating [19].

The appropriate tension must be applied and maintained in a straw for efficient, stable operation. Tension is applied through calibrated mechanical force but can be lost through relaxation mechanisms. Additionally, since the straws are primarily composed of Mylar, when under tension they will experience a lengthening over time (referred to as ”creep”). This creep effect will reduce the tension in the straw over time. Straw creep has been measured over a duration of approximately one year [19]. For the values of straw tension to be used in the modules, the amount of straw creep is negligible and not a concern.

Both wire and straw tension will be measured after assembly using vibrational resonance techniques appropriate to our short straws. Utilizing the experience of other experiments that have used straw tubes, testing procedures have been devised to measure both the straw tension and the wire tension after the straw tube modules have been assembled. These tests are detailed in Ref. [19] and Ref. [20]. Both tests are based on measuring the frequency of the induced EMF in a vibrating wire/straw in a magnetic field. Proof-of-prinicple tests have shown that the tension can be measured precisely enough for our needs. (Note that due to the short straws/wires and nearly vertical orientation, straw/wire sag is not a significant concern.)

All electronics components will be tested prior to installation on the tracking stations including a suitable burn-in period. The high voltage circuits will be tested for leakage current. The threshold characteristics of each channel will be tested with a threshold scan. A noise scan will be performed for various threshold settings to identify channels with large noise fractions. The FPGA TDCs will be validated by comparing their output to commercial TDC devices with higher resolution. After the final assembly of a straw module (including the electronics), all modules will undergo a system test utilizing cosmic rays or a radioactive source to verify the operation and performance of all channels.

## 19.9 Value Management

The tracker technology for Muon $g - 2$ is well established and has been implemented in other high energy and nuclear physics experiments. Value management principles have been applied over time during the development of the technology. Value management moving forward is mainly related to labor costs since the straw tracker assembly will be labor intensive. We have identified collaborating institutions with the technical capabilities to perform a large fraction of the assembly work at minimal cost.

We are subcontracting engineering to university engineering departments and using Fermilab engineering resources to perform independent design reviews before production or procurement. This keeps the overall engineering costs low while maintaing the standards of Fermilab engineering.

The back-end readout electronics and data acquisition for the tracker are equivalent to those used for the calorimeters. This simplifies the design and operation of the system. However, once the final specifications are known, we will investigate possible cost savings by using different system components. Current FPGA technology is sufficient to meet the needs of the tracker electronics. These will be purchased once they are no longer the most current devices which should lead to significant cost savings. Sufficient spares will be purchased to ensure the stock for the lifetime of the experiment.

The straw terminations require injection molded pieces. The cost of these pieces is almost entirely driven by the cost of the mold and so design iterations are costly. To mitigate this, we intend to first produce all injection molded pieces with a 3-D printer and construct straws with the printed pieces to validate the design before the molds are procured.

# Chapter 20

# Auxiliary detectors

## 20.1 Fiber beam monitors

### 20.1.1 Requirements

The fiber beam monitor system is designed to serve three purposes:

- As a commissioning instrument, to determine the position ($x$, $y$) and angle ($x'$, $y'$) of the beam at injection. (These coordinates are defined in Section 8.3.2.)
- To monitor the evolution of these beam properties during the kick and scraping phases.
- To observe and directly characterize periodic beam motion, notably the modulation of beam centroid position and width by coherent betatron oscillations, as described in Chapter 4.

In order to serve these purposes, the fiber beam monitor system is subject to the following requirements:

- The pulse width and deadtime must be much less than one cyclotron period of 150 ns, by at least one order of magnitude.
- The system must be able to characterize a muon beam whose intensity ranges from 5% to 200% of the expected number of muons ($1.8 \times 10^4$) that are captured in the storage ring.
- The spatial resolution of each detector must be sufficient to observe the transition from $x'$ to $x$ and from $y'$ to $y$ over a 90° phase advance. (The procedure for this analysis is described in Ref. [1].)
- The detector must be able to reside in a vacuum of $10^{-6}$ Torr, with a vacuum load of less than $5 \times 10^{-5}$ Torr L/s.
- The detector must be able to function in a 1.5 T magnetic field.
- The detector must not perturb the local magnetic field by more than 10 ppm. There must be no transient field perturbations of less than 1 ms duration except during special runs when the detector is activated and inserted.

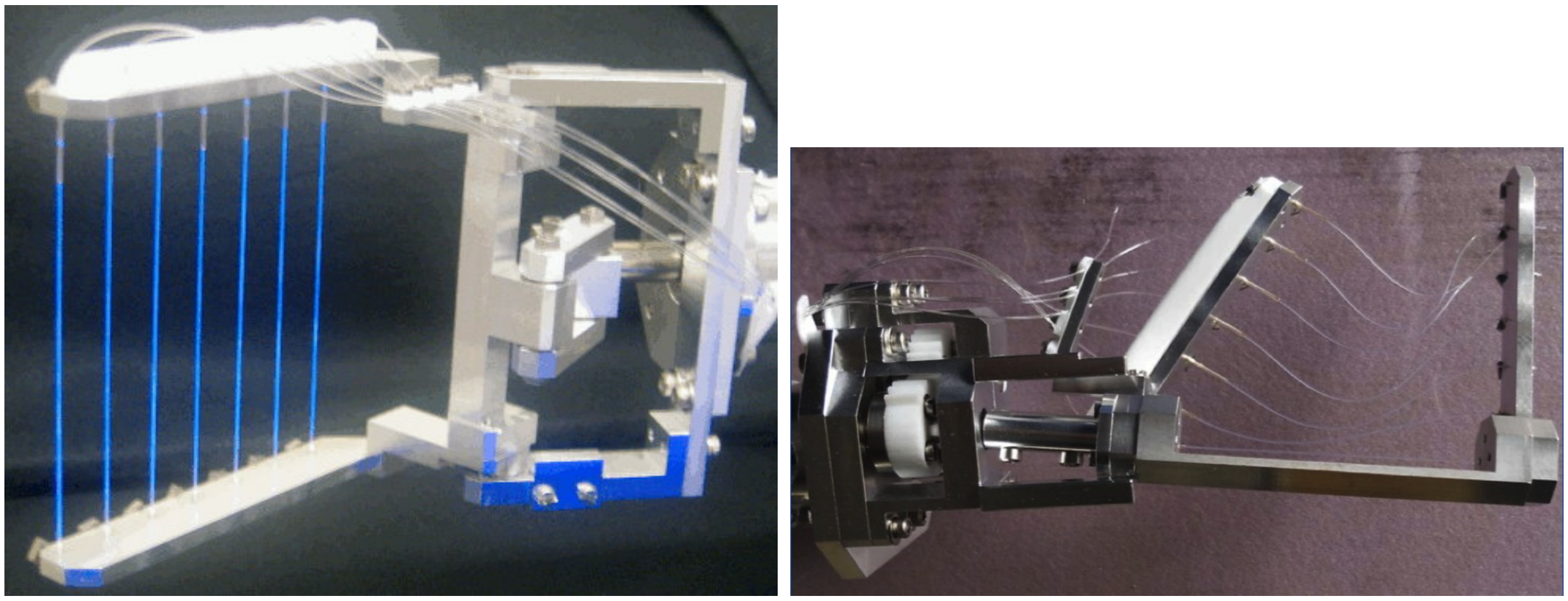

(a) (b)

Figure 20.1: (a) The 180° $x$-profile monitor, glowing under ultraviolet illumination in the laboratory. (b) The 270° $y$-profile monitor, which was found to be damaged when it was removed from the Brookhaven E821 storage ring.

### 20.1.2 Recommended Design

The fiber beam monitors were originally built for E821 by a group at KEK that is not part of the Fermilab collaboration [2, 3]. We intend to refurbish and reuse all components from the existing system that remain suitable.

Each fiber beam monitor holds a "harp" of seven scintillating fibers of 0.5 mm diameter, each 90 mm long and separated from its neighbors by 13 mm, as shown in Figure 20.1(a). Each scintillating fiber is bonded to a standard optical fiber that connects it to a vacuum feedthrough. There are a total of four devices, and they are deployed near the 180° and 270° positions in the ring. The two 180° fiber beam monitors should observe an image of the beam as it was injected at the inflector, while the two 270° fiber beam monitors should map $x'$ and $y'$ at the inflector into $x$ and $y$ at 270°. At each location, one fiber beam monitor suspends the fibers vertically to measure in $x$, and the other arranges them horizontally to measure in $y$. The fibers stay inside the vacuum, and they can be plunged into the beam path. As shown in Figure 20.2, they can be also rotated into a horizontal plane, where all fibers see the same beam, for calibration, or upright for measurement. Because ferromagnetic material cannot be placed this close to the precision magnetic field, aluminum motors and actuators driven by compressed air are used for this motion.

Three of the four fiber beam monitors recovered from E821 appear to be in mechanically good condition. One fiber beam monitor (the 270° $y$-profile monitor) was found to be significantly damaged, with a snapped fiber and bent frame components, as shown in Figure 20.1(b). This damage may have existed since early E821 runs; an unexpectedly high muon loss rate when the fiber harps were inserted suggests that there were unintended scattering sources in the beam. We have disassembled the affected part of the damaged harp and found that there is one aluminum part that will need to be replaced, along with a shaft that has already been straightened and a few bent screws. The work to fabricate the replacement

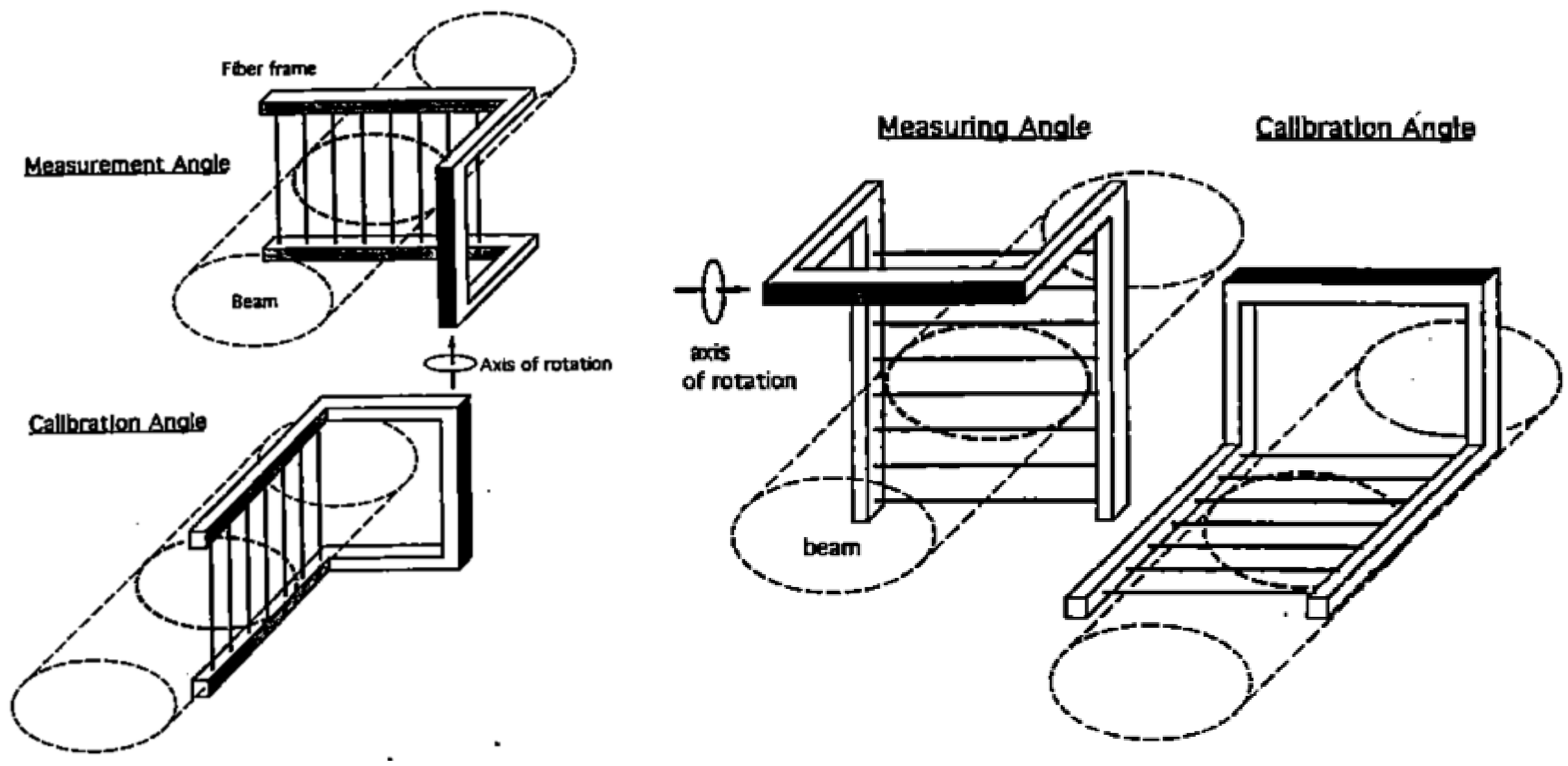

(a) (b)

Figure 20.2: Rotational motion corresponding to the calibration and measurement positions for (a) $x$ and (b) $y$. This figure is reproduced from [3].

part has already begun.

In addition to the fiber that is completely broken on this harp, another shows visible indications of strain. There are also two broken and one strained fibers on the 180° $x$-profile monitor, and two fibers on the 180° $y$-profile monitor are broken at the tips so that they are not supported rigidly. As laboratory testing has proceeded, we found other broken and damaged fibers; they break accidentally with very little force. Also, as shown in Figure 20.3, we have measured the photoelectron yield of an existing fiber taken from a harp and compared it with that of two new fiber samples provided by Kuraray. We can expect a factor of 2.7 more light (4.2 versus 1.6 photoelectrons) from a new fiber. Consequently, we are developing a reliable, reproducible technique to replace all of the fibers.

We initially tested a technique that yields a mechanically stable connection with no visible gap between two polished fibers: holding one of the fibers from below with a clamp, placing a 1 cm length of 24-gauge light-wall PTFE tubing over it, dispensing a drop of optical cement into the tubing with a needle, and inserting the second fiber from above. When the fibers were cleaned and polished properly before being joined, we observed transmission of 71% with this method. This agrees with the result of Ref. [4], which reports on a similar method, and with the 69% transmission that we measured for an existing glue joint that broke from a harp.

We attribute the light loss primarily to stripped cladding at the ends of the fibers, so we are now refining a technique that relies less on mechanical polishing. The scintillating and clear fibers will be embedded in a hole drilled in a block of PTFE, cemented in place with a vacuum-compatible epoxy. The blocks as a whole will be flycut with a polycrystalline diamond cutter; only minimal fine polishing of the block with 1 micron and 0.3 micron lapping paper will be needed as a final step. Alignment holes are drilled in the blocks to allow the fibers to be held together temporarily with M1 screws while an optical adhesive between them cures. Figure 20.4 shows a block with a single scintillating fiber for testing; the eventual intent is to embed all 7 fibers from one harp in a single connector block. A first test with this fiber has shown transmission of 77% from a scintillating to a clear fiber, and we are continuing to improve the technique.

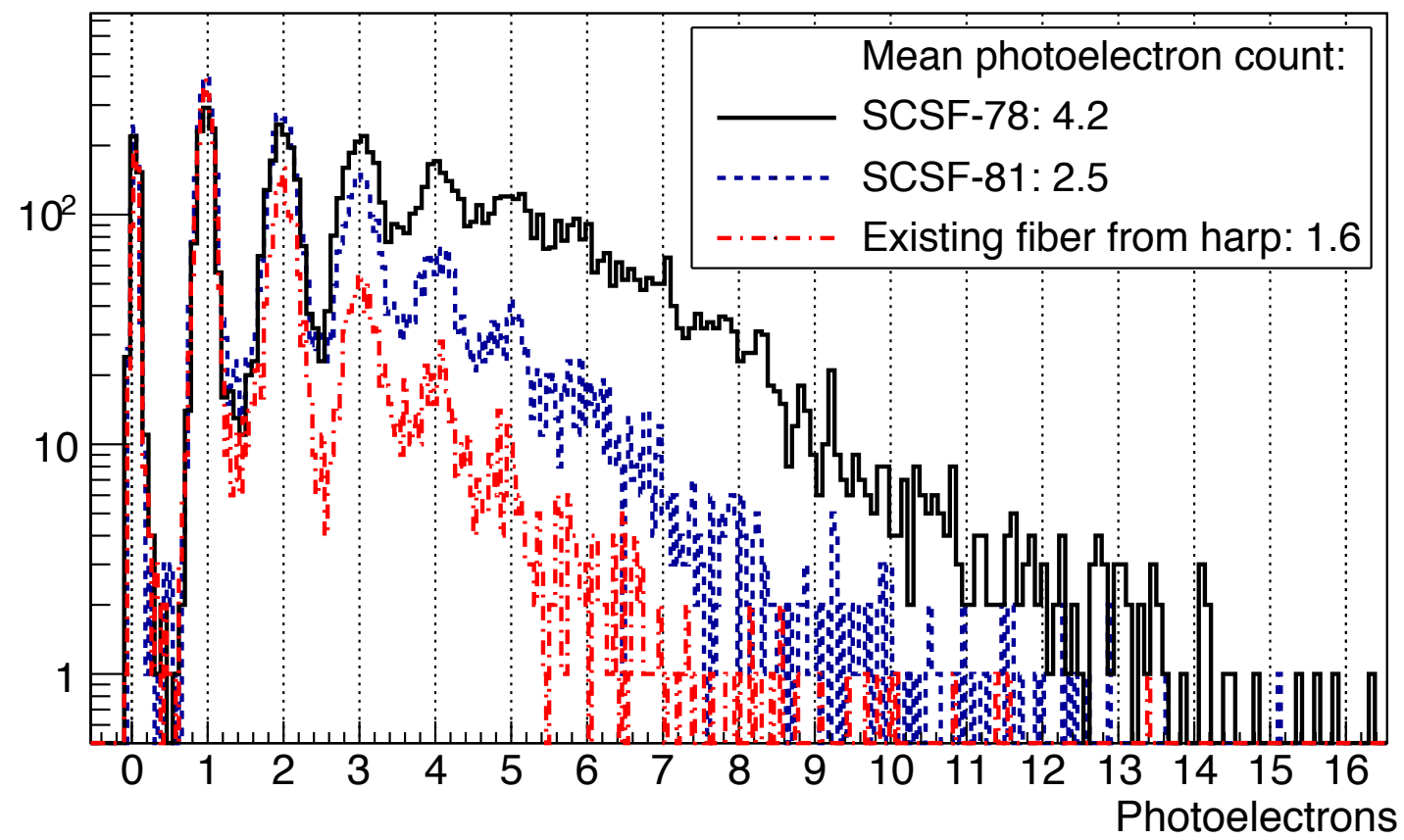

Figure 20.3: Number of photoelectrons per minimum ionizing event from a $^{90}$Sr beta source for two new fiber samples provided by Kuraray, in comparison to an existing fiber that broke from one of the harps.

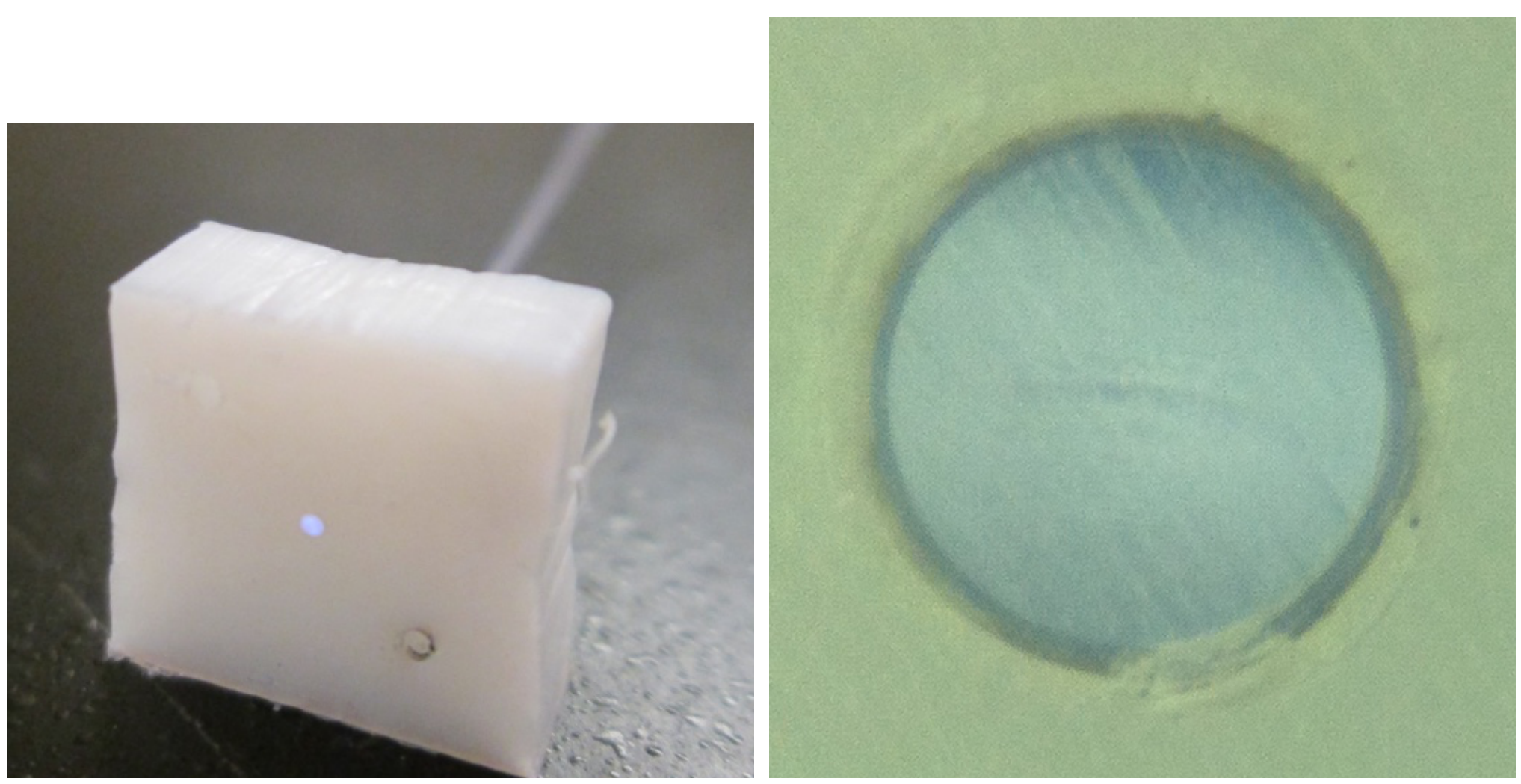

Figure 20.4: Two views (microscopic view on right) of a scintillating fiber that has been flycut inside a PTFE block. We are continuing to improve the procedure to minimize the remaining surface roughness and cladding separation.

We have cleaned the exterior parts of all of the fiber harps, and three of them have been tested through their full range of motion using a prototype of a new motion control system. After repairs are completed to the damaged harp, we expect to be able to do the same for it. Similarly, we will test and assure the vacuum integrity of the system before returning the fiber harps to Fermilab. We have already successfully tested one harp by observing the rate of pressure rise with a computer-interfaced Pirani gauge after a turbomolecular pump is valved out. After about a day of pumping, the rate was $2 \times 10^{-5}$ Torr L/s.

In E821, the fibers were read out with conventional photomultipliers in a remote location, at the end of a long fiber, where the magnetic field was reduced. Replacement by SiPMs mounted directly on the fiber harps will allow the long fiber to be eliminated. SiPMs also have higher photon detection efficiency than conventional photomultipliers. Initial SiPM tests have been conducted with the Hamamatsu S10362-11-050C, for which we have developed a readout board with a simple two-stage voltage preamplifier. It has a $1{\times}1$ mm$^2$ area that is suitable for fiber readout applications.

This SiPM is a reasonably appropriate match to the estimated number of photoelectrons, although a model with more pixels remains under consideration as an alternative. A GEANT4 simulation indicated that the most probable energy deposit is 0.06 MeV in each interaction, leading to approximately 6 photons at the SiPM. The proposed SiPM, with ∼65% quantum efficiency and 61.5% fill factor, would therefore yield 2.4 photoelectrons per interaction. This is less than the result shown in Figure 20.3 because it allows for some losses at the scintillating-to-clear fiber boundary and at the vacuum feedthrough. Approximately 1% of stored muons should interact with a central fiber in each turn. The rate calculation summarized in Table 5.1 shows that we can expect $1.8 \times 10^4$ on the initial turns around the ring. This would lead to 432 photoelectrons reaching the SiPM, which does have the potential to saturate the 400 available pixels; however, if too much light really becomes an issue, it would be straightforward to insert an attenuating filter. The maximum dark count rate of 800 kcps would give one photoelectron of noise every 8 cyclotron periods, which is very small compared to the expected amount of light. To fully cover the fibers on each of the four monitors, we need 28 SiPMs. We have acquired 30 Hamamatsu S10362-11-050C devices that were left unused by a project at Argonne National Laboratory.

The SiPM requires a preamplifier. For laboratory tests, we are currently using a design based on a circuit developed by the electronics group at the Paul Scherrer Institute and used in several experiments there [6, 7]. It uses two stages of Mini-Circuits MAR-series gain-block amplifiers separated by a pole-zero cancellation. A schematic of this circuit, appropriately adapted for the fiber harp application, appears in Figure 20.5. A prototype of this circuit has been constructed, and the full width of the pulse is ∼10 ns, which is more than sufficient for the required time resolution. Because the signal will be recorded by the new waveform digitizer modules that are being developed for the calorimeters, we plan to replace the second stage in this design with the same LMH6881 programmable-gain differential amplifier that is used on each crystal. Prior to installation, we will test the magnetic susceptibility of the new preamplifier, which is the only really new part of the system; the rest of the fiber harp was demonstrated in E821 to be compatible with the requirements of the precision field.

Waveforms from these SiPMs will be recorded by the same waveform digitizer that is being developed for the calorimeters; an additional MicroTCA crate will be located near the center of the storage ring for this purpose.

# 20.2 Entrance counters

## 20.2.1 Requirements

The time at which the muon bunch enters the ring must be subtracted from the time of each decay positron in order to align data from different fills properly. The relative intensity of each fill is also monitored. An entrance counter, positioned just outside the inflector, is needed to record the time and intensity of each fill. In E821, the cyclotron "fast rotation" structure, described in Section 4.3.1, was removed by adding a uniformly distributed pseudorandom number from zero up to the cyclotron period ($\sim$150 ns) to the entrance time of each bunch. This procedure, which will presumably be needed again in the new experiment, artificially degrades the time resolution of the entrance counter, setting the scale of the requirement on that parameter to a level that is easily met by SiPM technology:

- The counter must be able to determine the mean time of each muon bunch with a time resolution that is much less than the cyclotron period of 150 ns, by at least one order of magnitude.

- The counter must be able to adequately characterize a beam whose intensity ranges from 1% to 200% of the expected $7.3 \times 10^5$ muons per fill at the inflector entrance.

In E821, "flashlets" of beam that leaked from the AGS onto the target during the measuring period led to a potential systematic error. Muons produced by protons from a bunch that had not been cleanly kicked arrived at the experiment at the cyclotron period of the AGS. It is difficult to envision how this phenomenon could arise at Fermilab; any out-of-time muons would somehow need to be stored in the delivery ring without being kicked in. Nevertheless, it is worthwhile to be prepared with an extinction monitor to verify the absence of these out-of-time muons. Such a monitoring detector must satisfy this requirement:

- The counter must be able to detect a single isolated muon following a pulse of up to 200% of the expected $7.3 \times 10^5$ muons per fill, after a delay of 10 $\mu$s.

The 10 $\mu$s delay is set by the circumference of the recycler ring, which would give any "flashlets" a period of 10.9 $\mu$s. It is much longer than the recovery time constant of a SiPM, so it should be easily achievable.

## 20.2.2 Recommended design

In E821, the primary entrance ("T0") counter consisted of a 1 mm thick, 10 cm diameter volume of Lucite that produced Cerenkov light. It was coupled to a two-inch Hamamatsu

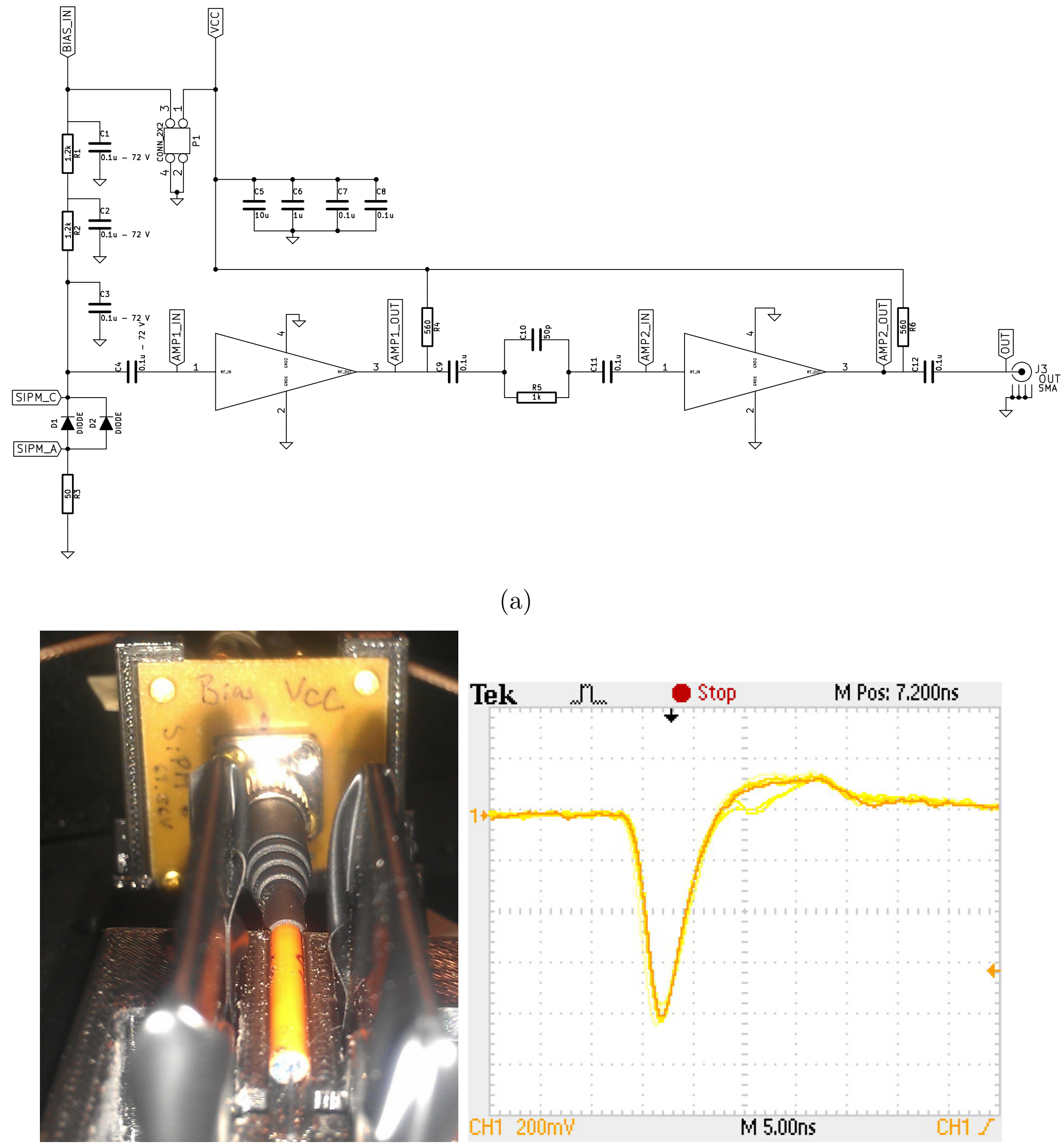

Figure 20.5: (a) Schematic of SiPM preamplifier circuit for fiber harps, based on a design from the Paul Scherrer Institute. (b) Photograph of a prototype board with a connectorized scintillating fiber sample attached. (c) Oscilloscope trace of a pulse from it.

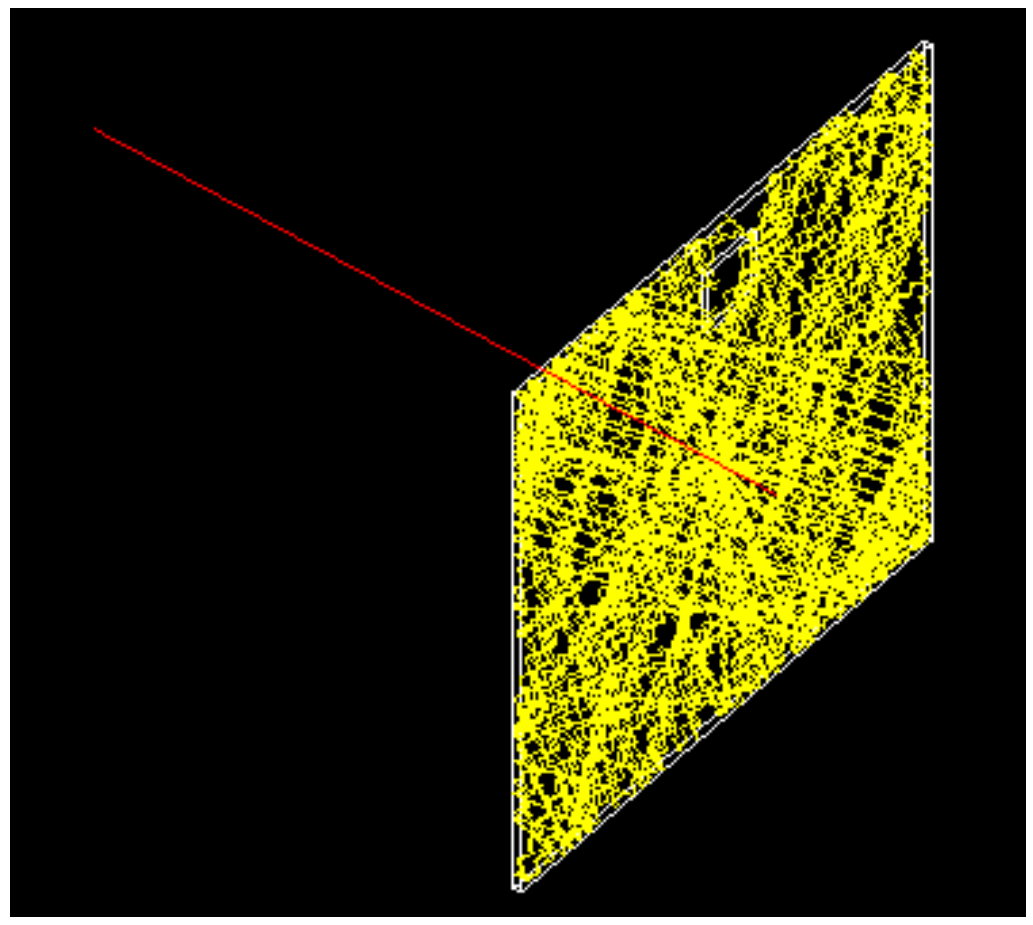

Figure 20.6: Tracking of photons produced by a minimum-ionizing muon in a 1.5 mm thick Lucite Cerenkov counter.

R1828 photomultiplier. It was initially believed that this device could be reused. However, when it was located after transportation to Fermilab, it was found to be damaged beyond convenient repair. Consequently, a new device is needed. The flashlet counter was a plastic scintillation detector that was only used in early runs of E821. The photomultiplier was configured to be gated off at the primary beam injection time by reversing the voltages on two dynodes. Consequently, the gain could be set to observe small amounts of beam entering at later times. Unfortunately, this device was also not found to be in usable condition.

A GEANT4 Monte Carlo simulation of two entrance counter materials was performed. In each case, a beam of 3.1 GeV muons was passed through the center of a $10 \times 10$ cm$^2$ area of a thin sheet of material with a SiPM of $12 \times 12$ mm$^2$ active area glued to the flat side near the edge, as shown in Figure 20.6. Photons from scintillation and Cerenkov processes were tracked, and the result was a Gaussian distribution centered at $N_{Lucite} = 22.4$ photons per muon reaching the SiPM for a 1.5 mm thick Lucite sheet, and $N_{Scint} = 277$ for a 0.5 mm thick Eljen EJ-212 scintillator sheet.

The spectrum of Cerenkov photons in the Lucite extends from 390 to 570 nm; the spectrum of scintillation light from EJ-212 extends from 400 to 520 nm but is peaked near 420 nm. The photon detection efficiency of typical SiPM devices for these wavelengths is approximately 30%. Consequently, approximately 7 detected photoelectrons per muon would be expected for the Lucite, compared to approximately 80 for the scintillator. Realistic indices of refraction were included in the simulation, but the surfaces of the plastic were assumed to be perfectly polished, so this study gives an upper limit on the potential performance of the detector. However, a prototype of this counter has been constructed and is currently being tested with cosmic rays in the laboratory, and the result will be compared to the light yield of a thin scintilator.

Muons deposit an average of 100 keV of ionization energy in a 0.5 mm plastic scintillator. If the full beam intensity of $7.3 \times 10^5$ per fill is spread over a 1 cm$^2$ area, the radiation dose would be about $2 \times 10^{-5}$ Gy per fill. Significant adverse effects in plastic scintillator are typically seen at the level of $10^4$ Gy [9], so the counter would be expected to last for $5 \times 10^8$ fills, or approximately 500 days of 12 Hz operation. This is well-matched to the expected duration of the experiment, although it may require replacement at some point.

We will therefore adopt as a baseline plan to build the entrance counter from a 0.5 mm thick EJ-212 plastic scintillator. It will have two positions to attach light detection devices, either photomultipliers or SiPMs. One of these devices will be used as on the calorimeter, coupled to the scintillator without attenuation; this will play the role of the flashlet counter. The other position will have a slot in front where a neutral density filter can be placed to attenuate the light to produce a linear response to the primary beam pulse; his will be the T0 counter. As the beam is tuned and the intensity increases, it will be possible to change the neutral density filter to match. The signals from each of these detectors will be routed to spare channels of the new calorimeter waveform digitizers.

## 20.3 Performance

The existing fiber beam monitors were used in E821. They fulfilled the requirements of that experiment, which were very similar to those proposed here.

The SiPM upgrade to the fiber beam monitors will further increase the number of detected photons and therefore improve the signal-to-noise ratio. However, the performance was already sufficient to directly observe and characterize the coherent betatron motion, which was published as Figure 21 in [5].

In E821, the fiber beam monitors were not prepared for the first day of the run. We will ensure that all of these detectors are ready for the first day of muon beam operation so that they can fulfill their requirements as beam commissioning devices. They can be used in this role to monitor the amplitude of the coherent betatron motion as a function of the injection and kick parameters.

## 20.4 Alternatives

Initially, we evaluated reusing the conventional photomultipliers that were used with the fiber beam monitors in E821. In that experiment, a $\sim$3 m long fiber connected each of the fibers from the feedthrough to an Amperex XP2202/B photomultiplier tube that was located in a cable tray above the storage ring in a location where the magnetic fringe field could be shielded with mu-metal. The signals were small enough that various models of LeCroy linear amplifiers in an adjacent NIM crate were needed to drive the long cables to the counting room. The photomultipliers and voltage divider bases that were used in E821 had already been reused from a previous project, and they are clearly aging devices that are in need of replacement. An initial inspection showed that they are in poor condition and unsuitable for future use. It would have been necessary to develop a new light conversion system, whether or not we moved to SiPM technology. Given the collaboration's familiarity with SiPMs from their extensive use in the calorimeter development (discussed in Section 17.4), their compactness, and their comparatively low cost, there was a clear choice.

The fiber beam monitor is not suited for a determination of the equilibrium radius of the stored beam. A GEANT4 simulation showed that energy loss in the fibers moves the average radius inward by $\sim$0.1 mm/$\mu$s, so the radius will be altered before equilibrium can be established. Even an order of magnitude less energy loss would still be unacceptable for this measurement, so it is not plausible that any system that intercepts the beam would be useful for it. We briefly considered diagnostic devices that would remain continuously deployed in the storage ring. However, any detector that intercepts the muons, even a low-mass wire chamber, would degrade the beam lifetime unacceptably, as shown in Figure 20.7. The E821 experience with a completely parasitic detector, the pickup electrodes, was also unsatisfactory; they were paralyzed by the pulsed high voltage devices.

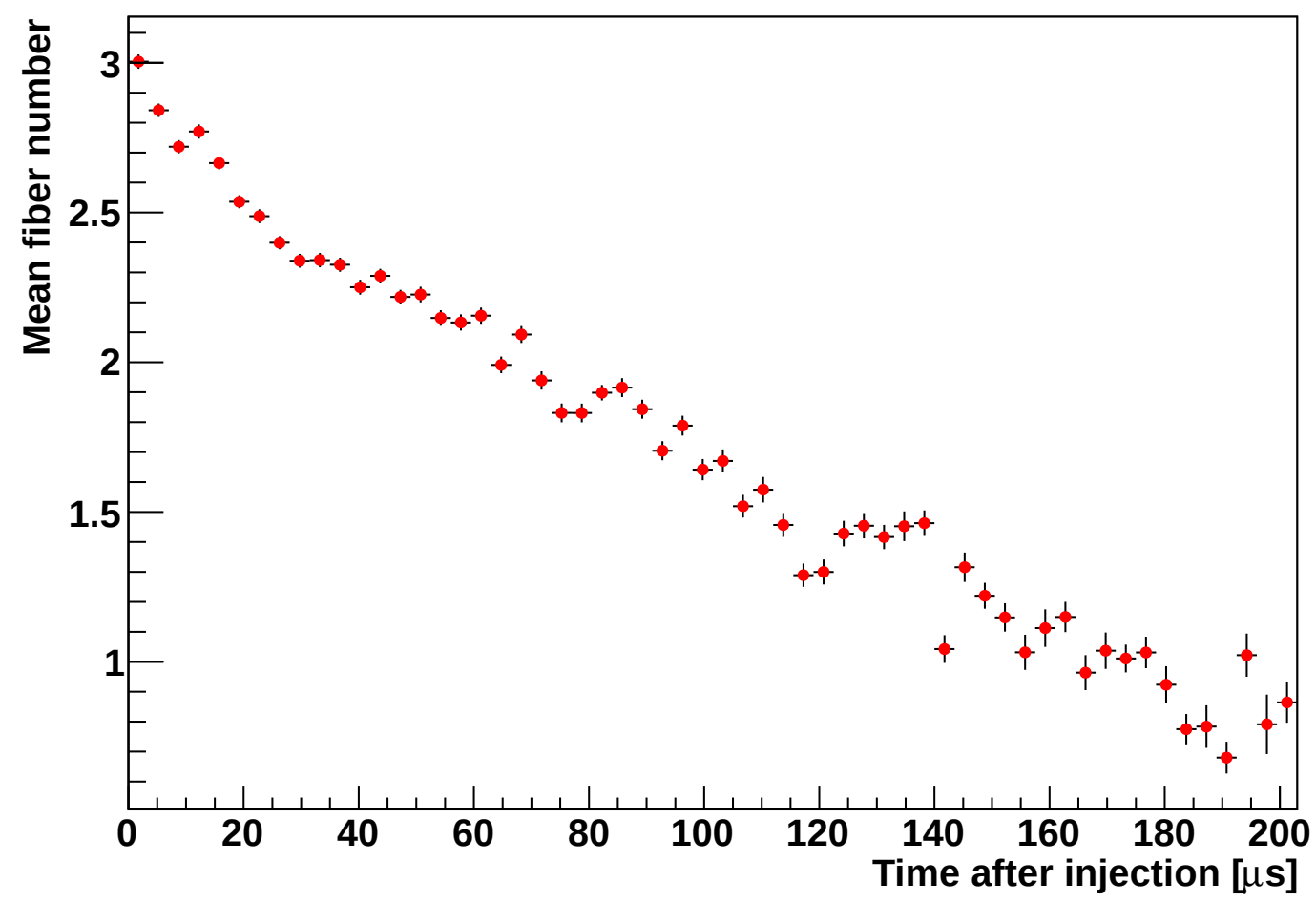

Figure 20.7: Simulated radial beam centroid position, in units of fiber number (7 mm), when the fiber harps are inserted. Energy loss in the fibers causes the beam to shift radially inward.

We had previously proposed to use waveform digitizers from the MuLan experiment [8] to read out the auxiliary detectors. They were existing devices that could have been obtained at no cost. However, we recognized that there would be significant value in simplifying the data acquisition system and the clock and control system distribution by using only one model of waveform digitizer. In the end, we determined that the time and effort that would have been required to install and support the MuLan system was not justified by the initial savings on the hardware.

Similarly, although we have acquired a set of SiPMs that could be used for the fiber readout, a new technological generation with significantly less dark current and greater photon detection efficiency is now available. This would also allow us to reconsider the most appropriate number of pixels in the device; as shown above, saturation of a 400 pixel SiPM may become a concern. We have recently obtained samples of Hamamatsu S12571-025C devices to test, which have 1600 pixels, and we are continuing to evaluate the value of purchasing this new model.

In E821, the primary method of monitoring the rate of flashlets was to suppress the firing of the electrostatic quadrupoles periodically, preventing the injected muon bunch from being stored. The number of suppressed fills could be varied, but it was typically one out of 25. Any signals that appeared in the calorimeters during these fills were presumed to be from flashlets, which was verified by observing the cyclotron period of the AGS in the time structure of the signals. While this method is effective, it unnecessarily discards a few percent of the data.

## 20.5 ES&H

The most significant hazards associated with the auxiliary detectors are electrical. The bias voltages needed for the SiPM readout of the fiber beam monitors will be approximately 70 V. To mitigate this hazard, a current-limited power supply will be used, with the current limit set to the lowest value that allows the devices to operate. All electrical devices will be subject to Fermilab's standard design review and operational readiness clearance processes.

The fiber beam monitors will be powered by compressed air at less than 150 psi. Requirements for appropriate personal protective equipment, such as eye protection, when working around compressed air lines will be determined in consultation with Fermilab ES&H experts. Similarly, the fiber beam monitors interface with the ultra high vacuum system, and they are within the large fringe field of the storage ring magnet. We will work with Fermilab ES&H to establish appropriate procedures to mitigate these hazards. The vacuum test stand to be used in laboratory tests before installation is small enough (less than 12 inch inner diameter and 35 cubic foot volume) to fall outside the scope of Fermilab's vessel certification requirements.

During the laboratory design and construction phase, other hazards will be relevant. Tools, including some with sharp blades as well as soldering irons and hot air stations, will be needed and will be treated with due caution. Two-part optical cement will be used to bond the fibers; it, and any other chemicals that are needed, will be handled in accordance with the requirements in the MSDS. Low-activity radiation sources will be used to test the light output of scintillating fibers, following the usual radiation safety precautions.

## 20.6 Risks

There is a risk of hidden damage, or degradation over time, to the fiber beam monitors that might require more repair work than anticipated. This damage may not be discovered until mechanical, vacuum, and light output tests are completed. There is also a risk that, after testing, we may find that the SiPMs that we were able to acquire at no cost are not suitable for the application and that we need to procure another model.

However, the only risk that would be expected to have a noticeable impact on the total project cost would be the destruction or damage beyond repair of one or more of the existing fiber harp devices by an accident in shipping, storage, or testing. Because much of the original knowledge of the system has been lost, to re-create a fiber beam monitor from scratch would require a significant level of unplanned engineering cost in addition to the precision machining work.

Because the fiber beam monitors interface with the beam vacuum system, any leak could cause downtime for the experiment. The motion control system could potentially fail in a way that would not allow them to be retracted, which would also cause downtime, requiring the entire storage ring to be brought up to atmospheric pressure to remove them. We intend to minimize these risks by careful testing before installation.

## 20.7 Quality Assurance

We will test the motion and vacuum integrity of the fiber beam monitors with extensive exercises in a test chamber in the laboratory before they are installed in the storage ring. We will also check the output of each fiber, and therefore the functionality of each SiPM channel, with a set of light pulsers and radiation sources.

## 20.8 Value Management

The auxiliary detectors represent a successful application of value management principles. All components that are suitable for reuse from Brookhaven E821 will be reused. The primary upgrade to the fiber beam monitor devices will be a SiPM readout system. For that installation, suitable unused SiPM devices (spares from a previous project) were identified and made available by Argonne National Laboratory.

## 20.9 R&D

The following studies are in progress or will begin very soon, in summer 2015:

- We are continuing to characterize techniques for bonding scintillating fibers to clear fibers and to check the transmission of light through the clear fibers and the vacuum feedthroughs.
- We will vacuum leak-test each of the fiber harp stations.
- We will repair the known mechanical damage to one fiber harp and proceed to verify that it moves correctly.
- We will build and test the final SiPM preamplifier design, along with a final version of the part that will hold it onto the fiber outside the vacuum feedthrough.
- We will develop a final version of the motion control system for the plunging of the fiber harps that interfaces with the experiment's MIDAS Slow Control Bus network.
- We will build prototypes of the new entrance counters with both 1.5 mm Lucite and 0.5 mm EJ-212 scintillator, and we will test them with cosmic rays to determine the light yield. This will allow the simulation to be validated. In the event that the performance of the prototype is sufficient to meet the requirements of the experiment, it will become the final entrance counter.

# Chapter 21

# Data Acquisition System

## 21.1 Physics Goals

The data acquisition system must read, process, monitor and store the data produced by the various detector systems. Most importantly, the DAQ must provide a distortion-free record of the detector signals resulting from the decay positrons during the 700 $\mu$s-long spills from the muon storage ring. Additionally, the system must record all data required to perform the corrections from effects such as pulse pileup, gain instabilities and beam dynamics. Furthermore, the system must allow the monitoring needed to guarantee the overall integrity of data taking and record-keeping needed to document the experimental conditions during data taking.

## 21.2 Overall Requirements

The overall requirements of the data acquisition system – from the detector sub-systems and the accelerator time structure to the data storage and the data monitoring – are summarized in Table 21.1.

The DAQ must handle the accelerator-defined time structure of the data readout from the detector systems. Under normal operations we anticipate a 12 Hz average rate of muon spills that comprises sequences of four consecutive 700 $\mu$s spills with 11 ms spill-separations for each booster batch received by the muon $g$-2 experiment. The procedures for reading, processing, monitoring and storing these data must not introduce time-dependent losses or time-dependent distortions of the detector signals.

The DAQ must handle the readout, processing, monitoring and storage of the data obtained from the 1296 channels of 800 MSPS, 12-bit, waveform digitizers instrumenting the individual $PbF_2$ crystals of the twenty four calorimeters. For each spill the raw data will consist of 1296 channels of 700 $\mu$s-long streams of continuously-digitized ADC samples. The DAQ must process these raw data into derived datasets including: T method data (*i.e.* individual islands of digitized pulses), Q method data (*i.e.* accumulated histograms of calorimeter spectra), and other calibration, diagnostic and systematic data. At a 12 Hz spill rate the readout (*i.e.* raw) data rate will be about 18 GB/s and the stored (*i.e.* derived) data rate will be about 80 MB/s.

| Feature | Requirement |
|---|---|
| spill time structure | 12 Hz spills in groups of 4 spills with 11 ms separations |
| readout electronics | AMC13-based, spill-async readout and other spill-sync readout |
| calorimeter raw data | 18 GB/s total from 1296 digitizer chans. in 24 $\mu$TCA crates |
| calorimeter derived data | 75 MB/s total of T, Q and other derived datasets |
| other detector data | 5 MB/s total of tracker, auxiliary detector raw data |
| event builder system | spill-based events assembled from ~30 fragments at ~80 MB/s |
| data monitoring system | online analyser, database to insure data quality / integrity |
| data storage | transfer to FNAL archive at $\leq$100 MB/s rate for ~2 PB total |
| run control | web-based local / remote control, monitoring, *etc* |

Table 21.1: Summary of the major requirements of the DAQ system.

The DAQ must also handle the readout, processing, monitoring and storage of the data obtained from the three positron tracking stations. This system consists of roughly 3000 channels of straw tubes with associated amplifier-discriminator-TDC electronics. The raw data – consisting of time stamps and spill numbers from individual straw tubes – is expected to yield a r oughly 3 MB/s time-averaged data rate.

Additionally, the DAQ must handle the readout, processing, monitoring and storage of data from the auxiliary detector systems. These systems include: the muon entrance counters, beam position monitors, fiber harp detectors, laser system monitors, electric quadrupole monitors and kicker monitors. The system involves both instrumentation that is operated during normal data taking (*e.g.* the muon entrance detector) and instrumentation that is operated during dedicated data taking (*e.g.* the fiber harp detectors). The read out for the laser calibration monitors will use spare channels of the 800 MSPS, 12-bit, $\mu$TCA-based waveform digitizers instrumenting the calorimeters stations. The read out for the muon entrance counters (2 channels), fiber harp detectors (28 channels), electric quadrupole monitors (4 channels) and kicker monitors (6-9 channels) will use a single dedicated crate of 800 MSPS, 12-bit, $\mu$TCA-based waveform digitizers. The readout electronics for the beam position monitors will use 1 GSPS CAEN VX1782 waveform digitizers with spill-asynchronous 1 GbE readout.

The expected data rates from auxiliary detector systems are: (i) 17 MB/s during dedicated data taking with the fiber harp detectors and (ii) 3 MB/s during normal data taking with the muon entrance counters, laser calibration monitors, electric quadrupole monitors and kicker monitors.

The DAQ must coordinate the acquisition of data by the frontend readout processes with the accelerator-defined spill cycles. This coordination is designed to incorporate both readout systems where data is transferred synchronously between spill cycles and readout electronics where data is transferred asynchronously with spill cycles (*e.g.* the TCPIP network packets of raw data from $\mu$TCA based digitizers and tracker TRMs).

The DAQ must assemble the individual fragments of spill-by-spill data from networked readout processes into complete, deadtime-free records of each muon spill. This includes assembling the data banks of T method and other datasets from the twenty four calorimeter stations as well as the data from the three tracker stations and the auxiliary detector systems. In total the event builder must match and assemble the fragments originating from roughly thirty frontend processes at an expected rate of about 80 MB/s. The resulting fully-assembled spill-by-spill events must be transferred to the Fermilab computing facilities.

The DAQ must provide the local / remote run control for data taking as well as facilities for configuration and readback of configuration parameters such as digitizer settings, multihit TDC settings *etc.* The system must provide the monitoring of data integrity and data quality and a comprehensive database of the experimental conditions and configuration parameters during data taking. The system must additionally provide for local storage of sufficient data for online analysis tasks.

The DAQ will require clean, uninterruptible power of roughly 50 kW total power with appropriate power distribution for roughly thirty rackmount computers and their associated network switches, mass storage devices, *etc.* The control room will require air circulation and cooling power for appropriate temperature and humidity control with temperature, humidity and air velocity sensors with digital readout.

The DAQ will require a reliable, fast network connection between the MC-1 computer room and the Fermilab data storage facilities that is capable of a sustained data rate of roughly 100 MB/s. Based on roughly one year of total running time, the experiment will require a permanent data storage capacity from Fermilab data storage facilities of $\sim$2 Petabytes.

## 21.3 Recommended Design

### 21.3.1 DAQ structure

The DAQ will acquire data in blocks that correspond to individual muon spills in the storage ring. Each event will represent a complete deadtime-free history of the entire activity in the detector systems for a complete spill – rather than events corresponding to individual positrons. This scheme will utilize the on-board memories in waveform digitizer and multi-hit TDCs to temporarily buffer the recorded data before its data transfer to the data acquisition. The design will be implemented as a modular, distributed computer system on a parallel, layered array of networked, commodity processors with graphical processing units (GPUs). The DAQ group has developed and operated very similar architectures [1] for the MuLan, MuCap and MuSun experiments at the Paul Scherrer Institute (these experiments involved high statistics, part-per-million measurements of the lifetimes of both free muon and muonic atoms).

The data acquisition system is depicted schematically in Fig. 21.1. It comprises a frontend processor layer responsible for readout and processing of waveform digitizer and multihit TDC data, a backend layer responsible for event assembly and data storage, a slow control layer responsible for control and read-back, and a data analysis layer responsible for monitor-ing data integrity. The DAQ hardware will comprise a networked cluster of high performance processors running Scientific Linux (4U rackmount eight-core processors with 10 GbE, PCIe3 and NVIDIA K40 GPUs). To maximize the bandwidth, point-to-point networks will handle the traffic between the readout electronics and the frontend process, another sub-network will handle the traffic between the frontend layer and the backend layer, and another network will handle the traffic between the backend layer and the analysis layer. A gateway machine will allow the data transfer between the g-2 private sub-networks and FNAL data storage.

The DAQ software will be based on the MIDAS data acquisition package [2], ROOT data analysis package [3], and NVIDIA's parallel computing platform CUDA [4]. Detailed information on the MIDAS data acquisition package – *e.g.* installation, documentation, user forum, and device drivers for various hardware – are available at Ref. [2]. The MIDAS software consists of function templates and library routines for processes that handle the read out from frontend electronics, event building, data logging DAQ operations. MIDAS also incorporates a fast online database for storing experimental configurations, and a web interface for local / remote control of data taking as well as integrated alarm and slow control systems.

The DAQ will be housed in the computer room in the MC-1 building. 10 GbE optical fiber links ($\sim$30 cables) will provide the point-to-point network connections between the detector sub-systems in the experimental hall and the data acquisition in the control room. for the data readout. A 1 GbE connection between a network switch in the computer room and another network switch in the experimental hall will provide the backbone for IPBus communications between readout electronics and frontend processors.

### 21.3.2 Asynchronous readout frontends for calorimeter stations

The calorimeter readout consists of one frontend processor per calorimeter station. Each frontend processor will read out the 54 waveform digitizer channels associated with the $6 \times 9$ $PbF_2$ crystals of a single calorimeter station (plus one additional channel for each calorimeter station's laser monitor). Each group of 54 waveform digitizer channels – together with a commercial MCH controller module [5] and a custom-built AMC13 controller module [6] – will occupy a single $\mu$TCA crate. The waveform digitizers and AMC13 controller module act as network hosts located on the MCH controller sub-network. The IPBus protocol [7] is used to perform the configuration and control of the waveform digitizers and the AMC13 controller. For read out, the calorimeter frontend process (a TCPIP client) receives digitizer data in 32 kByte blocks from the AMC13 controller (a TCPIP server) via a dedicated 10 GbE point-to-point network connection between the AMC13 controller and the frontend computer.

For each spill the raw calorimeter data will consist of $24 \times 54 = 1296$ channels of 700 $\mu$s-long streams of continuously-digitized, 800 MSPS, 12-bit, ADC samples – a total of 1.45 GBytes per spill or 18 GBytes per second (assuming the 12-bit samples each occupy 2 bytes). The frontend readout process comprises: a *TCP_thread* that receives and buffers the raw data blocks from the AMC13 controller, a *GPU_thread* that manages the GPU-based data processing, and a *MIDAS_thread* that handles the transfer of processed data as MIDAS-formatted events to the event builder. Mutual exclusion (Mutex) locks are used to synchronize the execution of the frontend threads. The arrangement was designed to provide the necessary performance to compress the continuously-digitized ADC samples into the T/Q method datasets at the software level. The T method datasets will consist of individual "islands" of above-threshold calorimeter signals and the Q method datasets will consist of sum histograms of consecutive spills of continuously digitized samples. Additionally – for activities such as detector commissioning and data monitoring – the design allows for storage of either all raw data (at low raw data rates) or pre-scaled raw data (at high raw data rates).

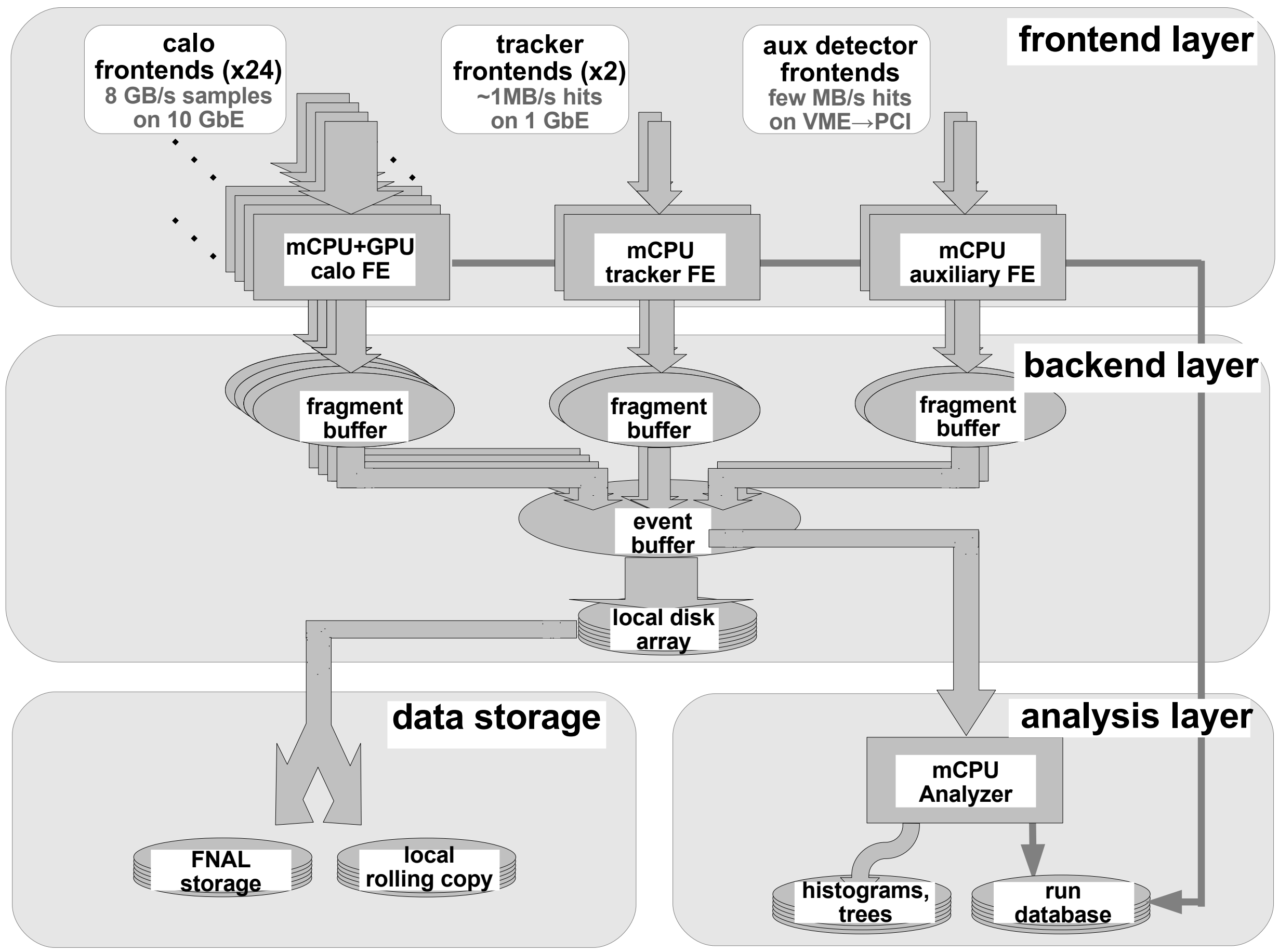

Figure 21.1: Conceptual design of the $g$-2 data acquisition. The figure shows: (i) the frontend layer for readout and processing of data from the calorimeter stations, tracking stations and auxiliary detector systems, (ii) the backend layer for event building and data migration, (iii) the analysis layer for monitoring data quality and recording run-by-run experimental conditions, configuration parameters, *etc.*, and (iv) data storage. The layers comprise arrays of networked commodity processors.

The algorithms for constructing the T/Q method datasets involve copying, masking and summing arrays and basic digital signal processing and therefore are ideally suited to GPU-based parallelization using standard algorithms and CUDA libraries. A major advantage of GPU-based processing of raw data into compressed datasets is the comparative ease of testing and implementing a variety of algorithms for processing sub-tasks such as digital filtering, pulse triggering, pulse clustering, *etc.* The scheme also offers the flexibility to implement other datasets – such as pile-up datasets (*e.g.* by summing fills before storing islands) and diagnostic datasets (*e.g.* by storing prescaled fills of continuously digitized samples) – as needed.

### 21.3.3 Asynchronous readout frontends for tracker stations

The positron tracking system consists of three tracking stations that each comprise about 1000 individual channels of straw tube detectors. The raw data from the frontend electronics of the three tracker stations are transferred via a serial link to custom-built tracker readout modules (TRMs). The TRMs – and their associated MCH and AMC13 controllers – are housed in a single $\mu$TCA crate. The TRM modules act as network hosts on the MCH controller sub-network and are configured and controlled using the IPBus protocol via the 1 GbE network connection between the MCH controller and the frontend computer. The tracker readout process (a TCPIP client) receives TRM data in 32 kByte blocks from the AMC13 controller (a TCPIP server) via a dedicated 10 GbE point-to-point network connection between the AMC13 controller and the frontend computer.

Each tracker frontend process will receive the TRM raw data – *i.e.* tracker hits defined by a channel number, spill number and time stamp – and pack and dispatch these data as MIDAS-formatted databanks over the frontend network to the event builder. Since the readout mechanisms for the $\mu$TCA-based WFD and TRM modules are the same, the same readout framework can handle the WFD data readout and TRM data readout. The TRM / WFD frontend software will only differ in details related to the device-dependent IPBus communication and associated online database entries and the presence (or the absence) of the GPU-based data processing for the WFD (TRM) data.

### 21.3.4 Synchronous readout frontends for auxiliary detectors

The auxiliary detector systems comprise the muon entrance counters, beam position monitors, fiber harp detectors, electric quadrupole monitors and kicker monitors. The beam position monitors will be readout by 1 GSPS CAEN VX1742 waveform digitizers. The other auxiliary detectors with be readout by a dedicated $\mu$TCA crate of 800 MSPS, 12-bit waveform digitizers. In both cases the data are trasferred over 11/10 GbE ethernet and readout using the same framework as the calorimeter frontend and the tracker frontend. The calorimiter, tracker and auxiliary readout frontends will differ only in the details of the IPBUS configuration and the corresponding ODB structure for the relevant WFD /TDC modules.

### 21.3.5 Master frontend and accelerator-DAQ synchronization

The DAQ design incorporates a master frontend process and hardware control logic in order to synchronize the data acquisition readout cycles with the accelerator-defined spill cycles. Importantly, the master frontend and control logic must accommodate both the spill-synchronous data readout from the VME crates via Struck SIS 3100/3300 interface modules and the spill-asynchronous data readout from the $\mu$TCA crates via 10 GbE network links. The hardware control logic design uses a *DAQ_ready* signal to authorize the recording of the data from the next spill by the WFD/TDC electronics. The software control logic uses remote procedure calls (RPCs) to provide status messages (*i.e. DAQ_ready*, *begin_of_spill*, *end_of_spill*) to readout processes and *FE_ready* messages from the synchronous frontends to the master frontend.

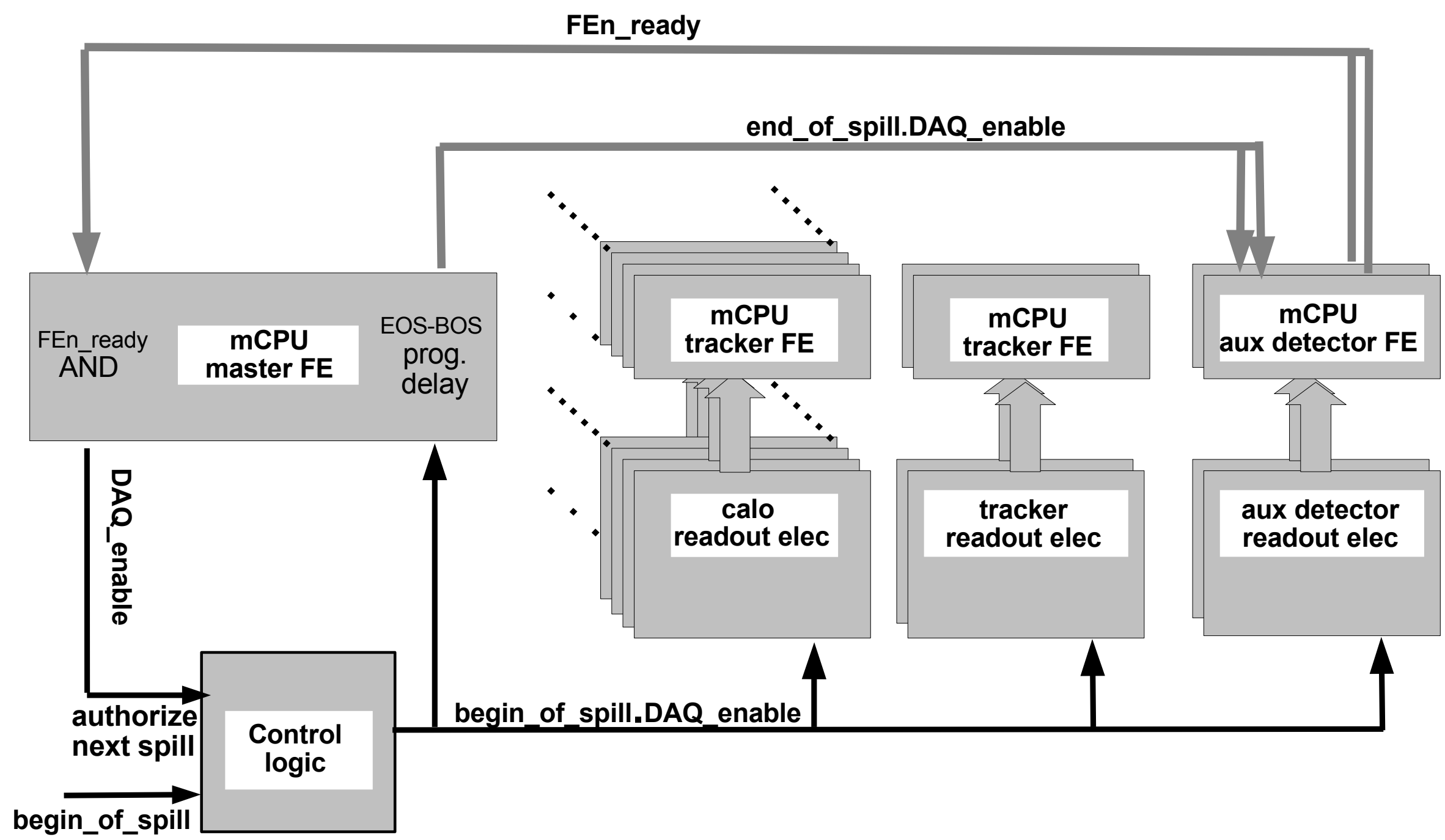

Figure 21.2: Conceptual design of the DAQ control logic. The logic is based upon an accelerator-defined *begin_of_spill* signal and a DAQ-defined *DAQ_enable signal*. When the DAQ is ready for data taking, it enables the distribution of the next *begin_of_spill* to the WFD/TDC electronics via a *DAQ_enable signal*. When the any synchronous readout processes have completed their required tasks between successive spills they report their readiness to the master frontend process via a *FE_ready* remote procedure call. The master frontend issues the next *DAQ_enable* signal when all synchronous frontends have reported their readiness.

The design of the synchronization logic that coordinates the spill cycles and readout cycles is shown in Fig. 21.2. The design involves two hardware control signals – an accelerator-derived *begin_of_spill* signal and DAQ-derived *DAQ_enable* signal. On starting a run, the master frontend processes issues a *DAQ_enable* signal to authorize recording the next spill. The subsequent authorized *begin_of_spill* signal, *i.e. begin_of_spill·DAQ_enable*, is then distributed to both the WFD/TDC electronics to enable data recording and the master frontend to generate *begin_of_spill* and time-delayed *end_of_spill* RPCs. The digitization time intervals for each detector system is a configuration parameter that is stored in the DAQ database and set in the readout electronics. The synchronous readout processes – which require the readout of the data size before the acquisition of the next spill – use remote procedure calls to report their readiness for data taking. This "readiness" report is made by a *FE_ready* remote procedure call from the synchronous readout processes to the master process. On receipt of all *FE_ready* RPCs from all synchronous readout processes, the master processes then issues another *DAQ_enable* signal to authorize recording the next spill. The asynchronous readout processes – that receive TCPIP data packets over 10GbE network links with AMC13 modules – are not required to report their readiness before acquiring another spill. [1]

Note we plan to use a PCI-based GPS synchronization card [9] to GPS timestamp the digitized spills in order to facilitate later coordination between the detector system readout and the magnetic field readout.

### 21.3.6 Event building and data logging

Each frontend readout process will transmit its spill-by-spill data fragments as MIDAS-format databanks across the frontend network to the backend processor. Initially, the data fragments from the twenty four calorimeter processes, three tracking system processes and various auxiliary detector processes, are transferred to shared memory segments on the backend machine (one memory segment per frontend process). After matching the MIDAS serial numbers and muon spill indexes of event fragments the event builder process assembles all data fragments into single events representing a complete record of each spill. The spill events are then written by the event builder process to a further memory segment known as the system memory segment and are then available for data storage tasks, data analysis tasks, *etc.*

The backend layer will use a two step procedure to permanently store a full copy of the data on the Fermilab Computing facilities and temporarily store a rolling copy of the recent data on our analysis layer. First, the data will be transferred from the system memory segment to a temporary disk file on a local redundant disk array on the backend processor. Next, the temporary data files on the backend processor will be asynchronously migrated to both the Fermilab computing facilities for permanent storage and the DAQ analysis layer for local analysis projects. This approach will minimize any delays in data taking due to latencies associated with the permanent archiving of the experimental data and make the current data available for local analysis and monitoring projects.

### 21.3.7 Online database and data monitoring

The DAQ analysis layer will provide both integrity checking and online histogramming. The online analyzer will receive events over the network from the system memory segment on the backend layer. These events will be received "as available" in order to avoid introducing any delays into the readout or the data storage. The online analysis will utilize the ROME framework generator for event-based data analysis [11] In ROME the analysis framework is automatically generated using a XML framework definition file that defines all necessary experiment-specific classes.

[1] The asynchronous readout processes use an independent readout thread with a ring buffer to maximize the TCPIP data transfer rate.

The analyzer will utilize a modular, multistage approach to analysis tasks where different analysis tasks will be implemented as individual analyzer modules and then switched on / off as needed. Each analysis module will have access to a global structure that contains both the raw MIDAS databanks from the readout processes and any derived MIDAS databanks from the preceding analysis modules. Low-level modules will be responsible for unpacking the databanks and checking their integrity. Intermediate-level modules will be responsible for various histogramming tasks to ensure the correct operations of detector systems. High-level modules will be responsible for online "physics" analysis such as fits to the precession signal.

The data acquisition system will incorporate database support to provide a comprehensive run-by-run record of the experimental conditions, configuration parameters, *etc.*, during the entire experiment.[2] The run-by-run database will store information derived from the MIDAS online database such as run start time, run stop time, operator run-time comments, the number of events, and hardware settings including the high voltage setting, digitizer configuration parameters, multihit-TDC configuration parameters, *etc.* In addition, the database will record such quantities as detector gains, pedestals, *etc.*, fitted frequencies, lifetimes, *etc.*, that are derived from the analysis layer. This information are the foundations for the offline data analysis.

A $g-2$ web-based interface will provide both local and remote access to run-by-run histograms, trend plots and the experimental database.

## 21.4 Design Performance

### 21.4.1 Test stands for prototyping and development

Several test stands are being used for the prototyping work on the data acquisition system. A test stand at UKy and another stand at Fermilab are being used for the development of the calorimeter readout system. A test stand at UCL is supporting the development work on the tracker readout system and another stand at Dubna is supporting the development of the ROME data monitor.

As an example, the UKy test stand that was used for several R&D projects on the calorimeter readout, is shown in Figs. 21.3 and 21.4. It comprises a network of three frontend processors and one backend processors. The backend processor hosts the MIDAS server process (MSERVER) that manages inter-process communications, the MIDAS web daemon (MHTTPD) that provides run control, the MIDAS event builder (MEVB), as well as data storage and analysis tools. Frontend processor FE01 comprised two quad-core Intel Xeon X5550 CPUs and a 2496-core, NVIDIA Kepler K20 GPU. As is planned for the full DAQ system, the test stand was arranged as a local sub-network of frontend processors with a backend file server.

[2]The MIDAS data acquisition package includes a central database called the "Online DataBase" (ODB). The ODB stores run parameters, frontend readout parameters, backend logging parameters, status / performance data, as well as other user-defined information. All processes participating in data taking have full access to the information in the MIDAS ODB.

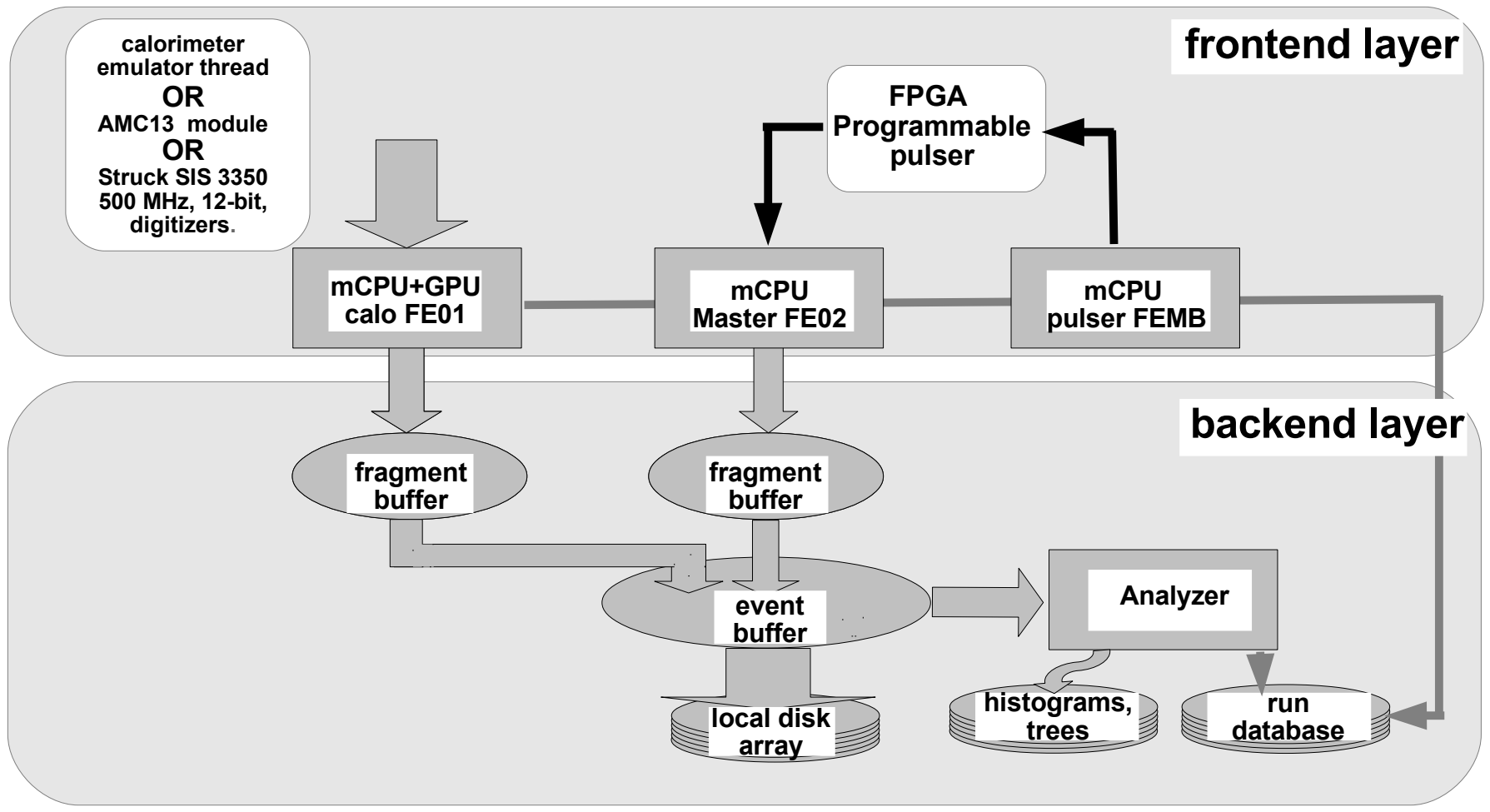

Figure 21.3: Layout of UKy MIDAS-based DAQ-platform for R&D projects. The frontend FE01 incorporates both readout of emulated continuously-digitized ADC samples from a calorimeter simulation process and real continuously-digitized ADC samples from a Struck SIS3350 digitizer system. The frontend FEMB emulates the Fermilab accelerator control signals and the frontend FE02 provides the software synchronization of spill-by-spill readout. The system also includes an event builder layer and data monitor layer.

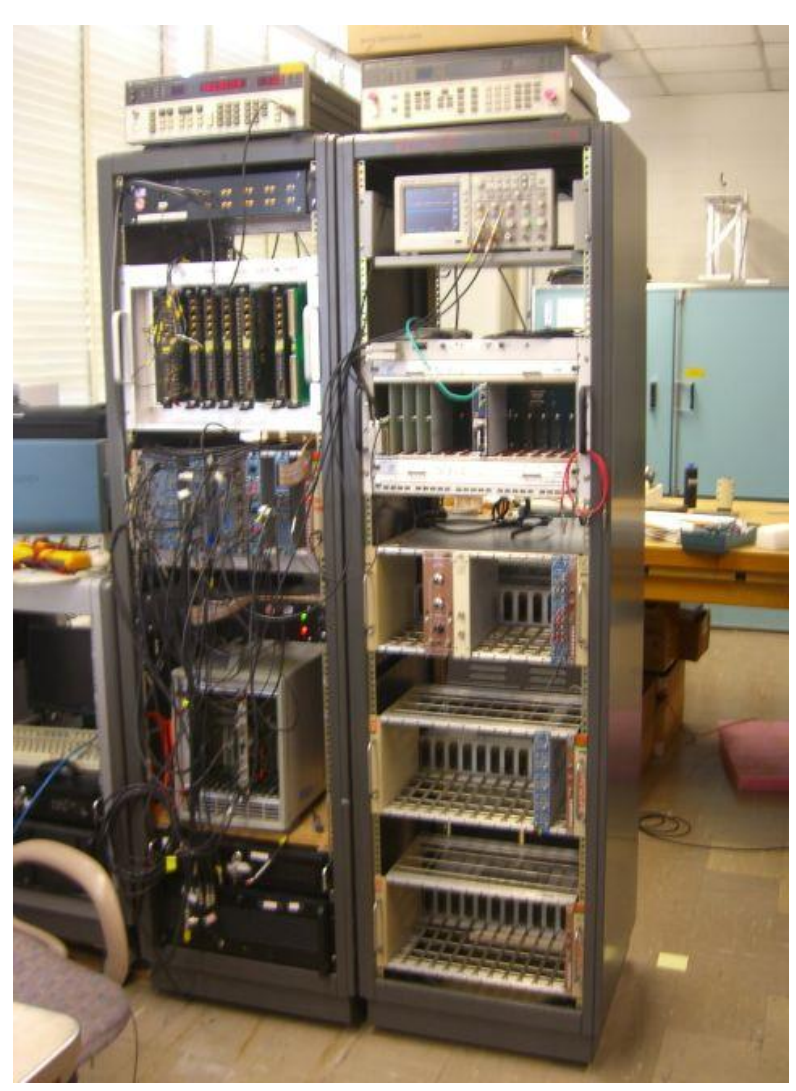

Figure 21.4: Photograph of the DAQ test stand including $\mu$TCA-based electronics and VME-based electronics at UKy.

The Fermilab, UCL and UKy test stands all incorporate both prototype readout electronics as well as either calorimeter station emulators (Fermilab / UKy) or tracker station emulators (UCL).

### 21.4.2 Development and prototyping of T / Q method data processing

Fig. 21.5 shows a typical spill of simulated data – *i.e.* $3.5\times10^5$ ADC samples of 700 $\mu$s continuous-digitization – that was Monte-Carlo generated by the calorimeter emulator. In building streams of ADC samples, the decay positrons were generated with the appropriate energy-time distributions for 3.094 GeV/c decays and the calorimeter hits were generated with appropriate x-y distributions and pulse shapes. The raw data were read out and processed into T/Q method datasets in the DAQ frontend layer and then analyzed and histogrammed in the DAQ analysis layer. Representative plots of T method energy and time distributions of decay positrons are shown in Fig. 21.6.

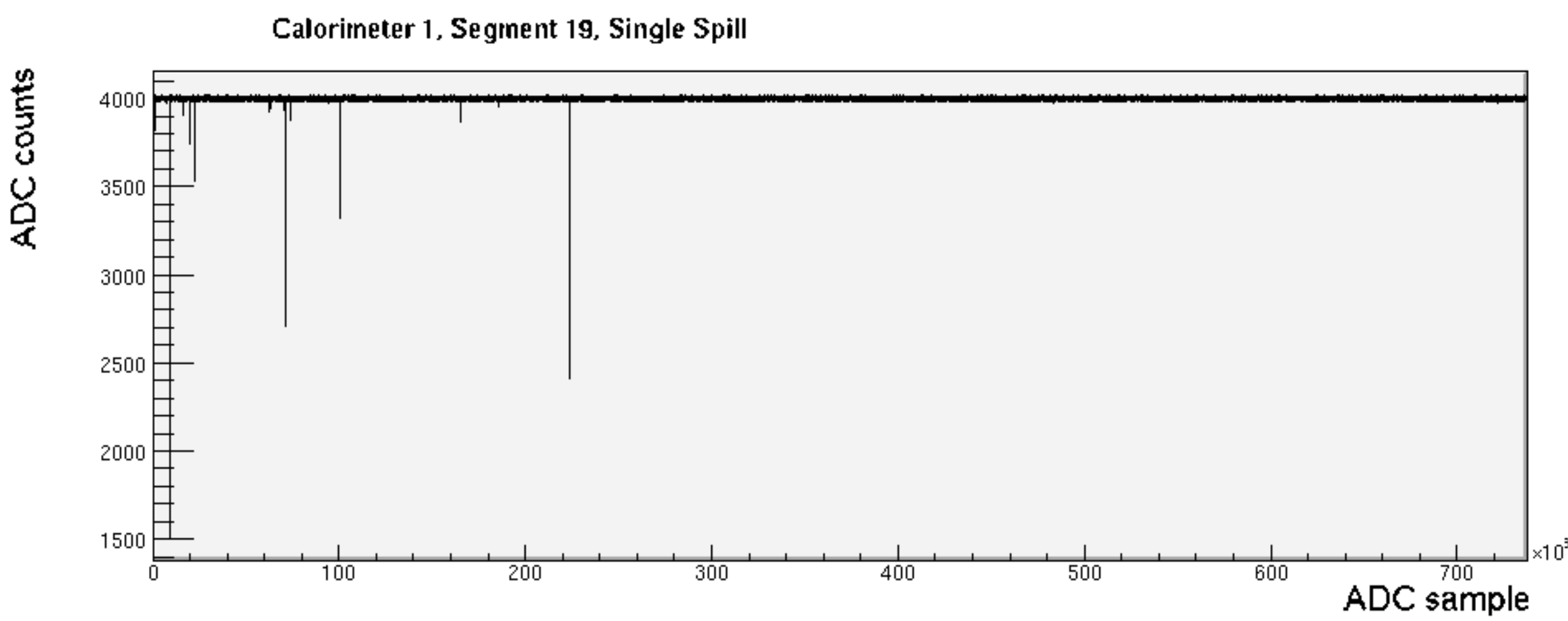

Figure 21.5: A representative single spill of simulated data generated by the calorimeter emulator. The data correspond to $3.5\times10^5$ ADC samples of 700 $\mu$s continuous-digitization for one segment of one calorimeter.

Results from frontend timing tests of the GPU-based, T/Q method processing of the simulated calorimeter data are shown in Fig. 21.7. After completing the readout of each spill of ADC samples the raw data are transferred from the CPU memory to the GPU memory. The GPU then initiates a sequence that involves: derivation of the segment-summed calorimeter samples from the individual crystal segment samples, identification of the T method above-threshold pulses in the summed calorimeter samples, assembly islands of the T method above-threshold islands with pre-/post-samples, and transfer of the resulting

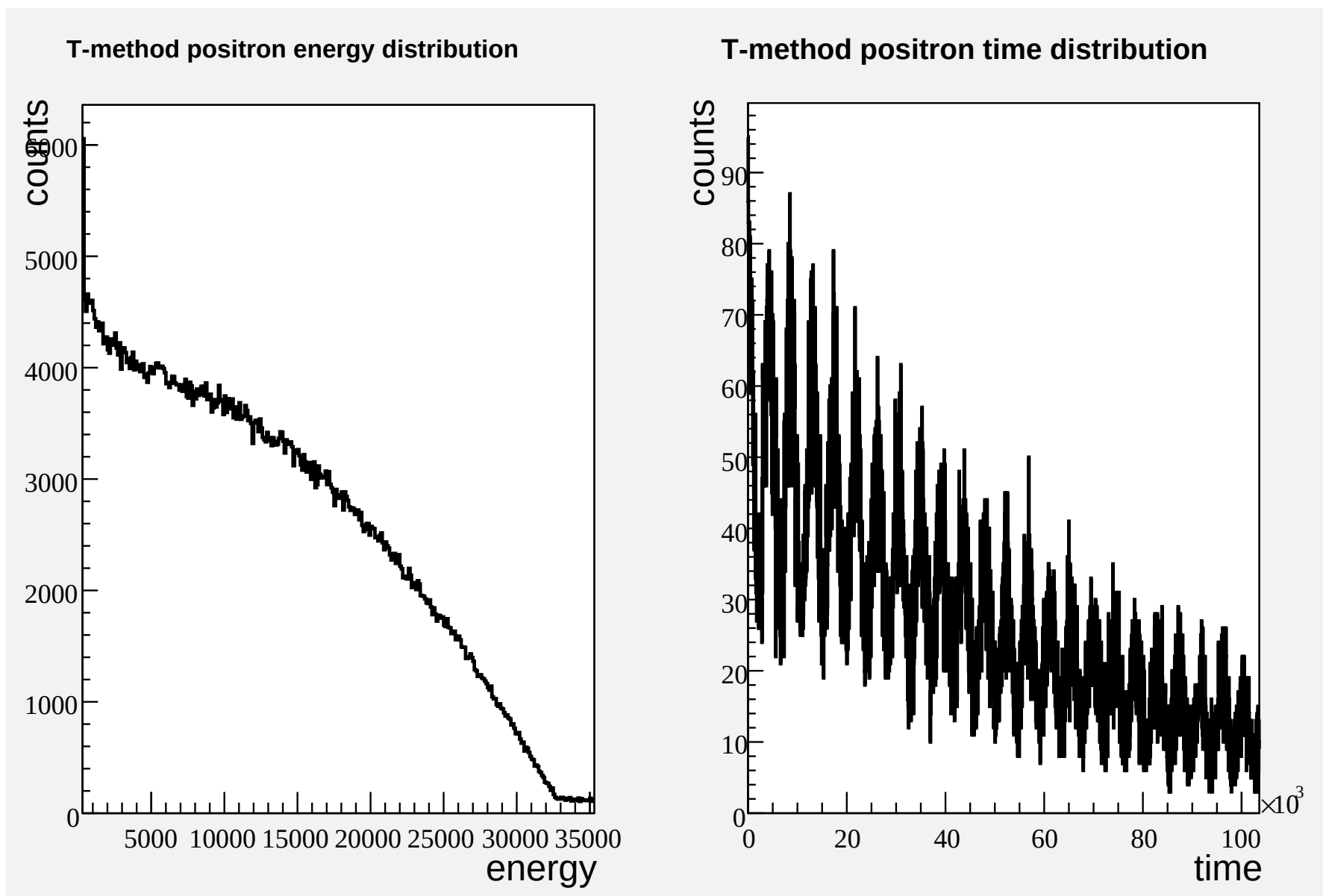

Figure 21.6: Energy distribution (lefthand plot) and time distribution (righthand plot) of energy / times of positron hits. The data were generated as continuous-digitization spills by the calorimeter emulator, were processed into T Method datasets in the calorimeter, and histogrammed in the analysis layer. The energy distribution shows the positron endpoint energy and the time distribution shows the anomalous precession frequency.

T method data from the GPU memory to the CPU memory. Additionally, a Q method dataset was constructed by summing consecutive blocks of 32 ADC samples of digitizer data and then copied from the GPU memory to the CPU memory. Finally, the T/Q method datasets are packaged into MIDAS databanks and transferred to the backend layer.

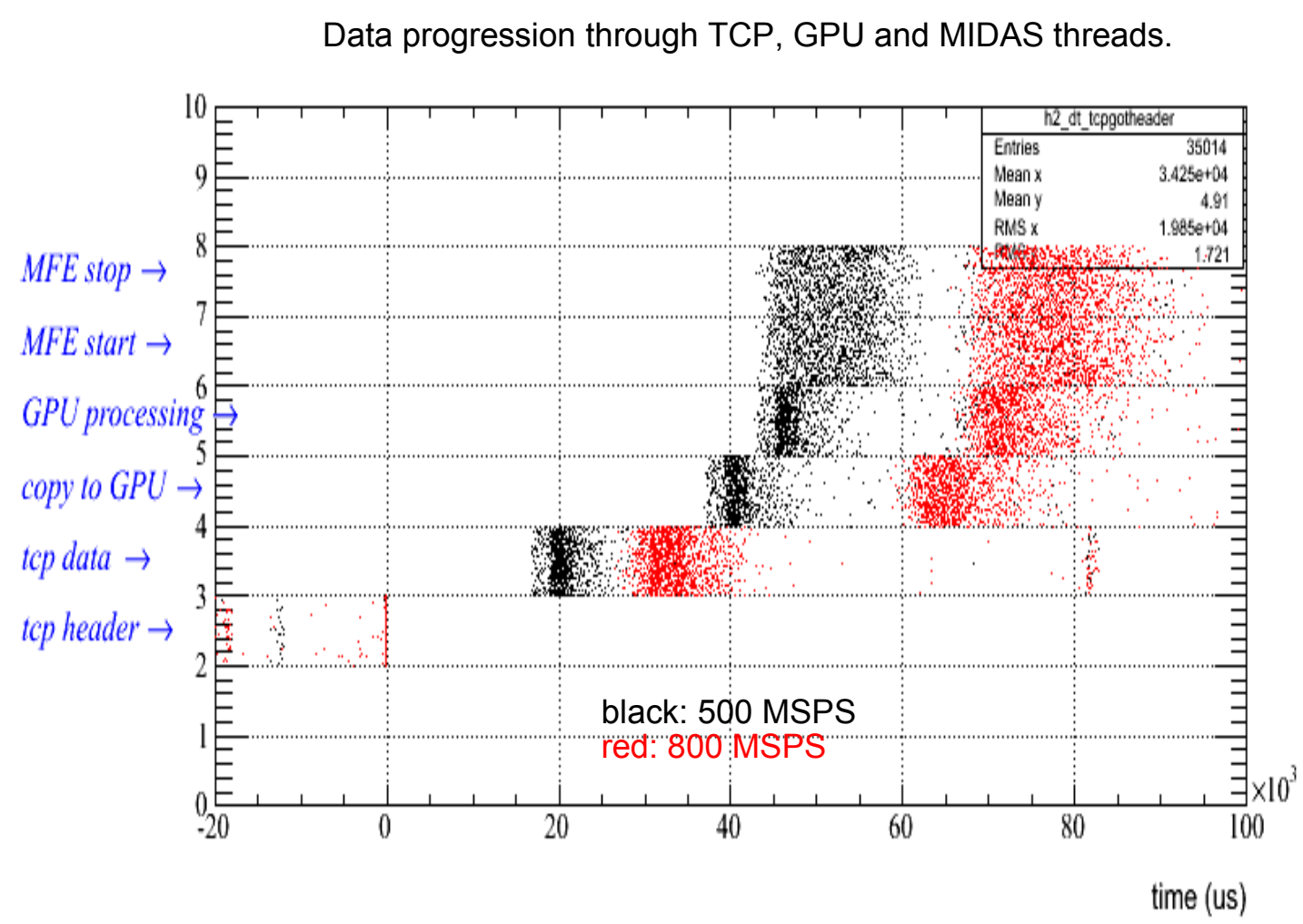

Figure 21.7: Timing plot of the readout, processing and transmission of simulated data through the TCP, GPU and MIDAS threads in the calorimeter readout frontend. The *TCP_thread* handles the receiving and unpacking of the 10 GbE network data, the *GPU_thread* manages the GPU processing into T, Q and other datasets, and *MFE_thread* manages the assemble and transfer of MIDAS-formatted events. The plot show the processing time for spills digitized at both 500 MSPS and 800 MSPS. The above results were obtained using the UKy test stand.

### 21.4.3 Development and prototyping of event building

Also conducted were timing tests of event building on simulated databanks in the backend layer of the data acquisition. The tests showed the rate limitations in event building were largely governed by memory copy operations during event fragment assembly. For the backend processor in the UKy test stand – a six core, Intel i7 processor with 8 GBytes of high-bandwidth DDR3 memory – the system was able to handle a data rate exceeding 100 MB/s without introducing time delays.

### 21.4.4 Development and prototyping of AMC13 readout

Both the UKy test stand and the Fermilab test stand include a $\mu$TCA crate, commercial MCH controller module and custom-built AMC13 module. Since April 2015 the Fermilab test stand also incorporates a prototype 5-channel waveform digitizer.

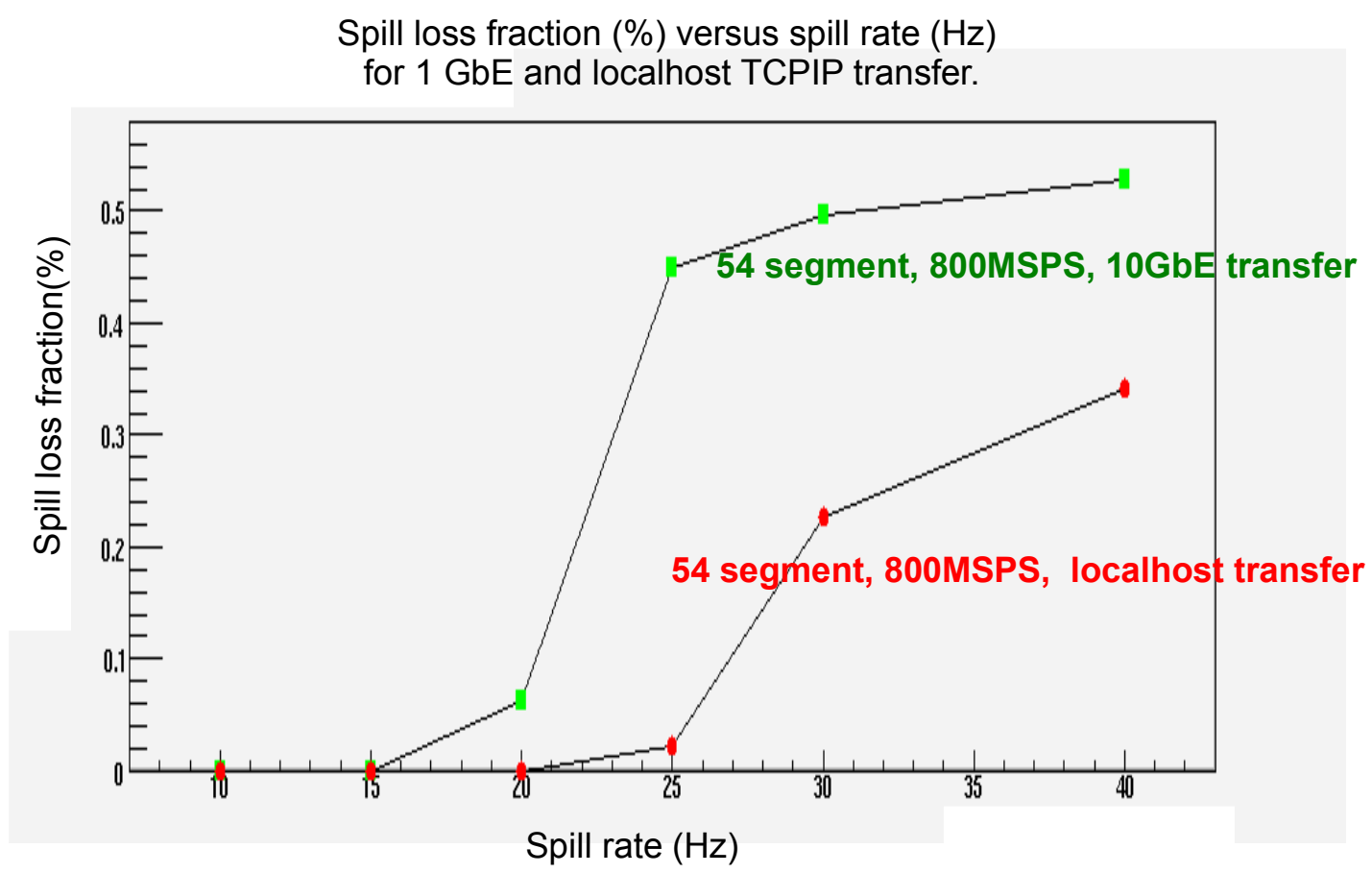

Figure 21.8: Spill lost fraction (%) versus spill rate (Hz) for 800 MSPS digitization with TCPIP data transfer over 10 GbE (green) and localhost (red).

This hardware has enabled the development work on the control, configuration and readout of AMC13 modules and more recently the control and configuration of the waveform digitiizer module. To date we've developed a *CaloReadoutAMC* MIDAS frontend for the readout of the AMC13 events and a *CaloTriggerAMC* MIDAS frontend to trigger the AMC13 to transmit events to the readout process. The new software incorporates both the IPbus protocol for AMC13 configuration / control and a TCPIP thread for AMC13 10 GbE readout. The TCPIP thread incorporates a time-efficient algorithm for unpacking the 32 kB blocks of AMC13 aggregated data into per-channel, per-module, time-ordered ADC samples. We have successfully demonstrated both the IPbus communication for AMC13 configuration / control and the 10 GbE readout of the AMC13 aggregated data.

## 21.5 Alternatives

The collaboration considered two alternative data acquisition frameworks: the CODA DAQ package [12] that is used at JLab and the artdaq DAQ package [13] that is under development at Fermilab. One advantage of using the MIDAS is that collaboration members have already developed very similar DAQ architectures with the MIDAS framework for other experiments. The $g$-2 DAQ can therefore profit from the software/hardware development for the earlier experiments. Another advantage of MIDAS is the availability of an extensive range of DAQ tools including an event builder, an analysis framework, a slow control system, a data alarm system, data storage and database tools, as well as large collections of device drivers for readout hardware. Its modularity makes MIDAS very convenient for both data acquisition development and detector system prototyping. MIDAS has an active community of software

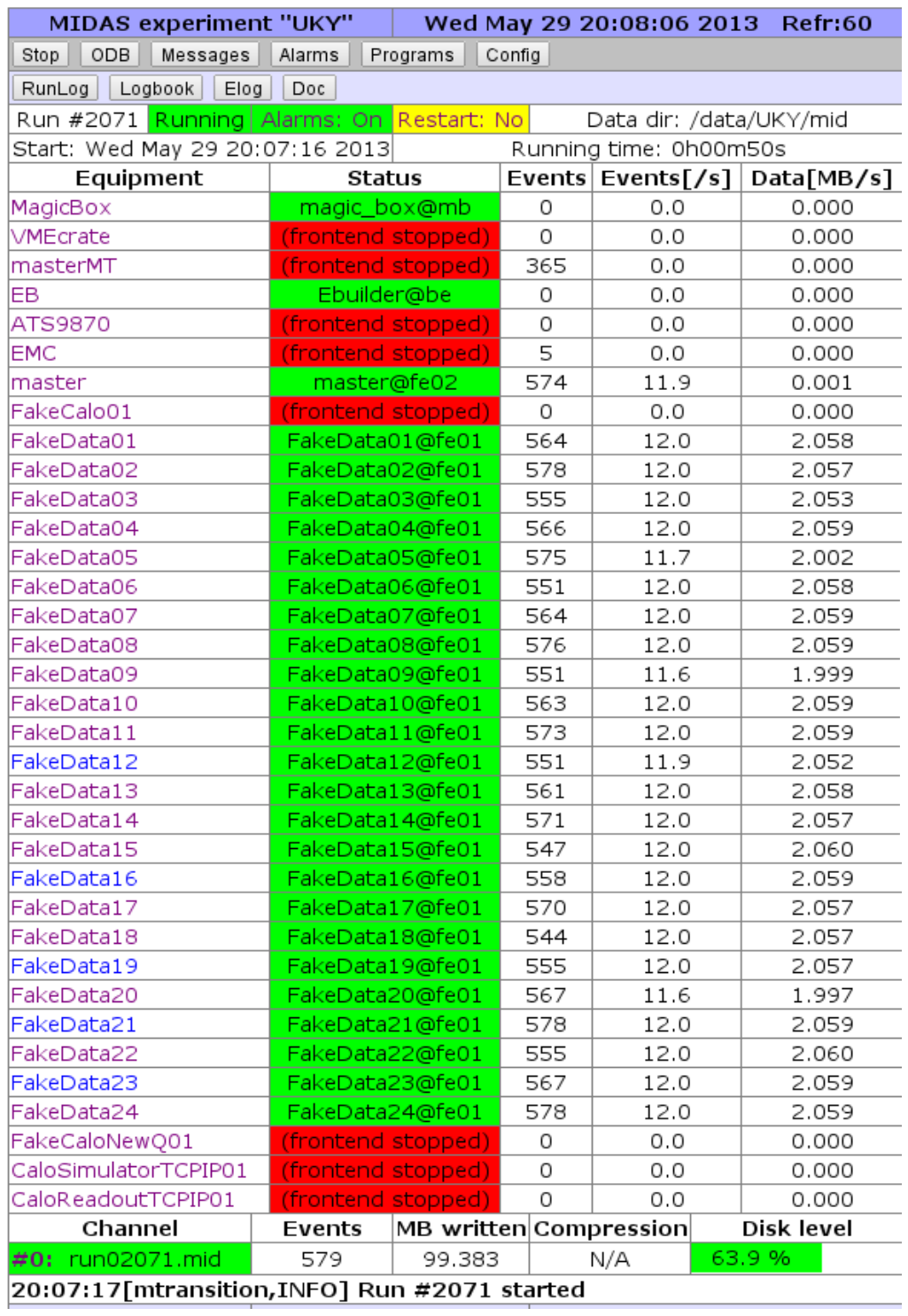

MIDAS experiment "UKY" | Wed May 29 20:08:06 2013 Refr:60

Stop ODB Messages Alarms Programs Config

RunLog Logbook Elog Doc

Run #2071 Running Alarms: On Restart: No Data dir: /data/UKY/mid

Start: Wed May 29 20:07:16 2013 Running time: 0h00m50s

| Equipment | Status | Events | Events[/s] | Data[MB/s] |
|---|---|---|---|---|
| MagicBox | magic_box@mb | 0 | 0.0 | 0.000 |
| VMEcrate | (frontend stopped) | 0 | 0.0 | 0.000 |
| masterMT | (frontend stopped) | 365 | 0.0 | 0.000 |
| EB | Ebuilder@be | 0 | 0.0 | 0.000 |
| ATS9870 | (frontend stopped) | 0 | 0.0 | 0.000 |
| EMC | (frontend stopped) | 5 | 0.0 | 0.000 |
| master | master@fe02 | 574 | 11.9 | 0.001 |
| FakeCalo01 | (frontend stopped) | 0 | 0.0 | 0.000 |
| FakeData01 | FakeData01@fe01 | 564 | 12.0 | 2.058 |
| FakeData02 | FakeData02@fe01 | 578 | 12.0 | 2.057 |
| FakeData03 | FakeData03@fe01 | 555 | 12.0 | 2.053 |
| FakeData04 | FakeData04@fe01 | 566 | 12.0 | 2.059 |
| FakeData05 | FakeData05@fe01 | 575 | 11.7 | 2.002 |
| FakeData06 | FakeData06@fe01 | 551 | 12.0 | 2.058 |
| FakeData07 | FakeData07@fe01 | 564 | 12.0 | 2.059 |
| FakeData08 | FakeData08@fe01 | 576 | 12.0 | 2.059 |
| FakeData09 | FakeData09@fe01 | 551 | 11.6 | 1.999 |
| FakeData10 | FakeData10@fe01 | 563 | 12.0 | 2.059 |
| FakeData11 | FakeData11@fe01 | 573 | 12.0 | 2.059 |
| FakeData12 | FakeData12@fe01 | 551 | 11.9 | 2.052 |
| FakeData13 | FakeData13@fe01 | 561 | 12.0 | 2.058 |
| FakeData14 | FakeData14@fe01 | 571 | 12.0 | 2.057 |
| FakeData15 | FakeData15@fe01 | 547 | 12.0 | 2.060 |
| FakeData16 | FakeData16@fe01 | 558 | 12.0 | 2.059 |
| FakeData17 | FakeData17@fe01 | 570 | 12.0 | 2.057 |
| FakeData18 | FakeData18@fe01 | 544 | 12.0 | 2.057 |
| FakeData19 | FakeData19@fe01 | 555 | 12.0 | 2.057 |
| FakeData20 | FakeData20@fe01 | 567 | 11.6 | 1.997 |
| FakeData21 | FakeData21@fe01 | 578 | 12.0 | 2.059 |
| FakeData22 | FakeData22@fe01 | 555 | 12.0 | 2.060 |
| FakeData23 | FakeData23@fe01 | 567 | 12.0 | 2.059 |
| FakeData24 | FakeData24@fe01 | 578 | 12.0 | 2.059 |
| FakeCaloNewQ01 | (frontend stopped) | 0 | 0.0 | 0.000 |
| CaloSimulatorTCPIP01 | (frontend stopped) | 0 | 0.0 | 0.000 |
| CaloReadoutTCPIP01 | (frontend stopped) | 0 | 0.0 | 0.000 |

| Channel | Events | MB written | Compression | Disk level |
|---|---|---|---|---|
| #0: run02071.mid | 579 | 99.383 | N/A | 63.9 % |

20:07:17[mtransition,INFO] Run #2071 started

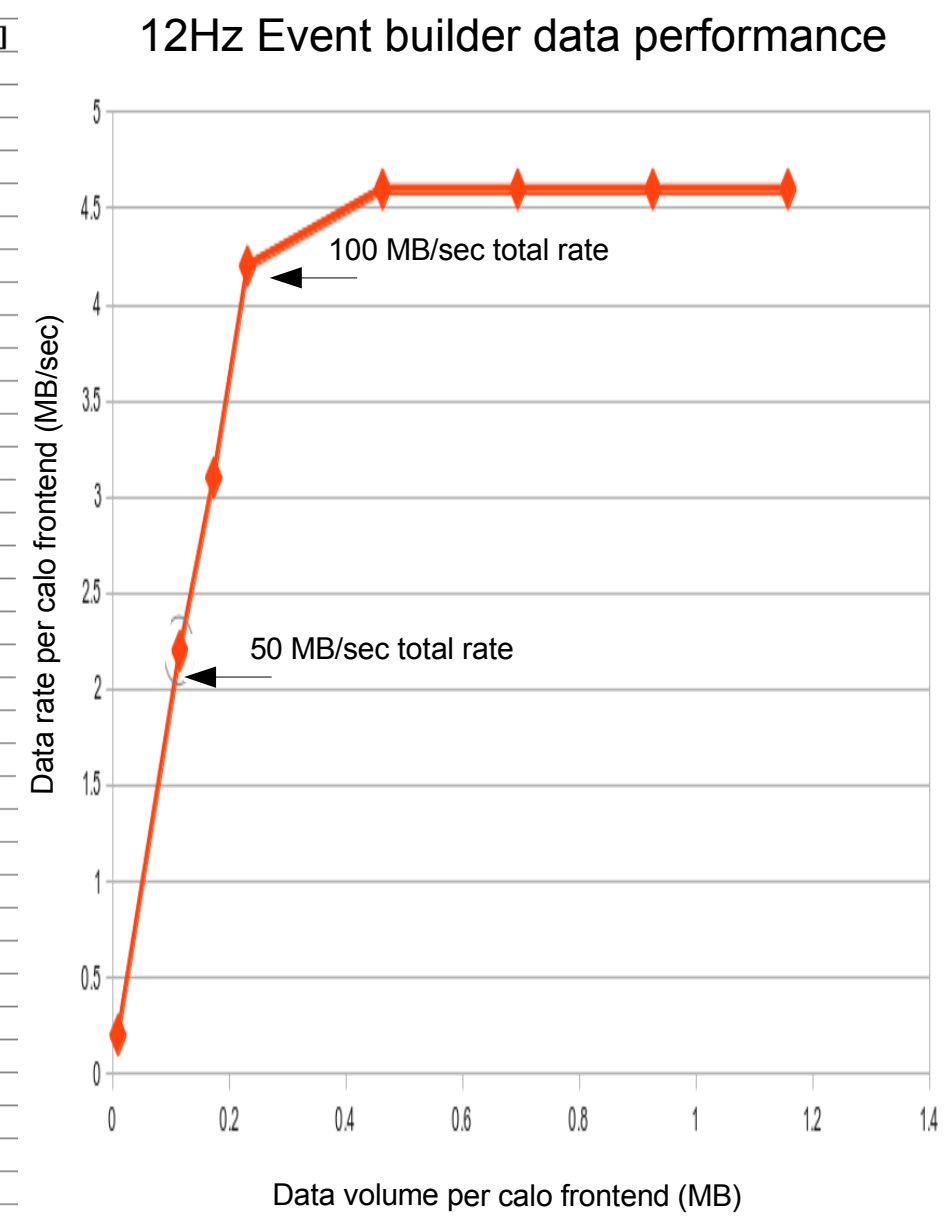

Figure 21.9: Test of event building for twenty four networked frontends. Lefthand side is screen shot of MIDAS control page show twenty four calorimeter frontends, master frontend (handling control logic), magic box frontend (generating spill cycles), and event builder and data logger. Righthand side show frontend data rate (MB/sec and total event builder rate (MB/sec) as function of data size of processed spills.

developers and is widely used at numerous nuclear and particle physics laboratories.

Several alternatives were considered for processing the raw calorimeter data ino T, Q and other method datasets. In particular, the collaboration considered the possibilities of deriving the T/Q method datasets at either the software level in the frontend processors or the firmware level in the digitizer FPGA electronics. One of the major advantages of the software implementation is the greater flexibility to modify or add new datasets or analysis algorithms as needed during the design, commissioning and running phases of the experiment. If necessary, one advantage of implementing some tasks in firmware is the lowering of the raw data rates from the readout electronics to the frontend processors.

Different architectures and parallelization schemes were also considered for the calorimeter frontend processors. In one parallelization scheme using multicore CPUs the frontends process individual spills in separate CPU threads in order to achieve the necessary data compression bandwidth. In another parallelization scheme using many core GPUs the frontends process individual samples in separate GPU threads in order to achieve the necessary data compression bandwidth. In tests, the GPU-based approach was found better matched to parallelizing the tasks involved in deriving T/Q method datasets, and takes advantage of general purpose algorithms and CUDA library functions for such operations.

One alternative under consideration for the data acquisition from the auxiliary detectors is the use of extra channels of $\mu$TCA-base waveform digitizers in place of existing channels of VME-based waveform digitizers. This alternative would reduce the complexity of the DAQ system through the use of a single readout interface.

## 21.6 ES&H

The components of the data acquisition do not involve either hazardous materials or unusual electrical / mechanical hazards. The system will comply with safety standards for power distribution, will require appropriate cooling power for the $\leq$50 kW computer system, and will require temperature, humidity and air velocity sensors.

## 21.7 Risks

Since the calorimeter readout involves the computation of derived T/Q method datasets from continuously digitized ADC samples, the largest risk in the data acquisition system is the corruption or the distortion of the positron time spectra. These risks are mitigated by DAQ integrity testing using DAQ test stands (see Sec. 21.8). They are further mitigated by pre-scaled collection of raw data and continuous monitoring of data integrity during experiment running.

A further risk is insufficient performance of the data acquisition – in particular the derivation of T/Q method datasets – that would impact rates of data accumulation during data taking. This risk is mitigated by the possibility of moving some tasks in the dataset derivation from the GPU hardware to the Kintex 7 FPGA in the AMC13 interface module. The risk is also mitigated by the flexibility of the GPU processing and the ease of a GPU upgrade. In addition, the combined memory of the digitizer modules and the AMC13 interface module, is able to buffer about 13 seconds of consecutive spills and therefore help to mitigate effects of rate variations due to the irregular fill cycle.

Another risk is delays in the software development of the data acquisition that would impact the schedules for detector installation, experiment commissioning and data taking. The DAQ is assembled from commodity computing hardware, so procurement and delivery is not likely a significant risk. The staged development and release approach to DAQ work should reduce the risk of delays in the availability of the software.

## 21.8 Quality Assurance

Fermilab, UCL and UKy have each established test stand for DAQ development, testing and quality assurance. The first stage of development and implementation for the calorimeter data readout, data processing and event building has been underway for two years using a calorimeter station emulator. The second stage of development and implementation was begun in Spring 2014 with the acquisition of a $\mu$TCA crate and MCH / AMC13 control modules. Other collaborators at Cornell and Washington are also establishing test stands for DAQ-related tasks on detector prototyping and electronics readout.

Quality assurance of the DAQ components will also be conducted using the test stands at various institutions. The major DAQ specifications requiring quality assurance are Such stress tests of DAQ components include: (i) rate, integrity and functionality tests of 10 GbE data transfer from the AMC13 controllerr and the frontend TCPIP readout thread, (ii) rate, integrity and functionality tests of PCIe data transfer and GPU data processing of the raw calorimeter samples into the derived data datasets in the frontend GPU thread, (iii) functionality and integrity tests of the IPBus control of the AMC13, Rider modules, and TRM modules, (iv) functionality and integrity tests of the data readout for the tracker stations and the auxiliary detectors, (v) functionality and integrity tests of the framework for the flexible readout of spill-asynchronous and spill-synchronous frontends, (vi) functionality and integrity tests of the framework for the DAQ / electronics control involving the accelerator generated begin-of-spill signal and the DAQ generated spill-authorization signal. (vii) rate, integrity and functionality tests of event builder system to handle approximately 30 event-fragments and a $\leq$100 MB/s rate, (viii) integrity and functionality tests of event builder system to handle both fragments for event building and raw / histogram data-types without event building. The data quality monitor is an integral part of the quality assurance for these parameters.

A phased build approach will be utilized for the DAQ implementation in the MC-1 computer room. As of May 2015 a system for readout and processing of two calorimter stations is installed in Fermilab MC-1. Our schedule is to complete a 6 calorimeter readout system during Summer 2015, a 12 calorimeter readout system during Fall / Winter 2015, and a 24 calorimeter readout system during Spring / Summer 2016. Mock data taking with the laser calibration system will begin Summer 2016.

## 21.9 Value Management

Significant value engineering has been employed in the design of the DAQ as is demonstrated in the alternatives section. In the future, we will continue to monitor developments in commodity electronics for data processing, networking, storage, *etc.* We plan to make our purchases at appropriate times to minimize costs and maximize performance in meeting the DAQ requirements for the experiment.

In addition to T/Q method data compression we plan to evaluate the lossless compression of digitizer data using standard libraries (*e.g.* ZLIB library [10]). For continuously-digitized ADC samples that consist of occasional pulses the loss-less algorithms should offer efficient compression. However, the sensitivity of compression to sources of noise, as well as the time-cost of the data compression, must be carefully studied.

## 21.10 R&D

Our R&D – discussed in Sec. 21.4 – will continue using the DAQ test stands at Fermilab, UCLm UKy and other institutions. Future R&D projects include:

- continuing development, testing and implementation of the data readout from the AMC13 controller with particular attention to optimizing performance and integrity

testing

- continuing development, testing and implementation of the GPU-based processing of the T, Q, and other datasets using emulator calorimeter data and waveform digitizer data
- further development, testing and implementation of the hardware / software logic for the synchronization of the accelerator and readout cycles for synchronous and asynchronous frontends
- further development, testing and implementation of the event building for synchronous / asynchronous data sources.

# Chapter 22

# Slow Controls

## 22.1 Overview and general requirements

The $g-2$ experiment is a complex system that involves many subsystems for which adequate sensoring and control during normal operation is required. The purpose of the slow controls and its associated data acquisition system is to set and monitor parameters such as voltages, currents, gas flows, temperatures, etc. These tasks are essential for the successful operation of the experiment over many months of data taking. Immediate online feedback allows the monitoring of the quality of the incoming data and opportunities to react to changes. For example, part-per-thousand gain stability for the silicon photo-multiplier readout of the electron calorimeter is required to meet the systematic uncertainty budget for $\omega_a$. While the gain stability of these photo-detectors will be monitored at the $10^{-3}$ level or better via a dedicated laser calibration system, immediate feedback on the two parameters (bias voltage and temperature) determining the gain of these devices is achieved via such continuous monitoring. There are many of other cases where such external parameters will be useful in this high precision measurement to establish a full understanding of all systematic uncertainties.

For the setting and read-back of parameters, the slow control system must provide sufficient sensors or control units which will either be directly integrated into the design of new subsystems or come as external devices. Most of these systems will connect to the slow control DAQ via the Midas Slow Control Bus (MSCB [1]) which is a cost-effective field bus developed at the Paul Scherrer Institute (PSI), Switzerland. This very mature system has been successfully employed in other similar experiments and allows for easy integration into the data acquisition framework MIDAS [2]. The slow control DAQ will also include communication interfaces to other external systems like the magnetic and cryogenic controls of the $g-2$ storage ring (iFix [3]) and the Fermilab accelerator (ACNet [4]). Other external devices like the $\mu$TCA crates for the readout electronics of the electron calorimeter will be interfaced and monitored.

The demand and read-back values for all parameters controlled by the slow control system will be stored in a PostgreSQL database for easy online access and wherever possible also in the MIDAS data stream for later analysis. While a local copy of the database will be available for online monitoring and analysis, a full copy will be transferred to a Fermilab database server for long-term storage. For efficient use of the read-backs during data taking, user friendly visualization tools will be provided in order to easily access the stored database information. A web browser based framework will be developed to display the large number of different channels monitored by the system.

Preventing unsafe running conditions will require special handling of some critical detector subsystems. Certain sensors will be connected to the experiment's Programmable Logic Controller (PLC) based safety system to provide interlocks and alarms for such situations. For example, the gas flow of the straw tracker will be monitored and shutdown if the flow read-backs are outside normal ranges. This will be the same PLC system used for the main critical experimental systems (like the cryogenic and vacuum controls of the $g-2$ ring) (see section 22.2.4).

## 22.2 Design

### 22.2.1 Software and hardware architecture

The slow control system will comprise a variety of sensors and control units described in more detail in the following section. Some of these systems will be purchased as single units (e.g. power supplies) and interfaced via common standard protocols (e.g. RS232, TCP/IP). Other subsystems will be custom-built and their design requires integration of an appropriate slow control interface. The usage of field buses like CAN, Profibus and LON are not justified as their integration requires significant effort. Instead, we will employ the Midas Slow Control Bus (MSCB [1]) which is a field bus developed at PSI. This system was optimized for the environment of a typical physics experiment and for cost-efficiency (typically $20 per node). In addition, it conveniently integrates into the MIDAS data acquisition system which is the basic design choice for the slow control computing infrastructure.

The MSCB is the default choice for all sensors and control units that are custom built for the $g-2$ experiment. The MSCB is based on the RS485 protocol which is similar to RS232 except for employing differential signals for superior noise immunity. RS485 is a multi-drop, half duplex communication standard so that many nodes can be connected to the same bus but only one can send data at a time. A single submaster can facilitate the communication between the MIDAS host computer and up to 256 individual MSCB nodes. In fact, by employing a layer of repeaters, up to 65,536 nodes can be operated on a single bus with up to a few km long cables. The MSCB requires two signal wires for the differential signal and a ground wire. Three additional lines provide power (+5 V, ±12 V). The usage of a 10-wire flat ribbon cable provides four additional digital lines for application specific usage.

The MSCB protocol is byte oriented and uses bit 9 from RS232 for addressing purposes. As this bit usually cannot be switched on and off quickly enough in the UART (universal asynchronous receiver/transmitter) of a PC, simply using RS232-RS485 converters is not sufficient. This can be overcome by employing a submaster on the computer side with a micro-controller to provide the handshake with the PC and enough memory to avoid data loss. In this scheme, bit rates of up to 42 kB/s are sustainable.

The development of the MSCB hardware at PSI had several iterations with increasingly sophisticated units. The latest generation is a general purpose unit, SCS2000, as shown in Fig. 22.1(a) and is successfully employed in the MEG experiment at PSI. The SCS2000 unit has an on-board programmable logic device (CLPD, Xilinx XC2C384) which communicates with the submaster via the MSCB on one side. On the other end, there are slots for 8 independent MSCB daughter cards which are each accessed by the CLPD via a 2-lane SPI and a parallel 8-bit bus. The available daughter cards come with a multitude of different functionality. Examples are shown in Fig. 22.1(b) and the complete set of these daughter cards comprises functions like digital I/O channels, 24-bit ADCs, DACs, current sources, valve controls, and many more. Each SCS2000 unit can carry daughter cards of different functionality so that we will be able to fill up each unit to meet the various applications for $g-2$. Because the MSCB protocol and communication is handled by the central programmable logic device in the SCS2000, the daughter cards only require a simple design and the whole package offers a relatively cost-efficient solution.

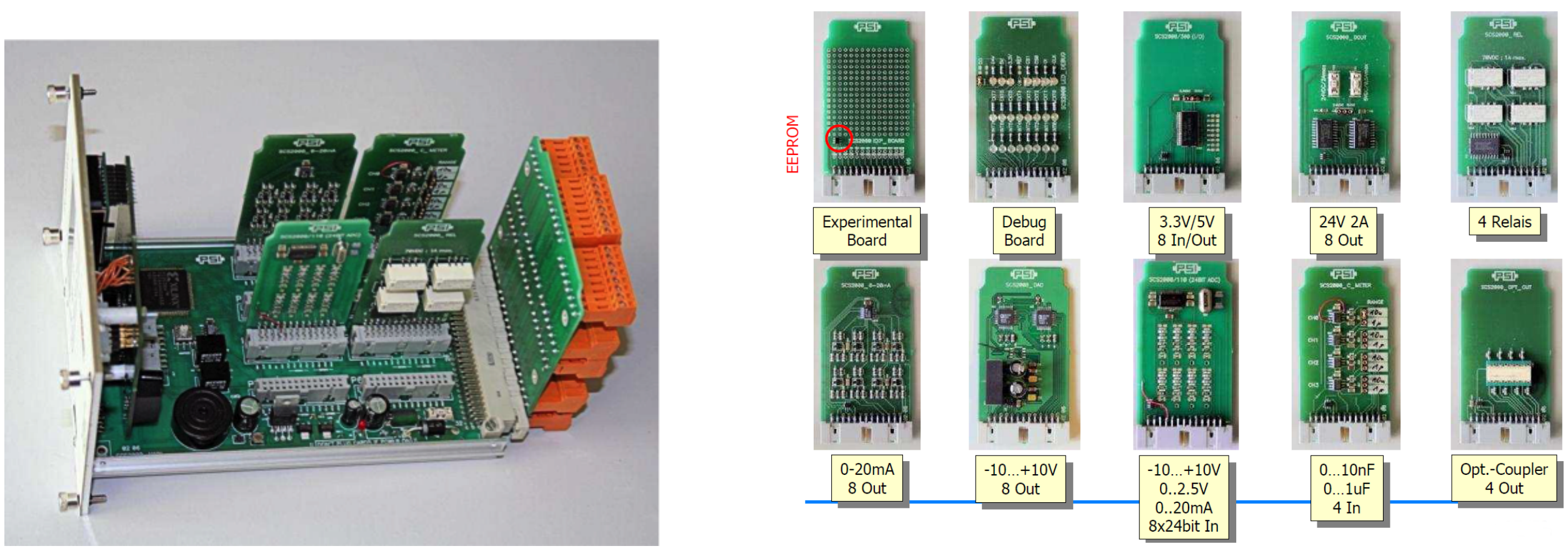

(a) (b)

Figure 22.1: a) SCS-2000 general purpose control unit. b) Examples of available SCS200 daughter cards.

As mentioned before, the MIDAS software framework will be used for the slow control data acquisition. Straightforward integration of MSCB-based hardware is already provided by appropriate drivers integrated into the software package. The end user has to provide an application specific frontend module to control the specific sensor or control unit, i.e. to set and readout parameters of the hardware system. Setting of the parameters such as detector voltages, amplifier gains for the SiPM readout of the calorimeter or the readout rates of sensors are handled by corresponding settings in the online database (ODB) on the slow control computer. Some of these values will be set based on the readback and subsequent online analysis of slow control parameters. A backend main server will handle the collection of the readout data with an adapted event builder provided in the MIDAS software. The assembled MIDAS events from all slow control subsystems are then handed off to a data logger module which will store the data in the MIDAS output stream and a PostgreSQL database locally as well as transfer it to the Fermilab long-term storage server.

Figure 22.2 summarizes the general components of the slow control system indicated by the solid colored boxes. A single slow control backend host (brown box) manages the communication with all MSCB nodes (blue boxes) via the MSCB submaster (green box). Non-MSCB based sensor and control nodes (purple boxes) will communicate directly with the backend server via appropriate interfaces (e.g. USB, serial port, ethernet, ...). During the development phase of the $g-2$ experiment, we expect several institutions to set up their own MIDAS and MSCB host computers for testing of individual components (e.g. the MSCB interface for the SiPM bias voltage control). Although a single main PC and submaster are sufficient to handle all MSCB nodes in the $g-2$ experiment, these additional available host computers with their MSCB submaster and nodes can be easily integrated into the slow control system. Therefore, the final implementation in E989 will involve additional MSCB frontend hosts to control special subsystems. Data exchange between a frontend computer and the slow control backend computer happens via ethernet network. This scheme adds redundancy to the system in case of maintenance or failure of one of the computers since MSCB nodes and their appropriate MIDAS software frontend can be easily ported. The system is completed by the stand-alone alarm system (red box) to provide appropriate actions in case of unsafe operating conditions of various detector subsystems.

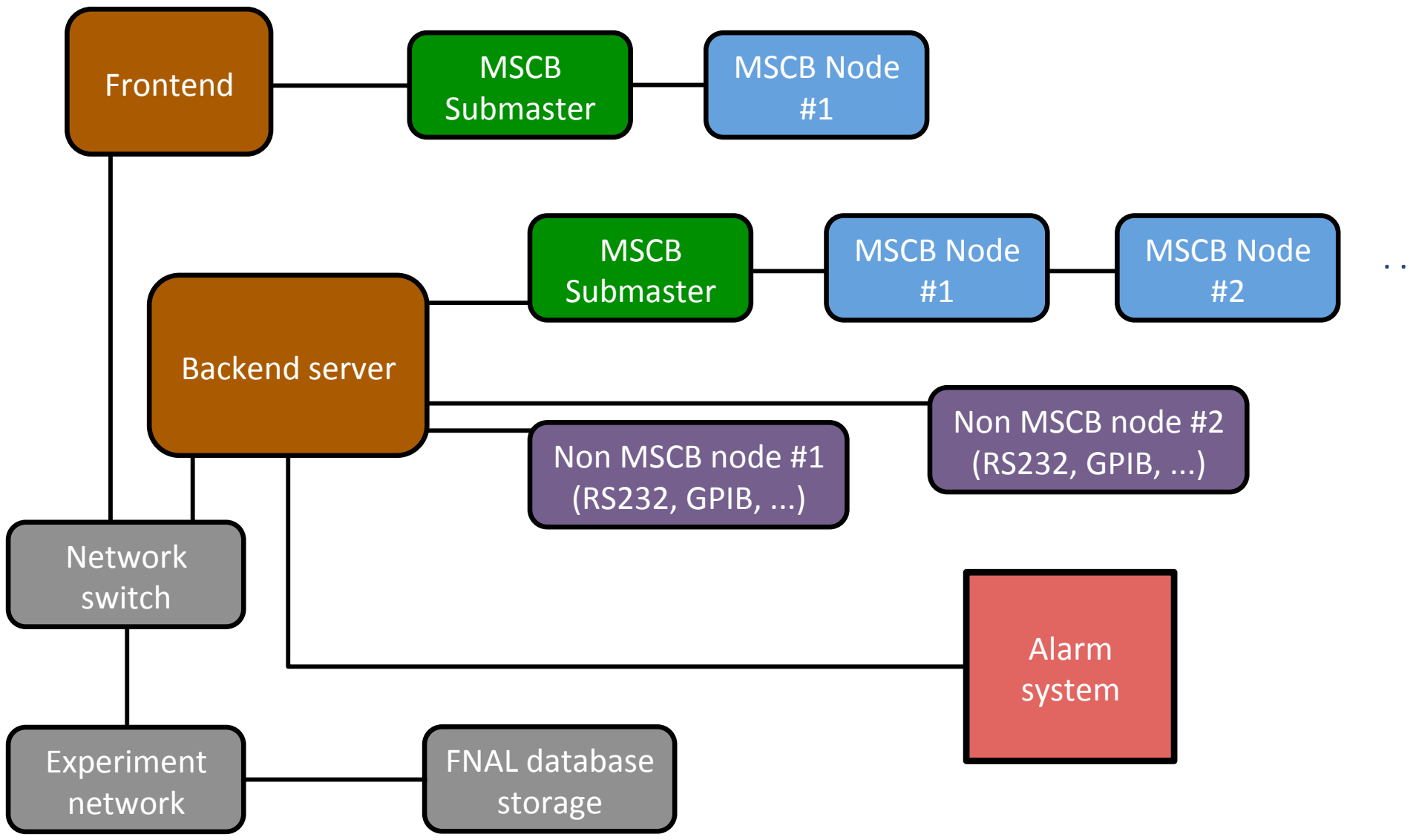

Figure 22.2: Slow control system for the $g-2$ experiment: The basic layout includes a backend host (brown box) which manages the communication with all MSCB nodes (blue boxes) via the MSCB submaster (green box). Non-MSCB nodes (purple boxes) directly connect to the backend via the appropriate interface (USB, serial port, ethernet, ...). Additinal frontend computer(s) with their own MSCB bus and nodes for dedicated applications communicate and exchange data with the backend server via the Ethernet network. The stand-alone alarm system (red box) will provide adequate measures to prevent unsafe running conditions.

In the following subsections, we will describe the sensors and control units, their requirements and the institutional responsibility. Thereafter, the design of the alarm system, the backend server and the data storage are provided.

### 22.2.2 Sensors and controls

The $g-2$ experiment will employ a variety of systems to facilitate the overall measurement of the muon anomalous magnetic moment. Figure 22.3 displays the current required functionality with respect to the slow control measurements broken down by various subsystems. The corresponding Table 22.1 lists the actual parameters set and monitored via the slow control data acquisition and the institutional responsibility for each component. The read-back precision are best estimates and are subject to change.

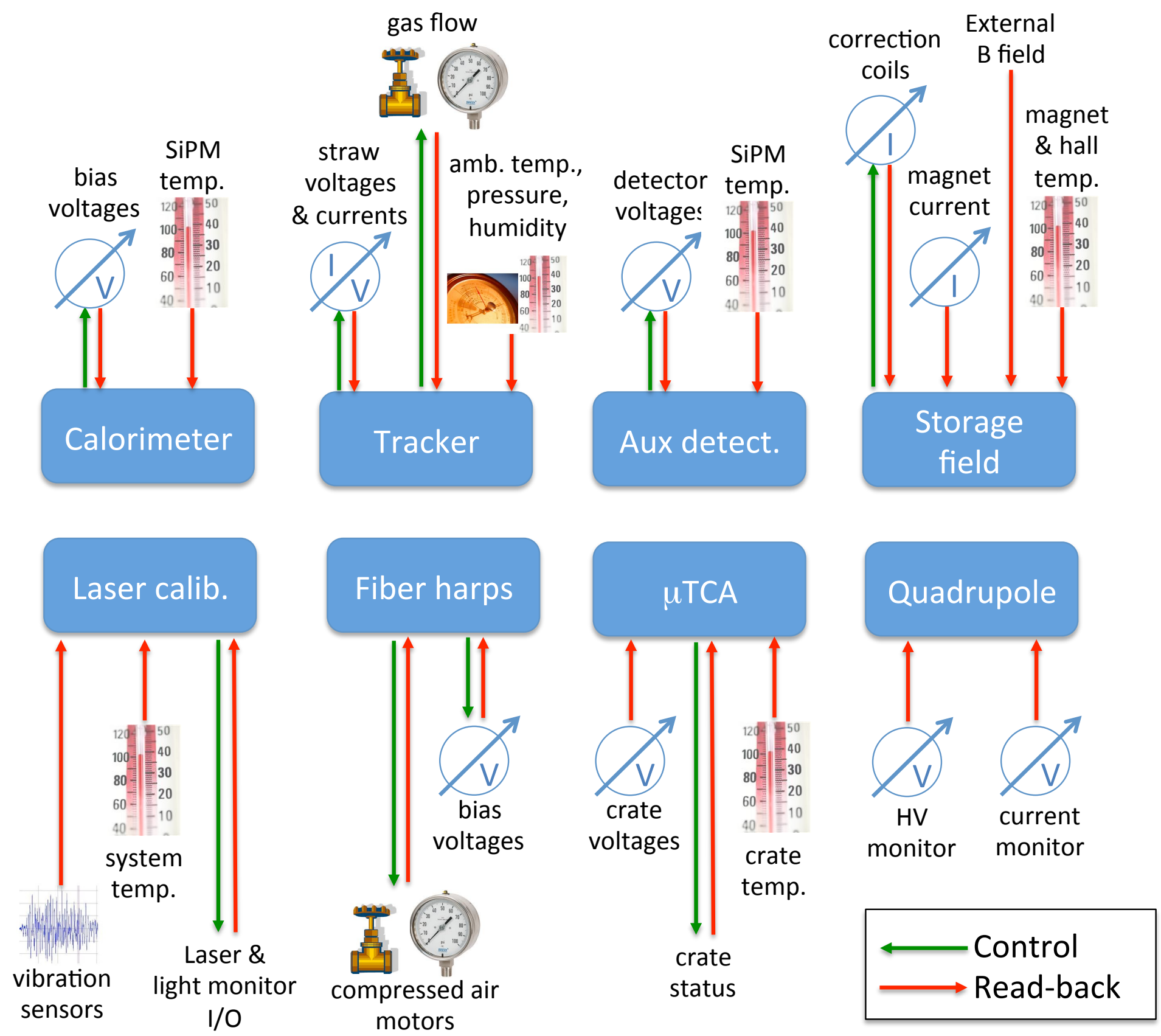

Figure 22.3: Schematic breakdown of the individual slow control nodes showing the controlled parameters and sensor read-backs.

The photo-readout of the electron calorimeter will be based on silicon photo-multipliers (see section 17). The design incorporates a surface mount SiPM on a readout board integrating the bias voltage supply and an amplification of the readout signal with adjustable gain. Since the experiment requires part-per-thousand gain stability, a stabilization and monitoring of the two external parameters that determine the SiPM gain, namely the bias voltage and temperature, is required. The bias voltage of each of the 1296 SiPM channels is set and monitored for each channel and the temperature sensors will be placed on each of the amplifier boards. Compensation of changes in the gain of each channel will be performed by adjustments to the variable gain setting of the differential amplifier stage.

The associated laser calibration system for the calorimeter allows us to monitor absolute gain changes in each calorimeter channel. For that purpose, we will need to monitor the laser intensity and the temperature of the light distribution system at several locations. The total number of channels required for the laser system is expected to be less than 100.

The tracker system comprises three stations of straws located inside the scallop regions of the vacuum chambers. The slow control will provide readings for ambient temperature, humidity, and pressure at those three locations. It will also monitor the gas flow and temperature as well as currents and voltages for both the straw high voltage and the electronics low voltage systems of each of the eight modules per station. The slow control will provide the mechanism to set the high voltage demand values in addition to the read-back of the actual values. The experiment's PLC safety system (see section 22.2.4) will provide interlocks for immediate shutdown of gas and HV in case of irregular running conditions.

Table 22.1: List of control and read-back parameters in the $g-2$ experiment handled by the slow control unit with anticipated read-back precision and rates, channel counts and the institutional responsibility for the implementation of the actual devices.

| **Parameter** | **Read-back precision** | **Channel count** | **Responsibility** |
|---|---|---|---|
| **Calorimeter** | | | |
| SiPM bias voltage | ∼mV | 1300 | UVa, JMU |
| SiPM amplifier gain | | 1300 | UW |
| SiPM temperature | 0.1° C | 1300 | UW |
| **Laser calibration** | | | |
| Laser temperature | < 0.5 °C | < 10 | INFN, NIU |
| Output signals (enable) | | < 48 | INFN |
| Input signals | | < 48 | INFN |
| Serial laser interface | – | < 10 | INFN |
| **Tracker** | | | |
| HV voltage | ∼ 1 V | 54 | FNAL |
| HV current | 0.1 $\mu$A | 54 | FNAL |
| HV status | – | 54 | FNAL |
| LV voltage | ∼ 0.1 V | 54 | UCL |
| LV current | ∼10 $\mu$A | 54 | UCL |
| Electronics temperature | ∼0.5 C° | 348 | FNAL, BU |
| Cooling temperature | ∼1 C° | 54 | FNAL, NIU |
| Amb. pressure | few mbar | 3 | NIU |
| Amb. temperature | < 0.5 °C | 3 | NIU |
| Amb. humidity | few % | 3 | NIU |
| Gas flow | | 48 | FNAL, NIU |
| **Electric quadrupole** | | | |
| Voltage (0-10 V) | 0.1 V | 5 | BNL |
| Current (0-10 V) | 0.1 V | 5 | BNL |
| HV disable / enable | – | 5 | BNL |
| **Aux. detector: Fiber harps** | | | |

Table 22.1 – *Continued from previous page*

| **Parameter** | **Read-back precision** | **Channel count** | **Responsibility** |
|---|---|---|---|
| SiPM bias voltage | few mV | 2 | Regis |
| SiPM temperature | 0.1° C | 4 | Regis |
| Motor control | - | 4 | Regis |
| **Aux. detector: Entrance counter** | | | |
| SiPM bias voltage | few mV | 2 | Regis |
| **Field** | | | |
| Main magnet current | | 1 | FNAL |
| Surface coil current | | 200 | FNAL |
| Yoke temperature | $< 0.5\,°C$ | $\sim 60$ | NIU |
| Hall temperature | $< 0.5\,°C$ | $\sim 5$ | NIU |

The quadrupoles are supplied by five power supplies which each have two low voltage (0-10 V) outputs for monitoring of the actual high voltage and the current, respectively. The slow control will incorporate 2 ADC daughter cards (±10 V range) for the SCS2000 units that will accomodate the 10 channels of this low voltage measurements. A remote HV enable (2.5–15 V) / disable (0-1.5 V) signal for each unit is handled by one 8 channel digital output card for the SCS2000 unit. The quadrupole power supplies will also be fast interlocked (potential free switch) by the PLC alarm system in case of bad vacuum, a storage ring magnet quench, or X-ray detection during access to the main experimental hall during operation.

The fiber harp detectors will be equipped with SiPMs as the photo-sensitive detectors. Their bias voltage power will be supplied through two additional channels of the calorimeter bias supply system. As the SiPMs for the readout of the fibers are grouped in 4 rows of 7, we anticipate monitoring the SiPM temperatures with one probe per row. As the fiber harps are rotated into the beam by compressed air actuators, 2 control channels and read-backs must be available. These are controlled by an Arduino board which will have an MSCB interface.

Beside the fiber harp, the auxiliary detectors also include the so-called $t_0$ entrance counter which is a Lucite Cerenkov sheet readout by two SiPMs. It requires two channels for the control and read-back of the bias voltage. If required, temperature monitoring can be added.

The communication between the slow control DAQ and the $\mu$TCA crates will be done via software (see Section 22.2.3). The $\mu$TCA crates already have an integrated on-shelf manager that can read the status of parameters provided by the crate such as voltages or temperature.

The field measurement in Fig. 22.3 includes readouts of the correction coil currents. It is most likely that these are directly interfaced by the DAQ for the fixed NMR probes. Current read-backs of the main storage ring magnet and other beamline magnets will not be directly interfaced by the slow control DAQ but the data will be received via software interfaces. The slow control will incorporate temperature sensors placed onto the magnet steel around the ring and hall temperature monitors. Since changes in the magnet temperature are the main driver for changes in the field homogeneity, monitoring will allow for detection of any irregular temperature trends which could be caused by a deterioration of the magnet insulation. Overall, we expect a total of <100 temperature probes for the entire experiment (mainly for the magnet steel) with a read-back precision of at most 0.1° C. Since we are mostly sensitive to temperature changes, the absolute accuracy is of less importance. For the implementation of these temperature sensors, we will use the above mentioned general purpose SCS2000 unit with existing 8-channel temperature daughter cards based on the Analog Device AD590 2-terminal temperature transducer. Since each channel senses the current in the AD590, long cables of more than 10 m can be used so that the SCS2000 unit(s) may be located at the center of the ring.

### 22.2.3 Communication with external systems

The slow control DAQ will not only retrieve data from the various sensors described above but also communicate with other systems in the $g-2$ experiment and the Fermilab accelerator infrastructure. As of now, there are a total of three such systems. Communication will need to be established with the main ring control system, the Fermilab accelerator complex, and the $\mu$TCA crates used for the readout of the electron calorimeter stations.

The ring control system for the cryogenics and vacuum is based on PLC interfaces which are accessed via the human machine interface iFix. The communication path (thick double arrow) between the iFix server and the slow control DAQ system will be facilitated via an Object Linking and Embedding for Process Control (OPC) server integrated into iFix. The communication on the slow control DAQ side is handled by an OPC client which is available as commercial or open-source products for the Linux based system. Alternatively, the OPC server might directly write into the PostgreSQL database.

During the $g-2$ operation, some parameters of the accelerator (like magnet currents, beam intensities, status of other beam elements) will be stored in the output datastream. This information can be retrieved via a data broker from the accelerator network (ACNet). Retrieval of accelerator related parameters is already implemented at Fermilab in the larger context of a beam database for the intensity frontier experiments (IFbeam) and we will be able to benefit from this existing implementation by adapting it to our needs and software infrastructure. The data is usually stored in PostgreSQL format and can be integrated into our experimental condition database.

A third system that we want to establish communication with is the $\mu$TCA crates for the readout of the electron calorimeters and possibly other electronics in the experiment. These crates typically provide internal status parameters (e.g. temperature, fan speeds, error indicators etc.) that are useful to monitor to quickly identify hardware problems or failures. System management and monitoring is achieved by means of software solutions based on the Intelligent Platform Management Interface (IPMI), a standardized computer system interface. An IPMI system manager connected to an application programming interface (API) over TCP sockets has been developed for the $\mu$TCA crates employed in the CMS experiment. We will adapt this software development for $g-2$ as it provides the functionality of monitoring various crate parameters.

### 22.2.4 Alarm system

While the slow control DAQ provides means to ramp down high voltages or close a valve, the availability of an additional fast hardware interrupt is preferable for the case of unsafe running conditions. An alarm system will serve the purpose of allowing quick and safe shutdown of certain elements of the $g-2$ detectors. This will be part of the PLC-based system handling the more critical components like the cryogens of the magnet as well as vacuum controls. There is plenty of capacity present within this PLC system and, if necessary, this alarm system can be implemented as a separate sub-master connected to the main system. The system described here will deal with detector components which are not critical in the sense of life threatening unsafe conditions. The interrupts provided by the slow control alarm system are mainly for protection of the detector components and electronics.

At this moment, we plan to provide hardware interlocks for the high voltages and the non-flammable gas for the straw detectors which are located inside the vacuum. Scenarios necessitating shutdown of voltages and gas flow could be vacuum leaks in the ring vacuum chambers, overheating or high fluctuations in the straw current that could indicate a developing problem. The quadrupole power supplies will also be connected to the system and interlocked in case of bad vacuum, a magnet quench or X-rays from sparking plates during person access to the $g-2$ hall. An interlock for the laser calibration system might be useful to protect the system in case of overheating or abnormal parameters. Similarly, hardware interlocks for the SiPM bias voltages could be provided in the same scheme if the request for it arises.

The PLC sits at the center of the system as shown in the schematic layout in Figure 22.4. Various input levels from other systems such as a good vacuum indicator or ring magnet status feed into the PLC. Those input levels are then used in the program running inside the PLC to determine the appropriate output levels of the interlocks for the various detectors. Our design also includes additional switches on the input level side as a measure to allow for bypassing of certain alarm channels. This can be very helpful during detector testing, maintenance or debugging where it is desirable to disable a specific input or output channel without interfering with all others. As the PLC is programmable, a hardcoded timeout interval could be added to automatically switch back on the bypassed channel.

The output channels of the PLC are then fed into the interlock channels of the detector electronics to shutdown high voltages, close gas valves or switch off other components. Additional alarm horns and sirens will be triggered for interlocks that require immediate intervention by shift personnel. A logical OR of all channels will be fed onto pin 10 of the parallel port of into the slow control computer where it can generate an interrupt to trigger software alarms and phone calls to appropriate system experts or the control room. Interpretation of the cause of the specific alarm happens by feeding an additional copy of each output channel onto the data pins of the parallel port card. A similar scheme of parallel port interrupts was already successfully applied in the test beams for the calorimeter and straw tracker and was used to monitor signals of the arriving beam to the DAQ. The adaption of the developed kernel drivers for the alarm system is therefore relatively easy.

### 22.2.5 Backend server

The backend server is the central computer in the slow control DAQ to communicate with the various control units and sensors and retrieve all read-backs. Since data rates on the slow control backend server are low (less than 1 MB/s), a standard modern Linux desktop is sufficient. It will be equipped with enough interfaces (RS232, USB, MSCB) for the ex-ternal devices. As mentioned above, we will work within the MIDAS software framework to coordinate the different tasks. The various sensors and controls can be accessed individually by independent frontend programs which run in parallel within the main MIDAS server. Each frontend has its specific functionality to set experimental parameters (like high voltages for each SiPM), read-back parameters, and to change read-back rates. Some hardware parameters might be set depending on the outcome of certain analyses routines. These analysis frontends can also be run on the backend server since MIDAS already provides for a convenient framework of an online analyzer.

For MSCB devices, necessary hardware drivers are provided by MIDAS so that the actual implementation of the interfacing software frontend is simplified. For other hardware connecting to the backend over RS232 or USB, MIDAS also includes software components that will make integration of these subsystems into the slow control easier. Such frontend code has been developed previously for experiments like MuLan [5] and MuCap [6] at PSI by some of the current E989 collaborators. Therefore, the implementation of the various frontends for all sensors and controls will not pose a major effort.

### 22.2.6 Data storage and access tools

For the data storage of slow control parameters, we will use a PostgreSQL databases. While MIDAS has already built in options for MySQL handling, Fermilab's preferred choice is PostgreSQL and so is the current anticipated choice for E989. Integration of PostgreSQL capabilities should be feasible with minimal effort and we have started adapting the appropriate parts of MIDAS. The backend server will have standard ethernet network connection(s) for the communication with external systems (see section 22.2.3) and synchronization of the local database with the remote long term storage at Fermilab. We will employ the automated script-based mechanisms developed at Fermilab for this purpose. Overall, the database handling and storage is expected to nicely integrate into the existing infrastructure.

From Table 22.1 one can deduce that the anticipated maximal channel count for the slow control is about 3000 readout channels with expected rates of $\sim 1\,\mathrm{s}^{-1}$. If we recorded for every single channel three float values (4 bytes) in form of a timestamp, demand, and current read-back value, we therefore can deduce a conservative upper limit of the expected data rate of 32 kB/s or 3 GB per day. Given the standard storage sizes of more than 1 TB today, the overall slow control data for the entire $g-2$ data taking period will be easily storable and does not pose any major challenge.

Any data acquisition requires a well-designed user interface for online monitoring and the offline analysis. For example, a user friendly visualization interface to inspect the large number of different channels (the calorimeter alone has 1300 channels) is essential during data taking. Based on past developments for muon precision experiments at PSI and other current Intensity Frontier experiments at Fermilab, we will have a variety of options to establish such tools. The IFbeam software tools incorporate the python based Web Server Gateway Interface and subsequent Google Charts to access and display database information in the web browser. The experiments at PSI, MuLan and MuCap, used custom developed web browser based tools to query and display the database information as well as standalone graphics displays within the ROOT framework [7]. The ROME software framework is well integrated into the MIDAS data acquisition framework and can be used for online monitoring of slow controls and other data [8]. At this point, we have started to collect software requirements across the entire experiment to coordinate and efficiently develop such tools that fit most of these applications. In general, usage of a single tool will increase user friendliness but it could be advantageous to have optimized tools for various different data streams. In any case, the specific implementation will profit from extensive former experience which will guide the collaboration in making the final decisions in the future.

## 22.3 Alternative Design Considerations

The information recorded by the slow digitization DAQ is quite independent from any other DAQ system in the $g-2$ experiment. Therefore, we have investigated the usage of alternative software packages like the ORCA system. The collaboration has used this system in the ongoing SiPM tests at UW in order to gain practical experience with this system. Another option is the EPICS software which is well supported at the Advanced Photon Source at ANL and at FNAL. However a careful comparison of the three systems has revealed that MIDAS is our best choice for the software framework for the slow control DAQ. Its major advantages are the fact that several of the $g-2$ collaborators have many years of experience with this system. It has been used successfully by a variety of experiments at PSI and other laboratories. We also have a good relationship with the main developers of MIDAS at PSI. Last but not least, synergies with the fast detector DAQ are obvious as it is based on the same framework. The amount of maintenance and debugging reduces and collaborators on shifts will only need to familiarize themselves with the subtleties of one system.

The default choice of the MSCB for hardware components is tightly connected to the decision for using the MIDAS framework as the latter has easy integration of MSCB components. In addition, the MSCB is optimized for cost efficiency. We have looked into the usage of more commercial products (e.g. National Instruments hardware with possible integration into LabVIEW) but such systems would simply increase the cost. In addition, some of our systems require custom built components (e.g. the extremely stable low voltage power supply for the SiPM) and therefore, we can profit from the simplicity of the MSCB protocol. Finally, the MIDAS and MSCB framework is very open and we have good connections to one of the experts of this system at the Paul Scherrer Institute, Switzerland. We are therefore confident, that development of new modules should be feasible with limited effort. It should also be noted, that we can still rely on non-MSCB off-the-shelf components if it turns out that they are an optimal choice to control or monitor some of our subsystems. Communication with such devices via typical standards of RS232 or USB is available within the MIDAS framework. Our default choice is therefore very modular and expandable but comes at a quite optimal cost.

## 22.4 ES&H

The slow control system will involve sensor and control units that mainly need low voltages and currents for operations. If high voltages (like for the SiPM bias voltage or the PMT voltage) are involved, adequate protection (shielded cables, enclosed and fused electronic components) will be employed to comply with Fermilab's safety rules. All SCS2000 units purchased from PSI come with a full enclosure and meet standard electrical safety requirements. The components for the slow control do not require any hazardous materials and there are no mechanical hazards since the components are typically small.

The alarm system included in the design of the slow control will interlock non-critical components to prevent direct damage to the hardware. It does not include any life-threating hazards.

## 22.5 Risks

The default design of slow control relies on the mature MSCB system that has been successfully employed in several experiments. Therefore, there is only a small risk that components will not work appropriately to the specified requirements. Sensors (like temperature, voltage, currents etc.) are readily available and there are no real indication that they would not meet the requirements in the E989 experiment. In the unlikely event that we cannot meet the requirements, a design of an appropriate component would require additional resources. Since the design of a new MSCB node is not too complex, the associated cost risk is rather small.

A failure in meeting the specified requirements for controlling devices and read back of performance parameters potentially causes an inability of detecting a loss in the data quality during the experiment. This could result in the necessity of dismissing data from the analysis and could result in the need of longer data taking to acquire the full statistics.

Any components installed close to the precision magnetic field (especially electronics circuits with time-varying currents) can cause a static or dynamic distortion to the homogeneity of the field and possibly decrease the precision in its measurement. Mitigation of this risk is achieved by using non-magnetic materials close to the field region and by testing all components for their magnetic properties in a 1.45 T test magnet and with specially designed pickup coils for transient fields.

## 22.6 Quality Assurance

The implementation of the slow control system relies on well established software in the form of the MIDAS framework. In addition, we will employ the very matured MSCB hardware whenever possible or purchase commercially available systems. Quality assurance measures are therefore mainly limited to verifying that custom-built sensors and control units meet the requirements that all systems work properly and comply with all safety regulations. We will extensively test individual components and the full system in dedicated bench tests before the final installation in the experimental hall. As outlined in the risks, these tests will include the verification of the stringent magnetic requirements for components installed in the vicinity of the precision magnetic field of the storage ring. In addition, several institutions across the collaboration will have their own small slow control system to develop individual components. This will help identifying any problems and debugging the system's functionality. At Argonne, we have already successfully tested one SCS2000 together with temperature and ADC readout cards. These tests have verified the easy integration and handling of those units within the MIDAS framework.

Since the slow control provides an online monitoring of the status of many systems in the $g-2$ experiment, care will be taken to properly design the appropriate visualization tools providing easy access to all parameters. This will be an important component in detecting any changes in the quality of the collected data during the experiment.

## 22.7 Value Management

The usage of the freely available open-source MIDAS software and the specifically cost-optimized MSCB hardware is key in keeping the slow control systems overall cost low. Some components that cannot be readily purchased (like the SiPM bias voltage supply with its stringent requirements, see section 17.4.2) need to be custom-built. Most of these will be designed and implemented by collaborators at universities and outside the US in order to keep the overall cost low. At the same time, the centralized integration of all components at Argonne will allow verification of the full system and detection of any interference of different sensors or control units.

## 22.8 R&D

Necessary R&D for cutom-built components that will be integrated in the slow control system is performed by some of the collaborating institutions and will be described in the appropriate sections in this document. Examples for these are the SiPM bias voltage supply (section 17.4.2) or the laser calibration system (section 17.4.3).

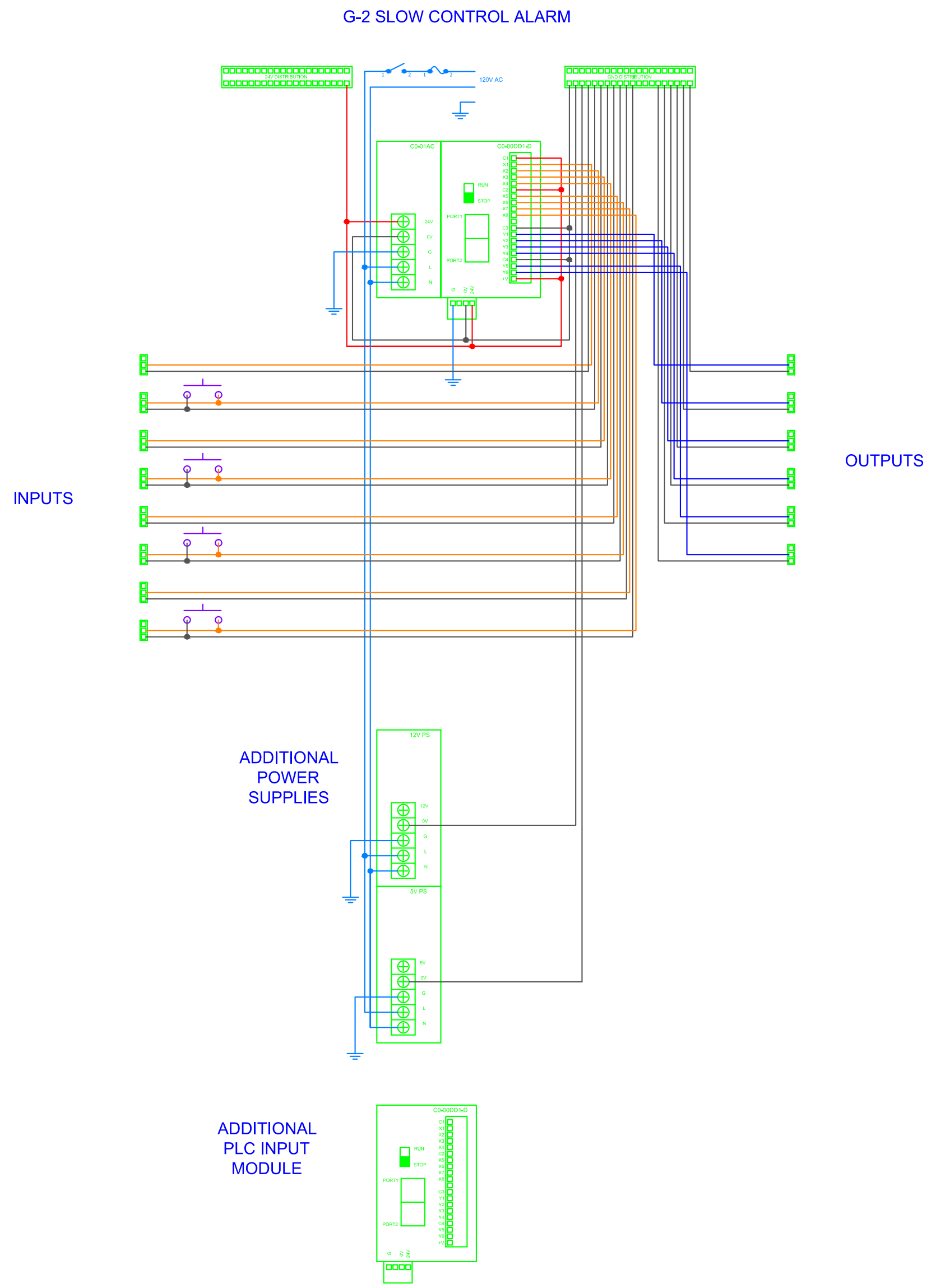

Figure 22.4: Layout for the stand-alone slow control alarm system based on a CLICK PLC.

# Appendix A

# Relocation of the E821 Storage Ring Magnet

## A.1 Reader's Note

This appendix documents the plan for the storage ring move, which occurred in the summer of 2013. It is left intact for the reader as an example of Value Management and Risk Mitigation. Images and videos documenting the move are available on a dedicated website [1].

## A.2 Introduction

The muon storage ring magnet consists of superconducting coils inside their cryostats and the steel yoke and pole pieces. The steel is easily disassembled and shipped by truck, i.e., the time reversal of the process we used twenty years ago. However, the 15 m-diameter coils were wound in Brookhaven Building 919. In order to maintain the exceptional magnetic field, the coils may not be disassembled to the degree that would allow conventional trucking. Special transportation for the very large load is required. Transporting the coils in their horizontal orientation is highly desired in order to prevent extraordinary forces and stresses on the coils.

The largest portion of the coil transport will occur by barge from Long Island, New York to Illinois via the Mississippi River system to the Illinois Waterway. Along the eastern seaboard the barge will travel through the Intracoastal Waterway keeping the barge near ports and in calm waters as compared to open sea travel. An ocean tug will be used from Long Island to New Orleans. A river tug will be used for the remainder of the trip to Lemont, Illinois. A back up plan could route the barge north via the St. Lawrence Seaway and Great Lakes to the Illinois Waterway.

A feasibility study in 2012 studied the best mode of transportation for the remaining short distance over land between the labs and ports in both Long Island and Illinois. The result of the study indicated that the use of a specialized truck and trailer is the best option. Some vendors in the heavy haul industry are capable of performing the work required with a specialized truck/trailer suitable for moving the $g-2$ coils. A transportation review based on the feasibility study was held at FNAL in September, 2012. One of the recommendations from this review was to be sure that we document the coil/cryostat system before the move. The documentation is given in Fermilab $g-2$ doc-db references [2], [3], [4], [5], and [6].

A Request for Proposal (RFP) was written at Fermilab and a meeting for the coil/cryostat transportation was held at BNL in November, 2012. Four companies replied to the RFP and attended the meeting; three of these submitted proposals. Emmert International was chosen to perform the work and the contract was signed.

The present plan is to truck the coil/cryostat from Brookhaven National Laboratory to Smith Point Marina in Suffolk County, Long Island. From this port the barge will travel to the Ozinga port on the Illinois Waterway. From the port in Lemont, the coils will travel via specialized truck/trailer to FNAL this summer.

An analysis has been performed by Emmert International to determine the deflection of the shipping fixture arms while supporting the coils. This has been determined for various support conditions that the fixture will undergo during the phases of the shipment. The results of the Emmert calculations have been verified at Fermilab. The expected forces and deflections have also been imposed on the coils in a Finite Element Analysis at Fermilab. The stresses imposed on the coils are seen to be low on the order of a few hundred psi. The coils and other internal components of the cryostats are not expected to be damaged as a result.

The shipment of the coils will be performed using a quality assurance plan. The plan will provide a means of assuring that the coils will not see stresses above those that we plan for. Severe storms will be avoided. Distant storms that cause higher than normal wave motion will be monitored. The shipment will be monitored with accelerometers capable of transmitting a signal. For wave motion approaching our limits, the barge will be called to safe harbor. A safe harbor plan will be a part of this quality assurance plan. The accelerometer readings will be recorded for later analysis as well.

## A.3 Preparations for Shipping

Figure A.1 shows a recent picture of the cryostats and the mostly disassembled steel. In this photo the upper yoke plates have been removed as well as much of the spacer plates. The coils will be removed for shipment before most of the lower yoke and the remaining spacer plates will be moved.

The following are the important activities occurring (or in process of occuring) in preparation for the move:

- Replacing all the G10 radial stops with Aluminum stops. The G10 stops do not touch the mandrel when warm, only when cold. The Aluminum stops are longer and are designed to touch the mandrel. This prevents the coil from moving side-to-side.

- For the outer coil, vertical bolts at the hangar locations will be inserted through the cryostat's top surface, and engage the mandrel. This is additional protection to prevent the mandrel from moving side-to-side. FEA simulations of this item and of the first item, show that these safeguards are sufficient for handling the worse case of 0.7 g side load.

- The exposed (unpainted) surfaces of the yoke steel was coated with Cosmoline to prevent rusting.
- A structure has been designed to support the interconnect and the hardware outside the outer cryostat (see reference [7]). This is to minimize the stress on both the coils and cryostat walls.
- A shrink wrap will cover the cryostats during the move.
- During the move, dry nitrogen will be flowing through the cryostat to keep it dry.

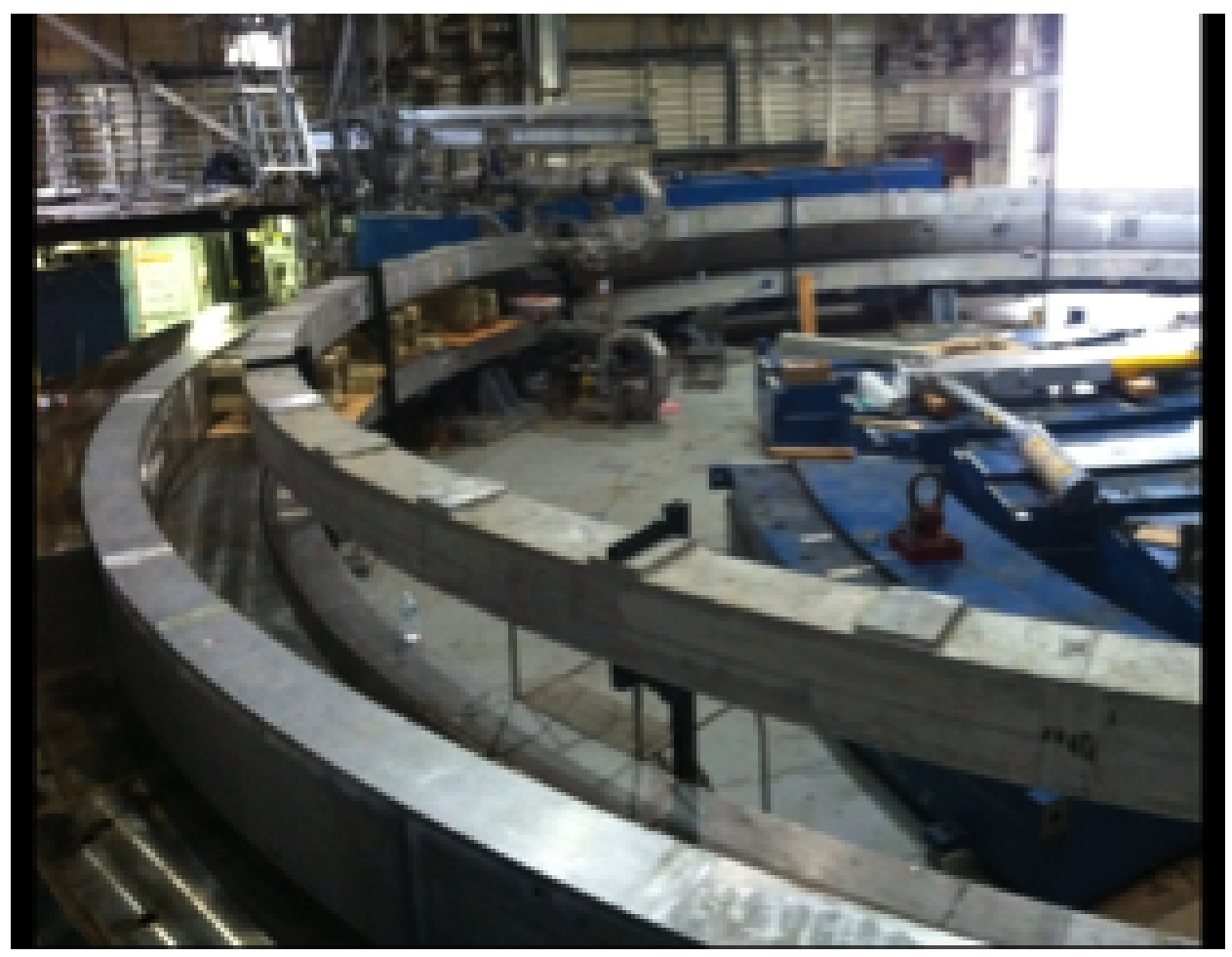

Figure A.1: Coils/Cryostats at BNL.

## A.4 The Coil Shipping Fixture and Transportation

Figure A.2 shows the shipping fixture as specified by Fermilab and designed and built by Emmert International per the criteria to carry the coils. The coils will remain very flat during the shipment to limit the stress imposed onto the coils.

Figure A.3 shows an engineering drawing of the mover and support fixture. The overall length of this rig is in excess of 117 feet. The trailer has three hydraulic zones to keep the load level and to distribute the weight to the wheels evenly. The truck will move slowly over the roadways ranging from walking speed to a maximum of 10 mph depending on the terrain and the proximity of obstacles along the path. The shipment will move over public roadways during night time hours to limit disruption to public traffic.

Figure A.4 shows a model of the mover and support fixture. The 50 foot diameter coils require roughly the width of four traffic lanes to traverse the roadways in Long Island and Illinois.

Figure A.5 shows a drawing of the shipping fixture with coils secured to the barge. The barge that we plan to utilize has a 54 foot width by 180 foot length. This barge size is chosen to limit the maximum roll, pitch, and heave the coils will experience over the water.

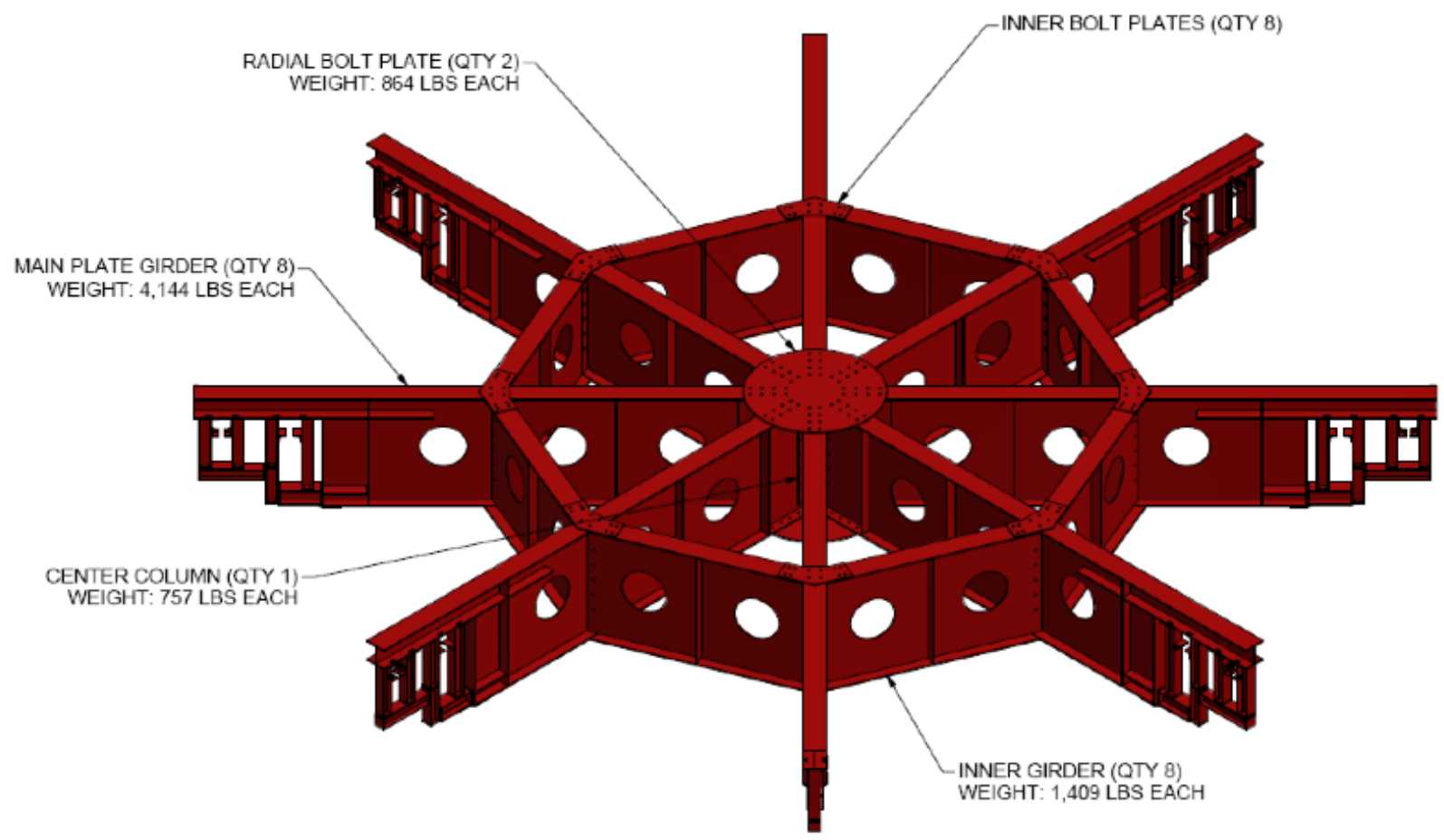

Figure A.2: The shipping fixture.

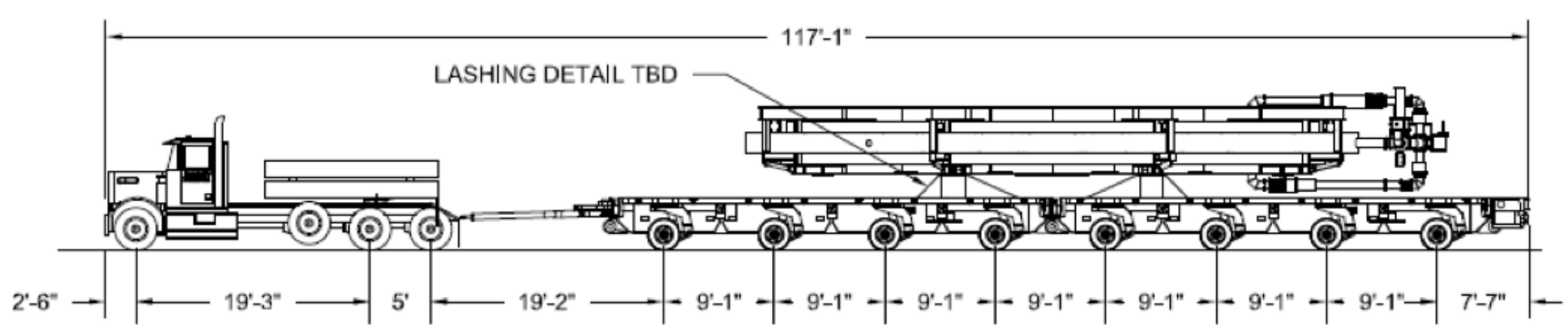

Figure A.3: Specialized Truck and Trailer for Coil Shipment.

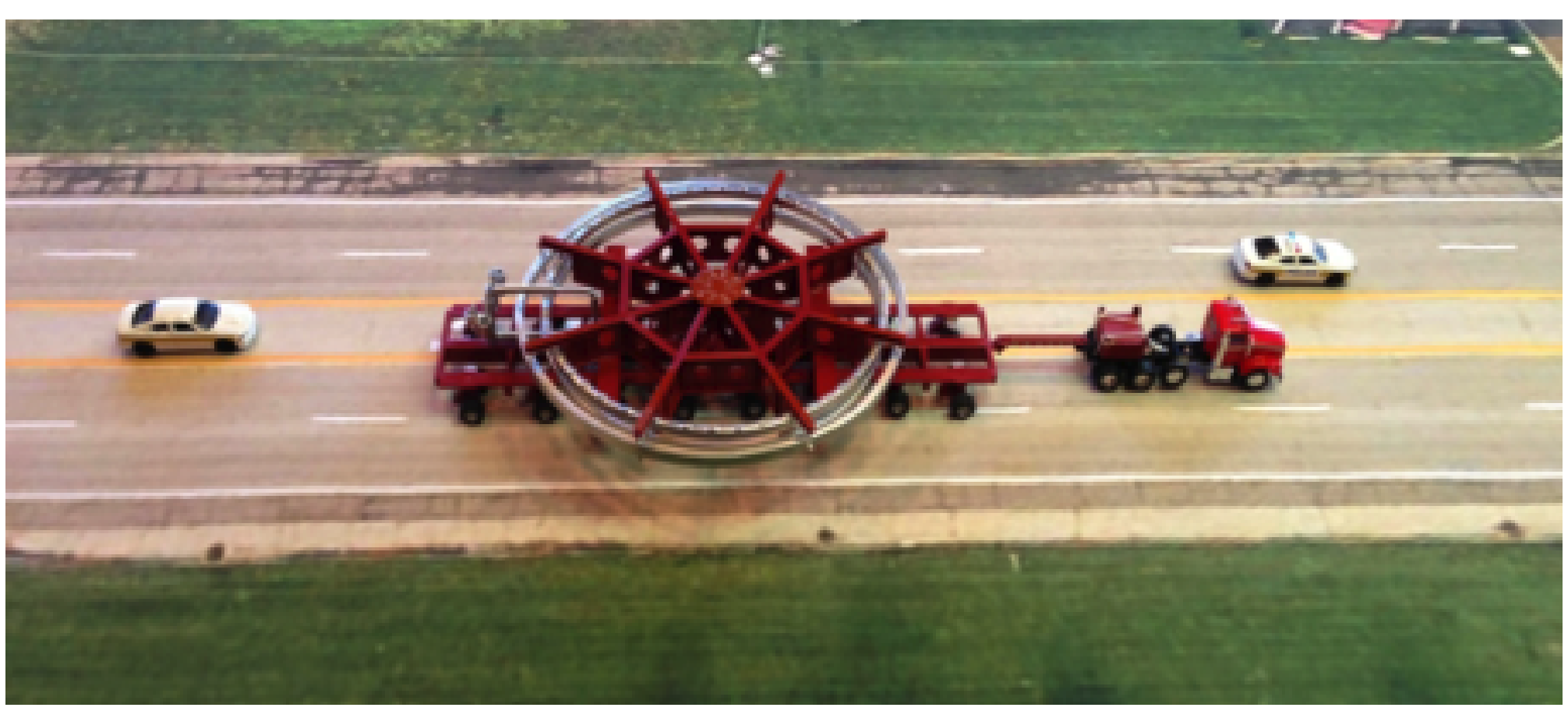

Figure A.4: Scaled model showing the specialized truck and trailer holding the coils.

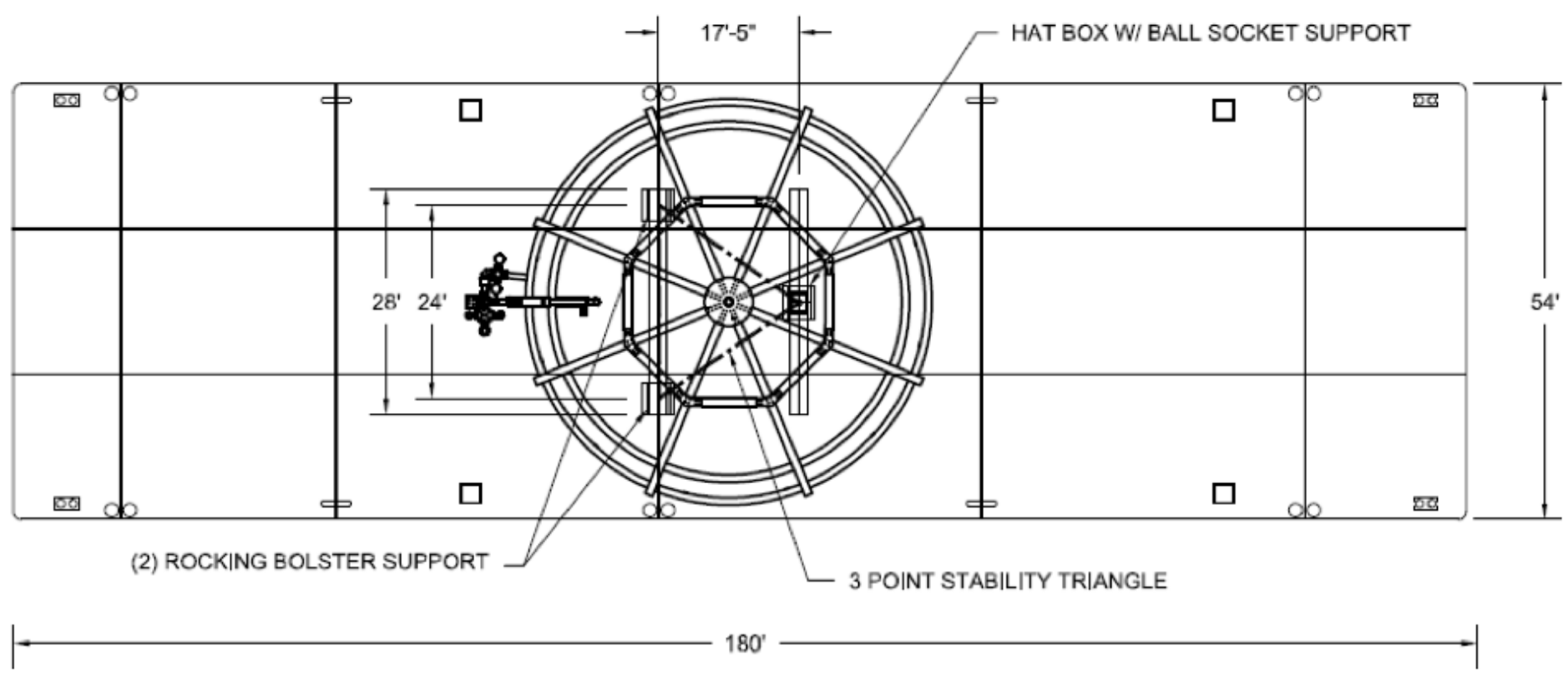

Figure A.5: Shipping fixture with coils shown secured to the barge.